%% file: dissertation.tex
\documentclass[10pt,twoside,fleqn,a4paper]{book}
\usepackage[german]{babel}
\usepackage{fancyheadings}
\usepackage{psboxit}
\usepackage{cite}        
\usepackage{version}
\usepackage{amsmath2000}
\usepackage{amscd2000}
\usepackage{amsfonts}
\usepackage{amssymb}
\usepackage{latexsym}
\usepackage{epsfig}
\usepackage{rotating}
\usepackage{dcolumn}
\usepackage{hhline}
\usepackage{alltt}
\usepackage{endnotes}
 \normalsize
\renewcommand{\chaptermark}[1]%
  {\markboth{#1}{}\plainheadrulewidth=0.0pt}
\renewcommand{\sectionmark}[1]%
  {\markright{\thesection\ #1}}
\newcommand{\clearemptydoublepage}{\newpage{\pagestyle{empty}\cleardoublepage}}
\includeversion{FINAL} \excludeversion{DRAFT}
\begin{FINAL}
\end{FINAL}
\begin{DRAFT}
\end{DRAFT}
\newcommand{\zet}{{z}}           
\newcommand{\bzet}{{\bar\zet}}   
\newcommand{\zbz}{{\zet\bzet}}   
\newcommand{\hh}{_{h \mskip-1mu h}}
\newcommand{\ellell}{_{\ell \mskip-2mu \ell}}
\newcommand{\ellp}{_{\ell \mskip-1mu p}}
\newcommand{\tTll}{\tilde{T}\ellell}      
\newcommand{\tsill}{\tilde{\si}\ellell}   
\newcommand{\pT}{T^{\D\bm{\prime}}}   
\let\Ga=\Gamma
\let\De=\Delta
\let\Th=\Theta
\let\La=\Lambda
\let\Si=\Sigma

\let\Ph=\Phi

\let\Om=\Omega
\let\al=\alpha
\let\be=\beta
\let\ga=\gamma
\let\de=\delta
\let\ep=\epsilon
\let\ze=\zeta
\let\et=\eta
\let\th=\theta

\let\ka=\kappa
\let\la=\lambda
\let\rh=\rho
\let\si=\sigma
\let\ta=\tau

\let\ph=\phi
\let\ch=\chi
\let\ps=\psi
\let\om=\omega
\let\vep=\varepsilon
\let\vth=\vartheta
\let\vka=\varkappa   
\let\vph=\varphi

\let\vrh=\varrho

\newcommand{\bm}[1]{\mbox{\boldmath${#1}$}}   
\newcommand{\bmG}[2][\large]{\sizeq{#1}{#2}}   

\def\bbbr{{\rm I\!R}}   

\def\bbbn{{\rm I\!N}}   

\def\bbbone{{\mathchoice {\rm 1\mskip-4mu l} {\rm 1\mskip-4mu l}
  {\rm 1\mskip-4.5mu l} {\rm 1\mskip-5mu l}}}
\def\bbbc{{\mathchoice {\setbox0=\hbox{$\displaystyle\rm C$}\hbox{\hbox
  to0pt{\kern0.4\wd0\vrule height0.9\ht0\hss}\box0}}
  {\setbox0=\hbox{$\textstyle\rm C$}\hbox{\hbox
  to0pt{\kern0.4\wd0\vrule height0.9\ht0\hss}\box0}}
  {\setbox0=\hbox{$\scriptstyle\rm C$}\hbox{\hbox
  to0pt{\kern0.4\wd0\vrule height0.9\ht0\hss}\box0}}
  {\setbox0=\hbox{$\scriptscriptstyle\rm C$}\hbox{\hbox
  to0pt{\kern0.4\wd0\vrule height0.9\ht0\hss}\box0}}}}
\def\bbbq{{\mathchoice {\setbox0=\hbox{$\displaystyle\rm
  Q$}\hbox{\raise
  0.15\ht0\hbox to0pt{\kern0.4\wd0\vrule height0.8\ht0\hss}\box0}}
  {\setbox0=\hbox{$\textstyle\rm Q$}\hbox{\raise
  0.15\ht0\hbox to0pt{\kern0.4\wd0\vrule height0.8\ht0\hss}\box0}}
  {\setbox0=\hbox{$\scriptstyle\rm Q$}\hbox{\raise
  0.15\ht0\hbox to0pt{\kern0.4\wd0\vrule height0.7\ht0\hss}\box0}}
  {\setbox0=\hbox{$\scriptscriptstyle\rm Q$}\hbox{\raise
  0.15\ht0\hbox to0pt{\kern0.4\wd0\vrule height0.7\ht0\hss}\box0}}}}
\def\bbbt{{\mathchoice {\setbox0=\hbox{$\displaystyle\rm
  T$}\hbox{\hbox to0pt{\kern0.3\wd0\vrule height0.9\ht0\hss}\box0}}
  {\setbox0=\hbox{$\textstyle\rm T$}\hbox{\hbox
  to0pt{\kern0.3\wd0\vrule height0.9\ht0\hss}\box0}}
  {\setbox0=\hbox{$\scriptstyle\rm T$}\hbox{\hbox
  to0pt{\kern0.3\wd0\vrule height0.9\ht0\hss}\box0}}
  {\setbox0=\hbox{$\scriptscriptstyle\rm T$}\hbox{\hbox
  to0pt{\kern0.3\wd0\vrule height0.9\ht0\hss}\box0}}}}
\def\bbbs{{\mathchoice
  {\setbox0=\hbox{$\displaystyle     \rm S$}\hbox{\raise0.5\ht0\hbox
  to0pt{\kern0.35\wd0\vrule height0.45\ht0\hss}\hbox
  to0pt{\kern0.55\wd0\vrule height0.5\ht0\hss}\box0}}
  {\setbox0=\hbox{$\textstyle        \rm S$}\hbox{\raise0.5\ht0\hbox
  to0pt{\kern0.35\wd0\vrule height0.45\ht0\hss}\hbox
  to0pt{\kern0.55\wd0\vrule height0.5\ht0\hss}\box0}}
  {\setbox0=\hbox{$\scriptstyle      \rm S$}\hbox{\raise0.5\ht0\hbox
  to0pt{\kern0.35\wd0\vrule height0.45\ht0\hss}\raise0.05\ht0\hbox
  to0pt{\kern0.5\wd0\vrule height0.45\ht0\hss}\box0}}
  {\setbox0=\hbox{$\scriptscriptstyle\rm S$}\hbox{\raise0.5\ht0\hbox
  to0pt{\kern0.4\wd0\vrule height0.45\ht0\hss}\raise0.05\ht0\hbox
  to0pt{\kern0.55\wd0\vrule height0.45\ht0\hss}\box0}}}}
\def\bbbz{{\mathchoice {\hbox{$\sf\textstyle Z\kern-0.4em Z$}}
  {\hbox{$\sf\textstyle Z\kern-0.4em Z$}}
  {\hbox{$\sf\scriptstyle Z\kern-0.3em Z$}}
  {\hbox{$\sf\scriptscriptstyle Z\kern-0.2em Z$}}}}
\newcommand{\grgl}{\:\hbox  to -0.2pt{\lower2.5pt\hbox{$\sim$}\hss}{\raise1pt\hbox{$>$}}\:}
\newcommand{\grglL}{\:\hbox to -0.2pt{\lower2.5pt\hbox{$\sim$}\hss}{\raise3pt\hbox{$>$}}\:}
\newcommand{\klgl}{\:\hbox  to -0.2pt{\lower2.5pt\hbox{$\sim$}\hss}{\raise1pt\hbox{$<$}}\:}
\newcommand{\klglL}{\:\hbox to -0.2pt{\lower2.5pt\hbox{$\sim$}\hss}{\raise3pt\hbox{$<$}}\:}
\newcommand{\nlongrightarrow}{\vv/\zzzz\zzzz\longrightarrow}
\newcommand{\nequiv}{\mskip7.5mu/\mskip-16.5mu\equiv}
\newcommand{\efn}[1]{\mbox{e}^{#1}}   

\newcommand{\hyperF}[2]{{}_{#1}\mskip-.75mu{\rm F}_{\mskip-.75mu{#2}}}   
\newcommand{\tr}[1][\,]{\mbox{tr}{#1}}
\newcommand{\Det}{\mbox{Det}\,}
\newcommand{\circint}{\circlearrowleft\mskip-23mu\int}
\newcommand{\derL}[1]{\overset{\leftharpoonup}{#1}}
\newcommand{\derR}[1]{\overset{\rightharpoonup}{#1}}

\newcommand{\pa}{\partial}
\newcommand{\paL}{\derL{\pa}}
\newcommand{\paR}{\derR{\pa}}

\newcommand{\del}[1]{\pa^{\mskip-.75mu{#1}}}
\newcommand{\delL}[1]{\paL{}^{\mskip-.75mu{#1}}}

\newcommand{\Del}[1]{D^{\mskip-.75mu{#1}}}

\newcommand{\DelR}[1]{D{}^{\mskip-.75mu{#1}}}

\newcommand{\paslash}{\pa\hspace*{-.5625em}/}
\newcommand{\paslashL}{\derL{\paslash}}
\newcommand{\paslashR}{\derR{\paslash}}

\newcommand{\delslashL}[1]{\paslashL{}^{\mskip-.75mu{#1}}}
\newcommand{\delslashR}[1]{\paslashR{}^{\mskip-.75mu{#1}}}

\newcommand{\Aslash}{A\mskip-9.5mu/}
\newcommand{\kslash}{k\mskip-9.5mu/}
\newcommand{\qslash}{q\mskip-8mu/}
\newcommand{\vepslash}{\vep\mskip-8mu/}
\newcommand{\bracketL}{\mbox{\boldmath$\langle$}}
\newcommand{\bracketM}{\mbox{\boldmath$|$}}
\newcommand{\bracketR}{\mbox{\boldmath$\rangle$}}
\newcommand{\bra}[1]{ \bracketL {#1} \bracketM }
\newcommand{\ket}[1]{ \bracketM {#1} \bracketR }
\newcommand{\bracket}[2]{ \bracketL {#1} \bracketM {#2} \bracketR }
\newcommand{\vev}[1]{\bracketL{#1}\bracketR}
\newcommand{\vacL}{\bracketL}
\newcommand{\vacR}{\bracketR_{\!A}}
\newcommand{\vac}[1]{ \vacL {#1} \vacR }
\newcommand{\cumL}{\mbox{\boldmath$\langle\!\langle$}}
\newcommand{\cumR}{\mbox{\boldmath$\rangle\!\rangle$}}
\newcommand{\cum}[1]{ \cumL {#1} \cumR }
\newcommand{\rbracketL}{ \mbox{\boldmath$($} }
\newcommand{\rbracketR}{ \mbox{\boldmath$)$} }
\newcommand{\rbra}[1]{\rbracketL {#1} \bracketM }
\newcommand{\rket}[1]{ \bracketM {#1} \rbracketR }
\newcommand{\rbracket}[2]{ \rbracketL {#1} \bracketM {#2} \rbracketR }
\newcommand{\permI}[1]{ \mbox{\boldmath$\lceil$} {#1} \mbox{\boldmath$\rceil$} }
\newcommand{\permII}[1]{ \mbox{\boldmath$\lceil\!\lceil$} {#1} \mbox{\boldmath$\rceil\!\rceil$} }
\newcommand{\Hss}[2]{\bm{|\mskip-2mu{#1}\mskip-2mu{#2}\mskip-2mu|}}
\newcommand{\Hpm}{\bm{|\mskip-6mu\pm\mskip-5mu\mp\mskip-2mu|}}
\newcommand{\Hmp}{\bm{|\mskip-6mu\mp\mskip-5mu\pm\mskip-2mu|}}
\newcommand{\Hpp}{\bm{|\mskip-6mu\pm\mskip-5mu\pm\mskip-2mu|}}
\newcommand{\D}{\displaystyle}   
\newcommand{\T}{\textstyle}
\newcommand{\mPl}{\mbox{$m_{\rm P\mskip-2mu l}$}}   
\newcommand{\lPl}{\mbox{$l_{\rm P\mskip-2mu l}$}}   
\newcommand{\BRST}{{\rm\scriptscriptstyle%
                     \mskip-1.5mu B\mskip-1.5mu R\mskip-1.5mu S\mskip-1.5mu T}}
\newcommand{\RT}{{\mskip-1.5mu R\mskip-1.5mu T}}
\newcommand{\Nc}{N_{\rm\!c}}
\newcommand{\SUNc}{SU\!(\Nc)}
\newcommand{\suNc}{su(\Nc)}
\newcommand{\Drst}[1]{{\mskip-1mu\mathfrak{#1}}}
\newcommand{\dimNc}{d_{\mskip-1mu SU\!(N_{\rm\!c})}}
\newcommand{\dimDrst}[1]{d_{\mskip-1mu\mathfrak{#1}}}
\newcommand{\trDrst}[1]{\mbox{tr}_{\mskip-1mu\mathfrak{#1}}\,}
\newcommand{\bbbOne}[1]{\bbbone_{\mskip-1mu\mathfrak{#1}}}   
\newcommand{\normDrst}[1]{n_\Drst{#1}}
\newcommand{\csDrst}[1]{c_2\!(\Drst{#1})}
\newcommand{\cssDrst}[1]{c(\Drst{#1})}
\newcommand{\HaarDmu}{\mathfrak{D}_{\T\!\mu}(A)}
\newcommand{\IN}{{\rm in}}
\newcommand{\OUT}{{\rm out}}
\newcommand{\deZ}{{\de\mskip-1mu Z\idx{2}}}
\newcommand{\sgnd}{{\rm sgn}_{\mskip-1mu d}}
\newcommand{\rb}[1]{{\rm\bf{#1}}}   
\newcommand{\rbb}[1]{\mbox{\boldmath${\rm\bar{#1}}$}}   
\newcommand{\rbG}[2][\large]{{\bmG[#1]{#2}}}   
\newcommand{\vecp}[1]{\vec{#1}^{\mskip3mu\prime}}
\newcommand{\AQ}{{A\!Q}}
\newcommand{\projNFAt}{\tilde\digamma}   
\newcommand{\projNAt}[2][\mskip2mu]{\projNFAt^{\mskip-2mu{#1}}\mskip-2mu[{#2}\mskip1mu]}
\newcommand{\projt}[3][\mskip2mu]{\projNFAt^{\mskip-2mu{#1}}\mskip-2mu[{#2}\mskip1mu]%
                                     \mskip-2mu(|{#3}|)}
\newcommand{\projtbig}[3][\mskip2mu]{\projNFAt^{\mskip-2mu{#1}}\mskip-2mu[{#2}\mskip1mu]%
                                     \mskip-2mu\big(\big|{#3}\big|\big)}
\newcommand{\xE}[1][x]{{#1}_{\mskip-.5mu\rm E\mskip-1mu}}   
\newcommand{\xM}[1][x]{{#1}_{\mskip-.5mu\rm M\mskip-1mu}}   
\newcommand{\eE}[1][e]{{#1}_{\mskip-.5mu\rm E\mskip-2mu}}   
\newcommand{\slE}[1][\mathbb{SL}]{{#1}_{\mskip-.25mu\rm E\mskip-1mu}}    
\newcommand{\iIM}{{\rm i}}
\newcommand{\Dstar}{{\D\star}}
\newcommand{\zem}[1][{m}]{\ze_{\mskip-1mu{#1}}}
\newcommand{\oE}{o.E.d.A.\@}
\newcommand{\Arg}{{\rm Arg}\,}
\newcommand{\intI}[3]{\mbox{$I^{I}_{{#1},{#2}}\mskip-3mu({#3})$}}   
\newcommand{\intIINA}[2][{\pm}]{\mbox{$I^{I\mskip-3mu I}_{#2,\mskip-1mu{#1}}$}}
\newcommand{\intII}[4][{\pm}]{\mbox{$I^{I\mskip-3mu I}_{#2,\mskip-1mu{#1}}\mskip-3mu({#3},{#4})$}}
\newcommand{\intMD}[4]{\mbox{$I^{({#1},{#2})}_{#3}\mskip-3mu({#4})$}}   
\newcommand{\intIII}[2]{\mbox{$I^{I\mskip-3mu I\mskip-3mu I}_{#1}\mskip-3mu({#2})$}}
\newcommand{\Dnbr}[1]{{\bm{\T\!{\D{#1}}\!}}}
\newcommand{\snbr}[1]{{\bm{\scriptscriptstyle\!{\scriptstyle{#1}}\!}}}

\newcommand{\Dbr}[1]{{\bm{\T(\!{\D{#1}}\!)}}}

\newcommand{\sbr}[1]{{\bm{\scriptscriptstyle(\!{\scriptstyle{#1}}\!)}}}
\newcommand{\rQ}[1]{\rb{r}_\sbr{#1}}

\newcommand{\bprll}{\rb{b}_{\|}}
\newcommand{\bperp}{\rb{b}_{\mskip-2mu\perp}}
\newcommand{\dsiP}[2][\mskip-2mu(\tilde{x}\Dmfp)]{{d\si\Dmfp^{#2}{#1}}}   
\newcommand{\dsiM}[2][\mskip-2mu(\tilde{x}\Dmfm)]{{d\si\Dmfm^{#2}{#1}}}   
\newcommand{\dsiI}[2][\mskip-2mu(\tilde{x}\Dimath)]{{d\si\Dimath^{#2}{#1}}}   
\newcommand{\dsiJ}[2][\mskip-2mu(\tilde{x}\Djmath)]{{d\si\Djmath^{#2}{#1}}}   
\newcommand{\dsiIxE}[2][\mskip-2mu(\tilde{x}\Dimath)]{{\xE[d\si]{}\Dimath^{#2}{#1}}}   
\newcommand{\dsiJxE}[2][\mskip-2mu(\tilde{x}\Djmath)]{{\xE[d\si]{}\Djmath^{#2}{#1}}}
\newcommand{\dsiPA}[3][\mskip-2mu(\tilde{x}\Dmfp)]{{\big[d\si_{#3}\big]%
                         ^{\mskip-1mu{#2}}{#1}}}
\newcommand{\dsiMA}[3][\mskip-2mu(\tilde{x}\Dmfm)]{{\big[d\si_{#3}\big]%
                         ^{\mskip-1mu{#2}}{#1}}}

\newcommand{\ddsiPAU}[3][\mskip-2mu(\tilde{x}\Dmfp)]{{\big[d\tilde\si_{#3}\big]%
                         _{\mskip-1mu{#2}}{#1}}}
\newcommand{\pDipol}[1][{\ }]{$\Dbr{+}$-Dipol{#1}}
\newcommand{\pDipols}[1][{\ }]{$\Dbr{+}$-Dipols{#1}}
\newcommand{\mDipol}[1][{\ }]{$\Dbr{-}$-Dipol{#1}}
\newcommand{\mDipols}[1][{\ }]{$\Dbr{-}$-Dipols{#1}}
\newcommand{\Gatot}{\Ga^{tot}}
\newcommand{\Gaee}{\Ga^{e^{\!+}\!e^{\!-}}}
\newcommand{\Gall}{\Ga^{l^{\mskip-1mu+}\mskip-1mu l^{\mskip-1mu-}}}
\newcommand{\GeV}[1][\;]{\mbox{$\rm{#1}Ge\mskip-3mu V$}}
\newcommand{\MeV}[1][\;]{\mbox{$\rm{#1}Me\mskip-3mu V$}}
\newcommand{\keV}[1][\;]{\mbox{$\rm{#1}ke\mskip-3mu V$}}
\newcommand{\fm}[1][\;]{{\rm{#1}fm}}
\newcommand{\mbarn}[1][\;]{{\rm{#1}mb}}
\newcommand{\microbarn}[1][\;]{{\rm{#1}\mu b}}
\newcommand{\nbarn}[1][\;]{{\rm{#1}nb}}
\newcolumntype{f}[1]{D{.}{.}{#1}}
\newcolumntype{g}[1]{D{,}{}{#1}}
\newcolumntype{h}[1]{D{/}{}{#1}}
\newcommand{\PM}{\mbox{$\pm$}}
\newcommand{\NON}{,\non}
\newcommand{\centi}{\mbox{$\times\!10^{-2}$}}
\newcommand{\milli}{\mbox{$\times\!10^{-3}$}}

\newcommand{\non}{\mbox{---}}
\newcommand{\overlap}[2][V]{\mbox{$\ps_{{#1}}{}^{\zz\D\dagger}{}_{\zz({#2})}\,%
                                \ps_{{\D\ga}(Q^2,{#2})}$}}
\newcommand{\roverlap}[2][V]{\mbox{$r\,\ps_{{#1}}{}^{\zz\D\dagger}{}_{\zz({#2})}\,%
                                \ps_{{\D\ga}(Q^2,{#2})}$}}
\newcommand{\gaga}[1]{\mbox{$\ps_{\D\ga}{}^{\zz\D\dagger}{}_{\zz(Q^{\prime2},{#1})}\,%
                                \ps_{{\D\ga}(Q^2,{#1})}$}}
\newcommand{\RLT}[1][]{R_{{#1}L\mskip-1.5mu T}}
\newcommand{\Rpi}{R_{\D\mskip-.75mu\pi}}
\newcommand{\Jps}{J\!/\!\ps}
\newcommand{\iJps}{{\mskip-3mu J \mskip-3mu/\mskip-3mu \ps}}
\newcommand{\bbB}[1][b]{|\mskip-2mu\rb{#1}\mskip-2mu|}   
\newcommand{\tfde}{{\de}t}                                
\newcommand{\tf}{\ta}                                
\newcommand{\tfQ}{\ta^{\mskip-1mu2}}                 
\newcommand{\tfb}{\rbG{\ta}}                          
\newcommand{\tfbQ}{\rbG{\ta}^{\mskip-2mu2}}           
\newcommand{\tfbB}{|\mskip-2mu\rbG{\ta}\mskip-2mu|}   
\newcommand{\psRest}[1][]{\mbox{$%
   \ps_{\mskip-1.5mu R\mskip-1.5mu e\mskip-1.5mu s\mskip-1.5mu t{#1}}$}}
\newcommand{\meff}[1][f\!,]{m_{#1{\rm eff}}}
\newcommand{\ip}{i^{\mskip-.5mu\prime}}
\newcommand{\ipp}{i^{\mskip-.5mu\prime\mskip-1.5mu\prime}}
\newcommand{\pmatrixZE}[2]{\begin{pmatrix}%
                            {#1}\\{#2}%
                                \end{pmatrix}}
\newcommand{\pmatrixZZ}[4]{\begin{pmatrix}%
                            {#1}&{#2}\\{#3}&{#4}%
                                \end{pmatrix}}
\newcommand{\pmatrixZD}[6]{\begin{pmatrix}%
                            {#1}&{#2}&{#3}\\{#4}&{#5}&{#6}%
                                \end{pmatrix}}
\newcommand{\pmatrixZV}[8]{\begin{pmatrix}%
                            {#1}&{#2}&{#3}&{#4}\\{#5}&{#6}&{#7}&{#8}%
                                \end{pmatrix}}
\newcommand{\pmatrixDD}[9]{\begin{pmatrix}%
                            {#1}&{#2}&{#3}\\{#4}&{#5}&{#6}\\{#7}&{#8}&{#9}%
                                \end{pmatrix}}
\newcommand{\ZWEI}[3][{\ }]{\mbox{{#2}$\mskip-1mu${#3}}}
\newcommand{\DREI}[4][{\ }]{\mbox{{#2}$\mskip-1mu${#3}$\mskip-1mu${#4}}}
\newcommand{\VIER}[5][{\ }]{\mbox{{#2}$\mskip-1mu${#3}$\mskip-1mu${#4}$\mskip-1mu${#5}}}
\newcommand{\FUNF}[6][{\ }]{%
  \mbox{{#2}$\mskip-1mu${#3}$\mskip-1mu${#4}$\mskip-1mu${#5}$\mskip-1mu${#6}}}
\newcommand{\ExpB}{\VIER[]{E}{6}{6}{5}/\!\citeUP{Adams97}}
\newcommand{\ExpC}{\DREI[]{N}{M}{C}/\!\citeUP{Arneodo94}}
\newcommand{\qqquad}{\quad\qquad}
\newcommand{\zz}{\!\!}   
\newcommand{\zzz}{\!\!\!}
\newcommand{\zzzz}{\!\!\!\!}
\newcommand{\vv}{\;\;}   
\newcommand{\vvv}{\;\;\;}
\newcommand{\vvvv}{\;\;\;\;}
\newcommand{\mf}[1]{\mathfrak{#1}}
\newcommand{\mfp}{\mathfrak{p}}
\newcommand{\mfm}{\mathfrak{m}}
\newcommand{\dbtilde}[1]{\Tilde{\Tilde{#1}}}   
\newcommand{\ixp}{x^{\mskip-2mu\prime}}
\newcommand{\ixpp}{x^{\mskip-2mu\prime\mskip-1.5mu\prime}}

\newcommand{\oC}{^{\mskip0mu C}}   
\newcommand{\oNC}{^{\mskip0mu N\mskip-2mu C}}
\newcommand{\uC}{_{\mskip-1.5mu C}}   
\newcommand{\uNC}{_{\mskip-1.5mu N\mskip-2mu C}}
\newcommand{\iES}{{1\mskip-2mu S}}   
\newcommand{\iZS}{{2\mskip-2mu S}}
\newcommand{\dbprime}{{\prime\mskip-1.5mu\prime}}   
\newcommand{\tlprime}{{\prime\mskip-1.5mu\prime\mskip-1.5mu\prime}}   
\newcommand{\ivrh}{_{\D\mskip-1.5mu \vrh}}
\newcommand{\isi}{_{\D\mskip-1.5mu \si}}
\newcommand{\iin}{_{\rm\mskip-1.5mu in}}
\newcommand{\iga}{{\D\ga}}
\newcommand{\irh}{{{\D\rh}(770)}}
\newcommand{\irhp}{{{\D\rh}(1450)}}
\newcommand{\irhpp}{{{\D\rh}(1700)}}
\newcommand{\idx}[2][]{_{\mskip-1mu\bm{\scriptstyle{#2}{#1}}}}
\newcommand{\oidx}[2][]{^{\mskip-1mu\bm{\scriptstyle{#2}{#1}}}}
\newcommand{\Dimath}[1][]{\idx[{#1}]{\imath}}
\newcommand{\Djmath}[1][]{\idx[{#1}]{\jmath}}
\newcommand{\Dmfp}[1][]{\idx[{#1}]{\mfp}}
\newcommand{\Dmfm}[1][]{\idx[{#1}]{\mfm}}
\newcommand{\Doperp}[1][]{\oidx[{#1}]{\perp}}
\newcommand{\bea}{\begin{eqnarray}}
\newcommand{\eea}{\end{eqnarray}}
\newcommand{\nn}{\nonumber}
\newcommand{\sizeq}[2]{{\mbox{#1${#2}$}}}   

\newcommand{\equalindent}{0em}   

\renewcommand{\textfraction}{0.1}
\newcommand{\xaxis}[2][]{\makebox[#1]{\hfill{#2}\hfill}}   
\newcommand{\yaxis}[2][]{\begin{sideways}\makebox[#1]{\hfill{#2}\hfill}\end{sideways}}
\renewcommand{\thefootnote}{\thechapter.\arabic{footnote}}
\newcommand{\FOOT}[1]{\footnote{#1}}
\renewcommand{\theendnotes}{}
\newcommand{\bffootnote}{{$^{\bf \thechapter.\addtocounter{footnote}{1}\arabic{footnote}}$}}
\newcommand{\rmfootnote}{{$^{\rm \thechapter.\addtocounter{footnote}{1}\arabic{footnote}}$}}
\newcommand{\citeFN}[1]{\mbox{$^{\mbox{\scriptsize\ref{#1}}}$}}   
\newcommand{\citeFNbf}[1]{$^{\bf\FN{#1}}$}   
\newcommand{\FN}[1]{\ref{#1}}
\newcommand{\FNg}[1]{\FN{#1} auf Seite~\pageref{#1}}
\newcommand{\refg}[1]{\ref{#1} auf Seite~\pageref{#1}}   
\newcommand{\Glg}[1]{Gl.~(\ref{#1}) auf Seite~\pageref{#1}}
\newcommand{\citeUP}[1]{\mbox{$^{\mbox{\scriptsize\cite{#1}}}$}}
\begin{document}
\PScommands

\include{P1}          
\clearemptydoublepage
%
%
\include{MNEMOSYNE}   
\clearemptydoublepage
\include{P3}          
\clearemptydoublepage
\include{P5}          
\include{NOTATION}    
\clearemptydoublepage

\pagenumbering{roman}

\renewcommand{\contentsname}{\huge Inhaltsverzeichnis}
\lhead[\fancyplain{}{\sc\thepage}]%
      {\fancyplain{}{\sc Inhaltsverzeichnis}}
\rhead[\fancyplain{}{\sc Inhaltsverzeichnis}]%
      {\fancyplain{}{\sc\thepage}}
\tableofcontents
\clearemptydoublepage
\renewcommand{\listfigurename}{\huge Abbildungsverzeichnis}
\lhead[\fancyplain{}{\sc\thepage}]%
      {\fancyplain{}{\sc Abbildungsverzeichnis}}
\rhead[\fancyplain{}{\sc Abbildungsverzeichnis}]%
      {\fancyplain{}{\sc\thepage}}
\listoffigures
\clearemptydoublepage
\renewcommand{\listtablename}{\huge Tabellenverzeichnis}
\lhead[\fancyplain{}{\sc\thepage}]%
      {\fancyplain{}{\sc Tabellenverzeichnis}}
\rhead[\fancyplain{}{\sc Tabellenverzeichnis}]%
      {\fancyplain{}{\sc\thepage}}
\listoftables
\clearemptydoublepage

\pagenumbering{arabic}

\renewcommand{\theequation}{\thechapter.\arabic{equation}}
\newcommand{\rightmarkQCD}{\rightmark}
\include{QCD}
\clearemptydoublepage

\renewcommand{\chaptername}{\Large Kapitel}
\renewcommand{\thefootnote}{\thechapter.\arabic{footnote}}

\begin{FINAL}
  \include{VAKUUM-F}
  \clearemptydoublepage
  \include{ANALYT-F}
  \clearemptydoublepage
  \include{GROUND-F}
  \clearemptydoublepage
  \include{EXCITED-F}
  \clearemptydoublepage
\end{FINAL}
\begin{DRAFT}
  \include{VAKUUM}
  \clearemptydoublepage
  \include{ANALYT}
  \clearemptydoublepage
  \include{GROUND}
  \clearemptydoublepage
  \include{EXCITED}
  \clearemptydoublepage
\end{DRAFT}

\include{RESUMEE}
\clearemptydoublepage

\renewcommand{\appendixname}{\huge Anhang}
\renewcommand{\thechapter}{\Alph{chapter}}
\renewcommand{\theequation}{\thechapter.\arabic{equation}}
\setcounter{chapter}{0}
\renewcommand{\chaptername}{\Large Anhang}

\setcounter{equation}{0}
\include{APP_ALGEBRA}
\clearemptydoublepage

\setcounter{equation}{0}
\include{APP_KINEMATIK}
\clearemptydoublepage

\setcounter{equation}{0}
\include{APP_STREUUNG}
\clearemptydoublepage
\begin{FINAL}
  \setcounter{equation}{0}
  \include{APP_BOOSTS-F}
  \clearemptydoublepage
\end{FINAL}
\begin{DRAFT}
  \setcounter{equation}{0}
  \include{APP_BOOSTS}
  \clearemptydoublepage
\end{DRAFT}
\begin{FINAL}
  \setcounter{equation}{0}
  \include{APP_INTEGRATE-F}
  \clearemptydoublepage
\end{FINAL}
\begin{DRAFT}
  \setcounter{equation}{0}
  \include{APP_INTEGRATE}
  \clearemptydoublepage
\end{DRAFT}

\setcounter{equation}{0}
\include{APP_CLTFN}
\clearemptydoublepage

\setcounter{equation}{0}
\include{APP_LCWFN}
\clearemptydoublepage
\begin{FINAL}
  \setcounter{equation}{0}
  \include{APP_TABLES-F}
  \clearemptydoublepage
\end{FINAL}
\begin{DRAFT}
  \setcounter{equation}{0}
  \include{APP_TABLES}
  \clearemptydoublepage
\end{DRAFT}
%

%

\bibliographystyle{plain}   
\renewcommand{\bibname}{\huge Literaturverzeichnis}
\addcontentsline{toc}{chapter}{\numberline{}Literaturverzeichnis}
\lhead[\fancyplain{}{\sc\thepage}]
      {\fancyplain{}{\sc{Literaturverzeichnis}}}
\rhead[\fancyplain{}{\sc{Literaturverzeichnis}}]
      {\fancyplain{}{\sc\thepage}}
\bibliography{dissertation}     
\clearemptydoublepage

\include{DANK}
%
%
%
%
%
%
%

\end{document}

%% file: P1.tex
\lhead[\fancyplain{}{}]%
      {\fancyplain{}{}}
\rhead[\fancyplain{}{}]%
      {\fancyplain{}{}}
\setlength{\parskip}{0.5cm}
\begin{center}
{\bf{INAUGURALDISSERTATION \\
zur\\
Erlangung der Doktorw"urde\\
der\\
Naturwissenschaftlich-Mathematischen \\
Gesamtfakult{\"a}t\\
der\\
Ruprecht-Karls-Universit"at\\
Heidelberg\vspace*{15cm}\\}}
vorgelegt von\\
Dipl.-Phys.\@ Gerhard Kulzinger\\
aus Heidelberg\\
\bigskip
Tag der m"undlichen Pr"ufung: 30.\@ Januar 2002
\end{center}

%% file: MNEMOSYNE.tex
\lhead[\fancyplain{}{}]%
      {\fancyplain{}{}}
\rhead[\fancyplain{}{}]%
      {\fancyplain{}{}}

\vspace*{\stretch{1}}
\begin{center}
\begin{minipage}{10.5cm}
\begin{verse}
{\footnotesize
{\bf 	Friedrich H"olderlin \\
	\vspace*{.6cm}
	Mnemosyne \\
	\vspace*{.3cm}
}
	Ein Zeichen sind wir, deutungslos \\
	Schmerzlos sind wir und haben fast \\
	Die Sprache in der Fremde verloren. \\
	Wenn nemlich "uber Menschen \\
	Ein Streit ist an dem Himmel und gewaltig \\
	Die Monde gehn, so redet \\
	Das Meer auch und Str"ome m"ussen \\
	Den Pfad sich suchen. Zweifellos \\
	Ist aber Einer. Der \\
	Kann t"aglich es "andern. Kaum bedarf er \\
	Gesetz. Und es t"onet das Blatt und Eichb"aume wehn dann neben \\
	Den Firnen. Denn nicht verm"ogen \\
	Die Himmlischen alles. Nemlich es reichen \\
	Die Sterblichen eh'an den Abgrund. Also wendet es sich, das Echo \\
	Mit diesen. Lang ist \\
	Die Zeit, es ereignet sich aber \\
	Das Wahre.
}
\end{verse}
\end{minipage}
\end{center}
\vspace*{\stretch{1.62}}

%% file: P3.tex
\lhead[\fancyplain{}{}]%
      {\fancyplain{}{}}
\rhead[\fancyplain{}{}]%
      {\fancyplain{}{}}
\setlength{\parskip}{0.5cm}
\bigskip
\bigskip
\vspace*{0.5cm}
\begin{center}
{\bf{\Large{%
Hochenergiestreuung im\\[.2cm]
  {\zzz}Nichtperturbativen Vakuum der Quantenchromodynamik{\zzz}}\vspace*{18cm}\\}}
%
%
\hspace*{-10mm} Gutachter: Prof.\@ Dr.\@ Hans G"unter Dosch\\
\hspace*{-1.5mm} Prof.\@ Dr.\@ Bogdan Povh
\end{center}
\setlength{\parskip}{0cm}

%% file: P5.tex
\lhead[\fancyplain{}{}]%
      {\fancyplain{}{}}
\rhead[\fancyplain{}{}]%
      {\fancyplain{}{}}
\addtocontents{toc}{\protect\contentsline {chapter}{\numberline {}{\rm Zusammenfassung/Abstract}}{}}

\begin{center}
{\bf{Zusammenfassung}}
\end{center}
\setlength{\unitlength}{1mm}
%
Wir untersuchen Streuung in nichtperturbativer Quantenchromodynamik~(\DREI{Q}{C}{D}) f"ur gro"se invariante Schwerpunktenergie~$\surd s$.
Unserer Analyse zugrunde liegt das Modell des Stochas\-tischen Vakuums~(\DREI{M}{S}{V}) von Dosch und Simonov und die nichtperturbative \mbox{\,$s \!\to\! \infty$-asymp}\-totische Formel Nachtmanns f"ur die Streuung zweier Wegner-Wilson-Loops.
Wir argumentieren, da"s die G"ultigkeit der Formel hin zu gro"sen aber endlichen Werten von~$s$ erweitert werden kann dadurch, da"s die urspr"unglichen nahezu lichtartigen Loops {\sl nicht\/} ersetzt werden durch ihre exakt lichtartigen Limites.
Die $T$-Amplitude h"angt daher ab von~$s$ "uber den kinematischen Faktor hinaus.
Diese \mbox{$s$-Ab}\-h"angigkeit wird explizit berechnet.
Wir f"uhren abschlie\-"send durch die analytische Fortsetzung in Euklidische Raumzeit und machen auf diese Weise Hochenergiestreuung zug"anglich Euklidischen Theorien wie \DREI{Q}{C}{D} als \vspace*{-.125ex}Gittereichtheorie. \\
\indent
Auf Basis der
   \vspace*{-.125ex}\mbox{\,$s \!\to\! \infty$-asymp}\-totischen Formel berechnen wir ferner exklusive Photo- und Leptoproduktion von Vektormesonen am Proton:%
  ~\mbox{$\ga^{\scriptscriptstyle({\D\ast})}p \!\to\! V p$},~-- f"ur%
  ~\mbox{$V \!\equiv\! \rh(770),\om(782)$}, \mbox{$\ph(1020),\Jps(3097)$} und%
  ~\mbox{\,$\rh(1450), \rh(1700)$}.
Wesentlich zugrunde liegt die zu  verschwindenden Virtualit"a\-ten~$Q$ hin universalisierte Photon-Lichtkegelwellenfunktion der Lichtkegelst"orungstheorie~(\VIER{L}{C}{P}{T}) und entsprechend modellierte Wellenfunktionen f"ur die Vektormesonen.
Wir verifizieren den \DREI{M}{S}{V}-spezifischen Mechanismus der Ausbildung wechselwirkender gluonischer Strings zwischen den Quark-Konstitueneten.
%
%

%
\vspace*{1cm}
\begin{center}
{\Large{\bf{High-energy scattering in the\vspace*{1mm}\\ nonperturbative vacuum of Quantum Chromodynamics}}}
\end{center}

\begin{center}
{\bf{Abstract}}
\end{center}
%
%
We consider scattering in nonperturbative Quantum Chromodynamics~(\DREI{Q}{C}{D}) for large invariant centre of mass energy~$\surd s$.
Our analysis is based upon the Stochastic Vacuum Model~(\DREI{S}{V}{M}) of Dosch and Simonov and on the nonperturbative \mbox{$s \!\to\! \infty$-asymp}\-totic formula of Nachtmann for the scattering of two Wegner Wilson loops.
We argue, that its validity can be expanded to large but finite values of~$s$ by {\sl not\/} substituting the initial near light-like loops for their exact light-like limits.
Thus, the \mbox{$T$-amp}\-litude depends on~$s$ beyond the kinematical factor.
This \mbox{$s$-de}\-pendence is explicitly calculated.
We conclude performing the analytical continuation to Euclidean spacetime and in that way make accessible high-energy scattering to Euclidean theories such as \DREI{Q}{C}{D} lattice \vspace*{-.125ex}gauge theory. \\
\indent
Based upon the
   \mbox{$s \!\to\! \infty$-asymp}\-totic formula, we further calculate exclusive photo- and leptoproduction of a vectormeson at the proton:%
  ~\vspace*{-.125ex}\mbox{$\ga^{\scriptscriptstyle({\D\ast})}p \!\to\! V p$},~-- for%
  ~\mbox{$V \!\equiv\! \rh(770), \om(782), \ph(1020)$}, $\Jps(3097)$ and%
  ~\mbox{\,$\rh(1450), \rh(1700)$}.
Essential entities are the photon light-cone wave function of light-cone perturbation theory~(\VIER{L}{C}{P}{T}) universalised to vanishing virtualities~$Q$ and wave functions for the vectormesons modelled accordingly.
We verify the \DREI{S}{V}{M}-specific mechanism of formation of interacting gluonic strings between the quark constituents.
%
%

%
%
%

%
%
%
%
%

%% file: NOTATION.tex
\lhead[\fancyplain{}{}]%
      {\fancyplain{}{}}
\rhead[\fancyplain{}{}]%
      {\fancyplain{}{}}
\addtocontents{toc}{\protect\contentsline {chapter}{\numberline {}\vspace*{-7.75ex}}{}}
\addtocontents{toc}{\protect\contentsline {chapter}{\numberline {}{\rm Notation}}{}}

\vspace*{\stretch{100}}
\begin{center}
{\bf{Notation}}
\end{center}

\noindent
Griechische Indizes~$\mu$,~$\nu$,~$\rh$,~$\si,\ldots$ beziehen sich auf die Raumzeit-Koordinaten, i.a.~$0$,~$1$,~$2$,~$3$.

\medskip\noindent
Lateinische Indizes~$i$,~$j$,~$k,\ldots$ beziehen sich auf die raumartigen Koordinaten, i.a.~$1$,~$2$,~$3$.

\medskip\noindent
Vektoren sind definiert als Spalten kontravarianter Komponenten.

\medskip\noindent
Vektoren der Raumzeit sind bezeichnet mit dem Buchstaben an sich:~$x$.

\medskip\noindent
Karthesische Dreier-Vektoren sind bezeichnet durch einen Pfeil:~$\vec{x}$.

\medskip\noindent
Kartesische (Zweier-)Vektoren, in denen gestrichen sind die longitudinalen Komponenten (i.a.\@ mit Indizes~0,~3), sind bezeichnet als transversal und notiert in geradem Fettdruck:~$\rb{x}$.

\medskip\noindent
Zeitartige Komponenten der Raumzeit gehen in die Signatur derer Metrik ein mit~$+1$, raumartige Komponenten mit~$-1$.

\medskip\noindent
Dirac-Indizes werden i.a.\@ nicht notiert, wenn doch durch griechische Buchstaben~$\al$,~$\be$,~$\ga$,~$\de,\ldots$

\medskip\noindent
Griechische Indizes~$\al$,~$\be$,~$\ga$,~$\de,\ldots$ bezeichnen i.a.\@ die Matrixindizes einer nicht spezifizierten, beliebigen Darstellung~$\Drst{R}$ der Eichgruppe.

\medskip\noindent
Lateinische Indizes~$a$,~$b$,~$c$,~$d,\ldots$ nummerieren die Generatoren einer beliebigen Darstellung der Eichgruppe oder bezeichnen die Matrixindizes der adjungierten Darstellung~$\Drst{A}$.

\medskip\noindent
Lateinische Indizes~$m$,~$n,\ldots$ bezeichnen die Matrixindizes der fundamentalen Darstellung~$\Drst{F}$ der Eichgruppe.

\medskip\noindent
"Uber doppelt auftretende Indizes wird summiert, wenn nicht ausdr"ucklich anders vereinbart.

\medskip\noindent
Wir arbeiten in Einheiten, in denen das Plancksche Wirkungsquantum geteilt durch~Zwei~Pi und die Vakuum-Lichtgeschwindigkeit identisch Eins sind:~$\hbar \!\equiv\! c \!\equiv\! 1$.
\vspace*{-.25ex}Es existiert nur eine Dimension, o.E.d.A.\@ die der Energie, gemessen in\GeV.
\vspace*{-.25ex}Mit\;~\mbox{$\hbar \!=\! 6.582\,118\,89\,(26) \!\cdot\! 10^{-25}\GeV\,{\rm s}$},\; \mbox{$c \!=\! 299\,792\,458\;{\rm m\, s}^{-1}$}\; und\; \mbox{${\rm e} \!=\! 1.602\,176\,462\,(63) \!\cdot\! 10^{-19}\;{\rm C}$}\; gelten die \vspace*{-.125ex}Umrechnungsfaktoren: \mbox{$(\mskip-2mu1\mskip-2mu\GeV)\!/\!c^2 \!=\! 1.782\,661\,731\,(70) \!\cdot\! 10^{\mskip-2mu-\mskip-2mu27}\;{\rm kg}$}, \mbox{$(\mskip-2mu1\mskip-2mu\GeV)^{\mskip-2mu-\mskip-2mu1} \hbar c \!=\! 0.197\,326\,960\,2\,(77)\fm$}.

\vspace*{\stretch{1}}

%% file: QCD.tex
\lhead[\fancyplain{}{\sc\thepage}]%
      {\fancyplain{}{\sc\rightmarkQCD}}
\rhead[\fancyplain{}{\sc{{\footnotesize Einf"uhrung:} Quantenchromodynamik}}]
      {\fancyplain{}{\sc\thepage}}
\psfull
%

%
%

%
\chapter*{\vspace*{-.671ex}{\Large Einf"uhrung:}
            {\huge Quantenchromodynamik}}
\chaptermark{}                                     
\addcontentsline{toc}{chapter}{\numberline{}{\footnotesize Einf"uhrung:} Quantenchromodynamik}

Nach heutigem Verst"andnis der Physik gibt es vier fundamentale Wechselwirkungen, "uber die Materie miteinander in Beziehung tritt.
Grundlegendes Verst"andnis dieser Wechselwirkungen auf Basis fundamentaler Prinzipe impliziert notwendig ihr Verst"andnis auf Quantenniveau: auf Niveau ihrer fundamentalen Elementarteilchen.
Mit dem {\it Standardmodell der Elementarteilchenphysik\/}~(SM) verf"ugt die Physik "uber eine Theorie in diesem Sinne, die alle beobachteten Ph"anomene~-- nicht eingeschlossen die Gravitationswechselwirkung%
\FOOT{
   Konsequenz der kleinen Newtonschen Gravitationskonstanten~\mbox{$G \!=\! 6.707 \!\cdot\! 10^{-39}\, \hbar c(\!\!\GeV\!\!/\!c^2\!)^{\!-2}$}~ist~die~schwa\-che Kopplung des Gravitons.   Quanteneffekte manifestieren sich signifikant auf der Skala der Planck-Masse \mbox{$\mPl \!=\! \surd\hbar c\!/\!G \!=\! 1.221 \!\cdot\! 10^{19}\GeV\!\!/\!c^2$}, die Strukturen der Planck-L"ange \mbox{$\lPl \!=\! \hbar c \!/\! \mPl c^2 \!=\! 1.616 \!\cdot\! 10^{-20}\fm$} aufl"ost, das hei"st siebzehn Gr"o"senordnungen~(!) entfernt von den heute im Beschleuniger-Labor zug"anglichen~Skalen.
}~--
in beeindruckender~Wei\-se beschreibt.
Wir verweisen allgemein auf Ref.~\cite{Nachtmann92} und wenden uns unmittelbar dem Sektor des SM zu, auf den im wesentlichen sich unsere Arbeit bezieht: dem Sektor der Starken Wechselwirkung, der Quantenchromodynamik~(QCD).

Die QCD des SM%
\FOOT{
  Wir gehen nicht auf supersymmetrische QCD und Formulierungen im Rahmen von (Super)\-String\-theorien ein~\cite{Green95}.   Beide Konzepte zusammen sind hochaktuell:
  Arbeiten von Seiberg und Witten haben eine Dualit"at zwischen stark und schwach gekoppeltem Regime von $({\cal N}\zz =\zz 2)$-supersymmetrischen (oder super\-konformen) $SU(N)$-invarianten Eichtheorien aufgezeigt~\cite{Seiberg94,Seiberg94a}, die auch f"ur $({\cal N}\zz =\zz 1)$-supersymmetrische und SM-QCD bedeutende Konsequenzen hat~\cite{Shifman97}; vgl.\@ auch die Refn.~\cite{AlvarezGaume97,AlvarezGaume97a,AlvarezGaume97b,AlvarezGaume97c,Lerche97}. Nichtperturbative Beitr"age sind also in der dualen Theorie perturbativ zug"anglich.
  Nach einer Vermutung von Maldacena~\cite{Maldacena97} existiert eine Dualit"at zwischen der {\sl large-N\/}-Entwicklung einer konformen Quantenfeldtheorie in $d$ Dimensionen und der perturbativen Entwicklung einer {\sl M(atrix)-Stringtheorie\/} im Hintergrund einer $AdS_{d+1}\!\otimes\!M$-Raumzeit, wobei $AdS_{d+1}$ der $(d\!+\!1)$-dimensionale Anti-deSitter- und $M$ ein beliebiger kompakter Raum sind; vgl.\@ auch die Refn.~\cite{Gross98,Gross98a}. Auf der Basis einer Dualit"at in diesem Sinne zwischen {\sl superkonformer\/} $SU(N)$-invarianter nichtabelscher Eichtheorie in 4 Dimensionen~-- dies die Klasse von Theorien eines supersymmetrisch erweiterten SM~-- und einer Typ-IIB-Stringtheorie in einem $AdS_5\!\otimes\!\mbox{\boldmath$S$}^5$-Hintergrund w"are in nat"urlicher Weise die Vereinheitlichung mit der letzten Wechselwirkung im Rahmen einer Superstring-Gravitation greifbar.
}
ist mathematisch in das Konzept der {\it Lagrange'schen Quantenfeldtheorie\/} gefa"st, Refn.~\cite{Itzykson88,Nachtmann92,ZinnJustin96,Kugo97,Weinberg96,Cheng94}.
In diesem Rahmen bedeutet das zugrundeliegende Bild von {\it Wechselwirkung vermittels des Austauschs von Teilchen\/} die Kopplung von Quantenfeldern untereinander und gegebenenfalls an sich selbst.
Eine renormierbare Quantenfeldtheorie ist~-- bei Angabe eines Regularisierungs- und Renormierungsschemas und von Zahlenwerten f"ur dessen Parameter~-- vollst"andig bestimmt durch ihr Wirkungs-Funktional, das gegeben ist als das \mbox{$d \zz\equiv\zz 4$-dimen}\-sionale Raumzeit-Integral ihrer Lagrangedichte~$\mf{L}(x)$, die Funktional ist der fundamentalen Quantenfelder der Theorie.
Die Lagrangedichte ist allgemein bereits dadurch wesentlich in ihrer Struktur eingeschr"ankt, da"s f"ur $dx\mf{L}(x)$ Symmetrien, cum~grano~salis {\it Invarianzen\/} gefordert werden, die in der Natur~\mbox{realisierten} "`Symmetrien"' entsprechen.
Zus"atzlich wird gefordert vor dem Hintergrund der Newtonschen Axiome, da"s nicht h"ohere als zweite Ableitungen auftreten.
Das Prinzip,~\mbox{ausgehend} von fundamentalen "`Symmetrien"' zu grundlegenden Theorien zu gelangen, ist wichtiges Konzept der~Elementar\-teilchenphysik.
Wir tragen seiner Bedeutung im besonderen f"ur unsere Arbeit Rechnung~und gehen zun"achst ein auf die Invarianzen, die der Lagrangedichte der QCD inh"arent sind.

\section*{\vspace*{-.5ex}Invarianzen}
\enlargethispage{1ex}
\renewcommand{\rightmarkQCD}{Invarianzen}
\addcontentsline{toc}{section}{\numberline{} \hspace*{-12pt}Invarianzen}

\label{T:diskreteSymmetrien}{\bf Invarianz unter den diskreten Transformationen \boldmath$\cal C$, \boldmath$\cal P$ und \boldmath$\cal T$, einzeln}%
\FOOT{
  \label{FN:axiomatischeQFT}Invarianz bez"uglich der kombinierten $\cal CPT$-Transformation folgt bereits aus sehr allgemeinen Prinzipen: hinreichend sind schon die Wightman-Axiome, die Forderungen der Art stellen wie Kausalit"at, oder da"s der Impulsoperator seine Eigenwerte auf dem Vorw"artslichtkegel annimmt.   Zu diesen Axiomen wie auch zum axiomatischen Zugang zu Quantenfeldtheorie im allgemeinen sei verwiesen auf die Refn.~\cite{Streater80,Bogolubov90}. \\
  Zum Kontext:   Seit den f"unfziger, sp"atestens in den sechziger Jahren wurde das Konzept der Quantenfeldtheorie auf der Basis einer fundamentalen mikroskopischen Lagrangedichte immer mehr zwiesp"altig, ja als das falsche Konzept angesehen.   Ursache war zum einen das bekannte {\sl Zero~charge\/}-Problem der Quantenelektrodynamik (QED), vgl.\@ Ref.~\cite{Landau55}, das sich zwar erst bei extrem hohen Energien zeigt und somit als {\sl akademisch\/} abgetan werden konnte, aber nichts desto trotz eine prinzipielle Inkonsistenz der Theorie aufzeigt (Landau hat dies 1955 mit der Bemerkung konstatiert, "`weak coupling electrodynamics is a theory, which is, fundamentally, logically incomplete"', siehe Ref.~\cite{Landau55a}). Im Falle der Starken Wechselwirkung kann dieses Problem nicht mehr ignoriert werden.   Zum anderen und noch schwererwiegend war die tiefe Skepsis gegen"uber dem Verfahren der Renormierung, wie sie seit den ersten Ans"atzen durch Dirac~-- schon von ihm selbst (vgl.\@ Ref.~\cite{Salam90}) wie auch von Wigner als dem zweiten Wegbereiter der Lagrange'schen Quantenfeldtheorie~-- allgemein empfunden und immer wieder ausgedr"uckt wurde: Renormierung~-- ein mathematischer Trick, ihre physikalische Bedeutung nicht wirklich verstanden.
  Diese Skepsis blieb bestehen trotz des Erfolges von Renormierung in der QED und wurde geteilt selbst von deren Begr"undern.   So erkl"arte Feynman 1961 lokale Felder, da nicht direkt observabel, f"ur bedeutungslos: "`I still hold to this belief and do not subscribe to the philosophy of renormalization"', siehe Ref.~\cite{Feynman61}.
  An die Stelle einer Lagrangedichte in Feldern fundamentaler Konstituenten wurden in der {\sl axiomatischen Quantenfeldtheorie\/} die observablen Elemente der $S(treu)$-Matrix gesetzt, auf der Grundlage derer und unter der Annahme allgemeiner Prinzipe wie Kausalit"at, Unitarit"at und Analytizit"at~-- aber unter Aufgabe des "`unphysikalischen"' Anspruchs, die mikroskopische Dynamik verstehen zu wollen~-- man glaubte, quasi in einem {\sl Bootstrap\/}-Verfahren zu einer eindeutigen Theorie zu gelangen.
  Heute ist bekannt, da"s es unendlich viele solcher konsistenten $S$-Matrix-Theorien gibt: jede nichtabelsche Eichtheorie bez"uglich einer beliebigen halbeinfachen Liegruppe ({\sl Yang-Mills-Theorie} nach~Ref.~\cite{Yang54}), mit beliebig vielen Multipletts~$N_{\rm\!F}$ von Fermionen, solange sie sich noch UV-asymptotisch wie eine freie Theorie verh"alt (d.h.\@ $N_{\rm\!F} \!\le\!16$ f"ur das Beispiel von $SU(N)$-Invarianz).   Dies ist Konsequenz der Beweise von zum einen ihrer {\sl Renormierbarkeit\/} durch 't~Hooft 1971 und zum anderen ihrer {\sl Asymptotischen Freiheit\/} durch Gross, Wilczek 1973, vgl.\@ die Refn.~\cite{tHooft71,tHooft71a} respektive~\cite{Gross73}.   Beides sind Meilensteine der Elementarteilchenphysik, auch insofern als sie {\sl konstruktiv\/} zur Lagrange'schen Formulierung zur"uckgef"uhrt haben.   Vgl.\@ die Refn.~\cite{Gross98b,tHooft98,tHooft98a}.
}\hspace*{8.5pt}
entspricht den drei separaten Forderungen:
da"s ein physikalischer Proze"s prinzipiell genauso real sein sollte bei Vertauschung von Teilchen und Antiteilchen (Ladungskonjugationsinvarianz,~$\cal C$),
da"s es prinzipiell nicht entscheidbar sein sollte, ob die Welt, die man "`direkt"', oder ob die, die man in einem Spiegel sieht, die reale ist (Parit"atsinvarianz,~$\cal P$)
und da"s auf {\it mikroskopischer\/}, das hei"st der Ebene von Elementarteilchen, ebenso aus prinzipiellen Gr"unden, nicht entscheidbar sein sollte, ob ein Proze"s "`vorw"arts"' oder "`r"uckw"arts"' in der Zeit abl"auft (Zeitumkehrinvarianz,~$\cal T$).
Es sei angemerkt, da"s hierzu in Kontrast die Schwache Wechselwirkung~$\cal P$ wie auch die Kombination~$\cal CP$ verletzt.
\vspace*{-2ex}

\paragraph{Poincar\'einvarianz} bedeutet im Sinne der Speziellen Relativit"atstheorie~(SRT), da"s die mathematische Struktur der Gleichungen die gleiche sein sollte, ob physikalische Ph"anomene von einem "`ruhenden"' Beobachter beschrieben werden oder von einem Beobachter, der sich relativ zum ersten gleichf"ormig bewegt.
Die Feldgleichungen nach Euler-Lagrange sollten daher Tensorgleichungen sein bez"uglich einer Transformation zwischen solchen {\it nichtbeschleunigten}, oder {\it Inertialsystemen\/} und in definierter Weise~-- {\it kovariant\/}~-- transformieren. \\
\indent
Formal gesprochen, leiten sich diese Transformationen her von der Poincar\'egruppe~$\vrh$, der direkten Summe der Translationen in der Raumzeit und der Gruppe der Lorentztransformationen $\cal L$.
Kovarianz oder Tensorverhalten der Gleichungen hei"st genau zu fordern, da"s die Lagrangedichte zum einen nicht explizit vom Weltpunkt $x$ abh"angt und zum anderen sich wie ein Skalar verh"alt unter Transformationen der \label{T:spez-orth-LorentzGruppe}\vspace*{-.125ex}speziellen orthochronen Lorentzgruppe~${\cal L}_+^{\bm{\scriptstyle\uparrow}}$, das ist die Eins-Zusammengangskomponente der Lorentzgruppe.
Sei verwiesen auf Ref.~\cite{Nachtmann92}, bzgl.\@ tiefergehender Fragen auf die Refn.~\cite{Bogolubov90,Sexl92}.
\vspace*{-2ex}

\paragraph{Eichinvarianz} bedeutet, da"s die observablen Aussagen einer Quantenfeldtheorie dieselben bleiben, wenn ihre Lagrange'schen Felder einer {\it Eichtransformation\/} unterworfen werden.
Dieses Prinzip leitet sich letztlich aus folgender "Uberlegung her:
Da die quantentheoretische (Materie-)Wellenfunktion~$\ph(x)\!\in\!\bbbc$ zu interpretieren ist als {\it Amplitude\/} einer Wahrscheinlichkeitsdichte, beschreibt sie noch dieselbe Physik nach einer beliebigen Phasen"anderung und ist in diesem Sinne nur bestimmt bis auf Transformationen unter einer unit"aren Matrix, das hei"st einer Matrix $U\!\in\!U(1)$.
Analog sollte ein Spinorfeld~$\ps(x)$ in der fundamentalen Darstellung der speziellen unit"aren Gruppe $SU(N)$~-- wie es in einer realistischen Quantenfeldtheorie Materie beschreibt~-- nur bis auf eine Transformation unter einer Matrix $U\!\in\!SU(N)$ festgelegt sein.
Mit anderen Worten ist eine Klasse~$[\ps]$ {\it global\/}-$SU(N)$-"aquivalenter Spinorfelder definiert durch die Transformation $\ps(x)\!\to\!U\ps(x)$, die bezeichnet wird als {\it globale Eichtransformation\/} bez"uglich der {\it Eich\/}- oder {\it Strukturgruppe\/} $\mf{G}\!=\!SU(N)$. {\it Global\/} deshalb, weil die Matrix $U$ nicht vom Weltpunkt $x$ abh"angt. \\
\indent
Zugrundeliegend dem Bild von Wechselwirkung vermittels Teilchenaustausch sind andererseits genau die Forderungen von {\it Lokalit"at\/} und {\it Kausalit"at}.
Lokalit"at hei"st, Felder koppeln punktf"ormig aneinander.
Kausalit"at im Sinne der SRT hei"st, da"s Ereignisse an zwei Weltpunkten nur dann in einen kausalen Zusammenhang gebracht werden k"onnen, wenn es eine zeitartige Trajektorie gibt, die diese verbindet.%
\FOOT{
  Dies ist letztlich Konsequenz der pseudo-Riemannschen Signatur der Minkowski-Raumzeit, die genau zu Begriffen f"uhrt wie {\sl Zeitartigkeit\/}, {\sl Vorw"artslichtkegel\/} etc.   Auf Basis dieser Raumzeit impliziert die Forderung von Invarianz unter speziell-relativistischen Koordinatentransformationen zun"achst die Existenz einer Geschwindigkeit~$c$, die {\sl fundamental\/} ist in dem Sinne, da"s sie invariant ist unter diesen Transformationen, das hei"st in einem beliebigen Inertialsystem dieselbe ist.   Dies wiederum impliziert, da"s {\sl kausal\/} nicht schneller Einflu"s genommen werden kann als mit~$c$.   Physikalisch ist~$c$ zu identifizieren mit der Vakuumlichtgeschwindigkeit.   Wir verweisen auf Ref.~\cite{Sexl92}.
}
In conclusio:
Austauschteilchen propagieren vom einen zum anderen {\it Wechselwirkungspunkt\/} und ben"otigen dazu {\it eine von Null verschiedene Zeit\/}.
Fernwechselwirkung existiert nicht. \\
\indent
Mit diesem Prinzip in Widerspruch steht, da"s bei einer globalen Eichtransformation an beliebig entfernten Weltpunkten unter der gleichen, auf der Raumzeit konstanten Matrix~$U$, also {\it instantan\/} transformiert wird.
Auf diesen Fernparallelismus hat als erster Weyl kritisch hingewiesen.
Seine L"osung ist genau das Konzept der {\it (lokalen) Eichtheorie\/}; siehe den "Ubersichtsartikel von Straumann~\cite{Straumann87} und die Originalrefn.~\cite{Weyl18,Weyl18a,Weyl18b,Weyl19}. \\
\indent
Wird nicht {\it globale\/}, sondern st"arker {\it lokale\/} Eichinvarianz gefordert, das hei"st die Transformationsmatrix darf von Weltpunkt zu Weltpunkt variieren, also $U\!\to\!U(x)$, die $\in\!SU(N)$ f"ur jedes $x$ (mathematisch: $[\ps]$ ist die Klasse {\it lokal\/}-$SU(N)$-"aquivalenter Spinorfelder), so impliziert dies zwangsl"aufig die Existenz eines Felds $A(x)$ des {\it Tangentialfaserb"undels\/}~-- das also f"ur jeden Weltpunkt $x$ der Raumzeit-Mannigfaltigkeit Element der in $x$ angehefteten Tangentialfaser der Gruppe $SU(N)$ ist, das hei"st eines in $x$ angebrachten isomorphen Bildes der Liealgebra~$su(N)$.%
\FOOT{
  Wir beschr"anken uns auch weiterhin auf $\mf{G}\!=\!SU(N)$ als prominenteste Klasse von Transformationsgruppen insofern, als sie dem SM zugrunde liegt. Wir merken aber an, da"s wesentlich zur Eichung einer Quantenfeldtheorie mit einer Gruppe $\mf G$ die Bijektivit"at der Exponentialabbildung ist, die das Feld~$A(x)$ des Tangentialfaserb"undels in die Gruppe selbst abbildet; diese Eigenschaft ist schon gegeben, wenn $\mf G$ eine zusammenh"angende kompakte Liegruppe ist.
}
Das Feld~$A(x)$ wird als {\it Eichfeld\/} bezeichnet und beschreibt physikalisch genau das Austauschteilchen, das die {\it Eichwechselwirkung\/} vermittelt. \\
\indent
\label{T:Eichtheorie}Weyls Konzept der Eichtheorie entwickelt sich weitgehend aus infinitesimalgeometrischen "Uberlegungen.
Wir wollen einen fundierten Abri"s in Hinblick auf Eichtheorien in der Physik geben.
Wir f"uhren ein in die Begriffe der zugrundeliegenden Mathematik, der {\it Differential\-geometrie auf Liealgebren\/}.
Vgl.\@ die~Refn.~\cite{Sexl95,Klingmann93} bzgl.\@ des Kalk"uls alternierender Differentialformen und Ref.~\cite{Sexl95} Kap.~11, und die Refn.~\cite{Klingmann94,Beiglboeck94} bzgl. der Anwendung auf Lie-Strukturen (die Begriffe \vspace*{-.25ex}{\it Fasermetrik\/}, {\it G-Struktur}). \\
\indent
Sei $M$ eine Mannigfaltigkeit und sei an jedem Punkt $x\!\in\!M$ eine "`Kopie"' $\mathbb{V}_x$ (d.i.\@ ein isomorphes Bild) irgendeines fest vorgegebenen Vektorraums $\mathbb{V}$ angebracht.
Ein derartiges Gebilde hei"st ein {\it Vektor(raum)b"undel\/} "uber der Basis $M$ mit Standardfaser $\mathbb{V}$, symbolisch $E(M,\mathbb{V})$; die Vektorr"aume $\mathbb{V}_x$ hei"sen seine {\it Fasern}.
Die Verallgemeinerung des Tangentialvektorfelds in diesem Rahmen ist der Begriff des {\it Schnittes\/} eines B"undels:
Ein Schnitt $\la$ ordnet jedem Punkt $x\!\in\!M$ einen Vektor $\la(x)\!\in\!\mathbb{V}_x$ aus der Faser "uber~$x$ zu.
Spezialf"alle sind das {\it Tangentialb"undel\/}~$T(M)$ mit $\mathbb{V}_x\!:=\!T_x$, das {\it Kotangentialb"undel\/}%
\FOOT{
  Die linearen Funktionale "uber~$\mathbb V$ bilden den zu $\mathbb V$ {\sl dualen Vektorraum\/}~$\mathbb{V}^{\D\ast}$, d.h.\@ f"ur~$a\!\in\!\mathbb{V}^{\D\ast}$,~$u,v\!\in\!\mathbb{V}$ und~$\al,\be\!\in\!\bbbr$ gilt:~$a(v)\!=$ reelle Zahl (Skalarprodukt von $a$ und $v$) und~$a(\al u\!+\!\be v)\!=\!\al a(u)\!+\!\be a(v)$.
}
$T^{\D\ast}(M)$ mit $\mathbb{V}_x \!:=\! T^{\D\ast}_x$ und allgemeiner {\it Tensorb"undel\/}~$T_b{}^a(M)$ mit~$\mathbb{V}_x\!:=\!T_x\! \otimes\!\ldots\!\otimes \!T^{\D\ast}_x\! \otimes\!\ldots$
Analog werden mit einem gegebenen Vektorb"undel $E(M,\mathbb{V})$ weitere B"undel $E^{\D\ast}(M,\mathbb{V}^{\D\ast})$ und%
  ~\vspace*{-.25ex}$E_b{}^a(M,\mathbb{V}_b{}^a)$ assoziiert. \\
\indent
Sei~$\{\pa_\mu\}$ eine holonome%
\FOOT{
  Eine Basis eines Tangentialraumes $T_x(M)$ hei"st {\sl holonom}, falls (lokal) Basisvektoren $\pa_\mu$ gefunden werden k"onnen, die Tangentialvektoren sind an die Koordinatenlinien eines geeigneten Koordinatensystems.
}
Basis des Tangentialb"undels~$T(M)$ und sei $\{dx^\mu\}$ die (dann ebenfalls holonome) Kobasis, das hei"st die dazu duale Basis des assoziierten Kob"undels $T^{\D\ast}(M)$; es gilt also \mbox{$dx^\mu(\pa_\nu)\!=\de^\mu{}_\nu$}.
Die Komponentenzerlegung einer beliebigen $p$-Form~-- d.i.\@ ein {\it Ko(tangential)tensorfeld\/}~$\in\!\La^p(T_x^{\D\ast})\;\forall x$, wo $\La$ die {\it "au"sere\/} (oder Gra"smann-) {\it Algebra\/} ist~-- lautet dann%
  ~\vspace*{-.25ex}\mbox{$a(x)\!=\!a_{\mu_{\!1}\!\cdots\!\mu_{\!p}}\!(x)\, dx^{\mu_{\!1}}\wedge\cdots\wedge dx^{\mu_{\!p}}$}. \\
\indent
Mit diesen Begriffen setzen wir an, wo die "Uberlegungen Weyls ihren Ausgangspunkt nehmen, n"amlich bei der Feststellung der Inkommensurabilit"at von Elementen aus Fasern, die an verschiedenen Punkten~$x$ der Mannigfaltigkeit~$M$ (hier: an verschiedenen Weltpunkten der speziell-relativistischen Minkowski-Raumzeit) angeheftet sind: \\
\indent
Formal ist das Eichfeld%
  ~\vspace*{-.125ex}$A(x)$, in o.g.\@ Basis $A(x)\!=\!A_\mu(x)dx^\mu$, eine $E(M,\mathbb{V})$-wertige 1-Form, wo $\mathbb{V}$ ein Dar\-stellungsraum der Eichgruppe $\mf{G}$ ist (hier: der Vektorraum der adjungierten Darstellung der~$SU(N)$), das hei"st f"ur jedes $x\!\in\!M$ ist es ein Element aus $\mathbb{V}_x\!\otimes\!T^{\D\ast}_x(M)$: Element der {\it an $x$ angehefteten Vektorfaser},~-- und daher {\it Menge von Elementen verschiedener R"aume\/}, die insofern%
  ~\vspace*{-.25ex}{\it inkommensurabel\/}, das hei"st nicht miteinander vergleichbar sind. \\
\indent
Kommensurabilit"at kann definiert werden nur f"ur Vektoren~$A$ derselben Faser (allgemein: f"ur Tensoren derselben Tensorfaser).
Um Vektoren vergleichen zu k"onnen, die zu Fasern zweier infinitesimal benachbarter Punkte $x$ und $x\!+\!dx$ geh"oren, bedarf es der Definition einer {\it "Ubertragung\/} (auch bezeichnet als: {\it Zusammenhang, Konnexion\/}).
Dies ist eine Vorschrift, wie sich ein Vektor der Faser "uber $x$ in die Faser "uber $x\!+\!dx$ "ubertr"agt.
Dabei wird per definitionem die "`Richtung"' von%
  ~\vspace*{-.25ex}$A$ nicht ge"andert. \\
\indent
"Aquivalent zur Definition einer bez"uglich des Vektorb"undels $E(M,\mathbb{V})$ {\it affinen torsionsfreien "Ubertragung} von beliebigen $E(M,\mathbb{V})$-wertigen $p$-Formen (insbes.\@ der 1-Form $A$)~-- dies ist eine "Ubertragung in obigem Sinne, die {\it linear\/} in der "`"Ubertragungsstrecke"' ist, und f"ur das Tangentialfaserb"undel $T(M)$ eine {\it holonome\/} Basis besitzt~-- ist die Definition einer {\it kovarianten Ableitung}, formal: die Definition einer {\it "au"seren kovarianten Ableitung~$D\wedge$ von~$E(M,\mathbb{V})$-wertigen $p$-Formen}.
Da Vektorb"undel-wertige $p$-Formen jedem Punkt~$x\!\in\!M$ ein Element aus $\mathbb{V}_x\otimes\La^p(T^{\D\ast}_x)$ zuordnen, sind sie Summen von Produkten $\la\otimes\al$, wo~$\la$ ein Schnitt des B"undels $E(M,\mathbb{V})$ und $\al$ eine gew"ohnliche%
  ~\vspace*{-.25ex}$p$-Form ist. \\
\indent\enlargethispage{1.25ex}
Es gen"ugt dabei, die Operation $D\wedge$ f"ur solche Produkte~$\la\!\otimes\!\al$ als~$E(M,\mathbb{V})$-wertige $(p\!+\!1)$-Form zu definieren, und Additivit"at zu fordern.
Man definiert also eine Leibniz-Regel $D\wedge(\la\!\otimes\!\al)\!:=\!(D\la)\wedge\al+\la\otimes d\wedge\al$, wobei $D$ entweder auf den Schnitt~$\la$ oder auf die gew"ohnliche $p$-Form~$\al$ wirkt.
F"ur die $p$-Form wird in nat"urlicher Weise%
  ~\vspace*{-.125ex}$D\wedge\al\!:=\!d\wedge\al$ gesetzt ($d$~der gew"ohnliche Differentialoperator f"ur $p$-Formen; in o.g.\@ Basis:~$d \!\equiv\! dx^\mu \pa_\mu$).

Eine kovariante Differentiation $D$ in einem Vektorb"undel $E(M,\mathbb{V})$ ist andererseits formal eine Vorschrift, die jedem Tangentialvektorfeld $u$ und jedem Schnitt $\la$ des B"undels einen Schnitt $D_u\la$ zuordnet, das hei"st seine kovariante Ableitung nach $u$.
Das {\it kovariante\/} (oder: {\it absolute\/}) Differential $D\la$ wird daher als {\it $E(M,\mathbb{V})$-wertige 1-Form\/} definiert, wobei $(D\la)(u)$ f"ur jedes Vektorfeld $u$ der Schnitt $(D\la)(u)\!:=\!D_u\la$ ist.
Explizit ist dieses $D$~-- und damit die "Ubertragung~-- definiert durch seine Wirkung auf eine {\it Schnittbasis\/} $\{b_I\}$, wobei $I$ eine Indexmenge von der Dimension der Faser~$\mathbb{V}$ ist, das hei"st, "uber die Gleichung $D b_I\!=\!\iIM\,g A^J{}_I\!\otimes\!b_J$, durch die Matrix~$(A^J{}_I)$ von 1-Formen. Dies ist die {\it "Ubertragungsmatrix\/}, $g$~wird als {\it Eichkopplung\/} bezeichnet. {\it Kovarianz\/} ist dabei garantiert durch definiertes Verhalten bei Eichtransformationen, das hei"st bei einem Basiswechsel%
  ~\vspace*{-.25ex}$\{b_I\!\to\!U^J{}_I b_J\}$. \\
\indent
Ohne neue Strukturen einf"uhren zu m"ussen, kann diese Operation von~$D$ auf die assoziierten Ko- und Tensorb"undel%
  ~\vspace*{-.125ex}$E^{\D\ast}(M,\mathbb{V}^{\D\ast})$ und~$E_b{}^a(M,\mathbb{V}_b{}^a)$ ausgedehnt werden, worauf wir aber nicht weiter \vspace*{-.125ex}eingehen wollen. \\
\indent
Jedenfalls ist damit die kovariante Ableitung~$D$ und damit die "Ubertragung vollst"andig bestimmt.
F"ur einen beliebigen Schnitt $\la$ mit Komponentenzerlegung $\la\!=\!\la^I b_I$, die $\la^I$ also gew"ohnliche Funktionen, ergibt sich $D\la\!=\!(D\la)^I\!\otimes b_I$ mit $(D\la)^I\!:=\!d\la^I\!+\!\iIM\,gA^I{}_J\la^J$.
\label{T:Eichtransformation}In der Notation der Elementarteilchenphysik:~$D\!=\!\bbbone\,d \!+\! \iIM\,gA$, mit dem Verhalten unter Eichtransformationen~$\ps \!\to\! \ps^U \!:=\! U\ps$ (mit~$U(x)$ beliebig $\!\in\!\mathbb{V}_x$,~$\forall x$) {\it per definitionem\/} wie%
  ~\vspace*{-.25ex}$D \!\to\! D^U \!:=\! UDU^{\D\dagger}$. \\
\indent
Das B"undel $E(M,\mathbb{V})$, aufgefa"st als Mannigfaltigkeit, hat im allgemeinen nichtverschwin\-dende Kr"ummung, was sich ausdr"uckt in der nichttrivialen $E(M,\mathbb{V})$-wertigen Kr"ummungs-2-Form\label{T:Kruemmungsform}: $F \!=\! (\iIM\,g)^{-1}D\wedge D \!=\! (d\wedge A) + \iIM\,g\,A\wedge A \!=\! (D\wedge A)$, in o.g.\@ Basis: $F\!=\!\frac{1}{2}F_{\mu\nu}dx^\mu\wedge dx^\nu$, so da"s%
  ~\vspace*{-.25ex}\mbox{$F_{\mu\nu}\!=\!\pa_\mu A_\nu\!-\!\pa_\nu A_\mu\!+\!\iIM\,g[A_\mu,A_\nu]$}, mit $[\,\cdot\,,\,\cdot\,]$ dem Kommutator ist. \\
\indent
Die durch $D$ vermittelte Richtungs"ubertragung in $E(M,\mathbb{V})$ ist also {\it nicht-integrabel\/}, das hei"st wegabh"angig und daher definiert nur, wenn bezogen auf eine Kurve ${\cal C}\!:\![0,1]\!\to\!M$:
Die~"`Rich\-tung"' eines Schnittes $\la$ (insbes.\@ des Vektors $A$), die sich bei seiner kovarianten "Uber\-tragung per definitionem nicht "andert, ist genau seine Richtung bez"uglich dieser Kurve; man spricht daher auch von {\it Parallelverschiebung\/}.
Ein Schnitt~$\la$, dessen Werte $\la(x)\!\in\!\mathbb{V}_x$,\; $\la(x\!+\!dx)\!\in\!\mathbb{V}_{x+dx},\;\ldots$ im Sinne der "Ubertragung {\it parallel} sind, erf"ullt $D\la\!=\!0$, was so viel hei"st wie:%
  ~\vspace*{-.25ex}\mbox{(Wert in~$x\!+\!dx$) $\equiv$ (Wert in~$x$, nach~$x\!+\!dx$ "ubertragen)}. \\
\indent
Wir schlie"sen diesen Paragraphen "uber das allgemeine Konzept der Eichtheorie mit einer Bemerkung zu Weyls urspr"unglicher Intention, der Konstruktion einer "`konsistenten"'%
\FOOT{
  \label{FN:Weyl-Einstein}Weyl im Briefwechsel mit Einstein, zitiert nach Straumann~\cite{Straumann87}, Originalrefn.\@ ebenda.
}
Allgemeinen Relativit"atstheorie~\vspace*{-.25ex}(\DREI{A}{R}{T}). \\
\indent\enlargethispage{.125ex}
Weyl stellte fest, da"s in der Riemannschen Geometrie, wie sie der ART Einsteins zugrunde liegt, "`ein letztes ferngeometrisches Element"'~\citeFN{FN:Weyl-Einstein} enthalten ist.
Die Riemannsche Metrik erlaubt es n"amlich, die {\it L"ange\/} von Vektoren in zwei voneinander beliebig entfernten Weltpunkten zu vergleichen, w"ahrend ja die {\it Richtung\/}s"ubertragung nicht-integrabel, also vom Weg der "Ubertragung abh"angig ist.
Diese "`Inkonsistenz"'~\citeFN{FN:Weyl-Einstein} zu beseitigen, gelangte er zu einer Formulierung der ART, die im ausgef"uhrten Sinne eine (lokale) Eichtheorie ist "uber der Strukturgruppe $\mf{G}$ der allgemeinen Koordinatentransformationen, formal: $\mf{G}\!=\!\vrh_+^{\bm{\scriptstyle\uparrow}}$, die spezielle orthochrone Poincar\'egruppe (d.i.\@ die Eins-Zusammenhangskomponente von $\vrh$).
Das darin auftretende Eichfeld $A$ identifizierte Weyl mit dem elektromagnetischen Feld, siehe die Refn.~\cite{Weyl29,Weyl31,Weyl31a}, und sah so "`Elektrizit"at und Gravitation aus einer gemeinsamen Quelle"'~\citeFN{FN:Weyl-Einstein} hergeleitet.
In Abwesenheit elektromagnetischer Felder~-- die L"angen"ubertragung wird integrabel und es gibt eine Eichung, in der $A$ verschwindet~-- geht seine Theorie in die Einsteinsche "uber. Sie "`gestattet sogar in einem gewissen Sinne zu begreifen, warum die Welt vierdimensional%
  ~\vspace*{-.25ex}ist"'~\citeFN{FN:Weyl-Einstein}. \\
\indent
Diese mathematisch konsequentere Weylsche~\DREI{A}{R}{T} hat weitreichende Konsequenzen.
So sollten beim bekannten Zwillingsparadoxon~-- wie es sich aus der Zeitdilatation der SRT erkl"art~-- zus"atzlich {\it Uhreneffekte zweiter Art\/} auftreten, das hei"st die wieder an einen Weltpunkt zusammengef"uhrten Uhren sich nicht nur in ihrem {\it Stand\/} sondern auch in ihrem~{\it Gang\/} unterscheiden.
Dies bedeutet, der Begriff "`Einheitsuhr"' ist nicht mehr definiert, da der Gang einer Uhr von ihrer Vorgeschichte abh"angt (analog der Begriff "`Einheitsma"sstab"'). Das invariante Linienelement $ds$ ist nicht mehr "uber Me"sprozesse mit unendlich kleinen Ma"sst"aben und Uhren zug"anglich, sondern Lichtstrahlen sind das einzige Mittel, metrische Verh"altnisse in der Umgebung eines Weltpunktes \vspace*{-.25ex}empirisch zu ermitteln. \\
\indent
Solche Uhreneffekte sind heute mit gro"ser Sicherheit ausgeschlossen, das hei"st die Weylsche~\DREI{A}{R}{T} ist in der Natur nicht realisiert.
Die Relativit"atstheorie beh"alt "uber $ds$ ihre empirische Basis.
Geblieben, und zwar beherrschend die Elementarteilchenphysik heute, ist~-- "`Genie-Streich ersten Ranges"'~,~"`grandiose Leistung des Gedankens"',~"`Nat"urlichkeit und Sch"onheit [des] Gedankenganges"'%
\FOOT{
  Einstein im Briefwechsel mit Weyl, zitiert nach Straumann~\cite{Straumann87}, Originalrefn.\@ ebenda.
}~--
Weyls Konzept der Eichtheorie.
\vspace*{-1ex}

\section*{\vspace*{-.5ex}"`Ph"anomene und Konzepte"'}
\renewcommand{\rightmarkQCD}{"`Ph"anomene und Konzepte"'}
\addcontentsline{toc}{section}{\numberline{} \hspace*{-12pt}"`Ph"anomene und Konzepte"'}

Die Invarianzen der Lagrangedichte der QCD und die Struktur nichtabelscher Eichtheorien sind ausf"uhrlich dargestellt.
Wir zeichnen den Weg der Ph"anomenologie der Starken Wechselwirkung nach hin zur QCD, einer Eichtheorie auf Basis der nichtabelschen Strukturgruppe%
  ~\vspace*{-.25ex}\mbox{\,$\mf{G} \!=\! \SUNc$}, mit~$\Nc \!=\! 3$ die Anzahl der {\it Colour\/}-Freiheitsgrade. \\
\indent
Im Anfang war die QED~$\ldots$
Die QCD ist konstruiert in Analogie zur QED, motiviert durch deren Erfolg.
Dieser dokumentiert sich im Postulat physikalischer Gr"o"sen mit in keiner anderen Theorie erreichten Pr"azision:
Beispielsweise folgt f"ur das {\it anomale magnetische Moment\/} des Myons,%
  ~\mbox{$a_\mu \!=\! (g_\mu\!-\!2)/2$}, theoretisch%
  ~\vspace*{-.125ex}\mbox{\,$a_\mu|_{\rm th.} \!=\! 116\,591\,596 \:(67) \times\!10^{\!-11}$} im Vergleich mit experimentell%
  ~\vspace*{-.125ex}\mbox{\,$a_\mu|_{\rm exp.} \!=\! 116\,592\,300 \:(840) \times\!10^{\!-11}$}; vgl.\@ die Refn.~\cite{Wilczek99,PDG00}, insbes.\@ Ref.~\cite{Davier98} bzgl.\@ der darin enthaltenen Korrekturen aufgrund der SM-Starken und Schwachen Wechselwirkung.
Sicherlich r"uhrt diese Pr"azision mit her von dem gl"ucklichen Umstand, da"s der Entwicklungsparameter der perturbativen Theorie auf {\it in praxi\/} relevanten Skalen numerisch klein ist.
Letztlich Ursache ist aber die Theorie: neben dem Renormierungs\-verfahren das {\it Prinzip der Eichinvarianz\/}, das der QED zugrundeliegt auf Basis der {\it abelschen\/} Strukturgruppe $U(1)$.~--
Der Erfolg der QED hat von Anfang an dazu hingef"uhrt, das Eichprinzip auf h"ohere, nichtabelsche Gruppen zu verallgemeinern, um in analogem Rahmen zu einer konsistenten Theorie der \vspace*{-.25ex}Starken Wechselwirkung zu gelangen. \\
\indent
So formulierten schon~1954 Yang und Mills in Ref.~\cite{Yang54} eine Eichtheorie f"ur die Wechselwirkung von Nukleonen, die formal basiert auf der Forderung lokaler $SU(2)$-Invarianz und sich bezieht auf den Raum des Heisenbergschen Isospins.%
\FOOT{
  Heisenberg in Ref.~\cite{Heisenberg32}:~"`und durch eine~$\ldots$ Zahl~$\vrh^\ze$, die der beiden Werte~$+1$ und~$-1$ f"ahig ist.~$\vrh^\ze \!=\! +1$ soll bedeuten, das Teilchen sei ein Neutron,~$\vrh^\ze \!=\! -1$ bedeutet, das Teilchen sei ein Proton"'.
}
Dies genau war die Konstruktion der neuen Klasse {\it nichtabelscher\/} Eichtheorien.%
\FOOT{
  Die Weylsche ART ist die "alteste Eichtheorie; mit der Strukturgruppe $\vrh_+^{\bm{\scriptstyle\uparrow}}$ auch die "alteste nichtabelsche.
}
In Referenz auf diese Arbeit werden Eichtheorien auf der Basis halbeinfacher Liegruppen $\mf{G}$~-- oft spezieller: auf Basis der~$SU(N)$~-- bezeichnet als \vspace*{-.25ex}{\it Yang-Mills-Theorien\/}. \\
\indent\enlargethispage{1.125ex}
Diese Klasse von Theorien geriet danach wieder in Vergessenheit.
Grund daf"ur waren zum einen Fragen prinzipieller Natur, die sich schon bei der QED gestellt hatten.
Wir haben in Fu"snote~\FN{FN:axiomatischeQFT} das {\it Zero~charge\/}-Problem und die tiefe Skepsis gegen"uber dem Verfahren der Renormierung als solchem angesprochen, und da"s dies bis hin zur generellen Ablehnung Lagrange'scher \vspace*{-.125ex}Quantenfeldtheorie gef"uhrt hat.
Beides, so kann man im nachhinein sagen, resultierte gro"senteils aus einem mangelnden Verst"andnis der Renormierungsgruppe.

Zum anderen koppeln nichtabelsche Eichfelder auch an sich selbst und induzieren so Nichtlinearit"aten, die die Strukturen gegen"uber der abelschen Theorie wesentlich ver"andern. \\
\indent
Ferner war ein massives Problem, \mbox{da"s Eichinvarianz einen {\it expliziten Massenterm} f"ur das} Eich-Vektorfeld~$A$ verbietet, Yang-Mills-Theorien es daher~-- zun"achst~-- als {\it masseloses\/} Feld postulieren.
Dies schien weder vereinbar zu sein mit der Vier-Fermion-Kopplung der wohl\-etablierten, wenn auch nicht renormierbaren Fermi-Theorie der Schwachen Wechselwirkung, noch waren masselose intermedi"are Vektorbosonen der Starken \vspace*{-.25ex}Wechselwirkung beobachtet. \\
\indent
Was speziell die Starken Wechselwirkung betraf, war nicht klar, auf welche der experimentell beobachteten Symmetrien die Strukturgruppe~$\mf{G}$ zu beziehen sei.
Das Quarkmodell~\cite{GellMann64b,Zweig65,Zweig64} war zwar anerkannt mit {\it Quarks\/} als Materie- und {\it Gluonen\/} als Eich-Vektorfeldern (die die Wechselwirkung vermitteln), mit der {\it Colour\/}-Symmetrie~-- allerdings als reines {\it Abstraktum\/}, reine {\it Arbeitshypothese\/}, die keinen direkten Bezug zur Wirklichkeit beanspruchte:
So fa"ste noch 1972 Gell-Mann (neben Zweig der Begr"under des Quarkmodells), fast entschuldigend, als den "`main point"' seinen vorausgegangenen Vortrag zusammen:
"`Since the entities [quarks and some kind of glue] we start with are fictitious, there is no need for any conflict with the bootstrap or conventional dual parton point of view"', vgl.\@ Ref.~\cite{GellMann72}.
Oder im Zusammenhang mit Stromalgebren, aber aus dem gleichen Denken heraus~\cite{GellMann64a}: "`We construct a mathematical theory of the strongly interacting particles, which may or may not have anything to do with reality, find suitable algebraic relations that hold in the model, postulate their validity, and then throw away the model"'.
Dieser Vorbehalt r"uhrte daher, da"s die Lagrange'schen Felder des Quarkmodells asymptotisch nicht beobachtet waren und ihr {\it Confinement\/} in Colour-Singletts als Mechanismus, der dies aus prinzipiellen Gr"unden verhindert, nicht vorstellbar schien.
Die Betonung lag daher lange auf der approximativen $SU(3)$-{\it Flavour\/}-Symmetrie~\cite{GellMann61,GellMann64,Neeman61,GellMann72}, die direkt observabel war. (Heute ist klar, da"s sie mehr oder weniger zuf"allig ist, dadurch da"s die drei leichtesten Quarks relativ leicht sind im Vergleich zu~$\La_{\rm QCD}$, \vspace*{-.25ex}der hadronischen Skala.) \\
\indent
Es war dies eine Reihe von Problemen, die gel"ost werden mu"sten. Wir wollen nur kurz die wichtigsten Stationen der Entwicklung nennen, die dahin gef"uhrt hat, da"s 1974 Quantenchromodynamik~-- in ihrer heutigen Gestalt als $\SUNc$-invariante Yang-Mills-Theorie~-- schlagartig anerkannt war; siehe auch die \vspace*{-.25ex}Darstellung in den Refn.~\cite{Gross98b,tHooft98}. \\
\indent
Diese Entwicklung wurde sehr konkret Mitte der sechziger Jahre, als Higgs und Kibble einen Mechanismus identifizierten, der Symmetrien der Lagrangedichte unter einer kontinuierliche Transformation spontan bricht; siehe die Refn.~\cite{Higgs64,Higgs64a,Higgs66,Guralnik65}, in Hinblick auf gruppentheoretische Aspekte Ref.~\cite{Kibble67}.
Diesem {\it Higgs-Kibble-Mechanismus\/} liegt wesentlich das {\it Goldstone-Theorem\/}~\cite{Goldstone61} zugrunde, das besagt, da"s Nicht-Invarianz des Vakuums unter einer kontinuierlichen Symmetrie der Lagrangedichte die Existenz masseloser Spin-0-Bosonen impliziert.
Angewandt auf lokale Eichtheorien werden Massenterme f"ur die Eich-Vektorfelder generiert:
Die {\it would~be\/} masselosen Goldstone-Bosonen kombinieren mit den {\it would~be\/} masselosen Eich-Vektorfeldern~-- bis auf jeweils eines~-- zu massiven Vektorfeldern; sie liefern deren longitudinalen Freiheitsgrad, die Anzahl der Freiheitsgrade \vspace*{-.25ex}insgesamt bleibt erhalten. \\
\indent\enlargethispage{.625ex}
Schon 1961 schlug Glashow~\cite{Glashow61} vor zur Vereinheitlichung der Schwachen mit der elektro\-magnetischen Wechselwirkung eine Yang-Mills-Theorie auf der Basis der nichtabelschen Strukturgruppe%
\FOOT{
  Die $U(1)$ hier bezieht sich auf {\sl Hyperladung\/} und ist nicht identisch mit der Eichgruppe der blo"sen QED.
}
  \vspace*{-.125ex}\mbox{\,$\mf{G} \!=\! SU(2) \!\otimes\! U(1)$}, mu"ste sie allerdings noch von Hand ab"andern, um der Masse der intermedi"aren Vektorbosonen Rechnung zu tragen.
Salam~\cite{Salam64,Salam68} und Weinberg~\cite{Weinberg67} gelangten zu derselben Theorie, bedienten sich zur Generierung von Massentermen f"ur die Eichfelder aber des \vspace*{-.125ex}Higgs-Kibble-Mechanismus'.
Unbefriedigend blieb noch der hadronische Sektor der Theorie.
Unbeantwortet war noch die Frage ihrer Renormierbarkeit~-- geschweige denn, da"s ein \vspace*{-.125ex}Renormierungsverfahren explizit formuliert gewesen w"are. \\
\indent
Dennoch fand diese vereinheitlichte Theorie augenblicklich Beachtung.
Veltman glaubte an eine nichtabelsche Eichstruktur der Schwachen Wechselwirkung und arbeitete seit 1963 daran, ein Renormierungsschema f"ur Yang-Mills-Theorien zu finden; vgl.\@ die entsprechenden Referenzen in~\cite{tHooft98}.
Allerdings lehnte er den Higgs-Kibble-Mechanismus ab, wie dieser auch allgemein als "`ugly"' empfunden wurde im Vergleich zum "`clean"' Konzept von Yang-Mills-Theorien.
Nicht so 't~Hooft, er pr"aferierte ihn gegen"uber der Alternative der Einf"uhrung expliziter Massenterme, da diese die Eichinvarianz preisgaben~-- und damit Unitarit"at und Kausalit"at~-- und daher versprachen, zu gravierenden Probleme bei der Renormierung zu f"uhren.
Als Kompromi"s untersuchte er in seiner Dissertation bei Veltman zun"achst die Renormierbarkeit {\it masseloser\/} Yang-Mills-Theorien.
Sein Beweis im Jahr~1971 geh"ort bis heute zu den grundlegendsten Arbeiten der theoretischen Physik "uberhaupt, Ref.~\cite{tHooft71}.
Renormierbarkeit~-- das hie"s im Sinne von {\it power counting}, das hei"st {\it perturbativ, zu allen Ordnungen\/}~-- basiert im wesentlichen auf der G"ultigkeit bestimmter verallgemeinerter {\it Ward-Identit"aten} (deren {\it Off-mass shell\/}-Versionen heute als {\it Slavnov-Taylor-Identit"aten\/} bekannt sind).
Da diese Identit"aten durch den Higgs-Kibble-Mechanismus nicht wesentlich modifiziert werden, konnte 't~Hooft direkt im Anschlu"s die Renormierbarkeit von~-- in diesem Sinne~-- \vspace*{-.25ex}{\it massiven\/} Yang-Mills-Theorien zeigen, Ref.~\cite{tHooft71a}. \\
\indent
Im darauf"|folgenden Jahr konnten 't~Hooft und Veltman im Rahmen des neuen Verfahrens der dimensionalen Regularisierung ein {\it Regularisierungs- und Renormierungsschema\/} f"ur Eichtheorien vom Yang-Mills-Typ explizit angeben~\cite{tHooft72}.
Das Feynmansche Funktionalintegral~\cite{Feynman50} war schon~1967, in Gross' Worten~\cite{Gross98b}: "`reemerged from obscurity"', als Faddeev und Popov mit seiner Hilfe {\it Feynman-Regeln\/} f"ur nichtabelsche Eichtheorien "uber einfachen Liegruppen $\mf{G}$ und {\it f"ur beliebige Diagramme\/} ableiteten~\cite{Faddeev67,Faddeev69} (f"ur {\it 1-Loop}-Diagramme hatte bereits Feynman die "Anderungen gegen"uber der QED aufgrund der Nicht-Abelizit"at beschrieben~\cite{Feynman63} und De~Witt im Detail ausgearbeitet~\cite{DeWitt65}).
In einer zweiten Arbeit~1972 gaben `t~Hooft und Veltman~\cite{tHooft72a} ein Verfahren an, wie man Feynman-Regeln in beliebigen Eichungen ableiten kann und gaben weiter~-- ausdr"uckend deren "Aquivalenz~-- einen kombinatorischen Beweis daf"ur, da"s die $S$-Matrix einer renormierbaren Eichtheorie, ohne Kompaktheit f"ur die zugrundeliegende Liegruppe $\mf{G}$ fordern zu m"ussen, unabh"angig ist von der \vspace*{-.125ex}Wahl der Eichung. \\
\indent
Zusammenfassend war das Konzept nichtabelscher Eichtheorien~1972 allgemein anerkannt von Seiten der Theoretischen Physik, und mit ihm die elektroschwache Theorie von Glashow--Weinberg--Salam; zur "Ubersicht verweisen wir auf die Refn.~\cite{Lee72,Abers72}.
Experimentell best"atigt wurden die von ihr postulierten {\it neutralen Str"ome\/} 1973 in Neutrino-Reaktionen, Ref.~\cite{Hasert73}, und die {\it Eichbosonen\/}~$W^\pm$ und~$Z^0$ 1983 am Proton-Antiproton-Speicherring $\rm Sp\bar{p}S$ im \vspace*{-.125ex}\VIER[]{C}{E}{R}{N}, Refn.~\cite{Arnison83,Banner83,Bagnaia83}. \\
\indent\enlargethispage{.375ex}
Mit dem Verst"andnis der "`exakten Regeln"' durch diese elektroschwache Theorie, konnte sie schnell auf Hadronen ausgedehnt werden.
Es konnten bald Vorhersagen gemacht werden, da"s etwa das gegenseitige K"urzen anomaler Beitr"age~-- die andernfalls die Renormierbarkeit der Theorie zerst"orten~-- erfordert dieselbe Anzahl von Quark{\it flavours\/} und Lepton{\it spezies\/}.
So wurde das {\it Charm\/}-Quark postuliert und gefunden; der Entdeckung der Familie von $\ta$-~und $\nu_{\!\ta}$-Lepton folgte die Entdeckung von~\vspace*{-.125ex}{\it Bottom\/}- und~{\it Top\/}-Quark. \\
\indent
Die Situation~1972/73 war also die folgende.
Es gab eine neue Klasse nichtabelscher Eichtheorien, die {\it theoretisch\/} weit ausgearbeitet waren: Allgemeine Feynman-Regeln konnten konsistent abgeleitet werden in beliebiger Eichung, die einander "aquivalent waren in dem Sinne, da"s sie zu derselben eichunabh"angigen $S$-Matrix f"uhrten.
Man verf"ugte "uber ein explizites Regularisierungs- und Renormierungsschema, dem zugrunde lag die G"ultigkeit verallgemeinerter Ward-Identit"aten, die Unitarit"at und Kausalit"at der Theorie implizierte.~--
Es waren dies notwendige Eigenschaften einer \vspace*{-.125ex}realistischen Theorie und insofern schon axiomatischer Ausgangspunkt des \vspace*{-.25ex}Bootstrap-Verfahrens gewesen; vgl.\@ Fu"sn.~\FN{FN:axiomatischeQFT}. \\
Andererseits war die Vereinheitlichung der Schwachen und elektromagnetischen Wechselwirkung als eine solche nichtabelsche Eichtheorie \vspace*{-.125ex}{\it experimentell\/} augenblicklich erfolgreich.

Der Gedanke lag nahe, analog k"onne eine konsistente Theorie der Starken Wechselwirkung formuliert werden.
Dahingehenden Fortschritten stand lange entgegen das Ph"anomen~{\it Con\-finement}.
Quarks und Gluonen waren zwar in gewissem Sinne, n"amlich um hadronische Symmetrien zusammenzufassen in ihrer fundamentalen Bedeutung akzeptiert, galten aber~-- da nicht asymptotisch, nicht direkt observabel~-- als {\it nicht real\/}, als \vspace*{-.25ex}{\it Abstraktum\/}. \\
Im Herbst~1968 postulierte {\it Bjorken\/} das nach ihm benannte {\it Skalenverhalten\/} tief-inelastischer Strukturfunktionen~\cite{Bjorken69}, das kurz danach in Experimenten am SLAC best"atigt wurde.
Es wurde von Feynman erkl"art in seinem bekannten {\it Partonbild\/}~\cite{Feynman69}, in dem sich Hadronen aus {\it Partonen\/} konstituieren, die sich bei immer kleineren Abst"anden mehr und mehr wie freie Teilchen verhalten.
Callan und Gross~\cite{Callan68} konnten bereits zuvor die experimentellen Daten dahin deuten, da"s diese Konstituenten im wesentlichen Spin-$1\!/\!2$ tragen; dar"uberhinaus zeigten die Experimente bald $1\!/\!3$ als deren Baryonquantenzahl.
Allerdings konnte man sich keine Quantenfeldtheorie vorstellen, deren Kopplung zwischen den Quarks einerseits so schwach sei, da"s {\it Asymptotische Freiheit\/} im genannten Sinne resultiert, und andererseits nicht erlaube, da"s Hadronen in ihre \vspace*{-.125ex}Quarkkonstituenten aufbrechen. \\
\indent\enlargethispage{.5ex}
Auf Basis der Renormierungsgruppe~-- durch Wilson wieder ins allgemeine Bewu"stsein zur"uckgerufen~\cite{Wilson71}~-- wurde argumentiert, da"s {\it Bjorken scaling\/} generell im Widerspruch stehe mit Quantenfeldtheorie:%
\FOOT{
  'T~Hooft macht dieses allgemeine Empfinden konkret mit den Worten Gross' aus dem Jahr~1971: "`no quantum field theory will ever explain Bjorken scaling"'; vgl.\@ die Refn.~\cite{tHooft98,tHooft98a}.
}
Aufgrund von Renormierung "andern sich die Dimensionen von Operatoren mit der (Energie)Skala~$\mu$, die mit den zugrundeliegenden Prozessen die betrachtete Physik bestimmt.
In gleicher Weise "andern sich dimensionslose Kopplungskonstanten.
Die Abh"angigkeit, das hei"st das "`Laufen"' der renormierten Eichkopplung~$g_{\rm ren.}\!(\mu)$ mit der Skala~$\mu$ wird beschrieben durch die entsprechende {\it Betafunktion\/}~$\be[g_{\rm ren.}\!(\mu)]$, die allgemein definiert ist als die logarithmische Ableitung von~$g_{\rm ren.}\!(\mu)$ nach~$\mu$, in Formeln~\mbox{$\be[g_{\rm ren.}\!(\mu)] \!:=\! d\!/\! d\ln\!\mu\; g_{\rm ren.}\!(\mu)  \!=\! \mu\, d\!/\! d\mu\; g_{\rm ren.}\!(\mu)$}.
Das U\!V-Verhalten einer Quantenfeldtheorie~-- explizit werden ihre {\it ein-Teilchen-irreduziblen renormierten ($n$-Punkt-)Green\-funktionen\/} diskutiert, vgl.\@ $G^{(n)}_{\rm 1-irred.}$ auf Seite~\pageref{T:1-irredGreenfnen}~-- ist bestimmt durch die Nullstellen der Betafunktion:~$\be[g^{\rm F}_{\rm ren.}\!(\mu)]\!=\!0$, das hei"st durch die Theorie am {\it Fixpunkt~{\rm F} der Renormierungsgruppengleichungen\/}~$g_{\rm ren.}\!(\mu) \!\to\! g^{\rm F}_{\rm ren.}\!(\mu)$.
Genau dann zeigt eine Theorie (bis auf Korrekturen $\ln\!\mu$) naives Skalenverhalten, das hei"st ist asymptotisch frei, wenn sie einen Fixpunkt am Ursprung $g^{\rm F}_{\rm ren.}\!(\mu) \!=\!0$ besitzt und dieser U\!V-stabil ist.
Tats"achlich besitzt jede Theorie trivialerweise einen Fixpunkt bei $g^{\rm F}_{\rm ren.}\!(\mu) \!=\!0$.
Ob er stabil ist, bestimmt in einer Entwicklung der Betafunktion~$\be[g_{\rm ren.}\!(\mu)]$ in Potenzen von~$g_{\rm ren.}\!(\mu)$ das Vorzeichen des f"uhrenden Koeffizienten.
In allen zu Anfang der siebziger Jahre bekannten Quantenfeldtheorien war dieser f"uhrende Koeffizient {\it positiv}, das hei"st der Fixpunkt verschwindender renormierter Kopplung {\it U\!V-instabil\/}:~$g_{\rm ren.}\!(\mu)$ w"achst an mit der Energie~$\mu$, das hei"st mit kleinerwerdendem Abstand~$1\!/\!\mu$.
Dies hielt man f"ur ein Charakteristikum von Quantenfeldtheorie im allgemeinen und sah von daher in Asymptotischer Freiheit das Argument f"ur deren konzeptuelle Mangelhaftigkeit.
Als Gross und Wilczek vor diesem Hintergrund der Renormierungsgruppe die Betafunktion nichtabelscher Eichtheorien auf der Basis halb\-einfacher Liegruppen $\mf{G}$ betrachteten, hatten sie tats"achlich die Absicht, f"ur die letzte bekannte Klasse von Theorien~-- quasi als Todessto"s f"ur eine Quantenfeldtheorie der Starken Wechselwirkung~-- die U\!V-Instabilit"at des Fixpunktes $g^{\rm F}_{\rm ren.}\!(\mu) \!=\!0$ zu zeigen.
Sie fanden im April~1973~\cite{Gross73} das Gegenteil: die Formel f"ur den relevanten Koeffizienten, die besagt, da"s er {\it negativ\/} ist, der Fixpunkt also {\it U\!V-stabil\/}~-- solange nur~$N_{\rm\!F}$, die Zahl der Quarkflavours, 16 nicht "uberschreitet;%
\FOOT{
  Bereits im Sommer zuvor w"ahrend eines Meetings in Marseille~\cite{tHooft72b} schrieb 't~Hooft dieselbe Formel an die Tafel, publizierte sie aber nicht.   Wohl angesichts des ungel"osten Confinement-Problems erkannte er nicht die Bedeutung seiner Rechnung,~-- insbesondere zur Erkl"arung des Bjorken-Skalenverhaltens.   Wir verweisen auf die Darstellungen in den Refn.~\cite{Gross98b,tHooft98}.
}
vgl.\@ auch Politzer, Ref.~\cite{Politzer73}.
Explizit zeigten Coleman und Gross~\cite{Coleman73}, da"s dieses Ergebnis~-- Asymptotische Freiheit~-- nicht m"oglich ist f"ur Theorien, die nur Fermionen und Skalare, aber kein (Eich-)Vektorfeld enthalten; vgl.\@ die \vspace*{-.25ex}Verallgemeinerung in Ref.~\cite{tHooft98}. \\
\indent
Das experimentell beobachtete Bjorken-Skalenverhalten war von {\it first principles\/} verstanden.
Gross und Wilczek berechneten die logarithmischen Abweichungen~\cite{Gross73a} und schlugen deren Messung als Pr"azisionstest vor; tats"achlich~beobachtet wurden sie ab Ende der siebziger Jahre und sind heute best"atigt auf \vspace*{-.25ex}besser als ein Prozent. \\
\indent
Die Anwendung Asymptotischer Freiheit konnte ausgedehnt werden auf viele neue Prozesse; sie wird heute als {\it perturbative\/} QCD bezeichnet.
Ihr Erfolg hat wesentlichen Anteil daran, da"s QCD im allgemeinen schnell als {\it die\/} Theorie der Starken Wechselwirkung anerkannt wurde.
Sehr auch durch die grundlegende "Anderung im Denken, zu dem der Erfolg perturbativer QCD f"uhrte:
Quarks und Gluonen wurden anerkannt als {\it physikalische Realit"at\/}.
Da"s sie in Colour-Singletts konfiniert sind, wurde verstanden als Konsequenz des entgegengesetzten Limes kleiner Skalen~$\mu$ (d.h.\@ gro"ser Abst"ande $1\!/\!\mu$):~-- unabh"angig von Asymptotischer Freiheit, und daher {\it nicht im Widerspruch\/} dazu.
Das Ph"anomen {\it Confinement\/} wurde damit nicht gel"ost, \vspace*{-.25ex}aber akzeptiert. \\
\indent
In der Folgezeit wurden im Rahmen von QCD gro"se Fortschritte gemacht, den Mechanismus, der zu Confinement f"uhrt, von {\it first principles\/} zu verstehen.
Wir verweisen an dieser Stelle auf den Abri"s 't~Hoofts in den Refn.~\cite{tHooft98,tHooft98a}, kommen aber zur"uck auf diesen Hintergrund {\it nichtperturbativer\/}~QCD, vor dem unsere Arbeit zu sehen ist.
\vspace*{-1ex}

\section*{\vspace*{-.5ex}Formale Darstellung}
\renewcommand{\rightmarkQCD}{Formale Darstellung}
\addcontentsline{toc}{section}{\numberline{} \hspace*{-12pt}Formale Darstellung}

Dies der Weg dahin, da"s QCD heute als {\it die\/} Theorie der Starken Wechselwirkung anerkannt ist.
In Gegen"uberstellung mit der QED, der Theorie der elektromagnetischen Wechselwirkung auf Basis einer Strukturgruppe~$U(1)$ gelangen wir~-- auch vor Augen unsere Einf"uhrung in die Mathematik nichtabelscher Eichtheorien auf S.~\pageref{T:Eichtheorie}\,ff.~-- zur QCD, der Theorie der Starken Wechselwirkung auf der Basis der~$\SUNc$,~$\Nc \!\equiv\! 3$ die {\it Colour\/}-Freiheitsgrade.
Dies hei"st konkret: \vspace*{-.25ex}zu ihrer Lagrangedichte. \\
\indent
Die wesentlichen Unterschiede sind Konsequenzen der {\it Nicht-Abelizit"at\/}.
Physikalisch manifestiert sie sich in der Selbstkopplung der Eichfelder.
Mathematisch resultiert sie daraus, da"s in der Kr"ummungs-2-Form~$F\!=\!d\wedge A+\iIM\,g\,A\wedge A$~, wie diskutiert auf S.~\pageref{T:Kruemmungsform}, das "au"sere Produkt des Eichfelds mit sich selbst~$A\wedge A$ nur in der abelschen Theorie verschwindet.
Ansonsten bleibt es im~$\suNc$-wertigen Feldst"arkentensor~$F_{\mu\nu}\!=\!\pa_\mu A_\nu\!-\!\pa_\nu A_\mu\!+\!\iIM\,g[A_\mu,A_\nu]$~als Kommutatorterm stehen.
Die Nichtlinearit"at, den dieser Term darstellt, verhindert letztlich eine simple Extrapolation der abelschen Theorie hin zur nichtabelschen.
Wir diskutieren, wie direkte Verallgemeinerung zu teilweise kontr"aren Konsequenzen f"uhrt. 
Zur diesbzgl.\@ umrissenen \vspace*{-.25ex}$\SUNc$-Eichalgebra vgl.\@ Anh.~\ref{APP:suNc-Eichalgebra}. \\
\indent\enlargethispage{1ex}
\label{T:SUNcAlgebra}Sei $\{T^a\}$ eine Basis der Liealgebra~$\suNc$, die mit der Eichgruppe~$\SUNc$ assoziiert ist.
Die~$T^a$ sind {\it spurlose hermitesche Operatoren\/} in einer Darstellung~$\Drst{R}$ der~$\suNc$, \vspace*{0ex}das hei"st~$\tr T^a \!=\!0$ und~$T^{a\D\dagger} \!=\! T^a$,~$a\!=\!1,2,\ldots,\dimNc$, mit~$\dimNc \!\equiv\! \Nc^2\!-\!1$ der \vspace*{-.25ex}Dimension der Gruppe.
Die $T^a$ werden bezeichnet als die {\it Erzeugenden\/} oder~{\it Generatoren\/} der Darstellung.
Darstellungsunabh"angig erf"ullt ihr Kommutator%
  ~\mbox{$[T^a,T^b]\!=\!\iIM\,f_{abc}T^c$} mit~$f_{abc}$ den Strukturkonstanten der~$\suNc$, die o.E.d.A.\@ als \vspace*{-.25ex}{\it voll antisymmetrisch\/} und {\it reell\/} definiert sind. \\
\indent
Die elementaren Lagrange'schen Felder der QCD sind die Materie-Spinorfelder~$\ps_n$ und Eich-Vektorfelder~$A_{\mu a}$, mit~$n \!=\! 1, 2,\ldots\!,\dimDrst{F}$ und~$a \!=\! 1,2,\ldots,\dimDrst{A}$.
Dabei ist allgemein~$\dimDrst{R}$ die Dimension des Raumes, in dem die Darstellung~$\Drst{R}$ wirkt,~-- explizit~$\dimDrst{F} \!=\! \Nc$ die Dimension der {\it fundamentalen Darstellung\/}~$\Drst{F}$ und~$\dimDrst{A} \!\equiv\! \dimNc \!=\! \Nc^2 \!-\!1$ die der {\it adjungierten Darstellung\/}~$\Drst{A}$.~--
Die {\it Quarkfelder\/}~$\ps_n$ transformieren unter der fundamentalen Darstellung.%
\FOOT{
  \label{FN:Drst_Fstar}Wir unterscheiden mit "`fundamental"' begrifflich nicht zwischen der Darstellung~$\Drst{F}$ der Quarks und der dazu komplex konjugierten (Stern-)Darstellung~$\Drst{F}^{\D\ast}$ der Antiquarks.
}
Als Normierung~$\tr T_\Drst{F}^aT_\Drst{F}^b \!=\! \normDrst{F}\,\de^{ab}$ w"ahlen wir standardm"asig~$\normDrst{F} \!=\! 1\!/\!2$ und f"ur~$\Nc \!=\!3 $ die $\la$-Matrizen Gell-Manns als Erzeugende:~$(T_\Drst{F}^a)_{mn} \!=\! 1\!/\!2 (\la^a)_{mn}$.
Die {\it Gluonfelder\/}~$A_{\mu a}$ transformieren unter der~adjungierten Darstellung, definiert durch%
  ~\vspace*{-.25ex}\mbox{$(T_\Drst{A}^a)_{bc} \!:=\! -\iIM\,f_{abc}$}. \\
\indent
Wir zerlegen ein Element~$B$ einer beliebigen Darstellung~$\Drst{R}$ der Liealgebra~$\suNc$~ge\-m"a"s~$B \!=:\! B_aT_\Drst{R}^a$ in Komponenten entlang der Erzeugenden.
Die so definierten~-- darstellungs\-abh"angigen~-- Komponenten~$B_a$ erh"alt man zur"uck durch~$B_a \!=\! \normDrst{R}{}^{\!-1}\,\tr BT^a_\Drst{R}$, mit~$\normDrst{R}$ der~{\it Nor\-mierung\/} der Generatoren gem"a"s~$\tr T^a_\Drst{R}T^b_\Drst{R} \!=\! \normDrst{R}\,\de^{ab}$, standardm"a"sig~$\normDrst{F} \!=\! 1\!/\!2$ und~$\normDrst{A} \!=\! \Nc$ f"ur die fundamentale beziehungsweise adjungierte Darstellung.
"Uber~$\normDrst{R} \dimNc \!=\! \csDrst{R} \dimDrst{R}$ ist diese Normierung verkn"upft mit dem {\it quadratischen Casimir-Operator\/} der \vspace*{-.25ex}Darstellung.%
\FOOT{
  Der quadratische Casimir-Operator~$\csDrst{R}$ einer Darstellung~$\Drst{R}$ ist allgemein definiert durch "`die aufsummierten Quadrate aller Generatoren"', $T_\Drst{R}^a T_\Drst{R}^a \!=:\! \csDrst{R}\, \bbbOne{R}$; die Eins der Darstellung ist dabei die~$\dimDrst{R}$-dimen\-sionale Einheitsmatrix,~-- d.h.\@ explizit $\csDrst{F} \!=\! \normDrst{F}\, \dimNc / \dimDrst{F} \!=\! (\Nc^2\!-\!1) \!/ 2\Nc$ und $\csDrst{A} \!=\! \normDrst{A} \!=\! \Nc$.
} \\
%
\indent
So haben wir f"ur das Eich-Vektorfeld die Zerlegung~$A_\mu \!=\! A_{\mu a}T^a$, wobei die Komponenten $A_{\mu a}$ Elemente aus~$\bbbr$ sind (Zahl der Freiheitsgrade!).
Die kovariante Ableitung\label{T:kovAbl} lautet allgemein~$(D_\mu)_{\al\be} \!=\! \pa_\mu (\bbbOne{R})_{\al\be} \!+\! \iIM\,gA_{\mu a}(T^a_\Drst{R})_{\al\be}$ in~$\Drst{R}$ der Darstellung des Felds, auf das~$D_\mu$ wirkt: 
Das hei"st f"ur die Quarkfelder~$(D_\mu \ps)_n \!=\! \pa_\mu \ps_n \!+\! \iIM\,gA_{\mu a}(T^a_\Drst{F})_{nm}\ps_m$ (fundamentale Darstellung)
und f"ur die Gluonfeldst"arken~$(D_\mu F_{\rh\si})_a \!=\! \pa_\mu F_{\rh\si a} \!-\! gf_{abc}A_{\mu b}F_{\rh\si c}$ (adjungierte Darstellung).
Wir haben dabei f"ur den Feldst"arkentensor gem"a"s der Zerlegung~$F_{\mu\nu} \!=\! F_{\mu\nu a}T^a$ die Komponenten~$F_{\mu\nu a} \!=\! \pa_\mu A_{\nu a} \!-\! \pa_\nu A_{\mu a} \!-\! gf_{abc}A_{\mu b}A_{\nu c}$ eingef"uhrt.
Sie ergeben sich in dieser Gestalt unabh"angig von der Darstellung: aus $F_{\mu\nu}\!=\!(\iIM\,g)^{-1}[D_\mu,D_\nu]$~mithilfe der~-- darstellungsunabh"angigen~-- Relation%
  ~\vspace*{-.25ex}\mbox{$[T^a,T^b] \!=\! \iIM\,f_{abc}T^c$}~f"ur den Kommutator der Erzeugenden. \\
\indent
Die {\it Lagrangedichte der\/}~QCD%
\FOOT{
  \label{FN:Flavour}Sei durch die gesamte Arbeit implizit summiert "uber $N_{\rm\!F}$~{\sl Flavours} von Fermion-Multipletts~$\ps$.
}
setzt sich zusammen aus der {\it Yang-Mills\/}- und der Materie- oder {\it Dirac-Lagrangedichte\/}
\vspace*{-.25ex}
\begin{align} \label{QCD:Lagrangedichte}
\mf{L}\;
  =\; \mf{L}_{\rm Y\!M}\;+\; \mf{L}_{\rm D}\;
  =\; -\frac{1}{2}\, \tr F_{\mu\nu} F^{\mu\nu}\;
         +\; \bar{\ps}\, (\iIM\,\ga^\mu D_\mu - m)\, \ps
    \\[-4ex]\nn
\end{align}
die Spur ist wieder bez"uglich der~$\suNc$ zu bilden, die Feldst"arken zu nehmen in der {\it fundamentalen Darstellung\/}:~$\tr F_{\mu\nu} F^{\mu\nu} \!=\! F_{\mu\nu a}F^{\mu\nu}{}_b\, \tr T^a_\Drst{F}T^b_\Drst{F} \!=\! \frac{1}{2}F_{\mu\nu a}F^{\mu\nu}{}_a$.
Die Lagrangedichte~$\mf{L}$ besitzt die eingangs diskutierten Invarianzen und steht~-- h"ohere als erste Ableitungen der Felder ausgeschlossen~-- {\it in eindeutiger Weise\/} f"ur die Theorie eines~$\Nc$-Multipletts~$\ps$ von Spin-$1\!/\!2$-Materiefeldern, das "uber ein entsprechendes Eich-Vektorfeld~\vspace*{-.25ex}$A$ wechselwirkt. \\
\indent
Wird die \label{T:JacobiBianchi}Jacobi-Identit"at f"ur die kovariante Ableitung,~$[D_\mu,[D_\rh,D_\si]]+\text{zykl.}=\!0$, kontrahiert mit dem Pseudotensor~$\vep^{\mu\nu\rh\si}$, so folgt mithilfe%
  ~\mbox{$F_{\rh\si}\!=\!(\iIM\,g)^{-1}[D_\rh,D_\si]$} die {\it Bianchi-Identit"at\/}%
  ~\mbox{\,$0 \!=\! \vep^{\mu\nu\rh\si}D_\mu F_{\rh\si} \!\equiv\! D_\mu(\vep^{\mu\nu\rh\si}F_{\rh\si})$}.
Sie ist die nichtabelsche Verallgemeinerung der homogenen Maxwellgleichungen und offenbar unmittelbare Konsequenz des {\it Eichprinzips\/}: des Begriffes nichtverschwindender {\it Kr"ummung\/} und daher {\it nicht-integrabler "Ubertragung im Eichraum\/}, die sich ihrerseits ausdr"uckt im Eich-Vektorfeld der \vspace*{-.25ex}{\it eichkovarianten Ableitung\/}. \\
\indent\enlargethispage{1ex}
Aus der Lagrangedichte~\vspace*{-.125ex}$\mf{L}$ resultieren die {\it Feldgleichungen der\/}~QCD nach Euler-La\-grange. Dies ist zum einen die {\it Dirac-Gleichung\/} im Hintergrund des Eichfelds%
  ~\vspace*{-.125ex}\mbox{\,$(\iIM\,\ga^\mu D_\mu \!-\! m)\ps\!=\!0$}.
Zum anderen sind dies die nichtabelsch verallgemeinerten inhomogenen Maxwellgleichungen $D_\mu F^{\mu\nu}\!=\!gJ^\nu$, in Komponenten~$(D_\mu F^{\mu\nu})_a \!=\! \pa_\mu F^{\mu\nu}{}_a \!-\! gf_{abc}A_{\mu b}F^{\mu\nu}{}_c \!=\! gJ^\nu{}_a$ (also die kovariante Ableitung auf die adjungierte Darstellung bezogen), mit dem {\it Dirac-Materiestrom\/}~$J$, dessen Komponenten durch%
  ~\vspace*{-.25ex}\mbox{$J^\nu{}_a \!:=\! \bar{\ps} \ga^\nu T^a \ps$} definiert sind. \\
\indent
Dieser Strom ist f"ur {\it freie Quarks\/} erhalten so da"s diese mit seiner Hilfe charakterisiert werden k"onnen bez"uglich der {\it Colour-Ladung\/}, die er assoziiert.
Kombinationen von Quarks mit definierter Colour k"onnen angegeben werden, insbesondere global colour-neutrale Had\-ronen als nichtlokale Objekte, zusammengesetzt aus Quarkkonstituenten.
Koppeln Quarks allerdings an Gluonen, so ist~$J$ nicht erhalten und gestattet nicht, Zust"ande bez"uglich Colour zu charakterisieren.
Dies spiegelt wider, da"s auch Gluonen Colour-Ladung tragen~-- eklatanter Kontrast zu Photonen in der QED, die nicht \vspace*{-.25ex}elektrisch geladen sind. \\
\indent
\label{T:Konstituentenquarks}Eine geeignete erhaltene Colour-Ladung in der QCD~-- und dadurch sind {\it Konstituentenquarks\/} erst definierbar~-- folgt aus der Invarianz ihrer Lagrangedichte unter Eichtransformationen.
Wir wollen diese daher rekapitulieren, folgen dabei der Darstellung von Lavelle und McMullan in Ref.~\cite{Lavelle97}; bez"uglich formalerer \vspace*{-.25ex}Aspekte verweisen wir auf Ref.~\cite{Itzykson88}. \\
\indent
Werde das Konzept der Eichtheorie auf das Spinorfeld~$\ps$ angewandt, das hei"st "uberf"uhren es lokale Eichtransformationen in physikalisch "aquivalente Kopien.
Sei~$U(x)$ die Transformationsmatrix, das hei"st f"ur jeden Weltpunkt~$x$ ein beliebiges Element der Eichgruppe~$\SUNc$~-- in geeigneter Darstellung.
Dann transformiert die kovariante Ableitung {\it per definitionem\/} wie~$D_\mu \!\to\! D^U_\mu \!:=\! U D_\mu U^{\D\dagger}$.
In Termen der Lagrange'schen Felder der Theorie sind lokale Eichtransformationen vollst"andig bestimmt durch
\vspace*{-.25ex}
\begin{alignat}{5}
&\ps\;& &\to\;& &\ps^U\;&
  &:=\;& &U\, \ps
    \label{QCD:Eichtransf} \\
&A_\mu\;& &\to\;& &A^U_\mu\;&
  &:=\;& &\frac{1}{\iIM\,g}\, U\, (D_\mu\, U^{\D\dagger})\;
    =\; U\, A_\mu\, U^{\D\dagger}\;
         +\; \frac{1}{\iIM\,g}\, U\, (\pa_\mu\, U^{\D\dagger})
    \tag{\ref{QCD:Eichtransf}$'$}
    \\[-4.75ex]\nn
\end{alignat}
Es gilt:
\vspace*{-1ex}
\begin{align} \label{QCD:Eichtransf_D}
D_\mu\;
   \to\; D^U_\mu\;
  =\; U\, D_\mu\, U^{\D\dagger}
\end{align}
{\it per~definitionem\/}.
Mit der Darstellung%
  ~\vspace*{-.25ex}\mbox{\,$F_{\mu\nu}\!=\!(\iIM\,g)^{-1}[D_\mu,D_\nu]$} und Unitarit"at von~$U(x)$ an~je\-dem Weltpunkt~$x$ der Raumzeit:%
  ~\mbox{\,$U^{\D\dagger}(x)U(x) \!=\! \bbbone$}, transformiert der Feldst"arkentensor wie
%
\begin{align} \label{QCD:Eichtransf_F}
F_{\mu\nu}\;
   \to\; D^U_{\mu\nu}\;
  =\; U\, F_{\mu\nu}\, U^{\D\dagger}
\end{align}
Damit ist evident die {\it Eichinvarianz\/} der Lagrangedichte%
  ~\vspace*{-.25ex}$\mf{L}$ der QCD, vgl.\@ Gl.~(\ref{QCD:Lagrangedichte}). \\
\indent
Schreibt man $U(x)\!=\!\efn{\iIM\,\th(x)}$, ist~$\th$ ein Element der Liealgebra~$\suNc$ mit der Komponen\-tenzerlegung~$\th\!=\!\th_aT^a$.
Es folgt der mit den Eichtransformationen assoziierte {\it Noether-Strom\/}~$j_\th$, wobei~$j_\th{}^\nu \!=\! -g^{-1} F^{\nu\mu}{}_a (D_\mu\th)_a \!+\! J^\nu{}_a \th_a \!=\! -g^{-1} F^{\nu\mu}{}_a \pa_\mu\th_a \!+\! (f_{abc}A_{\mu b}F^{\mu\nu}{}_c \!+\! J^\nu{}_a) \th_a$ und sich die kovariante Ableitung auf die adjungierte Darstellung bezieht.
F"ur {\it globale\/} Eichtransformationen, das hei"st~$U$ ist konstant und damit~$\th$, ergibt sich f"ur seine Komponente entlang~$\th_a$ der Ausdruck in den runden Klammern,~$j^\nu{}_a \!:=\! f_{abc} A_{\mu b}F^{\mu\nu}{}_c \!+\! J^\nu{}_a$.
Die Erhaltung dieses Stromes~$j$ ist Konsequenz des {\it Noetherschen Theorems\/},~-- folgt aber auch unmittelbar aus den Feldgleichungen $\pa_\mu F^{\mu\nu}{}_a \!-\! gf_{abc}A_{\mu b}F^{\mu\nu}{}_c \!=\! gJ^\nu{}_a$, indem man den Term mit den Strukturkonstanten auf die rechte Seite bringt:%
  ~\vspace*{-.25ex}\mbox{$\pa_\mu F^{\mu\nu}{}_a \!=\! g(f_{abc}A_{\mu b}F^{\mu\nu}{}_c \!+\! J^\nu{}_a) \!=\! gj^\nu{}_a$}. \\
\indent\enlargethispage{.75ex}
Der Generator der Eichtransformationen ist zu identifizieren mit der assoziierten erhaltenen {\it Noether-Ladung\/}~$G(\th) \!=\! \int d^3\!x\, j_\th{}^0\!(x) \!=:\! \int d^3\!x\, G_a\!(x)\, \th_a\!(x)$; dabei haben wir den Integranden zerlegt in Komponenten~$G_a \!:=\! -g^{-1}(D_iE^i)_a \!+\! J^0{}_a$ entlang~$\th_a$ und durch~$E^i_a\!:=\!F^{i0}{}_a$ das {\it chromo-elektrische Feld\/} definiert.
Die Ladung~$G(\th)$ generiert die Transformationen
\vspace*{-.5ex}
\begin{alignat}{3} \label{QCD:Eichtransf_inf}
&\de_\th A_{ia}\, (x)\;&
  &=\; [\, \iIM\,G(\th), A_{ia}(x)]\;&
  &=\; -\, \frac{1}{g}\, (D_i\th)_a\, (x) 
    \\
&\de_\th \ps\, (x)\;&
  &=\; [\, \iIM\,G(\th), \ps(x)]\;&
  &=\; \phantom{-\,}
       \iIM\, \th \ps\, (x)
    \tag{\ref{QCD:Eichtransf_inf}$'$}
\end{alignat}
Dies ist die infinitesimale Gestalt der Eichtransformationen der Felder~\pagebreak$A_i$ und~$\ps$.
Wir betonen, da"s wir nicht das Transformationsgesetz der {\it Zeitkomponenten~$A_0$ der Eich-Vektorfelder\/} wiedergefunden haben.
Dies ist Konsequenz daraus, da"s~$A_{0a}$ keine dynamische Variable ist, {\it kein physikalischer Freiheitsgrad\/},~-- sondern integrierender Faktor, {\it Lagrange-Multiplikator\/}.
Die zugeh"origen kanonisch konjugierten Felder~$E^0{}_a$ sind genau $F^{00}{}_a$, die identisch verschwinden; also $E^0{}_a\!(x)\!=\!0$.
Variation der Lagrangedichte~$\mf{L}$ nach~$A_{0a}\!(x)$ ergibt~$G_a\!(x)\!=\!0$: die {\it nicht\-abelsche Verallgemeinerung des Gau"s'schen Gesetzes},~$(D_iE^i)_a\!(x)\!=\!gJ^0{}_a\!(x)$.
Soweit halten wir fest:
Die QCD ist ein {\it System mit%
  ~\mbox{\,$2\!\cdot\!(\Nc^2 \!-\! 1)$} Zwangsbedingungen\/}.

Dies ist bereits f"ur die {\it klassische Chromodynamik\/} der Fall.
In der {\it quantisierten Theorie\/}~-- wir denken an die \VIER{B}{R}{S}{T}-Operatorformulierung, s.u.\@ -- werden die Felder~$A_{i}$ und~$\ps$ durch {\it Feldoperatoren\/} ersetzt; von diesen mit den entsprechenden kanonisch konjugierten Operatoren werden {\it f"ur gleiche Zeiten (Anti)Kommutatorrelationen\/} gefordert.
Mithilfe dieser Relationen haben wir f"ur infinitesimale Eichtransformationen Kommutatordarstellungen des entsprechenden Felds mit der Noether-Ladung~$G(\th)$ angegeben, die offensichtlich in Widerspruch stehen~-- und damit die fundamentalen (Anti)Kommutatorrelationen~-- mit der strikten Forderung, da"s~$G_a\!(x)\!=\!0$ {\it als Operatoren\/}.
Sie ist daher abzuschw"achen.
Dies geschieht dadurch, da"s~$G_a\!(x)\ket{\ps}=\!0$ nur f"ur eine Unterklasse von Zust"anden~$\ket{\ps}$ gefordert wird und genau diese als {\it physikalisch\/} erkl"art.
Sie bilden einen {\it linearen Unterraum\/}~-- das Superpositionsprinzip ist also unber"uhrt~--, und es verschwinden die auf sie bezogenen~Erwar\-tungswerte der Zwangsbedingungen,~$\bra{\ps}G_a\!(x)\ket{\ps}=\!0$.
Dies f"uhrt zu einem konsistenten \vspace*{-.125ex}Bild. \\
\indent
Hierin macht auch der Kommutator~$[G_a\!(x),G_b\!(x')] \!=\! -\iIM\,f_{abc}G_c\!(x)\de(x\!-\!x')$ Sinn, der die~$G_a$ als Erzeuger der Eichtransformationen ausweist, was allerdings suggeriert~-- bez"uglich der {\it Interpretation des Gau"s'schen Gesetzes\/}~$G_a\!(x)\ket{\ps}=\!0$~--, alle physikalischen Zust"ande seien eichinvariant.
Dies aber f"uhrt schon in der QED auf Probleme im Zusammenhang mit der Konstruktion {\it elektrisch geladener\/} physikalischer Zust"ande:
Bei einer globalen Eichtransformation, \mbox{$\th \!=\! \text{\sl konst.\/}$}, geht die abelsche Ladung in die elektrische Ladung "uber, so da"s physikalische Zust"ande, die {\it per definitionem\/} durch diese annihiliert werden, notwendig elektrisch ungeladen sind.
Wir haben im Rahmen des vorliegenden Abri"ses nicht den Raum, die subtile Argumentation von Lavelle und McMullan,~Ref.~\cite{Lavelle97}, nachzuzeichnen, die dieses Problem aufl"ost.
Ihre abschlie"sende Aussage ist aber:
Eine in der QCD ad"aquate Colour-Ladung kann konsistent definiert werden, allerdings nur f"ur eichinvariante Zust"ande; lokale Eichtransformationen m"ussen auf eine Unterklasse~$U(x) \!\in\! \SUNc$ eingeschr"ankt werden, so da"s globale Eichtransformationen nicht mehr aus diesen folgen durch Konstantsetzen von Parametern, sondern unabh"angig betrachtet werden m"ussen.~--
Die {\it Colour-Ladung}%
  ~\mbox{$Q \!=\! Q_aT^a$}, mit~$Q_a \!:=\! \int d^3\!x\, j^0{}_a\!(x) \!=\! \int d^3\!x\, (f_{abc}A_{\mu b}F^{\mu0}{}_c \!+\! J^0{}_a)$, ist per contructionem erhalten~$dQ\!/\!dt\!=\!0$, ist aber nicht eichinvariant; sie wird eichinvariant,~$Q \!\equiv\! Q^U$, wo~$Q \!\to\! Q^U \!:=\! U Q U^{\D\dagger}$, wenn~sie auf physikalische Zust"ande~$\ket{\ps}$ wirkt.
Dies folgt daraus, da"s sich weiter schreiben l"a"st~$Q_a \!=\! \int d^3\!x\, (g^{-1}\pa_iE^i{}_a \!+\! G_a)$, und so auf einen physikalischen Zustand~$\ket{\ps}$ bezogen effektiv~$j^0{}_a \!=\! g^{-1}\pa_iE^i{}_a$ zu setzen ist.
Anwendung des Gau"s'schen Satzes f"uhrt auf ein Oberfl"achenintegral, so da"s man dahin gef"uhrt ist, die zul"assigen Eichtransformationen auf Elemente der Eins-Zusammenhangskomponente der Eichgruppe~$\SUNc$ einzuschr"anken, die im r"aumlich Unendlichen in die Eins \vspace*{-.25ex}"ubergehen. \\
\indent\enlargethispage{1.125ex}
Mithilfe dieser Ladung~$Q$ k"onnen von first principles {\it Konstituentenquarks\/} konstruiert werden mit definierter {\it Colour-Ladung\/}~-- in Termen derer QCD alternativ formuliert werden kann, vgl.\@ Ref.~\cite{Lavelle97}.
Explizit meint dies die Konstruktion eines {\it dressed quarks\/}: ein Lagrange'sches Quark, dergestalt umgeben von einer Wolke virtueller Gluonen und (Anti)Quarks, da"s das gesamte System eichinvariant ist.%
\FOOT{
  Die Konstruktion dynamischer {\sl dressings\/} f"ur individuelle Quarks in Ref.~\cite{Lavelle97} ist a priori {\sl perturbativ}. Von hier kommend, spiegelt sich das Einsetzen von Confinement im Auftreten eines Mechanismus wider~-- formal aufgrund der {\sl Gribov ambiguities\/} der nichtabelschen Theorie, vgl. Fu"snote~\FN{FN:Gribov_ambiguities}~--, der {\sl nichtperturbative dressings\/} verhindert und dadurch, da"s (Konstituenten)Quarks asymptotisch werden.
}
Der Begriff des Konstituentenquarks spielt in der~vor\-liegenden Arbeit eine wichtige Rolle; er ist {\it wohldefiniert von first principles\/}.

Da"s die QCD eine Theorie mit {\it Zwangsbedingungen\/} ist, hat wesentliche Konsequenzen f"ur den Begriff der {\it Eichinvarianz\/} in der Standardformulierung der QCD als {\it manifest Poincar\'e-kovariante lokale Theorie\/}; vgl.\@ Ref.~\cite{Lavelle97}, \vspace*{-.25ex}bzgl.\@ tiefergehender Fragen Ref.~\cite{Itzykson88}. \\
\indent
Der Eichsektor der QCD konstituiert sich aus den Feldern~$A_{\mu a}$, das sind~$4\!\cdot\!(\Nc^2 \!-\! 1)$~Frei\-heitsgrade.
Wir haben gesehen, da"s Eichinvarianz~$2\!\cdot\!(\Nc^2 \!-\! 1)$ Zwangsbedingungen impliziert, n"amlich~$E^0{}_a\!=\!0$ und das Gau"s'sche Gesetz~$G_a\ket{\ps}=\!0$.
Dies ist analog zur QED, in der die vier Freiheitsgrade des Vektorpotentials auf die zwei transversalen Polarisationen des Photons eingeschr"ankt werden~-- und dabei (zun"achst) Poincar\'ekovarianz verlorengeht.
Diese ist aber von eminenter Bedeutung im perturbativen Regularisierungs- und Renormierungsprogramm, insofern als es bislang "uberhaupt erst f"ur manifest Poincar\'e-kovariante lokale Theorien erfolgreich hat durchgef"uhrt werden k"onnen; \vspace*{-.25ex}vgl.\@ die Refn.~\cite{tHooft71,tHooft71a,tHooft98a,tHooft72}. \\
\indent
Um die Nicht-Kovarianz zu vermeiden, die der Dynamik der QCD inh"arent ist, zerst"ort man ihre Eichinvarianz von Hand, indem man zur Lagrangedichte~$\mf{L}$ einen {\it eichfixierenden Term\/}~$\mf{L}_{\rm gf.}$ hinzuaddiert, der genau einen Repr"asentanten ausw"ahlt je "Aquivalenzklasse von~$A$ (die f"ur genau eine physikalische Feldst"arke steht).
Explizit ist%
  ~\mbox{$\mf{L}_{\rm gf.} \!=\! -\xi^{-1}\tr{\cal F}^2[A]$} f"ur die Eichfixierung~${\cal F}[A]\!=\!C$, wobei~${\cal F}$ ein {\it lokales Funktional\/} in~$A$ ist (d.h.\@ Funktion von~$A_\mu\!(x)$ und~$\pa_\mu A_\nu\!(x)$), das wie die Funktion~$C(x)$ seine Werte in der Liealgebra~$\suNc$ annimmt; der Parameter~$\xi$ wird induziert durch \vspace*{-.25ex}{\it gau"s'sche\/} Mittelung von~$C$ und ist frei w"ahlbar. \\
Konsequenz dieses Verfahrens ist, da"s wieder alle~$4\!\cdot\!(\Nc^2 \!-\! 1)$ Komponenten~$A_{\mu a}$ dynamische Feldvariablen sind.
Offensichtlich ist aber dadurch zum einen die Physik der Theorie nicht mehr dieselbe, und zum anderen hat ihr Zustandsraum {\it indefinite Metrik\/}. 
In der QED wird simultan das eine gel"ost, das andere umgangen~-- dadurch, da"s man nach Gupta und Bleuler~\cite{Gupta50,Bleuler50} physikalische Zust"ande auf den linearen Unterraum beschr"ankt, der definiert ist durch Restriktion der Lorentz-Eichung auf {\it positive Frequenzen\/}:%
  ~\vspace*{-.25ex}\mbox{\,$(\pa_\mu A^\mu)^+\ket{\ps}=\!0$}. \\
\indent
In der QCD ist dies nicht m"oglich.
Um den physikalischen Gehalt der Theorie nicht zu~"an\-dern, m"ussen {\it Faddeev-Popov-Geistfelder\/} eingef"uhrt werden, die als {\it fermionische (d.h.\@ anti\-kommutierende) Skalare\/} formal jeweils den Freiheitsgrad $(-1)$ beitragen, \mbox{vgl.\@ Ref.~\cite{Faddeev67,Faddeev69}}.
Hinzuf"ugen von je%
  ~\mbox{\,$(\Nc^2 \!-\! 1)$}~Geister~$c_a\!(x)$ und Antigeister~$\bar{c}_a\!(x)$ zum Eichfixierungsterm f"uhrt also wieder auf die korrekte Anzahl von Freiheitsgraden zur"uck.
Zur Lagrangedichte~$\mf{L}$ der QCD ist also neben dem Anteil~$\mf{L}_{\rm gf.}$ zur Eichfixierung ein Geist-Anteil~$\mf{L}_{\rm ghost}$ zu addieren.
Dieser h"angt allgemein von der Eichfixierung ab, das hei"st von dem Funktional~${\cal F}$; er lautet explizit%
  ~\vspace*{-.125ex}\mbox{$\mf{L}_{\rm ghost} \!=\! -\bar{c}_a (\de{\cal F}_a[A] /\de A_{\mu b}) (D_\mu c)_b$}, wobei sich gem"a"s
   \mbox{$(D_\mu c)_a \!=\! \pa_\mu c_a -gf_{abc}A_{\mu b}c_c$}~die kovariante Ableitung auf die \vspace*{-.25ex}adjungierte Darstellung bezieht.
So w"ahlt die Fixierung auf das {\it (Eich)Orbit\/}, das durch%
  ~\mbox{${\cal F}[A] \!\equiv\! \pa_\mu A^\mu \!=\!0$} gegeben ist, das hei"st
   \mbox{$\mf{L}_{\rm gf.} \!=\! -(2\xi)^{-1}\,(\pa_\mu A^\mu{}_a)^2$}, die Klasse der {\it Lorentz-Eichungen\/} aus; ihr Geistterm folgt zu%
  ~\mbox{$\mf{L}_{\rm ghost} \!=\! \bar{c}_a\, \pa^\mu (D_\mu c)_a$}.
Es  werden~$\xi\!=\!1$ als die {\it Feynman-\/} und%
  ~\vspace*{-.25ex}\mbox{$\xi\!=\!0$} als die {\it Landau-Eichung\/} bezeichnet. \\
\indent\enlargethispage{.75ex}
Mit der Lagrangedichte%
  ~\vspace*{-.125ex}\mbox{$\mf{L}\!=\!\mf{L}_{\rm Y\!M}\!+\!\mf{L}_{\rm D}$}, vgl.\@ Gl.~(\ref{QCD:Lagrangedichte}), ist in diesem Sinne die {\it effektive Lagrangedichte der\/}~QCD:
\vspace*{-.25ex}
\begin{align} \label{QCD:Lagrangedichte_eff}
\mf{L}_{\rm eff.}\;
  =\; \mf{L}\;
        +\; \mf{L}_{\rm gf.}\;
        +\; \mf{L}_{\rm ghost}
    \\[-5ex]\nn
\end{align}
mit
\vspace*{-1.5ex}
\begin{align}
&\mf{L}_{\rm gf.}\;
  \hspace*{9.5pt}
  =\; -\, \xi^{-1}\; \tr\, {\cal F}^2[A]\;
  \hspace*{27pt}
  =\; -\, \frac{1}{2\xi}\; (\pa_\mu A^\mu{}_a)^2 
    \tag{\ref{QCD:Lagrangedichte_eff}$'$} \\[-.5ex]
&\mf{L}_{\rm ghost}\;
  =\; -\, \bar{c}_a\vv \frac{\de{\cal F}_a[A]}{\de A_{\mu b}}\vv (D_\mu c)_b\;
  =\; -\, \bar{c}_a\; \pa^\mu\; (D_\mu c)_a
    \tag{\ref{QCD:Lagrangedichte_eff}$''$}
    \\[-4.625ex]\nn
\end{align}
und die zweite Darstellung den Fall der Lorentz-Eichungen illustriert.

Eichinvarianz ist von Hand zerst"ort, {\it Invarianz unter {\rm \VIER{B}{R}{S}{T}}-Transformationen\/} an ihre Stelle getreten~-- wie gezeigt wurde unabh"angig voneinander von Becchi, Rouet, Stora~\cite{Becchi74,Becchi76} und Tyutin~\cite{Tyutin75}.
F"ur die Felder~\mbox{\,$A_{\mu a}$} und~\mbox{\,$\ps$} haben sie {\it infinitesimal\/} die Gestalt
\vspace*{-.25ex}
\begin{alignat}{2} \label{QCD:BRST-Transf_inf}
&\de_\BRST A_{\mu a}\, (x)\;&
  &=\; -\, \frac{1}{g}\, (D_\mu c)_a\, (x)
    \\
&\de_\BRST \ps\, (x)\;&
  &=\; \phantom{-\,}
       \iIM\, c \ps\, (x)
    \tag{\ref{QCD:BRST-Transf_inf}$'$}
    \\[-4ex]\nn
\end{alignat}
haben; siehe auch die Refn.~\cite{Becchi96,Stora96,Green95,Kugo97}.
Vergleich mit den entsprechenden Ausdr"ucken f"ur Eichtransformationen~-- die Gln.~(\ref{QCD:Eichtransf_inf}),~(\ref{QCD:Eichtransf_inf}$'$)~-- \vspace*{-.25ex}unterstreicht den engen Zusammenhang. \\
\indent
Diese Invarianz der effektiven Lagrangedichte~$\mf{L}_{\rm eff.}$ wird durch die nilpotente {\it {\rm \VIER{B}{R}{S}{T}}-La\-dung\/}~\mbox{\,$Q_\BRST^{}$} generiert:~\mbox{\,$Q_\BRST^2 \!=\! 0$}.
{\it Physikalische Zust"ande\/} sind dadurch definiert, da"s sie gefordert werden als \VIER{B}{R}{S}{T}-{\it invariant\/};~dies zeichnet als physikalisch dieselben Zust"ande aus, wie die Bedingung oben: da"s sie annihiliert werden durch den Gau"s'schen Operator.
Die Forderung, da"s \VIER{B}{R}{S}{T}-invarianten Zust"anden eine definierte Colour-Ladung zugeordnet werden kann, "ubersetzt sich dahin, da"s die Geister im r"aumlich Unendlichen verschwinden.
\label{T:BRST-versusEichinvarianz}F"ur Zust"ande, die sich konstituieren aus Eich- und Materiefelder, ist aufgrund der Analogie der Transformationsgleichungen "aquivalent, entweder \VIER{B}{R}{S}{T}-{\it Invarianz\/} zu fordern oder~{\it Eichin\-varianz\/} unter den wie oben eingeschr"ankten Transformationen.
Wie allgemein "ublich sei im folgenden mit~"`Eichinvarianz"' sprachlich nicht mehr differenziert das eine und das \vspace*{-.25ex}anderen. \\
\indent
Hiermit haben wir vorgestellt die {\it Standardformulierung der\/}~QCD%
\FOOT{
  \label{FN:Gribov_ambiguities}Sei angemerkt, da"s der vorgestellten Quantisierung im Funktionalintegral-Formalismus nach Faddeev und Popov~\cite{Faddeev67,Faddeev69} das {\sl perturbative Konzept\/} bereits inh"arent ist:   Insofern, als sie fordert, da"s~${\cal F}[A]\!=\!C$ das Eichfeld auf {\sl genau ein\/} (Eich)Orbit fixiert.   In einer nichtabelschen Eichtheorie ist dies aber nicht der Fall f"ur gro"se Werte des Eichfelds, das hei"st auf nichtperturbativen Skalen; bzgl. dieser {\sl Gribov ambiguities\/}~\cite{Gribov78} vgl.\@ Ref.~\cite{Itzykson88}.   Der Formalismus ist {\sl nichtperturbativ\/} "`krank"' definiert.   Es gibt Versuche, ihn entsprechend zu modifizieren~\cite{Shabanov99}; solange dies nicht gelungen ist, bleibt nur das Problem als Dokument dessen zu ignorieren, da"s (nichtabelsche) Quantenfeldtheorie noch immer nicht mathematisch in aller Strenge formuliert ist.
},
deren wesentliche Charakteristika sind: {\it Lokalit"at}, {\it Poincar\'ekovarianz\/} und {\it Eich-\/} im Sinne von \vspace*{-.25ex}{\it \VIER{B}{R}{S}{T}-Invarianz\/}. \\
\indent
Aus ihrer Lagrangedichte hergeleitet werden Feynman-Regeln, die eine {\it lokale und Poincar\'e-kovariante Diagrammatik\/} darstellen, vgl.\@ Ref.~\cite{Veltman94}.
Auf dieser basisiert der Beweis `t~Hoofts ihrer Renormierung und sein explizites Regularisierungs- und Renormierungsverfahren, vgl.\@ unsere Ausf"uhrungen dazu; ebenso der Beweis Asymptotischer Freiheit, vgl.\@ unsere Ausf"uhrungen dazu.
Beide Beweise zusammen hei"sen Selbstkonsistenz der Theorie und ihre Konsistenz mit der Ph"anomenologie der Starken Wechselwirkung im Ultravioletten (perturbativen) Bereich.
Ihre Konsistenz auch im Infraroten (nichtperturbativen) Bereich ist akzeptiert, wenn auch keine Technik existiert, ihre zentralen \mbox{Ph"anomene wie sie Chirale~Sym}\-metriebrechung und Confinement formal-analytisch aus der effektiven Lagrangedichte~$\mf{L}_{\rm eff.}$ herzuleiten.
Die Diagrammatik {\it perturbativer\/}~\DREI{Q}{C}{D} scheint dazu ungeeignet zu sein.
\vspace*{-1ex}

\section*{\vspace*{-.5ex}Ausblick}
\renewcommand{\rightmarkQCD}{Ausblick}
\addcontentsline{toc}{section}{\numberline{} \hspace*{-12pt}Ausblick}

Die vorliegende Arbeit wendet sich zu der Struktur der Theorie im Infraroten, das hei"st auf die Struktur der~\DREI{Q}{C}{D} in ihrem {\it nichtperturbativ\/} dominierten Bereich, dem gleicherma"sen zugrunde liegt die vorgestellte lokale Poincar\'e-kovariante Formulierung der~\DREI{Q}{C}{D} als Lagrange'sche Quantenfeldtheorie.
Diese ist~-- wir rekapitulieren~-- vollst"andig bestimmt durch die formulierte effektive Lagrangedichte~$\mf{L}_{\rm eff.}$, wenn vorgegeben sind ein Regularisierungs- und Renormierungsverfahren und \vspace*{-.25ex}Zahlenwerte f"ur dessen Parameter. \\
\indent\enlargethispage{1ex}
Alle physikalischen Aussagen, das hei"st alle observablen Gr"o"sen einer renormierbaren Quantenfeldtheorie lassen sich ausdr"ucken durch ihre {\it ein-Teilchen-irreduziblen \vspace*{-.125ex}($n$-Punkt-)""Greenfunktionen\/}~$G^{(n)}_{\rm 1\!-\!irred.}$, die in diesem Sinne synonym stehen f"ur die Theorie an sich, s.u.\@ Seite~\pageref{T:1-irredGreenfnen}.
Diese Greenschen Funktionen sind im {\it Funktional- oder Pfadintegralzugang\/} definiert "uber normierte Funktionalintegrale eines entsprechenden Produkts von $n$ Quantenfeldern, gewichtet mit der exponentierten {\it effektiven Wirkung\/}~$\exp iS_{\rm eff.}$, die genau das Raumzeit-Integral der effektiven Lagrangedichte ist:~$S_{\rm eff.} \!=\! \int dx\mf{L}_{\rm eff.}(x)$.
"`Die Theorie l"osen"' besteht im eigentlichen darin, diese Funktionalintegrale auszuwerten.
Allerdings lassen sich Funktionalintegrale allgemein l"osen nur f"ur eine Wirkung, in der die Felder {\it quadratisch\/} auftreten; nur dann lassen sich die Integrationen auf {\it Gau"s'sche Integrationen\/} zur"uckf"uhren, nur diese lassen sich \vspace*{-.25ex}analytisch auswerten. \\
\indent
{\it Perturbative\/}~QCD ist  definiert genau durch eine perturbative Behandlung des Funktionalintegrals.
Dabei wird die effektive Wirkung formal in zwei Summanden aufgeteilt: in einen in diesem Sinne quadratischen Anteil, der analytisch ausgewertet wird, und in einen Rest, der als {\it kleine Korrektur\/} im Rahmen einer St"orungsreihe erfa"st wird.
Effektiv ist dies eine Entwicklung in dem Parameter~$g_{\rm ren.}^2\!/\!4\pi$ (mit~$g_{\rm ren.}$ die {\it renormierte Eichkopplung\/} der~QCD), und sie ist sinnvoll, nur solange diese klein ist, das hei"st Abbruch der Reihe nach einer endlichen Anzahl Glieder ein kleiner Fehler bedeutet.
Dies ist der Fall im ultravioletten Bereich der Theorie, der asymptotisch frei ist:~$g_{\rm ren.}(\mu)$ verschwindet mit anwachsender (Energie)Skala~$\mu$.
Dies ist nicht der Fall im Infraroten, wo die nichtabelsche Eichstruktur der Theorie~-- die Selbstkopplung des Eich-Vektorfelds durch den Kommutatorterm~$[A_\mu,A_\nu]$ im Feldst"arkentensor~-- zu einem Anwachsen der Kopplung bis hin zu ihrer formalen Divergenz am sogenannten \vspace*{-.25ex}{\it Landau-Pol\/} f"uhrt.
\vspace*{-.5ex}

\bigskip\noindent
{\bf Kapitel~\ref{Kap:VAKUUM}}\vspace*{-.125ex}\quad geht hierauf n"aher ein.
Es wird sich die dringende Frage nach einem Verfahren ergeben, mithilfe dessen das Funktionalintegral ausgewertet werden kann, ohne auf eine st"orungstheoretische Behandlung im angesprochenen Sinne zur"uckgreifen zu m"ussen.
Wir werden dieser Frage begegnen, indem wir das {\it Modell des Stochastischen Vakuums\/}~(\DREI[]{M}{S}{V}) von Dosch und Simonov motivieren und einf"uhren.
Die gef"uhrte Diskussion und der angegebene Formalismus liegen wesentlich zugrunde den nachfolgenden Kapitel.
\vspace*{-.5ex}

\bigskip\noindent
{\bf Kapitel~\ref{Kap:ANALYT}}\vspace*{-.125ex}\quad f"uhrt aus, wie das Modell des Stochastischen Vakuums bezogen wird auf Hochenergiestreuung und einer Analyse zug"anglich macht nichtperturbative Beitr"age.~--
Die eigentliche Zielrichtung unserer Arbeit: \vspace*{-.125ex}{\it Hochenergiestreuung im nichtperturbativen Vakuum der \vspace*{-.25ex}Quantenchromodynamik}. \\
\indent
Explizit geschieht dies mithilfe einer nichtperturbativen Formel f"ur die Streuung zweier Colour-Singuletts, die repr"asentiert sind in Termen (eichinvarianter) Wegner-Wilson-Loops, bei {\it kleinem\/} invarianten Quadrat des Impuls"ubertrags, typischerweise%
  ~\vspace*{-.125ex}\mbox{$-t \!<\! 1$~GeV$^2$}, {\it im Limes unendlichen\/} invarianten Quadrats der Schwerpunktenergie,%
  ~\vspace*{-.125ex}\mbox{$s \!\to\! \infty$}. \\
\indent
Wir geben an die Herleitung dieser Formel durch Nachtmann.
Vor dem Hintergrund dieser Herleitung argumentieren wir, in welcher Weise die Formel verallgemeinert werden kann zur Anwendung auf gro"se aber {\it endliche\/} Schwerpunktenergie.
Wir gelangen zu einem Ausdruck, der dominiert ist von dem Erwartungswert bez"uglich des nichtperturbativen QCD-Vakuums zweier Wegner-Wilson-Loops, deren zeitartige Linien {\it nahe\/} des Lichtkegels liegen~-- statt exakt darauf.
Die Streuamplitude kann daher {\it analytisch fortgesetzt\/} werden von der physikalischen {\it Minkowskischen Raumzeit\/} in deren analytische Fortsetzung ins {\it Euklidische\/}.
Wir arbeiten diese Fortsetzung \vspace*{-.25ex}explizit aus. \\
\indent\enlargethispage{2ex}
Zum einen f"uhrt die Rechnung zu einem tieferen Verst"andnis davon, wie sich vor dem Hintergrund des \DREI{M}{S}{V} nichtperturbative Streubeitr"age bei hohen Energien~$\surd s$ konstituieren im Zusammenspiel von longitudinaler und transversaler Dynamik.
Zum anderen machen wir die Streuamplitude, die {\it a priori\/} nur definiert ist in der physikalischen Minkowskischen Raumzeit, zug"anglich einer Auswertung in Euklidischer Raumzeit.
Dies ist von \vspace*{-.125ex}Bedeutung aus verschiedenen Gr"unden:
Hier ist die Argumentation sehr viel nat"urlicher, die zu den Annahmen f"uhrt, auf denen das \DREI{M}{S}{V} ruht;~-- mit anderen Worten, wir best"atigen \vspace*{-.125ex}{\it a posteriori\/} die naive Minkowskische Formulierung des \DREI[]{M}{S}{V}.
Und hier ist im Prinzip eine {\it modellunabh"angige\/} Auswertung m"oglich: durch numerische Simulationen im Rahmen von~\DREI{Q}{C}{D} als Gittereichtheorie.
\vspace*{-.5ex}

\bigskip\noindent
In den beiden folgenden Kapiteln wird der angegebene Formalismus bezogen auf explizite Teilchenreaktionen, die in besonderer Weise dominiert sind durch nichtperturbative~\vspace*{-.25ex}\DREI{Q}{C}{D}. \\
\indent
{\bf Kapitel~\ref{Kap:GROUND}}\vspace*{-.125ex}\quad diskutiert diffraktive Leptoproduktion eines Vektormesons~$V$ im Grundzustand an einem Nukleon~$N$:
Das Lepton strahlt ein virtuelles Photon~$\ga^{\D\ast}$ ab, das diffraktiv und bei hoher Schwerpunktenergie an~$N$ streut und im Sinne der Reaktion~\mbox{$\ga^{\D\ast}N \!\to\! VN$} exklusiv ein Vektormeson produziert; das Nukleon bleibt dabei vollst"andig intakt.
Diese Reaktion wird theoretisch zug"anglich im Rahmen des \DREI{M}{S}{V} mit zus"atzlicher Kenntnis der hadronischen Wellenfunktionen der \vspace*{-.25ex}involvierten Teilchen. \\
\indent
Zun"achst rechtfertigt der diskutierte kinematische Bereich gro"ser inverianter Schwerpunktenergie~$\surd s$, die Zust"ande zu repr"asentieren als Superposition ihres f"uhrenden Fock-Zustandes.
Auf der einen Seite der Streuung wird zugrunde gelegt f"ur das Nukleon der einfache Ansatz von Quark-Diquark-Paaren, deren beiden Konstituenten gleichverteilt sind bez"uglich ihres longitudinalen Impulses und Gau"s'sch verteilt bez"uglich ihres trensversalen Relativimpulses.
Auf der anderen Seite der Streuung wird das Photon repr"asentiert als Superposition von Quark-Antiquark-Zust"anden, deren Wellenfunktion bestimmt ist durch {\it Light-Cone Perturbation Theory\/}~(\VIER{L}{C}{P}{T}), das hei"st durch gew"ohnlichen St"orungstheorie auf dem Lichtkegel.
Diese Beschreibung ist wohldefiniert f"ur gro"se Virtualit"aten~$Q$ des Photons, bricht aber bei kleinen Skalen~$Q$ zusammen.
Dies geschieht bei zunehmender Manifestation der {\it Brechung der Chiralen Symmetrie\/}%
\FOOT{
  \label{FN:chiraleSymmetrie}Die Lagrangedichte der QCD besitzt zus"atzlich zu den Symmetrien unter ${\cal C}$-, ${\cal P}$- und ${\cal T}$-, lokale $\SUNc$-Eich- und Poincar\'e-Transformationen noch weitere Symmetrien: Es sind dies, im Limes verschwindender Fermionmassen, die in Fu"snote~\FN{FN:Flavour} angemerkte globale~$SU\!(N_{\rm\!F})$-Symmetrie und die {\it chirale Symmetrie\/}, d.h.\@ Invarianz unter globalen $SU\!(N_{\!F})_l\otimes SU\!(N_{\rm\!F})_r$-Transformationen, wobei die Skripte f"ur {\it Links\/}- und {\it Rechts-H"andigkeit\/} stehen.
}
und von {\it Confinement}; dann wenn diese Effekte die Quark-Antiquark-Dipole auf transversale Ausdehnungen beschr"anken, die sehr viel kleiner sind als~$1\!/\!Q$, die transversale Ausdehnung des perturbativen Photons.
In~praxi ist dies massiv zu erwarten f"ur Virtualit"aten kleiner als ein GeV,%
  ~\vspace*{-.25ex}\mbox{$Q^2 \!<\! 1\GeV^2$}. \\
\indent
In Analogie zu der Wellenfunktion des Photons modellieren wir eine Quark-Antiquark-Wellenfunktion des Vektormesons.
Dies hei"st im wesentlichen, die Helizit"atsstruktur des Quarks und Antiquarks von der Wellenfunktion des Photons zu "ubernehmen und den das Photon in der \VIER{L}{C}{P}{T} charakterisierende Energienenner%
  ~\mbox{$\big(z(1\!-\!z)\,Q^2 + m^2 + \rb{k}^2\big){}^{\!-1}$} zu ersetzen durch eine Funktion, die abh"angt von dem transversalen Impuls~$\rb{k}^2$, der verkn"upft ist mit dem transversalen Relativimpuls von Quark und Antiquark, und dem Anteil~$z$ des Quarks,~\mbox{$(1 \!-\! z)$} des Antiquarks am gesamten Lichtkegelimpuls.
F"ur diese Funktion wird ein Ansatz gemacht, der motiviert ist durch die Gau"s'schen Wellenfunktionen des transversalen Harmonischen Oszillators; seine zwei Parameter werden Fixiert durch Forderung von Normiertheit und Reproduktion der observablen elektronischen Zerfallsbreite des \vspace*{-.25ex}Vektormesons. \\
\indent
Auf Basis der in dieser Weise konstruierten hadronischen Wellenfunktionen werden berechnet charakteristische Observable der Reaktion~$\ga^{\D\ast} N\!\to\! V N$ und verglichen mit den experimentellen Daten~-- soweit diese vorhanden sind.
Wir untersuchen insbesondere die Produktion der Vektormesonen~$\rh(770)$, $\om(782)$, $\ph(1020)$ und $J\!/\!\ps(3097)$ und finden insgesamt gute "Ubereinstimmung mit dem Experiment. \\
\indent\enlargethispage{2.375ex}
Die~-- drei~-- Parameter des~\DREI[]{M}{S}{V} sind numerisch fixiert durch Prozesse, die allenfalls entfernt in Zusammenhang stehen mit den diskutierten Reaktionen.
Wir sehen \vspace*{-.125ex}daher best"atigt als {\it wesentlich\/} den grundlegenden Mechanismus des~\DREI[]{M}{S}{V}, der nichtperturbative Beitr"age zur Hochenergiestreuung erkl"art durch die {\it Ausbildung und Wechselwirkung gluonischer Strings\/} zwischen den (Anti)Quarkkonstituenten.
\vspace*{-.75ex}

\bigskip\noindent
{\bf Kapitel~\ref{Kap:EXCITED}}\vspace*{-.125ex}\quad arbeitet die Relevanz dieses Mechanismus weiter heraus in Reaktionen, die hierzu pr"adestiniert sind aufgrund der erheblich gr"o"seren transversalen Ausdehnung der involvierten~-- struenden~-- \vspace*{-.25ex}Teilchen. \\
\indent
Zum einen erweitern wir unsere \vspace*{-.125ex}Analyse, indem wir auch Photonen mit kleiner bis verschwindender Virtualit"at~-- das hei"st Photoproduktion:~\mbox{\,$Q^2 \!\equiv\! 0$}~-- in unsere Diskussion mit einbeziehen.
Dies, so kann argumentiert werden, ist m"oglich durch phenomenologische Extrapolation der renormierten Lagrange'schen Quarkmasse, wie sie auftritt in der \VIER{L}{C}{P}{T}, von $Q^2 \!\cong\! 1\GeV^2$ nach~$Q^2 \!=\! 0$ hin zu einer Konstituentenquarkmasse. \\
\indent
Zum anderen betrachten wir neben der exklusiven Produktion des \vspace*{-.125ex}$\rh$-Vektormesons auch die seiner ersten beiden angeregten Zust"ande~\vspace*{-.125ex}$\rh(1450)$ und~$\rh(1700)$.
Ben"otigt werden hierzu entsprechend die Quark-Antiquark-Wellenfunktionen der angeregten Vektormesonen.
Wir geben Argumente daf"ur an, die angeregten Zust"ande als Mischung eines Quark-Antiquark-$2S$-Zustandes und eines inerten Restes aufzufassen, der dadurch per definitionem charakterisiert ist, da"s seine Kopplung an das Photon unterdr"uckt ist.
Es sind dies also entweder Anregungen h"oheren Bahndrehimpulses, deren Kopplung differiert von Null erst durch relativistische Effekte, oder h"ohere Fock Zust"ande: im einfachsten Fall ein Quark-Antiquark-$2D$- beziehungsweise ein hybrider Quark-Antiquark-Gluon-Zustand.
Die explizite Konstruktion der Wellenfunktion geschieht analog zur Wellenfunktion des Grundzustandes~-- der unver"andert "ubernommen wird vom vorhergehenden Kapitel~\ref{Kap:GROUND}.
Die Parameter werden analog fixiert und zus"atzlich gefordert Orthogonalit"at zum \vspace*{-.25ex}Grundzustand. \\
\indent
Auf Basis dieser ph"anomenologischen Ans"atze werden berechnet charakteristische Observable und diese verglichen mit den experimentellen Daten~-- soweit diese vorhanden sind.
Abgesehen von einzelnen subtileren Fragen, erhalten wir en gros trotz der Einfachheit unserer Ans"atze gute bis sehr gute "Ubereinstimmung.
Dies trifft insbesondere zu f"ur Gr"o"sen, die wir postulieren, w"ahrend sie in der sonstigen Literatur nur parametrisiert \vspace*{-.25ex}sind. \\
\indent
Zusammenfassend die Kapitel~\ref{Kap:GROUND} und~\ref{Kap:EXCITED} sind zwei Aspekte zu betonen.
Angesichts dessen, da"s die Parameter des~\DREI{M}{S}{V} numerisch fixiert sind in Prozessen mit allenfalls entfernten Zusammenhang mit den diskutierten Reaktionen, unterstreicht die gute "Ubereinstimmung mit experimentellen Daten~-- soweit vorhanden~-- die Relevanz des Modells.
Das hei"st die allgemeine Relevanz seines grundlegenden Mechanismus der Ausbildung und Wechselwirkung gluonischer Strings f"ur nichtperturbative Quantenchromodynamik.
Andererseits geben wir in gro"sem Umfang numerische Postulate an f"ur Observable, die wir vorschlagen zur Messung~-- zur Verifizierung (oder Falsifizierung) dieses Mechanismus.
\vspace*{-.75ex}

\bigskip\noindent
Die wesentlichen Resultate unserer Arbeit von Kapitel~\ref{Kap:GROUND} und~\ref{Kap:EXCITED} sind ver"offentlicht in den beachteten Refn.~\cite{Dosch96} beziehungsweise~\cite{Kulzinger98,Kulzinger98a,Kulzinger99}.
Ihnen liegt zugrunde die \mbox{$s \!\to\! \infty$-asympto}\-tische \mbox{$T$-Amp}\-litude.
Die \mbox{$T$-Amp}\-litude f"ur gro"se aber endliche Werte von~$s$ ist Resultat der Analyse von Kapitel~\ref{Kap:ANALYT}, die chronologisch den Abschlu"s unserer Arbeit darstellt; ihre Ver"offentlichung ist in Vorbereitung in \vspace*{-.125ex}Ref.~\cite{Kulzinger99a}. \\
\indent
Wir weisen ferner hin auf Ref.~\cite{Kulzinger95}, unsere Diplomarbeit an der Ruprecht-Karls-Universi\-t"at Heidelberg, die mit Fragen aufgeworfen hat, die hier diskutiert und \vspace*{16ex}beantwortet werden.
\theendnotes

%% file: VAKUUM-F.tex
\lhead[\fancyplain{}{\sc\thepage}]
      {\fancyplain{}{\sc\rightmark}}
\rhead[\fancyplain{}{\sc\leftmark}]
      {\fancyplain{}{\sc\thepage}}
\chapter[Stochastisches Vakuum]{\huge Stochastisches Vakuum}
\label{Kap:VAKUUM}

Wir diskutieren in dieser Arbeit diffraktive Streuung bei hohen invarianten Schwerpunktenergien als Konsequenz {\it nichtperturbativer\/}~QCD.
Zu nichtperturbativer QCD existiert kein kanonischer Zugang aber die verschiedensten Ans"atze.
Wir werden daher zun"achst unseren Zugang zu diesem Bereich der QCD kleiner (Energie)Skalen formulieren: das {\it Modell des Stochastischen Vakuums\/}~(\DREI[]{M}{S}{V}) von Dosch und Simonov.
In diesem ersten Kapitel wollen wir dieses Modell in seinen Grundz"ugen vorstellen: die zugrundeliegende Idee~-- wie diese motiviert ist und die de facto Annahmen, die sie bedeutet~-- die Gr"o"sen, in Termen derer das Modell arbeitet, die notwendige Mathematik und den resultierenden Formalismus im allgemeinen.
Zwangsl"aufig, in Hinblick auf unsere Anwendung auf diffraktive Hochenergiestreuung, gehen wir ein auf die Beschreibung von Confinement.

\bigskip\noindent
{\it Nichtperturbative\/}~QCD steht synonym f"ur den Infrarotbereich, also den Bereich kleiner (Energie)Skalen~$\mu$ der QCD, die allgemein definiert ist durch die Lagrangedichte~$\mf{L}$ beziehungsweise die effektive Lagrangedichte~$\mf{L}_{\rm eff.}$, vgl.\@ Gl.~(\ref{QCD:Lagrangedichte}) bzw.~(\ref{QCD:Lagrangedichte_eff}).

Die Asymptotische Freiheit dieser Theorie auf {\it gro"sen\/} Skalen~$\mu$, das hei"st das Verschwinden der renormierten Eichkopplung~$g_{\rm ren.}\!(\mu)$ asymptotisch mit~$\mu\!\to\!\infty$ legt eine Entwicklung in diesem Parameter nahe und definiert durch die entsprechende St"orungsreihe {\it perturbative\/}~QCD:
Terme h"oherer Ordnung bedeuten eine nur "`kleine"' Korrektur und schon die ersten Terme vermitteln Aussagen von in praxi ausreichender Genauigkeit; eine "`kleine"' Anzahl h"oherer Ordnungen ist dann miteinzubeziehen, wenn eine h"ohere Genauigkeit gefordert wird.

Kontr"ar hierzu ist das Verhalten der Theorie auf {\it kleinen\/} Skalen~$\mu$.
(Diffraktion ist genau dominiert von gro"sen Abst"anden der Ordnung~$1\!/\!\mu$.)
So zeigt die Extrapolation der Eichkopplung~$g_{\rm ren.}\!(\mu)$ hin zu kleinen~$\mu$ mit Argumenten der Renormierungsgruppe (also das perturbative Konzept impliziert), da"s diese anw"achst wie~$(\ln\mu\!/\!\La_{\rm QCD})^{-1}$, das hei"st formal sogar divergiert f"ur~$\mu \!=\! \La_{\rm QCD}$, dem Landau-Pol der QCD~-- sie also weit davon entfernt ist, "`kleiner"' Parameter zu sein.
Auf kleinen Skalen konvergiert die St"orungsreihe also nicht, die herk"ommliche St"orungstheorie, mit anderen Worten: perturbative QCD, ist nicht definiert.

Da"s eine Entwicklung in~$g_{\rm ren.}\!(\mu)$ auf beliebigen Skalen~$\mu$ nicht definiert sein kann, folgt schon {\it a~priori}, seit bekannt ist, da"s Eichfeldkonfigurationen existieren~-- aufgrund einfacher "Uberlegungen sogar existieren m"ussen, vgl.\@ in Ref.~\cite{Nachtmann92}~--, die zur effektiven Lagrangedichte und damit zur effektiven Wirkung~$S_{\rm eff.} \!=\! \int dx\mf{L}_{\rm eff.}\!(x)$ der QCD proportional zu~$g_{\rm ren.}^{-2}$ beitragen.
Dies insofern, als sich observable Gr"o"sen~-- im Formalismus des Funktionalintegrals~-- herleiten aus dem Erzeugenden Funktional~$Z$ nach Gl.~(\ref{ErzFnl}), in das wesentlich eingeht das Exponential der effektiven Wirkung,~$\exp iS_{\rm eff.}$; f"ur verschwindende Kopplung~$g_{\rm ren.}$ besitzt~$Z$ also eine wesentliche Singularit"at und daher observable Gr"o"sen keine Entwicklung um diesen Punkt.
Auf der anderen Seite sind diese Beitr"age relevant, da f"ur sie die (Euklidische) effektive Wirkung ein Extremum annimmt, sie also L"osungen der klassischen Theorie sind, um die herum Quantenfluktuationen auftreten.

Wir k"onnen hier nicht n"aher auf diese sogenannten {\it Instanton(konfiguration)en\/} eingehen, die Mitte der siebziger Jahre explizit angegeben wurden, vgl.\@ die Refn.~\cite{Belavin75,tHooft76}; was ihre Konstruktion und Relevanz f"ur die QCD betrifft verweisen wir auf die Refn.~\cite{Callan78,Shuryak82,Dyakonov84}.
Wesentlich ist aber, da"s sie in Minkowskischer Raumzeit physikalisch zu interpretieren sind als Wahrscheinlichkeit f"ur den "Ubergang (im Raum des Eichfelds) zwischen zwei topologisch nicht-"aquivalenten Grundzust"anden~-- Vakua~-- der Theorie.
Oder: Das {\it physikalische\/} Vakuum der QCD ist zu verstehen als "Uberlagerung unendlich vieler topologisch nicht-"aquivalenter Vakua.
Es hat in diesem Sinne eine {\it nichttriviale\/} Struktur.

Ein "ahnlich ankn"upfendes Verst"andnis des physikalischen Vakuums liegt den {\it Summenregeln der\/}~QCD zugrunde, wie sie 1979 von Shifman, Vainshtein, Zakharov formuliert worden sind in Ref.~\cite{Shifman79}, vgl.\@ auch Ref.~\cite{Novikov84}:
Ausgehend von perturbativer QCD treten nichtperturbative Beitr"age auf als power corrections, die bestimmt sind durch Erwartungswerte bez"uglich des {\it physikalischen\/} Vakuums, die identisch verschwinden w"urden bez"uglich des {\it perturbativen\/} Vakuums.
Explizit wird die {\it Operatorproduktentwicklung\/} Wilsons, vgl.\@ Ref.~\cite{Wilson69}, ausgedehnt auf nichtperturbative Skalenbereiche und angewandt auf Korrelatoren von Vektorstr"omen, vgl.\@ Ref.~\cite{Novikov80}.
In der Entwicklung treten normalgeordnete Operatoren in den Lagrange'schen Quantenfeldern~${\cal O}_d$ auf mit beliebig hoher kanonischen Dimension~$d$.
Alle bis auf~${\cal O}_0\!\equiv\!\bbbone$ verschwinden per definitionem bez"uglich des perturbativen Vakuums; ihre Erwartungswerte bez"uglich des physikalischen Vakuums, parametrisieren nichtperturbative Beitr"age "uber eine nichttriviale Struktur des Vakuums der QCD.

In dieselbe Richtung einer nichttrivialen Struktur des Vakuums der QCD, gehen die Arbeiten `t~Hoofts von 1978/79, siehe die Refn.~\cite{tHooft78,tHooft79}.
Sie zeigen, da"s das Vakuum einer nichtabelschen Eichtheorie aufgefa"st werden kann als Kondensat magnetischer Monopole, das sich daher verh"alt wie ein dualer Supraleiter; Confinement das Ausbilden eines (chromo)elektrischen gluonischen Strings erkl"art sich durch einen dualen Mei"sner-Effekts.
Vgl.\@ auch die Refn.~\cite{tHooft98,Polyakov87}.

Das gemeinsame Bild aller dieser Ans"atze ist das folgende:
Im Sinne von Quantenfluktuationen werden aus dem Vakuum heraus permanent und, da masselos, in gro"ser Zahl niederenergetische (oder -frequenten) Gluonen erzeugt, treten aufgrund der Nicht-Abelizit"at der QCD miteinander in eine komplizierte, nichtlineare Wechselwirkung und werden wieder vernichtet.
Dies f"uhrt zu einem permanent fluktuierenden Eichfeld-Hintergrund.

Im Formalismus des Funktionalintegrals, der bekanntereren und intuitiveren Quantenformulierung nichtabelscher Feldtheorien (im Vergleich zum BRST-Operatorformalismus), wird dieser Hintergrund gerade vermittelt durch die Integration "uber alle Eichfeldkonfigurationen, die in komplizierter Weise gewichtet sind durch~$\exp iS_{\rm eff.} \!=\! \exp\iIM\! \int dx\mf{L}_{\rm eff.}\!(x)$.
Hier gehen ein die Yang-Mills- und die Dirac-Lagrangedichten,~$\mf{L}_{\rm Y\!M}$ bzw.\@ $\mf{L}_{\rm D}$, wie auch die Zwangsbedingungen der Theorie, die sich in einem Eichfixierungs- und Geistterm,~$\mf{L}_{\rm gf.}$ bzw.\@ $\mf{L}_{\rm ghost}$, manifestieren; vgl.\@ die Gln.~(\ref{QCD:Lagrangedichte_eff}$'$) bzw.~(\ref{QCD:Lagrangedichte_eff}$''$) bzgl.\@ deren expliziten Gestalt.

Es ergeben sich Schlu"sfolgerung f"ur die Formulierung des \DREI[]{M}{S}{V}, vgl. auch Ref.~\cite{Dosch94}.
Dosch und Simonov nehmen zun"achst an, das Eichfeldma"s des Funktionalintegrals aufspalten zu k"onnen in die Summe zweier Anteile, die von hoch- beziehungsweise niederfrequenten Gluon-Eichfeldern dominiert sind.
Der erste Anteil kann aufgrund von Asymptotischer Freiheit kanonisch behandelt werden im Rahmen von perturbativer QCD.
Der zweite, verantwortlich f"ur den nichtperturbativen Eichfeld-Hintergrund, wird durch einen expliziten Ansatz approximiert. \\
Es wird argumentiert:
Das Infrarotproblem der QCD ist Konsequenz des perturbativen Konzeptes, das auf beliebigen Skalen nicht geeignet ist~-- da nicht definiert~--, die Theorie auszuwerten; es ist insofern Artefakt des Zugangs zur Theorie und nicht ihr selbst inh"arent.
In der observablen Physik existiert kein Infrarotproblem.
Offensichtlich sind im Funktionalintegral die Eichfeldkonfigurationen in subtiler Weise genau so gegeneinander gewichtet, da"s Beitr"age, die f"ur sich genommen divergieren, einander aufheben.

Dies suggeriert die zwei Annahmen, die Grundannahmen des \DREI[]{M}{S}{V}:
Die Fluktuationen niederfrequenter Eichfeldkonfigurationen (die ihrer Komplexit"at wegen bislang nicht haben systematisch erfa"st werden k"onnen) k"onnen approximiert werden als {\it stochastisch\/} und sie f"uhren zu {\it infrarot-endlichen\/} Aussagen f"ur observable Gr"o"sen.

Formal hei"st dies:
F"ur (observable) Erwartungswerte im nichtperturbativen Vakuum der QCD~-- repr"asentiert im Formalismus des Funktionalintegrals, das hei"st als Integral des entsprechenden Eichfeld-Funktionals, das aufzufassen ist als dessen Mittelung bez"uglich eines komplizierten Integrationsma"ses "uber s"amtliche niederfrequenten Eichfeldkonfigurationen~-- wird angenommen, es existiert eine {\it konvergente Kumulanten- oder Clusterentwicklung\/}.
Darunter wiederum ist zu verstehen eine konvergente Entwicklung, deren $n$-ter Term gegeben ist durch die {\it Kumulante $n$-ter Ordnung\/}, eine nichtlokale Gr"o"se mit $n$~Raumzeit-Argumenten~$x_1,\ldots,x_n$, die definiert ist "uber ihre {\it Clustereigenschaft\/}, das hei"st sie verschwindet "`schnell"' (de facto exponentiell), wenn nur eines dieser Argumente den Cluster verl"a"st (den Bereich fest umrissener Ausdehnung), in dem die "ubrigen Argumente liegen.

Diese Entwicklung induziert als Parameter im wesentlichen die {\it Korrelationsl"ange\/} als Ma"s der Ausdehnung des korrelierten Bereiches und als Ma"s der Korrelationsst"arke die {\it Normierungen\/} der einzelnen Kumulanten~-- die aufgrund der Bianchi-Identit"aten~\mbox{$D_\rh F_{\mu\nu} \!+\text{zykl.} \!=\!0$} nicht unabh"angig voneinander sind, vgl.\@ die Refn.~\cite{Antonov95,Shevchenko97}.

Wesentlich ist, da"s der zugrundeliegende stochastische Proze"s nicht bez"uglich der {\it Eichfelder\/}~$A_\mu$, sondern bez"uglich der {\it Eichfeldst"arken\/}~$F_{\mu\nu}$ angenommen wird.
Die Kumulanten beziehen sich auf diese Felder und stehen f"ur die {\it Korrelation auf kleinen Abst"anden von Eichfeldst"arken}.
Dies impliziert dann~-- anstatt sie von vornherein auszuschlie"sen~-- die komplexe {\it Korrelation auf gro"sen Abst"anden von Eichfeldern\/}, mit anderen Worten deren Wechselwirkung bis hin "uber makroskopische Distanzen, wie sie gerade verantwortlich ist f"ur den von Aharonov und Bohm beschriebenen Effekt.

Das \DREI{M}{S}{V} in seiner allgemeinsten Form definiert sich genau "uber die Existenz einer konvergenten Kumulantenentwicklung in diesem Sinne.
Noch unabh"angig von dem zugrundeliegenden stochastischen Proze"s, das hei"st noch ohne die ihn definierenden Kumulanten explizit anzugeben~-- letztlich aufgrund deren Clustereigenschaft~--, f"uhrt es zu {\it linearem Confinement\/} f"ur ein statisches Quark-Antiquark-Paar. 

Wird dann der stochastische Proze"s spezifiziert, ist das dadurch definierte \DREI{M}{S}{V} im engeren Sinne ein voraussagekr"aftiges Instrument, indem es in die Lage versetzt, nichtperturbative Observable analytisch anzugehen und auszuwerten.

Es ist offensichtlich um so voraussagekr"aftiger, durch je weniger Parameter es bestimmt ist, je "`einfacher"' der zugrundeliegende Proze"s ist.
Der einfachste stochastische Proze"s ist ein {\it (zentrierter) Markov-\/} oder {\it Gau"s'scher Proze"s\/}, das hei"st nur die {\it Kumulante zweiter Ordnung\/} ist von Null verschieden.
Das \DREI{M}{S}{V} im engeren Sinne spezifiziert in genau dieser Weise den Proze"s, der die Fluktuationen der niederfrequenten Eichfeldkonfigurationen beschreibt.
Diese Annahme ist "`nat"urlich"' schon insofern, als die stochastische Interpretation der Fluktuationen nur dann sinnvoll ist, wenn die assoziierte Kumulantenentwicklung "`schnell"' konvergiert, wenn h"ohere Ordnungen also nur "`kleine"' Korrekturen darstellen.
Verschwinden die Kumulanten h"oherer Ordnung identisch und damit die Korrekturen, ist dies trivialerweise der Fall und die erste Approximation eines beliebigen sinnvollen Prozesses.

Impliziere das \DREI{M}{S}{V} zus"atzlich diese Gau"s'sche Annahme.
Es ist also vollst"andig bestimmt durch die {\it Kumulante zweiter Ordnung\/}, die nichtlokale Verallgemeinerung des {\it Gluonkondensats\/}~$\vac{g^2FF} \!\equiv\! \vac{g^2 F_{\mu\nu a}\!(0) F^{\mu\nu}{}_a\!(0)}$ der Summenregeln von Shifman, Vainshtein, Zakharov.
Sie ist schon weitgehend bestimmt durch die Invarianz der QCD unter ${\cal C}$-, ~${\cal P}$- und~${\cal T}$-Transformationen und ihrer (lokalen) Eich- und Poincar\'einvarianz.
Sieht man ab von ihrer expliziten funktionalen Form f"ur korrelierte Raumzeit-Argumente (die tats"achlich zweitrangig ist), so treten zwei Parameter auf: die Normierung der Kumulante zweiter Ordnung, die gegeben ist durch~$\vac{g^2FF}$, und die Korrelationsl"ange~$a$ der Eichfeldst"arken.
Beides sind Gr"o"sen von herausragender Bedeutung in nichtperturbativer QCD im allgemeinen, ihre Zahlenwerte daher innerhalb gewisser Bereiche festgelegt.

Wir wollen diese Gedanken formaler fassen und das \DREI{M}{S}{V} explizit formulieren.
\vspace*{-.5ex}

\section{Observable als Eichfeld-Funktionalintegrale}

Wie in der Einleitung ausgef"uhrt ist eine renormierbare Quantenfeldtheorie (bei Angabe eines Renormierungsschemas und Zahlenwerten f"ur dessen Parameter) vollst"andig bestimmt durch ihre Lagrangedichte. F"ur die QCD haben wir effektiv
\begin{samepage}
%
\begin{align} \label{Lagr_eff}
\mf{L}_{\rm eff.}\;
  =\; \mf{L}\;+\; \mf{L}_{\rm gf.}\; +\; \mf{L}_{\rm ghost}
\end{align}
Dabei ist
%
\begin{align} 
\mf{L}\;
  =\; \mf{L}_{\rm Y\!M}\;+\; \mf{L}_{\rm D}\;
  =\; -\frac{1}{2}\, \tr F_{\mu\nu} F^{\mu\nu}\;
        +\; \bar{\ps} \big(\iIM\,\ga^\mu D_\mu - m\big) \ps
\end{align}
die eigentliche Lagrangedichte der QCD, die sich aus dem Yang-Mills- und dem Dirac-Anteil zusammensetzt, vgl.\@ Gl.~(\ref{QCD:Lagrangedichte}); die Dichten
%
\begin{align} \label{Lagr_gfghost}
\mf{L}_{\rm gf.}\;
  =\; - \xi^{-1}\, \tr {\cal F}^2[A]\qquad
\mf{L}_{\rm ghost}\;
  =\; - \bar{c}_a\; \frac{\de{\cal F}_a[A]}{\de A_{\mu b}}\; (D_\mu c)_b
\end{align}
sind der die Eichung auf~${\cal F}[A]\!=\!C$ fixierende Term beziehungsweise der zugeh"orige Term der Faddeev-Popov-Geistfelder, vgl.\@ Gl.~(\ref{QCD:Lagrangedichte_eff}).
Wir erinnern an Eichfelder und -feldst"arken:
%
\begin{align} \label{Feldstrkn}
&F_{\mu\nu}\;
  =\; (\iIM\,g)^{-1}[D_\mu,D_\nu]\;
  =\; \pa_\mu A_\nu - \pa_\nu A_\mu + \iIM\,g[A_\mu,A_\nu]
    \\[.5ex]
&F_{\mu\nu}
  = F_{\mu\nu a}\, T^a_\Drst{R}\qquad
  A_\mu
    = A_{\mu a}\, T^a_\Drst{R}\qquad 
  F_{\mu\nu a}
    = \pa_\mu A_{\nu a} - \pa_\nu A_{\mu a} - gf_{abc}A_{\mu b}A_{\nu c}
    \tag{\ref{Feldstrkn}$'$}
\end{align}
und die eichkovariante Ableitung:
%
\begin{align} \label{kovAbl}
D_\mu\;
  =\; \pa_\mu + \iIM\,g A_\mu\qquad
(\Del{x}_\mu)_{\al\be}\;
  =\; \del{x}_\mu\, (\bbbOne{R})_{\al\be}
      + \iIM\,g\, A_{\mu a}\!(x)\, (T_\Drst{R}^a)_{\al\be}
\end{align}
Dabei sind~$T^a_\Drst{R}$ die Generatoren und~$\bbbOne{R}$ die Eins einer beliebigen Darstellung~$\mf{R}$ der Eichgruppe~$\SUNc$ und~$\del{x}_\mu \!\equiv\! \pa /\pa x^\mu$. Sei verwiesen auf Ref.~\cite{Itzykson88}.

Die effektive Lagrangedichte~$\mf{L}_{\rm eff.}$ ist also Funktional in den Quantenfeldern: in den Eich- und fundamentalen Materiefeldern,~$A$ bzw.\@ $\ps$,~$\bar{\ps}$, und den Geistern~$c$,~$\bar{c}$.
Zentrale Gr"o"se in der Funktionalintegral-Formulierung der QCD ist $Z$, das {\it Erzeugendes Funktional\/}
%
\begin{align} \label{ErzFnl}
Z[J, \et, \bar{\et}]\;
  =\; \mf{N}^{-1}\; \int\! \mf{D}\!(A, \ps, \bar{\ps}, c, \bar{c})\;
    \exp\,\iIM\!\int\zz dx
      \left( \mf{L}_{\rm eff.} + \tr J^\mu A_\mu + \bar{\et}\, \ps + \bar{\ps}\, \et
      \right)
\end{align}
wobei entsprechende {\it Quellenfelder\/}~$J$ und~$\et$,~$\bar{\et}$ eingef"uhrt sind.
Die Normierungskonstante~$\mf{N}^{-1}$ folgt aus der Forderung~$Z|_{J,\et,\bar{\et}=0} \equiv\!1$.
Die Integration~$dx$ ist zu verstehen als "uber die~vier\-dimensionale \pagebreak Minkowskische Raumzeit.
\end{samepage}

Eine Quantenfeldtheorie zu l"osen hei"st explizit, ihre {\it Greenfunktionen\/} zu bestimmen.
Diese sind formal {\it Vakuumerwartungswerte von zeitgeordneten Produkten wechselwirkender Quantenfelder\/} und bestimmen die $S$-Matrix der Theorie vollst"andig.
Verschieden definierte Greenfunktionen sind von Bedeutung: \\
Die {\it ($n$-Punkt-)Greenfunk\-tionen\/}~$G^{(n)}$ werden erzeugt durch das gerade definierte Funktional~$Z$ und zwar durch Anwendung~-- in definierter Reihenfolge~-- von Funktionalableitungen nach den Quellen~$J$ und~$\et$,~$\bar{\et}$ und deren anschlie"sendes Nullsetzen, generisch:
%
\begin{align} \label{Greenfnen}
&G^{(n=l+2m)}
  \!(x_1,\cdots,x_l; y_1,\cdots,y_m; \bar{y}_1,\cdots,\bar{y}_m)
    \nn \\[1ex]
&=\; \bra{\Om}\; T\vv
       {\T\prod}_{i=1}^l\, A_{\mu_i}\!(x_i)\vv
       {\T\prod}_{j=1}^m\, \ps(y_j)\vv
       {\T\prod}_{\bar{j}=1}^m\, \bar{\ps}(\bar{y}_{\bar{j}})\;
     \ket{\Om}
    \\[2ex]
&=\; \mf{N}^{-1}\; \int\! \mf{D}\!(A, \ps, \bar{\ps}, c, \bar{c})\;
    \left( T\;
      {\T\prod}_{i=1}^l\, A_{\mu_i}\!(x_i)\;
      {\T\prod}_{j=1}^m\, \ps(y_j)\;
      {\T\prod}_{\bar{j}=1}^m\, \bar{\ps}(\bar{y}_{\bar{j}})\,
    \right) \exp\,\iIM\!\int\zz dx \mf{L}_{\rm eff.}
    \nn \\[-2ex]
   &\tag{\ref{Greenfnen}$'$} \\[-1ex]
&=\; \left. \left( T\;
    {\T\prod}_{i=1}^l\;
       \frac{1}{\iIM}\frac{\de}{\de J^{\mu_i}\!(x_i)}\vv
    {\T\prod}_{j=1}^m\,
       \frac{1}{\iIM}\frac{\de}{\de \bar{\et}(y_j)}\vv
    {\T\prod}_{\bar{j}=1}^m\,
      -\frac{1}{\iIM}\frac{\de}{\de \et(\bar{y}_{\bar{j}})}\,
         \right)\;
    Z[J, \et, \bar{\et}]\vv \right|_{J, \et, \bar{\et} = 0}
    \tag{\ref{Greenfnen}$''$}
\end{align}
wobei~$T$ f"ur den Zeitordnungsoperator steht (die zweite Zeile suggestive Notation aus der Operatorformulierung mit~$\Om$ dem {\it physikalischen Vakuum}). Ableitungen nach Gra"smann-wertigen~-- d.h.\@ antikommutierenden~-- Feldern antikommutieren selbst und sind als linksseitige zu verstehen, Eichgruppenindizes sind weggelassen. \\
Die {\it zusammenh"angenden ($n$-Punkt-)Greenfunktionen\/}~$G^{(n)}_{\rm c}$ werden analog zu Gl.~(\ref{Greenfnen}) erzeugt durch das Funktional~$Z_{\rm c}\!=\!\ln Z$. \\
Die {\it ein-Teilchen-irreduziblen ($n$-Punkt-)Greenfunktionen\/}\label{T:1-irredGreenfnen}~$G^{(n)}_{\rm 1\!-\!irred.}$~-- dies sind die Greenfunktionen, die Feynman-Diagrammen entsprechen, die zusammenh"an\-gend bleiben, wenn eine beliebige interne Linie durchgeschnitten wird~-- folgen aus der Legendretransformierten~$Z_{\rm 1-irred}$ von~$Z_{\rm c}$ bez"uglich der Quellenfelder~$J$ und~$\et$,~$\bar{\et}$:
%
\begin{align} \label{1-irredErzFnl}
&\iIM\, Z_{\rm 1\!-\!irred.}[A, \ps, \bar{\ps}]\;
  =\; Z_{\rm c}[J, \et, \bar{\et}]\;
        -\;\iIM\!\int\zz dx
              \left( \tr J^\mu A_\mu + \bar{\et}\, \ps + \bar{\ps}\, \et \right)
    \\[1ex]
&\text{mit}\qquad
  A_\mu\;
    =\; \frac{1}{\iIM}\frac{\de Z_{\rm c}}{\de J^\mu}\qquad
  \ps\;
    =\; \frac{1}{\iIM}\frac{\de Z_{\rm c}}{\de \bar{\et}}\qquad
  \bar{\ps}\;
    =\; -\frac{1}{\iIM}\frac{\de Z_{\rm c}}{\de \et}
    \tag{\ref{1-irredErzFnl}$'$}
\end{align}
die konjugierten Felder definieren und bez"uglich der Quellen invertierbar sind.
      
Wir betrachten das Erzeugende Funktional~$Z$ nach Gl.~(\ref{ErzFnl}).
Die Materie\-felder~$\ps$,~$\bar{\ps}$~wie auch die Geistfelder~$c$,~$\bar{c}$ treten im Wirkung-Exponential nicht h"oher als quadratisch auf, vgl.\@ die Gln.~(\ref{Lagr_eff})-(\ref{Lagr_gfghost}); ihre Integrationen sind {\it Gau"s'sch\/} und k"onnen formal ausgef"uhrt werden:
%
\begin{align} \label{ErzFnl1}
&\hspace*{-16pt}
 \begin{aligned}[t]
  Z[J, \et, \bar{\et}]\;
  =\; \mf{N}^{-1}\; \int\! \mf{D}\!(A)\vv
        \Det\!\big(i{\cal O}_{\rm ghost}\big)\;
       &\Det\!\big(-\iIM\,{\cal O}_{\rm D}\big)\;
        \exp\, -\iIM \iint\zz dx\,dx'\;
        \bar{\et}\; {\cal O}_{\rm D}{}^{\zzz -1}\; \et
    \\ 
       &\times \exp\; \iIM \int\zz dx
         \left( \mf{L}_{\rm Y\!M} + \mf{L}_{\rm gf.} + \tr J^\mu A_\mu
         \right)
 \end{aligned}
    \\[1ex]
&\hspace*{-16pt}
 \text{mit}\qquad
  \begin{alignedat}[t]{2}
  &\big({\cal O}_{\rm ghost}\big)_{ab}(x,x')\;&
    &=\; \frac{\de{\cal F}_a[A]}{\de A_{\mu c}\!(x)}\;
           (\Del{x}_\mu)_{cb}\; \de(x-x')
    \\[.5ex]
  &\big({\cal O}_{\rm D}\big)_{nm}(x,x')\;&
    &=\; \big(\iIM\,\ga^\mu \Del{x}_\mu - m\big)_{nm}\, \de(x-x')\qquad
         {\cal O}_{\rm D}{}^{\zzz -1} = \big(\iIM\,\ga^\mu \Del{x}_\mu - m\big)^{-1}
  \end{alignedat}
    \tag{\ref{ErzFnl1}$'$}
\end{align}
Es sind~\mbox{\,${\cal O}_{\rm ghost}$} und~\mbox{\,${\cal O}_{\rm D}$},~\mbox{\,${\cal O}_{\rm D}{}^{\zzz -1}$} Funktionale im Eich-Vektorfeld~$A$; bei Bildung der {\it Faddeev-Popov-Geist-\/} und der {\it Fermion-Funktionaldeterminanten},~\mbox{$\Det\!(\iIM{\cal O}_{\rm ghost})$} bzw.~\mbox{$\Det\!(-\iIM{\cal O}_{\rm D})$}, sind sie aufzufassen als Matrizen mit (kontinuierlichen) Raumzeit-Indizes und Indizes bez"uglich der Eichgruppe in entsprechender Darstellung.
Die kovariante Ableitung versteht sich demgem"a"s als bez"uglich der {\it adjungierten Darstellung\/}~\mbox{\,$\mf{A}$}: \mbox{$(D_\mu)_{cb} \!=\! \pa_\mu \de_{cb} \!+\! gf_{cbd}A_{\mu d}$}, respektive der {\it fundamentalen Darstellung\/}~\mbox{\,$\mf{F}$}: \mbox{$(D_\mu)_{nm} \!=\! \pa_\mu \de_{nm} \!+\! \iIM\,g A_{\mu a}(T^a_\Drst{F})_{nm}$}; vgl.\@ Gl.~(\ref{kovAbl}).

Wir betrachten Greenfunktionen nach Gl.~(\ref{Greenfnen}) mit dem Erzeugenden Funktional~$Z$ in der Darstellung von Gl.~(\ref{ErzFnl1}).
Anwendung der~$2m$ Funktionalableitungen nach den Gra"smann-wertigen Quellen~$\bar{\et}(y_j)$,~$\et(\bar{y}_{\bar{j}})$ projiziert $m$-mal das Funktional~$i{\cal O}_{\rm D}{}^{\zzz -1}\![A]$ aus dem Exponenten herunter (bevor dieser gleich Null gesetzt wird) und f"uhrt auf die Summe dieser Monome "uber alle verschiedenen Paarungen der Raumzeit-Argumente (versehen mit dem Vorzeichen der entsprechenden Permutation~\mbox{\,$\si \!\in\! S_m$}, das hei"st der nat"urlichen Zahlen~$1,\ldots m$, und dividiert durch~$m!$, ihre Anzahl):
%
\begin{align} \label{Greenfnen1}
&G^{(n=l+2m)}
  \!(x_1,\cdots,x_l; y_1,\cdots,y_m; \bar{y}_1,\cdots,\bar{y}_m)
    \\[.5ex]
&=\; \vac{\; T\;
      {\T\prod}_{i=1}^l\, A_{\mu_i}\!(x_i)
      \cdot \frac{1}{m!}\, {\T\sum}_{\si\in S_m}\, {\rm sign}\!(\si)\;
       \left(\iIM\,{\cal O}_{\rm D}{}^{\zzz -1}\right)\!(y_1,\bar{y}_{\si\!(1)})\;\cdots
       \left(\iIM\,{\cal O}_{\rm D}{}^{\zzz -1}\right)\!(y_m,\bar{y}_{\si\!(m)})
      \;}
    \nn
\end{align}
Die Notation~$\vac{\;\cdot\;}$, bezogen auf ein beliebiges Funktional~${\cal G}$ im Eich-Vektorfeld~$A$, steht f"ur das Funktionalintegral
%
\begin{align} \label{vev}
&\vac{\,{\cal G}\,}\;
  =\; \int\! \HaarDmu\vv {\cal G}[A]
\end{align}
Es ist~\mbox{\,$\HaarDmu$} das Haarsche Eichfeldma"s, das definiert ist durch
%
\begin{align}
&\HaarDmu\;
  =\; \mf{D}(A)\vv \mu[A]
    \tag{\ref{vev}$'$} \\[1ex]
&\text{mit}\qquad
  \begin{aligned}[t]
  \mu[A]
  =\; \mf{N}^{-1}\vv
        \Det\!\big(\iIM\,{\cal O}_{\rm ghost}[A]\big)\vv
       &\Det\!\big(-\iIM\,{\cal O}_{\rm D}[A]\big)\;
    \\
       &\times
        \exp\;\iIM\!\int\zz dx
          \left( \mf{L}_{\rm Y\!M}[A] + \mf{L}_{\rm gf.}[A]
          \right)
  \end{aligned}
    \tag{\ref{vev}$''$}
\end{align}
Das implizite Auftreten des eichfixierenden Funktionals~${\cal F}$ suggeriert eine Abh"angigkeit von der Eichfixierung.
Tats"achlich ist dies nicht der Fall,~\mbox{\,$\HaarDmu$} ist eindeutig definiert und eichinvariant.
Die Eichinvarianz%
\FOOT{
  Cum grano salis nicht Eich- sondern BRST-Invarianz; bzgl.\@ unseres Sprachgebrauchs vgl.\@ Seite~\pageref{T:BRST-versusEichinvarianz}.
}
des Funktionalintegrals~$\vac{\;\cdot\;}$ eines eichinvarianten Funktionals in~$A$ "ubertr"agt sich unmittelbar von der urspr"unglichen Darstellung nach Gl.~(\ref{Greenfnen}).

Durch die Greenschen Funktionen~$G^{(n)}$ nach Gl.~(\ref{Greenfnen1}) l"a"st sich ein beliebiger {\it Erwartungswert bez"uglich des physikalischen Vakuums\/} formal darstellen als Integral~$\vac{\;\cdot\;}$ eines definierten Eichfeld-Funktionals.
Mit der Konsequenz der Eichinvarianz dieses Funktionals~-- wie auch der Lokalit"at aufgrund der Forderung von Kausalit"at (vgl.\@ die ausf"uhrliche Diskussion in Ref.~\cite{Lavelle97})~-- auch beliebige {\it Observable\/} und insbesondere {\it nichtperturbative Observable}~-- unter dem Caveat von Fu"snote~\FNg{FN:Gribov_ambiguities}.

Allgemein macht die Darstellung~$\vac{\,{\cal G}\,}$ des Erwartungswertes einer Gr"o"se bez"uglich des physikalischen Vakuums deutlich, vgl.\@ Gl.~(\ref{vev}), da"s dieser genau die Mittelung des entsprechenden Funktionals~${\cal G}[A]$ "uber alle Konfiguration des Eichfelds~$A$ ist, die jeweils gewichtet sind mit dem komplizierten Funktional~$\mu[A]$, vgl.\@ Gl.~(\ref{vev}). 

Wir wollen uns {\it anschaulich\/}~-- {\it formal\/} ist dies in eichinvarianter Weise nicht m"oglich~-- das Eichfeld aufgespalten denken in einen Anteil "`kleiner"' und einen "`gro"ser"' Impulse: einen Anteil {\it nieder-\/} beziehungsweise {\it hochfrequenter Gluonen\/}.
Nichtperturbative Observable sind dadurch definiert, da"s sie im wesentlichen bestimmt sind durch in diesem Sinne niederfrequente Eichfeld-Konfigurationen (f"ur genau die in der Darstellung~$\vac{\;\cdot\;}$ das \DREI{M}{S}{V} einen Ansatz macht); hochfrequente Konfigurationen bewirken {\it per definitionem\/} eine nur kleine Korrektur, die, wenn nicht ganz vernachl"assigt, berechnet wird im perturbativen Rahmen.

\section{Mathematische Hilfsmittel}

In diesem Abschnitt f"uhren wir mathematische Hilfsmittel ein, die essentiell in die Formulierung des \DREI{M}{S}{V} eingehen.

Dies sind zun"achst Konnektoren, mit deren Hilfe (im Sinne lokaler Eichkovarianz) paralleltransportierte Feldst"arken definiert werden, die wegen der Einfachheit ihres Verhaltens unter Eichtransformationen die geeigneten Gr"o"sen sind zur Formulierung manifest eichinvarianter Funktionale in~$A$. \\
Wir identifizieren Konnektoren bez"uglich geschlossener Kurven, sogenannte Wegner-Wilson-Loops, als die relevanten Eichfeld-Funktionale und geben mit dem nichtabelschen Stokes'schen Satz an, wie sie in Termen paralleltransportierter Feldst"arken geschrieben werden. \\
Wir geben an, wie der Vakuumerwartungswerte~$\vac{\;\cdot\;}$ eines Wegner-Wilson-Loops dargestellt wird als Entwicklung in Kumulanten.

\subsection{Manifest eichinvariante Funktionale durch Konnektoren}
\label{SubSect:ManifestEichinvariant}

Eichartefakte treten {\it a~priori\/} nicht auf, wenn eine Theorie in eichinvarianten Gr"o"sen formuliert ist.
Stehe~${\cal O}\![A]$ allgemein f"ur ein {\it eichinvariantes Funktional\/} in~$A$.
Zu seiner Konstruktion rufen wir uns {\it lokale Eichtransformationen\/} in Erinnerung, das hei"st Transformationen unter einer Matrix~$U(x)$, die in jedem Punkt~$x$ der Raumzeit Element der Eichgruppe~$\SUNc$ ist in geeigneter Darstellung~$\Drst{R}$, vgl.\@ die Gln.~(\ref{QCD:Eichtransf}),~(\ref{QCD:Eichtransf}$'$):
%
\begin{alignat}{5}
&\ps\;& &\to\;& &\ps^U\;&
  &:=\;& &U\, \ps
    \label{Eichtransf} \\
&A_\mu\;& &\to\;& &A^U_\mu\;&
  &:=\;& &\frac{1}{\iIM\,g}\, U\, (D_\mu\, U^{\D\dagger})\;
    =\; U\, A_\mu\, U^{\D\dagger}\;
         +\; \frac{1}{\iIM\,g}\, U\, (\pa_\mu\, U^{\D\dagger})
    \tag{\ref{Eichtransf}$'$}
    \\[-4.5ex]\nn
\intertext{\vspace*{-.5ex}Dabei transformiert die kovariante Ableitung {\it per definitionem\/} wie}
&D_\mu\;& &\to\;& &D^U_\mu\;&
  &=\;& &U\, D_\mu\, U^{\D\dagger}
    \label{Eichtransf_D}
    \\[-4.5ex]\nn
\intertext{\vspace*{-.5ex}vgl.\@ Gl.~(\ref{QCD:Eichtransf_D}).   Aus~$F_{\mu\nu}\!=\!(\iIM\,g)^{-1}[D_\mu,D_\nu]$, vgl.\@ Gl.~(\ref{Feldstrkn}), und der Unitarit"at der Transformation,~$U^{\D\dagger} U \!=\! \bbbOne{R}$, folgt damit f"ur den Feldst"arkentensor}
&F_{\mu\nu}\;& &\to\;& &D^U_{\mu\nu}\;&
  &=\;& &U\, F_{\mu\nu}\, U^{\D\dagger}
    \label{Eichtransf_F}
\end{alignat}
vgl.\@ Gl.~(\ref{QCD:Eichtransf_F}).
Dieses einfache Transformationsverhalten legt nahe, eichinvariante Funktionale~\mbox{\,${\cal O}\![A]$} in Termen von Feldst"arken~\mbox{\,$F$} statt von Feldern~\mbox{\,$A$} zu konstruieren.

Allerdings ist in den Gln.~(\ref{Eichtransf})-(\ref{Eichtransf_F}) die Transformationsmatrix am entsprechenden Weltpunkt zu nehmen.
F"ur nichtlokale Funktionale dargestellt durch Produkte von Feld\-st"arken an Weltpunkten~$x_1,x_2,\ldots$, die mithilfe der Matrizen~$U(x_1),U(x_2),\ldots$ transformieren, treten daher Produkten~$U^{\D\dagger}\!(x_1)U\!(x_2),\ldots$ auf, die im allgemeinen~$\!\neq\! \bbbOne{R}$ sind.
Wir definieren daher verallgemeinerte Feldst"arken mit geeigneterem Transformationsverhalten.

Hierzu f"uhren wir zun"achst den Eichgruppen-wertigen {\it Konnektor\/}~$\Ph$ ein, nichtabelsche Verallgemeinerung des Schwinger-Strings der QED:
%
\begin{align} \label{Konnektor}
\Ph(x_0,x; {\cal C})\;
  =\; P\; \exp -\iIM\,g \int_{\cal C} d{x'}^\mu\; A_\mu\!(x')
\end{align}
mit~$A_\mu\!=\!A_{\mu a}\, T^a_\Drst{R}$,~$\mf{R}$ eine beliebige Darstellung der~$\SUNc$, und~$P$ dem {\it Pfadordnungsoperator~$P \!\equiv\! P_{\cal C}$\/} bez"uglich der orientierten Kurve~${\cal C}$ von~$x$ nach~$x_0$:
%
\begin{align} 
{\cal C} \!\equiv\! {\cal C}_{x_{\!0}x}:\; t\in[0,1] \to x'(t)\qquad
  \text{mit}\qquad
  x'(0) = x\qquad
  x'(1) = x_0
\end{align}
das hei"st bez"uglich des Kurvenparameters~$t$.
Allgemein hat~$\Ph$ die Eigenschaften
\begin{align} \label{Konnektor_Eigenschaften}
&\Ph^{\D\dagger}(x_0,x; {\cal C}_{x_{\!0}x})
  = \Ph^{-1}(x_0,x; {\cal C}_{x_{\!0}x})
  = \Ph(x,x_0; {\cal C}_{x_{\!0}x}{}^{\zzz -1}) \\[.5ex]
&\Ph(x_0,x; {\cal C}_{x_{\!0}y}\circ{\cal C}_{yx})
  = \Ph(x_0,y; {\cal C}_{x_{\!0}y})\; \Ph(y,x; {\cal C}_{yx})
        \tag{\ref{Konnektor_Eigenschaften}$'$}
\end{align}
und ist L"osung der Differentialgleichung im Kurvenparameter~$t$ mit Randbedingung:
\bea \label{Konnektor_DGL}
\frac{d}{dt}\; \Ph(t) = -\iIM\,g\; \frac{d{x'}^\mu(t)}{dt}\; A_\mu\!(x'(t))\; \Ph(t) \qquad
\Ph(0) = \bbbOne{R}
\eea
Die Bedeutung des Konnektors~$\Ph$ folgt aus seinem Transformationsverhalten
\bea \label{Eichtransf_Ph}
\Ph(x_0,x;{\cal C}) \to \Ph^U\!(x_0,x;{\cal C}) := U\!(x_0)\; \Ph(x_0,x; {\cal C})\; U^{\D\dagger}\!(x)
\eea
das in Kontrast steht zu Gl.~(\ref{Eichtransf_D}) und~(\ref{Eichtransf_F}).

$\Ph$ ist aufzufassen als {\it Paralleltransporter\/}: Mithilfe von $\Ph(x_0,x; {\cal C})$ l"a"st sich der {\it Colour\/}-Gehalt einer Gr"o"se {\it in eichkovarianter Weise\/} entlang einer Kurve~${\cal C}$ von~$x$ nach~$x_0$ transportieren.
Beziehen wir in diesem Sinne~$\Ph(x_0,x; {\cal C})$ auf den Feldst"arkentensor, das hei"st definieren wir den {\it paralleltransportierten Feldst"arkentensor\/} durch
\bea \label{Fparallel}
F_{\mu\nu}\!(x; x_0,{\cal C})
  = \Ph(x_0,x; {\cal C})\; F_{\mu\nu}\!(x)\; \Ph^{\D\dagger}(x_0,x; {\cal C})
\eea
so transformiert dieser wie:
\bea \label{Eichtransf_Fparallel}
F_{\mu\nu}\!(x; x_0,{\cal C}) \to F_{\mu\nu}^U\!(x; x_0,{\cal C})
  := U\!(x_0)\; F_{\mu\nu}\!(x; x_0,{\cal C})\; U^{\D\dagger}\!(x_0)
\eea
vgl.\@ Gl.~(\ref{Eichtransf_F}) und~(\ref{Eichtransf_Ph}),~-- das hei"st wie eine bez"uglich ihres Colour-Gehalts in~$x_0$ lokale Gr"o"se.
Der paralleltransportierte Feldst"arkentensor~\mbox{\,$F_{\mu\nu}\!(x; x_0,{\cal C})$} ist der konventionelle Feldst"arkentensor~\mbox{\,$F_{\mu\nu}\!(x)$} modifiziert in dem Sinne, da"s dessen {\it Colour-Gehalt entlang der Kurve~${\cal C}$ paralleltransportiert ist an den Punkt~$x_0$\/}~-- der bezeichnet wird als {\it Referenzpunkt\/}.

In Termen paralleltransportierter Feldst"arkentensoren~$F_{\mu\nu}\!(x; x_0,{\cal C}_{x_{\!0}x})$ definieren wir die ~$n$-lokalen Eichfeld-Funktionale
\bea 
{\cal O}_n[A]
  = g^n F_{\mu_1\nu_1}\!(x_1; x_0,{\cal C}_{x_{\!0}x_{\!1}})\;
        F_{\mu_2\nu_2}\!(x_2; x_0,{\cal C}_{x_{\!0}x_{\!2}})\;
        \cdots\; F_{\mu_n\nu_n}\!(x_n; x_0,{\cal C}_{x_{\!0}x_{\!n}})
\eea
Dabei sei der Referenzpunkt~$x_0$ beliebig, aber fest gew"ahlt, ebenso die Kurven~${\cal C}_{x_{\!0}x_{\!i}}$ von Punkten~$x_i$ nach~$x_0$.
Aus Gl.~(\ref{Eichtransf_Fparallel}) und wegen der Eichinvarianz des physikalischen Vakuums~$\mf{D}\!\mu(A^U) \!=\! \HaarDmu$, oder in Operatornotation~$U\ket{\Om} \!=\! \ket{\Om}$, folgt die Eichinvarianz von
%
\begin{align} \label{vev_Fparallel}
&\vac{\;{\cal O}_n\;}
    \\
&=\; \vac{\; g^n F_{\mu_1\nu_1}\!(x_1; x_0,{\cal C}_{x_{\!0}x_{\!1}})\;
              F_{\mu_2\nu_2}\!(x_2; x_0,{\cal C}_{x_{\!0}x_{\!2}})\;
              \cdots\; F_{\mu_n\nu_n}\!(x_n; x_0,{\cal C}_{x_{\!0}x_{\!n}})
      \;} \nn
\end{align}
und daher Proportionalit"at zur Eins
%
\begin{align} \label{vev_Fparallel1}
\vac{\;{\cal O}_n\;}\;
  =\; \trDrst{R} \vac{\;{\cal O}_n\;}\cdot \bbbOne{R}
\end{align}
Die Spur bezieht sich auf eine beliebige Darstellung~$\mf{R}$ der Eichgruppe~$\SUNc$
und ist definiert als normiert auf die Dimension~$\dimDrst{R}$ der Darstellung:
%
\begin{align} \label{trDrst}
&\trDrst{R}\; =\; \frac{1}{\dimDrst{R}}\; \tr\qquad
  \Longrightarrow\qquad
  \trDrst{R}\, \bbbOne{R}\; \equiv\; 1\quad
    \forall \mf{R}
    \\[.5ex]
&\text{mit}\qquad
  \dimDrst{F}\; \equiv\; N_{\rm\!c},\quad
  \dimDrst{A}\; \equiv\; N_{\rm\!c}^2\!-\!1
    \tag{\ref{trDrst}$'$}
\end{align}
in der fundamentalen beziehungsweise adjungierten Darstellung und mit~$\bbbOne{R}$ der entsprechenden Eins (d.h.\@ der~$\dimDrst{R}$-dimensionalen Einheitsmatrix); vgl.\@ Seite~\pageref{T:SUNcAlgebra}.
Die Spur~$\tr$ und damit~$\trDrst{R}$ vertauscht mit der Bildung des Vakuumerwartungswertes~$\vac{\;\cdot\;}$ wegen dessen Linearit"at.

Um {\it manifeste Eichinvarianz\/} unserer Formulierung zu garantieren, werden wir statt mit Greenfunktionen in Termen von Eichfeldern~$A(x)$ nach Gl.~(\ref{Greenfnen1}) in der hierzu "aquivalenten Darstellung nach Gl.~(\ref{vev_Fparallel}) arbeiten: mit Vakuumerwartungswerten~$\vac{\;{\cal O}_n\;}$ in Termen~ent\-lang definierter Kurven~${\cal C}_{x_{\!0}x_{\!i}}$ zu einem festen (aber beliebigen) Referenzpunkt~$x_0$ paralleltransportierter Feldst"arken~$F_{\mu\nu}\!(x_i; x_0,{\cal C}_{x_{\!0}x_{\!i}})$.

\bigskip\noindent
Wir schlie"sen diesen Abschnitt mit einem Exkurs.
Zur Illustration geben wir die {\it Koordinateneichung bez"uglich des Referenzpunktes~$x_0$\/} an:
%
\begin{align} \label{KoordEichg}
(x \!-\! x_0)^\mu\; A_\mu\!(x)\;
  =\; 0\qquad
  A_\mu\!(x_0)\;
  =\; 0
\end{align}
sie ist offensichtlich {\it nicht kovariant\/}.

In ihr wird f"ur Paralleltransport nach~$x_0$ entlang {\it geradliniger\/} Kurven~${\cal C}_{x_{\!0}x}$ unsere Formulierung besonders einfach:
Die entsprechenden Konnektoren~$\Ph(x,x_0; {\cal C}_{x_{\!0}x})$ sich auf die Eins reduzieren und daher die mit ihrer Hilfe paralleltransportierten Feldst"arken mit den konventionellen identisch.
Wir zeigen~$F_{\mu\nu}\!(x; x_0,{\cal C}_{x_{\!0}x}) \!=\! F_{\mu\nu}\!(x)$.

Wir betrachten zun"achst die totale Ableitung von~$\;t^2 F_{\mu\nu}\!(y)$, mit~$y(t) \!=\! x_0 \!+\! t(x \!-\!x_0)$, nach dem Parameter~$t$.
Sie schreibt sich mithilfe der Eichbedingung, vgl.\@ Gl.~(\ref{KoordEichg}):
%
\begin{align} 
\frac{d}{dt} \Big[ t^2\, F_{\mu\nu}\!(y) \Big]\;
 =\; t\; \Big[
     2\, F_{\mu\nu}\!(y) + (x \!-\! x_0)^\rh\; \Del{x}_\rh\, F_{\mu\nu}\!(y)
   \Big]
\end{align}
und mithilfe der Jacobi-Identit"at f"ur die kovariante Ableitung~$D_\rh F_{\mu\nu} \!+\! D_\mu F_{\nu\rh} \!+\! D_\nu F_{\rh\mu} \!=\!0$, vgl.\@ Seite~\pageref{T:JacobiBianchi}
\vspace*{-.5ex}
\begin{align} 
\frac{d}{dt} \Big[ t^2\, F_{\mu\nu}\!(y) \Big]\;
  &=\; t\; \Big[
      F_{\mu\nu}\!(y) + (x \!-\! x_0)^\rh\; \Del{x}_{\mu}\; F_{\rh\nu}\!(y)
    \Big] \;-\; \{\mu\!\leftrightarrow\!\nu\}
  \\
  &=\; \Del{x}_{\mu} \Big[
         t\; (x \!-\! x_0)^\rh\; F_{\rh\nu}\!(y)
         \Big] \;-\; \{\mu\!\leftrightarrow\!\nu\}
    \nn
    \\[-4.5ex]\nn
\end{align}
Wir haben damit die Darstellung des Feldst"arkentensors als Integral
\vspace*{-.5ex}
\begin{align} \label{KoordEichg_F}
F_{\mu\nu}\!(x)\;
  &=\; \int_0^1\! dt\;
      \frac{d}{dt}\; \Big[ t^2\; F_{\mu\nu}\!(x_0 \!+\! t(x \!-\!x_0)) \Big]
    \\
  &=\; \Del{x}_{\mu}\; \Big[
      \int_0^1\! dt\; t\; (x \!-\! x_0)^\rh\; F_{\rh\nu}\!(x_0 \!+\! t(x \!-\!x_0))
    \Big] \;-\; \{\mu\!\leftrightarrow\!\nu\}
    \nn
    \\[-4.5ex]\nn
\end{align}
Andererseits gilt~$F_{\mu\nu} \!=\! (\iIM\,g)^{-1}[D_\mu,D_\nu] \!=\! \pa_\mu A_{\nu} \!-\! \pa_\nu A_{\mu} \!+\! \iIM\,g[A_{\mu},A_{\nu}]$, vgl.\@ Gl.~(\ref{Feldstrkn}), das hei"st $F_{\mu\nu}\!(x) \!=\! \Del{x}_\mu A_\nu\!(x) - \{\mu\!\leftrightarrow\!\nu\}$.
Gegen"uberstellung mit Gl.~(\ref{KoordEichg_F}) ergibt
\begin{samepage}
%
\begin{align} \label{KoordEichg_A}
A_\mu\!(x)\;
  =\; \int_0^1\! dt\; t\; (x \!-\! x_0)^\rh\; F_{\rh\mu}\!(x_0 \!+\! t(x \!-\!x_0))
\end{align}
als (nichtlokale) Darstellung von~$A$ durch~$F$.

F"ur das Linienintegral im Exponenten des Konnektors~\mbox{\,$\Ph(x_0,x; {\cal C})$}, vgl.\@ Gl.~(\ref{Konnektor}), gilt entlang der {\it geradlinigen\/} Verbindung~${\cal C}$ von einem Punkt~$x$ nach~$x_0$:
\vspace*{-.25ex}
\begin{align} \label{KoordEichg_Ph}
\int_{{\cal C}} d{x'}^\mu\; A_\mu\!(x')\;
  &=\; - \int_0^1\; dt'\! (x \!-\! x_0)^\mu\; A_\mu\!(x'(t'))
    \\
  &=\; - \int_0^1\! dt\; \int_0^1\! dt'\; t\,t'\; (x \!-\! x_0)^\rh\; (x \!-\! x_0)^\mu\;
      F_{\rh\mu}\!(x_0 \!+\! t\,t' (x \!-\!x_0))
    \tag{\ref{KoordEichg_Ph}$'$} \\
  &=\; 0
    \tag{\ref{KoordEichg_Ph}$''$}
    \\[-4.25ex]\nn
\end{align}
\end{samepage}%
mit der zweiten Identit"at wegen Gl.~(\ref{KoordEichg_A}) und der dritten wegen der Antisymmetrie des~Feld\-st"arkentensors bez"uglich der Lorentz-Indizes.

Entlang gerader Verbindungen zum Referenzpunkt~$x_0$ ist der Konnektor nach Gl.~(\ref{Konnektor}) in Koordinateneichung bez"uglich~$x_0$ die Exponentialfunktion an der Stelle Null, das hei"st die Eins.
Feldst"arken und paralleltransportierte Feldst"arken sind identisch:
%
\begin{align} 
\Ph(x,x_0; {\cal C}_{x_{\!0}x})\;
  \equiv\; \bbbOne{R}\qquad
  \text{so da"s}\qquad
  F_{\mu\nu}\!(x; x_0,{\cal C}_{x_{\!0}x})\;
  =\; F_{\mu\nu}\!(x)
\end{align}
dabei ist $\mf{R}$ die Darstellung von~$\Ph$, vgl.\@ Gl.~(\ref{Konnektor}).

Wir betonen, da"s wir uns weder auf diese Eichung noch auf Geradlinigkeit der Kurven~${\cal C}_{x_{\!0}x_{\!i}}$ beschr"anken.
Wir arbeiten in Termen allgemeiner paralleltransportierter Feldst"ar\-ken~$F_{\mu\nu}\!(x_i; x_0,{\cal C}_{x_{\!0}x_{\!i}})$.
Unsere Formulierung ist {\it manifest eich- und Poincar\'e-kovariant\/}.
\vspace*{-.5ex}

\subsection{Nichtabelscher Stokes'scher Satz}
\label{SubSect:Stokes}

Wir betrachten den Konnektor~$\Ph$ von~$x$ zur"uck nach~$x$ entlang~${\cal C} \!\equiv\! {\cal C}_{xx}$, einer beliebigen geschlossenen Kurve~\mbox{\,${\cal C} \!\equiv\! {\cal C}_{xx}$}
\begin{samepage}
\vspace*{-.5ex}
\begin{align} \label{Konnektor_LoopA}
&\Ph(x,x; {\cal C})\;
  =\; P\; \exp -\iIM\,g \int_{\cal C} d{x'}^\mu\; A_\mu\!(x')
    \\[.5ex]
&\text{mit}\qquad
  {\cal C} \!\equiv\! {\cal C}_{xx}:\; t\in[0,1] \to x'(t)\qquad
  x'(0) = x'(1) = x \nn
    \\[-4.5ex]\nn
\end{align}
Das Eich-Vektorfeld~$A$, vermittels~$A_\mu \!=\! A_{\mu a} T^a_\Drst{R}$, und damit~$\Ph$ sind aufzufassen als bez"uglich einer beliebigen Darstellung~$\mf{R}$ der Eichgruppe.

Die Spur von~$\Ph$ ist {\it eichinvariant\/} und wird bezeichnet als {\it Wegner-Wilson-Loop\/}
%
\begin{align} \label{WW-Loop}
W({\cal C})\;
  =\; \trDrst{R}\, \Ph(x,x; {\cal C})
\end{align}
Zur Definition der Spur~$\trDrst{R}$ vgl.\@ die Gln.~(\ref{trDrst}),~(\ref{trDrst}$'$).
Vakuumerwartungswerte~$\vac{\;\cdot\;}$ von Produkten von Wegner-Wilson-Loops sind von herausragender Bedeutung als Ordnungsparameter von Eichtheorien auf Skalen, die perturbativ nicht zug"anglich sind; vgl.\@ die Refn.~\mbox{\cite{Wegner71,Wilson74}}.~--
Wir werden sehen, inwiefern auch in der vorliegenden Arbeit.

Zur Ankn"upfung an den vorangehenden Abschnitt~\ref{SubSect:ManifestEichinvariant} ist in Gl.~(\ref{Konnektor_LoopA}) "uberzugehen von Eichfeldern~$A$ zu Eichfeldst"arken~$F$.

Der {\it Stokes'sche Satz\/} auf einer (pseudo-)Riemannschen Mannigfaltigkeit~$G$ ist f"ur das Integral der "au"seren Ableitung einer beliebigen {\it abelschen\/} $(s\!-\!1)$-Form~$\om$ "uber eine \mbox{$s$-dimen}\-sionale Untermannigfaltigkeit~$G_s \!\subset\! G$ genau folgende Identit"at:
\vspace*{-.25ex}
\begin{align} \label{Stokes_Diff'formen}
\int_{G_s}\; d\wedge\om\;
  =\; \int_{\pa G_s}\; \om
    \\[-4.25ex]\nn
\end{align}
vgl.\@ Gl.~(\ref{APP:Stokes_Diff'formen}) in Anhang~\ref{APP-Sect:Integration} und Ref.~\cite{Forster96}.

Seien~\mbox{\,$\om \!\equiv\! A \!=\! A_\mu dx^\mu$} und~\mbox{\,$F \!=\! \frac{1}{2} F_{\mu\nu} dx^\mu\wedge dx^\nu$} respektive Eichfeld und Eichfeldst"arke einer {\it abelschen\/} Theorie, etwa der QED.
Dann gilt~\mbox{\,$d\wedge\om \!\equiv\! d\wedge A \!=\! F$}.
Sei ferner \mbox{$G_s \!=\! {\cal S} \!\equiv\! {\cal S}({\cal C})$} eine beliebige Fl"ache mit Rand~\mbox{\,$\cal C$}, das hei"st~\mbox{\,$\pa G_s \!\equiv\! \pa{\cal S} \!\equiv\! {\cal C}$}.
Dann lautet Gl.~(\ref{Stokes_Diff'formen}):
\vspace*{-.25ex}
\begin{align} \label{StokesF-A}
&\int_{{\cal S}({\cal C})}\; F\;
  =\; \int_{\cal C}\; A
    \\[.5ex]
  &\Longleftrightarrow\qquad
  \frac{1}{2}\; \int_{{\cal S}({\cal C})} d\si^{\mu\nu}\!(x')\; F_{\mu\nu}\!(x')\;
    =\; \int_{\cal C} d{x'}^\mu\; A_\mu\!(x')
    \tag{\ref{StokesF-A}$'$}
    \\[-4.25ex]\nn
\end{align}
mit~\mbox{\,$d\si^{\mu\nu}\!(x') \!\equiv\! dV^{\mu\nu}\!(x') \!=\! d{x'}^\mu \wedge d{x'}^\nu$}, vgl.\@ die Gln.~(\ref{APP:dV_Gs_Def}),~(\ref{APP:Stokes}) und Ref.~\cite{Stephani91}.
Diese Identit"at ist zu verallgemeinern auf das {\it nichtabelsche\/} Eichfeld/die Eichfeldst"arke der QCD.
\end{samepage}

\begin{figure}
\begin{minipage}{\linewidth}
  \begin{center}
  \setlength{\unitlength}{1mm}\begin{picture}(120,61.3)   
    \put(0,0){\epsfxsize120mm \epsffile{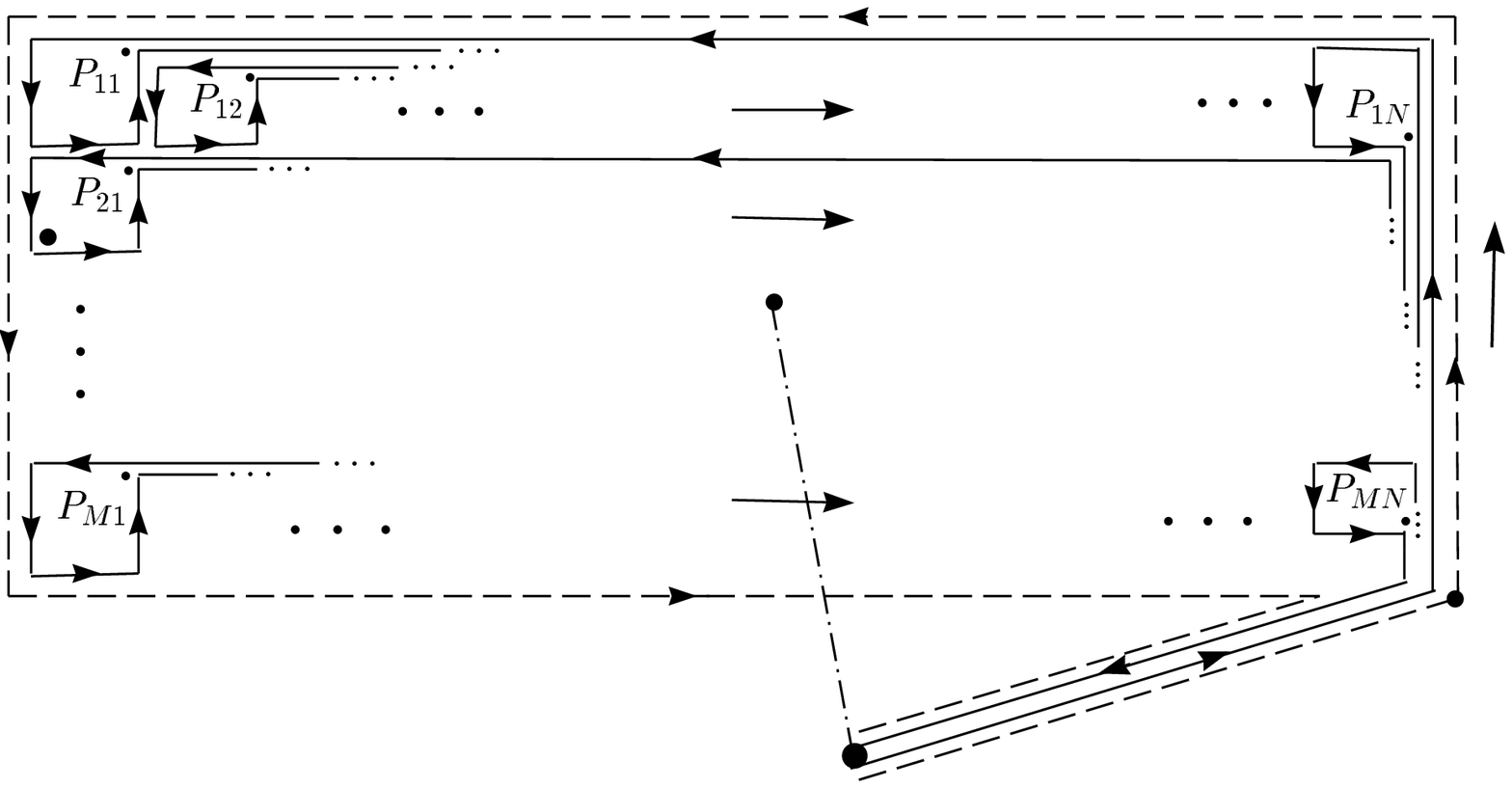}}
    \put( 98,62){\normalsize ${\cal C}$}
    \put(118,11){\normalsize $A$}
    \put( 62, 1){\normalsize $x_0$}
    \put( 57,37){\normalsize $x_0'$}
  \end{picture}
  \end{center}
\vspace*{-4ex}
\caption[Deformation der Kurve~\protect${\cal C}$ des Wegner-Wilson-Loops zu der Kurve~\protect$\tilde{\cal C}$, die infinitesimal plakettiert die Fl"ache~\protect${\cal S} \!\equiv\! {\cal S}(\tilde{\cal C})$ mit~\protect$\pa{\cal S} \!\equiv\! {\cal C}$]{
  Die Kurve~${\cal C}$ (gestrichelt) wird aufgeschnitten in~$A$ und durch Einschub von Einsen deformiert zu der Kurve~$\tilde{\cal C}$.   Diese f"achert pfadgeordnet mit infinitesimalen~Plaket\-ten~$P_{I\!J}$ eine (hier:\@ plane) Fl"ache~${\cal S} \!\equiv\! {\cal S}(\tilde{\cal C})$,~$\pa{\cal S} \!\equiv\! {\cal C}$, auf.   Im Grenzfall unendlich feiner~Pla\-kettierung treten auf:\@ an den Punkten~$x' \!\equiv\! x_{11}, x^\dbprime \!\equiv\! x_{12},\ldots \!\in\! {\cal S}$ die zu einem Referenspunkt $x_0 \!\in\! {\cal S}$ entlang Kurven~${\cal C}_{x_{\!0}\ixp}, {\cal C}_{x_{\!0}\ixpp},\ldots$ paralleltransportierte Feldst"arken.   Die Pfeile nach rechts illustrieren die Wirkung des Pfadordnungsoperators~$P_{\tilde{\cal C}}$ f"ur diese spezielle Auff"acherung.   Der Colour-Gehalt der Feldst"arken kann o.E.d.A.\@ weiter paralleltransportiert werden von~$x_0$ an einen beliebigen anderen Referenz-Punkt~$x_0'$.   Wir danken Edgar Berger f"ur die "Uberlassung dieser Abbildung; sie illustriere das im Text diskutierte Verfahren.
\vspace*{-0ex}
}
\label{Fig:Deformation}
\end{minipage}
\end{figure}

Der {\it Nichtabelsche Stokes'sche Satz\/} f"ur den Konnektor~$\Ph(x,x; {\cal C})$, vgl.\@ Gl.~(\ref{Konnektor_LoopA}), wird allgemein hergeleitet, indem die Kurve~$\cal C$ durch Einschub von Einsen~$\bbbOne{R} \!=\! \Ph^{\D\dagger}\Ph$ deformiert wird, vgl.\@ die Refn.~\cite{Arefeva80,Bralic80,Fishbane81,Diosi83,Simonov89,Shevchenko98}, besonders Ref.~\cite{Nachtmann96}.
Er lautet:
\vspace*{-.25ex}
\begin{align} \label{Konnektor_LoopF}
&\Ph(x,x; {\cal C})\;
  =\; P_{\cal C}\; \exp -\iIM\,g \int_{\cal C} d{x'}^\mu\; A_\mu\!(x')
    \\ 
&=\; \Ph(x,x_0; {\cal C}_{x_{\!0}x}{}^{\zzz -1})\vv
       \bigg( P_{\tilde{\cal C}}\; \exp -\frac{\iIM\,g}{2}
               \int_{{\cal S}(\tilde{\cal C})} d\si^{\mu\nu}\!(x')\;
               F_{\mu\nu}\!(x'; x_0,{\cal C}_{x_{\!0}\!\ixp})
       \bigg)\vv
     \Ph(x_0,x; {\cal C}_{x_{\!0}x})
    \nn
    \\[-4.5ex]\nn
\end{align}
Spurbildung~\mbox{\,$\trDrst{R}$} bez"uglich der Darstellung~\mbox{\,$\mf{R}$} von~\mbox{\,$\Ph$} ergibt, vgl.\@ die Gln.~(\ref{trDrst}),~(\ref{trDrst}$'$):
\vspace*{-.25ex}
\begin{align} \label{Konnektor_LoopFtr}
&W({\cal C})\;
  =\; \trDrst{R}\; \Ph(x,x; {\cal C})\;
  =\; \trDrst{R}\; P_{\cal C}\; \exp -\iIM\,g \int_{\cal C} d{x'}^\mu\; A_\mu\!(x')
    \\
&=\; \trDrst{R}\; P_{\tilde{\cal C}}\; \exp -\frac{\iIM\,g}{2} \int_{{\cal S}(\tilde{\cal C})}
       d\si^{\mu\nu}\!(x')\; F_{\mu\nu}\!(x'; x_0,{\cal C}_{x_{\!0}\!\ixp})
    \nn
    \\[-4.5ex]\nn
\end{align}
Diese Gleichungen, insbesondere das pfadgeordnete Exponential des Fl"achenintegrals, bed"ur\-fen gegen"uber dem konventionellen Stokes'schen Satz folgender Anmerkungen, vgl.\@ Abb.~\ref{Fig:Deformation}.

Es tritt ein {\it Referenzpunkt\/}~$x_0$ auf, der als Element der Integrationsfl"ache~${\cal S} \!\equiv\! {\cal S}(\tilde{\cal C})$ beliebig gew"ahlt werden kann, und {\it an~$x_0$ paralleltransportierte Feldst"arken\/}~$F_{\mu\nu}\!(x'; x_0,{\cal C}_{\ixp\!x_{\!0}})$. \\
Der~Paralleltransport geschieht entlang einer Kurve~$\tilde{\cal C}$, die die Fl"ache~$\cal S$ aufspannt ("`plakettiert"') und aus der Kurve~$\cal C$ formal hervorgeht, indem diese durch Einschub von Einsen $\bbbOne{R} \!=\! \Ph^{-1}\Ph \!=\! \Ph(x',x_0; {\cal C}_{x_{\!0}\ixp}{}^{\zzz -1}) \Ph(x_0,x'; {\cal C}_{x_{\!0}\ixp}) \!=\! \Ph(x^\dbprime,x_0; {\cal C}_{x_{\!0}\ixpp}{}^{\zzz -1}) \Ph(x_0,x^\dbprime; {\cal C}_{x_{\!0}\ixpp}) \!=\ldots$ deformiert wird
nach dem Schema: von~$x_0$ nach~$x'$~-- infinitesimale Plakette um~$x'$~-- von~$x'$ zur"uck nach~$x_0$~-- von~$x_0$ nach~$x^\dbprime$ (infinitesimal entfernt von~$x'$)~-- infinitesimale Plakette um~$x^\dbprime$~-- von~$x^\dbprime$ zur"uck nach~$x_0$~--~$\ldots$
{\it In conclusio\/}:
Erst die Wahl der Kurven~${\cal C}_{x_{\!0}\ixp},\; {\cal C}_{x_{\!0}\ixpp},\ldots$ zum Referenzpunkt~$x_0$ legt die Deformation~$\tilde{\cal C}$ von~$\cal C$ vollst"andig fest:
das hei"st zum einen die Pfadordnung entlang~$\tilde{\cal C}$ (angedeutet durch den Operator~$P_{\tilde{\cal C}}$\/), die konsistent ist mit der Pfadordnung des Kurvenintegrals entlang~$\cal C$,
und zum anderen die von~$\tilde{\cal C}$ aufgespannte Fl"ache~${\cal S} \!\equiv\! {\cal S}(\tilde{\cal C})$, von der nur zu fordern ist, da"s ihr Rand mit~$\cal C$ zusammenf"allt:~$\pa{\cal S} \!=\! {\cal C}$.
Umgekehrt hat man Freiheit in der Wahl der Pfadordnung auf der Fl"ache~${\cal S}$ wie auch von~${\cal S}$ selbst, so lange~${\cal S} \!\equiv\! {\cal S}(\tilde{\cal C})$ und~$\tilde{\cal C}$ eine Deformation von~${\cal C}$ ist im Sinne des Einschubs von Einsen.

Im abelschen Fall vereinfachen sich die Gln.~(\ref{Konnektor_LoopF}),~(\ref{Konnektor_LoopFtr}) zum konventionellen Stokes'\-schen Satz.

\subsection{Entwicklung in Kumulanten}

In diesem Abschnitt f"uhren wir ein das Konzept der Entwicklung in Kumulanten.
Wichtige fr"uhe Reviews, auf die wir uns wesentlich beziehen, sind die Arbeiten van Kampens, vgl.\@ die Refn.~\cite{VanKampen74,VanKampen76}.
Diese formulieren gleicherma"sen mathematisch streng und anschaulich anhand einer Reihe von Beispielen den Begriff der Kumulante zur L"osung stochastischer Differentialgleichungen.
Kumulanten wurde auf dieser Basis zuerst bezogen auf das Eichfeldma"s nichtperturbativer Quantenchromodynamik~-- in Zusammenhang mit der Beschreibung von Confinement~-- zun"achst von Dosch in Ref.~\cite{Dosch87}, dann von Dosch, Simonov in Ref.~\cite{Dosch88}.
Der Bezug auf Hochenergiestreuung ist ausf"uhrlich dargestellt von Kr"amer in Ref.~\cite{Kraemer91}; er impliziert insbesondere die Unterscheidung van Kampenscher und Waldenfels'scher Kumulanten wie die Frage der Faktorisierung auf Matrix- versus Matrixelementeniveau. \\
\indent
Unsere Darstellung steht vor dem Hintergrund dieser Arbeiten.
Sie h"alt sich formal an die systematische Diskussion Nachtmanns in Ref.~\cite{Nachtmann96}, geht aber "uber diese hinaus.
\vspace*{-.5ex}

\bigskip\noindent
Wir betrachten die Mittelung eines $t$-geordneten Exponentials, beispielsweise den Vakuumerwartungswert
\vspace*{-.5ex}
\begin{align} \label{wla}
w(\la)\;
  =\; \vac{ P\; \exp\; \la \int_0^1 dt\; B(t)\; }
\end{align}
Dabei sei~$B(t) \!=\! B_a(t)T^a_\Drst{R}$ zun"achst ein beliebiges Liealgebra-wertiges Funktional in~$A$ in einer beliebigen Darstellung~$\mf{R}$ der Eichgruppe; der (Pfad)Ordnungsoperator~$P \!\equiv\! P_t$ beziehe sich auf den Parameter~$t$.
Sei~$\la \!\in\! \bbbc$ beliebig.

Wir werden sp"ater~$w(\la)$ in Verbindung bringen mit dem Vakuumerwartungswert des geschlossenen Konnektors~$\Ph(x,x; {\cal C})$ aus Gl.~(\ref{Konnektor_LoopF}), das hei"st~$B$ mit der paralleltransportierten Feldst"arke~$F$ und~$t \!\in\! [0,1]$ identifizieren mit dem Kurvenparameter entlang~$\tilde{\cal C}$, einer definierten Deformation von~$\cal C$ im Sinne des vorangegangenen Unterabschnitts.

Der Erwartungswert~$w(\la)$ nach Gl.~(\ref{wla}) ist formal definiert durch die Reihenentwicklung der Exponentialfunktion:
%
\begin{align} \label{wla_Entw}
&w(\la)\;
  =\; 1 + {\T\sum}_{n=1}^{\infty}\; \frac{\la^n}{n!}\; b_n
    \\[.5ex]
&\text{mit}\qquad
  b_n\;
  =\; \int_0^1 dt_1 \cdots \int_0^1 dt_n\; B_n(t_1,\ldots t_n)
    \tag{\ref{wla_Entw}$'$} \\
&\hspace*{3.5em}
  B_n(t_1,\ldots t_n)\;
  \equiv\; \vac{ P\; B(t_1) \cdots B(t_n) }
    \tag{\ref{wla_Entw}$''$}
\end{align}
Die Funktion~$B_n$ ist wegen des Pfadordnungsoperators~$P$ {\it vollst"andig symmetrisch\/} in ihren~$n$ Argumenten.

Geschieht die Erwartungswertbildung zu~$w(\la)$ bez"uglich einer {\it stochastischen Variablen\/}~-- konkret: kann bei Bildung von~$\vac{\;\cdot\;}$ die Verteilung von Eichfeldkonfigurationen~$A(x)$ durch das Haarsche Ma"s~$\HaarDmu$ nach Gl.~(\ref{vev}) aufgefa"st werden als {\it stochastisch\/}~--, so existiert f"ur~$w(\la)$ eine {\it konvergente Entwicklung in Kumulanten\/}, das hei"st die Identit"at
\vspace*{-.5ex}
\begin{align} \label{lnwla_KumEntw}
&\ln w(\la)\;
  =\; {\T\sum}_{n=1}^\infty\; \frac{\la^n}{n!}\; k_n
    \\
&\text{mit}\qquad
  k_n\;
  =\; \int_0^1 dt_1 \cdots \int_0^1 dt_n\; K_n(t_1,\ldots,t_n)
    \tag{\ref{lnwla_KumEntw}$'$}
    \\[-4.5ex]\nn
\end{align}
Die Funktion~$K_n$ wird bezeichnet als {\it Kumulante $n$-ter Ordnung}; sie kann definiert werden als {\it vollst"andig symmetrisch\/} in ihren Argumenten, s.u.\@ die Diskussion von Gl.~(\ref{Kn_Monome}) und~(\ref{Kn}). 

Durch Gleichsetzen der Entwicklungen von Gl.~(\ref{wla_Entw}) und~(\ref{lnwla_KumEntw}) in der Form
\vspace*{-.5ex}
\begin{align} \label{bn_Gl}
&w(\la)\;
  =\; 1 + {\T\sum}_{n=1}^\infty\; \frac{\la^n}{n!}\; b_n
    \\
&\stackrel{\D!}{=}\; \exp\; [\ln w(\la)]\;
  =\; {\T\sum}_{m=0}^\infty\; \frac{1}{m!}\;
      \left[ {\T\sum}_{n=1}^\infty\; \frac{\la^n}{n!}\; k_n \right]^m\qquad
  \forall\la\in\bbbc
    \nn
    \\[-4.5ex]\nn
\end{align}
folgt:
\vspace*{-.5ex}
\begin{align} \label{bn}
b_n\;
  =\; {\T\sum}_{m=1}^n\; \frac{1}{m!}\;
        {\T\sum}_{ \T{}^{\vv i_1,\ldots i_m \in \bbbn}_{i_1+\ldots+i_m=n} }\;
        \frac{n!}{i_1!\cdots i_m!}\; \permI{ k_{i_1}\cdots k_{i_m} }
    \\[-4.5ex]\nn
\end{align}
Es treten s"amtliche {\it verschiedene Partitionen\/} des Index~$n$ in~$m$ Summanden~$i_1,\ldots i_m \!\in\! \bbbn$ auf.
Wir haben dar"uberhinaus die Notation~$\permI{\;\cdot\;}$ eingef"uhrt, mit
\vspace*{-.5ex}
\begin{align} \label{permI}
\permI{ k_{i_1}\cdots k_{i_m} }\;
  \equiv\; k_{i_1}\cdots k_{i_m}
             \;+\; \text{versch.\@ Reihenfolgen der $m$ Faktoren}
    \\[-4.5ex]\nn
\end{align}
die Summe s"amtlicher {\it verschiedenen Reihenfolgen der~$m$ Faktoren\/}~$k_i$, die im allgemeinen Liealgebra-wertig, das hei"st Matrizen sind.

Gl.(\ref{bn_Gl}) ist invertierbar, oder einfacher durch Logarithmieren von Gl.~(\ref{bn_Gl}):
\vspace*{-.5ex}
\begin{align} \label{kn_Gl}
&\ln w(\la)\;
  =\; {\T\sum}_{n=1}^\infty\; \frac{\la^n}{n!}\; k_n
    \\
&\stackrel{\D!}{=}\; \ln \left[ 1 + {\T\sum}_{n=1}^\infty\; \frac{\la^n}{n!}\; b_n \right]\;
  =\; {\T\sum}_{m=1}^\infty\; \frac{(-1)^{m\!-\!1}}{m}
      \left[ {\T\sum}_{n=1}^\infty\; \frac{\la^n}{n!}\; b_n \right]^m\qquad
  \forall\la\in\bbbc
    \nn
\end{align}
erh"alt man
\vspace*{-.5ex}
\begin{align} \label{kn}
k_n\;
  =\; {\T\sum}_{m=1}^n\; \frac{(-1)^{m\!-\!1}}{m}\;
        {\T\sum}_{ \T{}^{\vv i_1,\ldots i_m \in \bbbn}_{i_1+\ldots+i_m=n} }\;
        \frac{n!}{i_1!\cdots i_m!}\; \permI{ b_{i_1}\cdots b_{i_m} }
    \\[-4.5ex]\nn
\end{align}
analog zu Gl.~(\ref{bn}); bzgl.\@ der Notation~$\permI{\;\cdot\;}$ vgl.\@ Gl.~(\ref{permI}).
Wir veranschaulichen diese Gleichung durch explizite Angabe der Ausdr"ucke f"ur die ersten Indizes~$n$:
\vspace*{-.5ex}
\begin{alignat}{2} 
&k_1\;&
  &=\; b_1
    \\[.5ex]
&k_2\;&
  &=\; b_2 - \frac{1}{2}\cdot \frac{2!}{(1!)^2} b_1^2
    \nn \\[.5ex]
&k_3\;&
  &=\; b_3 - \frac{1}{2}\cdot \frac{3!}{1!2!} \permI{b_1 b_2}
      + \frac{1}{3}\cdot \frac{3!}{(1!)^3}
    \nn \\[.5ex]
&k_4\;&
  &=\; b_4 - \frac{1}{2}\cdot \left( \frac{4!}{1!3!} \permI{b_1 b_3}
                + \frac{4!}{(2!)^2} b_2^2 \right)  
      + \frac{1}{3}\cdot \frac{4!}{(1!)^2 2!} \permI{b_1^2 b_2}
      - \frac{1}{4}\cdot \frac{4!}{(1!)^4} b_1^4
    \nn \\[.5ex]
&k_5\;&
  &=\; b_5 - \frac{1}{2}\cdot \left( \frac{5!}{1!4!} \permI{b_1 b_4}
                + \frac{5!}{2!3!} \permI{b_2 b_3} \right)
      + \frac{1}{3}\cdot \left(\frac{5!}{(1!)^2 3!} \permI{b_1^2 b_3}
                + \frac{5!}{1!(2!)^2} \permI{b_1 b_2^2} \right)
    \nn \\
&&&\phantom{=\;}
      - \frac{1}{4}\cdot\frac{5!}{(1!)^3 2!} \permI{b_1^3 b_2}
      + \frac{1}{5}\cdot\frac{5!}{(1!)^5} b_1^5
    \nn
    \\[-4.5ex]\nn
\end{alignat}
\vspace*{-.5ex}
\begin{alignat}{2}
&k_6\;&
  &=\; b_6 - \frac{1}{2}\cdot \left( \frac{6!}{1!5!} \permI{b_1 b_5}
                + \frac{6!}{2!4!} \permI{b_2 b_3}
                + \frac{6!}{(3!)^2} b_3^2 \right)
    \nn \\
&&&\phantom{=\;}
      + \frac{1}{3}\cdot \left(\frac{6!}{(1!)^2 4!} \permI{b_1^2 b_4}
                + \frac{6!}{1!2!3!} \permI{b_1 b_2 b_3}
                + \frac{6!}{(2!)^3} b_2^3 \right) \nn \\
&&&\phantom{=\;}
      - \frac{1}{4}\cdot \left(\frac{6!}{(1!)^3 3!} \permI{b_1^3 b_3}
                + \frac{6!}{(1!)^2(2!)^2} \permI{b_1^2 b_2^2} \right)
      + \frac{1}{5}\cdot\frac{6!}{(1!)^4 2!} \permI{b_1^4 b_2}
      - \frac{1}{6}\cdot\frac{6!}{(1!)^6} b_1^6
    \nn \\[-.5ex]
&&&\ldots
    \nn
\end{alignat}
Gl.(\ref{kn}) bestimmt~$k_n$ als Summe von Monomen
\vspace*{-.5ex}
\begin{align} 
b_{i_1}\; b_{i_2} \cdots b_{i_m}
    \\[-4.5ex]\nn
\end{align}
wobei die Indizes~$i_1,\ldots i_m \!\in\! \bbbn$ s"amtliche verschiedenen Partitionen des Index~$n$ in~$m$ Summanden durchlaufen:~$i_1 \!+ \ldots +\! i_m \!=\! n$ ~($\forall m,\; 1 \!\leq\! m \!\leq\! n$), und~-- die Notation~$\permI{\;\cdot\;}$~-- die~$m$ Faktoren~$b_i$ in s"amtlichen verschiedenen Reihenfolgen stehen.

Die Kumulante~$n$-ter Ordnung~$K_n(t_1,\ldots t_n)$ folgt daher aufgrund ihrer Definition in Gl.~(\ref{lnwla_KumEntw}$'$) als Summe von Monomen
\begin{samepage}
\vspace*{-.5ex}
\begin{align} \label{Kn_Monome}
B_{i_1}(t_1,\ldots t_{i_1})\;
  B_{i_2}(t_{i_1+1},\ldots t_{i_1+i_2})\cdots
  B_{i_m}(t_{i_1+\ldots+i_{m-1}+1},\ldots t_n)
    \\[-4.5ex]\nn
\end{align}
Diese~$m$ Faktoren~($1 \!\leq\! m \!\leq\! n$) treten in s"amtlichen verschiedenen Reihenfolgen auf. \\
\indent
Da sich das Integral~$k_n$, als dessen Integrand~$K_n$ definiert ist, "uber ein {\it vollst"andig symmetrisches\/}~$n$-dimensionales Integrationsgebiet erstreckt, kann~$K_n(t_1,\ldots t_n)$ definiert werden als vollst"andig symmetrisch in seinen Argumenten.
Die Monome in Gl.~(\ref{Kn_Monome}) treten daher zus"atzlich auf als {\it vollst"andig symmetrisiert\/} in den~$t_i$.
Es gilt:
\vspace*{-.5ex}
\begin{align} \label{Kn}
&K_n(t_1,\ldots t_n)
    \\
&\equiv\;
  {\T\sum}_{m=1}^n\; \frac{(-1)^{m\!-\!1}}{m}\;
  {\T\sum}_{ \T{}^{\vv i_1,\ldots i_m \in \bbbn}_{i_1+\ldots+i_m=n} }\;
  \permII{ B_{i_1}\; B_{i_2}\cdots B_{i_m} }_{t_1,\ldots t_n}
    \nn
    \\[-4.5ex]\nn
\end{align}
Dabei haben wir die Notation~$\permII{\;\cdot\;}$ eingef"uhrt, mit
\vspace*{-.5ex}
\begin{align} \label{permII}
&\permII{ B_{i_1}\; B_{i_2}\cdots B_{i_m} }_{t_1,\ldots t_n}
    \\[.25ex]
&\equiv\;
  \permI{ B_{i_1}(t_1,\ldots t_{i_1})\;
          B_{i_2}(t_{i_1+1},\ldots t_{i_1+i_2})\cdots
          B_{i_m}(t_{i_1+\ldots+i_{m-1}+1},\ldots t_n) }
    \nn \\[.25ex]
&\phantom{\equiv\vvv}
  +\; \text{versch.\@ Partitionen der $n$ Argumente}
    \nn
    \\[-4.5ex]\nn
\end{align}
Sie impliziert also~$\permI{\;\cdot\;}$ nach Gl.~(\ref{permI}), das hei"st s"amtliche verschiedenen Reihenfolgen der~$m$ Faktoren, und bedeutet dar"uberhinaus Symmetrisierung der~$n$ Argumente~$t_i$, das hei"st Summation "uber s"amtlichen {\it verschiedenen Partitionen von~$t_1,\ldots t_n$ in Mengen von~$i_1,\ldots i_m$ Elementen\/}~-- was gerade dazu f"uhrt, da"s der Multinomialkoeffizient~$n!/i_1!\cdots i_m!$ von Gl.~(\ref{kn}) verschwunden ist in Gl.~(\ref{Kn}).

Der dem Erwartungswert zugrunde liegende stochastische Proze"s ist vollst"andig bestimmt durch die Kumulanten~$K_n$ in der Form von Gl.~(\ref{Kn}).

\bigskip\noindent
Wir haben in Gl.~(\ref{lnwla_KumEntw}) die Entwicklung in Kumulanten f"ur~$E(\la)$ definiert.
Die Konvergenz dieser Entwicklung impliziert~-- vgl.\@ die Refn.~\cite{VanKampen74,VanKampen76}~-- da"s die Kumulante~$n$-ter Ordnung~$K_n(t_1,\ldots t_n)$ f"ur alle~$n \!\geq\! 2$ nur dann wesentlich von Null verschieden ist, wenn s"amtliche ihrer~$n$ Argumente in einem Bereich des Integrationsintervalls~$[0,1]$ fest bestimmter Gr"o"se zusammengeclustert sind; anderenfalls verschwindet~$K_n$ exponentiell mit den Differenzen ihrer Argumente.
Diese Eigenschaft wird als {\it Cluster-Eigenschaft der Kumulanten\/} bezeichnet und definiert sie gleichsam.
\end{samepage}

Die Cluster-Eigenschaft ist dahingehend zu interpretieren, da"s die durch die stochastische Variable~$A$ bestimmten Gr"o"sen~$B(t)$ in Gl.~(\ref{wla}) an verschiedenen Stellen~$t$ nur dann {\it korreliert\/} fluktuiert, wenn alle diese Stellen in einem Cluster fest bestimmter Gr"o"se liegen.
In diesem Sinne zu verstehen ist die Standard-Notation der Kumulante:
%
\begin{align} \label{Kn_Korrelator}
\cum{B(t_1)\cdots B(t_n)}\;
  \equiv\; K_n(t_1,\ldots t_n)
\end{align}
und ihre Bezeichnung als {\it Korrelator\/}.

Die Konvergenz der Entwicklung in Kumulanten, vgl.\@ Gl.~(\ref{lnwla_KumEntw}), impliziert aber auch, da"s die Gr"o"sen~$k_n$ mit h"oherem Index~$n$ immer weniger zu~$\ln w(\la)$ beitragen.
Die Annahme, da"s s"amtliche Kumulanten h"oherer als zweiter Ordnung verschwinden,~\mbox{\,$K_n \!\equiv\! 0,\; \forall n\!>\!2$}, kann daher aufgefa"st werden als {\it grobe erste N"aherung\/} eines beliebig komplizierten stochastischen Prozesses.
Diese Annahme hei"st genau Annahme eines {\it Gau"s'schen stochastischen Prozesses\/} f"ur die Eichfeldkonfigurationen~$A(x)$ bei der Bildung des Vakuumerwartungswertes~$\vac{\;\cdot\;}$.

Wir werden~$B(t)$  in Verbindung bringen mit dem paralleltransportierten Feldst"arkentensor~$F_{\mu\nu}$.
Aufgrund der Poincar\'einvarianz des physikalischen Vakuums ist~$\vac{F_{\mu\nu}} \!\equiv\! 0$ und nach Gl.~(\ref{Kn}) daher~$K_1(t) \!\equiv\! B_1(t) \!\equiv\! \vac{B(t)} \!\equiv\! 0$.
Das hei"st der Gau"s'sche Proze"s, der in der vorliegenden Arbeit von Relevanz sein wird, ist~-- wie man sagt~-- {\it zentriert\/}.

Wir betrachten einen {\it zentrierten Gau"s'schen Proze"s\/}, der per definitionem vollst"andig bestimmt ist durch seine {\it Kumulante zweiter Ordnung\/}~$K_2(t_1,t_2)$.

Wir l"osen Gl.~(\ref{Kn}) auf nach dem Term auf der rechten Seite mit dem h"ochsten Index und erhalten
\vspace*{-.25ex}
\begin{align} 
&\permII{B_n}_{t_1,\ldots t_n}\;
  =\; B_n(t_1,\ldots t_n)
    \\
&=\; K_n(t_1,\ldots t_n)
  \;+\; {\T\sum}_{m=2}^n\; \frac{(-1)^m}{m}\;
  {\T\sum}_{ \T{}^{\vv i_1,\ldots i_m \in \bbbn}_{i_1+\ldots+i_m=n} }\;
  \permII{ B_{i_1}\; B_{i_2}\cdots B_{i_m} }_{t_1,\ldots t_n}
    \nn
    \\[-4.5ex]\nn
\end{align}
Das hei"st~$B_1 \!=\! K_1 \!\equiv\! 0$ und~$B_2 \!=\! K_2$.
F"ur Indizes~$n \!>\! 2$ steht wegen~$K_{n>2} \!\equiv\! 0$ und~$m \!\geq\! 2$ die Summe von Monomen mit mindestens zwei Faktoren, das hei"st h"ohere Erwartungswerte~$B_{n>2}$ {\it faktorisieren\/}~-- sukzessive in~$B_2 \!\equiv\! K_2$; da insbesondere alle Kumulanten ungerader Ordnung identisch Null sind, "ubertr"agt sich diese Eigenschaft auf die~$B_n$.
F"ur einen {\it zentrierten Gau"s'schen Proze"s\/} gilt zusammenfassend:
\begin{samepage}
\vspace*{-.5ex}
\begin{align} \label{zentrGauss0}
&B_2(t_1,t_2) = K_2(t_1,t_2)
  \\[.5ex]
&B_n(t_1,\ldots t_n)
    \nn \\[-.25ex]
&=\;
  \begin{cases}
    0 &
    \hspace*{1em}\text{$n$ ungerade} \\
    {\T\sum}_{m=2}^n\; {\D \frac{(-1)^m}{m} }\;
      {\T\sum}_{ \T{}^{\vvvv j_1,\ldots j_m \in \bbbn}_{2j_1+\ldots+2j_m=n} }\;
      \permII{ B_{2j_1}\; B_{2j_2}\cdots B_{2j_m} }_{t_1,\ldots t_n} &
    \hspace*{1em}\text{$n \!\geq\! 4$, gerade}
  \end{cases}
    \nn
    \\[-4.5ex]\nn
\end{align}
Die Faktorisierung ist dabei bestimmt durch die Summe "uber s"amtlicher {\it verschiedenen Partitionen\/} des Index~$n$ in~$m$ {\it geradzahlige\/} Summanden:~$2j_1 \!+ \ldots +\! 2j_m \!=\! n$ mit~$j_1,\ldots j_m \!\in\! \bbbn$.
Diese Gleichung vereinfacht sich zu
%
\begin{align} \label{zentrGauss1}
&B_n(t_1,\ldots t_n)\;
  =\; c_n\; \permII{B_2^{\;n\!/\!2}}_{t_1,\ldots t_n}
    \\[.5ex]
&\text{mit}\qquad
  c_n\;
  =\; {\T\sum}_{m=2}^{n\!/\!2}\; \frac{(-1)^m}{m}\;
        {\T\sum}_{ \T{}^{\vvvv j_1,\ldots j_m \in \bbbn}_{2j_1+\ldots+2j_m=n} }\;
        c_{2j_1} \; c_{2j_2}\cdots c_{2j_m} \quad\text{$n \!\geq\! 4$, gerade}
    \tag{\ref{zentrGauss1}$'$} \\
&\hspace*{3.5em}
  B_2(t_1,t_2)\;
  =\; K_2(t_1,t_2) \qquad c_2 \equiv 1
    \nn
\end{align}
\end{samepage}%
Der Ausdruck~$\permII{B_2^{\;n\!/\!2}}_{t_1,\ldots t_n}$ steht f"ur die Summe s"amtlicher $n!/(2!)^{n\!/\!2}$ Ausdr"ucke~$\permI{B_2^{\;n\!/\!2}}$ von Vertauschungen der Argumente~$t_1,\ldots t_n$, vgl.\@ Gl.~(\ref{permII}).
Die Gl.~(\ref{zentrGauss0}) implizite Rekursion reduziert sich hier auf den Koeffizienten~$c_n$; wir haben f"ur die ersten Indizes~\mbox{\,$c_4 \!=\! 1/2$}, \mbox{$c_6 \!=\! -1/12$},~\mbox{\,$c_8 \!=\! 1/6$},\;\ldots
Zur Illustration geben wir mit
%
\begin{align} \label{B4-Fak_Matrix}
&B_4(t_1,t_2,t_3,t_4)
    \\[.5ex]
&=\; \begin{aligned}[t]
  \frac{1}{2} \Big\{
      &B_2(t_1,t_2)\, B_2(t_3,t_4)
     + B_2(t_1,t_3)\, B_2(t_2,t_4)
     + B_2(t_1,t_4)\, B_2(t_2,t_3)
    \nn \\
    + &B_2(t_3,t_4)\, B_2(t_1,t_2)
     + B_2(t_2,t_4)\, B_2(t_1,t_3)
     + B_2(t_2,t_3)\, B_2(t_1,t_4)
        \Big\}
  \end{aligned}
    \nn
\end{align}
die Faktorisierung von~\mbox{\,$B_4(t_1,t_2,t_3,t_4) \!\equiv\! \vac{P\,B(t_1)B(t_2)B(t_3)B(t_4)}$} an.

\section{Annahmen und Konstituierung des \DREI[]{M}{S}{V}}
\label{Abschn:ANN-KONST}

Wir haben nun die Mathematik zur Hand, nichtperturbative Observable unter der Annahme eines stochastischen Prozesses f"ur~$\vac{\;\cdot\;}$ auszudr"ucken durch die ihn definierenden Kumulanten: das \DREI{M}{S}{V} explizit zu formulieren.

\bigskip\noindent
Wir betrachten einen Wegner-Wilson-Loop nach Gl.~(\ref{WW-Loop}).
Mithilfe des Nichtabelschen Stokes'schen Satzes nach Gl.~(\ref{Konnektor_LoopFtr}) gehen wir "uber vom Eich-Vektorfeld zum paralleltransportierten Eich-Feldst"arkentensor.
Sein Erwartungswert~$\vac{\;\cdot\;}$ lautet:
%
\begin{align} \label{WW-LoopF}
\vac{W({\cal C})}\;
  =\; \trDrst{R}\,
      \vac{ P_{\tilde{\cal C}}\; \exp
        -\frac{\iIM\,g}{2} \int_{{\cal S}(\tilde{\cal C})} d\si^{\mu\nu}\!(x')\;
        F_{\mu\nu}\!(x'; x_0,{\cal C}_{x_{\!0}\!\ixp})\; }
\end{align}
vgl.~$\trDrst{R}$ in den Gln.~(\ref{trDrst}),~(\ref{trDrst}$'$).
Hier bezieht sich die Pfadordnung~$P_{\tilde{\cal C}}$ auf die Kurve~$\tilde{\cal C}$, das hei"st die Deformation der Kurve~${\cal C}$, die die Fl"ache~${\cal S} \!\equiv\! {\cal S}(\tilde{\cal C})$ aufspannt; es gilt~$\pa{\cal S} \!=\! {\cal C}$.
Der Referenzpunkt~$x_0$ ist beliebig aber fest auf~${\cal S}$.
Die Kurven~${\cal C}_{x_{\!0}\!\ixp}$ verlaufen von den Punkten~$x'$ der Fl"ache nach~$x_0$.
Diese Kurven, mit ihren Inversen~${\cal C}_{x_{\!0}\!\ixp}{}^{\zzz -1} \!\equiv\! {\cal C}_{\ixp\!x_{\!0}}$ in definierter Weise verkettet, bilden genau die Deformation~$\tilde{\cal C}$ und damit~${\cal S}$.
Vgl.\@ die Diskussion im Anschlu"s an Gl.~(\ref{Konnektor_LoopFtr}).

Gegen"uberstellen von Gl.~(\ref{WW-LoopF}) und~(\ref{wla}) konkretisiert die Funktion~$w(\la)$.
Sei~$\la \!\equiv\! 1$ gesetzt, das Eichfeld-Funktional~$B(t)$ wird dann identifiziert "uber die Ersetzung
\begin{samepage}
%
\begin{align} \label{BnachF}
\int_0^1 dt\; B(t)
  \vv\longrightarrow\vv
  -\frac{\iIM\,g}{2}\; \int_{{\cal S}(\tilde{\cal C})} d\si^{\mu\nu}\!(x')\;
    F_{\mu\nu}\!(x'; x_0,{\cal C}_{x_{\!0}\!\ixp})
\end{align}
im wesentlichen mit der paralleltransportierten Feldst"arke~$F_{\mu\nu} \!=\! F_{\mu\nu a}T^a_\Drst{R}$ in einer Darstellung~$\mf{R}$ der Eichalgebra~$\suNc$~-- integriert "uber den zweiten, keiner definierten Ordnung unterworfene Parameter, der zusammen mit dem Parameter der Kurve~$\tilde{\cal C}$ lokal die Fl"ache~${\cal S}$ aufspannt.
Dieser wiederum ist mit~$t \!\in\! [0,1]$ in Gl.~(\ref{BnachF}) beziehungsweise~(\ref{wla}) zu identifizieren, und folglich~$P_{\tilde{\cal C}}$ mit dem Ordnungsoperator~$P \!\equiv\! P_t$ dort.

Konkretisiere Gl.~(\ref{BnachF}) den Erwartungswert~$w(\la \!\equiv\! 1)$, so hei"st die {\bf Annahme~\mbox{\boldmath$(1)$} einer konvergenten Entwicklung in Kumulanten} nach Gl.~(\ref{lnwla_KumEntw}) genau Konvergenz von
\vspace*{-.5ex}
\begin{align} \label{WW-Loop_KumEntw}
&\vac{W({\cal C})}\;
  =\; \trDrst{R}\, \exp\; \left[ {\T\sum}_{n=1}^\infty\; \frac{1}{n!}\; k_n \right]
                \\[1ex]
&\begin{aligned}[t]
  \text{mit}\qquad
  &k_n\; 
     =\; \left( -\frac{\iIM}{2}\right)^{\!n}\;
           \int_{{\cal S}(\tilde{\cal C})} d\si^{\mu_1\nu_1}\!(x_1)\vv \cdots
           \int_{{\cal S}(\tilde{\cal C})} d\si^{\mu_n\nu_n}\!(x_n)\vv K_n(x_1,\ldots,x_n)
                 \nn \\[.5ex]
  &K_n(x_1,\ldots,x_n)\;
     \equiv\; \cum{ g^n F^{(1)} \cdots F^{(n)} }
                \nn \\
  &F^{(i)}\;
     \equiv\; F_{\mu_i\nu_i}\!(x_i; x_0,{\cal C}_{x_{\!0}\!x_{\!i}})
                \nn
 \end{aligned}
    \nn
    \\[-4.5ex]\nn
\end{align}
vgl.\@ die Notation von Gl.~(\ref{Kn_Korrelator}). \\
\indent
Die Clustereigenschaft dieser Kumulanten~$K_n$ manifestiert sich darin,
\end{samepage}%
da"s~$K_n(x_1,\ldots x_n)$ f"ur~$n \!\geq\! 2$ nur dann wesentlich von Null verschieden ist, wenn s"amtliche ihrer Raumzeit-Argumente~$x_1,\ldots x_n$, das hei"st die paralleltransportierten Feldst"arken in einer Sph"are festen Radius' zusammengeclustert sind.%
\FOOT{
  \label{FN:EuklidMinkowski}An dieser Stelle ist eine grundlegende Bemerkung notwendig:   Die Clusterung, das hei"st die {\sl korrelierte\/} Fluktuation paralleltransportierter Feldst"arken bezogen auf Sph"aren macht a~priori nur Sinn bez"uglich Euklidischer Sph"aren:   Das \DREI{M}{S}{V} macht a~priori nur Sinn formuliert auf einer Raumzeit mit {\sl Euklidischer\/} (Riemannscher) Metrik.   Um die prinzipielle Argumentation dieses Abschnitts nicht unn"otigerweise zu komplizieren, geben wir zun"achst nur eine Formulierung an: die des \DREI{M}{S}{V} auf der physikalischen Raumzeit mit {\sl Minkowskischer\/} (pseudo-Riemannscher) Metrik~-- da diese unmittelbar anwendbar ist auf nichperturbative Hochenergiestreuung.   Wir argumentieren dessen ungeachtet auf der Basis einer Euklidischen Metrik.   Damit suggerieren wir Existenz und "Aquivalenz zweier Formulierungen: auf der physikalischen Minkowski-Raumzeit und ihrer Fortsetzung ins Euklidische.   Das folgende Kapitel~\ref{Kap:ANALYT} zeigt dies dann explizit durch Angabe der {\sl analytischen Fortsetzung\/} des \DREI[]{M}{S}{V}~-- und diskutiert die physikalischen Konsequenzen. 
}
Sei~$a$ der Radius der Sph"are (oder charakterisiere~$a$ die Gr"o"se eines anderen Gebietes in diesem Sinne), innerhalb derer(dessen) sich die stochastische Fluktuation des Eichfelds~$\vac{\;\cdot\;}$ manifestiert als {\it Korrelation\/} der paralleltransportierten Eichfeldst"arken.
$a$ wird allgemein als {\it Korrelationsl"ange\/} bezeichnet.

Schon die Existenz einer konvergenten Entwicklung in Kumulanten, vgl.\@ Gl.~(\ref{WW-Loop_KumEntw}), und zwar "uber deren Cluster-Eigenschaft, hat schon weitreichende Konsequenzen.

Wir betrachten einen Wegner-Wilson-Loop~$W({\cal C})$, dessen Ausdehnung sehr viel gr"o"ser ist als~$A$; Kriterium hierf"ur sei~${\cal S}_{\rm min.} \!\gg\! a^2$, wobei~${\cal S}_{\rm min.}$ definiert ist als die Fl"ache~${\cal S}(\tilde{\cal C})$, bestimmt durch eine Deformation~$\tilde{\cal C}$ im ausgef"uhrten Sinne, deren Betrag {\it minimal\/} ist. \\
Wir sch"atzen den Ausdruck~$k_n$ aus Gl.~(\ref{WW-Loop_KumEntw}) ab.
Sei dazu die Cluster-Eigenschaft der Kumulante~$K_n(x_1,\ldots x_n)$ dadurch approximiert, da"s~$K_n \!\equiv\!1$, falls s"amtliche ihrer~$n$ Argumente innerhalb einer Sph"are des Radius'~$a$ liegen, und ansonsten~$K_n \!\equiv\!0$ (d.h.\@ $\vth$-Funktions-artiger Abfall und Normierung der Korrelation auf 1):
Die ersten~$n\!-\!1$~Integrationen~$d\si_i \!\equiv\! d\si^{\mu_i\nu_i}\!(x_i)$,~$i \!=\! 1,\ldots n\!-\!1$, "uber die Fl"ache~${\cal S}$ ergeben dann den Faktor~$a^2$, n"amlich wenn die Argumente~$x_1,\ldots x_{n\!-\!1}$ von~$x_n$ nicht weiter entfernt sind als~$a$; die letzte Integration~$d\si_n$ schlie"slich ist frei und ergibt den Betrag von~${\cal S}$.
Letztlich ist die Minimalfl"ache~${\cal S}_{\rm min.}$ zu nehmen, vgl.\@ die Refn.~\cite{Dosch87,Dosch88}, und wir erhalten die Absch"atzung
%
\begin{align} \label{WW-Loop_arealaw}
\vac{W({\cal C})}\;
  \cong\; \trDrst{R}\, \exp\; - \si\, |{\cal S}_{\rm min.}|
\end{align}
mit~$\si$ einer Konstanten der Dimension~{\it Fl"ache$^{-1} =$ Energie pro L"ange\/}.

Dies ist genau das {\it area law\/}.
Es begr"undet die Bedeutung des Wegner-Wilson-Loops in nichtabelschen Eichtheorien als Ordnungsparameter auf nichtperturbativ dominierten Skalenbereichen.
Wegner-Wilson-Loops wurden 1971 erstmals definiert von Wegner in der statistischen Physik, Ref.~\cite{Wegner71}, bevor 1974 Wilson mit ihrer Hilfe {\it Confinement\/} von Quarks diskutierte im Rahmen von QCD als Gittereichtheorie, Ref.~\cite{Wilson74}.

Da das Verstehen des Confinements von Eichladungen urspr"ungliches Moment war zur Formulierung des \DREI{M}{S}{V} und der Confinement zugrundeliegende Mechanismus von herausragender Bedeutung ist f"ur unsere Arbeit, gehen wir n"aher darauf ein; vgl.\@ Ref.~\cite{Simonov96}.

Wir betrachten das eichinvariante Paar eines statischen Quarks und Antiquarks\label{T:statQ-AQ-Paar} mit r"aumlicher Separation~$R$.
Die klassische Weltlinie des Quarks verl"auft parallel der Zeitachse der Raumzeit, die des Antiquarks mit r"aumlichem Abstand~$R$ parallel zu dieser aber entgegengesetzt orientiert.
Der Propagation des Quarks und Antiquarks durch die Raumzeit entsprechen Konnektoren entlang dieser Kurven.
Wir betrachten das statische Paar "uber einen Zeitraum~$T$ (bzg.\@ seines Ruhesystems).
Eichinvarianz manifestiert sich darin, da"s die Konnektoren entlang der Weltlinien verbunden sind durch Konnektoren zu den Zeiten~0 und~$T$.
Resultat ist eine geschlossene orientierte Rechteckkurve, bezeichnet mit~${\cal C}_\RT$.
Das Paar ist daher repr"asentiert~-- vgl.\@ die Verkettungseigenschaft von Konnektoren nach Gl.~(\ref{Konnektor_Eigenschaften}$'$)~-- als Spur eines Konnektors entlang der geschlossenen Kurve~${\cal C}_\RT$, das hei"st als Wegner-Wilson-Loop~\mbox{\,$W({\cal C}_\RT)$}, der steht in der fundamentalen Darstellung~$\mf{F}$ der Quarks.

F"ur die potentielle Energie dieses Paares gilt
%
\begin{align} \label{QbarQ_Potential}
V(R)\;
  =\; {\T\lim}_{T\to\infty}\; -\frac{1}{T}\, \ln\, \vac{W({\cal C}_\RT)}
\end{align}
Da der Betrag der Minimalfl"ache mit Rand~${\cal C}_\RT$ gerade~$RT$ ist, ergibt Gl.~(\ref{WW-Loop_arealaw}) die Absch"atz\-ung~\mbox{\,$V(R) \!=\! \si\, R$}, vgl.\@ die Refn.~\cite{Dosch87,Dosch88}.
Die potentielle Energie eines statischen Quarks und Antiquarks w"achst also {\it linear\/} mit wachsender Separation~$R$.
Das resultierende physikalische Bild ist die Ausbildung eines {\it gluonischen Strings\/} zwischen dem Quark und dem Antiquark, das diese konfiniert mit der St"arke~$\si$, der Stringspannung.
Wir haben {\it lineares Confinement\/}.

Wir betonen, da"s das \DREI{M}{S}{V} dies folgert allein auf Basis der Annahme, da"s die Entwicklung in Kumulanten bez"uglich paralleltransportierter Feldst"arken konvergiert~-- Gl.~(\ref{WW-Loop_KumEntw}) sinnvoll in diesem Sinne existiert.
Der zugrundeliegende stochastische Proze"s, das hei"st die Kumulanten an sich sind ausdr"ucklich noch nicht spezifiziert.
\vspace*{-.5ex}

\bigskip\noindent
Um zu quantitativen Aussagen zu gelangen wird das \DREI{M}{S}{V} im engeren Sinne definiert.
Dies geschieht durch die zus"atzliche {\bf Annahme~\mbox{\boldmath$(2)$} eines Gau"s'schen Prozesses} im Sinne des vorangegangenen Abschnitts.

Wie dort schon angemerkt ist dieser wegen der Poincar\'einvarianz des Vakuums der QCD {\it zentriert\/}:~$K_1(x) \!\equiv\! \cum{ g F_{\mu\nu}\!(x; x_0,{\cal C}_{x_{\!0}\!x}) } \!\equiv\! \vac{ g F_{\mu\nu}\!(x; x_0,{\cal C}_{x_{\!0}\!x}) } \!\equiv\!0$.
Unter der Ersetzung nach Gl.~(\ref{BnachF}) haben wir Relationen, die Gl.~(\ref{zentrGauss0}) entsprechen, das hei"st Faktorisierung nach Gl.~(\ref{zentrGauss1}):
Vakuumerwartungswerte~$\vac{ P\, g^n F^{(1)} \cdots F^{(n)} }$ faktorisieren in die Kumulante zweiter Ordnung, das hei"st in~$K_2(x_1,x_2) \!\equiv\! \cum{ g^2 F^{(1)} F^{(2)} } \!\equiv\! \vac{ P\, g^2 F^{(1)} F^{(2)} }$.
Mit ihrer Kenntnis ist das \DREI{M}{S}{V} im engeren Sinne vollst"andig bestimmt.
Wir entwickeln im folgenden einen Ansatz.

Wir haben mit der Notation~$F^{(i)} \equiv F_{\mu_i\nu_i}\!(x_i; x_0,{\cal C}_{x_{\!0}\!x_{\!i}})$ aus Gl.~(\ref{WW-Loop_KumEntw}):
%
\begin{align} \label{K2_vac-g2FF}
&K_2\;
  =\; (K_2)_{a_1a_2}\, T^{a_1}T^{a_2}
    \\[1ex]
&=\; \vac{ P\, g^2 F^{(1)} F^{(2)} }\;
   =\; \vac{ P\, g^2 F^{(1)}{}_{\zz a_1} F^{(2)}{}_{\zz a_2}\, T^{a_1}T^{a_2} }\;
   =\; P\, \vac{ g^2 F^{(1)}{}_{\zz a_1} F^{(2)}{}_{\zz a_2} }\, T^{a_1}T^{a_2}
    \nn
\end{align}
mit der letzten Identit"at wegen der Linearit"at von~$\vac{\;\cdot\;}$ und dem Pfadordnungsoperator~$P$, der die Reihenfolge der~$ T^{a_1}$,~$T^{a_2}$ bestimmt, den Generatoren einer Darstellung~$\Drst{R}$ der~$\suNc$. \\
Dies ist ein Operator, der aufgrund der Eichinvarianz des physikalischen Vakuums, mit einem beliebigen anderen Operator vertauscht; er ist daher proportional zum quadratischen Casimir-Operator~$T^aT^a \!=\! \csDrst{R} \bbbOne{R}$, also zur Eins der Darstellung:
\begin{samepage}
%
\begin{align} \label{K2_Eins}
&K_2\;
  =\; (K_2)_{a_1a_2}\, T^{a_1}T^{a_2}\;
  =\; c_\Drst{R}\, (\csDrst{R})^{-1}\, T^a T^a\;
  =\; c_\Drst{R}\, \bbbOne{R}
    \\[1ex]
&\text{mit}\qquad
  c_\Drst{R}\;
  =\; \csDrst{R} / \dimNc\; (K_2)_{aa}\;
  =\; \normDrst{R} / \dimDrst{R}\; (K_2)_{aa}
    \nn
\end{align}
vgl.\@ Gl.~(\ref{vev_Fparallel1}).
Dabei folgt die Konstante~$c_\Drst{R}$ durch Spurbildung mithilfe~$\tr T^{a_1}T^{a_2} \!=\! \normDrst{R} \de^{a_1a_2}$; sie h"angt ab von der Darstellung~$\Drst{R}$ "uber den Faktor~$\csDrst{R}$ deren quadratischen Casimir-Operators beziehungsweise von deren Normierung~$\normDrst{R}$ und Dimension~$\dimDrst{R}$.

Unmittelbare Konsequenz der zweiten Identit"at in Gl.~(\ref{K2_Eins}) ist Proportionalit"at der Komponenete~$(K_2)_{a_1a_2}$ der Kumulante zweiter Ordnung zum Kroneckersymbol~$\de_{a_1a_2}$:
%
\begin{align} \label{K2Komp_Kronecker}
(K_2)_{a_1a_2}\;
  =\; \de_{a_1a_2} / \dimNc\; (K_2)_{aa}
\end{align}
\end{samepage}%
Die zu den Gln.~(\ref{K2_Eins}),~(\ref{K2Komp_Kronecker}) "aquivalenten Relationen in der suggestiven Notation von~Vaku\-umerwartungswerten, vgl.\@ Gl.~(\ref{K2_vac-g2FF}), lauten:
\vspace*{-.25ex}
\begin{alignat}{3}
&\vac{ g^2 F^{(1)} F^{(2)} }\;&
  &=\;& \csDrst{R} / \dimNc\vv
          &\vac{ g^2 F^{(1)}{}_{\zz a} F^{(2)}{}_{\zz a} }\cdot \bbbOne{R}
    \tag{\ref{K2_Eins}$'$} \\[1ex]
&\vac{ g^2 F^{(1)}{}_{a_1} F^{(2)}{}_{a_2} }\;&
  &=\;& \de_{a_1a_2} / \dimNc\vv
          &\vac{ g^2 F^{(1)}{}_{\zz a} F^{(2)}{}_{\zz a} }
    \tag{\ref{K2Komp_Kronecker}$'$}
    \\[-4.25ex]\nn
\end{alignat}
wobei der Pfadordnungsoperator~$P$ aufgrund der Symmetrie bez"uglich der Eichgruppenindizes "uberfl"ussig und daher weggelassen ist.
"Uber diese Relationen ist die Kumulante zweiter Ordnung vollst"andig bestimmt durch den bez"uglich der Eichgruppenindizes kontrahierten Vakuumerwartungswert~$\vac{ g^2 F^{(1)}{}_a F^{(2)}{}_a }$.

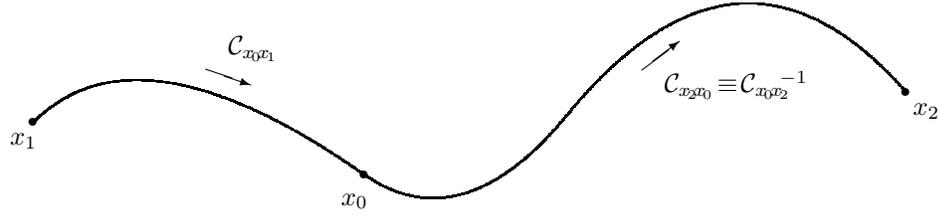
\begin{figure}
\begin{minipage}{\linewidth}
  \begin{center}
  \setlength{\unitlength}{1mm}\begin{picture}(120,30)   
    \linethickness{0.5pt}
    \put( -1,11){\normalsize $x_1$}
    \qbezier( 2,14)(16,28)( 46, 7)
        \put( 2,14){\normalsize \circle*{1}}
                      \put( 46, 7){\normalsize \circle*{1}}
    \put(25,21){\vector(3,-1){6}}
    \put(28,23){\normalsize ${\cal C}_{x_{\!0}\!x_{\!1}}$}
    \put( 43, 3){\normalsize $x_0$}
    \qbezier(46, 7)(59,-2)( 73,15)
    \qbezier(73,15)(96,43)(118,18)
                      \put(118,18){\normalsize \circle*{1}}
    \put(83,21){\vector(4,3){5}}
    \put(86,18){\normalsize ${\cal C}_{x_{\!2}\!x_{\!0}}
                   \!\equiv\! {\cal C}_{x_{\!0}\!x_{\!2}}{}^{\zzz -1}$}
    \put(119,15){\normalsize $x_2$}
  \end{picture}
  \end{center}
\vspace*{-4ex}
\caption[Korrelationstensor~\protect$\big(D_{\mu_1\nu_1\mu_2\nu_2}\!(x_1, x_2; x_0, {\cal C}_{x_{\!0}\!x_{\!1}}, {\cal C}_{x_{\!0}\!x_{\!2}})\big)$ "uber~\protect${\cal C}_{x_{\!2}\!x_{\!1}} \!\equiv\! {\cal C}_{x_{\!0}\!x_{\!2}}{}^{\zzz -1} \!\circ\! {\cal C}_{x_{\!0}\!x_{\!1}}$]{
  Die funktionale Abh"angigkeit des Tensor~$D \!=\! (D_{\mu_1\nu_1\mu_2\nu_2})$, vgl.\@ Gl.~(\ref{K2-g2FF_Dvier}), kann folgenderma"sen aufgefa"st werden: $D$ h"angt ab von den Positionen~$x_1$,~$x_2$ der paralleltransportierten Feldst"arken, von der diese Punkte verbindenden Kurve~${\cal C}_{x_{\!2}\!x_{\!1}} \!\equiv\! {\cal C}_{x_{\!0}\!x_{\!2}}{}^{\zzz -1} \!\circ\! {\cal C}_{x_{\!0}\!x_{\!1}}$ und von dem auf dieser Kurve gew"ahlten Referenzpunkt~$x_0$.
}
\label{Fig:Kurve-x1x2}
\end{minipage}
\end{figure}

Wir f"uhren ein die Notation
\vspace*{-.25ex}
\begin{align} \label{Gluonkondensat_Minkowski}
\vac{g^2 FF}\;
  \equiv\;
  \vac{ g^2 F_{\mu\nu     a}\!(x_1; x_0,{\cal C}_{x_{\!0}\!x_{\!1}})
            F^{\mu\nu}{}_{a}\!(x_2; x_0,{\cal C}_{x_{\!0}\!x_{\!2}})
      }\; \Big|_{x_1 \equiv x_2}
    \\[-4.25ex]\nn
\end{align}
f"ur die Minkowskische Version des {\it Gluonkondensats\/}, wie definiert in den Summenregeln der QCD von Shifman, Vainshtein, Zakharov.
Herausziehen dieses Faktors gem"a"s
\vspace*{-.25ex}
\begin{align} \label{K2-g2FF_Dvier}
&\vac{ g^2 F^{(1)}{}_{\zz a} F^{(2)}{}_{\zz a} }
    \\
  &=\; \vac{ P\, g^2
           F_{\mu_1\nu_1 a}\!(x_1; x_0, {\cal C}_{x_{\!0}\!x_{\!1}})
           F_{\mu_2\nu_2 a}\!(x_2; x_0, {\cal C}_{x_{\!0}\!x_{\!2}}) }
    \nn \\[.25ex]
&=:\; \vac{g^2 FF} \cdot
       D_{\mu_1\nu_1\mu_2\nu_2}
         \!(x_1, x_2; x_0, {\cal C}_{x_{\!0}\!x_{\!1}}, {\cal C}_{x_{\!0}\!x_{\!2}})\qquad
  \text{mit}\qquad
  D_{\mu\nu}{}^{\mu\nu} \Big|_{\D x_1 \!\equiv\! x_2}\;
  =\; 1
    \nn
    \\[-4.25ex]\nn
\end{align}
definiert "uber seine kovarianten Komponenten den dimensionslosen Lorentz-Tensor vierter Stufe~$D \!=\! (D_{\mu_1\nu_1\mu_2\nu_2})$, der den reinen Konfigurationsanteil der Kumulante zweiter Ordnung subsumiert, vgl.\@ die Gln.~(\ref{K2_Eins}$'$),~(\ref{K2Komp_Kronecker}$'$).
Er h"angt ab von den Positionen~$x_1$,~$x_2$ der beiden paralleltransportierten Feldst"arken, von dem Referenzpunkt~$x_0$, {\it an den\/}, und von den Kurven~${\cal C}_{x_{\!0}\!x_{\!1}}$,~${\cal C}_{x_{\!0}\!x_{\!2}}$, {\it entlang derer\/} diese paralleltransportiert werden, vgl. Abb.~\ref{Fig:Kurve-x1x2}.

F"ur diesen Tensor~$D$ wird im folgenden gemacht: die {\bf Annahme~\mbox{\boldmath$(3)$} der Unabh"angigkeit vom Referenzpunkt~\boldmath$x_0$ und von den Kurven~\boldmath${\cal C}_{x_{\!0}\!x_{\!1}}$ und \boldmath${\cal C}_{x_{\!0}\!x_{\!2}}$}.

Referenzpunkt wie Kurven sind mit denen zu identifizieren, die der Nichtabelsche Stokes'sche Satz einf"uhrt, vgl.\@ Gl.~(\ref{Konnektor_LoopFtr}).
Sie sind daher a~priori~-- innerhalb gewisser Randbedingungen~-- frei w"ahlbar:
Unabh"angigkeit von den Kurven ist gegeben, wenn die Integrationen "uber die Argumente~$x_1$ und~$x_2$ nach Gl.~(\ref{WW-Loop_KumEntw}), das hei"st insbesondere die Fl"ache~${\cal S} \!\equiv\! {\cal S}(\tilde{\cal C})$ und die Pfadordnung entlang~$\tilde{\cal C}$, kompatibel sind mit den Integrationen des Nichtabelschen Stokes'schen Satzes.
Beides werden wir respektieren.
Unabh"angigkeit vom Referenzpunkt~$x_0$ ist gegeben, wenn alle Kumulanten berechnet werden, die den stochastischen Proze"s definieren, der der Bildung~$\vac{\;\cdot\;}$ des Vakuumerwartungswerts zugrunde liegt.
Annahme~$(2)$ eines Gau"s'schen Prozesses~-- aufgefa"st als Weglassen der h"oheren Kumulanten des tats"achlichen komplizierteren Prozesses~-- ist daher mehr Approximation als Annahme.
In~praxi ist bei "`sinnvoller"' Wahl des Referenzpunktes die Abh"angigkeit numerisch nur schwach; wir kommen hierauf zur"uck, sei auch verwiesen auf die Refn.~\cite{Rueter94,Rueter94a,Dosch95}.

Poincar\'e- und Parit"atskovarianz von~$D$ als Tensor impliziert, da"s seine Tensorstruktur nur aus zwei Tensoren konstruiert sein kann: dem {\it metrischen Tensor\/}~$g \!=\! (g_{\mu\nu})$ und der Differenz~$\xi$ seiner Argumente, und seine skalare funktionale Abh"angigkeit nur aus~$\xi^2$, dem invarianten Differenzquadrat.
"Aquivalent zu~$\xi$ in dieser Hinsicht ist die partielle Ableitung~$\pa \!\equiv\! \pa/\pa\xi$, da sie auf solche Funktionen wie~$\pa \equiv\! 2\xi\, d/d\xi^2$ wirkt.
Wir definieren
%
\begin{align} \label{xi}
\xi\;
  =\; \big(x_{\!1} \!-\! x_{\!2}\big)/ a
\end{align}
Sei~\mbox{\,$a$} eine Konstante der Dimension einer L"ange, dann ist~\mbox{\,$\xi$} dimensionslos.
[Wir werden~\mbox{\,$a$} identifizieren als "`Korrelationsl"ange"'.]
Damit gilt f"ur~\mbox{\,$D \!\equiv\! \big(D_{\mu\nu\rh\si}\big)$} die Darstellung:
\vspace*{-.5ex}
\begin{align} \label{Dvier_DDxi0}
D_{\mu\nu\rh\si}\!(\xi)\;
  =\; \vka\; t\oC{}_{\zzzz \mu\nu\rh\si}\vv D\uC(\xi^2)\;
      +\; (1\!-\!\vka)\; t\oNC{}_{\zzzz \mu\nu\rh\si}\vv
        \frac{1}{8}\, \int_{\D-\infty}^{\D\xi^2} du\; D\uNC(u)
    \\[-5ex]\nn
\end{align}
mit
\vspace*{-1.25ex}
\begin{alignat}{2} \label{tC,tNC}
&t\oC{}_{\zzzz \mu\nu\rh\si}\;&
  &=\; \frac{1}{12}\;
         \big(
           g_{\mu\rh}\, g_{\nu\si} - g_{\mu\si}\, g_{\nu\rh}
         \big)\;
   =\; \frac{1}{12}\vv \de^{\al\be}_{\mu\nu}\vv g_{\al\rh}\, g_{\be\si}
    \\
&t\oNC{}_{\zzzz \mu\nu\rh\si}\;&
  &=\; \phantom{1}
       \frac{1}{6}\;
         \big(
             g_{\mu\rh}\, \pa_\nu\, \pa_\si - g_{\mu\si}\, \pa_\nu\, \pa_\rh
           + g_{\nu\si}\, \pa_\mu\, \pa_\rh - g_{\nu\rh}\, \pa_\mu\, \pa_\si
         \big)
    \tag{\ref{tC,tNC}$'$} \\[-.25ex]
&&&=\; \phantom{1}
       \frac{1}{6}\vv g_{\al\ga}\vv
         \de^{\al\be}_{\mu\nu}\, \pa_{\be}\vv
         \de^{\ga\de}_{\rh\si}\, \pa_{\de}
    \nn
    \\[-4.5ex]\nn
\end{alignat}
Der Lorentz-Tensor~\mbox{\,$D_{\mu\nu\rh\si}$} und damit~\mbox{\,$t\oC{}_{\zzzz \mu\nu\rh\si}$},~\mbox{\,$t\oNC{}_{\zzzz \mu\nu\rh\si}$} sind antisymmetrisch unter Vertauschung der Indizes~$\mu \!\leftrightarrow\! \nu$ oder~$\rh \!\leftrightarrow\! \si$.
Dies ist evident, geschrieben in Termen verallgemeinerter Kronecker-Symbole, die allgemein definiert sind als Determinante konventioneller Kronecker-Symbole, vgl.\@ Gl.~(\ref{APP:Kronecker_det-Def}):
\vspace*{-.5ex}
\begin{align} 
\de^{\mu_1\cdots\mu_s}_{\, \nu_1\cdots\nu_s}\;
  :=\; \det \big( \de^{\mu_i}_{\nu_j} \big)\quad
  =\; \de^{[\mu_1\cdots\mu_s]}_{\;\, \nu_1\cdots\nu_s}\;
  =\; \de^{\, \mu_1\cdots\mu_s}_{[\nu_1\cdots\nu_s]}\;
  =\; \pmatrixZV{\mu_1}{\mu_2}{\cdots}{\mu_s}
                {\nu_1}{\nu_2}{\cdots}{\nu_s}
    \\[-4.5ex]\nn
\end{align}
Sie sind folglich voll antisymmetrisch in den kontra- wie kovarianten Indizes und gleich dem Signum der Indexpermutation~\mbox{$\si^{\D\mskip-1mu\star}\!: \mu_i \!\to\! \nu_i,\, \forall i \!\in\! \{1,2,\ldots,s\}$}, falls~\mbox{\,$\{\mu_i\}$} und~\mbox{\,$\{\nu_i\}$} paarweise verschiedene Indexmengen sind,~-- und Null sonst; bzgl.\@ der Klammernotation vgl.\@ Gl.~(\ref{APP:SignumPermutation}). \\
\indent
Die Kontraktionen der Tensorstrukturen~$t\oC$ und~$t\oNC$ sind normiert gem"a"s
\begin{samepage}
\vspace*{-.25ex}
\begin{align} \label{tsvier_zkontr}
&t\oC{}_{\zzz \mu\nu}{}^{\mu\nu}\;
  =\; 1
    \\[.25ex]
&t\oNC{}_{\zzz \mu\nu}{}^{\mu\nu}\;
  =\; \pa^2
    \tag{\ref{tsvier_zkontr}$'$}
    \\[-4.25ex]\nn
\end{align}
mit~\mbox{\,$\pa^2 \!\equiv\! \pa_\mu\pa^\mu$} dem d'Alembert-Operator.

Die Darstellung Gl.~(\ref{Dvier_DDxi0}) leiten wir in Ref.~\cite{Kulzinger95} her.
Sie ist die geeignetste f"ur unsere Zwecke, differiert aber von der Standarddarstellung, die folgt durch Ausf"uhren einer der partiellen Ableitungen in~$t\oNC$ (angewandt auf das Integral bleibt wegen~$\pa \!\equiv\! 2\xi\; d/d\xi^2$ der Integrand an der oberen Grenze stehen) und Umbenennung~$D\uC \!\to\! D$ und~$D\uNC \!\to\! D_1$.

Gl.~(\ref{Dvier_DDxi0}) ist die Zerlegung\label{T:Zerlegung_tC,tNC} in zwei Tensorstrukturen~$t\oC$ und~$t\oNC$, die gewichtet sind mit einem Parameter~$\vka$ und multipliziert mit skalaren (Korrelations-)Funktionen~$D_C$ beziehungsweise~$D\uNC$ des invarianten Abstandquadrats~$\xi^2$.
Der Term mit~$t\oNC_{\mu\nu\rh\si}$ verschwindet identisch {\it per constructionem\/}, wenn einer der Differentialoperatoren
\vspace*{-.25ex}
\begin{align} \label{Bianchi-Op}
&\ep^{\al\be\mu\nu}\; \del{x_{\!1}}_\be
    \\[.25ex]
&\ep^{\al\be\rh\si}\; \del{x_{\!2}}_\be
    \tag{\ref{Bianchi-Op}$'$}
    \\[-4.25ex]\nn
\end{align}
\end{samepage}%
auf ihn wirkt; der Term mit~$t\oC$ ist a~priori definiert als der Rest.
Diese Zerlegung der Kumulante zweiter Ordnung in den Feldst"arken~-- vgl.\@ die Gln.~(\ref{K2_Eins}$'$),~(\ref{K2Komp_Kronecker}$'$) und~(\ref{K2-g2FF_Dvier})~-- resultiert also in {\it einen\/} Term~($N\!C$\/), wenn die Feldst"arken der Theorie die {\it abelsche\/} Bianchi-Identit"at erf"ullen:~$\del{x}_\mu F_{\nu\rh}\!(x) \!+\text{zykl.} \!=\! 0$, und in die {\it Summe zweier Terme\/}~($C$\/ und~$N\!C$\/), wenn sie diese nicht erf"ullen.
Die Funktion $D\uC$ ist ungleich Null also nur in einer {\it nicht-abelschen Theorie\/} wie der QCD oder in einer {\it abelschen Theorie mit magnetischen Monopolen\/}.
Die Indizes~$C$ und~$N\!C$ stehen f"ur {\it confining\/} (konfinierend) beziehungsweise {\it non-confining\/} (nicht konfinierend) und greifen ein zentrales Resultat des \DREI{M}{S}{V} auf:
F"ur die Stringspannung~$\si$ als Ma"s  f"ur die St"arke {\it linearen Confinements\/}~-- vgl.\@ Seite~\pageref{WW-Loop_arealaw}~-- wird berechnet:%
\FOOT{
  Die zweite Identit"at gilt auf Basis des unten in Gl.~(\ref{Dvier_DDkkontrAnsatz}) gemachten Ansatzes.
}
%
\vspace*{-.75ex}
\begin{align} \label{Stringspannung-si_DC}
&\si\;
   =\; \vka\, \vac{g^2FF}\, a^2\cdot
        \frac{\pi}{4\Nc}\, \int_0^\infty du^2\, \xE[D\uC](-u^2)
    \\[-1ex]
  &\hspace*{172pt}
   =\;  \vka\, \vac{g^2FF}\, (\la_\nu a)^2\cdot
        \frac{\pi}{4\Nc}\; 2(\nu \!-\! 3)
    \nn
    \\[-4.75ex]\nn
\end{align}
vgl.\@ die Ref.~\cite{Dosch94a}.
Ihr Nichtverschwinden setzt das Nichtverschwinden der Funktion~$D\uC$ voraus:
Das \DREI{M}{S}{V}  postuliert lineares Confinement~-- f"ur die QCD, nicht aber f"ur die QED.
\vspace*{-.25ex}

\bigskip\noindent
Wir geben die Ans"atze des \DREI{M}{S}{V} f"ur die Korrelationsfunktionen~$D\uC$ und~$D\uNC$ an.
Sei ver\-wiesen auf unserer Diplomarbeit, Ref.~\cite{Kulzinger95}. \\
\indent
Wir schreiben die rechte Seite von Gl.~(\ref{Dvier_DDxi0}) als ihr Fourier-Integral.
Sei~$k$ der zu~$\xi$ konjugierte dimensionslose Impuls [$k$ in Einheiten von~$a^{-1}$, da~$\xi$ in Einheiten von~$a$], seien~\mbox{$\tilde{D}\uC$},~\mbox{$\tilde{D}\uNC$} die Fourier-transformierten Korrelationsfunktionen.
Dann folgt
\begin{samepage}
\vspace*{-.5ex}
\begin{align} \label{Dvier_DDk}
D_{\mu\nu\rh\si}\!(\xi)\;
  =\; \int \frac{d^4k}{(2\pi)^4}\; \efn{\D-\iIM\, k \!\cdot\! \xi}\;
      \Big[\;
        \vka\; \tilde{t}\oNC{}_{\zzzz \mu\nu\rh\si}\vv \tilde{D}\uC(k^2)\;
        +\; (1\!-\!\vka)\; \frac{1}{2}\; \tilde{t}\oNC{}_{\zzzz \mu\nu\rh\si}\vv
               \tilde{D}^{\D\prime}\uNC(k^2)\; \Big]
    \\[-8.75ex]\nn
\end{align}
mit
\vspace*{-1.25ex}
\begin{alignat}{4} \label{tC,tNC-FT}
&\tilde{t}\oC{}_{\zzzz \mu\nu\rh\si}\;&
  &=\; t\oC{}_{\zzzz \mu\nu\rh\si}\;&
  &=\;& \frac{1}{12}&\vv \de^{\al\be}_{\mu\nu}\vv g_{\al\rh}\, g_{\be\si}
    \\
&\tilde{t}\oNC{}_{\zzzz \mu\nu\rh\si}\;&
  &=\; t\oNC{}_{\zzzz \mu\nu\rh\si} \Big|_{\T \pa\to -\iIM\, k}\;&
  &=\;& -\, \frac{1}{6}&\vv g_{\al\ga}\vv
         \de^{\al\be}_{\mu\nu}\, k_{\be}\vv
         \de^{\ga\de}_{\rh\si}\, k_{\de}
    \tag{\ref{tC,tNC-FT}$'$}
    \\[-4.5ex]\nn
\end{alignat}
Der Strich bezeichnet im Sinne~\mbox{\,$\tilde{D}^{\D\prime}\uNC(k^2) \!\equiv\! d/dk^2\, \tilde{D}\uNC(k^2)$} die totale Ableitung einer Funktion nach ihrem Argument; die Tensoren~\mbox{\,$\tilde{t}\oC$},~\mbox{\,$\tilde{t}\oNC$} kontrahieren zu:
\vspace*{-.25ex}
\begin{align} \label{tsvier_kkontr}
&\tilde{t}\oC{}_{\zzz \mu\nu}{}^{\mu\nu}\;
  =\; 1
    \\[.25ex]
&\tilde{t}\oNC{}_{\zzz \mu\nu}{}^{\mu\nu}\;
  =\; -\, k^2 
    \tag{\ref{tsvier_kkontr}$'$}
    \\[-4.25ex]\nn
\end{align}
vgl.\@ die Gln.~(\ref{tsvier_zkontr}),~(\ref{tsvier_zkontr}$'$).
Daraus folgt
\vspace*{-.5ex}
\begin{align} \label{Dvier_DDkkontr}
&D(\xi^2)\;
  \equiv\; D_{\mu\nu}{}^{\mu\nu}\!(\xi)
    \\
  &\phantom{D(\xi^2)\;}
   =\; \int \frac{d^4k}{(2\pi)^4}\; \efn{\D-\iIM\, k \!\cdot\! \xi}\;
         \Big[ \vka\; \tilde{D}\uC(k^2)
          \;+\; (1\!-\!\vka)\;
                \Big( -\frac{1}{2}\; k^2\; \tilde{D}^{\D\prime}\uNC(k^2) \Big) \Big]
    \nn
    \\[-4.5ex]\nn
\end{align}

Dies ist eine Funktion, die im Rahmen von QCD als $\SUNc$-Gittereichtheorie numerisch berechnet wird, vgl.\@ die Refn.~\cite{Campostrini89,DiGiacomo90,DiGiacomo92} und~\cite{DElia97,DElia97a,DElia97b} bez"uglich QCD ohne beziehungsweise mit dynamischen Fermionen.
Ref.~\cite{DiGiacomo92} zeigt f"ur explizit~$\Nc \!\equiv\! 3$, da"s beide Tensorstrukturen im (Gitter-)relevanten Bereich des Ortsraums {\it proportionales Abklingverhalten\/} zeigen.
Ref.~\cite{Kraemer91} macht daher, zun"achst f"ur beliebiges~$\vka$, den Ansatz:
\vspace*{-.5ex}
\begin{align} \label{Dvier_DDkkontrAnsatz}
&\tilde{D}(k^2)\;
  \equiv\; \vka\; \tilde{D}\uC(k^2)\;
       +\; (1\!-\!\vka)\;
         \Big( -\frac{1}{2}\, k^2\, \tilde{D}^{\D\prime}\uNC(k^2) \Big)
    \\
  &\phantom{D(\xi^2)\;}
   \stackrel{\D!}{=}\vv
       -6\iIM\, A_\nu\, \la_\nu^6\; \frac{k^2}{(\la_\nu^2 k^2 - 1 +\iIM\,\ep)^\nu}\qqquad
    \forall\, \vka,\vv
    \nu \in \bbbc,\vv {\rm Re}\, \nu > 3
    \nn
    \\[-4.5ex]\nn
\end{align}
\end{samepage}%
Dieser Ansatz ist die unmittelbare Verallgemeinerung des Feynman-Propagators%
  ~\vspace*{-.125ex}\mbox{\,$\tilde\De_{\mskip-1mu F}$}, wie wir explizit zeigen in \vspace*{-.125ex}Ref.~\cite{Kulzinger95}, Anh.~B.
Sie impliziert dessen Polstruktur in der komplexen $k^0$-Ebene, das hei"st die Existenz keiner weiteren Pole als%
  ~\mbox{\,$k^0 \!=\! \pm\surd m^2 \!+\! \vec{k}{}^2$}, mit%
  ~\mbox{\,$m^2 \!\equiv\! \la_\nu^{-2}$}, und die wohlbekannte mit diesen assoziierte Epsilon-Vorschrift zur Integration entlang der Achse~\mbox{\,${\rm Re}k^0$}.
Die Wick-Drehung in Euklidische Raumzeit geschieht daher vollst"andig analog; in Hinblick auf die Analytische Fortsetzung der $T$-Amplitude in Kapitel~\ref{Kap:ANALYT} f"uhren wir sie f"ur Definiertheit explizit durch in Anhang~\ref{APP-Subsect:Wick}.
Unter dieser Wick-Drehung gehen die Funktionen~\mbox{\,$D\uC$},~\mbox{\,$D\uNC$} "uber in die {\it Euklidischen Korrelationsfunktionen\/}~\mbox{\,$\xE[D\uC]$},~\mbox{\,$\xE[D\uNC]$}.
Diese h"angen ab von%
  ~\mbox{\,$-\xE[\xi]^2 \!=\! \xE[\xi]^\mu \xE[\xi]^\mu \!>\! 0$}, dem {\it negativ definiten\/} Euklidischen Vier-Skalarprodukt, vgl.\@ Def.\@ in Gl.~(\ref{APP:xcdoty-E}).
Die Funktionen~\mbox{\,$\xE[D]{}\uC$},~\mbox{\,$\xE[D]{}\uNC$} sind dort in der Weise konstruiert, da"s sie f"ur%
  ~\mbox{\,$\nu \!\in\! \bbbn \!>\! 3$} und gro"ses reell positives Argument%
  ~\mbox{\,$-\xE[\xi]^2$} {\it exponentielles Abfallverhalten\/} zeigen:%
  ~\mbox{\,$\sim\! \exp\sqrt{\!-\xE[\xi]^2}$}.
Sie sind daher zu interpretieren als {\it Korrelationsfunktionen\/}.
Diese Eigenschaft "ubertr"agt sich unmittelbar auf die {\it Minkowskischen Korrelationsfunktionen\/}~\mbox{\,$D\uC(\xi^2)$},~\mbox{\,$D\uNC(\xi^2)$} f"ur gro"ses negatives, das hei"st {\it raumartiges\/} Argument.

Die a~priori freien Konstanten~$A_n$ und~$\la_n$ in Gl.~(\ref{Dvier_DDkkontrAnsatz}) sind so zu bestimmen, da"s sie zum einen die durch Gl.~(\ref{K2-g2FF_Dvier}) festgelegte Normierung von~$D(\xi^2) \!=\! D_{\mu\nu}{}^{\mu\nu}\!(\xi)$, vgl.\@ Gl.~(\ref{Dvier_DDkkontrAnsatz}), garantieren und zum anderen der~$\xi$ impliziten Konstanten~$a$ die Bedeutung einer Korrelationsl"ange geben.
Dies geschieht "uber die Forderungen
\vspace*{-.5ex}
\begin{align} \label{Ala_Bedingg}
1\;
  \stackrel{\D!}{=}\; D \Big|_{\D\xi^2 \!\equiv\! 0}\qquad
  \text{und}\qquad
1\;
  \stackrel{\D!}{=}\; \int_0^{\infty} \! du\; D(-u^2)
    \\[-4.5ex]\nn
\end{align}
Durch die zweite Bedingung~-- an das Abfallverhalten von~\mbox{$D$} f"ur raumartiges Argument, ergo von~\mbox{$\xE[D]$}~-- erh"alt die $\xi$ implizite Skala~$a$ die Bedeutung einer Korrelationsl"ange.
Es folgt~-- vgl.\@ die Gln.~(\ref{APP:Anu-Minkowski}),~(\ref{APP:lanu-Minkowski}):%
\FOOT{
  \label{FN:AnhangCLTFN}Im Zuge der Herleitungen in Anhang~\ref{APP:CLTFN} werden verifiziert die Gln.~(\ref{Dvier_DDk}),\,(\ref{A,la-explizit}),\,(\ref{A,la-explizit}$'$),\,(\ref{Dvier_DDxi}); vgl.\@ dort.
}
%
\vspace*{-1.5ex}
\begin{alignat}{2} \label{A,la-explizit}
&A_\nu\;
  =\; -\, \efn{\T\nu\pi\iIM}\cdot
         \frac{4\pi^2}{3}\,
         \frac{\Ga(\nu)}{\Ga(\nu \!-\! 3)}\quad
  =\; -\, \efn{\T\nu\pi\iIM}\cdot 8\pi^2\, \pmatrixZE{\nu\!-\!1}{3}&&
    \\[-.25ex]
  &\la_\nu\;
  =\; \frac{4}{3}\,
      \frac{\Ga(\nu \!-\! 3)}{%
            \Ga\big(\frac{1}{2}\big)\,
              \Ga\big((\nu \!-\! 3) \!+\! \frac{1}{2}\big)}\quad
  =\; \frac{8}{3\pi}\,
      \frac{\Ga(\nu \!-\! 3)\!\cdot\! \Ga\big(\frac{3}{2}\big)}{%
            \Ga\big((\nu \!-\! 3) \!+\! \frac{1}{2}\big)}&&
    \tag{\ref{A,la-explizit}$'$} \\[-2.5ex]
  &&&\qqquad
    {\rm Re}\, \nu > 3
    \nn
    \\[-5.5ex]\nn
\end{alignat}
der Binomialkoeffizient f"ur Eintr"age~$\in\! \bbbc$ in Standard-Definition. \\
\indent
Forderung von Gl.~(\ref{Dvier_DDkkontrAnsatz}) f"ur beliebige Parameterwerte~$\vka$ impliziert:%
\FOOT{
  gesetzt~$\vka \!=\! 0$, beziehungsweise~$\vka \!=\! 1$
}
%
\vspace*{-.75ex}
\begin{align} \label{DDk}
\tilde{D}\uC(k^2)\;
  =\; -\, \frac{1}{2}\, k^2\, \tilde{D}^{\D\prime}\uNC(k^2)\;
  =\; 6\iIM\, A_\nu\cdot \la_\nu^2 (-k^2)\vv
        \la_\nu^4\; \frac{1}{(\la_\nu^2 k^2 - 1 +\iIM\,\ep)^\nu}
    \\[-4.75ex]\nn
\end{align}
Diese explizite Relation zwischen den Korrelationsfunktionen~\mbox{\,$\tilde{D}\uC$} und~\mbox{\,$\tilde{D}\uNC$}~\mbox{implementieren} wir in Ref.~\cite{Kulzinger95}, Anh.~B in der Weise, da"s folgt die Darstellung, die wir auch der vorliegenden Arbeit zugrunde legen~-- vgl.\@ die Gln.~(\ref{APP:FT_F,D-C}),~(\ref{APP:FT_F,D-NC}),~(\ref{APP:F^C,F^NC}), in~\mbox{\,$d \!\equiv\! 4$} Dimensionen:%
\citeFN{FN:AnhangCLTFN}
\vspace*{-.25ex}
\begin{align} \label{Dvier_DDxi}
&\hspace*{-8pt}
 D_{\mu\nu\rh\si}\!(\xi)\;
  =\; \vka\vv t\oC{}_{\zzzz \mu\nu\rh\si}\; \pa^2\vv F\oC(\xi^2)
        +\; (1\!-\!\vka)\vv t\oNC{}_{\zzzz \mu\nu\rh\si}\vv F\oNC(\xi^2)
    \\[.75ex]
  &\hspace*{-8pt}
   \begin{alignedat}[t]{2}
    \text{mit}\qquad
      F\oC(\xi^2)\; \equiv\; F\oNC(\xi^2)\;
        &=\; \la_\nu^2\;
               \frac{\nu \!-\! 1 \!-\! d\!/\!2}{d\!/\!2}\cdot
               {\cal K}_{\nu-\frac{d}{2}}\!(\ze)
             \;\Big|_{d\equiv4}\vv&&
    \\[-.5ex]
        &=\; \la_\nu^2\, \frac{\nu \!-\! 3}{2}\cdot {\cal K}_{\nu-2}(\ze)\vv&
     &\ze \equiv \sqrt{-\xi^2/\la_\nu^2 + \iIM\,\ep}
   \end{alignedat}
    \nn
    \\[-5.5ex]\nn
\end{align}
und
\vspace*{-1.25ex}
\begin{align} \label{calKmu-Def}
{\cal K}_\mu(\ze)\;
  \equiv\; \Big(\frac{1}{2}\Big)^{\!\mu-1}\! \frac{1}{\Ga(\mu)}\vv
         \ze^\mu\,
         {\rm K}_\mu(\ze)\qquad
     \underset{\ze\to0}{\longrightarrow}\vv 1,\quad
     \forall{\rm Re}\, \mu > 0
    \\[-5ex]\nn
\end{align}
Bzgl.\@ der Tensorkomponenten~\mbox{$t\oC{}_{\zzzz \mu\nu\rh\si}$},~\mbox{$t\oNC{}_{\zzzz \mu\nu\rh\si}$} vgl.\@ die Gln.~(\ref{Dvier_DDxi0}),\,(\ref{Dvier_DDxi0}$'$).
\mbox{Die Funktion~${\cal K}_\mu$ ist} definiert "uber~${\rm K}_\mu$, die modifizierte Besselfunktion zweiter Art mit Index~$\mu$ in der~\mbox{Konvention} von Ref.~\cite{Abramowitz84}; sie ist {\it per~constructionem\/} normiert auf Eins f"ur verschwindendes Argument. \\
\indent
Die modifizierte Besselfunktion%
  ~\vspace*{-.125ex}\mbox{\,${\rm K}_\mu(\ze),\, \forall\mu \!\in\! \bbbc$} geht gegen Null wie%
  ~\mbox{\,$\sim\! \exp\!-|\ze|$} f"ur~\mbox{\,$|\ze| \!\to\! \infty$} in der rechten Halbebene:%
  ~\mbox{$|\Arg\,\ze| \!<\! \frac{\pi}{2}$}, und divergiert wie~\mbox{\,$\sim\! \exp\!|\ze|$} \pagebreak\mbox{auf der imagin"aren Achse};
vgl.\@ Ref.~\cite{Kulzinger95}, Anh.~B und Ref.~\cite{Abramowitz84}.
Die Epsilon-Vorschrift~\mbox{$\ep \!\to\! 0\!+$} ist daher {\it essentiell\/} f"ur zeitartige und {\it redundant\/} f"ur raumartige~$\xi$.
In nichtperturbativer Streuung im Limes unendlicher Schwerpunktenergie,%
  ~\mbox{\,$\surd s \!\to\!\infty$}, zeigen wir in Ref.~\cite{Kulzinger95}~-- und f"ur den Fall gro"ser aber endlicher Werte~\mbox{\,$\surd s$} im folgenden Kapitel~\ref{Kap:ANALYT}~--, da"s nur {\it raumartige\/} Differenzvektoren~$\xi$ beitragen: das hei"st%
  ~\mbox{\,$\xi^2 \!<\! 0$} ergo%
  ~\mbox{\,$\ze \!\equiv\! \sqrt{-\xi^2/\la_\nu^2 + \iIM\,\ep} \!\cong\! \sqrt{-\xi^2}\big/\la_\nu \!\in\! \bbbr^+$}. \\
\indent
Wir zeigen in Ref.~\cite{Kulzinger95}~-- vgl.\@ auch Anh.~\ref{APP:CLTFN}~-- vgl.\@ auch, da"s die Darstellung von~\mbox{\,$D_{\mu\nu\rh\si}$} durch Gl.~(\ref{Dvier_DDxi}) g"ultig ist f"ur Indizes~\mbox{\,$\nu \!\in\! \bbbc,\, {\rm Re}\nu \!>\! 3$}.
Die beste Anpassung an die Gitterdaten ist gegeben durch~\mbox{\,$\nu$} als der kleinstm"oglichen nat"urlichen Zahl, vgl.\@ die Refn.~\cite{DiGiacomo92,Kraemer91,Dosch94a}.
Sei in diesem Sinne gesetzt~\mbox{\,$\nu \!\equiv\! 4$} und der Index im folgenden unterdr"uckt:
\vspace*{-.75ex}
\begin{align} \label{A,la-n=4explizit}
&A\;
  \equiv\; A_4\;
  =\; -\, 8\pi^2
    \\[-.5ex]
&\la\;
  \equiv\; \la_4\;
  =\; \frac{8}{3\pi}
    \tag{\ref{A,la-n=4explizit}$'$}
    \\[-4.5ex]\nn
\end{align}
vgl.\@ die Gln.~(\ref{A,la-explizit}),~(\ref{A,la-explizit}$'$). \\
\indent
Aus zitierter Ref.~\cite{DiGiacomo92} wird extrahiert die Gewichtung der Tensorstruktur~\mbox{$t\oC$} versus~\mbox{$t\oNC$}; es folgt~\mbox{$\vka \!\cong\! 0.74$}.
Dieser Zahlenwert schr"ankt ein die Werte f"ur Gluonkondensat~\mbox{$\vac{g^2FF}$}~und Korrelationsl"ange~{$a$}~-- der zwei Zahlen-Parameter des \DREI{M}{S}{V}; vgl.\@ unten auf Seite~\pageref{T:Parameter}.
\vspace*{-.5ex}

\bigskip\noindent
Dies die Annahmen und Ans"atze des \DREI[]{M}{S}{V}, die gestatten seine Anwendung auf den Vakuum\-erwartungswert {\it eines\/} Wegner-Wilson-Loops~\mbox{$\vac{W({\cal C})}$}.
Es kann Confinement eines statischen Quark-Antiquark-Paares quantifiziert werden:
die Stringspannung~$\si$, vgl.\@ Gl.~(\ref{Stringspannung-si_DC}) und die Refn.~\cite{Dosch87,Dosch88},
die spinabh"angigen Terme, Ref.~\cite{Marquard87},
das transversale Profil des konfinierenden gluonischen Strings, Refn.~\cite{Rueter94,Rueter94a,Dosch95},~-- als nur die wichtigsten Anwendungen in der Niederenergiephysik. \\
\indent
Wir betrachten Hochenergiephysik.
Im Limes unendlicher invarianter Schwerpunktenergie~$\surd s$ leitet Nachtmann in Ref.~\cite{Nachtmann91} die $T$-Amplitude her f"ur die Streuung von (Anti)Quarks und verallgemeinert sie in Ref.~\cite{Nachtmann96} auf Gluonen und eichinvariante Quark-Antiquark-Paare.
Sie ist wesentlich bestimmt durch den Vakuumerwartungswert zweier Wegner-Wilson-Linien entlang der klassischen im Limes~$s \!\to\! \infty$ lichtartigen Trajektorien der Partonen, in deren Darstellung der~$\SUNc$~-- f"ur eichinvariante Quark-Antiquark-Paare durch den Vakuumerwartungswert zweier lichtartiger Wegner-Wilson-Loops der fundamentalen Darstellung~$\mf{F}$.

Wir ben"otigen daher eine Formel zur Berechnung des Vakuumerwartungswertes {\it zweier\/} Wegner-Wilson-Loops~$\vac{W({\cal C}\Dmfp)W({\cal C}\Dmfm)}$.
In Ref.~\cite{Kraemer91} wird gezeigt f"ur den Vakuumerwartungswert {\it eines\/} Wegner-Wilson-Loops, da"s in sinnvoller Approximation dasselbe Resultat folgt aus der strengen Entwicklung in Kumulanten nach van Kampen, das hei"st auf Basis der Faktorisierung in die Kumulante zweiter Ordnung nach Gl.~(\ref{zentrGauss1}) auf {\it Matrixniveau\/}:
\begin{samepage}
\vspace*{-.25ex}
\begin{align} \label{Fak_Matrix}
&B_n(t_1,\ldots t_n)
  = c_n\; \permII{B_2^{\;n\!/\!2}}_{t_1,\ldots t_n}
    \\[1ex]
&\begin{alignedat}[t]{4}
  \text{mit}\quad
  &B_n(t_1,\ldots t_n)&\vv &\longrightarrow&\vv
    &\vac{ P\, g^n F^{(1)} \cdots F^{(n)} }&&
    \\
  &K_2(t_1,t_2)&\vv &\longrightarrow&\vv
    &K_2(x_1,x_2)&\qquad
    &\text{f"ur\vv $n \!=\!2$}
  \end{alignedat}
    \nn 
    \\[-4ex]\nn
\end{align}
und auf Basis der Faktorisierung auf {\it Matrixelementeniveau\/} wie folgt:
\vspace*{-.25ex}
\begin{align} \label{Fak_MatrixElement}
&B_{n,\,a_1\cdots a_n}(t_1,\ldots t_n)
  = c_n\; \permII{B_2^{\;n\!/\!2}}_{(t_1,a_1),\ldots (t_n,a_n)}
    \\[1ex]
&\begin{alignedat}[t]{4}
  \text{mit}\quad
  &B_{n,\,a_1\cdots a_n}(t_1,\ldots t_n)&\vv &\longrightarrow&\vv
    &\vac{ P\, g^n F^{(1)}{}_{\zz a_1} \cdots F^{(n)}{}_{\zz a_n} }&&
    \\
  &(K_2)_{a_1a_2}(t_1,t_2)&\vv &\longrightarrow&\vv
    &(K_2)_{a_1a_2}(x_1,x_2)&\qquad
    &\text{f"ur\vv $n \!=\!2$}
  \end{alignedat}
    \nn 
    \\[-4ex]\nn
\end{align}
\end{samepage}%
Es ist \mbox{$F^{(i)} \equiv F_{\mu_i\nu_i}\!(x_i; x_0,{\cal C}_{x_{\!0}\!x_{\!i}})$}, vgl.\@ Gl.~(\ref{WW-Loop_KumEntw}).
Bzgl.\@ der Notation~$\permII{\;\cdot\;}$ vgl.\@ Gl.~(\ref{permII}), bzgl.\@ der Ersetzung~$B\!\to\!F$ Gl.~(\ref{BnachF}).
Im Gegensatz zu Gl.~(\ref{Fak_Matrix}) ist Gl.~(\ref{Fak_MatrixElement}) unmittel\-bar~anwendbar auf den Erwartungswert zweier Loops.
Statt Annahme~$(2)$ eines Gau"s'schen stochastischen Prozesses f"ur~$\vac{\;\cdot\;}$ machen wir {\bf \label{Annahme(2')}Annahme~\bm{(2')}: Faktorisierung der Vaku\-umerwartungswerte}%
~\mbox{\bm{\vac{ P\, g^n F_{\mu_1\nu_1 a_1}\!(x_1; x_0,{\cal C}_{x_{\!0}\!x_{\!1}}) \cdots F_{\mu_n\nu_n a_n}\!(x_n; x_0,{\cal C}_{x_{\!0}\!x_{\!n}}) }}}; wir nehmen an Gl.~(\ref{Fak_MatrixElement}) statt Gl.~(\ref{Fak_Matrix}).
In Hinblick auf sp"ater halten wir fest
\vspace*{-.5ex}
\begin{align} \label{B4-Fak_MatrixElement}
&\vac{1,2,3,4}\;
  =\; \vac{1,2}\, \vac{3,4} + \vac{1,3}\, \vac{2,4} + \vac{1,4}\, \vac{2,3}
    \\[.25ex]
&\text{mit}\qquad
  \vac{i_1,\ldots i_n}\;
  \equiv\; \vac{ P\, g^n F^{(i_1)}{}_{\zz a_1} \cdots F^{(i_n)}{}_{\zz a_n} }
    \nn
    \\[-4.5ex]\nn
\end{align}
als das Pendant zu Gl.~(\ref{B4-Fak_Matrix}) auf Matrixelementeniveau.

Im Vakuumerwartungswert~$\vac{\;\cdot\;}$ zweier Wegner-Wilson-Loops%
~\mbox{$W({\cal C}\Dmfp) \!\equiv\! W({\cal S}(\tilde{\cal C}\Dmfp))$}
und \vspace*{-.25ex}\mbox{$W({\cal C}\Dmfm) \!\equiv\! W({\cal S}(\tilde{\cal C}\Dmfm))$} treten a~priori zwei Pfadordnungsoperatoren auf:%
~\vspace*{-.25ex}$P_{\tilde{\cal C}\Dmfp}$ bez"uglich der Deformation~$\tilde{\cal C}\Dmfp$
und~\vspace*{-.25ex}$P_{\tilde{\cal C}\Dmfm}$ bez"uglich~$\tilde{\cal C}\Dmfm$.
Entsprechend in den Vakuumerwartungswerten, die folgen aus der Entwicklung der Loop-Exponentiale.
Die Gln.~(\ref{Fak_Matrix}),~(\ref{Fak_MatrixElement}) sind weiterhin anwendbar auf Basis der folgenden "Uberlegung, die Nachtmann vorschl"agt in Ref.~\cite{Nachtmann96}:
Die Referenzpunkte der Loops, die Konsequenz sind des "Ubergang von Eichfeld zu Eichfeldst"arke mithilfe des Nichtabelschen Stokes'schen Satzes, werden identifiziert, so da"s sich die Integrationsfl"achen~${\cal S}(\tilde{\cal C}\Dmfp)$ und~${\cal S}(\tilde{\cal C}\Dmfm)$ ber"uhren und ein einziger Pfadordnungsoperator~$P \!\equiv\! P_{\tilde{\cal C}}$ definiert werden kann, der sich bezieht auf die Verkettung~$\tilde{\cal C} \!\equiv\! \tilde{\cal C}\Dmfp \circ \tilde{\cal C}\Dmfm$ der Deformationen.
\vspace*{-.5ex}

\bigskip\noindent
\label{T:ProduktKumulanten}Wir bemerken abschlie"send, da"s Berger, Nachtmann in Ref.~\cite{Berger98} angeben eine formale Entwicklung in Kumulanten f"ur den Vakuumerwartungswert zweier Wegner-Wilson-Loops, die wir skizzieren wie folgt.%
\FOOT{
  Besten Dank Edgar Berger und Timo Paulus f"ur die Diskussion.
}
Grundlegend ist zun"achst die Darstellung des Produkts zweier Loop-Exponentiale als ein einziges Loop-Exponential eines Produkt-Raumes [der Raum des direkten Produkts zweier fundamentaler~$\SUNc$-Darstellungen~\mbox{$T_\mf{F}^a \!\otimes\! T_\mf{F}^a$}]; die Spuren gehen "uber in eine "'Produkt-Spur"'~[\mbox{$\trDrst{F} \!\otimes\! \trDrst{F}$}].
Auf das eine "`Produkt-Exponential"' wird bezogen Annah\-me~(1), vgl.\@ Gl.~(\ref{WW-Loop_KumEntw}): die Annahme, da"s existiert seine Entwicklung in "`Produkt-Kumulan\-ten"'~$k_{\mskip-1.5mu\otimes\mskip-1mu n}$.
Die "`Produkt-Kumulante"' zweiter Ordnung~$k_{\mskip-1.5mu\otimes\mskip-1mu 2}$ wird ausgedr"uckt durch die Vakuumerwartungswerte~\mbox{$\vac{ P\, g^n F^{(1)}{}_{\zz a_1} \cdots F^{(n)}{}_{\zz a_n} }$} in den "`Faktor-R"aumen"' der initialen Loops.
F"ur diese wird wieder gemacht Annahme~(2'): angenommen Faktorisierung auf Matrixelementeniveau in den "`Faktor-R"aumen"', vgl.\@ Gl.~(\ref{Fak_MatrixElement}); die Vakuumerwartungswerte~\mbox{$\vac{ P\, g^2 F^{(1)}{}_{\zz a_1} F^{(2)}{}_{\zz a_2} }$} werden ausgedr"uckt durch den Korrelationstensor~$D$ nach Gl.~(\ref{K2-g2FF_Dvier}) und f"ur diesen gemacht wie hier Annahme~(3).
Der Korrelationstensor~$D$ tritt auf in Form derselben Funktion~$\ch$ wie hier.
Ausf"uhren der "`Produkt-Spur"' f"uhrt zur"uck in die "`Faktor-R"aume"' und stellt her den abschlie"senden funktionalen Zusammenhang von~$\ch$.

Berger, Nachtmann nehmen an dieselbe Faktorisierung, sie finden~$k_{\mskip-1.5mu\otimes\mskip-1mu 1} \!\equiv\! 0$ und vernachl"assigen h"ohere "`Produkt-Kumulanten"'~$k_{\mskip-1.5mu\otimes\mskip-1mu n}$,~\mbox{$n \!>\! 2$}.
Konsequenz ist, da"s die~\mbox{$T$-Ampli}\-tude der Streuung dort bestimmt ist in Form~\vspace*{-.25ex}\mbox{\,$\big(\frac{2}{3}\efn{-\iIM\frac{1}{3}\ch} \!+\! \frac{1}{3}\efn{\iIM\frac{2}{3}\ch}\big) \!-\! 1 \cong -\ch^2 + {\cal O}(\ch^3)$},~wohin\-gegen hier in Form~\mbox{\,$-\ch^2$}; vgl.\@ unten Fu"sn.\,\FNg{FN:ch-Berger,Nachtmann}.
Berger, Nachtmann summieren eine bestimmte Klasse von Beitr"agen auf; die Diskrepanz beider Formeln verschwindet f"ur gro"se transversale Sto"sparameter~$\rb{b}$, f"ur die~$\ch$ klein ist; vgl.\@ die Diskussion in Ref.~\cite{Berger98}.

Der eine wie der andere Zugang basiert auf Annahmen, die a~priori nicht mehr oder weniger gerechtfertigt sind.%
\FOOT{
  Die Annahme von Konvergenz der Entwicklung in "`Produkt-Kumulanten"'~$k_{\mskip-1.5mu\otimes\mskip-1mu n}$ ist gleichbedeutend mit der Existenz eines stochastischen Prozesses im Produkt-Raum; vgl.\@ van Kampen in den Refn.~\cite{VanKampen74,VanKampen76}.   Statt zu "ubernehmen die Ad~hoc-Hypothese von Faktorisierung auf Matrixelementeniveau in den "`Faktor-R"aumen"', scheint es uns logisch konsequenter, eine Annahme zu machen f"ur diesen stochastischen Proze"s.   Dies ist dann {\sl auch physikalisch\/} sinnvoll, wenn die Entwicklung selbst sinnvoll ist~-- im Sinne, da"s sie geschieht in einem kleinen Entwicklungsparameter.   Die Annahme eines (zentrierten) Gau"s'schen Prozesses sollte bereits gute Approximation sein und die einzige von Null verschiedene "`Produkt-Kumulante"'~$k_{\mskip-1.5mu\otimes\mskip-1mu 2}$ unmittelbar ausdr"uckbar durch den Korrelationstensor~$D$~-- ohne weitere Faktorisierungshypothese.
}
Wir arbeiten mit der ausf"uhrlich vorgestellten urspr"unglichen Formulierung des \DREI{M}{S}{V}.
\theendnotes

%% file: ANALYT-F.tex
\lhead[\fancyplain{}{\sc\thepage}]%
      {\fancyplain{}{\sc\rightmark}}
\rhead[\fancyplain{}{\sc{Analytische nahezu lichtartige \protect$T$-Amplitude}}]%
      {\fancyplain{}{\sc\thepage}}
%

%
\chapter[Analytische nahezu lichtartige \bm{T}-Amplitude]{%
   \huge Analytische\\ nahezu lichtartige \bm{T}-Amplitude}
\label{Kap:ANALYT}
\enlargethispage{1.05ex}

Wir betrachten nichtperturbative Hochenergiestreuung zweier Wegner-Wilson-Loops: bei gro"sem invarianten Quadrat~$s$ der Schwerpunktenergie und kleinem invarianten Quadrat~$-t$ des Impuls"ubertrags.
Wir zeigen {\it Existenz\/} und {\it "Aquivalenz\/} zweier Formulierungen des~\DREI[]{M}{S}{V}~-- der auf der physikalischen Minkowskischen Raumzeit und der auf deren analytischen Fortsetzung ins Euklidische.
Dies impliziert Aufl"osung des~-- scheinbaren~-- \mbox{Widerspruchs von a\,prio}\-ri nur im Euklidischen fundierten Annahmen aber im Minkowskischen konstituiertem~\DREI[]{M}{S}{V}.

Wir argumentieren, da"s die von Nachtmann im Limes~$s \!\to\! \infty$ hergeleitete nichtperturbative $T$-Amplitude f"ur die Parton-Parton-Streuung von (Anti)Quarks und Gluonen in ihrer G"ultigkeit verallgemeinert werden kann:
Sie gibt f"ur gro"se aber endliche Werte von~$s$ das f"uhrende Verhalten in~$s$ im Sinne einer {\it leading log\/}-Approximation korrekt wider.
Dies genau dadurch, da"s die klassischen Trajektorien der Partonen genommen werden als die {\it physikalischen mit kleiner zeitartigen Komponente\/} statt ersetzt werden durch deren f"ur~$s \!\to\! \infty$ {\it lichtartige\/} Limites.~--
Betrachtet von Nachtmanns Formel her: dadurch da"s die lichtartigen Trajektorien vom Lichtkegel "`ein wenig wegger"uckt"' werden.

Die Formel Nachtmanns ist im wesentlichen bestimmt durch den Vakuumerwartungswert~$\vac{\;\cdot\;}$ zweier {\it Wegner-Wilson-Linien\/}, das hei"st Konnektoren~$\Ph$ in der den Partonen entsprechenden Darstellung~$\Drst{R}$ der Eichgruppe entlang deren klassischen Trajektorien.
Die Streuung zweier Quark-Antiquark-Paaren ist im wesentlichen bestimmt durch den Vakuumerwartungswert zweier {\it Wegner-Wilson-Loops\/}, die zustande kommen durch orientiertes Verbinden der Trajektor-Konnektoren durch Konnektoren bei gro"sen positiven und negativen Zeiten, formal bei~$\pm\infty$.
Wir zeichnen Nachtmanns Herleitung nach bezogen auf physikalische klassische Trajektorien {\it nahe\/} des Lichtkegels~-- so explizit, wie die Klarheit der Darstellung erfordert~-- und gelangen zu einer Formel f"ur die $T$-Amplitude, die im wesentlichen bestimmt ist durch den Vakuumerwartungswert zweier Wegner-Wilson-Loops {\it nahe\/} des Lichtkegels.

Diesen Vakuumerwartungswert werten wir zum einen aus mithilfe des \DREI{M}{S}{V} in der Formulierung des vorangehenden Kapitels, das hei"st in der Minkowskischen Theorie.
Zum anderen kann er~-- da Funktional {\it zeit-\/} und nicht {\it lichtartiger\/} Trajektorien~-- analytisch fortgesetzt werden ins Euklidische.
Diese Fortsetzung werten wir aus mithilfe der Formulierung des \DREI{M}{S}{V} auf der ins Euklidische fortgesetzten Raumzeit, das hei"st in der Euklidischen Theorie.

Wir finden, da"s beide Darstellungen der $T$-Amplitude verkn"upft sind durch die analytische Fortsetzung eines {\it hyperbolischen Winkels\/}~\mbox{$\ps$} zu einem {\it Euklidischen Winkel\/}~\mbox{$\th$}, die~\mbox{definie}\-ren: \mbox{$\ps$}~den {\it aktiven Lorentz-Boost\/} der beiden Wegner-Wilson-Linien gegeneinander in der Minkowski-Raumzeit,~\mbox{$\th$} deren {\it$O(4)$-Verdrehung\/} gegeneinander in der ins Euklidische fortgesetzten Raumzeit.
Wir verallgemeinern damit ein f"ur Quark-Quark-Streuung ver"offentlichtes etabliertes  Resultat auf den Fall der Streuung eichinvarianter, physikalischer Objekte. \\
\indent
Die Formulierung der Euklidischen Theorie besitzt eine Reihe von Aspekten von Bedeutung.
So liegt der Argumentation hin zu den Annahmen des \DREI{M}{S}{V} wesentlich zugrunde eine Raumzeit mit {\it Euklidischer (Riemann\-scher)\/} statt {\it Minkowskischer (pseudo-Riemannscher)\/} Metrik:
Korrelation paralleltransportierter Feldst"arken~-- Observabler "uberhaupt~-- macht Sinn bez"uglich Euklidischer Sph"aren, nicht aber (wozu diese unter Fortsetzung werden) bez"uglich Minkowskischer Hyperboloide, die sich entlang des Lichtkegels bis ins Unendliche erstrecken, und so implizierten {\it maximale Korrelation\/} aller lichtartig separierten Weltpunkte unabh"ngig von deren r"aumlichen Separation.
Dar"uberhinaus ist die Formulierung der Euklidischen Theorie essentielle Voraussetzung daf"ur, Hochenergiestreuung auszuwerten im Rahmen numerischer Simulationen von QCD als Gittereichtheorie. \\
\indent
"Uberblickartig diskutieren wir das resultierende f"uhrende Verhalten der $T$-Amplitude f"ur gro"ses~$s$.
Dies geschieht in Referenz auf Quark-Quark-Streuung, in der das "`Wegr"u"cken"' der klassischen Partontrajektorien vom Lichtkegel zu verstehen ist als $s$-Regularisierung $\ln s$-divergenter Feynman-Loop-Diagramme.
Summation von $\ln s$-Termen aber mu"s auf die experimentell beobachtete $s$-Asymptotik f"uhren, die f"ur die $T$-Amplitude eine Abh"angigkeit wie~$s^{1 + \ep}$ hei"st und von der rein kinematischen durch ein "`kleines"' Epsilon~$\ep$ abweicht.%
\FOOT{
  Die nichtperturbative Wechselwirkung wird aufgefa"st als vermittels des Austauschs~des {\it Soft Pomeron\/}, dessen effektiver Intercept gegeben ist durch~$1 \!+\! \ep$ mit~$\ep \!\cong\! 0.0808$; vgl.\@ Gl.\,(\ref{si^tot_pom-ZW}) und die Refn.\,\cite{Donnachie92,Donnachie98}.
}
\vspace*{-.5ex}

\section{Nahezu versus exakt lichtartige Trajektorien}
\label{Sect:Nahezu_vs_exakt}

In den Refn.~\cite{Nachtmann91,Nachtmann96} leitet Nachtmann eine nichtperturbative Formel her f"ur Parton-Parton-Streuung bei gro"ser invarianter Schwerpunktenergie~$\surd s$.
Seiner Analyse liegt zugrunde der Formalismus des Funktionalintegrals.
Sie impliziert die Eikonalapproximation f"ur die L"osung der Dirac-Gleichung in Anwesenheit eines externen nichtabelschen Eichfelds.
Die n"achstf"uhrende Ordnung, unterdr"uckt durch Faktoren~$1\!/\!\surd s$, wird vernachl"assigt.
Dies ist de facto der Grenzwert~$s \!\to\! \infty$:
Die physikalischen klassischen Partontrajektorien werden {\it lichtartig\/}.
Nachtmanns Formel ist wesentlich bestimmt durch den Vakuumerwartungswert zweier Wegner-Wilson-Linien entlang dieser lichtartigen Trajektorien. \\
\indent
Im folgenden zeichnen wir Nachtmanns Analyse nach bezogen auf physikalische {\it nahe\-zu\/} lichtartige Partontrajektorien.
Hier definieren und begr"unden wir dieses Vorgehen.
\vspace*{-.5ex}

\bigskip\noindent
Wir betrachten die Streuung zweier Partonen in deren Schwerpunktsystem.
Sei gew"ahlt die $x^3$-Achse als Kollisionsrichtung und mit ihr verkn"upfte Gr"o"sen bezeichnet als {\it longitudinal\/}, mit~$x^i$, \mbox{$i \!\in\! \{1,2\}$}, verkn"upfte Gr"o"sen als {\it transversal\/} und \vspace*{-.25ex}zusammengefa"st als \mbox{(Spalten-)Vek}\-tor%
  ~\mbox{$\rb{x} \!\equiv\! (x^i)$},~\mbox{$i \!\in\! \{1,2\}$}:%
  ~\mbox{\,$\rb{x} \!=\! (x^1,x^2){}^{\T t}$} in geradem Fettdruck. \\
\begin{samepage}
\indent
F"ur einen beliebigen Lorentz-Vektor~$x$,
der definiert ist als \vspace*{-.25ex}(Spalten-)Vektor seiner kon\-travarianten Komponenten:%
  ~\mbox{$x \!\equiv\! \big(x^\mu\big)$},~\mbox{$\mu \!\in\! \{0,1,2,3\}$}:%
  ~\mbox{\,$x \!=\! \big(x^0,x^1,x^2,x^3\big){}^{\T t}$},
f"uhren wir \vspace*{-.25ex}Licht\-kegelkoordinaten ein durch%
  ~\mbox{$\bar{x} \!\equiv\! \big(x^{\bar\mu}\big)$},~\mbox{$\bar{\mu} \!\in\! \{+,-,1,2\}$}:%
  ~\mbox{\,$\bar{x} \!=\! \big(x^+,x^-,x^{\bar1},x^{\bar2}\big){}^{\T t}$}, das hei"st durch den (Spalten-)Vektor kontravarianter Komponenten, die definiert seien wie folgt:
\vspace*{-.5ex}
\begin{align} \label{LC-Koord}
&x^\pm\; :=\; \al\, \big(x^0 \pm x^3\big)\qquad
  x^{\bar i}\; :=\; x^i \quad i\in\{1,2\}
    \\[1ex]
&\text{d.h.}\qquad
\begin{aligned}[t]
 &\bar{x}\;
  =\; \mathbb{L}\, x\qquad
  x^{\bar\mu}\;
  =\; \mathbb{L}^{\bar\mu}{}_\nu\; x^\nu
    \nn \\[-.25ex]
 &\mathbb{L}
  \equiv \big(\mathbb{L}^{\bar\mu}{}_\nu\big)
  = \al\, \pmatrixZZ{1}{1}{1}{-1}\qquad
  -\, [\det\mathbb{L}]^{-1}
  = 1/(2\al^2)
  = g_{+-}
  = g_{-+}
 \end{aligned}
    \nn
    \\[-4.75ex]\nn
\end{align}
mit~\mbox{$g_{++} \!\equiv\! 0$},~\mbox{$g_{--} \!\equiv\! 0$},
explizit ausgeschrieben in der "Ubergangsmatrix~\mbox{$\mathbb{L} \!\equiv\! \big(\mathbb{L}^{\bar\mu}{}_\nu\big)$} nur die nichttrivialen longitudinalen Komponenten~\mbox{$(+,-) \!\leftrightarrow\! (0,3)$}
und definiert~\mbox{$\al \!\in\! \bbbr^+$} als noch offene Normierung, die wir in~praxi setzen:~\mbox{$\al \!=\! 1\!/\!\surd2$}; \vspace*{-.125ex}vgl.\@ Anh.~\ref{APP:LC-Koord}.
\end{samepage}

Seien analog%
\FOOT{
  Tensoren und Tensorkomponeten seien generell bezeichnet im Sinne~\mbox{$T \!\equiv\! (T^{\mu_1\cdots\mu_s})$} versus~\mbox{$\bar{T} \!\equiv\! (T^{{\bar\mu}_1\cdots{\bar\mu}_s})$} versus~\mbox{$\tilde{T} \!\equiv\! (T^{{\tilde\mu}_1\cdots{\tilde\mu}_s})$} mit~\mbox{$\mu \!\in\! \{0,1,2,3\}$},~\mbox{$\bar\mu \!\in\! \{+,-,1,2\}$} bzw.~\mbox{$\tilde\mu \!\in\! \{\mfp,\mfm,1,2\}$}.
}
definiert Koordinaten%
  ~\mbox{$\tilde{x} \!\equiv\! \big(x^{\tilde{\mu}}\big)$},~\mbox{$\tilde{\mu} \!\in\! \{\mfp,\mfm,1,2\}$}:%
  ~\mbox{\,$\tilde{x} \!=\! \big(x^\mfp,x^\mfm,x^{\tilde1},x^{\tilde2}\big){}^{\T t}$}.
F"ur ihre {\it longitudinalen\/} Komponenten gelte~$x^\mfp \!\to\! x^+$ und~$x^\mfm \!\to\! x^-$ im Limes~$s \!\to\! \infty$, und f"ur ihre {\it transversalen\/} Komponenten~$x^{\tilde i} \!\equiv\! x^{\bar i} \!\equiv\! x^i$.
Das hei"st die longitudinalen Komponenten seien definiert als die Richtungen der {\it physikalischen\/} klassischen Partontrajektorien~${\cal C}$~-- die angenommen werden als {\it geradlinig\/}.
So besitzt longitudinal {\it per definitionem\/} die Trajektorie des in positive $x^3$-Richtung propagierenden Partons nur eine $x^\mfp$-Komponente, die Trajektorie des in negative $x^3$-Richtung propagierenden Partons nur eine $x^\mfm$-Komponente,~-- im Limes~$s \!\to\! \infty$ nur eine $x^+$- beziehungsweise~$x^-$-Komponente.
Sei suggestiv notiert
\vspace*{-.5ex}
\begin{align} \label{Trajektorien_lim}
{\cal C}\Dmfp\;
  \underset{\text{$s \!\to\! \infty$}}{\longrightarrow}\; {\cal C}\idx{+}\qquad
  \text{und}\qquad
{\cal C}\Dmfm\;
  \underset{\text{$s \!\to\! \infty$}}{\longrightarrow}\; {\cal C}\idx{-}
    \\[-4.5ex]\nn
\end{align}
bez"uglich der physikalischen Trajektorien und deren Limites.
Wir haben vor Augen ein "`plus"'- und ein "`minus"'-Parton und indizieren allgemein die zugeh"ohrigen Gr"o"sen mit den Skripten~"`\bm{\mfp}"' beziehungsweise~"`\bm{\mfm}"', im Limes~$s\!\to\! \infty$ mit~"`\bm{+}"' beziehungsweise~"`\bm{-}"'.

\begin{figure}
\begin{minipage}{\linewidth}
  \begin{center}
  \setlength{\unitlength}{.8mm}\begin{picture}(100,88) 
    \put(0,0){\epsfxsize80mm \epsffile{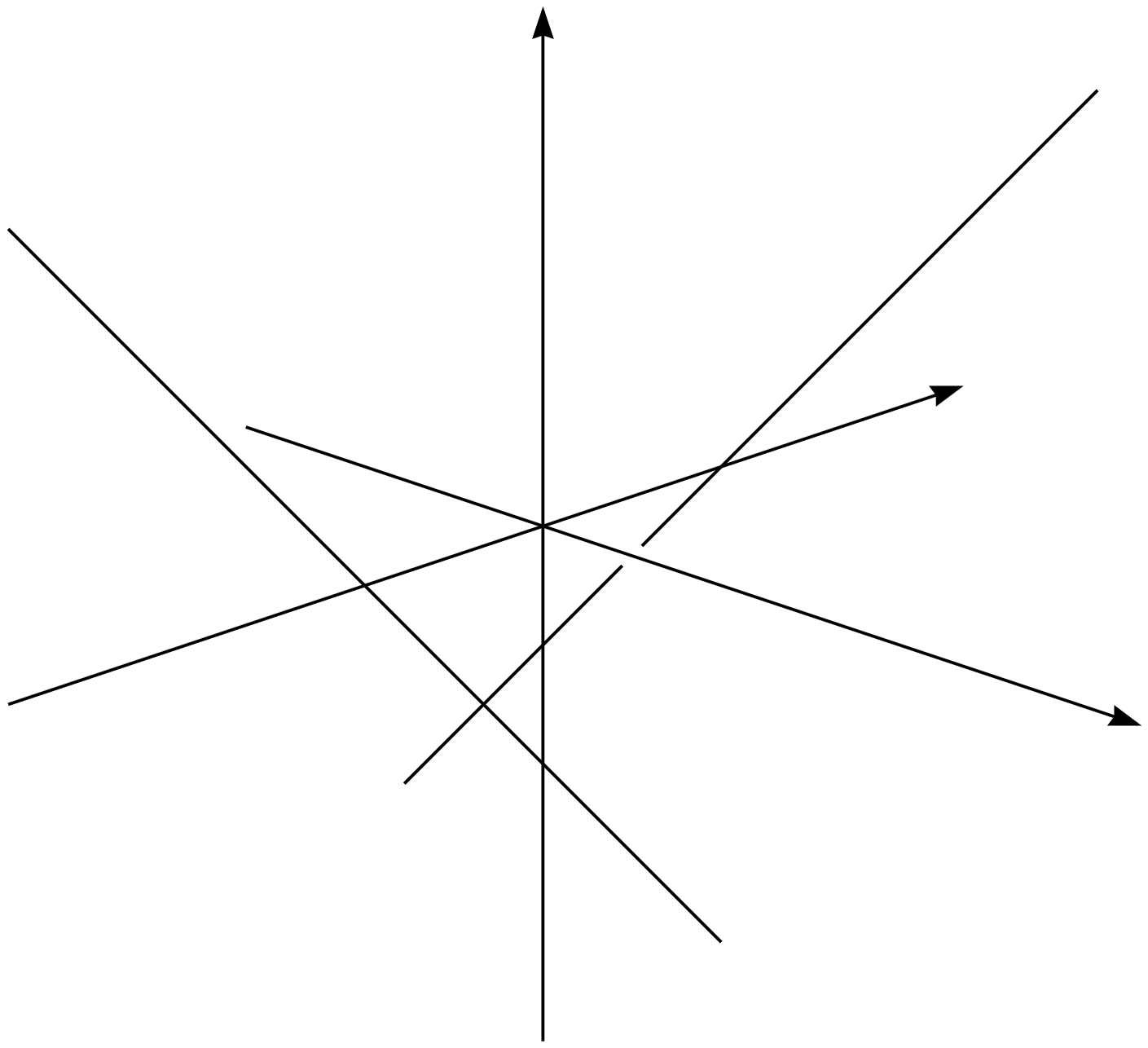}}
    \put(77,75){\normalsize$V\idx{+}$}
    \put(12,65){\normalsize$V\idx{-}$}
    \put(39,81){\normalsize$x^0$}
    \put(85,25){\normalsize$x^3$}
    \put(76,50){\normalsize$\rb{x}$}
  \end{picture}
  \end{center}
\vspace*{-4.5ex}
\caption[Streuung der Wegner-Wilson-Linien~\protect$V\idx{+} \!\equiv\! \Ph({\cal C}\idx{+})$,~\protect$V\idx{-} \!\equiv\! \Ph({\cal C}\idx{-})$]{
  Die Wegner-Wilson-Linien~$V\idx{+} \!\equiv\! \Ph({\cal C}\idx{+})$,~$V\idx{-} \!\equiv\! \Ph({\cal C}\idx{-})$~-- im Schwerpunktsystem der Streuung~-- entlang der klassischen Partontrajektorien, die im Limes~$s \!\to\! \infty$ "ubergehen in die lichtartigen Trajektorien~${\cal C}\idx{+}$,~${\cal C}\idx{-}$.   Ihr gemeinsamer Vakuumerwartungswert~$\vac{\;\cdot\;}$ bestimmt wesentlich die $T$-Amplitude der Streuung.
\vspace*{-.5ex}
}
\label{Fig:WW-Linien}
\end{minipage}
\end{figure}

Wir verweisen ausdr"ucklich auf Anhang~\ref{APP:Boosts}.
Dort ist entwickelt der formale wie interpretatorische Zusammenhang, der ein {\it tieferes Verst"andnis\/} vermittelt bereits an dieser Stelle, {\it essentiell\/} ist dann in den Abschnitten~\ref{Sect:T-Amplitude.Auswertung} und~\ref{Sect:T-Amplitude.Analytizit"at}. \\
\indent
Das anschauliche Bild ist das folgende; vgl.\@ Abb.~\ref{Fig:WW-Linien}.
Seien die beiden Partonen zun"achst in Ruhe.
Aus diesem System mit den longitudinalen Koordinaten~$x^0$ und~$x^3$ heraus werden sie "`langsam"' in positive beziehungsweise negative~$x^3$-Richtung aufeinander zu geboostet im Sinne aktiver Lorentz-Transformationen~$\La\Dmfp \!\equiv\! \La(\be\Dmfp)$ beziehungsweise~$\La\Dmfm \!\equiv\! \La(\be\Dmfm)$.
Dabei seien die {\it Beta-Parameter\/} definiert als positiv~$\in\![0,1)$, das hei"st das "`plus"'-Parton propagiere mit der Geschwindigkeit~$\be\Dmfp$, das "`minus"'-Parton mit der Geschwindigkeit~$-\be\Dmfm$ in die \mbox{$x^3$-Rich}\-tung.
{\pagebreak}H"ange~$\be\Dmfm$ dabei "`in jedem Moment"' des Boosts so von~$\be\Dmfp$ ab, da"s sich das System der aufeinander zulaufenden Partonen im "`momentanen"' Schwerpunktsystem befindet.
Dem Beta-Parameter "aquivalente Variable der Lorentz-Transformation ist die {\it Rapidit"at\/}~-- der {\it hyperbolische Winkel\/}~$\ps$, den in der $x^0,\!x^3$-Projektion des Minkowski-Diagramms die geboostete Zeitachse mit der urspr"unglichen einschlie"st; f"ur diesen gilt allgemein:
%
\vspace*{-.5ex}
\begin{align} 
\ps(\be)\;
  =\; \frac{1}{2}\, \ln \frac{1 \!+\! \be}{1 \!-\! \be}\;
  =\; {\rm artanh}\, \be
    \\[-4.5ex]\nn
\end{align}
das hei"st~\mbox{$\ps \!\in\! [0,+\infty)$} f"ur~\mbox{$\be \!\in\! [0,1)$}. \\
\indent
Die Trajektorien der so geboosteten Partonen verlaufen in der $x^0,\!x^3$-Projektion des Min\-kowski-Diagramms entlang der geboosteten Zeitachsen~$x^\mfp$ beziehungsweise entlang~$x^\mfm$ und schlie"sen mit der urspr"unglichen Zeitachse~$x^0$ die {\it endlichen\/} hyperbolischen Winkel~$\ps\Dmfp$ beziehungsweise~$-\ps\Dmfm$ ein, das hei"st zusammen den {\it endlichen\/} hyperbolischen Winkel~$\ps \!\equiv\! \ps\Dmfp \!+\! \ps\Dmfm$.
Die Beta- wie Gamma-Parameter sind in diesem Bild aktiver Lorentz-Boosts Funktionen des "`momentanen"' invarianten Quadrats der Schwerpunktenergie~$s$.
Der Limes~$s \!\to\! \infty$ wird realisiert durch~$\be\Dmfp$,~$\be\Dmfm \!\to\! 1$, das hei"st~$\ps\Dmfp$,~$\ps\Dmfm \!\to\! \infty$, also~$\ps \!\equiv\! \ps\Dmfp \!+\! \ps\Dmfm \!\to\! \infty$.
Vgl.\@ Anh.~\ref{APP:Boosts}.
Die Trajektorien~${\cal C}\idx{+}$,~${\cal C}\idx{-}$ fallen zusammen mit der ersten beziehungsweise zweiten Winkelhalbierenden zweier Schnitte der Minkowski-Raumzeit parallel zur $x^0\!x^3$-Ebene, vgl.\@ Abb.\ref{Fig:WW-Linien}.
Diesebez"uglich sind die physikalischen Trajektorien~${\cal C}\Dmfp$,~${\cal C}\Dmfm$ in den Lichtkegel hineingedreht.

Wir betrachten Wegner-Wilson-Linien bezogen genau auf diese Trajektorien, auf dieses geometrische Bild.
\vspace*{-.5ex}

\bigskip\noindent
Vor dieser Geometrie betrachten wir Nachtmanns Formel f"ur die $T$-Amplitude.
Sie ist im wesentlichen bestimmt durch den Vakuumerwartungswert~-- im Sinne der Definition von~$\vac{\;\cdot\;}$ durch Gl.~(\ref{vev})~-- zweier Wegner-Wilson-Linien entlang der klassischen Partontrajektorien, die zu nehmen sind in den Darstellungen der Partonen in der Eichgruppe~$\Drst{R}\idx{+}$ beziehungsweise~$\Drst{R}\idx{-}$.
Aufgrund des Limes~$s \!\to\! \infty$, den Nachtmann de facto ausf"uhrt, sind diese Trajektorien lichtartig und im Sinne und Notation von Gl.~(\ref{Trajektorien_lim}) genau~${\cal C}\idx{+}$ und~${\cal C}\idx{-}$:
%
\begin{align} \label{Parton-Parton_lim}
&\vac{\, Z\idx{2,+}^{-1}\, \big(V\idx{+} - 1\big)\cdot
         Z\idx{2,-}^{-1}\, \big(V\idx{-} - 1\big) \,}
    \\
&\text{mit,\vv \bm{\imath} $\,\equiv\,$ \bm{+,-}:}\qquad
  V\Dimath
    = P_{{\cal C}\Dimath}\; \exp -\iIM g \int_{\T{\cal C}\Dimath}  dx^\mu\; A_\mu\!(x)\qquad
  Z\idx{2,\imath}
    = \vac{\trDrst{R\Dimath} V\Dimath}
    \nn
    \\[-4.5ex]\nn
\end{align}
Es sind~$V\idx{+} \!\equiv\! \Ph({\cal C}\idx{+})$ und~$V\idx{-} \!\equiv\! \Ph({\cal C}\idx{-})$ die Wegner-Wilson-Linien; die Konnektoren~$\Ph$, definiert in Gl.~(\ref{Konnektor}), stehen in den Darstellungen~$\Drst{R}\idx{+}$ bzw.~$\Drst{R}\idx{-}$, bzgl.~$\trDrst{R}$ vgl.\@ die Gln.~(\ref{trDrst}),~(\ref{trDrst}$'$).

Dieser Vakuumerwartungswert ist zu interpretieren wie folgt.
Die Partonen propagieren durch die Raumzeit mit festem "au"serem Eichfeld~$A(x)$ und nehmen entlang ihrer klassischen Trajektorien nichtabelsche Phasen auf: die Eichgruppen-wertigen Exponentiale~\mbox{\,$V\Dimath$}.
Im Sinne von~$\vac{\;\cdot\;}$, das hei"st mit dem Haarschen Eichfeldma"s~$\HaarDmu$ nach Gl.~(\ref{vev}), ist dann zu mitteln "uber s"amtliche Konfigurationen~$A(x)$ des nichtperturbativen Vakuums der QCD.
Streuung geschieht vermittels korrelierter Fluktuation der Vakuum-Eichfeldkonfiguration.
\vspace*{-.5ex}

\bigskip\noindent
Die Formel Nachtmanns wurde sp"ater von Meggiolaro im Rahmen eines anderen Formalismus hergeleitet in den Refn.~\cite{Meggiolaro95,Meggiolaro97a}.
Dieser Zugang wurde in den fr"uhen sechziger Jahren entwickelt von Fradkin~\cite{Fradkin64} und basiert auf einer Funktionalintegral-Formulierung von Quantenfeldtheorien in erster Quantisierung, vgl.\@ hierzu Feynman in Ref.~\cite{Feynman48}.
Explizit werden darin Propagatoren (2-Punkt-Greenfunktionen) in Anwesenheit einer externen Met\-rik~$(g_{\mu\nu})$ und in einem externen nichtabelschen Eichfeld~$(A_{\mu a})$ ausgedr"uckt durch ein Funktionalintegral "uber Teilchentrajektorien und berechnet in einer Eikonal-Approximation.
Mit Kenntnis der Propagatoren~-- {\it on-mass~shell\/}, trunkiert und Beitr"age von Vakuum-Diagramme herausdividiert~-- extrahiert der von Lehmann, Symanzik, Zimmermann entwickelte \DREI{L}{S}{Z}-Formalismus die $S$- beziehungsweise $T$-Amplitude.
Venziano hat diesen Formalismus wieder aufgegriffen; in Ref.~\cite{Fabbrichesi93} diskutiert er Streuung durch Gravitationswechselwirkung in einer vierdimensionalen gekr"ummten Raumzeit bei Energien auf der Skala der Planckmasse.
Hierdurch motiviert betrachtet Meggiolaro im selben Formalismus zun"achst skalare QCD.
Die Verallgemeinerung auf "`reale"' fermionische QCD ist genau Nachtmanns Formel.

Meggiolaro geht im Sinne seines Zugangs einen Schritt weiter und ersetzen in dieser Formel die Grenzwerte der klassischen Partontrajektorien~${\cal C}\idx{+}$ und~${\cal C}\idx{-}$ durch die urspr"unglichen physikalischen Trajektorien~${\cal C}\Dmfp$ beziehungsweise~${\cal C}\Dmfm$.
Das hei"st Gl.~(\ref{Parton-Parton_lim}) geht "uber in
\begin{samepage}
%
\begin{align} \label{Parton-Parton}
&\vac{\, Z\idx{2,\mfp}^{-1}\, \big(V\Dmfp - 1\big)\cdot
         Z\idx{2,\mfm}^{-1}\, \big(V\Dmfm - 1\big) \,}
    \\
&\text{mit,\vv \bm{\imath} $\,\equiv\,$ \bm{\mfp,\mfm}:}\qquad
  V\Dimath
    = P_{{\cal C}\Dimath}\; \exp -\iIM g \int_{\T{\cal C}\Dimath}  dx^\mu\; A_\mu\!(x)\qquad
  Z\idx{2,\imath}
    = \vac{\trDrst{R\Dimath} V\Dimath}
    \nn
    \\[-4.5ex]\nn
\end{align}
Dabei stehen die Wegner-Wilson-Linien~$V\Dmfp \!\equiv\! \Ph({\cal C}\Dmfp)$ und~$V\Dmfm \!\equiv\! \Ph({\cal C}\Dmfm)$ in den Darstellungen der Partonen, respektive~$\mathfrak{R}\Dmfp$ und~$\mathfrak{R}\Dmfm$. \\
\indent
Meggiolaro f"uhrt Argumente daf"ur an, da"s diese Ersetzung sinnvoll und zumindest a~priori auch geboten ist.
Wir beziehen uns auf sie und zeichnen sie daher kurz nach.
\vspace*{-.5ex}

\bigskip\noindent
Verlinde, Verlinde zeigen in Ref.~\cite{Verlinde93} f"ur Quark-Quark-Streuung, da"s klassische Partontrajektorien {\it exakt lichtartig\/} ein {\it singul"arer\/} Grenzwert darstellen, der also Divergenzen impliziert.
Diese k"onnen verstanden werden als Infrarotdivergenzen infolge verschwindender Quarkmassen, denn dies genau  zwingt die Propagation der Quarks auf den Lichtkegel.
Eine M"oglichkeit diese Divergenzen zu regularisieren besteht darin, den Trajektorien eine kleine zeitartige Komponente zu geben: genau in der Art, da"s sie identisch sind den Trajektorien von Quarks mit nichtverschwindender Masse.

Dies induziert die Vorschrift, {\it a~priori\/} auszugehen von den physikalischen Trajektorien~${\cal C}\Dmfp$,~${\cal C}\Dmfm$ statt von ihren Grenzwerten~${\cal C}\idx{+}$,~${\cal C}\idx{-}$: ihnen ihre {\it physikalische\/} zeitartige Komponente zu belassen.
Es geht insofern nicht um "`Wegr"ucken"' vom Lichtkegel, sondern um "`Nicht-erst-Hinr"ucken"'.
Die $T$-Amplitude ist wohldefiniert und auszuwerten auf Basis nicht von Gl.~(\ref{Parton-Parton_lim}), sondern von Gl.~(\ref{Parton-Parton}):
Die Partonen bewegen sich respektive mit~$\be\Dmfp$ und~$-\be\Dmfm$ in~$x^3$-Richtung.
Die korrespondierenden Wegner-Wilson-Linien schlie"sen in der $x^0,\!x^3$-Projektion der Minkowski-Raumzeit den endlichen hyperbolischen Winkel~$\ps \!\equiv\! \ps\Dmfp \!+\! \ps\Dmfm$ ein.
Der Hochenergielimes~$s \!\to\! \infty$ ist erst als letzter Schritt zu nehmen in der Form~$\be\Dmfp$,~$\be\Dmfm \!\to\! 1$ beziehungsweise~$\ps \!\to\! \infty$.
Damit zusammen h"angt, da"s die Transformationsmatrix~\vspace*{-.25ex}\mbox{$\mathbb{L} \!\equiv\! \big(\mathbb{L}^{\bar\mu}{}_\nu\big)$}, die nach Gl.~(\ref{LC-Koord}) auf die Lichtkegelkoordinaten transformiert, zwar {\it Grenzwert\/} der Lorentz-Transformationen~\mbox{\,$\La\Dmfp$},~\mbox{\,$\La\Dmfm$} gegeneinander, selbst aber {\it keine Lorentz-Transformation\/} ist.

Diese Vorschrift wird {\it a~posteriori\/} legitimiert durch Resultate, die Meggiolaro auf ihrer Basis herleitet~-- wie auch andere Autoren, vgl.\@ die Refn.~\cite{Korchemsky94,Korchemskaya95,Arefeva94}.
So entwickelt er in Ref.~\cite{Meggiolaro96} f"ur Quark-Quark-Streuung die $T$-Amplitude auf der Basis von Gl.~(\ref{Parton-Parton}) in der renormierten Kopplungskonstanten~$g_{\rm ren.}$ bis zu Termen proportional~$g_{\rm ren.}^4$ und erh"alt genau das Resultat der konventionellen perturbativen QCD, vgl.\@ Cheng, Wu~\cite{Cheng87} und Lipatov~\cite{Lipatov89}.
Logarithmen in~$s$, die in perturbativer QCD auftreten aufgrund divergenter Feynman-Loop-Diagramme, treten in Meggiolaros Herleitung auf aufgrund des Zusammenhangs der invarianten Schwerpunktenergie~$\surd s$ und dem Parameter~$\ps$ des Lorentz-Boosts der Partonen gegeneinander:
%
\begin{align} \label{lns}
s\;
  =\; 2m^2\, (\cosh \ps + 1)\qquad
  \text{d.h.}\qquad
  \ps\; \underset{\text{$s \!\to\! \infty$}}{\sim}\; \ln s
\end{align}
Es skaliert also~$s$ mit~$m^2$, dem invarianten Quadrat des Quark-Vierer-Impulses.
\vspace*{-.5ex}
\end{samepage}

\bigskip\noindent
Die $T$-Amplitude auf Basis von Gl.~(\ref{Parton-Parton}) reproduziert nicht nur die korrekte f"uhrende Ordnung in~$s$ im Limes~$s\!\to\!\infty$, sondern erm"oglicht auch ihre analytische Fortsetzbarkeit ins Euklidische.
Meggiolaro gibt in Ref.~\cite{Meggiolaro96} den formalen Beweis bis Ordnung~${\cal O}(g_{\rm ren.}{}^{\zz4})$ und in Ref.~\cite{Meggiolaro97} zu jeder Ordnung.
Diesem Beweis liegt zugrunde~-- wir kommen hierauf zur"uck~-- im wesentlichen die Forderung der Fortsetzbarkeit der gluonischen Green-Funktionen auf der Minkowskischen Raumzeit zu den entsprechenden Schwinger-Funktionen auf deren Fortsetzung ins Euklidische.
Vgl.\@ hierzu Ref.~\cite{Bogolubov90}.\\
\indent
Dies er"offnet die M"oglichkeit, Quark-Quark-Streuung bei gro"ser aber endlicher invarianter Schwerpunktenergie~$\surd s$ zu betrachten auf einer Raumzeit mit Euklidischer (Riemannscher) Metrik und stellt den ersten Schritt dar hin zu ihrer Auswertung im Rahmen numerischer Simulationen auf einem Euklidischen Gitter.
\begin{figure}
\begin{minipage}{\linewidth}
  \begin{center}
  \setlength{\unitlength}{.8mm}\begin{picture}(100,86)   
    \put(0,0){\epsfxsize80mm \epsffile{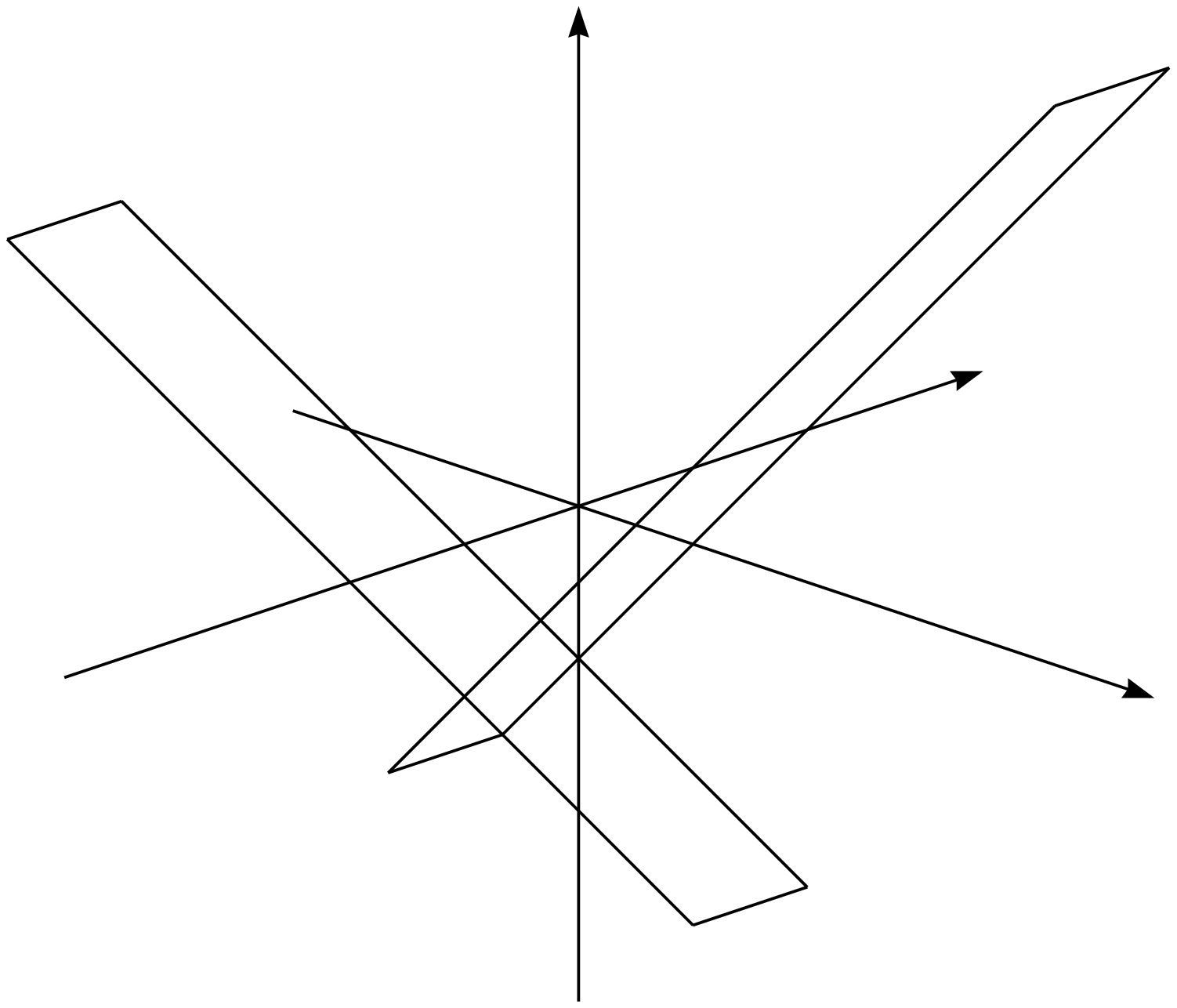}}
    \put(75,75){\normalsize$W\idx{+}$}
    \put(18,65){\normalsize$W\idx{-}$}
    \put(40,78){\normalsize$x^0$}
    \put(86,24){\normalsize$x^3$}
    \put(76,48){\normalsize$\rb{x}$}
  \end{picture}
  \end{center}
\vspace*{-4.5ex}
\caption[Streuung der Wegner-Wilson-Loops~\protect$W\idx{+} \!\equiv\! \trDrst{R} \Ph(\pa{\cal S}\idx{+})$,~\protect$W\idx{-} \!\equiv\! \trDrst{R} \Ph(\pa{\cal S}\idx{-})$]{
  Die Wegner-Wilson-Loops~$W\idx{+} \!\equiv\! \trDrst{R} \Ph(\pa{\cal S}\idx{+})$ und~$W\idx{-} \!\equiv\! \trDrst{R} \Ph(\pa{\cal S}\idx{-})$ im Schwerpunktsystem der Streuung.   Ihre longitudinalen Linien verlaufen entlang der klassischen im Limes~$s \!\to\! \infty$ lichtartigen Trajektorien der (Anti)Quarks, orientiert in positive (negative) Zeitrichtung.   Bei gro"sen positiven und negativen Zeiten, in praxi bei~$\pm\infty$, sind sie verbunden durch Konnektoren, die Eichinvarianz garantiert und in der Art orientiert sind, da"s resultieren Konnektoren bez"uglich geschlossener orientierter Kurven~$\pa{\cal S}\idx{+}$ beziehungsweise~$\pa{\cal S}\idx{-}$ der Raumzeit, das hei"st Wegner-Wilson-Loops.   Ihr gemeinsamer Vakuumerwartungswert~$\vac{\;\cdot\;}$ bestimmt wesentlich die $T$-Amplitude der Streuung.
\vspace{-.5ex}
}
\label{Fig:WW-Loops}
\end{minipage}
\end{figure}
\vspace*{-.5ex}

\bigskip\noindent
Unser Interesse gilt der Steuung eichinvarianter Obiekte~-- letztlich physikalischer Hadronen.
Nachtmann leitet in Ref.~\cite{Nachtmann96} im Limes~$s \!\to\! \infty$ die entsprechende $T$-Amplitude her und findet, da"s sie im wesentlichen bestimmt ist durch den Vakuumerwartungswert~$\vac{\;\cdot\;}$ zweier lichtartiger Wegner-Wilson-Loops im Sinne von Gl.~(\ref{Parton-Parton_lim}); vgl.\@ die Abbn.~\ref{Fig:WW-Linien} und~\ref{Fig:WW-Loops}.
Wir zeichnen diese Herleitung nach bezogen auf Wegner-Wilson-Loops, die im Sinne von Gl.~(\ref{Parton-Parton}) vom Lichtkegel wegger"uckt sind. \\
\indent
Zum einen ist dies aufzufassen als unmittelbare Verallgemeinerung der Arbeiten Meggiolaros auf den Fall eichinvarianter physikalischer Objekte.~--
In Hinblick auf dieselbe Fragenstellung: Untersuchung der f"uhrenden Ordnung gro"ser, aber endlicher~$s$ und Formulierung einer Euklidischen Theorie als erster Schritt hin zu einer numerischen Auswertung im Rahmen von QCD als Gittereichtheorie. \\
\indent
Zum anderen~-- so das Resultat des vorangegeangenen Kapitels~-- ist {\it a~priori\/} wohldefiniert nur eine Formulierung des \DREI{M}{S}{V} auf einer Raumzeit mit Euklidischer (Riemannscher) Met\-rik.
Wir verifizieren daher seine naive Minkowskische Formulierung {\it a~posteriori\/} dadurch, da"s wir sie herleiten als konsistente analytische Fortsetzung aus dem Euklidischen.

\section{Kinematik}
\label{Sect:Kinematik}

In den folgenden Abschnitten geben wir an Konstruktion, Auswertung und analytische Fortsetzung der $T$-Amplitude f"ur nahezu lichtartige Trajektorien.
Die dem zugrundeliegende Kinematik sei diskutiert hier und formal vertieft in Anhang~\ref{APP:Kinematik}. \\
\indent
Wir betrachten Streuung
\vspace*{-.5ex}
\begin{align} \label{2hto2h}
h^1(P_1) \;+\; h^2(P_2)\;
  \longrightarrow\;
  h^{1'}(P_{1'}) \;+\; h^{2'}(P_{2'})
    \\[-4.5ex]\nn
\end{align}
mit~$P_i$ den Vierer-Impulsen der Zust"ande~$h^i$.
Seien indiziert mit~\mbox{$i \!=\! 1,1'$} "`plus"'-, mit~\mbox{$i \!=\! 2,2'$} "`minus"'-Teilchen im Sinne von Propagation mit "`gro"ser"' Geschwindigkeit in positive respektive negative $x^3$-Richtung; bezeichnen ungestrichene Indizes ein-, gestrichene Indizes auslaufende Teilchen.
Ihre Lichtkegelkomponenten mit Normierung~$\al$ lauten dann:%
\FOOT{
  \label{FN-A:LC-Impuls}Bzgl.\@ Lichtkegelkoordinaten vgl.\@ Gl.~(\ref{LC-Koord}) und Anh.~\ref{APP:LC-Koord}: Definition~\mbox{$P^\pm \!=\! \al\, (P^0 \!\pm\! P^3)$},~\mbox{$\rb{P} \!=\! (P^1,P^2)^t$}, invertiert:~\mbox{$P^0 \!=\! (P^+ \!+\! P^-)\!/\!2\al$},~\mbox{$P^3 \!=\! (P^+ \!-\! P^-)\!/\!2\al$}, folgt f"ur das invariante Skalarprodukt zweier beliebiger Vektoren~$P$,~$Q$:~\mbox{$P \!\cdot\! Q \!=\! [P^0Q^0 \!-\! P^3Q^3] \!-\! \rb{P} \!\cdot\! \rb{Q} \!=\! [P^+Q^- \!+\! P^-Q^+]\!/\!2\al^2 \!-\! \rb{P} \!\cdot\! \rb{Q}$}, das hei"st:~\mbox{$g_{+-} \!=\! g_{-+} \!=\! 1\!/\!2\al^2$}.
}
%
\vspace*{-.5ex}
\begin{align} \label{h-Impuls}
\hspace*{-2pt}
\left(\begin{array}{l}
  P_i^+\\
  P_i^- \equiv\! \al^2 \rb{M}_i^2\!/P_i^+\\
  \rb{P}_i
  \end{array}\right)\quad
  \text{f"ur\;~$i \!=\! 1,1'$,}\qquad
\left(\begin{array}{l}
  P_i^+ \equiv\! \al^2\, \rb{M}_i^2\!/P_i^-\\
  P_i^-\\
  \rb{P}_i
  \end{array}\right)\quad
  \text{f"ur\;~$i \!=\! 2,2'$}
    \\[-4.5ex]\nn
\end{align}
\vspace*{-5ex}
\begin{alignat}{3} \label{h-Impuls+_lim}
\text{mit}\qquad
  P_1^+,\, P_{1'}^+\vv \text{und}\vv P_2^-,\, P_{2'}^-
    \vv\underset{\text{$s \!\to\! \infty$, $t$ fest}}{\longrightarrow}\vv \infty
    \\[-5ex]\nn
\end{alignat}
Es bezeichnen~$M_i^2$ und~$\rb{M}_i^2$, mit
\vspace*{-.5ex}
\begin{align} \label{transversaleMasse}
\rb{M}_i^2\;
  \stackrel{\D!}{=}\; M_i^2 + \rb{P}_i^2
    \\[-4.5ex]\nn
\end{align}
die Quadrate von {\it invarianter\/} beziehungsweise {\it transversaler Masse\/} des Zustands~$h^i$; diese k"onnen negativ sein, \vspace*{-.5ex}etwa f"ur ein Photon mit Virtualit"at~$Q$:~\mbox{$M^2{}_{\zzzz\D\ga^{\D\ast}} \!\equiv\! -Q^2$ mit $Q^2 \!\ge\! 0$}.
Die Abh"angigkeit in den Gl.~(\ref{h-Impuls}) de facto von nur einer longitudinalen Komponente ist Konsequenz der {\it on-mass~shell\/}-Forderung an die physikalischen Zust"ande:
\vspace*{-.5ex}
\begin{align} \label{OnMassShell}
P_i^2\; =\; M_i^2\qquad
  \Longleftrightarrow\qquad
  P_i^+ P_i^-\; =\; \al^2\, \rb{M}_i^2
    \\[-4.5ex]\nn
\end{align}
Wir betrachten diesbez"uglich als unabh"angig~-- vgl.\@ Gl.~(\ref{h-Impuls})~-- die Komponenten
\vspace*{-.5ex}
\begin{align} \label{h-Impuls-unabh"angig}
P_1^+,\, P_{1'}^+\vv \text{und}\vv P_2^-,\, P_{2'}^-
    \\[-4.5ex]\nn
\end{align}
In Termen dieser arbeiten wir im folgenden und sprechen sie an als die "`gro"sen"' Komponenten~-- in dem Sinne, da"s sie gegen Unendlich gehen im Limes~$s \!\to\! \infty$, $t$ fest; vgl.\@ Gl.~(\ref{h-Impuls+_lim}).
Von diesen abh"angig sind die "`kleinen"' Komponenten~\mbox{$P_1^-$, $P_{1'}^-$} und~\mbox{$P_2^+$, $P_{2'}^+$}. \\
\indent
Seien entsprechend "`kleine"' Gr"o"sen~$\vep_i$ f"ur~$i \!=\! 1,1',2,2'$ definiert durch:%
\FOOT{
  Die Definition geschieht zweckm"a"sigerweise in Termen von Masse{\sl quadraten\/} und {\sl nicht\/} dimensionslos.
}
%
\vspace*{-.75ex}
\begin{alignat}{3} \label{epsilons}
&\vep_i\;
  \stackrel{\D!}{=}\; \al\, \rb{M}_i^2\!/P_i^+&\qquad
  \Longleftrightarrow\vv
  &P_i^+ = \al\, \rb{M}_i^2\vv \vep_i^{-1}&
  \qquad\qquad&\text{f"ur\;~$i \!=\! 1,1'$}
    \\
&\vep_i\;
  \stackrel{\D!}{=}\; \al\, \rb{M}_i^2\!/P_i^-&\qquad
  \Longleftrightarrow\vv
  &P_i^- = \al\, \rb{M}_i^2\vv \vep_i^{-1}&
  \qquad\qquad&\text{f"ur\;~$i \!=\! 2,2'$}
    \tag{\ref{epsilons}$'$}
    \\[-5ex]\nn
\end{alignat}\nopagebreak
\vspace*{-4.5ex}
\begin{align} \label{epsilons_lim}
&\text{mit}\qquad
  \vep_i
    \vv\underset{\text{$s \!\to\! \infty$, $t$ fest}}{\longrightarrow}\vv 0
  \qquad\qquad\text{f"ur\;~$i \!=\! 1,1',2,2'$;\vv vgl.\@ Gl.~(\ref{h-Impuls+_lim})}
    \\[-4.5ex]\nn
\end{align}

Durch die~$\vep_i$ schreiben sich die "`kleinen"' Komponenten~\mbox{$P_1^-$, $P_{1'}^-$} und~\mbox{$P_2^+$, $P_{2'}^+$} suggestiv in der Form:
\vspace*{-.5ex}
\begin{alignat}{3} \label{h-Impuls-}
&P_i^-\;
  =\; \al\, \vep_i&
  \qquad\qquad&\text{f"ur\;~$i \!=\! 1,1'$}
    \\
&P_i^+\;
  =\; \al\, \vep_i&
  \qquad\qquad&\text{f"ur\;~$i \!=\! 2,2'$}
    \tag{\ref{h-Impuls-}$'$}
    \\[-4.5ex]\nn
\end{alignat}
\vspace*{-4.5ex}
\begin{alignat}{3} \label{h-Impuls-_lim}
\text{mit}\qquad
  P_1^-,\, P_{1'}^-\vv \text{und}\vv P_2^+,\, P_{2'}^+
    \vv\underset{\text{$s \!\to\! \infty$, $t$ fest}}{\longrightarrow}\vv 0
    \\[-5ex]\nn
\end{alignat}
Seien die "`gro"sen"' Komponenten verstanden als endlich, folglich die "`kleinen"' Komponenten und~$\vep_i$ als~{\it epsilon-wertig\/} im Sinne {\it nichtverschwindend infinitesimal\/}.
\vspace*{-.625ex}Mit Definition \mbox{$P^\pm \!\stackrel{\D!}{=}\! \al(P^0 \!\pm\! P^3)$} gilt f"ur Energie- und Impuls-Komponenete in~$x^3$-Richtung allgemein:
\begin{align} \label{P03}
P^0\; &=\; \frac{1}{2\al}\, \big(P^+ + P^-\big)
    \\
P^3\; &=\; \frac{1}{2\al}\, \big(P^+ - P^-\big)
    \tag{\ref{P03}$'$}
    \\[-4.5ex]\nn
\end{align}
in Termen der~\mbox{\,$\vep_i$}:
\vspace*{-.5ex}
\begin{alignat}{2} \label{P03_vep}
P_1^{\stackrel{\scriptstyle0}{3}}\;
  &=\; \big(\rb{M}_1^2\; \vep_1^{-1} \pm \vep_1 \big)\!\big/2\al&
  \qquad
  \text{und:}\qquad
  P_2^{\stackrel{\scriptstyle0}{3}}\;
  &=\; \big(\vep_2 \pm \rb{M}_2^2\; \vep_2^{-1}\big)\!\big/2\al
    \\[.5ex]
P_{1'}^{\stackrel{\scriptstyle0}{3}}\;
  &=\; \big(\rb{M}_{1'}^2\; \vep_{1'}^{-1} \pm \vep_{1'}\big)\!\big/2\al&
  \qquad
  P_{2'}^{\stackrel{\scriptstyle0}{3}}\;
  &=\; \big(\vep_{2'} \pm \rb{M}_{2'}^2\; \vep_{2'}^{-1}\big)\!\big/2\al
    \tag{\ref{P03_vep}$'$}
    \\[-4.5ex]\nn
\end{alignat}
Index~$0$ oberes/~$3$ unteres Zeichen.
Wir arbeiten in Termen der vier Drei-Tupel~\mbox{$[\vep_i,\rb{P}_i]$}.
Die Komponenten~$P_i^+$,~$P_i^-$ und~$P_i^0$,~$P_i^3$ folgen unmittelbar aus den angegebenen Gleichungen.
\begin{figure}
\begin{minipage}{\linewidth}
  \begin{center}
  \setlength{\unitlength}{0.8mm}\begin{picture}(100,43)  
    \thinlines
    \put( 5, 5){\vector(4, 1){36}}
    \put(59,14){\vector(4,-1){36}}
    \put(50,14){\vector(0, 1){17}}
    \put(59,31){\vector(4, 1){36}}
    \put( 5,40){\vector(4,-1){36}}
    \put(-18,40){$(P_1^+,\rb{P}_1)$}
    \put(-18, 3){$(P_2^-,\rb{P}_2)$}
    \put(102,40){$(P_{1'}^+,\rb{P}_{1'})$}
    \put(102, 3){$(P_{2'}^-,\rb{P}_{2'})$}
    \put( 0,20){$s$}
    \put(49,40){$t$}
  \end{picture}
  \end{center}
\vspace*{-3ex}
\caption[Streukinematik:~\protect$h^1(P_1) \!+\! h^2(P_2) \!\to\! h^{1'}(P_{1'}) \!+\! h^{2'}(P_{2'})$ allgemein]{
  Die Kinematik der Streuung, vgl.\@ Gl.~(\ref{2hto2h}), ist f"ur {\it on-mass~shell\/}-Zust"ande~$h^i(P_i)$ vollst"andig bestimmt durch jeweils drei Impulskomponenten: per definitionem die "`gro"se"' longitudinale Komponente~$P_i^\pm$ (die gegen Unendlich geht im Limes~$s \!\to\! \infty$,~$t$ fest) und der transversale Vektor~$\rb{P}_i$.}
\label{Fig:Kinematik}
\end{minipage}
\end{figure}
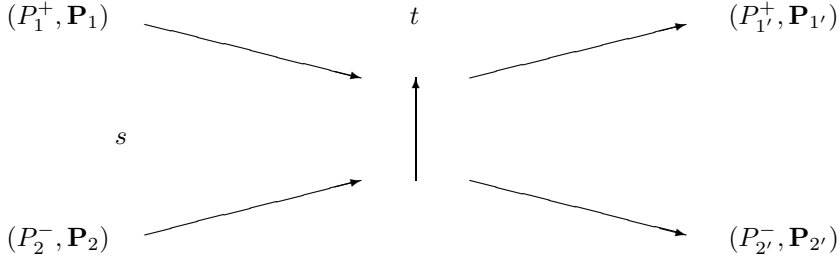
\\\indent
Es sind die Invarianten der Streuung nach Mandelstam; vgl.\@ Gl.~(\ref{2hto2h}) und Abb.~\ref{Fig:Kinematik}:
\vspace*{-.5ex}
\begin{alignat}{2} \label{Mandelstam_Def}
&s&\; &=\; (P_1\phantom{'} + P_2)^2
    \\[.25ex]
&t&\; &=\; (P_{1'} - P_1)^2
    \tag{\ref{Mandelstam_Def}$'$} \\[.25ex]
&u&\; &=\; (P_{2'} - P_1)^2
    \tag{\ref{Mandelstam_Def}$''$}
    \\[-4.5ex]\nn
\end{alignat}
Nur zwei davon~-- wir w"ahlen~$s$ und~$t$~-- sind unabh"angig, die dritte~-- also~$u$~-- folgt aus der Relation~$s + t + u = {\T\sum}_i\, M_i^2 + \frac{1}{2}\, P^2$; dabei ist 
\vspace*{-.5ex}
\begin{align} \label{ViererDifferenzimpuls}
P\; =\; (P_{1'} + P_{2'}) - (P_1 + P_2)
    \\[-4.5ex]\nn
\end{align}
der {\it Differenzimpuls zwischen den aus- und einlaufenden Zust"anden\/}.
Er verschwindet identisch,~\mbox{$P \!\equiv\! 0$}, wenn die Streu-Wechselwirkung den Gesamtimpuls erh"alt.%
\FOOT{
  \label{FN:Impulserhaltung}F"ur die im folgenden Abschnitt~\ref{Sect:T-Amplitude.Konstruktion} hergeleitete $T$-Amplitude der Streuung ist dies der Fall als Konsequenz der Impulserhaltung an jedem Vertex der Quantenchromodynamik.
}
Seien die Impulse~\mbox{$P_i \!=\! (P_i^+,P_i^-,\rb{P}_i)$}, vgl.\@ Gl.~(\ref{h-Impuls}), bezogen auf ein beliebiges, aber festes (inertiales) Koordinatensystem.
Dann schreiben sich die Gln.~(\ref{Mandelstam_Def}),~(\ref{Mandelstam_Def}$'$) in Termen der Tupel~$[\vep_i,\rb{P}_i]$ und Massequadrate~$\rb{M}_i^2$ wie folgt:
\vspace*{-.5ex}
\begin{align} \label{Mandelstam-st}
s\; &=\; - (\rb{P}_1 \!+\! \rb{P}_2)^2\;
        +\; \rb{M}_1^2 + \rb{M}_2^2\;
        +\; \big[ \rb{M}_1^2\, \rb{M}_2^2\cdot \big(\vep_1\, \vep_2\big)^{-1}
                 + (\vep_1\, \vep_2) \big]
    \\[.5ex] 
t\; &=\; - (\rb{P}_{1'} \!-\! \rb{P}_1)^2\;
        +\; \big[\rb{M}_1^2\,    \big(1 - \vep_1^{-1}\, \vep_{1'}\big)
               + \rb{M}_{1'}^2\, \big(1 - \vep_1\,      \vep_{1'}^{-1}\big)\big]\;
    \tag{\ref{Mandelstam-st}$'$}
    \\[-4.5ex]\nn
\end{align}
Physikalisch hat~$s$ die Bedeutung des Quadrats der invarianten Schwerpunktenergie und~$-t$ die des Quadrats des invarianten Impulstransfers der Streuung.

Sei aus formalen Gr"unden f"ur den Moment noch nicht impliziert Erhaltung des Gesamtimpulses, das hei"st~\mbox{$P \!\equiv\! 0$}.
Dann beziehen sich~$s$,~$t$ zun"achst auf die  {\it einlaufende\/} beziehungsweise {\it 1,1'-Seite\/} der Streuung.
Entsprechende Gr"o"sen~$s'$,~$t'$, die sich beziehen auf die {\it auslaufende\/} beziehungsweise {\it 2,2'-Seite\/} der Streuung sind definiert in Analogie zu den Gln.~(\ref{Mandelstam_Def}),~(\ref{Mandelstam_Def}$'$) durch:
\vspace*{-.5ex}
\begin{alignat}{2} \label{Mandelstam-prime_Def}
&s'&\; &=\; (P_{1'} + P_{2'})^2
    \\
&t'&\; &=\; (P_{2'} - P_2\phantom{'})^2
    \tag{\ref{Mandelstam-prime_Def}$'$}
    \\[-4.5ex]\nn
\end{alignat}
dabei gilt offensichtlich, vgl.\@ Gl.~(\ref{ViererDifferenzimpuls}):
\vspace*{-.5ex}
\begin{align} \label{P=0=>s'=s,t'=t}
P \equiv 0\qquad
  \Longrightarrow\qquad
  s' \equiv s\quad
  \text{und}\quad
  t' \equiv t
    \\[-4.5ex]\nn
\end{align}
Im folgenden werden explizit formuliert Relationen in Termen der ungestrichenen Variablen~$s$,~$t$.
Durch Indexsubstitution gem"a"s der Gln.~(\ref{Mandelstam_Def})\bm{\to}(\ref{Mandelstam-prime_Def}) und (\ref{Mandelstam_Def}$'$)\bm{\to}(\ref{Mandelstam-prime_Def}$'$) folgen unmittelbar analoge Relationen in Termen der gestrichenen Variablen~$s'$,~$t'$.

Sei analog zu~$P$, vgl.\@ Gl.~(\ref{ViererDifferenzimpuls}), definiert der Vierer-Vektor~$Q$ und in Gegen"uberstellung geschrieben:
\vspace*{-.5ex}
\begin{alignat}{2} \label{Vierer-Vektoren-PQ}
&P&\; &=\; (P_{1'} - P_1) + (P_{2'} - P_2)
    \\
&Q&\; &=\; (P_{1'} - P_1) - (P_{2'} - P_2)
    \tag{\ref{Vierer-Vektoren-PQ}$'$}
    \\[-4.5ex]\nn
\end{alignat}
Wir dr"ucken~$t$ aus symmetrisch durch gestrichene und ungestrichene Gr"o"sen gem"a"s
\vspace*{-.5ex}
\begin{align}
t\; =\; \frac{1}{2}\, \big(t + t'\big)\; +\; \frac{1}{2}\, P\cdot Q
    \\[-4.5ex]\nn
\end{align}
und finden durch Einsetzen von Gl.~(\ref{Mandelstam-st}$'$) und ihrem gestrichenen Pendant:
\vspace*{-.5ex}
\begin{align} \label{Mandelstam-t-symm}
\hspace*{-6pt}
t\;
&=\; -(\rb{P}_{1'} \!-\! \rb{P}_1)^2
    \\
&\phantom{=\;} +\; \frac{1}{2}\, \Big[
           \rb{M}_1^2\,    \big(1 - \vep_1^{-1}\, \vep_{1'}\big)
         + \rb{M}_{1'}^2\, \big(1 - \vep_1\,      \vep_{1'}^{-1}\big)
         + \rb{M}_2^2\,    \big(1 - \vep_2^{-1}\, \vep_{2'}\big)
         + \rb{M}_{2'}^2\, \big(1 - \vep_2\,      \vep_{2'}^{-1}\big)
         \Big]
    \nn \\
&\phantom{=\;} +\; \frac{1}{2}\, \Big[P^0\, Q^0 - P^3\, Q^3\Big]
    \nn
    \\[-4.5ex]\nn
\end{align}
Erhaltung des Gesamtimpulses impliziert:~\mbox{$P \!\equiv\! 0$}, verschwindet identisch die letzte Zeile und wir erhalten die bez"uglich der relevanten Terme symmetrisierte Darstellung von Gl.~(\ref{Mandelstam-t-symm}).

In Hinsicht auf die Abh"angigkeit von~$s$ sind zu bestimmen die Gr"o"sen Epsilon f"ur ungestrichene und gestrichene Indizes.
Daraus folgen die longitudinalen Impulskomponenten~$P^\pm$ beziehungsweise~$P^0$,~$P^3$ und das Quadrat~$-t$.
Wir tun dies im folgenden.

Zwischen den Elementen der vier Drei-Tupel~$[\vep_i,\rb{P}_i]$ bestehen Relationen, die~-- subsumiert in~$P$~-- widerspiegeln die Impulsstruktur der zugrundeliegenden Wechselwirkung.
F"ur definiertes~$P$ lauten diese:
\vspace*{-.5ex}
\begin{alignat}{2} \label{GesImpulsErhaltg}
&P^+&\;
  &=\; \al\, \big[
          \big(\rb{M}_{1'}^2\; \vep_{1'}^{-1}\; +\; \vep_{2'}\big)\;
      -\; \big(\rb{M}_1^2\;    \vep_1^{-1}\;    +\; \vep_2\big)
        \big]
    \\[.5ex]
&P^-&\;
  &=\; \al\, \big[
          \big(\vep_{1'}\; +\; \rb{M}_{2'}^2\; \vep_{2'}^{-1}\big)\;
      -\; \big(\vep_1\;    +\; \rb{M}_2^2\;    \vep_2^{-1}\big)
        \big]
    \tag{\ref{GesImpulsErhaltg}$'$} \\[.5ex]
&\rb{P}&\;
  &=\; \phantom{\al\,}
          \big(\rb{P_{1'}} + \rb{P_{2'}}\big)\;
      -\; \big(\rb{P_1}    + \rb{P_2}\big)
    \tag{\ref{GesImpulsErhaltg}$''$}
    \\[-4.5ex]\nn
\end{alignat}
Explizit, wenn die Wechselwirkung den Gesamtimpuls erh"alt:~\mbox{$P \!\equiv\! 0$}, verschwinden identisch die linken Seiten dieser Gleichungen.

Die Elemente der Drei-Tupel~$[\vep_i,\rb{P}_i]$ werden weiter verkn"upft durch explizite Wahl eines (inertialen) Bezugsystems.
Geschehe dies in analoger Form zu den Gln.~(\ref{GesImpulsErhaltg})-(\ref{GesImpulsErhaltg}$''$), durch Spezifizieren der einlaufenden Impulse auf der linken Seite der Gleichungen:
\vspace*{-.5ex}
\begin{align} \label{Bezugsystem}
&P_1^3 + P_2^3\;
  =\; \frac{1}{2}\ \Big[
        \big(\rb{M}_1^2\; \vep_1^{-1}\; -\; \vep_1\big)\;
    -\; \big(\rb{M}_2^2\; \vep_2^{-1}\; -\; \vep_2\big)
      \Big]
    \\
&\rb{P_1} + \rb{P_2}
    \tag{\ref{Bezugsystem}$'$}
    \\[-4.5ex]\nn
\end{align}
das hei"st durch Spezifizierung des Dreier-Vektors~\mbox{$\vec{P}_1 \!+\! \vec{P}_2$}.
Explizit f"ur das {\it Schwerpunktsystem\/} bez"uglich der \mbox{$x^3$-Rich}\-tung ist zu setzen~\mbox{$P_1^3 \!+\! P_2^3 \!\equiv\! 0$}, bez"uglich der transversalen \mbox{$\rb{x}$-Rich}\-tung~\mbox{$\rb{P}_1 \!+\! \rb{P}_2 \!\equiv\! \bf{0}$}.
Die Wahl bezieht sich allgemein zun"achst auf die einlaufenden Zust"an\-de, determiniert vollst"andig aber auch das auslaufende System.

Die Verkn"upfungen durch die Gln.~(\ref{GesImpulsErhaltg})-(\ref{GesImpulsErhaltg}$''$) und~(\ref{Bezugsystem}),~(\ref{Bezugsystem}$'$) f"uhren zu den folgenden Konsequenzen f"ur die longitudinalen Gr"o"sen~$\vep_i$.
Wir sind dabei in erster Linie interessiert an deren Abh"angigkeit von~$s$, aus der unmittelbar folgt die Abh"angigkeit der "`gro"sen"' Impulskomponeneten~$P_1^+$,~$P_{1'}^+$ und~$P_2^-$,~$P_{2'}^-$, daraus der "`kleinen"' Komponenten, und des Quadrats~$-t$ des invarianten Impulstransfers nach Gl.~(\ref{Mandelstam-st}$'$).

Sei  zun"achst bezeichnet mit~$\vep$ das Produkt~$\vep_1\vep_2$:
\vspace*{-.5ex}
\begin{align} \label{vep=vep1vep2}
\vep\; \equiv\; \vep_1 \vep_2
    \\[-4.5ex]\nn
\end{align}
Einerseits aus Gl.~(\ref{Bezugsystem}) folgt unmittelbar $\vep_2$ als Funktion von~$\vep_1$:
\vspace*{-.5ex}
\begin{align} \label{vep2_vrh_vep1}
&\vep_2(\vep_1)\;
  =\; - \vrh\; +\; \sqrt{\vrh^2 + \rb{M}_2^2}
    \\[.5ex]
  &\text{mit}\qquad
  \vrh(\vep_1)\; =\;
    \frac{1}{2}\, \big(\rb{M}_1^2\; \vep_1^{-1} - \vep_1\big)\;
      -\; (P_1^3 \!+\! P_2^3)^2
    \tag{\ref{vep2_vrh_vep1}$'$}
    \\[-4.5ex]\nn
\end{align}
Mithilfe dieser Relationen ist das Produkt~\mbox{$\vep \!\equiv\! \vep_1\vep_2$} Funktion von~$\vep_1$ allein:
\vspace*{-.5ex}
\begin{align} \label{vep1vep2-vrh}
&\vep|\ivrh\big(\vep_1\big)\;
  =\; \big(\vep_1\vep_2\big)\ivrh\big(\vep_1\big)
    \\
&\stackrel{\D!}{=}\; \vep_1\, \Big[\, -\vrh\; +\; \sqrt{\vrh^2 + \rb{M}_2^2}\, \Big]
    \nn
    \\[-4.5ex]\nn
\end{align}
entsprechend Gl.~(\ref{vep=vep1vep2}) und~$\vep$ in suggestiver Indizierung.

Andererseits nach Gl.~(\ref{Mandelstam-st}) h"angt~$s$ ab von den~$\vep_i$ allein "uber das Produkt~\mbox{$\vep \!\equiv\! \vep_1\vep_2$}; hiernach aufgel"ost Gl.~(\ref{Mandelstam-st}), folgt dessen Abh"angigkeit von~$s$:
\vspace*{-.5ex}
\begin{align} \label{vep1vep2_si_s}
&\vep|\isi(s)\;
  =\; \big(\vep_1\vep_2\big)\isi(s)\;
    \\
&\stackrel{\D!}{=}\; \big(\vep_1\vep_2\big)(s)\;
  =\; \si\; -\; \sqrt{\si^2 - \rb{M}_1^2\,\rb{M}_2^2}
    \nn \\[1ex]
  &\text{mit}\qquad
  \si(s)\; =\;
    \frac{1}{2}\, \Big[\, s\; -\;
      \big[\rb{M}_1^2 + \rb{M}_2^2 - \big(\rb{P}_1 \!+\! \rb{P}_2\big)^2\big]\,
      \Big]
    \tag{\ref{vep1vep2_si_s}$'$}
    \\[-4.5ex]\nn
\end{align}
entsprechend Gl.~(\ref{vep=vep1vep2}) und~$\vep$ in Indizierung im Sinne von Gl.~(\ref{vep1vep2-vrh}).

\begin{samepage}
Auf Basis dieser Relationen~-- durch Identifizieren~\mbox{$\vep|\ivrh \equiv \vep|\isi$}~-- werden in~\mbox{Anhang \ref{APP-Sect:vepi-s}} berechnet~$s$ als Funktion von~$\vep_1$ und umgekehrt~$\vep_1$ als Funktion von~$s$.
Wir finden:
\vspace*{-.5ex} 
\begin{align} \label{s_vep1}
&s(\vep_1)\;
  =\; \big[\rb{M}_1^2 + \rb{M}_2^2 - \big(\rb{P}_1 \!+\! \rb{P}_2\big)^2\big]\; +\; 2\,\si
    \\[1ex]
  &\text{mit}\qquad
  2\,\si(\vep_1)\;
    =\; \sqrt{\vrh^2 + \rb{M}_2^2}\cdot \big(\rb{M}_1^2\; \vep_1^{-1} + \vep_1\big)\;
                          +\; \vrh\cdot \big(\rb{M}_1^2\; \vep_1^{-1} - \vep_1\big)
    \tag{\ref{s_vep1}$'$}
    \\[-4.5ex]\nn
\end{align}
Dabei ist aufzufassen~$\vrh \!\equiv\! \vrh(\vep_1)$ entsprechend Gl.~(\ref{vep2_vrh_vep1}$'$) als Funktion von~$\vep_1$.
\end{samepage}\pagebreak
Weiter durch Einsetzen~\mbox{$\vep_1 \!\equiv\! \al\rb{M}_1^2\!/P_1^+$} folgt unmittelbar~$s$ als Funktion von~$P_1^+$,~-- $P_2^-$ ist eliminiert.
Umgekehrt finden wir:
\vspace*{-.5ex} 
\begin{align} \label{vep1_s}
&\vep_1(s)\;
  =\; \be\, \Big[-\big(P_1^3 \!+\! P_2^3\big)\;
        +\; \sqrt{\big(P_1^3 \!+\! P_2^3\big)^2 + \ga}\;\Big]
    \\[.5ex]
  &\text{mit}\qquad
  \be(\vep)\;
    =\; \vep \big/\; \big(\rb{M}_2^2 + \vep\big)
    \tag{\ref{vep1_s}$'$} \\
  &\phantom{\text{mit}\qquad}
  \ga(\vep)\;
    =\; \big(\rb{M}_1^2 + \vep\big) \big(\rb{M}_2^2 + \vep\big) \big/\; \vep
    \tag{\ref{vep1_s}$''$}
    \\[-4ex]\nn
\end{align}
Dabei ist aufzufassen~\mbox{$\vep \!\equiv\! \vep|\isi(s) \!=\! \big(\vep_1\vep_2\big)(s)$} entsprechend der Gln.~(\ref{vep1vep2_si_s}),~(\ref{vep1vep2_si_s}$'$).
Durch
\begin{align} \label{vep2_s}
\vep_2\;
  =\; \vep|\isi\; \big/\vv \vep_1
\end{align}
vgl.\@ Gl.~(\ref{vep=vep1vep2}), folgt~$\vep_2$ als Funktion von~$s$, durch Indexsubstitution~\mbox{$1 \!\to\! 1'$,~$2 \!\to\! 2'$} im Sinne\zz
\begin{align} \label{Subst1->1',2->2'}
\rb{M}_1^2\; \to\; \rb{M}_{1'}^2\,,\vv
  \rb{P}_1\; \to\; \rb{P}_{1'}\qquad
  \text{und}\qquad
  \rb{M}_2^2\; \to\; \rb{M}_{2'}^2\,,\vv
  \rb{P}_2\; \to\; \rb{P}_{2'}
\end{align}
schlie"slich Darstellungen f"ur die gestrichenen Gr"o"sen: f"ur~$s'(\vep_{1'})$ und~$\vep_{1'}(s')$,~$\vep_{2'}(s')$ mit~$s' \!\equiv\! s$ nach Gl.~(\ref{P=0=>s'=s,t'=t}) bei Erhaltung des Gesamtimpulses~\mbox{$P \!\equiv\! 0$}. \\
\indent
Mit den Epsilon~$\vep_i(s)$,~$i \!=\! 1,1',2,2'$, sind bekannt als Funktionen der invarianten Schwerpunktenergie~$\surd s$ alle relevanten Gr"o"sen:
die Lichtkegelkomponenten der Impulse~$P_i^+$,~$P_i^-$ durch die Gln.~(\ref{epsilons}),~(\ref{epsilons}$'$) und~(\ref{h-Impuls-}),~(\ref{h-Impuls-}$'$),
die Energie- und Drei-Komponenten~$P_i^0$,~$P_i^3$ durch die Gln.~(\ref{P03_vep}),~(\ref{P03_vep}$'$),
die Dreier-Impuls-Betr"age~$|\vec{P}_i| \!=\! \surd\rb{P}_i^2 \!+\! {P_i^3}^2$
und das Quadrat des invarianten Impulstransfers~$-t$, vgl.\@ die Gln.~(\ref{Mandelstam-st}$'$),~(\ref{Mandelstam-t-symm}). \\
\indent
In Anhang~\ref{APP-Sect:Entwicklungen} geben wir an deren Reihen-Entwicklungen f"ur gro"se~$s$, das hei"st f"ur kleine Parameter
\vspace*{-.5ex}
\begin{align} \label{kappa}
\ka\; \equiv\; \frac{1}{s}
    \\[-4.5ex]\nn
\end{align}
Wir diskutieren den allgemeinen Fall, der Erhaltung des Gesamtimpulses impliziert:~\mbox{$P \!\equiv\! 0$}, aber noch nicht spezifiziert das Bezugsystem:~\mbox{$\rb{P}_1 \!+\! \rb{P}_2$}, \mbox{$P_1^3 \!+\! P_2^3$},~-- und den Fall, der dieses konkretisiert als das Schwerpunktsystem bez"uglich der drei Raumrichtungen.
Diese Entwicklungen liegen zugrunde der Diskussion dieses wie der folgenden Kapitel.
Exemplarisch und in Hinblick auf sp"ater seien hier nur zitiert:
\vspace*{-.5ex}
\begin{align} \label{Pi^3-P3IN,PtrIN=0}
\big|\vec{P}_1(\ka)\big|\;
  &=\; \big|\vec{P}_2(\ka)\big|
    \\
&=\;
  {\frac{1}{2}}\; \frac{1}{\sqrt\ka}\vv
  -\; {\frac{1}{2}}\,
         (M_1^2 \!+\! M_2^2)\vv
         \sqrt\ka\vv
  -\; M_1^2\, M_2^2\vv
         \ka^{3\!/\!2}\vv
  +\; {\cal O}(\ka^{5\!/\!2})
    \nn \\[.5ex]
\big|\vec{P}_{1'}(\ka)\big|\;
  &=\; \big|\vec{P}_{2'}(\ka)\big|
    \tag{\ref{Pi^3-P3IN,PtrIN=0}$'$} \\
&=\;
  {\frac{1}{2}}\; \frac{1}{\sqrt\ka}\vv
  -\; {\frac{1}{2}}\,
         (M_{1'}^2 \!+\! M_{2'}^2)\vv
         \sqrt\ka\vv
  -\; M_{1'}^2\, M_{2'}^2\vv
         \ka^{3\!/\!2}\vv
  +\; {\cal O}(\ka^{5\!/\!2})
    \nn
    \\[-4.5ex]\nn
\end{align}
\begin{samepage}
vgl.\@ Gl.~(\ref{APP:P1vec-P3IN,PtrIN=0}$'$) wie auch die Gln.~(\ref{APP:P1^3-P3IN,PtrIN=0}),~(\ref{APP:P1vec-P3IN,PtrIN=0}),~-- und:%
\FOOT{
  \label{FN:t-fest}Es h"angt~$t$ ab von~$s$; sei insofern~\mbox{"`$t$ fest"'} generell verstanden bzgl.\@ des transversalen Anteils~\mbox{$-(\rb{P}_{1'} \!-\! \rb{P}_1)^2$} das hei"st als~\mbox{"`t fest bis auf~${\cal O}(\ka) \!\equiv\! {\cal O}(s^{-1})$"'} im Sinne kleiner effektiver Skalen~\mbox{$\rb{M}_i^2\,\ka \equiv \rb{M}_i^2\,s^{-1}$}.
}
%
\vspace*{-.5ex}
\begin{align} \label{tsymm-P3IN,PtrIN=0}
t(\ka)\;
&=\;
  -(\rb{P}_{1'} \!-\! \rb{P}_1)^2
    \\[.5ex]
&\phantom{=\;}
  - (\rb{M}_1^2 \!-\! \rb{M}_{1'}^2)\,(\rb{M}_2^2 \!-\! \rb{M}_{2'}^2)
    \vv \ka
    \nn \\[.5ex]
&\phantom{=\;}
  - {\frac{1}{2}}\,
  \Big\{
    (\rb{M}_1^2 \!-\! \rb{M}_{1'}^2) + (\rb{M}_2^2 \!-\! \rb{M}_{2'}^2)
  \Big\}
    \nn \\[-1ex]
&\phantom{=\; - {\frac{1}{2}}}
  \times \Big\{
    (\rb{M}_1^2 \!+\! \rb{M}_{1'}^2)\,(\rb{M}_2^2 \!-\! \rb{M}_{2'}^2) + 
        (\rb{M}_1^2 \!-\! \rb{M}_{1'}^2)\,(\rb{M}_2^2 \!+\! \rb{M}_{2'}^2)
  \Big\}
    \vv {{\ka}^2}
    \nn \\[.5ex]
&\phantom{=\;}
  + {{{\cal O}(\ka^3)}}
    \nn
    \\[-4.5ex]\nn
\end{align}
vgl.\@ Gl.~(\ref{APP:t-P3IN,PtrIN=0}).
Dabei ist gesetzt~\mbox{$\rb{P}_1 \!+\! \rb{P}_2 \!\equiv\! \bm{0}$} und~\mbox{$P_1^3 \!+\! P_2^3 \!\equiv\! 0$}, das hei"st Bezug genommen auf das Schwerpunktsystem bez"uglich der drei Raumrichtungen.
\end{samepage}

Die Asymmetrie zwischen~$\vep_1$ und~$\vep_2$, die unsere Diskussion und Herleitung suggeriert, ist selbstverst"andlich nur scheinbar; die Gln.~(\ref{Pi^3-P3IN,PtrIN=0}),~(\ref{Pi^3-P3IN,PtrIN=0}$'$) verifizieren dies a~posteriori.
Es folgt allgemein die~"`$1$-Gr"o"se"' aus der~"`$2$-Gr"o"se"' und die~"`$1'$-Gr"o"se"' aus der~"`$2'$-Gr"o"se"' durch Indexsubstitution~$1 \!\leftrightarrow\! 2$ beziehungsweise~$1' \!\leftrightarrow\! 2'$ im Sinne
\vspace*{-.25ex}
\begin{align} \label{Subst1<->2,1'<->2'}
\rb{M}_1^2\; \leftrightarrow\; \rb{M}_2^2\,,\vv
  \rb{P}_1\; \leftrightarrow\; \rb{P}_2\qquad
  \text{bzw.}\qquad
  \rb{M}_{1'}^2\; \leftrightarrow\; \rb{M}_{2'}^2\,,\vv
    \rb{P}_{1'}\; \leftrightarrow\; \rb{P}_{2'}
    \\[-4.25ex]\nn
\end{align}
und die "`gestrichenen"' Gr"o"sen durch die "`ungestrichenen"' durch~$1 \!\to\! 1'$,~$2 \!\to\! 2'$ im Sinne von Gl.~(\ref{Subst1->1',2->2'}).
\vspace*{-.5ex}

\section[Nahezu lichtartige~$T$-Amplitude: Konstruktion]{%
         Nahezu lichtartige~\bm{T}-Amplitude: Konstruktion}
\label{Sect:T-Amplitude.Konstruktion}

In Ref.~\cite{Nachtmann96} leitet Nachtmann die $T$-Amplitude von Hadron-Hadron-Streuung her bei festem kleinen invarianten Impulstransfer: typischerweise~\mbox{\,$\surd\!-t \!<\! 1\GeV^2$}, und im Limes unendlicher invarianter Schwerpunktenergie:~\mbox{\,$\surd s \!\to\! \infty$}.
Wir zeichnen diese Herleitung nach und verallgemeinern auf den Fall gro"ser, aber endlicher~\mbox{\,$\surd s$}.
\vspace*{-.5ex}

\subsection{Hadronniveau~I. Exposition}
\label{Subsect:HadronniveauI}

Seien~$\ket{h^i(P_i)}$ hadronische Zust"ande, die bei gro"sem, aber endlichem~$s$~-- im Sinne von Mesonen und deren f"uhrender Fock-Komponente~-- identifiziert seien mit Wellenpaketen von Quark-Antiquark-Paaren, oder: {\it Colour-Dipolen}.%
\FOOT{
  \label{FN:q(qq)-qqq}Diese Beschreibung schlie"st mit ein Baryonen, in denen zwei Quarks zusammengeclustert sind zu einem Diquark (das sich unter Eichtransformationen verh"alt wie ein Antiquark).   Die Beschreibung als Zustand dreier separierter Quarks ist prinzipiell identisch, nur aufwendiger; sei verwiesen auf die Refn.~\cite{Kraemer91,Nachtmann96}.
}
Seien definiert diese Zust"ande wie folgt:
%
\begin{align} \label{h-ket}
\ket{h^i(P_i)}\;
  =\; \frac{\de_{n_i{\bar n}_i}}{\sqrt{N_{\rm\!c}}}\vv
        \int \frac{d^2\rb{k}_i}{(2\pi)^2} \int_0^1 \frac{d\zet_i}{2\pi}\;
        {\tilde\vph}_{s_i\!{\bar s}_i}^i (\zet_i, \rb{k}_i)\vv
        \ket{ q_{s_i,n_i}(\zet_i, \rb{k}_i)\; 
                \bar{q}_{{\bar s}_i,{\bar n}_i}(\bzet_i, \rb{k}_i) }
\end{align}
Wir betrachten eichinvariante Zust"ande, so da"s wir a~priori herausziehen ein normiertes Kronecker-Symbol bez"uglich Eichgruppenindizes:
Die Indizes~$n_i$,~${\bar n}_i \!=\! 1,\ldots \dimDrst{F}$ beziehen sich auf die fundamentale Darstellung~$\mf{F}$ der Eichgruppe mit Dimension~\mbox{$\dimDrst{F} \!\equiv\! \Nc$}.
Bzgl.\@ der Differenzierung von~\mbox{\,$\mf{F}$} f"ur Quarks respektive~\mbox{\,$\mf{F}^\ast$} f"ur Antiquarks vgl.\@ Fu"sn.\,\FNg{FN:Drst_Fstar}.

Per constructionem durch Gl.~(\ref{h-ket}) ist~\mbox{\,${\tilde\vph}_{s_i\!{\bar s}_i}^i(\zet_i, \rb{k}_i)$} die quantentheoretische Wellenfunktion des Colour-Dipols~\mbox{\,$h^i(P_i)$}.
Das hei"st sie ist die Wahrscheinlichkeitsamplitude daf"ur, da"s das Quark~\mbox{\,$q$} in~\mbox{\,$\ket{h^i(P_i)}$} charakterisiert ist durch Spin~$s_i$, Colour~$n_i$, Flavour~$f_i$, das Antiquark~\mbox{\,$\bar{q}$} durch Spin~$\bar{s}_i$, Colour~$\bar{n}_i$, Flavour~$\bar{f}_i$~-- und entsprechende Raumzeit-Parameter.%
\FOOT{
  Seien weiterhin global unterdr"uckt Flavour-Indizes.
}%
Bewege sich der Colour-Dipol~\mbox{\,$h^i(P_i)$} ann"ahernd mit Lichtgeschwindigkeit, das hei"st~\mbox{\,$P_i^+ \!\gg\! P_i^-$} [oder umgekehrt] unter geeigneter Wahl der $x^3$-Achse.
Dann ist per definitionem~\mbox{\,${\tilde\vph}_{s_i\!{\bar s}_i}^i$} seine {\it Lichtkegelwellenfunktion\/} und die Raumzeit-Parameter sind genau~\mbox{\,$\zet_i$} und~\mbox{\,$\rb{k}_i$}.
Diese sind definiert wie folgt.

Seien~\mbox{\,$p_i$},~\mbox{\,${\bar p}_i$} f"ur~\mbox{\,$i \!\in\! 1\{,1',2,2'\}$} die Vierer-Impulse von Quarks respektive Antiquarks.
Seien diese definiert durch ihre gro"se Lichtkegel- und durch ihre transversale Vektorkomponente, vgl.~(\ref{h-Impuls}).
Es gilt f"ur die Quarks:
\begin{samepage}
%
\begin{alignat}{3} \label{Q_Impulse}
p_i^+\; &=\; \zet_i P_i^+ &\qquad
  \rb{p}_i\; =\; &\zet_i \rb{P}_i + \rb{k}_i &\qquad
  \text{f"ur}\quad &i \!=\! 1,1'
    \\[.5ex]
p_i^-\; &=\; \zet_i P_i^- &\qquad
                   &\sim                     &\qquad
  \text{f"ur}\quad &i \!=\! 2,2'
    \tag{\ref{Q_Impulse}$'$}
\end{alignat}
Die Komponenten der Antiquarks folgen durch Substitution
\end{samepage}
%
\begin{align} \label{Q-AQ-Substitution}
{\bar p}_i\;
  =\; p_i\, \big[\text{$\zet_i   \!\to\! \bzet_i$\vv
                    \& $\rb{k}_i \!\to\!    -\rb{k}_i$}\big]
    \\[-4ex]\nn
\end{align}
Dies ist unmittelbare Konsequenz der Forderungen:
\vspace*{-.5ex}
\begin{alignat}{3} \label{AQ_Impulse-Forderung}
p_i^+ + \bar{p}_i^+\;
   &\stackrel{\D!}{=}\; P_i^+ &\qquad
  \rb{p}_i\; +\; &\rbb{p}_i\;
    \stackrel{\D!}{=}\; \rb{P}_i &\qquad
  \text{f"ur}\quad &i \!=\! 1,1'
    \\[-.5ex]
p_i^- + \bar{p}_i^-\;
   &\stackrel{\D!}{=}\; P_i^- &\qquad
                 &\sim &\qquad
  \text{f"ur}\quad &i \!=\! 2,2'
    \tag{\ref{AQ_Impulse-Forderung}$'$}
\end{alignat}
die genau ausdr"ucken, da"s sich die gro"sen longitudinalen und die transversalen Komponenten von Quark und Antiquark~-- im Sinne von Impulserhaltung~-- addieren zu den entsprechenden Komponenten des Colour-Dipols~\mbox{\,$h^i(P_i)$}.
Quark und Antiquark auf dem Lichtkegel sind on mass-shell; infinitesimal wegger"uckt vom Lichtkegel seien sie angenommen als approximativ on mass-shell;~-- das hei"st die (Anti)Quarkimpulse besitzen Darstellungen "aquivalent denen in Gl.~(\ref{OnMassShell}) mit deren Implikationen f"ur die kleinen Lichtkegelkomponenten.

Wir weichen in den Gln.~(\ref{Q_Impulse}),~(\ref{Q_Impulse}$'$) bez"uglich des einen Punktes ab von Nachtmann in Ref.~\cite{Nachtmann96}, da"s der Anteil~\mbox{\,$\zet_i$} des Quarks beziehungsweise~\mbox{\,$\bzet_i \!\equiv\! 1 \!-\! \zet_i$} des Antiquarks am Gesamt-Lichtkegelimpuls~-- statt des Faktors~\mbox{\,$1\!/\!2$}~-- eingeht auch in den transversalen Komponenten.
Im allgemeinen ist~\mbox{\,$\zet_i \!\ne\! 1\!/\!2$}, so da"s unser Ansatz essentiell differiert von dem Nachtmanns.
F"ur den transversalen {\it Differenzimpuls\/} von Quark und Antiquark im Zustand~\mbox{\,$\ket{h^i(P_i)}$} folgt~\mbox{\,$(\rb{p}_i \!-\! \rbb{p}_i) \!=\! 2\rb{p} \!-\! \rb{P}$}, vgl.\@ die Gln.~(\ref{AQ_Impulse-Forderung}),(\ref{AQ_Impulse-Forderung}$'$), und mithilfe der Gln.~(\ref{Q_Impulse}),(\ref{Q_Impulse}$'$) unmittelbar:
\vspace*{-.25ex}
\begin{align} \label{relevant_relative}
\rb{k}_i\;
  =\; \frac{1}{2}\, \big(\rb{p}_i - \rbb{p}_i\big)
        + \Big(\frac{1}{2} \!-\! \zet_i\Big)\, \rb{P}_i
    \\[-4.25ex]\nn
\end{align}
Diskrepanz zu Nachtmann ist also, da"s%
  ~\mbox{\,$2\rb{k}$} und~\mbox{\,$(\rb{p}_i \!-\! \rbb{p}_i)$} in unserer Definition nicht identisch sind.
Dies zieht Konsequenzen nach sich in Bezug auf die Gr"o"se, die als relevanter~-- zum Impulstransfer konjugierter~-- Sto"sparameter der Streuung identifiziert wird. \\
\indent
In Anhang~\ref{APP-Subsect:LCWFN-Kovarianz} zeigen wir explizit, da"s bei Definition von%
  ~\mbox{\,$\zet_i$} als dem Anteil des Quarks am Gesamt-Lichtkegelimpuls:%
  ~\mbox{\,$p_i^\pm \!=\! \zet_iP^\pm$}~f"ur die gro"se Komponente,~-- Definition eines transversalen "`Relativimpulses"'~\mbox{\,$\rb{k}_i$} von Quark und Antiquark durch%
  ~\mbox{\,$\rb{p}_i \!=:\! \la'\rb{P}_i \!+\! \rb{k}_i$} notwendig impliziert%
  ~\mbox{\,$\la' \!\equiv\! \zet_i$}, wenn gefordert wird Kovarianz bez"uglich einer beliebigen Lorentz-Transformation%
  ~\mbox{\,$\La \!\equiv\! \big(\la^\mu{}_\nu\big)$} im weitesten Sinne, das hei"st formal:%
  ~\mbox{\,$p_i^\pm \!=\! \zet_iP^\pm \,\bm{\to}\, p_i^{\pm'} \!=\! \zet_iP^{\pm'}$} und%
  ~\mbox{\,$\rb{p}_i \!=\! \la'\rb{P}_i \!+\! \rb{k}_i \,\bm{\to}\, \rb{p}'_i
                     \!=\! \la'\rb{P}'_i \!+\! \rb{k}'_i$}.
Wir danken Dosch, Nachtmann, Michael R"uter f"ur die Diskussion, auf die hin der Beweis zustande gekommen ist.%
\FOOT{
  Forderung von Lorentz-Kovarianz l"a"st zu, den durch die Gln.~(\ref{Q_Impulse}),~(\ref{Q_Impulse}$'$) definierten Vektor~\mbox{\,$\rb{k}$} zu ersetzen durch~\mbox{\,$\rb{k}' \!=\! c\,\rb{k}, c \!\in\! \bbbr$}.   Es ist sinnvoll keinen Faktor einzuf"uhren zwischen~\mbox{\,$\rb{k}$} und den transversalen (Anti)Quarkimpulsen~\mbox{\,$\rb{p}_i$},~\mbox{\,$\rbb{p}_i$}.   Denn diese sind Fourier-konjugiert zu den (Anti)Quarkpositionen, ergo~-- vgl.\@ die Gln.~(\ref{Q_Impulse}),(\ref{Q_Impulse}$'$)~-- der Vektor~\mbox{\,$\rb{k}$} konjugiert zur Differenz dieser Positionen.   Zum einen kann die $T$-Amplitude nur abh"angen von Differenzvektoren, zum anderen ist der reine Abstand von Quark und Antiquark des Colour-Dipols Gr"o"se von suggestiver Anschauuung.
}

Seien die in Gl.~(\ref{h-ket}) definierten Zust"ande~$\ket{h^i(P_i)}$ normiert in Standardkonvention:
\begin{samepage}
%
\begin{align} \label{h_Norm}
\bracket{h^i(P)}{h^i(P')}\;
  =\; (2\pi)^3 2P_{0\!+}\, \de(\vec{P} \!-\! \vec{P}') \qquad
        \text{f"ur festes\vv$i \!\in\! \{1,1',2,2'\}$}
\end{align}
Dabei ist~$P_{0\!+} \!=\! {}\idx{+}\! \sqrt{\smash[b]{\vec{P}^2 \!+\! M^2}}$ der Betrag der Null-Komponente von~$P$; die Diracsche Delta-Distribution bezieht sich auf die Komponenten des Dreier-Vektors.
Wir zeigen in Anhang~\ref{APP:Skalarprodukt}, da"s hieraus f"ur die Lichtkegelwellenfunktionen folgt
%
\begin{align} \label{vph-k_Norm}
\int \frac{d^2\rb{k}}{(2\pi)^2}
  \int_0^1 \frac{d\zet}{2\pi}\vv 2\zet\bar{\zet}\vv
        {\tilde\vph}_{s\mskip-1mu\bar{s}}^{i\D\dagger}(\zet, \rb{k})\,
        {\tilde\vph}_{s\mskip-1mu\bar{s}}^i (\zet, \rb{k})\;
  =\; 1\qquad
  \text{f"ur festes\vv$i \!\in\! \{1,1',2,2'\}$}
\end{align}
Spinsumme impliziert.
Wir definieren mit
%
\begin{align} \label{vph-x_vph-k}
\vph_{s\mskip-1mu{\bar s}}^i(\zet,\rb{x})\;
 =\; \int \frac{d^2\rb{k}}{(2\pi)^2}\vv \efn{\D\iIM\,\rb{k} \!\cdot\! \rb{x}}\;
     \sqrt{2\zet\bzet}\vv {\tilde\vph}_{s\mskip-1mu{\bar s}}^i(\zet, \rb{k})
\end{align}
\end{samepage}%
die bez"uglich ihrer transversalen Koordinaten Fourier-transformierte Lichtkegelwellenfunktion; sei dabei~$\rb{x}$ die zu~$\rb{k}$ konjugierte Variable im (transversalen) Ortsraum.
Wir beachten den Faktor~$\sqrt{2\zet\bzet}$, in dem wir von der "ublichen Definition abweichen.
Normierung im der~${\tilde\vph}^i_{s_i\!{\bar s}_i}$ Impulsraum nach Gl.~(\ref{vph-k_Norm}) hei"st Normierung der~$\vph_{s \bar{s}}^i$ im Ortsraum wie
\vspace*{-.5ex}
\begin{align} \label{vph-x_Norm}
\int d^2\rb{x}
  \int_0^1 \frac{d\zet}{2\pi} \;
        \vph_{s \bar{s}}^{i\D\dagger}(\zet, \rb{x})\,
        \vph_{s \bar{s}}^i (\zet, \rb{x})\;
  =\; 1 \qquad
  \text{f"ur festes\vv$i \!\in\! \{1,1',2,2'\}$}
    \\[-4.5ex]\nn
\end{align}
Spinsumme impliziert.
In Anhang~\ref{APP:Skalarprodukt} zeigen wir f"ur das Skalarprokukt in diesem Sinne zweier "`plus"'- oder zweier "`minus"'-Zust"ande, das hei"st f"ur deren "Uberlapp die Relation
\vspace*{-.5ex}
\begin{alignat}{2} \label{h_bracket}
&\bracket{h^{i'}(P_{i'})}{h^i(P_i)}&&
    \\[-.5ex]
&\hspace*{\equalindent}
  =\; (2\pi)^3\, 2(P_i)_{0\!+}\, \de(\vec{P}_{i'} \!-\! \vec{P}_i)
        \cdot \int d^2&&\rb{x}\; \int_0^1 \frac{d\zet}{2\pi}\;
        \vph_{s\mskip-1mu{\bar s}}^{i'\D\dagger} (\zet, \rb{x})\,
        \vph_{s\mskip-1mu{\bar s}}^i (\zet, \rb{x})
    \nn \\[.5ex]
&&&\text{f"ur feste\vv$i,\,i'$, beide~$\in\! \{1,1'\}$ oder~$\in\! \{2,2'\}$}
    \nn
    \\[-4.5ex]\nn
\end{alignat}
F"ur~$i' \!\equiv\! i$ impliziert Gl.~(\ref{h_bracket}) die Normierung der Zust"ande~$\ket{h^i(P_i)}$ nach Gl.~(\ref{h_Norm}) dann, wenn die Lichtkegelwellenfunktionen normiert sind entsprechend Gl.~(\ref{vph-k_Norm}) respektive~(\ref{vph-x_Norm}).

Wir k"onnen bereits hier die Konsequenzen dessen ablesen, da"s wir abweichend von Nachtmann auch die transversalen (Anti)Quarkimpulse allgemein mit~$\zet_i$ und nicht mit dem konstanten Faktor~$1\!/\!2$ definieren, vgl.\@ die Gln.~(\ref{Q_Impulse}),~(\ref{Q_Impulse}$'$) und~(\ref{Q-AQ-Substitution}) bzw.~(\ref{relevant_relative}):

In der $T$-Amplitude der Hadron-Hadron-Streuung nach Gl.~(\ref{2hto2h}) werden analoge Integrationen auftreten wie in Gl.~(\ref{h_bracket}).
Sie werden sich beziehen auf den Vakuumerwartungswert zweier Wegner-Wilson-Loops, der nur noch von transversalen Vektoren abh"angt,~-- von der {\it Projektion der Loops in den Transversalraum\/}.
"Uber die Transversalvektoren wird dieser Vakuumerwartungswert aber sehr wohl noch abh"angen von den Faktoren~$\zet_i$, die daher nicht trivialerweise~-- im Sinne von Gl.~(\ref{vph-x_Norm}), das hei"st nur auf die Wellenfunktionen bezogen~-- ausintegriert werden k"onnen.
Das Bild ist zwar das getrennter (kleiner) transversalen Skala~$\surd-t$ und der (gro"ser) longitudinalen~$\surd s$, die Faktorisierung geschieht aber nur {\it effektiv\/}.

Im Limes gro"ser invarianter Schwerpunktenergie sind die streuenden Zust"ande zu {\it infinitesimal flachen\/} transversalen Scheiben Lorentz-kontrahiert, so da"s sie sich nur f"ur eine {\it infinitesimal kleine\/} Zeit durchdringen.
Ihre Wechselwirkung sollte beschrieben werden durch eine effektive zweidimensionale Quantenfeldtheorie im Transversalraum.
Da"s sie in der Tat {\it effektiv\/} ist in dem Sinne, da"s sie notwendigerweise longitudinale Dynamik subsummiert, zeigt die subtile Diskussion, die in diesem Rahmen Lipatovs Gluon-Emissions-Vertex verlangt, vgl.\@  Ref.~\cite{Verlinde93}.
\vspace*{-.5ex}

\subsection{Partonniveau. Durchf"uhrung}
\label{Subsect:Partonniveau}

Die Zust"ande~$h^i(P_i)$ in Definition nach Gl.~(\ref{h-ket}) sind Wellenpakete von Quark-Antiquark-Paaren.
Ihrer Streuung, vgl.\@ Gl.~(\ref{2hto2h}), liegt zugrunde auf Partonniveau der Proze"s
\begin{samepage}
\vspace*{-.25ex}
\begin{align} \label{4Qto4Q}
q(1) \;+\; \bar{q}(\bar1) \;+\; q(2) \;+\; \bar{q}(\bar2)\;
  \longrightarrow\;
q(1') \;+\; \bar{q}(\bar1') \;+\; q(2') \;+\; \bar{q}({\bar2'})
    \\[-4.25ex]\nn
\end{align}
wobei abk"urzend notiert sei
\vspace*{-.5ex}
\begin{align} \label{Q_abbrev}
q(i)\;
  &\equiv\; q_{s_i,n_i}(\zet_i, \rb{k}_i)\;
   \equiv\; q_{s_i,n_i}(p_i)
    \\[.5ex]
\bar{q}({\bar i})\;
  &\equiv\; \bar{q}_{{\bar s}_i,{\bar n}_i}(\bzet_i, \rb{k}_i)\;
   \equiv\; \bar{q}_{{\bar s}_i,{\bar n}_i}(\bar{p}_i)
    \tag{\ref{Q_abbrev}$'$}
    \\[-4.5ex]\nn
\end{align}
f"ur Quarks~$q$ beziehungsweise Antiquarks~$\bar{q}$, vgl.\@ die Gln.~(\ref{Q_Impulse}),~(\ref{Q_Impulse}$'$) bzw.~(\ref{Q-AQ-Substitution}).
Stehe generell~$i$ f"ur die Gesamtheit~$\{i\}$ der internen Quantenzahlen (Dirac, Colour, Flavour) und des Impulses mit Index~$i$; sei der Querstrich f"ur die Antiquarks unterdr"uckt, wenn er folgt aus dem Zusammenhang.
\end{samepage}

Der Partonproze"s nach Gl.~(\ref{4Qto4Q}) ist vollst"andig bestimmt durch das {\it $S$-Matrixelement\/}
%
\begin{align} \label{S-Element_in-in}
&\bracket{\, \bar{q}(\bar{2}')\, q(2')\, \bar{q}(\bar{1}')\, q(1'),\, \IN \,}{\,
             S\, \bracketM\,
             q(1)\, \bar{q}(\bar{1})\, q(2)\, \bar{q}(\bar{2}),\, \IN \,}
    \\[.5ex]
&=\; \bracket{\, 
       \bar{q}(\bar{2}')\, q(2')\, \bar{q}(\bar{1}')\, q(1'),\, \OUT \,}{\,
       q(1)\, \bar{q}(\bar{1})\, q(2)\, \bar{q}(\bar{2}),\, \IN \,}
    \nn
\end{align}
Dabei ist~$S$ der {\it $S$-Operator\/}, der {\it unit"ar\/}, das hei"st unter Normerhaltung die Darstellung eines beliebigen Zustands~$f$ bez"uglich der auslaufenden Basis abbildet auf seine Darstellung bez"uglich der einlaufenden:
\begin{align} \label{S-Operator}
\bra{f,\OUT\,} \;=:\; \bra{f,\IN\,}S \qquad
        \text{mit}\qquad
   SS^{\D\dagger}\; =\; S^{\D\dagger} S\; =\; \bbbone
\end{align}
Definition nach Ref.~\cite{Itzykson88}.

Die zweite Darstellung in Gl.~(\ref{S-Element_in-in}) bez"uglich der jeweils "`nat"urlichen"' Basis werde ausgeschrieben mithilfe entsprechender Erzeugungs- und Vernichtungsoperatoren, vgl.\@ Ref.~\cite{Itzykson88}:
%
\begin{align} \label{S-Element_ErzVern}
&\bracket{\, \bar{q}(\bar{2}')\, q(2')\, \bar{q}(\bar{1}')\, q(1'),\, \OUT \,}{\,
             q(1)\, \bar{q}(\bar{1})\, q(2)\, \bar{q}(\bar{2}),\, \IN \,}
    \\
&=\; \bra{\,\Om,\OUT\,}\;
        b_\OUT\!(\bar{2}')\, a_\OUT\!(2')\,
        b_\OUT\!(\bar{1}')\, a_\OUT\!(1')\,
        a^{\D\dagger}_\IN\!(1)\, b^{\D\dagger}_\IN\!(\bar{1})\,
        a^{\D\dagger}_\IN\!(2)\, b^{\D\dagger}_\IN\!(\bar{2})\;
     \ket{\,\Om,\IN\,}
    \nn
    \\[-4.5ex]\nn
\end{align}

Die Operatoren~$a^{\D\dagger}_\IN$,~$b^{\D\dagger}_\IN$ in Gl.~(\ref{S-Element_ErzVern}) erzeugen ein freies Quark, Antiquark im einlaufenden, die Operatoren~$a_\OUT$,~$b_\OUT$ vernichten ein (freies) Quark, Antiquark im auslaufenden Zustand, das hei"st bei gro"sen negativen beziehungsweise positiven Zeiten:~$x^0 \!\to\! \pm\infty$.

Es gelten~-- f"ur "`\IN"'- und "`\OUT"'-, das hei"st asymptotische Zust"ande~-- f"ur Quarks
%
\begin{alignat}{3}
&a^{\D\dagger}(i)\;&
  &=\; \int_{x^0} d^3\vec{x}\; \overline{\ps}(x)\, \ga^0\; \efn{-\iIM p_i x}\, u(i)\vv&
  &\equiv\; \int_{x^0} d^3\vec{x}\; \overline{\ps}(x)\, \ga^0\; \rbracket{x}{i}
    \label{ErzVernQ-Felder} \\
&a(i)\;&
  &=\; \int_{x^0} d^3\vec{x}\; \overline{u}(i)\, \efn{\iIM p_i x}\; \ga^0\, \ps(x)\vv&
  &\equiv\; \int_{x^0} d^3\vec{x}\; \rbracket{i}{x}\; \ga^0\, \ps(x)
    \tag{\ref{ErzVernQ-Felder}$'$}
    \\[-6.5ex]\nn
\intertext{\vspace*{-1.5ex}und f"ur Antiquarks}
&b^{\D\dagger}(\bar{i})\;&
  &=\; \int_{x^0} d^3\vec{x}\;
        \overline{v}(\bar{i})\, \efn{\iIM\bar{p}_i x}\;\ga^0\, \ps(x)\vv&
  &\equiv\; \int_{x^0} d^3\vec{x}\;
        \rbracket{\bar{i}}{x}\; \ga^0\, \ps(x)
    \label{ErzVernAQ-Felder} \\
&b(\bar{i})\;&
  &=\; \int_{x^0} d^3\vec{x}\;
        \overline{\ps}(x)\, \ga^0\; \efn{-\iIM\bar{p}_i x}\, v(\bar{i})\vv&
  &\equiv\; \int_{x^0} d^3\vec{x}\;
        \overline{\ps}(x)\, \ga^0\; \rbracket{x}{\bar{i}}
    \tag{\ref{ErzVernAQ-Felder}$'$}
\end{alignat}
verstanden als Integration "uber die Hyperfl"ache~\mbox{\,$x^0 \!=\! const.$}\;
Dabei sind
%
\begin{align} 
u(i)
  \equiv u_{s_i,n_i}(p_i)\qquad
  \text{und}\qquad
v(\bar{i})
  \equiv v_{{\bar s}_i,{\bar n}_i}(\bar{p}_i)
\end{align}
die Dirac-Spinoren f"ur Quarks beziehungsweise Antiquarks.
Die Indizes~$i$,~$\bar{i}$ stehen wieder f"ur die Gesamtheiten~$\{i\}$,~$\{\bar{i}\}$ der internen Quantenzahlen und der Impulsraumvariablen im Sinne von Gl.~(\ref{Q_abbrev}).
Zur Dirac-Algebra vgl.\@ auch Anhang~\ref{APP:Dirac-Algebra}.

Weiter treten in den Gln.~(\ref{ErzVernQ-Felder})-(\ref{ErzVernAQ-Felder}$'$) auf die freien (Anti)Quarkfelder~$\ps$,~$\overline{\ps}$; wir halten uns an die Standarddefinition
\begin{samepage}
\begin{align} \label{psFeld}
\ps(x)
  &= \int d^Rk\; {\T\sum}_s\; \Big(
        a_{s,n}(k)\, u_{s,n}(k)\, \efn{-\iIM k x}\,
      + b^{\D\dagger}_{s,n}(k)\, v_{s,n}(k)\, \efn{\iIM k x} \Big) \\
\overline{\ps}(x)
  &= \ps^{\D\dagger}(x)\, \ga^0 \tag{\ref{psFeld}$'$}
\end{align}
mit Integrationsma"s
\vspace*{-.5ex}
\begin{align} 
d^Rk\;
  =\; \frac{d^3\vec{k}}{(2\pi)^3\, 2k_{0\!+}}\;
  =\; \frac{dk}{(2\pi)^4}\vv 2\pi\, \de(k^2 - m^2)\, \th(k^0)\qquad
  k_{0\!+} = {}\idx{+}\! \sqrt{\smash[b]{\vec{k}^2 \!+\! m^2}}
    \\[-4.5ex]\nn
\end{align}
\end{samepage}%
Einschr"ankung der Vierer-Impulse auf die Vorw"arts-Masseschale~-- so das Ma"s~\mbox{\,$d^Rk$} selbst~-- ist {\it invariant\/} unter Transformationen der speziellen orthochronen Lorentz-Gruppe~\mbox{\,${\cal L}\idx{+}^{\bm{\scriptstyle\uparrow}}$}; vgl.\@ Anh.~\ref{APP-Sect:Lorentz-Boosts},~\ref{APP-Sect:aktiv-passiv}, insbes.\@ die Gln.~(\ref{APP:LT-La-Komponenten})-(\ref{APP:LT-La-Komponenten}$'''$) und~(\ref{APP:DrehungRij})-(\ref{APP:DrehungRij}$'''$) in~\ref{APP-Subsect:LCWFN-Kovarianz}.

Im Hinblick auf sp"ater haben wir durch die zweiten Identit"aten in den Gln.~(\ref{ErzVernQ-Felder})-(\ref{ErzVernAQ-Felder}$'$) die {\it Zustandsvektoren\/}~$\rket{\,\cdot\,}$ und~$\rbra{\,\cdot\,}$ f"ur Quark-Indizes~$i$ und Antiquark-Indizes~$\bar{i}$ eingef"uhrt.
Sie erf"ullen die {\it freie renormierte Dirac-Gleichung\/}, vgl.\@ unten Gl.~(\ref{rbra-rket_DiracGl}).
Definiert durch ihre Darstellung im Ortsraum haben wir f"ur ein- und auslaufend (Anti)Quarks
%
\begin{alignat}{5} \label{rbra-rket}
&\rbracket{x}{i}\;&
  &=\; \efn{-\iIM p_i x}\, u(i)& \qquad
        \text{bzw.}\qquad
&\rbracket{i}{x}\;&
  &=\; \overline{u}(i)\, \efn{\iIM p_i x}\;
  =\;& &\overline{\rbracket{x}{i}}
    \\
&\rbracket{\bar{i}}{x}\;&
  &=\; \overline{v}(\bar{i})\, \efn{-\iIM\bar{p}_i x}& \qquad
        \text{bzw.}\qquad
&\rbracket{x}{\bar{i}}\;&
  &=\; \efn{\iIM\bar{p}_i x}\, v(\bar{i})\;
  =\;& \ga^0\, &\rbracket{\bar{i}}{x}^{\D\dagger}
    \tag{\ref{rbra-rket}$'$}
\end{alignat}
Sie sind pr"agnante Bra-Ket-Notation im Diracschen Sinne.

%
\subsubsection{\DREI{L}{S}{Z}-Reduktionsformalismus}

Wir skizzieren den Reduktionsformalismus nach Lehmann, Symanzik, Zimmermann.
Wie er auf Basis der der Gln.~(\ref{ErzVernQ-Felder})-(\ref{ErzVernAQ-Felder}$'$) $S$-Matrixelemente von $n$ Partonen mit der entsprechenden $n$-{\it Punkt-Greenfunktion\/}~$G^{(n)}$ nach Gl.~(\ref{Greenfnen}) verkn"upft.

\vspace*{-.375ex}Im $n$-Teilchen-$S$-Matrixelement entsprechend den Gln.~(\ref{S-Element_in-in}),~(\ref{S-Element_ErzVern}) wird Gl.~(\ref{ErzVernQ-Felder}) angewandt zun"achst o.E.d.A.\@ auf einen Erzeuger~$a_\IN\!^{\D\dagger}(j)$ bez"uglich der einlaufenden "`\IN"'-Basis und mit den Quantenzahlen~$\{j\}$.
Das hei"st das Raumintegral bezieht sich auf die Hyperfl"ache konstanter {\it negativ unendlicher\/} Zeit~$x^0$; aufgefa"st als Funktion von~$x^0$ ist dies sein Grenzwert~\mbox{\,$x^0\!\to\!-\infty$}.
Mit der Identit"at
\vspace*{-.5ex}
\begin{align} \label{Int-Diff_LSZin}
\lim_{x^0\to-\infty}\;
  =\; \lim_{x^0\to\infty}\; -\; \int_{-\infty}^{\infty}dx^0\; \pa_0
    \\[-4.5ex]\nn
\end{align}
wird dieser "uberf"uhrt in die {\it Differenz\/} des Grenzwerts konstanter {\it positiv unendlicher\/} Zeit $x^0\!\to\!+\infty$ und des so generierten Ableitungsterm.
Dieser ist gleich dem vierdimensionalen Raumzeit-Integral
\vspace*{-.5ex}
\begin{align} \label{AbleitungstermQin}
\pa_{(j)}\;
  =\; \int dx\; \pa_0\; \big[\, \overline{\ps}(x)\, \ga^0\; \rbracket{x}{j}\, \big]
    \\[-4.5ex]\nn
\end{align}
\vspace*{-.375ex}Der Limes-Term ist~-- vgl.\@ Gl.~(\ref{ErzVernQ-Felder}) von rechts nach links~-- gerade der urspr"ungliche Erzeuger bez"uglich der auslaufenden "`\OUT"'-Basis:~$a_\OUT^{\D\dagger}\!(j)$. \\
\indent
Dieser wird nach links permutiert, wo er letztlich den Vakuumzustand annihiliert.%
\FOOT{
  \label{FN:OmInvarianz}Das Vakuum ist invariant unter dem $S$-Operator, das hei"st~$\ket{\,\Om\,} \equiv \ket{\,\Om,\IN\,} \equiv \ket{\,\Om,\OUT\,}$.
}
Bevor er dies tut, ergibt er an {\it genau einem\/} der Vernichter, o.E.d.A.\@ an~$a_\OUT\!(j')$ mit Index~$j'$, den Zusatzterm~$\de(j',j)$; dies aufgrund der nichttrivialen Antikommutatorrelationen
\begin{samepage}
%
\begin{alignat}{3} \label{Antikommutator}\;
&\big\{ a(i'),& a^{\D\dagger}(i) \big\}\;
   &=\; \de_{s_{i'}\!s_i} \de_{n_{i'}\!n_i}\;
        (2\pi)^3\, 2(p_i)_{0\!+}\;
        \de(\vec{p}_{i'} \!-\! \vec{p}_i)\;&
   &=\; \de(i',i)
    \\[.5ex]
&\big\{ b({\bar i}'),& b^{\D\dagger}({\bar i}) \big\}\;
   &=\; \de_{{\bar s}_{i'}\!{\bar s}_i} \de_{{\bar n}_{i'}\!{\bar n}_i}\;
        (2\pi)^3\, 2({\bar p}_i)_{0\!+}\;
        \de(\vec{\bar p}_{i'} \!-\! \vec{\bar p}_i)\;&
   &=\; \de({\bar i}',{\bar i}) \tag{\ref{Antikommutator}$'$}
\end{alignat}
\vspace*{-.375ex}symmetrisieren entsprechend~$(p_i)_{0\!+} \!=\! \surd(p_{i'})_{0\!+}(p_i)_{0\!+}$ und~$(\bar{p}_i)_{0\!+} \!=\! \surd(\bar{p}_{i'})_{0\!+}(\bar{p}_i)_{0\!+}$ unter der Diracschen Delta-Distribution.
Mit jeder Vertauschung nach links ergibt~$a_\OUT^{\D\dagger}\!(j)$ nur noch ein Vorzeichen und letztlich Null.
\vspace*{-.375ex}Exakt gesprochen tauscht die {\it Summe dreier Terme\/} nach links durch und sammelt in definierter Weise Vorzeichen auf: der Vernichter~$a_\OUT^{\D\dagger}\!(i)$, der Zusatzterm~$\de(j',j)$ und der (negative) Gra"smann-wertige Ableitungsterm. \\
\indent
Der Ableitungsterm~$\pa_{(j)}$, vgl.\@ Gl.~(\ref{AbleitungstermQin}), ist dabei multipliziert mit~$a_\OUT\!(j')$.
Dieser Vernichter wird seinerseits geschrieben als Grenzwert~$x^0\!\to\!+\infty$ eines Raumintegrals nach Gl.~(\ref{ErzVernQ-Felder}$'$).
Mit der Identit"at
\vspace*{-.5ex}
\begin{align} 
\lim_{x^0\to\infty}\;
  =\; \lim_{x^0\to-\infty}\; +\; \int_{-\infty}^{\infty}dx^0\; \pa_0 
    \\[-4.5ex]\nn
\end{align}
\end{samepage}%
vgl.\@ Gl.~(\ref{Int-Diff_LSZin}), wird dieser "uberf"uhrt in die {\it Summe\/} des Grenzwerts~$x^0\!\to\!+\infty$ und dem zu Gl.~(\ref{AbleitungstermQin}) analogen Ableitungsterm
\vspace*{-.5ex}
\begin{align} \label{AbleitungstermQout}
\pa_{(j')}\;
  =\; \int dx'\; \pa_0\; \big[\, \rbracket{j'}{x'}\; \ga^0\, \ps(x')\, \big]
    \\[-4.5ex]\nn
\end{align}
Der Limes-Term ist genau der Vernichter~$a_\IN\!(j')$, wird nach rechts durchpermutiert, sammelt dabei Vorzeichen auf und annihiliert schlie"slich den Vakuumzustand~\citeFN{FN:OmInvarianz}.
Dabei entstehen a~priori Zusatzterme~$\de(j',i)$ aufgrund der nichttrivialen Antikommutatorrelationen nach Gl.~(\ref{Antikommutator}).
Diese verschwinden alle identisch, da der Index~$i \!\equiv\! j$ nicht mehr vorkommt.
Der Limes-Term f"allt daher \vspace*{-.375ex}weg.

In conclusio sind die Operatoren~$a^{\D\dagger}_\IN\!(j)$ und~$a_\OUT\!(j')$ verschwunden und an ihre Stelle getreten, mit definiertem Vorzeichen, die Differenz
\vspace*{-.5ex}
\begin{align} \label{LSZ_entity}
\de(j',j) - \pa_{(j')} \cdot \pa_{(j)}
    \\[-4.5ex]\nn
\end{align}
mit expliziten Gestalt der~$\pa_{(j)}$,~$\pa_{(j')}$ wie in den Gln.~(\ref{AbleitungstermQin}),~(\ref{AbleitungstermQout}). \\
\indent
In den Ableitungstermen~$\pa_{(j)}$,~$\pa_{(j')}$ k"onnen die auf die Zustandsvektoren~$\rket{i}$,~$\rbra{j'}$ wirkenden Differentiationen~$\ga^0\del{x}_0$ beziehungsweise~$\ga^0\del{x'}_0$ mithilfe der freien renormierten Dirac-Gleichung~-- die~$\rket{j}$,~$\rbra{j'}$ l"osen, vgl.\@ unten Gl.~(\ref{rbra-rket_DiracGl})~-- ersetzt werden durch die r"aumlichen Differentiationen.
Diese k"onnen partiell integriert werden~-- die Randterme verschwinden~-- und wirken, von links beziehungsweise rechts, auf die Felder~$\ps(x)$,~$\overline{\ps}(x)$: zusammen mit~$\ga^0\del{x_j}_0$ beziehungsweise~$\ga^0\del{x_{j'}}_0$ genau als die \vspace*{-.375ex}{\it freien renormierten Dirac-Operatoren\/}. \\
\indent
Dieses Verfahren wird sukzessive auf s"amtliche Operatoren~$a^{\D\dagger}_\IN\!(i)$ und~$a_\OUT\!(i')$ angewandt.
Es treten a~priori s"amtliche verschiedenen Kontraktionen von Indexpaaren~$\de(i',i)$,~$\de(\bar{i}',\bar{i})$ auf im Sinne von Gl.~(\ref{LSZ_entity}).
Welche davon ungleich Null sind, folgt "uber die nichttrivialen Antikommutatorrelationen nach den Gln.~(\ref{Antikommutator}),~(\ref{Antikommutator}$'$) letztlich aus den internen Quantenzahlen und Impulsen der (Anti)Quarks,~-- der Kinematik der Streuung.

%
\subsubsection{(Anti)Quark-Propagation durch vorgegebenen Gluon-Hintergrund}

Wir kommen zur"uck auf das $S$-Matrixelement nach den Gln.~(\ref{S-Element_in-in}),~(\ref{S-Element_ErzVern}).
Resultat ist die folgende Reduktion:
Es wirkt f"ur jedes der acht Argumente der (Anti)Quarks der freie renormierte Dirac-Operator auf die {\it 8-Punkt-Greenfunktion\/}~$G^{(8)}$
\begin{align} 
&G^{(8)} \!(\bar{2}',2', \bar{1}',1', 1,\bar{1}, 2,\bar{2}) \\[.5ex]
&\hspace*{\equalindent} =\; \bra{\,\Om}\; T\;
        \overline{\ps}(\bar{2}')\, \ps(2')\,
        \overline{\ps}(\bar{1}')\, \ps(1')\,
        \overline{\ps}(1)\, \ps(\bar{1})\,
        \overline{\ps}(2)\, \ps(\bar{2})\;
                        \ket{\Om\,} \nn
\end{align}
vgl. Gl.~(\ref{Greenfnen}) und Fu"sn.\,\FN{FN:OmInvarianz}.
Die Integrationen~$dx$,~$dx'$ in~(\ref{LSZ_entity})~-- siehe auch die Gln.~(\ref{AbleitungstermQin}), (\ref{AbleitungstermQout})~-- stehen f"ur abschlie"sende Fourier-Transformation in den Impulsraum.

Im betrachteten~\mbox{\,$t$}-Kanal, vgl. die Gln.~(\ref{Q_Impulse}),~(\ref{Q_Impulse}$'$) und~(\ref{Q-AQ-Substitution}), folgt
%
\begin{align} \label{S-Element_calM}
&\bracket{\, \bar{q}(\bar{2}')\, q(2')\, \bar{q}(\bar{1}')\, q(1'),\,\IN \,}{\,
             S\, \bracketM\,
             q(1)\, \bar{q}(\bar{1})\, q(2)\, \bar{q}(\bar{2}),\,\IN \,}
    \\[1ex]
&\begin{aligned}[t]
 =\; \vacL\,
       \big( &\de(1',1) - \iIM\, Z\idx{2}^{-1}\, {\cal M}_{1'\mskip-1mu 1}^F[A] \big)\,
       \big( \de(\bar{1}',\bar{1})
               - \iIM\, Z\idx{2}^{-1}\, \bar{\cal M}_{\bar{1}'\mskip-1mu\bar{1}}^F[A] \big)
    \nn \\[.5ex]
      \times\big( &\de(2',2) - \iIM\, Z\idx{2}^{-1}\, {\cal M}_{2'\mskip-1mu 2}^F[A] \big)\,
        \big( \de(\bar{2}',\bar{2})
                - \iIM\, Z\idx{2}^{-1}\, \bar{\cal M}_{\bar{2}'\mskip-1mu\bar{2}}^F[A] \big)\, 
    \vacR \nn
 \end{aligned}
    \nn
\end{align}
wobei wir formal wieder zur"uckkehren von der \VIER{B}{R}{S}{T}-Operatornotation~$\bra{\,\Om} \cdot \ket{\Om\,}$ zur Funktionalintegraldarstellung~$\vac{\;\cdot\;}$, vgl.\@ Gl.~(\ref{vev}).
Wir merken an, da"s der Ausdruck nach Gl.~(\ref{S-Element_calM}) "uber~\mbox{\,$\de(i,i)$} und~\mbox{\,$\de(\bar{i},\bar{i})$} noch s"amtliche nichtzusammenh"angenden, {\it disconnected\/} Terme mit einschlie"st.

Diese Gleichung ist fundamentale Basis unserer Analyse.
Wir diskutieren wie folgt die auftretenden Gr"o"sen.

Die Konstante~$Z\idx{2}$ ist Element des reellen Einheitsintervalls~\mbox{$(0,1]$} und verkn"upft mit der Renormierung des Quark-Spinorfelds durch
\bea \label{Z2-renorm}
\ps(x) \;\underset{\text{$x^0 \!\to\! \pm\infty$}}{\longrightarrow}\;
  Z\idx{2}^{-1\!/\!2}\; \ps_{\overset{\scriptstyle\rm out}{\scriptstyle\rm in}}(x)
\eea
im schwachen Sinne, das hei"st als Erwartungswert.%
\FOOT{
  \label{FN:RenKonstanten}Vgl.\@ die Bemerkung in Anschlu"s an die Gln.~(\ref{SF(0)}),~(\ref{SF(0)}$'$).
}
Sie tritt auf im Zuge des "Ubergangs im \DREI{L}{S}{Z}-Formalismus von Lagrange'schen {\it nichtrenormierten\/} Quarkfeldern zu asymptotischen "`\IN"'- und "`\OUT"'-, das hei"st {\it renormierten\/} Feldern.

Die Gr"o"sen~${\cal M}_{i'\mskip-1mu i}^F$, und~$\bar{\cal M}_{\bar{i}'\mskip-1mu \bar{i}}^F$ sind bezeichnet und definiert nach Nachtmann:
\vspace*{-.5ex}
\begin{alignat}{3} \label{calM}
&{\cal M}_{i'\mskip-1mu i}^F[A]\;&
  &=\; \phantom{-\,}
       \rbra{i'}\vv&
         \big(\iIM\ga^\mu \paR_\mu - m_{\rm ren.}\big)\vv
        &S_F[A]\vv \big(\iIM\ga^\nu \paL_\nu + m_{\rm ren.}\big)\vv
         \rket{i}
    \\[.5ex]
&\bar{\cal M}_{\bar{i}'\mskip-1mu \bar{i}}^F[A]\;&
  &=\; -\, \rbra{\bar{i}}\vv&
         \big(\iIM\ga^\mu \paR_\mu - m_{\rm ren.}\big)\vv
        &S_F[A]\vv \big(\iIM\ga^\nu \paL_\nu + m_{\rm ren.}\big)\vv
         \rket{\bar{i}'}
    \tag{\ref{calM}$'$}
    \\[-4.5ex]\nn
\end{alignat}
Wir betonen die Vertauschung~\mbox{\,$\rbra{\bar{i}} \cdot \rket{\bar{i}'}$} f"ur Antiquarks gegen"uber~\mbox{\,$\rbra{i'} \cdot \rket{i}$} f"ur Quarks.
Es ist definiert~\mbox{\,$m_{\rm ren.}$} als die {\it renormierte\/} Quarkmasse, im Gegensatz zu~\mbox{\,$m$} der nichtrenormierten Lagrange'schen.
Sie sind verkn"upft "uber die Beziehung~\citeFN{FN:RenKonstanten}
%
\begin{align} \label{m-renorm}
m_{\rm ren.}\;
  =\; m\; +\; \de m
\end{align}
Die Gr"o"sen~\mbox{\,${\cal M}_{i'\mskip-1mu i}^F$},~\mbox{\,$\bar{\cal M}_{\bar{i}'\mskip-1mu \bar{i}}^F$} sind, wie notiert in den Gln.~(\ref{calM}),~(\ref{calM}$'$), Funktionale im Gluon-Eichfeld~\mbox{\,$A$}~-- "uber das Eichfeld-Funktional~\mbox{\,$S_F[A] \!\equiv\! S_F[A](x,x';m^2)$}.
Die Bedeutung von \mbox{\,${\cal M}_{i'\mskip-1mu i}^F$},~\mbox{\,$\bar{\cal M}_{\bar{i}'\mskip-1mu \bar{i}}^F$} erschlie"st sich "uber diese Gr"o"se.

Bevor wir hierauf n"aher eingehen, erinnern wir an die Definition der Zustandsvektoren~$\rket{\cdot}$,~$\rbra{\cdot}$ in den Gln.~(\ref{rbra-rket}),~(\ref{rbra-rket}$'$) und machen die folgende Anmerkung zur Schreibweise.
Wir arbeiten in der Bra-Ket-Notation Nachtmanns, in der die Gr"o"sen aufgefa"st sind als Vektoren beziehungsweise Matrizen bez"uglich des internen Eichgruppen- und Dirac-Raumes {\it und\/} des Konfigurationsraumes.
Multiplikation impliziert daher zus"atzlich zur Summation "uber Eichgruppen- und Dirac-Indizes Vierer-Integration der Ortsraum-Variable "uber den gesamten Raum.
Die Eins lautet ausgeschrieben~\mbox{\,$\bbbone \!=\! \de_{n'n} \de_{s's} \de(x' \!-\! x))$} und impliziert die Diracsche Delta-Distribution bez"uglich der vierdimensionalen Minkowski-Raumzeit.
So lauten die Gln.~(\ref{calM}),~(\ref{calM}$'$) ausgeschrieben:
%
\begin{alignat}{2}
&{\cal M}_{i'\mskip-1mu i}^F[A]&&
    \label{calM_explizit} \\
  &=\; \phantom{-\,}
         \int\!dx\, dx'\vv
         \rbracket{i'}{x}\vv
         (\iIM\delslashR{x} - m_{\rm ren.})\vv&
        &S_F[A](x,x'; m^2)\vv
         (\iIM\delslashL{x'} + m_{\rm ren.})\vv
         \rbracket{x'}{i}
    \nn \\[.5ex]
&\bar{\cal M}_{\bar{i}'\mskip-1mu \bar{i}}^F[A]&&
    \tag{\ref{calM_explizit}$'$} \\
  &=\, -\,
       \int\!dx\, dx'\vv
         \rbracket{\bar{i}}{x}\vv
         (\iIM\delslashR{x} - m_{\rm ren.})\vv&
        &S_F[A](x,x'; m^2)\vv
         (\iIM\delslashL{x'} + m_{\rm ren.})\vv
         \rbracket{x'}{\bar{i}'}
    \nn
    \\[-4.5ex]\nn
\end{alignat}
in Notation mit wohlbekanntem {\it Feynman-Slash\/}:~$\paslash \!\equiv\! \ga^\mu\pa_\mu$.

Wir kommen zur"uck auf das Funktional~$S_F[A]$.
Es ist genau die {\it Greenfunktion\/}~-- mit Polvorschrift nach {\it Feynman\/}~[Skript~"`$F$"'\/]~-- der {\it vollen nichtrenormierten Dirac-Gleichung\/}, das hei"st f"ur ein (Anti)Quarks in vorgegebener Konfiguration~$A(x)$ des Gluon-Eichfelds.%
~An\-wendung des vollen nichtrenormierten Dirac-Operators ergibt per~definitionem von links:%
\FOOT{
  \label{FN:Vierer-Integration}\vspace*{-.75ex}Die implizite Vierer-Integration ist bereits ausgef"uhrt:   Aufgrund der Diagonalit"at der~Differential\-operatoren steht zun"achst~\mbox{\,$\de(x \!-\! x')\, \big(\iIM\ga^\mu \DelR{x'}_\mu[A] - m\big)$} respektive~\mbox{\,$\big(\iIM\ga^\mu (-\delL{x}_\mu[A] + \iIM gA_\mu(x)) - m\big)\, \de(x \!-\! x')$}.
}
%
\begin{align} \label{SF_lim}
&\big(\iIM\ga^\mu D_\mu[A] - m\big)\vv
  S_F[A]\;
  =\; -\, \bbbone
    \\[.5ex]
&\text{d.h.}\qquad
  \big(\iIM\ga^\mu \DelR{x}_\mu[A] - m\big)\vv
  S_F[A](x,x';m^2)\;
  =\; -\, \de(x-x')
    \tag{\ref{SF_lim}$'$}
\end{align}
mit~$\Del{x}_\mu[A] = \del{x}_\mu \!+\! \iIM gA_\mu(x)$, vgl.\@ Gl.~(\ref{kovAbl}),~-- und von rechts:\citeFN{FN:Vierer-Integration}
\vspace*{-.5ex}
\begin{align} \label{SF_r}
S_F[A]\; \big(\iIM\ga^\mu (-\paL_\mu + \iIM gA_\mu) - m\big)\;
  =\; -\bbbone
    \\[-4.5ex]\nn
\end{align}
Differentiation bezieht sich generell auf s"amtliche Funktionen in Richtung des Pfeils.

In Hinblick auf die Differentialoperatoren in den Gl.~(\ref{calM}),~(\ref{calM}$'$) geben wir~-- entsprechend den Gln.~(\ref{SF_lim}),~(\ref{SF_r})~-- die Relationen an f"ur die {\it Feynmansche Greenfunktion der freien renormierten Diracgleichung\/}~\mbox{\,$S_F^{(0)}$}:
\vspace*{-.75ex}
\begin{align} \label{SF(0)}
&\big(\iIM\ga^\mu \paR_\mu - m_{\rm ren.}\big)\vv
  S_F^{(0)}\;
  =\; -\bbbone
    \\[.5ex]
&S_F^{(0)}\vv
  \big(\iIM\ga^\mu (-\paL_\mu) - m_{\rm ren.}\big)\;
  =\; -\bbbone
    \tag{\ref{SF(0)}$'$}
    \\[-4.5ex]\nn
\end{align}
Sie sind relevant f"ur die in den Gln.~(\ref{rbra-rket}),~(\ref{rbra-rket}$'$) definierten Zustandsvektoren~\mbox{\,$\rket{\,\cdot\,}$},~\mbox{\,$\rbra{\,\cdot\,}$}, die per definitionem L"osungen sind der freien renormierten Diracgleichung.
Die Gr"o"sen~\mbox{$m_{\rm ren.}$} und~\mbox{\,$Z\idx{2}$} sind definiert als der Pol von~$S_F^{(0)}$ im Impulsraum respektive das zugeh"orige Residuum; Gl.~(\ref{Z2-renorm}) und~(\ref{m-renorm}) sind unmittelbare Konsequenz von dieser Definition.

Physikalisch ist~$S_F$ genau der {\it volle nichtrenormierte Feynman-Propagator\/} eines Quarks im Gluon-Hintergrund~$A$ und tritt auf im $S$-Matrixelement nach Gl.~(\ref{S-Element_calM}) aufgrund der Relation mit den nichtrenormierten, das hei"st von $m$ abh"angenden Quarkfeldern:
\vspace*{-.75ex}
\begin{align} \label{SF-ps}
&\iIM\, S_F[A](x,x';m^2)\;
  =\; \ps(x)\, \overline{\ps}(x')\;
  =\; \iIM\, \big({\cal O}_{\rm D}{}^{\zzz-1}\big)[A](x,x';m^2)
\end{align}
Das hei"st~\mbox{\,$\vac{\iIM\,S_F[A]} \!=\! \vac{\ps\overline{\ps}} \!=\! \vac{\iIM\,{\cal O}_{\rm D}{}^{\!\!\!-1}[A]}$} ist der volle nichtrenormierte Feynman-Propa\-gator der QCD, geschrieben in Referenz auf Gl.~(\ref{Greenfnen}) mit~\mbox{\,${\cal O}_{\rm D}{}^{\zzz-1} \!=\! (\iIM\ga^\mu \Del{x}_\mu - m)^{-1}$}, wie definiert in Gl.~(\ref{ErzFnl1}).

Die physikalische Bedeutung der das $S$-Matrixelement bestimmenden Gr"o"sen~${\cal M}_{i'\mskip-1mu i}^F$,~$\bar{\cal M}_{\bar{i}'\mskip-1mu \bar{i}}^F$, vgl.\@ die Gln.~(\ref{calM}),~(\ref{calM}), erschlie"st sich dadurch, da"s wir sie mithilfe der Gluonfeld-abh"angigen Zustandsvektoren~$\rket{\ps_i^F}$ beziehungsweise~$\rbra{\bar{\ps}_{\bar i}^F}$ umschreiben.
Definiert durch
\begin{samepage}
\vspace*{-.75ex}
\begin{alignat}{2} \label{psiVekt}
&\rket{\ps_i^F[A]}\;&
  &=\; S_F[A]\vv
         \big(\iIM\ga^\mu \paL_\mu + m_{\rm ren.}\big)\vv
         \rket{i}
    \\
&\rbra{\bar{\ps}_{\bar i}^F[A]}\;&
  &=\; \rbra{\bar{i}}\vv
         \big(\iIM\ga^\mu \paR_\mu - m_{\rm ren.}\big)\vv
         S_F[A]
    \tag{\ref{psiVekt}$'$}
    \\[-4.5ex]\nn
\end{alignat}
l"osen sie die {\it volle nichtrenormierte Dirac-Gleichung\/}, das hei"st:
\vspace*{-.75ex}
\begin{align} \label{psiVekt_DiracGl}
&\big(\iIM\ga^\mu (\paR_\mu + igA_\mu) - m\big)\vv
  \rket{\ps_i^F}\;
  =\; 0
    \\
&\rbra{\bar{\ps}_{\bar i}^F}\vv
  \big(\iIM\ga^\mu (-\paL_\mu + igA_\mu) - m\big)\;
  =\; 0
    \tag{\ref{psiVekt_DiracGl}$'$}
    \\[-4.5ex]\nn
\end{align}
Dabei ist wesentlich, da"s die Zust"ande~$\rket{i}$,~$\rbra{\bar{i}}$ die {\it freie renormierte Dirac-Gleichung\/} l"osen, das hei"st:
\vspace*{-.75ex}
\begin{align} \label{rbra-rket_DiracGl}
&\big(\iIM\ga^\mu \paR_\mu - m\big)\vv
  \rket{i}\;
  =\; 0
    \\
&\rbra{\bar{i}}\vv
  \big(\iIM\ga^\mu (-\paL_\mu) - m\big)\;
  =\; 0
    \tag{\ref{rbra-rket_DiracGl}$'$}
    \\[-4.5ex]\nn
\end{align}
Mithilfe von~$\rket{\ps_i^F}$ beziehungsweise~\mbox{\;$\rbra{\bar{\ps}_{\bar i}^F}$} schreiben sich die Gln.~(\ref{calM}),~(\ref{calM}$'$) wie folgt:
\vspace*{-.75ex}
\begin{alignat}{5} \label{calM_psiVekt-F}
&{\cal M}_{i'\mskip-1mu i}^F[A]\;&
  &=\;  \rbra{i'}\; \big(\iIM\ga^\mu \paR_\mu - m_{\rm ren.}\big)\; \rket{\ps_i^F[A]}\;&
  &=\;  \rbra{i'}\; \big(g\ga^\mu A_\mu - \de m\big)\;           \rket{\ps_i^F[A]}
    \\[.5ex]
&\bar{\cal M}_{\bar{i}'\mskip-1mu \bar{i}}^F[A]\;&
  &=\;  \rbra{\bar{\ps}_{\bar i}^F[A]}\;
          \big(\iIM\ga^\mu \paL_\mu + m_{\rm ren.}\big)\; \rket{\bar{i}'}\;&
  &=\;  \rbra{\bar{\ps}_{\bar i}^F[A]}\;
          \big(g\ga^\mu A_\mu - \de m\big)\; \rket{\bar{i}'}
    \tag{\ref{calM_psiVekt-F}$'$}
    \\[-4.5ex]\nn
\end{alignat}
Die zweiten Gleichheitszeichen gelten aufgrund der Gln.~(\ref{psiVekt_DiracGl}),~(\ref{psiVekt_DiracGl}$'$).
Nachtmann findet diese Identit"aten "aquivalent, indem er benutzt:
%
\begin{align} \label{Lippmann-Schwinger}
S_F[A]\;
  &=\; S_F^{(0)}\vv
         -\vv S_F[A]\; \big(g\ga^\mu A_\mu - \de m\big)\; S_F^{(0)}
    \\[.5ex]
  &=\; {\T\sum}_{n=0}^\infty
         \big[ - S_F^{(0)}\; \big(g\ga^\mu A_\mu - \de m\big) \big]^n\vv
         S_F^{(0)}
    \tag{\ref{Lippmann-Schwinger}$'$}
\end{align}
\end{samepage}%
die {\it Lippmann-Schwinger-Gleichung\/} f"ur~\mbox{\,$S_F$}, vgl.\@ die Gln.~(\ref{SF(0)}),~(\ref{SF(0)}$'$). 

Auf Basis der Darstellungen durch die Gln.~(\ref{calM_psiVekt-F}),~(\ref{calM_psiVekt-F}$'$) halten wir fest:
Die Gr"o"se~\mbox{\,${\cal M}_{i'\mskip-1mu i}^F$} ist das Integral "uber eine vollst"andige einlaufende Wellenfunktion~\mbox{\,$\ps_i^F(x) \!=\! \rbracket{x}{\ps_i^F}$} eines Quarks~\mbox{\,$i$}, das an einem vorgegebenen (externen) Gluon-Potential \mbox{\,$(g\Aslash - \de m)$} streut und projiziert wird auf eine freie auslaufende Quark-Welle~\mbox{\,$\rbracket{x}{i'}$}.~--
Diese Darstellungen suggerieren daher die physikalische Interpretation von~\mbox{\,${\cal M}_{i'\mskip-1mu i}^F[A]$} als der entsprechenden {\it Streuamplitude\/}.
Die Gr"o"se~\mbox{\,$\bar{\cal M}_{\bar{i}'\mskip-1mu \bar{i}}^F[A]$} ist analog zu interpretieren als die Streuamplitude einer vollst"andigen einlaufenden Wellenfunktion~\mbox{\,$\bar{\ps}_{\bar i}^F(x) \!=\! \rbracket{\bar{\ps}_{\bar i}^F}{x}^{\D\ast}$} eines Antiquarks~\mbox{\,$\bar{i}$} an demselben (externen) Gluon-Potential.
\vspace*{-.5ex}

\bigskip\noindent
Die Amplituden~\mbox{\,${\cal M}_{i'\mskip-1mu i}^F$},~\mbox{\,$\bar{\cal M}_{\bar{i}'\mskip-1mu \bar{i}}^F[A]$}~-- vgl.\@ die Gln.~(\ref{calM}),~(\ref{calM}$'$) und~(\ref{calM_psiVekt-F}),~(\ref{calM_psiVekt-F}$'$)~-- sollten miteinander verkn"upft sein aufgrund der Invarianz der QCD unter Ladungskonjugation~\mbox{\,${\cal C}$}.
Wir verweisen auf die einleitende Diskussion auf Seite~\pageref{T:diskreteSymmetrien}.

Hierbei werden Teilchen mit Antiteilchen vertauscht, das hei"st Operatoren~$a \!\leftrightarrow\! b$,~$a^{\D\dagger} \!\leftrightarrow\! b^{\D\dagger}$ im $S$-Matrixelement.
Dies hei"st bez"uglich der (Anti)Quark- und der Gluon-Eichfelder:
\vspace*{-.75ex}
\begin{alignat}{5} \label{calC-Transf}
&\ps&\vv
  &\overset{{\cal C}}{\longrightarrow}\vv
    \ps^{\cal C}&\;
  &=&\; C\, \overline{\ps}^t&&
    &\text{d.h.}\qquad
  u^{\cal C}\; =\; C\, \overline{v}^t\qquad
  v^{\cal C}\; =\; C\, \overline{u}^t
    \\
&A&\vv
  &\overset{{\cal C}}{\longrightarrow}\vv
    A^{\cal C}&\;
  &=&\; C\, A\, C^{-1}&& = -A^t =
    -&A^{\D\ast}
    \tag{\ref{calC-Transf}$'$}
    \\[-4.5ex]\nn
\end{alignat}
Dabei ist~$C$ eine definierte Matrix, die nichttrivial wirkt im Dirac-Raum; sie ist \vspace*{-.25ex}antisym\-metrisch, antihermitesch und unit"ar:~$-C \!=\! C^t \!=\! C^{\D\dagger} \!=\! C^{-1}$.

Unter Ladungskonjugation~${\cal C}$ verhalten sich ein- respektive auslaufende Quarks wie
\begin{samepage}
\vspace*{-.75ex}
\begin{align} \label{calC_Q}
&\rket{i}\;
  \overset{{\cal C}}{\longrightarrow}\; \rket{i}^{\cal C}\;
  =\; C\, \rbra{\bar{i}}^t
    \\
&\rbra{i}\;
  \overset{{\cal C}}{\longrightarrow}\; \rbra{i}^{\cal C}\;
  =\; \rket{\bar{i}}^t\, C
    \tag{\ref{calC_Q}$'$}
    \\[-4.5ex]\nn
\end{align}
Dies impliziert f"ur die Greenfunktion:
\vspace*{-.25ex}
\begin{align} \label{calC_SF}
S_F[A]\;
  \overset{{\cal C}}{\longrightarrow}\;
  S_F[A^{\cal C}]\;
  =\; C^t\, \big( S_F[A] \big)^t\, C
    \\[-4.25ex]\nn
\end{align}
Die vollen einlaufenden Wellenfunktionen des (Anti)Quarks~\mbox{\,$\rket{\ps_i^F}$},~\mbox{\,$\rbra{\bar{\ps}_{\bar i}^F}$} in der Definition der Gln.~(\ref{psiVekt}),~(\ref{psiVekt}$'$) erf"ullen dann genau
\vspace*{-.25ex}
\begin{align} 
\rket{\ps_i^F[A]}\;
  \overset{{\cal C}}{\longrightarrow}\; \rket{\ps_i^F[A]}^{\cal C}\;
  =\; C^t\, \rbra{\bar{\ps}_{\bar i}^F[A^{\cal C}]}^t
    \\[-4.25ex]\nn
\end{align}
Aus den Gln.~(\ref{calC_Q}),~(\ref{calC_Q}$'$) und~(\ref{calC_SF}) zusammen folgt
\vspace*{-.25ex}
\begin{align} \label{calC_calM}
{\cal M}_{i'\mskip-1mu i}^F[A]\;
  \overset{{\cal C}}{\longrightarrow}\;
  \big( {\cal M}_{i'\mskip-1mu i}^F[A] \big)^{\cal C}\;
  =\; \bar{\cal M}_{\bar{i}'\mskip-1mu \bar{i}}^F[A^{\cal C}]
    \\[-4.25ex]\nn
\end{align}
Dies verifiziert genau, was wir a~priori erwarten:
Die Propagation eines Quarks durch die Konfiguration~$A$ des Gluon-Eichfelds ist "aquivalent zum ${\cal C}$-transformierten Bild: Das aus dem Quark durch~${\cal C}$-Transformation hervorgegangene Antiquark propagiert~-- zur"uck in der Zeit~-- durch die Konfiguration~$A^{\cal C}$. 

Ferner gen"ugt es, die Quark-Streuamplitude~\mbox{\,${\cal M}_{i'\mskip-1mu i}^F[A]$} zu betrachten.
Die Antiquark-Amplitude~\mbox{\,$\bar{\cal M}_{\bar{i}'\mskip-1mu \bar{i}}^F[A^{\cal C}]$} folgt unmittelbar durch Substitution, vgl.\@ die Gln.~(\ref{calC-Transf}),~(\ref{calC-Transf}$'$).

%
\subsubsection{(Anti)Quark-Streuamplituden%
    ~\bm{{\cal M}_{i'\mskip-1mu i}^F[A]},~\bm{\bar{\cal M}_{\bar{i}'\mskip-1mu \bar{i}}^F[A]}%
    ~als Wegner-Wilson-Linien}

Aufgabe ist es, die Streuamplituden des (Anti)Quarks an der vorgegebenen Konfiguration~$A$ des Gluon-Eichfelds~\mbox{\,${\cal M}_{i'\mskip-1mu i}^F[A]$},~\mbox{\,$\bar{\cal M}_{\bar{i}'\mskip-1mu \bar{i}}^F[A]$} einer Auswertung im Rahmen {\it nichtperturbativer\/}~QCD zug"anglich zu machen.
Das hei"st~-- wir beschr"anken uns auf das Quark~-- die vollen Wellenfunktionen~$\rket{\ps_i^F}$ zu bestimmen als L"osungen der vollen nichtrenormierten Dirac-Gleichung, vgl.\@ Gl.~(\ref{psiVekt_DiracGl}).
\end{samepage}

Unmittelbares Problem ist, da"s~$\rket{\ps_i^F}$ als einlaufender Zustand nicht den physikalischen Randbedingungen gen"ugt.
Insofern als f"ur negativ unendliche Zeiten~$\rket{\ps_i^F}$ {\it nicht\/} "ubergeht in den freien "`\IN"'-Zustand~$\rket{i}$:
\bea \label{psiVekt_AsymptF}
\rket{\ps_i^F} \;\underset{x^0\to-\infty}{\nlongrightarrow}\; \rket{i}
\eea
entsprechend~$\rbra{\bar{\ps}_{\bar i}^F}$ nicht in~$\rbra{\bar i}$.

Allerdings ist die Greenfunktion~\mbox{\,$S_F(x,x';m^2)[A]$} als L"osung von Gl.~(\ref{SF_lim}) nur bestimmt bis auf ein Funktional~\mbox{\,$\si(x,x';m^2)[A]$}, das identisch verschwindet unter der Wirkung des vollen nichtrenormierten Dirac-Operators:
%
\begin{align} 
\big(\iIM\ga^\mu D_\mu[A] - m\big)\vv \si[A]\;
  =\; 0
\end{align}
Neben~$S_F$ existieren bekanntlich weitere Greenfunktionen.~--
In dem Sinne, da"s diese eine Gl.~(\ref{SF_lim}) analoge Gleichung l"osen und von~$S_F$ genau in ihrer $\ep$-Vorschrift differieren: wie ihre Impulsraum-Darstellung bez"uglich~$k^0$- von~$-\infty$ nach~$+\infty$ Fourier-zu-integrieren ist in den Ortsraum, das hei"st wie in der komplexen $k^0$-Ebene der Integrationsweg an den Polen~$(\pm\surd\vec{k}^2 \!+\! m^2,\,0)$ vorbeizuf"uhren ist.

Nachtmann betrachtet die {\it retardierte\/} Greenfunktion~$S_r$.
Es gilt Gl.~(\ref{SF_lim}) unter der Ersetzung von~$S_F$ durch~$S_r$~-- wie auch alle weiteren "uber~$S_F$ definierten Gr"o"sen, analog definiert "uber~$S_r$, analoge Relationen erf"ullen.
Eingeschlossen~$\rket{\ps_i^r}$, vgl.\@ Gl.~(\ref{psiVekt}), die volle retardierte Wellenfunktion eines einlaufenden Quarks, die das Pendant zu Gl.~(\ref{psiVekt_DiracGl}) erf"ullt.
Wesentliche Beobachtung ist, da"s~$\rket{\ps_i^r}$ genau die geforderte Asymptotik besitzt:
%
\begin{align} \label{psiVekt_Asympt}
\rket{\ps_i^r}\;
  \underset{x^0\to-\infty}{\longrightarrow}\; \rket{i}
\end{align}
Gedanke ist daher genau der, generell~$S_F$ zu ersetzen durch~$S_r$.

Dies ist m"oglich, falls die longitudinalen (Anti)Quark-Impulse sehr viel gr"o"ser sind als die des Gluon-Eichfelds~$A$, dies ist gegeben im Grenzwert~\mbox{\,$s \!\to\! \infty$} f"ur~\mbox{\,$-t \!<\! 1\GeV^2$}, fest.
Es gilt die Asymptotik
%
\begin{align} 
{\cal M}_{i'\mskip-1mu i}^F[A]\; -\; {\cal M}_{i'\mskip-1mu i}^r[A]\;
  \underset{\text{$s \!\to\! \infty$, $-t \!<\! 1$~GeV$^2$, fest}}{\sim}\;
  {\cal O}(\vep_{i'},\vep_i)
\end{align}
mit~${\cal O}(\vep_{i'},\vep_i) \!\equiv\! {\cal O}(1\!/\!\surd s)$; vgl.\@ Abschn.~\ref{Sect:Kinematik}.
Nachtmann zeigt dies in Ref.~\cite{Nachtmann91} mithilfe der Lipp\-mann-Schwinger-Gleichungen f"ur~$S_F$ und~$S_r$, vgl.\@ Gl.~(\ref{Lippmann-Schwinger}$'$), durch vollst"andige Induktion nach $n$ bez"uglich ihrer Entwicklungen nach den freien Greenfunktionen.

In conclusio werden in der zentralen Gr"o"se unserer Betrachtung, dem $S$-Matrixelement nch Gl.~(\ref{S-Element_calM}), die Feynmanschen Streuamplituden~\mbox{\,${\cal M}_{i'\mskip-1mu i}^F$},~\mbox{\,$\bar{\cal M}_{\bar{i}'\mskip-1mu \bar{i}}^r$} ersetzt durch die {\it retardierten\/}
\begin{samepage}
%
\begin{alignat}{3} \label{calM_psiVekt-r}
&{\cal M}_{i'\mskip-1mu i}^r\;&
  &=\; \rbra{i'}\vv
         \big(g\ga^\mu &A_\mu - \de m\big)\vv
         \rket{\ps_i^r}
    \\[.5ex]
&\bar{\cal M}_{\bar{i}'\mskip-1mu \bar{i}}^r\;&
  &=\; \rbra{\bar{\ps}_{\bar i}^r}\vv
         \big(g\ga^\mu &A_\mu - \de m\big)\vv
         \rket{\bar{i}'}
    \tag{\ref{calM_psiVekt-r}$'$}
\end{alignat}
vgl.\@ die Gln.~(\ref{calM_psiVekt-F}),~(\ref{calM_psiVekt-F}$'$).
\vspace*{-.5ex}

\bigskip\noindent
Nachtmann leitet f"ur~\mbox{\,$\rket{\ps_i^r}$} als der L"osung der vollen nichtrenormierten Dirac-Gleichung, vgl.\@ Gl.~(\ref{psiVekt_DiracGl}), die Darstellung in der Eikonal-Appro\-ximation her, Ref.~\cite{Nachtmann91}.
Diese N"aherung basiert auf der Annahme, die longitudinalen (Anti)Quark-Impulse seien sehr viel gr"o"ser als die des Gluon-Eichfelds~$A$.
Dies ist realisiert im Grenzwert~\mbox{\,$s \!\to\! \infty$} und f"ur~\mbox{\,$-t \!<\! 1\GeV^2$}, fest.
Nachtmann arbeitet in diesem Grenzwert.
Er gibt an f"ur die einlaufenden Quarks:
\end{samepage}
%
\begin{alignat}{3} \label{psiVekt_WW-Vx}
&\ps^r_1(x)\;&
  &=\; \rbracket{x}{\ps_1^r}\;&
  &=\; V\idx{+}(x)\; \rbracket{x}{1}\;
        +\; {\cal O}(\vep_1)
    \\[.5ex]
&\ps^r_2(x)\;&
  &=\; \rbracket{x}{\ps_2^r}\;&
  &=\; V\idx{-}(x)\; \rbracket{x}{2}\;
        +\; {\cal O}(\vep_2)
    \tag{\ref{psiVekt_WW-Vx}$'$}
\end{alignat}
mit~\mbox{\,${\cal O}(\vep_i) \!=\! {\cal O}(1\!/\!\surd s)$}.
Dabei sind~$V\idx{+}(x)$,~$V\idx{-}(x)$ die Eichgruppen-, das hei"st {\it matrixwertigen\/} pfadgeordneten Phasenfaktoren:
\vspace*{-.5ex}
\begin{alignat}{2} \label{WW-Vx_lim}
&V\idx{+}[A](x)\;&
  &=\; P\; \exp -ig \int_{-\infty}^{x^+} d{x'}^+\; A_+\!({x'}^+,x^-,\rb{x})
    \\[-.5ex]
&V\idx{-}[A](x)\;&
  &=\; P\; \exp -ig \int_{-\infty}^{x^-} d{x'}^-\; A_-\!(x^+,{x'}^-,\rb{x})
    \tag{\ref{WW-Vx_lim}$'$}
    \\[-4.5ex]\nn
\end{alignat}
Wobei wir streng unterscheiden zwischen ko- und kontravarianten Komponenten, vgl.\@ Anhang~\ref{APP:LC-Koord}; vgl.\@ die Bem.\@ zu Gl.~(\ref{APP:gbar^-1}).
Diese Phasen  werden bezeichnet als {\it Wegner-Wilson-Linien\/} und sind nichts anderes als die Eichgruppen-Konnektoren~$\Ph$ entlang der klassischen Trajektorien der Quarks als Funktionen des Endpunktes~$x$ in der Raumzeit, vgl.\@ Gl.~(\ref{Konnektor}).
Aufgrund des Limes unendlicher invarianter Schwerpunktenergie~$\surd s$ sind sie~{\it lichtartig\/}.

F"ur die Limites bez"uglich ihrer oberen Integrationsgrenzen, das hei"st in Richtung ihrer "`gro"sen"' Impulse gilt
\vspace*{-.5ex}
\begin{alignat}{2}
&\lim_{x^+\to\infty} V\idx{+}(x^+,x^-,\rb{x})\;&
  &=\; V\idx{+}(\infty,x^-,\rb{x})
    \label{WW-V_lim} \\
&\lim_{x^-\to\infty} V\idx{-}(x^+,x^-,\rb{x})\;&
  &=\; V\idx{-}(x^+,\infty,\rb{x})
    \tag{\ref{WW-V_lim}$'$} \\
&\text{und}\qquad
  \lim_{x^+\to-\infty} V\idx{+}(x)\;&
  &=\; \bbbOne{F} 
    \label{WW-V_lim-One} \\
&\phantom{\text{und}\qquad}
  \lim_{x^-\to-\infty} V\idx{-}(x)\;&
  &=\; \bbbOne{F}
    \tag{\ref{WW-V_lim-One}$'$}
    \\[-4.5ex]\nn
\end{alignat}
mit den klassischen lichtartigen Trajektorien~\mbox{\,${\cal C}\idx{+}$},~\mbox{\,${\cal C}\idx{-}$} wie eingef"uhrt in Gl.~(\ref{Trajektorien_lim}) und den vollen Wegner-Wilson-Linien~\mbox{\,$V\idx{+} \!\equiv\! \Ph({\cal C}\idx{+})$},~\mbox{\,$V\idx{-} \!\equiv\! \Ph({\cal C}\idx{-})$} wie in Gl.~(\ref{Parton-Parton_lim}$'$).

Die Streuamplituden~\mbox{\,${\cal M}_{i'\mskip-1mu i}^r$},~\mbox{\,$\bar{\cal M}_{\bar{i}'\mskip-1mu \bar{i}}^r$} entsprechend den Gln.~(\ref{calM_psiVekt-r}),~(\ref{calM_psiVekt-r}$'$) folgen im wesentlichen durch Anwendung der definierenden Differentialgleichung f"ur Konnektoren, die nach Gl.~(\ref{Konnektor_DGL}) einen Ableitungsoperatior nach dem Kurvenparameter generiert, der wirkt auf die Wegner-Wilson-Linien; vgl.\@ die Gln.~(\ref{WW-V_lim}),~(\ref{WW-V_lim}$'$) und~(\ref{WW-V_lim-One}),~(\ref{WW-V_lim-One}$'$).
Nachtmann findet letztlich f"ur Quarks:%
\FOOT{
  \label{FN:de-m}unter Vernachl"assigung der Terme mit~$\de m$ im Limes~$s \!\to\! \infty$
}
%
\vspace*{-.5ex}
\begin{align} \label{calM-Q_WW-V}
{\cal M}_{1'\mskip-1mu 1}^r[A]\;
  =\; \iIM\, \de_{s_{1'}\mskip-1mu s_1}\,
        \int d_{1'\mskip-1mu 1}\!(x^-,\rb{x})\;
        \big[ V\idx{+}[A](\infty,x^-,\rb{x})
                - \bbbOne{F}\big]_{\!n_{1'}\mskip-1mu n_1}
    \\
{\cal M}_{2'\mskip-1mu 2}^r[A]\;
  =\; \iIM\, \de_{s_{2'}\mskip-1mu s_2}\,
        \int d_{2'\mskip-1mu 2}\!(y^+,\rb{y})\;
        \big[ V\idx{-}[A](y^+,\infty,\rb{y})
                - \bbbOne{F}\big]_{\!n_{2'}\mskip-1mu n_2}
    \tag{\ref{calM-Q_WW-V}$'$}
    \\[-4.5ex]\nn
\end{align}
Diese Amplituden~${\cal M}_{i'\mskip-1mu i}^r$ sind Funktionale im Gluon-Eichfeld~$A$ genau "uber~$V\idx{\pm}$, vgl.\@ die Gln.~(\ref{WW-V_lim}),~(\ref{WW-V_lim}$'$) und~(\ref{WW-V_lim-One}),~(\ref{WW-V_lim-One}$'$).
Aufgrund des Verhaltens unter Ladungskonjugation~\mbox{\,${\cal C}$} zum einen des Eichfelds \mbox{\,$A^{\cal C} \!\equiv\! -A^{\D\ast}$} und zum anderen dieser Amplituden, vgl.\@ Gl.~(\ref{calC-Transf}$'$) und~(\ref{calC_calM}), ist f"ur Antiquarks genau~\mbox{\,$V\idx{\pm}[A^{\cal C} \!\equiv\! -A^{\D\ast}] \equiv V^{\D\ast}\idx{\pm}[A]$} zu nehmen; es folgt
\vspace*{-.5ex}
\begin{align} \label{calM-AQ_WW-V}
\bar{\cal M}_{{\bar1}'\mskip-1mu {\bar1}}^r[A]\;
  =\; \iIM\, \de_{{\bar s}_{1'}\mskip-1mu {\bar s}_1}\,
        \int d_{{\bar1}'\mskip-1mu {\bar1}}\!({\bar x}^-,\rbb{x})\;
        \big[ V\idx{+}^{\D\ast}[A](\infty,{\bar x}^-,\rbb{x})
                - \bbbOne{F}\big]_{\!{\bar n}_{1'}\mskip-1mu {\bar n}_1}
    \\
\bar{\cal M}_{{\bar2}'\mskip-1mu {\bar2}}^r[A]\;
  =\; \iIM\, \de_{{\bar s}_{2'}\mskip-1mu {\bar s}_2}\,
        \int d_{{\bar2}'\mskip-1mu {\bar2}}\!({\bar y}^+,\rbb{y})\;
        \big[ V\idx{-}^{\D\ast}[A]({\bar y}^+,\infty,\rbb{y})
                - \bbbOne{F}\big]_{\!{\bar n}_{2'}\mskip-1mu {\bar n}_2}
    \tag{\ref{calM-AQ_WW-V}$'$}
    \\[-4.5ex]\nn
\end{align}
unter Definition der Integrationsma"se
\vspace*{-.5ex}
\begin{align} \label{di'i}
&d_{1'\mskip-1mu 1}\!(x^-,\rb{x})
    \\[-.5ex]
  &\phantom{d_{1'\mskip-1mu 1}\!}
   \equiv\; 2\sqrt{p_{1'}^+p_1^+}\vv
        g_{+-}\, dx^- d^2\rb{x}\vv
        \exp\, \iIM\big\{ g_{+-}(p_{1'}^+ \!-\! p_1^+) x^-
                - (\rb{p}_{1'} \!-\! \rb{p}_1) \!\cdot\! \rb{x} \big\}
    \nn \\
&d_{2'\mskip-1mu 2}\!(y^+,\rb{y})
    \tag{\ref{di'i}$'$} \\[-.5ex]
  &\phantom{d_{1'\mskip-1mu 1}\!}
   \equiv\; 2\sqrt{p_{2'}^- p_2^-}\vv
        g_{+-}\, dy^+ d^2\rb{y}\vv
        \exp\, \iIM\big\{ g_{+-}(p_{2'}^- \!-\! p_2^-) y^+
                - (\rb{p}_{2'} \!-\! \rb{p}_2) \!\cdot\! \rb{y} \big\}
    \nn
    \\[-4.25ex]\nn
\end{align}
und
\vspace*{-.5ex}
\begin{align} \label{di'i-bar}
&d_{\bar1'\mskip-1mu \bar1}\!(\bar{x}^-,\rbb{x})
    \\[-.5ex]
  &\phantom{d_{1'\mskip-1mu 1}\!}
   \equiv\; 2\sqrt{\bar{p}_{1'}^+\bar{p}_1^+}\vv
        g_{+-}\, d\bar{x}^- d^2\rbb{x}\vv
        \exp\, \iIM\big\{ g_{+-}(\bar{p}_{1'}^+ \!-\! \bar{p}_1^+) \bar{x}^-
                - (\rbb{p}_{1'} \!-\! \rbb{p}_1) \!\cdot\! \rbb{x} \big\}
    \nn \\
&d_{\bar2'\mskip-1mu \bar2}\!(\bar{y}^+,\rbb{y})
    \tag{\ref{di'i-bar}$'$} \\[-.5ex]
  &\phantom{d_{1'\mskip-1mu 1}\!}
   \equiv\; 2\sqrt{\bar{p}_{2'}^- \bar{p}_2^-}\vv
        g_{+-}\, d\bar{y}^+ d^2\rbb{y}\vv
        \exp\, \iIM\big\{ g_{+-}(\bar{p}_{2'}^- \!-\! \bar{p}_2^-) \bar{y}^+
                - (\rbb{p}_{2'} \!-\! \rbb{p}_2) \!\cdot\! \rbb{y} \big\}
    \nn
    \\[-4.5ex]\nn
\end{align}
mit Fourier-Transformation der "`kleinen"' longitudinalen und der transversalen Komponenten durch die Ausdr"ucke nach den Quadratwurzeln und mit~\mbox{$g_{+-} \!=\! (2\al^2)^{\!-1} \!=\! (-\det\mathbb{L})^{-1}$} nach Definition der Lichtkegelkoordinaten, vgl.\@ Gl.~(\ref{LC-Koord}) und Anhang~\ref{APP:LC-Koord}. \\
\indent
Im Limes~$s \!\to\! \infty$ gilt Helizit"atserhaltung im $s$-Kanal.
Dies ist Konsequenz dessen, da"s die f"uhrenden Spinor-Produkte~\mbox{$\overline{u}(i') \ga^{\bar\mu} u(i)$} und~\mbox{$\overline{v}(\bar{i}) \ga^{\bar\mu} v(\bar{i}')$}, mit~\mbox{$\bar{\mu} \!\in\! \{+,-,1,2\}$}, sind:
\vspace*{-.5ex}
\begin{align}
&\overline{u}(1')\, \ga^+\, u(1)\vv
  \underset{\text{$s \!\to\! \infty$}}{\sim}\vv
  \de_{s_{1'}\!s_1}\; 2\sqrt{p_{1'}^+p_1^+}
    \label{Helizitaetserhaltung-u} \\[-.375ex]
&\overline{u}(2')\, \ga^-\, u(2)\vv
  \underset{\text{$s \!\to\! \infty$}}{\sim}\vv
  \de_{s_{2'}\!s_2}\; 2\sqrt{p_{2'}^-p_2^-}
    \tag{\ref{Helizitaetserhaltung-u}$'$} \\
&\text{und}\qquad
  \overline{v}(\bar1)\, \ga^+\, v(\bar1')\vv
  \underset{\text{$s \!\to\! \infty$}}{\sim}\vv
  \de_{\bar{s}_{1'}\!\bar{s}_1}\; 2\sqrt{\bar{p}_{1'}^+\bar{p}_1^+}
    \label{Helizitaetserhaltung-v} \\[-.375ex]
&\phantom{\text{und}\qquad}
  \overline{v}(\bar2)\, \ga^-\, v(\bar2')\vv
  \underset{\text{$s \!\to\! \infty$}}{\sim}\vv
  \de_{\bar{s}_{2'}\!\bar{s}_2}\; 2\sqrt{\bar{p}_{2'}^-\bar{p}_2^-}
    \tag{\ref{Helizitaetserhaltung-v}$'$}
    \\[-4.5ex]\nn
\end{align}
unterdr"uckt Eichgruppenindizes, vgl.\@ Gl.~(\ref{APP:uv-Spinoren_ga^mu}).
Hieraus folgen die Kronecker-Symbole in den Spinindizes in den Gln.~(\ref{calM-Q_WW-V}),~(\ref{calM-Q_WW-V}$'$) und~(\ref{calM-AQ_WW-V}),~(\ref{calM-AQ_WW-V}$'$) und die Wurzelausdr"ucke der "`gro"sen"' longitudinalen Komponenten in den Gln.~(\ref{di'i}),~(\ref{di'i}$'$) und~(\ref{di'i-bar}),~(\ref{di'i-bar}$'$).

Im $S$-Matrixelement auf Partonniveau~-- vgl.\@ Gl.~(\ref{S-Element_calM})~-- treten die Amplituden~\mbox{\,${\cal M}_{i'\mskip-1mu i}^r$}, \mbox{\,${\cal M}_{{\bar i}'\mskip-1mu {\bar i}}^r$} auf zusammen mit {\it disconnected\/}, das hei"st nichtzusammenh"angenden Termen.
In An\-hang~\ref{APP:discon} stellen wir diese Terme~$\de(i',i)$,~$\de(\bar{i}',\bar{i})$ dar durch Integrale~\mbox{$\int d_{i'\mskip-1mu i}$},~\mbox{$\int d_{\bar{i}'\mskip-1mu \bar{i}}$} entsprechende den Gln.~(\ref{di'i}),~(\ref{di'i}$'$) beziehungsweise~(\ref{di'i-bar}),~(\ref{di'i-bar}$'$).
Dies erm"oglicht ihre gemeinsame Behandlung mit den Einsen~$\bbbOne{F}$ der Eichgruppe in den Ausdr"ucken f"ur~\mbox{\,${\cal M}_{i'\mskip-1mu i}^r$} und~\mbox{\,${\cal M}_{{\bar i}'\mskip-1mu {\bar i}}^r$} nach den Gln.~(\ref{calM-Q_WW-V}),~(\ref{calM-Q_WW-V}$'$) beziehungsweise~(\ref{calM-AQ_WW-V}$'$),~(\ref{calM-AQ_WW-V}$'$).
Zusammen mit den Gln.~(\ref{APP:Delta_diStrichi-Mass}),~(\ref{APP:Delta_diStrichi-Mass}$'$) und~(\ref{APP:Delta_diStrichi-Mass-bar}),~(\ref{APP:Delta_diStrichi-Mass-bar}$'$) folgt f"ur die relevanten Ausdr"ucke:
\begin{samepage}
\vspace*{-.25ex}
\begin{align} \label{de-calM}
\big( \de(1',1)
  &- \iIM\, Z\idx{2}^{-1}\, {\cal M}_{1'\mskip-1mu 1}^F[A] \big)\;
    \\[-.375ex]
  &=\; \de_{s_{1'}\!s_1}\;
       \int d_{1'\mskip-1mu 1}\!(x^-,\rb{x})\vv
       Z\idx{2}^{-1}
         \big( V\idx{+}[A](\infty,x^-,\rb{x})
               - \deZ\, \bbbOne{F}\big)_{\!n_{1'}\mskip-1mu n_1}
    \nn \\
\big( \de(2',2)
  &- \iIM\, Z\idx{2}^{-1}\, {\cal M}_{2'\mskip-1mu 2}^F[A] \big)
    \tag{\ref{de-calM}$'$} \\[-.375ex]
  &=\; \de_{s_{2'}\!s_2}\;
       \int d_{2'\mskip-1mu 2}\!(y^+,\rb{y})\vv
       Z\idx{2}^{-1}
         \big( V\idx{-}[A](y^+,\infty,\rb{y})
               - \deZ\, \bbbOne{F}\big)_{\!n_{2'}\mskip-1mu n_2}
    \nn
    \\[-4.75ex]\nn
\end{align}
und
\vspace*{-.75ex}
\begin{align} \label{de-calM-bar}
\big( \de(\bar1',\bar1)
  &- \iIM\, Z\idx{2}^{-1}\, \bar{\cal M}_{\bar1'\mskip-1mu \bar1}^F[A] \big)
    \\[-.375ex]
  &=\; \de_{\bar{s}_{1'}\!\bar{s}_1}\;
       \int d_{\bar1'\mskip-1mu \bar1}\!(\bar{x}^-,\rbb{x})\vv
       Z\idx{2}^{-1}
         \big( V\idx{+}^{\D\dagger}[A](\infty,\bar{x}^-,\rbb{x})
               - \deZ\, \bbbOne{F}\big)_{\!\bar{n}_1\mskip-1mu \bar{n}_{1'}}
    \nn \\
\big( \de(\bar2',\bar2)
  &- \iIM\, Z\idx{2}^{-1}\, \bar{\cal M}_{\bar2'\mskip-1mu \bar2}^F[A] \big)
    \tag{\ref{de-calM-bar}$'$} \\[-.375ex]
  &=\; \de_{\bar{s}_{2'}\!\bar{s}_2}\;
       \int d_{\bar2'\mskip-1mu \bar2}\!(\bar{y}^+,\rbb{y})\vv
       Z\idx{2}^{-1}
         \big( V\idx{-}^{\D\dagger}[A](\bar{y}^+,\infty,\rbb{y})
               - \deZ\, \bbbOne{F}\big)_{\!\bar{n}_2\mskip-1mu \bar{n}_{2'}}
    \nn
    \\[-4.5ex]\nn
\end{align}
Dabei ist benutzt~\mbox{\,$V^{\D\ast} \!\equiv\! V^{\D\dagger t}$} in den letzten Relationen mit den Antiquark-Amplituden~-- vgl.\@ die umgekehrte Reihenfolge der Eichgruppen-Indizes~$\bar{n}_i$,~$\bar{n}_{i'}$~-- und definiert
\vspace*{-.25ex}
\begin{align} \label{deZ}
\deZ\;
  \equiv\; 1 - Z\idx{2}
    \\[-4ex]\nn
\end{align}
als abk"urzende Notation
\end{samepage}

Mithilfe der Relationen der Gln.~(\ref{de-calM}),~(\ref{de-calM}$'$) und~(\ref{de-calM-bar}),~(\ref{de-calM-bar}$'$) folgt aus Gl.~(\ref{S-Element_calM}) unmittelbar
%
\begin{align} \label{S-Element_calM,V}
&\bracket{\, \bar{q}(\bar{2}')\, q(2')\, \bar{q}(\bar{1}')\, q(1'),\,\IN \,}{\,
             S\, \bracketM\,
             q(1)\, \bar{q}(\bar{1})\, q(2)\, \bar{q}(\bar{2}),\,\IN \,}
    \nn \\[.75ex]
  &=\;
  \begin{aligned}[t]
    \vacL\,
       \big( &\de(1',1) - \iIM\, Z\idx{2}^{-1}\, {\cal M}_{1'\mskip-1mu 1}^F[A] \big)\,
       \big( \de(\bar{1}',\bar{1})
               - \iIM\, Z\idx{2}^{-1}\, \bar{\cal M}_{\bar{1}'\mskip-1mu\bar{1}}^F[A] \big)
    \\[.5ex]
      \times\big( &\de(2',2) - \iIM\, Z\idx{2}^{-1}\, {\cal M}_{2'\mskip-1mu 2}^F[A] \big)\,
        \big( \de(\bar{2}',\bar{2})
                - \iIM\, Z\idx{2}^{-1}\, \bar{\cal M}_{\bar{2}'\mskip-1mu\bar{2}}^F[A] \big)\, 
    \vacR
  \end{aligned}
   \\[.75ex]
  &=\; \de_{s_{1'}\!s_1}\;
       \de_{\bar{s}_{1'}\!\bar{s}_1}\;
       \de_{s_{2'}\!s_2}\;
       \de_{\bar{s}_{2'}\!\bar{s}_2}
    \tag{\ref{S-Element_calM,V}$'$} \\
  &\phantom{=\;}
    \times\,
      \int d_{1'\mskip-1mu 1}\!(x^-,\rb{x})
      \int d_{{\bar1}'\mskip-1mu {\bar1}}\!({\bar x}^-,\rbb{x})
      \int d_{2'\mskip-1mu 2}\!(y^+,\rb{y})
      \int d_{{\bar2}'\mskip-1mu {\bar2}}\!({\bar y}^+,\rbb{y})
    \nn \\[-.25ex]
  &\phantom{=\;}
    \begin{alignedat}[t]{2}
    \times\, \vacL\,
      &Z\idx{2}^{-1}
         \big( V\idx{+}[A](\infty,x^-,\rb{x})
               - \deZ\, \bbbOne{F}\big)_{\!n_{1'}\mskip-1mu n_1}\;&
      &Z\idx{2}^{-1}
         \big( V\idx{+}^{\D\dagger}[A](\infty,\bar{x}^-,\rbb{x})
               - \deZ\, \bbbOne{F}\big)_{\!\bar{n}_1\mskip-1mu \bar{n}_{1'}}
    \\[-.5ex]
      &Z\idx{2}^{-1}
         \big( V\idx{-}[A](y^+,\infty,\rb{y})
               - \deZ\, \bbbOne{F}\big)_{\!n_{2'}\mskip-1mu n_2}\;&
      &Z\idx{2}^{-1}
         \big( V\idx{-}^{\D\dagger}[A](\bar{y}^+,\infty,\rbb{y})
               - \deZ\, \bbbOne{F}\big)_{\!\bar{n}_2\mskip-1mu \bar{n}_{2'}}
    \vacR
  \end{alignedat}
    \nn
\end{align}
f"ur das $S$-Matrixelement der Streuung auf Partonniveau.
\vspace*{-.5ex}

%
\subsection{Hadronniveau~II. Conclusio}
\label{Subsect:HadronniveauII}

Wir kehren zur"uck auf Hadronniveau.
Das volle $S$-Matrixelement der Hadron-Hadron-Streuung nach Gl.~(\ref{2hto2h}) wird ausgedr"uckt durch das $S$-Matrixelement des zugrundeliegenden Parton-Parton-Prozesses entsprechend Gl.~(\ref{4Qto4Q}).
Wir rekapitulieren die mesonartigen Zust"ande~$h^i(P_i)$ in der Form
\begin{samepage}
\vspace*{-.25ex}
\begin{align} \label{h-ketREPET}
&\ket{h^i(P_i)}\;
  =\; \frac{\de_{n_i\!{\bar n}_i}}{\sqrt{N_{\rm\!c}}}\vv
        \int d{\tilde\vph}_{s_i\!{\bar s}_i}^i (\zet_i, \rb{k}_i)\vv
        \ket{ q_{s_i,n_i}(\zet_i, \rb{k}_i)\; 
                \bar{q}_{{\bar s}_i,{\bar n}_i}(\bzet_i, \rb{k}_i) }
    \\[-4ex]\nn
\end{align}
vgl.\@ Gl.~(\ref{h-ket}), mit Integrationsma"s~\mbox{\,$d{\tilde\vph}_{s\mskip-1mu{\bar s}}^i (\zet, \rb{k})$} definiert durch:
%
\begin{align} \label{h-ket_Mass}
d{\tilde\vph}_{s\mskip-1mu{\bar s}}^i (\zet, \rb{k})\;
  \equiv\; \frac{d^2\rb{k}}{(2\pi)^2} \frac{d\zet}{2\pi}\;
      {\tilde\vph}_{s\mskip-1mu{\bar s}}^i (\zet, \rb{k})
\end{align}
Die Zustandvektoren~$\ket{h^i(P_i)}$ nach Gl.~(\ref{h-ket}) oder~(\ref{h-ketREPET}) vermitteln zwischen dem $S$-Matrixelement auf Hadron- und Partonniveau:
%
\begin{align} \label{S-Element2h_calM}
&\bracket{\, h^{2'}\!(P_{2'})\, h^{1'}\!(P_{1'}),\,\IN \,}{\,
             S\, \bracketM\,
             h^1(P_1)\, h^2(P_2),\,\IN \,}
    \\[.75ex]
  &=\; \frac{\de_{n_1\!{\bar n}_1}}{\sqrt{N_{\rm\!c}}}\,
         \frac{\de_{n_{1'}\!{\bar n}_{1'}}}{\sqrt{N_{\rm\!c}}}\,
         \frac{\de_{n_2\!{\bar n}_2}}{\sqrt{N_{\rm\!c}}}\,
         \frac{\de_{n_{2'}\!{\bar n}_{2'}}}{\sqrt{N_{\rm\!c}}}
    \nn \\
  &\phantom{=\;}
    \times\,
       \int d{\tilde\vph}_{s_1'\!{\bar s}_1'}^{1'\D\dagger} (\zet_{1'}, \rb{k}_{1'})
         \int d{\tilde\vph}_{s_1\!{\bar s}_1}^1 (\zet_1, \rb{k}_1)
         \int d{\tilde\vph}_{s_2'\!{\bar s}_2'}^{2'\D\dagger} (\zet_{2'}, \rb{k}_{2'})
         \int d{\tilde\vph}_{s_2\!{\bar s}_2}^2 (\zet_2, \rb{k}_2)
    \nn \\
  &\phantom{=\;}
    \times\,
       \bracket{\, \bar{q}(\bar{2}')\, q(2')\, \bar{q}(\bar{1}')\, q(1'),\,\IN \,}{\,
                     S\, \bracketM\,
                     q(1)\, \bar{q}(\bar{1})\, q(2)\, \bar{q}(\bar{2}),\,\IN \,}
    \nn
\end{align}
Eingesetzt den expliziten Ausdruck nach Gl.~(\ref{S-Element_calM,V}$'$), ergibt Kontraktion der Spin-Indizes die Spinsumme der Lichtkegelwellenfunktionen; Kontraktion mit den Kronecker-Symbolen bez"uglich der Eich-Indizes f"uhrt auf die Spuren~\mbox{$\trDrst{F} \!=\! 1\!/\!\Nc\, \tr$}, normiert wie~\mbox{$\trDrst{F}\bbbOne{F} \!=\! 1$}:
\end{samepage}
\vspace*{-.5ex}
\begin{align} \label{S-Element2h_WW-V}
&\bracket{\, h^{2'}\!(P_{2'})\, h^{1'}\!(P_{1'}),\,\IN \,}{\,
             S\, \bracketM\,
             h^1(P_1)\, h^2(P_2),\,\IN \,}
    \\[.75ex]
&\underset{\text{$s \!\to\! \infty$}}{\sim}
 \begin{aligned}[t]
   &\phantom{\times}
    \int d{\tilde\vph}_{s_1\!{\bar s}_1}^{1'\D\dagger} (\zet_{1'}, \rb{k}_{1'})
      \int d{\tilde\vph}_{s_1\!{\bar s}_1}^1 (\zet_1, \rb{k}_1)
      \int d{\tilde\vph}_{s_2\!{\bar s}_2}^{2'\D\dagger} (\zet_{2'}, \rb{k}_{2'})
      \int d{\tilde\vph}_{s_2\!{\bar s}_2}^2 (\zet_2, \rb{k}_2)
    \nn \\
   &\times
      \int d_{1'\mskip-1mu 1}\!(x^-,\rb{x})
      \int d_{{\bar1}'\mskip-1mu {\bar1}}\!({\bar x}^-,\rbb{x})
      \int d_{2'\mskip-1mu 2}\!(y^+,\rb{y})
      \int d_{{\bar2}'\mskip-1mu {\bar2}}\!({\bar y}^+,\rbb{y})
    \nn \\
   &\times
      \vacL\,
      \trDrst{F}\!\big[
      Z\idx{2}^{-1}
        \big( V\idx{+}[A](\infty,x^-,\rb{x})
              - \deZ\,\bbbOne{F} \big)\vv
      Z\idx{2}^{-1}
        \big( V\idx{+}^{\D\dagger}[A](\infty,{\bar x}^-,\rbb{x})
              - \deZ\,\bbbOne{F} \big)
    \big]
    \nn \\[-.5ex]
   &\phantom{\vacL\,} \times
      \trDrst{F}\!\big[
      Z\idx{2}^{-1}
        \big( V\idx{-}[A](y^+,       \infty, \rb{y})
              - \deZ\,\bbbOne{F} \big)\vv
      Z\idx{2}^{-1}
        \big( V\idx{-}^{\D\dagger}[A] ({\bar y}^+,\infty,\rbb{y})
              - \deZ\,\bbbOne{F} \big)
      \big]\,
    \vacR
    \nn
  \end{aligned}
    \nn
\end{align}

Der Vakuumerwartungswert der vier Wegner-Wilson-Linien und disconnected Terme h"angt ab nicht von s"amtlichen zw"olf Komponenten der (Anti)Quark-Positionen~$x$,~${\bar x}$ und~$y$,~${\bar y}$.
Zun"achst reduziert die Translationsinvarianz des physikalischen Vakuums die Zahl unabh"an\-giger Variablen um vier; dann k"onnen ausgef"uhrt werden die zwei verbleibenden longitudinalen Integrationen.
Konsequenz ist die Abh"angigkeit von sechs transversalen Komponenten. \\
\indent
Wir verweisen auf Anhang~\ref{APP:S-Element2h}, in dem wir diese Reduktion explizit durchf"uhren und identifizieren die relevanten Variablen.%
\FOOT{
  \label{FN:relevanteVariable}Unser Ansatz f"ur die transversalen Impulse der Quarks~(Antiquarks) impliziert eine Abh"angigkeit von deren Bruchteil~$\zet_i$~($\bzet_i$) am gesamten~-- longitudinalen~-- Lichtkegelimpuls, vgl.\@ die Gln.~(\ref{Q_Impulse}),~(\ref{Q_Impulse}$'$).   Infolge dessen identifizieren wir als relevant Variable, die differieren von denen Nachtmanns in Ref.~\cite{Nachtmann96}.
}
Wir zitieren f"ur das $S$-Matrixelement~(\ref{S-Element2h_WW-V}) die Darstellung~(\ref{APP:S-Element2h_WW-V}), die in Anhang~\ref{APP:S-Element2h} folgt als Resultat konsequenter Reduktion auf diese Variablen:
%
\begin{align} \label{S-Element2h_WW-V_APP}
&\hspace*{-10pt}
 \bracket{\, h^{2'}\!(P_{2'})\, h^{1'}\!(P_{1'}),\,\IN \,}{\,
             S\, \bracketM\,
             h^1(P_1)\, h^2(P_2),\,\IN \,}
    \\[.75ex]
&\hspace*{-10pt}
 \underset{\text{$s \!\to\! \infty$}}{\sim}\;
        \iIM\, (2\pi)^4\; \de(P)
    \nn \\[-.5ex]
&\hspace*{-10pt}
 \phantom{P_2}\times
        -\, 2\iIM\,s\; \int d^2\rb{b}\;
             \efn{\T-\iIM\,\tfb \!\cdot\! \rb{b}}\vv
             \int d\vph_{1',1} (\zet_1, \rb{X})\;
             \int d\vph_{2',2} (\zet_2, \rb{Y})
    \nn \\[-1ex]
&\hspace*{-10pt}
 \phantom{P_2}\times\,
 \hspace*{-5pt}
  \begin{aligned}[t]
  \vacL\,
    &\trDrst{F}\!\big[
      Z\idx{2}^{-1} \big(
        V\idx{+}             (\infty,0, \bzet_1 \rb{X} \!+\! \rb{b}\!/2)
        - \deZ\,\bbbOne{F} \big)\;
      Z\idx{2}^{-1} \big(
        V\idx{+}^{\D\dagger} (\infty,0, -\zet_1 \rb{X} \!+\! \rb{b}\!/2)
        - \deZ\,\bbbOne{F} \big)
    \big]
    \nn \\[-.5ex]
  \times
    &\trDrst{F}\!\big[
      Z\idx{2}^{-1} \big(
        V\idx{-}             (0,\infty, \bzet_2 \rb{Y} \!-\! \rb{b}\!/2)
        - \deZ\,\bbbOne{F} \big)\;
      Z\idx{2}^{-1} \big(
        V\idx{-}^{\D\dagger} (0,\infty, -\zet_2 \rb{Y} \!-\! \rb{b}\!/2)
        - \deZ\,\bbbOne{F} \big)
    \big]\,
  \vacR
    \nn
  \end{aligned}
    \nn
    \\[-3.5ex]\nn
\end{align}
und rekapitulieren im folgenden die explizit wie implizit auftretenden Gr"o"sen; vgl.\@ Anh.~\ref{APP:S-Element2h}. \\
\indent
Es ist definiert das Integrationsma"s, f"ur~\mbox{$i \!=\! 1,2$} und~\mbox{$(\zet, \rb{Z})$} f"ur~\mbox{$(\zet_1, \rb{X}),\, (\zet_2, \rb{Y})$}:
\begin{samepage}
\vspace*{-.5ex}
\begin{align} \label{dvphi'i}
d\vph_{i',i} (\zet, \rb{Z})\;
  \equiv\; \frac{d\zet}{2\pi}\; d^2\rb{Z}\vv
      \vph_{s\mskip-1mu{\bar s}}^{i'\D\dagger} (\zet, \rb{Z})\;
      \vph_{s\mskip-1mu{\bar s}}^i (\zet, \rb{Z})
    \\[-4.5ex]\nn
\end{align}
das also gewichtet entsprechend des "Uberlapps%
~\mbox{\,$\vph_{s\mskip-1mu{\bar s}}{}^{\zz i'\D\dagger} \vph_{s\mskip-1mu{\bar s}}^i$}~-- Spinsumme impliziert~-- der Lichtkegelwellenfunktionen; vgl.\@ Gl.~(\ref{APP:dvphi'i}). \\
\indent
Es ist~$P$ der Differenzvektor zwischen ein- und auslaufendem Vierer-Impuls in der Definition von Gl.~(\ref{ViererDifferenzimpuls}).
Die Erhaltung des Gesamtimpulses in Form~$\de(P)$ ist unmittelbare Konsequenz der Translationsinvarianz des Vakuumerwartungswertes~\mbox{\,$\vac{\;\cdot\;}$}, das Quadrat der invarianten Schwerpunktenergie~$s$ der Konsequenz der Kinematik. \\
\indent
Der Vierer-Vektor~$\tf$ bezeichnet den {\it Gesamt-Impuls"ubertrag\/}, der verkn"upft ist mit dem invarianten Impuls"ubertrag~$t$ nach Mandelstam, vgl.\@ Gl.~(\ref{Mandelstam_Def}):
%
\begin{align} \label{ViererImpulsuebertrag-ta}
\tf\;
  \equiv\;  P_{1'} - P_1 \qquad
  \text{mit}\qquad
  \tfQ\; =\; t
\end{align}
Zu~$\tf$ konjugiert ist der {\it Impakt\/}~$b$.
Relevant ist seine transversale Projektion:
%
\begin{align} \label{b_rb}
\rb{b}\;
  =\; \rb{X}_\zet - \rb{Y}_\zet\;
  =\; (\zet_1\, \rb{x} + \bzet_1\, \rbb{x})
    - (\zet_2\, \rb{y} + \bzet_2\, \rbb{y})
\end{align}
Mit~$X_\zet$,~$Y_\zet$ sind bezeichnet~-- f"ur das "`plus"'-~beziehungsweise "`minus"'-Paar~-- die {\it \mbox{$\zet_i$-gewich}\-tete Mitte\/} auf dem Quark-Antiquark-Abstandvektor:
%
\begin{alignat}{2} \label{gewMitten-XzetYzet}
&X_\zet\;&
  &=\; \zet_1\, x\; +\; \bzet_1\, {\bar x}
    \\
&Y_\zet\;&
  &=\; \zet_2\, y\; +\; \bzet_2\, {\bar y}
    \tag{\ref{gewMitten-XzetYzet}$'$}
\end{alignat}
Die Abstandvektoren selbst sind gegeben durch die Differenzen der (Anti)Quark-Positionen:
%
\begin{alignat}{2} \label{Diffvekt-XY}
&X\;&
  &=\; x\; -\; {\bar x}
    \\
&Y\;&
  &=\; y\; -\; {\bar y}
    \tag{\ref{Diffvekt-XY}$'$}
\end{alignat}
\end{samepage}%
Der Vakuumerwartungswert~\mbox{\,$\vac{\;\cdot\;}$} der vier Wegner-Wilson-Linien und disconnected Terme in Gl.~(\ref{S-Element2h_WW-V_APP}) h"angt ab nur von den transversalen Projektionen der (Anti)Quark-Positionen im Ortsraum, die gegeben sind f"ur die "`plus"'-Zust"ande~(Index 1) durch:
\vspace*{-.75ex}
\begin{align} \label{Q-AQ-Pos_rb1}
&\rb{x}\;
  =\; \phantom{-}\bzet_1\, \rb{X}\; +\; \rb{b}/2\; +\; \rbG{\om}
    \\[-.5ex]
&\rbb{x}\;
  =\; -\zet_1\,       \rb{X}\; +\; \rb{b}/2\; +\; \rbG{\om}
    \tag{\ref{Q-AQ-Pos_rb1}$'$}
    \\[-4.75ex]\nn
\end{align}
f"ur die "`minus"'-Zust"ande~(Index 2) durch:
\vspace*{-.75ex}
\begin{align} \label{Q-AQ-Pos_rb2}
&\rb{y}\;
  =\; \phantom{-}\bzet_2\, \rb{Y}\; -\; \rb{b}/2\; +\; \rbG{\om}
    \\[-.5ex]
&\rbb{y}\;
  =\; -\zet_2\,            \rb{Y}\; -\; \rb{b}/2\; +\; \rbG{\om}
    \tag{\ref{Q-AQ-Pos_rb2}$'$}
    \\[-6ex]\nn
\end{align}
bzgl.~$\om \!=\! (X_\zet \!+\! Y_\zet)\!/2 \!\stackrel{\D!}{=}\! 0$ vgl.\@ Fu"sn.\,\FN{APP-FN:om=0}.
Die Abh"angigkeit der transversalen Impulse der~(An\-ti)Quarks in unserem Ansatz~-- vgl.\@ die Gln.~(\ref{Q_Impulse}),~(\ref{Q_Impulse}$'$) und Fu"sn.\,\FN{FN:relevanteVariable}~-- "ubertr"agt sich unmittelbar auf ihre transversalen Positionen~$\rb{x}$,~$\rbb{x}$ und~$\rb{y}$,~$\rbb{y}$.
Dies impliziert, da"s der relevante Impakt der Streuung folgt als die Differenz der~$\zet_i$-abh"angigen Vektoren~$\rb{X}_\zet$,~$\rb{Y}_\zet$: als der Verbindungsvektor nicht der {\it blo"sen Mitten\/}, sondern der {\it $\zet_i$-gewichteten Mitten\/}~\mbox{Quark-An}\-tiquark-Dipole.
Dies genau der Unterschied zur Diskussion Nachtmanns in Ref.~\cite{Nachtmann96}.
\vspace*{-.5ex}

\bigskip\noindent
Unmittelbar streu-relevant ist nicht das $S$- sondern das $T$-Matrixelement.
Dabei ist der $S$-Operator~-- vgl.\@ Gl.~(\ref{S-Operator})~-- verkn"upft mit dem $T$-Operator "uber%
\FOOT{
  Definition nach Ref.~\cite{Itzykson88}
}
%
\begin{align} \label{T-Operator}
S\;
  =\; \bbbone\; +\; \iIM\, (2\pi)^4\; \de(P)\vv T
\end{align}
so da"s folgt f"ur das~$T$-Matrixelement:
%
\begin{align} \label{T-Element2h_V}
&\hspace*{-10pt}
 \bracket{\, h^{2'}\!(P_{2'})\, h^{1'}\!(P_{1'}),\,\IN \,}{\,
             T\, \bracketM\,
             h^1(P_1)\, h^2(P_2),\,\IN \,}
    \\[.75ex]
&\hspace*{-10pt}
 \underset{\text{$s \!\to\! \infty$}}{\sim}\;
        -\, 2\iIM\,s\; \int d^2\rb{b}\;
             \efn{\T-\iIM\,\tfb \!\cdot\! \rb{b}}\vv
             \int d\vph_{1',1} (\zet_1, \rb{X})\;
             \int d\vph_{2',2} (\zet_2, \rb{Y})
    \nn \\[-1ex]
&
 \phantom{P_2}\times
 \hspace*{-5pt}
  \begin{aligned}[t]
  \vacL\,
    &\trDrst{F}\!\big[
      Z\idx{2}^{-1} \big(
        V\idx{+}             (\infty,0, \bzet_1 \rb{X} \!+\! \rb{b}\!/2)
        - \deZ\,\bbbOne{F} \big)\;
      Z\idx{2}^{-1} \big(
        V\idx{+}^{\D\dagger} (\infty,0,      -\zet_1 \rb{X} \!+\! \rb{b}\!/2)
        - \deZ\,\bbbOne{F} \big)
    \big]
    \nn \\[-.5ex]
  \times\,
    &\trDrst{F}\!\big[
      Z\idx{2}^{-1} \big(
        V\idx{-}             (0,\infty, \bzet_2 \rb{Y} \!-\! \rb{b}\!/2)
        - \deZ\,\bbbOne{F} \big)\;
      Z\idx{2}^{-1} \big(
        V\idx{-}^{\D\dagger} (0,\infty,      -\zet_2 \rb{Y} \!-\! \rb{b}\!/2)
        - \deZ\,\bbbOne{F} \big)
    \big]
    \nn \\[-.5ex]
    &\phantom{\trDrst{F}\!}
     -\vv 1\,
  \vacR
    \nn
  \end{aligned}
    \nn
\end{align}
vgl.\@ Gl.~(\ref{S-Element2h_WW-V_APP}).
Die im Vakuumerwartungswert subtrahierte Eins ist unmittelbare Konsequenz des Einsoperators in Gl.~(\ref{T-Operator}).

Das $T$-Matrixelement nach Gl.~(\ref{T-Element2h_V}) wird im folgenden dargestellt in Termen von Wegner-Wilson-{\it Loops\/} statt {\it -Linien}, also in manifest eichinvarianter Form.

Dazu betrachten wir die Renormierungskonstante~\mbox{$Z\idx{2}$} des (Anti)Quark-Spinorfelds nach Gl.~(\ref{Z2-renorm}), die nach Nachtmann in Ref.~\cite{Nachtmann91} wie folgt verkn"upft ist im Limes~\mbox{\,$s \!\to\! \infty$} mit dem Vakuumerwartungswert~\mbox{\,$\vac{\;\cdot\;}$} der Spur einer einzelnen vollen%
\FOOT{
  im Sinne des exponentierten Integrals des Eichfeldes entlang der vollen Parton-Trajektorie
}
Wegner-Wilson-Linie:
\begin{samepage}
%
\begin{align} \label{Z2-renorm_WW-V_allg}
Z\idx{2}\vv
  \underset{\text{$s \!\to\! \infty$}}{\sim}\vv
  \vac{\,\trDrst{F} V\idx{+}(\infty,0,\rb{0})\,}
\end{align}
vgl.\@ die Gln.~(\ref{WW-V_lim}),~(\ref{WW-V_lim}$'$).
Poincar\'einvarianz des Vakuums der Quantenchromodynamik und/oder seine Invarianz unter Ladungskonjugation~${\cal C}$ f"uhrt dann unmittelbar auf
%
\begin{align} \label{Z2-renorm_WW-V}
Z\idx{2}\vv
&\underset{\text{$s \!\to\! \infty$}}{\sim}\vv
  \vac{\,\trDrst{F} V\idx{+}(\infty,0,  \phantom{-}\bzet_1 \rb{X} \!+\! \rb{b}\!/2)\,}
    \\[-.5ex]
&\underset{\text{$s \!\to\! \infty$}}{\sim}\vv
  \vac{\,\trDrst{F} V\idx{+}^{\D\dagger}(\infty,0, -\zet_1 \rb{X} \!+\! \rb{b}\!/2)\,}
    \tag{\ref{Z2-renorm_WW-V}$'$} \\[.25ex]
&\underset{\text{$s \!\to\! \infty$}}{\sim}\vv
  \vac{\,\trDrst{F} V\idx{-}(0,\infty,  \phantom{-}\bzet_2 \rb{Y} \!-\! \rb{b}\!/2)\,}
    \tag{\ref{Z2-renorm_WW-V}$''$} \\[-.5ex]
&\underset{\text{$s \!\to\! \infty$}}{\sim}\vv
  \vac{\,\trDrst{F} V\idx{-}^{\D\dagger}(0,\infty, -\zet_2 \rb{Y} \!-\! \rb{b}\!/2)\,}
    \tag{\ref{Z2-renorm_WW-V}$'''$}
\end{align}
\end{samepage}%
Die Umformungen des $T$-Matrixelements beziehen sich auf den Vakuumerwartungswert der vier Wegner-Wilson-Linien und disconnected Terme; wir bezeichnen diesen kurz mit~$\vac{4V}$, vgl.\@ die drei letzten Zeilen von Gl.~(\ref{T-Element2h_V}), und schreiben:%
\FOOT{
   \label{FN:arguments-suppressed}Seien f"ur Pr"agnanz der Notation im folgenden unterdr"uckt Argumente von Funktionalen und Funktionen.
}
%
\begin{alignat}{2} \label{vierVvev-0}
\vac{4V}\;
 =\; 
  \vacL\,
 &\vac{ \trDrst{F} V\idx{+} }^{\!-1}\,
  \vac{ \trDrst{F} V\idx{+}^{\D\dagger} }^{\!-1}\vv&
    &\trDrst{F}\!\big[
        \big( V\idx{+} - \deZ\,\bbbOne{F} \big)
        \big( V\idx{+}^{\D\dagger} - \deZ\,\bbbOne{F} \big)\,
    \big]
    \\
  \times\,
 &\vac{ \trDrst{F} V\idx{-} }^{\!-1}\,
  \vac{ \trDrst{F} V\idx{-}^{\D\dagger} }^{\!-1}\vv&
    &\trDrst{F}\!\big[
        \big( V\idx{-} - \deZ\,\bbbOne{F} \big)
        \big( V\idx{-}^{\D\dagger} - \deZ\,\bbbOne{F} \big)
    \big]\vv
    -\; 1\,
  \vacR
    \nn
\end{alignat}
unter Herausziehen der Konstanten~$Z\idx{2}$~-- in Form der Gln.~(\ref{Z2-renorm_WW-V})-(\ref{Z2-renorm_WW-V}$'''$) zur Verdeutlichung ihres Zusammenhangs mit den respektiven Wegner-Wilson-Linien~-- aus den Spuren. \\
\indent
F"ur gro"se invariante Schwerpunktenergie~$\surd s$ differiert die Renormierungskonstante des Quarkfelds~\mbox{\,$Z\idx{2}$} von Eins nur in Korrekturen der Ordnung~\mbox{\,$1\!/\!\surd s$}:
\begin{samepage}
%
\begin{align} \label{Z2,deZ-lim}
&Z\idx{2}\vv
  \underset{\text{$s \!\to\! \infty$}}{\sim}\vv 1
    \\
 &\Longrightarrow\qquad
  \deZ\;
  =\; 1 - Z\idx{2}\vv
  \underset{\text{$s \!\to\! \infty$}}{\sim}\vv 0
    \tag{\ref{Z2,deZ-lim}$'$}
\end{align}
Aufgefa"st in diesem Sinne~\mbox{\,$\deZ \!\equiv\! 0$}, folgt f"ur Gl.~(\ref{vierVvev-0}):
%
\begin{alignat}{2} \label{vierVvev-1}
\vac{4V}\;
 =\; 
  \vacL\,
 &\vac{\, \trDrst{F} V\idx{+} \,}^{\!-1}\,
  \vac{\, \trDrst{F} V\idx{+}^{\D\dagger} \,}^{\!-1}\vv&
    &\trDrst{F}\!\big[
      V\idx{+} V\idx{+}^{\D\dagger} \big]
    \\
  \times\,
 &\vac{\, \trDrst{F} V\idx{-} \,}^{\!-1}\,
  \vac{\, \trDrst{F} V\idx{-}^{\D\dagger} \,}^{\!-1}\vv&
    &\trDrst{F}\!\big[
      V\idx{-} V\idx{-}^{\D\dagger} \big]\vv
    -\; 1\,
  \vacR
    \nn
\end{alignat}
indem weiterhin explizit ausgeschrieben sind die Konstanten~\mbox{\,$Z\idx{2} \!\equiv\! 1$}. \\
\indent
Der "Ubergang zu Wegner-Wilson-Loops geschieht auf Basis der folgenden "Uberlegung.
Wir betrachten Streuung {\it eichinvarianter\/} Zust"ande~$h^i(P_i)$.
Dem tr"agt der Ansatz der Zustandvektoren~$\ket{h^i(P_i)}$, vgl.\@ die Gln.~(\ref{h-ket}),~(\ref{h-ketREPET}), bereits Rechnung durch die Kronecker-Symbole in den Eichgruppenindizes, durch die entsprechend verkn"upft ist die Colour von Quark und Antiquark des Zustands.
Diese induzieren die Spuren~$\trDrst{F}$ im $T$-Matrixelement nach Gl.~(\ref{T-Element2h_V}), vgl.\@ die Bem.\@ vorab Gl.~(\ref{S-Element2h_WW-V}).
Vollst"andig garantiert ist Eichinvarianz aber erst durch Einf"uhren von Schwinger-Strings: in definierter Weise orientierter Eichfeld-Konnektoren, die die Wegner-Wilson-Linien der (Anti)Quarks eines Zustands verbinden bei gro"sen positiven und negativen Zeiten bei \mbox{$x^0 \!=\! \pm T$}~-- in~praxi mit~\mbox{$T \!\to\! \infty$}~-- und so diese "uberf"uhren in {\it Wegner-Wilson-Loops\/}; vgl.\@ die Refn.~\cite{Kraemer91,Dosch94a,Nachtmann96}.
Formal hei"st dies:%
\FOOT{
  Unsere Konvention f"ur den Wegner-Wilson-Loop~$W$ impliziert Bildung der Spur bez"uglich der entsprechenden Darstellung~$\Drst{R}$.   Sei mit~$W'$ bezeichnet der Ausdruck {\sl vor\/} Spurbildung, das hei"st:~\mbox{\,$W \!=\! \trDrst{R}\! W' \!=\! W'_{\al\al}$} mit \mbox{$W' \!=\!  P_{\tilde{\cal C}} \exp\, -\iIM g\!/\!2 \iint_{{\cal S}(\tilde{\cal C})} d\si^{\mu\nu} F_{\mu\nu}$}, vgl.\@ Gl.~(\ref{Konnektor_LoopFtr}).   Mit~$W'$ dem {\sl eichinvarianten Pendant\/} zu~$V$ erh"alt die Konfrontation von Gr"o"sen in Termen von Wegner-Wilson-{\sl Loops\/} versus {\sl -Linien\/} suggestive Gestalt.
}
%
\vspace*{-.25ex}
\begin{alignat}{3} \label{Linien-zu-Loops}
&\trDrst{F} \big[ V\idx{+} V\idx{+}^{\D\dagger} \big]&\vv
  &\longrightarrow\quad&
  &\trDrst{F} W'\idx{+}\;
  \equiv\; W\idx{+}
    \\
&\trDrst{F} \big[ V\idx{-} V\idx{-}^{\D\dagger} \big]&\vv
  &\longrightarrow\quad&
  &\trDrst{F} W'\idx{-}\;
  \equiv\; W\idx{-}
    \tag{\ref{Linien-zu-Loops}$'$}
\end{alignat}
bzgl.\@ der Spur~\mbox{$\trDrst{R} \!=\! 1\!/\! \dimDrst{R}\tr$} vgl.\@ Gl.~(\ref{WW-Loop}). \\
\indent
In Ref.~\cite{Kulzinger95} zeigen wir f"ur den Vakuumerwartungswert~\mbox{\,$\vac{\;\cdot\;}$} eines einzelnen {\it lichtartigen\/} Wegner-Wilson-Loop:
%
\begin{align} \label{VVdagger->W}
&\vac{ \trDrst{F} W'\idx{+} }\;
  \equiv\; \vac{ W\idx{+} }\vv
  \underset{\text{$s \!\to\! \infty$}}{\sim}\vv 1
    \\[.25ex]
&\vac{ \trDrst{F} W'\idx{-} }\;
  \equiv\; \vac{ W\idx{-} }\vv
  \underset{\text{$s \!\to\! \infty$}}{\sim}\vv 1
    \tag{\ref{VVdagger->W}$'$}  
\end{align}
Dies suggeriert, auf Basis von Gl.~(\ref{Z2,deZ-lim}), das hei"st explizit von
\end{samepage}%
\vspace*{-.25ex}
\begin{align} \label{Z2^2-renorm_V}
&Z\idx{2}^2\vv
  \underset{\text{$s \!\to\! \infty$}}{\sim}\vv
      \vac{\, \trDrst{F} V\idx{+} \,}
        \vac{\, \trDrst{F} V\idx{+}^{\D\dagger} \,}\quad
  \underset{\text{$s \!\to\! \infty$}}{\sim}\vv 1
    \\
&Z\idx{2}^2\vv
  \underset{\text{$s \!\to\! \infty$}}{\sim}\vv
      \vac{\, \trDrst{F} V\idx{-} \,}
   \vac{\, \trDrst{F} V\idx{-}^{\D\dagger} \,}\quad
  \underset{\text{$s \!\to\! \infty$}}{\sim}\vv 1
    \tag{\ref{Z2^2-renorm_V}$'$}
\end{align}
zu identifizieren:
\vspace*{-1ex}
\begin{alignat}{2} \label{Z2^2-renorm_W}
&\vac{\, \trDrst{F} V\idx{+} \,}
    \vac{\, \trDrst{F} V\idx{+}^{\D\dagger} \,}\;
  =\;& &Z\idx{2}^2\quad
  \stackrel{\D!}{=}\; \vac{ \trDrst{F} W'\idx{+} }\;
    \equiv\; \vac{ W\idx{+} }
    \\
&\vac{\, \trDrst{F} V\idx{-} \,}
    \vac{\, \trDrst{F} V\idx{-}^{\D\dagger} \,}\;
  =\;& &Z\idx{2}^2\quad
  \stackrel{\D!}{=}\; \vac{ \trDrst{F} W'\idx{-} }\;
    \equiv\; \vac{ W\idx{-} }
    \tag{\ref{Z2^2-renorm_W}$'$}
    \\[-4ex]\nn
\end{alignat}
im Limes~\mbox{$s \!\to\! \infty$}, vgl.\@ die Gln.~(\ref{Linien-zu-Loops}),~(\ref{Linien-zu-Loops}$'$).
Es sind~$Z\idx{2}$ und~$Z\idx{2}^2$ zu interpretieren respektive als Renormierungskonstante der Wegner-Wilson-Linien und -Loops. \\
\indent
F"ur Gl.~(\ref{vierVvev-1}) folgt, da"s die Vakuumerwartungswerte der Spuren der einzelnen Linien~-- die $Z\idx{2}$-Fakotren~-- "uberf"uhrt werden in die Vakuumerwartungswerte der (Spuren der) respektiven Loops.
Da sie identisch Eins sind, kann ferner die subtrahierte Eins, die r"uhrt aus dem "Ubergang von $S$- zu $T$-Operator, symmetrisch absorbiert werden in die Spuren der Vakuumerwartungswerts beider Loops.
Es folgt:
\vspace*{-.25ex}
\begin{alignat}{2} \label{vierVvev-2}
\vac{4V}\;
  &=\; \vac{ \trDrst{F} W'\idx{+} }^{\!-1}\,
       \vac{ \trDrst{F} W'\idx{-} }^{\!-1}\vv
       \vacL\,
         \trDrst{F}\!\big[ W'\idx{+} - \bbbOne{F} \big]\;
         \trDrst{F}\!\big[ W'\idx{-} - \bbbOne{F} \big]\,
       \vacR
    \\[.5ex]
  &=\; \vac{ W\idx{+} }^{\!-1}\,
       \vac{ W\idx{-} }^{\!-1}\vv
       \vacL\,
         [ W\idx{+} - 1 ]
         [ W\idx{-} - 1 ]\,
       \vacR
    \tag{\ref{vierVvev-2}$'$}
    \\[-4.25ex]\nn
\end{alignat}
Dieser Formel wesentlich zugrunde liegt die Identifizierung von~$Z\idx{2}^2$ mit~\mbox{$\vac{ W\idx{+} }$},~\mbox{$\vac{ W\idx{-} }$}.
Formales Argument hierf"ur ist die Identit"at beider Gr"o"sen mit Eins im Limes~\mbox{$s \!\to\! \infty$}.
Argument ist ferner:
Da"s die $T$-Amplitude f"ur Streuung eichinvarianter Wegner-Wilson-Loops~$W\idx{+}$,~$W\idx{-}$ a~priori vollst"andig ausdr"uckbar sein sollte durch die relevanten eichinvarianten Gr"o"sen: das hei"st genau durch diese Loops, insbesondere $Z\idx{2}$ durch~\mbox{$\vac{ W\idx{+} }$},~\mbox{$\vac{ W\idx{-} }$}.
Und da"s sie von "aquivalenter Form sein sollte wie die $T$-Amplitude f"ur Streuung von Partonen.
Gl.~(\ref{vierVvev-2}) ist daher zu konfrontieren mit den Formeln, die Nachtmann herleitet f"ur die Streuung zweier Partonen, etwa f"ur zwei Quarks:
\vspace*{-.25ex}
\begin{alignat}{2} \label{Tqq_zweiVvev}
&\vac{2V}\;
  =\; \vac{ \trDrst{F} V\idx{+} }^{\!-1}\,
        \vac{ \trDrst{F} V\idx{-} }^{\!-1}\vv
      \vacL\,
        \big[ V\idx{+} - \bbbOne{F} \big]_{n_{1'}n_1}\;
        \big[ V\idx{-} - \bbbOne{F} \big]_{n_{2'}n_2}\,
      \vacR
    \\
&\text{wobei}\qquad
  T_{Q \mskip-1.5mu Q}^{(s,t)}
  \underset{\text{$s \!\to\! \infty$}}{\sim}\;
    -\, 2\iIM\,s\;
      \int d^2\rb{b}\;
      \efn{\T-\iIM\,\tfb \!\cdot\! \rb{b}}\vv
      \vac{2V}
    \tag{\ref{Tqq_zweiVvev}$'$}
    \\[-4.75ex]\nn
\end{alignat}
vgl.\@ Ref.~\cite{Nachtmann91}; f"ur Gluonen~-- vgl.\@ Ref.~\cite{Nachtmann96}~-- ist "uber zu gehen von der fundamentalen zur adjungierten Darstellung, das hei"st von~$\Drst{F}$ zu~$\Drst{A}$.
Wir verweisen auf die ausf"uhrliche Diskussion dieser Formel~-- vgl.\@ Gl.~(\ref{Parton-Parton_lim})~-- in Abschnitt~\ref{Sect:Nahezu_vs_exakt}.

F"ur das $T$-Matrixelement nach Gl.~(\ref{T-Element2h_V}) folgt mit Gl.~(\ref{vierVvev-2}$'$) abschlie"send:
\vspace*{-.25ex}
\begin{align} \label{T-Element2h_W}
T\hh^{(s,t)}\;
  &\equiv\;
  \bracket{\, h^{2'}\!(P_{2'})\, h^{1'}\!(P_{1'}),\,\IN \,}{\,
             T\, \bracketM\,
             h^1(P_1)\, h^2(P_2),\,\IN \,}
    \\[.25ex]
  \underset{\text{$s \!\to\! \infty$}}{\sim}\;
    &-\, 2\iIM\,s\;
  \begin{aligned}[t]
   &\int d^2\rb{b}\;
    \efn{\T-\iIM\,\tfb \!\cdot\! \rb{b}}\vv
    \int d\vph_{1',1} (\zet_1, \rb{X})\;
      \int d\vph_{2',2} (\zet_2, \rb{Y})
    \\[-.5ex]
   &\times\,
    \vac{ W\idx{+} }^{\!-1}\,
      \vac{ W\idx{-} }^{\!-1}\vv
    \vacL\,
      [ W\idx{+} - 1 ]
      [ W\idx{-} - 1 ]\,
    \vacR
  \end{aligned}
    \nn
    \\[-5ex]\nn
\end{align}
mit
\vspace*{-1.25ex}
\begin{alignat}{3} \label{WW-explizit}
&W\idx{+}\;&
  &\equiv\; W\idx{+}(\zet_1, \rb{X}, +\rb{b}\!/2)\quad&
  &\equiv\; W({\cal C}\idx{+})
    \\[.25ex]
&W\idx{-}\;&
  &\equiv\; W\idx{-}(\zet_2, \rb{Y}, -\rb{b}\!/2)\quad&
  &\equiv\; W({\cal C}\idx{-})
    \tag{\ref{WW-explizit}$'$}
    \\[-4.5ex]\nn
\end{alignat}
Es ist der Wegner-Wilson-Loop~\mbox{$W\idx{+}$} bez"uglich Index und Argumente zu lesen wie folgt~[entsprechend~\mbox{$W\idx{-}$}]:
Zun"achst ist~$W\idx{+}$ der Eichfeld-Konnektor entlang der geschlossenen orientierten Rechteckkurve~${\cal C}\idx{+}$ von Quark- und Antiquark-Trajek\-torie und deren Verbindungen bei~\mbox{$x^0 \!=\! \pm T\idx{+}$}~-- also $T\idx{+}$ die halbe "`L"ange"' dieser Trajektorien im Sinne der Eigenzeit der Propagation.
Longitudinal besitzt~${\cal C}\idx{+}$ nur eine~\mbox{$x^+$-Kom}\-ponente: liegt {\it exakt\/} in der durch~\mbox{$x^- \!\equiv\! 0$} charakterisierten Lichtkegel-Hyperfl"ache.%
\FOOT{
  Konsequenz dessen, da"s sich die Herleitung de~facto geschieht im Limes~\mbox{$s \!\to\! \infty$}
}
Transversal sind die~${\cal C}\idx{+}$ konstituierenden (Anti)Quarks separiert durch den Vektor~$\rb{X}$, der zeigt vom Antiquark zum Quark, vgl.\@ Gl.~(\ref{Diffvekt-XY}).
Nach Gl.~(\ref{Q_Impulse}) tr"agt das Quark den Bruchteil~$\zet_1$ am gesamten Lichtkegelimpuls des Zustands~$h^1(P_1)$.
Ausgehend vom Antiquark den Bruchteil~$\zet_1$ in Richtung Quark, das hei"st abgetragen den Vektor~\mbox{$\zet_1\rb{X}$}, sei definiert als die "`$\zet_1$-gewichtete Mitte"' von Quark und Antiquark.
Diese ist positioniert an der Stelle~$+\rb{b}\!/2$. \\
\indent
Der Diskussion dieses wie der folgenden Kapitel liegt wesentlich zugrunde diese Geometrie, auf die wir noch ausf"uhrlich eingehen.
\vspace*{-.5ex}

\bigskip\noindent
In Abschnitt~\ref{Sect:Nahezu_vs_exakt} ist argumentiert, da"s die G"ultigkeit des $T$-Matrixelements auf Basis {\it exakt lichtartiger\/} Partontrajektorien verallgemeinert wird, indem diese betrachtet werden als die unphysikalischen Limites der physikalischen {\it nahezu lichtartigen\/} Trajektorien und durch diese ersetzt werden; dies impliziert f"ur die geschlossenen Kurven der Wegner-Wilson-Loops:
\vspace*{-.25ex}
\begin{align} \label{Subst-Trajekt}
&{\cal C}\idx{+}\;
  \longrightarrow\; {\cal C}\Dmfp
    \\
&{\cal C}\idx{-}\;
  \longrightarrow\; {\cal C}\Dmfm
    \tag{\ref{Subst-Trajekt}$'$}
    \\[-4.25ex]\nn
\end{align}
mit~\mbox{${\cal C}\Dmfp \!\to\! {\cal C}\idx{+}$},~\mbox{${\cal C}\Dmfm \!\to\! {\cal C}\idx{-}$} im Limes~\mbox{$s \!\to\! \infty$}, vgl.\@ Gl.~(\ref{Trajektorien_lim}). \\
\indent
In Anhang~\ref{APP:Boosts} werden motiviert und definiert Koordinatenlinien~\mbox{$\tilde\mu \!\in\! \{\mfp,\mfm,1,2\}$} durch die Richtungen der physikalischen nahezu lichtartigen Weltlinien der Partonen, so da"s longitudinal die der~"`plus"'-Partonen nur eine~\bm{\mfp}- und die der~"`minus"'-Partonen nur eine~\bm{\mfm}- (kontravariante) Komponente besitzt.
Wir setzen im folgenden voraus den formalen wie interpretatorischen Zusammenhang dieser Koordinatenlinien, wie er ausf"uhrlich hergestellt und diskutiert wird in Anhang~\ref{APP-Sect:Lorentz-Boosts} bis~\ref{APP-Sect:Minkowski}.
In Termen dieser Koordinaten wird die Ersetzung der Trajektorien nach Gl.~(\ref{Subst-Trajekt}) formal realisiert durch Substitution der Lorentz-Indizes
\vspace*{-.25ex}
\begin{align} \label{Subst-LorentzInd}
&\bm{+}\;
  \longrightarrow\; \bm{\mfp}
    \\
&\bm{-}\;
  \longrightarrow\; \bm{\mfm}
    \tag{\ref{Subst-LorentzInd}$'$}
    \\[-4.25ex]\nn
\end{align}
mit~\mbox{$\bm{\mfp} \!\to\! \bm{+}$},~\mbox{$\bm{\mfm} \!\to\! \bm{-}$} im Limes~\mbox{$s \!\to\! \infty$}.
In diesem Sinne geht auf Hadronniveau%
\FOOT{
  Bzgl.\@ des "Ubergangs auf Partonniveau sei verwiesen auf Gl.~(\ref{Parton-Parton}) gegen"uber~(\ref{Parton-Parton_lim}).
}
das \mbox{$T$-Ma}\-trixelement~-- vgl.\@ Gl.~(\ref{T-Element2h_W}) und~(\ref{WW-explizit}),~(\ref{WW-explizit}$'$)~-- "uber in:
\vspace*{-.25ex}
\begin{align} \label{T-Element2h_W-mf}
T\hh^{(s,t)}\;
  &\equiv\;
   \bracket{\, h^{2'}\!(P_{2'})\, h^{1'}\!(P_{1'}),\,\IN \,}{\,
               T\, \bracketM\,
               h^1(P_1)\, h^2(P_2),\,\IN \,}
    \\[.5ex]
  =\; &-\, 2\iIM\,s\;
  \begin{aligned}[t]
   &\int d^2\rb{b}\;
    \efn{\T-\iIM\,\tfb \!\cdot\! \rb{b}}\vv
    \int d\vph_{1',1} (\zet_1, \rb{X})\;
      \int d\vph_{2',2} (\zet_2, \rb{Y})
    \\
   &\times\,
    \vac{ W\Dmfp }^{\!-1}\,
      \vac{ W\Dmfm }^{\!-1}\vv
    \vacL\,
      [ W\Dmfp - 1 ]
      [ W\Dmfm - 1 ]\,
    \vacR
  \end{aligned}
    \nn
    \\[-4.5ex]\nn
\end{align}
mit
\vspace*{-.5ex}
\begin{alignat}{3} \label{WW-explizit-mf}
&W\Dmfp\;&
  &\equiv\; W\Dmfp(\zet_1, \rb{X}, +\rb{b}\!/2)\quad&
  &\equiv\; W({\cal C}\Dmfp)
    \\[.5ex]
&W\Dmfm\;&
  &\equiv\; W\Dmfm(\zet_2, \rb{Y}, -\rb{b}\!/2)\quad&
  &\equiv\; W({\cal C}\Dmfm)
    \tag{\ref{WW-explizit-mf}$'$}
    \\[-4.5ex]\nn
\end{alignat}
Die Geometrie des Wegner-Wilson-Loops~$W\idx{+}$ ist in~extensio diskutiert in Anschlu"s an die Gln.~(\ref{WW-explizit}),~(\ref{WW-explizit}$'$).
Die des Loops~$W\Dmfp$ differiert dieser gegen"uber genau darin~[entsprechend~$W\Dmfm$ gegen"uber~$W\idx{-}$], da"s die longitudinalen Lorentz-Indizes zu substituieren sind entsprechend~\mbox{$\bm{+} \!\to\! \bm{\mfp}$},~\mbox{$\bm{-} \!\to\! \bm{\mfm}$}.
Die orientierte Loop-Rechteckkurve~${\cal C}\Dmfp$ besitzt nur eine~\mbox{$x^\mfp$-Kom}\-ponente: liegt {\it exakt\/} in der durch~\mbox{$x^\mfm \!\equiv\! 0$} charakterisierten Hyperfl"ache.
$W\Dmfp$,~$W\idx{+}$ sind {\it gegeneinander verdreht\/} longitudinal in der $x^0\!x^3$-Ebene und {\it identisch\/} transversal. \\
\indent
Die Auswertung der $T$-Amplitude geschieht mit Methoden der Differemtialgeometrie~auf (pseudo-)Riemannschen Mannigfaltigkeiten und ist formal identisch auf Basis der Kurven~${\cal C}\Dmfp$, ${\cal C}\Dmfm$ und deren Limites~${\cal C}\idx{+}$,~${\cal C}\idx{-}$ f"ur~\mbox{$s \!\to\! \infty$}~-- Gl.\,(\ref{T-Element2h_W-mf}) versus~(\ref{T-Element2h_W}):
\mbox{Die longitudinale Dy}\-namik ist vollst"andig subsumiert durch longitudinale Komponenten des~\pagebreak\mbox{metrischen Tensors} der respektiven "`nat"urlichen"' Koordinatenlinien~-- \mbox{$\mu \!\in\! \{+,-,1,2\}$} versus~\mbox{$\tilde\mu \!\in\! \{\mfp,\mfm,1,2\}$}.
Diskrepanz besteht daher genau darin, da"s die Au"serdiagonalelemente~$g_{+-}$,~$g_{-+}$ des einen metrischen Tensors zu substituieren sind durch die Elemente~$g_{\mfp\mfm}$,~$g_{\mfm\mfp}$ des anderen und da"s die Diagonalelemente~$g_{++}$,~$g_{--}$ identisch verschwinden, w"ahrend die Elemente~$g_{\mfp\mfp}$,~$g_{\mfm\mfm}$ f"ur endliche Werte von~$s$ ungleich Null sind und zus"atzliche nichtverschwindende Terme generieren.
Die unabh"angigen Elemente~$g_{\mfp\mfp}$,~$g_{\mfp\mfm}$ h"angen, definiert "uber die Richtungen der Weltlinien der Partonen, nichttrivial ab von deren invarianter Schwerpunktenergie~$\surd s$.
Folglich differiert die $s$-Abh"angigkeit der "`nahezu lichtartigen"' $T$-Amplitude von der rein kinematischen Abh"angigkeit proportional zu~$s$ der "`exakt lichtartigen"' $T$-Amplitude.
\vspace*{-.5ex}

\section[Nahezu lichtartige~$T$-Amplitude: Auswertung]{%
         Nahezu lichtartige~\bm{T}-Amplitude: Auswertung}
\label{Sect:T-Amplitude.Auswertung}

Wir betrachten die $T$-Amplitude f"ur die Streuung der hadronischen Zust"ande~$h^{1}\!(P_{1})$,~$h^{2}\!(P_{2})$ in~$h^{1'}\!(P_{1'})$,~$h^{2'}\!(P_{2'})$ nach Gl.~(\ref{T-Element2h_W-mf}).
Sei durch
\begin{samepage}
%
\begin{align} \label{T2h_T2ell-mf}
T\hh^{(s,t)}\;
  &\equiv\;
   \bracket{\, h^{2'}\!(P_{2'})\, h^{1'}\!(P_{1'}),\,\IN \,}{\,
               T\, \bracketM\,
               h^1(P_1)\, h^2(P_2),\,\IN \,}
    \\[.375ex]
  =\; &\int d^2\rb{b}\;
       \efn{\T-\iIM\,\tfb \!\cdot\! \rb{b}}\;
       \int d\vph_{1',1} (\zet_1, \rb{X})\;
         \int d\vph_{2',2} (\zet_2, \rb{Y})\vv
       \tTll^{(s,\rb{b})}(\zet_1, \rb{X}; \zet_2, \rb{Y}; \rb{b})
    \nn
    \\[-4.875ex]\nn
\end{align}
definiert, vgl.\@ Fu"sn.\,\FN{FN:arguments-suppressed}:
\vspace*{-.25ex}
\begin{align} \label{tTll_WW-mf}
\tTll^{(s,\rb{b})}\;
  &=\; -\, 2\iIM\,s\vv
         \vac{4V}
    \\[.5ex]
  &=\; -\, 2\iIM\,s\vv
         \vac{W\Dmfp}^{-1}\; \vac{W\Dmfm}^{-1}\vv
         \vac{\, [ W\Dmfp - 1 ]
                 [ W\Dmfm - 1 ] \,}
    \tag{\ref{tTll_WW-mf}$'$}
\end{align}
vgl.\@ Gl.~(\ref{vierVvev-2}$'$).
Wir rekapitulieren, da"s die Notation~$W\Dmfp$,~$W\Dmfm$ bereits impliziert normierte Spurbildung~\mbox{$\trDrst{F} \!=\! 1\!/\! \dimDrst{F} \tr$} bez"uglich der fundamentalen Darstellung~$\Drst{F}$ der Loops. \\
\indent
Die Funktion~\mbox{$\tTll \!\equiv\! \tTll^{(s,\rb{b})}$} ist zu interpretieren als die $T$-Amplitude der zugrundeliegenden Streuung der Wegner-Wilson-Loops~$W\Dmfp$,~$W\Dmfm$ "`bez"uglich des transversalen Sto"sparameters~$\rb{b}$"', das hei"st der bez"uglich~$\tfb$ Fourier-transformierten $T$-Amplitude.~--
Der Vierer-Impulstransfer~$\tf$ ist transversal bis auf Korrekturen~$\propto\! s^{-2}$, die Mandelstam-Variable daher f"ur gro"ses~$s$ approximativ gegeben durch~\mbox{$t \!=\! \tfQ \!\cong\! \tfbQ$}; vgl.\@ unten die Gln.~(\ref{tfbB_-t}),~(\ref{tfbB_-t}$'$).
Wir halten fest, da"s gilt bez"uglich der {\it Energiedimension},~\mbox{$[E] \!=\! +1$}:~\mbox{\,$[T\hh^{(s,t)}] \!=\! 0$}, folglich~\mbox{$[\tTll^{(s,\rb{b})}] \!=\! +2$}.

Die Amplitude~\mbox{$\tTll$} ist neben den Lichtkegelwellenfunktionen die eigentliche streu-relevan\-te Gr"o"se.
Sie ist hergeleitet in nichtperturbativ determinierter Quantenchromodynamik: f"ur kleinen invarianten Impuls"ubertrag, etwa~\mbox{$-t \!<\! 1\GeV^2$}, sie wird im folgenden ausgewertet im Rahmen des \DREI{M}{S}{V} und in der resultierenden Form \DREI{M}{S}{V}-spezifisch.

Die Streuung wird nach Gl.~(\ref{tTll_WW-mf}$'$) determiniert durch die Korrelation~\mbox{\,$\vac{\;\cdot\;}$} der Wegner-Wilson-Loops~$W\Dmfp$,~$W\Dmfm$~-- "uber die nichtperturbativen Fluktuation des Vakuum-Eichfelds~$A$~-- normiert mit den inversen Vakuumerwartungswerten der einzelnen Loops.
Die Loops selbst sind vollst"andig bestimmt als Funktionale von Fl"achen~\mbox{${\cal S}\Dmfp \!\equiv\! {\cal S}(\tilde{\cal C}\Dmfp)$},~\mbox{${\cal S}\Dmfm \!\equiv\! {\cal S}(\tilde{\cal C}\Dmfm)$}, f"ur~\mbox{$\imath \!=\! \mfp,\mfm$}:
%
\begin{align} \label{W_S-mf}
&W\Dimath\;
  \equiv\; W\big({\cal C}\Dimath\big)\;
  \equiv\; W\big({\cal S}\Dimath\big)
    \\[.5ex]
&\text{mit}\qquad
  \pa{\cal S}\Dimath
    \equiv {\cal C}\Dimath\qquad
  {\cal S}\Dimath
    \equiv {\cal S}(\tilde{\cal C}\Dimath)
    \tag{\ref{W_S-mf}$'$}
\end{align}
explizit, mit~\mbox{$\tilde\mu,\tilde\nu \!\in\! \{\mfp,\mfm,1,2\}$}:%
\FOOT{
  Die Auswertung der Funktionen~$\ch\idx{\imath\jmath}$ in den Abschnitten~\ref{Subsect:chNC} und~\ref{Subsect:chC} geschieht~-- zwar gr"o"stenteils nur argumentativ~-- durch Integration "uber Mannigfaltigkeiten und Untermannigfaltigkeiten verschiedener Dimension; sei dies im folgenden suggestiv verdeutlicht durch Integrationszeichen in entsprechender Zahl.
}
\end{samepage}%
%
\begin{align} \label{Konnektor_LoopFtr-mf}
W\Dimath\;
  =\; \trDrst{R\Dimath}\; P_{\tilde{\cal C}\Dimath}\vv
        \exp\; -\frac{\iIM\,g}{2}
        \iint_{{\cal S}(\tilde{\cal C}\Dimath)}
        \dsiI[\mskip-2mu(\tilde{x}\Dimath')]{\tilde\mu\tilde\nu}\;
        F_{\tilde\mu\tilde\nu}\!
          (\tilde{x}\Dimath'; x_0,{\cal C}_{x_{\!0}\!\tilde{x}\Dimath'})
    \\[-4ex]\nn
\end{align}
mit~$\Drst{R}\Dimath$ der Darstellung von~$W\Dimath$, die in~praxi zu identifizieren ist mit der fundamentalen Darstellung~$\Drst{F}$ der (Anti)Quarks die die hadronischen Zust"ande konstituieren. \\
\indent
Die Darstellung der Wegner-Wilson-Loops nach Gl.~(\ref{Konnektor_LoopFtr-mf}) impliziert bereits den "Ubergang mithilfe des Nichtabelschen Stokes'schen Satzes von der Kurve~${\cal C}\Dimath$ zu der Fl"ache~${\cal S}\Dimath$.
Diese ist frei w"ahlbar innerhalb der zwei Forderungen:
Da"s~${\cal S}\Dimath$ berandet ist durch die orientierte Kurve~${\cal C}\Dimath$.
Und da"s~${\cal S}\Dimath$ infinitesimal plakettiert ist durch die Deformation~$\tilde{\cal C}\Dimath$ von~${\cal C}\Dimath$.%
\FOOT{
  In Zusammenhang mit dem Nichtabelschen Stokes'schen Satz wird in Abschn.~\ref{SubSect:Stokes} dar"uberhinaus gefordert, da"s der Referenzpunkt~$x_0$, der induziert wird durch den zwingenden "Ubergang zu {\it paralleltransportierten\/} Feldst"arken, zu fordern ist als Element der respektiven Fl"ache.
Diese Forderung ist nicht wirklich notwendig, da der Colour-Gehalt der Feldst"arken o.E.d.A.\@ von~$x_0$ an einen belibigen anderen (Referenz)Punkt~$x'_0$ weiter paralleltransportiert werden kann; vgl.\@ insbes.\@ Abb.~\refg{Fig:Deformation}.
} \\
%
\indent
Die paralleltransportierte Feldst"arke~\mbox{$F_{\mu\nu}\!(\tilde{x}\Dimath'; x_0,{\cal C}_{x_{\!0}\!\tilde{x}\Dimath'})$} bez"uglich Referenzpunkt~$x_0$~-- wir rekapitulieren~-- ist definiert als die konventionelle Feldst"arke am Weltpunkt~$\tilde{x}\Dimath'$, deren Eichgehalt paralleltransportiert ist an~$x_0$.
Dies geschieht mithilfe von Konnektoren entlang der Kurve~${\cal C}_{x_{\!0}\!\tilde{x}\Dimath'}$ und ihrer Inversen~\mbox{${\cal C}_{\tilde{x}\Dimath'\!x_{\!0}} \!\equiv\! {\cal C}_{x_{\!0}\tilde{x}\Dimath'}{}^{\zzz -1}$} in der Weise, da"s die paralleltransformierte Feldst"arke bez"uglich Eichtransformationen transformiert mit~$U(x_0)$ wie eine Gr"o"se, angeheftet an der Raumzeit im Punkt~$x_0$.
Vgl.\@ Gl.~(\ref{Eichtransf_Fparallel}).

Die so zu lesende paralleltransportierte Feldst"arke ist nach Gl.~(\ref{Konnektor_LoopFtr-mf}) bez"uglich~$\tilde{x}\Dimath'$ zu integrieren "uber die Fl"ache \mbox{$S\Dimath \!\equiv\! {\cal S}(\tilde{\cal C}\Dimath)$}.
Diese Integration schlie"st ein \vspace*{-.125ex}Pfadordnung~$P_{\tilde{\cal C}\Dimath}$ entlang der Deformation~$\tilde{\cal C}\Dimath$ der Kurve~${\cal C}\Dimath$.
Explizit ist diese Deformation~$\tilde{\cal C}\Dimath$ gegeben als die Verkettung von Kurven~${\cal C}_{x_{\!0}\!\tilde{x}\Dimath'}$ und deren Inversen~\mbox{${\cal C}_{\tilde{x}\Dimath'\!x_{\!0}} \!\equiv\! {\cal C}_{x_{\!0}\tilde{x}\Dimath'}{}^{\zzz -1}$} bez"uglich infinitesimal benachbarter Weltpunkten~\mbox{$\tilde{x}\Dimath' \!=\! \tilde{x}\Dimath[1]', \tilde{x}\Dimath[2]', \tilde{x}\Dimath[3]', \ldots$}, die repr"asentieren die zu durchlaufende~-- die Fl"ache~${\cal S}\Dimath$ plakettierende~-- Deformation~$\tilde{\cal C}\Dimath$; vgl.\@ Abschn.~\ref{SubSect:Stokes}, zur Illustration Abb.~\ref{Fig:Deformation}.
\vspace*{-.5ex}

\subsection[Identifizierung von~\protect\mbox{$\tilde\ch\idx{\imath\jmath}
              = (1 \!-\! \vka)\, \tilde\ch\idx{\imath\jmath}\oNC
                    + \vka\, \tilde\ch\idx{\imath\jmath}\oC$},%
              ~\protect\mbox{\,$\imath,\jmath \!\in\! \{\mfp,\mfm\}$}]{%
            \vspace*{-.5ex}%
            Identifizierung von~\bm{\tilde\ch\idx{\imath\jmath}
              = (1 \!-\! \vka)\, \tilde\ch\idx{\imath\jmath}\oNC
                    + \vka\, \tilde\ch\idx{\imath\jmath}\oC},%
              ~\bm{\,\imath,\jmath \!\in\! \{\mfp,\mfm\}}}

Seien die Fl"achen~${\cal S}\Dmfp$,~${\cal S}\Dmfm$ fest gew"ahlt im Sinne der zu fordernden Einschr"ankungen.
Wir setzen die Wegner-Wilson-Loops~\mbox{$W\Dmfp \!\equiv\! W({\cal S}\Dmfp)$},~\mbox{$W\Dmfm \!\equiv\! W({\cal S}\Dmfp)$} in der Darstellung nach Gl.~(\ref{Konnektor_LoopFtr-mf}) ein in die Amplitude~\mbox{$\tTll \!\equiv\! \tTll^{(s,\rb{b})}$} nach Gl.~(\ref{tTll_WW-mf}) und werten diese aus wie folgt. \\
\indent
Die Loop-Exponentiale werden entwickelt bis zum ersten nichtverschwindenden Term und die Reihen danach abgebrochen.
Aufgrund der Subtraktion der Einsen und der Spurfreiheit der Generatoren~$T_{\Drst{R\Dimath}}^a$%
\FOOT{
  Bzgl.\@ der Darstellung~$\Drst{R}\Dimath$ des Loops~$W\Dimath$ vgl.\@ die Anmerkung in Anschlu"s an Gl.~(\ref{Konnektor_LoopFtr-mf}).
},
~\mbox{$a\!=\!1,2,\ldots,\dimNc$}, mit~\mbox{$\dimNc \!\equiv\! \Nc^2\!-\!1$} der Dimension der~$\SUNc$, verschwinden identisch der erste und zweite Term der Entwicklung.
Es resultiert der Vakuumerwartungswert von vier paralleltransportierten Feldst"arken~\mbox{$F_{\mu\nu}\!(\tilde{x}\Dimath; x_0,{\cal C}_{x_{\!0}\!x\Dimath})$}, die je zwei stammen von~$W\Dmfp$ und~$W\Dmfm$ und durchlaufen respektive auf~${\cal S}\Dmfp$ und~${\cal S}\Dmfm$. \\
\indent
Es wird gemacht Annahme~(2') des \DREI{M}{S}{V}~-- vgl.\@ Seite~\pageref{Annahme(2')}~-- das hei"st angenommen, ein Gau"s'scher stochastischer Proze"s liege zugrunde der nichtperturbativen Fluktuation des Vakuum-Eichfelds~$A$ in der Bildung des Vakuumerwartungswerts~$\vac{\;\cdot\;}$.
Dem approximativ "aquivalent~-- vgl.\@ die Diskussion dort~-- kann angenommen werden Faktorisierung der Vakuumerwartungswerte~\mbox{$%
  \vac{ P\, g^n F_{\mu_1\nu_1 a_1}\!(x_1; x_0,{\cal C}_{x_{\!0}\!x_{\!1}}) \cdots
                F_{\mu_n\nu_n a_n}\!(x_n; x_0,{\cal C}_{x_{\!0}\!x_{\!n}}) }$} f"ur gerade~$n$ in Summen von~$n\!/\!2$ Faktoren~\mbox{$%
  \vac{ P\, g^2 F_{\mu_1\nu_1 a_1}\!(x_1; x_0,{\cal C}_{x_{\!0}\!x_{\!1}})
                F_{\mu_2\nu_2 a_2}\!(x_2; x_0,{\cal C}_{x_{\!0}\!x_{\!2}}) }$}, Vakuumerwartungswerte einer ungeraden Zahl paralleltransportierte Feldst"arketensoren verschwinden identisch; vgl.\@ Gl.~(\ref{Fak_MatrixElement}).
F"ur vier Feldst"arken schreibt sich diese Faktorisierung explizit:
\begin{samepage}
\vspace*{-.5ex}
\begin{align} \label{4F-VEV_2F-VEV-mf}
&\vac{1,2,3,4}\;
  =\; \vac{1,2}\; \vac{3,4}\; +\; \vac{1,3}\; \vac{2,4}\; +\; \vac{1,4}\; \vac{2,3}
    \\[.25ex]
&\text{mit}\qquad
 \begin{aligned}[t]
  &\vac{i_1,\ldots i_n}\;
    \equiv\; \vac{ P\, g^n F^{(i_1)} \cdots F^{(i_n)} }
    \\[-.25ex]
  &F^{(i)}\;
    \equiv\; F_{\mu_i\nu_i a_i}\!(x_i; x_0,{\cal C}_{x_{\!0}\!x_{\!i}})
 \end{aligned}
    \nn
    \\[-4.5ex]\nn
\end{align}
vgl.\@ Gl.~(\ref{B4-Fak_MatrixElement}); insbes.\@ die abweichende Definition von~$F^{(i)}$ als {\it $a_i$-Matrixelement\/} hier.
\end{samepage}

Der Vier-Feldst"arken-Vakuumerwartungswert zerf"allt gem"a"s dieser Relation in die Summe dreier Terme von je dem Produkt zweier Zwei-Feldst"arken-Vakuumerwartungswerten.
Seien o.E.d.A.\@ die Indizes~1,2 bezogen auf den "`\bm{\mfp}"'-, die Indizes~3,4 auf den "`\bm{\mfm}"'-Loop, die Indizes der Raumzeit-Argumente bezogen auf die Koordinatenlinien~\mbox{$\mu \!\in\! \{\mfp,\mfm,1,2\}$}.
Dann korreliert der erste Term in Gl.~(\ref{4F-VEV_2F-VEV-mf}) je zwei Feldst"arken auf der "`\bm{\mfp}"'-Fl"ache~${\cal S}\Dmfp$ und je zwei auf der "`\bm{\mfm}"'-Fl"ache~${\cal S}\Dmfm$; \vspace*{-.25ex}die zwei verbleibenden Terme korrelieren in beiden Faktoren je eine Feldst"arke auf~${\cal S}\Dmfp$ mit einer auf~${\cal S}\Dmfm$.
F"ur~\mbox{$\tTll \!\equiv\! \tTll^{(s,\rb{b})}$} nach Gl.~(\ref{tTll_WW-mf}$'$) folgt:%
\FOOT{
  \label{FN:tTll-Vorfaktor}Der Vorfaktor ergibt sich wie folgt:
Zwei Faktoren~\vspace*{-.25ex}\mbox{$1 \!/\! 2!$} folgen~aus der Entwicklung der Exponentialfunktionen.
Als Normierung der Spuren~\vspace*{-.25ex}\mbox{$\trDrst{R\Dimath} \!=\! 1\!/\!\dimDrst{R\Dimath}\,\tr$}, vgl.\@ Gl.~(\ref{Konnektor_LoopFtr}), folgen die Faktoren~\mbox{$1 \!/\! \dimDrst{R\Dmfp}$},~\mbox{$1 \!/\! \dimDrst{R\Dmfm}$}.
Spurbildung~$\tr$ mit Normierung~\mbox{$\tr T_\Drst{R\Dimath}^aT_\Drst{R\Dimath}^b \!=\! \normDrst{R\Dimath}\,\de^{ab}$} der Generatoren ergibt die Faktoren~$\normDrst{R\Dmfp}$,~$\normDrst{R\Dmfm}$.
Die Eichgruppen-Faktoren aus Gl.~(\ref{K2Komp_Kronecker}$'$) kontrahieren wie~\mbox{$\de_{aa} \!/\! \dimNc \!\cdot\! \de_{bb} \!/\! \dimNc$} bzw.~\mbox{$\de_{ab} \!/\! \dimNc \!\cdot\! \de_{ab} \!/\! \dimNc$}.
}
%
\vspace*{-.25ex}
\begin{align} \label{tTll_WW_ch-mf}
\tTll\;
  =\; -\, 2\iIM\,s\vv
         \frac{\normDrst{R\Dmfp}\, \normDrst{R\Dmfm}}{%
                 (2!)^2\, \dimDrst{R\Dmfp}\, \dimDrst{R\Dmfm}}\cdot
         \vac{W\Dmfp}^{-1}\; \vac{W\Dmfm}^{-1}\;
         \Big[\,
           \ch\idx{\mfp\mskip-2mu\mfp}\, \ch\idx{\mfm\mskip-2mu\mfm}\;
           +\; \frac{2}{\dimNc}\;\, \ch\idx{\mfp\mskip-2mu\mfm}{}^{\zz2}
         \,\Big]
    \\[-4ex]\nn
\end{align}
Die Wegner-Wilson-Loops sind Konsequenz von (Anti)Quark-Konnektoren; ihre Darstellungen sind daher zu identifizieren mit der fundamentalen Darstellung:~\mbox{$R\Dmfp \!\equiv\! R\Dmfm \!\equiv\! \Drst{F}$}; f"ur diese gilt~\mbox{$\normDrst{F} \!\equiv\! 1\!/\!2$},~\mbox{$\dimDrst{F} \!\equiv\! \Nc$} f"ur Normierung und Dimension, so da"s folgt mit~\mbox{$\dimNc \!=\! \Nc^2 \!-\! 1$}:
\vspace*{-.25ex}
\begin{align}
\tTll\;
  =\; -\, 2\iIM\,s\vv
         \frac{1}{(4\Nc)^2}\cdot
         \vac{W\Dmfp}^{-1}\; \vac{W\Dmfm}^{-1}\;
         \Big[\,
           \ch\idx{\mfp\mskip-2mu\mfp}\, \ch\idx{\mfm\mskip-2mu\mfm}\;
           +\; \frac{2}{\Nc^2 \!-\! 1}\;\, \ch\idx{\mfp\mskip-2mu\mfm}{}^{\zz2}
         \,\Big]
    \tag{\ref{tTll_WW_ch-mf}$'$}
    \\[-4ex]\nn
\end{align}
\vspace*{-.25ex}Die Korrelation der paralleltransportierten Feldst"arken%
~\mbox{$F_{\mu\nu a}\!(\tilde{x}\Dimath'; x_0, {\cal C}_{x_{\!0}\!\tilde{x}\Dimath'})$},
 \mbox{$F_{\rh\si a}\!(\tilde{x}\Djmath'; x_0, {\cal C}_{x_{\!0}\!\tilde{x}\Djmath'})$}, die respektive integriert werden "uber die Fl"achen%
~\mbox{${\cal S}\Dimath \!\equiv\! {\cal S}(\tilde{\cal C}\Dimath)$},%
~\mbox{${\cal S}\Djmath \!\equiv\! {\cal S}(\tilde{\cal C}\Djmath)$}, mit~\mbox{$\imath,\jmath \!\in\! \{\mfp,\mfm\}$}, ist dabei subsumiert in den wie folgt definierten Funktionen~$\ch\idx{\imath\jmath}$ mit~\mbox{$\imath,\jmath \!\in\! \{\mfp,\mfm\}$}:%
\FOOT{
  \vspace*{-.375ex}Die Operatoren~$P_{\tilde{\cal C}\Dimath}$,~$P_{\tilde{\cal C}\Djmath}$ f"ur Fl"achenordnung~-- vgl.\@ Gl.~(\ref{Konnektor_LoopFtr-mf})~--  sind redundant f"ur zwei Feldst"arken unter Spurbildung und daher weggelassen im folgenden.
}
%
\vspace*{-.5ex}
\begin{align} \label{ch_FF-mf}
\ch\idx{\imath\jmath}\;
  =\; \Big(-\frac{\iIM\,g}{2}\Big)^{\!2}\,
        \iint_{{\cal S}(\tilde{\cal C}\Dimath)}
         &\dsiI[\mskip-2mu(\tilde{x}\Dimath')]{\tilde\mu\tilde\nu}\;
        \iint_{{\cal S}(\tilde{\cal C}\Djmath)}
          \dsiJ[\mskip-2mu(\tilde{x}\Djmath')]{\tilde\rh\tilde\si}
    \\[-.25ex]
     &\phantom{d\si\Dimath}
      \times \vac{\;
       F_{\tilde\mu\tilde\nu a}\!
         (\tilde{x}\Dimath'; x_0, {\cal C}_{x_{\!0}\!\tilde{x}\Dimath'})\,
       F_{\tilde\rh\tilde\si a}\!
         (\tilde{x}\Djmath'; x_0, {\cal C}_{x_{\!0}\!\tilde{x}\Djmath'})
      \;}
    \nn
    \\[-4.5ex]\nn
\end{align}
auf die ferner zur"uckgef"uhrt werden die Vakuumerwartungswerte der einzelnen Wegner-Wil\-son-Loops~$\vac{W\Dmfp}$,~$\vac{W\Dmfm}$~-- und die insofern die eigentlich relevanten Funktionen sind in Hinblick auf Auswertung und analytische Fortsetzung der $T$-Amplitude.

Als zweifaches Fl"achenintegral der Gr"o"se~\mbox{$\vac{\, g^2%
  F_{\tilde\mu\tilde\nu a}\!(\tilde{x}\Dimath'; x_0, {\cal C}_{x_{\!0}\!\tilde{x}\Dimath'})\,
  F_{\tilde\rh\tilde\si a}\!(\tilde{x}\Djmath'; x_0, {\cal C}_{x_{\!0}\!\tilde{x}\Djmath'}) \,}$} mit Energie-Dimension~$+4$ sind die Funktionen~$\ch\idx{\imath\jmath}$ dimensionslos.
Wir gehen von diesen "uber zu den Funktionen~$\tilde\ch\idx{\imath\jmath}$ durch Definition:
\begin{samepage}
\vspace*{-.5ex}
\begin{align} \label{ch_ch-tilde-mf}
\ch\idx{\imath\jmath}\;
  =:\; -\iIM\vv \vac{g^2 FF}a^4\; \cdot\; \tilde\ch\idx{\imath\jmath}\qqquad
  \imath,\jmath \in \{\mfp,\mfm\}
    \\[-4ex]\nn
\end{align}
Herausziehen des Faktors~$-\iIM$ garantiert, da"s die Funktionen~$\tilde\ch\idx{\imath\jmath}$ positiv reell sind "f"ur {\it maximale Korrelation\/}:~\mbox{$0 \!=\! \tilde{x}'{}^2 \!=\! (\tilde{x}\Dmfp' \!-\! \tilde{x}\Dmfm')^2$}, vgl.\@ unten Gl.~(\ref{ch-tilde_FF-mf}).
Herausziehen des dimensionslosen Faktors~\mbox{$\vac{g^2 FF}a^4$},~-- mit~$\vac{g^2 FF}$ dem Minkowskischen Gluonkondensat und~$a$ der Korrelationsl"ange~--, eliminiert in den Funktionen~$\tilde\ch\idx{\imath\jmath}$ die zwei Zahlenparameter des \DREI{M}{S}{V} im eigentlichen Sinne.%
\FOOT{
  Die Unabh"anggikeit von diesen Parametern ist unmittelbar klar f"ur~$\vac{g^2 FF}$; sie wird offensichtlich f"ur~$a$ nach Skalierung der Vierer-Vektoren~$\tilde{x}\Dimath'$ in Einheiten der Korrelationsl"ange im Sinne von Gl.~(\ref{x=ximath-xjmath}$'$).
}
Es gilt, mit~\mbox{$\imath,\jmath \!\in\! \{\mfp,\mfm\}$}:
\vspace*{-.5ex}
\begin{align}
&\tilde\ch\idx{\imath\jmath}\;
  =\; \iIM\, \Big(-\frac{\iIM}{2}\Big)^{\!2}\,
        \iint_{{\cal S}(\tilde{\cal C}\Dimath)} \dsiI{\tilde\mu\tilde\nu}\;
        \iint_{{\cal S}(\tilde{\cal C}\Djmath)} \dsiJ{\tilde\rh\tilde\si}\vv
        D_{\tilde\mu\tilde\nu\tilde\rh\tilde\si}\!(\tilde{x})
    \label{ch-tilde_FF-mf} \\
&\text{unter Definition\rmfootnote}\hspace*{16pt}\qquad
  \tilde{x}\;
     \equiv\; \tilde{x}\Dimath - \tilde{x}\Djmath
    \label{x=ximath-xjmath} \\[-.125ex]
&\phantom{unter}\text{und Reskalierung}\qquad
  \tilde{x}\Dimath
    \equiv \tilde{x}\Dimath'/a
    \tag{\ref{x=ximath-xjmath}$'$}
    \\[-4.25ex]\nn
\end{align}%
\end{samepage}%
\footnotetext{
  \label{FN:x=ximath-xjmath,Qmfp}Der Zusammenhang klar stellt, ob~-- vgl.\@ Gl.~(\ref{x=ximath-xjmath}) vs.~(\ref{Q-AQ-Pos_rb1})~--~$\rb{x}$ gleich~\mbox{$\rb{x}\Dimath \!-\! \rb{x}\Djmath$}, dem Differenzvektor der Feldst"arken auf~${\cal S}(\tilde{\cal C}\Dimath)$,~${\cal S}(\tilde{\cal C}\Djmath)$, oder gleich~\vspace*{.125ex}\mbox{$\bzet_1\rb{X} \!+\! \rb{b}/2 \!+\! \rbG{\om}$}, der Position des mit~${\cal S}(\tilde{\cal C}\Dmfp)$ assoziierten Quarks.
}
Mit dieser Reskalierung sind L"angen implizit~$\tilde\ch\idx{\imath\jmath}$ gegeben durch ungestrichene Raumzeit-Argu\-mente in "`Einheiten von~$a$"'.
Vgl.\@ die Gln.~(\ref{Gluonkondensat_Minkowski})-(\ref{Dvier_DDxi0}). \\
\indent
Im Sinne der Tensorstrukturen%
  ~$t\oC{}_{\zzzz \tilde\mu\tilde\nu\tilde\rh\tilde\si}$ und%
  ~$t\oNC{}_{\zzzz \tilde\mu\tilde\nu\tilde\rh\tilde\si}$ des Korrelationstensors%
  ~\mbox{$\tilde{D} \!\equiv\! \big(D_{\tilde\mu\tilde\nu\tilde\rh\tilde\si}\big)$} sind unmittelbar definiert durch
\vspace*{-.25ex}
\begin{align} \label{ch_vka,chC,chNC-tilde}
\tilde\ch\idx{\imath\jmath}\;
  =\; \vka\;  \tilde\ch\idx{\imath\jmath}\oC\;
      +\; (1 \!-\! \vka)\; \tilde\ch\idx{\imath\jmath}\oNC\qqquad
  \imath,\jmath \in \{\mfp,\mfm\}
    \\[-4ex]\nn
\end{align}
die konfinierende Funktion%
  ~\vspace*{-.125ex}$\tilde\ch\idx{\imath\jmath}\oC$ und die nicht-konfinierende Funktion~$\tilde\ch\oNC\idx{\imath\jmath}$. \\
\indent
Im folgenden Abschnitt~\ref{Subsect:surfacesSDmf} wird getroffen eine Wahl der Fl"achen~${\cal S}\Dmfp$,~${\cal S}\Dmfm$ und angegeben explizite Parametrisierungen, auf Basis derer in den \vspace*{-.125ex}Abschnitten~\ref{Subsect:chNC},~\ref{Subsect:chC} ausgewertet werden die Funktionen%
  ~\vspace*{-.125ex}$\tilde\ch\idx{\imath\jmath}\oC$,~$\tilde\ch\idx{\imath\jmath}\oNC$.
Wesentliches Resultat ist die Faktorisierung der Funktion%
  ~\vspace*{-.125ex}$\tilde\ch\idx{\imath\jmath}$ in einen rein longitudinal bestimmten Anteil~\mbox{$-\det\mathbb{SL}\, g_{\imath\jmath}$} und ein~-- mit dem Gro"sbuchstaben bezeichnetes~-- Funktional~$\tilde{X}\idx{\imath\jmath}$, das bestimmt ist durch die Geometrie in der \mbox{$x^1\!x^2$-Trans}\-versalebene, die wiederum effektiv abh"angt von den~$\zet_i$, das hei"st von Gr"o"sen longitudinalen Ursprungs:%
\FOOT{
  \label{FN:eff-transv-QFT}Dies ist insofern a~priori klar, als Hochenergiestreuung beschreibbar sein sollte durch eine {\sl effektive Quantenfeldtheorie der Transversalebene\/}; vgl.\@ etwa Verlinde, Verlinde in Ref.~\cite{Verlinde93}.
}
\begin{samepage}
\vspace*{-.25ex}
\begin{align} \label{ch=gXi}
\tilde\ch\idx{\imath\jmath}\;
  =\; -\, \det\mathbb{SL}\vv g_{\imath\jmath}\;\cdot\;
        \tilde{X}\idx{\imath\jmath}\qqquad
  \imath,\jmath \in \{\mfp,\mfm\}
    \\[-4ex]\nn
\end{align}
In Anhang~\ref{APP-Sect:Minkowski} wird hergeleitet~\mbox{\,$g_{\mfp\mfp} \!=\! \vrh^2$},~\mbox{\,$g_{\mfp\mfm} \!=\! \vrh^2 \cosh\ps$} f"ur die unabh"angigen Komponenten des metrischen Tensors und \mbox{\,$\vrh^2 \!=\! (-\, \det\mathbb{SL}\cdot \sinh\ps)^{-1}$} f"ur die Normierung~$\vrh$ der Koordinaten in Richtung der Teilchen-Weltlinien, vgl.\@ die Gln.~(\ref{APP:gtilde-kov}),~(\ref{APP:gtilde-kov}$'$) bzw.~(\ref{APP:detS,detS^-1t}).
Es folgt:
\vspace*{-.25ex}
\begin{alignat}{2} \label{-detSLgmf_s}
&-\det\mathbb{SL}\vv g_{\mfp\mfp}\;&
  &=\; \sinh^{-1}\!\ps
    \\[-.25ex]
&-\det\mathbb{SL}\vv g_{\mfp\mfm}\;&
  &=\; \tanh^{-1}\!\ps\quad
   =\; 1\; +\; \efn{\D-\ps}\, \sinh^{-1}\!\ps
    \tag{\ref{-detSLgmf_s}$'$}
    \\[-4ex]\nn
\end{alignat}
die letzte Identit"at mit~\vspace*{-.25ex}\mbox{$\cosh\ps \!=\! \sinh\ps \!+\! \efn{\D-\ps}$}. \\
\indent
Die Faktoren~\mbox{$-\det\mathbb{SL}\, g_{\imath\jmath}$} h"angen ab vom Quadrat~$s$ der invarianten Schwerpunktenergie "uber~\mbox{$\ps \!=\! \ps\Dmfp \!+\! \ps\Dmfm$}, mit~\mbox{$\be\Dmfp \!=\! \tanh\ps\Dmfp$},~\mbox{$\be\Dmfm \!=\! \tanh\ps\Dmfm$} den Beta-Parametern der aktiven Boosts der Teilchen head-to-head.
Die Normierung~$\vrh$ der Koordinaten~\mbox{$\tilde\mu \!\in\! \{\mfp,\mfm,1,2\}$} ist bestimmt durch die Forderung von {\it L"angentreue\/} der Abbildung, die vermittelt mit den Lichtkegelkoordinaten~\mbox{$\bar\mu \!\in\! \{+,-,1,2\}$}, vgl.\@ Anh.~\ref{APP-Sect:Minkowski} auf Seite~\pageref{APP:vrh}.
Dies hei"st formal~\mbox{$\det\mathbb{S} \!\equiv\! 1$} und impliziert im Limes%
  ~\mbox{$s \!\to\! \infty \Leftrightarrow \ps \!\to\! \infty$} unmittelbar%
  ~\mbox{$g_{\mfp\mfm} \!\to\! g_{+-}$}%
  ~\mbox{[und~\mbox{$g_{\mfp\mfp} \!\to\! g_{++} \!\equiv\! 0$}]; ergo gilt}:
   \mbox{$-\det\mathbb{SL}\, g_{\mfp\mfm} \!\equiv\! -\det\mathbb{L}\, g_{\mfp\mfm} \!\to\! -\det\mathbb{L}\, g_{+-} \!\equiv\! 1$}%
  ~[und~\mbox{$-\det\mathbb{SL} g_{\mfp\mfp} \!\to\! -\det\mathbb{L} g_{++} \!\equiv\! 0$}]~-- {\it unabh"an\-gig\/} von der Normierung~$\al$ der Lichtkegelkoordinaten, von der abh"angt~$g_{+-}$ und~$-\det\mathbb{L}$.%
\FOOT{
  \label{FN:ch-Berger,Nachtmann}In Kapitel~\ref{Kap:VAKUUM} auf Seite~\pageref{T:ProduktKumulanten} ist diskutiert die Entwicklung des Vakuumerwartungswerts~$\vac{\;\cdot\;}$ zweier Weg\-ner-Wilson-Loops in "`Produkt-Kumulanten"'; vgl.\@ Berger, Nachtmann in Ref.~\cite{Berger98}.   Die dort angesprochene Funktion~$\ch$ ist genau~$\ch\idx{+-}$, das hei"st~$\ch\idx{\mfp\mskip-2mu\mfm}$ im Limes~\mbox{$s \!\to\! \infty$}.   Aus Gl.~(\ref{tTll_WW_ch-mf}) resultiert in der Tat~\mbox{$\tTll \!\propto\! \ch\idx{\mfp\mskip-2mu\mfm}^2 \!\to\! \ch\idx{+-}^2$} im Limes~\mbox{$s \!\to\! \infty$}, da Gl.~(\ref{ch=gXi}) impliziert~\mbox{$\ch\idx{\mfp\mskip-2mu\mfp} \!\propto\! g_{\mfp\mfp}$},~\mbox{$\ch\idx{\mfm\mskip-2mu\mfm} \!\propto\! g_{\mfm\mfm}$} und gilt \mbox{$g_{\mfp\mfp} \!\to\! g_{++} \!\equiv\! 0$},~\mbox{$g_{\mfm\mfm} \!\to\! g_{--} \!\equiv\! 0$}.
} \\
%
\indent
Explizit sind die Funktionale%
  ~\vspace*{-.125ex}$\tilde{X}\idx{\imath\jmath}$ Integrale der paralleltransportierten Feldst"arken "uber ${\cal S}\Dimath\Doperp$,~${\cal S}\Djmath\Doperp$, das hei"st "uber die Projektionen in den Transversalraum der initialen Fl"achen~${\cal S}\Dimath$,~${\cal S}\Djmath$.
Mithilfe des Stokes'schen Satzes k"onnen die so verbleibenden zwei Integrale reduziert werden:
F"ur die respektive konfinierende Funktion%
  ~\vspace*{-.125ex}$\tilde\ch\idx{\imath\jmath}\oC$ ein Integral, so da"s korreliert sind die Endpunkte von~${\cal S}\Dimath\Doperp$ mit jedem Punkt von~${\cal S}\Djmath\Doperp$~-- und umgekehrt.
F"ur die respektive nicht-konfinierende Funktion~$\tilde\ch\idx{\imath\jmath}\oNC$ beide Integrale, so da"s korreliert sind die Endpunkte von~${\cal S}\Dimath\Doperp$ mit denen von~${\cal S}\Djmath\Doperp$.
Die Wechselwirkung resultiert als {\it nichtlokal\/} beziehungsweise {\it lokal\/}. \\
\indent
Bez"uglich der funktionalen Abh"anggikeit der Gr"o"sen mit Indizes%
  ~\vspace*{-.125ex}\mbox{\,$\imath,\jmath,\ldots \!\in\! \{\mfp,\mfm\}$} sei festgehalten, da"s diese im Sinne
\vspace*{-.25ex}
\begin{align} 
\bm{\mfp}\,
    \leftrightarrow\,
    \big\{ \zet_1,\, \rb{X},\, +\rb{b}\!/\!2 \big\}\qquad
  \text{und/oder}\qquad
  \bm{\mfm}\,
    \leftrightarrow\,
    \big\{ \zet_2,\, \rb{Y},\, -\rb{b}\!/\!2 \big\}
    \\[-4ex]\nn
\end{align}
abh"angen von den hierdurch \vspace*{-.125ex}definierten "`\bm{\mfp}"'- und "`\bm{\mfm}"'-Variablens"atzen.
Dies folgt aus der Abh"anggikeit der Wegner-Wilson-Loops%
  ~\vspace*{-.375ex}$W\Dmfp$,~$W\Dmfm$, vgl.\@ die Gln.~(\ref{WW-explizit-mf}),~(\ref{WW-explizit-mf}$'$).
\end{samepage}

\subsection[Explizite Raumzeit-Parametrisierung von%
              ~\protect${\cal S}\Dimath \!\equiv\! {\cal S}(\tilde{\cal C}\Dimath)$,%
              ~\protect\mbox{\,$\imath \!=\! \mfp,\mfm$}]{%
            \vspace*{-.25ex}
            Explizite Raumzeit-Parametrisierung von%
              ~\bm{{\cal S}\Dimath \!\equiv\! {\cal S}(\tilde{\cal C}\Dimath)},%
              ~\bm{\,\imath \!=\! \mfp,\mfm}}
\label{Subsect:surfacesSDmf}

\begin{figure}
\begin{minipage}{\linewidth}
  \begin{center}
  \setlength{\unitlength}{.9mm}\begin{picture}(120,85.5)   
    \put(0,0){\epsfxsize108mm \epsffile{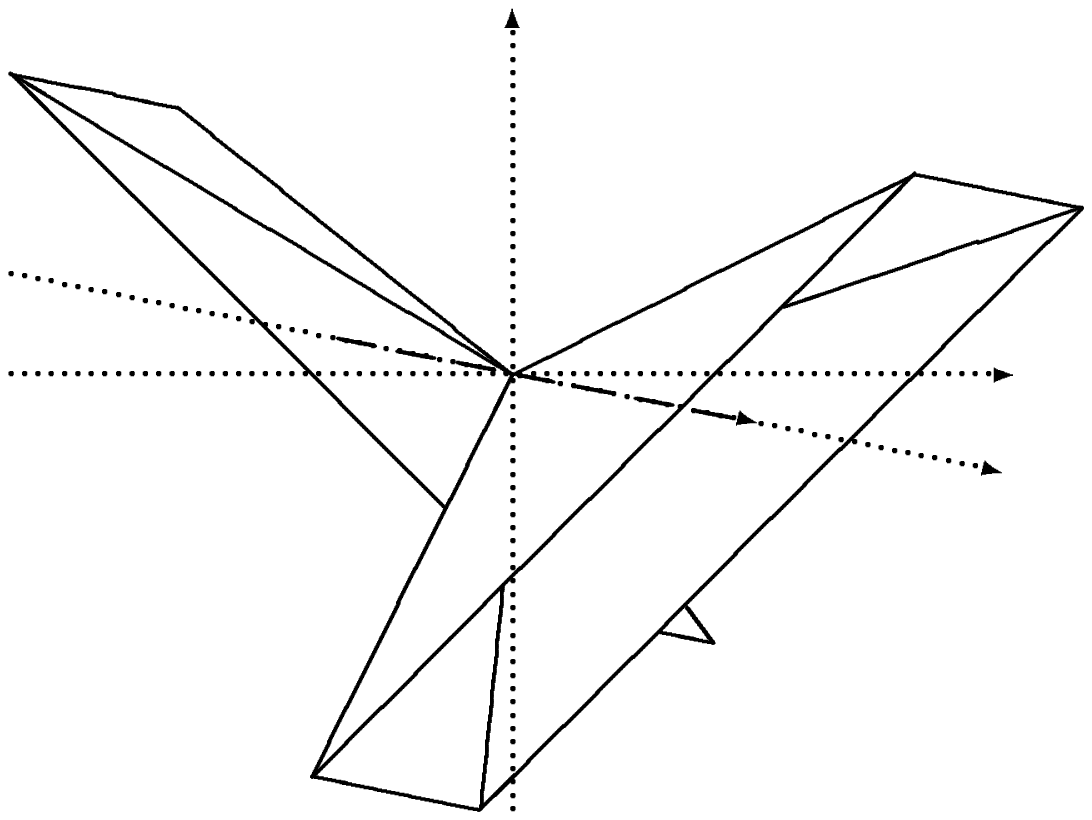}}
    \linethickness{0.5pt}
    \put( 53,82){\normalsize$x^0$}
    \put(112,43){\normalsize$x^3$}
    \put(107,32){\normalsize$\rb{x}$}
    \put( 80,38){\normalsize$\rb{b}$}
    \put( 56,51){\normalsize$x_0 \!\equiv\! \om$}
    \put(59.8,47.2){\circle*{1.2}}
    \put( 85,24){\normalsize$W\Dmfp$}
    \put( 85,64){\normalsize${\cal S}\Dmfp$}
    \put( 63,30){\vector(1,1){10}}
    \put( 91,37){\vector(-1,-1){10}}
    \put(103,70){\normalsize$Q$}
    \put(121,66){\normalsize$\AQ$}
    \put(  5,69){\normalsize$W\Dmfm$}
    \put( 34,69){\normalsize${\cal S}\Dmfm$}
    \put( 48,36){\vector(-1,1){10}}
    \put( 20,78){\normalsize$\AQ$}
    \put(  2,81){\normalsize$Q$}
  \end{picture}
  \end{center}
\vspace*{-3.75ex}
\caption[Raumzeit-Darstellung der Pyramiden~\protect\mbox{$P\Dmfp$},~\protect\mbox{$P\Dmfm$}]{
  Raumzeit-Darstellung der Pyramiden~$P\Dmfp$,~$P\Dmfm$ mit Mantelfl"achen%
~\mbox{${\cal S}\Dmfp \!\equiv\! {\cal S}(\tilde{\cal C}\Dmfp)$},
 \mbox{${\cal S}\Dmfm \!\equiv\! {\cal S}(\tilde{\cal C}\Dmfm)$} "uber den Wegner-Wilson-Loops%
~\mbox{$W\Dmfp \!\equiv\! W({\cal S}\Dmfp)$},%
~\mbox{$W\Dmfm \!\equiv\! W({\cal S}\Dmfm)$}.   Die Integrationen der paralleltransportierten Feldst"arken beziehen sich auf~\mbox{${\cal S}\Dmfp$},~\mbox{${\cal S}\Dmfm$}.   Wir betonen, da"s die transversalen Projektionen der Loops nicht parallel sind.
\vspace*{-.5ex}
}
\label{Fig:pyramids}
\end{minipage}
\end{figure}

F"ur Definiertheit%
\FOOT{
  Unserer Auswertung ist zug"anglich einer allgemeineren Klasse von Fl"achen; vgl.\@ die Anmerkungen dort.
}%
~-- vgl.\@ Abb.~\ref{Fig:pyramids}~-- spezifizieren wir die Integrationsfl"achen~${\cal S}\Dmfp$,~${\cal S}\Dmfm$ als die Mantelfl"achen (planer) Pyramiden~$P\Dmfp$,~$P\Dmfm$, deren Basen gegeben sind als die von den~Kurven ${\cal C}\Dmfp$,~${\cal C}\Dmfm$ der Wegner-Wilson-Loops%
  ~\mbox{$W\Dmfp \!\equiv\! W({\cal C}\Dmfp)$},~\mbox{$W\Dmfm \!\equiv\! W({\cal C}\Dmfm)$} berandeten (planen) Rechteck\-fl"achen und deren Apex%
  ~\mbox{$\om \!\equiv\! \om^{\tilde\mu} \tilde{e}_{(\tilde\mu)}$} gew"ahlt wird identisch dem gemeinsamen Referenzpunkt $x_0$~-- im Sinne h"ochster Symmetrie%
\FOOT{
  \label{FN:h"ochsteSymmetrie}Wir kommen ausf"uhrlich zur"uck auf diese Wahl "`im Sinne h"ochster Symmetrie"'; vgl.\@ Kap.~\ref{Subsect:TransversaleKonfiguration}.
}
als die Mitte des transversalen Impaktvektors~$\rb{b}$.
\begin{figure}
\vspace*{-.75ex}
\begin{minipage}{\linewidth}
  \begin{center}
  \setlength{\unitlength}{.9mm}\begin{picture}(120,71)   
    \put(0,0){\epsfxsize108mm \epsffile{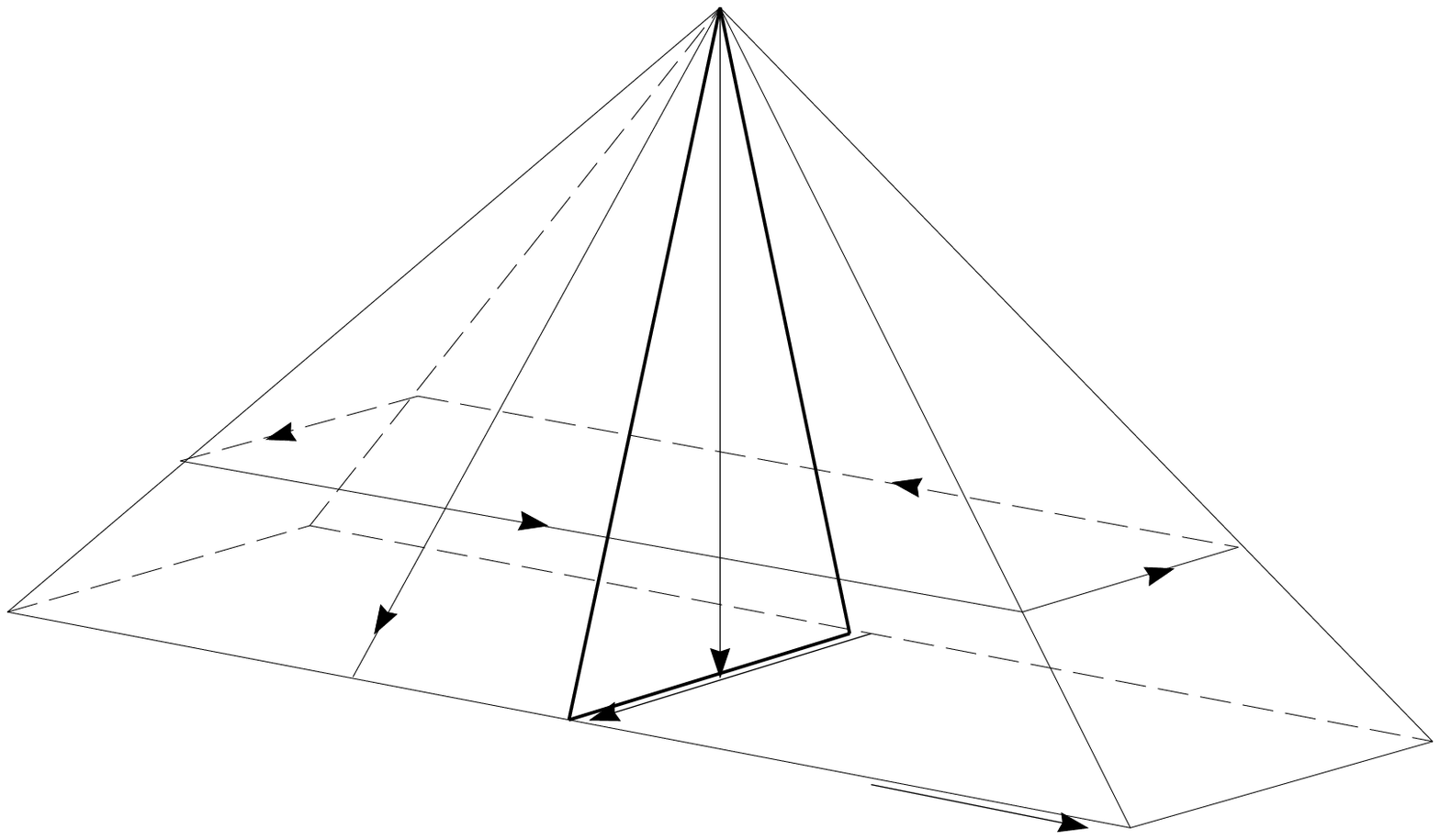}}
    \put(-18,69){\normalsize Parametrisierungen}
    \put(-18,64.5){\normalsize ${\cal S}\Dmfp$,~$V\Dmfp$ und~$L\Dmfp$:}
    \put(63,69){\normalsize$x_0 \!\equiv\! \om$}
    \put(18,10){\normalsize$W\Dmfp$}
    \put(122, 6){\normalsize$\AQ$}
    \put( 97,-2){\normalsize$Q$}
    \put(43.0, 5.0){\normalsize$\rb{X}$}
    \put(63.0,11.5){\normalsize$\zet_1\rb{X}$}
    \put(54.5, 9.0){\normalsize$\bzet_1\rb{X}$}
      \put(47.5,10){\circle*{1}}
      \put(71  ,17.2){\circle*{1}}
    \put(56.5,38){\normalsize$\rb{b}$}
    \put(78,-1){\normalsize$u\Dmfp\, \tilde{e}_{(\mfp)}$}
    \put(26,18){\normalsize$s\Dmfp$}
    \put(94,18){\normalsize$\ta\Dmfp$}
  \end{picture}
  \end{center}
\vspace*{-3ex}
\caption[Raumzeit-Parametrisierung der Pyramide~\protect\mbox{$P\Dmfp$}]{
  Raumzeit-Parametrisierung~-- durch Parameter~$s\Dmfp$,~$\ta\Dmfp$,~$u\Dmfp$~-- der planen Pyramide~$P\Dmfp$ "uber dem Wegner-Wilson-Loop~\mbox{$W\Dmfp \!\equiv\! W({\cal S}\Dmfp)$}.   Es ist~\mbox{${\cal S}\Dmfp \!\equiv\! {\cal S}(\tilde{\cal C}\Dmfp)$} mit Deformation~$\tilde{\cal C}\Dmfp$ von~${\cal C}\Dmfp$ ihre Mantelfl"ache, $V\Dmfp$~ ihr Volumen,~$L\Dmfp$ ihre Grundfl"ache.   Die Integration der pa\-ralleltransportierten Feldst"arken bezieht sich zun"achst auf~${\cal S}\Dmfp$; sie reduziert sich f"ur die konfinierende Funktion~$\tilde\ch\oC$ auf die transversale Projektion~${\cal S}\Dmfp\Doperp$, f"ur die nicht-kon\-finierende~$\tilde\ch\oNC$ auf die Endpunkte von~${\cal S}\Dmfp\Doperp$ [Schenkel des Dreiecks "uber~$\rb{X}$ bzw.\@ Punkte].
}
\label{Fig:pyramid-mfp}
\end{minipage}
\end{figure}
Explizite Parametrisierungen~-- vgl.\@ Abb.~\ref{Fig:pyramid-mfp}~-- geben wir an wie folgt, wobei wir uns beschr"anken auf die "`$\mfp$"'-Seite der Streuung: \vspace*{-.125ex}die Pyramide~$P\Dmfp$; bzgl.~$P\Dmfm$ vgl.\@ die Gln.~(\ref{Subst:mfp->mfm}), (\ref{Subst:mfp->mfm}$'$).
Wir f"uhren ein Parameter~$s\Dmfp$,~$\ta\Dmfp$,~$u\Dmfp$, die laufen in den Intervallen
%
\begin{align} \label{s,ta,u-Lauf}
s\Dmfp:\;
  0 \to 1\qquad
\ta\Dmfp:\;
  0 \to 1\qquad
u\Dmfp:\;
  -T\Dmfp/2 \to T\Dmfp/2
\end{align}
Bzgl.~$T\Dmfp$~-- der "`L"ange"' des Loops~$W\Dmfp$ in $x^\mfp$-Richtung~-- vgl.\@ Gl.~(\ref{APP:Tmfp_zeDmfp^mfp-pre}$'$). \\
\indent
\vspace*{-.125ex}Die Mantelfl"ache~\mbox{${\cal S}\Dmfp \!\equiv\! {\cal S}(\tilde{\cal C}\Dmfp)$} von~$P\Dmfp$~-- mit~$\tilde{\cal C}\Dmfp$ der Deformation von~${\cal C}\Dmfp$, die~${\cal S}\Dmfp$ infinitesimal plakettiert~-- ist dann gegeben als vier (plane) Dreiecke, parametrisiert:
\FOOT{
  Diese repr"asentieren~-- dem Umlaufsinn nach von oben nach unten~-- den Quark-, den Quark-Antiquark-, den Antiquark- und den Antiquark-Quark-Konnektor des Wegner-Wilson-Loops~$W\Dmfp$.
}
%
\vspace*{-.75ex}
\begin{align} \label{Smfp-Parametr}
&{\cal S}\Dmfp\;
    \\[-.75ex]
 &\begin{alignedat}[t]{6}
  =\vvv
  &\Big\{\, \tilde{x}\Dmfp \!\equiv\! (x\Dmfp^{\tilde\mu})\Big|\;
         \tilde{x}\Dmfp(s\Dmfp,u\Dmfp)&
      &=\, \om\, + s\Dmfp \big[&\, u\Dmfp\;
           &\vrh^{-1}\, \tilde{e}_{(\mfp)}&
           &+ \big(&(1 \!-\! \zet_1) &X^i
                     + b^i/2\big)\, \tilde{e}_{(i)}&
           &\,\big]
   \,\Big\}
    \nn \\[-.25ex]
  {\T\bigcup}\;
  &\Big\{\, \tilde{x}\Dmfp \!\equiv\! (x\Dmfp^{\tilde\mu})\Big|\;
         \tilde{x}\Dmfp(s\Dmfp,\ta\Dmfp)&
      &=\, \om\, + s\Dmfp \big[&\,
           &\vrh^{-1}\, \tilde{e}_{(\mfp)}&
           &+ \big(&([1\!-\!\ta\Dmfp] \!-\! \zet_1) &X^i
                     + b^i/2\big)\, \tilde{e}_{(i)}&
           &\,\big]
   \,\Big\}
    \nn \\[-.25ex]
  {\T\bigcup}\;
  &\Big\{\, \tilde{x}\Dmfp \!\equiv\! (x\Dmfp^{\tilde\mu})\Big|\;
         \tilde{x}\Dmfp(s\Dmfp,u\Dmfp)&
      &=\, \om\, + s\Dmfp \big[&\, -u\Dmfp\;
           &\vrh^{-1}\, \tilde{e}_{(\mfp)}&
           &+ \big(&(0 \!-\! \zet_1) &X^i
                     + b^i/2\big)\, \tilde{e}_{(i)}&
           &\,\big]
   \,\Big\}
    \nn \\[-.25ex]
  {\T\bigcup}\;
  &\Big\{\, \tilde{x}\Dmfp \!\equiv\! (x\Dmfp^{\tilde\mu})\Big|\;
         \tilde{x}\Dmfp(s\Dmfp,\ta\Dmfp)&
      &=\, \om\, + s\Dmfp \big[&\,
          -&\vrh^{-1}\, \tilde{e}_{(\mfp)}&
           &+ \big(&(\ta\Dmfp \!-\! \zet_1) &X^i
                      + b^i/2\big)\, \tilde{e}_{(i)}&
           &\,\big]
   \,\Big\}
    \nn
  \end{alignedat}
    \\[-4.25ex]\nn
\end{align}
Ruhe zun"achst das Quark-Antiquark-Paar~"`\bm\mfp"' im System mit karthesischen Koordinatenlinien~\mbox{$\mu \!\in\! \{0,1,2,3\}$}.
Unter dem aktiven Lorentz-Boost mit Geschwindigkeit~$\be\Dmfp$ in \mbox{$+x^3$-Rich}\-tung transformiert die Richtung der (Anti)Quark-Weltlinien~-- des Loops~$W\Dmfp$~-- in der%
  ~\vspace*{-.125ex}\mbox{$x^0\!x^3$-Ebene} wie%
  ~\vspace*{-.125ex}\mbox{$e_{(0)} \!\to\! \ga\Dmfp \big(e_{(0)} + \be\Dmfp e_{(3)}\big)$}.
Durch%
  ~\mbox{$\tilde{x} \!=\! x^{\tilde\mu} \tilde{e}_{(\tilde\mu)}$} mit%
  ~\mbox{$\tilde\mu \!\in\! \{\mfp,\mfm,1,2\}$ sind verkn"upft der Vektor}
   \vspace*{-.125ex}\mbox{$\tilde{e}_{(\mfp)} \!=\! \vrh\, \ga\Dmfp \big(e_{(0)} + \be\Dmfp e_{(3)}\big)$} und die kontravariante Komponente~$x^\mfp$ der Weltlinien, vgl.\@ die Gln.~(\ref{APP:etilde_e}),~(\ref{APP:xtilde-Zerlegung-kontra}).
Ihre Endpunkte~[bzgl.~$\tilde{e}_{(\mfp)}$] sind%
  ~\vspace*{-.25ex}\mbox{$\big(\ze\Dmfp^{\tilde\mu}\big) \!\equiv\! \big(\!\pm\ze\Dmfp^\mfp,0,\rb{z}^{\T t}\big){}^{\T t}$} mit~\mbox{$\rb{z} \!=\! \rb{x},\rbb{x}$ und}
   \mbox{$\ze\Dmfp^\mfp \!=\! g_{\mfp\mfp}^{\vv-1\!/\!2} \!\cdot\! T\Dmfp\!/\!2 \!=\! \vrh^{-1} \!\cdot\! T\Dmfp\!/\!2$}, vgl.\@ die Gln.~(\ref{APP:Tmfp_zeDmfp^mfp}),\,(\ref{APP:g_mfpmfp})~-- \mbox{abh"angig von der Normierung~$\vrh$}, die fixiert ist in Gl.~(\ref{APP:vrh}$'$) durch Forderung von L"angentreue des Zusammenhangs~\vspace*{-.125ex}\mbox{$\tilde\mu \!\leftrightarrow\! \bar\mu$}. \\
\indent
Das Volumen~$V\Dmfp$ ist parametrisiert analog zu~${\cal S}\Dmfp$:
\begin{samepage}
\vspace*{-1.125ex}
\begin{align} \label{Vmfp-Parametr}
&V\Dmfp\;
    \\[-.5ex]
  &=\;
   \Big\{\, \tilde{x}\Dmfp \!\equiv\! (x\Dmfp^{\tilde\mu})\Big|\;
         \tilde{x}\Dmfp(s\Dmfp,\ta\Dmfp,u\Dmfp)
      =\, \om\, + s\Dmfp \big[\, u\Dmfp\;
           \vrh^{-1}\, \tilde{e}_{(\mfp)}
           + \big((\ta\Dmfp \!-\! \zet_1) X^i
              + b^i/2\big)\, \tilde{e}_{(i)}
        \,\big]
   \,\Big\}
    \nn
    \\[-4.5ex]\nn
\end{align}
die Grundfl"ache~$L\Dmfp$:
\vspace*{-1ex}
\begin{align} \label{Lmfp-Parametr}
&L\Dmfp\;
  =\; V\Dmfp\Big|_{s\Dmfp\equiv1}
    \\[-.5ex]
  &=\;
   \Big\{\, \tilde{x}\Dmfp \!\equiv\! (x\Dmfp^{\tilde\mu})\Big|\;
         \tilde{x}\Dmfp(\ta\Dmfp,u\Dmfp)
      =\, \om\, + u\Dmfp\;
           \vrh^{-1}\, \tilde{e}_{(\mfp)}
           + \big((\ta\Dmfp \!-\! \zet_1) X^i
              + b^i/2\big)\, \tilde{e}_{(i)}
   \,\Big\}
    \nn
    \\[-4.5ex]\nn
\end{align}
Resultat der longitudinalen $u\Dmfp$-Integration ist "`transversale Projektion"' von~${\cal S}\Dmfp$,~$V\Dmfp$,~$L\Dmfp$ auf deren eigentlich  relevanten Untermannigfaltigkeiten~${\cal S}\Dmfp\Doperp$,~$V\Dmfp\Doperp$,~$L\Dmfp\Doperp$, charakterisiert durch
\vspace*{-.25ex}
\begin{align} 
\tilde{x}\Dmfp^\mfp\;
  \equiv\; 0
    \\[-4ex]\nn
\end{align}
das hei"st durch identisch verschwindende \vspace*{-.25ex}longitudinalen Komponente. \\
\indent
F"ur die projizierte Mantelfla"che~${\cal S}\Dmfp\Doperp$~-- mit%
  ~\mbox{$\rbG{\om} \!\equiv\! \om^i \tilde{e}_{(i)} \!=\! (0,0,\om^1,\om^2)^{\T t}$}~-- gilt explizit:%
\FOOT{
  Vgl.\@ Gl.~(\ref{Smfp-Parametr}): den ersten und dritten Abschnitt der Parametrisierung von~${\cal S}\Dmfp$.
}
%
\vspace*{-.5ex}
\begin{alignat}{3} \label{Smfp^perp-Parametr}
{\cal S}\Dmfp\Doperp\;
  =\vvv
  &\Big\{\, \tilde{x}\Dmfp \!\equiv\! (x\Dmfp^{\tilde\mu})\Big|\;
         \tilde{x}\Dmfp(s\Dmfp)&
      &=\, \rbG{\om}\, + s\Dmfp \big(& \bzet_1 &X^i
              + b^i/2\big)\, \tilde{e}_{(i)}
   \,\Big\}
    \\[-.25ex]
  {\T\bigcup}\;
  &\Big\{\, \tilde{x}\Dmfp \!\equiv\! (x\Dmfp^{\tilde\mu})\Big|\;
         \tilde{x}\Dmfp(s\Dmfp)&
      &=\, \rbG{\om}\, + s\Dmfp \big(&- \zet_1 &X^i
              + b^i/2\big)\, \tilde{e}_{(i)}
   \,\Big\}
    \nn
    \\[-7.5ex]\nn
\end{alignat}
\end{samepage}

Das projizierte Volumen~$V\Dmfp\Doperp$ ist parametrisiert:
\vspace*{-.25ex}
\begin{align} \label{Vmfp^perp-Parametr}
&V\Dmfp\Doperp\;
  =\;
   \Big\{\, \tilde{x}\Dmfp \!\equiv\! (x\Dmfp^{\tilde\mu})\Big|\;
         \tilde{x}\Dmfp(s\Dmfp,\ta\Dmfp)
      =\, \rbG{\om}\, + s\Dmfp \big((\ta\Dmfp \!-\! \zet_1) X^i
              + b^i/2\big)\, \tilde{e}_{(i)}
   \,\Big\}
    \\[-4ex]\nn
\end{align}
die projizierte Grundfl"ache~$L\Dmfp\Doperp$:
\vspace*{-.25ex}
\begin{align} \label{Lmfp^perp-Parametr}
&L\Dmfp\Doperp\;
  =\; V\Dmfp\Doperp\Big|_{s\Dmfp\equiv1}\;
  =\;
   \Big\{\, \tilde{x}\Dmfp \!\equiv\! (x\Dmfp^{\tilde\mu})\Big|\;
         \tilde{x}\Dmfp(\ta\Dmfp)
      =\, \rbG{\om}\, + \big((\ta\Dmfp \!-\! \zet_1) X^i
              + b^i/2\big)\, \tilde{e}_{(i)}
   \,\Big\}
    \\[-4ex]\nn
\end{align}
Die Parameter~$s\Dmfp$,~$\ta\Dmfp$,~$u\Dmfp$ in den Gln.~(\ref{Smfp-Parametr})-(\ref{Lmfp-Parametr}) und~(\ref{Smfp^perp-Parametr})-(\ref{Lmfp^perp-Parametr}) laufen einheitlich in den Intervallen angegeben in Gl.~(\ref{s,ta,u-Lauf}). \\
\indent
Mit diesen Parametrisierungen sind gegeben explizite Fl"achenelemente, die wir angeben wie folgt.
F"ur das Volumen~$V\Dmfp$ und seine transversale Projektion~$V\Dmfp\Doperp$ gilt~-- f"ur die dualen Elemente:
\begin{samepage}
\vspace*{-.5ex}
\begin{align} \label{ddsi-Vmfp}
\ddsiPAU{\tilde\mu}{V\Dmfp}\;
  &=\; \frac{1}{1!}\; \ep_{\tilde\mu\tilde\nu\tilde\rh\tilde\si}\;
         \frac{\pa x^{\tilde\nu}}{\pa s\Dmfp}\,
         \frac{\pa x^{\tilde\rh}}{\pa \ta\Dmfp}\,
         \frac{\pa x^{\tilde\si}}{\pa u\Dmfp}\;
         ds\Dmfp\, d\ta\Dmfp\, du\Dmfp
    \\
  &=\; 2\, g_{\tilde\mu}{}^\mfm\;
         \frac{\pa x^\mfp}{\pa u\Dmfp}\, du\Dmfp\vv
       \otimes\;
         \ddsiPAU{\mfm\mfp}{V\Dmfp\Doperp}
    \tag{\ref{ddsi-Vmfp}$'$}
    \\[-4.5ex]\nn
\end{align}
mit
\vspace*{-.5ex}
\begin{align} 
\ddsiPAU{\tilde\mu\tilde\nu}{V\Dmfp\Doperp}\;
  =\; \frac{1}{2!}\; \ep_{\tilde\mu\tilde\nu\tilde\rh\tilde\si}\;
         \frac{\pa x^{\tilde\rh}}{\pa s\Dmfp}\,
         \frac{\pa x^{\tilde\si}}{\pa \ta\Dmfp}\;
         ds\Dmfp\, d\ta\Dmfp
    \\[-4.5ex]\nn
\end{align}
und f"ur die nicht-verschwindenden nicht dualen Elemente:
\vspace*{-.5ex}
\begin{align} \label{dsi-Vmfp}
&\dsiPA{\mfp ij}{V\Dmfp}
    \nn \\[.5ex]
  &=\; \frac{\pa\big(x^\mfp, x^i, x^j\big)}{\pa\big(s\Dmfp, \ta\Dmfp, u\Dmfp\big)}\;
         ds\Dmfp\, d\ta\Dmfp\, du\Dmfp
    \\
  &=\; \bigg\{\,
         \frac{\pa x^\mfp}{\pa s\Dmfp}\,
           \frac{\pa\big(x^i, x^j\big)}{\pa\big(\ta\Dmfp, u\Dmfp\big)}\;
     -\; \frac{\pa x^\mfp}{\pa \ta\Dmfp}\,
           \frac{\pa\big(x^i, x^j\big)}{\pa\big(s\Dmfp, u\Dmfp\big)}\;
     +\; \frac{\pa x^\mfp}{\pa u\Dmfp}\,
           \frac{\pa\big(x^i, x^j\big)}{\pa\big(s\Dmfp, \ta\Dmfp\big)} 
       \,\bigg\}\;
         ds\Dmfp\, d\ta\Dmfp\, du\Dmfp
    \tag{\ref{dsi-Vmfp}$'$}
    \\[-4.5ex]\nn
\end{align}
Ergo
\vspace*{-.5ex}
\begin{align} \label{dsi-Vmfp-Faktor}
&\dsiPA{\mfp ij}{V\Dmfp}\;
  =\; \frac{\pa x^\mfp}{\pa u\Dmfp}\, du\Dmfp\vv
       \otimes\;
         \dsiPA{ij}{V\Dmfp\Doperp}
    \\
  &\text{mit}\qquad
  \dsiPA{ij}{V\Dmfp\Doperp}\;
  =\; \frac{\pa\big(x^i, x^j\big)}{\pa\big(s\Dmfp, \ta\Dmfp\big)}\;
        ds\Dmfp\, d\ta\Dmfp
    \tag{\ref{dsi-Vmfp-Faktor}$'$}
    \\[-4.5ex]\nn
\end{align}
wegen
\vspace*{-.5ex}
\begin{align} 
\frac{\pa\big(x^i, x^j\big)}{\pa\big(\ta\Dmfp, u\Dmfp\big)}\;
  \equiv\; 0\qqquad
\frac{\pa x^\mfp}{\pa \ta\Dmfp}\;
  \equiv\; 0
    \\[-4.5ex]\nn
\end{align}
vgl.\@ Gl.~(\ref{Vmfp-Parametr}). \\
\indent
F"ur die Grundfl"ache~$L\Dmfp$ und ihre transversale Projektion~$L\Dmfp\Doperp$ gilt analog f"ur die nichtverschwindenden Fl"achenelemente:
\vspace*{-.5ex}
\begin{align} \label{dsi-Lmfp}
\dsiPA{i\mfp}{L\Dmfp}\;
  &=\; \frac{\pa\big(x^i, x^\mfp\big)}{\pa\big(\ta\Dmfp, u\Dmfp\big)}\;
         d\ta\Dmfp\, du\Dmfp
    \\
  &=\; \bigg\{\,
         \frac{\pa x^i}{\pa \ta\Dmfp}\, \frac{\pa x^\mfp}{\pa u\Dmfp}\;
     -\; \frac{\pa x^\mfp}{\pa \ta\Dmfp}\, \frac{\pa x^i}{\pa u\Dmfp}
       \,\bigg\}\;
         d\ta\Dmfp\, du\Dmfp
    \tag{\ref{dsi-Lmfp}$'$}
    \\[-4.5ex]\nn
\end{align}
Ergo
\end{samepage}
\vspace*{-.5ex}
\begin{align} \label{dsi-Lmfp-Faktor}
&\dsiPA{i\mfp}{L\Dmfp}\;
  =\; \frac{\pa x^\mfp}{\pa u\Dmfp}\, du\Dmfp\vv
       \otimes\;
         \dsiPA{i}{L\Dmfp\Doperp}
    \\
  &\text{mit}\qquad
  \dsiPA{i}{L\Dmfp\Doperp}\;
  =\; \frac{\pa x^i}{\pa \ta\Dmfp}\;
        d\ta\Dmfp
    \tag{\ref{dsi-Lmfp-Faktor}$'$}
    \\[-4.5ex]\nn
\end{align}
wegen
\vspace*{-.5ex}
\begin{align} \label{U'Det-Null-Lmfp}
\frac{\pa x^\mfp}{\pa \ta\Dmfp}\;
  \equiv\; 0\qqquad
\frac{\pa x^i}{\pa u\Dmfp}\;
  \equiv\; 0
    \\[-4.5ex]\nn
\end{align}
vgl.\@ Gl.~(\ref{Lmfp-Parametr}). \\
\indent
F"ur die Auswertung in den folgenden Abschnitten werden diese Darstellungen nicht wirklich explizit ben"otigt, sondern allein die {\it Faktorisierung\/} des $V\Dmfp$- und $L\Dmfp$-Elements bez"uglich longitudinalen und transversalen Komponenten, vgl.\@ Gl.~(\ref{dsi-Vmfp-Faktor}) bzw.~(\ref{dsi-Lmfp-Faktor}).
Um diese zu garantieren, ist f"ur die Mantelfl"ache~${\cal S}\Dmfp$ zu fordern {\it Planarit"at\/} in $\tilde{e}_{(\mfp)}$-Richtung; vgl.\@ Gl.~(\ref{Smfp-Parametr}): den ersten und dritten Abschnitt von~${\cal S}\Dmfp$. \\
\indent
Ferner werden explizit ben"otigt die Fl"achenelemente der Transversalprojektion~${\cal S}\Dmfp\Doperp$ der Mantelfl"ache~${\cal S}\Dmfp$; bez"uglich deren zwei Abschnitte gilt, vgl.\@ Gl.~(\ref{Smfp^perp-Parametr}):
\begin{samepage}
\vspace*{-.5ex}
\begin{alignat}{2}
\dsiPA{i}{{\cal S}\Dmfp\Doperp}\;
  &=\; \frac{\pa x^i}{\pa s\Dmfp}\;
         ds\Dmfp&
  &=\; \big(r\Dmfp^Q\big)^i\, ds\Dmfp
    \label{dsi-Smfp^perp_rQ} \\
&\qquad\text{bzw.}&
  &=\; \big(r\Dmfp^{\AQ}\big)^i\, ds\Dmfp
    \tag{\ref{dsi-Smfp^perp_rQ}$'$}
    \\[-4.5ex]\nn
\end{alignat}
mit
\vspace*{-.5ex}
\begin{alignat}{4} \label{r_Q,AQ-mfp}
&r\Dmfp^Q\;&
  &\equiv\;&  &\bzet_1 X + b/2&\quad
  &=\; x - \om
    \\
&r\Dmfp^{\AQ}\;&
  &\equiv\;& - &\zet_1 X + b/2&\quad
  &=\; \bar{x} - \om
    \tag{\ref{r_Q,AQ-mfp}$'$}
    \\[-4.5ex]\nn
\end{alignat}
als systematische Notation f"ur die Differenzvektoren der Positionen von Quark~$Q$ und Anti\-quark~$\AQ$ der Pyramide~$P\Dmfp$ und deren Apex~$\om$. \\
\indent
Explizite Parametrisierungen f"ur~${\cal S}\Dmfm$,~$V\Dmfm$,~$L\Dmfm$ und~${\cal S}\Dmfm\Doperp$,~$V\Dmfm\Doperp$,~$L\Dmfm\Doperp$ bez"uglich der Pyramide~$P\Dmfm$ folgen aus den Parametrisierungen bez"uglich~$P\Dmfp$ durch Substitution
\vspace*{-.5ex}
\begin{align} \label{Subst:mfp->mfm}
&\tilde{e}_{(\mfp)}\;
  \longrightarrow\; \tilde{e}_{(\mfm)}
    \\[.25ex]
&\text{und}\qquad
  \zet_1\;
    \longrightarrow\; \zet_2\,,\qquad
  X\;
    \longrightarrow\; Y\,,\qquad
  b\;
    \longrightarrow\; -b
    \tag{\ref{Subst:mfp->mfm}$'$}
    \\[-4.5ex]\nn
\end{align}
\vspace*{-.125ex}Es ist~\mbox{$\tilde{e}_{(\mfm)} \!=\! \vrh\, \ga\Dmfm \big(e_{(0)} - \be\Dmfm e_{(3)}\big)$}, vgl.\@ Gl.~(\ref{APP:etilde_e}$'$),
die Richtung der Weltlinie des mit~$\be\Dmfm$ in~\mbox{$-x^3$-Rich}\-tung aktiv geboosteten Teilchens und ihre Endpunkte~\mbox{$(\ze\Dmfm^{\tilde\mu}) \!\equiv\! \big(0,\pm\ze\Dmfm^\mfm,\rb{z}^{\T t}\big){}^{\T t}$} mit \mbox{$\rb{z} \!=\! \rb{y},\rbb{y}$} und~\mbox{$\ze\Dmfm^\mfm \!=\! 1\!/\!\surd g_{\mfm\mfm}\cdot T\Dmfm/2 \!=\! \vrh^{-1}\cdot T\Dmfm/2$}, vgl.\@ die Gln.~(\ref{APP:Tmfp_zeDmfp^mfp}$'$),~(\ref{APP:g_mfpmfp}). \\
\indent
Die korrespondierenden Fl"achenelemente folgen entsprechend.
So gilt f"ur das Element des Volumens~$V\Dmfm$ die Faktorisierung
\vspace*{-.5ex}
\begin{align} \label{dsi-Vmfm-Faktor}
&\dsiMA{\mfm ij}{V\Dmfm}\;
   =\; \frac{\pa x^\mfm}{\pa u\Dmfm}\, du\Dmfm\vv
       \otimes\;
         \dsiMA{ij}{V\Dmfm\Doperp}
    \\[-.375ex]
  &\text{mit}\qquad
  \dsiMA{ij}{V\Dmfm\Doperp}\;
  =\; \frac{\pa\big(x^i, x^j\big)}{\pa\big(s\Dmfm, \ta\Dmfp\big)}\;
        ds\Dmfm\, d\ta\Dmfm
    \tag{\ref{dsi-Vmfm-Faktor}$'$}
    \\[-4.5ex]\nn
\end{align}
und f"ur das Element der Grundfl"ache~$L\Dmfm$
\vspace*{-.5ex}
\begin{align} \label{dsi-Lmfm-Faktor}
&\dsiMA{i\mfm}{L\Dmfm}\;
  =\; \frac{\pa x^\mfm}{\pa u\Dmfm}\, du\Dmfm\vv
       \otimes\;
         \dsiMA{i}{L\Dmfm\Doperp}
    \\[-.375ex]
  &\text{mit}\qquad
  \dsiMA{i}{L\Dmfm\Doperp}\;
  =\; \frac{\pa x^i}{\pa \ta\Dmfm}\;
        d\ta\Dmfm
    \tag{\ref{dsi-Lmfm-Faktor}$'$}
    \\[-4.5ex]\nn
\end{align}
angenommen Planarit"at der Mantelfl"ache~${\cal S}\Dmfm$ der Pyramide~$P\Dmfm$ in $\tilde{e}_{(\mfm)}$-Richtung; vgl.\@ die Gln.~(\ref{dsi-Vmfp-Faktor}),~(\ref{dsi-Vmfp-Faktor}$'$) bzw.~(\ref{dsi-Lmfp-Faktor}),~(\ref{dsi-Lmfp-Faktor}$'$). \\
\indent
F"ur die transversale Projektion~${\cal S}\Dmfm\Doperp$ von~${\cal S}\Dmfm$ sind die Fl"achenelemente abschnittweise gegeben durch
\end{samepage}
\vspace*{-.5ex}
\begin{alignat}{2} \label{dsi-Smfm^perp_rQ}
\dsiMA{i}{{\cal S}\Dmfm\Doperp}\;
  &=\; \frac{\pa x^i}{\pa s\Dmfm}\;
         ds\Dmfm&
  &=\; \big(r\Dmfm^Q\big)^i\, ds\Dmfm
    \\
&\qquad\text{bzw.}\qquad&
  &=\; \big(r\Dmfm^{\AQ}\big)^i\, ds\Dmfm
    \tag{\ref{dsi-Smfm^perp_rQ}$'$}
    \\[-4.5ex]\nn
\end{alignat}
mit
\vspace*{-.5ex}
\begin{alignat}{4} \label{r_Q,AQ-mfm}
&r\Dmfm^Q\;&
  &\equiv\;&  &\bzet_2 Y - b/2&\quad
  &=\; y - \om
    \\[.25ex]
&r\Dmfm^{\AQ}\;&
  &\equiv\;& - &\zet_2 Y - b/2&\quad
  &=\; \bar{y} - \om
    \tag{\ref{r_Q,AQ-mfm}$'$}
    \\[-4.5ex]\nn
\end{alignat}
als systematische Notation f"ur die Differenzvektoren der Positionen von Quark und Antiquark der Pyramide~$P\Dmfm$ und deren Apex.
Vgl.\@ die Gln.~(\ref{dsi-Smfp^perp_rQ}), (\ref{dsi-Smfp^perp_rQ}$'$) und~(\ref{r_Q,AQ-mfp}),~(\ref{r_Q,AQ-mfp}$'$).

Es ist hiermit der technische Apparat gegeben, die $\tilde\ch$-Funktionen nach Gl.~(\ref{ch-tilde_FF-mf}) auszuwerten.
Eingesetzt der Korrelationstensor~\mbox{$\tilde{D} \!\equiv\! \big(D_{\tilde\mu\tilde\nu\tilde\rh\tilde\si}\big)$} in Darstellung nach Gl.~(\ref{Dvier_DDxi}), mit Lorentz-Indizes~\mbox{$\tilde\mu,\tilde\nu,\tilde\rh,\tilde\si \!\in\! \{\mfp,\mfm,1,2\}$} und~\mbox{$n \!\equiv\! 4$}, folgen im Sinne von Gl.~(\ref{ch_vka,chC,chNC-tilde}) die Funktionen~$\tilde\ch\oNC\idx{\imath\jmath}$ und~$\tilde\ch\oC\idx{\imath\jmath}$~-- mit~\mbox{$\imath,\jmath \!\in\! \{\mfp,\mfm\}$}.
Die nicht-konfinierenden Funktionen~$\tilde\ch\oNC\idx{\imath\jmath}$ werden ausgewertet im folgenden Abschnitt~\ref{Subsect:chNC}, die konfinierenden Funktionen~$\tilde\ch\oC\idx{\imath\jmath}$ in~\ref{Subsect:chC}.

\subsection[Nicht-konfinierende Funktionen~\protect$\tilde\ch\idx{\imath\jmath}\oNC$,%
              ~\protect\mbox{\,$\imath,\jmath \!\in\! \{\mfp,\mfm\}$}]{%
            Nicht-konfinierende Funktionen~\bm{\tilde\ch\idx{\imath\jmath}\oNC},%
              ~\bm{\,\imath,\jmath \!\in\! \{\mfp,\mfm\}}}
\label{Subsect:chNC}

Wir betrachten die nicht-konfinierenden $\tilde\ch$-Funktionen, zun"achst~\mbox{$\tilde\ch\oNC \!\equiv\! \tilde\ch\idx{\mfp\mskip-2mu\mfm}\oNC$}.
Es gilt~-- vgl.\@ die Gln.~(\ref{ch-tilde_FF-mf}),~(\ref{ch_vka,chC,chNC-tilde}) und~(\ref{Dvier_DDxi})~-- mit~\mbox{$\tilde\mu,\tilde\nu,\tilde\rh,\tilde\si \!\in\! \{\mfp,\mfm,1,2\}$}:
%
\begin{align} \label{chNC-0}
\tilde\ch\oNC\;
  =\; \iIM\, \Big(-\frac{\iIM}{2}\Big)^{\!2}\,
        \iint_{S\Dmfp} \dsiP{\tilde\mu\tilde\nu}\;
        \iint_{S\Dmfm} \dsiM{\tilde\rh\tilde\si}\vv
        t\oNC{}_{\zzzz \tilde\mu\tilde\nu\tilde\rh\tilde\si}\;
        F\oNC(\tilde{x}^2)
\end{align}
dabei ist definiert, vgl.\@ Gl.~(\ref{x=ximath-xjmath}):
%
\begin{align} \label{x=xmfp-xmfm}
\tilde{x}\;
  =\; \tilde{x}\Dmfp
        - \tilde{x}\Dmfm\qquad
  \tilde{x}\Dmfp \in S\Dmfp,\vv
  \tilde{x}\Dmfm \in S\Dmfm
\end{align}
mit~\mbox{$\tilde{x} \!\equiv\! (x^{\tilde\mu})$} und~\mbox{$\tilde{x}\Dmfp \!\equiv\! (x\Dmfp^{\tilde\mu})$},~\mbox{$\tilde{x}\Dmfm \!\equiv\! (x\Dmfm^{\tilde\mu})$}.
Die nicht-konfinierenden Tensorstruktur~\mbox{$t\oNC{}_{\zzzz \tilde\mu\tilde\nu\tilde\rh\tilde\si}$} ist explizit angegeben in Gl.~(\ref{tC,tNC}$'$); wir rekapitulieren:
\begin{samepage}
\vspace*{-.5ex}
\begin{align} \label{tNC}
t\oNC{}_{\zzzz \tilde\mu\tilde\nu\tilde\rh\tilde\si}\;
  =\; \frac{1}{6}\vv g_{\tilde\al\tilde\ga}\vv
        \de^{\tilde\al\tilde\be}_{\tilde\mu\tilde\nu}\, \pa_{\tilde\be}\vv
        \de^{\tilde\ga\tilde\de}_{\tilde\rh\tilde\si}\, \pa_{\tilde\de}
    \\[-4.5ex]\nn
\end{align}
sie kontrahiert wie
\vspace*{-.5ex}
\begin{align} \label{tNC-kontrahiert}
t\oNC{}_{\zzzz \tilde\mu\tilde\nu}{}^{\tilde\mu\tilde\nu}\;
  =\; \tilde\pa^2
  \equiv \pa^{\tilde\mu}\pa_{\tilde\mu}
    \\[-4.5ex]\nn
\end{align}
Die nicht-konfinierende Korrelationsfunktion~$F\oNC$ ist definiert durch
\vspace*{-.5ex}
\begin{align} \label{F^NC}
F\oNC(\tilde{x}^2)\;
   =\; \frac{1}{2}\, \la^2\cdot {\cal K}_2(\ze)
    \\[-4.5ex]\nn
\end{align}
unter Definition
\vspace*{-.25ex}
\begin{align} \label{calK_mu}
&{\cal K}_\mu(\ze)\;
  =\; \Big(\frac{1}{2}\Big)^{\!\mu-1}\!
             \frac{1}{\Ga(\mu)}\;
             \ze^{\mu}\, {\rm K}_{\mu}\!(\ze)\qqquad
  \ze \equiv \sqrt{-\tilde{x}^2/\la^2 + \iIM\, \ep}
       \cong \sqrt{-\tilde{x}^2}\big/\la
    \\[-.25ex]
  &\text{mit}\qquad
  {\cal K}_\mu(\ze) \to 1,\vv
    \forall{\rm Re}\mu > 0\qquad
  \text{f"ur}\quad
  \ze \to 0
    \tag{\ref{calK_mu}$'$}
    \\[-4ex]\nn
\end{align}
vgl.\@ Gl.~(\ref{calKmu-Def})~-- und~\mbox{${\rm K}_\mu$} der modifizierten Besselfunktion zweiter Art nach Ref.~\cite{Abramowitz84}.
Diese Relationen basieren auf Gl.~(\ref{Dvier_DDxi}) mit Index~\mbox{$\nu \!\equiv\! 4$}, ergo:~\mbox{$\la \!\equiv\! \la_4 \!=\! 8\!\big/\!3\pi$}, vgl.\@ Gl.~(\ref{A,la-n=4explizit}$'$)~-- der physikalisch suggerierten Darstellung.
F"ur Vollst"andigkeit sei angegeben mit
\vspace*{-.5ex}
\begin{align} \label{F,D-NC}
F\oNC(\tilde{x}^2)\;
   =\; \frac{1}{8}\, \int_{-\infty}^{\tilde{x}^2} du\; D\uNC(u)
    \\[-4.5ex]\nn
\end{align}
der Zusammenhang mit der $D$-Korrelationsfunktion nach Gl.~(\ref{Dvier_DDxi0}). \\
\indent
Wir fassen die Mantelfl"ache der Pyramide~$P\Dimath$, mit~\mbox{$\imath \!=\! \mfp,\mfm$}, in orientierter Weise auf als vollst"andigen Rand ihres Volumens~$\pa V\Dimath$ subtrahiert ihre Grundfl"ache~$L\Dimath$:
\vspace*{-.25ex}
\begin{align} \label{S=paV-L}
S\Dimath\;
  =\; \pa V\Dimath - L\Dimath
    \\[-4.5ex]\nn
\end{align}
ergo f"ur die Fl"achenintegrale:
\end{samepage}%
\vspace*{-.5ex}
\begin{align} \label{S=paV-L-Int}
\iint_{S\Dimath}\;
  \dsiI{\tilde\mu\tilde\nu}\vv
  T_{\tilde\mu\tilde\nu}
 &=\; \bigg[\iint_{\pa V\Dimath}\;
          - \iint_{L\Dimath}
      \bigg]\vv
      \dsiI{\tilde\mu\tilde\nu}\vv
      T_{\tilde\mu\tilde\nu}
    \\[-4.5ex]\nn
\end{align}
und mithilfe des Stokes'schen Satzes:
%
\begin{align} \label{paV<->V-Stokes}
\iint_{\pa V\Dimath}\;
  \dsiI{\tilde\mu\tilde\nu}\vv
  T_{\tilde\mu\tilde\nu}
  =\; \iiint_{V\Dimath}\;
        \dsiI{\tilde\mu\tilde\nu\tilde\rh}\;
        \del{\tilde{x}\Dimath}_{\tilde\mu}\vv
        T_{\tilde\mu\tilde\nu}
    \\[-4ex]\nn
\end{align}
Dabei ist~\mbox{\,$\tilde{T}\!\equiv\! \big(T_{\tilde\mu\tilde\nu}(\tilde{x}\Dimath)\big)$} ein beliebiger Fourier-integrabler~$(0,2)$-Lorentz-Tensor und~$\dsiI{\tilde\mu\tilde\nu}$ zu identifizieren mit~$d\tilde{V}^{\tilde\mu\tilde\nu}$ in den Gln.~(\ref{APP:dVtilde_dV-Konv}),~(\ref{APP:Stokes_dual-Konv}).
F"ur~$\tilde\ch\oNC$ folgt, vgl.\@ Gl.~(\ref{chNC-0}):
%
\begin{align} \label{chNC-1}
\hspace*{-10pt}
\tilde\ch\oNC\;
  =\; \iIM\, \Big(-\frac{\iIM}{2}\Big)^{\!2}\;
        &\bigg[\iiint_{V\Dmfp}
                 \dsiP{\tilde\mu\tilde\nu\tilde\nu'}\;
                 \pa_{\tilde\nu'}\;
          -\; \iint_{L\Dmfp}
                 \dsiP{\tilde\mu\tilde\nu}
         \bigg]
    \\[-.25ex]
      \times\; &\bigg[\iiint_{V\Dmfm}
                 \dsiM{\tilde\rh\tilde\si\tilde\si'}\;
                 (-\pa_{\tilde\si'})\;
          -\; \iint_{L\Dmfm}
                 \dsiM{\tilde\rh\tilde\si}
         \bigg]\vv
        t\oNC{}_{\zzzz \tilde\mu\tilde\nu\tilde\rh\tilde\si}\;
        F\oNC(\tilde{x}^2)
    \nn
\end{align}
mithilfe der Identit"at
\vspace*{-.25ex}
\begin{align} \label{pa=del(x)}
\pa_{\tilde\mu}\;
  \equiv\; \del{\tilde{x}}_{\tilde\mu}\;
  =\; \del{\tilde{x}\Dmfp}_{\tilde\mu}\;
  =\; -\, \del{\tilde{x}\Dmfm}_{\tilde\mu}
    \\[-4ex]\nn
\end{align}
die unmittelbar folgt aus Gl.~(\ref{x=xmfp-xmfm}). \\
\indent
Unter Kontraktion mit dem vollst"andig antisymmetrischen Lorentz-Tensor \mbox{\,$\tilde{T} \!\equiv\! \big(T^{\tilde\mu_1\cdots\tilde\mu_s}\big)$} faktorisiert das verallgemeinerte Kronecker-Symbol:%
\FOOT{
  Dies ist unmittelbar klar, gelesen von rechts nach links; vgl.\@ Gl.~(\ref{APP:ProjectAntisymm}).
}
%
\begin{align} \label{Kronecker-Faktor}
\de^{\tilde\mu_1\tilde\mu_2\cdots\tilde\mu_s}_{\tilde\nu_1\tilde\nu_2\cdots\tilde\nu_s}\vv
  T^{\tilde\nu_1\tilde\nu_2\cdots\tilde\nu_s}\;
  =\; s!\vv T^{\tilde\mu_1\tilde\mu_2\cdots\tilde\mu_s}\;
  =\; s!\vv \de^{\tilde\mu_1}_{\tilde\nu_1}
            \de^{\tilde\mu_2}_{\tilde\nu_2}\cdots
            \de^{\tilde\mu_s}_{\tilde\nu_s}\vv
       T^{\tilde\nu_1\tilde\nu_2\cdots\tilde\nu_s}
\end{align}
Unter Kontraktion mit den Fl"achenelementen gilt daher die "Aquivalenz:
\vspace*{-.5ex}
\begin{align} \label{tNC-antisymm}
t\oNC{}_{\zzzz \tilde\mu\tilde\nu\tilde\rh\tilde\si}\;
  &=\; \frac{1}{6}\vv g_{\tilde\al\tilde\ga}\vv
        \de^{\tilde\al\tilde\be}_{\tilde\mu\tilde\nu}\, \pa_{\tilde\be}\vv
        \de^{\tilde\ga\tilde\de}_{\tilde\rh\tilde\si}\, \pa_{\tilde\de}
    \\
  &\cong\; \frac{1}{6}\vv g_{\tilde\al\tilde\ga}\vv
        2\, \de^{\tilde\al}_{\tilde\mu}\de^{\tilde\be}_{\tilde\nu}\, \pa_{\tilde\be}\vv
        2\, \de^{\tilde\ga}_{\tilde\rh}\de^{\tilde\de}_{\tilde\si}\, \pa_{\tilde\de}\;
   =\; \frac{2}{3}\vv g_{\tilde\mu\tilde\nu}\;
        \pa_{\tilde\rh}\pa_{\tilde\si}
    \nn
    \\[-4.5ex]\nn
\end{align}
vgl.\@ Gl.~(\ref{tNC}).
Aufgrund von~\vspace*{-.25ex}\mbox{\,$\dsiP{\tilde\mu\tilde\nu\tilde\nu'} \pa_{\tilde\nu'} \pa_{\tilde\nu} \!\equiv\! 0$} und~\mbox{\,$\dsiM{\tilde\rh\tilde\si\tilde\si'} \pa_{\tilde\si'} \pa_{\tilde\si} \!\equiv\! 0$} verschwindet identisch das $V\Dmfp$- respektive das $V\Dmfm$-Integral. \\
\indent
Die Grundfl"ache~$L\Dmfp$ besitzt~-- unabh"angig von der Parametrisierung~-- eine longitudinale Komponente~\bm{\mfp} und eine transversale~$i$.
Das Fl"achenelement~\mbox{$\dsiP{\tilde\mu\tilde\nu}$} ist daher ungleich Null nur f"ur~\mbox{\,$\tilde\mu \!\equiv\! i$},~\mbox{$\tilde\nu \!\equiv\! \mfp$} und umgekehrt; es folgen zwei identische Terme.
Entsprechend das $L\Dmfm$-Integral.
Der Faktor Vier k"urzt den Nenner des Vorfaktors.
Es folgt, mit~\mbox{$i,j \!\in\! \{1,2\}$}:
%
\begin{alignat}{3} \label{chNC-2}
\tilde\ch\oNC\;
  =\; -\, \iIM\;
        \iint_{L\Dmfp}
         &\dsiP{i\mfp}\;
        \iint_{L\Dmfm}
          \dsiM{j\mfm}\vv
        t\oNC{}_{\zzzz i\mfp j\mfm}\vv
        F\oNC(\tilde{x}^2)
\end{alignat}
dabei ist mit Gl.~(\ref{tNC}):
\vspace*{-.25ex}
\begin{align} \label{tNC_imfpjmfm}
t\oNC{}_{\zzzz i\mfp j\mfm}\;
  =\; \frac{1}{6}\vv g_{\tilde\al\tilde\ga}\vv
        \de^{\tilde\al\tilde\be}_{i\,\mfp}\, \pa_{\tilde\be}\vv
        \de^{\tilde\ga\tilde\de}_{j\mfm}\, \pa_{\tilde\de}\;
  =\; \frac{1}{6}\;
        \big(g_{ij}\, \pa_\mfp\, \pa_\mfm\;
         +\; g_{\mfp\mfm}\, \pa_i\, \pa_j\big)
    \\[-4ex]\nn
\end{align}
aufgrund von~\mbox{\,$g_{\mfp j} \!\equiv\! 0$},~\mbox{$g_{i\mfm} \!\equiv\! 0$}, das hei"st aufgrund der Faktorisierung des metrischen Tensors%
  ~\mbox{\,$\tilde{g} \!\equiv\! \big(g_{\tilde\mu\tilde\nu}\big)$} bez"ug\-lich longitudinalen und transversalen Komponenten. \\
\indent
F"ur die Mantelfl"ache~${\cal S}\Dmfp$ ist gefordert {\it Planarit"at\/} in $\tilde{e}_{(\mfp)}$-, f"ur~${\cal S}\Dmfm$ in $\tilde{e}_{(\mfm)}$-Richtung.
Dies induziert {\it Faktorisierung\/} der Fl"achenelemente bez"uglich longitudinaler und transversalen Komponenten und impliziert f"ur die {\it Grundfl"achen\/}~$L\Dmfp$,~$L\Dmfm$~-- vgl.\@ die Gln.~(\ref{dsi-Lmfp-Faktor}),~(\ref{dsi-Lmfm-Faktor}):
\vspace*{-.5ex}
\begin{align} \label{Lmfp-Faktor}
&\dsiP{i\mfp}\;
  =\; \dsiP{i}\;
        \otimes\; \dsiP{\mfp}
    \\
&\text{d.h.}\qquad
  \iint_{L\Dmfp} \dsiP{i\mfp}\;
  =\; \int_{L\Dmfp\Doperp} dx\Dmfp^{i}\;
        \otimes\; \int dx\Dmfp^{\mfp}
    \tag{\ref{Lmfp-Faktor}$'$}
    \\[-5ex]\nn
\end{align}
und
\vspace*{-1.5ex}
\begin{align} \label{Lmfm-Faktor}
&\dsiM{j\mfm}\;
  =\; \dsiM{j}\;
        \otimes\; \dsiM{\mfm}
    \\
&\text{d.h.}\qquad
  \iint_{L\Dmfm} \dsiM{j\mfm}\;
  =\; \int_{L\Dmfm\Doperp} dx\Dmfm^{j}\;
        \otimes\; \int dx\Dmfm^{\mfm}
    \tag{\ref{Lmfm-Faktor}$'$}
    \\[-5.5ex]\nn
\end{align}
mit Notation
\vspace*{-.5ex}
\begin{align} \label{dx^mu=dsi^mu(x)}
dx\Dmfp^{\tilde\mu}\;
  \equiv\; \dsiP{\tilde\mu}\qquad
dx\Dmfm^{\tilde\mu}\;
  \equiv\; \dsiM{\tilde\mu}\qquad
  \tilde\mu \in \{\mfp,\mfm,1,2\}
    \\[-4.5ex]\nn
\end{align}
Die Integration der longitudinalen Komponente bezieht sich dabei auf die volle L"ange des respektiven Wegner-Wilson-Loops~-- das hei"st der diesem zugrunde liegenden Parton-Trajektorien~${\cal C}\Dmfp$,~${\cal C}\Dmfm$, deren Endpunkte~\mbox{$\big(\pm\ze\Dmfp^\mfp,0,\rb{z}^{\T t}\big){}^{\T t}$},~\mbox{$\big(0,\pm\ze\Dmfm^\mfm,\rb{z}^{\T t}\big){}^{\T t}$} bestimmt sind durch die Eigenzeiten~$T\Dmfp$,~$T\Dmfm$, "uber die diese durch das Vakuum-Eichfeld~$A$ propagieren und h"angen ab von den Richtungen~\bm{\mfp},~\bm{\mfm}.
Vgl.\@ die Bemn.\@ zu Gl.~(\ref{Smfp-Parametr}) und~(\ref{Subst:mfp->mfm}) wie \mbox{Anh.~\ref{APP-Sect:Minkowski}, S.\,\pageref{APP-T:Tmfp,Tmfm}}. \\
\indent
Auf Basis dieser Faktorisierung gilt f"ur~$\tilde\ch\oNC$:
\vspace*{-.25ex}
\begin{align} \label{chNC_IL[F]}
\tilde\ch\oNC\;
  =\; -\, \iIM\vv
        \int_{L\Dmfp\Doperp} &dx\Dmfp^{i}\;
        \int_{L\Dmfm\Doperp}  dx\Dmfm^{j}\vv
        t\oNC{}_{\zzzz i\mfp j\mfm}\vv
        I_L[F\oNC]
    \\[-5.5ex]\nn
\end{align}
unter Definition
\vspace*{-.5ex}
\begin{align} \label{IL[F]-Def}
I_L[F]\;
  &=\; \int dx\Dmfp^{\mfp}\;
       \int dx\Dmfm^{\mfm}\vv
         F(\tilde{x}^2)
    \\
  &=\; \int_{-T\Dmfp\!/\!2}^{T\Dmfp\!/\!2}\; \frac{\pa x\Dmfp^\mfp}{\pa u\Dmfp}\, du\Dmfp\;
       \int_{-T\Dmfm\!/\!2}^{T\Dmfm\!/\!2}\; \frac{\pa x\Dmfm^\mfm}{\pa u\Dmfm}\, du\Dmfm\;
         F(\tilde{x}^2)
    \tag{\ref{IL[F]-Def}$'$}
    \\[-4.5ex]\nn
\end{align}
Wir betrachten das Funktional~$I_L$ f"ur eine allgemeine Funktion~$F$ der Raumzeit mit  wohldefinierter Fourier-Transformierten~$\tilde{F}$; {\it in~praxi\/} ist zu identifizieren
\vspace*{-.25ex}
\begin{align} \label{F=F^NC}
F\;
  \equiv\; F\oNC
    \\[-4.25ex]\nn
\end{align}
vgl.\@ Gl.~(\ref{chNC-2}) und die Gln.~(\ref{chNC_IL[F]}),~(\ref{IL[F]-Def}),~(\ref{IL[F]-Def}$'$). \\
\indent
In Anhang~\ref{APP-Subsect:umfp,umfm-Int} wird~$I_L$ geeignet vereinfacht f"ur allgemeine Funktion~$F$:
In Gl.~(\ref{IL[F]-Def}$'$) werden explizit ausintegriert die longitudinalen Parameter~$u\Dmfp$,~$u\Dmfm$; es folgt:
\begin{samepage}
\vspace*{-.5ex}
\begin{align} \label{IL[F]}
&I_L[F]\;
  =\; -\, \det\mathbb{SL}\;\cdot\; \projt{F}{\rb{x}}
    \\[.5ex]
&\text{mit}\qquad
  \projt{F}{\rb{x}}\;
    =\; \iint_{-\infty}^\infty \frac{d^2\rb{k}}{(2\pi)^2}\vv
          \efn{\D\iIM\, \rb{k} \!\cdot\! \rb{x}}\vv
          \tilde{F}(-\rb{k}^2)
    \tag{\ref{IL[F]}$'$}
    \\[-5.5ex]\nn
\end{align}
vgl.\@ Gl.~(\ref{APP:longInt}). \\
\indent
Nach Gl.~(\ref{chNC_IL[F]}) f"ur~\mbox{$\tilde\ch\oNC$} ist auf~$I_L[F]$~-- das hei"st auf die Funktion~\mbox{$\projNAt{F}$}~-- zu beziehen die Tensorstruktur~\mbox{$t\oNC{}_{\zzzz i\mfp j\mfm}$}, vgl.\@ Gl.~(\ref{tNC_imfpjmfm}).
Da~\mbox{$\projNAt{F}$} abh"angt nur vom Betrag des transversalen Vektors~\mbox{\,$\rb{x} \!\equiv\! (x^1,x^2)^{\T t}$}, ergibt identisch Null der Term von~\mbox{$t\oNC{}_{\zzzz i\mfp j\mfm}$} mit den longitudinalen Ableitungen:~\mbox{$g_{ij}\pa_\mfp\pa_\mfm$}.
Es folgt aus dem Term~\mbox{$g_{\mfp\mfm}\pa_i\pa_j$}, mit~\mbox{$i,j \!\in\! \{1,2\}$}:
\vspace*{-.25ex}
\begin{align} \label{chNC-4}
\tilde\ch\oNC\;
  &=\; -\, \det\mathbb{SL}\; g_{\mfp\mfm}\;\cdot\;
        \iIM\; \frac{1}{6}\vv
        \int_{L\Dmfp\Doperp} dx\Dmfp^{i}\; \del{\tilde{x}\Dmfp}_i\;
        \int_{L\Dmfm\Doperp} dx\Dmfm^{j}\; \del{\tilde{x}\Dmfm}_j\vv
        \projt{F\oNC}{\rb{x}}
    \\
  &=\; -\, \det\mathbb{SL}\; g_{\mfp\mfm}\;\cdot\;
        \iIM\; \frac{1}{6}\vv
        \int_{{\cal S}\Dmfp\Doperp} dx\Dmfp^{i}\; \del{\tilde{x}\Dmfp}_i\;
        \int_{{\cal S}\Dmfm\Doperp} dx\Dmfm^{j}\; \del{\tilde{x}\Dmfm}_j\vv
        \projt{F\oNC}{\rb{x}}
    \tag{\ref{chNC-4}$'$}
    \\[-5.5ex]\nn
\end{align}
\vspace*{-.125ex}dabei ist ein Signum absorbiert im "Ubergang~\mbox{$\pa_i \!\to\! \del{\tilde{x}\Dmfp}_i$},~\mbox{$\pa_j \!\to\! -\del{\tilde{x}\Dmfm}_j$}, vgl.\@ Gl.~(\ref{pa=del(x)}). \\
\indent
Der Integrand ist vollst"andiges Differential der Koordinatenlinien, derentlang integriert wird "uber die respektiven transversalen Projektionen.
Mithilfe des Stokes'schen Satzes reduzieren sich die Integrationen unmittelbar auf Differenzen bez"uglich der Endpunkte der Linien~$L\Dmfp\Doperp$,~$L\Dmfm\Doperp$ oder alternativ~${\cal S}\Dmfp\Doperp$,~${\cal S}\Dmfm\Doperp$~-- vgl.\@ Gl.~(\ref{chNC-4}) bzw.~(\ref{chNC-4}$'$).
\end{samepage}

Bezugnehmend auf die explizite Wahl von~${\cal S}\Dmfp$,~${\cal S}\Dmfm$ und~$L\Dmfp$,~$L\Dmfm$ als die Mantel- beziehungsweise Grundfl"achen planer Pyramiden~$P\Dmfp$,~$P\Dmfm$~-- vgl.\@ die Gln.~(\ref{Smfp-Parametr}),~(\ref{Lmfp-Parametr}) und Gl.~(\ref{Subst:mfp->mfm}$'$)~-- sind die transversalen Projektionen~$L\Dmfp\Doperp$,~$L\Dmfm\Doperp$ und~${\cal S}\Dmfp\Doperp$,~${\cal S}\Dmfm\Doperp$ {\it geradlinig\/}%
\FOOT{
  Es ist {\sl nicht\/} notwendig zu fordern {\sl transversale Planarit"at\/} der Fl"achen~${\cal S}\Dmfp$,~${\cal S}\Dmfm$ und~$L\Dmfp$,~$L\Dmfm$, das hei"st~{\sl Geradlinigkeit\/} der transversalen Projektionen~${\cal S}\Dmfp\Doperp$,~${\cal S}\Dmfm\Doperp$ und~$L\Dmfp\Doperp$,~$L\Dmfm\Doperp$.
Dies geschieht hier einzig f"ur Definiertheit.
}.
Wir werten~$\tilde\ch\oNC$ aus o.E.d.A.\@ auf Basis von Gl.~(\ref{chNC-4})~-- als Integral "uber~$L\Dmfp\Doperp$,~$L\Dmfm\Doperp$, orientiert~$\tilde{x}\Dmfp(\ta\Dmfp)$,~$\tilde{x}\Dmfm(\ta\Dmfm)$ f"ur~\mbox{$\ta\Dmfp, \ta\Dmfm\!: 0 \!\to\! 1$} vom respektiven Antiquark~$\AQ$ zum Quark~$Q$:
\vspace*{-.25ex}
\begin{alignat}{7} \label{APP:L_mf^perp}
&\hspace*{-0pt}
 L\Dmfp\Doperp(\ta\Dmfp\!:\, 0 \!\to\! 1):&\quad\vv
  &\AQ\!:&\vv
    &x\Dmfp(0)&\,
    &=\, \om \!+\! r\Dmfp^\AQ&\quad\vvv
  \longrightarrow\quad\vv
  &Q\!:&\vv
    &\tilde{x}\Dmfp(1)&\,
    &=\, \om \!+\! r\Dmfp^Q\zzzz\zzzz
    \\[.5ex]
&\hspace*{-0pt}
 L\Dmfm\Doperp(\ta\Dmfm\!:\, 0 \!\to\! 1):&\quad\vv
  &\AQ\!:&\vv
    &x\Dmfm(0)&\,
    &=\, \om \!+\! r\Dmfm^\AQ&\quad\vvv
  \longrightarrow\quad\vv
  &Q\!:&\vv
    &\tilde{x}\Dmfm(1)&\,
    &=\, \om \!+\! r\Dmfm^Q\zzzz\zzzz
    \tag{\ref{APP:L_mf^perp}$'$}
    \\[-4ex]\nn
\end{alignat}
mit deren Positionen in Termen der systematisch notierten Differenzvektoren~$r\Dmfp^Q$,~$r\Dmfp^{\AQ}$ und $r\Dmfm^Q$,~$r\Dmfm^{\AQ}$ der (Anti)Quarks und des gemeinsamen Apex~$\om$ der~-- o.E.d.A.\@ generalisierten~-- Pyramiden, vgl.\@ die Gln.~(\ref{r_Q,AQ-mfp}),~(\ref{r_Q,AQ-mfp}$'$) bzw.~(\ref{r_Q,AQ-mfm}),~(\ref{r_Q,AQ-mfm}$'$). \\
\indent
Im Sinne der expliziten Parametrisierungen~\mbox{$\tilde{x}\Dmfp(\ta\Dmfp)$},~\mbox{$\tilde{x}\Dmfm(\ta\Dmfm)$} gilt
%
\begin{alignat}{3} \label{dsi-pa}
&dx\Dimath^{i}\; \del{\tilde{x}\Dimath}_i\;&
  &=\; d\ta\Dimath\vv \frac{d x\Dimath^i}{d\ta\Dimath}
       \frac{\pa}{\pa x\Dimath^i}&
  &=\; d\ta\Dimath\; \frac{d}{d\ta\Dimath}\qqquad
    \imath = \mfp,\mfm
\end{alignat}
und f"ur die Funktion~$\tilde\ch\oNC$ nach Gl.~(\ref{chNC-4}):
\vspace*{-.5ex}
\begin{align} \label{chNC-5}
&\tilde\ch\oNC\;
   =\; -\, \det\mathbb{SL}\; g_{\mfp\mfm}\;\cdot\;
         \iIM\; \frac{1}{6}\vv
         \int_0^1 d\ta\Dmfp\, \frac{d}{d\ta\Dmfp}\vv
         \int_0^1 d\ta\Dmfm\, \frac{d}{d\ta\Dmfm}\vv
         \projt{F\oNC}{\rb{x}}
    \\[-4.5ex]\nn
\end{align}
Abgek"urzt~\mbox{\,$f(\rb{x}) \!\equiv\! \projt{F\oNC}{\rb{x}}$}, folgt f"ur den Integralausausdruck:
\vspace*{-.5ex}
\begin{align} \label{chNC-5_f}
&\int_0^1 d\ta\Dmfp\, \frac{d}{d\ta\Dmfp}\vv
   \int_0^1 d\ta\Dmfm\, \frac{d}{d\ta\Dmfm}\vv
     f\big( \rb{x}\Dmfp(\ta\Dmfp) \!-\! \rb{x}\Dmfm(\ta\Dmfm) \big)
    \\
&\phantom{\tilde\ch\oNC\;}
  =\; \bigg[\;
        \bigg[\vv
          f\big( \rb{x}\Dmfp(\ta\Dmfp) \!-\! \rb{x}\Dmfm(\ta\Dmfm) \big)
        \vv\bigg]\Big.^{\T\ta\Dmfm \!\equiv\! 1}_{\T\ta\Dmfm \!\equiv\! 0}
      \vv\bigg]\Big.^{\T\ta\Dmfp \!\equiv\! 1}_{\T\ta\Dmfp \!\equiv\! 0}
    \tag{\ref{chNC-5_f}$'$} \\[.5ex]
&\phantom{\tilde\ch\oNC\;}
  =\; \sum\big._{\T \ta\Dmfp,\ta\Dmfm \!\equiv\! 0,1}\vv
        (-1)^{\T \ta\Dmfp \!+\! \ta\Dmfm}\vv
        f\big( \rb{x}\Dmfp(\ta\Dmfp) \!-\! \rb{x}\Dmfm(\ta\Dmfm) \big)
    \tag{\ref{chNC-5_f}$''$}
    \\[-4.5ex]\nn
\end{align}
F"ur~\mbox{$\tilde\ch\oNC$} nach Gl.~(\ref{chNC-5}) folgt
\vspace*{-.5ex}
\begin{align} \label{chNC_IJ}
\tilde\ch\oNC\;
  =\; -\, \det\mathbb{SL}\; g_{\mfp\mfm}\;\cdot\;
        \iIM\; \frac{1}{6}\vv
        \sum\Big._{\!I,J\equiv Q,\AQ}\vv
          {\rm sign}_{I,J}\vv
          \projtbig{F\oNC}{\rb{r}\Dmfp^I \!-\! \rb{r}\Dmfm^J}
    \\[-4.5ex]\nn
\end{align}
in Termen von~$r\Dmfp^Q$,~$r\Dmfp^{\AQ}$ und~$r\Dmfm^Q$,~$r\Dmfm^{\AQ}$;
dabei ist f"ur Suggestivit"at der Schreibweise summiert "uber die (Anti)Quarks:~\mbox{$I,J \!\in\! \{Q,\AQ\}$}, und definiert
\vspace*{-.25ex}
\begin{align} \label{signIJ}
{\rm sign}_{I,J} \!=\! (-1)^{I+J}\qquad
  I,J \in \{Q,\AQ\},\quad
  Q \equiv 0,\;
  \AQ \equiv 1
    \\[-4ex]\nn
\end{align}
das hei"st durch Zuweisen der Zahlenwerte~\mbox{$Q \!\equiv\! 0$},~\mbox{$\AQ \!\equiv\! 1$}; s.u.\@ Fu"sn.\,\FN{APP-FN:AQ-negativ} auf Seite~\pageref{APP-FN:AQ-negativ}. \\
\indent
Die nicht-konfinierende Funktion~$\tilde\ch\oNC$ ist nach Gl.~(\ref{chNC_IJ})~-- da~$\projt{F\oNC}{\rb{x}}$ allgemein abh"angt nur von~$|\rb{x}|$~-- bestimmt durch die Betr"age der transversal projizierten Abstandvektoren der paralleltransportierten Feldst"arken an den (Anti)Quark-Positionen.
Deren Wechselwirkung ist in diesem Sinne {\it lokal\/}.
Die konfinierende Funktion~$\tilde\ch\oC$ basiert dazu in Kontrast auf einer {\it nicht-lokalen\/} Wechselwirkung der paralleltransportierten Feldst"arken im Sinne, da"s diese eingehen bez"uglich "`s"amtlicher Weltpunkte der Raumzeit zwischen den (Anti)Quarks"'.
Die Herleitung von~$\tilde\ch\oC$ im folgenden Abschnitt pr"azisiert diese Aussage. \\
\indent
Die Funktionen%
  ~\vspace*{-.25ex}\mbox{$\tilde\ch\oNC\idx{\mfp\mskip-2mu\mfp}$},~\mbox{$\tilde\ch\oNC\idx{\mfm\mskip-2mu\mfm}$} differieren von%
  ~\mbox{$\tilde\ch\oNC \!\equiv\! \tilde\ch\oNC\idx{\mfp\mskip-2mu\mfm} (\equiv\! \tilde\ch\oNC\idx{\mfm\mfp})$} genau darin, da"s beide paralleltransportierten Feldst"arken Elemente sind {\it derselben\/} Pyramiden-Mantelfl"ache~-- \mbox{entweder~${\cal S}\Dmfp$} oder~${\cal S}\Dmfm$.
So gilt nach Gl.~(\ref{chNC_IJ}) f"ur%
  ~\vspace*{-.375ex}\mbox{\,$\tilde\ch\idx{\mfp\mskip-2mu\mfp}\oNC$}~-- mit%
  ~\mbox{\,$\rb{r}\Dmfp^Q \!-\! \rb{r}\Dmfp^\AQ \!\equiv\! \rb{X}$} der (orientierten) transversalen Ausdehnung des Wegner-Wilson-Loops~$W\Dmfp$:
\vspace*{-.625ex}
\begin{align} \label{chNC_IJ-mfpmfp}
\tilde\ch\oNC\idx{\mfp\mskip-2mu\mfp}\;
  =\; -\, \det\mathbb{SL}\; g_{\mfp\mfp}\;\cdot\;
        \iIM\; \frac{1}{3}\,
        \Big\{\;
          \projt{F\oNC}{\bm{0}}\;
      -\; \projt{F\oNC}{\rb{X}}
        \;\Big\}
    \\[-4.625ex]\nn
\end{align}
und f"ur%
  ~\vspace*{-.125ex}\mbox{\,$\tilde\ch\idx{\mfm\mskip-2mu\mfm}\oNC$} analog:%
  ~\mbox{\,$g_{\mfp\mfp} \!\to\! g_{\mfm\mfm}$} und~\mbox{\,$r\Dmfp \!\to\! r\Dmfm$}, ergo~\mbox{\,$X \!\to\! Y$}. \\
\indent
Wir verweisen auf Anhang~\ref{APP:CLTFN}.
Dort leiten wir her~-- vgl.\@ Gl.~(\ref{APP:F^C,NC-Minkowski}$'$), mit~\mbox{\,$\nu \!\equiv\! 4$}:
\FOOT{
  In den Gr"o"sen in (geradem) Fettdruck sind gestrichen die ausintegrierten longitudinalen Komponenten.
}
%
\vspace*{-.5ex}
\begin{align} \label{projt[F]NC-Ansatz}
&\projt{F\oNC}{\rb{x}}\;
  =\; -\, \iIM\, 4\pi\cdot
         \la^4\vv {\cal K}_3(\bm\ze)
    \\[-.5ex]
  &\hspace*{140pt}
    \la \equiv \la_4 = 8\!\big/\!3\pi,\quad
    \bm\ze \equiv {\T\sqrt{|\rb{x}|^2 \!/\! \la^2 + \iIM\, \ep}}
            \cong  |\rb{x}| \!\big/\! \la
    \nn
    \\[-4.25ex]\nn
\end{align}
auf Basis unseres Ansatzes%
  ~\vspace*{-.125ex}\mbox{$F\oNC\!(\tilde{x}^2) =\! \frac{1}{2}\la^2\, {\cal K}_2(\ze)$},%
  ~\mbox{\,$\ze \equiv\! \sqrt{-\tilde{x}^2 \!/\! \la^2 \!+\! \iIM\, \ep} \cong \sqrt{-\tilde{x}^2}\big/\la$} f"ur die nicht-konfinierende \mbox{$F$-Korrelations}\-funktion.
Vgl.\@ Gl.~(\ref{F^NC})ff.\@, ferner Kap.~\ref{Abschn:ANN-KONST}, Gl.~(\ref{Dvier_DDxi})ff.\@ und Anh.~\ref{APP:CLTFN}, die Gln.~(\ref{APP:DDk-Ansatz}),~(\ref{APP:FT_F,D-NC}).
\vspace*{-.5ex}

\subsection[Konfinierende Funktionen~\protect$\tilde\ch\idx{\imath\jmath}\oC$,%
              ~\protect\mbox{\,$\imath,\jmath \!\in\! \{\mfp,\mfm\}$}]{%
            Konfinierende Funktionen~\bm{\tilde\ch\idx{\imath\jmath}\oC},%
              ~\bm{\,\imath,\jmath \!\in\! \{\mfp,\mfm\}}}
\label{Subsect:chC}

Wir betrachten die konfinierenden $\ch$-Funktionen, zun"achst~\mbox{$\tilde\ch\oC \!\equiv\! \tilde\ch\idx{\mfp\mskip-2mu\mfm}\oC$}.
Analog zu Gl.~(\ref{chNC-0}) gilt~-- vgl.\@ wieder die Gln.~(\ref{ch-tilde_FF-mf}),~(\ref{ch_vka,chC,chNC-tilde}) und~(\ref{Dvier_DDxi})~-- mit~\mbox{$\tilde\mu,\tilde\nu,\tilde\rh,\tilde\si \!\in\! \{\mfp,\mfm,1,2\}$}:
\begin{samepage}
\vspace*{-.5ex}
\begin{align} \label{chC-0}
\tilde\ch\oC\;
  =\; \iIM\, \Big(-\frac{\iIM}{2}\Big)^{\!2}\,
        \iint_{S\Dmfp} \dsiP{\tilde\mu\tilde\nu}\;
        \iint_{S\Dmfm} \dsiM{\tilde\rh\tilde\si}\vv
        t\oC{}_{\zzzz \tilde\mu\tilde\nu\tilde\rh\tilde\si}\;
        \tilde\pa^2\, F\oC(\tilde{x}^2)
    \\[-4.5ex]\nn
\end{align}
dabei ist~\mbox{$\tilde{x} \!=\! \tilde{x}\Dmfp \!-\! \tilde{x}\Dmfm$} mit~\mbox{\,$\tilde{x}\Dmfp \!\in\! S\Dmfp$},~\mbox{$\tilde{x}\Dmfm \!\in\! S\Dmfm$} nach Gl.~(\ref{x=xmfp-xmfm}).
Die konfinierende Tensorstruktur~\mbox{$t\oC{}_{\zzzz \tilde\mu\tilde\nu\tilde\rh\tilde\si}$} ist explizit gegeben in Gl.~(\ref{tC,tNC}); wir rekapitulieren:
\vspace*{-.75ex}
\begin{align} \label{tC}
t\oC{}_{\zzzz \tilde\mu\tilde\nu\tilde\rh\tilde\si}\;
  =\; \frac{1}{12}\vv
        \de^{\tilde\al\tilde\be}_{\tilde\mu\tilde\nu}\vv
        g_{\tilde\al\tilde\rh}\, g_{\tilde\be\tilde\si}
    \\[-4.75ex]\nn
\end{align}
sie kontrahiert wie
\vspace*{-.5ex}
\begin{align} \label{tC-kontrahiert}
t\oC{}_{\zzzz \tilde\mu\tilde\nu}{}^{\tilde\mu\tilde\nu} = 1
    \\[-4.5ex]\nn
\end{align}
Die konfinierende Korrelationsfunktion~$F\oC$ ist definiert durch
\vspace*{-.75ex}
\begin{align} \label{F^C}
F\oC(\tilde{x}^2)\;
   =\; \frac{1}{2}\, \la^2\cdot {\cal K}_2(\ze)
    \\[-4.75ex]\nn
\end{align}
mit~\vspace*{-.125ex}\mbox{\,${\cal K}_\mu(\ze)
     = (\frac{1}{2})^{\mu-1} \frac{1}{\Ga(\mu)}\, \ze^\mu {\rm K}_\mu(\ze)$},%
  ~\mbox{\,$\ze \!\equiv\! \sqrt{-\tilde{x}^2/\la^2 + \iIM\, \ep} \!\cong\! \sqrt{-\tilde{x}^2}\big/\la$} und%
  ~\mbox{\,${\cal K}_\mu(\ze) \!\to\! 1,\, \forall{\rm Re}\mu \!>\! 0$} f"ur~\mbox{$\ze \!\to\! 0$}; vgl.\@ die Gln.~(\ref{calKmu-Def}) und~(\ref{calK_mu}), (\ref{calK_mu}$'$).
Diese Relationen basieren auf Gl.~(\ref{Dvier_DDxi}) mit Index~\mbox{$\nu \!\equiv\! 4$}, ergo~\mbox{$\la \!\equiv\! \la_4 \!=\! 8\!\big/\!3\pi$}.
F"ur Vollst"andigkeit sei angegeben mit
\vspace*{-.25ex}
\begin{align} \label{F,D-C}
\tilde\pa^2\, F\oC(\tilde{x}^2)\;
  =\; D\uC(\tilde{x}^2)
    \\[-4ex]\nn
\end{align}
der Zusammenhang mit der $D$-Korrelationsfunktion nach Gl.~(\ref{Dvier_DDxi0}). \\
\indent
Die Gln.~(\ref{F^NC}),~(\ref{F^C}) implizieren
\vspace*{-.5ex}
\begin{align} \label{F^C<->F^NC}
F\oC\;
  \equiv\; F\oNC
    \\[-4.5ex]\nn
\end{align}
Die Existenz einer Relation zwischen nicht-konfinierender und konfinierender Korrelationsfunktion folgt aus dem Ansatz nach Gl.~(\ref{Dvier_DDkkontrAnsatz}).
Sie wird am sinnvollsten ausgedr"uckt durch die eine Korrelationsfunktion~${\cal K}_2$, die Diskrepanz beider Terme vollst"andig subsumiert in den Lorentz-Tensorstrukturen:~\mbox{\,$t\oNC{}_{\zzzz \tilde\mu\tilde\nu\tilde\rh\tilde\si}$} versus~\mbox{\,$t\oC{}_{\zzzz \tilde\mu\tilde\nu\tilde\rh\tilde\si} \tilde\pa^2$}; vgl.\@ auch Gl.~(\ref{tNC-kontrahiert}) versus~(\ref{tC-kontrahiert}). \\
\indent
Im Sinne~\mbox{$S\Dimath \!=\! \pa V\Dimath \!-\! L\Dimath$} f"ur~\mbox{$\imath \!=\! \mfp,\mfm$}, vgl.\@ Gl.~(\ref{S=paV-L}), seien die Mantelfl"achen der~\mbox{Pyramiden} wieder aufgefa"st als vollst"andiger Rand ihres Volumens, subtrahiert ihre Grundfl"ache:
\vspace*{-.25ex}
\begin{align} \label{chC-1}
\tilde\ch\oC\;
  =\; \iIM\, \Big(-\frac{\iIM}{2}\Big)^{\!2}\;
        &\bigg[\iiint_{V\Dmfp}
                 \dsiP{\tilde\mu\tilde\nu\tilde\nu'}\;
                 \pa_{\tilde\nu'}\;
          -\; \iint_{L\Dmfp}
                 \dsiP{\tilde\mu\tilde\nu}
         \bigg]
    \\[-.25ex]
      \times\; &\bigg[\iiint_{V\Dmfm}
                 \dsiM{\tilde\rh\tilde\si\tilde\si'}\;
                 (-\pa_{\tilde\si'})\;
          -\; \iint_{L\Dmfm}
                 \dsiM{\tilde\rh\tilde\si}
         \bigg]\vv
        t\oC{}_{\zzzz \tilde\mu\tilde\nu\tilde\rh\tilde\si}\;
        \tilde\pa^2\, F\oC(\tilde{x}^2)
    \nn
    \\[-4.75ex]\nn
\end{align}
\end{samepage}%
das konfinierende Pendant zu Gl.~(\ref{chNC-1}).

Nach Gl.~(\ref{Kronecker-Faktor}) sind "aquivalent unter der Kontraktion mit den Fl"achenelementen aufgrund deren vollst"andigen Antisymmetrie:
\begin{samepage}
\vspace*{-.25ex}
\begin{align} \label{tC-antisymm}
t\oC{}_{\zzzz \tilde\mu\tilde\nu\tilde\rh\tilde\si}\;
  &=\; \frac{1}{12}\vv
        \de^{\tilde\al\tilde\be}_{\tilde\mu\tilde\nu}\vv
        g_{\tilde\al\tilde\rh}\, g_{\tilde\be\tilde\si}
    \\
  &\cong\; \frac{1}{12}\vv
        2\, \de^{\tilde\al}_{\tilde\mu}\de^{\tilde\be}_{\tilde\nu}\vv
        g_{\tilde\al\tilde\rh}\, g_{\tilde\be\tilde\si}\;
   =\; \frac{1}{6}\;
        g_{\tilde\mu\tilde\rh}\, g_{\tilde\nu\tilde\si}
    \nn
    \\[-4.25ex]\nn
\end{align}
vgl.\@ Gl.~(\ref{tC}). \\
\indent
Grundfl"ache~$L\Dmfp$ wie Volumen~$V\Dmfp$ besitzen~-- unabh"angig von der Parametrisierung~-- eine longitudinale Komponente:~\bm{\mfp}, sonst transversale Komponente(n):~$i$ bzw.~\mbox{$i,i'$}.
Aus den Fl"achenelementen~\mbox{$\dsiP[]{\tilde\mu\tilde\nu\tilde\nu'}$} und~\mbox{$\dsiP[]{\tilde\mu\tilde\nu}$} folgen jeweils zwei identische Terme f"ur~\mbox{\,$\tilde\mu \!\equiv\! i$},~\mbox{$\tilde\nu \!\equiv\! \mfp$} und umgekehrt.
Zus"atzlich tritt auf im $V\Dmfp$-Integral f"ur~\mbox{\,$\tilde\mu \!\equiv\! i$},~\mbox{$\tilde\nu \!\equiv\! i'$},~\mbox{$\tilde\nu' \!\equiv\! \mfp$} der Term~\mbox{$\dsiP[]{ii'\mfp} \pa_\mfp$} mit longitudinaler Ableitung~$\pa_\mfp$.
Entsprechend das $V\Dmfm$-Integral.
Der Faktor Vier k"urzt~-- wie im nicht-konfinierenden Fall~-- den Nenner des Vorfaktors.
Es folgt, mit~\mbox{$i,i',j,j' \!\in\! \{1,2\}$}:
\vspace*{-.5ex}
\begin{align} \label{chC-2}
\tilde\ch\oC\;
  =\; -\, \iIM\;
        &\bigg[\iiint_{V\Dmfp}
                 \dsiP{i\mfp i'}\;
                 \pa_{i'}\;
          -\; \iint_{L\Dmfp}
                 \dsiP{i\mfp}
         \bigg]
    \\[-.25ex]
      \times\; &\bigg[\iiint_{V\Dmfp}
                 \dsiM{j\mfm j'}\;
                 (-\pa_{j'})\;
          -\; \iint_{L\Dmfm}
                 \dsiM{j\mfm}
         \bigg]\vv
        t\oC{}_{\zzzz i\mfp j\mfm}\;
        \tilde\pa^2\, F\oC(\tilde{x}^2)
    \nn \\[-.5ex]
  &\phantom{\bigg[\;}
   +\; \big[\text{Terme mit\vv $\pa_\mfp$
                   und/oder\vv $\pa_\mfm$}\big]
    \nn
    \\[-4.75ex]\nn
\end{align}
und mit Gl.~(\ref{tC}):
\vspace*{-.75ex}
\begin{align} \label{tC_imfpjmfm}
t\oC{}_{\zzzz i\mfp j\mfm}\;
  =\; \frac{1}{12}\vv
        \de^{\tilde\al\tilde\be}_{i\mfp}\vv
        g_{\tilde\al j}\, g_{\tilde\be \mfm}\;
  =\; \frac{1}{12}\;
        g_{ij}\, g_{\mfp\mfm}
    \\[-4.5ex]\nn
\end{align}
aufgrund von~\mbox{$g_{\mfp j},g_{i\mfm} \!\equiv\! 0$}, vgl.\@ Gl.~(\ref{tNC_imfpjmfm}). \\
\indent
Faktorisierung der Pyramiden-Mantelfl"achen~$S\Dmfp$,~$S\Dmfm$ bez"uglich longitudinaler und trans\-versaler Komponenten impliziert f"ur die {\it Volumina\/}~$V\Dmfp$,~$V\Dmfm$~-- vgl.\@ die Gln.~(\ref{dsi-Vmfp-Faktor}),~(\ref{dsi-Vmfm-Faktor}):\zz
\vspace*{-.5ex}
\begin{align} \label{Vmfp-Faktor}
&\dsiP{ii'\!\mfp}\;
  =\; \dsiP{ii'}\;
        \otimes\; \dsiP{\mfp}
    \\[.5ex]
&\text{d.h.}\qquad
  \iiint_{V\Dmfp} \dsiP{i\mfp i'}\;
  =\; -\, \iint_{V\Dmfp\Doperp} \dsiP{ii'}\;
        \otimes\; \int dx\Dmfp^{\mfp}
    \tag{\ref{Vmfp-Faktor}$'$}
    \\[-4.5ex]\nn
\end{align}
und
\vspace*{-.5ex}
\begin{align} \label{Vmfm-Faktor}
&\dsiM{jj'\!\mfm}\;
  =\; \dsiM{jj'}\;
        \otimes\; \dsiM{\mfm}
    \\[.5ex]
&\text{d.h.}\qquad
  \iiint_{V\Dmfm} \dsiM{j\mfm j'}\;
  =\; -\, \iint_{V\Dmfm\Doperp} \dsiM{jj'}\;
        \otimes\; \int dx\Dmfm^{\mfm}
    \tag{\ref{Vmfm-Faktor}$'$}
    \\[-4.5ex]\nn
\end{align}
mit~$V\Dmfp\Doperp$,~$V\Dmfm\Doperp$ den {\it transversalen Projektionen\/} der Pyramiden-Volumina im Sinne der expliziten Parametrisierung nach Gl.~(\ref{Vmfp^perp-Parametr}).
Bez"uglich der Grundfl"achen~$L\Dmfp$,~$L\Dmfm$ sei verwiesen auf die Gln.~(\ref{Lmfp-Faktor}),~(\ref{Lmfp-Faktor}$'$) und~(\ref{Lmfp-Faktor}),~(\ref{Lmfp-Faktor}$'$) und die Bemerkungen dort. \\
\indent
Eingesetzt die Faktorisierung der Fl"achenelemente von~$V\Dmfp$,~$V\Dmfm$ und~$L\Dmfp$,~$L\Dmfm$, folgt aus Gl.~(\ref{chC-2}) unmittelbar:
\end{samepage}%
\vspace*{-.5ex}
\begin{align} \label{chC_IL[F]}
\tilde\ch\oC\;
  =\; -\, \iIM\;
        &\bigg[-\,\iint_{V\Dmfp\Doperp}
                 \dsiP{ii'}\;
                 \pa_{i'}\;
          -\; \int_{L\Dmfp\Doperp}
                 dx\Dmfp^{i}
         \bigg]
    \\[-.25ex]
      \times\; &\bigg[-\,\iint_{V\Dmfm\Doperp}
                 \dsiM{jj'}\;
                 (-\pa_{j'})\;
          -\; \int_{L\Dmfm\Doperp}
                 dx\Dmfm^{j}
         \bigg]\vv
        t\oC{}_{\zzzz i\mfp j\mfm}\;
        \tilde\pa^2\vv
        I_L[F\oC]
    \nn \\
  &\phantom{\bigg[\;}
   +\; \big[\text{Terme mit\vv $\pa_\mfp$
                   und/oder\vv $\pa_\mfm$}\big]
    \nn
    \\[-4.5ex]\nn
\end{align}
mit
\begin{samepage}
\vspace*{-1.25ex}
\begin{align} \label{IL[F]-DefII}
  I_L[F]\;
  =\; \int dx\Dmfp^{\mfp}\;
      \int dx\Dmfm^{\mfm}\vv
        F(\tilde{x}^2)
    \\[-4.75ex]\nn
\end{align}
Analog zum nicht-konfinierenden Fall~-- vgl.\@ Gl.~(\ref{chNC_IL[F]})~-- f"uhrt diese Faktorisierung unmittelbar auf das longitudinal bestimmte Funktional~$I_L$, wie definiert in den Gl.~(\ref{IL[F]-Def}),~(\ref{IL[F]-Def}$'$).
Die Integration der longitudinalen Parameter~$u\Dmfp$,~$u\Dmfm$ ist f"ur allgemeine Funktion~$F$ explizit durchgef"uhrt in Anhang~\ref{APP-Subsect:umfp,umfm-Int}, das Resultat f"ur~$I_L[F]$ zitiert in den Gln.~(\ref{IL[F]}),~(\ref{IL[F]}$'$).
F"ur~$\tilde\ch\oC$ nach~(\ref{chC_IL[F]}) folgt auf dieser Basis:
\vspace*{-.25ex}
\begin{align} \label{chC-4}
\tilde\ch\oC\;
  =\; &-\, \det\mathbb{SL}
    \\[-.5ex]
  &\begin{aligned}[t]
   \times\, (-\,\iIM)\;
        &\bigg[-\,\iint_{V\Dmfp\Doperp}
                 \dsiP{ii'}\;
                 \del{\tilde{x}\Dmfp}_{i'}\;
          -\; \int_{L\Dmfp\Doperp}
                 dx\Dmfp^{i}
         \bigg]
    \\[-.5ex]
      \times\; &\bigg[-\,\iint_{V\Dmfm\Doperp}
                 \dsiM{jj'}\;
                 \del{\tilde{x}\Dmfm}_{j'}\;
          -\; \int_{L\Dmfm\Doperp}
                 dx\Dmfm^{j}
         \bigg]\vv
        t\oC{}_{\zzzz i\mfp j\mfm}\;
        \tilde\pa^2\vv
        \projt{F\oC}{\rb{x}}
   \end{aligned}
    \nn
    \\[-4.25ex]\nn
\end{align}
gesetzt%
  ~\vspace*{-.25ex}\mbox{\,$\pa_{i'} \!=\! \del{\tilde{x}\Dmfp}_{i'}$},%
  ~\mbox{\,$-\pa_{j'} \!=\! \del{\tilde{x}\Dmfm}_{j'}$}, vgl.\@ Gl.~(\ref{pa=del(x)}).
Da~\mbox{$\projt{F\oC}{\rb{x}}$} nur abh"angt von~$|\rb{x}|$, also nicht von den longitudinalen Komponenten von~$\tilde{x}$, verschwinden identisch die Terme mit~$\pa_\mfp$ und/oder~$\pa_\mfm$, vgl.\@ Gl.~(\ref{chC_IL[F]}), und es gilt~\mbox{$\tilde\pa^2 \!\equiv\! g^{\mfp\mfp}(\pa_\mfp\pa_\mfp \!+\! \pa_\mfm\pa_\mfm) \!+\! 2g^{\mfp\mfm}\pa_\mfp\pa_\mfm \!-\! \pa_i\pa_i \!=\! -\pa_i\pa_i$}. \\
\indent
In Gl.~(\ref{chC-4}) sind gegen"uber Gl.~(\ref{chC-1}) im wesentlichen die Integrationsgebiete "ubergegangen in ihre transversalen Projektionen.
Eine Dimension reduziert, gehen wir die Schritte zur"uck, die uns gef"uhrt haben von Gl.~(\ref{chC-0}) zu Gl.~(\ref{chC-1}):
Wir gehen "uber von~\mbox{$V\Dimath\Doperp$} zu~\mbox{$\pa V\Dimath\Doperp$} f"ur~\mbox{$\imath \!=\! \mfp,\mfm$} mithilfe des Stokes'schen Satzes:
\vspace*{-.5ex}
\begin{align} 
\iint_{V\Dimath\Doperp}\;
  \dsiI{\tilde\mu\tilde\nu}\;
  \del{\tilde{x}\Dimath}_{\tilde\nu}\vv
  T_{\tilde\mu}\;
  =\; -\, \int_{\pa V\Dmfp\Doperp}\;
        \dsiI{\tilde\mu}\vv
        T_{\tilde\mu}
    \\[-4.75ex]\nn
\end{align}
f"ur~\mbox{$\tilde{T}\!\equiv\! \big(T_{\tilde\mu}\big)$} ein beliebiger Fourier-integrabler~$(0,1)$-Lorentz-Tensor, das hei"st -Vektor, vgl.\@ Gl.~(\ref{paV<->V-Stokes}).~--
Und fassen zusammen die Fl"achenintegrale "uber~$\pa V\Dimath\Doperp$ und~$L\Dimath\Doperp$ im Sinne%
\FOOT{
  Es folgt~\mbox{$S\Dimath\Doperp \!=\! \pa V\Dimath\Doperp \!-\! L\Dimath\Doperp$} aus~\mbox{$S\Dimath \!=\! \pa V\Dimath \!-\! L\Dimath$} f"ur~\mbox{$\imath \!=\! \mfp,\mfm$}:  Seien die Orientierungen repr"asentiert durch "au"sere Normalenvektoren; dann sind transversal projizierte Differenzen identisch den respektiven Differenzen der transversal Projizierten.   Die longitudinale Integration hat zur Konsequenz transversale Projektion des Integrtaionsgebiets, so da"s a~priori "ubergeht $S\Dimath$ in $S\Dimath\Doperp$~-- ungeachtet der intermedi"aren, formalen Aufteilung.
}
%
\vspace*{-.25ex}
\begin{align} 
&S\Dimath\Doperp\;
  =\; \pa V\Dimath\Doperp - L\Dimath\Doperp
    \\[-.25ex]
&\text{ergo:}\qquad
  \int_{S\Dimath\Doperp}\;
  \dsiI{\tilde\mu}\vv
  T_{\tilde\mu}
  =\; \bigg[\int_{\pa V\Dimath\Doperp}\;
        -\; \int_{L\Dimath\Doperp}
      \bigg]\vv
      \dsiI{\tilde\mu}\vv
      T_{\tilde\mu}
    \\[-4.75ex]\nn
\end{align}
vgl.\@ Gl.~(\ref{S=paV-L}) bzw.~(\ref{S=paV-L-Int}). \\
\indent
Aus Gl.~(\ref{chC-4}) folgt unmittelbar~-- mit~\mbox{$dx\Dimath^{\tilde\mu} \!\equiv\! \dsiI{\tilde\mu}$}, vgl.\@ Gl.~(\ref{dx^mu=dsi^mu(x)}):
\vspace*{-.25ex}
\begin{align} 
\tilde\ch\oC\;
  =\; -\, \det\mathbb{SL}\;\cdot\; (-\,\iIM)\;
        \int_{{\cal S}\Dmfp\Doperp}
                 dx\Dmfp^{i}\vv
        \int_{{\cal S}\Dmfm\Doperp}
                 dx\Dmfm^{j}\vv
        t\oC{}_{\zzzz i\mfp j\mfm}\;
        \tilde\pa^2\, \projt{F\oC}{\rb{x}}
    \\[-5ex]\nn
\end{align}
und mit~{\,$t\oC{}_{\zzzz i\mfp j\mfm} \!=\! -\frac{1}{12}\, \de_{ij}\, g_{\mfp\mfm}$}, vgl.\@ Gl.~(\ref{tC_imfpjmfm}):
\vspace*{-.25ex}
\begin{align} \label{chC-6}
\tilde\ch\oC\;
  =\; -\, \det\mathbb{SL}\; g_{\mfp\mfm}\;\cdot\;
        \iIM\; \frac{1}{12}\vv
        \int_{{\cal S}\Dmfp\Doperp}
                 dx\Dmfp^{i}\vv
        \int_{{\cal S}\Dmfm\Doperp}
                 dx\Dmfm^{i}\vv
        \tilde\pa^2\, \projt{F\oC}{\rb{x}}
    \\[-5ex]\nn
\end{align}
Diese Darstellung von~\vspace*{-.25ex}$\tilde\ch\oC$ ist unmittelbares Pendant zu der von~$\tilde\ch\oNC$ nach Gl.~(\ref{chNC-4}). \\
\indent
In Anhang~\ref{APP-Subsect:S_mf^perp-Int} wird diese Darstellung ausgewertet in Form
\vspace*{-.5ex}
\begin{align} \label{chC_IT[F]}
&\tilde\ch\oC\;
  =\; -\, \det\mathbb{SL}\; g_{\mfp\mfm}\;\cdot\;
         \iIM\; \frac{1}{12}\vv
         I_T[F]
    \\[-4.5ex]\nn
\end{align}
unter Definition
\vspace*{-.5ex}
\begin{align} \label{IT[F]-Def}
I_T[F]\;
  =\; \int_{{\cal S}\Dmfp\Doperp} dx\Dmfp^{i}\vv
      \int_{{\cal S}\Dmfm\Doperp} dx\Dmfm^{i}\vv
        \pa_j \left[x^j F\right]
    \\[-5ex]\nn
\end{align}
\end{samepage}%
Und zwar auf Basis der speziellen Wahl der verallgemeinerten Pyramiden~$P\Dmfp$,~$P\Dmfm$ wie in Abschnitt~\ref{Subsect:surfacesSDmf}, Gl.~(\ref{Smfp-Parametr}):
Deren Mantelfl"achen~${\cal S}\Dmfp$,~${\cal S}\Dmfm$~-- "uber die ab~initio die paralleltransportierten Feldst"arken zu integrieren sind~-- werden angenommen als {\it planar\/} bez"uglich der transversalen Komponenten, das hei"st deren Projektionen~${\cal S}\Dmfp\Doperp$,~${\cal S}\Dmfm\Doperp$ als abschnittweise {\it geradlinig\/}.
Anhang~\ref{APP-Subsect:S_mf^perp-Int} abschlie"send wird argumentiert, da"s Geradlinigkeit beziehungsweise Planarit"at in diesem Sinne nicht notwendig ist. \\
\indent
Relevant f"ur die Auswertung in Anhang~\ref{APP-Subsect:S_mf^perp-Int} ist der Differentialoperator~$\pa_{\tilde\mu}$, der r"uhrt aus dem d'Alembert-Operator%
  ~\mbox{\,$\tilde\pa^2 \!\equiv\! \pa_{\tilde\mu} \pa_{\tilde\mu}$}%
\FOOT{
  {\sl cum~grano~salis\/} der negative transversale Laplace-Operator~\vspace*{-.25ex}\mbox{\,$-\pa_i \pa_i \!\equiv\! -\bm\nabla^2$}, da~\mbox{\,$\projNAt{F\oC}$}~-- vgl.\@ Gl.~(\ref{chC-6})~-- abh"angt nur von den Komponenten des transversalen Vektors~\vspace*{-.125ex}\mbox{$\rb{x} \!\equiv\! (x^1,x^2)^{\T t}$}
}.
Relevant ist nicht die explizite Darstellung der konfinierenden Korrelationsfunktion~$F\oC$ beziehungsweise deren Projektion~$\projNAt{F\oC}$, so da"s wir~-- auch f"ur K"urze der Notation~-- arbeiten mit der allgemeinen Funktion~\mbox{$F \!\equiv\! F(|\rb{x}|)$}, die aber in~praxi zu identifizieren auf Basis der Relation
%
\begin{align} 
x^j F\;
  \equiv\; \pa^j \projNAt{F\oC}
\end{align}
vgl.\@ die Gln.~(\ref{chC-6})-(\ref{IT[F]-Def}).
Da~\vspace*{-.125ex}\mbox{$\projNAt{F\oC}$} abh"angt nur von%
  ~\mbox{\,$|\rb{x}| \!\equiv\! \surd\rb{x}^2$},%
  ~\mbox{\,$\rb{x}^2 \!\equiv\! x^ix^i \!\equiv\! -g_{ij}x^ix^j$}, gilt f"ur den kontravarianten Ableitungsoperator~$\pa^j$, der wirkt auf~$\projNAt{F\oC}$:
\begin{samepage}
\vspace*{-.25ex}
\begin{align} \label{pa^j->dBetrag}
\pa^j\;
  =\; -\, \pa_j\;
  \equiv\; -\, \frac{\pa}{\pa x^j}\;
  =\; -\, \frac{\pa\rb{x}^2}{\pa x^j}\;
        \frac{d\surd\rb{x}^2}{d\rb{x}^2}\;
        \frac{d}{d|\rb{x}|}\;
  =\; -\, x^j\, \frac{1}{|\rb{x}|}\, \frac{d}{d|\rb{x}|}
    \\[-4.25ex]\nn
\end{align}
Folglich ist zu identifizieren:
\vspace*{-.25ex}
\begin{align} \label{F_projt(1)}
F(|\rb{x}|)\;
  \equiv\; \projt[(1)]{F\oC}{\rb{x}}\;
  \equiv\; -\, \frac{1}{|\rb{x}|} \frac{d}{d|\rb{x}|}\vv
             \projt{F\oC}{\rb{x}}
    \\[-4.25ex]\nn
\end{align}
indem definiert seien durch
\vspace*{-.25ex}
\begin{align} \label{projt(n)-Def}
\projt[(n)]{F\oC}{\rb{x}}\;
  \equiv\; \bigg(\! -\, \frac{1}{|\rb{x}|} \frac{d}{d|\rb{x}|}\bigg)\Big.^{\zz\T n}\vv
             \projt{F\oC}{\rb{x}}
    \\[-4.25ex]\nn
\end{align}
geeignete $n$-fache Ableitungen~\mbox{\,$\projNAt[(n)]{F\oC}$} der Projektion~\mbox{\,$\projNAt{F\oC}$}.%
\FOOT{
  formal: Ableitungen nach~\vspace*{-.125ex}\mbox{\,$-\frac{1}{2}|\rb{x}|^2$}
} \\
%
\indent
Die Auswertung des Funktionals~$I_T[F]$ nach Gl.~(\ref{IT[F]-Def}) in Anhang~\ref{APP-Subsect:S_mf^perp-Int} geschieht wie folgt:
Der Integralausdruck wird aufgefa"st als {\it direktes Produkt von Linienintegralen\/} entlang der Mannigfaltigkeiten%
  ~${\cal S}\Dmfp\Doperp\big|_{\scriptstyle I}$,~${\cal S}\Dmfm\Doperp\big|_{\scriptstyle J}$ und dargestellt als {\it Fl"achenintegral "uber das nichttriviale direkte Produkt\/}%
  ~\mbox{${\cal S}\Dmfp\Doperp\big|_{\scriptstyle I} \!\otimes\! {\cal S}\Dmfm\Doperp\big|_{\scriptstyle J}$} der Mannigfaltigkeiten; dabei wird zerlegt in Summen geradliniger Abschnitte:%
  ~\mbox{\,$I,J \!\in\! \{Q,\AQ\}$} die respektiven Quark- und Antiquark-Abschnitte bei Planarit"at der Pyramiden~$P\Dmfp$,~$P\Dmfm$ im diskutierten Sinne.
Die Fl"achenintegrale "uber%
  ~\mbox{${\cal S}\Dmfp\Doperp\big|_{\scriptstyle I} \!\otimes\! {\cal S}\Dmfm\Doperp\big|_{\scriptstyle J}$} werden "uberf"uhrt in Linienintegrale entlang%
  ~\mbox{$\pa\big({\cal S}\Dmfp\Doperp\big|_I \!\otimes\! {\cal S}\Dmfm\Doperp\big|_J\big)$} mithilfe des {\it dualen \mbox{Stokes'schen} Satzes\/}~-- da der Integrand {\it Divergenz\/}~$\pa_j$ ist einer Funktion.
Resultat ist das Linienintegrale entlang des {\it "au"seren\/} Randes~\mbox{$\pa\big({\cal S}\Dmfp\Doperp \!\otimes\! {\cal S}\Dmfm\Doperp\big)$}, in parametrisierter Form~-- vgl.\@ Gl.~(\ref{APP:IT[F]}):
\vspace*{-.25ex}
\begin{align} \label{IT[F]}
&I_T[F]
    \\[-.75ex]
  &\hspace*{11pt}
   =\vv \sum\Big._{\!I,J\equiv Q,\AQ}\vv
       {\rm sign}_{I,J}\quad
       \rb{r}\Dmfp^I \cdot \rb{r}\Dmfm^J\vv
       \int_0^1 ds\vv
         \Big\{\;
           F\big(\big|s\, \rb{r}\Dmfp^I - \rb{r}\Dmfm^J\big|\big)\;
       +\; F\big(\big|\rb{r}\Dmfp^I - s\, \rb{r}\Dmfm^J\big|\big)\;
         \Big\}
    \nn
    \\[-4ex]\nn
\end{align}
ergo~-- vgl.\@ Gl.~(\ref{APP:chC_IJ}):
\vspace*{-.25ex}
\begin{align} \label{chC_IJ}
&\tilde\ch\oC\;
   =\; -\, \det\mathbb{SL}\; g_{\mfp\mfm}\;\cdot\;
         \iIM\; \frac{1}{12}\vv
     \sum\Big._{\!I,J\equiv Q,\AQ}\vv
     {\rm sign}_{I,J}
    \\[-.25ex]
  &\phantom{\tilde\ch\oC\; =\; -\, \det}\times\,
     \rb{r}\Dmfp^I\cdot \rb{r}\Dmfm^J\vv
       \int_0^1 ds\vv
         \Big\{\;
           \projtbig[(1)]{F\oC}{s\, \rb{r}\Dmfp^I - \rb{r}\Dmfm^J}\;
       +\; \projtbig[(1)]{F\oC}{\rb{r}\Dmfp^I - s\, \rb{r}\Dmfm^J}
         \;\Big\}
    \nn
    \\[-4ex]\nn
\end{align}
\end{samepage}%
mit~\mbox{\,$F \!\equiv\! \projNAt[(1)]{F\oC}$} nach Gl.~(\ref{F_projt(1)}).
Es ist~\mbox{\,$\rb{r}\Dmfp^I \!\cdot\! \rb{r}\Dmfm^J
  \equiv +\de_{ij} r\Dmfp^{Ii} r\Dmfm^{Jj}$} mit~\mbox{$i,j \!\in\! \{1,2\}$} das Skalarprodukt der transversalen Vektoren.
Der Faktor~\mbox{${\rm sign}_{I,J} \!=\! (-1)^{I+J}$}\nopagebreak
ist definiert in Gl.~(\ref{signIJ}) durch Zuweisen der Zahlenwerte%
  ~\mbox{$Q \!\equiv\! 0$},~\mbox{$\AQ \!\equiv\! 1$} und~-- vgl.\@ die \mbox{nicht-konfinierende Struktur}
nach Gl.~(\ref{chNC_IJ})~-- tr"agt Rechnung dem relativen Signum von Beitr"agen mit%
  ~\vspace*{-.25ex}\mbox{$I \!\nequiv\! J$}: wenn genau eine paralleltransportierte Feldst"arke Element ist des Antiquark-Abschnitts.%
\FOOT{
  \label{APP-FN:AQ-negativ}Dies ist Konsequenz der induzierten Orientierung:~\mbox{$\text{\sl Apex} \!\to\! Q,\, \text{\sl Apex} \!\to\! \AQ$} [statt~\mbox{$\AQ \!\to\! \text{\sl Apex} \!\to\! Q$}], so da"s die Antiquarks eingehen mit entgegengesetztem Vorzeichen, die sich im~\mbox{$(\AQ,\AQ)$-Bei}\-trag kompensieren.
} \\
%
\indent
Diese Reduktion der transversalen Integrationen auf eine einzige beruht wesentlich auf der Nichttrivialit"at des direkten Produkts%
  ~\mbox{${\cal S}\Dmfp\Doperp\big|_{\scriptstyle I} \!\otimes\! {\cal S}\Dmfm\Doperp\big|_{\scriptstyle J}$} der Faktor-Mannigfaltigkeiten%
  ~${\cal S}\Dmfp\Doperp\big|_{\scriptstyle I}$ und~${\cal S}\Dmfm\Doperp\big|_{\scriptstyle J}$.
Sie kann~-- in dieser Form~-- {\it nicht\/} durchgef"uhrt werden f"ur die Funktionen~$\tilde\ch\idx{\mfp\mskip-2mu\mfp}\oC$ und~$\tilde\ch\idx{\mfm\mskip-2mu\mfm}\oC$, da diese determiniert sind durch die Integration zweier paralleltransportierter Feldst"arken "uber {\it dieselbe\/} Mannigfaltigkeit:~${\cal S}\Dmfp\Doperp$ beziehungsweis~${\cal S}\Dmfm\Doperp$.
Es ist daher f"ur diese Funktionen auszugehen von der urspr"unglichen Darstellung des Funktionals~$I_T[F]$ nach Gl.~(\ref{IT[F]-Def}), f"ur die in Anhang~\ref{APP-Subsect:S_mf^perp-Int} angegeben wird eine geeignete Parametrisierung.
Der Differentialoperator~$\pa_j$~-- nicht explizit ben"otigt in Hinblick auf den Stokes'schen Satz~-- wird "uberf"uhrt in eine Ableitung nach dem eigentlich relevanten Raumzeit-Argument~$|\rb{x}|$ und entsprechend umgeformt der Integrand:
\vspace*{-.5ex}
\begin{alignat}{2} \label{paj-F-explizit}
\pa_j \left[x^j F\right]\;
  &=\; g^j{}_j\, F\;
         +\; x^j\, \pa_j\, F&
   &\vv\text{mit\rmfootnote}\quad\vv
    g^j{}_j \equiv d \!-\! 2
                     \;\big|_{d\equiv4}
            = 2
    \\[-.5ex]
  &=\; 2\; F
        -\; |\rb{x}|^2 \bigg(\! -\, \frac{1}{|\rb{x}|} \frac{d}{d|\rb{x}|}\bigg)\; F&&
    \tag{\ref{paj-F-explizit}$'$} \\
  &=\; 2\; \projNAt[(1)]{F\oC}\;
        -\; |\rb{x}|^2\; \projNAt[(2)]{F\oC}&&
    \tag{\ref{paj-F-explizit}$''$}
    \\[-4.5ex]\nn
\end{alignat}
\footnotetext{
  Anhang~\ref{APP:CLTFN} nimmt Bezug auf den verallgemeinerten Minkowski-Raums~${\cal M}_{d,\si}$ mit Dimension~$d$ und Sig\-natur~$\si$, f"ur den hier steht~$d$ subtrahiert die zwei ausintegrierten~-- "`longitudinalen"'~-- Dimensionen.
}%
%
vgl.\@ die Gln.~(\ref{pa^j->dBetrag}),~(\ref{projt(n)-Def}).
Es folgt~-- vgl.\@ Gl.~(\ref{APP:IT[F]IJ-mfpmfp}):
\vspace*{-.5ex}
\begin{align} \label{chC_IJ-mfpmfp}
&\tilde\ch\idx{\mfp\mskip-2mu\mfp}\oC\;
   =\; -\, \det\mathbb{SL}\; g_{\mfp\mfp}\;\cdot\;
         \iIM\; \frac{1}{12}\vv
     \sum\Big._{\!I,J\equiv Q,\AQ}\vv
     {\rm sign}_{I,J} 
    \\[-.25ex]
  &\phantom{\tilde\ch\oC\; =\; - \det}\times\,
     \rb{r}\Dmfp^I\cdot \rb{r}\Dmfp^J\vv
        \int_0^1 ds\vv
        \int_0^1 ds'\vv
         \Big\{\;
           2\;
             \projt[(1)]{F\oC}{\rb{x}}\;
       -\; |\rb{x}|^2\;
             \projt[(2)]{F\oC}{\rb{x}}
         \;\Big\}
        \vv\bigg|_{(I,J)}
    \nn \\[-.25ex]
  &\text{mit}\qquad
    \rb{x}\big|_{(I,J)}\;
      \equiv\; \rb{x}\mskip-2mu(s, s')\big|_{(I,J)}\;
      \equiv\; s\, \rb{r}\Dmfp^I - s'\, \rb{r}\Dmfp^J
    \nn
    \\[-4.5ex]\nn
\end{align}
und daraus%
  ~\vspace*{-.125ex}\mbox{\,$\tilde\ch\idx{\mfm\mskip-2mu\mfm}\oC$} durch%
  ~\mbox{\,$g_{\mfp\mfp} \!\to\! g_{\mfm\mfm}$} und~\mbox{\,$\rb{r}\Dmfp \!\to\! \rb{r}\Dmfm$}. \\
\indent
Bez"uglich der streu-relevanten Projektionen%
  ~\mbox{\,$\projNAt[(1)]{F\oC},\, \projNAt[(1)]{F\oC}$} verweisen wir auf Anhang~\ref{APP:CLTFN}.
Dort leiten wir her~-- vgl.\@ die Gln.~(\ref{APP:F^C,NC-Minkowski}$''$),~(\ref{APP:F^C,NC-Minkowski}$'''$) mit~\mbox{\,$\nu \!\equiv\! 4$}:
%
\begin{align} \label{projt[F]C-Ansatz}
&\projt[(1)]{F\oC}{\rb{x}}\;
  =\, -\, \iIM\, \pi\cdot
         \la^2\vv
         {\cal K}_2(\bm\ze)
    \\[.25ex]
  &\projt[(2)]{F\oC}{\rb{x}}\;
  =\, -\, \iIM\, \frac{\pi}{2}\cdot
         {\cal K}_1(\bm\ze)
    \tag{\ref{projt[F]C-Ansatz}$'$} \\[-1ex]
  &\hspace*{140pt}
    \la \equiv \la_4 = 8\!\big/\!3\pi,\quad
    \bm\ze \equiv {\T\sqrt{|\rb{x}|^2 \!/\! \la^2 + \iIM\, \ep}}
            \cong  |\rb{x}| \!\big/\! \la
    \nn
    \\[-4.25ex]\nn
\end{align}
durch ein- beziehungsweise zweifache Differentiation%
  ~\vspace*{-.375ex}\mbox{$-\frac{1}{|\rb{x}|} \frac{d}{d|\rb{x}|}$} aus der Projektion%
  ~\mbox{$\projt{F\oC}{\rb{x}}$}
  ~\vspace*{-.125ex}\mbox{$\equiv\! \projt{F\oNC}{\rb{x}} = -\iIM\,4\pi \!\cdot\! \la^4\, {\cal K}_3(\bm\ze)$}, vgl.\@ Gl.\,(\ref{projt[F]NC-Ansatz}).
Diese Darstellung wiederum ist Konsequenz unseres Ansatzes%
  ~\vspace*{-.25ex}\mbox{$F\oC\!(\tilde{x}^2) \equiv F\oNC\!(\tilde{x}^2) = \frac{1}{2}\la^2\, {\cal K}_2(\ze)$},%
  ~\mbox{$\ze \equiv\! \sqrt{-\tilde{x}^2 \!/\! \la^2 \!+\! \iIM\, \ep} \cong \sqrt{-\tilde{x}^2}\big/\la$} f"ur die~konfi\-nierende \mbox{$F$-Korrelations}\-funktion.
Vgl.\@ Gl.\,(\ref{F^C})ff.\@, ferner Kap.\,\ref{Abschn:ANN-KONST}, Gl.\,(\ref{Dvier_DDxi})ff.\@ und Anh.\,\ref{APP:CLTFN}, die Gln.~(\ref{APP:DDk-Ansatz}),~(\ref{APP:FT_F,D-C}).
\vspace*{-.75ex}

\section[Nahezu lichtartige~\protect$T$-Amplitude: Diskussion]{%
         Nahezu lichtartige~\bm{T}-Amplitude: Diskussion}
\label{Sect:T-Amplitude.Diskussion}

Die konfinierenden und nicht-konfinierenden Funktionen konstituieren vollst"andig die nahezu lichtartige $T$-Amplitude, die wir zusammenfassend darstellen durch die hergeleiteten expliziten Ausdr"ucke und diskutieren vor diesem Hintergrund.
\vspace*{-.5ex}

\bigskip\noindent
Die $T$-Amplitude%
  ~\vspace*{-.25ex}\mbox{$T\hh^{(s,t)}\mskip-2mu$} f"ur Streuung der hadronischen Zust"ande%
  ~$h^{1}\!(P_{1})$,~$h^{2}\!(P_{2})$ in~$h^{1'}\!(P_{1'})$, $h^{2'}\!(P_{2'})$ ist gegeben in Gl.~(\ref{T2h_T2ell-mf}) in Termen der $T$-Amplitude%
  ~\vspace*{-.375ex}\mbox{$\tTll^{(s,\rb{b})}$} f"ur die zugrundeliegende Loop-Loop-Streuung.
Diese ist gegeben in Termen der Funktionen%
  ~\mbox{$\ch\idx{\imath\jmath}$},~\mbox{$\imath,\jmath \!\in\! \{\mfp,\mfm\}$}:
\vspace*{-.25ex}
\begin{align} \label{tTll_WW_ch-mf_ALL}
\hspace*{-12pt}
\tTll\;
  &=\; -\, 2\iIM\,s\vv
         \frac{\normDrst{R\Dmfp}\, \normDrst{R\Dmfm}}{%
                 (2!)^2\, \dimDrst{R\Dmfp}\, \dimDrst{R\Dmfm}}
    \\[-.75ex]
  &\hspace*{9.5pt}\hspace*{64.5pt}
   \times\,
        \Big[\,
          \ch\idx{\mfp\mskip-2mu\mfp}\, \ch\idx{\mfm\mskip-2mu\mfm}\;
          +\; \frac{2}{\dimNc}\;\, \ch\idx{\mfp\mskip-2mu\mfm}{}^{\zz2}
        \,\Big]\vv
        \exp\, -\frac{1}{2!}
          \Big(
            \frac{\normDrst{R\Dmfp}}{\dimDrst{R\Dmfp}}\;
              \ch\idx{\mfp\mskip-2mu\mfp}
          + \frac{\normDrst{R\Dmfm}}{\dimDrst{R\Dmfm}}\;
              \ch\idx{\mfm\mskip-2mu\mfm}
          \Big)
    \nn \\[.5ex]
  &\hspace*{-6pt}
   \underset{\Drst{R\Dimath} \equiv \Drst{F}}{=}\vv
       -\, 2\iIM\,s\vv
         \frac{1}{(4\Nc)^2}\vv
        \Big[\,
          \ch\idx{\mfp\mskip-2mu\mfp}\, \ch\idx{\mfm\mskip-2mu\mfm}\;
          +\; \frac{2}{\Nc^2 \!-\! 1}\;\, \ch\idx{\mfp\mskip-2mu\mfm}{}^{\zz2}
        \,\Big]\vv
        \exp\, -\frac{1}{4\Nc}
          \big(
            \ch\idx{\mfp\mskip-2mu\mfp}
          + \ch\idx{\mfm\mskip-2mu\mfm}
          \big)
    \tag{\ref{tTll_WW_ch-mf_ALL}$'$}
    \\[-4.25ex]\nn
\end{align}
vgl.\@ die Gln.~(\ref{tTll_WW_ch-mf}),~(\ref{tTll_WW_ch-mf}$'$),~-- dort eingesetzt in diesem Sinne:%
\FOOT{
  Die erste Identit"at in Gl.~(\ref{vevW_ch-mf}) folgt auf Basis der expliziten Darstellung von~$W\Dimath$ nach Gl.~(\ref{Konnektor_LoopFtr-mf})~unmittelbar aus Annahme~(1): der {\sl Existenz\/} einer konvergenten Entwicklung in Kumulanten~$K_n$,~\mbox{und aus Annah}\-me~(2) eines {\sl Gau"s'schen\/} zugrundeliegenden stochastischen Prozesses:~\mbox{$K_n \!\equiv\! 0,\, \forall n \!\ne\! 2$}; vgl.\@ Gl.~(\ref{WW-Loop_KumEntw}) bzw.\@ die Ausf"uhrungen auf Seite~\pageref{K2_vac-g2FF}.   F"ur den Vorfaktor gilt: $K_2$ ist multipliziert mit~$1\!/\!2!$; es ist%
  ~\mbox{\,$K_2 \equiv \vac{g^2 F^{(1)} F^{(2)}}$}
   \mbox{$= \csDrst{R} \!/\! \dimNc\, \vac{g^2 F^{(1)}{}_{\zz a} F^{(2)}{}_{\zz a}} \!\cdot\! \bbbOne{R}$} und%
  ~\mbox{\,$\vac{g^2 F^{(1)}{}_{\zz a} F^{(2)}{}_{\zz a}}
     = \vac{g^2 FF} \!\cdot\! D_{\mu_1\nu_1\mu_2\nu_2}$} mit~\mbox{\,$F^{(i)} \!\equiv\! F_{\mu_i\nu_i}$}, vgl.\@ die Gln.~(\ref{K2_Eins}$'$),~(\ref{K2-g2FF_Dvier}), ferner~\mbox{\,$\trDrst{R}\exp\al\bbbOne{R} = \exp\al$} f"ur~\mbox{$\bbbc$-Zahlen~$\al$}.   \mbox{Die zweite Identit"at in Gl.~(\ref{tTll_WW_ch-mf_ALL}) folgt} mit%
  ~\mbox{\,$\csDrst{R} \dimDrst{R} \!=\! \normDrst{R} \dimNc$}, vgl.\@ Gl.~(\ref{APP:csDrst*dimDrst_nDrst*dimNc}).   Gl.~(\ref{tTll_WW_ch-mf_ALL}$'$) folgt mit~\mbox{$\dimNc \!=\! \Nc^2 \!-\! 1$}, explizit eingesetzt%
  ~\mbox{\,$\dimDrst{F} \!\equiv\! \Nc$}, \mbox{$\normDrst{F} \!\equiv\! 1\!/\!2$}~-- vgl.\@ die Gln.~(\ref{APP:Quarks/Gluonen-DrstF/A}),~(\ref{APP:nDrstF-Konv})~-- im Fall~\mbox{\,$\Drst{R\Dimath} \!\equiv\! \Drst{F}$} von Quark-Loops.   Vgl.\@ Fu"sn.\,\FN{FN:tTll-Vorfaktor}
}
\begin{samepage}
\vspace*{-.5ex}
\begin{align} \label{vevW_ch-mf}
\vac{W\Dimath}\;
  &=\; \exp\; \frac{1}{2!}\, \frac{\csDrst{R\Dimath}}{\dimNc}\;
         \ch\idx{\imath\imath}\quad
   =\; \exp\; \frac{1}{2!}\, \frac{\normDrst{R\Dimath}}{\dimDrst{R\Dimath}}\;
         \ch\idx{\imath\imath}
    \\[-.5ex]
  &\hspace*{-6pt}
   \underset{\Drst{R\Dimath} \equiv \Drst{F}}{=}\vv
       \exp\; \frac{1}{4\Nc}\;
         \ch\idx{\imath\imath}
    \tag{\ref{vevW_ch-mf}$'$}
    \\[-4.5ex]\nn
\end{align}
f"ur die Vakuumerwartungswerte der einzelnen Wegner-Wilson-Loops~$\vac{W\Dimath}$,~\mbox{\,$\imath = \mfp,\mfm$}. \\
\indent
Durch die Funktion%
  ~$\ch\idx{\imath\jmath}$ ist definiert die reellwertige Funktion
  ~$\tilde\ch\idx{\imath\jmath}$:%
\FOOT{
  Die \mbox{$\tilde\ch\idx{\imath\jmath}$-im}pliziten Vektoren sind {\sl reskaliert} "`in Einheiten von~$a$"'~-- vgl.\@ Gl.~(\ref{x=ximath-xjmath}$'$)~-- folglich~$\tilde\ch\idx{\imath\jmath}$ selbst.
}
%
\vspace*{-.25ex}
\begin{align} \label{ch_ch-tilde-mf-REP}
&\ch\idx{\imath\jmath}\;
  =\; -\iIM\vv \vac{g^2 FF}a^4\; \cdot\; \tilde\ch\idx{\imath\jmath}
    \\[-3.75ex]
 &\hspace*{153pt}
  \imath,\jmath \in \{\mfp,\mfm\}
    \nn
    \\[-4.25ex]\nn
\end{align}
vgl.\@ Gl.~(\ref{ch_ch-tilde-mf}),~-- und diese im Sinne der Tensorstrukturen%
  ~\vspace*{-.125ex}\mbox{$t\oC{}_{\zzzz \tilde\mu\tilde\nu\tilde\rh\tilde\si}$} und%
  ~\mbox{$t\oNC{}_{\zzzz \tilde\mu\tilde\nu\tilde\rh\tilde\si}$} des Korrelationstensors als Summe eines konfinierenden und eines nicht-konfinierenden Anteils:
\vspace*{-.25ex}
\begin{align} \label{ch_vka,chC,chNC-tilde-REP}
&\tilde\ch\idx{\imath\jmath}\;
  =\; \vka\;  \tilde\ch\idx{\imath\jmath}\oC\;
      +\; (1 \!-\! \vka)\; \tilde\ch\idx{\imath\jmath}\oNC
    \\[-3.75ex]
 &\hspace*{153pt}
  \imath,\jmath \in \{\mfp,\mfm\}
    \nn
    \\[-4.25ex]\nn
\end{align}
vgl.\@ Gl.~(\ref{ch_vka,chC,chNC-tilde}).
Die Funktionen%
  ~$\tilde\ch\idx{\imath\jmath}\oC$,~$\tilde\ch\idx{\imath\jmath}\oNC$ sind unabh"angig f"ur die Indexpaare~$\mfp\mfp$,~$\mfm\mfm$ und~$\mfp\mfm$.
Sie sind definiert und ausgewertet in den vorangehenden Abschnitten~\ref{Subsect:chNC} und~\ref{Subsect:chC} und seien zusammenfassend angegeben in systematischer Form~-- vgl.\@ Gl.~(\ref{ch=gXi}):
\vspace*{-.25ex}
\begin{align} \label{ch=gXi-REP}
&\tilde\ch\idx{\imath\jmath}\;
  =\; -\, \det\mathbb{SL}\vv g_{\imath\jmath}\;\cdot\;
        \tilde{X}\idx{\imath\jmath}
    \\[-3.75ex]
 &\hspace*{153pt}
  \imath,\jmath \in \{\mfp,\mfm\}
    \nn
    \\[-4.25ex]\nn
\end{align}
unter Separation des rein longitudinal determinierten Faktors%
  ~\vspace*{-.125ex}\mbox{\,$-\det\mathbb{SL}\, g_{\imath\jmath}$} von Komponenten des metrischen Tensors.
Die Funktionen%
  ~$\tilde{X}\idx{\imath\jmath}$ sind vollst"andig bestimmt durch die Geometrie der Streuung in der~\mbox{$x^1\!x^2$-Transversal}\-ebene, in die aber~-- wir rekapitulieren Fu"sn.\,\FN{FN:eff-transv-QFT}~--pro\-jiziert ist die longitudinale Dynamik der Streuung in Form der Abh"angigkeit des relevanten Impaktvektors~$\rb{b}$ von den Quark-Anteilen~$\zet_1$,~$\zet_2$ am Lichtkegelimpuls der Loops~$W\Dmfp$,~$W\Dmfm$. \\
\indent
Bez"uglich der nicht-konfinierenden $N\!C$-Lorentz-Tensorstruktur folgt f"ur die~\mbox{$\bm\mfp\bm\mfm$-Funk}\-tion~-- vgl.\@ Gl.~(\ref{chNC_IJ}):
\vspace*{-1ex}
\begin{align} \label{ChNC-pm}
\tilde{X}\idx{\mfp\mskip-2mu\mfm}\oNC\;
  &=\; \iIM\; \frac{1}{6}\vv
         \sum\Big._{\!I,J\equiv Q,\AQ}\vv
           {\rm sign}_{I,J}\vv
           \projtbig{F\oNC}{\rb{r}\Dmfp^I \!-\! \rb{r}\Dmfm^J}
    \\[-.25ex]
  &=\; 8\cdot \la^4\,
         \frac{\pi}{12}\vv
         \sum\Big._{\!I,J\equiv Q,\AQ}\vv
           {\rm sign}_{I,J}\vv
           {\cal K}_3\big(\big|\rb{r}\Dmfp^I \!-\! \rb{r}\Dmfm^J\big| \!\big/\!\la\big)
    \tag{\ref{ChNC-pm}$'$}
    \\[-4.5ex]\nn
\end{align}
und f"ur die~\mbox{$\bm\mfp\bm\mfp$-Funk}\-tion~-- vgl.\@ Gl.~(\ref{chNC_IJ-mfpmfp}):
\vspace*{-.5ex}
\begin{align} \label{ChNC-pp}
\tilde{X}\idx{\mfp\mskip-2mu\mfp}\oNC\;
  &=\; \iIM\; \frac{1}{3}\,
         \Big\{\;
           \projt{F\oNC}{\bm{0}}\;
       -\; \projt{F\oNC}{\rb{X}}
         \;\Big\}
    \\[-.25ex]
  &=\; 16\cdot \la^4\,
         \frac{\pi}{12}\,
         \Big\{\,
           1\; -\; {\cal K}_3(|\rb{X}| \!\big/\!\la)
         \,\Big\}
    \tag{\ref{ChNC-pp}$'$}
    \\[-4.5ex]\nn
\end{align}
\end{samepage}%
die zweiten Identit"aten mit%
  ~\mbox{\,$\projt{F\oNC}{\rb{x}} \!=\! -\iIM\,4\pi \!\cdot\! \la^4\, {\cal K}_3(|\rb{x}| \!\big/\!\la)$} nach Gl.~(\ref{projt[F]NC-Ansatz}), mit%
  ~\mbox{\,${\cal K}_\nu\!(\ze) \!\to\! 1$} f"ur~\mbox{\,$\ze \!\to\! 0$}, vgl.\@ Gl.~(\ref{calK_mu}$'$).
Die~\mbox{$\bm\mfm\bm\mfm$-Funk}\-tion~$\tilde{X}\idx{\mfm\mskip-2mu\mfm}\oNC$ folgt aus~$\tilde{X}\idx{\mfp\mskip-2mu\mfp}\oNC$ durch "Ubergang~\mbox{\,$\rb{X} \!\to\! \rb{Y}$}. \\
\indent
Bez"uglich der konfinierenden $C$-Lorentz-Tensorstruktur folgt f"ur die~\mbox{$\bm\mfp\bm\mfm$-Funk}\-tion~-- vgl.\@ Gl.~(\ref{chC_IJ}):
\vspace*{-1.25ex}
\begin{align} \label{ChC-pm}
&\tilde{X}\idx{\mfp\mskip-2mu\mfm}\oC\;
   =\; \iIM\; \frac{1}{12}\vv
         \sum\Big._{\!I,J\equiv Q,\AQ}\vv
         {\rm sign}_{I,J}
    \\[-.25ex]
  &\phantom{\tilde\ch\oC\; =\; - \det}\times\,
     \rb{r}\Dmfp^I\cdot \rb{r}\Dmfm^J\vv
       \int_0^1 ds\;
         \Big\{\;
           \projtbig[(1)]{F\oC}{s\, \rb{r}\Dmfp^I - \rb{r}\Dmfm^J}\;
       +\; \projtbig[(1)]{F\oC}{\rb{r}\Dmfp^I - s\, \rb{r}\Dmfm^J}
         \;\Big\}
    \nn \\[.75ex]
  &\phantom{\tilde{X}\idx{\mfp\mskip-2mu\mfm}\oC\;}
   =\; \la^4\,
         \frac{\pi}{12}\vv
         \sum\Big._{\!I,J\equiv Q,\AQ}\vv
         {\rm sign}_{I,J}
    \tag{\ref{ChC-pm}$'$} \\[-.25ex]
  &\hspace*{21pt}
   \phantom{\tilde\ch\oC\; =\; - \det}\times\,
     \rb{r}\Dmfp^I\cdot \rb{r}\Dmfp^J \!\big/\! \la^2\vv
       \int_0^1 ds\;
         \Big\{\;
           {\cal K}_2\big(\big|s\, \rb{r}\Dmfp^I - \rb{r}\Dmfm^J\big| \!\big/\!\la\big)
       +\; {\cal K}_2\big(\big|\rb{r}\Dmfp^I - s\, \rb{r}\Dmfm^J\big| \!\big/\!\la\big)
         \;\Big\}
    \nn
    \\[-5ex]\nn
\end{align}
und f"ur die~\mbox{$\bm\mfp\bm\mfp$-Funk}\-tion~-- vgl.\@ Gl.~(\ref{chNC_IJ-mfpmfp}):
\vspace*{-1.25ex}
\begin{align} \label{ChC-pp}
&\tilde{X}\idx{\mfp\mskip-2mu\mfp}\oC\;
   =\; \iIM\; \frac{1}{12}\vv
      \sum\Big._{\!I,J\equiv Q,\AQ}\vv
      {\rm sign}_{I,J}
    \\[-.25ex]
  &\hspace*{-8pt}
   \phantom{\tilde\ch\oC\; =\; -\, \det}\times\,
     \rb{r}\Dmfp^I\cdot \rb{r}\Dmfp^J\vv
        \int_0^1 ds\vv
        \int_0^1 ds'\vv
         \Big\{\;
           2\;
             \projt[(1)]{F\oC}{\rb{x}}\;
       -\; |\rb{x}|^2\;
             \projt[(2)]{F\oC}{\rb{x}}
         \;\Big\}
        \vv\Big|_{(I,J)}
    \nn \\[.75ex]
  &\phantom{\tilde{X}\idx{\mfp\mskip-2mu\mfp}\oC\;}
   =\; 2\cdot \la^4\,
         \frac{\pi}{12}\vv
      \sum\Big._{\!I,J\equiv Q,\AQ}\vv
      {\rm sign}_{I,J}
    \tag{\ref{ChC-pp}$'$} \\[-.25ex]
  &\hspace*{-8pt}
   \phantom{\tilde\ch\oC\; =\; -\, \det}\times\,
     \rb{r}\Dmfp^I\cdot \rb{r}\Dmfp^J \!\big/\! \la^2\vv
        \int_0^1 ds\vv
        \int_0^1 ds'\vv
         \Big\{\;
             {\cal K}_2(|\rb{x}| \!\big/\!\la)\;
       -\; \frac{1}{4}\, \big(|\rb{x}| \!\big/\! \la\big)^2\;
             {\cal K}_1(|\rb{x}| \!\big/\!\la)
         \;\Big\}
        \vv\Big|_{(I,J)}
    \nn \\[.75ex]
  &\hspace*{0pt}
   \text{mit}\qquad
    \rb{x}\big|_{(I,J)}\;
      \equiv\; \rb{x}\mskip-2mu(s, s')\big|_{(I,J)}\;
      \equiv\; s\, \rb{r}\Dmfp^I - s'\, \rb{r}\Dmfp^J
    \nn
    \\[-4.25ex]\nn
\end{align}
die zweiten Identit"aten mit%
  ~\mbox{$\projt[(1)]{F\oC}{\rb{x}} \!=\! -\iIM\,\pi \!\cdot\! \la^2\, {\cal K}_2(|\rb{x}| \!\big/\!\la)$} nach Gl.~(\ref{projt[F]C-Ansatz}) beziehungsweise%
  ~\mbox{$\projt[(2)]{F\oC}{\rb{x}} \!=\! -\iIM\,\frac{\pi}{2} \!\cdot\! {\cal K}_1(|\rb{x}| \!\big/\!\la)$} nach Gl.~(\ref{projt[F]C-Ansatz}$'$).
Die~\mbox{$\bm\mfm\bm\mfm$-Funk}\-tion~$\tilde{X}\idx{\mfm\mskip-2mu\mfm}\oC$ folgt aus~$\tilde{X}\idx{\mfp\mskip-2mu\mfp}\oNC$ durch "Ubergang%
  ~\mbox{\,$\rb{r}\Dmfp \!\to\! \rb{r}\Dmfm$}, ergo%
  ~\mbox{\,$\zet_1 \!\to\! \zet_2$},~\mbox{$\rb{X} \!\to\! \rb{Y}$},~\mbox{$\rb{b} \!\to\! -\rb{b}$}. \\
\indent
Wir rekapitulieren die Positionen der \mbox{$\bm\mfp$-(Anti)Quarks}:%
\FOOT{
  relativ zum gemeinsamen Apex~\mbox{$\rbG{\om} \!\equiv\! \om^i \tilde{e}_{(i)}$} der verallgemeinerten Pyramiden~$P\Dmfp$,~$P\Dmfm$
}
%
\vspace*{-1ex}
\begin{alignat}{2} \label{r_Q,AQ-mfp-REP}
&\rb{r}\Dmfp^Q\;&
  &\equiv\; \phantom{-\,} \bzet_1\, \rb{X}\; +\; \rb{b} \!\big/\! 2
    \\[-.25ex]
&\rb{r}\Dmfp^{\AQ}\;&
  &\equiv\;           -\, \zet_1\, \rb{X}\; +\; \rb{b} \!\big/\! 2
    \tag{\ref{r_Q,AQ-mfp-REP}$'$}
    \\[-4.75ex]\nn
\end{alignat}
vgl.\@ die Gln.~(\ref{r_Q,AQ-mfp}),~(\ref{r_Q,AQ-mfp}$'$),~-- und invertiert:
\vspace*{-1ex}
\begin{alignat}{2} \label{X,b-r_Q,AQ-mfp}
&\rb{X}\;&
  &=\; \rb{r}\Dmfp^Q\; -\; \rb{r}\Dmfp^\AQ
    \\[-.25ex]
&\rb{b} \!\big/\! 2\;&
  &=\; \zet_1\, \rb{r}\Dmfp^Q\; +\; \bzet_1\, \rb{r}\Dmfp^\AQ
    \tag{\ref{X,b-r_Q,AQ-mfp}$'$}
    \\[-4.75ex]\nn
\end{alignat}
f"ur Ausdehnung beziehungsweise Impakt.
Die \mbox{$\bm\mfp$-Gr"o}\-"sen folgen durch "Ubergang
\vspace*{-.75ex}
\begin{align} \label{Subst:mfp->mfm-REP}
\zet_1\;
    \longrightarrow\; \zet_2\,,\qquad
  \rb{X}\;
    \longrightarrow\; \rb{Y}\,,\qquad
  \rb{b}\;
    \longrightarrow\; -\rb{b}
    \\[-4.75ex]\nn
\end{align}
vgl.\@ Gl.~(\ref{Subst:mfp->mfm}$'$).
Vor dem Hintergrund dieser Notation diskutieren wir wie folgt die Funktionen~$\tilde{X}\idx{\imath\jmath}\oC$,~$\tilde{X}\idx{\imath\jmath}\oNC$ f"ur Indexpaare~\mbox{$\imath\jmath \!\equiv\! \mfp\mfp$},~$\mfm\mfm$ und~$\mfp\mfm$.
\vspace*{-.5ex}

\bigskip\noindent
Wir stellen fest {\it "Ubereinstimmung\/} der \mbox{$\bm\mfp\bm\mfm$-Funk}\-tionen mit den entsprechenden Ausdr"ucken, die wir herleiten in Ref.~\cite{Kulzinger95}: vgl.\@ die Gln.~(\ref{ChNC-pm}$'$),~(\ref{ChC-pm}$'$) hier mit Gl.~(4.33) dort%
\FOOT{
  Wir weisen hin auf auf den Faktor~$12$, um den die \mbox{$\tilde\ch$-Funk}\-tionen dort gr"o"ser definiert sind als hier.
},~-- 
da {\it per~constructionem\/} der \mbox{$\mfp$-,$\mfm$-Koor}\-dinatenlinien der separierte Faktor%
  ~\mbox{\,$-\det\mathbb{SL}\, g_{\mfp\mfm}$} "ubergeht in%
  ~\mbox{\,$-\det\mathbb{L}\, g_{+-} \!\equiv\! 1$} im Limes~\mbox{$s \!\to\! \infty$}; vgl.\@ die Diskussion der Gln.~(\ref{-detSLgmf_s}),~(\ref{-detSLgmf_s}$'$). \\
\indent
Wir stellen fest \vspace*{-.125ex}in den gestrichenen Darstellungen~-- die Gln.~(\ref{ChNC-pm}$'$),~(\ref{ChNC-pp}$'$),~(\ref{ChC-pm}$'$) und (\ref{ChC-pp}$'$)~-- den {\it globalen Faktor\/}~$\la^4$, wenn~-- wie geschehen~-- die Vektoren%
  ~\vspace*{-.125ex}$\rb{r}\Dmfp^Q$,~$\rb{r}\Dmfp^\AQ$ und $\rb{r}\Dmfm^Q$,~$\rb{r}\Dmfm^\AQ$ konsistent skaliert sind mit~$\la$.
Diese Vektoren selbst sind bereits implizit skaliert mit der Korrelationsl"ange~$a$, vgl.\@ Gl.~(\ref{x=ximath-xjmath}$'$),~-- "uber deren Bedeutung als {\it Korrelations\/}-L"ange genau~$\la$ fixiert wird, vgl.\@ die zweite Relation in Gl.~(\ref{Ala_Bedingg}) bzw.~(\ref{APP:Ala_Bedingg}) und explizit Anh.~\ref{APP-Subsect:la_nu}.
Diese Skalierung~$a^{-1}$ der Vektoren~-- ergo die Skalierung~$a^{-4}$ der $\ch$- und \mbox{$\tilde\ch$-Funk}\-tionen~-- impliziert Skalierung mit~$a^4$ des faktorisierten Gluonkondensats~$\vac{g^2 FF}$; zusammen mit dem Faktor~$\la^4$ also~$(a\la)^4$.
Es ist folglich das Produkt~$(a\la)$ die eigentliche Korrelationsl"ange des stochastischen Proze"ses, der zugrunde liedt dem Eichfeld-Vakuumerwartungswert der Quantenchromodynamik;~--~$a$ ist numerisch modifiziert durch den Faktor~\mbox{\,$\la \!=\! 8\!\big/\!3\pi \!\cong\! 0.849$}. \\
\indent
Die ungestrichenen Darstellungen~-- die Gln.~(\ref{ChNC-pm}),~(\ref{ChNC-pp}),~(\ref{ChC-pm}) und (\ref{ChC-pp})~-- sind allgemein g"ultig f"ur nicht weiter spezifizierte Korrelationsfunktionen~$F\oC$,~$F\oNC$, die gestrichenen Darstellungen ist g"ultig auf Basis unseres speziellen Ansatzes f"ur diese.
Analytische L"osbarkeit der \mbox{${\cal K}_\mu$-Inte}\-grale ist daher denkbar und sei diskutiert bez"uglich der Funktion%
  ~\vspace*{-.125ex}$\tilde{X}\idx{\mfp\mskip-2mu\mfp}\oC$ nach Gl.~(\ref{ChC-pp}$'$).
Es gilt~-- mit%
  ~\vspace*{-.125ex}\mbox{\,$I,J \!\in\! \{Q,\AQ\}$}:
\vspace*{-.5ex}
\begin{align} \label{|rbx|_IJ-explizit}
\hspace{-4pt}
 \big|\rb{x}\mskip-2mu(s, s')\big|\, \Big._{(I,J)}\;
      \equiv\; \big|s\, \rb{r}\Dmfp^I - s'\, \rb{r}\Dmfp^J\big|\;
  =\; \left\{
  \begin{alignedat}{2}
    \vv &|s - s'|\cdot |\rb{r}\Dmfp^I|&
       &\qquad I \equiv J
    \\[.5ex]
    \vv &\big|(s \!-\! s') \rb{b} \!\big/\! 2
               + (\zet_1\, s \!+\! \bzet_1\, s') \rb{X}\big|&
       &\qquad I \nequiv J
  \end{alignedat}
      \right.
    \\[-4.5ex]\nn
\end{align}
Wir betrachten daher:%
\FOOT{
  Die Parameter~\mbox{$n \!\in\! \bbbn_0$},~\mbox{${\rm Re}\mu \!>\! 0$} und~$\rb{a}$,~$\rb{a}'$ seien in~praxi gesetzt entsprechend den Gln.~(\ref{ChC-pp}$'$),~(\ref{|rbx|_IJ-explizit}).
}
%
\vspace*{-.75ex}
\begin{align} \label{intPARA-s,s'-ze,ze'}
&\hspace*{-4pt}
 I_{n,\mu}\mskip-3mu(\rb{a},\rb{a}')\;
   \equiv\; \int_0^1 ds\vv
           \int_0^1ds'\vv
             \big(|\rb{x}| \!\big/\! \la\big)^n\;
             {\cal K}_\mu(|\rb{x}| \!\big/\!\la)
    \\[-.25ex]
  &\hspace*{-4pt}
   \phantom{I_{n,\mu}\mskip-3mu(\rb{a},\rb{a}')\;}
   =\; \int_{-1}^1 d\ze'\vv
           \int_0^{|\ze'|}d\ze\vv
             \big(|\rb{x}| \!\big/\! \la\big)^n\;
             {\cal K}_\mu(|\rb{x}| \!\big/\!\la)\quad
   \equiv\; \int_{-1}^1 d\ze'\vv I'_{n,\mu}\mskip-3mu(\rb{a},\rb{a}';\ze')
    \tag{\ref{intPARA-s,s'-ze,ze'}$'$} \\
  &\phantom{I_{n,\mu}}\text{mit}\quad
   |\rb{x}|
     \equiv \big|(s \!-\! s')\rb{a} + (\zet_1\, s \!+\! \bzet_1\, s')\rb{a}'\big|\qquad
   \text{bzw.}\quad
   |\rb{x}|
     \equiv |\ze\, \rb{a} + \ze' \rb{a}'|
    \nn
    \\[-4.5ex]\nn
\end{align}
die zweite Darstellung unter Substitution%
  ~\vspace*{-.125ex}\mbox{$(s,s') \!\to\! (\ze,\ze')$} mit%
  ~\mbox{$\ze \!\equiv\! s \!-\! s'$},~{$\ze' \!\equiv\! \zet_1s \!+\! \bzet_1s'$} und Jacobi-Determinante~$+1$.
Die Funktion%
  ~\vspace*{-.125ex}$\tilde{X}\idx{\mfp\mskip-2mu\mfm}\oC$ nach Gl.~(\ref{ChC-pm}$'$) ist bestimmt durch ein Integral vom Typ des inneren \mbox{$\ze$-Inte}\-grals%
  ~\vspace*{-.375ex}\mbox{\,$I'_{n,\mu}\mskip-3mu(\rb{a},\rb{a}';\ze')$} mit~\mbox{\,$\ze' \!\equiv\! 1$}.
Es gilt
\vspace*{-.5ex}
\begin{align} \label{|rbx|_al,al'}
&|x| \!\big/\! \la\;
  \equiv\; a \!\big/\! \la \!\cdot\! \sqrt{(\ze \!+\! \al)^2 \!-\! \al'{}^2}
    \\
  &\phantom{I_{n,\mu}}\text{mit}\qquad
   \al \equiv \rb{a} \!\cdot\! (\ze'\rb{a'}) \big/ a^2\,,\quad
   \al' \equiv \al^2 - (\ze'{a}')^2 \!\big/\! a^2\qquad
   \text{und}\quad
     a \equiv |\rb{a}|\,,\vv
     a' \equiv |\rb{a}'|
    \nn
    \\[-4.5ex]\nn
\end{align}
und f"ur das \mbox{$\ze$-Inte}\-gral unter Substitution~\mbox{$\ze \!\to\! z \!\equiv\! |x| \!\big/\! \la$}:
\vspace*{-.75ex}
\begin{align} \label{intPARA'-ze,ze'}
&I'_{n,\mu}\mskip-3mu(\rb{a},\rb{a}';\ze')\;
   =\; \int_0^{|\ze'|}\; d\ze\vv
             \big(|\rb{x}| \!\big/\! \la\big)^{\!n}\;
             {\cal K}_\mu(|\rb{x}| \!\big/\!\la)\qqquad\hspace*{4pt}
   |\rb{x}|\; \equiv\; |\ze\, \rb{a} + \ze' \rb{a}'|
    \\[-.25ex]
  &\phantom{I_{n,\mu}\mskip-3mu(\rb{a},\rb{a}';\ze')\;}
   =\; \big(a \!\big/\! \la\big)^{\!-3\!/\!2}\;
         \int_{z_\star}^{z^\star} dz\vv
         \frac{z^{n+1}\, {\cal K}_\mu\!(z)}{%
               \sqrt{z^2 \!+\! \al''{}^2}}\qqquad
  \al'' \equiv a \!\big/\! \la \!\cdot\! \al'
    \tag{\ref{intPARA'-ze,ze'}$'$} \\[.5ex]
  &\phantom{I_{n,\mu}}\text{mit}\qquad
    z_\star \equiv a \!\big/\! \la \!\cdot\!
                        \sqrt{\al^2 \!-\! {\al'}^2}\qquad
    z^\star \equiv a \!\big/\! \la \!\cdot\!
                        \sqrt{(|\ze'| \!+\! \al)^2 \!-\! \al'{}^2}\qquad
    \nn
    \\[-4.75ex]\nn
\end{align}
Daraus folgt f"ur~\mbox{\,$\rb{a}' \!\equiv\! \bm0 \Rightarrow \al,\al',\al'' \!\equiv\! 0$}:
\vspace*{-.75ex}
\begin{align} \label{intPARA'-ze,ze'-rba'=0}
I'_{n,\mu}\mskip-3mu(\rb{a},\bm0;\ze')\;
   =\; \big(a \!\big/\! \la\big)^{\!-3\!/\!2}\;
         \int_0^{a \!/\! \la \cdot |\ze'|}\; dz\vv
         z^n\, {\cal K}_\mu\!(z)
    \\[-4.75ex]\nn
\end{align}
In Ref.~\cite{Gradstein81}, Gradstein,~Ryshik~5.52.1 ist angegeben ein Integral%
\FOOT{
  eigentliche crux ist, da"s wir die Integrale ben"otigen als {\sl unbestimmte\/} Integrale 
},
aus dem unmittelbar folgt das \mbox{$\rb{a}' \!\equiv\! \bm0$-Inte}\-gral nach Gl.~(\ref{intPARA'-ze,ze'-rba'=0}) f"ur Indizes%
  ~\mbox{$n \!\equiv\! 1,\, \forall\mu \!\in\! \bbbc$}; ungeradzahlige Indizes~$n$ folgen durch partielle Integration.
F"ur geradzahlige Indizes~$n$ kann analytisch kein Ausdruck angegeben werden,~-- ergo nicht f"ur die $(Q,Q)$- und \mbox{$(\AQ,\AQ)$-Ter}\-me der Funktion%
  ~\vspace*{-.125ex}$\tilde{X}\idx{\mfp\mskip-2mu\mfp}\oC$~[$\tilde{X}\idx{\mfm\mskip-2mu\mfm}\oC$] nach Gl.~(\ref{ChC-pp}$'$).
Desgleichen kann analytisch kein Ausdruck angegeben werden%
\nopagebreak~f"ur das allgemeinere \mbox{$\rb{a}' \!\nequiv\! \bm0$-Inte}\-gral nach Gl.~(\ref{intPARA'-ze,ze'}$'$),~-- \mbox{ergo nicht f"ur die Integrale der $(Q,\AQ)$- und} \mbox{$(\AQ,Q)$-Terme} der Funktion%
  ~\vspace*{-.25ex}$\tilde{X}\idx{\mfp\mskip-2mu\mfp}\oC$~[$\tilde{X}\idx{\mfm\mskip-2mu\mfm}\oC$] und die der Funktion%
  ~$\tilde{X}\idx{\mfp\mskip-2mu\mfm}\oC$ nach Gl.~(\ref{ChC-pm}$'$). \\
\indent
Die Darstellung durch das Integral%
  ~\vspace*{-.125ex}\mbox{\,$I'_{n,\mu}\mskip-3mu(\rb{a},\rb{a}')$} nach Gl.~(\ref{intPARA-s,s'-ze,ze'}$'$) mit dem inneren \mbox{$\ze$-Inte}\-gral%
  ~\vspace*{-.125ex}\mbox{\,$I'_{n,\mu}\mskip-3mu(\rb{a},\rb{a}';\ze')$} nach Gl.~(\ref{intPARA'-ze,ze'}$'$) beziehungsweise~(\ref{intPARA'-ze,ze'-rba'=0}) kann aber zweckm"a"sig zugrunde gelegt werden einer numerischen Analyse.
\vspace*{-.5ex}

\bigskip\noindent
Qualitativ k"onnen die Ausdr"ucke der \vspace*{-.125ex}$\bm\mfp\bm\mfm$-, $\bm\mfp\bm\mfp$- und~\mbox{$\bm\mfm\bm\mfm$-Funk}\-tionen~-- einschlie"slich der sie wesentlich konstituierenden Integalausdr"ucke~-- leicht in bekannten Zusammenhang gebracht werden.
Streuung im Limes~\mbox{$s \!\to\! \infty$} ist bestimmt wegen%
  ~\vspace*{-.125ex}\mbox{\,$-\det\mathbb{SL}\, g_{\mfp\mfp} \!\to\! -\det\mathbb{L}\, g_{++} \!\equiv\! 0$} und%
  ~\vspace*{-.125ex}\mbox{\,$-\det\mathbb{SL}\, g_{\mfm\mfm} \!\to\! -\det\mathbb{L}\, g_{--} \!\equiv\! 0$} vollst"andig bestimmt durch die Funktionen%
  ~$\tilde{X}\idx{\mfp\mskip-2mu\mfm}\oNC$,~$\tilde{X}\idx{\mfp\mskip-2mu\mfm}\oC$ nach den Gln.~(\ref{ChNC-pm}$'$),~(\ref{ChC-pm}$'$).
Diese sind bekannt aus fr"uheren Arbeiten; Referenzen und eine ausf"uhrliche Diskussion folgen in Kapitel~\ref{Kap:GROUND}, vgl.\@ insbes.\@ die Abschnitte~\ref{Subsect:TransversaleKonfiguration} und~\ref{Subsect:Loop-Loop-Streuung}. \\
\indent
An dieser \vspace*{-.125ex}Stelle seien qualitativ diskutiert die (Integral)Ausdr"ucke, die konstituieren die \mbox{$\bm\mfp\bm\mfp$-Funk}\-tionen%
  ~\vspace*{-.25ex}$\tilde{X}\idx{\mfp\mskip-2mu\mfp}\oNC$,~$\tilde{X}\idx{\mfp\mskip-2mu\mfp}\oC$ nach den Gln.~(\ref{ChNC-pm}$'$),~(\ref{ChC-pm}$'$)~-- und analog die \mbox{$\bm\mfm\bm\mfm$-Funk}\-tionen~$\tilde{X}\idx{\mfm\mskip-2mu\mfm}\oNC$,~$\tilde{X}\idx{\mfm\mskip-2mu\mfm}\oC$.
Die durch sie vermittelten Beitr"age verschwinden erst im Limes~\mbox{$s \!\to\! \infty$} aufgrund des verschwindenden Komponente~$g_{\mfp\mfp}$ des metrischen Tensors als Vorfaktor.
Sie sind Konsequenz dessen, da"s f"ur endliches~$s$ der Vakuumerwartungswert eines einzelnen Wegner-Wilson-Loops nicht identisch Eins ist, da"s hei"st der nicht-verschwindenden internen Wechselwirkung der (Anti)Quarks des Loops.
Die Funktionen%
  ~\vspace*{-.25ex}$\tilde{X}\idx{\mfp\mskip-2mu\mfp}\oNC$,~$\tilde{X}\idx{\mfp\mskip-2mu\mfp}\oC$ sollten in Zusammenhang stehen mit der entsprechenden Gr"o"se eines statischen, das hei"st eines Quark-Antiquark-Paares im entgegengesetzten Limes verschwindender kinetischer Schwerpunktenergie: mit der (statischen) Stringspannung~$\si$, die dessen Confinement subsumiert.
Sie sollten repr"asentieren die Stringspannung eines schnell bewegten Quark-Antiquark-Paares.
Sei verwiesen auf die Diskussion in Zusammenhang der Gln.~(\ref{WW-Loop_arealaw}) und~(\ref{QbarQ_Potential}) auf Seite~\pageref{WW-Loop_arealaw}f.\\
\indent
Die \mbox{$\bm\mfp\bm\mfp$-Funk}\-tionen repr"asentieren die Integration zweier paralleltransportierter Feldst"ar\-ken bez"uglich desselben Wegner-Wilson-Loops~$W\Dmfp$, die {\it korreliert\/} sind als Konsequenz der Cluster-Eigenschaft der Kumulante Zweiter Ordnung, durch die wesentlich bestimmt ist der Korrelations-Lorentztensor und daher die Funktionen%
  ~\vspace*{-.25ex}$\tilde{X}\idx{\mfp\mskip-2mu\mfp}\oNC$,~$\tilde{X}\idx{\mfp\mskip-2mu\mfp}\oC$.
Korrelationsl"ange ist~$(a\la)$, St"arke bei maximaler Korrelation ist Eins per definitionem.
Die integrierte Funktion kann daher approximiert werden als Eins, falls ihre Argumente nicht weiter auseinander liegen als~$(a\la)$, \vspace*{-.125ex}und Null sonst. \\
\indent
F"ur die konfinierende Funktion%
  ~\vspace*{-.125ex}$\tilde{X}\idx{\mfp\mskip-2mu\mfp}\oC$ nach Gl.~(\ref{ChC-pp}$'$) folgt bei festgehaltener ersten Feldst"arke f"ur das Integral der zweiten genau%
  ~\vspace*{-.125ex}\mbox{\,$(a\la) \!\cdot\! 1$}, dann f"ur das Integral der ersten~Feld\-st"arke genau~$\big|{\cal S}\Dmfp\Doperp\big|$, der Betrag des Integrationsgebietes:
%
\begin{align} \label{ChC-pp-app}
\tilde{X}\idx{\mfp\mskip-2mu\mfp}\oC\;
  \propto\; \big|{\cal S}\Dmfp\Doperp\big|\quad\vv
  \underset{|\rb{X}| \!/\!\la \gg 1}{\sim}\vv
            \big|{\cal S}\Dmfp\Doperp\big._{\rm min.}\big|\;       
              \equiv\; |\rb{X}| \!\big/\!\la
    \\[-4.25ex]\nn
\end{align}
Dazu in Kontrast gilt f"ur die nicht-konfinierende Funktion%
  ~\vspace*{-.125ex}$\tilde{X}\idx{\mfp\mskip-2mu\mfp}\oNC$ nach Gl.~(\ref{ChNC-pp}$'$):
\vspace*{-.125ex}
\begin{align} \label{ChNC-pp-app}
\tilde{X}\idx{\mfp\mskip-2mu\mfp}\oNC\;
  \propto\; \Big\{\;
              1\; -\; \efn{\D-|\rb{X}| \!\big/\!\la}\vv P_3(|\rb{X}| \!\big/\!\la)
            \Big\}\quad\vv
  \underset{|\rb{X}| \!/\!\la \gg 1}{\sim}\vv
            \Big\{\;
              1\; -\; \efn{\D-|\rb{X}| \!\big/\!\la}
            \Big\}\;
    \\[-4.5ex]\nn
\end{align}
da sich%
  ~\vspace*{-.125ex}\mbox{${\cal K}_\mu\!(x)$} f"ur~\mbox{\,${\rm Re}\mu \!>\! 0,\, x \!\in\! \bbbr^+$} verh"alt wie~\mbox{$\sim\!\efn{\D-x} \!\cdot\! P_\mu\!(x)$}, mit~$P_\mu$ einer Potenzreihe, die normiert ist auf Eins \vspace*{-.125ex}f"ur verschwindendes Argument. \\
\indent
Die konventionelle Stringspannung des Systems des $\mfp$-Quarks und -Antiquarks~-- durchgef"uhrt der Limes~\mbox{\,$T\Dmfp \!\to\! \infty$}~-- ist genau der Koeffizient in dessen Potential vor deren minimaler transversaler Separation~$|\rb{X}|$.
Die Gln.~(\ref{ChC-pp-app}),~(\ref{ChNC-pp-app}) dokumentieren einen Beitrag durch die konfinierende $C$-, nicht aber durch die nicht-konfinierende \mbox{$N\!C$-Tensor}\-struktur.
Dieses Resultat ist notwendig f"ur Konsistenz damit, da"s die statische Stringspannung abh"angt von der $C$-, nicht aber von der $N\!C$-Korrelationsfunktion, vgl.\@ Gl.~(\ref{Stringspannung-si_DC}) auf Seite~\pageref{Stringspannung-si_DC}.
Und es legitimiert die Bezeichnung "`konfinierend"' und "`nicht-konfinierend"' auch f"ur die%
  ~\vspace*{-.125ex}$\bm\mfp\bm\mfp$- und \mbox{$\bm\mfm\bm\mfm$-Funk}\-tionen der entsprechenden Struktur%
  ~\mbox{$t\oC{}_{\zzzz \tilde\mu\tilde\nu\tilde\rh\tilde\si}$},%
  ~\mbox{$t\oNC{}_{\zzzz \tilde\mu\tilde\nu\tilde\rh\tilde\si}$} des Korrelationstensors.

Wir konstatieren, da"s die Funktionen%
  ~$\tilde\ch\idx{\imath\jmath}\oC$,~$\tilde\ch\idx{\imath\jmath}\oNC$ mit unabh"angigen Indexpaaren~$\mfp\mfp$,~$\mfm\mfm$ und~$\mfp\mfm$ bestimmt sind durch die Geometrie in der {$x^1\!x^2$-Trans}\-versalebene, in die aber projiziert ist die longitudinale Dynamik der Streuung in Form der Abh"angigkeit des relevanten Impaktvektors~$\rb{b}$~-- vgl.\@ Gl.~(\ref{X,b-r_Q,AQ-mfp}$'$) und~(\ref{Subst:mfp->mfm-REP})~-- von~$\zet_1$,~$\zet_2$, den Anteilen der respektiven Quarks am gesamten Lichtkegelimpuls der Wegner-Wilson-Loops~$W\Dmfp$,~$W\Dmfm$.
Diese Anteile~$\zet_1$,~$\zet_2$ sind Lorentz-invariant, wie wir zeigen in Anhang~\ref{APP-Subsect:LCWFN-Kovarianz}.
Sie h"angen folglich nicht ab von den aktiven Lorentz-Boosts der Quark-Antiquark-Systeme~"`\bm\mfp"' und~"`\bm\mfm"' gegen einander, insbesondere nicht von~$s$, dem durch diese Boosts kontrollierten Quadrat ihrer invarianten Schwerpunktenergie.
Folglich h"angt auch nicht ab von~$s$: die Geometrie in der {$x^1\!x^2$-Trans}\-versalebene~-- und die Funktionen%
  ~$\tilde\ch\idx{\imath\jmath}\oC$,~$\tilde\ch\idx{\imath\jmath}\oNC$. \\
\indent
Die $s$-Abh"angigkeit Funktionen%
  ~\vspace*{-.125ex}$\tilde\ch\idx{\imath\jmath}\oC$,~$\tilde\ch\idx{\imath\jmath}\oNC$ f"ur festes Indexpaar~$\imath\jmath$ ist vollst"andig subsumiert in dem separierten Faktor%
  ~\vspace*{-.125ex}\mbox{\,$-\det\mathbb{SL}\, g_{\imath\jmath}$}, der bestimmt ist durch die korrespondierende Komponente%
  ~\vspace*{-.125ex}$g_{\imath\jmath}$ des metrischen Tensors bez"uglich der Koordinatenlinien~$\mfp$,~$\mfm$.
Diese Faktoren sind vorab der Auswertung der \mbox{$\ch$-Funk}\-tionen angegeben in den Gln.~(\ref{-detSLgmf_s}),~(\ref{-detSLgmf_s}$'$); wir rekapitulieren diese in Form
\begin{samepage}
\vspace*{-.5ex}
\begin{alignat}{3} \label{-detSLgmf_s-simplified}
&-\det\mathbb{SL}\vv g_{\mfp\mfp}\;&
   &=\; -\det\mathbb{L}\vv g_{\mfp\mfp}\;&
   &=\; g_{\mfp\mfp} \big/ g_{+-}
    \\[-.25ex]
 &&&=\; \sinh^{-1}\!\ps\;&&
    \nn \\[.5ex]
&-\det\mathbb{SL}\vv g_{\mfp\mfm}\;&
   &=\; -\det\mathbb{L}\vv g_{\mfp\mfm}\;&
   &=\; g_{\mfp\mfm} \big/ g_{+-}
    \tag{\ref{-detSLgmf_s-simplified}$'$} \\[-.25ex]
 &&&=\; \tanh^{-1}\!\ps\;&&\hspace*{-12pt}
    =\; 1\; +\; \efn{\D-\ps}\, \sinh^{-1}\!\ps
    \nn
    \\[-4.5ex]\nn
\end{alignat}
und diskutieren sie wie folgt. \\
\indent
Die Koordinatenlinien~$\mfp$,~$\mfm$ sind definiert als die Richtungen der physikalischen nahezu lichtartigen Weltlinien der (Anti)Quarks, die exakt auf dem Lichtkegel verlaufen im Limes%
  ~\mbox{\,$s \!\to\! \infty$}.
Dies suggeriert zu fordern, da"s die Koordinaten%
  ~\mbox{$\tilde\mu \!\in\! \{\mfp,\mfm,1,2\}$} "ubergehen in Lichtkegelkoordinaten%
  ~\mbox{\,$\bar\mu \!\in\! \{+,-,1,2\}$} und entsprechend die longitudinalen Komponenten des metrischen Tensors:%
  ~\vspace*{-.125ex}\mbox{\,$g_{\mfp\mfm} \!\to\! g_{+-}$},~\mbox{\,$g_{\mfp\mfp} \!\to\! g_{++} \!\equiv\! 0$} bez"uglich der unabh"angigen Komponenten.
Dies ist simultan gegeben bei Forderung von {\it L"angentreue\/} der vermittelnden Abbildung, das hei"st, da"s ihre Determinante identisch Eins ist oder "aquivalent die Determinante des metrischen Tensors invariant:
\vspace*{-.5ex}
\begin{align} \label{vrh-Forderung}
&\hspace*{-8pt}
 \det \tilde{g}\;
    \equiv\; \det\! \big(g_{\tilde\mu\tilde\nu}\big)\vv
  \stackrel{\D!}{=}\vv
  \det \bar{g}\;
    \equiv\; \det\! \big(g_{\bar\mu\bar\nu}\big)
    \\[-.5ex]
&\hspace*{-8pt}
 \Longleftrightarrow\qquad
 1\vv
    \stackrel{\D!}{=}\vv \det \mathbb{S}\;
    \equiv\; \det\! \big(\mathbb{S}^{\tilde\mu}{}_{\bar\nu}\big)\;
    =\; \frac{\sqrt{-\det \bar{g}}}{\sqrt{-\det \tilde{g}}}
 \qqquad\text{f"ur alle\vv $\ps \in (0,\infty)$}
    \tag{\ref{vrh-Forderung}$'$}
    \\[-4.75ex]\nn
\end{align}
Durch diese Forderung wird fixiert in Anhang~\vspace*{-.125ex}\ref{APP-Sect:Minkowski}~-- vgl.\@ Gl.~(\ref{APP:vrh-Forderung})ff.\@ auf Seite~\pageref{APP:vrh-Forderung}~-- die Normierungskonstante%
  ~\mbox{$\vrh \!\in\! \bbbr^+$} der Koordinaten~\mbox{$\tilde\mu \!\in\! \{\mfp,\mfm,1,2\}$}. \\
\indent
Die ersten Identit"aten der Gln.~(\ref{-detSLgmf_s-simplified}),~(\ref{-detSLgmf_s-simplified}$'$) sind unmittelbar die Forderung%
  ~\vspace*{-.125ex}\mbox{\,$\det\mathbb{S} \!\equiv\! 1$} nach Gl.~(\ref{vrh-Forderung}$'$).
Die zweiten Identit"aten folgen aus%
  ~\vspace*{-.125ex}\mbox{$-\det\mathbb{L} \!\equiv\! (g_{+-})^{-1}$}, dem Zusammenhang der Determinanten der Transformation von konventionellen Koordinaten%
  ~\mbox{\,$\mu \!\in\! \{0,1,2,3\}$} auf Lichtkegelkoordinaten und der einen unabh"angigen Komponente~\vspace*{-.125ex}$g_{+-}$ deren metrischen Tensors~-- unabh"angig von deren Normierung~$\al$.
Die Faktoren%
  ~\vspace*{-.125ex}\mbox{$-\det\mathbb{SL}\, g_{\imath\jmath}$} sind in der Tat unabh"angig von der Konvention bez"uglich der Lichtkegelkoordinaten, da {\it per~constructionem\/}%
  ~\vspace*{-.125ex}\mbox{\,$g_{\mfp\mfm} \!\to\! g_{+-}$} und%
  ~\mbox{\,$g_{\mfp\mfp} \!\to\! g_{++} \!\equiv\! 0$} und daher im Grenzwert der \mbox{$\al$-ab}\-h"angige Nenner~$g_{+-}$ gek"urzt wird beziehungsweise multipliziert wird mit Null.
F"ur endliche Werte von~$s$, das hei"st~$\ps$ ist diese Unabh"angigkeit dokumentiert in den expliziten Ausdr"ucken~-- vgl.\@ die zweiten Zeilen der Gln.~(\ref{-detSLgmf_s-simplified}),~(\ref{-detSLgmf_s-simplified}$'$)~-- die folgen auf Basis der gestellten Forderung. \\
\indent
Die Faktoren%
  ~\vspace*{-.125ex}\mbox{\,$-\det\mathbb{SL}\, g_{\imath\jmath}$} sind gegeben als einfache Funktionen des {\it hyperbolischen Winkels\/}~$\ps$.
Dieser ist bestimmt als die Summe
\end{samepage}%
\vspace*{-.25ex}
\begin{align} \label{ps_ps-mf}
&\ps\;
  =\; \ps\Dmfp\; +\; \ps\Dmfm
    \\[-4.5ex]\nn
\end{align}
mit
\begin{samepage}
\vspace*{-1.25ex}
\begin{align} \label{ps-mf_be-mf}
\be\Dmfp\;
  =\; \tanh\ps\Dmfp\qquad
 \text{und}\qquad
 \be\Dmfm\;
  =\; \tanh\ps\Dmfm
    \\[-4.125ex]\nn
\end{align}
den Beta-Parametern, das hei"st den Geschwindigkeiten der aktiven Lorentz-Boost der Weg\-ner-Wilson-Loops~$W\Dmfp$,~$W\Dmfm$ gegen einander.
Durch diese wird unmittelbar kontrolliert das Quadrat~$s$ der invarianten Schwerpunktenergie des Loop-Loop-Systems.
Sei~$\be\Dmfm$~[$\ps\Dmfm$] gew"ahlt als Funktion von~$\be\Dmfp$~[$\ps\Dmfp$], so da"s der Schwerpunkt des Systems (longitudinal) ruht.
Dann ist der Limes~\mbox{\,$s \!\to\! \infty$} realisiert durch~\mbox{\,$\be\Dmfp \!\to\! 1$}, ergo~\mbox{\,$\ps\Dmfp \!\to\! \infty$}, das hei"st~\mbox{\,$\ps \!\to\! \infty$}.
Wir betrachten Werte dieser Parameter {\it nahe\/} dieser Grenzwerte, aber von diesen verschieden:%
  ~\mbox{\,$\be\Dmfp \!<\! 1$}, ergo~\mbox{\,$\ps\Dmfp \!<\! \infty$}.
Endlich Werte von~$s$ sind realisiert durch endliche Werte von~$\ps$.
 \\
\indent
Seien~$P\Dmfp$,~$P\Dmfm$ die Vierer-Impulsvektoren der Wegner-Wilson-Loops~$W\Dmfp$,~$W\Dmfm$.
Dann sind mit diesen assoziiert die Lorentz-invarianten (Ruhe)Massen%
  ~\vspace*{-.125ex}\mbox{$M\Dmfp \!\equiv\! \surd P\Dmfp^2$},%
  ~\mbox{$M\Dmfm \!\equiv\! \surd P\Dmfm^2$} mit~\oE~\mbox{$M\Dmfp,M\Dmfm \!\gtrless\! 0$}.
Diese sind aufzufassen als Parameter der Funktion~$\ps$ von~$s$:
   \vspace*{-.125ex}\mbox{\,$\ps \!\equiv\! \ps_{M\Dmfp,M\Dmfm}\!(s)$}.
Wir verweisen auf Anhang~\ref{APP-Sect:Streuung_aktiveBoosts}, in dem~-- f"ur allgemeine~$M\Dmfp$,~$M\Dmfm$~-- hergeleitet werden explizite Relationen zwischen~$\ps$, den Beta-Parametern~$\be\Dmfp$,~$\be\Dmfm$ und dem Quadrat~$s$ der invarianten Schwerpunktenergie.
Auf Basis dieser Relationen folgen unmittelbar aus den Gln.~(\ref{-detSLgmf_s-simplified}),~(\ref{-detSLgmf_s-simplified}$'$) die Faktoren~\mbox{\,$-\det\mathbb{SL}\, g_{\imath\jmath}$}, durch die vollst"andig bestimmt ist die \mbox{$s$-Ab}\-h"angigkeit der $\tilde\ch$-Funktionen und ergo der Loop-Loop-Streuung.

\begin{figure}
\begin{minipage}{\linewidth}
  \begin{center}
  \vspace*{3ex}
  \setlength{\unitlength}{1mm}\begin{picture}(130,55.7528)   
    \put(10,0){\epsfxsize120mm \epsffile{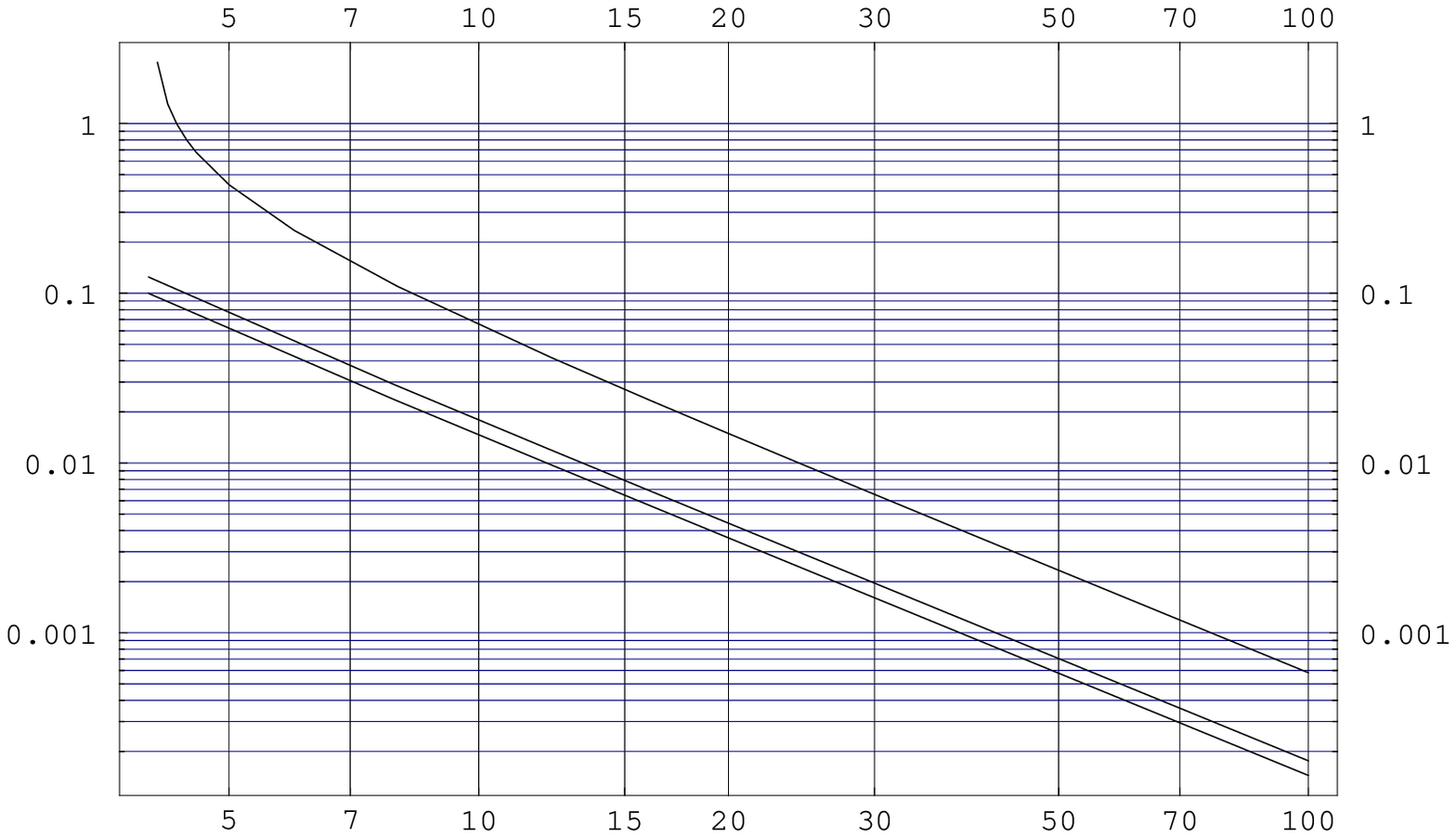}}
    \put(-5,58){\normalsize(a)\quad diagonal:}
    \put(90,49.5){\normalsize$\rh,p,\Jps \bm{-} p$}
    \put(113,0.5){\normalsize$\surd s\;[\GeV[]]$}
    \put(  9,0  ){\yaxis[55.7528mm]{\normalsize%
                    ${g_{\mfp\mfp}\big/g_{+-}}$}}
    \put( 15,0  ){\yaxis[55.7528mm]{\normalsize%
                    $\equiv\; -\det\mathbb{SL}\; g_{\mfp\mfp}\;
                       =\; \sinh^{\D-1}\!\ps$}}
  \end{picture}\\[6ex]
  \setlength{\unitlength}{1mm}\begin{picture}(130,102.308)   
    \put(10,0){\epsfxsize120mm \epsffile{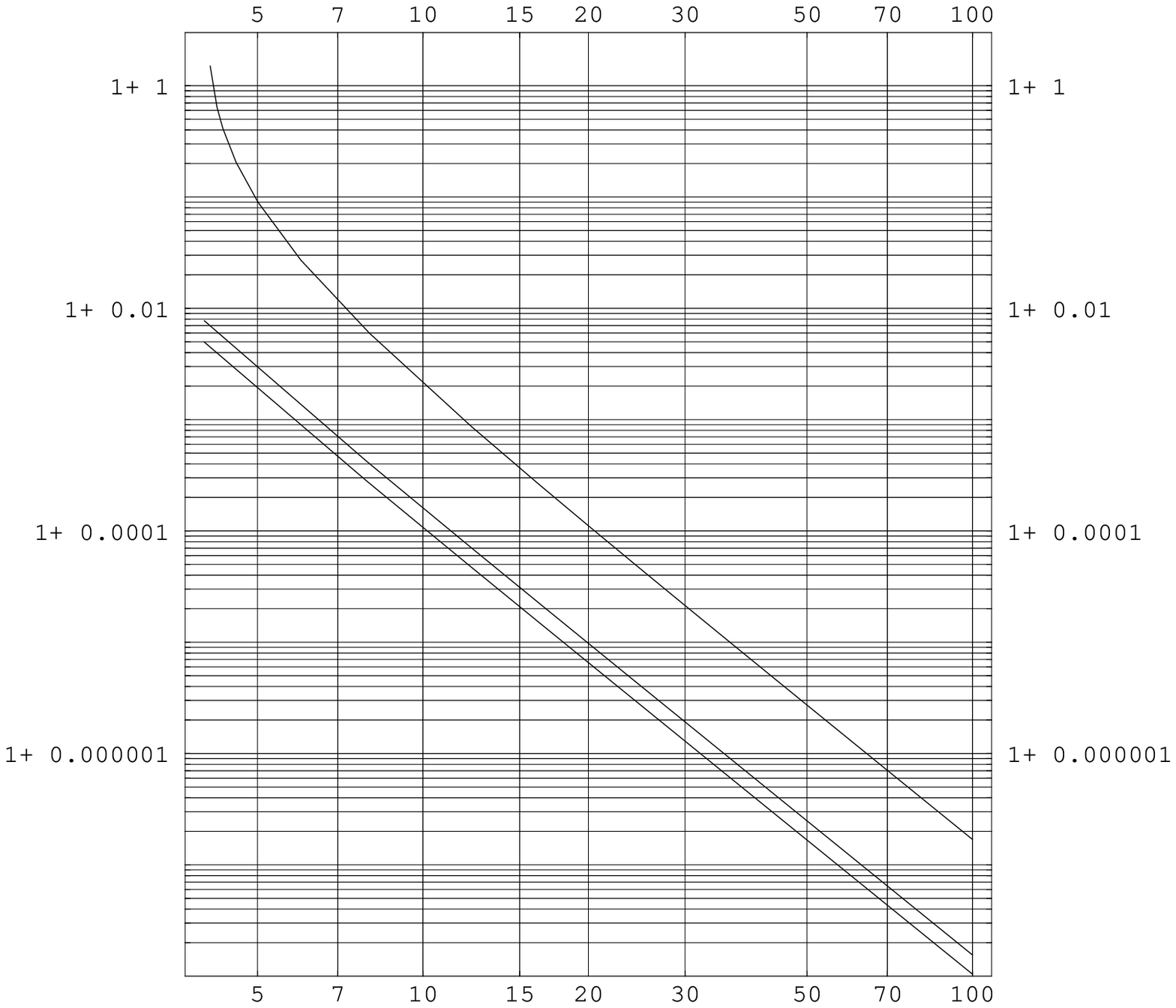}}
    \put(-5,104){\normalsize(b)\quad au"serdiagonal:}
    \put(90,95.5){\normalsize$\rh,p,\Jps \bm{-} p$}
    \put(113,0.5){\normalsize$\surd s\;[\GeV[]]$}
    \put( -1,0  ){\yaxis[102.308mm]{\normalsize%
                    ${g_{\mfp\mfm}\big/g_{+-}}$}}
    \put(  5,0  ){\yaxis[102.308mm]{\normalsize%
                    $\equiv\; -\det\mathbb{SL}\; g_{\mfp\mfm}\;
                       =\; 1 + \efn{\D-\ps} \sinh^{\D-1}\!\ps$}}
  \end{picture}
  \end{center}
\vspace*{-3ex}
\caption[\protect$\surd s$-Abh"angigkeit der Komponenten%
  ~\protect\mbox{$g_{\mfp\mfp},\, g_{\mfp\mfm}$} des metrischen Tensors]{
  Komponenten~$g_{\mfp\mfp}$,~$g_{\mfp\mfm}$ des metrischen Tensors als Funktionen der invarianten Schwerpunktenergie~$\surd s$.   (a) diagonal: die Gr"o"se~\mbox{$g_{\mfp\mfp}\big/g_{+-} \!\equiv\! \sinh^{-1}\!\ps$, die abweicht} von Null f"ur endliche~$\surd s$;   \vspace*{-.125ex}(b) au"serdiagonal: die Gr"o"se~\mbox{$g_{\mfp\mfm}\big/g_{+-} \!\equiv\! 1 + \efn{\D-\ps} \sinh^{-1}\!\ps$}, deren Abweichung von Eins zus"atzlich unterdr"uckt ist durch den Faktor~$\efn{\D-\ps}$.   Es h"angt~\mbox{$\ps \!=\! \ps\Dmfp \!+\! \ps\Dmfm$}~-- mit~\mbox{$\be\Dmfp \!=\! \tanh\ps\Dmfp$},~\mbox{$\be\Dmfm \!=\! \tanh\ps\Dmfm$}~-- ab von~$\surd s$ und den Massen des "`$\mfp$"'- und "`$\mfm$"'-Teilchens.   Die Kurven beziehen auf die F"alle, die~$M\Dmfp$ identifizieren mit der Ruhemasse des Protons und~$M\Dmfm$ mit der des~$\rh(770)$, des Protons oder des~$\Jps(3097)$~-- mit zunehmender Diskrepanz zum asymptotischen Wert.   Bzgl.\@ der zugrundeliegenden Zahlenwerte vgl.\@ Ref.~\cite{PDG00}, Particle Data Group: \mbox{\,$M_p \!=\! 0.938\,271\,998(38)\GeV$}; bzgl.\@ \mbox{$\rh(770),\,\Jps(3097)$} vgl.\@ Tabl.~\ref{Tabl:Charakt_rh,om,ph,Jps}.
}
\label{Fig:lglg-gpp,gpm}
\end{minipage}
\end{figure}
Zur Illustration ist dargestellt in Abbildung~\ref{Fig:lglg-gpp,gpm} die \mbox{$s$-Ab}\-h"angigkeit der relevanten Faktoren%
  ~\vspace*{-.125ex}\mbox{\,$-\det\mathbb{SL}\, g_{\mfp\mfp} \!\equiv\! g_{\mfp\mfp} \!\big/\! g_{+-} \!=\! \sinh^{-1}\!\ps$} und%
  ~\mbox{\,$-\det\mathbb{SL}\, g_{\mfp\mfm} \!\equiv\! g_{\mfp\mfm} \!\big/\! g_{+-} \!=\! 1 \!+\! \efn{\D-\ps} \sinh^{-1}\!\ps$}~-- der diagonalen und au"serdiagonalen Komponente des metrischen Tensors bez"uglich der Richtungen~$\mfp$,~$\mfm$ der Weltlinien der Wegner-Wilson-Loops, die normiert sind in der Weise, da"s%
  ~\vspace*{-.125ex}$g_{\mfp\mfp}$,~$g_{\mfp\mfm}$ f"ur~\mbox{\,$s \!\to\! \infty$} "ubergehen respektive in die Komponenten%
  ~\vspace*{-.125ex}$g_{++}$,~$g_{+-}$ bez"uglich Lichtkegelkoordinaten {\it unabh"angig\/} von deren Normierung.
Wir geben an die F"alle, in denen die Masse~$M\Dmfp$ identifiziert ist mit der Ruhemasse des Protons und~$M\Dmfm$ mit der des~$\rh(770)$, des Protons oder des~$\Jps(3097)$, die entsprechen der $s$-Abh"angigkeit von Proton-$\rh(770)$-, von Proton-Proton- und Proton-$\Jps(3097)$-Streuung; dabei sind nicht erfa"st Effekte, die r"uhren von der $s$-Abh"angigkeit der hadronischen (Quark-Antiquark-)Wellenfunktionen.%
\FOOT{
  \label{FN:WFN_s}So ist die Darstellung eines Hadrons als Superposition von Quark-Antiquark-Paaren~-- durch den niedrigsten Fock-Zustand allein~-- {\sl streng\/} nicht zul"assig f"ur endliche Werte von~$s$, {\sl approximativ\/} zul"assig nur~ober\-halb eines Schwellenwerts, wo Abstrahlungs- und Rekombinationsprozesse {\sl numerisch\/} vernachl"assigbar sind.
} \\
%
\indent
Wir finden, da"s die Abweichung des normierte Diagonalelements%
  ~\vspace*{-.125ex}\mbox{$g_{\mfp\mfp} \!\big/\! g_{+-}$} von seinem asymptotischen Wert Null gr"o"ser ist als die Abweichung des normierten Au"serdiagonalelements%
  ~\vspace*{-.125ex}\mbox{$g_{\mfp\mfm} \!\big/\! g_{+-}$} von dessen asymptotischen Wert Eins.
Dies ist unmittelbare Konsequenz dessen, da"s die eine Abweichung gegeben ist durch die Funktion%
  ~\vspace*{-.125ex}\mbox{\,$\sinh^{-1}\!\ps$}, die andere durch
  ~\vspace*{-.125ex}\mbox{\,$\efn{\D-\ps} \sinh^{-1}\!\ps$}, das hei"st zus"atzlich unterdr"uckt ist durch den Faktor~$\efn{\D-\ps}$.
Absolut betr"agt die Abweichung von~\mbox{$g_{\mfp\mfp} \!\big/\! g_{+-}$} f"ur~$\Jps(3097)$ etwa ein Prozent im Bereich von~$\surd s$ zwischen $20$ und~$30\GeV$.
Dies erscheint zun"achst numerisch eher klein und \vspace*{-.125ex}vernachl"assigbar zu sein. \\
\indent
Wir betrachten, wie die longitudinalen Metrikkomponenten%
  ~\vspace*{-.25ex}\mbox{$g_{\mfp\mfp} \!\big/\! g_{+-} \!\equiv\mskip-4mu g_{\mfm\mfm} \!\big/\! g_{+-} \!=\! \sinh^{-1}\!\ps$} und%
  ~\mbox{$g_{\mfp\mfm} \!\big/\! g_{+-} \!=\! 1 \!+\! \efn{\D-\ps} \sinh^{-1}\!\ps$} explizit bestimmen die $T$-Amplitude%
  ~\vspace*{-.125ex}\mbox{$\tTll^{(s,\rb{b})}$} f"ur die Streuung der Wegner-Wilson-Loops%
  ~\vspace*{-.25ex}$W\Dmfp$,~$W\Dmfm$:
\vspace*{-.5ex}
\begin{align} \label{tTll_WW_ch-mf_ALL-gmf}
\tTll\;
  &=\; 2\iIM\,s\vv
         \frac{1}{(4\Nc)^2}\vv
         \big(\vac{g^2 FF}a^4\big)^2
    \\[-.25ex]
  &\phantom{=\;}\qqquad\times
        \Big[\,
          \Big(\frac{g_{\mfp\mfp}}{g_{+-}}\Big)^{\zz2}\;
            \tilde{X}\idx{\mfp\mskip-2mu\mfp}\, \tilde{X}\idx{\mfm\mskip-2mu\mfm}\;
     +\; \frac{2}{\Nc^2 \!-\! 1}\;\,
            \Big(\frac{g_{\mfp\mfm}}{g_{+-}}\Big)^{\zz2}\;
            \tilde{X}\idx{\mfp\mskip-2mu\mfm}{}^{\zz2}
        \,\Big]
    \nn \\[-.25ex]
  &\phantom{=\;}\qqquad\times\,
        \exp\, -\frac{1}{4\Nc}\cdot
          \vac{g^2 FF}a^4\cdot
          \frac{g_{\mfp\mfp}}{g_{+-}}\,
          \big(
            \tilde{X}\idx{\mfp\mskip-2mu\mfp}
          + \tilde{X}\idx{\mfm\mskip-2mu\mfm}
          \big)
    \nn
    \\[-4.25ex]\nn
\end{align}
\end{samepage}%
eingesetzt in Gl.~(\ref{tTll_WW_ch-mf_ALL}$'$) die Gln.~(\ref{ch_ch-tilde-mf-REP}),~(\ref{ch=gXi-REP}).
Es ist%
  ~\vspace*{-.125ex}\mbox{$\tilde{X}\idx{\imath\jmath} \!=\!
    \vka \tilde{X}\idx{\imath\jmath}\oC \!+\! (1 \!-\! \vka) \tilde{X}\idx{\imath\jmath}\oNC$}~\mbox{im Sinne} von Gl.\,(\ref{ch_vka,chC,chNC-tilde-REP}); bzgl.%
  ~\mbox{$\tilde{X}\idx{\imath\jmath}\oC$},%
  ~\mbox{$\tilde{X}\idx{\imath\jmath}\oNC$} explizit vgl.\@ die Gln.\,(\ref{ChNC-pm}$'$),~(\ref{ChNC-pp}$'$),~(\ref{ChC-pm}$'$),~(\ref{ChC-pp}$'$).

Die Metrik-Komponenten%
  ~\vspace*{-.125ex}$g_{\mfp\mfp}$,~$g_{\mfp\mfm}$ sind nicht unabh"angig.
Induziert durch die Forderung der Gln.~\vspace*{-.75ex}(\ref{vrh-Forderung}),~(\ref{vrh-Forderung}$'$) besteht f"ur die Determinanten der met\-rischen Tensoren der Zusammenhang%
  ~\mbox{\,$\det\tilde{g} \!=\! (g_{\mfp\mfp})^2 \!-\! (g_{\mfp\mfm})^2 \!\stackrel{\T!}{=}\! \det\bar{g} \!=\! -(g_{+-})^2$}, ergo:
\begin{samepage}
%
\begin{align} \label{tilde-gpm_tilde-gpp}
\big(g_{\mfp\mfm} \!\big/\! g_{+-}\big)^2\;
  =\; 1\; +\; \big(g_{\mfp\mfp} \!\big/\! g_{+-}\big)^2
    \\[-4ex]\nn
\end{align}
Die Loop-Loop-Amplitude%
  ~\vspace*{-.125ex}\mbox{$\tTll^{(s,\rb{b})}$} folgt in Form:
\vspace*{-.5ex}
\begin{align} \label{tTll_WW_ch-mf_ALL-gmfpmfp}
\tTll\;
  &=\; 2\iIM\,s\vv
         \frac{1}{(4\Nc)^2}\vv
         \big(\vac{g^2 FF}a^4\big)^2
    \\[-.25ex]
  &\phantom{=\;}\qqquad\times
        \bigg[\,
          \frac{2}{\Nc^2 \!-\! 1}\; \tilde{X}\idx{\mfp\mskip-2mu\mfm}{}^{\zz2}\;
    +\; \Big(\frac{g_{\mfp\mfp}}{g_{+-}}\Big)^{\zz2}\cdot
          \Big(
            \tilde{X}\idx{\mfp\mskip-2mu\mfp}\, \tilde{X}\idx{\mfm\mskip-2mu\mfm}
          + \frac{2}{\Nc^2 \!-\! 1}\; \tilde{X}\idx{\mfp\mskip-2mu\mfm}{}^{\zz2}
          \Big)
        \,\bigg]
    \nn \\[-.25ex]
  &\phantom{=\;}\qqquad\times\,
        \exp\, -\frac{1}{4\Nc}\cdot
          \vac{g^2 FF}a^4\cdot
          \frac{g_{\mfp\mfp}}{g_{+-}}\,
          \big(
            \tilde{X}\idx{\mfp\mskip-2mu\mfp}
          + \tilde{X}\idx{\mfm\mskip-2mu\mfm}
          \big)
    \nn
    \\[-4.25ex]\nn
\end{align}
vgl.\@ die Gl.~(\ref{tTll_WW_ch-mf_ALL-gmf}).
Im Limes~\mbox{\,$s \!\to\! \infty$} verschwindet%
  ~\vspace*{-.125ex}\mbox{\,$g_{\mfp\mfp} \!\big/\! g_{+-}$}, ergo der zweite Summand in der eckigen Klammer und das Argument der Exponentialfunktion:
Es folgt die bekannte asymptotische Formel von Dosch, Ferreira, Kr"amer in Ref.~\cite{Dosch94a}; vgl.\@ auch die Refn.~\cite{Nachtmann96,Kulzinger95}. \\
\indent
F"ur endliche Werte von~$s$ ist%
  ~\vspace*{-.125ex}\mbox{\,$g_{\mfp\mfp} \!\big/\! g_{+-}$} nichtverschwindender kleiner Parameter, dessen absolute Gr"o"se vollst"andig bestimmt ist durch~$s$ und~$M\Dmfp$,~$M\Dmfm$.
Er tritt auf {\it quadratisch\/} in der eckigen Klammer, {\it linear\/} im Argument der Exponentialfunktion, so da"s durch diese bestimmt ist die f"uhrende Korrektur f"ur endliches~$s$ im Vergleich zum Grenzwert~\mbox{\,$s \!\to\! \infty$}.
Die Funktionen~$\tilde{X}\idx{\mfp\mskip-2mu\mfp}$,~$\tilde{X}\idx{\mfm\mskip-2mu\mfm}$ sind reell-positiv, ergo die f"uhrende Korrektur reell-negativ.
Der Absolutbetrag der $T$-Amplitude%
  ~\vspace*{-.125ex}\mbox{$\tTll^{(s,\rb{b})}$} f"ur die Streuung der Wegner-Wilson-Loops%
  ~$W\Dmfp$,~$W\Dmfm$ ist monoton leicht ansteigende Funktion von~$s$.
Dies ist konsistent mit der experimentellen Beobachtung, da"s der Beitrag nichtperturbativer Quantenchromodynamik~-- des {\sl Soft Pomeron\/}~-- zu totalen Wirkungsquerschnitten leicht ansteigt mit~$s$: bezogen auf die $T$-Amplitude wie~$s^{1+\ep}$ mit der Eins rein kinematischen Ursprungs und zu identifizieren mit dem expliziten Faktor~$s$ unserer Formel und~\mbox{\,$\ep \!\cong\! 0.0808$} dem effektiven Intercept des Soft Pomeron. \\
\indent
Wir halten fest, da"s die eckige Klammer darstellt den Vakuumerwartungswert beider Wegner-Wilson-Loops%
  ~\vspace*{-.125ex}\mbox{\,$\vac{\, [ W\Dmfp \!-\! 1 ] [ W\Dmfm \!-\! 1 ] \,}$} und die Exponentialfunktion dessen Normierung%
  ~\vspace*{-.125ex}\mbox{\,$\vac{W\Dmfp}^{-1}\; \vac{W\Dmfm}^{-1}$}~-- vgl.\@ Gl.~(\ref{tTll_WW-mf}$'$).
Die f"uhrende Korrektur%
  ~\mbox{\,$\propto\! g_{\mfp\mfp} \!\big/\! g_{+-}$} der Loop-Loop-Amplitude f"ur endliche Werte von~$s$ r"uhrt daher von den Vakuumerwartungswerten der einzelnen Loops, das hei"st von~\mbox{\,$Z\idx{2}$} der Renormierungskonstante des Quarkfelds; sie ist Konsequenz der nichtverschwindenden Selbst-Wechselwirkung des Loops f"ur endliches~$s$, das hei"st der Wechselwirkung des ihn konstituierenden Quarks und Antiquarks.
Erst die n"achst-f"uhrende Korrektur~\mbox{\,$\propto\! (g_{\mfp\mfp} \!\big/\! g_{+-})^2$} ist beeinflu"st durch die blo"se Loop-Loop-Wechselwirkung. \\
\indent
Angesichts der absolute Gr"o"se des Parameters%
  ~\mbox{\,$g_{\mfp\mfp} \!\big/\! g_{+-}$}~-- vgl.\@ Abb.~\ref{Fig:lglg-gpp,gpm}(a)~-- ist gerechtfertigt, die Exponentialfunktion zu entwickeln.
Es folgt:
\end{samepage}%
\vspace*{-.5ex}
\begin{align} \label{tTll_WW_ch-mf_ALL-gmfpmfp-expansion}
\hspace*{-8pt}
\tTll\;
  &=\; 2\iIM\,s\vv
       \Big( \frac{1}{4\Nc}\; \vac{g^2 FF}a^4 \Big)\big.^{\zz2}
    \\
  &\phantom{=\;}\qqquad\times
        \bigg[\,
          \frac{2}{\Nc^2 \!-\! 1}\; \tilde{X}\idx{\mfp\mskip-2mu\mfm}{}^{\zz2}
    \nn \\[-.25ex]
  &\phantom{=\;\qqquad\qquad}
   -\; (g_{\mfp\mfp} \!\big/\! g_{+-})\cdot
         \frac{2}{\Nc^2 \!-\! 1}
           \Big( \frac{1}{4\Nc}\; \vac{g^2 FF}a^4 \Big)\cdot
         \tilde{X}\idx{\mfp\mskip-2mu\mfm}{}^{\zz2}\,
           \big( \tilde{X}\idx{\mfp\mskip-2mu\mfp} + \tilde{X}\idx{\mfm\mskip-2mu\mfm} \big)
    \nn \\[-.25ex]
  &\begin{aligned}[t]\!
   \phantom{=\;\qqquad\qquad}
   +\; (g_{\mfp\mfp} \!\big/\! g_{+-})^2\cdot
         \Big\{
          &\Big(
             \tilde{X}\idx{\mfp\mskip-2mu\mfp}\, \tilde{X}\idx{\mfm\mskip-2mu\mfm}
           + \frac{2}{\Nc^2 \!-\! 1}\; \tilde{X}\idx{\mfp\mskip-2mu\mfm}{}^{\zz2}
           \Big)
    \nn \\[-.75ex]
          &+\; \frac{1}{\Nc^2 \!-\! 1}
                 \Big( \frac{1}{4\Nc}\; \vac{g^2 FF}a^4 \Big)\big.^{\zz2}\cdot
               \tilde{X}\idx{\mfp\mskip-2mu\mfm}{}^{\zz2}\,
                 \big( \tilde{X}\idx{\mfp\mskip-2mu\mfp}
                     + \tilde{X}\idx{\mfm\mskip-2mu\mfm} \big)^{\!2}
         \Big\}
   \end{aligned}
    \nn \\[-.5ex]
  &\phantom{=\;\qqquad\qquad}
   +\; {\cal O}\big((g_{\mfp\mfp} \!\big/\! g_{+-})^3\big)
        \,\bigg]
    \nn
    \\[-4.5ex]\nn
\end{align}
vgl.\@ Gl.~(\ref{tTll_WW_ch-mf_ALL-gmfpmfp}).
Die korrelierten Loop-Exponentiale im Ausdruck%
  ~\vspace*{-.25ex}\mbox{\,$\vac{\, [ W\Dmfp \!-\! 1 ] [ W\Dmfm \!-\! 1 ] \,}$} sind bereits entwickelt bis zur Ordnung%
  ~\vspace*{-.25ex}\mbox{\,${\cal O}\big((F_{\mu\nu})^6\big) \!=\!
    {\cal O}\big((\tilde\ch\idx{\imath\jmath})^3\big) \!=\!
    {\cal O}\big((g_{\imath\jmath} \!\big/\! g_{+-})^3\, (\tilde{X}\idx{\imath\jmath})^3\big)$}, so da"s diese \vspace*{-.25ex}Formel die konsequentere ist~-- da {\it exakt\/} bis zu dieser Ordnung. \\
\indent
Die Darstellungen von%
  ~\vspace*{-.25ex}$\tTll^{(s,\rb{b})}$~-- die Gln.~(\ref{tTll_WW_ch-mf_ALL-gmfpmfp}),~(\ref{tTll_WW_ch-mf_ALL-gmfpmfp-expansion})~-- dokumentieren Unterdr"uckung der $\bm\mfp\bm\mfm$- gegen"uber den \mbox{$\bm\mfp\bm\mfp$-,$\bm\mfm\bm\mfm$-Ter}\-men durch den Faktor%
  ~\mbox{\,$2 \!\big/\! (\Nc^2 \!-\! 1)$} von~$1\!\big/\!4$ in der physikalischen Quantenchromodynamik mit drei Colour-Freiheisgraden%
  ~\mbox{$\Nc \!\equiv\! 3$}:%
\FOOT{
  Die~$2$ repr"asentiert die zwei M"oglichkeiten, je zwei paralleltransportierte Feldst"arken von~$W\Dmfp$,~$W\Dmfm$ zu kontrahieren in Paare bez"uglich verschiedener Loops.   Der Colour-Faktor~\mbox{$1 \!\big/\! \dimNc \!\equiv\! 1 \!\big/\! (\Nc^2 \!-\! 1)$} folgt aus differierender Kontraktion der Eichgruppen-Tensorstruktur der $\bm\mfp\bm\mfm$- gegen"uber der \mbox{$\bm\mfp\bm\mfp$-,$\bm\mfm\bm\mfm$-Terme.   Vgl.\@ Fu"sn.\FN{FN:tTll-Vorfaktor}}.
}
%
\vspace*{-.5ex}
\begin{align} \label{Unterdr"uckung2/dimNc}
\tilde{X}\idx{\mfp\mskip-2mu\mfp},\,
    \tilde{X}\idx{\mfm\mskip-2mu\mfm}\vv
  \longleftrightarrow\vv
  \frac{\surd2}{\sqrt{\Nc^2 \!-\! 1}}\; \tilde{X}\idx{\mfp\mskip-2mu\mfm}
    \\[-4.5ex]\nn
\end{align}
Die Funktionen%
  ~$\tilde{X}\idx{\mfp\mskip-2mu\mfp}$,~$\tilde{X}\idx{\mfm\mskip-2mu\mfm}$ selbst erwartem wir von derselben Gr"o"senordnung wie%
  ~$\tilde{X}\idx{\mfp\mskip-2mu\mfm}$~-- eher gr"o"ser, da sie subsumieren Korrelation zweier paralleltransportierter Feldst"arken {\it desselben\/} Wegner-Wilson-Loops%
  ~\vspace*{-.125ex}$W\Dmfp$ oder~$W\Dmfm$.
Absolut etwa f"ur%
  ~\mbox{\,$\tilde{X}\idx{\mfp\mskip-2mu\mfp}\oC$}:~"`Stringspannung"'~$\si$ multipliziert mit der transversalen~\mbox{(Minimal-)Aus}\-dehnung~$\rb{X}$ von~$W\Dmfp$, mit~$\si$ von der Gr"o"senordnung der konventionellen Stringspannung  eines statischen Quark-Antiquark-Paares; vgl.\@ die Gln.~(\ref{ChC-pp-app}),~(\ref{ChNC-pp-app}). \\
\indent
Es sind zu identifizieren als die effektiven Entwicklungsparameter
\vspace*{-.5ex}
\begin{alignat}{2} \label{effEntwicklungsparameter}
&\al\idx{\mfp\mskip-2mu\mfm}\;&
  &\equiv\; \frac{1}{4\Nc}
               \vac{g^2 FF}a^4
    \\[-.375ex]
&\al\idx{\mfp\mskip-2mu\mfp}\;&
  &\equiv\; (g_{\mfp\mfp} \!\big/\! g_{+-})\cdot
               \frac{1}{4\Nc}\;
               \vac{g^2 FF}a^4
    \tag{\ref{effEntwicklungsparameter}$'$}
    \\[-4.5ex]\nn
\end{alignat}
Der Parameter%
  ~\vspace*{-.125ex}$\al\idx{\mfp\mskip-2mu\mfp}$ ist dominiert durch die Metrik-Komponente%
  ~\mbox{\,$g_{\mfp\mfp} \!\big/\! g_{+-}$}, die numerisch klein ist f"ur gro"se Werte von~$s$, ergo~$\al\idx{\mfp\mskip-2mu\mfp}$.
Im Limes~\mbox{\,$s \!\to\! \infty$} gilt%
  ~\vspace*{-.125ex}\mbox{\,$g_{\mfp\mfp} \!\big/\! g_{+-} \!\to\! 0$} und%
  ~\mbox{\,$g_{\mfp\mfm} \!\big/\! g_{+-} \!\to\! 1$}, so da"s die Loop-Loop-Amplitude in diesem Grenzwert vollst"andig gegeben ist als Entwicklung in dem Parameter%
  ~$\al\idx{\mfp\mskip-2mu\mfm}$, den wir versehen mit den Indizes des \mbox{$\bm\mfp\bm\mfm$-Terms}, da er auftritt nur multipliziert mit der~Colour-unterdr"uckten Funktion%
  ~\mbox{$\tilde{X}\idx{\mfp\mskip-2mu\mfm}$}, vgl.\@ Gl.~(\ref{Unterdr"uckung2/dimNc}).
Dieses Produkt ist in~praxi numerisch klein; vgl.\@ auch die Diskussion in Ref.~\cite{Kraemer91}.
Zusammen ist a~posteriori gerechtfertigt die Entwicklung in Kumulanten.
\vspace*{-.5ex}

%
%
%
\begin{sidewaysfigure}
\begin{minipage}{\linewidth}
\renewcommand{\thefootnote}{\thempfootnote}
\setlength{\unitlength}{1mm}
\begin{center}
  \begin{tabular}{|g{10}||f{19}|f{19}||f{10}|f{9}|f{15}|} \hline
  \multicolumn{6}{|c|}{Parameter~$\al,\al'$
                  des~-- doppel\/logarithmisch linearen~-- Zusammenhangs\vv%
                  $g_{\mfp\mfp}\!(s) \!\big/\! g_{+-} = \al \!\cdot\! s^{-(1+\al')}$}
    \\[.125ex] \hhline{:=:t:==:t:===:}
  \multicolumn{1}{|c||}{}
    & \multicolumn{1}{c|}{$g_{\mfp\mfp}\!(s_\star) \!\big/\! g_{+-}$}
    & \multicolumn{1}{c||}{$g_{\mfp\mfp}\!(s^\star) \!\big/\! g_{+-}$}
    & \multicolumn{1}{c|}{}
    & \multicolumn{1}{c|}{}
    & \multicolumn{1}{c|}{}
    \\[-1ex]
  \multicolumn{1}{|c||}{}
    & \multicolumn{1}{c|}{f"ur~$\surd s_\star \equiv 10^3\GeV$}
    & \multicolumn{1}{c||}{f"ur~$\surd s^\star \equiv 10^4\GeV$}
    & \multicolumn{1}{c|}{$\big.^{\D\ln\al}$}
    & \multicolumn{1}{c|}{$\big.^{\D\al}$}
    & \multicolumn{1}{c|}{$\big.^{\D\al'}$}
    \\[.125ex] \hhline{|-||--||---|}
  \mbox{$\vv p-$},\mbox{$\rh(770) \vv$}\;
    &\; 1.4436\,2742\,140 \times\!10^{-6}
    &\; 1.4436\,2531\,738 \times\!10^{-8}
    &\vv 0.3671\,6336 \vv
    &\vv 1.4436\,337  \vv
    &\; 0.3164\,8281 \times\!10^{-6}\; \\[.125ex]
  \mbox{$\vv p-$},\mbox{$p \vv$}\;
    &1.7607\,1178\,457 \times\!10^{-6}
    &1.7607\,0871\,546 \times\!10^{-8}
    &0.5657\,2338
    &1.7607\,210
    &0.3785\,1038 \times\!10^{-6} \\[.125ex]
  \mbox{$\vv p-$},\mbox{$\Jps(3097) \vv$}\;
    &5.8114\,7365\,669 \times\!10^{-6}
    &5.8114\,1341\,340 \times\!10^{-8}
    &1.7598\,653
    &5.8116\,544
    &2.2510\,178\phantom{4} \times\!10^{-6} \\
  \hhline{:=:b:==:b:===:}
  \end{tabular}
  \makebox(220,95){
  \setlength{\unitlength}{1mm}\begin{picture}(220,95)   
    \put(20,0){\epsfxsize180mm \epsfysize95mm \epsffile{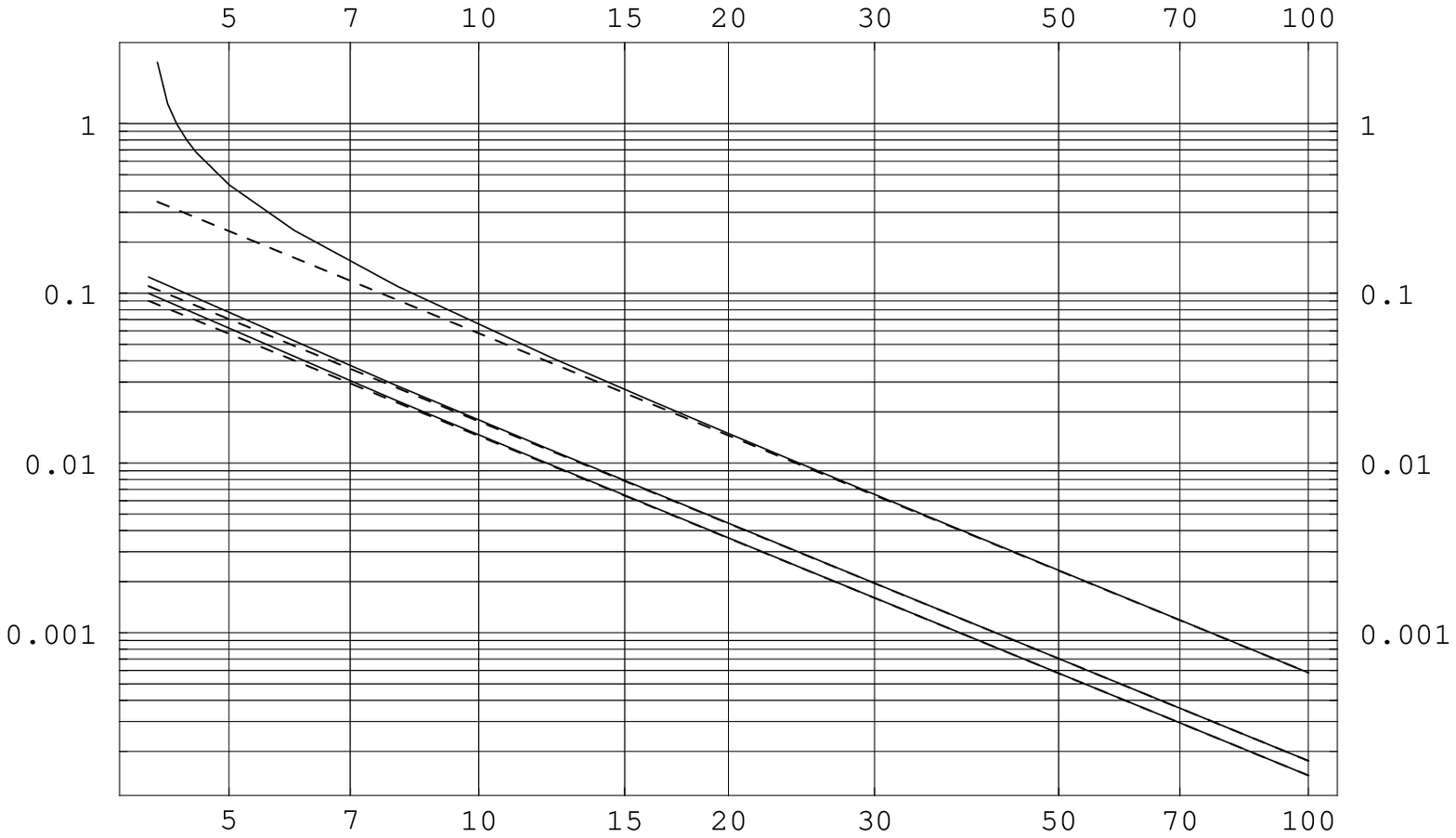}}
    \put(163, 86){\normalsize$\rh,p,\Jps \bm{-} p$}
    \put(188, .5){\normalsize$\surd s\;[\GeV[]]$}
    \put( 13,  0){\yaxis[95mm]{\normalsize%
                    ${g_{\mfp\mfp}\big/g_{+-}}\quad
                       \equiv\; -\det\mathbb{SL}\; g_{\mfp\mfp}\;
                       =\; \sinh^{\D-1}\!\ps$}}
  \end{picture}}
\vspace*{-2.5ex}
\refstepcounter{table}\label{Tabl:lglg-gpp-Parameter}
\refstepcounter{figure}\label{Fig:lglg-gpp-Parameter}
\addcontentsline{lot}{table}{\ref{Tabl:lglg-gpp-Parameter}\quad
  Parameter des asymptotischen Zusammenhangs%
  ~\protect\mbox{$g_{\mfp\mfp}\!(s) \!\big/\! g_{+-} = \al \!\cdot\! s^{-(1+\al')}$}}
\addcontentsline{lof}{figure}{\ref{Fig:lglg-gpp-Parameter}\quad
  \protect$\surd s$-Abh"angigkeit von~\protect$g_{\mfp\mfp}$
  versus \protect\mbox{$\surd s \!\to\! \infty$-Asymptotik}}
\begin{minipage}{\linewidth}
  Tabelle~\ref{Tabl:lglg-gpp-Parameter}/Abbildung~\ref{Fig:lglg-gpp-Parameter}: Doppel\/Logarithmische Auftragung der \vspace*{-.25ex}\mbox{$\surd s$-Ab}\-h"angigkeit der Metrik-Komponente~$g_{\mfp\mfp}$ zeigt asymp\-totisch linearen Zusammenhang:~\vspace*{-.25ex}\mbox{$\ln (g_{\mfp\mfp}\!(s) \!\big/\! g_{+-}) \!=\! \ln\al \!-\! (1\!+\!\al') \!\cdot\! \ln s \Leftrightarrow g_{\mfp\mfp}\!(s) \!\big/\! g_{+-}  \!=\! \al \!\cdot\! s^{-(1+\al')}$}.   Angegeben sind Parameter und Auftragung approximierender~Gera\-den~-- \mbox{konstruiert als Sekanten der Punkte f"ur~\mbox{$\surd s \!=\! 10^3$} und~$10^4\GeV$.   Es ist~\mbox{$M\Dmfp \!\equiv\! M_p$} und~\mbox{$M\Dmfm \!\equiv\! M_{\rh(770)},M_p, M_{\Jps(3097)}$} wie in Abb.~\ref{Fig:lglg-gpp,gpm}(a)}.
\end{minipage}
\end{center}
\end{minipage}
\end{sidewaysfigure}%
\renewcommand{\thefootnote}{\thechapter.\arabic{footnote}}
\bigskip\noindent
In Tabelle~\ref{Tabl:lglg-gpp-Parameter}/Abbildung~\ref{Fig:lglg-gpp-Parameter} ist quantitativ diskutiert die \mbox{$\surd s$-Ab}\-h"angigkeit der Komponente%
  ~\vspace*{-.25ex}\mbox{$g_{\mfp\mfp} \!\big/\! g_{+-}$} des metrischen Tensors.
Durch diese ist vollst"andig bestimmt die \mbox{$\surd s$-Ab}\-h"angigkeit der Loop-Loop-Amplitude%
  ~\vspace*{-.25ex}$\tTll^{(s,\rb{b})}$, vgl.\@ die Gln.~(\ref{tTll_WW_ch-mf_ALL-gmfpmfp}),~(\ref{tTll_WW_ch-mf_ALL-gmfpmfp-expansion}).
Bereits Abbildung~\ref{Fig:lglg-gpp,gpm}(a) dokumentiert asymptotisch f"ur~\mbox{\,$s \!\to\! \infty$} {\it lineare Abh"angigkeit\/} der doppel\/logarithmischen Auftragung.
Wir betrachten daher als funktionalen Ansatz:
\vspace*{-.25ex}
\begin{align} \label{log-log-Ansatz}
&\ln\, (g_{\mfp\mfp}\!(s) \!\big/\! g_{+-})\;
  =\; \ln\al - (1\!+\!\al')\cdot \ln s
    \\[-.5ex]
&\Longleftrightarrow\qquad
 g_{\mfp\mfp}\!(s) \!\big/\! g_{+-}\;
  =\; \al\cdot s^{\D-(1 \!+\! \al')}
    \tag{\ref{log-log-Ansatz}$'$}
    \\[-4ex]\nn
\end{align}
und bestimmen die Parameter~\vspace*{-.25ex}$\al$,~$\al'$. \\
\indent
Diese Parameter h"angen ab~-- wenn auch nur sehr schwach~-- davon, wie hoch der Bereich von~$\surd s$ gew"ahlt ist, in dem die exakte Kurve%
  ~\mbox{$g_{\mfp\mfp}\!(s) \!\big/\! g_{+-}$} approximiert wird; so sollte~$\al'$ identisch verschwinden im Limes~\mbox{$s \!\to\! \infty$}, in dem erwartet werden \mbox{exakt-$1\!\big/\!s$-Korrek}\-turen.
F"ur praktische Anwendung w"ahlen wir den Bereich%
  ~\mbox{\,$\surd s \!=\! 10^3 \!-\! 10^4\GeV$}, der zum einen gen"ugend hoch ist, als "`asymptotisch"' bezeichnet werden zu k"onnen, und zum anderen physikalisch zug"anglich zu sein.
F"ur Einfachheit des Verfahrens konstruieren wir die asymptotische Gerade~-- im Sinne doppel\/logarithmischer Auftragung~-- als \vspace*{-.125ex}Sekante durch die Punkte bez"uglich der Intervall-Endpunkte:%
  ~\vspace*{-.125ex}\mbox{\,$(s_\star,g_{\mfp\mfp}\!(s_\star) \!\big/\! g_{+-})$},~\mbox{\,$(s^\star,g_{\mfp\mfp}\!(s^\star) \!\big/\! g_{+-})$} mit%
  ~\mbox{$s_\star \!\equiv\! 10^3\GeV$},~\mbox{\,$s^\star \!\equiv\! 10^4\GeV$}.
In Tabelle~\ref{Tabl:lglg-gpp-Parameter} sind~-- f"ur die drei F"alle%
  ~\vspace*{-.125ex}\mbox{$M\Dmfp \!\equiv\! M_p$} und
   \vspace*{-.25ex}\mbox{$M\Dmfm \!\equiv\! M_{\rh(770)},M_p$} oder~$M_{\Jps(3097)}$ von Abb.~\ref{Fig:lglg-gpp,gpm}(a)~-- angegeben numerisch die Funktionswerte%
  ~\pagebreak\mbox{\,$g_{\mfp\mfp}\!(s_\star) \!\big/\! g_{+-}$},%
  ~\mbox{\,$g_{\mfp\mfp}\!(s^\star) \!\big/\! g_{+-}$} und die resultierenden Zahlenwerten f"ur%
  ~\vspace*{-.25ex}$\ln\al$,~$\al$ und~$\al'$.
Zur Illustration sind in Abbildung~\ref{Fig:lglg-gpp-Parameter} dargestellt in doppel\/logarithmischer Auftragung die durch sie bestimmten Geraden~-- gestrichelte versus durchgezogene Linien~-- gegen die exakten Kurven. \\
\indent
Wir fassen zusammen wie folgt Tabelle~\ref{Tabl:lglg-gpp-Parameter} und Abbildung~\ref{Fig:lglg-gpp-Parameter}.
Die Steigung der approximierenden Geraden differiert von der asymptotischen Geraden durch Werte von~\vspace*{-.125ex}$\al'$ von der Gr"o"senordnung einiger~$10^{-7}$ bis~$10^{-6}$.
Die Parameter~$\al$,~$\al'$ sind fixiert weit au"serhalb des \vspace*{-.125ex}\mbox{$\surd s$-Inter}\-valls, "uber das wir auftragen die approximierende Gerade und die exakte Kurve f"ur
  \vspace*{-.125ex}\mbox{$g_{\mfp\mfp}\!(s) \!\big/\! g_{+-}$}.
Signifikante Diskrepanz besteht dennoch nur f"ur kleine~\mbox{Werte von~$\surd s$}; oberhalb von~\vspace*{-.125ex}\mbox{\,$\surd s \!\cong\! 15,\,20$} beziehungsweise~$50\GeV$ ist in~praxi die Asymptotik "`perfekt"' realisiert. \\
\indent
Dies unterstreicht, da"s der absolute Effekt endlicher Werte von~\vspace*{-.125ex}$s$ gegen"uber dem Limes%
  ~\vspace*{-.125ex}\mbox{\,$s \!\to\! \infty$} nur klein ist.
Dies genau l"a"st erwarten die experimentelle Beobachtung, da"s der $s$-Abh"angigkeit der $T$-Amplitude entspricht ein leichter Anstieg wie~$s^{1+\ep}$, die abweicht von der kinematischen Abh"angigkeit durch ein numerisch kleines Epsilon von%
  ~\vspace*{-.125ex}\mbox{$\ep \!\cong\! 0.0808$}.
Die quantitative Verifizierung dieses subtilen Anstiegs mit~$s$ gegen den~Grenz\-wert f"ur%
  ~\vspace*{-.125ex}\mbox{\,$s \!\to\! \infty$} kann nur erfolgen auf Basis {\it expliziter\/} Auswertung der \mbox{$T$-Amp}\-litude%
  ~\vspace*{-.125ex}$\tTll^{(s,\rb{b})}$ entsprechend den Gln.~(\ref{tTll_WW_ch-mf_ALL-gmfpmfp}),~(\ref{tTll_WW_ch-mf_ALL-gmfpmfp-expansion}) f"ur die Streuung der Wegner-Wilson-Loops~\vspace*{-.125ex}$W\Dmfp$,~$W\Dmfm$.
Diese wiederum schlie"st ein die explizite Auswertung der Funktionen%
  ~\vspace*{-.125ex}$\tilde{X}\idx{\mfp\mskip-2mu\mfp}$,~$\tilde{X}\idx{\mfm\mskip-2mu\mfm}$, die ausdr"ucken die Selbstwechselwirkung der Loops und nicht auftreten im Limes%
  ~\vspace*{-.125ex}\mbox{\,$s \!\to\! \infty$}.
Gegen\-"uber der Auswertung der bekannten%
  ~\vspace*{-.125ex}\mbox{\,$s \!\to\! \infty$-asymp}\-totischen Loop-Loop-Amplitude ist daher die Auswertung der Loop-Loop-Amplitude f"ur endliche Werte von~$s$ technisch aufwendiger.
Sie ist aber straight-forward, so da"s wir sie f"ur diese Arbeit zur"uckstellen.%
\FOOT{
  Chronologisch steht die Analyse des vorliegenden Kapitels am Ende unserer Arbeit; wir ziehen sie vor, da sie auch darstellt die bekannte asymptotische Amplitude, die noch zugrunde liegt den folgenden Kapiteln.
}
\vspace*{-.5ex}

\bigskip\noindent
Es besteht berechtigte Annahme, da"s die hergeleitete nahezu lichtartige \mbox{$T$-Ampli}\-tude f"ur die Streuung zweier Wegner-Wilson-Loops die beobachtete \mbox{$s$-Ab}\-h"angigkeit reproduziert.~--
Unter der Einschr"ankung, da"s sie per~constructionem erfa"st nicht-observable Loop-Loop-, nicht aber observable Hadron-Hadron-Streuung, das hei"st den Einflu"s hadronischer Wellenfunktionen; vgl.\@ Fu"sn.\FN{FN:WFN_s}.
Wir kn"upfen an an die Diskussion in Abschnitt~\ref{Sect:Nahezu_vs_exakt}. \\
\indent
Die $T$-Amplitude f"ur Parton-Parton-Streuung f"ur gro"se invariante Schwerpunktenergie%
  ~\vspace*{-.125ex}$\surd s$ und kleinen invarianten Impulstransfer~$\surd\!-t$ ist gegeben in Termen der Vakuumerwartungswerte~$\vac{\;\cdot\;}$ beider und der einzelnen korrespondierenden Wegner-Wilson-Linien.
Diese beziehen sich auf die {\it physikalischen\/} Parton-Parton-Trajektorien, die nahezu lichtartig sind mit einer kleinen zeitartigem Komponente.
Die Herleitung Nachtmanns~-- vgl.\@ Ref.~\cite{Nachtmann91} und unseren Abri"s in Abschn.~\ref{Subsect:Partonniveau}~-- ber"ucksichtigt konsequent die f"uhrende Ordnung bez"uglich~$s$.
Dies involviert formal den "Ubergang der physikalischen nahezu lichtartigen Trajektorien zu deren lichtartigen Limites unter~\vspace*{-.125ex}\mbox{\,$s \!\to\! \infty$}. \\
\indent
Verlinde, Verlinde zeigen in Ref.~\cite{Verlinde93} f"ur Quark-Quark-Streuung, da"s dieser Limes singul"ar ist, wie etwa folgt aus der perturbativen Entwicklung.
Diese Singularit"at kann verstanden werden als Konsequenz infraroter Divergenzen, die daher r"uhren, da"s die im Limes~\vspace*{-.125ex}\mbox{\,$s \!\to\! \infty$} auf den Lichtkegel gezwungene Propagation der Quarks impliziert das Verschwinden ihrer Masse~$m$, und die dadurch regularisiert werden, da"s den Quark-Trajektorien eine "`kleine zeitartige Komponente"' gegeben wird, die genau entspricht der von Quarks mit endlicher Masse~$m$.
Dies impliziert, da"s die $T$-Amplitude~-- "uber den kinematische Faktor~\vspace*{-.125ex}$s$ hinaus~-- abh"angt von~$s$ vermittels der Richtungen der physikalischen Parton-Trajektorien. \\
\indent
Unser Zugang besteht genau darin, den Parton-Trajektorien in diesem Sinne ihre zeitartige Komponente zu belassen: \vspace*{-.125ex}zu arbeiten mit den initialen physikalischen Trajektorien und mit den diese induzierenden nahezu lichtartigen \vspace*{-.125ex}Wegner-Wilson-Linien und -Loops. \\
\indent
Auf derselben Basis diskutiert Meggiolaro in Ref.~\cite{Meggiolaro96} Quark-Quark-Streuung, indem er im Sinne Verlindes, Verlindes die Wegner-Wilson-Linien, die konstituieren die \mbox{$T$-Am}\-plitude, betrachtet als Funktionale der physikalischen Trajektorien.
Meggiolaro entwickelt die \vspace*{-.125ex}\mbox{$T$-Am}\-plitude konsequent bis zu Termen proportional~$g_{\rm ren.}^4$~-- mit~$g_{\rm ren.}$ der renormierten Kopplungskonstanten der QCD~-- und findet exakte "Ubereinstimmung mit dem Standard-Resultat der konventionellen perturbativen Theorie nach Cheng, Wu, vgl.\@ Ref.~\cite{Cheng87}.
Die identifizierte \mbox{$s$-Ab}\-h"angigkeit manifestiert sich explizit in Form
\vspace*{-.25ex}
\begin{align} \label{Meggiolaro-s}
g_{\rm ren.}^2\cdot \tanh^{-1}\!\ps\qquad
  \text{und}\quad\vv
  g_{\rm ren.}^4\cdot \tanh^{-2}\!\ps\,,\quad\vv
  g_{\rm ren.}^4\cdot \ps\, \tanh^{-2}\!\ps
    \\[-4.25ex]\nn
\end{align}
mit~$\ps$ wie in unserer Definition dem hyperbolischen Winkel, den die physikalischen Trajektorien einschlie"sen in der Boost-Ebene: mit%
  ~\mbox{\,$\be\Dimath \!=\! \tanh\ps\Dimath$},~\mbox{\,$\imath \!=\! \mfp,\mfm$} , den Beta-Parametern der aktiven Lorentz-Boosts gegeneinander und%
  ~\mbox{\,$\ps \!=\! \ps\Dmfp \!+\! \ps\Dmfm$}.
Im Rahmen unserer differentialgeometrischen Formulierung~-- die in nat"urlicher Weise definiert Koordinatenlinien~$\mfp$,~$\mfm$ als die Richtungen dieser Trajektorien~-- verstehen wir in Gl.~(\ref{Meggiolaro-s}) den inversen hyperbolischen Tangens als die eine unabh"angige Komponente%
  ~\vspace*{-.125ex}\mbox{\,$g_{\mfp\mfm} \big/ g_{+-} \!=\! \tanh^{-1}\!\ps$} des induzierten metrischen Tensors%
\FOOT{
  cum grano salis als deren Anteil {\sl unabh"angig\/} von der Konvention bez"uglich der Lichtkegelkoordinaten
};
vgl.%
  ~\vspace*{-.125ex}\mbox{$g_{\mfp\mfp} \!\big/\! g_{+-} \!=\!
      \surd1 \!-\! (g_{\mfp\mfm} \!\big/\! g_{+-})^2$} nach Gl.~(\ref{tilde-gpm_tilde-gpp}) und
  ~\mbox{$g_{\mfm\mfp} \!\equiv\! g_{\mfp\mfm}$},%
  ~\mbox{$g_{\mfm\mfm} \!\equiv\! g_{\mfp\mfp}$}. \\
\indent
Der letzte Term in Gl.~(\ref{Meggiolaro-s}) mit explizitem Faktor~$\ps$ folgt in Meggiolaros Analyse aus der%
  ~\vspace*{-.25ex}\mbox{\,$T_{(1)}^a \!\otimes\! T_{(2)}^a$-Colour-Struk}\-tur der Summe der zwei Feynman-Diagramme~-- dem {\it ladder\/}- und {\it cross\/}-Diagramm~-- f"ur den Austausch zweier Gluonen.
F"ur Streuung zweier \vspace*{-.125ex}Quarks derselben Masse~$m$ gilt%
  ~\mbox{\,$s \!=\! 2m^2\, (\cosh \ps + 1)$}, so da"s folgt%
  ~\vspace*{-.125ex}\mbox{\,$\ps \!\sim\! \ln s$} f"ur~\mbox{\,$s \!\to\! \infty$~-- vgl.\@ Gl.~(\ref{lns})~-- und} auf diese Weise generiert werden in Quark-Quark-Streuung die bekannten Logarithmen von~$s$. \\
\indent
Meggiolaro gelangt ferner zu denselben Ausdr"ucken%
  ~\vspace*{-.125ex}\mbox{\,$I(t) \!\propto
      \int\! d^2\rb{z}\, \efn{\D\iIM\rb{q} \!\cdot\! \rb{z}} \big({\rm K}_0(\la|\rb{z}|)\big){}^2$}, mit \vspace*{-.125ex}\mbox{\,$t \!=\! q^2 \!\cong\! -\rb{q}^2$}, wie die konventionelle perturbative Theorie; diese entsprechen Gluon-Loop-Diagrammen und sind wie dort zu regularisieren~-- vgl.%
  ~\vspace*{-.125ex}\mbox{\,$K_0(z) \!\sim\! -\ln z$} f"ur~\mbox{\,$z \!\to\! 0$}~-- durch eine endliche Gluonmasse~$\la$ [die am Ende:~\mbox{\,$\la \!\to\! 0\!+$}, identisch Null zu setzen ist]. \\
\indent
Diese Logarithmen~-- wie auch die Ausdr"ucke~\vspace*{-.25ex}$I(t)$~-- erscheinen uns Artefakte zu sein der perturbativen Entwicklung und des diese induzierenden Renormierungsverfahren.
Wir erwarten weniger, da"s ihre Aufsummierung zu dem "`korrekten Potenzverhalten"'~$s^{1+\ep}$ der \mbox{$T$-Am}\-plitude f"uhrt als vielmehr ein wesentlich nichtperturbativer Mechanismus, etwa im Sinne des Regge-Pol-Mechanismus auf Basis ihrer Analytizit"at.
Meggiolaro gibt in Ref.~\cite{Meggiolaro96} an die explizite Herleitung des Regge-Pol-Modells aus der $T$-Amplitude f"ur Quark-Quark-Streuung, die genau betrachtet die sie konstituierenden Wegner-Wilson-Linien als Funktionale der physikalischen Parton-Trajektorien.
Wir verweisen ausdr"ucklich auf diese Herleitung%
\FOOT{
  Wir skizzieren ihre wesentlichen Schritte:   Partialwellen-Zerlegung der $T$-Amplitude.   Annahme analytischer Fortsetzbarkeit der Partial-Amplituden~$A_l(t)$ mit Drehimpulsen~\mbox{\,$l \!\in\! \bbbn$} zu komplexen Werten~\mbox{\,$l \!\in\! \bbbc$}.   Umformung der Partialsumme~-- auf Basis der Produkt-Entwicklung von~\mbox{\,$\sin(\pi l)$}~-- in ein Integral, dessen Kontur in der komplexen \mbox{$l$-Ebe}\-ne genau die nicht-negativen nat"urlichen~$l$ einschlie"st.   "Uberf"uhren dieser Kontur in die Gerade~\mbox{\,${\rm Re}\,l \!=\! -\frac{1}{2}$} parallel zur imagin"aren \mbox{$l$-Ach}\-se, indem aufgrund des Residuensatzes explizit auftritt die Summe "uber die Residuen des Integranden rechts der Geraden:~\mbox{\,${\rm Re}\,l \!>\! -\frac{1}{2}$}.   Im Limes~\mbox{\,$s \!\to\! \infty$} verschwindet das Integral mindestens wie~$1\!\big/\!\surd s$ und die Amplitude ist gegeben durch die Summe "uber die Residuen.   Die $T$-Amplitude im Limes~\mbox{\,$s \!\to\! \infty$} ist gegeben als Summe von Termen, die abh"angen von~$s$ wie~\mbox{\,$s^{1+\al_n(t)}$} mit~$\al_n(t)$,~\mbox{\,$n \!=\! 1,2,\ldots$}, den Positionen der~-- f"ur Einfachheit angenommenen~-- {\sl einfachen\/} Pole der Partial-Amplituden~$A_l(t)$ in der komplexen \mbox{$l$-Ebe}\-ne.   Der Pol mit dem gr"o"sten Realteil,~\mbox{\,${\rm Re}\,\al_n(t) \!>\! -\frac{1}{2}$} dominiert die Summe~-- und wird konventionell bezeichnet als {\sl Regge-Pol\/}.   Bzgl.\@ der~-- sehr lesenswerten~-- Originalarbeiten Regges vgl.\@ die Refn.~\cite{Regge59,Regge60}.
}~--
zum einen wegen ihrer \vspace*{-.125ex}"Asthetik, zum anderen in Hinblick auf ihre Relevanz in Bezug auf die von uns diskutierte \mbox{$T$-Am}\-plitude%
  ~\vspace*{-.125ex}\vspace*{-.25ex}$\tTll^{(s,\rb{b})}$~-- vgl.\@ die Gln.~(\ref{tTll_WW_ch-mf_ALL-gmfpmfp}),~(\ref{tTll_WW_ch-mf_ALL-gmfpmfp-expansion})~-- f"ur die Streuung der Wegner-Wilson-Loops~\vspace*{-.125ex}$W\Dmfp$,~$W\Dmfm$ im Sinne eichinvarianter Objekte. \\
\indent
Diese Herleitung Meggiolaro geschieht in Euklidischer Raumzeit auf Basis der Feststellung, da"s die Analytische Fortsetzung der \vspace*{-.25ex}\mbox{$T$-Am}\-plitude sich reduziert auf die Analytische Fortsetzung des {\it hyperbolischen Winkels\/}%
  ~\vspace*{-.125ex}$\ps$ zwischen den Parton-Trajektorien in der Boost-Ebene in Minkowskischer Raumzeit zu dem {\it Winkel\/}%
  ~\vspace*{-.125ex}$\th$, den diese einschlie"sen in der entsprechenden Ebene in Euklidischer Raumzeit.
Meggiolaro zeigt diesen Zusammenhang zun"achst in Ref.~\cite{Meggiolaro96} bis zu Termen proportional~\vspace*{-.125ex}$g_{\rm ren.}^4$, dann in Ref.~\cite{Meggiolaro97} zu jeder Ordnung.
Wir gelangen zu demselben Resultat f"ur die Loop-Loop-Amplitude%
  ~\vspace*{-.25ex}$\tTll^{(s,\rb{b})}$, wie abschlie"send dieses Kapitel explizit gezeigt wird im folgenden Abschnitt.
\vspace*{-.5ex}

\section[Nahezu lichtartige~\protect$T$-Amplitude: Analytizit"at]{%
         Nahezu lichtartige~\bm{T}-Amplitude: Analytizit"at}
\label{Sect:T-Amplitude.Analytizit"at}

In Anhang~\ref{APP-Sect:Euklid} ist allgemein diskutiert der "Ubergang von Komponenten%
  ~\vspace*{-.125ex}$x^\mu$ bez"uglich Koordinatenlinien%
  ~\mbox{\,$\mu \!\in\! \{0,1,2,3\}$} in der Minkowski-Raumzeit zu Komponenten%
  ~\vspace*{-.125ex}$\xE^{\xE[\mu]}$ bez"uglich Koordinatenlinien%
  ~\vspace*{-.125ex}\mbox{\,$\xE[\mu] \!\in\! \{1,2,3,4\}$} in deren Fortsetzung ins Euklidische.
Es ist ferner~ausf"uhrlich dargestellt der formale wie interpretatorische Zusammenhang spezieller Koordinatenlinien
   \vspace*{-.125ex}\mbox{\,$\xE[\tilde\mu] \!\in\! \{\mfp,\mfm,1,2\}$} in dieser Euklidischen Raumzeit, deren longitudinale Komponenten~$\mfp$,~$\mfm$~-- analog zum Minkowskischen~-- definiert sind als die Richtungen physikalischer (klassischer) Teilchen-Trajektorien.
In Anhang~\ref{APP-Subsect:Wick} ist explizit ausgef"uhrt die Wick-Drehung, die vermittelt zwischen den Korrelationsfunktionen der einen und der anderen Raumzeit. \\
\indent
Die Formulierung von Anhang~\ref{APP-Sect:Euklid} f"ur diese Euklidische Raumzeit parallel und in voll\-st"andiger Analogie zu Anhang~\ref{APP-Sect:Minkowski} f"ur die \vspace*{-.125ex}Minkowski-Raumzeit~-- vgl.\@ auch Abb.~\ref{Fig:EuklidGENERAL} versus Abb.~\ref{Fig:MinkowskiGENERAL}~-- suggeriert bereits weitgehend die analytische Fortsetzung der \mbox{$T$-Am}\-plitude%
  ~\vspace*{-.125ex}\vspace*{-.25ex}$\tTll^{(s,\rb{b})}$ f"ur die Streuung der Wegner-Wilson-Loops%
  ~\vspace*{-.25ex}$W\Dmfp$,~$W\Dmfm$~-- vgl.\@ die Gln.~(\ref{tTll_WW_ch-mf_ALL-gmfpmfp}),~(\ref{tTll_WW_ch-mf_ALL-gmfpmfp-expansion}).
Wir setzen daher im folgenden Anhang~\ref{APP:Boosts} voraus, so da"s gen"ugt die Fortsetzung von%
  ~\vspace*{-.125ex}$\tTll^{(s,\rb{b})}$ zu skizzieren.
\vspace*{-.5ex}

\bigskip\noindent
In Anhang~\ref{Subsect:AktiveO(4)Drehungen} sind ausf"uhrlich formuliert die aktiven \mbox{$O(4)$-Drehung}\-en, die unmittelbar entsprechen den aktiven Lorentz-Boosts im Minkowskischen.
Es ist diskutiert \vspace*{-.125ex}der Zusammenhang des {\it Winkels\/}~$\th$ in der
  \vspace*{-.125ex}\mbox{$\xE^4\!\xE^3$-Ebe}\-ne der Teilchen-Trajektorien gegeneinander im Euklidischen mit dem {\it hyperbolischen Winkel\/}~$\ps$, den diese einschlie"sen in der \mbox{$x^0\!x^3$-Boost-Ebe}\-ne im Minkowskischen~-- vgl.\@ Gl.~(\ref{APP:thp,thm-psp,psm}):%
\FOOT{
  \label{FN:th-Meggiolaro}Meggiolaro definiert in den Refn.~\cite{Meggiolaro96,Meggiolaro97} den Winkel~$\th$ mit {\sl umgekehrtem\/} Vorzeichen.
}
\begin{samepage}
\vspace*{-.25ex}
\begin{align} \label{-iIMth<->ps}
&-\, \iIM\,\th\vv
  \longleftrightarrow\vv \ps\qqquad
  \th \in (0,\pi \!\big/\! 2)\,,\quad
  \ps \in (0,\infty)
    \\
 &\text{mit}\qquad
  \th \equiv \th\Dmfp + \th\Dmfm\,,\quad\vv
  \ps \equiv \ps\Dmfp + \ps\Dmfm
    \tag{\ref{-iIMth<->ps}$'$}
    \\[-4ex]\nn
\end{align}
Es sind~\vspace*{-.125ex}$\th\Dmfp$,~$\th\Dmfm$ und~$\ps\Dmfp$,~$\ps\Dmfm$ die {\it Winkel\/} und {\it hyperbolischen Winkel\/} der~Tra\-jektorien gegen die $\xE^4$- und \mbox{$x^0$-Ach}\-se, vgl.\@ Gl.~(\ref{APP:ps=psp+psm}) bzw.~(\ref{APP:th=thp+thm}).
Der Limes~\mbox{$s \!\to\! \infty$}~ist realisiert durch
%
\begin{align} 
\th\vv \longrightarrow\vv \pi\!\big/\!2\qquad
  \Longleftrightarrow\qquad
  \ps\vv \longrightarrow\vv \infty
\end{align}
Die Normierungskonstante%
  ~\vspace*{-.125ex}$\xE[\vrh]$ der Koordinatenlinien~$\mfp$,~$\mfm$ im Euklidischen ist~-- analog zu~$\vrh$ im Minkowskischen~-- bestimmt genau durch die \vspace*{-.125ex}Forderung von L"angentreue der Abbildung zwischen den relevanten Koordinaten%
  ~\vspace*{-.125ex}$\xE^\mu$,~\mbox{\,$\mu \!\in\! \{1,2,3,4\}$} und%
  ~$\xE^{\tilde\mu}$,~\mbox{\,$\mu \!\in\! \{\mfp,\mfm,1,2\}$}%
\FOOT{
  Sei im folgenden~-- wie in Anhang~\ref{APP:Boosts}~-- unterdr"uckt das Skript~"`E"' an den Raumzeit-Indizes~\mbox{\,$\xE[\mu],\ldots,\xE[\tilde\mu],\ldots$}
}
oder "aquivalent der Forderung von Invarianz der Determinante des metrischen Tensors.
Dies~impli\-ziert explizite Ausdr"ucke f"ur die Komponenten des metrischen Tensors%
  ~\vspace*{-.125ex}\mbox{\,$\xE[\tilde{g}] \!\equiv\! \big(\xE[g]{}_{\tilde\mu\tilde\nu}\big)$} und da"s diese "ubergehen in die Komponenten des metrischen Tensors%
  ~\vspace*{-.125ex}\mbox{\,$\xE[g] \!\equiv\! \big(\xE[g]{}_{\mu\nu}\big)$} im Limes~\mbox{\,$s \!\to\! \infty$}:
\vspace*{-.25ex}
\begin{align} \label{gE_pp-expl}
&\xE[g]{}_{\mfp\mfp}\;
  =\; \xE[g]{}_{\mfm\mfm}\;
  =\; -\, \xE[\vrh]^2\quad
  =\; -\, \sin^{-1}\!\th
    \\[-.25ex]
  &\hspace*{203pt}
   \longrightarrow\vv -\, 1\qquad
  \text{f"ur\vv $\th \to \pi\!/\!2$}
    \tag{\ref{gE_pp-expl}$'$} \\[.75ex]
&\xE[g]{}_{\mfp\mfm}\;
  =\; \xE[g]{}_{\mfm\mfp}\;
  =\; -\, \xE[\vrh]^2\, \cos\th\quad
  =\; -\, \tan^{-1}\!\th
    \label{gE_pm-expl} \\[-.25ex]
  &\hspace*{213pt}
   \longrightarrow\vv 0\qquad
  \text{f"ur\vv $\th \to \pi\!/\!2$}
    \tag{\ref{gE_pm-expl}$'$}
    \\[-4.25ex]\nn
\end{align}
\end{samepage}%
vgl.\@ die Gln.~(\ref{APP:gE_pp-expl}),~(\ref{APP:gE_pp-expl}$'$) und~(\ref{APP:gE_pm-expl}),~(\ref{APP:gE_pm-expl}$'$).

Die Analyse im Minkowskischen~-- vgl.\@ Abschn.~\ref{Sect:T-Amplitude.Auswertung}~-- ist zun"achst vollst"andig neu durchzuf"uhren im Euklidischen.
De~facto ist dies offensichtlich nicht notwendig, sondern sie reduziert sich genau auf die folgenden Substitutionen:
Es sind zu ersetzen zum einen die Lorentz-Komponenten der Tensoren durch ihre Euklidischen, zum anderen die Korrelationsfunktionen durch ihre analytischen Fortsetzungen ins Euklidische. \\
\indent
Die differentialgeometrische Auswertung bleibt unver"andert; es ist lediglich anzubringen das Skript~"`E"' zur Indizierung, da"s sich die entsprechenden geometrischen Objekte beziehen auf die Euklidische, statt auf die Minkowskische Raumzeit.
In der abschlie"senden \vspace*{-.125ex}Formel f"ur die Euklidische \mbox{$T$-Am}\-plitude%
  ~\vspace*{-.125ex}\vspace*{-.25ex}$\xE[(\tTll^{(s,\rb{b})})]$ f"ur die Streuung der Euklidischen Wegner-Wilson-Loops%
  ~\vspace*{-.25ex}$\xE[W]{}\Dmfp$,~$\xE[W]{}\Dmfm$ ist insofern in Bezug auf die Minkowskische \mbox{$T$-Am}\-plitude%
  ~$\tTll^{(s,\rb{b})}$ entsprechend den Gln.~(\ref{tTll_WW_ch-mf_ALL-gmfpmfp}),~(\ref{tTll_WW_ch-mf_ALL-gmfpmfp-expansion})~-- ist zu ersetzen die unabh"angige Komponente des metrischen Tensors wie folgt:
\begin{samepage}
\vspace*{-.5ex}
\begin{align} \label{relevanteMetrikkomp}
&-\, \det\mathbb{SL}\vv g_{\mfp\mfp}\;
  =\; g_{\mfp\mfp} \!\big/\! g_{+-}\;
  =\; \sinh^{-1}\!\ps\;
  =\; \iIM\, \sin^{-1}\!(\iIM\ps)
    \\
&\phantom{{\iIM} x^0\;}
  \longrightarrow\vv
  +\, \det \slE\vv \xE[g]{}_{\mfp\mfp}\;
  =\; \xE[g]{}_{\mfp\mfp}\;
  =\; -\, \sin^{-1}\!\th
    \tag{\ref{relevanteMetrikkomp}$'$}
    \\[-4ex]\nn
\end{align}
bzgl.\@ der expliziten Ausdr"ucke vgl.\@ Gl.~(\ref{-detSLgmf_s-simplified}) bzw.\@ die Gln.~(\ref{gE_pp-expl}),~(\ref{gE_pp-expl}$'$); der "Ubergang zum Argument~\mbox{\,$\iIM\ps$} ist angegeben in Hinblick auf den Zusammenhang%
  ~\mbox{\,$\iIM\ps \!\leftrightarrow\! \th$} nach Gl.~(\ref{-iIMth<->ps}). \\
\indent
Im "Ubergang von Minkowskischen Null- zu Euklidischen Vier-Komponenten treten auf explizite Faktoren der imagin"aren Einheit.
Daraus folgt%
  ~\mbox{\,$F(\xi^2) \!\to\! \xE[F](\xE[\xi]^2) \!=\! -\iIM\!\cdot\! F\big(\xi^2 \!\to\! \xE[\xi]^2\big)$}, vgl.\@ Gl.\,(\ref{APP:F(xi)-FE(xiE)}).
Der Faktor~$-\iIM$ zusammen mit dem Signum in Gl.(\ref{relevanteMetrikkomp}$'$) ist genau der explizite Faktor~$\iIM$ in Gl.(\ref{relevanteMetrikkomp}).
Mit Beweis der Gln.~(\ref{relevanteMetrikkomp}),~(\ref{relevanteMetrikkomp}$'$) ist daher im wesentlichen gezeigt der Zusammenhang der respektiven $T$-Amplituden durch den Zusammenhang%
  ~\mbox{\,$\iIM\ps \!\leftrightarrow\! \th$} allein.
Dieser Beweis erfordert bei aller gew"unschten Pr"agnanz eine gewisse Explizitheit der Darstellung.
In diesem Sinne f"uhren wir wie folgt duch die Analyse im Euklidischen.
\vspace*{-.5ex}

\bigskip\noindent
Seien zun"achst rekapituliert\,-- vgl.\,Anh.\,\ref{APP-Sect:Euklid}\,-- unsere allgemeinen Definitionen und~Konventio\-nen.
Analytische Fortsetzung der Minkowski-Raumzeit in eine Euklidische sei verstanden als formale Substitution der pseudo-Riemannschen Metrik durch eine Riemannsche im Sinne:
\vspace*{-.5ex}
\begin{align} \label{g->deE}
&g\;
  \equiv\; (g_{\mu\nu})\;
  =\; {\rm diag}[+1,-1,-1,-1]
    \\[-.75ex]
&\phantom{{\iIM} x^0\;}
  \stackrel{\D!}{\longrightarrow}\;
      {\rm diag}[-1,-1,-1,-1]\;
  =\; -(\xE[\de]{}_{\mu\nu})\;
  \equiv\; -\xE[\de]\;
  \equiv\; \xE[g]
    \nn
    \\[-4.25ex]\nn
\end{align}
Bezogen auf Punkte~$x$ der Raumzeit impliziert formale Substitution der Metrik~\mbox{$g \!\to\! \xE[g] \!=\! -\xE[\de]$} bei Forderung von Invarianz des Skalarprodukts~-- vgl.\@ die Gln.~(\ref{APP:g->deE}),~(\ref{APP:xy->xyE}):
\vspace*{-.5ex}
\begin{alignat}{2} \label{x->xE}
&{\iIM} x^0\vv&
  &\longrightarrow\vv \xE^4
    \\
&x^i\vv&
  &\longrightarrow\vv \xE^i\qquad
  i \in \{1,2,3\}
    \tag{\ref{x->xE}$'$}
    \\[-4.25ex]\nn
\end{alignat}
das hei"st Identifizierung von%
  ~\vspace*{-.25ex}$\xE^4$ mit~$\iIM x^0$ und trivial von~$\xE^i$ mit~$x^i$ f"ur~$i \!\in\! \{1,2,3\}$. \\
\indent
Das Vierer-Skalarprodukt ist definiert f"ur Vektoren~$\xE[a]$,~$\xE[b]$ durch%
  ~\vspace*{-.125ex}\mbox{\,$\xE[a] \!\cdot\! \xE[b] \!\equiv\! \xE[g]{}_{\mu\nu} \xE[a]^\mu\, \xE[b]^\nu$} voll\-st"andig analog wie im Minkowskischen~-- es ist daher~{\it negativ definit\/}:
\vspace{-.5ex}
\begin{align} \label{xcdoty-E}
\xE[a]\cdot \xE[a]\;
  =\; -\, \xE[a]^\mu\, \xE[a]^\mu\vv
  \le\vv 0
    \\[-4.25ex]\nn
\end{align}
entsprechend kovariante Komponenten durch:
%
\begin{align} \label{xE-kov}
\xE[a]{}_\mu\;
  :=\; \xE[g]{}_{\mu\nu}\, \xE[a]^\nu\;
   =\; -\xE[\de]{}_{\mu\nu}\, \xE[a]^\nu
    \\[-4ex]\nn
\end{align}
Sie differieren durch ein {\it Signum\/} von den kontravarianten Komponenten:~\vspace*{-.25ex}\mbox{$\xE[a]{}_\mu \!=\! -\xE[a]^\mu$}. \\
\indent
\vspace*{-.25ex}Der Euklidische Epsilon-Pseudotensor%
  ~\vspace*{-.125ex}\mbox{$\xE[\ep] \!\equiv\! \big(\xE[\ep]^{\mu\nu\rh\si}\big)$} sei definiert als das Signum der Permutation seiner Indizes~-- durch seine kontravrianten Komponenten wie folgt:
\vspace*{-.5ex}
\begin{align} \label{epE-Def}
\xE[\ep]^{\mu_1\cdots\mu_4}\;
  \equiv\; \pmatrixZD{\mu_1}{\cdots}{\mu_4}{1}{\cdots}{4}\; \ep^{1234}\qquad
 \text{\sl per def.:}\qqquad
  \xE[\ep]^{1234}\;
  \equiv\; +1
    \\[-4.75ex]\nn
\end{align}
\end{samepage}%
folglich:%
~\mbox{$\xE[\ep]{}_{1234} \!=\!
  \xE[g]{}_{1\mu}\, \xE[g]{}_{2\nu}\, \xE[g]{}_{3\rh}\, \xE[g]{}_{4\si}\,
    \xE[\ep]^{\mu\nu\rh\si} \!=\!
  (-1)^4\, \xE[\ep]^{1234} \!\equiv\! +1$};
vgl.\@ die Gln.~(\ref{APP:epTensor-kontrav}),~(\ref{APP:SignumPermutation}). \\
\indent
Lorentz-Tensorgleichungen involvieren o.E.\@ Vektorfunktionen%
  ~\vspace*{-.125ex}\mbox{$a^\mu\!(x)$}, den metrischen Tensor~$g$ und den Epsilon-Pseudotensor~$\ep$; es folgt als allgemeine Substitutionsvorschrift f"ur den "Ubergang von Minkowskischer zu Euklidischer Raumzeit~-- vgl.\@ die Gln.~(\ref{APP:xy->xyE}),~(\ref{APP:ep->epE}):
\vspace*{-.5ex}
\begin{alignat}{4} \label{TensorGln:M->E}
&a^\mu\!(x)\;&
  &\longrightarrow\vv
      \phantom{-\iIM\; } \xE[a]^\mu\!(\xE)&&&&
    \\
&g_{\mu\nu}\;&
  &\longrightarrow\vv
      \phantom{-\iIM\; } \xE[g]{}_{\mu\nu}\;&
     &=\; -\, \xE[\de]{}_{\mu\nu}&&
    \tag{\ref{TensorGln:M->E}$'$} \\
&\ep_{\mu\nu\rh\si}\;&
  &\longrightarrow\vv
      -\iIM\; \xE[\ep]{}_{\mu\nu\rh\si}&&&&\qqquad
  \text{s.u.\@ Fu"sn.\,\FN{FN:tildeF}}
    \tag{\ref{TensorGln:M->E}$''$}
    \\[-4ex]\nn
\end{alignat}
mit Euklidischen Vektor-Komponenten~\vspace*{-.25ex}\mbox{\,$\xE[a]^\mu$} im Sinne der Gln.~(\ref{x->xE}),~(\ref{x->xE}$'$).
\vspace*{-.25ex}

\bigskip\noindent
Auf dieser Basis werden explizt angegeben die relevanten Gr"o"sen bez"uglich der Euklidischen Raumzeit aus denen bez"uglich der Minkowskischen. \\
\indent
Der (paralleltransportierte) Feldst"arkentensor ist antisymmetrisch in den Raumzeit-Indi\-zes und im Minkowskischen dargestellt wie folgt durch die colour-elektrischen und -magneti\-schen Felder~-- Konvention nach Ref.~\cite{Itzykson88}:%
\FOOT{
  \label{FN:E,B-Matrix}Zu lesen als \vspace*{-.125ex}\mbox{$\dimDrst{F} \!\times\! \dimDrst{F}$}-{\sl Matrizen\/}%
  ~\mbox{$E^i \!\equiv\! E^{ia} T_\Drst{F}^a$},~\mbox{$B^i \!\equiv\! B^{ia} T_\Drst{F}^a$}, mit%
  ~\mbox{$a \!=\! 1,2,\ldots\dimNc$},%
  ~\mbox{$\dimDrst{F} \!\equiv\! \Nc$},~\mbox{$\dimNc \!\equiv\! \Nc^2 \!-\! 1$}, oder {\sl unterdr"uckt\/} der Colour-Index~$a$; seien suggestiv ausgeschrieben nur die unabh"angigen Komponenten.
}
\begin{samepage}
\vspace*{-.75ex}
\begin{align} \label{F_E,B-Matrix}
F\;
  \equiv\; \big(F^{\mu\nu}\big)\;
  =\; \begin{pmatrix}
        0& -E^1& -E^2& -E^3\\
         &    0& -B^3& \phantom{-}B^2\\
         &     &    0& -B^1\\
         &     &     &    0
       \end{pmatrix}\quad
  =\; -\, F{}^{\D\,t}
    \\[-4.75ex]\nn
\end{align}
in kompakter Notation%
  ~-- mit~\mbox{\,$i,j,k,\ldots \!\in\! \{1,2,3\}$},~\mbox{\,$\ep_{123} \!\equiv\! 1$}:%
\FOOT{
  \label{FN:ep-Raum}Der Epsilon-Pseudotensor bez"uglich der drei {\sl Raum\/}koordinaten sei~-- entgegen Gl.~(\ref{APP:epTensor-kontrav})~-- konsequent~no\-tiert mit {\sl unteren\/} Indizes und in dieser Form gleicherma"sen benutzt im Minkowskischen {\sl und\/} Euklidischen.
}
%
\vspace*{-.25ex}
\begin{align} \label{F<->E,B}
&F^{i0}\;
  \equiv\; E^i,\qquad
  F^{ij}\;
   \equiv\; -\, \ep_{ijk}\, B^k
    \\[-.375ex]
 &\Longleftrightarrow\qquad
  E^i\;
   \equiv\; F^{i0},\qquad
  B^i\;
   \equiv\; -\, \frac{1}{2}\; \ep_{ijk}\, F^{jk}
    \tag{\ref{F<->E,B}$'$}
    \\[-4.5ex]\nn
\end{align}
dabei sind%
  ~\vspace*{-.125ex}\mbox{$\vec{E} \!\equiv\! (E^i)^{\D t}$},%
  ~\mbox{$\vec{B} \!\equiv\! (B^i)^{\D t}$} die konventionellen Dreier-Vektoren und~\mbox{\,$\ep_{ikl}\ep_{jkl} \!=\! 2\de_{ij}$} benutzt f"ur die zweite Zeile.
Der duale (paralleltransportierte) Feldst"arkentensor%
\FOOT{
  \label{FN:tildeF}Die Pseudo-Tensoren werden hier nicht explizit ben"otigt und angegeben nur f"ur Vollst"andigkeit.
}
ist definiert durch:%
\FOOT{
  \label{FN:StandardDualit"at}In Standard-Konvention nach Ref.~\cite{Itzykson88}~-- abweichend von der Konvention bez"uglich {\sl Dualit"at\/}, die wir zweckm"a"sig einf"uhren f"ur Integration auf {\sl allgemeinen\/} (pseudo-)Riemannschen Mannigfaltigkeiten~$G_{d,\si}$; vgl.\@ Anh.~\ref{APP-Sect:Integration}, insbes.\@ Gl.~(\ref{APP:Ttilde_T-Konv}).
}
%
\vspace*{-.5ex}
\begin{align} \label{tildeF-Def}
\tilde{F}^{\mu\nu}\;
  \equiv\; \frac{1}{2}\, \ep^{\mu\nu\rh\si}\, F_{\rh\si}
    \\[-4.5ex]\nn
\end{align}
mit~\mbox{\,$\ep^{0123} \!\equiv\! 1$}.
Daraus folgt unmittelbar:
\vspace*{-.25ex}
\begin{align} \label{tildeF<->E,B}
&\tilde{F}^{i0}\;
  \equiv\; B^i,\qquad
  \tilde{F}^{ij}\;
   \equiv\; \ep_{ijk}\, E^k\;
   \equiv\; -\, \ep_{ijk}\, (-E^k)
    \\[-.375ex]
 &\Longleftrightarrow\qquad
  B^i\;
   \equiv\; \tilde{F}^{i0},\qquad
  E^i\;
   \equiv\; \frac{1}{2}\; \ep_{ijk}\, \tilde{F}^{jk}
    \tag{\ref{tildeF<->E,B}$'$}
    \\[-4.5ex]\nn
\end{align}
ergo~\mbox{\,$\tilde{F} \!\equiv\! F[E^i \!\to\! B^i, B^i \!\to\! -E^i]$}, vgl.\@ die Gln.~(\ref{F<->E,B}),~(\ref{F<->E,B}$'$).
In Matrixform gilt:%
\citeFN{FN:E,B-Matrix}
\vspace*{-.75ex}
\begin{align} \label{tildeF_E,B-Matrix}
\tilde{F}\;
  \equiv\; \big(\tilde{F}^{\mu\nu}\big)\;
  =\; \begin{pmatrix}
        0& -B^1&           -B^2&           -B^3\\
         &    0& \phantom{-}E^3&           -E^2\\
         &     &              0& \phantom{-}E^1\\
         &     &               &              0
       \end{pmatrix}\quad
  =\; -\, \tilde{F}{}^{\D\,t}
    \\[-4.75ex]\nn
\end{align}
analog zu Gl.~(\ref{F_E,B-Matrix}).
Es folgt
\vspace*{-.5ex}
\begin{alignat}{3} \label{FF,FtildeF}
&F_{\mu\nu}\, F^{\mu\nu}\;&
  &=\vv 2\, \big( -\zz E^i E^i\, +\, B^i B^i \big)\quad&
  &\equiv\vv 2\, \big( -\zz \vec{E}^2\; +\, \vec{B}^2 \big)
    \\[.25ex]
&F_{\mu\nu}\, \tilde{F}^{\mu\nu}\;&
  &=\; -\, 4\, E^i B^i\quad&
  &\equiv\; -\, 4\, \vec{E} \!\cdot\! \vec{B}
    \tag{\ref{FF,FtildeF}$'$}
    \\[-4.5ex]\nn
\end{alignat}
\end{samepage}%
f"ur die beiden Lorentz-Invarianten.

Der (paralleltransportierte) Feldst"arkentensor im Euklidischen sei dargestellt vollst"andig analog wie im Minkowskischen~-- vgl.\@ die Gln.~(\ref{F<->E,B}),~(\ref{F<->E,B}$'$):%
\citeFN{FN:ep-Raum}
\vspace*{-.25ex}
\begin{align} \label{F<->E,B-Euklid}
&\xE[F]^{i4}\;
   \equiv:\; \xE[E]^i,\qquad
  \xE[F]^{ij}\;
   \equiv:\; -\, \ep_{ijk}\, \xE[B]^k
    \\[-.25ex]
 &\Longleftrightarrow\qquad
  \xE[E]^i\;
   \equiv\; \xE[F]^{i4},\qquad
  \xE[B]^i\;
   \equiv\; -\, \frac{1}{2}\; \ep_{ijk}\, \xE[F]^{jk}
    \tag{\ref{F<->E,B-Euklid}$'$}
    \\[-4.75ex]\nn
\end{align}
Dadurch definiert sind Euklidischen colour-elekrische und -magnetische Felder mit Dreier-Vektoren%
  ~\vspace*{-.125ex}\mbox{$\xE[\vec{E}] \!\equiv\! (\xE[E]^i)^{\D t}$},%
  ~\mbox{$\xE[\vec{B}] \!\equiv\! (\xE[B]^i)^{\D t}$}.
Die Definition nach Gl.~(\ref{F<->E,B-Euklid}) impliziert
\vspace*{-.25ex}
\begin{alignat}{4} \label{E,B->Euklid}
&\iIM E^i\;&
  &\equiv\;& \iIM F^{i0}\vv
  &\longrightarrow\vv&
   \xE[F]{}^{i4}\;
  &\equiv\; \xE[E]^i
    \\[-.25ex]
&B^i\;&
  &\equiv\;& -\, \frac{1}{2}\; \ep_{ijk}\, F^{jk}\vv
  &\longrightarrow\vv&
   -\, \frac{1}{2}\; \ep_{ijk}\, \xE[F]^{jk}\;
  &\equiv\; \xE[B]^i
    \tag{\ref{E,B->Euklid}$'$}
    \\[-4ex]\nn
\end{alignat}
im Sinne der Gln.~(\ref{x->xE}),~(\ref{x->xE}$'$).
Der duale (paralleltransportierte) Feldst"arkentensor%
\citeFN{FN:tildeF}
im Euklidischen ist analog zu Gl.~(\ref{tildeF-Def}) definiert durch:%
\citeFN{FN:StandardDualit"at}
\vspace*{-.5ex}
\begin{align} \label{tildeF-Def-Euklid}
\xE[\tilde{F}]^{\mu\nu}\;
  \equiv\; \frac{1}{2}\, \xE[\ep]^{\mu\nu\rh\si}\, \xE[F]{}_{\rh\si}
    \\[-4.5ex]\nn
\end{align}
mit~\mbox{\,$\xE[\ep]^{1234} \!\equiv\! 1$}.
Daraus folgt unmittelbar~-- vgl.\@ die Gln.~(\ref{tildeF<->E,B}),~(\ref{tildeF<->E,B}$'$):
\vspace*{-.25ex}
\begin{align} \label{tildeF<->E,B-Euklid}
&\xE[\tilde{F}]^{i4}\;
  \equiv\; -\, \xE[B]^i,\qquad
  \xE[\tilde{F}]^{ij}\;
   \equiv\; \ep_{ijk}\, \xE[E]^k\;
   \equiv\; -\, \ep_{ijk}\, (-\xE[E]^k)
    \\[-.25ex]
 &\Longleftrightarrow\qquad
  \xE[B]^i\;
   \equiv\; -\, \xE[\tilde{F}]^{i4},\qquad
  \xE[E]^i\;
   \equiv\; \frac{1}{2}\; \ep_{ijk}\, \xE[\tilde{F}]^{jk}
    \tag{\ref{tildeF<->E,B-Euklid}$'$}
    \\[-4.5ex]\nn
\end{align}
ergo~\mbox{\,$\xE[\tilde{F}]^i \!\equiv\! \xE[F][\xE[E]^i \!\to\! -\xE[B]^i, \xE[B]^i \!\to\! -\xE[E]^i]$}, vgl.\@ die Gln.~(\ref{F<->E,B-Euklid}),~(\ref{F<->E,B-Euklid}$'$).
F"ur Vollst"andigkeit~sei\-en angegeben%
  ~\mbox{\,$\xE[F] \!\equiv\! \big(\xE[F]^{\mu\nu}\big)$},%
  ~\mbox{\,$\xE[\tilde{F}] \!\equiv\! \big(\xE[\tilde{F}]^{\mu\nu}\big)$} in Matrixform:%
\citeFN{FN:E,B-Matrix}
\vspace*{-.5ex}
\begin{align} 
\hspace*{-12pt}
\big(\xE[F]^{\mu\nu}\big)\;
  =\; \begin{pmatrix}
        0& -\xE[B]^3& \phantom{-}\xE[B]^2& \xE[E]^1\\
         &         0&           -\xE[B]^1& \xE[E]^2\\
         &          &                   0& \xE[E]^3\\
         &          &                    &        0
       \end{pmatrix},\quad\vv
  \big(\xE[\tilde{F}]^{\mu\nu}\big)\;
  =\; \begin{pmatrix}
        0& \phantom{-}\xE[E]^3&           -\xE[E]^2& -\xE[B]^1\\
         &                   0& \phantom{-}\xE[E]^1& -\xE[B]^2\\
         &                    &                   0& -\xE[B]^3\\
         &                    &                    &         0
       \end{pmatrix}
    \\[-4.5ex]\nn
\end{align}
vgl.\@ die Gln.~(\ref{F_E,B-Matrix}),~(\ref{tildeF_E,B-Matrix}).
Es folgt~-- vgl.\@ die Gln.~(\ref{FF,FtildeF}),~(\ref{FF,FtildeF}$'$):
\vspace*{-.25ex}
\begin{alignat}{3} \label{FF,FtildeF-Euklid}
&\xE[F]{}_{\mu\nu}\, \xE[F]^{\mu\nu}\;&
  &=\vv 2\, \big( +\zz\xE[E]^i \xE[E]^i\, +\, \xE[B]^i \xE[B]^i\big)\quad&
  &\equiv\vv 2\, \big( +\zz\xE[\vec{E}]^2\; +\, \xE[\vec{B}]^2 \big)
    \\[.25ex]
&\xE[F]{}_{\mu\nu}\, \xE[\tilde{F}]^{\mu\nu}\;&
  &=\; -\, 4\, \xE[E]^i \xE[B]^i\quad&
  &\equiv\; -\, 4\, \xE[\vec{E}] \!\cdot\! \xE[\vec{B}]
    \tag{\ref{FF,FtildeF-Euklid}$'$}
    \\[-4.25ex]\nn
\end{alignat}
f"ur die beiden $O(4)$-Invarianten. \\
\indent
Der fundamentale Korrelations-Lorentztensor~\mbox{$D \!\equiv\! \big(D_{\mu\nu\rh\si}\big)$} ist zerlegt in die Summe seiner Anteile bez"uglich der Strukturen%
  ~\mbox{\,$t\oNC{}_{\zzzz \tilde\mu\tilde\nu\tilde\rh\tilde\si}$} und%
  ~\mbox{\,$t\oC{}_{\zzzz \tilde\mu\tilde\nu\tilde\rh\tilde\si}$}, die explizit gegeben sind durch die Gln.~(\ref{tNC}),~(\ref{tC}); es folgt:%
\FOOT{
  Die verallgemeinerten Kronecker-{\sl Symbole\/} sind definiert~-- \vspace*{-.125ex}unabh"angig von der Raumzeit~-- als Determinante konventioneller Kronecker-{\sl Sympole\/}:~\mbox{\,$\de^{\mu_1\cdots\mu_s}_{\, \nu_1\cdots\nu_s} \!=\! \det \big( \de^{\mu_i}_{\nu_j} \big)$}; vgl.\@ Gl.~(\ref{APP:Kronecker_det-Def}).
}
\begin{samepage}
\vspace*{-.5ex}
\begin{align} \label{tC,NC-Euklid}
&t\oC{}_{\zzzz \mu\nu\rh\si}\;
  =\; \frac{1}{12}\vv
        \de^{\al\be}_{\mu\nu}\vv
        g_{\al\rh}\, g_{\be\si}
    \\[-.5ex]
  &\hspace*{89pt}
   \longrightarrow\vv
  \xE[t]\oC{}_{\mu\nu\rh\si}\;
  =\; \frac{1}{12}\vv
        \de^{\al\be}_{\mu\nu}\vv
        \xE[g]{}_{\al\rh}\, \xE[g]{}_{\be\si}
    \nn
    \\[-4.5ex]\nn
\end{align}
und
\vspace*{-.5ex}
\begin{align}
&t\oNC{}_{\zzzz \mu\nu\rh\si}\;
  =\; \frac{1}{6}\vv g_{\al\ga}\vv
        \de^{\al\be}_{\mu\nu}\, \pa_\be\vv
        \de^{\ga\de}_{\rh\si}\, \pa_\de
    \tag{\ref{tC,NC-Euklid}$'$} \\[-.5ex]
  &\hspace*{89pt}
   \longrightarrow\vv
  \xE[t]\oNC{}_{\zzzz \mu\nu\rh\si}\;
  =\; \frac{1}{6}\vv \xE[g]{}_{\al\ga}\vv
        \de^{\al\be}_{\mu\nu}\, \xE[\pa]{}_\be\vv
        \de^{\ga\de}_{\rh\si}\, \xE[\pa]{}_\de
    \nn
    \\[-4.5ex]\nn
\end{align}
mit Euklidischen Komponenten im Sinne der \vspace*{-.25ex}Gln.~(\ref{x->xE}),~(\ref{x->xE}$'$). \\
\indent
Die konfinierenden und nicht-konfinierenden%
  ~\vspace*{-.125ex}\mbox{$\tilde\ch\idx{\imath\jmath}$-Funk}\-tionen im Minkowskischen sind ausgewertet in den Abschnitten~\ref{Subsect:chNC} und~\ref{Subsect:chC}.
Explizite Darstellungen \vspace*{-.125ex}sind angegeben f"ur%
  ~\mbox{\,$\imath \!\equiv\! \mfp, \jmath \!\equiv\! \mfm$} in den Gln.~(\ref{chNC-0}),~(\ref{chC-0}).
Unter analytischer Fortsetzung gehen sie "uber respektive in
\end{samepage}%
\vspace*{-.5ex}
\begin{align} \label{chC,NC-Euklid}
\xE[\tilde\ch]{}\oC\idx{\imath\jmath}\;
  \phantom{\vv}
  =\; \iIM\, \Big(-\frac{\iIM}{2}\Big)^{\!2}\,
        \iint_{S\Dimath} \dsiIxE{\tilde\mu\tilde\nu}\;
        \iint_{S\Djmath} \dsiJxE{\tilde\rh\tilde\si}\vv
        \xE[t]\oC{}_{\tilde\mu\tilde\nu\tilde\rh\tilde\si}\;
        \xE[\tilde\pa]^2\, \xE[F]\oC(\tilde{x}^2)
    \\[-4.75ex]\nn
\end{align}
und
\vspace*{-.75ex}
\begin{alignat}{2}
\xE[\tilde\ch]{}\oNC\idx{\imath\jmath}\;
  =\; \iIM\, \Big(-\frac{\iIM}{2}\Big)^{\!2}\,
        \iint_{S\Dimath} \dsiIxE{\tilde\mu\tilde\nu}\;
        \iint_{S\Djmath} \dsiJxE{\tilde\rh\tilde\si}\vv
        \xE[t]\oNC{}_{\zzzz \tilde\mu\tilde\nu\tilde\rh\tilde\si}\;
        \xE[F]\oNC(\tilde{x}^2)
    \tag{\ref{chC,NC-Euklid}$'$}
    \\[-4.5ex]\nn
\end{alignat}
Es ist%
  ~\mbox{\,$\tilde{x} \!\equiv\! \big(x^{\tilde\mu}\big)$} und%
  ~\mbox{\,$\tilde{x} \!\equiv\! \tilde{x}\Dimath \!-\! \tilde{x}\Djmath$},%
  ~\mbox{\,$\imath,\jmath \!\in\! \{\mfp,\mfm\}$}; ferner sind%
  ~\mbox{\,$\tilde\mu,\ldots \!\in\! \{\mfp,\mfm,1,2\}$} Koordinatenlinien in der Euklidischen Raumzeit.
Sei verwiesen auf Anhang~\ref{APP-Sect:Euklid}, in dem diese explizit definiert sind durch die Richtungen der klassischen Teilchen-Trajektorien und ausf"uhrlich dargestellt ist ihr Zusammenhang mit den entsprechenden Koordinatenlinien der Minkowski-Raumzeit. \\
\indent
Die Auswertung der Funktionen%
  ~\vspace*{-.125ex}$\xE[\tilde\ch]{}\oC\idx{\imath\jmath}$,%
  ~$\xE[\tilde\ch]{}\oNC\idx{\imath\jmath}$~-- vgl.\@ die Gln.~(\ref{chC,NC-Euklid}),~(\ref{chC,NC-Euklid}$'$)~-- im Euklidischen geschieht vollst"andig analog zu der Auswertung der Funktionen%
  ~\vspace*{-.125ex}$\tilde\ch\idx{\imath\jmath}\oC$,%
  ~$\tilde\ch\idx{\imath\jmath}\oNC$ im~Minkow\-skischen in den Abschnitten~\ref{Subsect:chNC} und~\ref{Subsect:chC}.
Es folgt in faktorisierter Form:
\begin{samepage}
\vspace*{-.25ex}
\begin{align} \label{chC_X-Euklid}
&\tilde\ch\oC\idx{\imath\jmath}\;
  =\; -\, \det\mathbb{SL}\vv g_{\mfp\mfp}\;
        \cdot\; \tilde{X}\oC\idx{\imath\jmath}\;
  =\; g_{\mfp\mfp} \!\big/\! g_{+-}\;
        \cdot\; \tilde{X}\oC\idx{\imath\jmath}
    \\[.25ex]
&\text{analog:}\qquad
 \xE[\tilde\ch]{}\oC\idx{\imath\jmath}\;
   =\; +\, \det \slE\vv \xE[g]{}_{\mfp\mfp}\;
        \cdot\; \xE[\tilde{X}]{}\oC\idx{\imath\jmath}\;
   =\; \xE[g]{}_{\imath\jmath}\;
        \cdot\; \tilde{X}\oC\idx{\imath\jmath}
    \tag{\ref{chC_X-Euklid}$'$}
    \\[-4ex]\nn
\end{align}
entsprechend f"ur die nicht-konfinierenden Funktionen~\mbox{\,$\tilde{X}\oNC\idx{\imath\jmath}$}.
Die Funktionen%
  ~\vspace*{-.25ex}\mbox{\,$\tilde{X}\oC\idx{\imath\jmath}$},%
  ~\mbox{\,$\tilde{X}\oNC\idx{\imath\jmath}$} in den zweiten Darstellungen sind die Funktionen im Minkowskischen, die vollst"andig bestimmt sind durch die Geometrie in der \mbox{$x^1\!x^2$-Trans}\-versalebene.
Explizite Darstellungen sind angegeben in den Gln.~(\ref{ChNC-pm})-(\ref{ChC-pp}) und in deren gestrichenen Pendants\vspace*{-.125ex}.
Von diesen Funktionen differieren die Funktionen%
  ~\vspace*{-.125ex}\mbox{\,$\xE[\tilde{X}]{}\oC\idx{\imath\jmath}$},%
  ~\mbox{\,$\xE[\tilde{X}]{}\oNC\idx{\imath\jmath}$} der ersten Darstellung in Gl.~(\ref{chC_X-Euklid}$'$) genau darin, da"s in ihnen auftritt die respektive ins Euklidische fortgesetzte Korrelationsfunktion%
  ~\vspace*{-.125ex}$\xE[F]\oC$,~$\xE[F]\oNC$ an Stelle der urspr"unglichen%
  ~\vspace*{-.125ex}$F\oC$,~$F\oNC$.
Die zweite Darstellung in Gl.~(\ref{chC_X-Euklid}$'$) ist daher unmittelbare Konsequenz deren Zusammenhangs
\vspace*{.25ex}
\begin{align} \label{F(xi)-FE(xiE)}
\xE[F](\xE[\xi]^2)\;
  =\; -\iIM\cdot F\big(\xi^2 \!\to\! \xE[\xi]^2\big)\qquad\qquad
  F \equiv F\oC, F\oNC, \ldots
\end{align}
vgl.\@ Anh.~\ref{APP-Subsect:Wick}, \vspace*{-.125ex}Gl.~(\ref{APP:F(xi)-FE(xiE)}). \\
\indent
Zun"achst nichttrivial~-- vgl.\@ die Gln.~(\ref{chC_X-Euklid}),~(\ref{chC_X-Euklid}$'$)~-- ist das {\it positive\/} Vorzeichen im Euklidischen gegen"uber dem {\it negativen\/} Vorzeichen im Minkowskischen, mit dem multipliziert ist die respektive Determinante der Transformationsmatrix%
  ~\mbox{\,$\slE$} und~\mbox{\,$\mathbb{SL}$}.
Bzgl.~\mbox{\,$\det\slE \!\equiv\! 1$} vgl.\@ die Fixierung der Euklidischen Koordinatenlinien%
  ~\mbox{\,$\mu \!\in\! \{\mfp,\mfm,1,2\}$} in Anh.~\ref{Subsect:AktiveO(4)Drehungen}, Gl.~(\ref{APP:vrhE-Forderung}$'$).
Dieses Signum ist die einzige Diskrepanz der Euklidischen Auswertung gegen"uber der Minkowskischen; es folgt im Zuge der Integration der longitudinalen Parameter~$u\Dmfp$,~$u\Dmfm$~-- vgl.\@ Anh.~\ref{APP-Subsect:umfp,umfm-Int}~--, da das Volumenelement im Euklidischen Impulsraum gegeben ist durch
\vspace*{.25ex}
\begin{align} 
d\xE[k]\;
  \equiv\; d\xE[k]^4\, d\xE[k]^3\, d^2\xE[\rb{k}]\;
  =\, +\det\slE\vv d\xE[k]{}_\mfp\, d\xE[k]{}_\mfm\, d^2\xE[\rb{k}]
    \\[-3.5ex]\nn
\end{align}
gegen"uber%
  ~\vspace*{-.125ex}\mbox{\,$dk \equiv\mskip-1.5mu dk^0 dk^3 d^2\rb{k} =\,
    -\det\mathbb{SL}\; dk_\mfp dk_\mfm d^2\rb{k}$} im Minkowskischen, vgl.\@ Gl.~(\ref{APP:dk,d4k}):
Herunterziehen der longitudinalen Raumzeit-Indizes ergibt zwei Vorzeichen statt nur eines. \\
\indent
Es gelten die expliziten Darstellungen:
%
\begin{alignat}{3} \label{relevanteMetrikkomp-explizit}
&\xE[g]{}_{\mfp\mfp}\;&
  &=\; -\, \sin^{-1}\!\th\quad&
  &\equiv\; \iIM\cdot \sinh^{-1}\!(-\iIM\th)
    \\[.25ex]
&\xE[g]{}_{\mfp\mfm}\;&
  &=\; -\, \tan^{-1}\!\th\quad&
  &\equiv\; \iIM\cdot \tanh^{-1}\!(-\iIM\th)
    \tag{\ref{relevanteMetrikkomp-explizit}$'$}
    \\[-4ex]\nn
\end{alignat}
vgl.\@ die Gln.~(\ref{gE_pp-expl}),~(\ref{gE_pp-expl}$'$) und~(\ref{gE_pm-expl}),~(\ref{gE_pm-expl}$'$).
Es gilt f"ur die unabh"angigen konfinierenden Funktionen:
\end{samepage}%
%
\begin{alignat}{4} \label{chC_X-Euklid-Analyt}
&\xE[\tilde\ch]{}\oC\idx{\mfp\mskip-2mu\mfp}[\th]\;&
  &=\; \sinh^{-1}\!(-\iIM\th)\;
        \cdot\; \tilde{X}\oC\idx{\imath\jmath}\qquad&
  &=\;& \tilde\ch\oC\idx{\mfp\mskip-2mu\mfp}&[\ps \!\to\! -\iIM\th]
    \\[.375ex]
 &&&\hspace*{28pt}
    \text{analog:}\qquad
  \xE[\tilde\ch]{}\oC\idx{\mfm\mskip-2mu\mfm}[\th]\;&
  &=\;& \tilde\ch\oC\idx{\mfm\mskip-2mu\mfm}&[\ps \!\to\! -\iIM\th]
    \tag{\ref{chC_X-Euklid-Analyt}$'$} \\[.5ex]
 &\xE[\tilde\ch]{}\oC\idx{\mfp\mskip-2mu\mfm}[\th]\;&
  &=\; \tanh^{-1}\!(-\iIM\th)\;
        \cdot\; \tilde{X}\oC\idx{\imath\jmath}\qquad&
  &=\;& \tilde\ch\oC\idx{\mfp\mskip-2mu\mfm}&[\ps \!\to\! -\iIM\th]
    \tag{\ref{chC_X-Euklid-Analyt}$''$}
    \\[-4.5ex]\nn
\end{alignat}
entsprechend f"ur die unabh"angigen nicht-konfinierenden Funktionen%
  ~\vspace{-.25ex}$\xE[\tilde\ch]{}\oNC\idx{\mfp\mskip-2mu\mfp}$,~$\xE[\tilde\ch]{}\oNC\idx{\mfm\mskip-2mu\mfm}$,%
  ~$\xE[\tilde\ch]{}\oNC\idx{\mfp\mskip-2mu\mfm}$.

F"ur die~\mbox{$T$-Am}\-plituden%
  ~\vspace*{-.125ex}\mbox{\,$\xE[(\tTll)] \!\equiv\! \xE[(\tTll^{(s,\rb{b})})]$} und%
  ~\mbox{\,$\tTll \!\equiv\! \tTll^{(s,\rb{b})}$} f"ur die Streuung der respektiven Wegner-Wilson-Loops in Euklidischer und Minkowskischer Raumzeit besteht folglich der Zusammenhang:
\vspace*{.25ex}
\begin{align} \label{TE_T-Analyt}
&\xE[(\tTll)][\th]\;
  =\; \tTll[\ps \!\to\! -\iIM\th]
    \\[1ex]
 &\text{umgekehrt:}\qquad
  \tTll[\ps]\;
  =\; \xE[(\tTll)]{}[\th \!\to\! \iIM\ps]
    \tag{\ref{TE_T-Analyt}$'$}
    \\[-3.5ex]\nn
\end{align}
Explizite Darstellungen folgen auf Basis von \vspace*{-.25ex}Gl.~(\ref{tTll_WW_ch-mf_ALL-gmfpmfp}) und~(\ref{tTll_WW_ch-mf_ALL-gmfpmfp-expansion}). \\
\indent
Zusammenfassend ist der Zusammenhang der%
  ~\vspace*{-.125ex}\mbox{$T$-Am}\-plituden f"ur die Streuung zweier Wegner-Wilson-Loops in Euklidischer und Minkowskischer Raumzeit vollst"andig gegeben
durch den Zusammenhang~-- vgl.\@ Gl.~(\ref{-iIMth<->ps}) und Fu"sn.\,\FN{FN:th-Meggiolaro}:
%
\begin{align} 
-\, \iIM\,\th\vv
  \longleftrightarrow\vv \ps
\end{align}
mit~$\th$ dem {\it Winkel\/} und~$\ps$ dem {\it hyperbolischen Winkel\/}, den respektive einschlie"sen die Loops
   $\xE[W]{}\Dmfp$,~$\xE[W]{}\Dmfp$ im Euklidischen und die Loops%
  ~\vspace*{-.125ex}$W\Dmfp$,~$W\Dmfm$ im Minkowskischen. \\
\indent
Die \mbox{$s$-Ab}\-h"angigkeit der Loop-Loop-Amplituden%
  ~\vspace*{-.125ex}\mbox{\,$\tTll$} und~\mbox{\,$\xE[(\tTll)]$} ist vollst"andig subsumiert in den respektiven longitudinalen Metrik-Komponenten%
  ~$g_{\mfp\mfm}$,~$g_{\mfm\mfp}$ und%
  ~$\xE[g]{}_{\mfp\mfm}$,~$\xE[g]{}_{\mfm\mfp}$.
Unsere Herleitung zeigt, da"s der Zusammenhang der \mbox{$T$-Amp}\-lituden gem"a"s der Gln.~(\ref{TE_T-Analyt}),~(\ref{TE_T-Analyt}$'$) daher {\it exakt\/} ist: zu jeder Ordnung einer Entwicklung in Kumulanten~-- \vspace*{-.125ex}oder etwa in der renormierten Eichkopplung~$g_{\rm ren.}^4$ der Quantenchromodynamik, in Termen derer Meggiolaro~-- vgl.\@ die Refn.~\cite{Meggiolaro96,Meggiolaro97}~-- seine Beweise f"uhrt f"ur die Quark-Quark-Amplitude. \\
\indent
Mit der Euklidischen Amplitude%
  ~\vspace*{-.125ex}$\xE[(\tTll)]$ als Funktion von~$\th$ ist daher die Minkowskische Amplitude%
  ~\vspace*{-.125ex}$\tTll$ gegeben durch analytische Fortsetzung von~$\th$ zu imagin"aren Werten:%
  ~\mbox{\,$\th \!\to\! \iIM\ps$}.
Auf diese Weise k"onnen auf Basis einer Euklidischen Theorie~-- etwa Quantenchromodynamik als Gittereichtheorie oder das Modell des Stochastischen Vakuums in seiner urspr"unglichen Formulierung, die im eigentlichen Sinne erst seine Annahmen rechtfertigt~-- physikalische Wirkungsquerschnitte f"ur hohe Energien berechnet werden; mit "Ubergang~\mbox{\,$\ps \!\to\! \infty$} in den abschlie"senden Ausdr"ucken ist gegeben der Limes~\mbox{\,$s \!\to\! \infty$} lichtartiger Wegner-Wilson-Loops. \\
\indent
Es ist uns wohl bewu"st, da"s dieser Weg in~praxi nicht trivial durchzuf"uhren ist:
Die Euklidische \mbox{$T$-Am}\-plitude ist zu {\it bestimmen\/} als Funktion des Winkels~$\th$, sie ist dann {\it analytisch fortzusetzen\/} zu imagin"aren Werten dieses Winkels:~\mbox{\,$\th \!\to\! \iIM\ps$} und abschlie"send zu {\it extrapolieren\/} zu Unendlichen Werten des hyperbolischen Winkels~$\ps$.
F"ur~\DREI{Q}{C}{D} als Gittereichtheorie ist dieser Weg noch nicht angegangen und erst von prinzipieller Bedeutung.
F"ur das~\DREI{M}{S}{V} ist er explizit durchgef"uhrt im vorliegenden \vspace*{41.5ex}Kapitel dieser Arbeit.
\theendnotes

%% file: GROUND-F.tex
\lhead[\fancyplain{}{\sc\thepage}]%
      {\fancyplain{}{\sc\rightmark}}
\rhead[\fancyplain{}{\sc{Leptoproduktion von~\protect$\rh(770)$,
                           \protect$\om(782)$,~\protect$\ph(1020)$ und~\protect$\Jps(3097)$}}]%
      {\fancyplain{}{\sc\thepage}}
\chapter[Leptoproduktion von~\protect\bm{\rh(770)},
           \protect\bm{\om(782)},~\protect\bm{\ph(1020)} und~\protect\bm{\Jps(3097)}]{%
   \huge Leptoproduktion von~\vspace*{-.125ex}\protect\bm{\rh(770)},
           \protect\bm{\om(782)},~\protect\bm{\ph(1020)} und~\protect\bm{\Jps(3097)}~\bffootnote}
\label{Kap:GROUND}
\footnotetext{
  Kapitel~\ref{Kap:ANALYT} ist verfa"st {\sl nach\/} Kapitel~\ref{Kap:GROUND} und~\ref{Kap:EXCITED}; es bestehen Redundanzen, die bewu"st {\sl nicht\/} ausger"aumt sind.
}
%

In den Kapiteln~\ref{Kap:GROUND} und~\ref{Kap:EXCITED} wenden wir uns weg von der Theorie hin zu der Ph"anomenologie des \DREI{M}{S}{V} und konfrontieren diese mit dem Experiment.
Wir pr"asentieren explizite Anwendungen des \DREI{M}{S}{V} auf Diffraktion, die wir in weiterem Rahmen betrachten wollen als Referenz auf nichtperturbativ dominierter QCD im allgemeinen.
Zu der impliziten Forderung gen"ugend kleinen invarianten Impuls"ubertrags tritt, als Voraussetzung f"ur die Skalentrennung im Sinne des vorangegangenen Kapitels~\ref{Kap:ANALYT}, die Forderung gen"ugend gro"ser invarianter Schwerpunktsenergie.
Konkret diskutieren wir als Proze"s der Hochenergiestreuung die Transmutation eines realen oder virtuellen Photons~$\ga^{\scriptscriptstyle({\D\ast})}$ in ein Vektormeson~$V$~--
kurz: die exklusive Photo- beziehungsweise Leptoproduktion eines Vektormesons~--
diffraktiv an einem Nukleon~$N$, das selbst also in diesem Proze"s intakt bleibt:~\mbox{\,$\ga^{\scriptscriptstyle({\D\ast})}N \!\to\! VN$}.%
\FOOT{
  \label{FN:Nukleon-Proton}Das Pr"afix "`Lepto-"' greift auf das physikalische Lepton, das im Experiment das einlaufende virtuelle wie auch das reale Photon abstrahlt; "`Photo-"' steht bei realem, physikalischem Photon. Experimentelle Daten existieren gro"senteils f"ur Protonen, wir sprechen daher h"aufig von~"`Proton"' statt von~"`Nukleon"'.
}
Photon und Vektormeson sind beides~\mbox{\,$J^{P\!C} \!=\! 1^{--}$}-Teilchen, so da"s die Wechselwirkung aufgefa"st werden kann als der Austausch eines Quasiteilchens mit den Quantenzahlen des Vakuums: des (Weichen) Pomerons.
Wir berechnen in~$t$ differentielle Wirkungsquerschnitte f"ur~$\surd-t \!<\! 1$~GeV und hoher aber fester Schwerpunktenergie~$\surd s \!>\! 10$~GeV, f"ur Definiertheit~$\surd s \!=\! 20$~GeV. \\
\indent
Kapitel~\ref{Kap:GROUND}, ver"offentlicht in Ref.~\cite{Dosch96}, leitet ein mit der Diskussion des zugrundeliegenden Bildes, im Rahmen dessen~-- in Gegen"uberstellung mit konkurrierenden Zug"angen~-- wir allgemein unseren Zugang herstellen.
Dieser Zugang induziert zun"achst die Einschr"ankung auf Photonen nicht beliebig kleiner Virtualit"aten, {\it a~posteriori\/}~$Q^2 \!>\! 1$~GeV$^2$.
Wir beziehen uns weiter zun"achst auf Vektormesonen im Grundzustand.
Der die Observablen wesentlich determinierende Mechanismus nichtperturbativer QCD wird im Rahmen des \DREI{M}{S}{V} identifiziert und diskutiert: ein String-String-Mechanismus, der besteht in der {\it Ausbildung und Wechselwirkung gluonischer Strings\/}.
Die Diskussion konkret der Resonanzen~$\rh(770)$,~$\om(782)$,~$\ph(1020)$ und~$\Jps(3097)$~-- in der Notation der Particle Data Group, Ref.~\cite{PDG98}~-- zeigt in Gegen"uberstellung mit den verf"ugbaren experimentellen Daten die Relevanz dieses Mechanismus'.
Sie suggeriert weiter die Analyse des nachfolgenden Kapitels. \\
\indent
Kapitel~\ref{Kap:EXCITED}, ver"offentlicht in den Refn.~\cite{Kulzinger98,Kulzinger98a,Kulzinger99}, arbeitet die Bedeutung dieses Mechanismus' weiter heraus.
Es dehnt die Analyse aus tiefer in einen Bereich nichtperturbativ dominierter QCD, der noch wesentlicher bestimmt sein {\it sollte\/}~-- und im \vspace*{-.125ex}Resultat {\it ist}~-- durch den identifizierten String-String-Mechanismus: m"oglichst gro"se Ausdehnung der involvierten Strings, oder gleichbedeutend (transversal) m"oglichst ausgedehnte hadronische Zust"ande.
Bezogen auf den konkreten Proze"s, der wesentlich variiert werden kann bez"uglich seiner Photon-Vektormeson-Komponente, hei"st dies
zum einen, unseren Zugang auszudehnen auf {\it virtuelle Photonen beliebig kleiner\/} bis {\it reale Photonen verschwindender Virtualit"at\/},
und zum anderen, zus"atzlich zu der Grundzustand-\mbox{$\rh(770)$-Reso}\-nanz deren {\it (transversal) ausgedehntere Anregungen\/} zu diskutieren, also zun"achst~-- wieder in der Notation von Ref.~\cite{PDG98}~-- die Zust"ande~$\rh(1450)$ und~$\rh(1700)$.
Zielrichtung ist auch hier die Konfrontation unserer Postulate mit den verf"ugbaren experimentellen Daten.
\vspace*{-.5ex}

\section{Zugrundeliegendes Bild und Zusammenhang}

Exklusive Photo- und Leptoproduktion eines Vektormesons~$\ga^{\scriptscriptstyle({\D\ast})}N \!\to\! VN$ ist in besonderer Weise geeignet, die Physik zu untersuchen, die Diffraktion im speziellen und nichtperturbativ dominierter QCD im allgemeinen zugrundeliegt.

Ist die Koh"arenzl"ange des (virtuellen) Photons gr"o"ser als der Radius des Hadrons, mit dem es zur Kollision kommt, kann dieses Photon~-- im Ruhesystem des Hadrons oder im Schwerpunktsystem der Kollision~-- aufgefa"st werden als hadronisches System aus Partonen im Feynmanschen Sinne.
Die Photon-Hadron-Wechselwirkung ist dann grundlegend vergleichbar einer konventionellen Hadron-Hadron-Wechselwirkung. \\
Der Proze"s ist in zweierlei Hinsicht determiniert durch die Physik gro"ser Abst"ande, das hei"st durch nichtperturbativ dominierte QCD:
Zum ersten aufgrund der Beschr"ankung auf Diffraktion, also {\it per definitionem\/} auf einen gen"ugend kleinen invarianten Impuls"uber\-trag~$\surd-t$.
Und zum zweiten aufgrund der gro"sen (transversalen) Ausdehnungen der involvierten Zust"ande. \\
Er ist in besonderer Weise geeignet, diese Physik zu untersuchen, insofern als er insbesondere bez"uglich seiner Photon-Vektormeson-Komponente eine Anzahl Parameter besitzt, die theoretisch {\it variiert\/} und im Experiment als unterscheidbare Signale {\it detektiert\/} werden k"onnen, und er daher eine breit angelegte Untersuchung gestattet.
Unter Betonung dieser Parameter stellt sich der Proze"s exklusiver Photo- und Leptoproduktion eines Vektormesons in folgender Weise dar:
Durch Variation der Wellenfunktion des hochenergetischen Photons in Bezug auf seine Virtualit"at~$Q$ und Polarisation~$\la$ wird ein hadronischer Zustand "`nahezu beliebiger"' (transversaler) Ausdehnung generiert.
Dieser transmutiert diffraktiv am Nukleon in ein Vektormeson mit entsprechender Hochenergiewellenfunktion, die ihrerseits, in Abh"angigkeit von invarianter Masse~$M_V$ und Polarisation~$\la'$, dessen (transversale) Ausdehnung bestimmt.
Der f"ur den String-String-Mechanismus wesentliche Parameter: (transversale) Ausdehnung der involvierten hadronischen Zust"ande, wird so mittelbar variiert und der Mechanismus und dessen Relevanz f"ur die Determinierung nichtperturbativer QCD untersucht.
\vspace*{-.5ex}

\bigskip\noindent
Hadronische Zust"ande in nichtperturbativ dominierter QCD k"onnen anschaulich%
\FOOT{
  \label{FN:pure-gauge-theory}Wir beziehen uns in dieser Anschauung auf die Approximation einer reinen Gluodynamik im Sinne der {\it quenched approximation\/}:   Mit letztlich dem Argument der Dominanz der Skala des Gluonkondensates~$\vac{g^2FF}$ "uber die des Quark\-kondensates~$\vac{\psi\bar{\psi}}$ vernachl"assigt sie konsequent Quarkloopdiagramme,~-- und kann so formal aufgefa"st werden als der der Limes unendlicher Quarkmassen.
}
aufgefa"st werden als Zusammenspiel zweier Komponenten: der diese Zust"ande konstituierenden Partonen, im allgemeinen Quarks, und der sich zwischen diesen ausbildenden gluonischen Strings.
Die Ausdehnung hadronischer Zust"ande wird bestimmt durch die (axiale wie radiale) Ausgedehnung der Strings zwischen den konstituierenden Partonen,~-- die auf diese Weise erst zu Hadronen fester Radii konfiniert werden.
Vor dem Hintergrund dieser Anschauung macht das \DREI{M}{S}{V} einen quantitativen Mechanismus identifizierbar, der nichtperturbativ dominierte QCD im allgemeinen aus derselben Ursache heraus erkl"art wie deren prominentestes Ph"anomen: das Ph"anomen des Confinements. \\
Dieser String-Mechanismus wurde zun"achst eingehend untersucht f"ur das Ph"anomen des Confinements selbst, das hei"st, in unserer Anschauung, dem {\it Ausbilden\/} gluonischer Strings zwischen den konstituierenden Partonen und deren dadurch vermittelten Einschlu"s zu eichinvarianten Objekten, den Hadronen.
F"ur die einfachste Darstellung eines statischen~-- in praxi: schweren~-- Hadrons als eichinvariantes Paar eines statischen Quarks und Antiquarks wurde der verantwortliche String buchst"ablich ausgemessen und unter anderem bez"uglich seines Energiegehaltes Konsistenz gezeigt mit einem bekannten Niederenergietheorem, der Michael-sum rule, vgl.\@ die Refn~\cite{Rueter94a,Dosch95}. \\
Der hierf"ur relevante String-Mechanismus in der Niederenergiephysik f"uhrt unmittelbar auf einen String-String-Mechanismus in der Hochenergiestreuung.
Dieser erkl"art die Observablen genau dadurch, da"s die gluonischen Strings der streuenden Hadronen miteinander in {\it Wechselwirkung\/} treten~-- letztlich durch deren transversale Ausdehnung und implizierend die Aussage, da"s Observable der Hadron-Hadron-Streuung nicht konstruiert werden k"onnen aus Quark-Quark-Streuung im Sinne von Quark\-additivit"at allein, vgl.\@ Ref.~\cite{Kraemer91}.
Konkret postuliert das \DREI{M}{S}{V} Hochenergie-Wirkungsquerschnitte, differentiell im invarianten Quadrat des Impulstransfers~$-t$, f"ur~$-t \!<\! 1$~GeV$^2$, und erkl"art die Korrelation zwischen slope-Parameter und totalem Wirkungsquerschnitt in der elastischen Proton-Proton-Streuung, vgl.\@ Ref.~\cite{Dosch94a}, bzgl.\@ der ph"anomenologischen Beobachtung dieser Korrelation Ref.~\cite{Povh87}.

Wir untersuchen diesen zugrundeliegenden String-String-Mechanismus des \DREI[]{M}{S}{V}.
Die Verifizierung und Demonstration seiner Relevanz mu"s f"ur Prozesse geschehen, die gro"se Effekte~-- auch und gerade im Vergleich zu konkurrierenden Beschreibungen~-- erwarten lassen.
Wir sind daher a~priori interessiert an m"oglichst ausgedehnten Strings, oder gleichbedeutend an m"oglichst ausgedehnten Zust"anden.
\vspace*{-.75ex}

\bigskip\noindent
Im vorangegangenen Kapitel haben wir Nachtmanns Zugang zu weiche Hochenergiestreuung nachgezeichnet.
Seine $T$-Amplitude ist dominiert durch einen Ausdruck, der dem diffraktiven, in diesem Sinne nichtperturbativen Charakter des Prozesses Rechnung tr"agt: der Vakuumerwartungswert zweier (nahezu) lichtartiger Wegner-Wilson-Loops, der ausgewertet werden kann im Rahmen des \DREI[]{M}{S}{V}.
Dar"uberhinaus ist aber wesentlich die Kenntnis der (Nahezu)Lichtkegelwellenfunktionen der in der Kollision involvierten hadronischen Zust"ande.

Diese hadronischen Zust"ande werden aufgefa"st als Wellenpakete von Colour-Dipolen~-- Mesonen von Quark-Antiquark-Dipolen, Baryonen von Quark-Diquark-Dipolen%
\FOOT{
  Der besseren Lesbarkeit wegen sei hierzwischen im folgenden sprachlich nicht mehr differenziert.
}~--
die bestimmt sind durch zum einen die Aufteilung des Lichtkegel-Gesamtimpulses auf die beiden Colourladungen:~$z$ der Anteil des Quarks, und zum anderen deren transversalen Abstandsvektor~$\rb{r}$.
Die (Nahezu)Lichtkegelwellenfunktionen beziehen sich im "ublichen Sinne einer Wellenfunktion auf diese Variablen~$z$ und~$\rb{r}$ und daher auf das ihnen zugrundeliegende Bild mit (nahezu) Lichtgeschwindigkeit propagierender hadronischer Zust"ande.

Zusammenfassend dieses Bild, fluktuiert also das hochenergetische einlaufende Photon~$\ga^{\scriptscriptstyle({\D\ast})}$ gem"a"s einer Wellenfunktion in Wellenpakete von Colour-Dipolen (Quark-Antiquark-Paaren), das jedes einzelne charakterisiert ist durch das Variablenpaar~$(z,\rb{r})$.
Diese Dipole $\{z,\rb{r}\}$ streuen diffraktiv am Nukleon und werden gebunden als definierter Zustand eines Vektormesons~$V$,~-- formalisiert wiederum durch {\it dessen\/} (Nahezu)Lichtkegelwellenfunktion.
Das Nukleon wird dabei der Einfachheit halber betrachtet als Wellenpakete von Colour-Dipolen (Quark-Diquark-Paaren), deren variable Abh"angigkeit~-- wir kommen hierauf zu\-r"uck~-- betrachtet werden kann als reduziert auf~$|\rb{r'}|$, den Betrag es transversalen Abstands der dar"uberhinaus gau"s'sch verteilt ist.
Dem so resultierenden Bild einer Dipol-Dipol-Streuung liegt die Beobachtung zugrunde, da"s die Zeitskala, auf der die beiden Colour-Dipole miteinander in Wechselwirkung treten sehr viel kleiner ist als die, auf der die Fluktuation aus beziehungsweise die Bindung in die asymptotischen hadronischen Zust"ande stattfindet.
Wir erinnern an Nachtmanns Diskussion dieser Zeitskalen und seinen Begriff "`Femtouniversum"', vgl.\@ Ref.~\cite{Nachtmann91}.

Exklusive Photo- und Leptoproduktion eines Vektormesons an einem Nukleon kann allgemein in vollst"andiger Analogie zu Hadron-Nukleon-Streuung berechnet werden: indem das Photon repr"asentiert wird durch seine hadronische Wellenfunktion, in dem von uns betrachteten Hochenergieproze"s durch seine (Nahezu)Lichtkegelwellenfunktion.
Diese kann unter bestimmten Bedingungen perturbativ berechnet werden, im Rahmen der Lichtkegelst"orungstheorie (light cone perturbation theory, LCPT).

F"ur kleine Virtualit"aten, a~posteriori~$Q^2 \!<\! 1$~GeV$^2$, ist das Photon von der Gr"o"se typischer Hadronen und wird dominiert durch die~-- nichtperturbative~-- Physik gro"ser Distanzen: seine perturbative Beschreibung im Rahmen der LCPT bricht zusammen.
Unser Ansatz f"ur diesen Bereich von~$Q^2$ war zun"achst der des Verallgemeinerten Vektormeson-Dominanz-Modells~(GVDM).
Dieses repr"asentiert das Photon als "Uberlagerung nicht von Colour-Dipolen aus (Anti)Quarks, sondern im Sinne der Quark-Hadron-Dualit"at als "Uberlagerung der Grundzustand-Vektormesonen~$\rh(770)$ und $\om(782)$,~$\ph(1020)$, $\Jps(3097)$ und deren immer h"oheren Anregungen.
Einerseits m"ussen f"ur eine Approximation des Photons auf diese Weise m"oglichst viele Anregungen ber"ucksichtigt werden, vgl.\@ Ref.~\cite{Dosch97}.
Andererseits werden f"ur jedes in die Diskussion mit einbezogene h"oher angeregte Vektormeson~$V$ de facto unbekannte Parameter eingef"uhrt (etwa deren mit einer komplexen Phase versehene Kopplung an den elektromagnetischen Strom~$f_V$).
Im Resultat f"uhrt dies zu einer Inflation der Parameter und zur totalen Deflation der Aussagekraft des Modells~-- und damit den Ansatz {\it ad absurdum\/}.
Wir haben ihn daher verworfen und werden im folgenden Kapitel~\ref{Kap:EXCITED} eine aussagekr"aftige Analyse f"ur kleine bis verschwindende Virtualit"aten des Photons vorstellen, die basiert auf der konsequenten Fortf"uhrung des im folgenden formulierten Zugangs f"ur mittlere Virtualit"aten.

Mit wachsender Virtualit"at~$Q$ des Photons, und abh"angig von einem subtilen Wechselspiel zwischen seiner Virtualit"at und Polarisation, verringert sich die transversale Ausdehnung~$|\rb{r}|$ der Colour-Dipole, deren Wellenpakete es konstituieren.
Wir werden im folgenden Kapitel diskutieren, wie sich in Hinhlick auf die Ph"anomenologie der Prozesse dieser "Ubergang der Dominanz gro"ser zu der kleiner Abst"ande vollzieht.
F"ur mittlere Virtualit"aten im Bereich von~$Q^2 \!=\! 1 - 10$~GeV$^2$ kann so die perturbative Beschreibung des Photons durch seine im Rahmen der LCPT berechneten Lichtkegelwellenfunktion verl"a"slich beschrieben werden.
Allerdings behalten die Wirkungsquerschnitte transversal polarisierter Photonen gro"se nichtperturbative Beitr"age, die daher r"uhren, da"s die Endpunkte des Impulsanteils in der Wellenfunktion, das hei"st~$z \!=\!0$ und~$z \!=\!1$, {\it nicht\/} auf Colour-Dipole kleiner transversaler Ausdehnung einschr"anken.
Dies erkl"art sich daraus, da"s die Skala, auf der sich die transversale Ausdehnung der Dipole mit wachsendem~$Q^2$ verringert im wesentlichen das Produkt~$z(1\!-\!z)Q^2$ ist.
Der String-String-Mechanismus des \DREI[]{M}{S}{V}, der genau ankn"upft an den gluonischen Strings zwischen den Valenzquark-Konstituenten und umso wichtiger ist, je ausgedehnter diese Strings sind, kommt auch hier wesentlich zum tragen. \\
Wir betonen, da"s diese Stringausdehnung~-- der (transversale) Abstand der Quarkkonstituenten des Dipols~-- abh"angt von der expliziten Form der Wellenfunktion.
Andererseits ist nur sehr wenig "uber die Physik bekannt, die die Lichtkegel-Hamiltonfunktion nichtperturbativer QCD bestimmt.
Die Wellenfunktionen in der Gestalt, wie wir sie angeben, und deren integrierte Verteilungen erf"ullen daher in erster Linie die ph"anomenologische Aufgabe, den Konstituentengehalt des hadronischen Zustandes zu paramtrisieren.
Den mit diesen Wellenfunktionen berechneten Wirkungsquerschnitten ist dies Unsicherheit inh"arent; die exakten Werte h"angen ab von der detaiilierten Form der Wellenfunktionen.
Das $Q^2$-Verhalten anderer Observable, wie des Verh"altnisses der Produktion longitudinaler zu transversaler Vektormesonen oder des slope-Parameters, versprechen dagegen, ein guter Test f"ur das String-Bild des \DREI{M}{S}{V} zu sein.

\bigskip\noindent
Bevor wir unseren Zugang explizit darstellen, gehen wir in aller K"urze auf konkurierende Zug"ange und Modelle ein.

Exklusive Elektroproduktion von Vektormesonen wird im Rahmen eines Weichen Pomerons von Donnachie und Landshoff in den Refn.~\cite{Donnachie87,Donnachie94} diskutiert.
In diesem Modell spielt die transversale Ausdehnung der streuenden hadronischen Zust"ande eine nur marginale Rolle.
Zum einen kann Hadron-Hadron-Streuung aufgrund der dem Modell inh"arenten Quarkadditivit"at systematisch reduziert werden auf Quark-Quark-Streuung.
Zum anderen wird angenommen, der transversale Abstand der Quarks im virtuellen Photon sei sehr viel kleiner als die Wellenfunktion des Vektormesons, so da"s diese ersetzt wird durch ihren Wert am Ursprung.
Diese Annahmen stehen nicht au"serhalb jeder Diskussion, umso mehr als zus"atzlich angenommen wird die Kopplung des Weichen Pomerons an Quarks h"ange ab von dessen Flavour.
F"ur mittlere~$Q^2$ im Bereich von~$1 - 10$~GeV$^2$, vgl.\@ Ref.~\cite{Donnachie87}, wird ein Pomeron-Formfaktor f"ur {\it far-off-shell\/}-Quarklinien eingef"uhrt.
F"ur gr"o"sere Werte von~$Q^2$, vgl.\@ Ref.~\cite{Donnachie94}, wird der Austausch zweier nichtperturbativer Gluon (mit entsprechenden nichtperturbativen Propagatoren) angewandt; dies f"uhrt dazu, da"s sich bei kleiner Ausdehnung des Quark-Antiquark-Dipols {\it colour singlet\/}-Beitr"age gegeneinander wegheben.

In den Ver"offentlichungen Refn.~\cite{Kopeliovich93,Nemchick94} wird ein Modell, das basiert auf perturbativem Zwei-Gluon-Austausch, zur Beschreibung nichtperturbativer Effekte in der Weise ausgeweitet, da"s die Gluon-Verteilung im Proton explizit mit in den Ansatz einbezogen wird, diese quasi durch Vergleich mit bestimmten Beispielen exklusiver Vektormesonproduktion parametrisiert und in dieser Form zur Postulation verwandter Prozesse benutzt wird.
Dieses Modell wird weiter verfeinert durch eine \VIER{B}{F}{K}{L}-artige Entwicklung, um Aussagen machen zu k"onnen "uber die $s$- wie auch $Q^2$-Abh"angingkeit der Observablen.
Zusammenfassend ist hier zugrundeliegendes Bild die Streuung von Colour-Dipolen.
Also "ahnlicher Ausgangspunkt wie die von uns im folgenden vorgestellte Analyse.
Wesentlicher Unterschied beider Zug"ange liegt in beider Reaktionsmechanismen.

Die Wichtigkeit der Gluon-Verteilung im Proton wird als notwendiges Ingredient auch f"ur harte Diffraktion betont.
Wird sie in straight-forward perturbative Rechnung miteinbezogen und deren G"ultigkeit eingeschr"ankt auf den Bereich, in dem eine gro"se Skala invarianten Impulstransfers gegeben ist, k"onnen exklusive Produktion von $\Jps(3097)$, vgl.\@ Ref.~\cite{Ryskin93}, oder Prozesse bei gro"sen Virtualit"aten von~$Q^2 \!>\! 10$~GeV$^2$ postuliert werden, vgl.\@ Ref.~\cite{Brodsky94}.

Mit dem \VIER[]{D}{E}{S}{Y}-Beschleuniger \VIER{H}{E}{R}{A} sind neue M"oglichkeiten geschaffen, den zug"anglichen $s$- und $Q^2$-Bereich exklusiver Vektormesonproduktion zu erweitern.
Daten, die von den \VIER[]{Z}{E}{U}{S}- und \ZWEI[]{H}{1}-Kollaborationen ver"offentlicht wurden, sprechen daf"ur, da"s bei gro"sen~$Q^2$ der~Wirkungsquerschnitt steiler ansteigt, als von dem Austausch des Weichen Pomerons her zu erwarten ist.
Eine m"ogliche Erkl"arung in der Sprache von Colour-Dipolen ist, da"s die Evolution der Wellenfunktion zu h"oheren~$s$ und~$Q^2$ darin resultiert, da"s sich immer mehr solcher Dipole im Hadron finden, vgl.\@ die Refn.~\cite{Mueller94,Nikolaev93}.
Die Kenntnis dieser Evolution w"are dann Voraussetzung, die $s$-Abh"angigkeit des Wirkungsquerschnitts eines virtuellen Photons bei steigender Aufl"osung zu diskutieren.

Diesbez"uglich verweisen wir auf Arbeiten von Dosch und R"uter, die den Zugang des \DREI{M}{S}{V} inzwischen dahingehend erweitern, da"s auch die $s$-Abh"angigkeit postuliert wird, vgl.\@ Ref.~\cite{Rueter98}.
Dies geschieht, indem im Sinne des ph"anomenologischen Donnachie-Landshoff-Modells, vgl.\@ Ref.~\cite{Donnachie98}, der Austausch eines Weichen Pomerons (motiviert aus der Regge-Theorie) und eines Harten Pomerons (aus perturbativen \VIER{B}{F}{K}{L}-Rechnungen) zugrundegelegt wird, die jeweils als einfache Pole in der komplexen Drehimpulsebene angenommen werden.
An das Weiche Pomeron koppeln Dipole von der Gr"o"se typischer Hadronen, an das Harte Pomeron sehr viel kleinere Dipole~-- behandelt im Rahmen des \DREI{M}{S}{V} herk"ommlich beziehungsweise als perturbativer Zwei-Gluon-Austausch.
Dieser erweiterte Zugang reproduziert die Resultate bei~\mbox{\,$\surd s \!=\! 20\GeV$} und erkl"art die~$s$-Abh"angigkeit des gesamten~\mbox{\,$Q^2$}- und~\mbox{\,$M_V$}-Spektrums von \VIER[]{H}{E}{R}{A}-Daten~-- insbesondere den "Ubergang der Dominanz des Weichen zu der des Harten Pomerons.

Der Frage der~$s$-Abh"angigkeit und dem Ph"anomen des Austauschs eines Harten Pomerons wenden wir uns im folgenden aber {\it nicht\/} zu; wir beschr"anken uns auf die feste Energie von~$\surd s \!=\! 20$~GeV und auf Physik des Weichen Pomerons.
\vspace*{-.5ex}

\bigskip\noindent
Wir konkretisieren die in Abschnitt~\ref{Sect:Kinematik} gef"uhrte allgemeine Diskussion der zugrundeliegenden Kinematik auf den Fall exklusiver Produktion von Vektormesonen.

Seien der einlaufende Vierer-Impuls des Photons mit~$q$, der des einlaufenden Nukleons mit~$p$ bezeichnet, mit~$q'$ und~$p'$ die entsprechenden auslaufenden Impulse, vgl.\@ Abb.~\ref{Fig-G:Kinematik} in Gegen"uberstellung zu Abb.~\refg{Fig:Kinematik}.
\begin{figure}
\begin{minipage}{\linewidth}
  \begin{center}
  \setlength{\unitlength}{0.8mm}\begin{picture}(100,44.3)  
    \thinlines
    \put( 5, 5){\vector(4, 1){36}}
    \put(59,14){\vector(4,-1){36}}
    \put(50,14){\vector(0, 1){17}}
    \put(59,31){\vector(4, 1){36}}
    \put( 5,40){\vector(4,-1){36}}
    \put(-28,40){\normalsize $\ga^{\scriptscriptstyle({\D\ast})}\zz:$}
    \put(-15,40){$(q^+,\rb{q})$}
    \put(-28, 3){\normalsize $N\zz:$}
    \put(-15, 3){$(p^-,\rb{p})$}
    \put(102,40){\normalsize $V\zz:$}
    \put(115,40){$({q'}^+,\rb{q'})$}
    \put(102, 3){\normalsize $N\zz:$}
    \put(115, 3){$({p'}^-,\rb{p'})$}
    \put( 0,20){$s$}
    \put(49,40){$t$}
    \put(54,20){$\tf$}
  \end{picture}
  \end{center}
\vspace*{-4ex}
\caption[Streukinematik:~\protect$\ga^{\scriptscriptstyle({\D\ast})} \!+\! N \!\to\! V \!+\! N$ speziell]{
  Die Kinematik der Streuung~$\ga^{\scriptscriptstyle({\D\ast})}N \!\to\! VN$.
Im Limes~$s \!\to\! \infty$ ist ein Teilchenimpuls~$P$ vollst"andig bestimmt durch die zwei transversalen~$\rb{P}$ und {\it eine\/} der longitudinale Impulskomponenten~$P^\pm$; wir arbeiten mit de "`gro"sen"', die "`kleine"' ist dann gegeben durch~$P^\pm \!=\! (\rb{P}^2 \!+\! M^2) \!/\! 2P^\mp \!=\! \rb{M}^2 \!/\! 2P^\mp$, mit~$M$ der Masse,~$\rb{M}$ der transversalen Masse des Teilchens. Vgl.\@ Abb.~\refg{Fig:Kinematik}.
\vspace*{-.5ex}
}
\label{Fig-G:Kinematik}
\end{minipage}
\end{figure}
Dann ist~$\tf \!=\! q' \!-\! q$ der Vierer-Gesamt-Impulstransfer und die unabh"angigen Lorentz-invarianten Variablen sind gegeben durch
\vspace*{-.5ex}
\begin{alignat}{3}
&s\;&
  &=\;& (p\; &+\, q)^2
    \label{s_pq} \\
&t\;& 
  &=\;& \tf^{\!2}\;
   =\; (q'  &-\, q)^2
    \label{t_ta_q'q} \\
&Q^2\;&
  &=\;& &-\, q^2
    \label{Q2_q2}
    \\[-4.5ex]\nn
\end{alignat}
vgl.\@ die Gln.~(\ref{Mandelstam_Def}),~(\ref{Mandelstam_Def}$'$).
Wir betrachten {\it weiche\/} Streuung, das hei"st \mbox{$-t \!<\! 1\GeV^2$}, bei {\it hohen\/} Energien, das hei"st~\mbox{$s \!\gg\! Q^2$} und~\mbox{$s \!\gg\! M_{\!N}^2$,~$M_V^2$}, etwa~\mbox{$s \!>\! 100$~GeV$^2$}.
Die Bjorken-Skalenvariable~\mbox{$x_{\!B\!j} \!=\! Q^2 \!/\! 2p \!\cdot\! q \!=\! Q^2 \!/\! (s \!+\! Q^2 \!-\! M_{\!N}^2)$} ist in diesem kinematischen Bereich klein von der Gr"o\-"senordnung~\mbox{$x_{\!Bj\!} \!<\! 0.1$}.

Weiter arbeiten wir im Schwerpunktsystem, in dem der Photon-Dreier-Impuls~$\vec{q}$ in positive~$x^3$-Richtung zeigt.
Die Betr"age der Dreier-Impulse sind dann gegeben durch
\vspace*{-.5ex}
\begin{alignat}{5} \label{Betr"age_3erImpulse}
&|\vec{p}\,|&\;
  &=\; |\vec{q}\,|&\;
  &=\; \frac{\surd s}{2}\,
         \bigg[\; 1\;
                -\; \frac{M_{\!N}^2 \!-\! Q^2}{s}&\;
               &+\; O(s^{\,-2})\;
         \bigg]
    \\
&|\vecp{p}|&\;
  &=\; |\vecp{q}|&\;
  &=\; \frac{\surd s}{2}\,
         \bigg[\; 1\;
                -\; \frac{M_{\!N}^2 \!+\! M_V^2}{s}&\;
               &+\; O(s^{\,-2})\;
         \bigg]
    \tag{\ref{Betr"age_3erImpulse}$'$}
    \\[-4.5ex]\nn
\end{alignat}
vgl.\@ die Gln.~(\ref{Pi^3-P3IN,PtrIN=0}),~(\ref{Pi^3-P3IN,PtrIN=0}$'$).
In diesem Bezugsystem erh"alt das Vektormeson einen kleinen transversalen Impuls vom Betrag~$|\rb{q'}| \!=\! \tfbB \!\cong\! \sqrt{s}\, \vth \!/\!2 \!<\!1$~GeV.
Im Unterschied zu elastischer Hadron-Hadron-Streuung, wegen~$q^2 \!=\! -Q^2 \!\le\! 0$, existiert kein Bezugsystem, in dem die Differenz der Dreier-Impuls-Betr"age auf der Nukleonseite identisch verschwindet; wir finden die Abh"angigkeit
\vspace*{-.5ex}
\begin{align} \label{delta}
\de\;
  =\; |\vec{p}\,| - |\vecp{p}|\;
  \cong\; \frac{Q^2 + M_V^2}{2\surd s}\;
  >\; 0
    \\[-4.5ex]\nn
\end{align}
Dies impliziert eine nichtverschwindende Zeitkomponente des Gesamt-Impulstransfers~%
\mbox{$\tf^{\mskip-1mu0} \!\cong\! \de$} zus"atzlich zu den Raumkomponeneten~%
\mbox{$\tf^{\mskip-1mu3} \!\cong\! -\de \!-\! \sqrt{s}\, \vth^2 \!/\!4$} und~%
\mbox{$\tfbB \!\cong\! \sqrt{s}\, \vth \!/\!2$}.
Und folglich:
\vspace*{-.5ex}
\begin{align} \label{tfbB_-t}
&t\;
  =\; \tfQ\;
  =\; -\tfbQ\; +\; \tfde
    \\
&\text{mit}\qquad
  \tfde\;
    =\; -\, \frac{M_{\!N}^2 (Q^2 \!+\! M_V^2)^2}{s^2}\vv +\; O(s^{-3})
    \tag{\ref{tfbB_-t}$'$}
    \\[-4.5ex]\nn
\end{align}
vgl.\@ Gl.~(\ref{tsymm-P3IN,PtrIN=0}).
Im Hochenergielimes~$s \!\to\! \infty$ dominieren die Raumkomponenten; fordern wir im Rahmen dieses Limes einen endlichen invarianten Impulstransfer~$\surd-t$, das hei"st einen kleinen Streuwinkel~$\vth \!/\!2 \!=\! O(1\!/\!\surd s)$, so ist der transversale Vektor~$\tfb$ der f"uhrende Eintrag in~$\tf$.
Wir werden daher im folgenden alle Komponenten bis auf~$\tfb$ vernachl"assigen.
\vspace*{-.5ex}

\section{Weiche Hochenergiestreuung im \DREI[]{M}{S}{V}}

Das \DREI{M}{S}{V} wurde in Kapitel~\ref{Kap:VAKUUM}, seine Anwendung auf weiche Hochenergiestreuung in Kapitel~\ref{Kap:ANALYT} diskutiert.
Wir rekapitulieren in aller K"urze diese Diskussion unter Betonung von Aspekten, die uns ein tieferes Verst"andnis in Hinblick auf unsere Untersuchung vermitteln.

Die $T$-Amplitude f"ur Hadron-Hadron-Streuung im Limes~$s \!\to\! \infty$, vgl.\@ Gl.~(\ref{T2h_T2ell-mf}), lautet:
\begin{samepage}
%
\begin{align} \label{GROUND:Tconn-Element2h_WW-W_lim}
&T\hh^{(s,t)}\;
  \equiv\;
  \bracket{\, h^{2'}\!(P_{2'})\, h^{1'}\!(P_{1'}),\,\IN \,}{\,
              T\, \bracketM\,
              h^1(P_1)\, h^2(P_2),\,\IN \,}
    \\
&=\; \int d^2\rb{b}\;
              \efn{\T-\iIM\,\tfb \!\cdot\! \rb{b}}\;
              \int d\vph_{1',1} (\zet_1, \rb{X})\;
              \int d\vph_{2',2} (\zet_2, \rb{Y})\vv
              \tTll^{(s,\rb{b})}(\zet_1, \rb{X}; \zet_2, \rb{Y}; \rb{b})
    \nn
\end{align}
mit den Lichtkegelwellenfunktionen absorbiert im Integrationsma"s
%
\begin{align} \label{GROUND:dvphi'i}
d\vph_{i',i} (\zet, \rb{Z})\;
  \equiv\; \frac{d\zet}{2\pi}\; d^2\rb{Z}\vv
      \vph_{s\mskip-1mu{\bar s}}^{i'\D\dagger} (\zet, \rb{Z})\;
      \vph_{s\mskip-1mu{\bar s}}^i             (\zet, \rb{Z})
\end{align}
vgl.\@ Gl.~(\ref{dvphi'i}).
Die Funktion~$\tTll \!\equiv\! \tTll^{(s,\rb{b})}$ ist definiert als die bez"uglich~$\tfb$, mit~$t \!=\! \tfQ$, Fourier-trans\-formierte $T$-Amplitude der zugrundeliegenden Streuung zweier \vspace*{-.125ex}Wegner-Wilson-Loops; sie ist explizit gegeben durch~-- vgl.\@ Gl.~(\ref{tTll_WW-mf}$'$):
%
\begin{align} \label{tTll_WW-ohneArgument}
&\begin{aligned}[t]
 \tTll\;
  =\; -\, 2\iIM\,s\vv
         \vac{W\Dmfp}^{-1}\; \vac{W\Dmfm}^{-1}\vv
        &\vac{\, [ W\Dmfp - 1 ]
                 [ W\Dmfm - 1 ] \,}
    \\
   \underset{\text{$s \!\to\! \infty$}}{\longrightarrow}\vv
      -\, 2\iIM\,s\vv
        &\vac{\, W\idx{+} \!\cdot\! W\idx{-}\, -\, 1 \,}
 \end{aligned}
    \\[1ex]
&\text{wegen}\qquad
    W\Dmfp \underset{\text{$s \!\to\! \infty$}}{\longrightarrow} W\idx{+},\vv
    W\Dmfm \underset{\text{$s \!\to\! \infty$}}{\longrightarrow} W\idx{-},\qquad
    \vac{W\idx{+}} = \vac{W\idx{-}} = 1
    \nn
\end{align}
\end{samepage}%
das hei"st im wesentlichen durch den Vakuumerwartungswert~$\vac{\;\cdot\;}$ des Produkts~$W\idx{+} \! W\idx{-}$.
Wir merken an, da"s unsere Notation~$W\idx{+}$,~$W\idx{-}$ normierte Spurbildung~\mbox{$\trDrst{F} \!=\! 1\!/\! \dimDrst{F} \tr$} bez"uglich der fundamentalen Darstellung~$\Drst{F}$ der Loops bereits impliziert.

{\it Longitudinal\/}, in der $x^0,\!x^3$-Ebene f"allt der Loop~$W\idx{+}$ [bzw.\@$W\idx{-}$] mit deren erster [bzw.\@ zweiter] Winkelhalbierenden, das hei"st mit der $x^+$-Achse [bzw.\@ $x^-$-Achse] zusammen.
Diese Seiten sind lichtartig und verlaufen entlang der physikalische klassischen Trajektorien eines Quarks und Antiquarks, die parallel zueinander mit (nahezu) Lichtgeschwindigkeit in positive [bzw.\@ negative] $x^3$-Richtung propagieren, also $x^3\!(x^0) \!=\! x^0$ [bzw.\@ $x^3\!(x^0) \!=\! -x^0$], und gemeinsam einen Quark-Antiquark-/Colour-Dipol darstellen. \\
{\it Transversal projiziert\/} ist die Loop-Loop-Konfiguration insofern von besonderer Relevanz, als durch sie~-- in dem wie folgt ausgef"uhrten Sinne~-- die Streuung vollst"andig bestimmt ist.

%
\subsection{Transversale Konfiguration}
\label{Subsect:TransversaleKonfiguration}

Wir geben die funktionale Abh"angigkeit der Wegner-Wilson-Loops~$W\idx{+}$,~$W\idx{-}$ und damit der Funktion~$\tTll$ aus Gl.~(\ref{tTll_WW-ohneArgument}) wie folgt an:
\begin{samepage}
\begin{align}
&\tTll\;
  =\; -2\iIM\,s\; \vac{\, W\idx{+} \!\cdot\! W\idx{-}\; -\; 1 \,}
    \tag{\ref{tTll_WW-ohneArgument}$'$} \\
&=\; \tTll(\zet_1, \rb{x}, \rbb{x}; \zet_2, \rb{y}, \rbb{y})\;
  =\; -2\iIM\,s\; \vac{\, W\idx{+}(\zet_1, \rb{x}, \rbb{x})
               \cdot W\idx{-}(\zet_2, \rb{y}, \rbb{y})\; -\; 1 \,}
    \label{tTll_WW} \\[.5ex]
&=\; \tTll(\zet_1, \rb{X}; \zet_2, \rb{Y}; \rb{b})\;\hspace*{.5em}
  =\; -2\iIM\,s\; \vac{\, W\idx{+}(\zet_1, \rb{X}, +\rb{b}\!/2)
               \cdot W\idx{-}(\zet_2, \rb{Y}, -\rb{b}\!/2)\; -\; 1 \,}
    \tag{\ref{tTll_WW}$'$}
\end{align}
Als Variable {\it longitudinalen Ursprungs\/} treten auf: die Anteile~$\zet_i$ der Quarks am jeweiligen gesamten Lichtkegelimpuls, als {\it transversale\/} Variable: die (Anti)Quarkposi\-tionen~$\rb{x}$,~$\rbb{x}$ und~$\rb{y}$,~$\rbb{y}$ und die Ausdehnungen der Colour-Dipole~$\rb{X}$,~$\rb{Y}$ und deren Impakt~$\rb{b}$.

Diese zwei alternativen Variablens"atze~\mbox{$\{\zet_1, \zet_2, \rb{x}, \rbb{x}, \rb{y}, \rbb{y}\}$} und~\mbox{$\{\zet_1, \zet_2, \rb{X}, \rb{Y}, \rb{b}\}$} parametrisieren die Projektion der Konfiguration der Wegner-Wilson-Loops in den Transversalraum: durch die Positionen der (Anti)Quarks oder durch die Konstellation der Quark-Anti\-\mbox{quark-/Co}lour-Dipole.
Allein von dieser Transversalprojektion h"angt die $T$-Amplitude der Loop-Loop-Streuung~$\tTll$ und folglich der Hadron-Hadron-Streuung~$T_{h\mskip-1mu h}$ ab.

Grund daf"ur ist, da"s die Definition der relevanten transversalen Vektoren die longitudinale Dynamik subsumiert:
Die $\rb{b}$-Integration mit dem Exponential~\mbox{$\exp\{ -\iIM\, \tfb \!\cdot\! \rb{b} \}$} in~$T_{h\mskip-1mu h}$, vgl.\@ Gl.~(\ref{GROUND:Tconn-Element2h_WW-W_lim}), ist Reminiszenz daran, da"s der relevante Impakt der Streuung~-- im Sinne von Fourier-konjugiert zu~$\tf$, dem Vierer-Gesamt-Impuls"ubertrag~-- genau der Vierer-Vektor~$b$ ist.
In Kapitel~\ref{Subsect:HadronniveauII} beziehungsweise Anhang~\ref{APP:S-Element2h} wurde gezeigt, da"s aus dieser Definition der Konjugation zu~$\tf$ folgt, da"s~$\rb{b}$ (in unserem Bezugsystem ist~$b$ rein transversal) die mit den $\zet_i$-Faktoren gewichteten Mitten der projizierten Dipole~$\rb{X}$,~$\rb{Y}$ verbindet.
Der Impakt~\mbox{\,$\rb{b}$} setzt die longitudinalen Gr"o"sen~$\zet_1$,~$\zet_2$ voraus, involviert "uber diese die Abh"angigkeit von der longitudinalen Dynamik~-- und "ubersetzt sie in den Transversalraum.

%
\bigskip\noindent
Wir rekapitulieren die formalen Definitionen und Zusammenh"ange dieser Variablen, in Termen derer wir die transversale Konfiguration parametrisieren; vgl.\@ Seite~\pageref{b_rb} und Abb.~\ref{Fig:Loop-Loop-transvKonfig}.

\begin{figure}
\begin{minipage}{\linewidth}
  \begin{center}
  \setlength{\unitlength}{.9mm}\begin{picture}(120,74)   
    \put(1.125,0){\epsfxsize108mm \epsffile{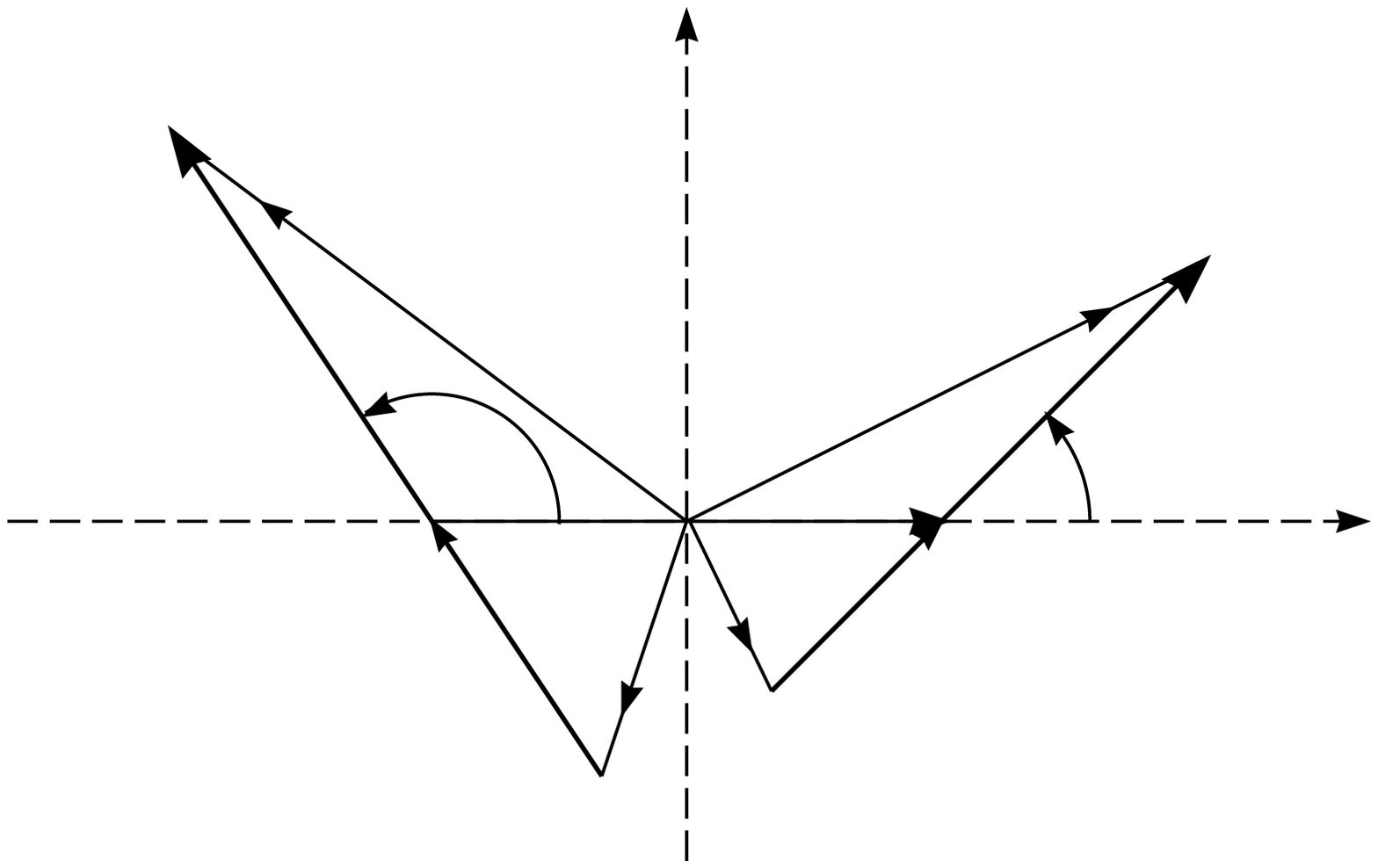}}
    \put(106,54){\normalsize $\rb{x}$}   
    \put( 66,11){\normalsize$\rbb{x}$}
    \put( 10,65){\normalsize $\rb{y}$}
    \put( 50, 3){\normalsize$\rbb{y}$}
    \put(97,42){\normalsize$\bzet_1\rb{X}$}   
    \put(79,23){\normalsize      $\zet_1\rb{X}$}
    \put(14,50){\normalsize$\bzet_2\rb{Y}$}
    \put(32,23){\normalsize      $\zet_2\rb{Y}$}
    \put(74,32){\normalsize$\rb{b}$}
    \put(61,33){\normalsize$\rbG{\om}$}   
    \put(110,25){\normalsize$\hat{e}_\rb{b}$}
    \put( 63,68){\normalsize$\hat{e}_{\perp\!\rb{b}}$}
  \end{picture}
  \end{center}
\vspace*{-3ex}
\caption[Transversale Konfiguration der Loop-Loop-Streuung durch~\protect$\rb{b},\rbG{\om}$]{
  Die Konfiguration der wechselwirkenden Wegner-Wilson-Loops~$W\idx{+}$,~$W\idx{-}$ im Transversalraum.
Die Vektoren~$\rb{x}$,~$\rbb{x}$ und~$\rb{y}$,~$\rbb{y}$ sind die Positionen der (Anti)\-Quarks des $\Dbr{+}$- bzw.\@ $\Dbr{-}$-Loops.
Die Quark-Antiquark-Differenzvektoren~$\rb{X}$ bzw.\@ $\rb{Y}$ sind die Ausdehnungen der entsprechenden Dipole.
Der Impaktvektor~$\rb{b}$, dessen Mitte bezeichnet ist mit~$\rbG{\om}$ (Schnittpunkt des Achsenkreuzes), ist orientiert in Richtung \pDipol und verbindet die $\zet_i$-gewichteten Mitten von~$\rb{X}$ und~$\rb{Y}$.
}
\label{Fig:Loop-Loop-transvKonfig}
\end{minipage}
\end{figure}
Die Paare~$\{\zet_1, \rb{x}\}$,~$\{\bzet_1, \rbb{x}\}$ und~$\{\zet_2, \rb{y}\}$,~$\{\bzet_2, \rbb{y}\}$ charakterisieren die (Anti)Quarks des $\Dbr{+}$-~beziehungsweise des \mDipols vollst"andig durch ihre Anteile am Lichtkegelimpuls und ihre transversalen Positionen; dabei schreiben wir abk"urzend:~$\bzet_i \!=\! (1 \!-\! \zet_i)$. \\
Diese Variablen sind nicht unabh"angig, wohl aber der Satz:
die Paare~$\{\zet_1, \rb{X}\}$ und~$\{\zet_2, \rb{Y}\}$ f"ur den $\Dbr{+}$-~beziehungsweise \mDipol und~$\rb{b}$ deren Impakt.
Konkret sind~$\rb{X}$ beziehungsweise~$\rb{Y}$ die vom Antiquark zum Quark hin orientierten transversalen Abstandvektoren, und es verbindet~$\rb{b}$, vom $\Dbr{-}$- zum \pDipol hin orientiert, deren $\zet_i$-gewichteten Mitten. \\
\indent
F"ur~$\rb{x}$,~$\rbb{x}$, die (Anti)Quark-Positionen des~\pDipols[], haben wir den Zusammenhang:
\end{samepage}
\vspace*{-.5ex}
\begin{alignat}{2} \label{GROUND:Q-AQ-Pos_rb1}
&\rb{x}\;
  =&\;  &\bzet_1\, \rb{X}\; +\; \rb{b}/2\; +\; \rbG{\om}
    \\[-.125ex]
&\rbb{x}\;
  =&\; -&\zet_1\,  \rb{X}\; +\; \rb{b}/2\; +\; \rbG{\om}
    \tag{\ref{GROUND:Q-AQ-Pos_rb1}$'$}
    \\[-4.75ex]\nn
\end{alignat}
und f"ur~$\rb{y}$,~$\rbb{y}$, die des \mDipols[]:
\begin{samepage}
\vspace*{-1ex}
\begin{alignat}{2} \label{GROUND:Q-AQ-Pos_rb2}
&\rb{y}\;
  =&\;  &\bzet_2\, \rb{Y}\; -\; \rb{b}/2\; +\; \rbG{\om}
    \\[-.25ex]
&\rbb{y}\;
  =&\; -&\zet_2\, \rb{Y}\; -\; \rb{b}/2\; +\; \rbG{\om}
    \tag{\ref{GROUND:Q-AQ-Pos_rb2}$'$}
    \\[-5ex]\nn
\end{alignat}
Bis auf den mit~$\rbG{\om}$ bezeichneten Punkt, den wir als Koordinatenursprung, das hei"st identisch Null w"ahlen werden~-- sind dies die Gln.~(\ref{Q-AQ-Pos_rb1}),~(\ref{Q-AQ-Pos_rb1}$'$) beziehungsweise~(\ref{Q-AQ-Pos_rb2}),~(\ref{Q-AQ-Pos_rb2}$'$).

Dieser Zusammenhang invertiert, gilt f"ur den Impaktvektor~$\rb{b}$ und f"ur dessen zun"achst noch allgemein gehaltene Mitte~$\rbG{\om}$:
\vspace*{-1ex}
\begin{alignat}{3}
&\rb{b}&\; 
  &=&\; &\rb{X}_\zet - \rb{Y}_\zet\;
    \label{GROUND:b_XzetYzet} \\[-.25ex]
&\rbG{\om}&\;
  &=&\; (&\rb{X}_\zet + \rb{Y}_\zet)/2
    \label{GROUND:om_XzetYzet}
    \\[-5ex]\nn
\end{alignat}
Anschaulich verbindet dieser Impaktvektor die $\zet_i$-gewichteten Dipol-Mitten~$\rb{X}_\zet$,~$\rb{Y}_\zet$, die wie folgt definiert sind:
\vspace*{-1ex}
\begin{alignat}{4} \label{GROUND:gewMitten-XzetYzet}
&\rb{X}_\zet&\;
  &=\; \zet_1\, \rb{x}&\; &+\; \bzet_1\, \rbb{x}&\;
  &=\; \phantom{-\,}
       \rb{b}/2\; +\; \rbG{\om}
    \\[-.25ex]
&\rb{Y}_\zet&\;
  &=\; \zet_2\, \rb{y}&\; &+\; \bzet_2\, \rbb{y}&\;
  &=\; -\, \rb{b}/2\; +\; \rbG{\om}
    \tag{\ref{GROUND:gewMitten-XzetYzet}$'$}
    \\[-5ex]\nn
\end{alignat}
F"ur die Quark-Antiquark-Abstandvektoren schlie"slich gilt:
\vspace*{-.5ex}
\begin{alignat}{3} \label{GROUND:Diffvekt-XY}
&\rb{X}&\;
  &=&\; \rb{x}\; &-\; \rbb{x}
    \\[-.25ex]
&\rb{Y}&\;
  &=&\; \rb{y}\; &-\; \rbb{y}
    \tag{\ref{GROUND:Diffvekt-XY}$'$}
    \\[-4.5ex]\nn
\end{alignat}
Vgl.\@ Gl.~(\ref{b_rb}), die Gln.~(\ref{gewMitten-XzetYzet}),~(\ref{gewMitten-XzetYzet}$'$) und~(\ref{Diffvekt-XY}),~(\ref{Diffvekt-XY}$'$). \\
\indent
Im folgenden arbeiten wir je nach Zweckm"a"sigkeit entweder mit dem einen oder dem anderen dieser Variablens"atze, ohne bez"uglich der {\it Notation\/}~-- vgl.\@ die Gl.~(\ref{tTll_WW}),~(\ref{tTll_WW}$'$)~-- zwischen den entsprechenden Funktionen zu unterscheiden.
\end{samepage}
\vspace*{-.5ex}

\bigskip\noindent
\begin{samepage}%
Wir unterbrechen unsere Diskussion mit einem Exkurs, der sich von wesentlich gr"o"serer Bedeutung herausstellen wird als die Koordinatentransformation, als die er zun"achst erscheint.

Die transversalen Abstandvektoren der Quark-Antiquark-Paare sind~$\rb{X}$,~$\rb{Y}$.
Sei der Vektor~$\rb{b}'$ mit Mitte~$\rbG{\om}'$ dadurch definiert, da"s er ihre {\it blo"sen Mitten\/} verbindet.
Es gilt
\vspace*{-1ex}
\begin{align} \label{bb'-omom'-aequiv}
\rb{b}'\; \equiv\; \rb{b}\qquad
\rbG{\om}'\; \equiv\; \rbG{\om}\qquad
  \Longleftrightarrow\qquad
\zet_1\; \equiv\; \zet_2\; \equiv\; 1\!/\!2
    \\[-5ex]\nn
\end{align}
in Gegen"uberstellung mit dem Vektor~$\rb{b}$ mit Mitte~$\rbG{\om}$, der die {\it $\zet_i$-gewichteten Mitten\/} von~$\rb{X}$ und~$\rb{Y}$ verbindet. 
Vgl.\@ Abb.~\ref{Fig':Loop-Loop-transvKonfig} in Kontrast zu Abb.~\ref{Fig:Loop-Loop-transvKonfig}. \\
%
%
%
%
An die Stelle der Gln.~(\ref{GROUND:Q-AQ-Pos_rb1}),~(\ref{GROUND:Q-AQ-Pos_rb1}$'$) treten f"ur den \pDipol[]:
\vspace*{-1ex}
\begin{alignat}{2} \label{GROUND':Q-AQ-Pos_rb1}
&\rb{x}\;
  =&\;  &\rb{X}/2\; +\; \rb{b}'/2\; +\; \rbG{\om}'
    \\[-.25ex]
&\rbb{x}\;
  =&\; -&\rb{X}/2\; +\; \rb{b}'/2\; +\; \rbG{\om}'
    \tag{\ref{GROUND':Q-AQ-Pos_rb1}$'$}
    \\[-5ex]\nn
\end{alignat}
und an die Stelle der Gln.~(\ref{GROUND:Q-AQ-Pos_rb2}),~(\ref{GROUND:Q-AQ-Pos_rb2}$'$) f"ur den \mDipol[]:
\vspace*{-1ex}
\begin{alignat}{2} \label{GROUND':Q-AQ-Pos_rb2}
&\rb{y}\;
  =&\;  &\rb{Y}/2\; -\; \rb{b}'/2\; +\; \rbG{\om}'
    \\[-.25ex]
&\rbb{y}\;
  =&\; -&\rb{Y}/2\; -\; \rb{b}'/2\; +\; \rbG{\om}'
    \tag{\ref{GROUND':Q-AQ-Pos_rb2}$'$}
    \\[-6ex]\nn
\end{alignat}
Addition und Gleichsetzen der (Anti)Quark-Positionen
nach~\mbox{(\ref{GROUND:Q-AQ-Pos_rb1}) $\zz+\zz$ (\ref{GROUND:Q-AQ-Pos_rb1}$'$) $\zz\stackrel{\T!}{=}\zz$ (\ref{GROUND':Q-AQ-Pos_rb1}) $\zz+\zz$ (\ref{GROUND':Q-AQ-Pos_rb1}$'$)}\vspace*{-.5ex}
beziehungsweise~\mbox{(\ref{GROUND:Q-AQ-Pos_rb2}) $\zz+\zz$ (\ref{GROUND:Q-AQ-Pos_rb2}$'$) $\zz\stackrel{\T!}{=}\zz$ (\ref{GROUND':Q-AQ-Pos_rb2}) $\zz+\zz$ (\ref{GROUND':Q-AQ-Pos_rb2}$'$)}
f"uhrt auf
\vspace*{-1ex}
\begin{align} \label{bb'-omom'}
\phantom{-\,}
\rb{b}'\; +\; 2 \rbG{\om}'\; 
  =\; (\bzet_1 - \zet_1)\, \rb{X}\; +\; \rb{b}\; +\; 2 \rbG{\om}
    \\[-.25ex]
-\, \rb{b}'\; +\; 2 \rbG{\om}'\;
  =\; (\bzet_2 - \zet_2)\, \rb{Y}\; -\; \rb{b}\; +\; 2 \rbG{\om}
    \tag{\ref{bb'-omom'}$'$}
    \\[-5ex]\nn
\end{align}
Addition und Subtraktion dieser Gleichungen f"uhrt respektive auf
\vspace*{-1ex}
\begin{alignat}{3}
&\rbG{\om}'&\;
  &=\; \rbG{\om}\; +\; 1\!/\!2&\,
        &[\, (1\!/\!2 - \zet_1)\, \rb{X} - (1\!/\!2 - \zet_2)\, \rb{Y}\, ]
    \label{om'_om} \\[-.25ex]
&\rb{b}'&\;
  &=\; \rb{b}\; +\;&
        &[\, (1\!/\!2 - \zet_1)\, \rb{X} - (1\!/\!2 - \zet_2)\, \rb{Y}\, ]
    \label{b'_b}
    \\[-5ex]\nn
\end{alignat}
vgl.~\mbox{\,$\bzet_i \!-\! \zet_i \!=\! 1 \!-\! 2\zet_i$}, als Relationen zwischen den gestrichenen und ungestrichenen Gr"o"sen~-- insbesondere in "Ubereinstimmung mit~(\ref{bb'-omom'-aequiv}):
Eine Diskrepanz existiert nur f"ur Konfigurationen mit nicht gleichverteilten Lichtkegelimpulsen. \\
\indent
Die Loop-Loop-Streuamplitude~$\tTll$ h"angt allein ab von der transversalen Konfiguration der Streuung.
Sei diese Konfiguration statt durch die Vektoren~$\rb{X}$,~$\rb{Y}$ und~$\rb{b}$ wie in Abbildung~\ref{Fig:Loop-Loop-transvKonfig} parametrisiert durch die Vektoren~$\rb{X}$,~$\rb{Y}$ und~$\rb{b}'$ wie in Abbildung~\ref{Fig':Loop-Loop-transvKonfig}~-- also auch implizit {\it unabh"angig\/} von den Anteilen~$\zet_i$ der Quarks am Lichtkegelimpuls; vgl.\@ Diskussion oben bzgl.\@ der impliziten Abh"angigkeit des Impakts~$\rb{b}$ von den~$\zet_i$.
In der Formel f"ur die $T$-Amplitude der Hadron-Hadron-Streuung~$T_{h\mskip-1mu h}$, vgl.\@ Gl.~(\ref{GROUND:Tconn-Element2h_WW-W_lim}), geht im Rahmen dieses Variablen"ubergangs~$\rb{b} \!\to\! \rb{b}'$, Jacobi-Determinante~$\equiv\! 1$,
die $\rb{b}$- "uber in eine $\rb{b}'$-Integration
und, vgl.\@ Gl.~(\ref{b'_b}), die Exponentialfunktion~\mbox{$\exp\{ -\iIM\, \tfb \!\cdot\! \rb{b} \}$}
"uber in einen Faktor~\mbox{$\exp\{ -\iIM\, \tfb \!\cdot\! \rb{b}' \}$}
und in einen zweiten Faktor~\mbox{$\exp\{ +i \tfb \!\cdot\! [(1\!/\!2 \!-\! \zet_1)\rb{X} - (1\!/\!2 \!-\! \zet_2)\rb{Y}] \}$}.
Die~$\zet_1,\!\zet_2$-Abh"angigkeit ist aus der Loop-Loop-Amplitude~$\tTll$ faktoriell herausgezogen.

\begin{figure}
\begin{minipage}{\linewidth}
  \begin{center}
  \setlength{\unitlength}{.9mm}\begin{picture}(120,74)   
    \put(1.125,0){\epsfxsize108mm \epsffile{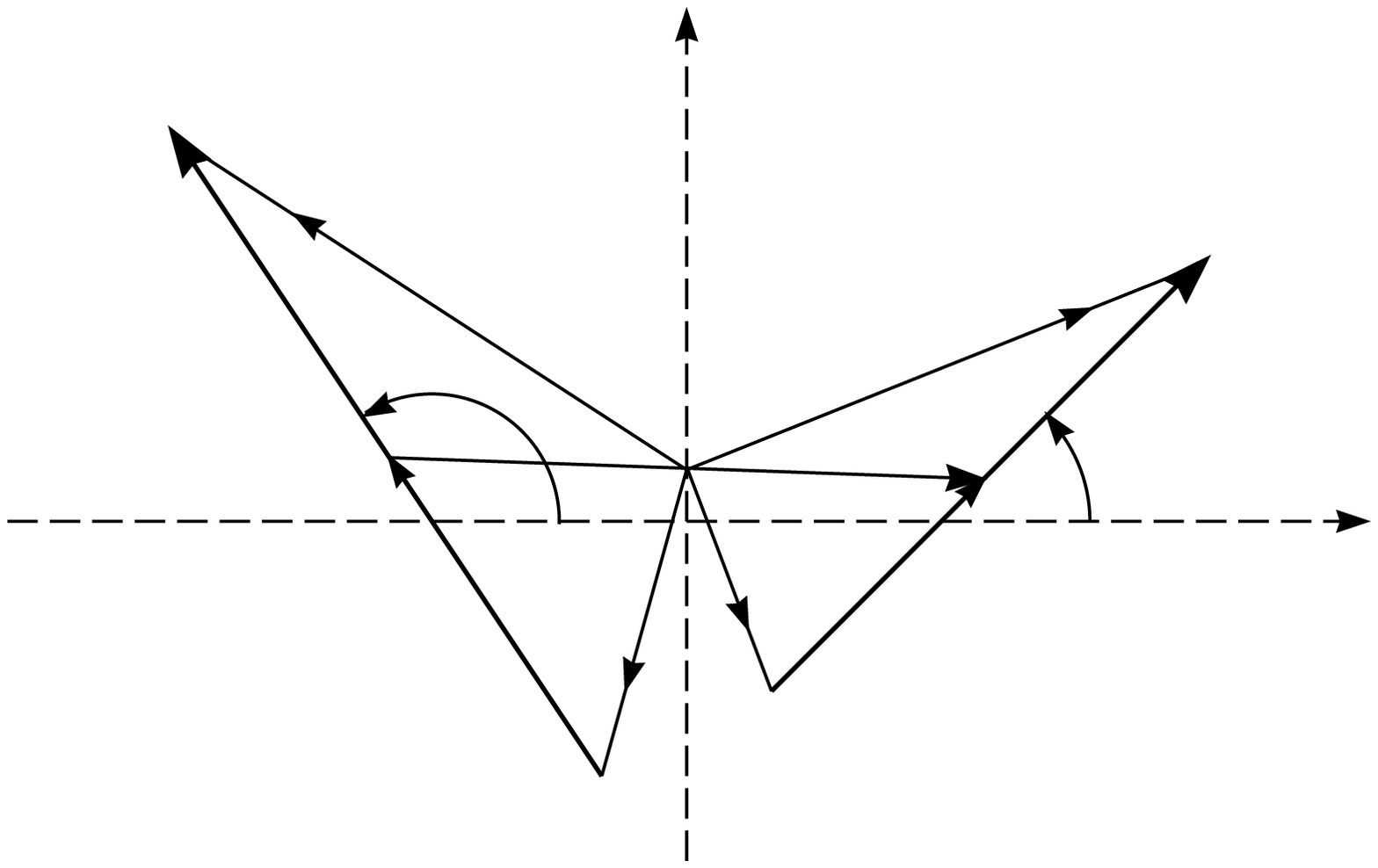}}
    \put(106,54){\normalsize $\rb{x}$}   
    \put( 66,11){\normalsize$\rbb{x}$}
    \put( 10,65){\normalsize $\rb{y}$}
    \put( 50, 3){\normalsize$\rbb{y}$}
    \put(97,42){\normalsize$\rb{X}$}   
    \put(14,50){\normalsize$\rb{Y}$}
    \put(77,36){\normalsize$\rb{b}'$}
    \put(61,38){\normalsize$\rbG{\om}'$}
    \put(110,25){\normalsize$\hat{e}_\rb{b}$}
    \put( 63,68){\normalsize$\hat{e}_{\perp\!\rb{b}}$}
  \end{picture}
  \end{center}
\vspace*{-3ex}
\caption[Transversale Konfiguration der Loop-Loop-Streuung durch~\protect$\rb{b}',\rbG{\om}'$]{
  In Kontrast zu dem Vektor~$\rb{b}$ mit Mitte~$\rbG{\om}$, vgl.\@ Abb.~\ref{Fig:Loop-Loop-transvKonfig}, verbindet der Vektor~$\rb{b}'$ mit Mitte~$\rbG{\om}'$ nicht die {\it $\zet_i$-gewichteten}, sondern die {\it blo"sen Mitten\/} der Dipole.
}
\label{Fig':Loop-Loop-transvKonfig}
\end{minipage}
\end{figure}

Bei genauerer "Uberlegung konstatieren wir folgendes:
Dokumentiert durch Gl.~(\ref{om'_om}) bewegen sich~$\rbG{\om}'$ und~$\rbG{\om}$ relativ zueinander, wenn die $\zet_1$-,~$\zet_2$- und die $\rb{X}$-,~$\rb{Y}$-Integrationen ausgef"uhrt werden.
Da wir den gemeinsamen Referenzpunkt~$x_0$ der Wegner-Wilson-Loops~-- im Sinne einer "`m"oglichst symmetrischen Wahl"'~-- w"ahlen werden als die {\it Mitte des Impaktvektors\/}, hat dies zur Konsequenz, da"s sich dieser in der einen "`Parametrisierung"' relativ zur anderen bewegt.
Es handelt sich daher nicht mehr um verschiedene Parametrisierungen derselben Geometrie.
Genau diese Beobachtung machen wir uns zunutze und untersuchen, wie eine solche permanente Zitterbewegung des Referenzpunktes, mit der korreliert ist eine entsprechende Bewegung der Fl"achen, "uber die die paralleltransportierten Feldst"arken integriert werden, sich quantitativ auf die postulierten Wirkungsquerschnitte auswirkt.
\end{samepage}

Wir finden vernachl"assigbare numerische Effekte von kleiner als einem halben Prozent.
Dies ist die A~posteriori-Rechtfertigung der Annahme einer nur geringen Abh"angigkeit unseres Zugangs von der Wahl des Referenzpunktes, wenn diese nur "`vern"unftig"' ist, im Sinne von "`m"oglichst symmetrisch"'.
Diese Annahme ist eingegangen in den nicht explizit vom Referenzpunkt abh"angenden Ansatz f"ur die Kumulante zweiter Ordnung in den paralleltransportierten Feldst"arken.%
\FOOT{
  \label{FN:K2-Ansatz}Vgl.\@ Kapitel~\ref{Kap:VAKUUM} die Gln.~(\ref{K2_Eins}),~(\ref{K2_Eins}$'$),~(\ref{K2Komp_Kronecker})~(\ref{K2Komp_Kronecker}$'$), Gl.~(\ref{K2-g2FF_Dvier}) und Gl.~(\ref{Dvier_DDxi0}).
}
Wir schlie"sen diesen Exkurs und fahren fort mit unserer Diskussion in Termen der ungestrichenen Vektoren wie veranschaulicht in Abbildung~\ref{Fig:Loop-Loop-transvKonfig}.

\subsection[\protect\vspace*{-.25ex}%
            Loop-Loop-Streuamplitude~\protect$\tTll^{(s,\rb{b})}$]{%
            Loop-Loop-Streuamplitude~\protect\bm{\tTll^{(s,\rb{b})}}}
\label{Subsect:Loop-Loop-Streuung}

Wir rekapitulieren die Umformungen der Loop-Loop-Streuamplitude~$\tTll$ als des Vakuumerwartungswerts~$\vac{\;\cdot\;}$ im wesentlichen des Produkts der Wegner-Wilson-Loops~$W\idx{+}$,~$W\idx{-}$, vgl.\@ Gl.~(\ref{tTll_WW}$'$),~-- in eine Gestalt, die in einfacher und anschaulicher Weise allein von der gerade diskutierten transversalen Konfiguration abh"angt.
Sei verwiesen auf Kapitel~\ref{Sect:T-Amplitude.Auswertung} und~\ref{Sect:T-Amplitude.Diskussion}. \\
\indent
Sei Abbildung~\ref{Fig:Loop-Loop-transvKonfig} formal aufgefa"st als transversaler Schnitt bei~\mbox{$x^+ \!\equiv\! \om^+, x^- \!\equiv\! \om^-$} bez"ug\-lich eines Punktes~$\om$, dessen transversale Komponente~$\rbG{\om}$ in dieser Ebene gerade die Mitte des Impaktvektors~$\rb{b}$ bezeichnet.
Sei dieser Punkt~$\om$ gew"ahlt als der Koordinatenur\-sprung~-- ohne unmittelbar~$\om \!\equiv\! \bm{0}$ zu setzen.
Die funktionale Abh"angigkeit der Loops~$W\idx{+}$,~$W\idx{-}$ ist dann zu lesen:
Die lange, longitudinale Seite des~$\Dbr{+}$-Loops $W\idx{+}(\zet_1, \rb{X}, +\rb{b}\!/\!2)$ verl"auft auf der~$x^+$-Achse; seine kurze, transversale Seite ist der Vektor~$\rb{X}$, dessen~$\zet_1$-gewichtete Mitte transversal bei~$+\rb{b}\!/\!2 \!+\! \rbG{\om}$ positioniert ist.
Analog der~$\Dbr{-}$-Loop $W\idx{-}(\zet_2, \rb{Y}, -\rb{b}\!/\!2)$. \\
\indent
Der Wegner-Wilson-Loop selbst sei in der Gestalt von Gl.~(\ref{W_S-mf}) konkretisiert:
\begin{samepage}
%
\begin{align} \label{GROUND:Konnektor_LoopFtr}
W({\cal C})\;
  =\; \trDrst{R}\; P_{\tilde{\cal C}}\vv
        \exp\; -\frac{\iIM\,g}{2}
        \int_{{\cal S}(\tilde{\cal C})} d\si^{\mu\nu}\!(x')\;
        F_{\mu\nu}\!(x'; x_0,{\cal C}_{x_{\!0}\!\ixp})
\end{align}
\end{samepage}%
In dieser Formel ist die Feldst"arke~$F(x'; x_0,{\cal C}_{x_{\!0}\!\ixp})$, deren Eichanteil mithilfe Konnektoren entlang einer Kurve~${\cal C}_{x_{\!0}\!\ixp}$ hin und entlang deren Inversen~${\cal C}_{\ixp\!x_{\!0}} \!\equiv\! {\cal C}_{x_{\!0}\ixp}{}^{\zzz -1}$ zur"uck an den Referenzpunkt~$x_0$ paralleltransportiert ist, bez"uglich ihres Arguments~$x'$ in folgendem Sinne zu integrieren:
In definierter Ordnung~$P_{\tilde{\cal C}}$ "uber eine Fl"ache~${\cal S}(\tilde{\cal C})$, die unendlich fein plakettiert zu denken ist durch die Deformation~$\tilde{\cal C}$, die zusammengesetzt ist aus den Kurven~${\cal C}_{x_{\!0}\!\ixp}$,~${\cal C}_{x_{\!0}\ixp}{}^{\zzz -1}$ und weiter den Kurven f"ur~$x^\dbprime$,~$x^\tlprime$,\ldots
Diese Fl"ache~${\cal S}(\tilde{\cal C})$ hat dabei den Bedingungen zu gen"ugen: da"s ihr Rand mit dem orientierten Loop zusammenf"allt und da"s sie den Referenzpunkt~$x_0$ als Element enth"alt,~-- sie ist dadurch aber noch nicht vollst"andig bestimmt.
Vgl.\@ die Diskussion zum Nichtabelschen Stokes'schen Satz \Glg{Konnektor_LoopFtr}. \\
\indent
Um auf die Amplitude~$\tTll$ in der Gestalt des Vakuumerwartungswerts zweier Wegner-Wilson-Loops die Kumulantenentwicklung anwenden zu k"onnen, vgl.\@ Abschnitt~\ref{Abschn:ANN-KONST}, m"ussen die Referenzpunkte von~$W\idx{+}$,~$W\idx{-}$ identisch sein, das hei"st es m"ussen die Integrationsfl"achen ${\cal S}(\tilde{\cal C}_\snbr{+})$,~${\cal S}(\tilde{\cal C}_\snbr{-})$ in diesem Punkt geometrisch verkn"upft sein.
Der Ansatz f"ur die Kumulante zweiter Ordnung in den paralleltransportierten Feldst"ar\-ken, vgl.\@ Fu"snote~\FN{FN:K2-Ansatz}, tr"agt Rechnung einer funktionalen Abh"angigkeit von der Vierer-Differenz~$\xi$ der Feldst"arkenpositionen allein und nicht auch vom gemeinsamen Referenzpunkt~$x_0$.
Wir haben argumentiert und gerade verifiziert, da"s dies dann gerechtfertigt ist, wenn in m"oglichst~symmetrischer Weise~$x_0$ und~-- damit in Zusammenhang~-- die Integrationsfl"achen gew"ahlt werden. \\
\indent
In diesem Sinne identifizieren wir die Referenzpunkte der Wegner-Wilson-Loops~$W\idx{+}$,~$W\idx{-}$ mit einem gemeinsamen Referenzpunkt~$x_0$.
Wir w"ahlen diesen identisch der Mitte~$\rbG{\om}$ des Impaktvektors~$\rb{b}$, das hei"st~$\rb{x_0} \!\equiv\! \rbG{\om}$.
Wir konkretisieren die Integrationsfl"achen als die orientierten Mantelfl"achen zweier Pyramiden, deren gemeinsamer Apex~\mbox{gerade der Referenzpunkt~$x_0$} und deren Basen der jeweilige plane Loop ist.
Wir verweisen auf Abbildung~\refg{Fig:pyramids}. \\
\indent
Die Geraden, die in dem in Abbildung~\ref{Fig:Loop-Loop-transvKonfig} dargestellten transversalen Schnitt von~$\rbG{\om}$ zu den Positionen~$\rb{x}$,~$\rbb{x}$ und~$\rb{y}$,~$\rbb{y}$ der (Anti)Quarks verlaufen, sind die Projektionen dieser Pyramiden-Mantelfl"achen in den Transversalraum.
Wir rekapitulieren im folgenden, wie sich die Integrationen der paralleltransportierten Feldst"arken "uber die Pyramiden-Mantelfl"achen reduzieren auf ihre Integration "uber diese Projektionen im Transversalraum. \\
\indent
Wir verweisen auf die ausf"uhrliche Darstellung dieser Geometrie in Kapitel~\vspace*{-.125ex}\ref{Subsect:surfacesSDmf}, die einschlie"st explizite Parametrisierungen der Pyramiden-Mantelfl"achen~${\cal S}(\tilde{\cal C}\Dmfp)$ und~${\cal S}(\tilde{\cal C}\Dmfm)$ "uber den nahezu lichtartigen Wegner-Wilson-Loops~$W\Dmfp$,~$W\Dmfm$ und deren transversale Projektionen.
Diese gehen "uber im hier betrachteten Limes~\mbox{\,$s \!\to\! \infty$} in die Fl"achen~${\cal S}(\tilde{\cal C}_\snbr{+})$ und~${\cal S}(\tilde{\cal C}_\snbr{-})$ "uber den lichtartigen Wegner-Wilson-Loops~$W\idx{+}$,~$W\idx{-}$.
\vspace*{-.5ex}

\begin{figure}
\begin{minipage}{\linewidth}
  \begin{center}
  \setlength{\unitlength}{.9mm}\begin{picture}(120,74)   
    \put(1.125,0){\epsfxsize108mm \epsffile{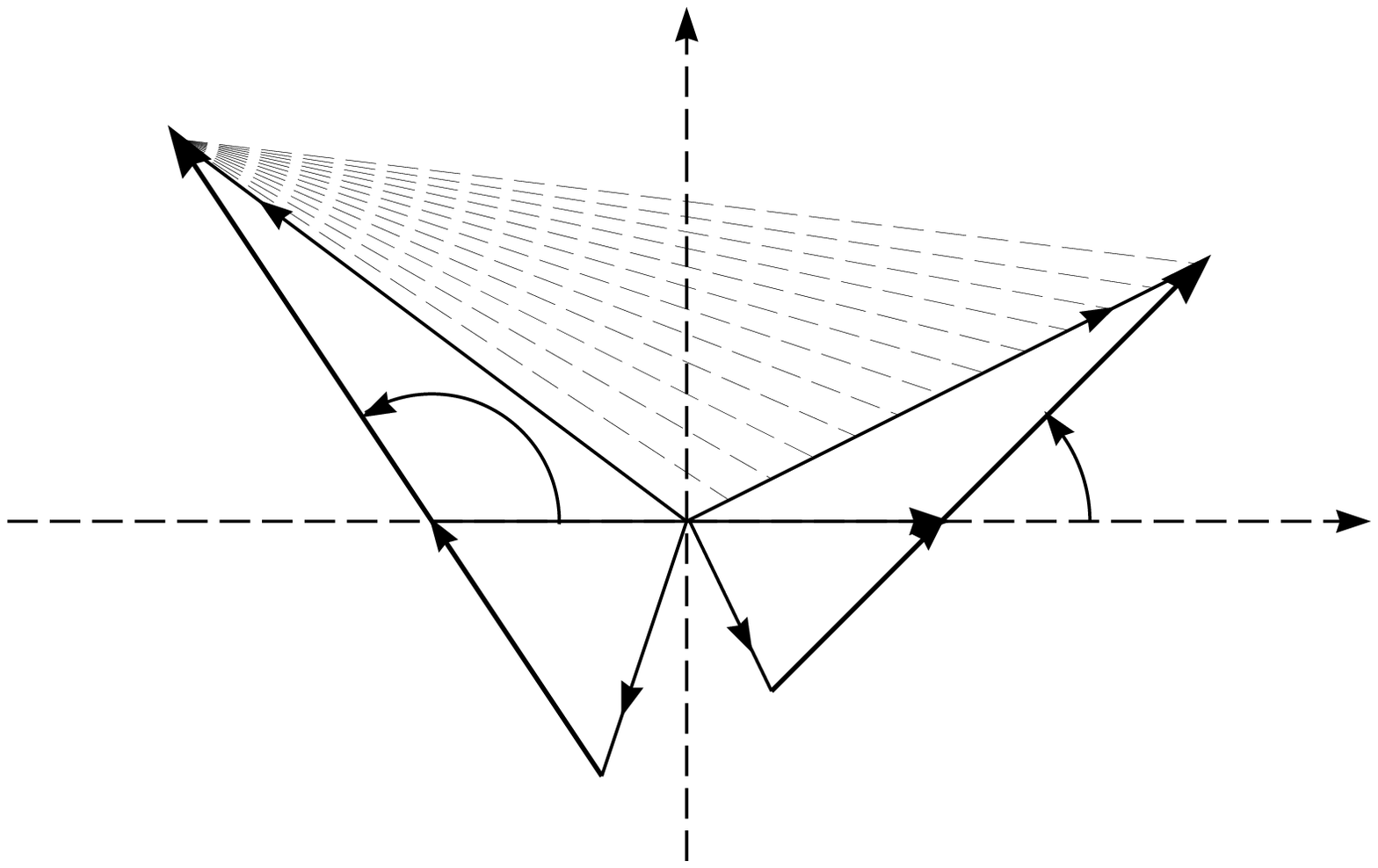}}
    \put(100,56){\normalsize$\rQ{+}^Q     \cdot\!(\la a) \equiv  \rb{x}$}
    \put( 64,10){\normalsize$\rQ{+}^{A\!Q}\cdot\!(\la a) \equiv \rbb{x}$}
    \put(  6,66){\normalsize$\rQ{-}^Q     \cdot\!(\la a) \equiv  \rb{y}$}
    \put( 35, 4){\normalsize$\rQ{-}^{A\!Q}\cdot\!(\la a) \equiv \rbb{y}$}
    \put(42,34){\normalsize$\th_\snbr{-}$}
    \put(95,34){\normalsize$\th_\snbr{+}$}
    \put(97,42){\normalsize$\rb{X}$}
    \put(14,50){\normalsize$\rb{Y}$}
    \put(74,32){\normalsize$\rb{b}$}
    \put(56,34){\normalsize$\rb{x}_0 \!\equiv\! \rbG{\om}$}
    \put(110,25){\normalsize$\hat{e}_\rb{b}$}
    \put( 63,68){\normalsize$\hat{e}_{\perp\!\rb{b}}$}
  \end{picture}
  \end{center}
\vspace*{-4ex}
\caption[Transversale Konfiguration der nichtlokalen Wechselwirkung des~\protect$C$-Terms]{
  In Kontrast zur Quark-Quark-Wechselwirkung, inh"arent der Funktion~$\tilde{X}\idx{\mfp\mskip-2mu\mfm}\oNC$, liegt~$\tilde{X}\idx{\mfp\mskip-2mu\mfm}\oC$ eine nichtlokale Wechselwirkung zugrunde.   Sei betrachtet die Kombination $\{ \rQ{+}^I, \rQ{-}^J \}$ f"ur feste~\mbox{$I,J \!\in\! \{Q,\AQ\}$}.   Nach Gl.~(\ref{X-C,X-NC-tilde_final}) treten zwei \vspace*{-.125ex}Kurvenintegrale auf entlang der {\sl Verbindungslinie\/} vom Referenzpunkt zu den \vspace*{-.25ex}Positionen~$\rQ{+}^I$,~$\rQ{-}^J$: entlang dem entsprechenden orientierten Abschnitt~\mbox{${\cal S}_\snbr{+}\Doperp\big|_I\!(s)$},~\mbox{${\cal S}_\snbr{-}\Doperp\big|_J\!(s)$}, der Transversalprojektionen der Pyramiden-Mantelfl"achen~${\cal S}_\snbr{+}$,~${\cal S}_\snbr{-}$~-- {\sl Geraden\/} f"ur plane Fl"achen.   Das erste Integral integriert die Korrelationen bez"uglich~${\cal K}_2$ auf: \vspace*{-.25ex}jedes Punktes auf~\mbox{${\cal S}_\snbr{+}\Doperp\big|_I$} mit~$\rQ{-}^J$, das zweite Integral die Korrelationen \vspace*{-.25ex}jedes Punktes auf~\mbox{${\cal S}_\snbr{-}\Doperp\big|_J$} mit~$\rQ{+}^I$.   Dargestellt ist~\mbox{$I \!\equiv\! J \!\equiv\! Q$}.
\vspace*{-1ex}
}
\label{Fig:chC-tilde_nichtlokal}
\end{minipage}
\end{figure}

\bigskip\noindent
Vor dieser Geometrie wird in Kapitel~\ref{Sect:T-Amplitude.Auswertung} und~\ref{Sect:T-Amplitude.Diskussion} die Streuamplitude%
  ~\vspace*{-.125ex}\mbox{\,$\tTll \!\equiv\! \tTll^{(s,\rb{b})}$} f"ur gro"se aber endliche Werte des Quadrats~$s$ der invarianten Schwerpunktenergie dargestellt wie folgt~-- vgl.\@ Gl.~(\ref{tTll_WW_ch-mf_ALL-gmfpmfp}):
\begin{samepage}
\vspace*{-.25ex}
\begin{align} \label{tTll_WW_ch-mf_ALL-gmfpmfp-REP}
\tTll\;
  &=\; 2\iIM\,s\vv
         \frac{1}{(4\Nc)^2}\vv
         \big(\vac{g^2 FF}a^4\big)^2
    \\[-.25ex]
  &\phantom{=\;}\qqquad\times
        \bigg[\,
          \frac{2}{\Nc^2 \!-\! 1}\; \tilde{X}\idx{\mfp\mskip-2mu\mfm}{}^{\zz2}\;
    +\; \Big(\frac{g_{\mfp\mfp}}{g_{+-}}\Big)^{\zz2}\cdot
          \Big(
            \tilde{X}\idx{\mfp\mskip-2mu\mfp}\, \tilde{X}\idx{\mfm\mskip-2mu\mfm}
          + \frac{2}{\Nc^2 \!-\! 1}\; \tilde{X}\idx{\mfp\mskip-2mu\mfm}{}^{\zz2}
          \Big)
        \,\bigg]
    \nn \\[-.25ex]
  &\phantom{=\;}\qqquad\times\,
        \exp\, -\frac{1}{4\Nc}\cdot
          \vac{g^2 FF}a^4\cdot
          \frac{g_{\mfp\mfp}}{g_{+-}}\,
          \big(
            \tilde{X}\idx{\mfp\mskip-2mu\mfp}
          + \tilde{X}\idx{\mfm\mskip-2mu\mfm}
          \big)
    \nn
    \\[-4ex]\nn
\end{align}
Im Limes~\mbox{\,$s \!\to\! \infty$} geht die Komponente%
  ~\mbox{\,$g_{\mfp\mfp}$} des metrischen Tensors bez"uglich der Koordinatenlinien%
  ~\mbox{\,$\tilde\mu \!\in\! \{\mfp,\mfm,1,2\}$} "uber in die~-- identisch verschwindende~-- Komponente%
  ~\mbox{\,$g_{++}$} des metrischen tensors bez"uglich der Lichtkegel-Koordinatenlinien%
  ~\mbox{\,$\bar\mu \!\in\! \{+,-,1,2\}$}.
Aus Gl.~(\ref{tTll_WW_ch-mf_ALL-gmfpmfp-REP}) folgt die bekannte \mbox{$s \!\to\! \infty$-asymp}\-totische Formel:
\vspace*{-.25ex}
\begin{align} \label{tTll_WW_ch}
&\tTll\;
  =\; 2\iIM\,s\vv
         \frac{1}{(4\Nc)^2}\,
         \frac{2}{\Nc^2 \!-\! 1}\vv
         \big(\vac{g^2 FF}a^4\big)^2\vv
         \tilde{X}\idx{\mfp\mskip-2mu\mfm}{}^{\zz2}
    \\[-4ex]\nn
\end{align}
\end{samepage}%
Die Funktion%
  ~\mbox{$\tilde{X}\idx{\mfp\mskip-2mu\mfm} \!\equiv\!
    \tilde{X}\idx{\mfp\mskip-2mu\mfm}(\zet_1, \rb{X}; \zet_2, \rb{Y}; \rb{b})$} h"angt ab allein von der Geometrie der Streuung in der \mbox{$x^1\!x^2$-Trans}\-versalebene und ist insbesondere unabh"angig von den der Energie~$\surd s$ der Streuung.
Sie ist Summe eines Anteils%
  ~\mbox{$\tilde{X}\idx{\mfp\mskip-2mu\mfm}\oC$} bez"uglich der konfinierenden Lorentz-Tensorstruktur~\mbox{$t\oC{}_{\zzzz \tilde\mu\tilde\nu\tilde\rh\tilde\si}$} des Korrelationstensors und eines Anteils%
  ~\mbox{$\tilde{X}\idx{\mfp\mskip-2mu\mfm}\oC$} bez"uglich der nicht-konfinierenden Struktur%
  ~\mbox{$t\oNC{}_{\zzzz \tilde\mu\tilde\nu\tilde\rh\tilde\si}$}~-- vgl.\@ die Gln.~(\ref{ch_vka,chC,chNC-tilde-REP}),~(\ref{ch=gXi-REP}):%
%
\begin{align} \label{X_vka,X-C,X-NC-tilde}
&\tilde{X}\idx{\mfp\mskip-2mu\mfm}\;
  =\; \vka\;  \tilde{X}\idx{\mfp\mskip-2mu\mfm}\oC\;
      +\; (1 \!-\! \vka)\; \tilde{X}\idx{\mfp\mskip-2mu\mfm}\oNC
\end{align}
Die nicht-konfinierende Funktion ist explizit angegeben in den Gln.~(\ref{ChNC-pm}),~(\ref{ChNC-pm}$'$):
\begin{samepage}
\vspace*{-.25ex}
\begin{align} \label{ChNC-pm-REP}
\tilde{X}\idx{\mfp\mskip-2mu\mfm}\oNC\;
  &=\; \iIM\; \frac{1}{6}\vv
         \sum\Big._{\!I,J\equiv Q,\AQ}\vv
           {\rm sign}_{I,J}\vv
           \projtbig{F\oNC}{\rb{r}\Dmfp^I \!-\! \rb{r}\Dmfm^J}
    \\[-.25ex]
  &=\; 8\cdot \la^4\,
         \frac{\pi}{12}\vv
         \sum\Big._{\!I,J\equiv Q,\AQ}\vv
           {\rm sign}_{I,J}\vv
           {\cal K}_3\big(\big|\rb{r}\Dmfp^I \!-\! \rb{r}\Dmfm^J\big| \!\big/\!\la\big)
    \tag{\ref{ChNC-pm-REP}$'$}
    \\[-4ex]\nn
\end{align}
f"ur die konfinierende in den Gln.~(\ref{ChC-pm}),~(\ref{ChC-pm}$'$):
\vspace*{-.25ex}
\begin{align} \label{ChC-pm-REP}
&\tilde{X}\idx{\mfp\mskip-2mu\mfm}\oC\;
   =\; \iIM\; \frac{1}{12}\vv
         \sum\Big._{\!I,J\equiv Q,\AQ}\vv
         {\rm sign}_{I,J}
    \\[-.25ex]
  &\phantom{\tilde{X}\idx{\mfp\mskip-2mu\mfm}\oC\; =\; - \det}\times\,
     \rb{r}\Dmfp^I\cdot \rb{r}\Dmfm^J\vv
       \int_0^1 ds\;
         \Big\{\;
           \projtbig[(1)]{F\oC}{s\, \rb{r}\Dmfp^I - \rb{r}\Dmfm^J}\;
       +\; \projtbig[(1)]{F\oC}{\rb{r}\Dmfp^I - s\, \rb{r}\Dmfm^J}
         \;\Big\}
    \nn \\[.75ex]
  &\phantom{\tilde{X}\idx{\mfp\mskip-2mu\mfm}\oC\;}
   =\; \la^4\,
         \frac{\pi}{12}\vv
         \sum\Big._{\!I,J\equiv Q,\AQ}\vv
         {\rm sign}_{I,J}
    \tag{\ref{ChC-pm-REP}$'$} \\[-.25ex]
  &\hspace*{21pt}
   \phantom{\tilde{X}\idx{\mfp\mskip-2mu\mfm}\oC\; =\; - \det}\times\,
     \rb{r}\Dmfp^I\cdot \rb{r}\Dmfp^J \!\big/\! \la^2\vv
       \int_0^1 ds\;
         \Big\{\;
           {\cal K}_2\big(\big|s\, \rb{r}\Dmfp^I - \rb{r}\Dmfm^J\big| \!\big/\!\la\big)
       +\; {\cal K}_2\big(\big|\rb{r}\Dmfp^I - s\, \rb{r}\Dmfm^J\big| \!\big/\!\la\big)
         \;\Big\}
    \nn
    \\[-4ex]\nn
\end{align}
Dabei folgen die gestrichenen Darstellungen auf Basis unseres expliziten Ansatzes f"ur die Korrelationsfunktionen; vgl.\@ Anh.~\ref{APP:CLTFN}.
Die Funktionen~\mbox{\,${\cal K}_\mu(\ze)$} sind definiert durch
\vspace*{-.25ex}
\begin{align} \label{calK_mu-REP}
&{\cal K}_\mu(\ze)\;
  =\; \Big(\frac{1}{2}\Big)^{\!\mu-1}\!
             \frac{1}{\Ga(\mu)}\;
             \ze^{\mu}\, {\rm K}_{\mu}\!(\ze)
    \\[-.25ex]
  &\text{mit}\qquad
  {\cal K}_\mu(\ze) \to 1,\vv
    \forall{\rm Re}\mu > 0\qquad
  \text{f"ur}\quad
  \ze \to 0
    \tag{\ref{calK_mu-REP}$'$}
    \\[-4ex]\nn
\end{align}
vgl.\@ Gl.~(\ref{calK_mu})~-- und~\mbox{${\rm K}_\mu$} der modifizierten Besselfunktion zweiter Art mit Index~$\mu$ in der Konvention von Ref.~\cite{Abramowitz84}.
Sie sind normiert auf Eins f"ur verschwindendes Argument. \\
\indent
Die nicht-konfinierende Funktion%
  ~\mbox{\,$\tilde{X}\idx{\mfp\mskip-2mu\mfm}\oNC$} ist zun"achst gegeben in Form des Integrals:
\vspace*{-.25ex}
\begin{align} \label{X-NC}
\tilde{X}\idx{\mfp\mskip-2mu\mfm}\oNC\;
  &=\; \iIM\; \frac{1}{6}\vv
        \int_{{\cal S}\Dmfp\Doperp} dx\Dmfp^{i}\; \del{\tilde{x}\Dmfp}_i\;
        \int_{{\cal S}\Dmfm\Doperp} dx\Dmfm^{j}\; \del{\tilde{x}\Dmfm}_j\vv
        \projt{F\oNC}{\rb{x}}
    \\[-4ex]\nn
\end{align}
vgl.\@ Gl.~(\ref{chNC-4}$'$) mit%
  ~\mbox{\,$\tilde\ch\idx{\mfp\mskip-2mu\mfm}\oNC \!=\!
    -\det\mathbb{SL}\; g_{\mfp\mfp}\cdot \tilde{X}\idx{\mfp\mskip-2mu\mfm}\oNC$} nach Gl.~(\ref{ch=gXi-REP}). \\
\indent
Die partiellen Ableitungen stammen aus der nicht-konfinierenden Tensorstruktur%
  ~\mbox{\,$t\oNC{}_{\zzzz \tilde\mu\tilde\nu\tilde\rh\tilde\si}$}.
Der Integrand ist daher vollst"andiges Differential beider Koordinatenlinien, so da"s das Integral folgt als Differenzen bez"uglich der Endpunkte der transversalen Projektionen~${\cal S}\Dmfp\Doperp$,~${\cal S}\Dmfm\Doperp$ der Pyramiden-Mantelfl"achen: den transversalen Projektionen der (Anti)Quark-Positionen.
Der resultierende Ausdruck~-- vgl.\@ Gln.~(\ref{ChNC-pm-REP})~-- ist dominiert durch die entsprechende Korrelationsfunktion, die f"ur gro"ses Argument, das hei"st gro"se Separation der (Anti)Quarks allgemein exponentiell abf"allt.
Die Funktion~$\tilde{X}\idx{\mfp\mskip-2mu\mfm}\oNC$ ist in der Tat nicht konfinierend. \\
\indent
Die konfinierende Funktion%
  ~\mbox{\,$\tilde{X}\idx{\mfp\mskip-2mu\mfm}\oC$} ist zun"achst gegeben in Form des Integrals~-- vgl.\@ die Gln.~(\ref{chC_IT[F]}),~(\ref{IT[F]-Def}$'$):
\vspace*{-.25ex}
\begin{align} \label{X-C}
&\tilde{X}\idx{\mfp\mskip-2mu\mfm}\oC\;
  =\; \iIM\; \frac{1}{12}\vv
         \int_{{\cal S}\Dmfp\Doperp} dx\Dmfp^{i}\vv
      \int_{{\cal S}\Dmfm\Doperp} dx\Dmfm^{i}\vv
        \pa_j \left[x^j F\right]
    \\[-4ex]\nn
\end{align}
\end{samepage}%
mit%
  ~\mbox{\,$\tilde\ch\idx{\mfp\mskip-2mu\mfm}\oC \!=\!
    -\det\mathbb{SL}\; g_{\mfp\mfp}\cdot \tilde{X}\idx{\mfp\mskip-2mu\mfm}\oC$} nach Gl.~(\ref{ch=gXi-REP}) und~\mbox{\,$F(|\rb{x}|) \!\equiv\! \projt[(1)]{F\oC}{\rb{x}}$} nach Gl.~(\ref{F_projt(1)}). \\
\indent
Der Integrand in diesem Ausdruck ist {\it nicht\/} vollst"andiges Differential beider Koordinatenlinien.
Es kann allgemein nur eine Integration ausgef"uhrt werden.
Die Funktion%
  ~\mbox{\,$\tilde{X}\idx{\mfp\mskip-2mu\mfm}\oC$} ist {\it nichtlokal\/} in dem Sinne, da"s sie abh"angt nicht nur von den Differenzen der~-- transversal projizierten~-- Positionen der (Anti)Quarks, sondern auch davon, wie diese verbunden sind durch die Projektionen~${\cal S}\Dmfp\Doperp$,~${\cal S}\Dmfm\Doperp$ der Pyramiden-Mantelfl"achen.
Wir argumentieren, da"s diese "`Verbindungen"' zu interpretieren sind als die~-- transverslen Projektionen~-- gluonischer Strings, die sich ausbilden zwischen den \vspace*{-.125ex}(Anti)Quarks. \\
\indent
Seien zun"achst die Funktionen%
  ~\vspace*{-.25ex}\mbox{\,$\tilde{X}\idx{\mfp\mskip-2mu\mfm}\oNC$} und%
  ~\mbox{\,$\tilde{X}\idx{\mfp\mskip-2mu\mfm}\oC$}~-- vgl.\@ Gl.~(\ref{ChNC-pm-REP}$'$) bzw.~(\ref{ChC-pm-REP}$'$)~-- abschlie"send angegeben in Form:
\vspace*{-.75ex}
\begin{alignat}{3} \label{X-C,X-NC-tilde_final}
&\tilde{X}\idx{\mfp\mskip-2mu\mfm}\oNC\;&
  &=\vv& 8\cdot &\la^4\,
         \frac{\pi}{12}\vv
         \sum\Big._{\!I,J\equiv Q,\AQ}\vv
           {\rm sign}_{I,J}\vv
           {\cal K}_3\big(\big|\rQ{+}^I \!-\! \rQ{-}^J\big|\big)
    \\[.5ex]
&\tilde{X}\idx{\mfp\mskip-2mu\mfm}\oC\;&
  &=\vv& &\la^4\,
         \frac{\pi}{12}\vv
         \sum\Big._{\!I,J\equiv Q,\AQ}\vv
         {\rm sign}_{I,J}
    \tag{\ref{X-C,X-NC-tilde_final}$'$} \\
  &&&&&\hspace*{16pt}\times\,
     \rQ{+}^I\cdot \rQ{-}^J\vv
       \int_0^1 ds\;
         \Big\{\;
           {\cal K}_2\big(\big|s\, \rQ{+}^I - \rQ{-}^J\big|\big)
       +\; {\cal K}_2\big(\big|\rQ{+}^I - s\, \rQ{-}^J\big|\big)
         \;\Big\}
    \nn
    \\[-5.75ex]\nn
\end{alignat}
mit Notation%
\FOOT{
  Die Konstanten~$\la$,~$a$ treten auf nur als Produkt~$(\la a)$~-- vgl.\@ die Gln.~(\ref{tTll_WW_ch-mf_ALL-gmfpmfp-REP}),~(\ref{tTll_WW_ch})~-- das daher die {\sl effektive\/} Korrelationsl"ange ist; mit~\mbox{\,$\la \!=\! 8\!\big/\!3\pi \!\cong\! 0.849$} ist diese um~$15\%$ keiner als~$a$ allein.
}
%
\vspace*{-.5ex}
\begin{align} \label{r^Q/AQ_(+)(-)}
&\rQ{+}^Q\;
   \equiv\;  \rb{x}/(\la a)\qquad
  \rQ{+}^{A\!Q}\;
   \equiv\; \rbb{x}/(\la a)\qquad
  \rQ{-}^Q\;
   \equiv\;  \rb{y}/(\la a)\qquad
  \rQ{-}^{A\!Q}\;
   \equiv\; \rbb{y}/(\la a)
    \\[-4.5ex]\nn
\end{align}
f"ur die skalierten Positionen der (Anti)Quarks des $\Dbr{+}$- und \mDipols[],~bzgl.~$\rb{x}$,~$\rbb{x}$ und~$\rb{y}$,~$\rbb{y}$, vgl.\@ die Gln.~(\ref{GROUND:Q-AQ-Pos_rb1}),~(\ref{GROUND:Q-AQ-Pos_rb1}$'$) bzw.~(\ref{GROUND:Q-AQ-Pos_rb2}),~(\ref{GROUND:Q-AQ-Pos_rb2}$'$).
Wir verweisen auf \vspace*{-.125ex}Abbildung~\ref{Fig:chC-tilde_nichtlokal}.
\\\indent
Die Funktion%
  ~$\tilde{X}\idx{\mfp\mskip-2mu\mfm}\oNC$ subsumiert Wechselwirkung der Dipol-Endpunkte oder (Anti)Quarks; sie induziert daher Quarkadditivit"at.
Die Funktion%
  ~$\tilde{X}\idx{\mfp\mskip-2mu\mfm}\oNC$ h"angt ab von der Ausdehnung der wechselwirkenden Colour-Dipole.
Die $T$-Amplitude des \DREI{M}{S}{V} bricht Quarkadditivit"at aufgrund des~$C$-Terms.
Das Verh"altnis der totalen Wirkungsquerschnitte Pion-Proton zu Proton-Proton folgt aber auch im \DREI{M}{S}{V} ungef"ahr zu Zwei Drittel:~$\si_{{\pi}p}^{\rm tot} \!/\!\si_{pp}^{\rm tot} \!\cong\! 2\!/\!3$; allerdings nicht aufgrund Quarkadditivit"at, sondern aufgrund dessen, da"s das Verh"altnis der Ausdehnungen von Pion zu Proton im Quadrat ungef"ahr~$2\!/\!3$ betr"agt.
Vgl.\@ die Bem.\@ zu Gl.~(\ref{transversaleWfn-allg}). \\
\indent
In Kapitel~\ref{Kap:VAKUUM} sind eingef"uhrt die Indizes~$C$ und~$N\!C$ als {\it confining\/} (Colour-Ladungen konfinierend in eichinvariante Zust"ande) und {\it non-confining\/} (nicht konfinierend).
Dies geschah in Konsequenz des Resultats des \DREI{M}{S}{V} f"ur ein statisches Quark-Antiquark-Paar~-- mathematisch repr"asentiert durch den Vakuumerwartungswert~$\vac{\;\cdot\;}$ eines einzelnen Wegner-Wilson-Loops.
So setzt das Nichtverschwinden der Stringspannung~$\si$~-- das hei"st (lineares) Confinement~-- notwendig voraus das Nichtverschwinden der Lorentz-Tensorstruktur~$t\oC$; sie ist insbesondere unabh"angig von der Struktur~$t\oNC$; vgl.\@ Gl.~(\ref{Stringspannung-si_DC}).
%

Wir diskutieren, da"s in diesem Sinne auch in weicher Hochenergiestreuung~-- mathematisch repr"asentiert durch den Vakuumerwartungswert~$\vac{\;\cdot\;}$ zweier, nahezu lichtartiger Wegner-Wilson-Loops~-- der~$C$-Term anzusprechen ist als konfinierend, der~$N\!C$-Term demgegen"uber als nicht konfinierend.
Ursache  hierf"ur ist genau die Nicht-Lokalit"at der Wechselwirkung immanent der Funktion~$\tilde{X}\idx{\mfp\mskip-2mu\mfm}\oC$ gegen"uber der blo"sen Quark-Quark-Wechselwirkung der Funktion~$\tilde{X}\idx{\mfp\mskip-2mu\mfm}\oNC$.
Diese Nicht-Lokalit"at f"uhrt auf einen Mechanismus, der sensitiv ist auf die Gr"osse der wechselwirkenden Colour-Dipole, in dem Sinne, da"s die "`Quark-Antiquark-Verbindungslinien"' wesentlich an der Streuung beteiligt ist.
Wir interpretieren dies als {\it String-String-Mechanismus\/}: das Ausbilden und Wechselwirken gluonischer Strings zwischen den Quarkkonstituenten des Dipols.

Die Abbildungen~\ref{Fig:int-ampl_NC,C}(a),(b) und~\ref{Fig:int-ampl-collinear_NC,C} suggerieren und veranschaulichen diese Interpretation.
Dargestellt ist die dimensionslose Funktion
\bea \label{tsill}
\tsill(\rb{b})\;
  =\; \frac{1}{s}\, {\rm Im}\vv \int_0^{2\pi}\! \frac{d\th_\snbr{-}}{2\pi}\vv
      \tTll(\ldots,\th_\snbr{-},\ldots;\rb{b})
\eea
wobei durch Integral und Punkte eine Dipol-Dipol-Konfiguration mit Impaktvektor~$\rb{b}$ angedeutet sei.
Die Funktion~$\tsill$ ist definiert durch~$\tTll$: die bez"uglich des transversalen Impulstransfers~$\tfb$, mit~$t \!=\! \tfQ$, Fourier-transformierte $T$-Amplitude f"ur die Wechselwirkung dieser Konfiguration; bzgl.~$\tTll$ vgl.\@ die Gln.~(\ref{tTll_WW_ch}),~(\ref{X-C,X-NC-tilde_final}),~(\ref{X-C,X-NC-tilde_final}$'$).
Aufgrund des {\it optischen Theorems\/} ist~$\si\ellell^{\rm tot} \!\equiv\! \int d^2\rb{b}\, \tsill(\rb{b})$ der entsprechende {\it totale Wirkungsquerschnitt\/},~$\tsill(\rb{b})$ selbst daher aufzufassen als Wechselwirkungsamplitude der zugrundeliegenden Dipol-Dipol-Konfiguration mit Impakt~$\rb{b}$, die wir wie folgt definieren.

Seien systematische Notation und polare Koordinaten eingef"uhrt durch:
\bea \label{r_+-}
\rb{r}_\snbr{+} \equiv \rb{X}, \quad
\rb{r}_\snbr{-} \equiv \rb{Y} \qquad
  \text{und}\qquad
  r_\snbr{\,\pm} \equiv |\rb{r}_\snbr{\,\pm}|, \quad
  \text{$\th_\snbr{\,\pm} \equiv$~Azimut von~$\rb{r}_\snbr{\,\pm}$ bzgl.~$\rb{b}$}
\eea
mit~$r_\snbr{+}$,~$r_\snbr{-}$ den transversalen Ausdehnungen des~$\Dbr{+}$-, \mDipols und~$\th_\snbr{+}$,~$\th_\snbr{-}$ den Azimutwinkeln um ihre~$\zet_\snbr{+}$-,~$\zet_\snbr{-}$-gewichteten Mitten in der Transversalebene, abgetragen vom~Impaktvektor~$\rb{b}$ so, da"s~$\th_\snbr{+},\th_\snbr{-} \!=\! 0$~[$\pi$], falls~$\rb{X},\rb{Y}$ parallel [antiparallel] zu~$\rb{b}$, vgl.\@ Abb.~\ref{Fig:chC-tilde_nichtlokal}. \\
\indent
Besitze der eine Dipol, o.E.d.A.\@ der \pDipol die Ausdehnung~$r_\snbr{+} \!=\! 12\,a$, sei~$\zet_\snbr{+} \!=\! 1\!/\!2$, das hei"st der Lichtkegellimpuls ist gleichverteilt auf Quark und Antiquark: die~$\zet_\snbr{+}$-gewichtete Mitte identisch mit seiner blo"sen Mitte.
Besitze der andere, \mDipol die Ausdehnung~$r_\snbr{-} \!=\! 1\,a$, sei~$\zet_\snbr{-} \!=\! 1\!/\!2$.
Er werde gemittelt "uber s"amtliche Orientierungen~$\th_\snbr{-}$ und beliebig verschoben relativ zum \pDipol[], der fest positioniert sei.
Wir betrachten~$\tsill$ als Funktion des entsprechenden Impakts, den wir gem"a"s~$\rb{b} \!=\! \bprll \!+\! \bperp$ zerlegen in einen Anteil parallel und in einen orthogonal zu der Richtung, die der \pDipol definiert.%
\FOOT{
  \label{FN:Impakt-Betrag}Wahl eines Koordinatensystems hei"st Wahl des Azimutwinkels~$\vph_\rb{b}$ des Impaktvektors bzgl.\@ dessen \mbox{$x^1$-Ach}\-se;~$\tsill$ ist unabh"angig von~$\vph_\rb{b}$, relevant ist~$\th_\snbr{+}$ als Azimut von~$\rb{b}$ zur Richtung des \pDipols[].
}

Wir betrachten im Vergleich die Abbildungen~\ref{Fig:int-ampl_NC,C}(a) und~(b).\@
\begin{figure}
\begin{minipage}{\linewidth}
  \begin{center}
  \setlength{\unitlength}{.9mm}
  \begin{picture}(120,81)   
    \put(0,0){\epsfxsize108mm \epsffile{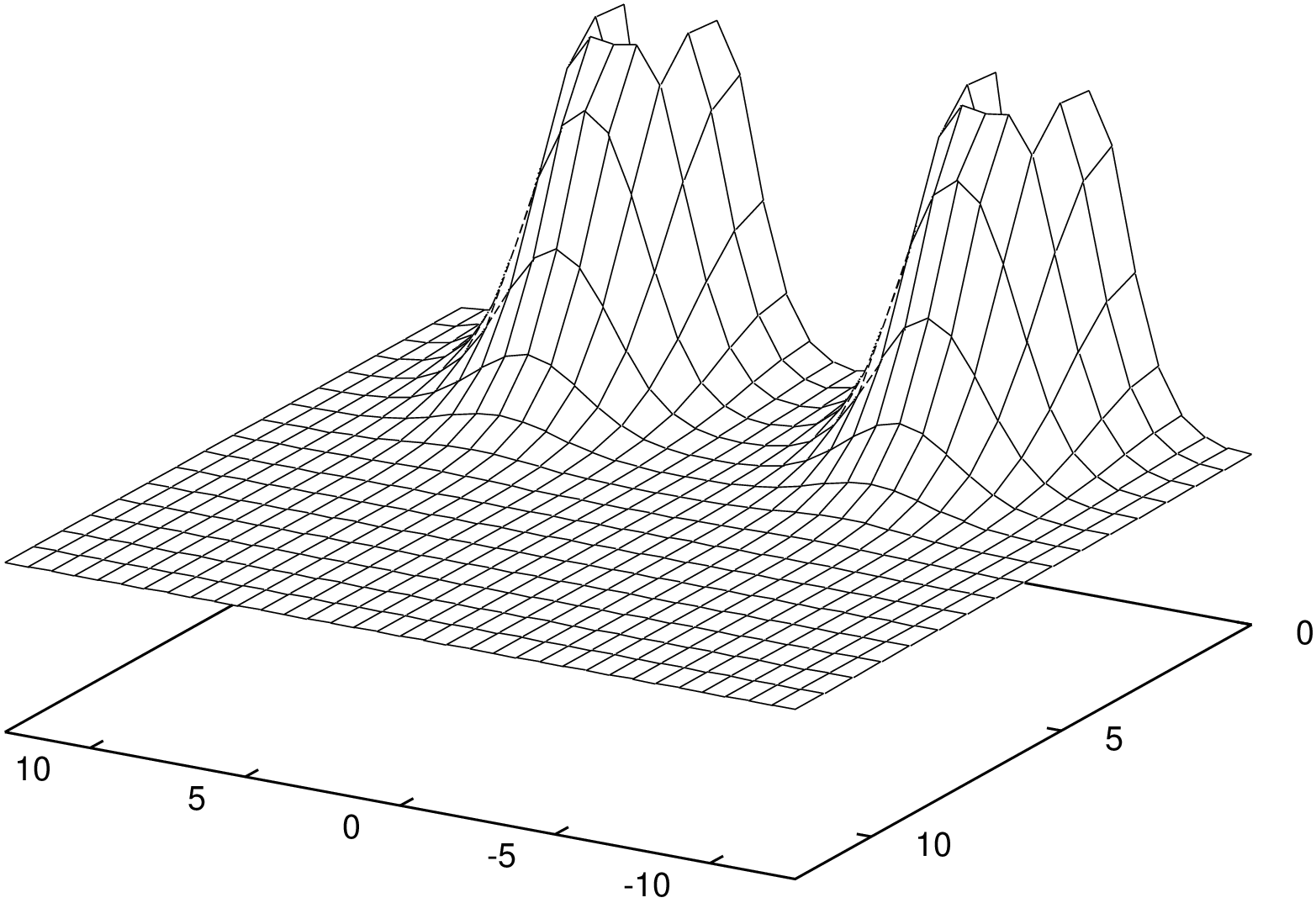}}
    \put( 6,70){\normalsize (a)\quad$N\!C$-Term}
    \put(93, 4){\normalsize $\bperp /\!a$}
    \put( 5, 4){\normalsize $\bprll /\!a$}
  \end{picture}\\[-37.5mm]
  \begin{picture}(120,41)   
    \linethickness{0.3pt}
    \qbezier(42,39)(77.75,32)(113.5,25)
    \qbezier(1,16)(21.5,27.5)(42,39)
    \linethickness{1.5pt} \qbezier(94,29)(77.75,32)(61.5,35)      
    \thicklines           \put(77.75,32){\vector(-4,-1){47.75}}   
    \linethickness{1.5pt} \qbezier(31.25,19)(30,20)(28.75,21)     
    \linethickness{0.2pt} \qbezier(17,13)(30,20)(58,36)
    \linethickness{0.2pt} \qbezier(85, 9)(30,20)(13.5,23)
  \end{picture} \\[-1ex]
  \begin{picture}(120,81)   
    \put(0,0){\epsfxsize108mm \epsffile{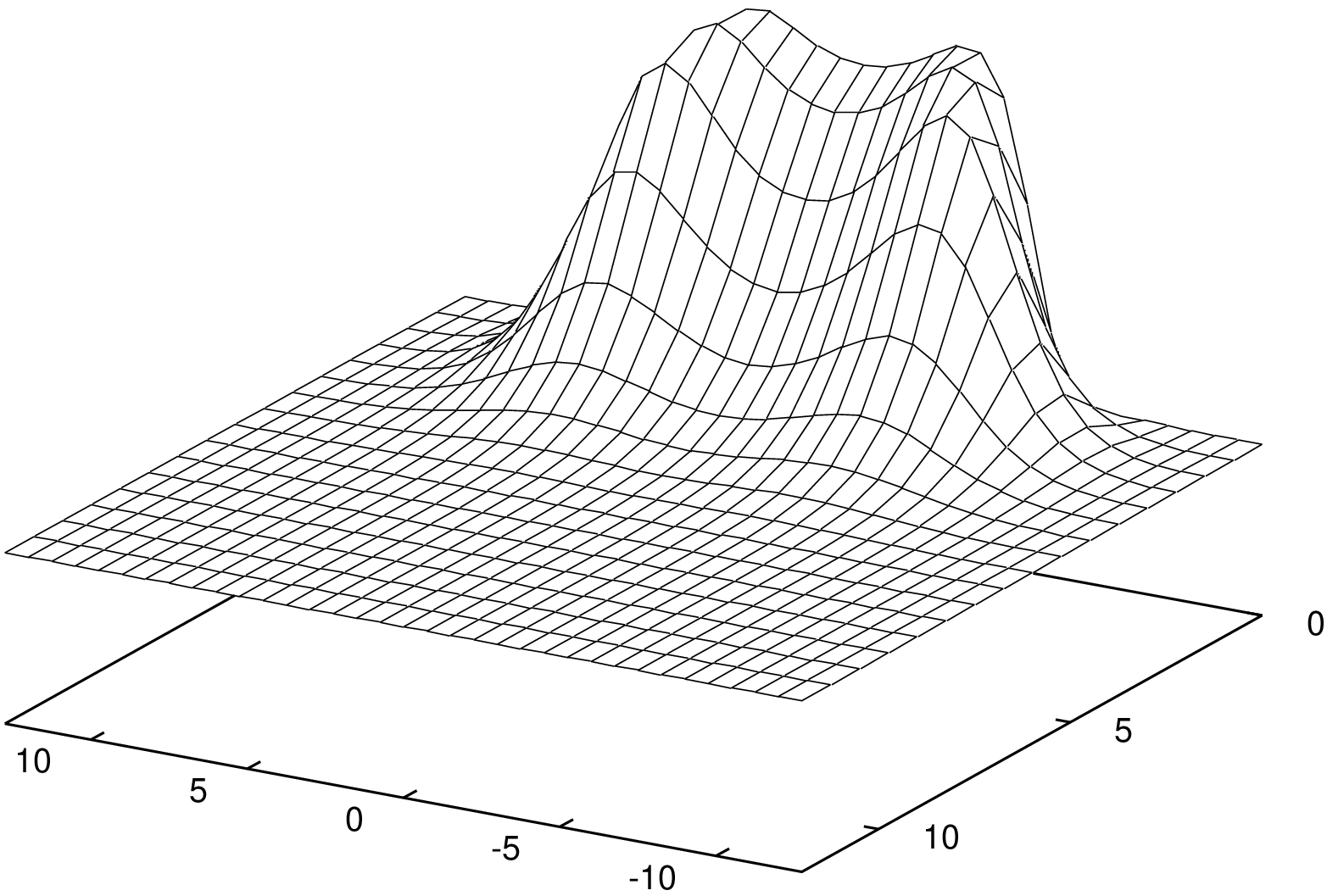}}
    \put( 6,70){\normalsize (b)\quad$C$-Term}
    \put(93, 4){\normalsize $\bperp /\!a$}
    \put( 5, 4){\normalsize $\bprll /\!a$}
  \end{picture}\\[-37.5mm]   
  \begin{picture}(120,41)   
    \linethickness{0.3pt}
    \qbezier(42,39)(77.75,32)(113.5,25)
    \qbezier(1,16)(21.5,27.5)(42,39)
    \linethickness{1.5pt} \qbezier(94,29)(77.75,32)(61.5,35)      
    \thicklines           \put(77.75,32){\vector(-4,-1){47.75}}   
    \linethickness{1.5pt} \qbezier(31.25,19)(30,20)(28.75,21)     
    \linethickness{0.2pt} \qbezier(17,13)(30,20)(58,36)
    \linethickness{0.2pt} \qbezier(85, 9)(30,20)(13.5,23)
  \end{picture}
  \end{center}
\vspace*{-3ex}
\caption[Wechselwirkung zweier Dipole,~\protect$N\!C$- und~\protect$C$-Term]{
  Amplitude~$\tsill$ f"ur die Wechselwirkung zweier Colour-Dipole als Funktion deren Impakts~$\rb{b}$, vgl.\@ Gl.~(\ref{tsill}).   Der \pDipol hat Ausdehnung~$r_\snbr{+} \!=\! 12\,a$ ($\zet_\snbr{+} \!=\! 1\!/\!2$) und ist fest positioniert.   Der \mDipol hat Ausdehnung~$r_\snbr{-} \!=\! 1\,a$ ($\zet_\snbr{-} \!=\! 1\!/\!2$), er werde gemittelt "uber seine Orientierungen und verschoben relativ zum \pDipol[].   (a) Nicht-konfinierender Fall,~$\vka \!\equiv\! 0$.   Wesentliche Beitr"age zur Wechselwirkung treten auf nur f"ur kleine Abst"ande der Dipol-Endpunkte, das hei"st (Anti)Quarks.   (b) Konfinierender Fall,~$\vka \!\equiv\! 1$.   Wesentliche Beitr"age treten auf f"ur kleine Abst"ande der Dipole an sich: es gen"ugt ein kleiner Abstand des kleinen Dipols zu der geraden Verbindung der Quarkkonstituenten des gro"sen, auch wenn er von diesen selbst weit separiert ist.
\vspace*{-1.875ex}
}
\label{Fig:int-ampl_NC,C}
\end{minipage}
\end{figure}
Dargestellt ist die Amplitude~$\tsill$ der Wechselwirkung zweier Colour-Dipole als Funktion deren Impakts~$\rb{b}$, vgl.\@ Gl.~(\ref{tsill}): in Abbildung~\ref{Fig:int-ampl_NC,C}(a) der~$N\!C$-Term und in~\ref{Fig:int-ampl_NC,C}(b) der~$C$-Term, die folgen durch Setzen~$\vka \!\equiv\! 0$ beziehungsweise~$\vka \!\equiv\! 1$, vgl.\@ Gl.~(\ref{X_vka,X-C,X-NC-tilde}).
F"ur jede Verschiebung des~\pDipols relativ zum~\mDipol sind diese Terme berechnet als Funktion des Impakts~$\rb{b} \!=\! \bprll \!+\! \bperp$ und aufgetragen "uber der~$\bprll,\! \bperp$-Ebene. \\
\indent
Wesentliche Beitr"age treten aufgrund des exponentiellen Abfalls der~$\tilde{X}\idx{\mfp\mskip-2mu\mfm}\oC$-,~$\tilde{X}\idx{\mfp\mskip-2mu\mfm}\oNC$-inh"arenten Korrelationsfunktionen~${\cal K}_2$,~${\cal K}_3$ nur auf f"ur kleine Abst"ande.
Dabei bezieht sich "`Abstand"' f"ur den~$N\!C$-Term auf die Separation der Endpunkte der Dipole, das hei"st der Separation ihrer Quarkkonstituenten.
F"ur den~$C$-Term dagegen bezieht sich "`Abstand"' auf die Separation der geraden Verbindungslinie der Quarkkonstituenten des einen zu der des anderen Dipols.
Dies ist insbsondere im dargestellten Falle eines gro"sen und eines kleinen Dipols von eklatantem Unterschied:
Bei Positionierung des kleinen Dipols auf der Verbindungslinie der Quarkkonstituenten des gro"sen, hat er zu dieser Linie minimalen, zu dessen Endpunkte mitunter sehr gro"sen Abstand.
Dies ist dahin zu interpretieren, da"s zwischen den Quarkkonstituenten eines Dipols ein {\it gluonischer String\/} ausgebildet ist, und diese wesentlich zur Amplitude~$\tTll$ der Dipol-Dipol-Wechselwirkung beitragen:
Der postulierte String-String-Mechanismus. \\
\indent
\begin{figure}
\begin{minipage}{\linewidth}
  \begin{center}
  \setlength{\unitlength}{.8mm}\begin{picture}(120,82)   
    \put(0,0){\epsfxsize96mm \epsffile{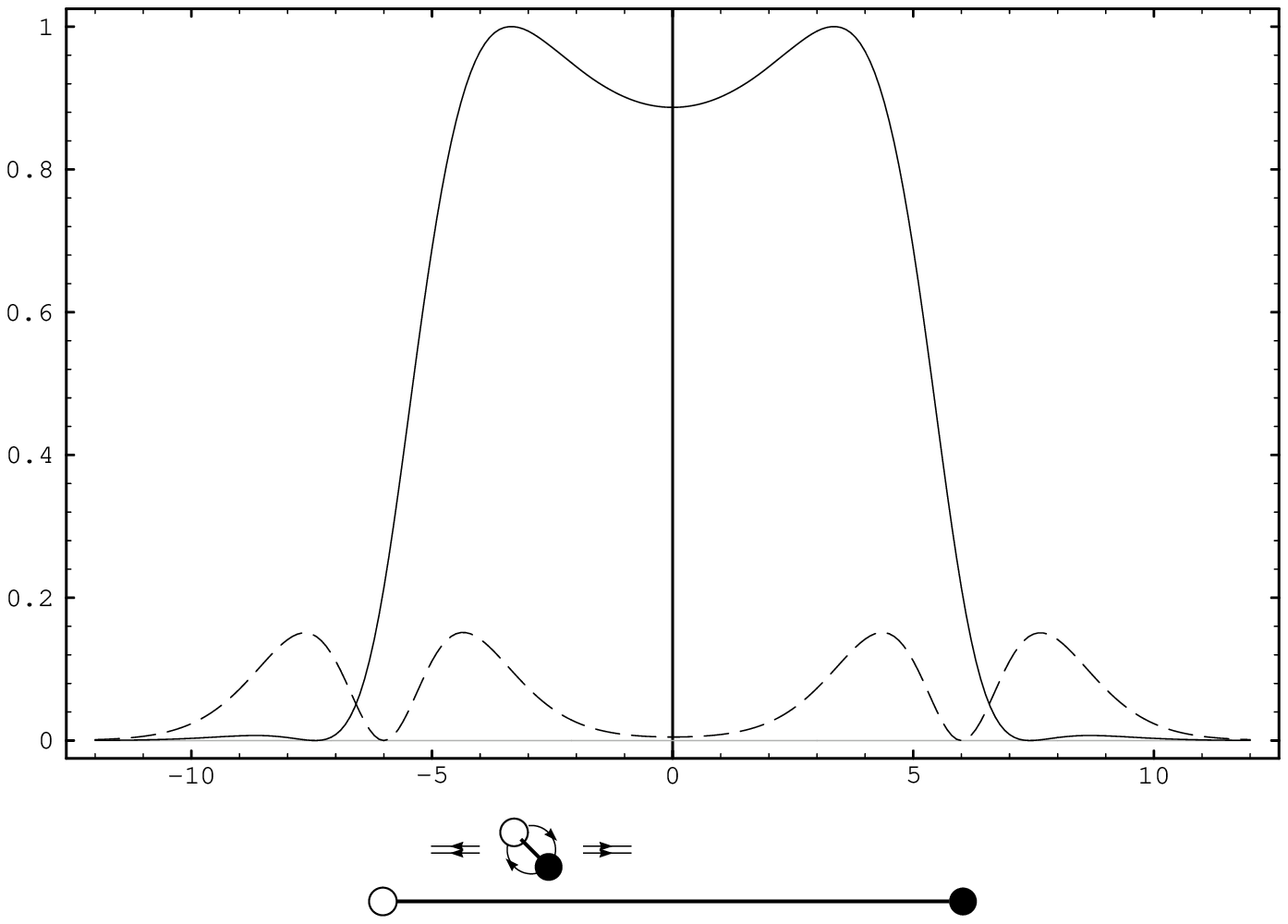}}
    \put(21,55){\normalsize $C$-Term}
    \put( 7,27){\normalsize $N\!C$-Term}
    \put(92,1){\normalsize \pDipol}
    \put(92,6){\normalsize \mDipol}
    \put(116,16){\normalsize $\bprll /\!a$}
  \end{picture}
  \end{center}
\vspace*{-2.5ex}
\caption[Wechselwirkung zweier kollinearer Dipole,~\protect$N\!C$- und~\protect$C$-Term]{
  Amplitude~$\tsill$ f"ur die Wechselwirkung zweier kollinearer Colour-Dipole als Funktion deren Impakts~$\bprll$, vgl.\@ Gl.~(\ref{tsill}).   $N\!C$- gegen $C$-Term, normiert auf das Maximum des $C$-Terms.   Der~$\Dbr{+}$- und~\mDipol sind identisch dimensioniert wie in Abbildung~\ref{Fig:int-ampl_NC,C} und identisch konfiguriert mit dem einzigen Unterschied, da"s die Position des~\mDipols eingeschr"ankt ist auf die gerade Verbindungslinie der Quarkkonstituenten des~\pDipols[].   Es ist~$\bperp \!\equiv\! 0$ per definitionem und~$\tsill$ aufgetragen "uber~\mbox{$\rb{b} \!\equiv\! \bprll$}: das hei"st genau die Profile "uber der $\bperp \!\equiv\! 0$-Linie der Abbildungen~\ref{Fig:int-ampl_NC,C}(a),(b).   W"ahrend sich der~$N\!C$-Term konstituiert aus Konfigurationen mit kleinen Abst"anden der Dipol-Endpunkte, treten im~$C$-Term wesentliche Beitr"age auch auf bei kleinen Abst"anden der Dipole an sich.   Die physikalische Kurve f"ur~\mbox{$\vka \!=\! 0.74$}, s.u.\@ Gl.~(\ref{Parameter}), folgt durch Multiplikation des~$N\!C$-Terms mit~\mbox{$(1 \!-\! \vka)\!/\vka \cong 1\!/3$}~und Addition zum~$C$-Term, das hei"st der konfinierende Term dominiert die Wechselwirkung.
}
\label{Fig:int-ampl-collinear_NC,C}
\end{minipage}
\end{figure}
Dieser Mechanismus wird noch deutlicher herauspr"apariert in Abbildung~\ref{Fig:int-ampl-collinear_NC,C}, in der die Profile der~\mbox{$\bperp \!\equiv\! 0$-Linien} der Abbildungen~\ref{Fig:int-ampl_NC,C}(a) und (b) gegeneinander aufgetragen sind:
Der~\mbox{\mDipol[]} wird verschoben nur auf der geraden Verbindungslinie der Quarkkonstituenten des des~\pDipols[], $\Dbr{+}$- und~\mDipol sind kollinear.

Wir merken an zu den Abbildungen~\ref{Fig:int-ampl_NC,C} und~\ref{Fig:int-ampl-collinear_NC,C}:
Erstens.\@
Von den Funktionen~$\tilde{X}\oC$,~$\tilde{X}\oNC$, vgl.\@ die Gln.~(\ref{X-C,X-NC-tilde_final}),~(\ref{X-C,X-NC-tilde_final}$'$), "ubertragen sich auf die $T$-Amplitude~$\tTll$ die Symmetrien:
\vspace*{-.5ex}
\begin{align} \label{Symm_tTll}
&\tTll(\zet_\snbr{+}, \th_\snbr{+}, \zet_\snbr{-}, \th_\snbr{-},\ldots)
   \\[.75ex]
  &=\; \tTll(1\!-\! \zet_\snbr{+}, \pi\!+\! \th_\snbr{+}, \zet_\snbr{-}, \th_\snbr{-},\ldots)\;
   =\; \tTll(\zet_\snbr{+}, \th_\snbr{+}, 1\!-\! \zet_\snbr{-}, \pi\!+\! \th_\snbr{-},\ldots)
    \nn
\end{align}
das hei"st sie ist symmetrisch bez"uglich gleichzeitiger Ersetzung%
  ~\vspace*{-.125ex}\mbox{$\th_\snbr{\,\pm} \!\to\! (\pi \!+\! \th_\snbr{\,\pm})$} und~$\zet_\snbr{\,\pm} \!\to\! \bzet_\snbr{\,\pm}$ im~$\Dbr{+}$- und/oder~\mbox{\mDipol[]}.
Zweitens.\@
In Wirklichkeit sind die Quarkkonstituenten weniger deutlich voneinander getrennt als in den Abbildungen~\ref{Fig:int-ampl_NC,C}(b),~\ref{Fig:int-ampl-collinear_NC,C}($N\!C$)~-- und entsprechend f"allt der $C$-Term weniger stark ab auf ihrer Verbindungslinie als in den Abbildungen~\ref{Fig:int-ampl_NC,C}(a),~\ref{Fig:int-ampl-collinear_NC,C}($C$):
Der Wert der Korrelationsl"ange~$a$ betr"agt ungef"ahr~$0.35\fm$, so da"s ein physikalischer Colour-Dipol eine transversale Ausdehnung von nur~$4 \!-\! 5\,a$ besitzt.%
\FOOT{
  Effektive Korrelationsl"ange ist nicht~$a$, sondern~$\la a$, vgl.\@ die Bem.\@ zu Gl.~(\ref{r^Q/AQ_(+)(-)}).   Wir werden dies im folgenden sprachlich nicht mehr differenzieren.
}
\vspace*{-.5ex}

\subsection[Hadron-Hadron-Streuamplitude~\protect$T\hh^{(s,t)}$]{Hadron-Hadron-Streuamplitude~\protect\bm{T\hh^{(s,t)}}}

Wir gehen "uber von der Fourier-transformierten $T$-Amplitude~$\tTll^{(s,\rb{b})}$ f"ur \vspace*{-.25ex}Loop-Loop-Streuung zur vollen~$T$-Amplitude~$T\hh^{(s,t)}$ f"ur Hadron-Hadron-Streuung.
Wir rekapitulieren:
\begin{align} 
T\hh^{(s,t)}\; &\equiv\;
  \bracket{\, h^{2'}\!(P_{2'})\, h^{1'}\!(P_{1'}),\,\IN \,}{\,
              T\, \bracketM\,
              h^1(P_1)\, h^2(P_2),\,\IN \,} \\[.5ex]
  &=\; \int d^2\rb{b}\;
         \efn{\T-\iIM\,\tfb \!\cdot\! \rb{b}}\;
         \int d\vph_{1',1} (\zet_\snbr{+}, \rb{r}_\snbr{+})\;
         \int d\vph_{2',2} (\zet_\snbr{-}, \rb{r}_\snbr{-})\vv
         \tTll^{(s,\rb{b})}(\zet_\snbr{+}, \rb{r}_\snbr{+};
                            \zet_\snbr{-}, \rb{r}_\snbr{-}; \rb{b}) \nn
\end{align}
Bzgl.\@ der von Gl.~(\ref{GROUND:Tconn-Element2h_WW-W_lim}) differierenden Notation~$\zet_\snbr{\,\pm}$,~$\rb{r}_\snbr{\,\pm}$ vgl.~(\ref{r_+-}).

Wir betrachten~$\tTll$ als ausintegriert bez"uglich der Variablen eines Dipols; sei dies o.E.d.A.\@ der~\mDipols[], das hei"st ausintegriert mit dem Ma"s~$d\vph_{2',2}$, das die Spinsumme der Lichtkegelwellenfunktionen~$\vph^{2'}_{s\bar s}{}^{\!\D\dagger}$,~$\vph^2_{s\bar s}$ impliziert, die Variablen~$\zet_\snbr{-}$,~$\rb{r}_\snbr{-}$, vgl.\@ Gl.~(\ref{GROUND:dvphi'i}).
Aufgrund derselben Argumentation wie oben, vgl.\@ Fu"snote~\FN{FN:Impakt-Betrag}, h"angt die resultierende Funktion vom Impaktvektor~$\rb{b}$ nur ab "uber dessen Betrag~$\bbB$ und nicht von dessen Azimut~$\vph_\rb{b}$, der Verdrehung relativ einer beliebig gew"ahlten~$x^1$-Achse.
Die~$\vph_\rb{b}$-Winkelintegration betrifft daher nur die Exponentialfunktion und wird trivial bei Wahl der~$x^1$-Achse in Richtung~$\tfb$; wir erhalten:
\bea \label{Thh_Tellp-allg}
T\hh\;
  =\; \int d\vph_{1',1} (\zet_\snbr{+}, \rb{r}_\snbr{+})\vv
        T\ellp(\zet_\snbr{+}, \rb{r}_\snbr{+}; \tfbB)
\eea
dabei gilt~$\tfbB \!=\! \sqrt{-t \!+\! \tfde} \!\cong\! \surd\!-t$, vgl.\@ Gl.~(\ref{tfbB_-t}), und ist definiert:
\bea \label{Tellp_tTll-allg}
T\ellp(\zet_\snbr{+}, \rb{r}_\snbr{+}; \tfbB)\;
  =\; 2\pi\, \int_0^{\infty} \bbB\, d\bbB\, J_0(\tfbB {\mskip-.5mu} \bbB)\;
        \int d\vph_{2',2} (\zet_\snbr{-}, \rb{r}_\snbr{-})\vv
          \tTll(\zet_\snbr{+}, \rb{r}_\snbr{+};
                \zet_\snbr{-}, \rb{r}_\snbr{-}; \bbB)
\eea
mit~$J_0$ der Besselfunktion erster Art in der Definition von Ref.~\cite{Abramowitz84}: $J_0(z) \!\to\! 1$ f"ur~$z \!\to\! 0$.

In der vorliegenden Arbeit diskutieren wir Photo- und Leptoproduktion verschiedener Vektormesonen~$V$ unter verschiedenen kinematischen Bedingungen~$(s,t)$ an einem festen intakt bleibenden Proton-Target, vgl.\@ Fu"snote~\FN{FN:Nukleon-Proton}.
Mit Bezug des Index~"`1"', des \pDipols auf die Vektormeson-Seite der Streuung und des Index~"`2"' des \mDipols auf die des Protons hei"st dies die Identifizierung:~$h^{1} \!\equiv\! \ga^{\scriptscriptstyle({\D\ast})}$,~$h^{1'} \!\equiv\! V$ und~$h^{2} \!\equiv\! h^{2'} \!\equiv\! p$.
Wir haben daher den unver"anderlichen Teil der Streuung, die Proton-Seite absorbiert in die Funktion~$T\ellp \!\equiv\! T\ellp^{(s,t)}$, die genau die $T$-Amplitude ist f"ur die Streuung eines Wegner-Wilson-Loops, das hei"st Colour-Dipols an dem festen Proton-Target.
Au"ser von der Kinematik~$(s,t)$, vgl.~\mbox{$\tfbB \!\cong\! \surd\!-t$}, h"angt sie ab von dem Paar~$\{ \zet_\snbr{+}, \rb{r}_\snbr{+} \}$, das den als variabel zu betrachtenden~\pDipol[]~charakterisiert.
Sie ist von unmittelbarer physikalischer Bedeutung.
\vspace*{-1.5ex}

\paragraph{\label{T:Parameter}Parameter.} Wir unterbrechen die Diskussion mit der Bestimmung der Parameter, die unserer Beschreibung weicher Hochenergiestreuung zugrunde liegen.
Dies sind zum einen die Parameter des \DREI{M}{S}{V} an sich: Gluonkondensat~$\vac{g^2FF}$, Korrelationsl"ange~$a$ und die Konstante~$\vka$, die~$N\!C$- und~$C$-Term gegeneinander gewichtet.
Zum anderen ist ein Ansatz zu machen f"ur die Wellenfunktion des Protons und anzugeben deren \vspace*{-.25ex}Parameter. \\
\indent
Diese Bestimmung geschieht bez"uglich weicher Proton-Proton-Streuung bei hoher invarianter Schwerpunktenergie~$\surd s$ wie bez"uglich Confinements eines statischen Quark-Antiquark-Paares.~--
Sie garantiert daher Konsistenz des vorgestellten Zugangs zu Streuph"anomenen des Hochenergiebereichs nichtperturbativer Quantenchromodynamik mit der Beschreibung im Rahmen des \DREI{M}{S}{V} von Ph"anomenen ihres Niederenergiebereich.

Wir gehen in derselben Weise vor wie Ref.~\cite{Dosch94a}, vgl.\@ auch Ref.~\cite{Kraemer91}:
Die Beschreibung eines statischen Quark-Antiquark-Paares reduziert sich unter sehr allgemeinen Annahmen auf die fundamentale Kumulante zweiter Ordnung in paralleltransportierten Feldst"arken: \mbox{$(K_2)_{a_1a_2} \!\equiv\! \vac{ g^2 F^{(1)}{}_{\zz a_1} F^{(2)}{}_{\zz a_2} }$}, mit~\mbox{$F^{(i)} \equiv F_{\mu_i\nu_i}\!(x_i; x_0,{\cal C}_{x_{\!0}\!x_{\!i}})$}.
Vgl.\@ die Gln.~(\ref{K2Komp_Kronecker}),~(\ref{K2Komp_Kronecker}$'$),~bzgl.\@ der Relation von~$(K_2)_{a_1a_2}$ mit dem Korrelations-Lorentz-Tensor~$D \!=\! (D_{\mu_1\nu_1\mu_2\nu_2})$~Gl.~(\ref{K2-g2FF_Dvier}) und bzgl.\@ dessen Gestalt Gl.~(\ref{Dvier_DDxi}) bzw.~(\ref{Dvier_DDxi}).
Diese Kumulante~$(K_2)_{a_1a_2}$ wird berechnet in den Refn.~\cite{Campostrini89,DiGiacomo90,DiGiacomo92} im Rahmen einer {\it pure\/} $SU\!(\Nc\!\equiv\!3)$-Gittereichtheorie, das hei"st einer reinen (Gitter)Eichtheorie ohne dynamische Fermionen.%
\FOOT{
  Neuere Gittersimulationen existieren, die versuchen perturbative Beitr"age pr"aziser von den nichtperturbativen Gluonfluktuationen zu isolieren, auch die dynamische Fermionen miteinbeziehen, vgl.\@ Ref.~\cite{DiGiacomo96} bzw.\@ die Refn.\,\mbox{\cite{DElia97,DElia97a,DElia97b}}; angesichts der nur geringen Diskrepanz in Hinblick auf numerische wie konzeptionelle Unsicherheit hier wie dort erscheint es gegenw"artig nicht als gerechtfertigt, die etablierte Formulierung des \DREI{M}{S}{V} umzuwerfen~-- etwa durch verfeinerte Anpassungen der Korrelationsfunktionen.
}
%
Ref.~\cite{DiGiacomo92} gelangt unmittelbar zu~$\vka \!\cong\! 0.74$, das hei"st bei maximaler Korrelation betr"agt der $N\!C$-Term nur ein Drittel des $C$-Terms.
Sie postuliert innerhalb der numerischen Unsicherheit im (Gitter-)relevanten Bereich des Ortsraums Proportionalit"at der entsprechenden~$D_{\mu_1\nu_1\mu_2\nu_2}$-inh"arenten Korrelationsfunktionen; dies induziert die Verkn"upfung der Korrelationsfunktionen im Impulsraum~$\widetilde{D}\uC$ und~$\widetilde{D}^{\D\prime}\uNC$ wie formuliert in den Gln.~(\ref{Dvier_DDkkontrAnsatz}),~(\ref{DDk}).
Bez"uglich des in diesen Gleichungen angegebenen Ansatz folgt weiter beste Anpassung an die auf dem Gitter "`gemessenen"' Korrelationsfunktionen f"ur den Index~$n \!\equiv\! 4$ (der auch im folgenden unterdr"uckt sei).

Im Rahmen des dargestellten Zugangs des \DREI{M}{S}{V} zu weicher Hochenergiestreuung wird in Ref.~\cite{Dosch94a} zun"achst die Streuung zweier gleicher hadronischer Zust"ande diskutiert und festgestellt die dominierende Sensitivit"at auf deren transversale Ausdehnung.
Bezogen auf die Streuung zweier Proton existiert das umfangreichste und genaueste Material experimenteller Daten.
Zum einen ist aber nur wenig bekannt "uber die exakte Gestalt der Proton-Lichtkegelwellenfunktion, zum anderen sollte sie in Hinblick auf die Aussagekraft des Modells nur wenige Parameter enthalten.
F"ur das Absolutquadrat der Proton-Wellenfunktion wird daher der einfache Ansatz gemacht:
\begin{samepage}
\vspace*{-.75ex}
\begin{align} \label{Proton-wfn2_vph}
\vph_{s\mskip-1mu{\bar s}}^{p\D\dagger} (\zet_\snbr{-}, \rb{r}_\snbr{-})\;\,
     \vph_{s\mskip-1mu{\bar s}}^p       (\zet_\snbr{-}, \rb{r}_\snbr{-})\;
  \equiv\; \big| \vph_p (\zet_\snbr{-}, &\rb{r}_\snbr{-}) \big|^2
    \\[-.75ex]
  &\stackrel{\D!}{=}\vv 2\, \om_{\!p}^2\; \de(\zet_\snbr{-} \!-\! 1\!/\!2)\;
           \efn{\D -\om_{\!p}^2 r_\snbr{-}^2}
    \nn
    \\[-4.5ex]\nn
\end{align}
mit~$r_\snbr{-} = |\rb{r}_\snbr{-}|$, vgl.\@ Gl.~(\ref{r_+-}), normiert wie in Gl.~(\ref{vph-x_Norm}).
Das Proton aufgefa"st als~$1S$-Zustand in einem Harmonischen-Oszillator-Potential, besitzt genau diesen Transversalanteil.
Mit~$\om_p$ dem Oszillatorparameter sind%
~\mbox{$R_p \!\equiv\! \vev{r_\snbr{-}^{\,2}}^{1\!/\!2} \!=\! 1/2\om_{\!p}$} und%
~\mbox{$R_{3,p} \!=\! \vev{\vec{r}_\snbr{-}^{\,2}}^{1\!/\!2} \!=\! \smash{\sqrt{3\!/\!2}} /2\om_{\!p}$}
der \vspace*{-.25ex}transversale beziehungsweise volle {\it root mean square-Radius\/} (rmsq-Radius) dieses Zustands.
Bez"uglich des Lichtkegelimpulses wird Gleichverteilung auf Quark und Antiquark angenommen:~$\zet_\snbr{-} \!=\! 1\!/\!2$.
Wir kommen darauf zur"uck, wie gerechtfertigt diese Approximationen sind. \\
\indent
Auf Basis dieses Absolutquadrats der Proton-Wellenfunktion werden in Ref~\cite{Dosch94a} f"ur Proton-Proton-Streuung unter anderem berechnet der {\it totaler Wirkungsquerschnitt\/}:
\vspace*{-.25ex}
\begin{align} \label{si^tot_Thh}
\si^{\rm tot}\;
  =\; \frac{1}{s}\, {\rm Im}\vv T\hh^{(s,t=0)}
    \\[-4.25ex]\nn
\end{align}
\end{samepage}%
auf Basis des optischen Theorems, und der {\it slope-Parameter}, der zusammenh"angt mit dem \mbox{$t$-dif}\-ferentiellen elastischen Wirkungsquerschnitt in Vorw"artsrichtung:
\vspace*{-.5ex}
\begin{align} \label{slope-Parameter_Thh}
B_0\hspace*{1.125em}
  =\; \frac{d}{dt}\; \ln\, \frac{d\si^{\rm el}}{dt}\; \Big|_{t=0} \qquad
  \text{mit}\qquad
  \frac{d\si^{\rm el}}{dt}(t)\;
  =\; \frac{1}{16\pi}\, \frac{1}{s^2}\, \big| T\hh^{(s,t)} \big|^2
    \\[-4.5ex]\nn
\end{align}
anschaulich der Abfall mit~$\tfbQ \!\cong\! -t$, dem invarianten Quadrat des Impulstransfers, des in~$t$ differentiellen Wirkungsquerschnitts an der Stelle~$t \!=\! 0$.
Durch Eliminierung der Abh"angigkeit vom Protonradius~$R_p$ in den resultierenden Funktionen~$\si^{\rm tot}(R_p/\!a)$ und ~$B_0(R_p/\!a)$ und durch Anbinden an die experimentellen Zahlenwerte f"ur~$\surd s \!=\! 20\GeV$,~$s'$ in\GeV$^2$:
\begin{align}
&\si_{pom.}^{\rm tot}(pp,p{\bar p})\;
  =\; 21.70\;{\rm mb}\cdot (s')^{0.0808}\;
      \big|_{\surd s = 20\GeV}\;
  \cong\; 35.2\;{\rm mb}&
    \label{si^tot_pom-ZW} \\[.5ex]
&B_0\;  \big|_{\surd s = 20\GeV}\;
  \cong\; 11.5\GeV^{-2}&
    \label{B_pp-ZW}
\end{align}
vgl.\@ Ref.~\cite{Donnachie92} bzw.~\cite{Amaldi80},%
\FOOT{
  Amaldi, Schubert geben an in ihrem "`Review aller ISR-Daten f"ur Proton-Proton-Streuuung"', vgl.\@ Ref.~\cite{Amaldi80}:~$B \!=\! 11.8 \!\pm\! 0.3\GeV^{-2}$ bez"uglich~$-t \!=\! 0.04\GeV^2$ und der niedrigsten ISR-Energie von~$\surd s \!=\! 23.5\GeV$; im Hinblick auf die h"oheren ISR-Energien folgt als {\sl educated guess\/} f"ur~$\surd s \!=\! 20\GeV$,~$t \!=\! 0$ der Wert von Gl.~(\ref{B_pp-ZW}).   Besten Dank an Timo Paulus, der uns auf diese Referenz aufmerksam gemacht hat.
}
folgt~$\vac{g^2FF}'$, das Gluonkondensat in\GeV$^4$, als Funktion von~$a'$, der Korrelationl"ange~$a$ in\fm, wir schreiben symbolisch:~\mbox{$\vac{g^2FF}^{\prime(1)}(a')$}.
Eine zweite Funktion \mbox{$\vac{g^2FF}^{\prime(2)}(a')$} folgt aus der Berechnung der Stringspannung~$\si$ im Rahmen des \DREI[]{M}{S}{V}:
\vspace*{-.5ex}
\begin{align} 
\si = \frac{8}{81\pi}\,a^2\,\vka\vac{g^2FF} \qquad
  \si \cong 0.18\GeV^2
    \\[-4.5ex]\nn
\end{align}
vgl.\@ Gl.~(\ref{Stringspannung-si_DC}) und die Refn.~\cite{Dosch87,Dosch88,Dosch94a}.
Eine dritte Funktion letztlich folgt aus Ref.~\cite{DiGiacomo92},~aus der Anbindung der QCD-Gitterskala~$\La_{latt.}$ an physikalische Einheiten:
\vspace*{-.5ex}
\begin{align} 
\La_{latt.}\; \cong\; 5\MeV
    \\[-4.5ex]\nn
\end{align}
Aufgrund der Unsicherheit, die den experimentellen Werten f"ur~$\si^{\rm tot}$,$B_0$ und den Werten von~$\si$,~$\La_{latt.}$ anhaften sind diese drei Kurven effektiv B"ander in der~$\vac{g^2FF},\!a$-Ebene.
In deren Schnittbereich werden die Parameter~$\vac{g^2FF}$ und~$a$ gew"ahlt, ~$R_p$ folgt unmittelbar quasi als Kurvenparameter von~$\vac{g^2FF}^{\prime(1)}(a')$.
Wir geben zusammenfassend an als die unserer Arbeit zugrundeliegenden Zahlenwerte:
\vspace*{-.5ex}
\bea \label{Parameter}
       \vka\; =\; 0.74       \qquad
\vac{g^2FF}\; =\; 2.49\GeV^4 \qquad
          a\; =\; 0.346\fm   \qquad
        R_p\; =\; 1.51\, a
    \\[-4.5ex]\nn
\eea
folglich~$R_p \!\cong\! 0.522\fm$ und~$\vac{g^2FF}a^4 \!\cong\! 23.5$ f"ur den dimensionslosen Parameter, in der Kombination nur das Gluonkondensat auftritt, vgl.\@ die Gln.~(\ref{tTll_WW_ch-mf_ALL-gmfpmfp-REP}),~(\ref{tTll_WW_ch}). \\
\indent
Die Diskrepanz zu den Werten aus Ref.~\cite{Dosch94a} r"uhrt daher, da"s dort~$\vka \!=\! 0.74$ gesetzt, aber im Sinne einer ersten Approximation der~$N\!C$-Term komplett vernachl"asiigt wird.
Wir berechnen beide Terme.
Nur der Vollst"andigkeit wegen sei angemerkt, da"s wir den experimentellen Wert f"ur den totalen Wirkungsquerschnitt~(\ref{si^tot_pom-ZW}) w"ahlen f"ur~$\surd s$ "`exakt"'~$20\GeV$ und nicht als "`exakt"'~$35.0\mbarn$ (f"ur kleineres~$\surd s \!\cong\! 19.2\GeV$).
Wesentlich dagegen, da"s der experimentelle Wert f"ur den slope-Parameter auf den Zahlenwert~(\ref{B_pp-ZW}) aktualisiert ist.

Der Zahlenwert f"ur das Gluonkondensat in~(\ref{Parameter}) ist um einen Faktor F"unf gr"o"ser als der Wert, zu dem Shifman, Vainshtein, Zakharov mithilfe QCD-Summenregeln in Ref.~\cite{Shifman79} gelangen.
Dies ist insofern nicht unerwartet, als zum einen deren Wert allgemein als unterste zul"assige Schranke betrachtet wird, und zum anderen wir in einer {\it pure gauge theory\/} arbeiten, vgl.\@ Fu"snote~\FN{FN:pure-gauge-theory}, deren Gluonkondensat um einen Faktor Zwei bis Drei gr"o"ser generiert wird; sei verwiesen auf Ref.~\cite{Dosch94a}, Anh.~C wie auf Ref.~\cite{Novikov81}.

Der transversale rmsq-Radius des Protons ist mit~$R_p \!\cong\! 0.52\fm$ kleiner als sein experimenteller transversaler {\it rmsq-charge-Radius\/}~$R^{\rm ch.}_p \!\cong\! 0.70\fm$, vgl.\@ Ref.~\cite{Simon80}.
Dies kann in der Weise interpretiert werden, da"s {\it sea quarks\/} zum Formfaktor beitragen und so den elektromagnetischen Ladungs-Radius im Vergleich zu dem der Valenzquark-Konstituenten anwachsen l"ast.
Dieselben sea quarks, in Gestalt zus"atzlicher Colour-Dipole, k"onnten beitragen zur Streuung~-- umso wichtiger, je gr"o"ser ihre Anzahl, insbesondere mit Anwachsen des invarianten Quadrats der Gesamtenergie~$s$.
Dies k"onnte erkl"aren das Anwachsen des totalen Wirkungsquerschnitts mit~$s$, vgl.\@ Gl.~(\ref{si^tot_pom-ZW}), das in unserem Ansatz genau dadurch simuliert werden kann, da"s der Protonradius~$R_p$ aufgefa"st wird als monoton steigende Funktion von~$s$; vgl.\@ die expliziten Ausdr"ucke in Ref.~\cite{Dosch94a}.

\bigskip\noindent
Wir kehren zur"uck zu der Diskussion der $T$-Amplitude~$T\hh$ f"ur Hadron-Hadron-Streuung in der Gestalt der Gln.~(\ref{Thh_Tellp-allg}),~(\ref{Tellp_tTll-allg}).
Die Lichtkegel-Wellenfunktionen der partizipierenden Zust"ande sind absorbiert im Integrationsma"s~$d\vph_{i',i}$, vgl.\@ die Gln.~(\ref{vph-x_Norm}),~(\ref{h_bracket}) und~(\ref{GROUND:dvphi'i}):
\begin{align} \label{GROUND:h_bracket}
&\bracket{h^{i'}(P_{i'})}{h^i(P_i)}\;
  =\; (2\pi)^3\, 2(P_i)_{0\!+}\, \de(\vec{P}_{i'} \!-\! \vec{P}_i)
         \cdot \int\! d\vph_{i',i} (\zet, \rb{r})
    \\
  &\text{mit}\qquad
  1\;
  =\; \int\! d\vph_{i,i} (\zet, \rb{r})\;
  =\; \int_0^1 \frac{d\zet}{2\pi}\; \int d^2\rb{r}\;
        \vph_{s\mskip-1mu{\bar s}}^{i\D\dagger} (\zet, \rb{r})\,
        \vph_{s\mskip-1mu{\bar s}}^i            (\zet, \rb{r})
    \tag{\ref{GROUND:h_bracket}$'$}
\end{align}
Gl.~(\ref{GROUND:h_bracket}) f"ur feste~$i,i'$, beide~$\in\! \{1,1'\}$ oder~$\in\! \{2,2'\}$, bzw.\@ Gl.~(\ref{GROUND:h_bracket}$'$) f"ur festes~$i \!\in\! \{1,1',2,2'\}$.

Eine Notation in "Ubereinstimmung mit der Literatur, auf die wir wesentlich Bezug nehmen insbesondere unsere Ver"offentlichungen, wird erreicht, indem wir umbenennen:
\bea \label{Umbenennung-vph-ps}
\ps_{s\mskip-1mu{\bar s}, i}\; 
    \equiv\; \surd2\cdot \vph_{s\mskip-1mu{\bar s}}^i \qquad
  \text{und}\qquad
  d\ps_{i',i}\;
    \equiv\; d\vph_{i',i}
\eea
indem die Notation f"ur den "Uberlapp~\mbox{$\ps_{i'}^{\D\dagger} \ps_{i} \equiv 2\cdot \vph_{s\mskip-1mu{\bar s}}^{i'\D\dagger} \vph_{s\mskip-1mu{\bar s}}^i$} die Spinsumme impliziere (wie bereits die Flavour-Summe) und unterdr"uckt sei der Index~"`$\Dnbr{+}$"' der Variablen der~\mbox{Photon-/Vek}\-tormeson-Seite.
Die Gln.~(\ref{Thh_Tellp-allg}),~(\ref{Tellp_tTll-allg}) schreiben sich dann:
\begin{align} %
&T\hh\;
  =\; \int \frac{d\zet d^2\rb{r}}{4\pi}\; \vv
        \ps_V^{\D\dagger} \ps_\iga (\zet, \rb{r})\vv
        T\ellp(\zet, \rb{r}; \tfbB)
    \label{Thh_Tellp} \\[1ex]
&T\ellp
  =\; 2\pi\, \int_0^{\infty} \bbB\, d\bbB\, J_0(\tfbB\, \bbB)\;
        \int \frac{d\zet_\snbr{-} d^2\rb{r}_\snbr{-}}{4\pi} \vv
        \big| \ps_p (\zet_\snbr{-}, \rb{r}_\snbr{-}) \big|^2 \vv
          \tTll(\zet, \rb{r}; \zet_\snbr{-}, \rb{r}_\snbr{-}; \rb{b})
    \label{Tellp_tTll}
\end{align}
Dabei ist entsprechend Gl.~(\ref{Proton-wfn2_vph}) zu setzen
\bea \label{Proton-wfn2_ps}
\ps_p^{\D\dagger} (\zet_\snbr{-}, \rb{r}_\snbr{-})\,
            \ps_p (\zet_\snbr{-}, \rb{r}_\snbr{-})\;
  =\; \big| \ps_p (\zet_\snbr{-}, \rb{r}_\snbr{-}) \big|^2
  =\; 4\, \om_{\!p}^2\; \de(\zet_\snbr{-} \!-\! 1\!/\!2)\;
          \efn{\D -\om_{\!p}^2 r_\snbr{-}^2}
\eea
Wir betrachten~$T\ellp$ als ausintegriert.
Aus den Gln.~(\ref{GROUND:h_bracket}$'$),~(\ref{Umbenennung-vph-ps}) folgt
\bea \label{Normierung}
1\; =\; \int d\ps_{i,i} (\zet, \rb{r})\;
  =\; \int \frac{d\zet d^2\rb{r}}{4\pi}\; \vv
        \big| \ps_i (\zet, \rb{r}) \big|^2
\eea
als Normierung.
Diese bezieht sich auf die Zust"ande~$i \!=\! 1' \!=\! V$ und~$2 \!=\! 2' \!=\! p$; die Photon-Lichtkegelwellenfunktion ist nicht normiert, vgl.\@ ihre Konstruktion unten auf Seite~\pageref{Subsect:Photon-Wfn}.

In Zusammenhang der Gln.~(\ref{Thh_Tellp-allg}),~(\ref{Tellp_tTll-allg}) greifen wir die Bemerkung von Fu"snote~\FNg{FN:q(qq)-qqq} auf:
Der Einfachheit halber fassen wir das Proton auf als Quark-Diquark-Zustand.
Aufgefa"st als Drei-Quark-Zustand, und in diesem Bild fixiert die Parameter, vgl.\@ Ref.~\cite{Dosch94a}, folgen prinzipiell die gleichen Vorhersagen,~-- allein der Rechenaufwand ist gr"o"ser, da der Zustand formal repr"asentiert ist durch drei Wegner-Wilson-Loops, das hei"st Colour-Dipole, die verkn"upft sind bez"uglich ihres Eichgehalts.
Die Berechnung des {\it Odderon\/}-Austausches~\mbox{($C \!\equiv\! P \!\equiv\! -1$)} zeigt, da"s das Quark-Diquark-Bild insofern zu favorisieren ist, als es diesen Beitrag unterdr"uckt gegen"uber dem des Pomerons~\mbox{($C \!\equiv\! P \!\equiv\! +1$)}; vgl.\@ Ref.~\cite{Rueter96}.

Es ist definiert~$T\ellp \!\stackrel{\D!}{=}\! T\ellp^{(s,t)}$ als die $T$-Amplitude f"ur die Streuung eines Wegner-Wilson-Loops, das hei"st Colour-Dipols an dem festen Proton-Target, vgl.\@ die Gln.~(\ref{Tellp_tTll-allg}),(\ref{Tellp_tTll-allg}).
Der totale Wirkungsquerschnitt f"ur die Streuung des \pDipols[]~$\{\zet,\rb{r}\} \!\equiv\! \{\zet,r,\th\}$, der Einfachheit halber gemittelt "uber s"amtliche Orientierungen~$\th$, an dem festen Proton-Target ist gegeben durch
\vspace*{-1.5ex}
\begin{align}
&\si^{\rm tot}\ellp(\zet, r^2)\;
  =\; \frac{1}{s}\, {\rm Im}\vv
      \pT\ellp(\zet, r, \tfbB \!\equiv\! 0)
    \label{sitot_ellp}
    \\[.25ex]
&\text{mit}\qquad
 \pT\ellp(\zet, r, \tfbB)\;
  =\; \int_0^{2\pi}\! \frac{d\th}{2\pi}\vv
        T\ellp(\zet, r, \th, \tfbB)
    \label{pT_ellp}
    \\[-5ex]\nn
\end{align}
aufgrund des optischen Theorems, vgl.\@ die Gln.~(\ref{tsill}),~(\ref{si^tot_Thh}).

\begin{figure}
\begin{minipage}{\linewidth}
  \begin{center}
  \setlength{\unitlength}{0.9mm}\begin{picture}(120,73.7)   
    \put(0,0){\epsfxsize108mm \epsffile{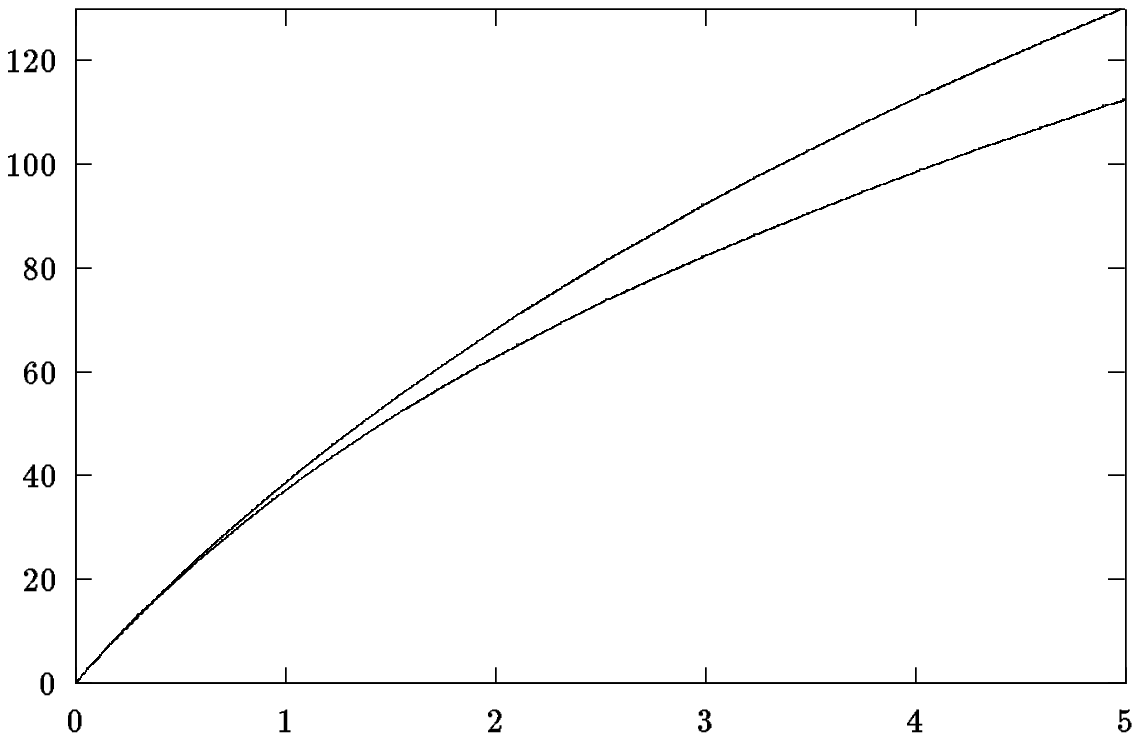}}
    \put(119,0.5){\normalsize$r^2\;[\fm^2]$}
    \put( -6,0  ){\yaxis[66.3mm]{\normalsize$\si\ellp^{\rm tot}(\zet,r^2)\;[\mbarn[]]$}}
    \put(52,50){\normalsize$\zet \!\equiv\! 0$}
    \put(85,50){\normalsize$\zet \!\equiv\! 1\!/\!2$}
  \end{picture}
  \end{center}
\vspace*{-3ex}
\caption[Dipol-Proton-totaler Wirkungsquerschnitt~\protect$\si\ellp^{\rm tot}(\zet,r^2)$ f"ur~\protect$\zet \!\equiv\! 0 \;{\rm vs.}\; 1\!/\!2$]{
  Totaler Wirkungsquerschnitt~$\si\ellp^{\rm tot}(\zet, r^2)$ f"ur die Streuung eines Colour-Dipols~$\{\zet,\rb{r}\}$, der Einfachheit halber gemittelt "uber s"amtliche Orientierungen~$\th$, als Funktion von~$r^2 \!\equiv\! \rb{r}^2$.   Die obere Kurve bezieht sich auf~$\zet \!=\! 0$, die untere auf~$\zet \!=\! 1\!/\!2$; sie veranschaulichen seine nur marginale $\zet$-Abh"angigkeit.   Bez"uglich der Dipolausdehnung~$r$ verh"alt er sich wie~$\si\ellp^{\rm tot}(\zet, r^2) \!\propto\! r^{2n}$, wobei~$n \!\equiv\! 1$ im Limes verschwindender Ausdehnung, vgl.\@ Ref.~\cite{Kulzinger95}, und leicht abfallend f"ur gr"o"sere Dipole.   Vgl.\@ auch~$\si\ellp^{\rm tot}(\zet\zz\equiv\zz1\!/\!2,r)$ in Abb.~\refg{Fig:si-ellp-tot_vka}.
}
\label{Fig:si-ellp-tot_zet}
\end{minipage}
\end{figure}

In Abbildung~\ref{Fig:si-ellp-tot_zet} ist er~auf\-getragen "uber~$r^2$ f"ur~$\zet \!\equiv\! 0$ und~$\zet \!\equiv\! 1\!/\!2$.
F"ur anwachsende Dipolausdehnung~$r$ w"achst der Dipol-Proton-totale Wirkungsquerschnitt~$\si\ellp^{\rm tot}(\zet, r^2)$ zun"achst quadratisch mit~$r$ bis zu einer Gr"o"se von etwa~$r \!\cong\! 1 - 2\,a$.
Danach steigt er weiter an, jedoch mit langsam kleinerwerdender Potenz.
Dieses stetige Ansteigen ist ganz wesentlich Effekt der Nicht-Lokalit"at des~$C$-Terms des Korrelationstensors~$D \!=\! (D_{\mu\nu\mu'\!\nu'})$.
Der~$N\!C$-Term allein saturiert.
Perturbativer Gluonaustausch tr"agt nur zum~$N\!C$-Term bei.
Hierin genau unterscheidet sich unser Zugang grundlegend von Modellen auf Basis perturbativen Gluonaustauschs.
So saturiert der entsprechende Dipol-Proton-Wirkungsquerschnitt in Ref.~\cite{Nikolaev91} bei etwa dem doppelten Protonradius.

\label{T:slope}Wir schlie"sen den Abschnitt "uber den allgemeinen Zugang des \DREI{M}{S}{V} zu weicher Hochener\-giestreuung mit der Diskussion eines weiteren Charaktristikums, in dem er sich grundlegend unterscheidet von perturbativen Modellen auf Basis lokaler Quark-Quark-Wechselwirkung mit inh"arenter Quark-Additivit"at.
Wir betrachten den \vspace*{-.375ex}Abfall der $T$-Amplituden~$T\ellp$,~$\pT\ellp$ f"ur Loop-Proton-Streuung mit~\mbox{$\tfbQ \!\cong\! -t$}, vgl.\@ Gl.~(\ref{tfbB_-t}).
Es gilt\,~$T\ellp,\pT\ellp \!\propto\! \exp -B_0\tfbQ /\!2$\, f"ur kleine~$\tfbB$, vgl.\@ Gl.~(\ref{slope-Parameter_Thh}).
In Ref.~\cite{Dosch94a} wird der slope-Parameter~$B_0$ in folgender Weise numerisch parametrisiert:%
\FOOT{
  In Ref.~\cite{Dosch94a}, Gl.~(76)~-- vgl.\@ dort besser Gl.~(97)~-- ist bez"uglich des konstanten Terms ein Schreibfehler~un\-terlaufen, der sich in unsere Ver"offentlichung Ref.~\cite{Dosch96} "ubertragen hat.
Weiter arbeiten wir mit den~\mbox{rmsq-Ra}\-dien~$R_\snbr{\,\pm}$, die um~$\sqrt{3}\!\big/\!2 \!\cong\! 0.866$ kleiner sind als die Ausdehnungsparameter dort:~$R_\snbr{\,\pm} \!=\! \sqrt{3}\!\big/\!2\; R_\snbr{\,\pm}{}^{\mskip-5mu\mbox{\scriptsize Ref.\zz\cite{Dosch94a}}}$.
}
%
\vspace*{-.5ex}
\begin{align} \label{slope_R1,R2}
B_0\; \cong\; 1.858\, a^2\; +\; 0.244\; (R_\snbr{+}^{\,2} + R_\snbr{-}^{\,2})
  \\[-4.5ex]\nn
\end{align}
Der konstante Term tr"agt mit~$\cong\! 5.71\GeV^{-2}$ ungef"ahr die H"alfte bei zu typischen Zahlenwerten~$B_0 \!\cong\! 11.5\GeV^{-2}$ f"ur den slope-Parameter, vgl.\@ Gl.~(\ref{B_pp-ZW}).
In einem perturbativen Modell auf Basis von Quarkadditivit"at kann unter bestimmten Bedingungen auch ein konstanter Term generiert werden.
Allerdings kommt dieser zustande als Quark-Formfaktor und ist daher wesentlich kleiner als der zweite Term~-- typischerweise um das Verh"altnis der elektromagnetischen Ladungsradien von Quark und Hadronen im Quadrat:~$(R^{\rm ch.}_q \!/\! R^{\rm ch.}_\snbr{\,\pm})^2$.
Vgl.\@ die Diskussion des {\it modifizierten Chou-Yang-Modells\/} in Ref.~\cite{Dosch94a}.

\section{Hadronische (Lichtkegel-)Wellenfunktionen}

In die $T$-Amplitude f"ur Hadron-Hadron-Streuung~$T\hh^{(s,t)}$ gehen wesentlich ein die hadronischen Quark-Antiquark-(Lichtkegel-)Wellenfunktionen der involvierten Zust"anden.
Diese treten neben die im letzten Abschnitt diskutierte \DREI[]{M}{S}{V}-spezifische Funktion~$\tTll^{(s,\rm{b})}$, die bez"uglich~$\tfb$ Fourier-transformierte $T$-Amplitude f"ur die Streuung zweier Wegner-Wilson-Loops, das hei"st zweier Colour-Dipole.
Der "Ubergang zu der $T$-Amplitude~$T\ellp^{(s,t)}$ f"ur die Loop-Proton-Streuung schlie"st bereits ein das Absolutquadrat der Proton-Wellenfunktion.

Wenig ist bekannt "uber die wirkliche Dynamik von Hadronen, die diesbez"ugliche Unsicherheit selbst auf qualitativem Niveau ist gro"s.
Es ist sicher eine Hauptaufgabe gegenw"artiger wie zuk"unftiger Experimente, grundlegende Tatsachen hadronischer Wellenfunktionen in Daten zu dokumentieren und so beizutragen, die Zusammenh"ange zu erhellen, die wesentlich r"uhren an das Verhalten der Quantenchromodynamik auf gro"sen L"angenskalen.
Wir zeigen, da"s elastische Photo- und Leptoproduktion von Vektormesonen hochsensitiv ist auf bestimmte Aspekte und Details der involvierten hadronischen Wellenfunktionen.
Ihre experimentelle wie theoretische Untersuchung verspricht daher, einen wichtigen Beitrag zu leisten zur Beantwortung grundlegender Fragen bez"uglich der Dynamik von Hadronen.

Gegenw"artig m"ussen aber teils erhebliche Annahmen und Approximationen gemacht werden.
Ziel ist, auf die Hadron-Hadron-Amplitude~$T\hh$, m"oglichst wenig aufgeweicht durch komplizierte Wellenfunktionen, den \DREI[]{M}{S}{V}-spezifischen String-String-Mechanismus zu projizieren, der inh"arent ist den Amplituden~$\tTll$,~$T\ellp$ und so zu verifizier- wie falsifizierbaren Postulaten dieses Mechanismus, zu gelangen.
Wesentliche Charakteristika der dynamischer Hadronen sollen zug"anglich gemacht werden durch Formulierung einfacher Ans"atze f"ur deren Wellenfunktionen mit m"oglichst wenigen sensitiven Parametern.

Nachdem in Ref.~\cite{Kraemer91} die hohe Sensitivit"at der \DREI[]{M}{S}{V}-Amplitude~$\tTll$ auf die transversalen Hadronausdehnungen festgestellt ist, wird in diesem Sinne in Ref.~\cite{Dosch94a} f"ur die transversalen Wellenfunktionen einer Vielzahl von Hadronen~$h$ der einfache Ansatz gemacht:
\vspace*{-.5ex}
\bea \label{transversaleWfn-allg}
\ps_h(r)\;
  =\; 2\, \om_h\; \efn{\D -\om_h^2 r^2\!/2}
    \\[-4.5ex]\nn
\eea
Die Oszillatorparameter~$\om_h$ sind korreliert mit den transversalen rmsq-Radien~$R_h$, die berechnet werden aus dem des Protons durch Multiplikation mit dem entsprechenden experimentellen Wert f"ur das Verh"altnis der elektromagnetischen Ladungsradien, etwa f"ur das \mbox{$\pi$-Meson}:~$R_\pi \!\cong\! 0.766\, R_p$, vgl.\@ die Diskussion in Anschlu"s an Gl.\,(\ref{Proton-wfn2_vph}), bzgl.~$R_p$ vgl.\,(\ref{Parameter}).
Dies f"uhrt auf~$\si^{\rm tot}_{{\pi}p} \!/\! \si^{\rm tot}_{pp} \!\cong\! 0.66$ f"ur das Verh"altnis der totalen Wirkungsquerschnitte~-- aufgrund des~Ver\-h"alnisses der Hadronradien, nicht aufgrund Quarkadditivit"at. \\
\indent
Die Abh"angigkeit von~$\zet$, dem Anteil des Quarks am gesamten Lichtkegelimpuls, wurde in Ref.~\cite{Dosch94a} nicht betrachtet.
Auf Basis des angegebenen Zahlenwertes f"ur~$R_\pi$ haben wir die Berechnung wiederholt und "ahnliche Resultate gefunden f"ur eine $\pi$-Wellenfunktion mit "`vern"unftiger"' $\zet$-Abh"angigkeit.
Die Hadron-Hadron-Amplitude bleibt also wesentlich determiniert durch die transversalen Ausdehnungen der streuenden Hadronen.
Wir behalten daher bei f"ur das Absolutquadrat der Proton-Wellenfunktion den einfachen Ansatz gegeben in den Gln.~(\ref{Proton-wfn2_vph}),~(\ref{Proton-wfn2_ps}), konform mit der transversalen Wellenfunktion nach Gl.~(\ref{transversaleWfn-allg}).

Aufgrund der Differenzierung verschiedener Helizit"atszust"ande ist die Photon-/Vektor\-meson-Seite der Streuung sehr viel sensitiver auf~$\zet$.
Wir formulieren im folgenden die Lichtkegelwellenfunktionen von Photon und Vektormeson und diskutieren, wie deren $\zet$-Abh"angigkeit teils erheblich mit Einflu"s nimmt auf die effektive transversale Ausdehnung~-- und so auf die Hadron-Hadron-Am\-plitude~$T\hh$ und die postulierten Observablen.

\vspace*{-1ex}
\bigskip\noindent
Streuung bei hohen Energien impliziert, da"s die einlaufenden, in unserem Fall elastischer Vektormesonproduktion auch die auslaufenden Teilchen auf Trajektorien nahe des Lichtkegels propagieren.
Es liegt nahe, ihre Wellenfunktionen zu konstruieren als {\it Lichtkegelwellenfunktionen\/}; seien im Bild von Valenzquark-Konstituenten mit~$\ps$ bezeichnet ihre Quark-Antiquark-Lichtkegelwellenfunktionen.
Die $T$-Amplitude~$T\hh$ f"ur Hadron-Hadron-Streuung folgt dann allgemein in der Gestalt der Gln.~(\ref{Thh_Tellp}),~(\ref{Tellp_tTll}) mit~$\tTll$ der zun"achst nicht n"aher bestimmten Amplitude f"ur die Streuung der beiden Quark-Antiquark-(Colour)Dipole, vgl.\@ neben Kapitel~\ref{Kap:ANALYT} bzw.\@ Ref.~\cite{Nachtmann91} auch Ref.~\cite{Gunion77}.

Die Funktionen~$\ps(\zet,\rb{r})$ in ihrer Eigenschaft als Quark-Antiquark-Lichtkegelwellenfunk\-tionen sind folgenderma"sen aufzufassen, vgl.\@ Kapitel~\ref{Subsect:HadronniveauI}:
Sie sind zun"achst definiert im Impulsraum,~$\tilde{\ps}(\zet,\rb{k})$ stellt die Wahrscheinlichkeitsamplitude daf"ur dar, in einem Hadron mit Impuls~\mbox{\,$\{P^+,\rb{P}\}$}%
\FOOT{
  \label{FN:LC-Impuls}Bzgl.\@ der Lichtkegelkoordinaten, vgl.\@ Anh.~\ref{APP:LC-Koord}; hier kurz:~$P^\pm \!=\! \al\, (P^0 \!\pm\! P^3)$ und~$\rb{P} \!=\! (P^1,P^2)^t$.   Ein Teilchenimpuls ist vollst"andig bestimmt durch das Paar~$\{P^+,\rb{P}\}$, aufgrund~$P^+P^- \!=\! \al^2 (M^2 \!+\! \rb{P}^2) \!=\! \al^2 \rb{M}^2$, mit~$M$ und~$\rb{M}$ der Masse bzw.\@ transversalen Masse des Teilchens.   Sei ab sofort~$\al \!=\! 1\!/\!\surd2$, vgl.\@ Anh.~\ref{APPSect:Photon-Wfn}.
}
und wohldefiniertem Drehimpuls- und Flavour-Gehalt ein Quark und Antiquark anzutreffen mit Impulsen~\mbox{$\{\zet P^+,\rb{k} \!+\! \zet\rb{P}\}$} beziehungsweise~\mbox{$\{\bzet P^+,-\rb{k} \!+\! \bzet\rb{P}\}$}, mit~$\bzet \!=\! 1 \!-\! \zet$.
Eine wesentliche Eigenschaft ist ihre Abh"angigkeit allein von den Variablen~$\zet$ und~$\rb{k}$, vgl.\@ die Refn.~\cite{Bjorken71,Brodsky81}.
Fourier-transformiert in den transversalen Ortsraum ist~$\rb{r}$ der zu~$k$ konjugierte Vektor von Quark zu Antiquark. 
\vspace*{-1ex}

\subsection{Lichtkegelwellenfunktion des Photons}
\label{Subsect:Photon-Wfn}

Die Quark-Antiquark-Wellenfunktion des Photons~$\ps$ h"angt ab von dessen Virtualit"at~$Q$ und Helizit"at~$\la$:~$\ps(\zet,\rb{r}) \!\equiv\! \ps_{\iga(Q^2,\la)}(\zet,\rb{r})$.
Sie ist die Wahrscheinlichkeitsamplitude daf"ur, im Photon ein Quark-Antiquark-(Colour)Dipol anzutreffen mit Anteilen am Gesamt-Lichtkegel\-impuls~$(\zet,\bzet)$ und transversaler Separation~\mbox{\,$\rb{r} \!=\! (r \cos\vph, r \sin\vph)^t$}.%
\FOOT{
  Es ist~$\vph \!\equiv\! \vph_\snbr{+}$, und~$\vph_\snbr{\,\pm} \!=\! \th_\snbr{\,\pm} \!+\! \vph_\rb{b}$ in Zusammenhang mit Gl.~(\ref{r_+-}) und Fu"snote~\FN{FN:Impakt-Betrag}.
}
Quark und Antiquark sind definierte Zust"ande bez"uglich Flavour~$(f,\bar{f})$ und Helizit"aten~$(h,\bar{h})$ [bzw.\@ Spins~$(s,\bar s)$]; ihr Colour-Gehalt induziert den globalen Faktor~$\surd\Nc$, vgl.\@ unten Fu"snote~\FN{FN:Wfns_Colour}.
Das Photon koppelt an die elektrische Ladung wie~$e_f \de_{\!f\!\bar f} \!\equiv\! e_f {\T\sum}_{f_i}\, \de_{\!f\!,f_i} \de_{\!\bar{f}\!,\bar{f}_i}$, mit der Summe "uber die Flavour~$f_i \!\equiv\!$~{\it down(d),\, up(u),\, strange(s),\, charm(c),\, bottom(b),\, top(t)\/} und~$e_f \!=\! +2\!/\!3 e$ und~$-1\!/\!3 e$ deren Ladung.
Die explizite Raum-Konfiguration der Photon-Lichtkegelwellenfunktion, in Termen der Variablen~$\zet$ und~$\rb{r}$ h"angt ab von den Quark-Helizit"aten~$h$,~$\bar{h}$ und der Photon-Helizit"at~$\la$.
Wir berechnen sie im Rahmen von Lichtkegelst"orungstheorie (light cone perturbation theory, LCPT) zu niedrigster Ordnung in Anhang~\ref{APPSect:Photon-Wfn}:
\bea \label{Photon-Wfn}
\ps_{\iga(Q^2,\la)}^{h,\bar h}(\zet,\rb{r})\;
  =\; \surd\Nc\vv e_f \de_{\!f\!\bar f}\vv \ch_{\iga(Q^2,\la)}^{h,\bar h}(\zet,\rb{r})
\eea
wobei%
\FOOT{
  \label{FN:ArgumenteIndizes}Seien Indizes und Argumente der Funktionen nur soweit explizit notiert, wie der Zusammenhang fordert.
}
%
\vspace*{-1.25ex}
\begin{alignat}{2}
&\hspace*{0em}
 \ch_{\iga(Q^2,\la\equiv0)}&\;
  &=\; -\, 2 \zbz\vv Q\; \de_{h,-\bar h}\vv
             \frac{{\rm K}_0(\vep r)}{2\pi}
    \tag{\ref{Photon-Wfn}$'$} \\
&\hspace*{0em}
 \ch_{\iga(Q^2,\la\equiv+1)}&\;
  &=\; \surd2\, \bigg[\,
         \iIM\,\vep \, \efn{\T +\iIM\,\vph}\;
           \big( \zet\, \de_{h+,\bar h-} - \bzet\, \de_{h-,\bar h+} \big)
           \frac{{\rm K}_1(\vep r)}{2\pi}\;
     +\; m_f\; \de_{h+,\bar h+}\,
           \frac{{\rm K}_0(\vep r)}{2\pi}\,
                \bigg] \nn \\
&\hspace*{0em}
 \ch_{\iga(Q^2,\la\equiv-1)}&\;
  &=\; \surd2\, \bigg[\,
         \iIM\,\vep \, \efn{\T -\iIM\,\vph}\;
           \big( \bzet\, \de_{h+,\bar h-} - \zet\, \de_{h-,\bar h+} \big)
           \frac{{\rm K}_1(\vep r)}{2\pi}\;
     +\; m_f\; \de_{h-,\bar h-}\,
           \frac{{\rm K}_0(\vep r)}{2\pi}\,
                \bigg] \nn
\end{alignat}
mit abk"urzend~$\de_{h+,\bar h-} \!\equiv\! \de_{h,+1\!/\!2} \de_{\bar h,-1\!/\!2}$,\ldots\,
Es steht~$m_f$ f"ur die laufende Quarkmasse mit Flavour~$f$, bzgl.\@ Zahlenwerten vgl.\@ unten Tabelle~\ref{Tabl:Charakt_rh,om,ph,Jps}, Fu"snote~\mbox{\FN{FN:LaufendeQuarkmassen}}; und es ist~$\vep$ definiert durch:
\bea \label{epsilon}
\vep = \sqrt{\zbz\, Q^2 + m_f^2}
\eea
Die Funktionen~${\rm K}_0$,~${\rm K}_1$ sind die vertrauten modifizierten Besselfunktioinen zweiter Art in der Definition von Ref.~\cite{Abramowitz84}.

Diese Funktionen~${\rm K}_0$,~${\rm K}_1$ fallen exponentiell ab mit ihrem Argument~$\vep r$.
Dies induziert ein subtiles Wechselspiel der Variablen~$\zet$ und~$Q$, abh"angig von der Quarkmasse~$m_f$, vgl.\@ Gl.~(\ref{epsilon}), das bestimmt, bis zu welcher Ausdehnung:~$r \klgl \vep^{-1}$, (Colour-)Dipole effektiv zu den Funktionsausdr"ucken~${\rm K}_0(\vep r)$,~${\rm K}_1(\vep r)$ beitragen.
Die Wellenfunktionen~$\ch_{\iga(Q^2,\la)}$ wiederum, vgl.~(\ref{Photon-Wfn}$'$), sind komplizierte zus"atzlich subtil von $\zet$ und $Q$ abh"angende Funktionale dieser ${\rm K}_0$-,${\rm K}_1$-Ausdr"ucke~-- und wesentlich verschieden f"ur longitudinale und transversale Polarisation des Photons,~$\la \!\equiv\! 0$ beziehungsweise~$\la \!\equiv\! \pm1$.
Im Resultat tragen zur transversalen Ausdehnung von longitudinal und transversal polarisiertem Photon wesentlich bei (Colour)Dipole verschiedenster transversaler Ausdehnungen~-- je nach Virtualit"at~$Q$ und je nach dem, ob mittlere Werte von~$\zet$ betrachtet werden oder $\zet$-Werte nahe der Intervall-Endpunkte~$0$ und~$1$.

\label{T:Photon-Wfn}Die longitudinale Wellenfunktion~$\ch_{\iga(Q^2,\la\equiv0)}$ ist gepeakt um~$\zet \!=\! 1\!/\!2$, so da"s sich das longitudinale Photon f"ur gro"ses~$Q^2$ wie ein kleiner Dipol der Ausdehnung~$r \!\sim\! 1\!/\!Q$ verh"alt.
Dazu in Kontrast sind die transversalen Wellenfunktionen~$\ch_{\iga(Q^2,\la\equiv\pm1)}$ fast flach bez"uglich~$\zet$ und die relevante Gr"o"se nicht~$Q^2$ sondern~$Q_{\rm eff}^2 \!\equiv\! 4\zbz Q^2$, so da"s sich ein transversales Photonen f"ur gro"ses~$Q^2$ komplizierter verh"alt:
Teils, n"amlich f"ur mittlere~$\zet$, wechselwirkt es als kleiner Dipol mit einer Ausdehnung von~$r \klgl 2\!/\!Q$.
Teils, n"amlich f"ur~$\zet \!\cong\! m_f\!/\!Q$ nahe dem Intervall-Endpunkt,~-- falls leichte Quarks mit Flavour~\mbox{\it d, u, s\/}~in\-volviert sind~-- als gro"ser Dipol mit einer Ausdehnung von bis zu~$r \!\sim\! 1\!/\!\sqrt{m_f Q}$; dagegen f"ur schwere Quarks mit Flavour \mbox{\it c, b, t\/} begrenzt allgemein die inverse Quarkmasse die Ausdehnung des Photons.\nopagebreak

Die Photon-Wellenfunktion~$\ps_{\iga(Q^2,\la)}$, vgl.\@ die Gln.~(\ref{Photon-Wfn}),~(\ref{Photon-Wfn}$'$), ist perturbativ konstruiert, nichtperturbativen Effekten wie {\it Chirale Symmetriebrechung\/} oder {\it Confinement\/} tr"agt sie nicht Rechnung.
Mit kleinerwerdendem~$Q^2$ werden die Ausdehnungen der wesentlich beitragenden Dipole, insbesondere f"ur transversale Polarisation des Photons, schnell gr"o"ser und "ubersteigen typische Hadronausdehnungen, dies widerspricht der wirklichen Physik.
In Photo- und Leptoproduktion von Vektormesonen wird die Ausdehnung der effektiv beitragenden Dipole bestimmt durch den "Uberlapp der Wellenfunktionen von Photon und Vektormeson.
Die Form der Wellenfunktionen der in diesem Kapitel diskutierten~$1S$-Vektormesonen unterdr"uckt den Bereich der $\zet$-Endpunkte und damit die gro"sen Dipole transversal polarisierter Dipole.
Auf diese Weise testen Photon-Virtualit"aten im Bereich~$Q^2 \grgl 1 \!-\! 2\GeV^2$ effektiv Dipole von transversaler Ausdehnung bis ungef"ahr~$1\fm$.
Dies l"a"st erwarten, da"s unsere Photon-Wellenfunktion in diesem~$Q^2$-Bereich verl"a"slich die f"ur uns relevanten Aspekte der wirklichen Photon-Wellenfunktion approximiert.
Die gute "Ubereinstimmung unseres Postulats mit den experimentellen Daten f"ur das Verh"altnis Longitudinaler zu transversaler Produktion des $\rh$-Mesons untermauert dies a~posteriori, vgl.\@ unten Abbildungen~\refg{Fig-G:R_LT} und mehr noch Abbildung~\refg{Fig:R_LT}. \\
Unser Zugang im Rahmen des \DREI{M}{S}{V} ist nichtperturbativ, er ist ad"aquater Zugang zu Effekten auf L"angenskalen typischer Hadronausdehnungen; Dominanz wesentlich kleinerer Dipole fordert, den Austausch perturbativer Gluonen miteinzubeziehen.
Unser Interesse gilt vornehmlich kleinen Virtualit"aten, so da"s wir statt dessen den betrachteten $Q^2$-Bereich auch nach oben einschr"anken, in diesem Kapitel konservativ auf~$1 \!-\! 2$ bis~$10\GeV^2$.

Es zeigt sich a~posteriori G"ultigkeit der in den Gln.~(\ref{Photon-Wfn}),~(\ref{Photon-Wfn}$'$) angegebene Photon-Wellenfunktion~$\ps_{\iga(Q^2,\la)}$ auch f"ur h"ohere~$Q^2$; wir geben daher im folgenden Kapitel Zahlenwerte bis~$Q^2 \!\cong\! 20\GeV^2$ an.
Wir modifizieren dort die Wellenfunktion~$\ps_{\iga(Q^2,\la)}$ in der Weise, da"s sich ihre G"ultigkeit erweitert hin zu kleineren bis verschwindenden Virtualit"aten.
Dies setzt uns in die Lage, auch ausgedehntere angeregte Vektormesonen zu diskutieren.
So testet die Produktion des~$2S$-Zustandes durch transversal polarisierte Photonen Dipole von transversaler Ausdehnung bis ungef"ahr~$2\fm$ f"ur~$Q^2 \!\equiv\! 1\GeV^2$ und~$2.5 \!-\! 2.8\fm$ f"ur~$Q^2 \!\equiv\! 0$.

\subsection[Lichtkegelwellenfunktionen der \protect$1S$-Vektormesonen
          ]{Lichtkegelwellenfunktionen der \protect\bm{1S}-Vektormesonen}
\label{Subsect:1S-Vektormeson-Wfn}

Die Lichtkegelwellenfunktionen der~$1S$-Vektormesonen seien definiert in Notation analog zu der des Photons und zun"achst separiert ihr Colour- und Flavour-Gehalt.
Dies sind der globale Faktor~$1\!/\!\surd\Nc$ bez"uglich Colour;%
\FOOT{
  \label{FN:Wfns_Colour}Die Colour-Freiheitsgrade sind bereits vollst"andig absorbiert in den Wegner-Wilson-Loops durch die Normierung der Spuren gem"a"s~$\trDrst{F} \!=\! 1\!/\!\dimDrst{F} \tr \!=\! 1\!/\!\Nc \tr$.   Unsere Konvention f"ur die Lichtkegelwellenfunktion, vgl.\@ Gl.~(\ref{Photon-Wfn}), schlie"st einen Faktor~$\surd\Nc$ ein, dieser ist zu k"urzen durch einen Faktor~$1\!/\!\surd\Nc$ im Vektormeson.
}
bez"uglich ihres Flavour-Gehaltes fassen wir~$\rh(770)$ und~$\om(782)$ auf als reinen Isospin-1- beziehungsweise Isospin-0-Zustand und~$\ph(1020)$ und $\Jps(3097)$ als reine Quarkonia.
Wir schreiben analog zu~Gl.~(\ref{Photon-Wfn}):
\vspace*{-.5ex}
\begin{align} \label{Vektormeson-Wfn}
\ps_{V(\la)}^{h,\bar h}(\zet,\rb{r})\;
  =\; 1\!/\!\surd\Nc\vv i_{V,f}\vv \ch_{V(\la)}^{h,\bar h}(\zet,\rb{r})
    \\[-4.5ex]\nn
\end{align}
mit symbolisch~-- f"ur~\mbox{\,$\rh \!=\! 1^+(1^{--})$} und~\mbox{\,$\om,\ph,\Jps \!=\! 0^-(1^{--})$}%
\FOOT{
  \label{FN:Parity}Notation~$I^G(J^{P\!C})$ nach Ref.~\cite{PDG00}.   F"ur einen~$q\bar{q}$-Zustand~$\ket{N^{2S+1}L_J}$ gilt, mit~\mbox{$S \!=\! 0,1$}:~\mbox{$G \!=\! (-1)^{L+S+I}$}, \mbox{~$P \!=\! (-1)^{L+1}$}, \mbox{$C \!=\! (-1)^{L+S}$}.   F"ur~$L \!=\! 0$ sind~$J^P \!=\! 0^-$ Pseudoskalare,~$1^-$ Vektoren.
}~--:
%
\vspace*{-.5ex}
\bea \label{Flavour-Gehalt}
i_{\rh,f} = (u\bar{u} \!-\! d\bar{d})/\surd2 \qquad
i_{\om,f} = (u\bar{u} \!+\! d\bar{d})/\surd2 \qquad
i_{\ph,f} = s\bar{s} \qquad
i_{\Jps,f} = c\bar{c}
  \\[-4.5ex]\nn
\eea
f"ur~$i_{\rh,f} \!=\! (u\bar{u} \!-\! d\bar{d})\!/\!\surd2 \!\equiv\! (\de_{\!f\!,u} \de_{\!\bar{f}\!,\bar{u}} \!-\! \de_{\!f\!,d} \de_{\!\bar{f}\!,\bar{d}})\!/\!\surd2$,\ldots mit%
~${\T\sum}_f\, i_{V,f}\, i_{V,f} \!=\! 1$,%
~${\T\sum}_f\, e_f \de_{\!f\!\bar f}\, i_{V,f} \!=\! e_V \!\equiv\! e\hat{e}_V$; dabei sind~$e_V$,~$\hat{e}_V$ die effektive Quark-Ladung im Vektormeson~$V$ in nat"urlichen beziehungsweise in Einheiten der Proton-Ladung~$e$; bzgl.\@ expliziter Werte von~$\hat{e}_V$ vgl.\@ Tabelle~\ref{Tabl:Charakt_rh,om,ph,Jps}. \\
\indent
Die Abh"angigkeit von der Raum-Konfiguration ist subsumiert in den Funktionen~$\ch_{V(\la)}$, die diese beschreiben in Termen der Variablen~$\zet$ und~$\rb{r}$, je nach Helizit"aten~$(h,\bar h)$ der Quarks und Helizit"at~$\la$ des Vektormesons.
Vgl.\@ die Photon-Funktionen~$\ch_{\iga(Q^2,\la)}$, Gl.~(\ref{Photon-Wfn}$'$). \\
%
\indent
Bez"uglich dieser Funktionen~$\ch_{V(\la)}$, allgemein der Raum-Konfiguration der Quark-Anti\-quark-Lichtkegelwellenfunktion eines (Vektor)Mesons, existiert nur sehr punktuelle Information.
Von dem Standpunkt der Theorie her ist gegenw"artig selbst die Form des Lichtkegel-Hamiltonoperators f"ur Zust"ande aus Valenzquark-Konstituenten nicht bekannt.
Theoretische Untersuchungen stehen erst am Anfang.
So werden Versuche unternommen, Lichtkegelwellenfunktionen f"ur Vektormesonen zu konstruieren als die {\it Melosh-Transformierten\/} von L"osungen eines relativistisch erweiterten Konstituentenquark-Modells, vgl.\@ Ref.~\cite{Cardelli95}.
Es wird eine String-Gleichung f"ur ein Meson auf dem Lichtkegel gel"ost, vgl.\@ Ref.~\cite{Morgunov97}.
Doch liefern diese "Uberlegungen (noch) keine L"osungen f"ur~$\ch_{V(\la)}$ in parametrisierter Form.
Wir haben im Einleitungskapitel auf Seite~\pageref{T:Konstituentenquarks} geschildert, wie theoretisch komplex die Formulierung des Konstituentenbildes von {\it first principles\/} ist.
Sie geschieht meist zun"achst f"ur QED, eine Verallgemeinerung auf QCD ist schwierig; vgl.\@ den bereits zitierten Review von Lavelle und McMullan, Ref.~\cite{Lavelle97}, und die dort angegeben formaleren Referenzen.

Wir werden daher auf Basis der punktuellen Information, {\it die\/} existiert, {\it ph"anomenologische\/} Ans"atze modellieren analog den Funktionen~$\ch_{\iga(Q^2,\la)}$ des Photons.
Im Bereich der Quantenchromodynamik auf gro"sen L"angenskalen werden Hadronen erfolgreich beschrieben im Rahmen von Quark-Modellen: statische Hadronen durch Gau"s'sche Wellenfunktionen, das hei"st aufgefa"st als Systeme von Konstituentenquarks in einem Harmonischen-Oszillator-Potential.
Die Aufgabe aber, diese nichtrelativistischen Wellenfunktionen in ein schnell-bewegtes System Lorentz-zu tranformieren, ist hochgradig nichttrivial.
Das Zusammenspiel von longitudinaler und transversaler Dynamik in der Physik nahe und auf dem Lichtkegel ist nicht wirklich verstanden, genausowenig, wie die Spin-Freiheitsgrade zu behandeln sind.
Das \DREI{M}{S}{V} andererseits gelangt auf Basis einfacher Gau"s'scher Wellenfunktionen, vgl.\@ Gl.~(\ref{transversaleWfn-allg}) bzw.\@ die Gln.~(\ref{Proton-wfn2_vph}),~(\ref{Proton-wfn2_ps}), in denen Spin- und $\zet$-Abh"angigkeit komplett vernachl"assigt sind, zu einer "uberzeugenden quantitativen Beschreibung von weicher Streuung bei hohen Energien.
Wir halten daher bei Einbeziehung einer Spin- und~$\zet$-Abh"angigkeit fest an dem Bild eines (transversalen) Harmonischen Oszillators.
Im Bereich kleiner L"angenskalen, werden mithilfe perturbativer Quantenchromodynamik unterst"utzt durch Summenregeln Eigenschaften der Wellenfunktionen im Limes verschwindender transversaler Separation~$r$ extrahiert, insbesondere bez"uglich der Endpunkte des $\zet$-Intervalls~$[0,1]$.
In die Streuamplitude f"ur harte exklusive Prozesse geht in guter Approximation nur der Wert der Wellenfunktion am Ursprung ein, vgl.\@ Ref.~\cite{Lepage80}, der innerhalb derselben Approximation gegeben ist durch die leptonische Zerfallsbreite des Mesons.

Vor diesem Hintergrund modellieren wir die Funktionen~$\ch_{V(\la\equiv0)}$; in Anhang~\ref{APPSect:Vektormeson-Wfn} gelangen wir zu, vgl.\@ Fu"snote~\FN{FN:ArgumenteIndizes}:
\vspace*{-.5ex}
\begin{alignat}{2} \label{1S-Vektormeson-Wfn}
&\hspace*{0em}
 \ch_{V(\la\equiv0)}&\;
  &=\; 4 \zbz\vv \om_{V,L}\vv \de_{h,-\bar h}\cdot
             g_{V,L}(r)\; h_{V,L}(\zet)
    \\[.5ex]
&\hspace*{0em}
 \ch_{V(\la\equiv+1)}&\;
  &=\; \Big[\,
         \iIM\, \om_{V,T}^2r\, \efn{\T +\iIM\,\vph}\;
           \big( \zet\, \de_{h+,\bar h-} - \bzet\, \de_{h-,\bar h+} \big)\;
     +\; m_f\; \de_{h+,\bar h+}\,
                \Big]\cdot g_{V,T}(r)\; h_{V,T}(\zet) \nn \\[.5ex]
&\hspace*{0em}
 \ch_{V(\la\equiv-1)}&\;
  &=\; \Big[\,
         \iIM\, \om_{V,T}^2r\, \efn{\T -\iIM\,\vph}\;
           \big( \bzet\, \de_{h+,\bar h-} - \zet\, \de_{h-,\bar h+} \big)\;
     +\; m_f\; \de_{h-,\bar h-}\,
                \Big]\cdot g_{V,T}(r)\; h_{V,T}(\zet) \nn
\end{alignat}
dabei sind definiert f"ur longitudinale und transversale Polarisation,~$\la \!\equiv\! 0,L$ beziehungsweise~$\la \!\equiv\! \pm1,T$, die Funktionen:
\vspace*{-.5ex}
\begin{alignat}{2}
&g_{V,\la}(r)&\;
  &=\;   \exp \bigg[ -\frac{1}{2}\, \om_{V,\la}^2\, r^2 \bigg]
    \label{g-Wfn} \\
&h_{V,\la}(\zet)&\;
  &=\; {\cal N}_{V,\la}\vv \sqrt{\zbz}\vv
         \exp \bigg[ -\frac{1}{2}\,
                     \frac{M_V^2 (\zet \!-\! 1\!/\!2)^2}{\om_{V,\la}^2} \bigg]
    \label{h-Wfn}
    \\[-4.5ex]\nn
\end{alignat}
Diese Darstellung h"alt sich an unsere Notation von Ref.~\cite{Kulzinger98}.
Sie differiert nur in dem einen Punkt, da"s in Hinblick auf Systematik mit den Photon-Funktionen~$\ch_{\iga(Q^2,\la)}$ auch in der Definition der Vektormeson-Funktionen~$\ch_{V(\la)}$ der Colour-Gehalt~-- in Form des Faktors~$1\!/\!\surd\Nc$~-- separiert ist, vgl. die Gln.~(\ref{Photon-Wfn}),(\ref{Vektormeson-Wfn}); die dimensonslosen Normierungskonstanten~${\cal N}_{V,\la}$ in den Funktionen~$h_{V,\la}(z)$ nach Gl.~(\ref{h-Wfn}) sind so um einen Faktor~${\cal N}_{V,\la}$,~$\Nc \!\equiv\! 3$, gr"o"ser als die entsprechenden Konstanten dort.

Die Argumentation hin zu diesem Ansatz ist angegeben im Zuge seiner Herleitung in Anhang~\ref{APPSect:Vektormeson-Wfn}.
Wir merken hier nur folgendes an:
Lichtkegelst"orungstheorie~(LCPT) liefert zun"achst Ausdr"ucke f"ur die Photon-Funktionen~$\tilde{\ch}_{\iga(Q^2,\la)}(\zet,\rb{k})$; die Funktionen~$\ch_{\iga(Q^2,\la)}(\zet,\rb{r})$ in Gl.~(\ref{Photon-Wfn}) sind definiert als deren Fourier-Transformierte bez"uglich~$\rb{k}$.
Die Funktionen~$\tilde{\ch}_{\iga(Q^2,\la)}$ faktorisieren in einen Term, dessen $\zet$- und~$\rb{k}$-Abh"angigkeit in komplizierter Weise abh"angt von dem Paar~$(h,\bar h)$ der Quark-Helizit"aten, und in einen zweiten Term, den {\it Energienenner des Photons\/}~\mbox{$1/\!(\vep^2 \!+\! \rb{k}^2)$}, mit~$\vep$ abh"angig von~$\zet$ nach Gl.~(\ref{epsilon}).
Der erste Term kann aufgefa"st werden als allgemeine Konsequenz der Quark-Antiquark-Valenz-Struktur, der zweite Term als Photon-spezifisch.
Als Ansatz f"ur die Vektormeson-Funktionen~$\tilde{\ch}_{V(\la)}(\zet,\rb{k})$ wird der erste Term im wesentlichen unver"andert "ubernommen, der zweite, der Photon-Energienenner, wird ersetzt durch eine Funktion, die analog abh"angt von~$\zet$ und~$k \!\equiv\! |\rb{k}|$ und die das spezifische Vektormeson~$V$ modelliert.
F"ur diese wird angesetzt~-- bis auf konstante Faktoren:
\bea \label{Ansatz-k_g,h-Wfn}
\tilde{g}_{V,\la}(k)\cdot h_{V,\la}(\zet)\;
  \equiv\;
  2\pi\, \om_{V,\la}^{-2}\vv
  g_{V,\la}\big(\om_{V,\la}^{-2}k\big)\cdot h_{V,\la}(\zet)
\eea
Die so konstruierten Funktionen~$\tilde{\ch}_{V(\la)}$ ergeben Fourier-transformiert bez"uglich~$\rb{k}$ die Funktionen~$\ch_{V(\la)}(\zet,\rb{r})$ von Gl.~(\ref{Vektormeson-Wfn}).

Die explizite funktionale Form von~$h_{V,\la}$,~$g_{V,\la}$ nach den Gln.~(\ref{h-Wfn}),~(\ref{g-Wfn}) wird vorgeschlagen von Wirbel, Stech, Bauer in Ref.~\cite{Wirbel85} in Zusammenhang exklusiver semileptonischer Zerf"alle der neutralen ~$D$- und $B$-Mesonen.
Die~$r$-Abh"angigkeit ist parametrisiert durch die Funktionen~$g_{V,\la}(r)$, das hei"st das Quark-Antiquark-System des Vektormesons wird aufgefa"st als konfiniert in einem Harmonischen-Oszillator-Potential mit Parameter~$\om_{V,\la}$.
Mit diesem unmittelbar verkn"upft "uber~$R_{V,\la} \!=\! \vev{r^2}^{1/2} \!=\! 1/\!2\om_{V,\la}$ und weiter "uber~$R_{V,\la;3} \!=\! \sqrt{3\!/\!2} R_{V,\la}$ sind der zweidimensionale beziehungsweise der dreidimensionale rmsq-Radius des Vektormesons mit Polarisation~$\la$.
Diese erwarten wir vergleichbar den entsprechenden elektromagnetischen Ladungsradien.
Elektromagnetische Ladungsradien f"ur Vektormesonen sind allerdings noch nicht experimentell bekannt.
Die~$\zet$-Abh"angigkeit der Vektormeson-Wellenfunktionen ist parametrisiert durch die Funktionen~$h_{V,\la}(z)$.
Diese sind invariant bez"uglich der Vertauschung von Quark und Antiquark:~$\zet \!\leftrightarrow\! \bzet$, vgl.~$(\zet \!-\! 1\!/\!2)^2 \!=\! 1\!/\!4 \!-\! \zbz$, und sind gepeakt f"ur gleichverteilten Lichtkegelimpuls:
global durch den Faktor~$\surd\zbz$ und durch die Exponentialfunktion um so st"arker, je gr"o"ser der Oszillatorparameter~$\om_{V,\la}$, das hei"st je kleiner das Vektormeson ist.

Wir machen f"ur alle betrachteten $1S$-Vektormesonenen {\it universell\/} den Ansatz dieser funktionalen Form.
F"ur ein spezielles Vektormeson~$V$ h"angt er ab, separat f"ur longitudinale und transversale Polarisation, nur von den beiden Parametern~$\om_{V,\la}$,~${\cal N}_{V,\la}$.
Diese werden durch die Forderungen fixiert, da"s die Wellenfunktionen zum einen normiert sind, und zum anderen den experimentellen Zahlenwert reproduzieren f"ur~$f_V$, die Kopplung von~$V$ an~den elektromagnetischen Strom~$J_{\rm em} \!=\! (e_{\!F}\bar{\ps}_F\ga^\mu\ps_F)$, impliziert Summation "uber alle (geladenen) Fermionen~$F$; vgl.\@ Anh.~\ref{APPSubsect:1S-Vektormeson-Wfn} auf Seite~\pageref{APP-T:1S-ParameterFix}.
Sei diese Kopplung~$f_V$ definiert durch
\vspace*{-.5ex}
\begin{align} \label{fV-Def}
\bra{\Om}J_{\rm em}^{\mu}(0)\ket{V(q,\la)}\;
  =\; e f_V\, M_V\, \vep^{\mu}(q,\la)
    \\[-3.5ex]\nn
\end{align}
dann ist sie durch
\vspace*{-.5ex}
\bea \label{fV_Gall}
\Gall_V\;
  =\; \frac{4\pi\al_{\rm em}^2}{3} \frac{f_V^2}{M_V} \cdot
        \bigg[ 1 \!+\! \frac{2m_l^2}{M_V^2} \bigg]
         \sqrt{1 \!-\! \frac{4m_l^2}{M_V^2}}
    \\[-3.5ex]\nn
\eea
verkn"upft mit der Breite des elektromagnetischen Zerfalls von~$V$ in das Leptonpaar~$l^+l^-$, mit~$m_l$ der Masse des Leptons.
Unserer Arbeit zugrunde liegen die experimentellen Werte f"ur die Zerfallsbreiten in~$e^+e^-$; aus Gl.~(\ref{fV_Gall}) folgen~-- unter Vernachl"assigung der Elektronmasse~$m_e$ wird der Faktor nach dem Punkt zu Eins~-- Zahlenwerte f"ur~$f_V$; vgl.\@ Tabl.~\ref{Tabl:Charakt_rh,om,ph,Jps}. \\
\indent
\label{T:1S-Parameter}F"ur festes Vektormeson~$V$ und feste Polarisation~$\la \!\equiv\! L,T$ sind die beiden Parameter $\om_{V,\la}$,~${\cal N}_{V,\la}$ zu bestimmen in Abh"angigkeit von~$M_V$,~$f_V$.
Die genannten Forderungen sind formal ein "uber~${\cal N}_{V,\la}$~gekop\-peltes System impliziter Gleichungen von Integralfunktionen in~$\om_{V,\la}$.
Ihre Formulierung und L"osung geschieht in Anhang~\ref{APPSubsect:1S-Vektormeson-Wfn} auf Seite~\pageref{APP-T:1S-ParameterFix}; die expliziten Zahlenwerte sind festgehalten in Tabelle~\ref{Tabl:Charakt_rh,om,ph,Jps}.
\begin{table}
\begin{minipage}{\linewidth}
\renewcommand{\thefootnote}{\thempfootnote}
\begin{center}
  \begin{tabular}{|h{8}||g{7}|g{7}|g{7}|g{7}|} \hline
  \multicolumn{5}{|c|}{Charakteristik der Grundzustand-$1^+(1^{--})$-%
                       ~und -$0^-(1^--)$-Vektormesonen}
    \\ \hhline{:=:t:====:}
  \multicolumn{1}{|c||}{}
    & \multicolumn{1}{c|}{$\rh$}
    & \multicolumn{1}{c|}{$\om$}
    & \multicolumn{1}{c|}{$\ph$}
    & \multicolumn{1}{c|}{$\Jps$}
    \\
    & \multicolumn{1}{c|}{$\!\equiv\! \rh(770)$}
    & \multicolumn{1}{c|}{$\!\equiv\! \om(782)$}
    & \multicolumn{1}{c|}{$\!\equiv\! \ph(1020)$}
    & \multicolumn{1}{c|}{$\!\equiv\! \Jps(3097)$}
    \\ \hhline{|-||----|}
  \mbox{$M_V$}/[\MeV[]]
    &  769.3  ,\PM 0.8
    &  782.57 ,\PM 0.12
    & 1019.417,\PM 0.014
    & 3096.87 ,\PM 0.04  \\
  \mbox{$\Gaee_V$}/[\keV[]]
    & 6.77,\PM 0.32
    & 0.60,\PM 0.02
    & 1.37,\PM 0.06
    & 5.26,\PM 0.37 \\
  \mbox{$f_V$}/[\GeV[]]
    & 0.,152\,6 & 0.,045\,8 & 0.,079\,1 & 0.,270 \\[.5ex]
  \multicolumn{1}{|c||}{Flavour\mbox{\footnote{
    \label{FN:LaufendeQuarkmassen}Als laufende Quarkmassen legen wir zugrunde~\mbox{$m_u \!=\! m_d \!=\! 0$}, \mbox{$m_s \!=\! 0.15\GeV$} und~\mbox{$m_c \!=\! 1.3\GeV$}.
}:\vv$i_{V,f}$}}
    & \multicolumn{1}{c|}{$(u\bar{u} \!-\! d\bar{d})/\surd2$}
    & \multicolumn{1}{c|}{$(u\bar{u} \!+\! d\bar{d})/\surd2$}
    & \multicolumn{1}{c|}{$s\bar{s}$}
    & \multicolumn{1}{c|}{$c\bar{c}$} \\
  \multicolumn{1}{|c||}{$\hat{e}_V$}
    & \multicolumn{1}{c|}{$1/\surd2$}
    & \multicolumn{1}{c|}{$1/(3\surd2)$}
    & \multicolumn{1}{c|}{$1/3$}
    & \multicolumn{1}{c|}{$2/3$} \\[.5ex]
  \mbox{$\om_{V,L}$}/[\GeV[]]
    & 0.,330 & 0.,300 & 0.,368 & 0.,682 \\
  \mbox{${\cal N}_{V,L}$}/\mbox{$[\surd\Nc\!\equiv\!3]$}
    & 4.,48 & 4.,55 & 4.,59 & 5.,13 \\
  \mbox{$R_{V,L}$}/[\fm[]]
    & 0.,299 & 0.,329 & 0.,267 & 0.,144 \\[.5ex]
  \mbox{$\om_{V,T}$}/[\GeV[]]
    & 0.,217 & 0.,212 & 0.,270 & 0.,573 \\
  \mbox{${\cal N}_{V,T}$}/\mbox{$[\surd\Nc\!\equiv\!3]$}
    & 5.,85 & 5.,95 & 4.,80 & 2.,15 \\
  \mbox{$R_{V,T}$}/[\fm[]]
    & 0.,527 & 0.,538 & 0.,439 & 0.,230 \\
  \hhline{:=:b:====:}
  \end{tabular}
  \end{center}
\vspace*{-3ex}
\caption[\protect$1S$-/Grundzustand-Vektormesonen~\protect$\rh(770)$,~\protect$\om(782)$,~\protect$\ph(1020)$,~\protect$\Jps(3097)$: Charakteristik und Parameter]{
  Charakteristik der Vektormesonen~$\rh(770)$,~$\om(782)$,~$\ph(1020)$ und~$\Jps(3097)$.   Die Zahlenwerte f"ur~$M_V$,~$\Gaee_V$ sind i.w.\@ unver"andert in der neuesten Ausgabe der Particle Data Group~\cite{PDG00}, daher dieser entnommen~-- au"ser des zentralen Werts von~$\Gaee_\ph$, der gegenw"artig mit~$1.32\keV$ angegben wird.   Der Effekt auf unsere Resultate ist unwesentlich, genauso das Ersetzen der Werte f"ur~$M_V$ durch die Ordnungszahlen in den runden Klammern.   Die Kopplungen~$f_V$ sind berechnet aus den Zerfallsbreiten~$\Gaee_V$.   Die Normierungskonstanten~${\cal N}_{V,\la}$ sind angegeben~in~Einheiten des Colour-Faktors~$\surd\Nc$,~$\Nc \!\equiv\! 3$.
\vspace*{-1ex}Abweichend von Ref.~\cite{Dosch96}~sind die Radien hier definiert durch~$R_{V,\la} \!\stackrel{\T!}{=}\! \surd\vev{\rb{r}^2}_{V,\la}$, vgl.\@ Anh.~\ref{APPSubsect:1S-Vektormeson-Wfn}, die Gln.~(\ref{APP:rmsq-Radius-1S}),~(\ref{APP:rmsq-Radius-1S}$'$).
\vspace*{-1ex}
}
\label{Tabl:Charakt_rh,om,ph,Jps}
\end{minipage}
\end{table} 
\renewcommand{\thefootnote}{\thechapter.\arabic{footnote}}
\bigskip\noindent
Mit den so definierten Lichtkegelwellenfunktionen f"ur Photon und Vektormesonen sind alle Funktionen angegeben, die eingehen in die $T$-Amplitude~$T\hh$ f"ur die exklusive Leptoproduktion dieser Vektormesonen elastisch am Proton.
Wir stellen die Formeln zusammen, mit denen wir im folgenden Abschnitt Observablen explizit berechnen werden.

Wir rekapitulieren~$T\hh$, Gl.~(\ref{Thh_Tellp}):
\vspace*{-.5ex}
\begin{align} \label{Thh_Tellp-coll}
T\hh\;
  =\; \int \frac{d\zet d^2\rb{r}}{4\pi}\vv
        \ps_V^{\D\dagger} \ps_\iga (\zet, \rb{r})\vv
        T\ellp(\zet, \rb{r}; \tfbB)
    \\[-4.5ex]\nn
\end{align}
mit~$T\ellp$ der $T$-Amplitude der zugrundeliegenden Loop(Dipol)-Proton-Streuung, Gl.~(\ref{Tellp_tTll}).
Wir betrachten~$T\ellp$ als zerlegt in Eigenfunktionen bez"uglich des Drehimpulsoperators $L_3$:
\vspace*{-.5ex}
\begin{align} \label{Tellp-L3-Zerlegung}
T\ellp(\zet,r,\vph,\tfbB)\;
  =\; {\T\sum}_{m\,{\rm gerade}}\vv \efn{\D \iIM m\vph}\vv T\ellp^{(m)}(\zet,r,\tfbB)
    \\[-4.5ex]\nn
\end{align}
Aufgrund der Periodizit"at von~$T\ellp$, vgl.\@ Gl.~(\ref{Symm_tTll}), treten nur gerade~$m$ auf.
F"ur die betrachtete Reaktion impliziert dies, da"s nur "Uberg"ange auftreten eines longitudinal polarisierten Photons zu einem longitudinal polarisierten Vektormeson und eines transversal polarisierten Photon zu einem transversal polarisierten Vektormeson.
Aus der expliziten Gestalt der Photon- und Vektormeson-Wellenfunktionen, vgl.\@ die Gln.~(\ref{Photon-Wfn}$'$),~(\ref{1S-Vektormeson-Wfn}), folgt die M"oglichkeit einer "Anderung der Helizit"at um zwei Einheiten:~$\la \!\equiv\! \pm1 \!\to\! \mp1$.
Wir haben diese "Uberg"ange berechnet und gefunden, da"s sie im gesamten betrachteten $Q^2$-Bereich weniger als~$2\%$ zum jeweiligen Wirkungsquerschnitt beitragen.
Wir vernachl"assigen sie daher.

In die Amplitude~$T\hh$ gehen die Lichtkegelwellenfunktionen von Photon und Vektormeson ein als der "Uberlapp:
\begin{align} 
\overlap{\la}\;
  \equiv\; {\T\sum}_{f,\bar f}\; {\T\sum}_{h,\bar h}\;
             \ps_{\iga(Q^2,\la)}^{h,\bar h\;{\D\dagger}}
             \ps_{V(\la)}^{h,\bar h}\;
  =\; e\hat{e}_V\; {\T\sum}_{h,\bar h}\;
             \ch_{\iga(Q^2,\la)}^{h,\bar h\;{\D\dagger}}
             \ch_{V(\la)}^{h,\bar h}
\end{align}
Die Kontraktion ist zu verstehen f"ur feste Helizit"at~$\la \!\equiv\! 0$,~$+1$ oder~$-1$; sie h"angt nur noch ab von~$\zet$,~$r$ und ist unabh"angig vom Azimutwinkel~$\vph$.
Der "Ubergang zu den Funktionen~$\ch$, vgl.\@ die Gln.~(\ref{Photon-Wfn}),~(\ref{Vektormeson-Wfn}), impliziert die Kontraktion der Flavour-Anteile; vgl.\@ Gl.~(\ref{Flavour-Gehalt}), bzgl.\@ expliziter Werte f"ur~$\hat{e}_V$ vgl.\@ Tabl.~\ref{Tabl:Charakt_rh,om,ph,Jps}.
Kontraktion der Funktionen~$\ch_{\iga(Q^2,\la)}$,~$\ch_{V(\la)}$ in der expliziten Gestalt der Gln.~(\ref{Photon-Wfn}$'$),~(\ref{1S-Vektormeson-Wfn}) f"uhrt auf den "Uberlapp:
\begin{align} \label{"Uberlapp}
&\overlap{\la\equiv L}(\zet,r)
    \\[-.5ex]
  &=\; -\, e \hat{e}_V\; 16 (\zbz)^2\vv
             \om_{V,L}\vv g_{V,L}(r)\; h_{V,L}(\zet)\vv 
             Q\vv \frac{{\rm K}_0(\vep r)}{2\pi}\vv
    \nn \\[1ex]
&\overlap{\la\equiv T}(\zet,r)
    \tag{\ref{"Uberlapp}$'$} \\[-.5ex]
  &=\; \phantom{-\,}
           e \hat{e}_V\; \surd2\vv
             g_{V,T}(r)\; h_{V,T}(\zet)\vv  
             \bigg[\, \om_{V,T}^2 r \vep \big(\zet^2 + \bzet^2 \big)
                        \frac{{\rm K}_1(\vep r)}{2\pi}\;
                  +\; m_f^2\; \frac{{\rm K}_0(\vep r)}{2\pi}\,
             \bigg] \nn
\end{align}
Der "Uberlapp ist identisch f"ur~$\la \!\equiv\! +1$ und $-1$; wir gehen daher "uber von {\it Helizit"aten\/} zu {\it Polarisationen}, mit systematisch~$\la \!\equiv\! L,T$ f"ur longitudinal, transversal. \\
\indent
Die im Quadrat des invarianten Impulstransfers~$-t$ differentiellen Wirkungsquerschnitte f"ur elastische Vektormeson-Leptoproduktion folgen damit zu, vgl.\@ die Gln.~(\ref{slope-Parameter_Thh}),~(\ref{Thh_Tellp-coll}):
\begin{align} \label{dsigmadt-L,T}
\frac{d\si_\la}{dt}(\tfbQ)\;
  =\; \D \frac{1}{16\pi} \frac{1}{s^2}\;
        \left|\; \int_0^{1\!/\!2}\zz d\zet\; \int_0^{\infty}\zz rdr\vv
               \overlap{\la}(\zet,r)\vv \pT\ellp(\zet,r,\tfbB)\;
        \right|^2
\end{align}
f"ur~$\la \!\equiv\! L,T$; bzgl.~$\tfbB \!\cong\! \surd\!-t$ vgl.\@ Gl.~(\ref{tfbB_-t}).
Der "Uberlapp der Wellenfunktionen h"angt~nicht ab von~$\vph$; dessen Integration bezieht sich allein auf die Amplitude~$T\ellp$ und ergibt deren Mittelung~$\pT\ellp$ "uber die azimutalen Orientierungen des Dipols~$\{\zet,\rb{r}\}$, vgl.\@ Gl.~(\ref{pT_ellp}).
Aufgrund der Symmetrie des Integranden, vgl.\@ Gl.~(\ref{Symm_tTll}), wird~$\zet$ bezogen auf das halbe Intervall.

Die Amplituden~$T\ellp$,~$\pT\ellp$ sind {\it rein imagin"ar\/} und h"angen ab von der invarianten Gesamtenergie~$\surd s$ nur {\it kinematisch\/}, das hei"st durch einen globalen Faktor~$s$.
Aus dem Verschwinden des Realteils folgt, da"s~$\,\frac{1}{\T s}\pT\ellp$ f"ur~$\tfbB \!\equiv\! 0$ identisch gleich ist der Funktion~$\si^{\rm tot}\ellp$, dem entsprechenden totalen Wirkungsquerschnitt, vgl.\@ Gl.~(\ref{sitot_ellp}) und Abb.~\ref{Fig:si-ellp-tot_zet}.
Aufgrund der nur kinematischen Energieabh"angigkeit von~$\pT\ellp$, n"amlich durch den globalen Faktor~$s$ folgt f"ur die differentiellen Wirkungsquerschnitte Konstanz bez"uglich~$s$, vgl.\@ Gl.~(\ref{dsigmadt-L,T}).
Sie beziehen sich auf den Wert~$\surd s \!\cong\! 20\GeV$, f"ur den unsere Parameter fixiert sind, vgl.\@ Seite~\pageref{T:Parameter}, insbes.\@ die Gln.~(\ref{si^tot_pom-ZW}),~(\ref{B_pp-ZW}).
Sie h"angen ab von~$Q^2$ "uber den "Uberlapp der~Wellenfunktio\-nen von Photon und Vektormeson und von~$\tfbQ$ "uber die Amplitude~$\pT\ellp$.
F"ur diese gilt, vgl.\@ Seite~\pageref{T:slope}:~$\pT\ellp \!\propto\! \exp -B_0\tfbQ /\!2$\, f"ur kleine~$\tfbB \!\cong\! \surd-t$, mit~$B_0$ dem slope-Parameter.

Experimentell sind diese Wirkungsquerschnitte schwierig zu trennen bez"uglich Polarisationen.
Daten in Abh"angigkeit von~$\tfbQ$ werden angegeben nur f"ur
\begin{align} \label{dsigmadt-exp}
\frac{d\si}{dt}\;
  =\; \ep\, \frac{d\si_L}{dt}\; +\; \frac{d\si_T}{dt}
\end{align}
meist f"ur festes~$Q^2$.
Diese Daten sind rar.
Die meisten und pr"azisere Daten finden sich f"ur die $\tfbQ$-integrierten Wirkungsquerschnitte
\begin{align} \label{sigma-exp}
\si\; =\; \ep\, \si_L\; +\; \si_T
\end{align}
als Funktion von~$Q^2$.

Dabei ist in den Gln.~(\ref{dsigmadt-exp}),~(\ref{sigma-exp}) mit~$\ep$ die Rate longitudinal polarisierter Photonen bezeichnet.
Sie h"angt ab vom Streuwinkel~$\th_l$ des Leptons, von dem das (virtuelle) Photon emittiert wird, und von dessen Energie~$\nu$; so gilt
\begin{align} \label{Epsilon}
\ep\;
  =\; \big[\, 1 + \big[ 1 \!+\! \nu^2 \!/\! Q^2 \big]\!\cdot 2 \tan^2(\th_l\!/2)\, \big]^{-1}
\end{align}
im Ruhesystem des Protons.

In die Herleitung unserer $T$-Amplitude~$T\hh$ ist die Annahme eingegangen, da"s im Limes~$s \!\to\! \infty$ Helizit"atserhaltung im~$s$-Kanal gilt, vgl.\@ die Gln.~(\ref{Helizitaetserhaltung-u}),~(\ref{Helizitaetserhaltung-u}$'$) und~(\ref{Helizitaetserhaltung-v}), (\ref{Helizitaetserhaltung-v}$'$).
Ihre G"ultigkeit kann untersucht werden in Zerf"allen von Vektormesonen.
Umgekehrt k"onnen wir unter ihrer Annahme berechnen
\begin{align} \label{R_LT}
\RLT\; =\; \si_L / \si_T
\end{align}
in Abh"angigkeit von~$Q^2$, das hei"st das Verh"altnis der integrierten Wirkungsquerschnitts bez"uglich longitudinaler und transversaler Polarisation als Funktion der Photon-Virtualit"at.

\section{Numerische Analyse}

In diesem Abschnitt analysieren wir detailiert die numerischen Konsequenzen, die sich aus diesen Formeln ergeben f"ur die observablen der Vektormesonen~$\rh(770)$,~$\om(782)$,~$\ph(1020)$ und~$\Jps(3097)$.
Wir stellen dieser speziellen Analyse voran die allgemeine Diskussion dieser Formeln.

Zun"achst bezieht sich unsere Analyse zu einem wichtigen Teil auf die Abh"angigkeit von der Photon-Virtualit"at in dem betrachteten Bereich von~$Q^2 \!=\! 1 \!-\! 2$ bis~$10\GeV^2$.
Wir diskutieren daher qualitativ die allgemeine $Q^2$-Abh"angigkeit der differentiellen Wirkungsquerschnitte f"ur longitudinale und transversale Polarisation.
Wir betrachten Gl.~(\ref{dsigmadt-L,T}) f"ur~$\la \!\equiv\! L,T$ mithilfe der Formeln f"ur den entsprechenden "Uberlapp der Wellenfunktionen, die Gln.~(\ref{"Uberlapp}),~(\ref{"Uberlapp}$'$).
Die $T$-Amplitude~$\pT\ellp(\zet, \rb{r}; \tfbB)$ f"ur Loop(Dipol)-Proton-Streuung zeigt Potenzverhalten bez"uglich~$r$ und f"allt f"ur kleines~$\tfbB \!\cong\!\surd-t $ exponentiell ab mit~$\tfbQ$.
Sie h"angt nur marginal ab von~$\zet$, vgl.\@ Abb.~\ref{Fig:si-ellp-tot_zet}.%
\FOOT{
  \label{FN:pT_ellp-si_ellp}Bzgl.\@ der Verkn"upfung mit der dort aufgetragenen Funktion~$\si^{\rm tot}\ellp(\zet,r^2)$ vgl.\@ die Diskussion zu Gl.~(\ref{dsigmadt-L,T}).
}
F"ur festes~$\tfbB$ gilt~$\pT\ellp \!\propto\! r^{2n}$, wobei~$n \!\equiv\! 1$ analytisch f"ur verschwindendes~$r$ langsam kleiner wird f"ur~$r \grgl a$, vgl.\@ Abb.~\ref{Fig:si-ellp-tot_zet}.\citeFN{FN:pT_ellp-si_ellp}
Wir interessieren uns daf"ur f"ur welche Werte von~$r$, abh"angig von~$Q^2$, die Amplitude~$\pT\ellp$ effektiv auszuwerten ist.
Da die Abh"angigkeit von~$\pT\ellp$ von~$\zet$ nur sehr schwach ist, vernachl"assigen wir sie f"ur den Moment komplett, so da"s sich die~$\zet$-Integration im differentiellen Wirkungsquerschnitt bezieht nur auf den "Uberlapp der Wellenfunktionen von Photon und Vektormeson.
Dieser bestimmt dann die effektiven Dipol-Ausdehnungen und diese die effektive Potenz~$n'$ in~$\pT\ellp \!\propto\! r^{2n'}$.
Durch Reskalieren~$r \!\to\! Qr$ in der Amplitude~$\pT\ellp$ wird diese dimensionslos und herausprojiziert entsprechende Faktoren von~$Q$; diese zusammen mit dem~$Q^2$-Verhalten des "Uberlapps bestimmen wesentlich das~$Q^2$-Verhalten des Wirkungsquerschnitts.
Wir unterscheiden auf diese Weise zwei verschiedene Bereich:
F"ur kleine~$Q^2$ ist die effektive Ausdehnung dominiert durch die Vektormeson-Wellenfunktion, wohingegen sie f"ur gro"se~$Q^2$ gegeben ist allein durch die des Photons; wir finden asymptotisch:
%
\begin{alignat}{7} \label{Q2-Asymptotik}
&d\si_L&\vv &\propto&\; &Q^2& \qquad
&d\si_T&\vv &=&\;       &\text{konst.}& \qquad\qquad
  &\text{f"ur\vvv $Q^2 \to 0$}
    \\
&d\si_L&\vv &\propto&\; &Q^{-6}& \qquad
&d\si_T&\vv &\propto&\;  &Q^{-8}& \qquad\qquad
  &\text{f"ur\vvv $Q^2 \to \infty$}
    \tag{\ref{Q2-Asymptotik}$'$}
    \\[-4.5ex]\nn
\end{alignat}
Diese Asymptotik der Wirkungsquerschnitte h"angt stark ab von dem zugrundegelegten Modell.
Die meisten gegenw"artigen Experimente beziehen sich auf Werte von~$Q^2$, die diese Asymptotik gro"ser $Q^2$ noch nicht erreichen.
Auch der von uns betrachtete Bereich von \mbox{$Q^2 \!=\! 1 \!-\! 2$} bis~$10\GeV^2$ ist zu beziehen weder auf die Asymptotik kleiner noch der gro"ser~$Q^2$:
\begin{figure}
\begin{minipage}{\linewidth}
  \begin{center}
  \setlength{\unitlength}{.9mm}\begin{picture}(120,71)   
    \put(0,0){\epsfxsize108mm \epsffile{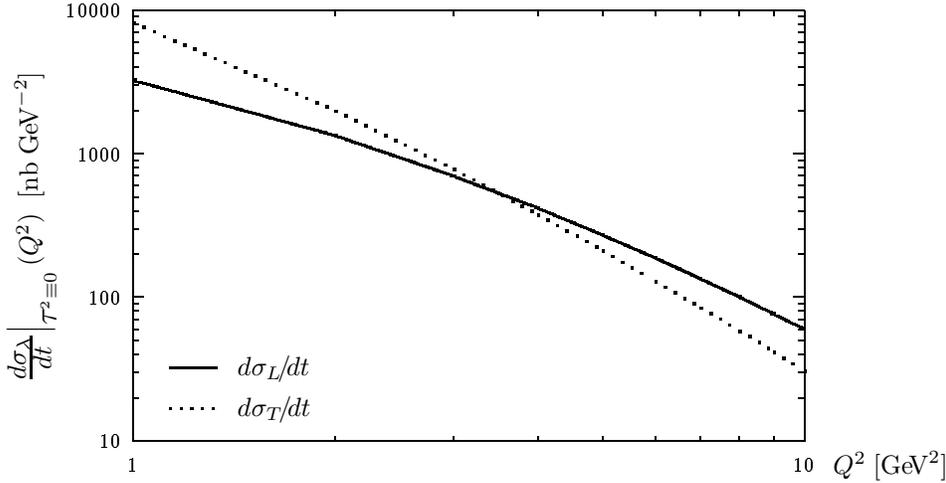}}
    \put(114,0.5){\normalsize$Q^2\;[\GeV[]^2]$}
    \put( -8,0  ){\yaxis[66.3mm]{\normalsize$%
                                 \left.\frac{\D d\si_\la}{\D dt}\right|_{\tfbQ\equiv0}(Q^2)%
                                 \vv[\nbarn[]\GeV^{-2}]$}}
    \put(     16,16){\rule{18pt}{0.8pt}}   
      \put(   26,15){\normalsize$d\si_L \!/\! dt$}
    \multiput(16,10)(1.56,0){5}{\rule{0.8pt}{0.8pt}}   
      \put(   26, 9){\normalsize$d\si_T \!/\! dt$}
  \end{picture}
  \end{center}
\vspace*{-4ex}
\caption[\protect$Q^2$-Abh"angigkeit von~\protect$d\si_\la \!/\! dt|_{\tfbQ\equiv0}$ f"ur~\protect$\rh$-Produktion,~\protect$\la \!\equiv\! L,T$]{
  $Q^2$-Verhalten von~$d\si_\la \!/\! dt(\tfbQ\!\equiv\!0)$ f"ur $\rh$-Produktion.   Die durchgezogene Linie bezieht sich auf longitudinale Polarisation, die gepunktete auf transversale.   Die effektive Potenz des Abfalls mit~$Q$ steigt an um zwei Einheiten im dargestellten Bereich von~$Q^2$.
}
\label{Fig:dsigma-dt_rh}
\end{minipage}
\end{figure}
Abbildung~\ref{Fig:dsigma-dt_rh} zeigt effektiv~$\log\,d\si_\la\!/\!dt$, den Logarithmus des differentiellen~Wirkungsquerschnitts f"ur das~$\rh$-Produktion bei~$\tfb \!\equiv\! 0$ in Gegen"uberstellung f"ur~\mbox{longitudinale} und transversale Polarisation~$\la \!\equiv\! L,T$.
Die effektive Potenz~$m'$ f"ur den Abfall~\mbox{$\log\,d\si_\la\!/\!dt \!\propto\! Q^{-m'}$} ist ungef"ahr~$2.5$ f"ur longitudinale und~$4$ f"ur transversale Polarisation bei~\mbox{$Q^2 \!\equiv\! 1\GeV^2$}.
Diese Potenz steigt an auf ungef"ahr~$4.5$ beziehungsweise~$6$ bei~$Q^2 \!\equiv\! 10\GeV^2$.
Die Asymptotik gro"ser~$Q^2$ beginnt also erst jenseits dieses Wertes, wir erwarten im Bereich~$Q^2 \!=\! 10 \!-\! 100\GeV^2$.

\begin{figure}
\begin{minipage}{\linewidth}
  \begin{center}
  \vspace*{1ex}
  \setlength{\unitlength}{.9mm}\begin{picture}(120,75.3)   
    \put(0,0){\epsfxsize108mm \epsffile{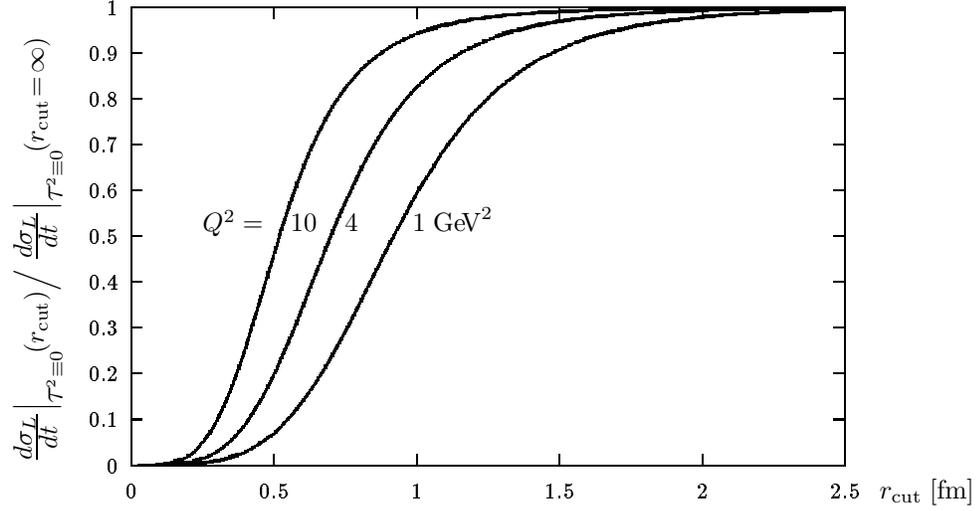}}
    \put(-18,76){\normalsize (a)\quad Longitudinal,~$L$}
    \put(118,0.5){\normalsize$r_{\rm cut}\;[\fm[]]$}
    \put(-10,0  ){\yaxis[66.3mm]{\normalsize%
            $\left.\left.\frac{\D d\si_L}{\D dt}%
                     \right|_{\tfbQ\equiv0}\zz(r_{\rm cut})\right/%
                   \left.\frac{\D d\si_L}{\D dt}%
                     \right|_{\tfbQ\equiv0}\zz(r_{\rm cut} \!=\!\infty)$}}
    \put(18,40){\normalsize$Q^2 =$}
    \put(31,40){\normalsize$10$}
    \put(39,40){\normalsize$4$}
    \put(49,40){\normalsize$1\GeV^2$}
  \end{picture}\\[5ex]   
  \setlength{\unitlength}{.9mm}\begin{picture}(120,75.3)   
    \put(0,0){\epsfxsize108mm \epsffile{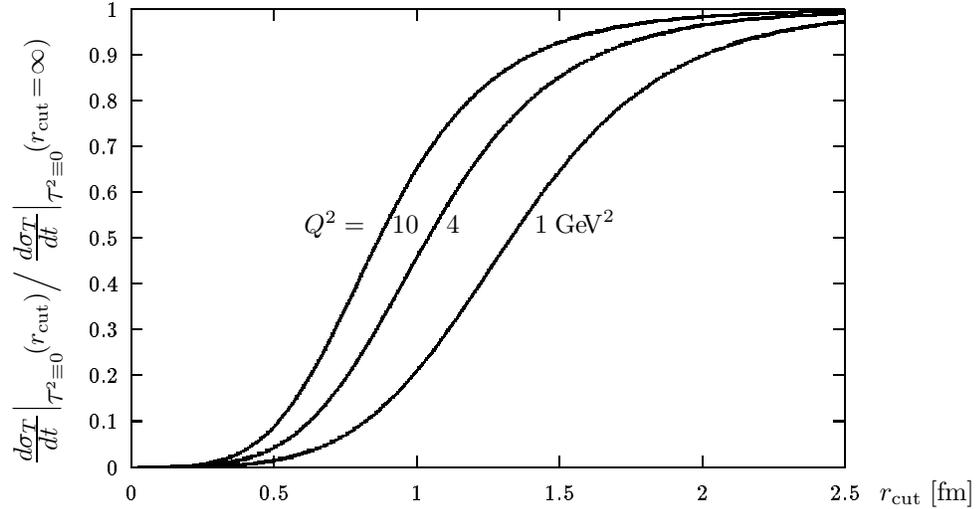}}
    \put(-18,76){\normalsize (b)\quad Transversal,~$T$}
    \put(118,0.5){\normalsize$r_{\rm cut}\;[\fm[]]$}
    \put(-10,0  ){\yaxis[66.3mm]{\normalsize%
            $\left.\left.\frac{\D d\si_T}{\D dt}%
                     \right|_{\tfbQ\equiv0}\zz(r_{\rm cut})\right/%
                   \left.\frac{\D d\si_T}{\D dt}%
                     \right|_{\tfbQ\equiv0}\zz(r_{\rm cut} \!=\!\infty)$}}
    \put(33,40){\normalsize$Q^2 =$}
    \put(46,40){\normalsize$10$}
    \put(54,40){\normalsize$4$}
    \put(67,40){\normalsize$1\GeV^2$}
  \end{picture}
  \end{center}
\vspace*{-4ex}
\caption[Verh"altnis
  \protect$d\sigma_\la \!/\! dt|_{\tfbQ\equiv0}(r_{\rm cut})/%
           d\sigma_\la \!/\! dt|_{\tfbQ\equiv0}(\infty)$ f"ur~\protect$\rh$-Produktion,~\protect$\la \!\equiv\! L,T$]{
  Differentieller Wirkungsquerschnitt f"ur $\rh$-Produktion bei~$\tfbB \!\equiv\! 0$ als Funktion des Cut-off~$r_{\rm cut}$ f"ur die transversale Ausdehnung der Dipole, normiert auf den vollen Wert.   Die Kurven beziehen sich auf~$Q^2 \!\equiv\! 1,\,4,\,10\GeV^2$, in~(a) auf longitudinale, in~(b) auf transversale Polarisation.   Mit wachsendem~$Q^2$ dominieren kleinere Dipole.   Der Wirkungsquerschnitt ist erst sp"ater auf\/integriert f"ur transversale Polarisation aufgrund des $\zet$-Endpunkte-Beitrags.
\vspace*{-1.5ex}
}
\label{Fig:dsdtLT-rcut}
\end{minipage}
\end{figure}
In Abbildung~\ref{Fig:dsdtLT-rcut} illustrieren wir die Bedeutung der transversalen Ausdehnung des Photons f"ur die Wirkungsquerschnitte.
Wir f"uhren als obere Grenze in der~$r$-Integration des differentiellen Wirkungsquerschnitts nach Gl.~(\ref{dsigmadt-L,T}) einen {\it Cut-Off\/}~$r_{\rm cut}$ ein, das hei"st wir beschr"anken die transversale Ausdehnung der beitragenden Dipole von Hand auf Werte~$r \!\leq\! r_{\rm cut}$.
Wir normieren diese Gr"o"se auf den vollen Wert mit entferntem Cut-Off, formal~$r_{\rm cut} \!=\! \infty$, und tragen ihn auf als Funktion von~$r_{\rm cut}$.
Wie erwartet ist die Quark-Antiquark-Lichtkegelwellenfunktion des Photons transversal ausgedehnter f"ur transversale als f"ur longitudinale Polarisation.
So r"uhrt f"ur~$Q^2 \!\equiv\! 4\GeV^2$ bei transversaler Polarisation mehr als~$50\%$ des Wirkungsquerschnitts von Dipolen transversal gr"o"ser als~$1\fm$, wohingegen bei longitudinaler Polarisation dies nur noch~$15\%$ sind.
Dies ist Konsequenz dessen, da"s die transversale Wellenfunktion fast flach bez"uglich~$\zet$ ist und ihre Ausdehnung grob gegeben ist als~$r \!\sim\! \vep^{-1}$; aufgrund der Proportionalit"at von~$\vep$ f"ur gro"se Photon-Virtualit"aten zu~$\surd\zbz Q^2$ und nicht zu~$\surd Q^2$, vgl.\@ Gl.~(\ref{epsilon}), f"uhren gro"se~$Q^2$ nicht notwendig zu Dominanz kleiner Dipole, sondern treten f"ur die $\zet$-Endpunkte weiterhin gro"se Dipole auf. \\
\indent
Unsere Resultate h"angen andererseits ab von den Lichtkegelwellenfunktionen der Vektormesonen.
Noch ohne ihre funktionale Form zu ver"andern zeigt sich dies durch Variation des (Oszillator)Parameters~$\om_{V,\la}$.
Wir halten den Wert der Vektormeson-Wellenfunktion~am Ursprung konstant.
Ein um~$5\%$ verkleinerter Wert von~$\om_{V,\la}$ ergibt eine Erh"ohung um~unge\-f"ahr~$20\%$ f"ur den Wirkungsquerschnitt bei~$Q^2 \!\equiv\! 1\GeV^2$.
Dieser Effekt verringert sich~auf $3\%$ bei~$Q^2 \!\equiv\! 10\GeV^2$.
Dies ist Konsequenz dessen, da"s der kleine Dipol in der Photon-Wellenfunktion bei gro"sen Virtualit"ten~$Q$ nur einen transversal engen Bereich der Vektor\-meson-Wellenfunktion um deren Ursprung testet. \\
\label{T:h_Vla-mod}Eine weitere M"oglichkeit die Abh"angigkeit von der Lichtkegelwellenfunktion des Vektormesons zu untersuchen ist, ihre funktionale Abh"angigkeit zu ver"andern.
So f"uhrt die Ersetzung~$\surd\zbz \!\to\! \zbz \surd\zbz$ in der Funktion~$h_{V,\la}(\zet)$, vgl.\@ Gl.~(\ref{h-Wfn}), zu einer sch"arferen Konzentrierung der Wellenfunktionen hin zu mittleren~$\zet$.
Dies f"uhrt zu einer Verringerung des Wirkungsquerschnitts um ungef"ahr~$30\%$ im gesamten~$Q^2$-Bereich.
Wie erwartet aufgrund der unterschiedlichen Bedeutung des $\zet$-Endpunkte-Beitrags ist der Effekt ausgepr"agter f"ur transversale als f"ur longitudinale Polarisation.

\subsection[Leptoproduktion von~\protect$\rh(770)$,~\protect$\om(782)$ und~\protect$\ph(1020)$
          ]{Leptoproduktion von~\protect\bm{\rh(770)},~\protect\bm{\om(782)} und~\protect\bm{\ph(1020)}}

Wir vergleichen unsere Resultate mit den experimentellen Daten im Bereich~$Q^2 \!=\! 1 \!-\! 10\GeV^2$.
Wir betonen vorab, da"s unser Modell im Rahmen des \DREI{M}{S}{V} keine neuen Parameter einf"uhrt, die in irgendeinem Sinne bezeichnet werden k"onnten als spezifisch f"ur Photo- und Leptoproduktion.
Die Parameter, die unserem Zugang auf Niveau der Quark-Gluon-Wechselwirkung zugrundeliegen, sind fixiert in Konsistenz mit Quark-Confinement in der Nieder- und elastischer Proton-Proton-Streuung in der Hochenergiephysik, vgl.\@ Seite~\pageref{T:Parameter}.

\vspace*{-1ex}
\paragraph{\bm{\rh}-Produktion.}\DREI{E}{M}{C} und \DREI{N}{M}{C} beobachten eine Skalierung des integrierten Wirkungsquerschnitts ungef"ahr wie~$1\!/\!Q^4$; wir reproduzieren dieses Verhalten sehr gut.
\begin{figure}
\begin{minipage}{\linewidth}
  \begin{center}
  \setlength{\unitlength}{.9mm}\begin{picture}(120,73.5)   
    \put(0,0){\epsfxsize108mm \epsffile{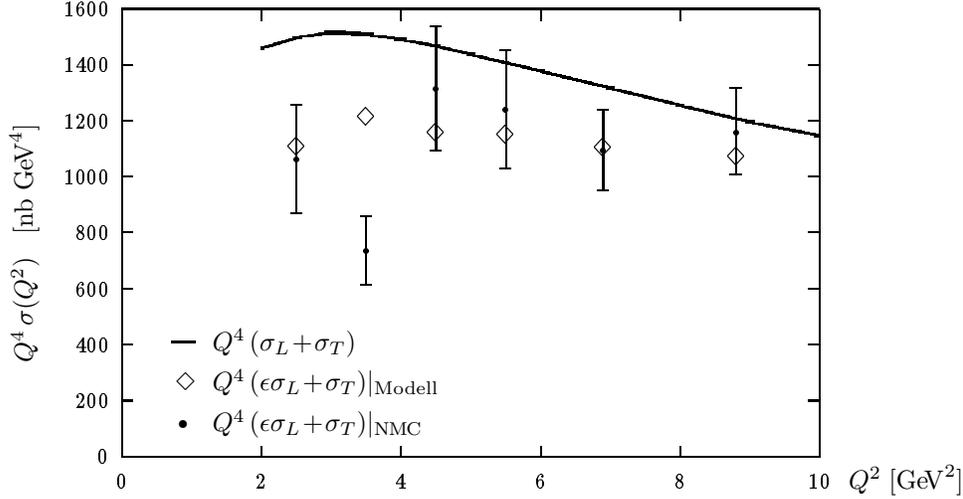}}
    \put(116,0.5){\normalsize$Q^2\;[\GeV[]^2]$}
    \put( -8,0  ){\yaxis[66.3mm]{\normalsize$%
                                 Q^4\,\si(Q^2)\vvv[\nbarn[]\GeV^4]$}}
    \put(  16  ,22){\rule{9pt}{0.8pt}}   
      \put(22  ,21){\normalsize$Q^4\,(\si_L \!+\! \si_T)$}
    \put(  16.5,15){$\Diamond$}   
      \put(22  ,15){\normalsize$Q^4\,(\ep\si_L \!+\! \si_T)|_{\rm Modell}$}
    \put(  17.7,10){\circle*{1}}
      \put(22  , 9){\normalsize$Q^4\,(\ep\si_L \!+\! \si_T)%
                                     |_{\mbox{\scriptsize\DREI[]{N}{M}{C}}}$}
  \end{picture}
  \end{center}
\vspace*{-3ex}
\caption[Skalierter integrierter~\protect$\rh$-Wirkungsquerschnitt~\protect$Q^4\,\si(Q^2)$]{
  Skalierter integrierte Wirkungsquerschnitt~$Q^4\,\si(Q^2)$ f"ur $\rh$-Produktion.   Als Punkte sind aufgetragen die \DREI[]{N}{M}{C}-Daten, vgl.\@ Ref.~\cite{Arneodo94}, als Rauten, unter Verwendung der entsprechenden \DREI[]{N}{M}{C}-Polarisationsrate~$\ep(Q^2)$, unser Postulat f"ur~$Q^4\,(\ep\si_L \!+\! \si_T)$.
\vspace*{-1ex}
}
\label{Fig:Q4si_rh}
\end{minipage}
\end{figure}
In Abbildung~\ref{Fig:Q4si_rh} sind aufgetragen die \DREI[]{N}{M}{C}-Daten f"ur Deuteron, vgl.\@ Ref.~\cite{Arneodo94}, und unser Postulat f"ur~\mbox{$Q^4\,(\si_L \!+\! \si_T)$}.
Diese Gr"o"se ist nicht gleich der gemessenen, da die \DREI[]{N}{M}{C}-Rate longitudinal polarisierter Photonen ungleich Eins ist:~$\ep \!\neq\! 1$, vgl.\@ Gl.~(\ref{sigma-exp}); wir tragen daher "uber den \DREI[]{N}{M}{C}-Werten f"ur~$Q^2$ die Gr"o"se~\mbox{$Q^4(\ep\si_L \!+\! \si_T)$} auf, indem wir deren Werte f"ur~$\ep(Q^2)$ verwenden.
Wir verweisen auch auf Tabelle~\refg{Tabl:Q>0-sigma_rh770}, die Daten von \DREI{E}{M}{C} und \DREI{N}{M}{C} f"ur integrierte elastische Wirkungsquerschnitte unseren Postulaten gegen"uberstellt.

In unserem Zugang ist das Verhalten des Wirkungsquerschnitts~$\si \!=\! (\si_L \!+\! \si_T)$ ungef"ahr wie~$1 \!/\! Q^4$ Resultat der Kombination des unterschiedlichen Verhaltens von~$\si_L$ und~$\si_T$, vgl.\@ die Gln.~(\ref{Q2-Asymptotik}),~(\ref{Q2-Asymptotik}$'$), das seinerseits Resultat ist des subtilen Wechselspiels zwischen der Abh"angigkeit der Dipol-Proton-Amplitude~$\pT\ellp$ von der transversalen Ausdehnung des Dipols mit dem effektiven "Uberlapp der Photon- und Vektormeson-Wellenfunktion.
Dieser Ursprung differiert wesentlich von der Dynamik, die auftritt bei Quark-Quark-Streuung, die aber dennoch zu "ahnlichen Resultaten f"uhrt.
Wie bereits diskutiert, ist demgegen"uber das asymptotische Verhalten f"ur gro"se~$Q^2$ beherrscht von der Ausdehnung der Photon-inh"arenten Dipolen und daher weitgehend unabh"angig von der funktionalen Form der Vektormeson-Wellenfunktion und dem Modell, wie diese Dipole wechselwirken mit dem Proton. \\
\indent
Ein m"oglicher Weg beide Zug"ange zu unterscheiden: Quark-Quark-Wechselwirkung von der String-String-Wechselwirkung des \DREI[]{M}{S}{V}, besteht in der pr"azisen Messung als Funktion von~$Q^2$ des Verh"altnisses des integrierten elastischen Wirkungsquerschnitts f"ur longitudinale zu dem f"ur transversale Polarisation:~$\RLT \!=\! \si_L \!/\! \si_T$, vgl.\@ Gl.~(\ref{R_LT}).
\begin{figure}
\begin{minipage}{\linewidth}
  \begin{center}
  \setlength{\unitlength}{.9mm}\begin{picture}(120,75.3)   
    \put(0,0){\epsfxsize108mm \epsffile{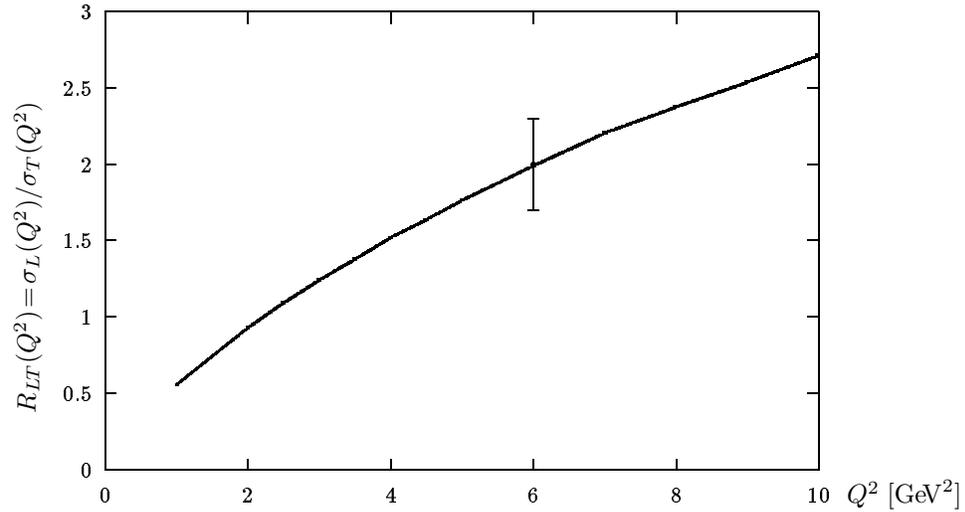}}
    \put(117,0.5){\normalsize$Q^2\;[\GeV[]^2]$}
    \put( -6,0  ){\yaxis[66.3mm]{\normalsize$%
                                 \RLT(Q^2) \!=\! \si_L(Q^2)/\si_T(Q^2)$}}
  \end{picture}
  \end{center}
\vspace*{-3ex}
\caption[\protect$Q^2$-Abh"angigkeit von~\protect$\RLT \!=\! \si_L \!/\! \si_T$ f"ur~\protect$\rh$-Produktion]{
  Integrierter elastischer Wirkungsquerschnitt f"ur $\rh$-Produktion, Verh"altnis longitudinale zu transversale Polarisation als Funktion von~$Q^2$.   Originalabbildung aus unserer Ver"offentlichung Ref.~\cite{Dosch96}: zum Zeitpunkt ihres Entstehens war der aufgetragene \DREI[]{N}{M}{C}-Datenpunkt, vgl.\@ Ref.~\cite{Arneodo94}, der einzige verl"a"sliche Wert im betrachteten~$Q^2$-Bereich; unsere Analyse war in guter "Ubereinstimmung mit weiteren Daten, doch lagen diese weit oberhalb des~$Q^2$-Bereichs von~$10 \!-\! 20\GeV^2$ oder waren mit gro"sen Fehlern behaftet.   Bzgl.\@ einer aktuellen Gegen"uberstellung vgl.\@ Abb.~\refg{Fig:R_LT}.
\vspace*{-.5ex}
}
\label{Fig-G:R_LT}
\end{minipage}
\end{figure}
Unser Postulat f"ur~$\RLT(Q^2)$ f"ur $\rh$-Produktion ist aufgetragen in Abbildung~\ref{Fig-G:R_LT}.
F"ur gro"se~$Q^2$ wird asymptotisch ein Verhalten erwartet wie~$\RLT \!\propto\! Q^2$, vgl.\@ die Gln.~(\ref{Q2-Asymptotik}),~(\ref{Q2-Asymptotik}$'$); dieses Verhalten ist in dem betrachteten~$Q^2$-Bereich von~$1 \!-\! 2$ bis~$10\GeV^2$ noch nicht erreicht, wir stellen ein Ansteigen fest weniger stark als~$Q^2$.
In diesem Bereich mittlerer~$Q^2$ erwarten wir signifikante Diskrepanz mit den konkurrierenden Modellen, die basieren auf blo"ser Quark-Quark-Wechselwirkung. \\
\indent
Abbildung~\ref{Fig-G:R_LT} ist die Originalabbildung aus unserer Ver"offentlichung Ref.~\cite{Dosch96}.
Inzwi\-schen sind mehr verl"assliche experimentelle Werte bekannt als der dort eingetragene \DREI[]{N}{M}{C}-Datenpunkt, vgl.\@ Ref.~\cite{Arneodo94}.
Im Rahmen der Erweiterung unseres Zugangs im folgenden Kapitel werden auch kleinere bis verschwindende Virtualit"aten zug"anglich, wobei unsere Vorhersage f"ur~$Q^2 \grgl 1\GeV^2$ unver"andert bestehen bleiben.
Mit Abbildung~\refg{Fig:R_LT} geben wir eine aktuelle Gegen"uberstellung mit dem Experiment an.
Wir finden dokumentieren dort in doppeltlogarithmischer Auftragung sehr gute "Ubereinstimmung im gesamten Bereich von~$Q^2 \!=\! 20\GeV^2$ bis hinunter zu den kleinsten bekannten experimentellen Datenpunkten.
Wir erwarten, da"s zuk"unftige Experimente das bereits angedeutete steilere Abfallen hin zu~$\RLT(Q^2 \!\equiv\! 0) \!=\! 0$, vgl.\@ die Gln.~(\ref{Q2-Asymptotik}),~(\ref{Q2-Asymptotik}$'$), mit kleineren~$Q^2$~noch deutlicher herauspr"aparieren werden.

Eine weitere M"oglichkeit zum Test der Wechselwirkung auf Quark-Gluon-Niveau ist die Untersuchung des differentiellen Wirkungsquerschnitts als Funktionen von~$\tfbQ \!\cong\! -t$, vgl.\@ Gl.~(\ref{tfbB_-t}).
\begin{figure}
\begin{minipage}{\linewidth}
  \begin{center}
  \setlength{\unitlength}{.9mm}\begin{picture}(120,73.7)   
    \put(0,0){\epsfxsize108mm \epsffile{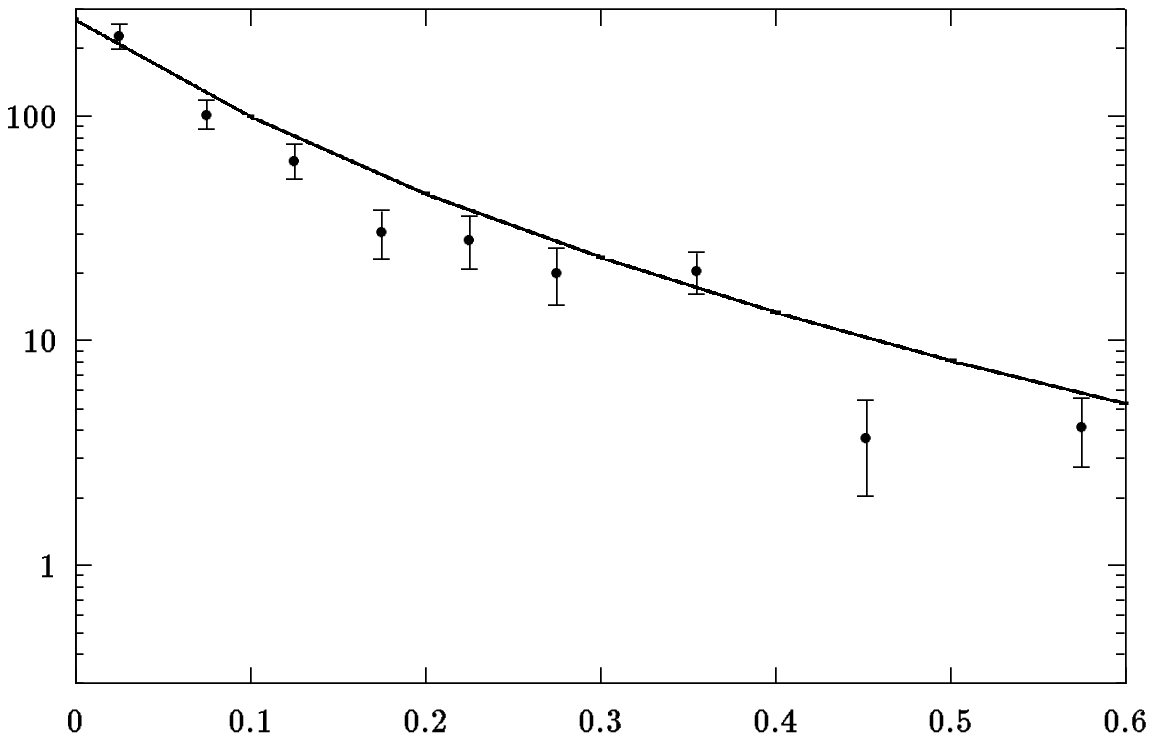}}
    \put(117,0.5){\normalsize$\tfbQ\;[\GeV[]^2]$}
    \put( -8,0  ){\yaxis[66.3mm]{\normalsize%
                    $\left.\left[\ep\,\frac{\D d\si_L}{\D dt}\!+\!%
                                   \frac{\D d\si_T}{\D dt}%
                           \right]\right|_{Q^2\equiv6{\rm\;Ge\mskip-3mu V}^2}(\tfbQ)%
                           \vvv[\nbarn[]\GeV^{-2}]$}}
  \end{picture}
  \end{center}
\vspace*{-4ex}
\caption[\protect$\tfbQ$-Verhalten von~\protect\mbox{$[\ep_{\mbox{\scriptsize\DREI[]{N}{M}{C}}} d\si_L\!/\!dt \!+\! d\si_T\!/\!dt]$} f"ur $\rh$-Produktion bei~\protect$Q^2 \!\equiv\! 6\GeV^2$\!]{
  Differentieller Wirkungsquerschnitt~$\ep_{\mbox{\scriptsize\DREI[]{N}{M}{C}}} d\si_L\!/\!dt \!+\! d\si_T\!/\!dt$ f"ur ~$\rh$-Produktion bei $Q^2 \!\equiv\! 6\GeV^2$ als Funktion von~$\tfbQ$.   Die Daten sind entnommen Ref.~\cite{Arneodo94}, vgl.\@ Tabl.~\ref{Tabl:Q>0-sigma_rh770}.
\vspace*{-2ex}
}
\label{Fig:dsigmadt-rhData}
\end{minipage}
\end{figure}
In Abbildung~\ref{Fig:dsigmadt-rhData} tragen wir auf, als Funktion von~$\tfbQ$, unser Resultat f"ur die Gr"o"se~$\ep d\si_L\!/\!dt \!+\! d\si_T\!/\!dt$ f"ur~$\rh$-Produktion bei~$Q^2 \!\equiv\! 6\GeV^2$.
Wir stellen diesem gegen"uber die \DREI[]{N}{M}{C}-Datenpunkte f"ur~$\rh$-Produktion elastisch an Deuteron au"serhalb der Region, in der koh"arente Produktion stattfindet, vgl.\@ Ref.~\cite{Arneodo94}.
Der Abfall ist nicht rein exponentiell, sondern gegeben durch eine konkave Kurve.
Wir finden~$\ep d\si_L\!/\!dt \!+\! d\si_T\!/\!dt \!\propto\! \exp\, -B(\tfbQ)\tfbQ /\!2$ mit~$B(0) \!\equiv\! B_0$, dem in Gl.~(\ref{slope-Parameter_Thh}) definierten slope-Parameter f"ur verschwindenden invarianten Impulstransfer und~$B(\tfbQ)$ langsam abfallend mit~$\tfbQ$ f"ur~$\tfbQ \grgl 0.1\GeV^2$.
Dieses Verhalten des effektiven slope-Parameters~$B(\tfbQ)$ ist modellabh"angig und ist Konsequenz des \DREI[]{M}{S}{V}-spezifischen String-String-Mechanismus inh"arent der Loop(Dipol)-Proton-Amplitude~$T\ellp$, vgl.\@ die Refn.~\cite{Dosch94a,Kulzinger95}.
Da die in $\tfbQ$ differentiellen Wirkungsquerschnitte schwer zu messen sind, existieren gegenw"artig kaum experimentelle Daten, aus denen~$B(\tfbQ)$ extrahiert werden k"onnte; wir m"ussen uns daher bez"uglich einer Konfrontation unseres Postulats mit dem Experiment begn"ugen mit dieser Abbildung.

\begin{figure}
  \begin{center}
  \setlength{\unitlength}{.9mm}\begin{picture}(120,72.2)   
    \put(0,0){\epsfxsize108mm \epsffile{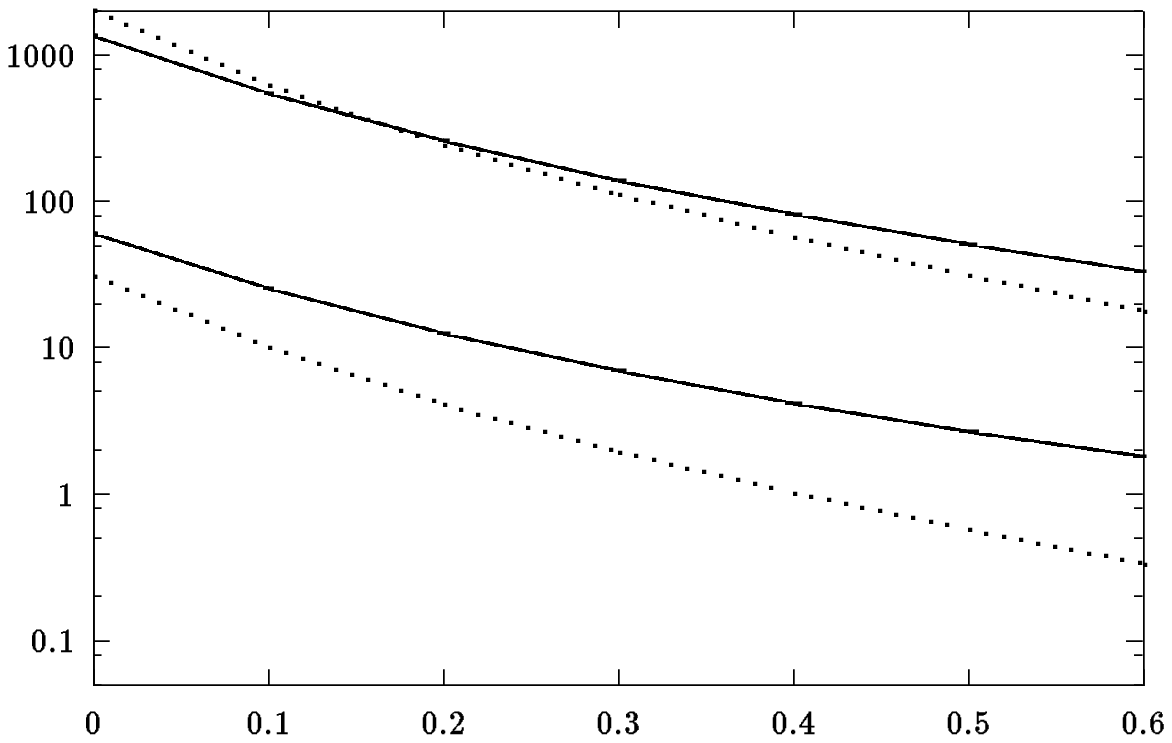}}
    \put(117,0.5){\normalsize$\tfbQ\;[\GeV[]^2]$}
    \put( -8,0  ){\yaxis[66.3mm]{\normalsize%
                    $\left.\frac{\D d\si_\la}{\D dt}%
                     \right|_{Q^2}%
                           (\tfbQ)\vvv[\nbarn[]\GeV^{-2}]$}}
    \put(     16,16){\rule{18pt}{0.8pt}}   
      \put(   26,15){\normalsize$d\si_L \!/\! dt$}
    \multiput(16,10)(1.56,0){5}{\rule{0.8pt}{0.8pt}}   
      \put(   26, 9){\normalsize$d\si_T \!/\! dt$}
    \put(70,54){\normalsize $\;\,Q^2 \!\equiv\!  2\GeV^2$}
    \put(70,36){\normalsize     $Q^2 \!\equiv\! 10\GeV^2$}
  \end{picture}
  \end{center}
\vspace*{-4ex}
\caption[\protect$\tfbQ$-Verhalten von \protect$d\si_\la\!/\!dt$ f"ur \protect$\rh$-Produktion bei~\protect$Q^2 \!\equiv\! 2,\,10\GeV^2$,~$\la \!\equiv\! L,T$]{
  Differentieller Wirkungsquerschnitt f"ur $\rh$-Produktion als Funktion von~$\tfbQ$ f"ur festes~$Q^2 \!\equiv\! 2$ und~$10\GeV^2$; die durchgezogene Linie bezieht sich auf longitudinale Polarisation, die gepunktete auf transversale.
\vspace*{-.75ex}
}
\label{Fig:dsigmadt-LT}
\end{figure}
\label{T:slope-dsigmadt_L,T}In Hinblick auf zuk"unftige Experimente zeigen wir in Abbildung~\ref{Fig:dsigmadt-LT} unsere Vorhersage f"ur die differentiellen Wirkungsquerschnitte als Funktion von~$\tfbQ$ f"ur longitudinale und transversale Polarisation getrennt; wir w"ahlen~$Q^2 \!\equiv\! 2$ und $10\GeV^2$ als einen niedrigeren und einen h"oheren Wert unseres~$Q^2$-Bereichs.
Wir finden wieder konkaves Abfallverhalten mit~$\tfbQ$.
Der Abfall f"ur kleine~$\tfbQ$ ist jedoch steiler f"ur transversale als f"ur longitudinale Polarisation; dieser Effekt wird schw"acher f"ur anwachsendes~$\tfbQ$.
Das hei"st der effektive slope-Parameter~$B_\la$ ist f"ur transversale Polarisation gr"o"ser als f"ur longitudinale:~$B_T \!>\! B_L$ f"ur festes~$\tfbQ$, und zwar umso mehr, je kleiner~$\tfbQ$ ist.
Wir erwarten dies aus folgendem~Grund:
Je kleiner~$\tfbQ$, desto mehr sollte~$B_\la$ in der Weise von den transversalen Ausdehnung der streuenden Hadronen bestimmt sein wie~$B_0$ nach Gl.~(\ref{slope_R1,R2}).
Andererseits ist die transversale Ausdehnung der effektiven Dipole gr"o"ser f"ur transversale als f"ur longitudinale Polarisation.

Wir vergleichen die Kurven in Abbildung~\ref{Fig:dsigmadt-LT} bez"uglich fester Polarisation aber unterschiedlichem~$Q^2$.
Sie verlaufen nicht ganz parallel und dokumentieren eine nichttriviale Abh"angigkeit des effektiven slope-Parameters auch von~$Q^2$.
Diese r"uhrt daher, da"s die Photon-Wellenfunktion mit wachsendem~$Q^2$ langsam transversal kleiner wird und so einen transversal kleineren Ausschnitt der Vektormeson-Wellenfunktion testet.

Zusammenfassend sehen wir gute Signatur, in einer pr"azisen Messung der differentiellen Wirkungsquerschnitte als Funktion von~$\tfbQ$ zu unterscheiden zwischen unserem Zugang und konkurrierenden.
Im folgenden Kapitel f"uhren wir in weiterem Rahmen unsere Diskussion der $\rh$-Produktion fort; bez"uglich differentieller Wirkungsquerschnitte verweisen wir auf Abbildung~\ref{Fig:dsdt_V,la}(a),(b) auf Seite~\pageref{Fig:dsdt_V,la} und die Tabellen~\ref{Tabl:dsigmadtQ2_0}-\ref{Tabl:dsigmadtQ2_20T} in Anhang~\ref{APP:TABLES}.

\vspace*{-1ex}
\paragraph{\bm{\om}-Produktion.}Das $\om$-Meson unterscheidet sich von dem $\rh$-Meson im wesentlichen durch seinen Flavour-Gehalt, genauer: seinen Isospin, vgl.\@ Gl.~(\ref{Flavour-Gehalt}) bzw.\@ Tab.~\ref{Tabl:Charakt_rh,om,ph,Jps}.
Wir erwarten daher, da"s die Lichtkegelwellenfunktionen beider Teilchen sehr "ahnlich sind.
Da diese proportional der entsprechenden Kopplung~$f_V$ an den elektromagnetischen Strom sind, sollte sich das Verh"altnis von Wirkungsquerschnitten verhalten wie, vgl.\@ Tabl.~\ref{Tabl:Charakt_rh,om,ph,Jps}:
\vspace*{-.5ex}
\begin{align} \label{fV2ratio-om/rh}
f_{{\D\om}(782)}^{\,2}\; \Big/\; f_\irh^{\,2}\;
  \cong\; 9.01 /\; 100
    \\[-4.5ex]\nn
\end{align}
Dies beobachten wir in der Tat bis auf etwa~$2\,\%$, vgl.\@ Tabl.~\refg{Tabl:Q=0_rh,om,ph,Jps},~\mbox{Fu"sn.\,\FN{FN:ratio-om/rh}}.%
~Die~theo\-retische Diskussion bleibt daher unver"andert gegen"uber dem $\rh$-Meson.
Andererseits liegt nur sehr begrenzt experimentelles Datenmaterial vor, mit dem wir unsere Vorhersagen konfrontieren k"onnten, so da"s wir unmittelbar "ubergehen zu den schwereren Vektormesonen.

\bigskip\noindent
\label{T:mf-ungleich-0}Unserer Analyse liegen zugrunde f"ur die laufenden Quarkmassen Zahlenwerte, wie angegeben in Tabelle\,\ref{Tabl:Charakt_rh,om,ph,Jps}, Fu"snote~\mbox{\FN{FN:LaufendeQuarkmassen}}:
Die up- und down-Massen werden identisch Null gesetzt:~\mbox{$m_u \!=\! m_d \!=\! 0$}, die strange-Masse betr"agt~\mbox{$m_s \!=\! 0.15\GeV$}, die charm-Masse~\mbox{$m_c \!=\! 1.3\GeV$}.
Das~$\ph$-Meson konstituiert sich also aus Quarks des ersten Flavours mit nichtverschwindender Masse; diese hat weitreichende Konsequenzen.
Wir diskutieren zun"acht allgemein die Konsequenzen einer {\it nichtverschwindenden\/} Quarkmasse.
Diese Diskussion bezieht sich bereits auf die unmittelbar folgende Analyse des~$\ph$-Mesons.
Sie wird fortgesetzt im n"achsten Unterabschnitt in der Analyse des~$\Jps$-Mesons, f"ur den Fall, da"s diese Quarkmasse {\it gro"s\/} ist.

Es tritt zun"achst ein zus"atzlicher Term auf im "Uberlapp der Lichtkegelwellenfunktionen von Photon und Vektormeson, der bislang identisch Null war, vgl.\@ den Term proportional~${\rm K}_0$ in Gl.~(\ref{"Uberlapp}$'$).
Bereits das Photon allein h"angt ab von der Masse der Quarkkonstituenten, in die es sich aufspaltet; das Epsilon, das deren transversale Separation, das hei"st die transversale Ausdehnung der Photon-Wellenfunktion determiniert, h"angt explizit ab von der Quarkmasse:~$\vep \!=\! \surd \zbz Q^2 \!+\! m_f^2$, vgl.\@ Gl.~(\ref{epsilon}).
Andererseits nimmt auch auf Seiten des Vektormesons dessen Quark-Gehalt Einflu"s: zum einen auf die transversale Ausdehnung der Vektormeson-Wellenfunktion und zum anderen auf die Verteilung des Parameters~$\zet$, das hei"st auf die Aufteilung des Lichtkegel-Gesamtimpulses auf Quark und Antiquark.
Der Einflu"s auf die transversale Ausdehnung ist qualitativ gut verstanden und auch quantitativ kontrollierbar.
Hingegen bez"uglich des Einflusses auf die~$\zet$-Verteilung~$h_{V,\la}(\zet)$, vgl.\@ Gl.~(\ref{h-Wfn}), ist nur bekannt, da"s bei gr"o"serer Quarkmasse die Verteilung mehr gepeakt sein sollte um den zentralen Wert~$\zet \!=\! 1\!/\!2$, der den Lichtkegelimpuls gleichverteilt.

\vspace*{-1ex}
\paragraph{\bm{\ph}-Produktion.}So wird f"ur die Lichtkegelwellenfunktion des~$\ph$-Mesons eine transversal kleinere Ausdehnung erwartet.
Dies zusammen mit der st"arkeren Unterdr"uckung der Endpunkte des~$\zet$-Intervalls l"a"st erwarten, da"s vor allem bei kleinen~$Q^2$ und st"arker f"ur longitudinale Polarisation als f"ur transversale die Wirkungsquerschnitte gegen"uber denen des $\rh$-Mesons mehr reduziert sein sollten als durch den Flavour-Faktor, vgl.\@ Tabl.~\ref{Tabl:Charakt_rh,om,ph,Jps}:
\vspace*{-.5ex}
\begin{align} 
f_{{\D\ph}(1020)}^{\,2}\; \Big/\; f_\irh^{\,2}\;
  \cong\; 26.9 /\; 100
    \\[-4.5ex]\nn
\end{align}
Ein "ahnlicher Effekt wird im Rahmen des \DREI{M}{S}{V} beobachtet im Verh"altnis von Pion-Proton- zu Kaon-Proton-Streuung, vgl.\@ hierzu Ref.~\cite{Dosch94a}.
Der Effekt sollte sich verringern mit steigender Virtualit"at des Photons, da f"ur steigende Werte von~$Q^2$ tendenziell die Photon-Wellenfunktion transversal kleiner wird und nur einen immer engeren Bereich der Vektormesonwellenfunktion um deren Ursprung testet:
Die Wirkungsquerschnitte sollten weniger abh"angen von der transversalen Ausdehnung des produzierten Vektormesons als von der des Photons.
Es gibt Hinweise daf"ur, da"s dieser Effekt in den \VIER[]{Z}{E}{U}{S}-Daten beobachtet wird, vgl.\@ Ref.~\cite{Derrick96a}.
Weiter k"onnte eine ver"anderte $\zet$-Verteilung~$h_{V,\la}(\zet)$ des~$\ph$- gegen"uber dem~$\rh$-Vektormeson die Wirkungsquerschnitte beeinflussen unabh"angig von~$Q^2$.

\begin{figure}
  \begin{center}
  \setlength{\unitlength}{.9mm}\begin{picture}(120,75)   
    \put(0,0){\epsfxsize108mm \epsffile{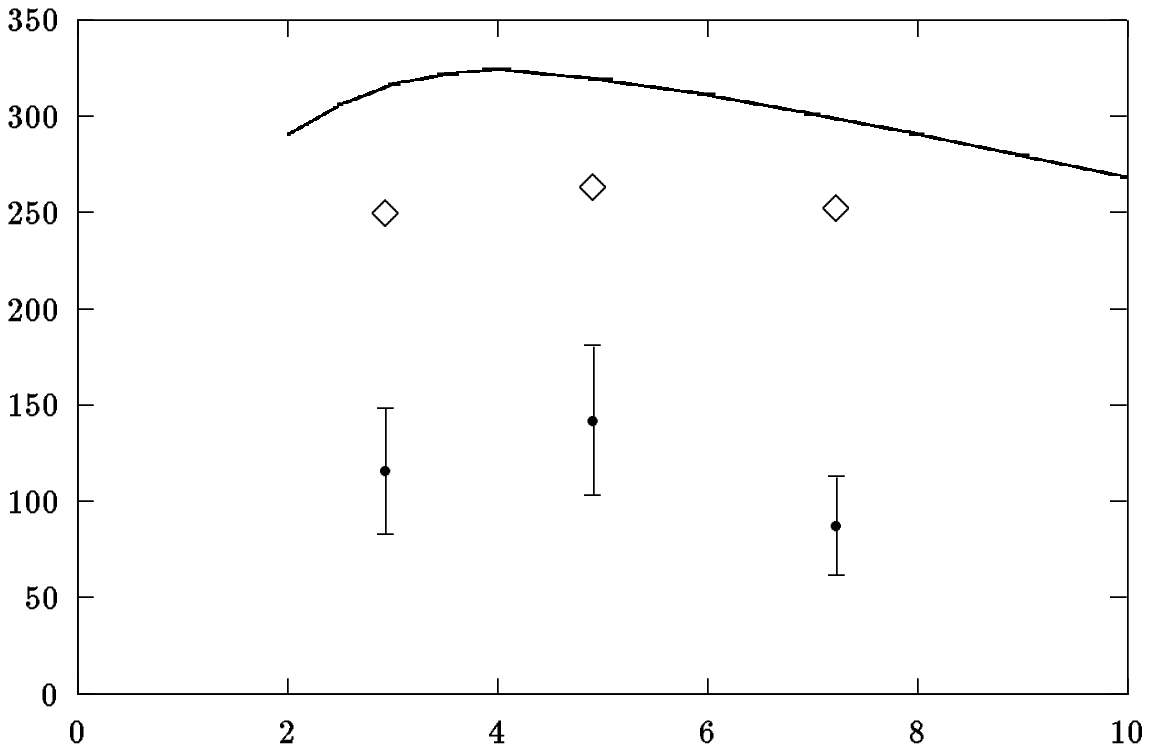}}
    \put(117,0){\normalsize$Q^2\;[\GeV[]^2]$}
    \put( -7,0){\yaxis[66.3mm]{\normalsize$Q^4\,\si(Q^2)\vvv[\nbarn[]\GeV^4]$}}
    \put(   9  ,21){\rule{9pt}{0.8pt}}   
      \put(15  ,20){\normalsize$Q^4\,(\si_L \!+\! \si_T)$}
    \put(   9.5,14){$\Diamond$}
      \put(15  ,14){\normalsize$Q^4\,(\ep\si_L \!+\! \si_T)|_{\rm Modell}$}
    \put(  10.7, 9){\circle*{1}}
      \put(15  , 8){\normalsize$Q^4\,(\ep\si_L \!+\! \si_T)%
                                     |_{\mbox{\scriptsize\DREI[]{N}{M}{C}}}$}
  \end{picture}
  \end{center}
\vspace*{-4ex}
\caption[Skalierter integrierter~\protect$\ph$-Wirkungsquerschnitt~\protect$Q^4\,\si(Q^2)$]{
  Skalierter integrierter Wirkungsquerschnitt f"ur~$\ph$-Produktion konfrontiert mit den experimentellen Daten von \DREI[]{N}{M}{C}, vgl.\@ Ref.~\cite{Arneodo94}.   Wir sehen die Ursache der Diskre\-panz in der funktionalen Form der $\ph$-Lichtkegelwellenfunktion.
\vspace*{-1ex}
}
\label{Fig:Q4si_ph}
\end{figure}

Wir legen unserer Berechnung f"ur das~$\ph$-Meson dieselbe funktionale Form zugrunde wie dem~$\rh$-Meson.
Wir reproduzieren damit die~$Q^2$-Abh"angigkeit des integrierten Wirkungsquerschnitts wie~$1\!/\!Q^4$, wie sie von \DREI{N}{M}{C} beobachtet wird.
Unsere Vorhersage ist allerdings absolut um einen Faktor Zwei zu gro"s; vgl.\@ Abbildun~\ref{Fig:Q4si_ph}.
Die $\ph$-Lichtkegelwellenfunktion sollte st"arker gepeakt sein um~$\zet \!=\! 1\!/\!2$.
Dies kann erreicht werden durch einen zus"atzlichen Faktor~$\zbz$ in ihrer Verteilung~$h_{\ph,\la}$, f"ur~$\la \!\equiv\! L,T$.
Wir haben diskutiert, da"s dies eine Verringerung der Wirkungsquerschnitte mit sich bringt von ungef"ahr~$30\%$ unabh"angig von~$Q^2$, vgl.\@ oben auf Seite~\pageref{T:h_Vla-mod}.
Dieser Effekt geht in die richtige Richtung und kann verst"arkt durch entsprechend st"arkere Akzentuierung der mittleren~$\zet$-Werte.
Es liegt aber nicht in unsere Absicht durch {\it fine tuning\/} Konsistenz mit allen Observablen zu erreichen.
Zumal gerade die zu Abbildung~\ref{Fig:Q4si_ph} korrespondierende Auftragung der \DREI[]{N}{M}{C}-Daten f"ur~$\rh$-Produktion, vgl.\@ Abb.~\ref{Fig:Q4si_rh}, suggeriert, da"s diese mit gr"o"seren Fehlern behaftet sein k"onnten, als ihre Fehlerbalken angeben.

F"ur das Verh"altnis der integrierten Wirkungsquerschnitte f"ur longitudinale und transversale Polarisation finden wir einen "ahnlichen Anstieg wie f"ur das~$\rh$-Meson; sein absoluter Wert liegt ungef"ahr~$20\%$ darunter.
Die~$t$-Abh"angigkeit der differentiellen Wirkungsquerschnitte zeigt ein "ahnliches Muster wie die f"ur~$\rh$-Produktion, wobei der $\tfbB \!\equiv\! 0$-Diffraktions-Peak~et\-was breiter ausl"auft.
\vspace*{-1ex}

\subsection[Photo- und Leptoproduktion von~\protect$\Jps(3097)$
          ]{Photo- und Leptoproduktion von~\protect\bm{\Jps(3097)}}

Eine nichtverschwindende Quarkmasse f"uhrt zu drastischen Ver"anderungen in den Lichtkegelwellenfunktionen von Photon und Vektormeson und folglich in deren "Uberlapp, vgl.\@ die Diskussion unmittelbar vor der Analyse des $\ph$-Mesons auf Seite~\pageref{T:mf-ungleich-0}.
Ist die Quarkmasse zus"atzlich {\it gro"s}, so gehen die ver"anderungen erheblich weiter.
Zun"achst ist zu beachten, da"s unser Zugang annimmt, da"s die Quarks auf (nahezu) lichtartigen Trajektorien propagieren.
Dies ist erst der Fall f"ur Energien weit oberhalb~$2m_f$.
Daher sind f"ur Leptoproduktion schwererer Vektormesonen wie des~$\Jps(3097)$ gr"o"sere invariante Schwerpunktenergien~$\surd s$ notwendig als die bisher diskutierten~$10\GeV$.
Bei entsprechend h"oheren Energien wird wiederum eine nichttriviale $s$-Abh"angigkeit beobachtet, die unser Modell nicht ber"ucksichtigt; vgl.\@ diesbzgl.\@ Ref.~\cite{Rueter98}.
Andererseits stellt die gro"se Quarkmasse eine {\it harte Skala\/} dar, die es erm"oglicht innerhalb unserer perturbativen Behandlung des Photons auch kleinere~$Q^2$ zu betrachten bis hin zu Photoproduktion:~$Q^2 \!\equiv\! 0$.
Weiter induziert die Differenz~$\tfde \!=\! -M_p^2(Q^2\!+\!M_V^2)^2\!/s^2$ von~$\tfbQ \!\cong\! -t \!+\! \tfde$ und~$-t$, vgl.\@ Gl.~(\ref{tfbB_-t}), formal eine untere Schranke in der~$\tfb$-Phasenraumintegration und f"uhrt f"ur Wirkungsquerschnitte zu dem zus"atzlichen Faktor~$\exp B\tfde$.
Bezogen auf~$\Jps$-Produktion und den von uns zu betrachtenden Energiebereich von~$\surd s \grgl 15\GeV$ finden wir, da"s dieser Effekt erst f"ur gro"se~$Q^2 \grgl 10\GeV^2$ deutlich wird.
So ergibt~$\surd s \!=\! 15\GeV$,~$Q^2 \!\equiv\! 10\GeV^2$ und der slope-Parameter im typischen Bereich~$B \!=\! 5 \!-\! 10\GeV^{-2}$ explizit~$\exp B\tfde \!=\! 0.97 \!-\! 0.94$.
Wir werden diesen Faktor im folgenden vernachl"assigen. \\
\indent
Eine gro"se Quarkmasse hat ferner Konsequenzen f"ur das Asymptotische Verhalten der Wirkungsquerschnitte f"ur kleine Abst"ande im Sinne gro"ser~$Q^2$.
Wir untersuchen die Modifikationen gegen"uber Gl.~(\ref{Q2-Asymptotik}$'$), die der zus"atzliche Massenterm im "Uberlapp von Photon- und $\Jps$-Lichtkegelwel\-lenfunktion generiert, vgl.\@ die Gln.~(\ref{"Uberlapp}),~(\ref{"Uberlapp}$'$).
Wir nehmen dazu f"ur den Moment an, die $\zet$-Verteilung sei nichtrelativistisch, das hei"st~\mbox{$h_{\iJps,\la}(\zet) \!\propto\! \de(\zet \!-\! 1\!/\!2)$} f"ur~\mbox{$\la \!\equiv\! L,T$},~vgl.\@ Gl.~(\ref{h-Wfn}), und f"ur Konsistenz~$m_c \!=\! M_\iJps\!/2$.
Wir betrachten so gro"se~$Q^2$, da"s das Epsilon der Photon-Wellenfunktion~\mbox{$\vep^2 \!=\! m_f^2 \!+\! \zbz Q^2 \!=\! (Q^2 \!+\! M_\iJps^2)\!/4$}~hinrei\-chend gro"s ist, \vspace*{-.375ex}um~\mbox{$g_{\iJps,\la}(r) \!=\! \exp\{-\om_{\iJps,\la}^2r^2\!/2\} \!\cong\! 1$} f"ur $\la \!\equiv\! L,T$, vgl.\@ Gl.~(\ref{g-Wfn}),~\mbox{approximieren} zu k"onnen; dann gilt~\mbox{\,$\pT\ellp(\zet \!\equiv\! 1\!/\!2,r,\tfbB \!\equiv\! 0) \cong \iIM s\, C\,r^2$}, mit~\mbox{\,$C \!\cong\! 4.3$}, und wir erhalten:
\vspace*{-.5ex}
\begin{alignat}{2} \label{Q2-Asymptotik_Jps}
&\left.\frac{d\si_L}{dt}\right|_{\tfbQ\equiv0}\zz(Q^2)&\;
  &\underset{\text{$Q^2 \!\to\! \infty$}}{\cong}\;
       \al_{\rm em}\cdot \big[8f_\iJps C\big]^2\,
       \frac{Q^2}{(Q^2 \!+\! M_\iJps^2)^4}
    \\[0ex]
&\left.\frac{d\si_T}{dt}\right|_{\tfbQ\equiv0}\zz(Q^2)&\;
  &\underset{\text{$Q^2 \!\to\! \infty$}}{\cong}\;
       \al_{\rm em}\cdot \big[8f_\iJps C\big]^2\,
       \frac{M_\iJps^2}{(Q^2 \!+\! M_\iJps^2)^4}\,
       \left[ \frac{M_\iJps^2 \!+\! 8\om_{\iJps,T}^2}{%
                    M_\iJps^2 \!+\! 2\om_{\iJps,T}^2} \right]^2
    \tag{\ref{Q2-Asymptotik_Jps}$'$}
    \\[-4.5ex]\nn
\end{alignat}
Zun"achst zeigen diese Ausdr"ucke, da"s die relevante Skala, mit der die $\Jps$-Produktionsquer\-schnitte abfallen~$Q^2 \!+\! M_\iJps^2$ ist und nicht~$Q^2$; erst f"ur~$Q^2 \!\gg\! M_\iJps^2$~verliert dieser Unterschied an Relevanz und laufen die Ausdr"ucke dieser Gleichungen in die Asymptotik von Gl.~(\ref{Q2-Asymptotik}$'$).
Wie f"ur den Fall leichter Quarks dominiert f"ur gro"se~$Q^2$ der Wirkungsquerschnitt f"ur longitudinale Polarisation "uber den f"ur transversale; mit dem Zahlenwert~$\om_{\iJps,T} \!\cong\! 0.57\GeV$, vgl.\@ Tabl.~\ref{Tabl:Charakt_rh,om,ph,Jps}, ist~\mbox{$(M_\iJps^2 \!+\! 8\om_{\iJps,T}^2)^2\!/\!(M_\iJps^2 \!+\! 2\om_{\iJps,T}^2)^2 \!\cong\! 1.42 \!\cong\! \surd2$} und wir finden:
\vspace{-0ex}
\begin{align} 
\RLT[\iJps,](Q^2)\vv
  \underset{\text{$Q^2 \!\to\! \infty$}}{\cong}\; 1/\surd2\cdot\vv Q^2/M_\iJps^2
    \\[-5ex]\nn
\end{align}
Ausgedr"uckt durch diese Funktion~$\RLT(Q^2)$ sollte der differentielle Wirkungsquerschnitt in Vorw"artsrichtung abfallen wie~$(\ep d\si_L\!/\!dt + d\si_T\!/\!dt)|_{\tfbQ\equiv0} \!\propto\! [\ep\RLT \!+\! 1]\,(Q^2 \!+\! M_\iJps^2)^{-4}$.%
~Experimen\-tell wird eine Skalierung beobachtet wie~$(Q^2 \!+\! M_\iJps^2)^{-m}$\!, mit~$m \!\cong\! 2$.
Dies k"onnte widerspiegeln, da"s~$\RLT$ selbst von~$Q^2$ abh"angt und~-- ansteigend wie~$Q^2$~-- den vollst"andig asymptotischen Abfall mit~$m \!=\! 4$ konterkariert:
Der beobachtete Abfall wird gut gefittet durch eine kleinere Potenz~$m$; zumal die Daten nicht ausreichend genau sind,~$m \!=\! 4$ auszuschlie"sen.
Wir betonen, da"s eine genaue Bestimmung des Potenzverhaltens bez"uglich der Skala~\mbox{$Q^2 \!+\! M_\iJps^2$} unverzichtbare Voraussetzung ist f"ur das Verst"andnis der zugrundeliegenden Physik.

\begin{figure}
  \begin{center}
  \setlength{\unitlength}{.9mm}\begin{picture}(120,75.2)   
    \put(0,0){\epsfxsize108mm \epsffile{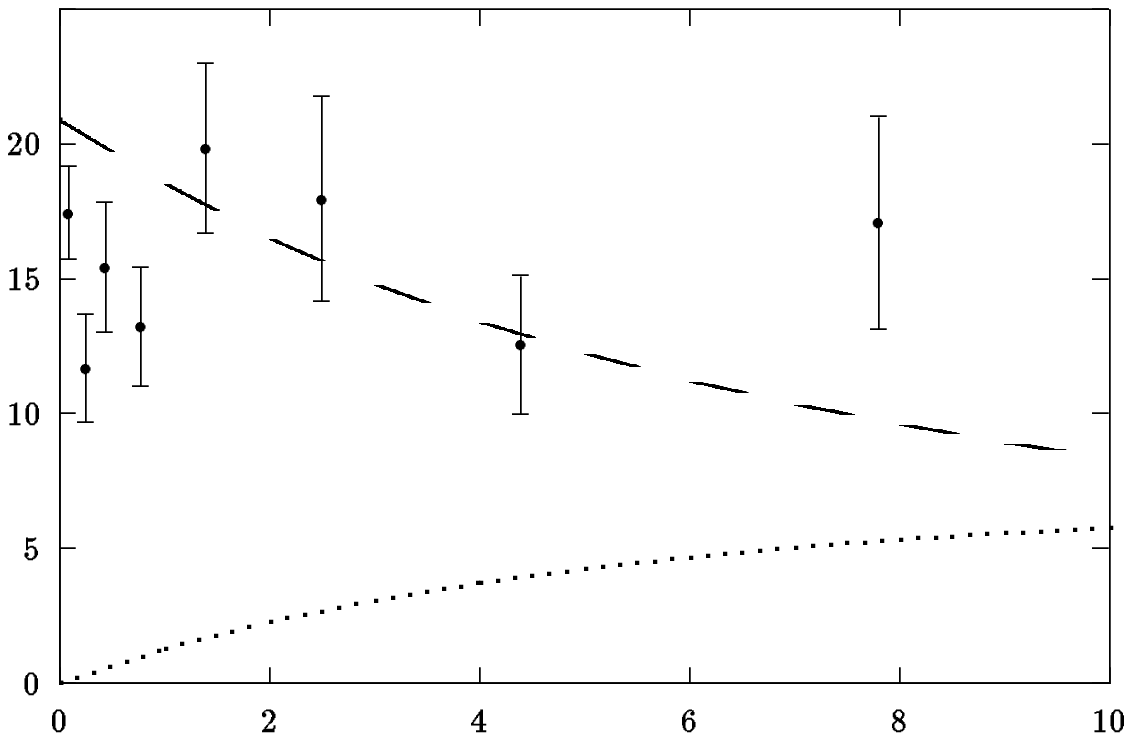}}
    \put(117,.5){\normalsize$Q^2\;[\GeV[]^2]$}
    \put( -7,0){\yaxis[66.3mm]{\normalsize$%
                     \big[1 \!+\! Q^2 \!/\! M_\iJps^2\big]^2\,\si_\la(Q^2)\vvv[\nbarn[]]$}}
    \multiput(10,26)(1.56,0){5}{\rule{0.8pt}{0.8pt}}   
      \put(   20,25){\normalsize$\la \!\equiv\! L$}
    \multiput(10,20)(2.6,0){3}{\rule{4.2pt}{0.8pt}}   
      \put(   20,19){\normalsize$\la \!\equiv\! T$}
  \end{picture}
  \end{center}
\vspace*{-3ex}
\caption[Skalierter integrierter~\protect$\Jps$-Wirkungsquerschnitt\,\protect\mbox{$[1 \!+\! Q^2 \!/\! M_\iJps^2]^2\,\si_{\la \equiv L,T}(Q^2)$}\protect\vspace*{-.375ex}]{
  Skalierter integrierter Wirkungsquerschnitt~$[1 \!+\! Q^2 \!/\! M_\iJps^2]^2\,\si_\la(Q^2)$ als Funktion von~$Q^2$: abweichend zu vorherigen Abbildungen bezieht sich die gepunktete Kurve auf longitudinale, die (fast) durchgezogene auf transversale.   Um mit den aufgetragenen \DREI[]{E}{M}{C}-Daten, vgl.\@ Ref.~\cite{Aubert83}, zu vergleichen, sind unsere Kurven entsprechend~$\ep \si_L \!+\! \si_T$ zu addieren, mit~$\ep \!\cong\! 0.7$ der gemessenen \DREI[]{E}{M}{C}-Polarisationsrate.
}
\label{Fig:sigma_Jps}
\end{figure}
In Abbildung~\ref{Fig:sigma_Jps} ist angegeben unser Postulat f"ur~$[1 \!+\! Q^2 \!/\! M_\iJps^2]^2\,\si_\la(Q^2)$ f"ur~$\la \!\equiv\! L,T$, die entsprechend skalierten integrierten Wirkungsquerschnitte f"ur longitudinale und transversale Polarisation.
Wir stellen ihnen gegen"uber die genannten Experimentellen Daten, die von \DREI{E}{M}{C} im Energiebereich~$\surd s \!=\! 10 \!-\! 20\GeV$ genommen sind, vgl.\@ Ref.~\cite{Aubert83}.
F"ur kleine~$Q^2$ nahe Photoproduktion sind im Energiebereich~$\surd s \!=\! 10 \!-\! 20\GeV$ mehrere experimentelle Werte angegeben, die aber nur grob einen Bereich von~$10 \!-\! 20\nbarn$ bestimmen.
Unsere Untersuchung der Asymptotik f"ur gro"se~$Q^2$ l"a"st bereits vermuten, da"s die Wirkungsquerschnitte wesentlich determiniert sind durch die Form der~$\zet$-Verteilumg~$h_{\iJps,\la}(\zet)$ und den Wert der charm-Quark-Masse~$m_u$.
Numerisch finden wir f"ur eine "Anderung von~$m_u$ um~$5\%$ f"ur die Wirkungsquerschnitte eine "Anderung von~$20\%$ bei~$Q^2 \!\equiv\! 0$ und von~$10\%$ bei~$Q^2 \!\equiv\! 10\GeV^2$. \\
F"ur ansteigendes~$Q^2$ folgt unsere Vorhersage qualitativ mehr und mehr dem Muster der Asymptotik kleiner Abst"ande, wie oben diskutiert; quantitativ tragen Dipole mittlerer transversaler Ausdehnung in der Weise effektiv bei, da"s der Abfall flacher resultiert als die Asymptotik allein erwarten lie"se.

\begin{figure}
  \begin{center}
  \setlength{\unitlength}{.9mm}\begin{picture}(120,73.7)   
    \put(0,0){\epsfxsize108mm \epsffile{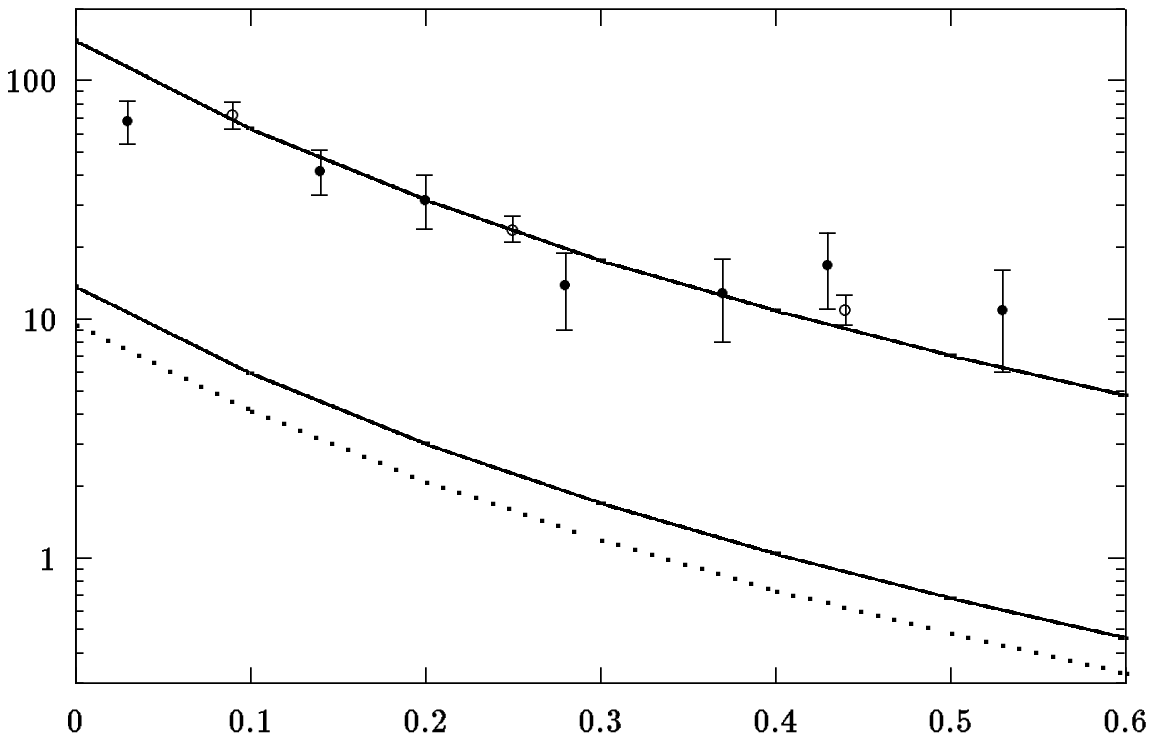}}
    \put(117,0.5){\normalsize$\tfbQ\;[\GeV[]^2]$}
    \put( -7,0  ){\yaxis[66.3mm]{\normalsize%
                    $\left.\frac{\D d\si_\la}{\D dt}%
                     \right|_{Q^2}%
                           (\tfbQ)\vvv[\nbarn[]]$}}
    \multiput(10,15)(1.56,0){5}{\rule{0.8pt}{0.8pt}}
      \put(   20,14){\normalsize$d\sigma_L/dt$}
    \put(     10, 9){\rule{18pt}{0.8pt}}
      \put(   20, 8){\normalsize$d\sigma_T \!/\! dt$}
    \put(50,54){\normalsize $\;\,Q^2 \!\equiv\!  0$}
    \put(50,29){\normalsize     $Q^2 \!\equiv\! 10\GeV^2$}
  \end{picture}
  \end{center}
\vspace*{-3ex}
\caption[\protect$\tfbQ$-Verhalten von \protect$d\si_\la\!/\!dt$ f"ur \protect$\Jps$-Produktion bei~\protect$Q^2 \!\equiv\! 0,\,10\GeV^2\!$,~$\la \!\equiv\! L,T$]{
  Differentiellen Wirkungsquerschnitte~$d\si_\la\!/\!dt$,~$\la \!\equiv\! L,T$, f"ur $\Jps$-Produktion in Abh"angigkeit von~$\tfbQ$ bei~$Q^2 \!\equiv\! 0,\,10\GeV^2\!$.   Die obere Kurve ist unsere Vorhersage f"ur Photoproduktion im Vergleich zu experimentellen Daten: der Messung von Ref.~\cite{Binkley82}~(ausgef"ullte Punkte) und der Extrapolation nach~$Q^2 \!\equiv\! 0$~der \DREI[]{E}{M}{C}-Daten von Ref.~\cite{Aubert83}~(offene Punkte).   "Ahnliche Daten sind gemessen von \DREI{N}{M}{C} bei~$Q^2 \!\equiv\! 1.5\GeV^2$, vgl.\@ Ref.~\cite{Arneodo94a}.   Die unteren Kurven zeigen unser Postulat f"ur~$Q^2 \!\equiv\! 10\GeV^2\!$.   Die Abh"angigkeit des slope-Parameters von der Polarisation, wie von~$Q^2$ sind nur marginal.
}
\label{Fig:dsigmadt_Jps}
\end{figure}
In Abbildung~\ref{Fig:dsigmadt_Jps} betrachten wir die Abh"angigkeit der differentiellen Wirkungsquerschnitte~$d\si_\la \!/\! dt$ f"ur~$\la \!\equiv\! L,T$ und festes~$Q^2$ in Abh"angigkeit von~$\tfbQ \!\cong\! -t$.
Unseren Vorhersagen stehen gegen"uber die experimentellen Daten f"ur~$Q^2 \!\equiv\! 0$ von Ref.~\cite{Binkley82}~(ausgef"ullte Punkte)~und die Extrapolation nach~$Q^2 \!\equiv\! 0$ der \DREI[]{E}{M}{C}-Daten von Ref.~\cite{Aubert83}~(offene Punkte).
Wir finden gute "Ubereinstimmung.
Zus"atzlich tragen wir unsere Postulate f"ur~$\Jps$-Leptoproduktion bei~$Q^2 \!\equiv\! 10\GeV^2$ auf.
In Kontrast zum Fall "`gro"ser"' Hadronen, vgl.\@ Abb.~\ref{Fig:dsigmadt-LT}, sind die Kurven weitgehend parallel, das hei"st wir beobachten nur sehr marginale Abh"angigkeit zum einen von der Polarisation, zum anderen von~$Q^2$.

\bigskip\noindent
Zusammenfassend berechnen wir differentielle Wirkungsquerschnitte~$d\si_\la\!/\!dt|_{Q^2}(\tfbQ)$ f"ur Leptoproduktion der Vektormesonen~$\rh(770)$,~$\om(782)$ und~$\ph(1020)$ elastisch am Proton und entsprechende Querschnitte f"ur Photo- und Leptoproduktion des~$\Jps(3097)$.
Wir postulieren diese Gr"o"sen separat f"ur longitudinale und transversale Polarisation, die sich Weise bezieht auf das einlaufende (virtuelle) Photon und das produzierte Vektormeson; Prozesse mit Helizit"ats"anderung um zwei Einheiten sind in unserem Zugang m"oglich, aber numerisch klein und daher vernachl"assigt.
Wir betrachten diese und daraus berechnete Observablen wie der integrierte Wirkungsquerschnitt~$\si_\la(Q^2)$ und das Verh"altnis~$\RLT(Q^2) \!=\! \si_L(Q^2) \!/\! \si_T(Q^2)$, sprechen an den slope-Parameter.

Das zugrundeliegende Modell auf Niveau der Quark-Gluon-Wechselwirkung ist das Modell des Stochastischen Vakuums.
Sein wesentliches Charakteristikum ist ein String-String-Mechanismus: das Ausbilden und die Wechselwirkung gluonischer Strings.
Seine Parameter sind vollst"andig fixiert durch Quark-Confinement bei niedrigen Energien und weicher Proton-Proton-Streuung bei hohen Energien; sie sind vollst"andig unabh"angig von Photo- und Leptoproduktion.

Die hadronischen Wellenfunktionen der involvierten Teilchen sind berechnet in Lichtkegelst"orungstheorie [Photon] beziehungsweise danach modelliert in wohl-etablierter Weise [Vektormeson] und einfachster Weise [Proton].

Wichtiges Spezifikum von Photo- und Leptoproduktion ist die nichttriviale Abh"angigkeit ihrer Wirkungsquerschnitte von der Photon-Virtualit"at.
Wir diskutieren die Asymptotik f"ur kleine und gro"se~$Q^2$.
Unser Zugang schr"ankt ein auf~$Q^2 \!=\! 1 \!-\! 2$ bis~$10\GeV^2$ als konservativen Bereich.
Dieser liegt weit ab von der Asymptotik, insbesondere der modellunabh"angigen gro"ser~$Q^2$.
Unsere Postulate beziehen sich auf mittlere~$Q^2$, die ganz wesentlich abh"angen vom zugrundeliegenden Quark-Gluon-Modell.
So korrespondiert eine Photon-Virtualit"at im betrachteten Bereich mit einer transversalen Ausdehnung seiner konstituierenden Quark-Antiquark-Dipole von~$0.5$ bis~$1.2\fm$, je nach Polarisation; entsprechend ausgedehnte transversale Struktur des Vektormesons wird getestet, vgl.\@ Abb.~\ref{Fig:dsdtLT-rcut}.
Unsere Vorhersagen reproduzieren nahezu perfekt das Verhalten bez"uglich der betrachteten intermedi"aren~$Q^2$; wir sehen darin den String-String-Mechanismus des \DREI{M}{S}{V} in wesentlichen Z"ugen best"atigt.
Wir geben weitere Observable an, wie die $\tfbQ$-Abh"angigkeit der differentiellen Wirkungsquerschnitte und die $Q^2$-Abh"angigkeit des Verh"altnisses~$\RLT$ der integrierten Wirkungsquerschnitte, die noch spezifischer abh"angen vom \DREI[]{M}{S}{V}-Mechanismus, gegenw"artig aber noch nicht ausreichend genau experimentell bestimmt sind.

Wir reproduzieren die absoluten Wirkungsquerschnitte f"ur~$\rh$,~$\om$ und $\Jps$; unser Resultat f"ur~$\ph$ ist um einen Faktor Zwei gr"o"ser.
Wir erkl"aren dies damit, da"s die Abh"angigkeit der~$\zet$-Verteilung von einer nichtverschwindenden Quarkmasse zu wenig verstanden ist (bevor sie f"ur die gro"se Masse des charm-Quarks wieder einfacher wird) und eine Definitive Aussage bez"uglich ihrer funktionalen Form nicht gemacht werden kann.
Der Faktor Zwei ist ein globaler; die $Q^2$-Abh"angigkeit des~$\ph$ wird korrekt reproduziert, wie wir finden im Vergleich mit den \VIER[]{Z}{E}{U}{S}-Daten f"ur das Verh"altnis von~$\ph$- zu $\rh$-Produktion, vgl.\@ Ref.~\cite{Derrick96a}.

Die Ausdehnung unseres Zugangs im folgenden Kapitel geschieht dahingehend, da"s noch spezifischer der String-String-Mechanismus des \DREI{M}{S}{V} getestet wird:
Wir erweitern den G"ultigkeitsbereich der Photon-Lichtkegelwellenfunktion hin zu kleiner bis verschwindender Virtualit"at.
Und wir betrachten h"oher angeregte Vektormesonen.
Es sind daher noch im "Uberlapp beider Wellenfunktionen Dipole involviert mit wesentlich gr"o"seren transversalen Ausdehnungen als bei den betrachteten $1S$-Vektormesonen; effektiv finden wir f"ur transversale Polarisation sp"urbare Beitr"age bis zu transversal~$2.5 \!-\! 2.8\fm$. \\
\indent
In Tabelle~\refg{Tabl:Q=0_rh,om,ph,Jps}, quasi als Nachtrag zu diesem Kapitel, geben wir Wirkungsquerschnitte der hier diskutierten Grundzustand-Vektormesonen an f"ur Photoproduktion.
Wir finden gute "Ubereinstimmung~-- auch f"ur das~$\ph(1020)$.
\theendnotes

%% file: EXCITED-F.tex
\lhead[\fancyplain{}{\sc\thepage}]
      {\fancyplain{}{\sc\rightmark}}
\rhead[\fancyplain{}{\sc{Photo- und Leptoproduktion von%
                         ~\protect$\rh(770)$,~\protect$\rh(1450)$ und~\protect$\rh(1700)$}}]
      {\fancyplain{}{\sc\thepage}}
\chapter[Photo- und Leptoproduktion von~\protect\bm{\rh(770)},~\protect\bm{\rh(1450)} und~\protect\bm{\rh(1700)}]{\vspace*{-.395ex}
   \huge Photo- und Leptoproduktion\\ von~\protect\bm{\rh(770)},~\protect\bm{\rh(1450)} und~\protect\bm{\rh(1700)}
}
\label{Kap:EXCITED}

Zentrales Moment der Untersuchungen dieses Kapitels ist, uns weiter "uber die grundlegenden Mechanismen klar zu werden, die in nichtperturbativer Quantenchromodynamik auf Quark-Gluon-Niveau agieren.
Wir betrachten als ganz wesentlich in dieser Hinsicht den String-String-Mechanismus, wie er Konsequenz ist der einfachen Annahmen des Modells des Stochastischen Vakuums: die Ausbildung gluonischer Strings zwischen den fundamentalen Partonen, aus denen wir uns einen hadronischen Zustand aufgebaut vorstellen, und deren Wechselwirkung.
Ziel ist es, die Relevanz dieses Mechanismus' weiter herauszupr"aparieren, als dies bereits geschehen ist.
Wir betrachten dazu Zust"ande gr"o"serer transversaler Ausdehnung, mit ausgedehnteren Strings. \\
\indent
In Termen der definierten Funktionen ist dies folgenderma"sen motiviert.
Die Funktion~$\pT\ellp(\zet,r,\tfbB)$ ist die~$T$-Amplitude f"ur die zugrundeliegende Streuung eines Loops, das hei"st Colour-Dipols~$\{\zet,\rb{r}\}$ an dem festen Protontarget.
F"ur festes~$\tfbB \!\cong\! \surd-t$ und~$\zet$ steigt sie mit der transversalen Dipolausdehnung~$r$ an wie~$r^{2n}$; dabei ist~$n \!\equiv\! 1$ an der Stelle~$r \!\equiv\! 0$ und f"allt langsam ab f"ur~$r \grgl 1 \!-\! 2\,a$ auf typische Werte~$n \!=\! 0.9 \!-\! 0.85$ f"ur mittlere~$r$.
Wesentliche Konsequenz des String-String-Mechanismus des \DREI{M}{S}{V} ist, da"s diese effektive Potenz zwar weiter~-- langsam~-- kleiner wird, aber~$\pT\ellp$ weiter ansteigt.
Dies ist wesentlicher Effekt des konfinierenden~$C$-Terms der eichinvarianten Zwei-Feldst"arken-Kumulante, vgl.\@ den Korrelationstensor~$D \!=\! (D_{\mu\nu\mu'\!\nu'})$, und nicht zu erreichen durch den~$N\!C$-Term allein.
Konkurrierende Modelle auf Basis perturbativer Quark-Quark-Wechselwirkung generieren Dipol-Proton-Amplituden entsprechend~$\pT\ellp$, die aber Quark-additiv sind und folglich von der Struktur unseres~$N\!C$-Terms; sie saturieren typischerweise f"ur~$r \!=\! 1 \!-\! 2\fm$, vgl.\@ etwa Ref.~\cite{Nikolaev91}.
Je gr"o"ser daher die transversale Ausdehnung~$r$ der Dipole ist, f"ur die~-- nicht unterdr"uckt durch den "Uberlapp der Photon- und Vektormeson-Wellenfunktionen~-- die Amplitude~$\pT\ellp$ aufgerufen wird, desto relevanter der String-String-Mechanismus und desto gr"o"ser die Diskrepanz der Postulate solcher Modelle zu unseren im Rahmen des \DREI[]{M}{S}{V}. \\
\indent
Hadronische Zust"ande gr"o"serer transversaler Ausdehnung sind im Rahmen von Photo- und Leptoproduktion von Vektormesonen elastisch am Proton involviert, wenn beider Wellenfunktionen "`breit"' sind:
Das einlaufende Photon mu"s tendenziell von m"oglichst kleiner Virtualit"at~$Q$ sein.
Das erzeugte Vektormeson mu"s in gleicher Weise eine "`breite"'\vspace*{-.5ex} Wellenfunktion besitzen.
Nur dann l"a"st ihr "Uberlapp~$\overlap{\la}$ Dipole gro"ser transversaler Ausdehnung~$r$ zu.
Das eine werden wir realisieren, indem wir die Lichtkegelwellenfunktion des Photons, wie sie zugrundeliegt der Analyse des vorangegangenen Kapitels, in ihrer G"ultigkeit ausdehnen auf kleinere bis verschwindende Virtualit"aten.
Das andere, indem wir h"ohere, "`breitere"' Resonanzen betrachten.
Wir nehmen vorweg, da"s in dem System mit den Quantenzahlen des~$\rh(770)$ als n"achsth"ohere Resonanz de~facto (mindestens) zwei Zust"ande so dicht beieinander liegen, da"s eine sie nur zusammen diskutiert werden k"onnen; wir bezeichnen diese mit~$\rh'$ und~$\rh^\dbprime$ und beziehen uns auf die Zust"ande~$\rh(1450)$ und~$\rh(1700)$ der Particle Data Group, Refn.~\cite{PDG98,PDG00}.
Nach Definition der Photon-Wellenfunktion f"ur kleine bis verschwindende Virtualit"aten, f"ur Definiertheit~$Q^2 \klgl 2\GeV^2$, im folgenden Abschnitt gehen wir ein auf das Rho-System im daran anschlie"senden, bevor wir uns zuwenden der numerischen Analyse ihrer Photo- und Leptoproduktion elastisch am Proton.
\vspace*{-2ex}

\section[Universelle Photon-Lichtkegelwellenfunktion
       ]{Universelle Photon-Lichtkegelwellenfunktion}
\label{Sect:Photon-Wfn_Q2klgl2}

Dosch, Gousset, Pirner gehen in Ref.~\cite{Dosch97} einen in gleicher Weise einfachen und effektiven Weg, die Lichtkegelwellenfunktion des Photons in ihrer G"ultigkeit auszudehnen auf kleinere bis verschwindende Virtualit"aten.
Wir geben einen Abri"s ihrer "Uberlegungen, die in Zusammenhang stehen mit den konventionellen Zug"angen zu den asymptotischen Bereichen gro"ser und kleiner~$Q^2$.
Sie f"uhren auf die konsistente Interpolation dieser Asymptotiken.
Auf eine universelle Photon-Lichtkekelwellenfunktion f"ur Virtualit"aten auch im Bereich~$Q^2 \klgl 2\GeV^2$.
\vspace*{-2ex}

\subsection{Zusammenhang}

Aufgrund Asymptotischer Freiheit wird die Asymptotik kleiner (Konstituenten-)Abst"ande, tendenziell gro"ser~$Q^2$ beschrieben durch die perturbative Theorie; die Photon-Lichtkegelwel\-lenfunktion des vorangegangenen Kapitels ist hergeleitet im Rahmen von LCPT.

Demgegen"uber wird die Asymptotik kleiner~$Q^2$ wesentlich determiniert durch Colour-Dipole von der Ausdehnung typischer Hadronen.
Dies suggeriert, die Photon-Wellenfunktion zu entwickeln in einer hadronischen Basis; f"ur transversale Polarisation:
\bea \label{Photon-Wfn_1S+Rest}
\ps_{\iga(Q^2,T)}\;
  =\; \sum_{V=\rh,\om,\ph} \frac{ef_VM_V}{M_V^2 \!+\! Q^2}\; \ps_{V(T)}\vv
        +\vv \psRest[(T)]
\eea
und f"ur longitudinale entsprechend durch Ersetzen~$ef_VM_V \!\to\! ef_VQ$ und~$\ps_{\#\!(T)} \!\to\! \ps_{\#\!(L)}$. \\
Dies ist die Entwicklung in $J^{P\!C} \!=\! 1^{--}$-hadronische Zust"ande.
Dabei sind die Grundzustand-Vektormesonen der drei leichten Flavour up, down, strange explizit ausgeschrieben im Sinne der Hypothese des Vektormeson-Dominanz-Modells~(VDM), da"s das Photon bei kleiner Virtualit"at im wesentlichen bestimmt ist durch die Resonanzen~$\rh(770)$,~$\om(782)$,~$\ph(1020)$.
Mit~$\psRest[(\la)]$ ist symbolisch bezeichnet die Summe der Wellenfunktionen s"amtlicher anderer $1^{--}$-Residuen: die h"oheren Anregungen von~$\rh$,~$\om$,~$\ph$ bez"uglich radialer und orbitalen Quantenzahlen und nicht-resonante Mehrteilchen-Zust"ande wie Multiquarkzust"ande und Hybride der Art Quark-Antiquark-Glue~$q\bar{q}g$, Quark-Antiquark-Glue-Glue~$q\bar{q}gg$~\ldots 

In Zusammenhang mit der Ausdehnung dieser Darstellung hin zu gr"o"seren~$Q^2$ ist von zentraler Bedeutung die Feststellung, da"s die Terme~$\psRest$ wesentlich das $Q^2$-Verhalten von Observablen mitbestimmen.
Dies wird exemplifiziert f"ur Leptoproduktion von Vektormesonen im Rahmen des vorangegangenen Kapitels und f"ur Compton-Streuung; die entsprechenden $T$-Amplituden schreiben sich, vgl.\@ die Gln.~(\ref{Thh_Tellp-coll}),~(\ref{dsigmadt-L,T}):
\begin{alignat}{3} \label{T_Lepto,Compton}
&T_\la\![\ga^{\scriptscriptstyle({\D\ast})}p \!\to\! Vp]&\;
  &=\; \int_0^{1\!/\!2}\zz d\zet\; \int_0^{\infty}\zz rdr&\vv
         \big[\overlap{\la}\big]\!(\zet,r)\vv
        &\pT\ellp(\zet,r,\tfbB)
    \\[.5ex]
&T_\la\![\ga^{\scriptscriptstyle({\D\ast})}p \!\to\! \ga^{\scriptscriptstyle({\D\ast})}p]&\;
  &=\; \int_0^{1\!/\!2}\zz d\zet\; \int_0^{\infty}\zz rdr&\vv
         \big[\gaga{\la}\big]\!(\zet,r)\vv
        &\pT\ellp(\zet,r,\tfbB)
    \tag{\ref{T_Lepto,Compton}$'$}
\end{alignat}
Dabei folgt die zweite Amplitude aus der ersten durch Ersetzen des auslaufenden Vektormesons durch ein Photon, dessen Virtualit"at bezeichnet sei mit~$Q'$.

Zum einen wird diskutiert die Asymptotik gro"ser~$Q^2$ bez"uglich des totalen Wirkungsquerschnitts von Photon-Proton-Streuung~$\si^{\rm tot} \!\equiv\! \si^{\rm tot}[\ga^{\D\ast}p]$ mit~$\si^{\rm tot} \!=\! \si^{\rm tot}_L \!+\! \si^{\rm tot}_T$.
Diese Asymptotik wird beschrieben durch die perturbative Theorie, die eine $T$-Amplitude der Form~(\ref{T_Lepto,Compton}$'$) generiert mit~$\pT\ellp$ der nicht n"aher bestimmten $T$-Amplitude f"ur den Austausch des Pomerons in der Streuung eines Loops(Colour-Dipols) am Proton.
Der totale Wirkungsquerschnitt~$\si_{(\la)}^{\rm tot}$ f"ur Polarisation~$\la \!\equiv\! L,T$ ist aufgrund des optischen Theorems genau der Integralausdruck von~(\ref{T_Lepto,Compton}$'$) in Vorw"artsrichtung~$\tfbB \!=\! 0$ f"ur~$Q^{\prime2} \!\equiv\! Q^2$.
Als Asymptotik gro"ser~$Q^2$ folgt dann mithilfe der universellen Relation~$\pT\ellp \!\propto\! r^2$ f"ur~$r \!\to\! 0$ f"ur den totalen Wirkungsquerschnitt~$\si^{\rm tot} \!\propto\! Q^{-2}$ bei Dominanz transversaler Polarisation.
Dem gegen"uber steht das VDM auf Basis der drei Grundzustand-Resonanzen~$\rh$,~$\om$,~$\ph$ allein:~$\si^{\rm tot}_L \!\propto\! Q^{-2}$ und~$\si^{\rm tot}_T \!\propto\! Q^{-4}$, das hei"st Dominanz longitudinaler Polarisation.
Der fehlende transversale inklusive Wirkungsquerschnitt mu"s daher resultieren aus~$\psRest[(T)]$.

Zum anderen wird diskutiert der integrierte Wirkungsquerschnitt f"ur die Leptoproduktion von Vektormesonen elastisch am Proton:~$\si^{\rm el} \!\equiv\! \si^{\rm el}[\ga^{\D\ast}p \!\to\! Vp]$ mit~$\si^{\rm el} \!=\! \si^{\rm el}_L \!+\! \si^{\rm el}_T$, vgl.\@ Gl.~(\ref{slope-Parameter_Thh}).
F"ur Virtualit"aten im Bereich~$Q^2 \!=\! 1\!-\! 2$ bis~$10\GeV^2$ wurde f"ur die Produktion der Grundzustand-Vektormesonen der drei leichten Flavour im vorangegangenen Kapitel auf Basis der Amplitude~(\ref{T_Lepto,Compton}) ein Verhalten wie~$\si^{\rm el} \!\propto\! Q^{-4}$ postuliert und experimentell best"atigt gefunden, vgl.\@ die Abbn.~\ref{Fig:Q4si_rh} und~\ref{Fig:Q4si_ph}; dabei stellten wir f"ur gr"o"sere~$Q^2$ fest eine zunehmende Dominanz longitudinaler Polarisation, vgl.\@ Abb.~\ref{Fig:dsigma-dt_rh}.
Hierzu in Kontrast folgt im Rahmen des VDM auf Basis der drei Grundzustand-Resonanzen~$\rh$,~$\om$,~$\ph$ allein:~$\si^{\rm el}_L \!\propto\! Q^{-2}$; wir stellen fest ein "uberschie"sen f"ur longitudinale Polarisation.
Diese Beitr"age m"ussen also gek"urzt werden durch~$\psRest[(L)]$.

Bei gr"o"seren~$Q^2$ k"onnen die Beitr"age von~$\psRest[(L)]$,~$\psRest[(T)]$ also nicht vernachl"assigt werden, de facto bereits nicht bei~$Q^2 \!\klgl\! 1\GeV^2$.
Einerseits mu"s der beobachtete subtile K"urzungsmechanismus der Art konsistent implementiert werden, da"s Resultat ist eine universelle Beschreibung des Photons in hadronischer Hochenergiestreuung.
Andererseits ist "uber die Zust"ande, die eingehen in~$\psRest$, quantitativ wenig bis "uberhaupt nichts bekannt.

Dies suggeriert zun"achst in einer Art gemischten Basis zu arbeiten: in der hadronischen f"ur kleine und in der Quark-Gluon-Basis f"ur mittlere bis gro"se~$Q^2$.
Dies ist prinzipiell m"oglich aber nicht dadurch erreicht,~$\psRest$ in Gl.~(\ref{Photon-Wfn_1S+Rest}) zu identifizieren mit der perturbativen Photon-Wellenfunktion.~--
Dies reproduzierte nicht die beobachtete (und in Ansatz gerade diskutierte) Ph"anomenologie, auch aus dem Grund, weil es zu unkontrolliertem {\it double counting\/} f"uhrte, das hei"st Quark-Gluon-Konfigurationen, die bereits impliziert sind in den Termen f"ur~$\rh$,~$\om$,~$\ph$, werden im perturbativen Anteil ein zweites Mal gez"ahlt.

Dosch, Gousset, Pirner gehen in Ref.~\cite{Dosch97} den dritten Weg, indem sie in einer reinen Quark-Gluon-Basis arbeiten.
Sie geben an die konsistente Fortsetzung der perturbativen Lichtkegelwellenfunktion des Photons zu kleinen bis verschwindenden Virtualit"aten; f"ur diese wird gefordert deren "Ubereinstimmung mit den drei expliziten Termen von Gl.~(\ref{Photon-Wfn_1S+Rest}) f"ur die Vektormesonen~$\rh(770)$,~$\om(782)$,~$\ph(1020)$.
Zentraler Punkt ihrer Herleitung ist, da"s die laufenden Quarkmassen der leichten Flavour der perturbativen Theorie:~\mbox{$m_u \!=\! m_d \!=\! 0$} und~\mbox{$m_s \!=\! 0.15\GeV$}, vgl.\@ Tabl.~\ref{Tabl:Charakt_rh,om,ph,Jps}, Fu"sn.~\mbox{\FN{FN:LaufendeQuarkmassen}}, ersetzt werden durch effektive Massen~$\meff(Q^2)$, $f \!=\! u,d,s$, die abh"angen von~$Q^2$ als der effektiven Skala, die bestimmt die Aufl"osung des Photons in der Streureaktion.
F"ur kleinerwerdende~$Q^2$ unterhalb~$Q^2 \!=\! 1 \!-\! 2\GeV^2$ steigen $\meff(Q^2)$ an von den laufenden, {\it bare\/} Werten~$m_f$ auf typische Konstituentenquark-Massen.

Eine effektive~$Q^2$-Abh"angigkeit der Quarkmassen der leichten Flavour in diesem Sinne wird in Ref.~\cite{Politzer76} nicht-perturbativ mittels einer Operator-Produkt-Entwicklung diskutiert und berechnet die Beitr"age aufgrund dynamischer, spontaner Brechung und aufgrund expliziter Brechung der~$SU\!(2) \otimes SU\!(2)$-{\it chiralen Symmetrie}.
Zus"atzliche Beitr"age werden erwartet aufgrund von {\it Confinement}.

%
\subsection{Transversaler Harmonischer Oszillator. Konstruktion}

Diese Ersetzung der laufenden Quarkmassen durch effektive,~$Q^2$-abh"angige ist motiviert aus der Analogie des transversalen (zweidimensionalen) Harmonischen Oszillators mit der Pho\-ton-Lichtkegelwellenfunktion in wesentlichen Charakteristika.
Wir zeichnen die Argumen\-tation von Dosch, Gousset, Pirner in Ref.~\cite{Dosch97} nach%
\FOOT{
  Wesentlich sind die grundlegende Analogie und die daraus motivierte Idee.   Wir stellen sie daher relativ ausf"uhrlich dar, auch auf die Gefahr hin einer gewissen Parallelit"at zu der pr"agnanten Darstellung dort.
}
hin zu deren Resultat f"ur~$\meff(Q^2)$ in parametrisierter Form.

Wir rekapitulieren, da"s unser Zugang zu Hochenergiestreuung im Rahmen des \DREI{M}{S}{V} we\-sentlich determiniert wird durch die transversalen Ausdehnungen der involvierten Colour-Dipole.
In Ref.~\cite{Dosch94} wurde weiche Hochenergiestreuung, in erster Linie \mbox{Proton-Proton-Streu}\-ung diskutiert auf Basis von Wellenfunktionen mit nur einem Parameter: n"amlich f"ur ihre transversale Ausdehnung, Spin-Freiheitsgrade wurden vollst"andig vernachl"assigt; vgl.\@ Gl.~(\ref{transversaleWfn-allg}).
Es wurde dennoch gute "Ubereinstimmung mit dem Experiment gefunden.

In diesem Sinne: als approximative Untersuchung wesentlicher Charakteristika betrachten wir~-- unter Vernachl"assigung der Spin-Freiheitsgrade~-- den transversalen Konfigurationsanteil der perturbativen Photon-Lichtkegelwellenfunktion:%
\FOOT{
  Die Diskussion bleibt qualitativ dieselbe bei Einbeziehung des Terms~$\propto\! {\rm K}_1$ f"ur transversale Polarisation.
}
%
\bea \label{Photon-Wfn_transv}
\ps_\iga(\zet,\rb{r})\;
  \propto\; \int \frac{d^2\rb{k}}{(2\pi)^2}\,
                 \frac{\efn{\D \iIM\,\rb{k}\rb{r}}}{\rb{k}^2 \!+\! \vep^2}\;
  =\;       \frac{{\rm K}_0(\vep r)}{2\pi} \qquad
  \text{mit}\quad
  \vep = \sqrt{\zbz\, Q^2 + m_f^2} \quad
  r = |\rb{r}|
\eea
vgl.\@ die Gln.~(\ref{Photon-Wfn}),~(\ref{Photon-Wfn}$'$) bzw.\@ Anh.~\ref{APPSect:Photon-Wfn}.
Ihre transversle Ausdehnung ist also beschr"ankt auf~$r \klgl \vep^{-1}$ und sie ist ad"aquate Beschreibung f"ur gro"se~$\vep$.
F"ur kleine~$\vep$ wirken Brechung der chiralen Symmetrie und Confinement zu gro"sen Ausdehnungen~$r$ entgegen.
Dies kann anschaulich so betrachtet werden:
Bei hohen Energien, weit entfernt vom Target, sollte die perturbative Theorie freier Quark-Antiquark-Paare g"ultig sein; aber f"ur kleine~$Q^2$ ist die Zeit ausreichend, da"s sich diese im Photon formieren zu gebundenen hadronischen Zust"anden.

Diesem Bild eines Quark-Antiquark-(Colour)Dipols in der Quantenchromodynamik entspricht das zweier Teilchen im Harmonischen Oszillatorpotential der Quantenmechanik in zweierlei Hinsicht.
Zum einen treten nicht auf gro"se transversale Separationen aufgrund des Potentials, das stark mit diesen ansteigt; zum anderen verschwindet es f"ur kurze Zeiten und verschwindende Separation, und ist die Dynamik vollst"andig bestimmt durch die kinetische Energie.~--
Dies ist genau Confinement beziehungsweise Asymptotische Freiheit.
In diesem Sinne interpretiert wird die Greensche Funktion des transversalen Harmonischen Oszillators:
\bea \label{GreenFn_HarmOsz}
G(\rb{r},\bm{0},-E)\;
  =\; \sum_{\rb{n}=(n^1,n^2)}
        \frac{\ps_\rb{n}^{\D\ast}(\rb{r}) \ps_\rb{n}(\bm{0})}{E_\rb{n} \!-\! E}\qquad
  \text{mit}\quad
  E_\rb{n} = (n^1 \!+\! n^2 \!+\! 1)\, \om
\eea
Die Wellenfunktion~$\ps_\rb{n}(\bm{0})$ steht f"ur den harten Proze"s der Produktion des Quark-Antiquark-Paares weit entfernt vom Target und die Wellenfunktion~$\ps_\rb{n}^{\D\ast}(\rb{r})$ bestimmt dessen transversale Separation.
Die Beschr"ankung auf kleine Zeiten ist realisiert durch gro"se negative Energien~\mbox{$E \!=\! -M$}; gro"se positive~$M$ entsprechen dabei dem tief-Euklidischen Bereich der QCD.
F"ur kleine negative Werte~$E \!=\! -M$ geht~\mbox{$G(\rb{r},\bm{0},-E) \!\to\! G(\rb{r},\bm{0},M)$} "uber in die freie zweidimensionale Greensche Funktion der Quantenmechanik:
\bea \label{GreenFn_frei}
G_0(\rb{r},\bm{0},M)\;
  =\; \int \frac{d^2\rb{k}}{(2\pi)^2}\,
           \frac{\efn{\D \iIM\,\rb{k}\rb{r}}}{\rb{k}^2\!/(2m) \!+\! M}\;
  =\;      2m\cdot \frac{{\rm K}_0(\sqrt{2mM} r)}{2\pi}      
\eea
Die Analogie zu der Photon-Lichtkegelwellenfunktion~$\ps_\iga(\zet,\rb{r})$, vgl.\@ Gl.\,(\ref{Photon-Wfn_transv}), ist frappant.
Die Korrespondenz~\mbox{$\surd 2mM \!\leftrightarrow\! \vep \!=\! \surd\zbz Q^2 \!+\! m_f^2$} verdeutlicht den Zusammenhang der Skalen~$M$ und~$Q^2$.

Auf Basis dieser Analogie ist es m"oglich sich im Studium der Photon-Lichtkegelwellen\-funktion zunutze zu machen, da"s f"ur den Harmonischen Oszillator eine Vielzahl von Resultaten analytisch zur Verf"ugung stehen oder in beliebiger Genauigkeit einfach berechnet werden k"onnen, vgl.\@ etwa Ref.~\cite{Grawert89}.
"`Approximationen"' k"onnen dahingehend untersucht werden, ob sie die exakten Resultate tats"achlich approximieren oder deren wesentliche Charakteristika nicht reproduzieren.

Auf diese Weise gelangen Dosch, Gousset, Pirner zun"achst zu zwei bedeutenden Feststellungen f"ur den Harmonischen Oszillator:
Sie finden zum einen, da"s die Determinierung von Momenten der exakten Greenfunktion mit der Grundzustand-Wellenfunktion~-- die in analoger Form f"ur die Lichtkegelwellenfunktion des Photons wesentlich eingehen in die Streuamplitude~-- einem subtilen K"urzungsmechanismus von Beitr"agen unterworfen ist, die r"uhren von verschiedenen h"oheren Anregungen.
Um die exakte Greenfunktion~$G(\rb{r},\bm{0},M)$ in der Darstellung der unendlichen Reihe nach Gl.~(\ref{GreenFn_HarmOsz}) zu approximieren durch eine entsprechende endliche Summe werden sehr viele Terme ben"otigt.
Dosch, Gousset, Pirner finden zum anderen, da"s zur Approximation von~$G$ sehr viel effektiver ist, auszugehen von der freien Greensche Funktion~$G_0(\rb{r},\bm{0},M)$, vgl.\@ Gl.~(\ref{GreenFn_frei}), und diese zu modifizieren durch Verschieben ihres Arguments entsprechend~\mbox{$\surd2mM \!\to\! s_0 \!+\! \surd2mM$}, mit~$s_0 \!\equiv\! s_0(M)$.

Beide Feststellungen lauten in Gegen"uberstellung:
Schon bei mittleren Werten von~$M$ wie~$M \!\cong\! 5\om$ werden mehr als zwanzig Terme in der Entwicklung~(\ref{GreenFn_HarmOsz}) ben"otigt, um eine bessere Approximation von~$G$ zu erreichen, als die blo"se freie Greensche Funktion~$G_0$ darstellt~-- noch ohne Verschieben ihres Arguments.
Verschiebung des Arguments mit eingeschlossen, gelingt eine sehr gute Approximation bis zu verschwindenden~$M$.

Aufgrund dieser Feststellungen wird ein Verfahren angegeben, das in konsistenter Weise die Verschiebung~$s_0$ bestimmt als Funktion von~$M$.
Das Verfahren wird so eingerichtet~-- die exakte Greensche Funktion~$G$ ist ja bekannt f"ur den Harmonischen Oszillator der Quantenmechanik, nicht aber~$\ps_\iga$ f"ur die QCD, vgl.\@ die Gln.~(\ref{Photon-Wfn_transv}),~(\ref{GreenFn_HarmOsz})~--, da"s es in gleicher Weise anwendbar ist auf den einen, wie auf den anderen Fall.
Es wird daher die Approximation: freie Greensche Funktion mit verschobenem Argument, nicht verglichen mit der exakten Greenschen Funktion.
Sondern mit einem ph"anomenologischen Ansatz f"ur diese, der in gleicher Weise gemacht werden kann f"ur den Harmonischen Oszillator und das Quark-Antiquark-Photon der QCD.
Dieser ist motiviert aus den QCD-Summenregeln nach Shifman, Vainshtein, Zakharov, vgl.\@ Ref.~\cite{Shifman79}, und lautet:~"`eine Resonanz plus perturbatives Kontinuum"'.
Er bezieht sich auf Funktionen, die in gleicher Weise definiert werden k"onnen f"ur den einen wie f"ur den anderen Fall.
F"ur den Harmonischen Oszillator geben wir diese nur an:
\bea \label{Pi_HarmOsz}
\Pi^{(n)}(M)\;
  \equiv\; \frac{d^n}{d(-M)^n}\, \Pi(M)\qquad
  \text{mit}\quad
  \Pi(M)\; \stackrel{\D!}{=}\; G(\bm{0},\bm{0},M)
\eea
dabei garantiert der Zusammenhang mit der exakten Greenschen Funktion~$G$ Anbindung an das "`exakte"' Resultat.

\bigskip\noindent
Wir gehen unmittelbar "uber zum Fall der Photon-Lichtkegelwellenfunktion der Quantenchromodynamik, den wir im folgenden skizzieren.

Die~$\Pi(M)$ entsprechende Funktion der QCD ist die skalare {\it Vakuum-Polarisationsfunktion\/} $\Pi(q^2,m_f^2)$; vgl.\@ generell auch Ref.~\cite{Shifman79}.
Diese ist definiert "uber den Polarisationstensor des Vektorstroms~$J_{\!(f)} \!=\! (\bar{\ps}_f \ga^\mu \ps_f)$ eines Quarks mit leichtem Flavour~$f \!=\! u,d,s$ und laufender Masse~$m_f$ (nicht impliziert in dieser Definition ist dessen Ladung~$e_f \!=\! +2\!/\!3 e, -1\!/\!3 e$):
\vspace*{-1ex}
\begin{align}
\Pi^{\mu\nu}(q,m_f^2)\;
  &=\; \iIM\; \int d^4x\, \efn{\D iqx}\;
           \bra{\Om} T \big\{ J_{\!(f)}^\mu\!(x), J_{\!(f)}^\nu\!(0) \big\} \ket{\Om}
    \\
  &=\; (q^\mu q^\nu \!-\! q^2 g^{\mu\nu})\vv \Pi(q^2,m_f^2)
    \nn
    \\[-4ex]\nn
\end{align}
Der konstante Term der Funktion~$\Pi(q^2,m_f^2)$ ist nicht definierbar, s.u.\@ Gl.~(\ref{I0xi_la-explizit}).
F"ur~gro\-"se~$q^2$ wird ihr Imagin"arteil~${\rm Im} \Pi_0(s,m_f^2)$ in niedrigster Ordnung der perturbativen Theorie repr"asentiert und berechnet durch einen einzigen Feynman-Graphen, der steht f"ur die Erzeugung, {\it freie\/} Propagation und anschlie"sende Vernichtung des Quark-Antiquark-Paares mit Flavour~$f$.%
\FOOT{
  Bzgl.\@ des n"achstf"uhrenden Terms~${\rm Im} \Pi_1(s,m_f^2)$ proportional~$\al_s$ auf Basis zweier zus"atzlicher Graphen, die den Austausch jeweils eines Gluons beinhalten, vgl.\@ etwa Ref.~\cite{Shifman79}.
}
Die erste partielle Ableitung von~$\Pi(-Q^2,m_f^2)$ nach~$-Q^2 \!=\! q^2$, s.u.\@ Gl.~(\ref{Pi(n)}), wird "ublicherweise angegeben in Form der Dispersionsrelation:
\vspace*{-.5ex}
\begin{align} \label{Pi0(1)-DispersRel}
&\Pi_0^{(1)}(Q^2,m_f^2)\;
   =\; \frac{1}{\pi}\;
         \int_0^\infty ds\vv
         \frac{{\rm Im}\, \Pi_0(s,m_f^2)}{(s \!+\! Q^2)^2}
    \\[1ex]
&\text{mit}\qquad
  \begin{aligned}[t] \label{ImPi0}
  {\rm Im}\, \Pi_0(s,m_f^2)\;
    &=\; \frac{\Nc}{12\pi}\, \frac{v(3 \!-\! v^2)}{2}\; \th(s \!-\! 4m_f^2) \qquad
           v = \sqrt{1 \!-\! 4m_f^2 \!/s}
    \\[-1ex]
    &=\; \frac{\Nc}{12\pi}\,
           \Big[ 1 \!+\! \frac{2m_f^2}{s} \Big] \sqrt{1 \!-\! \frac{4m_f^2}{s}}\vv
           \th(s \!-\! 4m_f^2)
  \end{aligned}
    \\[-4.5ex]\nn
\end{align}
Dabei ist~$v$ die Geschwindigkeit des Quarks; die Theta-Funktion schr"ankt ein auf Werte oberhalb der kinematischen Schwelle.
Der Quark-Vektorstrom~$J_{\!(f)}$ ist Teil des elektromagnetischen Stroms~$J_{\rm em}$; der hier auftretende Phasenraumfaktor entspricht daher dem in Gl.~(\ref{fV_Gall}).
Wir schreiben dimensionslos:%
\FOOT{
  Die gestrichenen Gleichungen im folgenden sind Darstellungen in Termen dimensionsloser Variablen; Verdeutlichung in Hinsicht auf formale Umformungen, seien sie ignoriert beim weniger strengen Lesen.
}
%
\vspace*{-.5ex}
\begin{align}
\phantom{\text{mit}\qquad }
{\rm Im}\, \tilde{\Pi}_0(\la,m_f^2)\;
  &=\; \frac{\Nc}{12\pi}\,
         \Big[ 1 \!+\! \frac{1}{2\la} \Big] \sqrt{1 \!-\! \frac{1}{\la}}\vv
         \th(\la \!-\! 1)
  \qquad
  \text{mit}\quad
  \la \equiv s\!/ 4m_f^2
    \tag{\ref{ImPi0}$'$} \\
  &=\; {\rm Im}\, \Pi_0(4m_f^2 \!\cdot\! \la,m_f^2)
    \nn
    \\[-4.5ex]\nn
\end{align}
Die Dispersionsrelation~(\ref{Pi0(1)-DispersRel}) gilt zu beliebiger Ordnung in~$\al_s$ und kann verallgemeinert werden auf h"ohere Ableitungen.
Wir definieren die $n$-ten partiellen Ableitungen nach~$-Q^2$ beziehungsweise~$-\xi \!\equiv\! -Q^2 \!/\! 4m_f^2$:
\begin{alignat}{4} \label{Pi(n)}
&\Pi^{(n)}(Q^2,m_f^2)&\;
  &\equiv&\; \Big[-\frac{\pa}{\pa Q^2}\Big]^n\, &\Pi(Q^2,m_f^2)&
    &\\
&\tilde{\Pi}^{(n)}(\xi,m_f^2)&\;
  &\equiv&\; \Big[-\frac{\pa}{\pa\xi}\Big]^n\, &\tilde{\Pi}(\xi,m_f^2)&
  &=\; \big(4m_f^2\big)^{\!n}\cdot\;
         \Pi^{(n)}(4m_f \!\cdot\! \xi,m_f^2)
    \tag{\ref{Pi(n)}$'$}
\end{alignat}
und formulieren allgemein die Dispersionsrelationen:
\begin{alignat}{4} \label{Pi(n)_DispersRel}
&\Pi^{(n)}(Q^2,m_f^2)&\;
   &=&\; \frac{n!}{\pi}\;
         \int ds\vv
         &\frac{{\rm Im}\, \Pi(s,m_f^2)}{(s \!+\! Q^2)^{n+1}}&
  \quad &n \!\ge\! 1
    \\[1ex]
&\tilde{\Pi}^{(n)}(\xi,m_f^2)&\;
   &=&\; \frac{n!}{\pi}\;
         \int d\la\vv
         &\frac{{\rm Im}\,
          \tilde{\Pi}(\la,m_f^2)}{(\la \!+\! \xi)^{n+1}}&
  \qquad &n \!\ge\! 1
  \qquad \text{mit}\quad
  \la \equiv s\!/ 4m_f^2
    \tag{\ref{Pi(n)_DispersRel}$'$}
\end{alignat}
Aufgrund Asymptotischer Freiheit besitzen die Funktionen~$\Pi^{(n)}$ f"ur~$n \!\ge\! 0$ im Euklidischen Bereich eine sehr gute phenomenologische Darstellung der Art "`eine Resonanz plus perturbatives Kontinuum"', vgl.\@ Ref.~\cite{Shifman79}:
\begin{align} \label{Pi-ph(n)}
\Pi_{\rm ph}^{(n)}(Q^2,m_f^2)\;
  =\; \frac{n! f_V^2}{(M_V^2 \!+\! Q^2)^{n+1}}\;
        +\; \frac{n!}{\pi}\;
            \int_{{\D s}_{\!t}}^\infty ds\vv \frac{{\rm Im}\, \Pi_0(s,m_f^2)}{(s \!+\! Q^2)^{n+1}} 
  \qquad n \!\ge\! 1
\end{align}
mit~$s_t$ der Schwelle, an der das perturbative Kontinuum einsetzt.
Als Resonanz tritt auf das Grundzustand-Vektormesonen~$V \!\equiv\! \rh(770)$, $\om(782)$ oder~$\ph(1020)$, das dem Quark bez"uglich Isospin und Flavour-Gehalt entspricht; vgl.\@ Gl.~(\ref{Flavour-Gehalt}) und Tabl.~\ref{Tabl:Charakt_rh,om,ph,Jps}.
Dabei bezeichnen~$M_V$ seine Masse und~$f_V$ seine Kopplung an den elektromagnetischen Strom, vgl.\@ Gl.~(\ref{fV-Def}).

In niedrigster Ordnung der perturbativen Theorie ist die Kopplung~$f_V$ proportional zu der Wellenfunktion des Vektormesons~$V$ am Ursprung; sie tritt daher in Gl.~(\ref{Pi-ph(n)}) auf als Residuum des Resonanz-Pols.
Aufgrund Quark-Hadron-Dualit"at besteht f"ur dieses Residuum die Relation:
%
\bea \label{fV_st}
f_V^2\;
  =\; \frac{1}{\pi}\vv \int_{0}^{{\D s}_{\!t}} ds\vv {\rm Im}\, \Pi_0(s,m_f^2)
    \\[-3.5ex]\nn
\eea
und mithilfe von Gl.~(\ref{ImPi0}$'$):
\vspace*{-.5ex}
\begin{align}
\frac{\pi f_V^2}{4m_f^2}\;
  &=\; \int_0^{\la_t} d\la\vv
        {\rm Im}\, \Pi_0(4m_f^2 \!\cdot\! \la,m_f^2) \qquad
  \text{mit}\qquad
  \la_t \equiv s_t\!/ 4m_f^2
    \tag{\ref{fV_st}$'$} \\[-.5ex]
  &=\; \frac{\Nc}{12\pi}\;
         \int_1^{\la_t} d\la\;
         \Big[ 1 \!+\! \frac{1}{2\la} \Big] \sqrt{1 \!-\! \frac{1}{\la}}\vv
  =\vv \frac{\Nc}{12\pi}\;
         \la\, \sqrt{1 \!-\! \frac{1}{\la}}\vv
         \bigg|_1^{\la_t}
    \nn
    \\[-4.5ex]\nn
\end{align}
Hierdurch folgt die Schwelle~$s_t$, an der das \vspace*{-.375ex}perturbative Kontinuum einsetzt.
Die Ableitungen~$\Pi_{\rm ph}^{(n)}$,~$n \!\ge\! 1$, sind damit wohldefiniert durch Gl.~(\ref{Pi-ph(n)}); sie stellen die phenomenologische, "`exakte"' Seite der Gleichung dar, durch die $\meff$ bestimmt wird.

\bigskip\noindent
Die "`approximierte"' Seite der Gleichung ist gegeben durch die approximierten Funktionen~$\Pi_{\rm app}^{(n)}$; diese werden identifiziert mit den {\it freien Funktionen~$\Pi_0^{(n)}$ mit verschobenem Argument\/}, vgl.\@ die Gln.~(\ref{ImPi0}),~(\ref{Pi(n)_DispersRel}) bzw.~(\ref{ImPi0}$'$),~(\ref{Pi(n)_DispersRel}$'$).

In welcher Weise explizit diese Verschiebung des Arguments durchzuf"uhren ist, folgt einerseits aus der entsprechenden Verschiebung~\mbox{$\surd 2mM \!\to\! s_0(M) \!+\! \surd 2mM$} im Fall des Harmonischen Oszillators, andererseits durch Identifizierung und Zusammenhang der relevanten Skalen~$M$ und~$Q$ aufgrund der Korrespondenz~\mbox{$\surd 2mM \!\leftrightarrow\! \vep \!=\! \surd\zbz Q^2 \!+\! m_f^2$} zwischen dem Harmonischen Oszillator der Quantenmechanik und der Photon-Lichtkegelwellenfunktion der QCD.
Vgl.\@ die Diskussion in Anschlu"s an Gl.~(\ref{GreenFn_frei}).
Die Verschiebung des Arguments in diesem Sinne kann subsumiert werden in
\begin{alignat}{4} \label{vep_m-to-meff}
&\ps_\iga(\zet,\rb{r})\;
  \propto&\; {\rm K}_0(\sqrt{\zbz Q^2 \!+\! m_f^2}&&\; |\rb{r}|)\vv
  &\longrightarrow&\vv
             {\rm K}_0(\sqrt{\zbz Q^2 \!+\! \meff^2}&\; |\rb{r}|)
    \\[-.5ex]
&&\qquad     m_f&&\vv
  &\longrightarrow&\vv
             \meff&
    \tag{\ref{vep_m-to-meff}$'$}
\end{alignat}
das hei"st im Ersetzen der laufenden durch eine effektive Quarkmasse.
Zwei F"alle werden unterschieden:
\vspace*{-.5ex}
\begin{alignat}{4} \label{meff-cases}
&\text{(i)}&\qquad
  &\meff&\; &=&\; &\meff(Q^2)
    \\
&\text{(ii)}&\qquad
  &\meff&\; &=&\; &\meff(Q_{\rm eff}^2 \!\equiv\! 4\zbz Q^2)
    \tag{\ref{meff-cases}$'$}
\end{alignat}
Das hei"st~$\meff$ wird aufgefa"st als Funktion von~$Q^2$ oder von~$Q_{\rm eff}^2 \!\equiv\! 4\zbz Q^2$, dem Quadrat der effektiven Virtualit"at des Photons, die abh"angt von~$\zet$, dem Anteil des Quarks am Gesamt-Lichtkegelimpuls; vgl.\@ die Diskussion der transversalen Photon-Wellenfunktion in Anschlu"s an Gl.~(\ref{epsilon}) auf Seite~\pageref{T:Photon-Wfn}.

\noindent
Auf dieser Basis folgt als die "`approximierte"' Seite der Gleichung:
\bea \label{Pi-app(n)}
\Pi_{\rm app}^{(n)}(Q^2,\meff^2)\;
  \equiv\; \Pi_0^{(n)}(Q^2,m_f^2 \!\to\! \meff^2)
\eea
Und die Gleichung selbst, durch die~$\meff$ bestimmt wird in Abh"angigkeit der rein hadronischen Gr"o"sen~$M_V$ und~$f_V$:
\begin{align} \label{meff-Best.Gl}
&\Pi_{\rm ph}^{(n)}(Q^2,m_f^2)\;
  =\; \Pi_{\rm app}^{(n)}(Q^2,\meff^2)
    \\[.5ex]
&\frac{n! f_V^2}{(M_V^2 \!+\! Q^2)^{n+1}}\;
    +\; \frac{n!}{\pi}\;
        \int_{{\D s}_{\!t}}^\infty ds\vv
        \frac{{\rm Im}\, \Pi_0(s,m_f^2)}{(s \!+\! Q^2)^{n+1}}
    \tag{\ref{meff-Best.Gl}$'$} \\
&\phantom{\Pi_{\rm ph}^{(n)}(Q^2,m_f^2)\;}  
  =\; \frac{n!}{\pi}\;
        \int_{4\meff^2}^\infty ds\vv
        \frac{{\rm Im}\, \Pi_0(s,\meff^2)}{(s \!+\! Q^2)^{n+1}} \qquad
  n \ge 1
    \nn
\end{align}
vgl.\@ die Gln.~(\ref{Pi(n)_DispersRel}),~(\ref{Pi-ph(n)}); auf der rechten Seite ist die untere Integrationsgrenze retundant ausgeschrieben, insofern sie bereits impliziert ist in~${\rm Im}\,\Pi_0$, vgl.\@ Gl.~(\ref{ImPi0}).
Wir betonen, da"s die Ableitungen nach Definition in Gl.~(\ref{Pi(n)}) {\it partielle\/} Ableitungen sind und daher nicht wirken auf die~$Q^2$-Abh"angigkeit implizit der effektiven Quarkmasse~$\meff$.

\hspace*{-2pt}Wir betrachten die Dispersionsintegrale in Gl.\,(\ref{meff-Best.Gl}$'$) und identifizieren mithilfe~Gl.\,(\ref{Pi(n)}$'$) als relevantes Integral bez"uglich der $n$-ten Ableitungen:
\vspace*{-.5ex}
\begin{align} \label{I0xi_la-Def}
&\big(4m^2\big)^{\!n}\cdot\; \Pi_\#^{(n)}(4m \!\cdot\! \xi,m^2)\; \Big|_{\rm Integral}
  =\; \tilde{\Pi}_\#^{(n)}(\xi,m^2)\; \Big|_{\rm Integral}
    \\
&=\vv \Big[-\frac{\pa}{\pa\xi}\Big]^n\, \tilde{\Pi}_\#(\xi,m^2)\; \Big|_{\rm Integral}
    \nn \\
&=\vv \frac{\Nc}{12\pi}\frac{n!}{\pi}\cdot\,
        \int_{\la_t,1}^\infty d\la\; \frac{1}{(\la \!+\! \xi)^{n+1}}\,
        \left[1 \!+\! \frac{1}{2\la}\right]\, \sqrt{1 \!-\! \frac{1}{\la}}\vv
  \stackrel{\D!}{=}\vv
    \frac{\Nc}{12\pi^2}\cdot\,
    I_n(\xi,\la)\; \Big|_{\la_t,1}^\infty
    \nn
    \\[-4.5ex]\nn
\end{align}
mit~\mbox{$\xi \!=\! Q^2 \!/\! 4m^2$} und, f"ur die linke Seite von Gl.~(\ref{meff-Best.Gl}$'$),~$\# \!\equiv\! {\rm ph}$~[bzw.\@ rechte Seite,~\mbox{$\# \!\equiv\! {\rm app}$}]: \mbox{$m \!\equiv\! m_f$\,[bzw.~$\meff$]} und die untere Integrationsgrenze gleich~\mbox{$\la_t \!\equiv\! s_t\!/\!4m_f^2$~[bzw.~$1$]}.
Die Funktion~$I_n(\xi,\la)$ ist definiert als die Stammfunktion des Integranden.%
\FOOT{
  Das in Gl.~(\ref{fV_st}$'$) angegebene Integral kann formal aufgefa"st werden als~$I_{-1}$.
}
Wir berechnen~$I_n(\xi,\la)$ f"ur m"oglichst kleinen Index~$n \!\in\! \bbbn_0$; alle h"oheren Dispersionsintegrale, das hei"st bez"uglich indizes~$n' \!>\! n$ folgen dann durch $(n' \!-\! n)$-malige Differentiation.
Der Index~$n \!\equiv\! 0$ ist zul"assig:
Zwar divergiert das Integral logarithmisch an der oberen Grenze, wie folgt aus der Absch"atzung des Integranden f"ur gro"se Argumente~$\la$; es kann aber regularisiert und renormiert werden.
Wir tun dies im folgenden.
Zun"achst finden wir:
\vspace*{-.5ex}
\begin{align} \label{I0xi_la-explizit}
I_0(\xi,\la)\;
&=\;
  \ln\,\la\vv
  +\vv 2\, \ln\Big[ 1 + \sqrt{1 \!-\! \frac{1}{\la}} \Big]\vv
    \\[-1ex]
&\quad
  -\vv \frac{1}{\xi}\, \sqrt{1 \!-\! \frac{1}{\la}}
    \nn \\
&\quad
  -\vv \left[1 \!-\! \frac{1}{2\xi}\right]\, \sqrt{1 \!+\! \frac{1}{\xi}}\vv
       \Bigg\{
         \ln\left[ \left.
              \!\left(\sqrt{1 \!+\! \frac{1}{\xi}} + \sqrt{1 \!-\! \frac{1}{\la}}\right)
              \zz\right/\zz
              \left(\sqrt{1 \!+\! \frac{1}{\xi}} - \sqrt{1 \!-\! \frac{1}{\la}}\right)\!
            \right]
    \nn \\
&\phantom{\quad \vv \left[1 \!+\! \frac{1}{2\xi}\right]\, \sqrt{1 \!-\! \frac{1}{\xi}}\vv \Bigg\{}
  +\; \ln\left[ \xi^2\, \left[1 \!-\! \frac{1}{2\xi}\right]\,
                       \left(1 \!+\! \frac{1}{\xi}\right)^{\zz3\!/\!2}
         \right]
       \Bigg\}
    \nn
    \\[-4.5ex]\nn
\end{align}

Der erste Term~$\ln\la$ ist verantwortlich f"ur die Divergenz des Integrals an der oberen Integrationsgrenze, Regularisierung geschieht durch seine Subtraktion.
Da er nicht abh"angt von~$\xi$, "andert diese Regularisierung nicht den Wert der h"oheren Ableitungen nach~$\xi$, das hei"st deren Renormierung ist trivial.
F"ur die nicht-abgeleitete Funktion selbst ist aufgrund seiner Unabh"angigkeit von~$\xi$ die Subtraktion dieselbe auf der rechten und linken Seite von Gl.~(\ref{meff-Best.Gl}$'$); und somit irrelevant auch in diesem Fall.
Aufgrund derselben Argumentation kann der zweite Term\;~$2\ln[1 \!+\! \surd1 \!-\! 1\!/\!\la]$ als konstant subtrahiert werden.
Schlie"slich der Term in der letzten Zeile ist unabh"angig von~$\la$ und verschwindet bei Differenzbildung bez"uglich der oberen und unteren Integrationsgrenze; er kann daher o.E.d.A.\@ vorab subtrahiert werden.
Sei in diesem Sinne definiert das renormierte Integral~$I_0^{\rm ren}(\xi,\la)$ als die verbleibende {\it zweite und dritte Zeile von\/} Gl.~(\ref{I0xi_la-explizit}).

Das Dispersionsintegral folgt dann als die Differenz~$I_0^{\rm ren}(\xi,\la \!\equiv\! \infty) \!-\! I_0^{\rm ren}(\xi,\la \!\equiv\! \la_t~\text{bzw.}~1)$, vgl.\@ Gl.~(\ref{I0xi_la-Def}).
Auf der "`approximierten"' Seite von Gl.~(\ref{meff-Best.Gl}$'$) verschwindet die Funktion~$I_0^{\rm ren}$ identisch an der untere Integrationsgrenze~$\la \!\equiv\! 1$ und wir gelangen zu dem Ausdruck, den Dosch, Gousset, Pirner in Ref.~\cite{Dosch97} angeben und ihrer Analyse zugrunde legen:
\vspace*{-.5ex}
\begin{align} \label{nPi0-DGP}
&\big(4\meff^2\big)^{\!n}\cdot\; \Pi_{\rm app}^{(n)}(4\meff \!\cdot\! \xi,\meff^2)\;
  =\; \tilde{\Pi}_{\rm app}^{(n)}(\xi,\meff^2)
    \\
&=\; \Big[-\frac{\pa}{\pa \xi}\Big]^n\vv
       \frac{\Nc}{12\pi^2}\cdot\, I_0^{\rm ren}(\xi,\infty)
    \nn \\
&=\; \Big[-\frac{\pa}{\pa \xi}\Big]^n\vv
       \frac{\Nc}{12\pi^2}\, \left\{
         - \frac{1}{\xi}\;
         -\; \Big[1 \!-\! \frac{1}{2\xi}\Big]\, \sqrt{1 \!+\! \frac{1}{\xi}}\vv
               \ln\left[\! \left. \left(\sqrt{1 \!+\! \frac{1}{\xi}} + 1\right) \zz\right/\zz
                                  \left(\sqrt{1 \!+\! \frac{1}{\xi}} - 1\right) \!\right]
      \right\}
    \nn
    \\[-4.5ex]\nn
\end{align}
mit~$\xi \!\equiv\! Q^2 \!/\! 4\meff^2$.

Auf dieser Basis bestimmen sie f"ur die leichten Flavour up/down und strange, die funktionale Abh"angigkeit der effektiven Quarkmassen~$\meff[u\!/\!d,]$ und~$\meff[s,]$, indem sie annehmen: eine Abh"angigkeit (i)~von~$Q^2$ oder (ii)~von~\mbox{$Q_{\rm eff}^2 \!\equiv\! 4\zbz Q^2$}, vgl.\@ die Gln.~(\ref{meff-cases}),~(\ref{meff-cases}$'$).

\begin{figure}
\begin{minipage}{\linewidth}
  \begin{center}
  \setlength{\unitlength}{1mm}\begin{picture}(140,43)
  \put(7,1){   
  \makebox(66,38.5){
  \begin{picture}(66,38.5)
    \put(0,0){\epsfxsize66mm \epsffile{FIGURES-F/Figure-4-1a.eps}}
    \put(-6, 0  ){\yaxis[38.5mm]{\normalsize $\meff(Q^2)\;[\MeV[]]$}}
    \put(47,-3.5){\normalsize$Q^2\;[\GeV[]^2]$}
    \put(0,40){(a)\vv up/down\vv ($m_{u\!/\!d} \!=\! 0$)}
    \put(38,20){\normalsize$\meff[u\!/\!d,](Q^2)$}
  \end{picture}
  }}
  \put(71,1){
  \makebox(66,38.5){
  \begin{picture}(66,38.5)
    \put(0,0){\epsfxsize66mm \epsffile{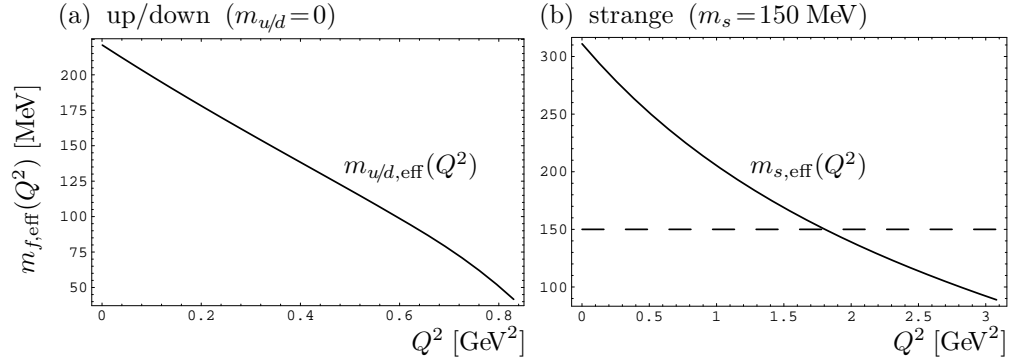}}
    \put(47,-3.5){\normalsize$Q^2\;[\GeV[]^2]$}
    \put(28,20){\normalsize$\meff[s,](Q^2)$}
    \put(0,40){(b)\vv strange\vv ($m_s \!=\! 150\MeV$)}
  \end{picture}
  }}
  \end{picture}
  \end{center}
\vspace*{-3ex}
\caption[Ref.~\cite{Dosch97}: Effektive Quarkmasse~\protect\mbox{$\meff(Q^2)$} f"ur~\protect$f \!\equiv\! up/down$ und~\protect$strange$]{
  Effektive Quarkmassen~\mbox{$\meff[u\!/\!d,]$} und~\mbox{$\meff[s,]$} als Funktionen von~$Q^2$, Fall~(i) nach Gl.~(\ref{meff-cases}).   Dieses Resultat ist entnommen Ref.~\cite{Dosch97}; wir danken den Autoren Dosch, Gousset, Pirner f"ur die freundliche "Uberlassung dieser Abbildungen.
}
\label{Fig:meff-DGP}
\end{minipage}
\end{figure}
Ihre Resultate f"ur den Fall~(i) sind angegeben in den Abbildungen~\ref{Fig:meff-DGP}(a),(b).
Sie stellen dar die funktionale Abh"angigkeit der effektiven Quarkmassen~$\meff[u\!/\!d,](Q^2)$ und~$\meff[s,](Q^2)$, die folgt aus den Gln.~(\ref{meff-Best.Gl}),~(\ref{meff-Best.Gl}$'$) f"ur die zweiten Ableitungen:~$n \!\equiv\! 2$.
Eine Abh"angigkeit von der Anzahl der Differentiationen ist gegeben, jedoch wird diese referiert als schwach.
Die $Q^2$-Abh"angigkeit von~$\meff(Q^2)$ ist in erster Approximation linear:
F"ur gro"se~$Q^2$ ist die effektive Quarkmasse~$\meff(Q^2)$ identisch der laufenden Quarkmasse~$m_f$, mit~\mbox{$m_{u\!/\!d} \!=\! 0$} und~\mbox{$m_s \!=\! 150\MeV$}, vgl.\@ Tabelle~\ref{Tabl:Charakt_rh,om,ph,Jps}, Fu"snote~\mbox{\FN{FN:LaufendeQuarkmassen}}; f"ur Virtualit"aten unterhalb einer Schwelle~$Q_{f\!,0}^2$ steigt sie linear an mit kleinerwerdendem~$Q^2$ auf den Wert~\mbox{$m_{f\!,0} \!\equiv\! \meff(Q^2 \zz\equiv\zz 0)$}.
Numerisch sind diese Massen~$m_{f\!,0}$ von der Gr"o"senordnung typischer Konstituentenquark-Massen:~\mbox{$m_{u\!/\!d,0} \!=\! 220\MeV$} und~\mbox{$m_{s,0} \!=\! 310\MeV$}, und die Schwellen f"ur die Photon-Virtualit"at liegen bei:~\mbox{$Q_{u\!/\!d,0}^2 \!=\! 1.05\GeV^2$} und~$Q_{s,0}^2 \!=\! 1.6\GeV^2$. \\
\noindent
Unsere Resultate f"ur die leichten Vektormesonen~$\rh(770)$,~$\om(782)$, $\ph(1020)$ im vorangegange\-nen Kapitel folgen daher im Rahmen einer universellen Photon-Lichtkegelwellenfunktion mit effektiver Quarkmasse unver"andert f"ur Virtualit"aten oberhalb dieser Werte:~\mbox{$Q^2 \!\ge\! 1.05\GeV^2$} beziehungsweise~\mbox{$Q^2 \!\ge\! 1.6\GeV^2$}, und werden erst modifiziert f"ur Werte darunter. \\
\indent
Dosch, Gousset, Pirner geben mit besser als~$5\%$ die Genauigkeit an, mit der die approximierte Polarisationsfunktionen~$\Pi_{\rm app}^{(n)}(Q^2,\meff^2)$ auf Basis dieser \vspace*{-.375ex}effektiven Quarkmassen die phenomenologischen, "`exakten"' Polarisationsfunktionen~$\Pi_{\rm ph}^{(n)}(Q^2,m_f^2)$ reproduzieren.

\bigskip\noindent
Dosch, Gousset, Pirner diskutieren ferner den Fall~(ii) nach Gl.~(\ref{meff-cases}$'$), \vspace*{-.25ex}in dem die effektive Quarkmasse~$\meff$ angenommen wird als Funktion von~$Q_{\rm eff}^2 \!\equiv\! 4\zbz Q^2$.
Wir machen diesbez"uglich nur die folgenden Bemerkungen.
Die Abh"angigkeit von~$Q_{\rm eff}^2$ impliziert effektiv eine Abh"anggikeit von~$\zet$, dem Anteil des Quarks am gesamten Lichtkegelimpuls.
Dem wird Rechnung getragen dadurch, da"s~${\rm Im}\,\Pi_0(s,m^2)$ nach Gl.~(\ref{ImPi0}) ersetzt wird durch:
\vspace*{-.5ex}
\begin{align} \label{ImPi0-zet}
&{\rm Im}\, \Pi_{0,\zet}(s,m^2)\;
  =\; \frac{\Nc}{12\pi}\; \int_0^1 d\zet\; 
        \bigg[ 1 \!+\! \frac{2m^2}{s} \bigg]\vv
        \th(s \!-\! m^2\!/(\zbz))
    \\[-4.5ex]\nn
\end{align}
Das hei"st der Phasenraum wird entsprechend des Werts von~$\zet$ eingeschr"ankt;~$m$ steht wieder f"ur~$m_f$ beziehungsweise~$\meff$.
Formal ist~${\rm Im}\Pi_{0,\zet}(s,m^2)$ eine Distribution bez"uglich~$s$ wie~${\rm Im}\Pi_0(s,m^2)$ nach Gl.~(\ref{ImPi0}); sie geht in diese "uber durch Ausf"uhren der $\zet$-Integration [Substitution~$\mu \!=\! 4\zbz$], das hei"st bei Bezug auf eine~$s$-unabh"angige Funktion.
Eingesetzt im Dispersionsintegral nach Gl.~(\ref{Pi(n)_DispersRel}) f"uhrt sie zu Dispersionsrelationen f"ur~$\Pi_{0,\zet}^{(n)}(Q^2,m^2)$ analog denen f"ur~$\Pi_0^{(n)}(Q^2,m^2)$,~$n \!\ge\! 1$.
Diese Modifikation besteht darin, da"s die $s$-Integration des Dispersionsintegrals "uber die untere Grenze~$m^2\!/\!(\zbz)$ abh"angt von~$\zet$ und "uber diese~\mbox{$\zet$-Kon}\-figurationen gemittelt wird. \\
\indent 
Resultat ist wie erwartet im wesentlichen eine glattere Interpolation zwischen laufender Quarkmasse~$m_f$ und Konstituentenquark-Masse~$m_{f\!,0}$.
Im Rahmen der Genauigkeit der Approximation und der Abh"angigkeit von der Anzahl~$n$ von Differentiationen kann~$m_{f\!,0}$ betrachtet werden als konstant; dann "andert sich allein die Schwelle~$Q_{f\!,0}^2$: und zwar zu den kleineren Werten~$0.69\GeV^2$ f"ur die Flavour up/down und~$1.16\GeV^2$ f"ur strange.

\vspace{-1ex}
\paragraph{\label{T:Univ-Photon-Wfn}Die universelle Photon-Lichtkegelwellenfunktion} ist auf Basis expliziter Parametrisierungen von~$\meff$ gegeben durch~-- wir fassen zusammen mithilfe der Gln.~(\ref{Photon-Wfn}),(\ref{Photon-Wfn}$'$):
\begin{align} \label{E:Photon-Wfn}
\ps_{\iga(Q^2,\la)}^{h,\bar h}(\zet,\rb{r})\;
  =\; \surd\Nc\vv e_f \de_{\!f\!\bar f}\vv \ch_{\iga(Q^2,\la)}^{h,\bar h}(\zet,\rb{r})
\end{align}
wobei, mit~$\vep \!=\! \surd\zbz Q^2 \!+\! \meff^2$, vgl.\@ Gl.~(\ref{epsilon}):
%
\begin{alignat}{2}
&\hspace*{-5pt}
 \ch_{\iga(Q^2,\la\equiv0)}&\;
  &=\; -\, 2 \zbz\vv Q\; \de_{h,-\bar h}\vv
             \frac{{\rm K}_0(\vep r)}{2\pi}
    \tag{\ref{E:Photon-Wfn}$'$} \\[-.5ex]
&\hspace*{-5pt}
 \ch_{\iga(Q^2,\la\equiv+1)}&\;
  &=\; \surd2\, \bigg[\,
         \iIM\,\vep \, \efn{\T +\iIM\,\vph}\;
           \big( \zet\, \de_{h+,\bar h-} - \bzet\, \de_{h-,\bar h+} \big)
           \frac{{\rm K}_1(\vep r)}{2\pi}\;
     +\; \meff\; \de_{h+,\bar h+}\,
           \frac{{\rm K}_0(\vep r)}{2\pi}\,
                \bigg] \nn \\
&\hspace*{-5pt}
 \ch_{\iga(Q^2,\la\equiv-1)}&\;
  &=\; \surd2\, \bigg[\,
         \iIM\,\vep \, \efn{\T -\iIM\,\vph}\;
           \big( \bzet\, \de_{h+,\bar h-} - \zet\, \de_{h-,\bar h+} \big)
           \frac{{\rm K}_1(\vep r)}{2\pi}\;
     +\; \meff\; \de_{h-,\bar h-}\,
           \frac{{\rm K}_0(\vep r)}{2\pi}\,
                \bigg] \nn
\end{alignat}
In den folgenden Abschnitten diskutieren wir ein System h"oher angeregter Resonanzen mit den Quantenzahlen des~$\rh(770)$.
F"ur Definiertheit geben wir an dieser Stelle die explizite Parametrisierung der effektiven Quarkmasse~$\meff[u\!/\!d,]$ f"ur Flavour up/down an, die wir dieser Diskussion zugrundelegen, mit~$m_{u\!/\!d,0} \!=\! 220\MeV$, vgl.\@ Ref.~\cite{Dosch97}:
\begin{align} \label{meff-explizit}
\meff[u\!/\!d,](Q^2)\;
  =\; \begin{cases}
  \vv m_{u\!/\!d,0}\cdot \big[1 - Q^2 \!/ Q_{u\!/\!d,0}^2\big]
    &\quad\text{f"ur\quad $Q^2\; <\; Q_{u\!/\!d,0}^2 \!=\! 1.05\GeV^2$} \\
  \vv m_{u\!/\!d} = 0
    &\quad\text{f"ur\quad $Q^2\; \ge\; Q_{u\!/\!d,0}^2$}
         \end{cases}
\end{align}
Dies ist die $Q^2$-abh"angige effektive Quarkmasse nach Fall~(i), Gl.~(\ref{meff-cases}) wie dargestellt in Ab\-bildung~\ref{Fig:meff-DGP}(a).
Das Ansteigen von~$\meff$ hin zu kleinen~$Q^2$ verhindert zu gro"se transversale Separationen~$r$ des Quark-Antiquark-Paares und ist insofern zu interpretieren als Mimikry von Effekten
, die diese unterdr"ucken: chirale Symmetriebrechung und Confinement.

\begin{figure}
\vspace*{-1ex}
\begin{minipage}{\linewidth}
  \begin{center}
  \setlength{\unitlength}{.8mm}\begin{picture}(120,85.5)
  \put(0,0){
  \makebox(120,85.5){
  \begin{picture}(120,85.5)
    \put(0,0){\epsfxsize96mm \epsffile{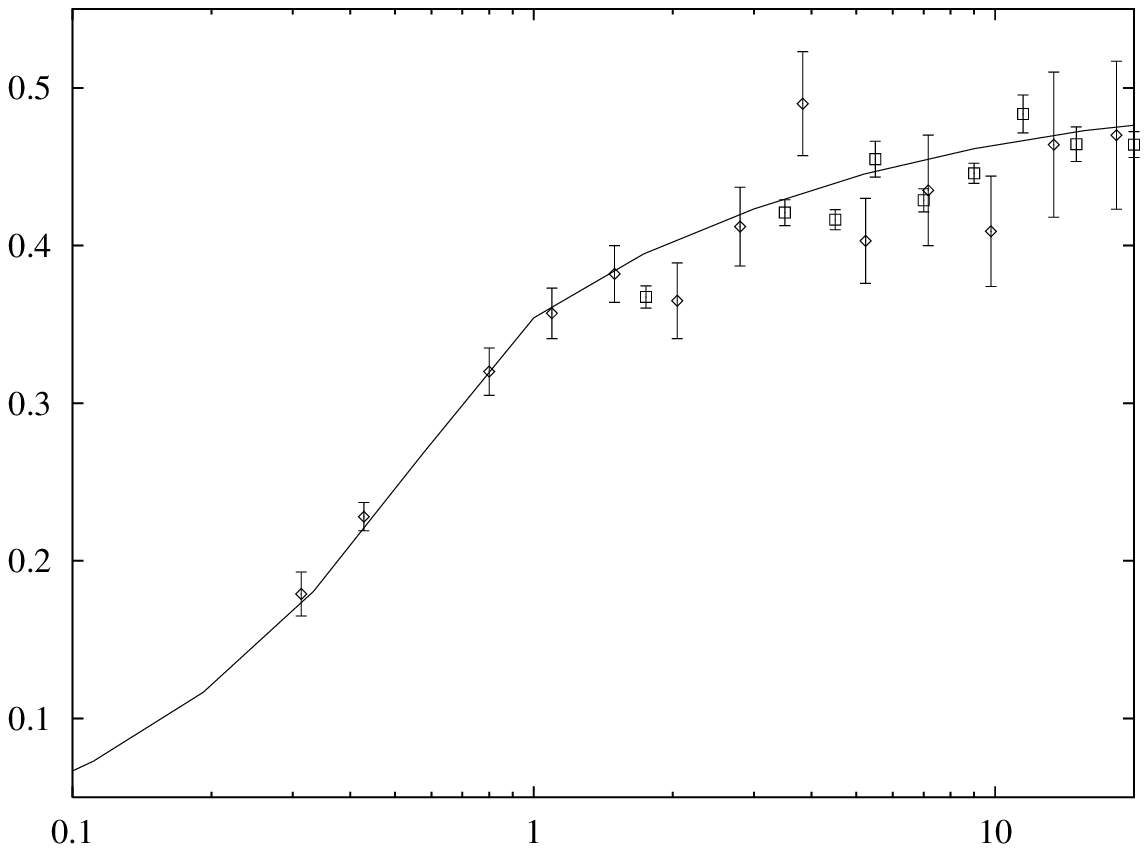}}
    \put(116,0.5){\normalsize$Q^2\;[\GeV[]^2]$}
    \put(-11,0  ){\yaxis[72mm]{\normalsize%
                    $0.9\times\, F_2(Q^2) \!=\! \frac{\D Q^2}{\D \pi e^2} %
                                     \left[\si_L^{\rm tot}[\ga^{\D\ast}p] %
                                    \!+\! \si_T^{\rm tot}[\ga^{\D\ast}p]%
                                     \right]$}}
    \end{picture}
    }}
  \end{picture}
  \end{center}
\vspace*{-4.25ex}
\caption[Ref.~\cite{Dosch97}: Strukturfunktion~\protect$F_2(Q^2)$ f"ur~\protect$\surd s \!=\! 20\GeV$]{
  $Q^2$-Abh"angigkeit des Beitrags der leichten Flavour up, down und strange zur Proton-Strukturfunktion~$F_2$ f"ur festes~$\surd s \!=\! 20\GeV$.   Die Daten stammen von \DREI[]{N}{M}{C}~[Quadrate] und \VIER[]{E}{6}{6}{5}~[Rauten], vgl.\@ die Refn.~\cite{Arneodo97,Adams96}.   Die Kurve ist berechnet in Ref.~\cite{Dosch97} als Resultat auf Basis der Photon-Lichtkegelwellenfunktion mit effektiven Quarkmassen~$\meff[u\!/\!d,]$ und~$\meff[s,]$, wie gezeigt in den Abbildungen~\ref{Fig:meff-DGP}(a),(b).   Sie ist reskaliert um einen Faktor~$0.9$, der folgen w"urde etwa aus der Verringerung des Wertes f"ur das Gluonkondensat um~$5\%$.   Wir danken Dosch, Gousset, Pirner f"ur die freundliche "Uberlassung dieser Abbildung.
\vspace*{-.5ex}
}
\label{Fig:F2-DGP}
\end{minipage}
\end{figure}

\bigskip\noindent
Der Erfolg des Verfahrens einer effektiven Quarkmasse, die abh"angt von der Skala des einlaufenden Impulstransfers, sei abschlie"send exemplarisch gezeigt an einem der Resultate von Dosch, Gousset, Pirner in Ref.~\cite{Dosch97}; er ist eindrucksvoll dokumentiert in Abbildung~\ref{Fig:F2-DGP}.
\\
\indent
Dargestellt ist ein Resultat in Zusammenhang mit {\it Compton-Streuung\/}, das berechnet ist auf Basis der universellen Photon-Lichtkegelwellenfunktion nach den Gln.~(\ref{E:Photon-Wfn}),~(\ref{E:Photon-Wfn}$'$) mit den effektiven Quarkmassen~$\meff[u\!/\!d,]$ und~$\meff[s,]$ nach Abbildung~\ref{Fig:meff-DGP}(a),(b).
Die zugrundeliegende $T$-Amplitude~$T_\la\![\ga^{\scriptscriptstyle({\D\ast})}p \!\to\! \ga^{\scriptscriptstyle({\D\ast})}p]$ f"ur longitudinale und transversale Polarisation~$\la \!\equiv\! L,T$ ist bereits angegeben in Gl.~(\ref{T_Lepto,Compton}$'$).
In Abbildung~\ref{Fig:F2-DGP} ist dargestellt f"ur feste invariante Schwerpunktenergie~$\surd s \!\cong\! 20\GeV$ und als Funktion von~$Q^2$ die {\it Proton-Strukturfunktion\/}~\mbox{$F_2(Q^2) \!=\! Q^2 \!/\! (\pi e^2) \!\cdot\! \si^{\rm tot}[\ga^{\scriptscriptstyle({\D\ast})}p]$}, mit~$\si^{\rm tot}[\ga^{\scriptscriptstyle({\D\ast})}p] \!=\! \si_L^{\rm tot}[\ga^{\scriptscriptstyle({\D\ast})}p] \!+\! \si_T^{\rm tot}[\ga^{\scriptscriptstyle({\D\ast})}p]$.
Der experimentell beobachtet Verlauf von~$F_2$ mit~$Q^2$ ist hochgradig nichttrivial und wird perfekt reproduziert bis hinunter zu den kleinsten gemessenen Werten von~$Q^2$.
Der Knick bei~$1 \!-\! 2\GeV^2$, der zustande kommt durch das Anwachsen der Quarkmassen f"ur Virtualit"aten unterhalb der Schwellen~$Q_{f\!,0}^2$, ist vorhanden in den experimentellen Daten.
Dies suggeriert, da"s die Brechung der chiralen Symmetrie bereits in gegenw"artigen Experimenten gemessen ist, und sie durch genaue Messungen der Proton-Strukturfunktionen in Leptoproduktion bei Virtualit"at~$Q$ besser zug"anglich sein k"onnte als in Schwerionen-Kollisionen bei endlicher Temperatur~$T \!=\! Q\!/(2\pi)$.
Vgl.\@ die Diskussion in Ref.~\cite{Dosch97}.
Die absolute Skalierung der postulierten Kurve auf neun Zehntel entspricht einer Reskalierung des Gluonkondensats auf~$95\%$ des zugrundegelegten Wertes,~-- und ist insofern nicht wirklich von Bedeutung.

\section[System h"oherer~\protect$1^+(1^{--})$-/Rho-Anregungen]{%
         System h"oherer \bm{1^+(1^{--})}-/Rho-Anregungen}
\label{Sect:Rho-Anregungen}

Auf Basis der so definierten universellen Photon-Lichtkegelwellenfunktion werden untersucht in Photo- und Leptoproduktion h"ohere Anregungen mit den Quantenzahlen des~$\rh(770)$, das hei"st mit~$I^G(J^{P\!C}) \!=\! 1^+(1^{--})$, vgl.\@ Fu"sn.~\refg{FN:Parity}.
Wir betrachten zun"achst den Zusammenhang, insbesondere die experimentelle Situation.
\vspace*{-1ex}

\subsection{Zusammenhang}
\label{Subsect:Zushang}

\begin{figure}
\begin{minipage}{\linewidth}
  \begin{center}
  \setlength{\unitlength}{.9mm}\begin{picture}(120,71.3)   
    \put(0,0){\epsfxsize108mm \epsffile{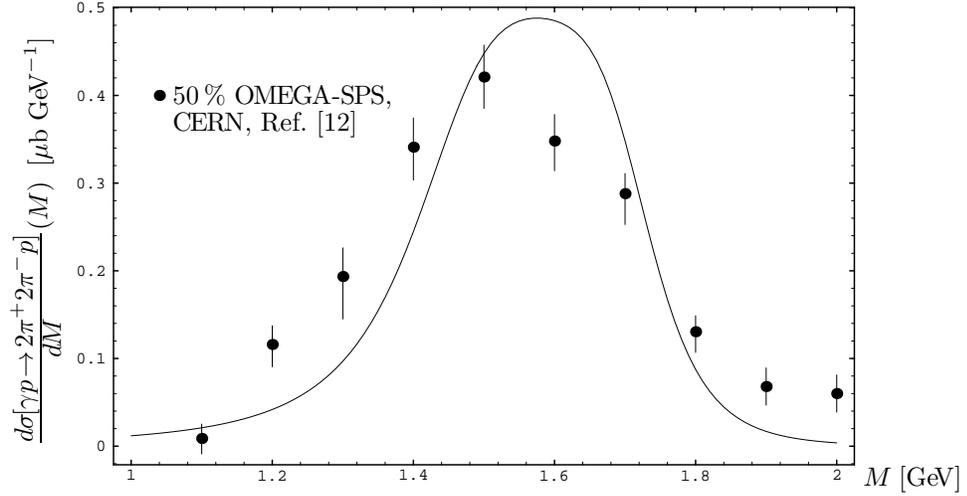}}
    \put(116,-0.5){\normalsize$M\;[\GeV[]]$}
    \put(-10, 0  ){\yaxis[64.17mm]{\normalsize
                                 $\frac{\D d\mskip-.5mu \si\mskip-2mu%
                                                           [\ga p \!\to\! 2\pi^+2\pi^-p]}{%
                                        \D d\mskip-1.5mu M}(M)\vv%
                                  [\microbarn[]\GeV^{-1}]$}}
    \put(14,56){\normalsize$50\,\%$~\FUNF[]{O}{M}{E}{G}{A}-\DREI[]{S}{P}{S},}
    \put(14,52){\normalsize\VIER[]{C}{E}{R}{N}, Ref.~\cite{Aston81}}
  \end{picture}
  \end{center}
\vspace*{-4.25ex}
\caption[Massespektrum f"ur Photoproduktion von~\protect$2\pi^+2\pi^-$ elastisch am Proton]{
  Massespektrum f"ur $2\pi^+2\pi^-$-Photoproduktion elastisch am Proton.   Die Kurve ist Resultat unserer Parametrisierung.   Vgl.\@ Text.    Die Daten stammen vom~\FUNF[]{O}{M}{E}{G}{A}-Spektrometer am \DREI[]{S}{P}{S}, \VIER[]{C}{E}{R}{N}, vgl.\@ Ref.~\cite{Aston81}; sie sind skaliert auf~$50\,\%$ ihresr Werte.
\vspace*{-.5ex}
}
\label{Fig:photo4pis}
\end{minipage}
\end{figure}

Wir sprechen von einem "`System von Anregungen"' und setzen damit bereits implizit voraus die erste im folgenden diskutierte Feststellung.
Die Spektren f"ur die Produktion von zwei und vier geladenen Pionen,~$\pi^+\pi^-$ und~$2\pi^+2\pi^-$, bez"uglich derer invarianten Masse~$M$ zeigen im Bereich von~$1 \!-\! 2\GeV$ ein komplexes Bild, das hindeutet auf ein Wechselspiel subtil interferierender Zust"ande.
\begin{figure}
\vspace*{-.5ex}
\begin{minipage}{\linewidth}
  \begin{center}
  \setlength{\unitlength}{.87694mm}\begin{picture}(120,67.9)   
    \put(0,0){\epsfxsize105.233mm \epsffile{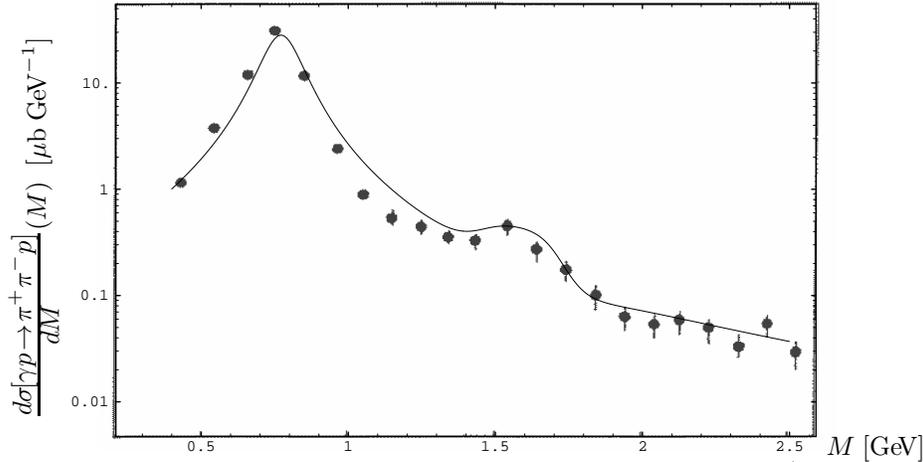}}
    \put(115,-0.5){\normalsize$M\;[\GeV[]]$}
    \put( -9, 0  ){\yaxis[59.5443mm]{\normalsize
                                 $\frac{\D d\mskip-.5mu \si\mskip-2mu%
                                                           [\ga p \!\to\! \pi^+\pi^-p]}{%
                                        \D d\mskip-1.5mu M}(M)\vv%
                                  [\microbarn[]\GeV^{-1}]$}}
  \end{picture}
  \end{center}
\vspace*{-5ex}
\caption[\protect{\vspace*{-.25ex}}Massespektrum f"ur Photoproduktion von~\protect$\pi^+\pi^-$ elastisch am Proton]{
  Massespektrum f"ur Hochenergie-Photoproduktion von~$\pi^+\pi^-$ elastisch am Proton.   Im Bereich~$M \!\cong\! 1.6\GeV$ zeigt sich konstruktive Interferenz.   Die experimentellen Daten stammen aus Ref.~\cite{Aston80}, mit einem Beitrag von~$50 \!\pm\! 20\nbarn$ des~$3^{--}$-Vektormesons~$\rh_3(1690)$ subtrahiert, vgl.\@ Ref.~\cite{Atkinson86}.   Die Auftragung geschieht entsprechend Donnachie, Mirzaie, Ref.~\cite{Donnachie87a}: nicht normiert auf absolute Zahlenwerte.   Die durchgezogene Kurve ist Resultat unserer Analyse, die basiert auf den im Text parametrisierten Ans"atzen f"ur~$\rh$,~$\rh'$,~$\rh^\dbprime$~und deren einfache Breit-Wigner-Distribution; vgl.\@ Anh.~\ref{APPSubsect:Photo,l+l-Annih}.
\vspace*{-.5ex}
}
\label{Fig:photo2pis}
\end{minipage}
\end{figure}
Zum einen in {\it Hochenergie-Photoproduktion elastisch am Proton\/} wird im $\pi^+\pi^-$-Spektrum auf der $\rh(770)$-Schulter bei~\mbox{$M \!\cong\! 1.6\GeV$} ein breites Maximum beobachtet; vgl.\@ Abb.~\ref{Fig:photo2pis}, und Ref.~\cite{Aston80} bzgl.\@ der experimentellen Daten.
Ein Anstieg an derselben Stelle findet sich auch im~\mbox{$2\pi^+2\pi^-$-Massen}\-spektrum, vgl.\@ die Refn.~\cite{Aston81,Aston82} und Abb.~\refg{Fig:photo4pis}.
\begin{figure}
\vspace*{-.5ex}
\begin{minipage}{\linewidth}
  \begin{center}
  \setlength{\unitlength}{.9mm}\begin{picture}(120,67)   
    \put(0,0){\epsfxsize108mm \epsffile{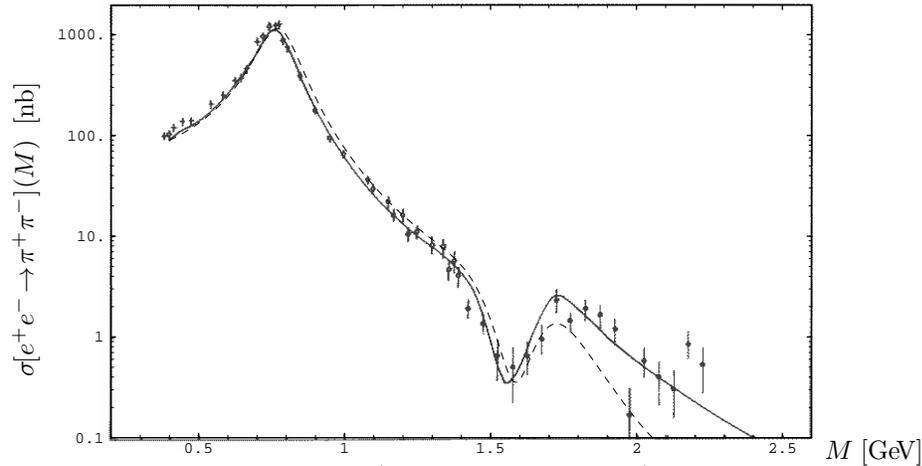}}
    \put(114,-0.5){\normalsize$M\;[\GeV[]]$}
    \put( -6, 0  ){\yaxis[60.3mm]{\normalsize%
                                    $\si\mskip-2mu[e^+e^- \!\to\! \pi^+\pi^-](M)\vv%
                                     [\nbarn[]]$}}
  \end{picture}
  \end{center}
\vspace*{-5ex}
\caption[Massespektrum f"ur~\protect$e^+e^-$-Annihilation in~\protect$\pi^+\pi^-$]{
  Massespektrum f"ur~$e^+e^-$-Annihilation in~$\pi^+\pi^-$.   Die Interferenz im Bereich~$M \!\cong\! 1.6\GeV$ ist destruktiv; sie bestimmt die relativen Vorzeichen der Kopplungen an den elektromagnetischen Strom~$f_V$,~$V \!\equiv\! \rh,\rh',\rh^\dbprime$, zu~$[+,-,+]$.   Die durchgezogene Kurve ist die Anpassung von Donnachie, Mirzaie auf Basis zweier interferierender Resonanzen~"`$\rh_1$"'$\!(1465)$ und~"`$\rh_2$"'$\!(1700)$, vgl.\@ Ref.~\cite{Donnachie87a}.   Die gestrichelte Kurve stellt dar das Resultat f"ur die Parametrisierungen von~$\rh'$,~$\rh^\dbprime$ im Text, vgl.\@ Tabl.~\ref{Tabl:Charakt_rh,rh',rh''} und Anh.~\ref{APPSubsect:Photo,l+l-Annih}.
\vspace*{-.5ex}
}
\label{Fig:eebar2pis}
\end{minipage}
\end{figure}
Zum~an\-deren in {\it Elektron-Positron-Annihilation\/} wird im $\pi^+\pi^-$-Spektrum bei~\mbox{$M \!\cong\! 1.2 \!-\! 2\GeV$} ein Muster beobachtet, das wesentlich in Kontrast steht zu dem in der Photoproduktion, vgl.\@ Abb.~\ref{Fig:eebar2pis}, und die Refn.~\cite{Quenzer78,Barkov85,Bisello85} bzgl.\@ der experimentellen Daten.
Bzgl.\@ des \mbox{$e^+e^-$-Anni}\-hilationspektrums in $2\pi^+2\pi^-$ vgl.\@ Abb.~\refg{Fig:eebar4pis}.

Dieser Unterschied suggeriert die Interpretation als zum einen {\it konstruktive\/} und zum anderen {\it destruktive Interferenz} und impliziert, da"s im Massebereich von~$1.2 \!-\! 2\GeV$ nicht nur ein Zustand mit den Quantenzahlen des~$\rh(770)$ produziert wird.
Evidenz f"ur die Existenz zweier Resonanzen wird seit langem diskutiert.
So kommen Donnachie, Mirzaie in Ref.~\cite{Donnachie87a}, vgl.\@ auch Ref.~\cite{Clegg94}, in einer Analyse der experimentellen Massespektren zu dem Schlu"s, da"s eine konsistente Erkl"arung auf Basis nur einer Resonanz nicht, wohl aber m"oglich ist auf Basis zweier Resonanzen~$\rh_1$,~$\rh_2$ mit Massen~$1465$ beziehungsweise~$1700\MeV$, die in etwa gleichstark koppeln an den elektromagnetischen Strom.
\label{T:rh(1600)}Der alte eine Eintrag f"ur~$\rh(1600)$ der Particle Data Group ist ersetzt durch zwei Eintr"age f"ur~$\rh(1450)$,~$\rh(1700)$; vgl.\@ die Diskussion unter letzterem ebenda, Ref.~\cite{PDG00}.
Ferner wird diskutiert die Existenz eines Hybrid-Zustands~$h(1450)$ mit den Quanten\-zahlen des~$\rh(770)$:~$1^+(1^{--})$, der dominant zerf"allt in~$\pi\,a_1(1260)$, vgl.\@ die Refn.~\cite{Clegg94,Close94,Close97}.
Wir legen unserer Analyse zwei Zust"ande zugrunde, die wir bezeichnen mit~$\rh'$,~$\rh^\dbprime$ und verstehen als~$\rh(1450)$,~$\rh(1700)$.
Diese sind zu pr"azisieren.

Mit Gesamtdrehimpuls~$J \!\equiv\! 1$ und aufgefa"st als "Uberlagerung von Paaren eines Quarks und eines Antiquarks, deren Spins ausgerichtet sein k"onnen parallel oder antiparallel, besitzt die Wellenfunktion, gem"a"s der die Paare verteilt sind, Bahndrehimpuls Null oder Zwei; dies suggeriert den Ansatz als $2S$- und $2D$-Zustand [kurz~$\ket{NL}$, f"ur systematisch~$\ket{N{}^{2S+1}L_{JM}}$].
Die Wellenfunktion der $2D$-Welle verschwindet am Ursprung, so da"s erst in relativistisch erweiterten Modellen eine nichtverschwindende Kopplung an den elektromagnetischen Strom besteht.
Diese relativistisch induzierte Kopplung~$f_V$ ist insofern von h"oherer Ordnung und numerisch kleiner als die der~$2S$-Welle, entsprechend~$\Gall_V$, die leptonische Zerfallsbreite.%
\FOOT{
  Donnachie, Mirzaie extrahieren aus den experimentellen Daten f"ur die zweite Anregung~$\rh_2$ mit Masse~$1.7\GeV$ eine leptonische Zerfallebreite von~$1.5 \!-\! 3.0\keV$, vgl.\@ Ref.~\cite{Donnachie87a}.   Als typisches relativistisch erweitertes Quarkmodell postuliert das von Godfrey, Isgur, vgl.\@ Ref.~\cite{Godfrey85}, f"ur einen $2D$-Zustand mit vergleichbarer Masse von~$1.66\GeV$ Zahlenwerte von~$0.15 \!-\! 0.45\keV$.
Vorwegnehmend resultiert unser Ansatz f"ur~$V \!\equiv\! \rh',\rh^\dbprime$ in Zahlenwerte f"ur die Kopplung~$f_V$ von~$-103$ und~$90.3\MeV$; damit ist zu konfrontieren der Wert~$32.5\MeV$ auf Basis einer~$2D$-Welle in Ref.~\cite{Godfrey85}. 
}
Dies war auch unser urspr"unglicher Ansatz:~$\rh^\dbprime$ aufzufassen als~$2D$-Zustand.
Wir argumentieren, weswegen wir ihn verworfen haben zugunsten eines anderen Ansatzes.
\vspace*{-.5ex}

\bigskip\noindent
Wir rekapitulieren Gl.~(\ref{fV-Def}):
\vspace*{-.5ex}
\begin{align} \label{E:fV-Def}
\bra{\Om}J_{\rm em}^{\mu}(0)\ket{V(q,\la)}\;
  =\; e f_V\, M_V\, \vep^{\mu}(q,\la)
    \\[-4.5ex]\nn
\end{align}
Durch sie ist~$f_V$ definiert als die Kopplung des Vektormesons~$V$ an den elektromagnetischen Strom~$J_{\rm em} \!=\! (e_{\!F}\bar{\ps}_{\!F}\ga^\mu\ps_{\!F})$, Summation "uber alle (geladenen) Fermionen~$F$ impliziert.

Der Zustand~$\ket{V(q,\la)}$ des Vektormesons mit Viererimpuls~$q$ und Helizit"at~$\la \!\equiv\! 0,\pm1$ wird aufgefa"st als Superposition von Quark-Antiquark-Paare, die verteilt sind bez"uglich einer {\it nicht-relativistischen\/} Wellenfunktion.
Auswertung der Formel~(\ref{E:fV-Def}) im Rahmen konventioneller relativistischer Quantenfeldtheorie ergibt dann, vgl.\@ Ref.~\cite{Bergstroem79}:%
\FOOT{
  Die Phase in diesen Gleichungen ist nicht physikalisch.   Die Diskrepanz f"ur die $D$-Welle im Vergleich zu Ref.~\cite{Bergstroem79} ist Konsequenz dessen, da"s wir in den Konventionen von Landau, Lifschitz arbeiten, deren Kugelfl"achenfunktionen~${\rm Y}_{l,l_3}$ die zus"atzliche Phase~$i^l$ implizieren, vgl.\@ Ref.~\cite{Landau86}.
}
%
\begin{align} \label{f_V-nS,nD}
f_{V\!,nS}\;
  &=\; \hat{e}_V\surd\Nc\vv M_{V\!,nS}^{-1/2}\cdot\;
        \La_{\la,S}\;\cdot
        2\sqrt{4\pi}\;
        \int_0^\infty \frac{k^2dk}{(2\pi)^3}\; \tilde{R}_{n,0}(k)\;\cdot
        \frac{2}{3} \Big[1 \!+\! \frac{1}{2}\frac{m}{k^0}\Big]
    \\
f_{V\!,nD}\;
  &=\; \hat{e}_V\surd\Nc\vv M_{V\!,nD}^{-1/2}\cdot\;
        \La_{\la,D}\;\cdot
        2\sqrt{4\pi}\;
        \int_0^\infty \frac{k^2dk}{(2\pi)^3}\; \tilde{R}_{n,2}(k)\;\cdot
        \frac{\surd2}{3} \Big[1 \!-\! \frac{m}{k^0}\Big]
    \tag{\ref{f_V-nS,nD}$'$}
\end{align}
mit~$k \!=\! |\vec{k}|$ dem Betrag des Dreier-Relativimpuls' der Quarks,~$m$ deren Konstituentenmasse, \mbox{$k^0 \!=\! \surd k^2 \!+\! m^2$} deren Energie und~$\tilde{R}_{nl}$ dem Radialteil der Impulsraum-Wellenfunktion:
\begin{align} 
1\; =\; \int \frac{d^3\vec{k}}{(2\pi)^3}\;
        \Big|\tilde{\ph}_{n,l,l_3}(\vec{k})
             \equiv
             \tilde{R}_{n,l}(k)\, {\rm Y}_{l,l_3}(\hat{k})
        \Big|^2\;
    =\; \int \frac{k^2dk}{(2\pi)^3}\; \big|\tilde{R}_{n,l}(k)\big|^2\;
\end{align}
Die Faktoren~$\La_{\la,S},\La_{\la,D}$ in den Gln.~(\ref{f_V-nS,nD}),~(\ref{f_V-nS,nD}$'$) sind f"ur feste Helizit"at~$\la \!\equiv\! 0,\pm1$ identisch Eins; sie geben die Beitr"age an bez"uglich der Spin-Konfigurationen~$(s,{\bar s})$ der Quarks beziehungsweise der dritten Komponente des Bahndrehimpulses~$l_3$:
\begin{align} \label{La_la,S/D}
1\;
  &=\; \La_{\la,S}\;
  =\; {\T\sum}_{s,{\bar s}}\;
       \Big\{
       \de_\la^0\cdot   \frac{1}{2}\big[\de_{s+}\de_{\bar s-} \!+\! \de_{s-}\de_{\bar s+}\big]
     + \de_\la^\pm\cdot \de_{s\pm}\de_{\bar s\pm}
       \Big\}
    \\
1\;
  &=\; \La_{\la,D}\;
    \tag{\ref{La_la,S/D}$'$} \\[-.5ex]
  &=\; {\T\sum}_{l_3}\;
       \Big\{
       \de_\la^0\cdot
         \Big[\frac{2}{5}\de_{l_3,0}
              \!+\! \frac{3}{10}\de_{l_3,+1}
              \!+\! \frac{3}{10}\de_{l_3,-1}
         \Big]
     + \de_\la^\pm\cdot
         \Big[\frac{1}{10}\de_{l_3,0}
              \!+\! \frac{3}{10}\de_{l_3,\pm1}
              \!+\! \frac{3}{5}\de_{l_3,\pm2}
         \Big]
       \Big\}
    \nn
\end{align}
F"ur den (nicht-relativistischen) Harmonischen Oszillator mit Parameter~$\om$ gilt:
\vspace*{-.5ex}
\begin{alignat}{3} \label{tildeR_S,D}
&\tilde{R}_{00}(k)&\;
  &=\; \sqrt{4\pi}\vv 2^3\! \left(\frac{\surd\pi}{2\om}\right)\Big.^{\zz3\!/\!2}\vv
        \efn{\D-\om^{-2}k^2\!/2}&&
    \\[-.25ex]
&\tilde{R}_{20}(k)&\;
  &=\; \sqrt{4\pi}\vv 2^3\! \left(\frac{\surd\pi}{2\om}\right)\Big.^{\zz3\!/\!2}\vv
        \efn{\D-\om^{-2}k^2\!/2}&
        \vv\cdot\vv
       &\sqrt{\frac{3}{2}}\Big[\frac{2}{3}\, \om^{-2}k^2 - 1\Big]
    \tag{\ref{tildeR_S,D}$'$} \\[-.25ex]
&\tilde{R}_{22}(k)&\;
  &=\; \sqrt{4\pi}\vv 2^3\! \left(\frac{\surd\pi}{2\om}\right)\Big.^{\zz3\!/\!2}\vv
        \efn{\D-\om^{-2}k^2\!/2}&
        \vv\cdot\vv
       &\frac{2}{\surd15}\; \om^{-2}k^2
    \tag{\ref{tildeR_S,D}$'$}
    \\[-4.5ex]\nn
\end{alignat}
mit weiter nicht spezifizierter Phase.
Die Integrale der Gln.~(\ref{f_V-nS,nD}),~(\ref{f_V-nS,nD}$'$) auf Basis dieser Ausdr"ucke k"onnen analytisch gel"ost werden.
Wir k"urzen ab den Anteil der Kopplung~$f_V$, der nicht abh"angt von den Quantenzahlen von~$V$:
\vspace*{-.5ex}
\begin{align} 
\hat{f}_{nl}\;
  :=\; \frac{f_{V\!,nl}}{%
          \hat{e}_V\surd\Nc\vv M_{V\!,nl}^{-1\!/\!2}}\qqquad
  l \equiv S,D
    \\[-4.5ex]\nn
\end{align}
und finden, mit~$\ze \!\equiv\! m^2\!/\!4\om^2$ und~${\rm K}_0$,~${\rm K}_1$ den modifizierten Besselfunktionen zweiter Art:
\vspace*{-.5ex}
\begin{align} \label{f_V-1S,2S,2D}
\hspace*{-2.5ex}
\hat{f}_{1S}\;
  &=\; \frac{1}{\surd3}\! \left(\frac{2\om}{\surd\pi}\right)\Big.^{\zz3\!/\!4}\cdot
       \sqrt{\frac{2}{3}}\;
       \bigg[
         1 + \ze^{3\!/\!2}\, \efn{\D \ze}\; \sqrt{\frac{2}{\pi}}\;
             \big\{
             {\rm K}_1(\ze) - {\rm K}_0(\ze)
             \big\}
       \bigg]
    \\
\hspace*{-2.5ex}
\hat{f}_{2S}\;
  &=\; \frac{1}{\surd3}\! \left(\frac{2\om}{\surd\pi}\right)\Big.^{\zz3\!/\!4}
       \phantom{\sqrt{\frac{2}{3}}\; }\cdot
       \bigg[
         1 + \ze^{3\!/\!2}\, \efn{\D \ze}\; \sqrt{\frac{2}{\pi}}\;
             \Big\{
             \Big[\frac{1}{3} \!-\! \frac{8}{3}\ze\Big]\, {\rm K}_1(\ze)
           + \Big[1           \!+\! \frac{8}{3}\ze\Big]\, {\rm K}_0(\ze)
             \Big\}
       \bigg]
    \tag{\ref{f_V-1S,2S,2D}$'$} \\
\hspace*{-2.5ex}
\hat{f}_{2D}\;
  &=\; \frac{1}{\surd3}\! \left(\frac{2\om}{\surd\pi}\right)\Big.^{\zz3\!/\!4}\cdot
       \,\frac{2}{\surd5}\;
       \bigg[
         1 - \ze^{3\!/\!2}\, \efn{\D \ze}\; \sqrt{\frac{2}{\pi}}\;
             \Big\{
             \Big[\frac{4}{3} \!-\! \frac{8}{3}\ze\Big]\, {\rm K}_1(\ze)
                                  + \frac{8}{3}\ze\,      {\rm K}_0(\ze)
             \Big\}
       \bigg]
    \tag{\ref{f_V-1S,2S,2D}$''$}
    \\[-4.5ex]\nn
\end{align}
In Abbildung~\ref{Fig:Couplings_m,omPH} sind aufgetragen auf Basis dieser Ausdr"ucke~$f_{V\!,nl}(m,\om)$ f"ur~\mbox{$V \!\equiv\! \rh,\rh',\rh^\dbprime$} als Funktionen der Quarkmasse~$m$ beziehungsweise des Oszillatorparameters~$\om$.
\begin{sidewaysfigure}
\begin{minipage}{\linewidth}
\setlength{\unitlength}{1mm}
\makebox(220,101){
\begin{picture}(220,101)
\put(-2,51){
  \makebox{
  \begin{picture}(72, 45)
    \put(0,0){\epsfxsize72mm \epsffile{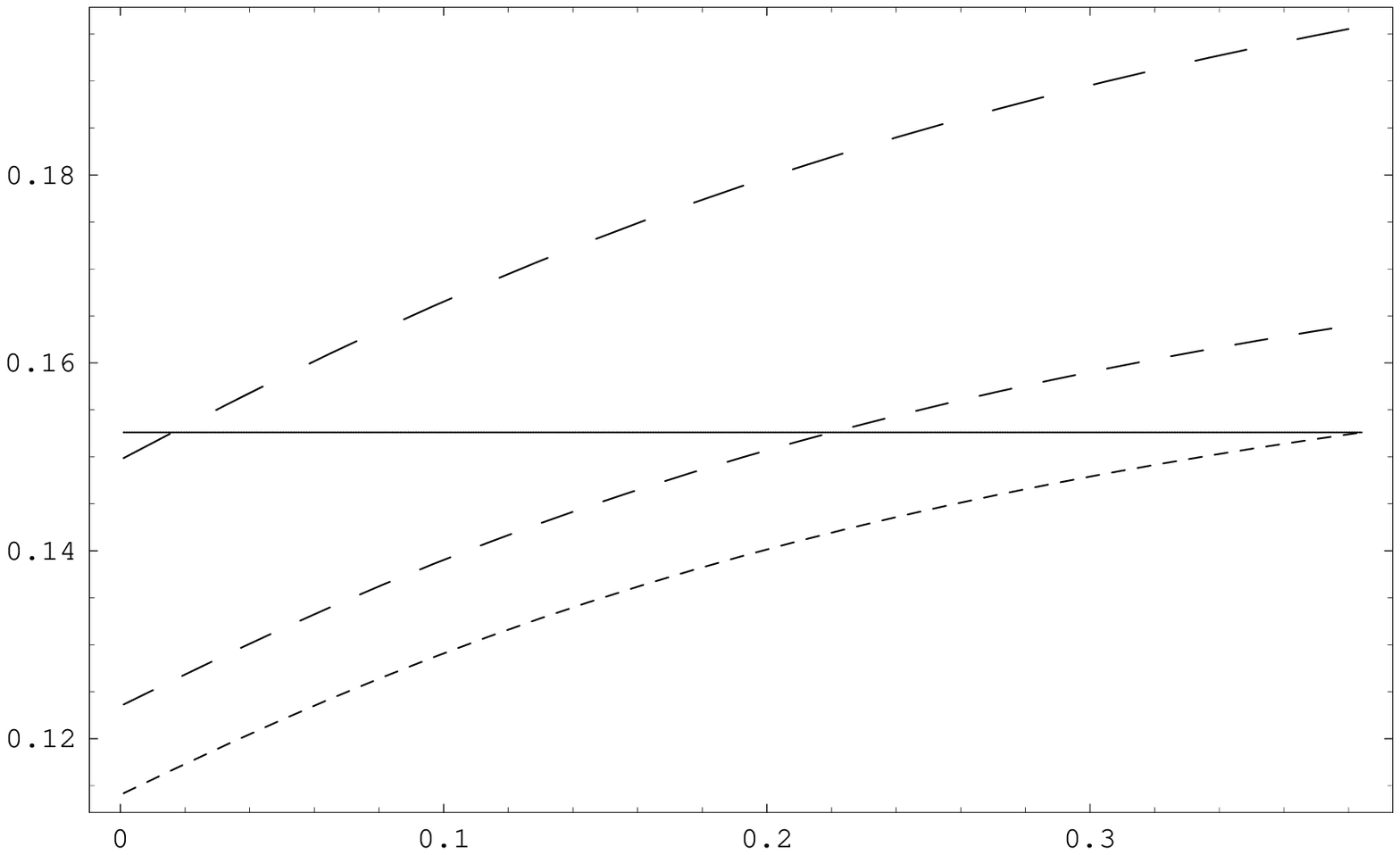}}
    \put(59,-2.5){\normalsize$m\;[\GeV[]]$}
    \put(0,47){\xaxis[72mm]{\normalsize $f_{\rh,1S}(m,\om)\;[\GeV[]]$}}
  \end{picture}
  }}
\put(-2,1){
  \makebox{
  \begin{picture}(72, 45)
    \put(0,0){\epsfxsize72mm \epsffile{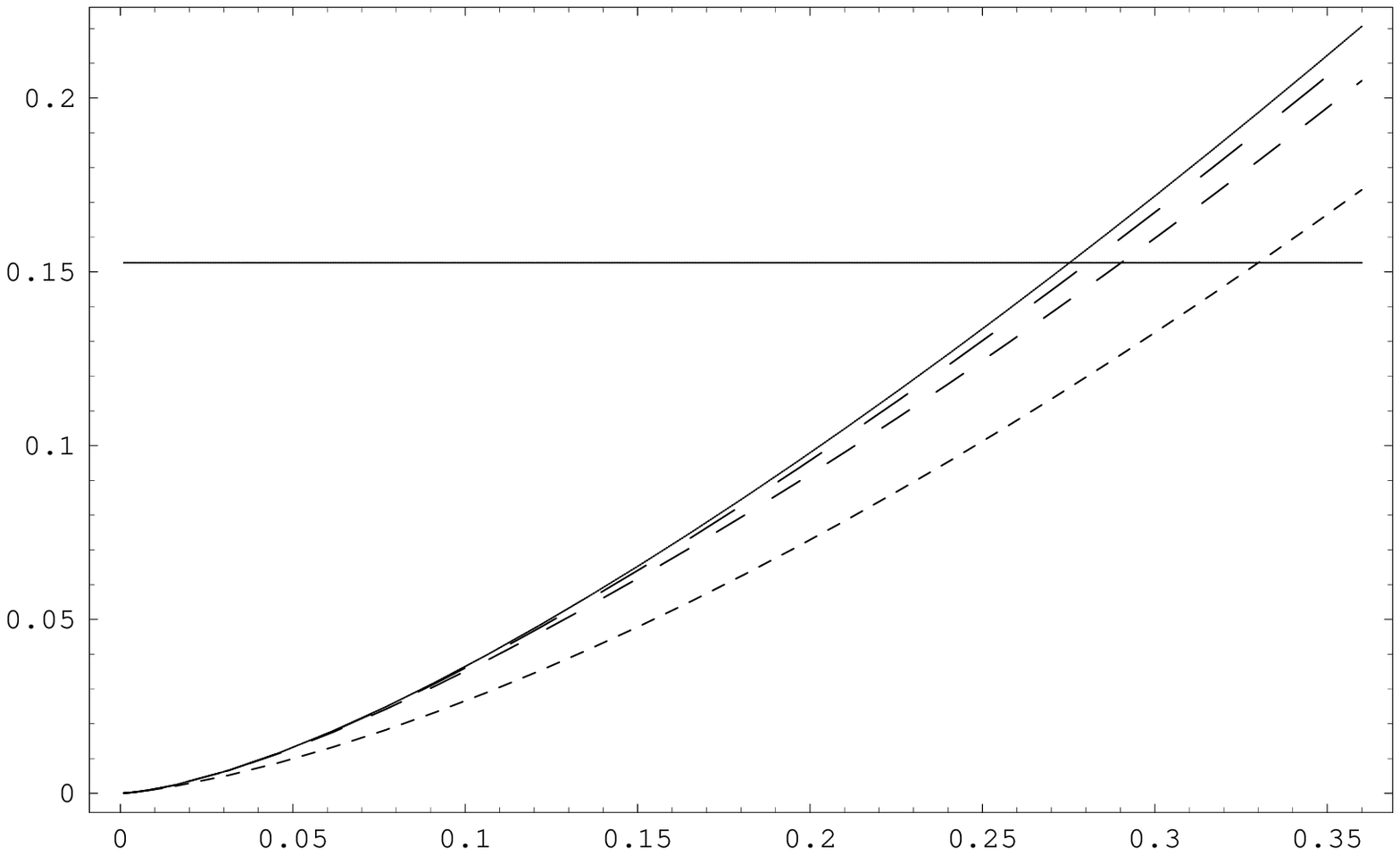}}
    \put(59,-3.5){\normalsize$\om\;[\GeV[]]$}
  \end{picture}
  }}
\put(71,51){
  \makebox{
  \begin{picture}(72, 45)
    \put(0,0){\epsfxsize72mm \epsffile{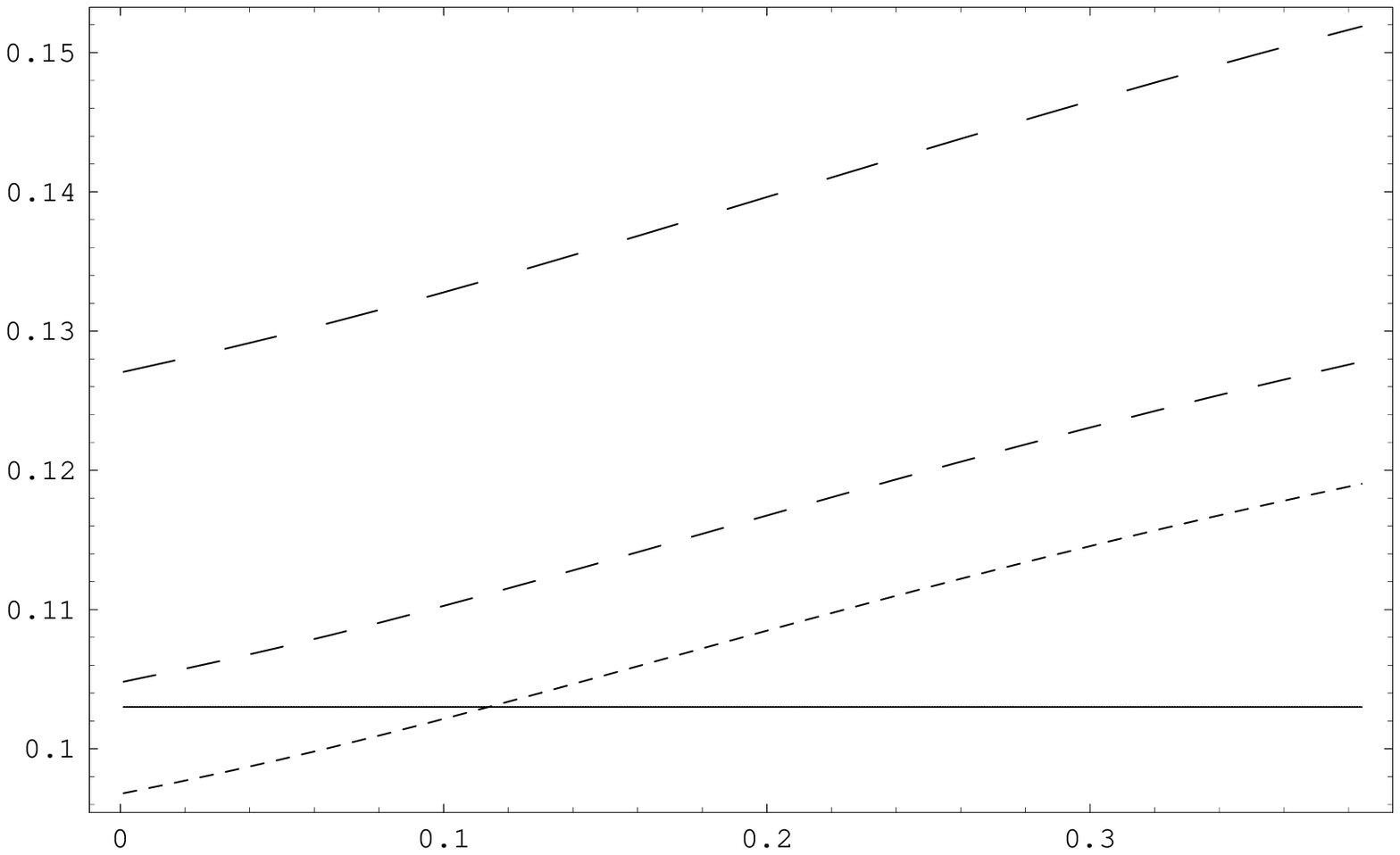}}
    \put(59,-2.5){\normalsize$m\;[\GeV[]]$}
    \put(0,47){\xaxis[72mm]{\normalsize $f_{\rh',2S}(m,\om)\;[\GeV[]]$}}
  \end{picture}
  }}
\put(71,1){
  \makebox{
  \begin{picture}(72, 45)
    \put(0,0){\epsfxsize72mm \epsffile{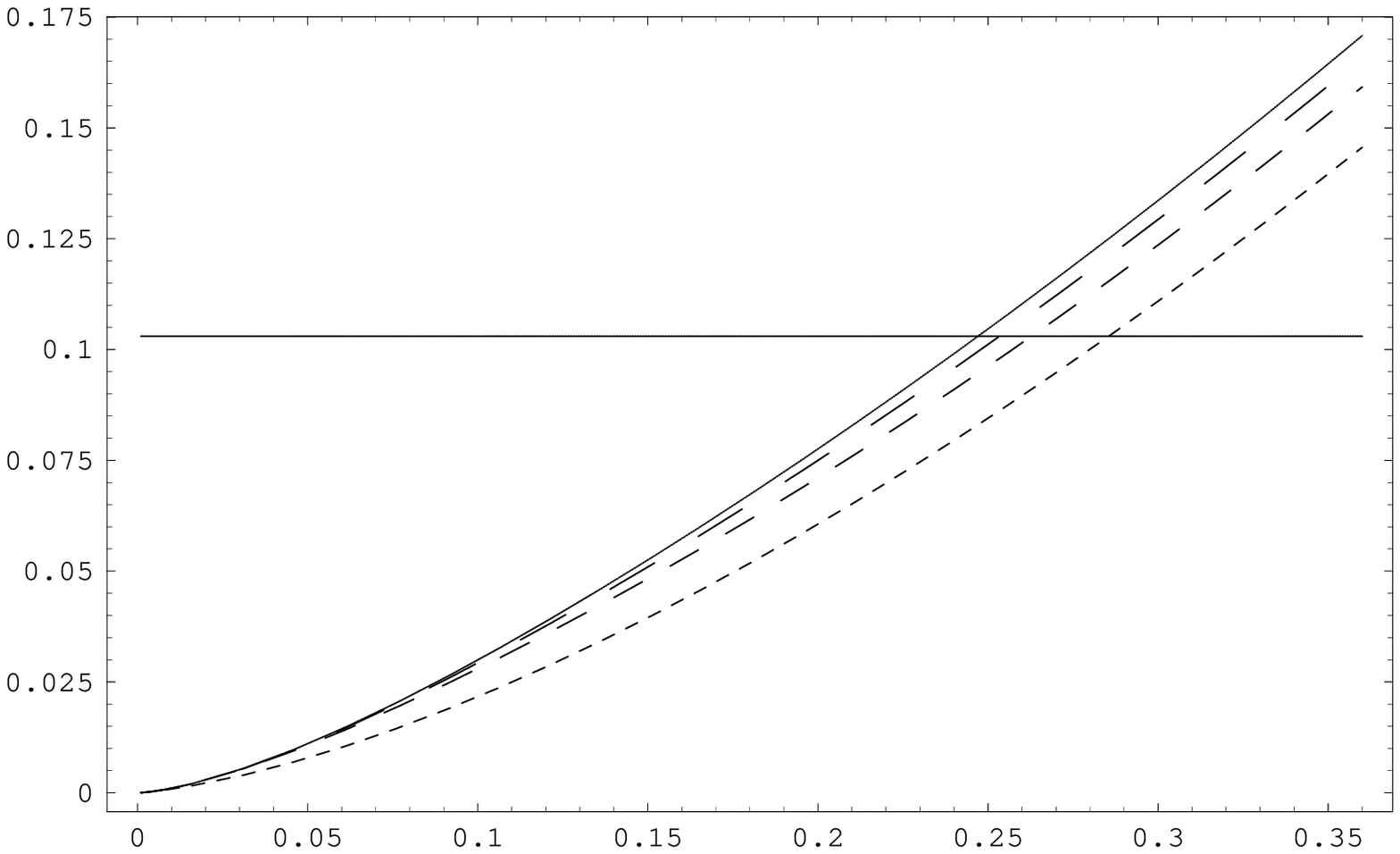}}
    \put(59,-3.5){\normalsize$\om\;[\GeV[]]$}
  \end{picture}
  }}
\put(144,51){
  \makebox{
  \begin{picture}(72, 45)
    \put(0,0){\epsfxsize72mm \epsffile{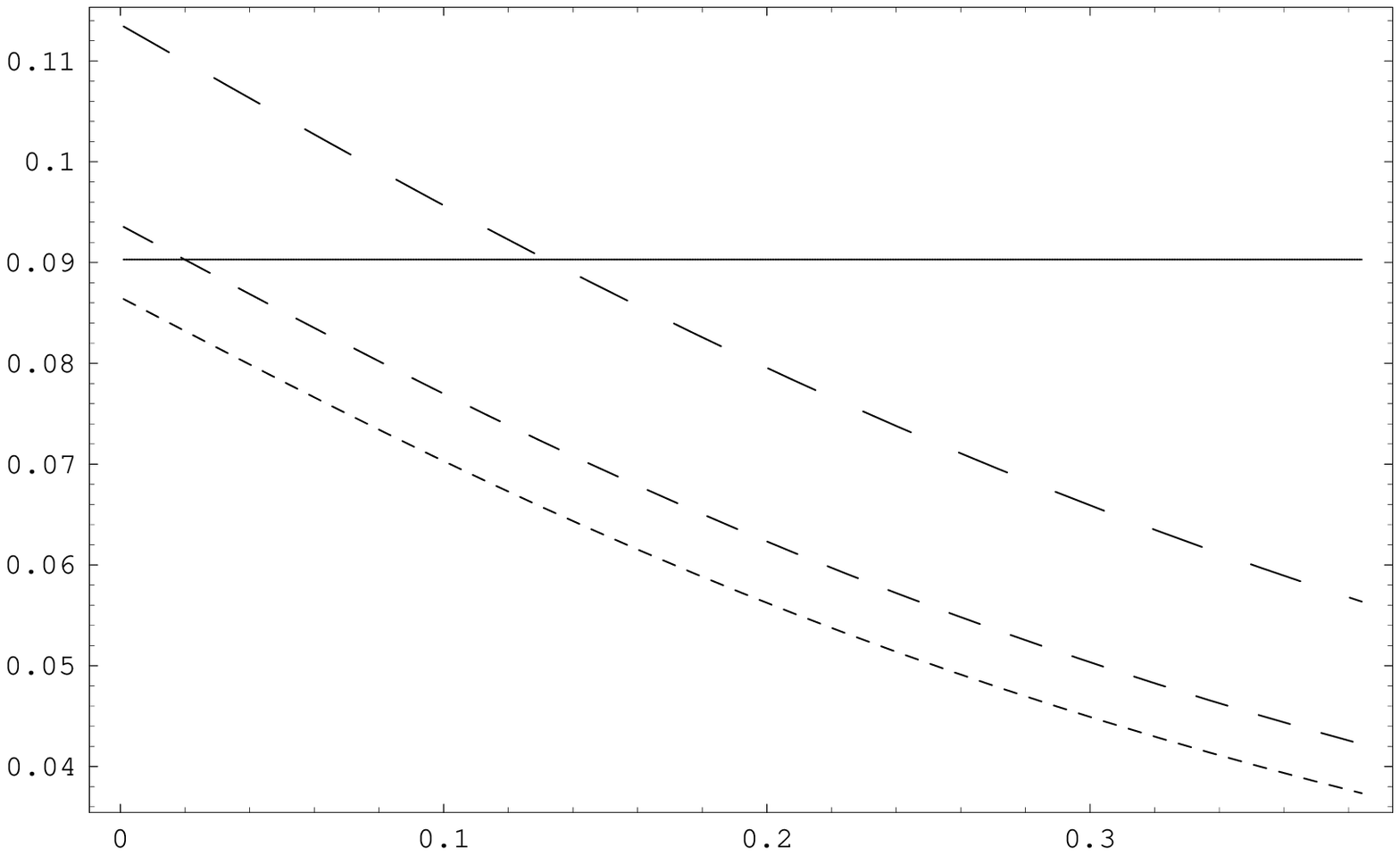}}
    \put(59,-2.5){\normalsize$m\;[\GeV[]]$}
    \put(0,47){\xaxis[72mm]{\normalsize $f_{\rh^\dbprime,2D}(m,\om)\;[\GeV[]]$}}
  \end{picture}
  }}
\put(144,1){
  \makebox{
  \begin{picture}(72, 45)
    \put(0,0){\epsfxsize72mm \epsffile{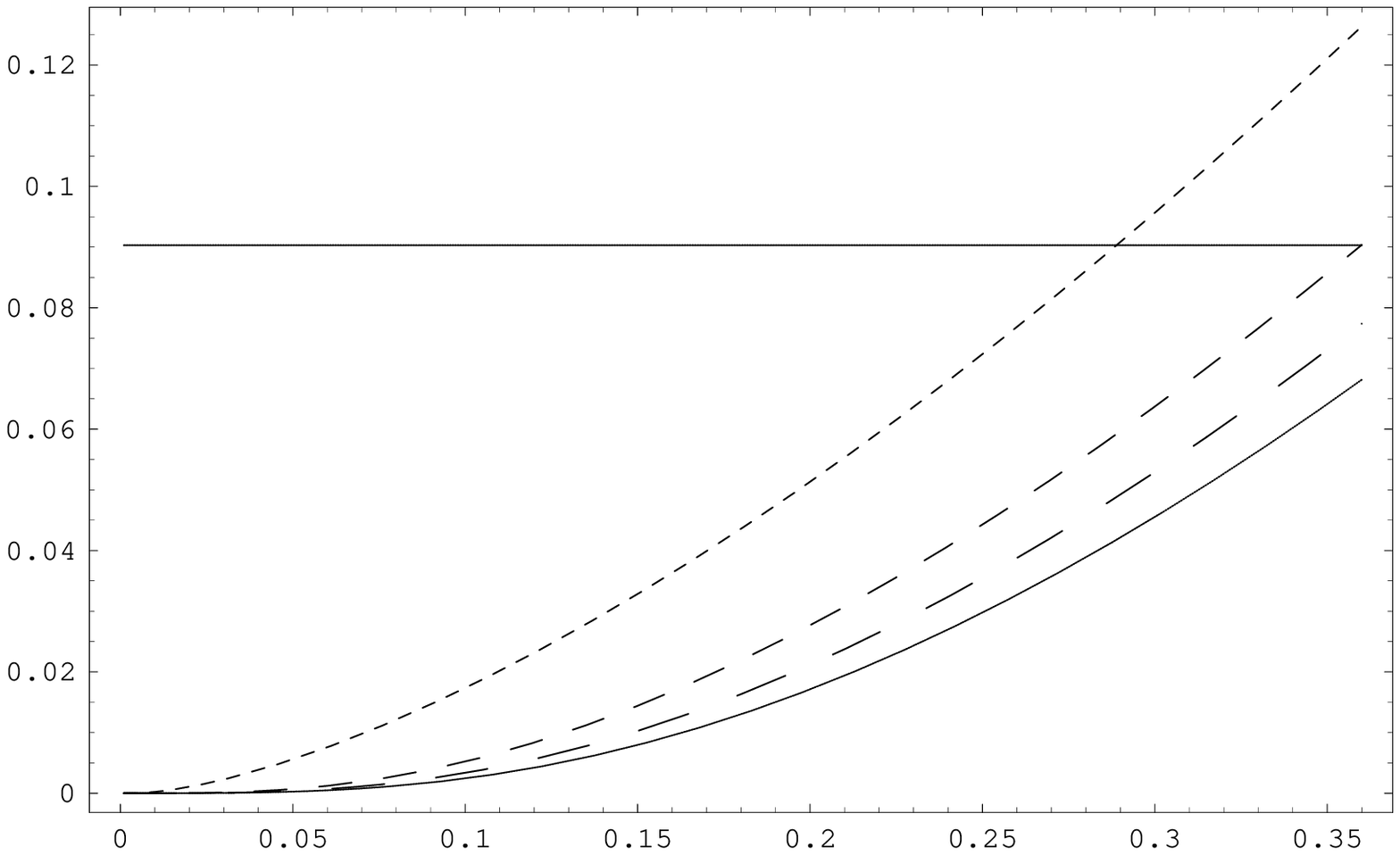}}
    \put(59,-3.5){\normalsize$\om\;[\GeV[]]$}
  \end{picture}
  }}
\end{picture}}
\caption[Quarkmodell-Kopplungen~\protect$f_V$ f"ur~\protect$V \!\equiv\! \rh,\rh',\rh^\dbprime\!$ als reine~\protect$1S$-,\protect$2S$-,\protect$2D$-Wellen\!]{
 Kopplungen~$f_V(m,\om)$ in einem relativistisch erweiterten Quarkmodell auf Basis des Harmonischen Oszillators.   Die physikalischen Zust"ande~$\rh(770)$,~$\rh(1450)$,~$\rh(1700)$ sind aufgefa"st als reine $1S$-,$2S$-,$2D$-Welle.   Horizontele Linien: f"ur~$\rh$ experimenteller Wert~$f_{\rh,{\rm exp}} \!=\! 0.1526\GeV$, f"ur~$\rh'$,~$\rh^\dbprime$ die Werte, die Resultat sind unseres Ansatzes im Text und die experimentellen~$\pi^+\pi^-$-,~$2\pi^+2\pi^-$-Spektren reproduziert, vgl.\@ Tabl.~\ref{Tabl:Charakt_rh,rh',rh''}.   Erste Zeile:   $f_V$ als Funktion der Quark-Konstituentenmasse~$m$ f"ur feste~$\om \!=\! 0.275, 0.290, 0.330\GeV$~[zunehmende Strichl"ange]; die ersten beiden Werte reproduzieren~$f_{\rh,{\rm exp}}$ f"ur die Quarkmasse~$M_\rh\!/\!2$ bzw.~$0.220\GeV$, der dritte Wert ist der Parameter~$\om_{\rh,T}$ unseres Ansatzes, vgl.\@ die Tabln.~\ref{Tabl:Charakt_rh,om,ph,Jps},~\ref{Tabl:Wfn-Parameter}.   Zweite Zeile:   $f_V$ als Funktion des Oszillatorparameters~$\om$ f"ur feste~$m \!=\! 0.0161, 0.220, 0.310\GeV, M_\rh\!/\!2$~[zunehmende Strichl"ange, durchgezogene Linie]; der erste Wert reproduziert~$f_{\rh,{\rm exp}}$ f"ur~$\om_{\rh,T} \!=\! 0.330\GeV$, der zweite, dritte Wert ist~$\meff[u\!/\!d,](Q^2\zz\equiv\zz0)$ bzw.~$\meff[s,](Q^2\zz\equiv\zz0)$ als typische Konstituentenmassen.
}
\label{Fig:Couplings_m,omPH}
\end{minipage}
\end{sidewaysfigure}
Wir setzen~\mbox{$\Nc \!\equiv\! 3$},\,$\hat{e}_V \!\equiv\! 1\!/\!\surd2$ und fassen f"ur den Moment die physikalischen Zust"ande~$\rh(770)$,\,$\rh(1450)$, $\rh(1700)$, vgl.\@ Tabl.~\ref{Tabl:Charakt_rh,rh',rh''}, auf als reine $1S$-,$2S$-,$2D$-Welle mit Massen~$M_{\rh\!,1S}$,~$M_{\rh'\!,2S}$, $M_{\rh^\dbprime\!,2D}$.
Wir finden aus den Kurven f"ur $f_{\rh^\dbprime,2D}(m,\om)$~[Spalte rechts]: da"s der Ansatz von~$\rh^\dbprime \!\equiv\! \rh(1700)$ als reine $2D$-Welle~-- ausgeklammert extreme Werte~f"ur~$m$,~$\om$~-- wesentlich kleinere Kopplungen~$f_{\rh^\dbprime}$ generieren, als notwendig sind, die experimentellen $\pi^+\pi^-$-,~$2\pi^+2\pi^-$-Massespektren zu reproduzieren; vgl.\@ unseren Ansatz, den wir nun fortfahren zu formulieren.

\bigskip\noindent
Die Beobachtung, da"s die Anregungen~$\rh(1450)$,~$\rh(1700)$ an das Photon etwa gleich stark und von der Gr"o"senordnung typischer $2S$-Wellen koppeln, verbietet den Ansatz eines der Zust"ande als reine $2D$-Welle.
Im vorangegangenen Kapitel wurde festgestellt, da"s Diffraktion im wesentlichen geschieht ohne Drehimpuls-"Ubertrag.
Beides zusammen ist starkes Argument daf"ur, da"s der~$2D$-Zustand stark unterdr"uckt sein sollte.

Die Notation~$nS$,~$nD$,\ldots r"uhrt her von nicht-relativistischen Wellenfunktionen, wie gerade diskutiert; sie sei benutzt als abk"urzende Bezeichnung f"ur die {\it Lichtkegelwellen\-funktionen\/} von Zust"anden, die im nicht-relativistischen Limes entsprechend charakterisiert sind.
Wir formulieren als unseren Ansatz f"ur die physikalischen Vektormeson-Zust"ande:
\vspace*{-.5ex}
\begin{alignat}{3} \label{Ansatz}
&\ket{\rh(770)}&\;
  &=\; 
       \ket{1S}
    \\
&\ket{\rh(1450)}&\;
  &=\;     \ket{2S}\cdot \cos\Th\; \phantom{(-)}
       +\; \ket{\mbox{\sl rest}}\cdot \sin\Th
    \tag{\ref{Ansatz}$'$} \\
&\ket{\rh(1700)}&\;
  &=\;     \ket{2S}\cdot (-\sin\Th)\; 
       +\; \ket{\mbox{\sl rest}}\cdot \cos\Th
    \tag{\ref{Ansatz}$''$}
    \\[-4.5ex]\nn
\end{alignat}
Dabei stehe summarisch~$\ket{\mbox{\sl rest}}$ f"ur Hybride~$\ket{\mbox{\sl h}}$ und den Zustand~$\ket{2D}$, deren jeweilige Kopplung an den elektromagnetischen Strom~-- an das Photon~-- unterdr"uckt und numerisch klein ist; wir vernachl"assigen daher diese Zust"ande.
Der Mischungswinkel~$\Th$ wird berechnet in einer Analyse der $\pi^+\pi^-$-,~$2\pi^+2\pi^-$-Massespektren, vgl.\@ Anh.~\ref{APPSubsect:Mischungswinkel}.
Zun"achst aber konstruieren wir die Lichtkegelwellenfunktionen der Zust"ande~$\ket{1S}$,~$\ket{2S}$.
\vspace*{-.5ex}

\subsection[\protect$1S$- und~\protect$2S$-Lichtkegelwellenfunktion. Konstruktion]{%
            \protect\bm{1S}- und~\protect\bm{2S}-Lichtkegelwellenfunktion. Konstruktion}

In~\ref{Subsect:1S-Vektormeson-Wfn} sind die Lichtkegelwellenfunktionen der $1S$-Vektormesonen~$\rh(770)$,~$\om(782)$, $\ph(1020)$ und~$\Jps(3097)$ angegeben.
Der Zustand~$\ket{1S}$ wird identifiziert mit der Wellenfunktion f"ur das~$\rh$-Meson dort, der Zustand~$\ket{2S}$ in Analogie konstruiert; wir schreiben, vgl.\@ Gl.~(\ref{Vektormeson-Wfn}):
\vspace*{.25ex}
\begin{align} \label{E:Vektormeson-Wfn}
\ps_{V(\la)}^{h,\bar h}(\zet,\rb{r})\;
  =\; 1\!/\!\surd\Nc\vv i_{V,f}\vv \ch_{V(\la)}^{h,\bar h}(\zet,\rb{r})
  \qquad\text{f"ur}\quad
  V \!\equiv\! 1S,2S\quad
  \la \!\equiv\! L,T
    \\[-3.5ex]\nn
\end{align}
mit~$i_{V,f} \!=\! (u\bar{u} \!-\! d\bar{d})\!/\surd2$ f"ur~$V \!\equiv\! \rh$-$1S,\rh$-$2S$,\ldots (Isospin~1), vgl.\@ Gl.~(\ref{Flavour-Gehalt}).
F"ur~\mbox{$V \!\equiv\! 1S$}~reka\-pitulieren wir, vgl.\@ Gl.~(\ref{1S-Vektormeson-Wfn}) und Fu"sn.~\ref{FN:ArgumenteIndizes}:%
\FOOT{
  Bzgl.\@ der effektiven Quarkmasse~$\meff[] \!\equiv\! \meff[u\!/\!d,]$, vgl.\@ unten die Diskussion in Anschlu"s an Gl.~(\ref{2S-Vektormeson-Wfn}).
}
%
\begin{alignat}{2} \label{E:1S-Vektormeson-Wfn}
&\hspace*{-5pt}
 \ch_{\iES(\la\equiv0)}&\;
  &=\; 4 \zbz\vv \om_{\iES,L}\vv \de_{h,-\bar h}\cdot
             g_{\iES,L}(r)\; h_{\iES,L}(\zet)
    \\[1ex]
&\hspace*{-5pt}
 \ch_{\iES(\la\equiv+1)}&\;
  &=\; \Big[\,
         \iIM\, \om_{\iES,T}^2r\, \efn{\T +\iIM\,\vph}\;
           \big( \zet\, \de_{h+,\bar h-} - \bzet\, \de_{h-,\bar h+} \big)\;
     +\; \meff[]\; \de_{h+,\bar h+}\,
                \Big]\cdot g_{\iES,T}(r)\; h_{\iES,T}(\zet) \nn \\[1ex]
&\hspace*{-5pt}
 \ch_{\iES(\la\equiv-1)}&\;
  &=\; \Big[\,
         \iIM\, \om_{\iES,T}^2r\, \efn{\T -\iIM\,\vph}\;
           \big( \bzet\, \de_{h+,\bar h-} - \zet\, \de_{h-,\bar h+} \big)\;
     +\; \meff[]\; \de_{h-,\bar h-}\,
                \Big]\cdot g_{\iES,T}(r)\; h_{\iES,T}(\zet) \nn
\end{alignat}
mit~$r \!\equiv\! |\rb{r}|$ und Definition von%
~$g_{V,\la}(r)$,~$h_{V,\la}(\zet)$ durch, vgl.\@ die Gln.~(\ref{g-Wfn}),~(\ref{h-Wfn}):%
\FOOT{
  Bzgl.\@ der Vektormeson-Masse~$M \!\equiv\! M_\iES$, vgl.\@ unten die Bem.\@ in Anschlu"s an Gl.~(\ref{h_to_h-excited}).
}
%
\vspace{-.5ex}
\begin{alignat}{3}
&g_{V,\la}(r)&\;
  &=\;   \exp \bigg[ -\frac{1}{2}\, \om_{V,\la}^2\, r^2 \bigg]&&
    \label{g-Wfn1S,2S} \\[-.5ex]
&h_{V,\la}(\zet)&\;
  &=\; {\cal N}_{V,\la}\vv \sqrt{\zbz}\vv
         \exp \bigg[ -\frac{1}{2}\,
                     \frac{M^2 (\zet \!-\! 1\!/\!2)^2}{\om_{V,\la}^2} \bigg]&&
    \label{h-Wfn1S,2S} \\[-3ex]
  &&&&&\qquad\text{f"ur}\quad
  V \!\equiv\! 1S,2S\quad
  \la \!\equiv\! L,T
    \nn
    \\[-4.5ex]\nn
\end{alignat}
Die Funktionen~$g_{V,\la}(r)$,~$h_{V,\la}(z)$ sind im wesentlichen die $1S$-Wellenfunktion des transversalen Harmonischen Oszillators und der Ansatz nach Wirbel, Stech, Bauer f"ur die Verteilung der longitudinalen Anteils~$\zet$ des Quarks am Gesamt-Lichtkegelimpuls.

Durch den Index~$V \!\equiv\! 2S$ seien Funktionen definiert, in Termen derer wir analog die Lichtkegelwellenfunktion~$\ket{2S}$ konstruieren.
Diese besitzen somit dieselbe funktionale Abh"angigkeit differieren aber in ihren Parametern~$\om_{V,\la}$,~${\cal N}_{V,\la}$.
Die Bezeichnung als~$2S$-, das hei"st {\it radiale\/} Anregung bezieht sich auf die Freiheitsgrade des {\it drei\/}dimensionalen Konfigurationsraums: transversal wie longitudinal.
Anschaulich gesprochen, sollte die Wellenfunktion des Zustands~$\ket{2S}$ einen "`Knoten"' bez"uglich~$r$ und einen "`Knoten"' bez"uglich~$\zet$ besitzen.
Wir konstruieren diese explizit in Anhang~\ref{APPSect:Vektormeson-Wfn}; hier beschr"anken wir uns darauf, diese Konstruktion zu skizzieren.
Zun"achst ist wesentlich, da"s sie geschieht im {\it Impulsraum}.

F"ur die zwei transversalen Anregungsmoden wird die~$1S$-Wellenfunktion des transversalen Harmonischen Oszillators~$\tilde{g}_{V,\la}(k) \!\propto\! g_{V,\la}\big(\om_{V,\la}^{-2}k\big)$,~$k \!\equiv\! |\rb{k}|$, vgl.\@ Gl.~(\ref{Ansatz-k_g,h-Wfn}), ersetzt durch die~$2S$-Wellenfunktion; de facto wird~$\tilde{g}_{\iZS,\la}(k)$ multipliziert mit dem entsprechenden Polynom ersten Grades in~$\om_{V,\la}^{-2}k^2$~-- und dem Faktor~$\surd2$ f"ur die {\it zwei\/} Freiheitsgrade:
\vspace*{-.5ex}
\begin{align} 
\tilde{g}_{\iES,\la}(k)\;
  \vv\underset{\text{$V \!\equiv\! 1S \!\to\! 2S$}}{\longrightarrow}\vv
  \tilde{g}_{\iZS,\la}(k)\cdot \surd2\cdot \big(1 - \om_{\iZS,\la}^{-2}k^2\big)
    \\[-4.5ex]\nn
\end{align}
F"ur die eine longitudinale Anregungsmode wird~$h_{\iZS,\la}(\zet)$ entsprechend multipliziert~--~in~Hin\-blick auf manifeste Invarianz bez"uglich des Austauschs von Quark und Antiquark:~\mbox{$\zet \!\leftrightarrow\! \bzet$}~--~mit einem Polynom in~$\zbz$, der Einfachheit halber mit einem Polynom ersten Grades:%
\FOOT{
  Die Analyse auf Basis eines Polynoms zweiten Grades f"uhrt auf vergleichbare Resultate, vgl.\@ Anh.~\ref{APPSubsect:2S-Vektormeson-Wfn}.
}
%
\vspace*{-.5ex}
\begin{align} \label{h_to_h-excited}
h_{\iES,\la}(z)\;
  \vv\underset{\text{$V \!\equiv\! 1S \!\to\! 2S$}}{\longrightarrow}\vv
      h_{\iZS,\la}(z)\cdot
      \big(\zbz - A_\la\big) \qquad\qquad
  \text{mit}\quad
  A_\la \in \bbbr
    \\[-4.5ex]\nn
\end{align}
Der Ausdruck im Impulsraum impliziert winkelabh"angige Terme mit Faktoren~$k$, die unter Fourier-Transformation "ubergehen in Ableitungen~$\pa\!/\pa r$; diese wirken auf das aus der Transformation resultierende Polynom in~$\om_{\iZS,\la}^2r^2$, das auftritt differenziert und nicht-differenziert. \\
\indent
Im Sinne eines konsistenten Modells sollte ein Zusammenhang zwischen den einzelnen Zust"anden bestehen:
Formal sollte in~$g_{\iZS,\la}$~[transversales Exponential] f"ur festes~$\la \!\equiv\! L,T$ der Oszillatorparameter~$\om_{\iZS,\la}$ gleich~$\om_{\iES,\la}$ sein; wir lassen ein leichtes Abweichen zu, vgl.\@ unten~\ref{Subsect:2SParameter}, Schritt Zwei in der Fixierung der $2S$-Parameter.
In demselben Sinne ist in~$h_{\iZS,\la}$~[longitudinales Exponential] universell~$M \!\equiv\! M_\iES$ gesetzt.

In Anhang~\ref{APPSubsect:2S-Vektormeson-Wfn} leiten wir her f"ur~$V \!\equiv\! 2S$, vgl.\@ Fu"sn.~\ref{FN:ArgumenteIndizes}:
\vspace*{-.5ex}
\begin{alignat}{2} \label{2S-Vektormeson-Wfn}
&\hspace*{-0pt}
 \ch_{\iZS(\la\equiv0)}&\,
  &=\; 4 \zbz\vv \om_{\iZS,L}\vv \de_{h,-\bar h} \\
  &&&\phantom{=\;\vv +\; \meff[]\; \de_{h+,\bar h+}\! }\times
         \big\{
           \big(\zbz \!-\! A_\la\big)
         + \surd2\, \big(\om_{\iZS,\la}^2r^2 \!-\! 1\big)
         \big\}\cdot
         g_{\iZS,L}(r)\; h_{\iZS,L}(\zet)
    \nn \\[.5ex]
&\hspace*{-0pt}
 \ch_{\iZS(\la\equiv+1)}&\,
  &=\; \Big[
         \iIM\, \om_{\iZS,T}^2r\, \efn{\T +\iIM\,\vph}\;
           \big( \zet\, \de_{h+,\bar h-} - \bzet\, \de_{h-,\bar h+} \big)
    \nn \\
  &&&\phantom{=\;\vv +\; \meff[]\; \de_{h+,\bar h+}\! }\times
         \big\{
           \big(\zbz \!-\! A_\la\big)
         + \surd2\, \big(\om_{\iZS,\la}^2r^2 \!-\! 3\big)
         \big\}
    \nn \\
  &&&\phantom{=\;\vv}
      +\; \meff[]\; \de_{h+,\bar h+}\,\cdot
          \big\{
            \big(\zbz \!-\! A_\la\big)
          + \surd2\, \big(\om_{\iZS,\la}^2r^2 \!-\! 1\big)
          \big\}
       \Big]\cdot
       g_{\iZS,T}(r)\; h_{\iZS,T}(\zet)
    \nn \\[.5ex]
&\hspace*{-0pt}
 \ch_{\iZS(\la\equiv-1)}&\,
  &=\; \Big[
         \iIM\, \om_{\iZS,T}^2r\, \efn{\T -\iIM\,\vph}\;
           \big( \bzet\, \de_{h+,\bar h-} - \zet\, \de_{h-,\bar h+} \big)
    \nn \\
  &&&\phantom{=\;\vv +\; \meff[]\; \de_{h+,\bar h+}\! }\times
         \big\{
           \big(\zbz \!-\! A_\la\big)
         + \surd2\, \big(\om_{\iZS,\la}^2r^2 \!-\! 3\big)
         \big\}
    \nn \\
  &&&\phantom{=\;\vv}
      +\; \meff[]\; \de_{h-,\bar h-}\,\cdot
          \big\{
            \big(\zbz \!-\! A_\la\big)
          + \surd2\, \big(\om_{\iZS,\la}^2r^2 \!-\! 1\big)
          \big\}
       \Big]\cdot
       g_{\iZS,T}(r)\; h_{\iZS,T}(\zet)
    \nn
    \\[-4.5ex]\nn
\end{alignat}
mit~$g_{\iZS,\la}(r)$,~$h_{\iZS,\la}(\zet)$ nach den Gln.~(\ref{g-Wfn1S,2S}),~(\ref{h-Wfn1S,2S}). \\
\indent
Wir betonen, da"s in den Gln.~(\ref{E:1S-Vektormeson-Wfn}),~(\ref{2S-Vektormeson-Wfn}) wie in der Photon-Lichtkegelwellenfunk\-tion, vgl.\@ die Gln.~(\ref{E:Photon-Wfn}),~(\ref{E:Photon-Wfn}$'$), die laufende Quarkmasse~$m_{u\!/\!d}$ ersetzt ist durch die {\it effektive Quarkmasse\/}~$\meff[](Q^2)$, die nach Gl.~(\ref{meff-explizit}) f"ur~$Q^2 \!<\! Q_0^2 \!=\! 1.05\GeV^2$ in nichttrivialer Weise abh"angt von~$Q^2$.%
\FOOT{
  Sei hier und im folgenden abk"urzend notiert:~$\meff[] \!\equiv\! \meff[u\!/\!d,]$ und~$Q^2 \!\equiv\! Q_{u\!/\!d,0}^2$.
}
Dies verwundert f"ur einen Augenblick, da~$\meff[]$ bestimmt ist f"ur die Lichtkegelwellenfunktion des Photons und abh"angt von der Gr"o"se~$Q^2$, die das Photon charakterisiert.
Hier genau liegt aber auch die Konsistenz:
Die Einf"uhrung der effektiven Quarkmasse~$\meff[]$ im vorangehenden Abschnitt geschieht in der "ublichen Interpretation, vgl.\@ etwa Ref.~\cite{Politzer76}: als die "`dem einlaufenden Impulstransfer~$Q^2$ entsprechende Parton-Masse"'.
Sie imitiert chirale Symmetriebrechung und Confinement entsprechend der effektiven Skala~$Q^2$ und ist insofern die relevante Quarkmasse auch in der Lichtkegelwellenfunktion des Vektormesons, das produziert wird durch ein Photon, charakterisiert durch diese Skala.
\vspace*{-1.75ex}

\subsection[\protect$1S$- und~\protect$2S$-Parameter. Fixierung]{%
            \protect\bm{1S}- und~\protect\bm{2S}-Parameter. Fixierung}
\label{Subsect:2SParameter}

Wir diskutieren im folgenden die Fixierung der $1S$- und $2S$-Parameter.
Sei diesbez"uglich ausdr"ucklich verwiesen auf Anhang~\ref{APPSect:Vektormeson-Wfn}, insbesondere auf Seite~\ref{APP-T:1S-ParameterFix} und~\pageref{APP-T:2S-ParameterFix}.
Es werden dort explizit angegeben und diskutiert die formalen Relationen, die zugrunde liegen der nicht-formalen Diskussion hier. \\
\indent
Aufgrund der unterschiedlichen Helizit"atenstruktur der Lichtkegelwellenfunktionen~$\ket{1S}$, $\ket{2S}$: Quark-Helizit"aten~$h,\bar h \!\equiv\! \pm1\!/\!2$ in Abh"angigket der Vektormeson-Helizit"at~$\la \!\equiv\! 0,\pm1$, sind diese formalen Relationen von unterschiedlicher Struktur f"ur longitudinale und transversale Polarisation, vgl.\@ die Gln.~(\ref{E:1S-Vektormeson-Wfn}),~(\ref{2S-Vektormeson-Wfn}).
F"ur transversale Polarisation h"angen sie weiter generell ab von der Quarkmasse, f"ur longitudinale Polarisation dagegen generell nicht.
Ersetzen der laufenden durch die effektive Quarkmasse~$m_f \!\to\! \meff$ induziert daher "uber diese eine nichttriviale~$Q^2$-Abh"angigkeit der transversalen Relationen f"ur~$Q^2 \!<\! Q_0^2 \!=\! 1.05\GeV^2$, vgl.\@ Gl.~(\ref{meff-explizit}).
Fixierung der Parameter und die folgende Diskussion bezieht sich daher a~priori auf feste, wenn auch beliebige Werte von~$Q^2$. \\
\indent
Die Fixierung der~$2S$-Parameter geschieht in Schritten.
{\bf Schritt Null}\label{T:SchrittNull}\vv ist die Fixierung der $1S$-Parameter~$\om_{\iES,\la}$,~${\cal N}_{\iES,\la}$.
Das prinzipielle Vorgehen ist bereits geschildert im vorangehenden Kapitel, vgl.\@ Seite~\pageref{T:1S-Parameter}; wir rekapitulieren:
Die Forderungen von {\it Normierung\/} der Lichtkegelwellenfunktion~$\ket{2S}$ und {\it Reproduktion von\/}~$f_\iES$, der Kopplung an den elektromagnetischen Strom, stellen f"ur feste Polarisation~$\la \!\equiv\! L,T$ und festes~$Q^2$ formal ein System zweier gekoppelter Gleichungen dar, das implizit "uber Funktionen, die Integrationen implizieren, bestimmt die Parameter~${\cal N}_{\iES,\la}$,~$\om_{\iES,\la}$ in Abh"angigkeit von den experimentell wohl-determinierten Zahlenwerten~$M_\iES \!\equiv\! M_\irh$ und~$f_{\iES,\la} \!\equiv\! f_\irh$ f"ur das~$\rh(770)$-Vektormeson.
Ein solches System von Gleichungen ist zu l"osen f"ur transversale und longitudinale Polarisation und s"amtliche Werte des betrachteten~$Q^2$-Intervalls.
Die Parameter~$\om_{\iES,L}$,~${\cal N}_{\iES,L}$ folgen als konstant, die Parameter~$\om_{\iES,T}$,~${\cal N}_{\iES,T}$~-- aufgrund der~$\meff[]$-Abh"angigkeit der Relationen~-- als nichttrivial abh"angig von~$Q^2$ f"ur Werte unterhalb~$Q_0^2 \!=\! 1.05\GeV^2$. \\
\indent
Zahlenwerte f"ur die~$1S$-Parameter~$\om_{\iES,\la}$,~${\cal N}_{\iES,\la}$ f"ur~$\la \!\equiv\! L,T$ und~$Q^2\!\equiv\! 0$ sind angegeben in Tabelle~\ref{Tabl:Wfn-Parameter}, linke Spalten; die $Q^2$-Abh"angigkeit f"ur transversale Polarisation ist graphisch dargestellt in Abbildung~\ref{Fig:1S-ParameterT_N,om}. \\
\indent
Dabei betonen wir zun"achst, da"s sich die Zahlenwerte von Tabelle~\ref{Tabl:Wfn-Parameter} beziehen auf verschwindende Photon-Virtualit"at~$Q^2 \!\equiv\! 0$.
Aufgrund der bei diesem Wert nichtverschwindenden effektiven Quarkmasse~\mbox{$\meff[u\!/\!d,](Q^2\zz\equiv\zz0) \!=\! 0.220\GeV$} sind die Zahlenwerte f"ur~$\rh(770)$ dort nicht gleich den entsprechenden Zahlen in Tabelle~\refg{Tabl:Charakt_rh,om,ph,Jps}; diese basieren auf einer verschwindenden laufenden Quarkmasse~$m_{u\!/\!d} \!=\! 0$.
"Ubereinstimmung ist in der Tat gegeben f"ur Werte~$Q^2 \!\ge\! Q_{u\!/\!d,0}^2 \!=\! 1.05\GeV^2$, da f"ur diese auch~$\meff[u\!/\!d,](Q^2)$ verschwindet.
Daher sind unsere Resultate f"ur~$\rh(770)$ und~$\om(782)$ aus Kapitel~\ref{Kap:GROUND} vollst"andig g"ultig oberhalb dieser Schwelle~[bzw.\@ die Resultate f"ur~$\ph(1020)$ oberhalb der Schwelle~$Q_{s,0}^2 \!=\! 1.6\GeV^2$].
\begin{table}
\begin{minipage}{\linewidth}
\renewcommand{\thefootnote}{\thempfootnote}
  \begin{center}
  \begin{tabular}{|h{-1}||g{7}|g{7}||f{7}|f{7}|} \hline
  \multicolumn{5}{|c|}{Parameter der $1S$- und $2S$-Lichtkegelwellenfunktionen%
                       ~f"ur~$Q^2 \!\equiv\! 0$}
  \\ \hhline{:=====:}
  \multicolumn{5}{|c|}{{\bf approximativ orthogonal}~(a.o.), Ref.~\cite{Kulzinger98}:}
  \\ \hhline{:=:t:==:t:==:}
  \multicolumn{1}{|c||}{}
    & \multicolumn{2}{c||}{$1S$}
    & \multicolumn{2}{c|}{$2S$}
  \\
    & \multicolumn{1}{c|}{\; Longitudinal\;}
    & \multicolumn{1}{c||}{\;Transversal\;}
    & \multicolumn{1}{c|}{\;Longitudinal\;}
    & \multicolumn{1}{c|}{\;Transversal\;}
  \\ \hhline{|-||--||--|}
  \mbox{$\om_{V,\la}$}/[\GeV[]]
    &      0.,330     &      0.,213
    &      0.\mbox{296\,53}\mbox{\footnote{
  \label{FN:app-orth}\protect\vspace*{-0ex}entspricht Verr"uckung~\bm{\De^{\!\rm a.o.} \!\cong\! 10.1\,\%} und~\bm{\De \!\cong\! 9.76\,\%} f"ur approximative bzw.\@ exakte Orthogonalit"at
}}
    &      0.234\,50\footnotemark[\value{mpfootnote}] \\
  \mbox{${\cal N}_{V,\la}$}/[\surd\Nc\!\equiv\!3]\footnote{
  \label{FN:calN-in-Nc}\vspace*{-2ex}vgl.\@ Unterschrift zu Tabl.~\refg{Tabl:Charakt_rh,om,ph,Jps}
}
    &      4.,48          &      3.,44          &             3.21\,16  &             1.96\,44  \\
  \mbox{$A_\la$}/  
    &       \NON          &       \NON          &             0.228\,46 & \mbox{$-$}\,0.328\,38 \\
  \mbox{$f_{V,\la}$}/[\GeV[]]
    & \bm{0.},\bm{152\,6} & \bm{0.},\bm{152\,6} & \mbox{$-$}\,0.137\,23 & \mbox{$-$}\,0.137\,23 \\
  \mbox{$R_{V,\la}$}/[\fm[]]
    &      0.,299         &     0.,536          &             0.572\,8  &             0.725\,0  \\
  \hhline{:=:b:==:b:==:}
  \multicolumn{5}{|c|}{{\bf orthogonal}, Referenz-Parametersatz:}
  \\ \hhline{:=:t:==:t:==:}
  \multicolumn{1}{|c||}{}
    & \multicolumn{2}{c||}{$1S$}
    & \multicolumn{2}{c|}{$2S$}
  \\
    & \multicolumn{1}{c|}{Longitudinal}
    & \multicolumn{1}{c||}{Transversal}
    & \multicolumn{1}{c|}{Longitudinal}
    & \multicolumn{1}{c|}{Transversal}
  \\ \hhline{|-||--||--|}
  \mbox{$\om_{V,\la}$}/[\GeV[]]
    &      0.,330     &      0.,213
    &      0.297\,72\citeFN{FN:app-orth}
    &      0.233\,74\citeFN{FN:app-orth} \\
  \mbox{${\cal N}_{V,\la}$}/[\surd\Nc\!\equiv\!3]\citeFN{FN:calN-in-Nc}
    &      4.,48          &      3.,44          &             3.20\,98  &             1.96\,34  \\
  \mbox{$A_\la$}/
    &       \NON          &       \NON          &             0.227\,94 & \mbox{$-$}\,0.291\,84 \\
  \mbox{$\, f_{V,\la}$}/[\GeV[]]
    & \bm{0.},\bm{152\,6} & \bm{0.},\bm{152\,6} & \mbox{$-$}\,0.137\,78 & \mbox{$-$}\,0.137\,78 \\
  \mbox{$R_{V,\la}$}/[\fm[]]
    &      0.,299         &      0.,536         &             0.570\,6  &             0.723\,0  \\
  \hhline{:=:b:==:b:==:}
  \end{tabular}
  \end{center}
\vspace*{-3ex}
\caption[\protect$1S$- und~\protect$2S$-Parameter f"ur~\protect$Q^2 \!\equiv\! 0$,~\protect\mbox{$\meff[](0) \!\equiv\! 0.220\GeV$}]{
  Parameter der Wellenfunktionen~$\ket{1S}$,~$\ket{2S}$ f"ur~$Q^2 \!\equiv\! 0$,~\mbox{$\meff[](0) \!\equiv\! 0.220\GeV$}.   Input in Fettdruck:~$f_\iES \!\equiv\! f_\irh$ und~$\De$; zus"atzlich:~$M_\iES \!\equiv\! M_\irh$, vgl.\@ die Tabln.~\ref{Tabl:Charakt_rh,om,ph,Jps},~\ref{Tabl:Charakt_rh,rh',rh''}, und, zur Berechnung von~$f_{\iZS,T}$, die Absch"atzung~$M_\iZS \!\cong\! 1.6\GeV$ f"ur den nichtphysikalischen~\mbox{$2S$-Zu}\-stand.   Die Parameters"atze beziehen sich auf approximative und exakte Orthogonalit"at von~$\ket{2S}$ auf~$\ket{1S}$, vgl.\@ Text und Anh.~\ref{APPSubsect:2S-Vektormeson-Wfn}.   Allein im Zahlenwert f"ur~$A_T$ geht die Diskrepanz "uber die Genauigkeit hinaus, die wir in Ref.~\cite{Kulzinger98} angegeben haben,~-- so da"s f"ur Kommensurabilit"at hier die Anzahl angegebener Dezimalstellen erh"oht wurde.
\vspace*{-.5ex}
}
\label{Tabl:Wfn-Parameter}
\renewcommand{\thefootnote}{\thechapter.\arabic{footnote}}
\end{minipage}
\end{table}
\begin{figure}
\begin{minipage}{\linewidth}
  \begin{center}
  \setlength{\unitlength}{.9mm}\begin{picture}(120,72.31)   
    \put(0,0){\epsfxsize108mm \epsffile{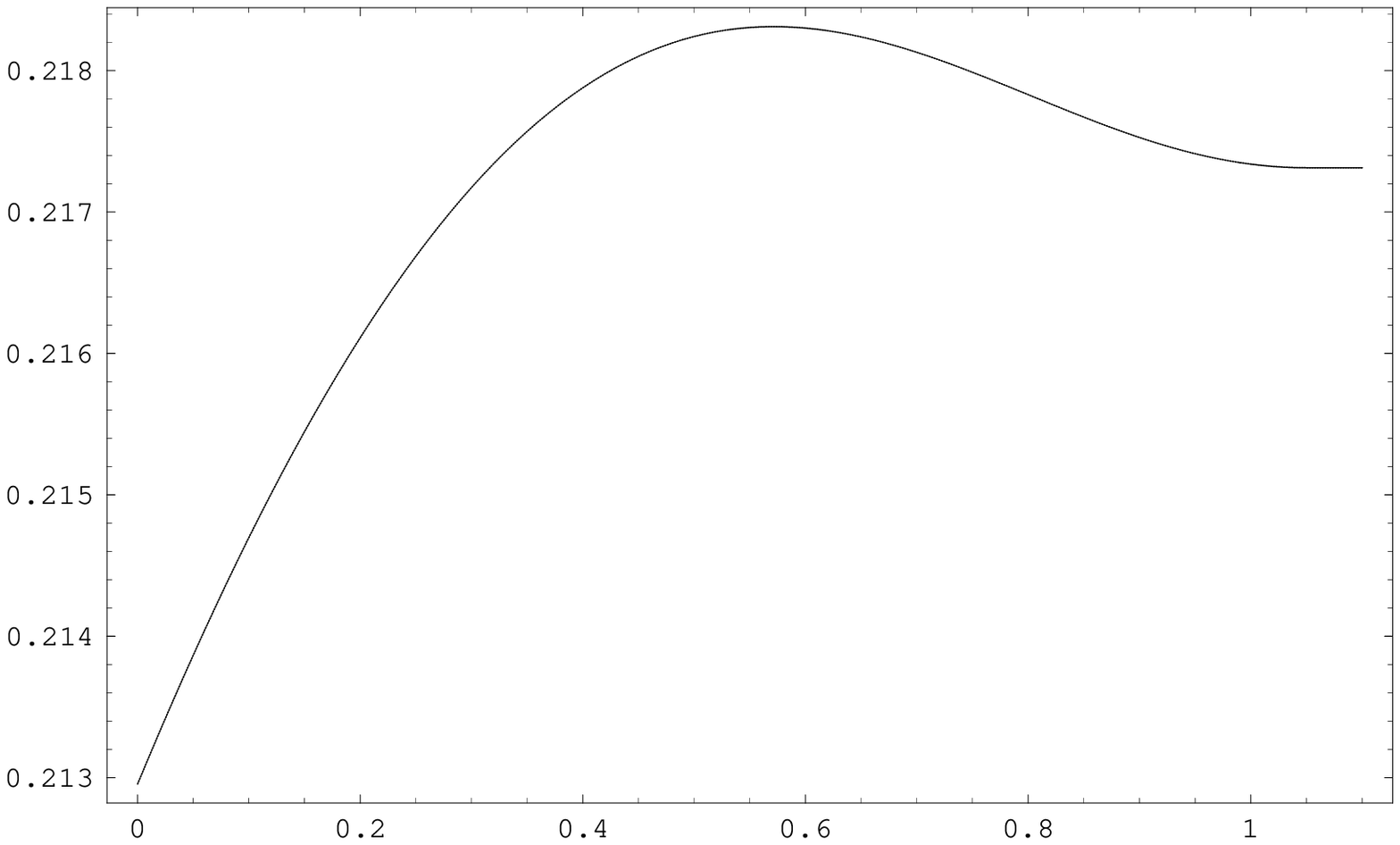}}
    \put(118,0){\normalsize$Q^2\;[\GeV[]^2]$}
    \put( -6,0){\yaxis[65.08mm]{\normalsize$\om_{\iES,T}(Q^2)\vv[\GeV[]]$}}
    \put(32,35){\normalsize$\om_{\iES,T}$}
  \end{picture}\\[.5ex]
  \setlength{\unitlength}{.9mm}\begin{picture}(120,71.97)   
    \put(0,0){\epsfxsize108mm \epsffile{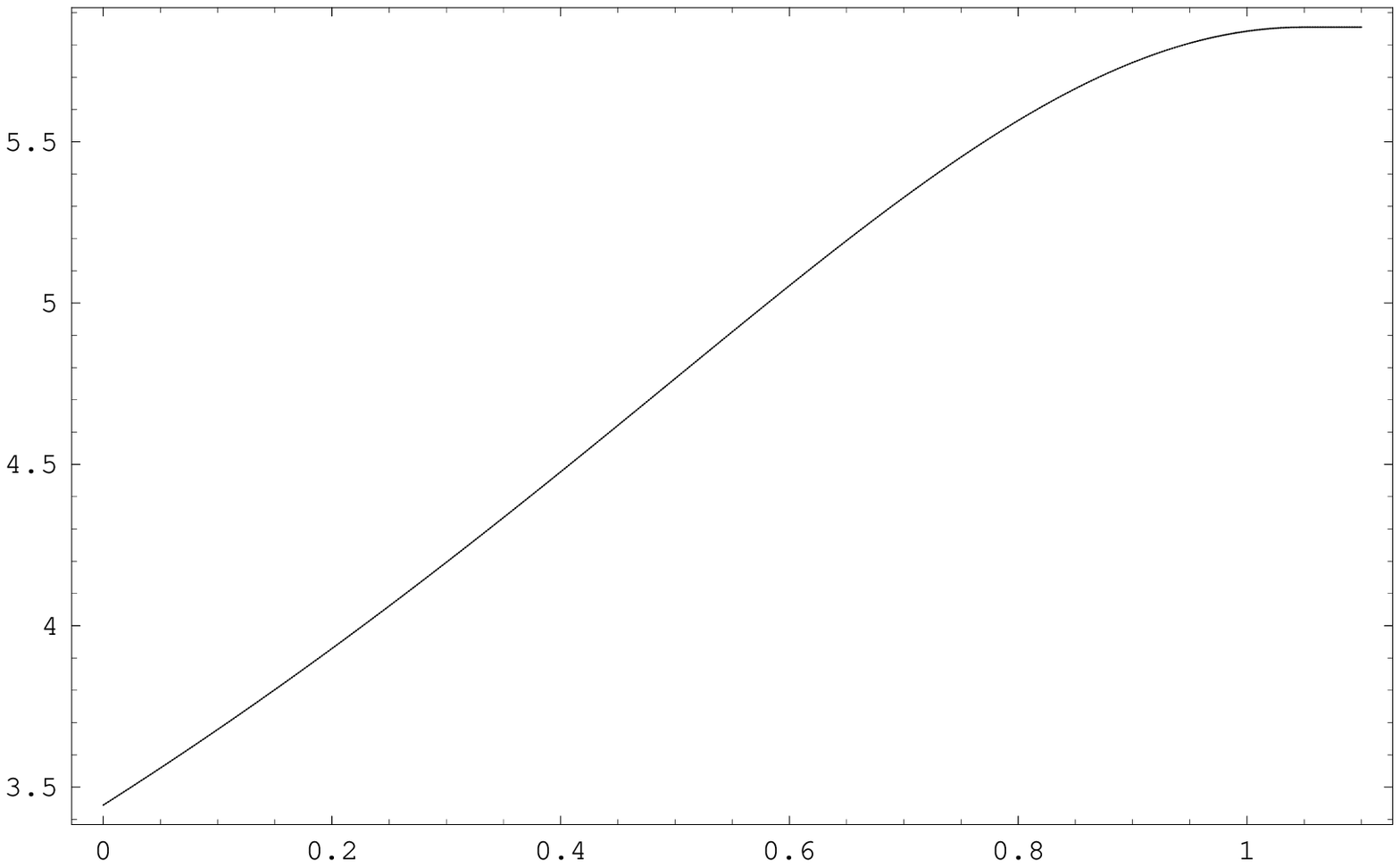}}
    \put(118,0){\normalsize$Q^2\;[\GeV[]^2]$}
    \put( -6,0){\yaxis[64.77mm]{\normalsize${\cal N}_{\iES,T}(Q^2)$}}
    \put(60,35){\normalsize${\cal N}_{\iES,T}$}
  \end{picture}
  \end{center}
\vspace*{-4.5ex}
\caption[\protect$1S$-Parameter~\protect$\om_{\iES,T}\!(\!Q^2\!)$,~\protect${\cal N}_{\iES,T}\!(\!Q^2\!)$, transversal]{
  Verhalten mit~$Q^2$ der Parameter~$\om_{\iES,T}$,~${\cal N}_{\iES,T}$.   Die Abh"angigkeit von~$Q^2$ wird induziert durch die effektive Quarkmasse~$\meff[](Q^2)$.   Diese betr"agt~\mbox{$\meff[](0) \!=\! 0.220\GeV$} f"ur~$Q^2 \!\equiv\! 0$, f"allt linear ab mit~$Q^2$ und verschwindet f"ur~$Q^2 \!\ge\! Q_0^2 \!=\! 1.05\GeV^2$; das hei"st f"ur gr"o"sere~$Q^2$ sind die Parameter~$\om_{\iES,T}$,~${\cal N}_{\iES,T}$ eingefroren auf ihre Werte f"ur~$Q^2 \!\equiv\! 1.05$ und diese identisch denen des vorangehenden Kapitels.
\vspace*{-1.75ex}
}
\label{Fig:1S-ParameterT_N,om}
\end{minipage}
\end{figure}
\\
\indent
Die Bestimmung von~$\om_{\iES,T}$ als Funktion von~$Q^2$ ist komplex; sie geschieht in~praxi f"ur~$\la \!\equiv\! L,T$ und jedes~$Q^2$ durch die numerische Bestimmung im Rahmen des Newtonschen Approximationsverfahrens des Fixpunktes einer komplizierten nicht-aufl"osbaren Gleichung von Integral-Funktionen in~$\om_{\iES,T}$.
Andererseits ist~$\om_{\iES,T}(Q^2)$ von zentraler Bedeutung f"ur die Berechnung aller "ubrigen Parameter.
Wir geben daher in Anhang~\ref{APPSubsect:2S-Vektormeson-Wfn} auf Seite~\pageref{Tabl-APP:om1ST,De,f2S}, Tabelle~\ref{Tabl-APP:om1ST,De,f2S} explizite Zahlenwerte an in ausreichender Genauigkeit und f"ur das gesamte relevante Intervall von~$Q^2$; siehe dort auch den bez"uglich~$Q^2$ konstanten Zahlenwert f"ur~$\om_{\iES,L}$. \\
\indent
Die Lichtkegelwellenfunktion~$\ket{2S}$ des $2S$-Zustands besitzt gegen"uber~$\ket{1S}$~-- f"ur feste Polarisation~\mbox{$\la \!\equiv\! L,T$}~-- neben~$\om_{\iZS,\la}$,~${\cal N}_{\iZS,\la}$ einen dritten Parameter, n"amlich~$A_\la$.%
~\mbox{Dieser} parametrisiert die Position des Knotens der longitudinalen Anregungsmode und stellt sich heraus als der Parameter, der allein die Forderung subsumiert von Orthogonalit"at von~$\ket{2S}$ auf~$\ket{1S}$ und durch diese bereits vollst"andig bestimmt ist, vgl.\@ die Gln.~(\ref{APP:OrthoGon_skalar-zet}),~(\ref{APP:Ala_ij}) bzw.~(\ref{APP:OrthoGon_skalar-zet}$'$),~(\ref{APP:Ala_ij}$'$).
Die Fixierung der Parameter~$\om_{\iZS,\la}$,~${\cal N}_{\iZS,\la}$ und~$A_\la$ kann nicht analog durchtgef"uhrt werden wie die des~$1S$-Zustands, und zwar aus dem folgenden Grund. \\
\indent
Der $2S$-Zustand ist nicht physikalisch und die~$M_\irh$ und~$f_\irh$ entsprechenden Parameter~$M_\iZS$ und~$f_\iZS$ daher experimentell nicht unmittelbar zug"anglich.
Seine Masse kann zwar abgesch"atzt werden durch die experimentell bestimmten Massen der physikalischen Zust"ande~$\rh(1450)$,~$\rh(1700)$, vgl.\@ die Gln.~(\ref{Ansatz}$'$),~(\ref{Ansatz}$''$).
Die Kopplungen~$f_\irhp$,~$f_\irhpp$ aber sind experimentell nicht bekannt, so da"s unser Zahlenwert f"ur~$f_\iZS$ reines Postulat sein wird.
In die Relationen, auf Basis derer wir die $2S$-Parameter bestimmen, gehen die Kopplung de facto nur ein als~$f_{\iZS,L}$ und als das Produkt~$f_{\iZS,T}\!\cdot\!M_\iZS$, vgl.\@ die Gln.~(\ref{fV_LCWfn}),~(\ref{fV_LCWfn}$'$).
Wir stellen andererseits die physikalische Forderung von Identit"at der Kopplungen f"ur~longi\-tudinale und transversale Polarisation:~\mbox{$f_\iZS \!\equiv\! f_{\iZS,L} \!\equiv\! f_{\iZS,T}$}; vgl.\@ etwa die Gln.~(\ref{f_V-nS,nD}),~(\ref{f_V-nS,nD}$'$) und~(\ref{La_la,S/D}),~(\ref{La_la,S/D}$'$) in dem vorgestellten relativistisch erweiterten Quarkmodell.
Damit folgt prinzipiell~$M_\iZS$.
Wir gehen hiervon abweichend den folgenden Weg:
Wir sch"atzen ab~\mbox{\,$M_\iZS \!\cong\! 1.6\GeV$}%
\FOOT{
  Im Sinne des alten Eintrags~$\rh(1600)$ der Particle Data Group, vgl.~\ref{Subsect:Zushang} auf Seite~\pageref{T:rh(1600)}.
}
und fordern Identit"at der Kopplungen f"ur longitudinale und transversale Polarisation.
Dies wird erreicht, indem wir zulassen, da"s f"ur feste Polarisation der Oszillatorparameter~$\om_{\iZS,\la}$ abweicht von~$\om_{\iES,\la}$, und fordern, da"s die relative Verr"uckung {\it minimal\/} und {\it identisch\/} ist.
Wir pr"azisieren dies im folgenden.

Sei betrachtet feste Polarisation~\mbox{$\la \!\equiv\! L,T$}.
{\bf Schritt~Eins}\vv setzt an die Oszillatorparameter des~$2S$-Zustands als identisch gleich denen des~$1S$-Zustands:
\vspace*{-1ex}
\begin{align} 
\hspace*{-1em}
  \text{Schritt~Eins:}\qquad
  \om_{\iZS,\la}\;
  \stackrel{\D!}{=}\; \om_{\iES,\la} \qquad
  \text{$\la \!\equiv\! L,T$,\vv fest}
    \\[-4.25ex]\nn
\end{align}
vgl.\@ Gl.~(\ref{APP:Parameter1.}).
{\bf Schritt~Eins$^{\bm{\prime}}$}\vv bstimmt dann sukzessive~$A_\la$,~${\cal N}_{\iZS,\la}$ und~$f_{\iZS,\la}$:
Aus der~ers\-ten Forderung an den Zustand~$\ket{2S}$, von {\it Orthogonalit"at\/} auf~$\ket{1S}$, folgt der Parameter~$A_\la$ in Abh"angigkeit von~$\om_{\iES,\la}$ und~$M_\iES$,%
\FOOT{
  \label{FN:M,f-rho-dependence}d.h.\@ ein expliziter Zahlenwert, der effektiv abh"angt von~$M_\iES \!\equiv\! M_\irh$ und~$f_\iES \!\equiv\! f_\irh$
}
vgl.\@ die Gln.~(\ref{APP:OrthoGon_skalar-zet}),~(\ref{APP:OrthoGon_skalar-zet}$'$) und~(\ref{APP:Ala_ij}),~(\ref{APP:Ala_ij}$'$).
Aus der zweiten Forderung an~$\ket{2S}$, seiner {\it Normiertheit\/}, folgt weiter mithilfe~$A_\la$ der Parameter~${\cal N}_{\iZS,\la}$\citeFN{FN:M,f-rho-dependence}, vgl.\@ die Gln.~(\ref{APP:Norm_skalar2S-zet}),~(\ref{APP:Norm_skalar2S-zet}$'$).
Aus der dritten Forderung an~$\ket{2S}$, von {\it Reproduktion der Kopplung an den elektromagnetischen Strom\/}~$J_{\rm em}$, folgt schlie"slich mithilfe~$A_\la$,~${\cal N}_{\iZS,\la}$ ein expliziter Zahlenwert f"ur~$f_{\iZS,\la}$, vgl.\@ die Gln.~(\ref{APP:fV_LCWfn2S-zet}),~(\ref{APP:fV_LCWfn2S-zet}$'$):
\vspace*{-1ex}
\begin{alignat}{3} \label{Parameter1'.}
\hspace*{-1em}
  \text{Schritt~Eins$^{\bm{\prime}}$:}\qquad
 &\om_{\iZS,\la}\qquad
  \text{$\la \!\equiv\! L,T$,\vv fest}&&
    \\
 &\to\quad
  \ket{2S}\vv \text{!\vv orthogonal~$\ket{1S}$}&\quad
 &\Longrightarrow\quad
  A_\la
    \nn \\
 &\to\quad
  \ket{2S}\vv \text{!\vv normiert}&\quad
 &\Longrightarrow\quad
  {\cal N}_{\iZS,\la}
    \nn \\
 &\to\quad
  \ket{2S}\vv \text{!\vv koppelt an~$J_{\rm em}$}&\quad
 &\Longrightarrow\quad
  f_{\iZS,\la}
    \nn
    \\[-4.25ex]\nn
\end{alignat}
Dies geschieht separat und unabh"angig voneinander f"ur longitudinale und transversale Polarisation.
Offensichtlich differieren die resultierenden Kopplungen:~\vspace*{-.25ex}$f_{\iZS,L} \!\ne\! f_{\iZS,T}$; wir fordern dagegen physikalisch ihre Identit"at.
{\bf Schritt Zwei}\vv ist daher, simultan wegzur"ucken~$\om_{\iZS,L}$ von~$\om_{\iES,L}$ und~$\om_{\iZS,T}$ von~$\om_{\iES,T}$:
\vspace*{-1ex}
\begin{align} \label{Parameter2.}
\hspace*{-1em}
  \text{Schritt~Zwei:}\qquad
 &f_{\iZS,L}\;
    \stackrel{\D!}{=}\; f_{\iZS,T}\quad
  \Longrightarrow\quad
  \om_{\iZS,\la}\;
    \ne\; \om_{\iES,\la} \qquad
  \text{$\la \!\equiv\! L,T$,\vv fest}
    \\
 &\om_{\iZS,L} \stackrel{\D!}{=} \om_{\iES,L} (1\!-\! \De)
    \tag{\ref{Parameter2.}$'$} \\[-.5ex]
 &\text{\&}\quad
  \om_{\iZS,T} \stackrel{\D!}{=} \om_{\iES,T} (1\!+\! \De) \qquad
  \text{$|\mskip-2mu\De\mskip-2mu|$\vv minimal}
    \nn
    \\[-4.25ex]\nn
\end{align}
vgl.\@ die Gln.~(\ref{APP:Parameter2.}),~(\ref{APP:Parameter2.}$'$).
Identit"at der Kopplungen~$f_{\iZS,L}$,~$f_{\iZS,T}$ kann in dieser Weise erreicht werden f"ur~$\De \!>\! 0$, das hei"st~$\om_{\iZS,L}$ ist kleiner zu w"ahlen als~$\om_{\iES,L}$ und~$\om_{\iZS,T}$ gr"o"ser als~$\om_{\iES,L}$.
Effektiv ist Identit"at der Kopplungen erreicht f"ur relative Verr"uckungen um etwa~$10\,\%$.
Explizit ist~$\De \!\equiv\! \De(Q^2)$ eine Funktion von~$Q^2$.

In~praxi gehen wir vor wie folgt:
Wir betrachten festes~$Q^2$, f"ur Definiertheit~$Q^2 \!\equiv\! 0$.
Wir setzen~$\om_{\iZS,\la} \!\equiv\! \om_{\iES,\la}$ als Schritt Eins.
Wir f"uhren durch Schritt~Eins$^{\bm{\prime}}$, das hei"st wir bestimmen sukzessive~$A_\la$, dann~${\cal N}_{\iZS,\la}$, dann~$f_{\iZS,\la}$~-- separat f"ur longitudinale und transversale Polarisation~$\la \!\equiv\! L,T$.
Vergleich von~$f_{\iZS,L}$,~$f_{\iZS,T}$ suggeriert einen Ansatz f"ur~$\De$; der gem"a"s Gl.~(\ref{Parameter2.}$'$) Werte f"ur~$\om_{\iZS,\la}$ induziert.
Mit diesen Oszillatorparameter wird wiederholt Schritt Eins$^{\bm{\prime}}$, das hei"st sukzessive bestimmt~$A_\la$,~${\cal N}_{\iZS,\la}$, und~$f_{\iZS,\la}$.
Vergleich von~$f_{\iZS,L}$,~$f_{\iZS,T}$ suggeriert eine Modifikation von~$\De$.
Mit den korrespondierenden~$\om_{\iZS,\la}$ wird wieder durchgef"uhrt Schritt Eins$^{\bm{\prime}}$.
Und so weiter.
Resultat dieser Iteration von Hand ist ein Wert~$\De(Q^2\zz\equiv\zz0)$, der garantiert neben Identit"at von~$f_{\iZS,L}$,~$f_{\iZS,T}$ die Forderungen von Orthogonalit"at und Normiertheit der Lichtkegelwellefunktionen~$\ket{1S}$,~$\ket{2S}$ f"ur feste Polarisation.

Dieses Iterationsverfahren ist durchzuf"uhren f"ur s"amtliche Werte von~$Q^2$ in dem Intervall von~$0$ bis~$Q_0^2 \!=\! 1.05\GeV^2$, in dem die Forderungen "uber die effektive Quarkmasse~$\meff[]$ in nichttrivialer Weise abh"angen von~$Q^2$.
Dieses Vorgehen ist offensichtlich sehr aufwendig.
Prinzipiell kann es formalisiert und automatisiert werden, etwa auf Basis des Newtonschen Approximationsverfahrens zur Bestimmung von Fixpunkten impliziter Funktionen.
Dieses legen wir zugrunde der Berechnung des~$1S$-Oszillatorparameters f"ur transversale Polarisation:~$\om_{\iES,T}$ in Schritt Null; vgl.\@ explizit Anh.~\ref{APPSubsect:1S-Vektormeson-Wfn}, die Gln.~(\ref{APP:NewtonApprox0}),~(\ref{APP:NewtonApprox}).
Aufgrund der Komplexit"at der Gleichungen zur Bestimmung von~$\De(Q^2)$ ist dies in diesem Fall nicht praktikabel; wir verweisen auf Anh.~\ref{APPSubsect:1S-Vektormeson-Wfn}, die Gln.~(\ref{APP:fV_LCWfn2S-zet}),~(\ref{APP:fV_LCWfn2S-zet}$'$),~(\ref{APP:Norm_skalar2S-zet}),~(\ref{APP:Norm_skalar2S-zet}$'$) und~(\ref{APP:OrthoGon_skalar-zet}),~(\ref{APP:OrthoGon_skalar-zet}$'$).
Wir sind dazu gezwungen,~$\De(Q^2)$ zu bestimmen~{\it von Hand}.

\begin{figure}
\begin{minipage}{\linewidth}
  \begin{center}
  \setlength{\unitlength}{.9mm}\begin{picture}(120,71)   
    \put(0,0){\epsfxsize108mm \epsffile{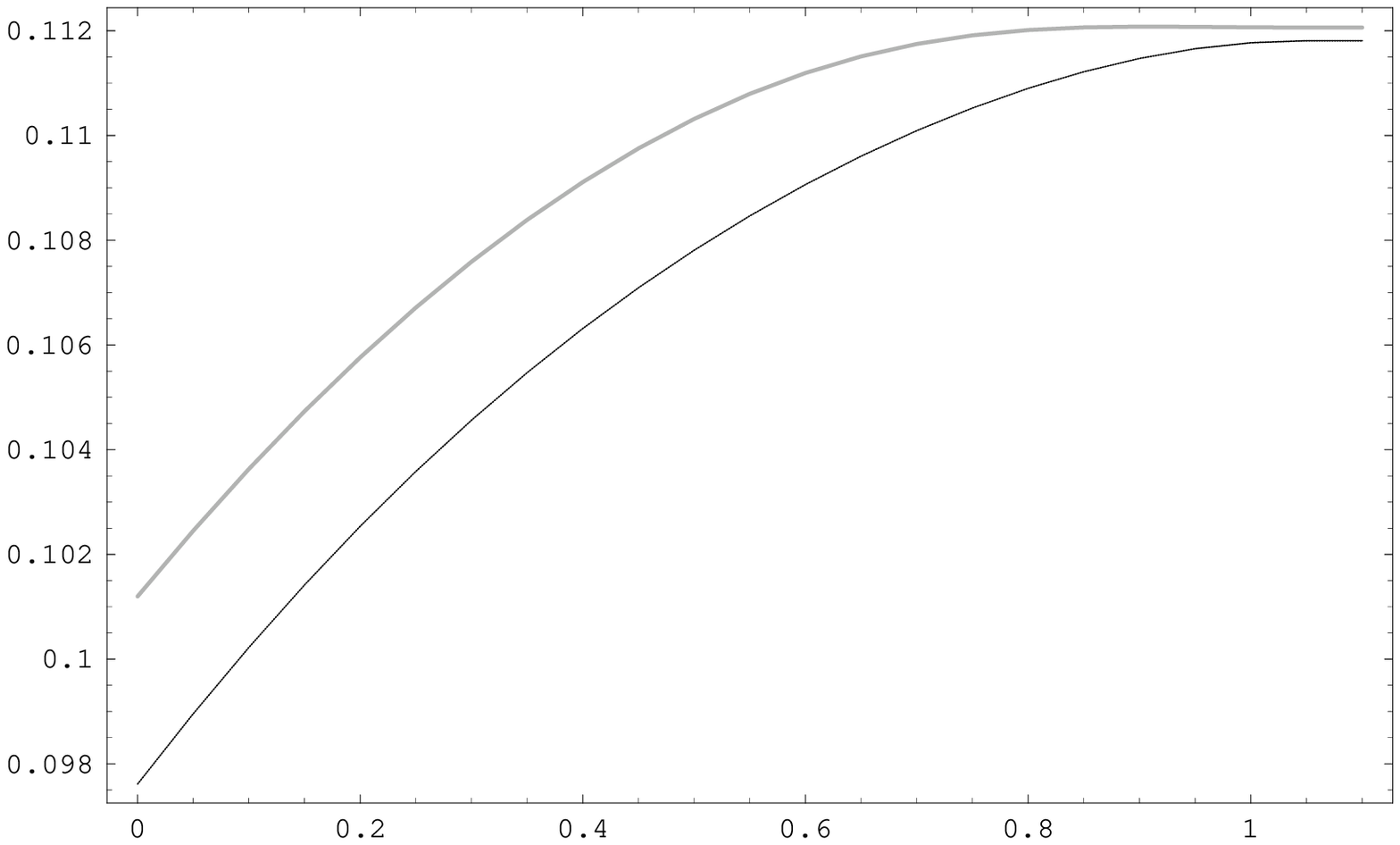}}
    \put(118,0){\normalsize$Q^2\;[\GeV[]^2]$}
    \put( -6,0){\yaxis[63.9mm]{\normalsize$\De^{\!\rm a.o.}(Q^2),\,\De(Q^2)$}}
    \put(29,47){\normalsize$\De^{\!\rm a.o.}$}
    \put(64,47){\normalsize$\De$}
  \end{picture}
  \end{center}
\vspace*{-4ex}
\caption[Parameter~\protect$\De^{\!\rm a.o.}\!(\!Q^2\!)$,~\protect$\De\!(\!Q^2\!)$, approximativ und exakt orthogonal]{
  Verhalten mit~$Q^2$ der Parameter~$\De^{\!\rm a.o.}$,~$\De$.   Diese bestimmen die Verr"uckung der Oszillatorparameter~$\om_{\iZS,\la}$ relativ zu~$\om_{\iES,\la}$ in der Weise, da"s garantiert ist Identit"at der Kopplungen~$f_{\iZS,L}$,~$f_{\iZS,T}$ bei minimaler und identischer relativen Verr"uckung.   Graue Kurve f"ur approximierte, schwarze f"ur exakte Orthogonalit"at von~$\ket{2S}$ auf~$\ket{1S}$, vgl.\@ Text.
\vspace*{-.5ex}
}
\label{Fig:Delta}
\end{minipage}
\end{figure}

In Abbildung~\ref{Fig:Delta} stellt die schwarze (durchgezogene) Kurve graphisch dar das Resultat f"ur~$\De$ als Funktion von~$Q^2$.
Zum einen ist die Bestimmung von~$\De(Q^2)$ sehr aufwendig.
Zum anderen determiniert~$\De(Q^2)$ gem"a"s Gl.~(\ref{Parameter2.}$'$) eindeutig die Oszillatorparameter~$\om_{\iZS,L}$,~$\om_{\iZS,T}$.
Diese sind~-- als Schritt Eins$^{\bm{\prime}}$~-- fundamentaler Ausgangspunkt der sukzessiven Bestimmung der Parameter~$A_\la$,~${\cal N}_{\iZS,\la}$, und~$f_{\iZS,\la}$.
Neben den $1S$-Oszillatorparametern, insbesondere~$\om_{\iES,T}(Q^2)$ ist daher die Funktion~$\De(Q^2)$ von zentraler Bedeutung, so da"s wir explizite Zahlenwerte in ausreichender Genauigkeit und f"ur das gesamte relevante Intervall von~$Q^2$ hinzuf"ugen der Tabelle~\refg{Tabl-APP:om1ST,De,f2S}.

Aus~$\De(Q^2)$ folgen unmittelbar die $2S$-Oszillatorparameter~$\om_{\iZS,\la}$.
Nach Gl.~(\ref{Parameter2.}$'$) {\it induziert\/} ihre Abh"angigkeit von~$Q^2$ eine $Q^2$-Abh"angigkeit des longitudinalen $2S$-Oszillatorpara\-meters~$\om_{\iZS,L}$ gegen"uber dem konstanten $1S$-Parame\-ter~$\om_{\iES,L}$~-- und {\it modifiziert\/} ihre Abh"ang\-igkeit die $Q^2$-Abh"angigkeit des transversalen $2S$-Oszillatorpara\-meters~$\om_{\iZS,T}$, die herr"uhrt von dem $1S$-Parameter~$\om_{\iES,T}$.
Konsequenz des Fixierungsverfahrens ist daher die nichttriviale Abh"angigkeit von~$Q^2$ f"ur Werte unterhalb~$Q_0^2 \!=\! 1.05\GeV^2$ {\it beider Polarisationen\/} aufgrund Schritt Zwei und {\it s"amtlicher\/}~$2S$-Parameter aufgrund Schritt Eins$^{\bm{\prime}}$.

\begin{figure}
\begin{minipage}{\linewidth}
  \begin{center}
  \setlength{\unitlength}{.9mm}\begin{picture}(120,71.54)   
    \put(0,0){\epsfxsize108mm \epsffile{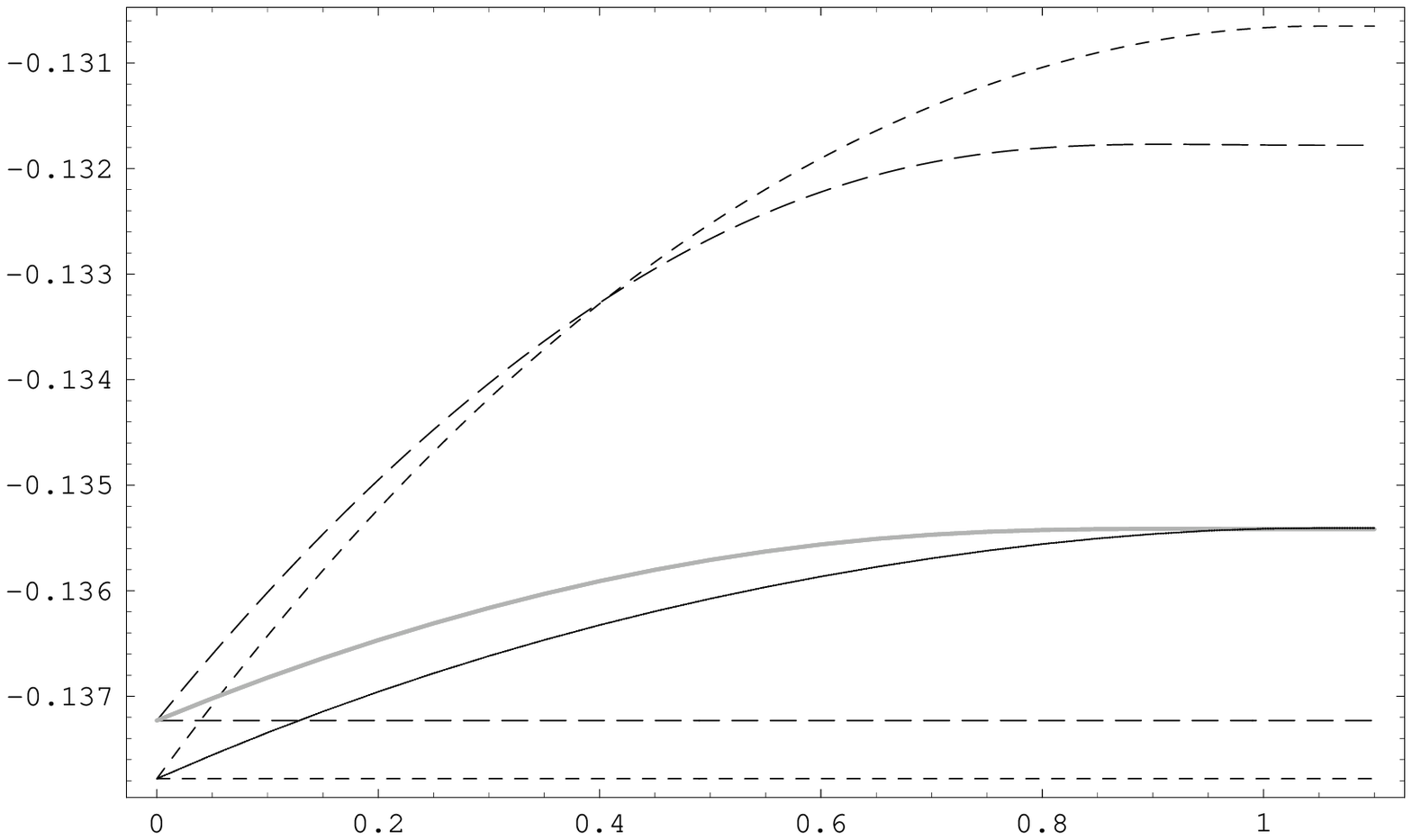}}
    \put(118,0){\normalsize$Q^2\;[\GeV[]^2]$}
    \put( -6,0){\yaxis[64.08mm]{\normalsize$f_{\iZS,\la}(Q^2)\;[\GeV[]]$}}
    \put(92,54){\normalsize$f_{\iZS,T}$}
    \put(92,30){\normalsize$f_{\iZS,L} \!\equiv\! f_{\iZS,T}$}
    \put(92,13){\normalsize$f_{\iZS,L}$}
  \end{picture}
  \end{center}
\vspace*{-4ex}
\caption[\protect$Q^2$-Verhalten von~\protect$f_{\iZS,\la}$, approximativ und exakt orthogonal]{
  Verhalten mit~$Q^2$ der Kopplungen~$f_{\iZS,\la}$.   Lang gestrichelt: Die approximativ orthogonale Parametrisierung von Ref.~\cite{Kulzinger98}; die Oszillatorparameter~$\om_{\iZS,\la}$ werden f"ur~$Q^2 \!\equiv\! 0$ minimal in der Weise verr"uckt von~$\om_{\iES,\la}$, da"s die Kopplungen~$f_{\iZS,\la}$ identisch sind, und eingefroren auf diese Werte~$\om_{\iZS,\la}(Q^2\zz\equiv\zz0)$.   F"ur nichtverschwindende~$Q^2$ ist Identit"at von~$f_{\iZS,L}$,~$f_{\iZS,T}$ nicht mehr erf"ullt, da die transversalen Relationen "uber~$\meff[]$ abh"angen von~$Q^2$, vgl.\@ Schritt Eins$^{\bm{\prime}}$.   Kurz gestrichelt: analog f"ur exakte Orthogonalit"at.   Die durchgezogenen Kurven zeigen den Effekt, wenn die~$\om_{\iZS,\la}$~-- abh"angig von~$Q^2$~-- so verschoben werden, da"s Identit"at der Kopplungen f"ur jedes~$Q^2$ garantiert ist: grau approximative, schwarz exakte Orthogonalit"at auf Basis von~$\De^{\!\rm a.o.}(Q^2)$ beziehungsweise~$\De(Q^2)$.
\vspace*{-1.5ex}
}
\label{Fig:f2S}
\end{minipage}
\end{figure}

Die Parameter~$A_\la$,~${\cal N}_{\iZS,\la}$ und die Kopplung~\mbox{$f_\iZS \!\equiv\! f_{\iZS,L} \!\equiv\! f_{\iZS,T}$} folgen aus~$\om_{\iZS,\la}$ sukzessive durch Schritt Eins$^{\bm{\prime}}$.
Zahlenwerte f"ur~$Q^2 \!\equiv\! 0$ sind angegeben im unteren Teil von Tabelle~\ref{Tabl:Wfn-Parameter}.
Die Abh"angigkeit von~$Q^2$ der Kopplung~$f_\iZS$ ist graphisch dargestellt in Abbildung~\ref{Fig:f2S}, die der Parameter~$\om_{\iZS,\la}$,~${\cal N}_{\iZS,\la}$,~$A_\la$ f"ur longitudinale Polarisation in Abbildung~\ref{Fig:2S-ParameterL_N,om,A}, f"ur transversale Polarisation in Abbildungen~\ref{Fig:2S-ParameterT_N,om,A}:
Wir verweisen auf die schwarzen durchgezogenen Kurven und betrachten die entsprechenden Parameter, die resultieren aus dem geschilderten Verfahren auf Basis von~$\om_{\iES,\la}$,~$\De(Q^2)$, vgl.\@ Tabl.~\refg{Tabl-APP:om1ST,De,f2S}, als {\it Referenz-Parametersatz}.%
~Wir fassen zusammen, da"s er charakterisiert: Lichtkegelwellenfunktionen~$\ket{1S}$,~$\ket{2S}$, die f"ur jedes~$Q^2$ normiert sind, orthogonal aufeinander stehen und eine f"ur longitudinale und transversale Polarisation identische Kopplungen~$f_\iZS \!\equiv\! f_{\iZS,L} \!\equiv\! f_{\iZS,T}$ an den elektromagnetischen Strom generieren.
Dabei ist die relative Verr"uckung der $2S$- von den $1S$-Oszillatorparameter minimal und identisch.
%
%

%
%
\begin{figure}
\begin{minipage}{\linewidth}
  \begin{center}
  \setlength{\unitlength}{.9mm}\begin{picture}(120,71)   
    \put(0,0){\epsfxsize108mm \epsffile{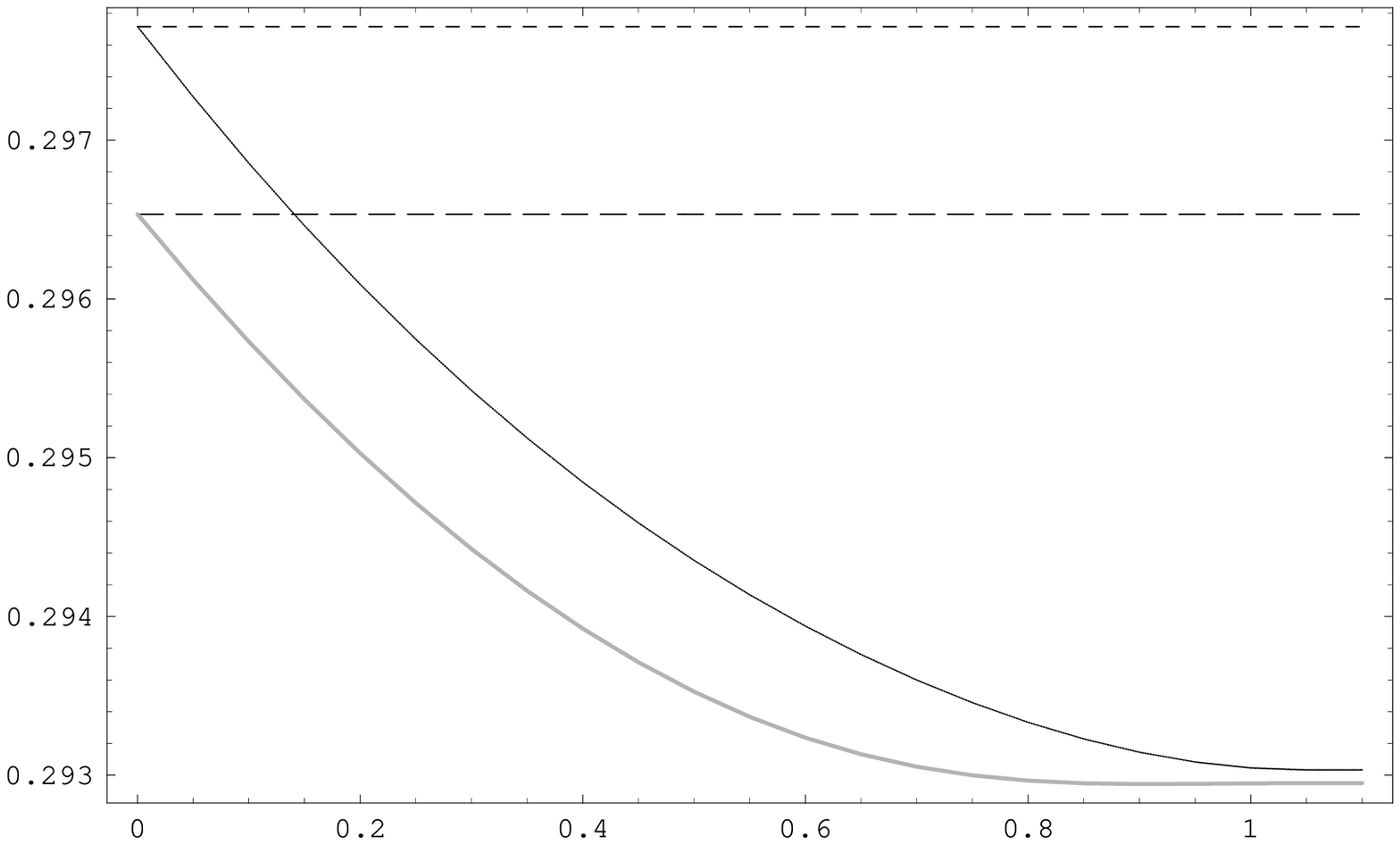}}
    \put(118,0){\normalsize$Q^2\;[\GeV[]^2]$}
    \put( -6,0){\yaxis[63.9mm]{\normalsize$\om_{\iZS,L}(Q^2)$}}
  \end{picture}\\[.5ex]
  \setlength{\unitlength}{.9mm}\begin{picture}(120,71)   
    \put(0,0){\epsfxsize108mm \epsffile{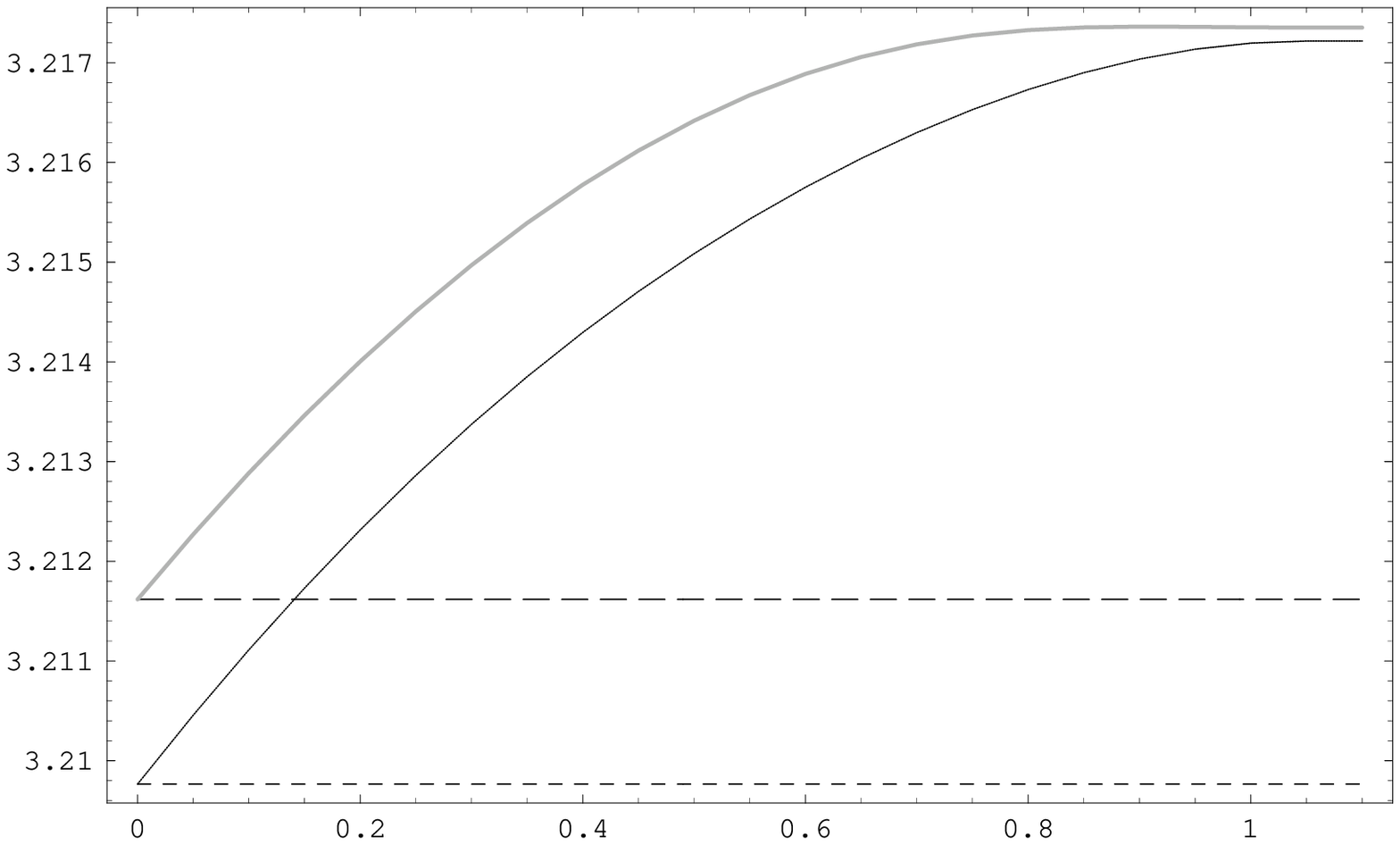}}
    \put(118,0){\normalsize$Q^2\;[\GeV[]^2]$}
    \put( -6,0){\yaxis[63.9mm]{\normalsize${\cal N}_{\iZS,L}(Q^2)$}}
  \end{picture}\\[.5ex]
  \setlength{\unitlength}{.9mm}\begin{picture}(120,71.28)   
    \put(0,0){\epsfxsize108mm \epsffile{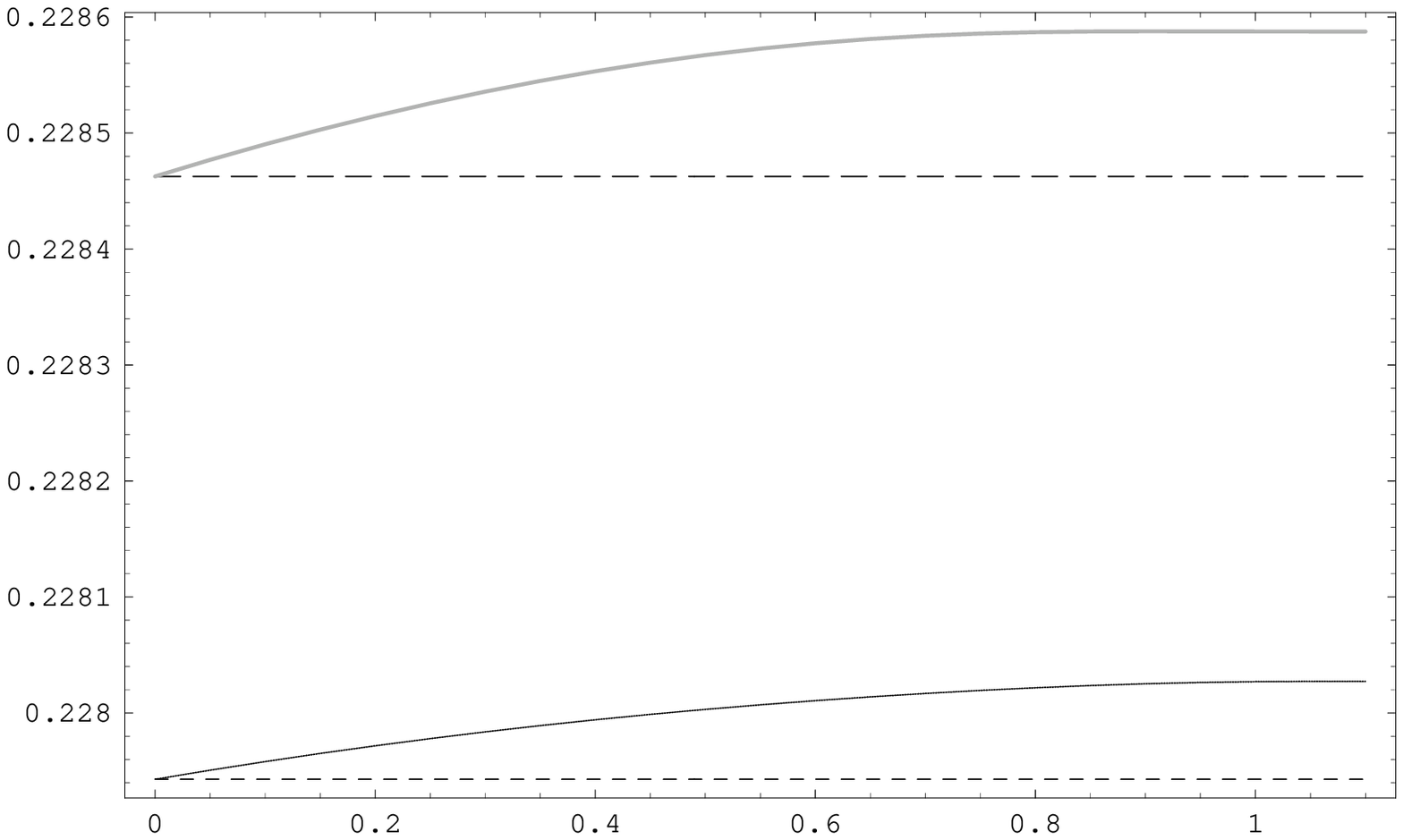}}
    \put(118,0){\normalsize$Q^2\;[\GeV[]^2]$}
    \put( -6,0){\yaxis[64.15mm]{\normalsize$A_L(Q^2)$}}
  \end{picture}
  \end{center}
\vspace*{-4ex}
\caption[\protect$2S$-Parameter~\protect$\om_{\iZS,L}\!(\!Q^2\!)$,~\protect${\cal N}_{\iZS,L}\!(\!Q^2\!)$,~\protect$A_L\!(\!Q^2\!)$, longitudinal]{
  Verhalten mit~$Q^2$ der~$2S$-Parameter~$\om_{\iZS,L}$,~${\cal N}_{\iZS,L}$,~$A_L$ f"ur longitudinale Polarisation,~$\la \!\equiv\! L$.   Gestrichelte und durchgezogene Kurven entsprechend Abbildung~\ref{Fig:f2S} und Text.   Die~$Q^2$-Abh"angigkeit der longitudinalen Parameter ist induziert~\mbox{durch die Forderung} von Identit"at der Kopplungen~$f_{\iZS,L}$,~$f_{\iZS,T}$ f"ur jedes~$Q^2$; sie ist insgesamt nur sehr schwach.
}
\label{Fig:2S-ParameterL_N,om,A}
\end{minipage}
\end{figure}
%
%
%
%
\begin{figure}
\begin{minipage}{\linewidth}
  \begin{center}
  \setlength{\unitlength}{.9mm}\begin{picture}(120,71)   
    \put(0,0){\epsfxsize108mm \epsffile{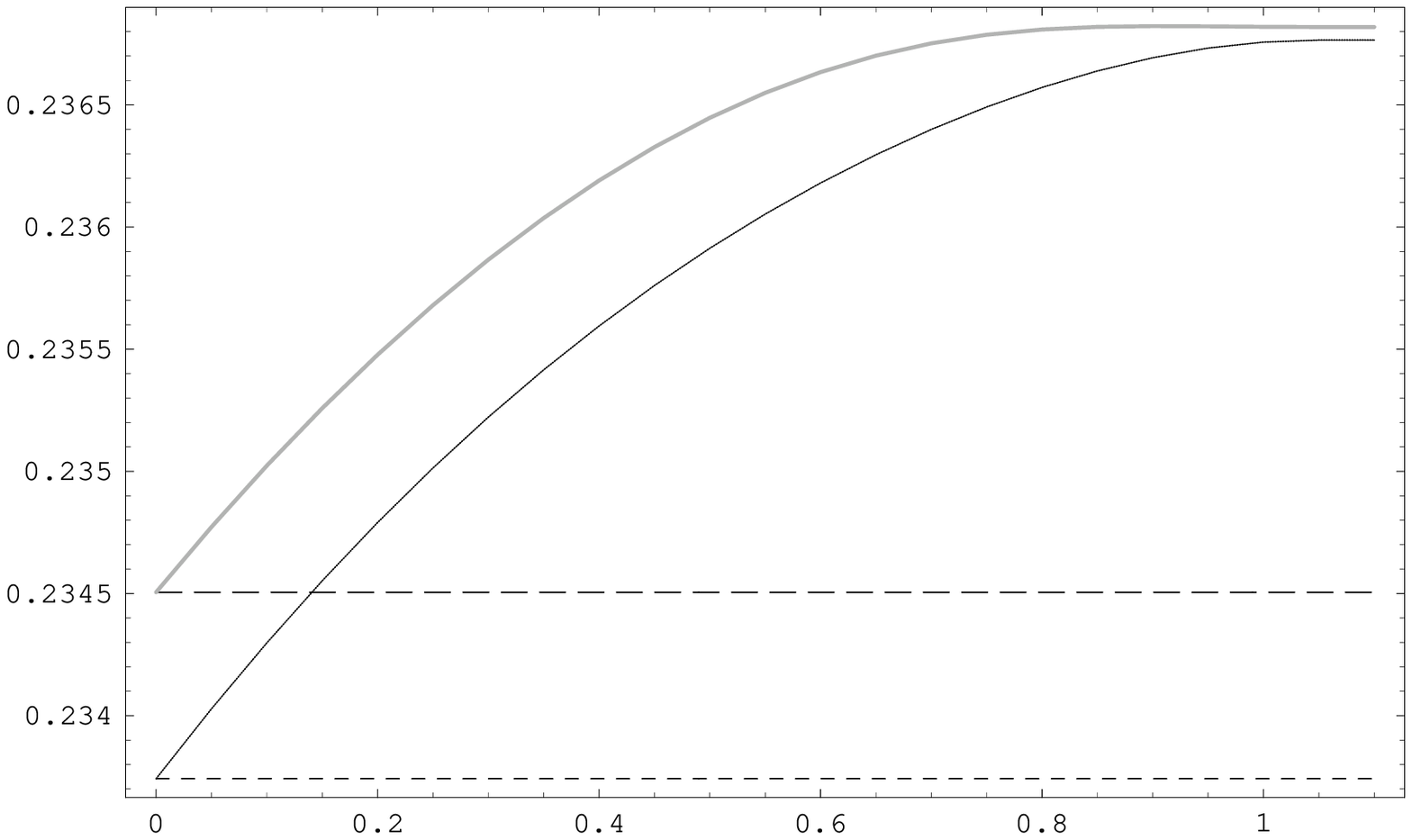}}
    \put(118,0){\normalsize$Q^2\;[\GeV[]^2]$}
    \put( -6,0){\yaxis[63.9mm]{\normalsize$\om_{\iZS,T}(Q^2)$}}
  \end{picture}\\[.5ex]
  \setlength{\unitlength}{.9mm}\begin{picture}(120,71)   
    \put(0,0){\epsfxsize108mm \epsffile{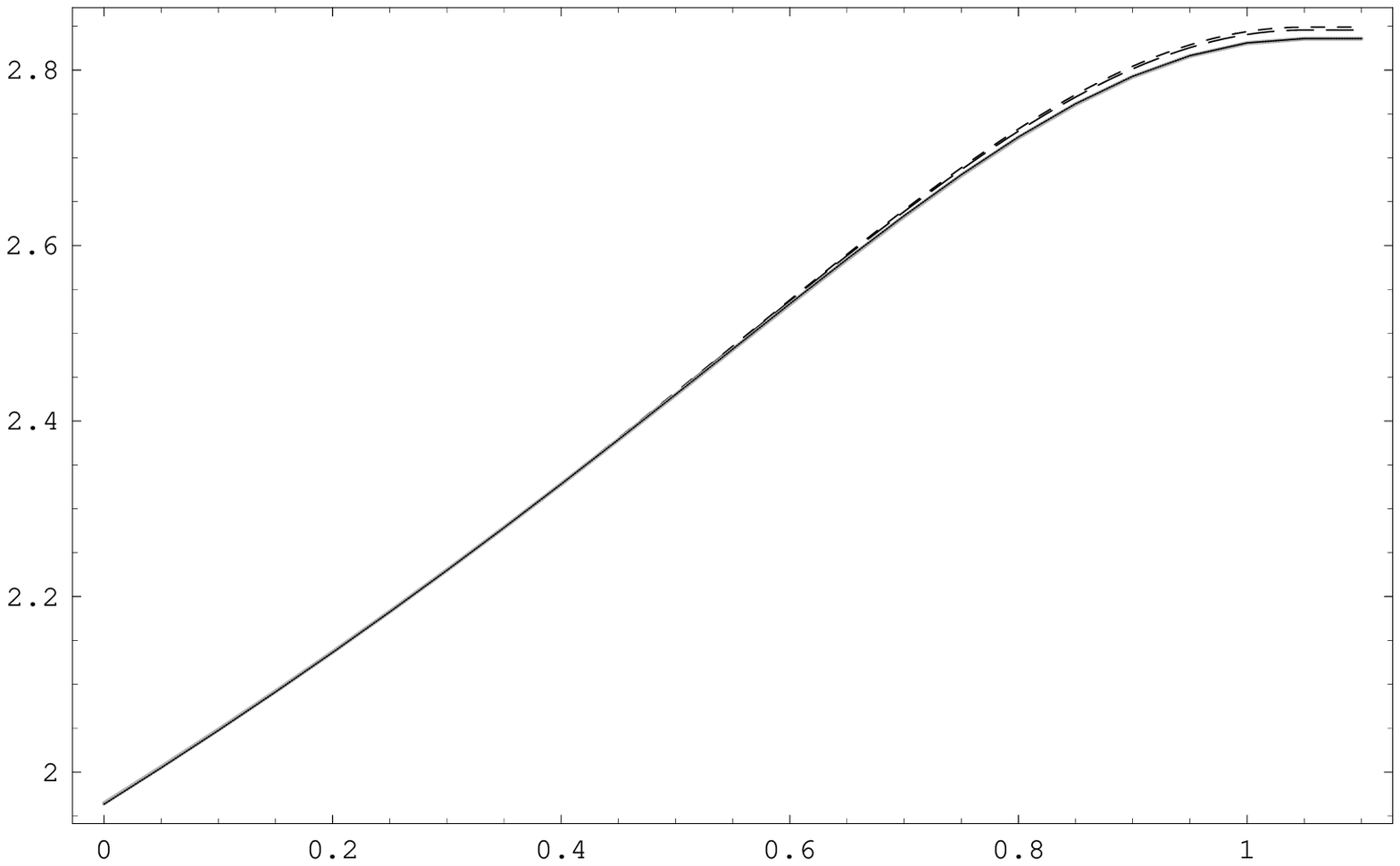}}
    \put(118,0){\normalsize$Q^2\;[\GeV[]^2]$}
    \put( -6,0){\yaxis[63.9mm]{\normalsize${\cal N}_{\iZS,T}(Q^2)$}}
    \put(58.75, 3.75){\epsfxsize54mm \epsffile{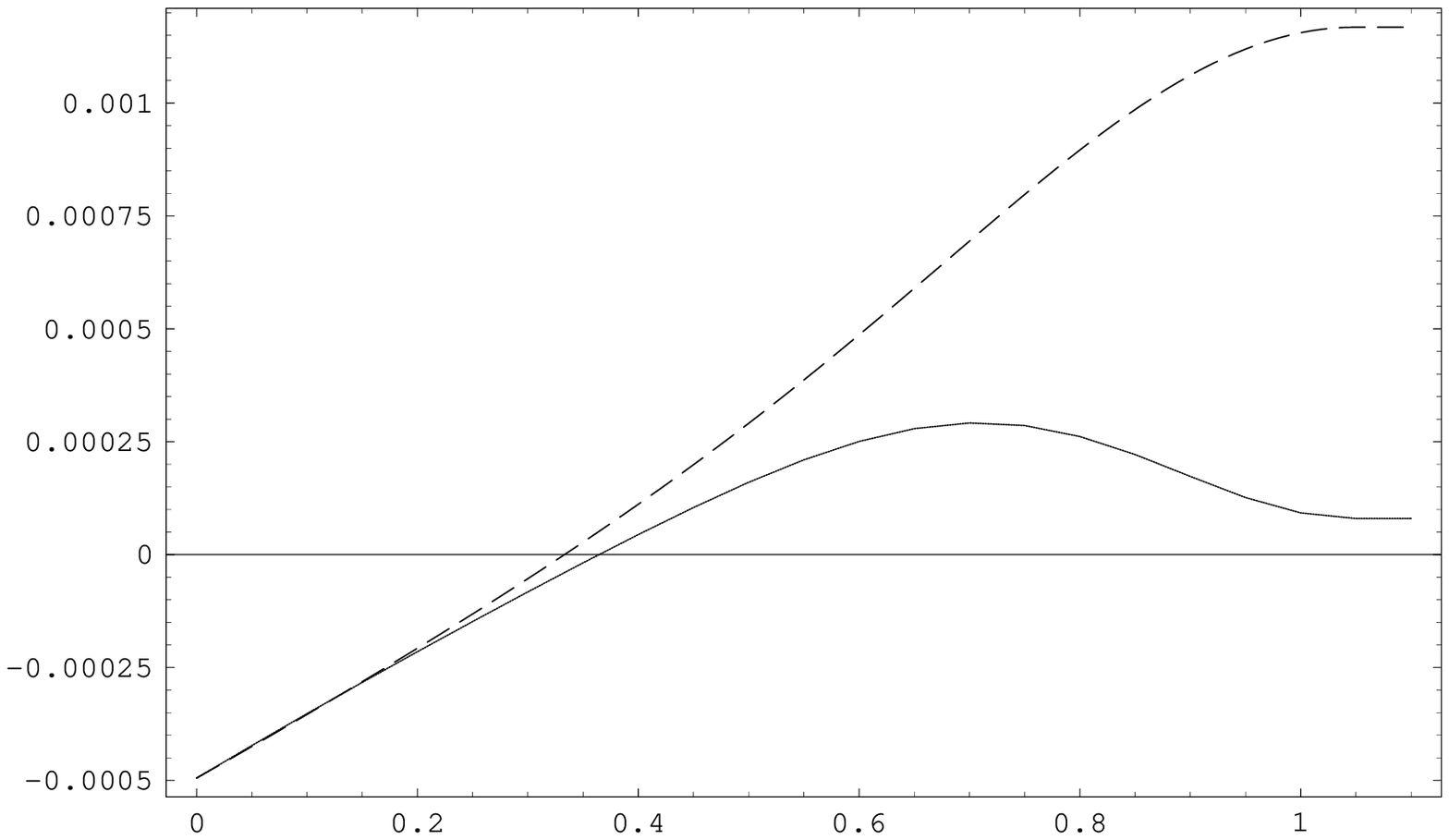}}
    \put(83,7){\small%
                 $({\cal N}_{\iZS,T} \!-\! {\cal N}_{\iZS,T}^{\rm a.o.})\!/ {\cal N}_{\iZS,T}$}
  \end{picture}\\[.5ex]
  \setlength{\unitlength}{.9mm}\begin{picture}(120,71.28)   
    \put(0,0){\epsfxsize108mm \epsffile{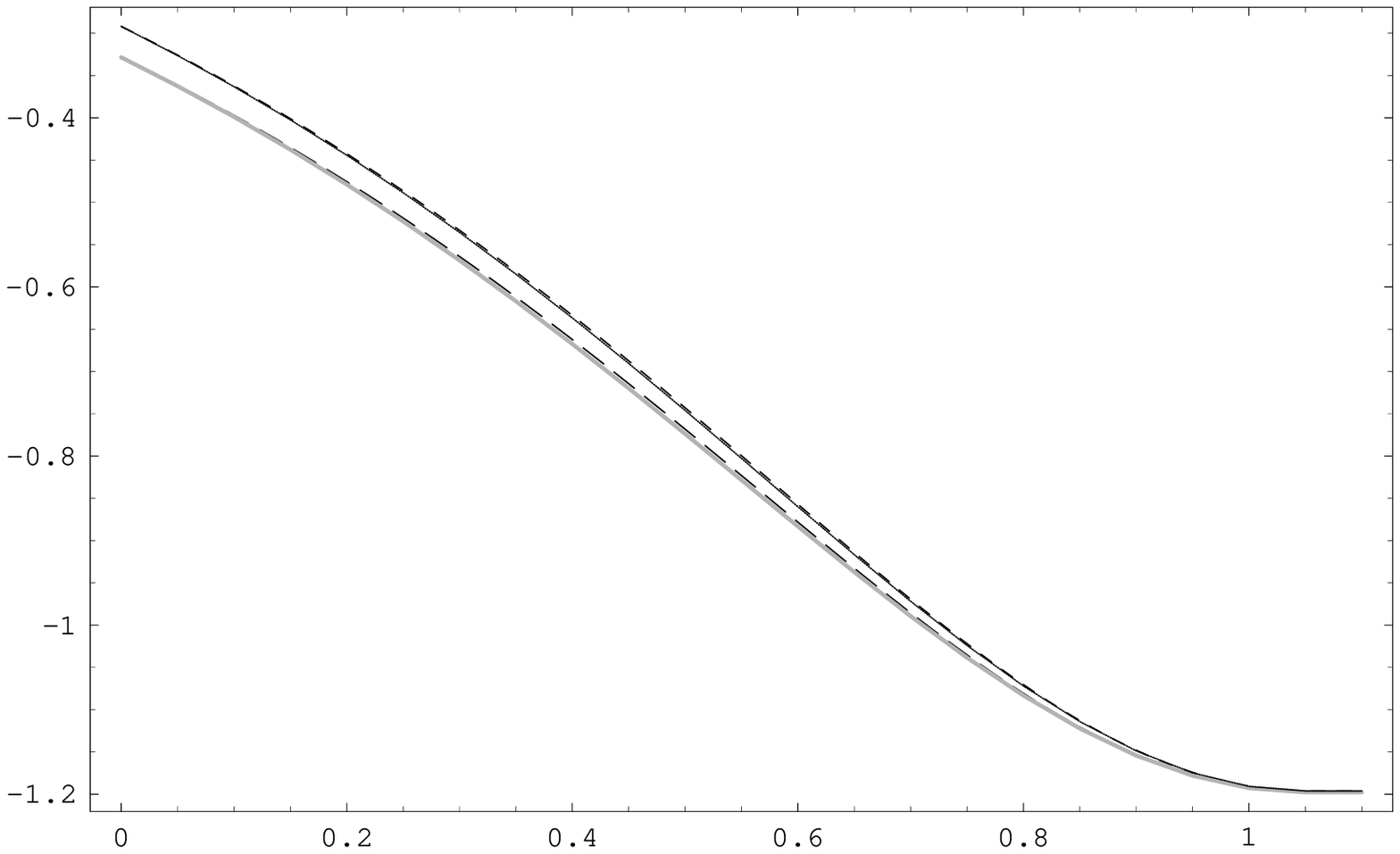}}
    \put(118,0){\normalsize$Q^2\;[\GeV[]^2]$}
    \put( -6,0){\yaxis[64.15mm]{\normalsize$A_T(Q^2)$}}
    \put(58.75,35){\epsfxsize54mm \epsffile{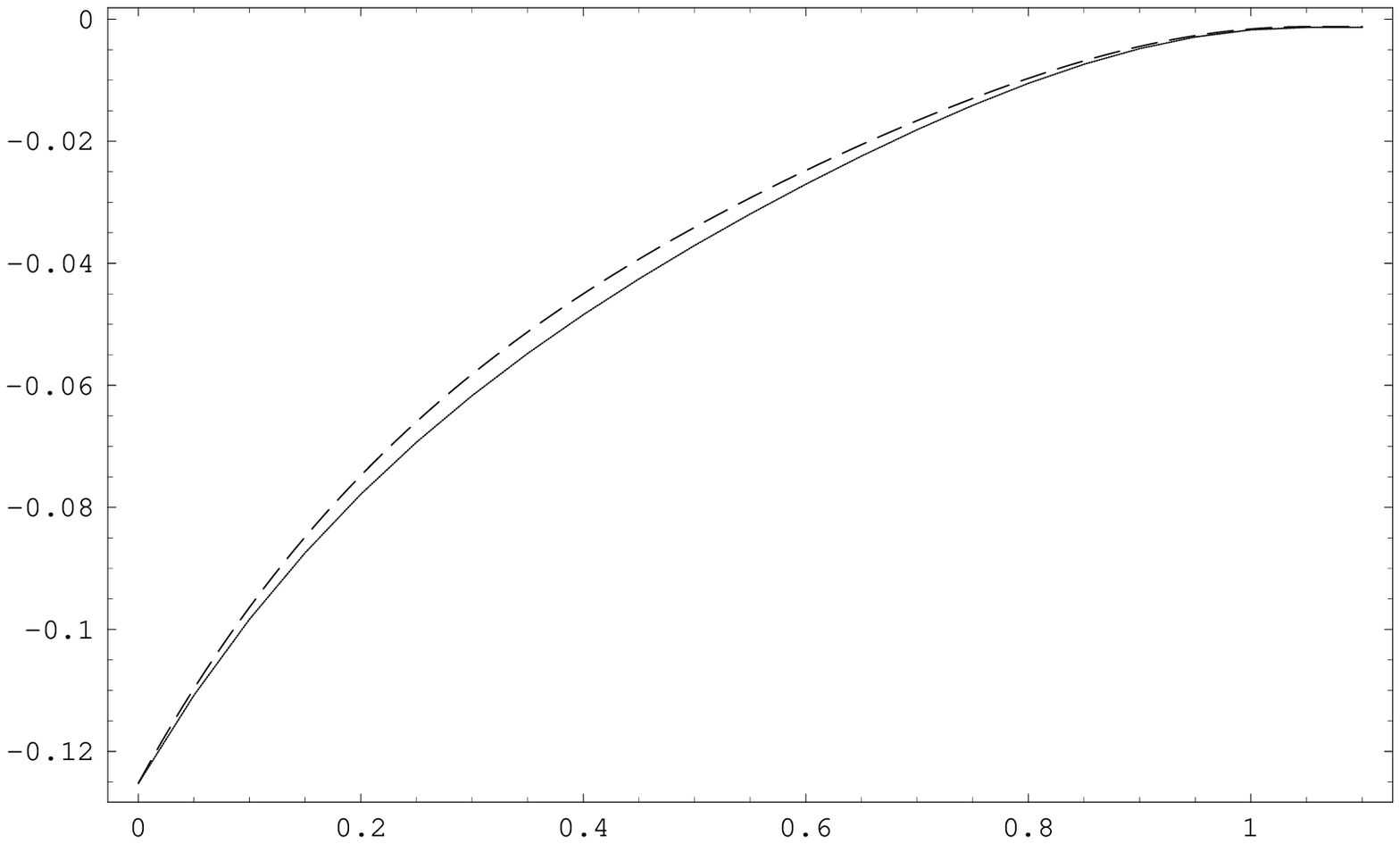}}
    \put(93,   38.25){\small$(A_T \!-\! A_T^{\rm a.o.})\!/ A_T$}
  \end{picture}
  \end{center}
\vspace*{-4ex}
\caption[\protect$2S$-Parameter~\protect$\om_{\iZS,T}\!(\!Q^2\!)$,~\protect${\cal N}_{\iZS,T}\!(\!Q^2\!)$,~\protect$A_T\!(\!Q^2\!)$, transversal]{
  Verhalten mit~$Q^2$ der~$2S$-Parameter~$\om_{\iZS,T}$,~${\cal N}_{\iZS,T}$,~$A_T$ f"ur transversale~Pola\-risation,~$\la \!\equiv\! T$.   Kurven entsprechend Abbildung~\ref{Fig:f2S} und Text.   Die relative Abweichung f"ur approximative Orthogonalit"at bezogen auf exakte ist gezeigt in den kleinen Abbildungen:   F"ur~${\cal N}_{\iZS,T}$ ist sie~$<\! 0.12\,\%$ und~$<\! 0.05\,\%$ f"ur Identit"at der Kopplungen nur f"ur~$Q^2 \!\equiv\! 0$ bzw.\@ f"ur alle~$Q^2$~[volle Linie], also irrelevant; f"ur~$A_T$ betr"agt sie in beiden F"allen bis zu $12\,\%$.
}
\label{Fig:2S-ParameterT_N,om,A}
\end{minipage}
\end{figure}
\vspace*{-1ex}
\bigskip\noindent
Im oberen Teil von Tabelle~\ref{Tabl:Wfn-Parameter} sind angegeben Zahlenwerte, die abweichen von denen des Referenz-Satzes; in den Abbildungen~\ref{Fig:Delta}-\ref{Fig:2S-ParameterT_N,om,A} sind neben den schwarzen durchgezogenen Kurven weitere Kurven aufgetragen.
In den folgenden zwei Bemerkungen diskutieren wir die argumentative Basis, auf der diese Zahlenwerte und Kurven stehen, und vergleichen sie kritisch mit denen des Referenz-Satzes. \\
\indent
{\bf Bemerkung~Eins.}\vv
In Anhang~\ref{APPSect:Vektormeson-Wfn}, vgl.\@ die Seiten~\pageref{APP-T:1S-ParameterFix}ff.\@ und~\pageref{APP-T:2S-ParameterFix}ff., ist geschildert, da"s wir Relationen, die eingehen in die dargestellte Fixierung der $2S$-Parameter, analytisch hergeleitet haben erst nach Ver"offentlichung von Ref.~\cite{Kulzinger98}.
Im Zuge ihrer numerischen Approximation wurde dort eine N"aherung angenommen, die dazu f"uhrt, da"s die Lichtkegelwellenfunktion~$\ket{2S}$ nur approximativ, nicht exakt orthogonal ist auf~$\ket{1S}$.
Die N"aherung bezieht sich allein auf Orthogonalit"at, nicht auf Normiertheit und Reproduktion der Kopplung; daher ist unmittelbar affektiert nur~$A_\la$.
Sie ist sensitiver auf eine nichtverschwindende Quarkmasse, so da"s sie zu deutlicheren Effekten f"uhren sollte f"ur transversale Polarisation und kleinen~$Q^2$.
Aufgrund des dargestellten Verfahrens zur sukzessiven Bestimmung der Parameter~-- Iteration von Schritt Zwei und Schritt Eins$^{\bm{\prime}}$~-- werden mittelbar affektiert s"amtliche Gr"o"sen:~$\om_{\iZS,\la}$,~${\cal N}_{\iZS,\la}$,~$A_\la$ und~$f_{\iZS,\la}$, ungeachtet der Polarisation. \\
\indent
In Tabelle~\ref{Tabl:Wfn-Parameter} sind angegeben die expliziten Zahlenwerte f"ur~$Q^2 \!\equiv\! 0$.
Wir finden, vgl.\@ oben und unten, die beiden rechten Spalten, da"s die numerische Diskrepanz bereits vollst"andig verschwindet in der Genauigkeit, die angegeben war in Ref.~\cite{Kulzinger98}; so betr"agt f"ur~$f_\iZS$ der Zahlenwert~$-0.137\,23$ gegen"uber~$-0.137\,78$ im Referenz-Satz.~--
Ausgenommen der transversale Parameter~$A_T$: f"ur diesen folgt~$-0.328$ gegen"uber~$-0.291$.
Die Diskrepanz betr"agt bis zu~$12\,\%$ und ist damit zun"achst signifikant, vgl.\@ die untere Abbildung in~\ref{Fig:2S-ParameterT_N,om,A}:
Die Kurve in langen Strichen ist zu vergleichen mit der Kurve in kurzen Strichen, die korrespondiiert mit dem Referenz-Satz.
Wir werden diesen Punkt ausf"uhrlich diskutieren. \\
\indent
Der Parametersatz von Ref.~\cite{Kulzinger98} ist in den Abbildungen~\ref{Fig:f2S}-\ref{Fig:2S-ParameterT_N,om,A} konsequent dargestellt durch die Kurven in langen Strichen.
Wir verglichen hier mit den Kurven in kurzen Strichen, statt mit den schwarzen durchgezogenen Linien~-- die konsequent darstellen den Referenz-Parametersatz~-- aus dem Grund der folgenden Bemerkung. \\
\indent
{\bf Bemerkung~Zwei.}\vv
Das Iterationsverfahren zur Bestimmung der Funktion~$\De(Q^2)$ ist sehr aufwendig und mu"s von Hand durchgef"uhrt werden.
Bereits dies ist in praktikablem Rahmen erst m"oglich, seit wir s"amtliche Relationen, die entsprechen den Forderungen von Orthonormalit"at und Reproduktion der Kopplung an denn elektromagnetischen Strom, analytisch hergeleitet haben.
Zum Zeitpunkt unserer Ver"offentlichung Ref.~\cite{Kulzinger98} war dies noch nicht der Fall.
Dort wird~$\De$ bestimmt f"ur~$Q^2 \!\equiv\! 0$ und die Oszillatorparameter~$\om_{\iZS,\la}$ entsprechend verr"uckt von den Werten~$\om_{\iES,\la}$.
Statt das Verfahren durchzuf"uhren f"ur s"amtliche Werte von~$Q^2$ im Intervall nichttrivialer Abh"angigkeit:~\mbox{$Q^2 \!<\! Q_0^2 \!=\! 1.05\GeV$}, werden die Oszillatorparameter eingefroren auf diese Werte~$\om_{\iZS,\la}(Q^2\zz\equiv\zz0)$ f"ur nichtverschwindende~$Q^2$.
Die Oszillatorparameter~$\om_{\iZS,\la}$ sind von Hand konstant gesetzt, vgl.\@ die Abbn.~\ref{Fig:2S-ParameterL_N,om,A},~\ref{Fig:2S-ParameterT_N,om,A}; mit diesen konstanten Zahlenwerten wird f"ur jedes~$Q^2$ wieder Schritt Eins$^{\bm{\prime}}$ durchgef"uhrt, das hei"st sukzessive bestimmt~$A_\la$,~${\cal N}_{\iZS,\la}$,~$f_{\iZS,\la}$.
Konsequenz ist, da"s f"ur feste Polarisation die Lichtkegelwellenfunktion~$\ket{2S}$ normiert ist und [approximativ]%
\FOOT{
  \label{FN:BemEins}im Sinne von Bemerkung~Eins
}
orthogonal steht auf~$\ket{1S}$ {\it f"ur jedes\/}~$Q^2$, identische Kopplungen~$f_{\iZS,L}$,~$f_{\iZS,T}$ liefert aber {\it nur f"ur\/}~$Q^2 \!\equiv\! 0$.

Mit Kenntnis aller Relationen {\it analytisch\/} ist das Iterationsverfahren zur Bestimmung der Verr"uckung~$\De$ praktikabler, obwohl es weiterhin von Hand durchgef"uhrt werden mu"s.
Dies setzt uns dennoch in die Lage, analog zu~$\De(Q^2)$ die Funktion~$\De^{\!\rm a.o.}(Q^2)$ zu bestimmen.
Dabei stehe der Index f"ur "`approximativ orthogonal"' und bestimme~$\De^{\!\rm a.o.}$ in vollst"andiger Analogie zu~$\De$ die Verr"uckung von~$\om_{\iZS,\la}$ relativ zu~$\om_{\iES,\la}$~-- impliziere aber dieselbe N"aherung, die gemacht ist in Ref.~\cite{Kulzinger98} und die f"uhrt auf bez"uglich~$\ket{1S}$ nur approximativ orthogonale Lichtkegelwellenfunktionen~$\ket{2S}$.
In Abbildung~\ref{Fig:Delta} ist als graue (durchgezogene) Kurve die Funktion~$\De^{\!\rm a.o.}(Q^2)$ aufgetragen in Gegen"uberstellung mit~$\De(Q^2)$.

\vspace*{-1ex}
\bigskip\noindent
Die Kurven der Abbildungen~\ref{Fig:f2S}-\ref{Fig:2S-ParameterT_N,om,A} beziehen sich konsequent auf Lichtkegelwellenfunktione~$\ket{1S}$,~$\ket{2S}$, die folgenderma"sen charakterisiert sind:
\renewcommand{\labelitemi}{$\circ$}
\begin{itemize}
\item \vspace*{-.5ex}durchgezogen schwarz, Referenz-Parametersatz:
  Orthogonalit"at,
  Normiertheit, Identit"at der Kopplungen~$f_{\iZS,L}$,~$f_{\iZS,T}$
    \vspace*{-.75ex}f"ur s"amtliche~$Q^2$
\item \vspace*{-.5ex}durchgezogen grau:
  approximative Orthogonalit"at\citeFN{FN:BemEins},
  Normiertheit, Identit"at der Kopplungen~$f_{\iZS,L}$,~$f_{\iZS,T}$
    \vspace*{-.75ex}f"ur s"amtliche~$Q^2$
\item \vspace*{-.5ex}kurze Striche:
  Orthogonalit"at, Normiertheit f"ur s"amtliche~$Q^2$;
  Identit"at der Kopplungen~$f_{\iZS,L}$,~$f_{\iZS,T}$
    \vspace*{-.75ex}nur f"ur~$Q^2 \!\equiv\! 0$%
\FOOT{
  \label{FN:BemZwei}im Sinne von Bemerkung~Zwei
}
\item \vspace*{-.5ex}lange Striche, Ref.~\cite{Kulzinger98}:
  approximative Orthogonalit"at\citeFN{FN:BemEins};
  Identit"at der Kopplungen~$f_{\iZS,L}$,~$f_{\iZS,T}$
    \vspace*{-.75ex}nur f"ur~$Q^2 \!\equiv\! 0$\citeFN{FN:BemZwei}
\end{itemize}
Zu Aussagen bez"uglich approximativer versus exakter Orthogonalit"at von~$\ket{2S}$ auf~$\ket{1S}$ im Sinne von Bemerkung Eins gelangen wir durch Vergleich~\mbox{der gestrichelten beziehungsweise} durchgezogenen Kurven.
Zu Aussagen bez"uglich Identit"at der Kopplungen~$f_{\iZS,L}$,~$f_{\iZS,T}$ f"ur s"amtliche~$Q^2$ versus nur f"ur~$Q^2 \!\equiv\! 0$ im Sinne von Bemerkung Zwei durch Vergleich der durchgezogenen grauen Kurve mit der in langen Strichen beziehungsweise durch Vergleich der durchgezogenen schwarzen Kurve mit der in kurzen Strichen. \\
\indent
Auf dieser Basis betrachten wir Abbildung~\ref{Fig:Delta}.
Die relativen Verr"uckungen~$\De^{\!\rm a.o.}$,~$\De$~der Oszillatorparameter~$\om_{\iZS,\la}$ von~$\om_{\iES,\la}$ steigen monoton an mit~$Q^2$ und variieren im Bereich von zehn bis elf Prozent.
Dabei ist f"ur approximative Orthogonalit"at die Verr"uckunfg zun"achst gr"o"ser, gleicht sich aber f"ur gr"o"sere~$Q^2$ der an f"ur exakte Orthogonalit"at.
Die Kurven flachen ab, das hei"st, wie erwartet, wird der Effekt kleiner f"ur kleinere Quarkmasse. \\
\indent
Dies wird noch deutlicher im Verhalten der Kopplungen in Abbildung~\ref{Fig:f2S}:
Die durchgezogenen Kurven sind nicht unterscheidbar f"ur~$Q^2 \!\ge\! 1\GeV^2$.
Wenn die Oszillatorparameter nicht angepa"st werden f"ur jedes~$Q^2$, sondern eingefroren sind auf die Werte f"ur~$Q^2 \!\equiv\! 0$, gestrichelte Kurven, ist~$f_{\iZS,L}$ konstant und steigt~$f_{\iZS,T}$ an mit~$Q^2$ aufgrund dessen, da"s die Schritt~Eins$^{\bm{\prime}}$ zugrundeliegenden Relationen unabh"angig sind von~$\meff[](Q^2)$ beziehungsweise davon abh"angen.
Die Aufspaltung von~$f_{\iZS,T}$ relativ zu~$f_{\iZS,L}$ f"ur gr"o"sere~$Q^2$ ist weniger stark f"ur approximative als f"ur exakte Orthogonalit"at, steigt aber an f"ur steigendes~$Q^2$ aufgrund des Anstiegs von~$\De^{\!\rm a.o.}(Q^2)$ wie auch von~$\De(Q^2)$.
Die Aufspaltung wird flacher f"ur gr"o"sere~$Q^2$ parallel zum Verlauf von~$\De^{\!\rm a.o.}$,~$\De$.
Absolut gesehen, bezieht sich die Varianz der Kopplungen auf einen Bereich, der nur klein ist.
Auf Niveau von Streuquerschnitten erwarten wir daher keinen signifikanten Effekt, der herr"uhrt von der absoluten Gr"o"se der Kopplung.
F"ur Photoproduktion,~$Q^2 \!\equiv\! 0$ ist exakte Identit"at der Kopplungen~$f_{\iZS,L}$,~$f_{\iZS,T}$ gegeben; die Diskrepanz zwischen approximativer und exakter Orthogonalit"at,~$-0.137\,23$ gegen"uber~$-0.131\,78$, ist von keiner praktischen Relevanz.

Das Verhalten mit~$Q^2$ der Parameter~$\om_{\iZS,\la}$,~${\cal N}_{\iZS,\la}$,~$A_\la$ f"ur longitudinale Polarisation~$\la \!\equiv\! L$ ist dargestellt in Abbildung~\ref{Fig:2S-ParameterL_N,om,A} und f"ur transversale Polarisation~$\la \!\equiv\! T$ in Abbildung~\ref{Fig:2S-ParameterT_N,om,A}.
Die gestrichelten Kurven stellen wieder dar, die F"alle~-- approximativer und exakter Orthogonalit"at~--, wenn die $2S$-Oszillatorparameter~$\om_{\iZS,\la}$ nicht angepa"st werden f"ur jedes~$Q^2$, sondern eingefroren sind auf die Werte f"ur~$Q^2 \!\equiv\! 0$.
Ihre Konstanz in den oberen Abbildungen ist daher von Hand.
Da die transversalen Relationen "uber~$\meff[]$ abh"angen von~$Q^2$, die longitudinalen hingegen nicht, folgt unmittelbar die Abh"angigkeit von~$Q^2$ der transversalen Parameter~${\cal N}_{\iZS,T}$,~$A_T$ und die Konstanz der longitudinalen Parameter~${\cal N}_{\iZS,L}$,~$A_L$.
Die durchgezogenen Kurven stellen wieder dar die F"alle identischer Kopplungen~$f_{\iZS,L}$,~$f_{\iZS,T}$ f"ur jedes~$Q^2$.
Das Abfallen von~$\om_{\iZS,L}(Q^2)$ gegen"uber dem Wert bei~$Q^2 \!\equiv\! 0$ und das entsprechende Ansteigen von~$\om_{\iZS,T}(Q^2)$ betr"agt kaum mehr als ein Prozent.
Der Effekt ist noch deutlich kleiner bei den Parametern~${\cal N}_{\iZS,\la}$,~$A_\la$: durchgezogene versus entsprechende gestrichelte Kurve.
Die absolute Skala der Ordinate ist generell nur sehr klein f"ur~${\cal N}_{\iZS,L}$,~$A_L$, beziehungsweise die Kurven nur kaum unterscheidbar f"ur~${\cal N}_{\iZS,T}$,~$A_T$.
Insofern sollten die Effekte auf Niveau der Streuquerschnitte nicht von Relevanz sein. \\
\indent\enlargethispage{1.0325ex}
Ausnahme ist der Parameter~$A_T$; er zeigt f"ur approximative Orthogonalit"at bezogen auf exakte eine Abweichung bis zu~$12\,\%$  f"ur kleine Werte von~$Q^2$.
Er parametrisiert die Position des Knotens der longitudinalen Anregungsmode f"ur transversale Polarisation.
Observable, die sensitiv sind auf die Position dieses Knotens, sollten abh"angen von approximativer versus exakter Orthogonalit"at.
Dies aber ist zu sehen in Zusammenhang mit dem generellen Modell-Charakter der~$2S$-Wellenfunktion:
Die Konstruktion der Knoten geschieht im Sinne einer {\it Modellierung} und soll heranf"uhren an ein Verst"andnis von Effekten, die bestimmt sind durch Beitr"age "`rechts und links des Knotens"', die im Falle eines Nulldurchgangs auftreten mit entgegengesetztem Vorzeichen und so f"uhren k"onnen zu signifikanten K"urzungen.
Sie geschieht nicht im Sinne eines definitiven Postulats der Position der Knoten.

Diesen Punkt ausgeklammert in diesem Sinne, erwarten wir nur marginale numerische Diskrepanz auf Niveau von Streuquerschnitten.
Wir werden daher im folgenden die Abbildungen unserer Ver"offentlichung, Ref.~\cite{Kulzinger98},~-- die basieren auf Tabelle~\ref{Fig:1S-ParameterT_N,om} oben und den Kurven in langen Strichen~-- unver"andert "ubernehmen, sie jedoch diskutieren auch in Hinblick auf Unterschiede, die zu erwarten sind auf Basis des Referenz-Parametersatzes. \\
\indent
Dieser ist der  konsequenteste im theoretischen Sinne, wir schlagen ihn vor f"ur zuk"unftige Untersuchungen.
Vgl.\@ diesbzgl.: Tabl.~\ref{Fig:1S-ParameterT_N,om} unten, die durchgezogenen schwarzen Kurven in den Abbn.~\ref{Fig:Delta}-\ref{Fig:2S-ParameterT_N,om,A} und Tabl.~\refg{Tabl-APP:om1ST,De,f2S} zu seiner numerischen Reproduktion.
\vspace*{-1ex}

\subsection[Charakteristik der physikalischen%
              ~\protect$\rh$-,~\protect$\rh'$,~\protect$\rh^\dbprime$ Zust"ande]{%
            Charakteristik der physikalischen%
              ~\protect\bm{\rh}-,~\protect\bm{\rh'}-,~\protect\bm{\rh^\dbprime}-Zust"ande}

Das Grundzustand-Vektormeson~$\rh$~[$\equiv\! \rh(770)$] wird entsprechend Gl.~(\ref{Ansatz}) identifiziert mit dem $1S$-Zustand und repr"asentiert durch die Lichtkegelwellenfunktion~$\ket{1S}$.
Die angeregten Vektormesonen $\rh'$,~$\rh^\dbprime$~[$\equiv\! \rh(1450),\rh(1700)$] sind entsprechend den Gln.~(\ref{Ansatz}$'$),~(\ref{Ansatz}$''$) verkn"upft mit dem nichtphysikalischen $2S$-Zustand "uber den Mischungswinkel~$\Th$ und in elastischer Photo- und Leptoproduktion im wesentlichen repr"asentiert entsprechend durch~$\ket{2S}$. \\
\indent
Sei durch~$B_{V\to f} \!\equiv\! \Ga_{V\to f}\!/\Gatot_V$ das Verzweigungsverh"altnis bezeichnet f"ur den Zerfall des Vektormesons~$V$ in den Zustand~$f$; wir definieren Gr"o"sen~$X_{V\!,i}$ f"ur~$V \!\equiv\! \rh,\rh',\rh^\dbprime$ wie folgt:
\vspace*{-.5ex}
\begin{align} \label{branchings}
X_1\; &=\; B_{e^+e^-}\;   \cdot\; B_{\pi^+\pi^-}
    \\
X_2\; &=\; B_{2\pi^+2\pi^-}\; /\; B_{\pi^+\pi^-}
    \tag{\ref{branchings}$'$} \\
X_3\; &=\; B_{\pi^+\pi^-}\;   +\; B_{2\pi^+2\pi^-}
    \tag{\ref{branchings}$''$}
    \\[-4.5ex]\nn
\end{align}
Sei der Index~$V$ f"ur das Vektormeson auch im folgenden unterdr"uckt, wenn er folgt aus dem Zusammenhang. \\
\indent
Der Mischungswinkel~$\Th$ wird definiert durch den Zusammenhang der physikalischen Zust"ande~$\ket{\rh(1450)}$,~$\ket{\rh(1700)}$ mit~$\ket{2S}$, vgl.\@ die Gln.~(\ref{Ansatz}$'$),~(\ref{Ansatz}$''$).
Er kann ausgedr"uckt werden durch~$M_V$,~$\Gatot_V$ und die Gr"o"sen~$X_{V\!,i}$ f"ur~$V \!\equiv\! \rh',\rh^\dbprime$:
\vspace*{-.5ex}
\begin{align} \label{Th}
\tan^2\Th\;
  =\; M_{\rh'} \Gatot_{\rh'}\, x_{\rh'}\;
      \big/\;
      M_{\rh^\dbprime} \Gatot_{\rh^\dbprime}\, x_{\rh^\dbprime} \qquad
  \text{mit}\quad
  x_V\; \equiv\; X_{V\!,1}(1\!+\!X_{V\!,2})\big / X_{V\!,3}
\end{align}
vgl.\@ Anhang~\ref{APPSubsect:Mischungswinkel}, Gl.~(\ref{APP:Th}).

\renewcommand{\thefootnote}{\thempfootnote}
\begin{table}
\begin{minipage}{\linewidth}
\begin{center}
  \begin{tabular}{|h{-1}||g{-1}|g{-1}|g{-1}|} \hline
  \multicolumn{4}{|c|}{Charakteristik der $1^+(1^{--})$-Vektormesonen%
                       ~$\rh(770)$,~$\rh(1450)$,~$\rh(1700)$}
    \\ \hhline{:=:t:===:}
  \multicolumn{1}{|c||}{}
    & \multicolumn{1}{c|}{$\rh$}
    & \multicolumn{1}{c|}{$\rh'$}
    & \multicolumn{1}{c|}{$\rh^\dbprime$}
    \\
    & \multicolumn{1}{c|}{$\phantom{1}\!\equiv\! \rh(770)$}
    & \multicolumn{1}{c|}{$\!\equiv\!           \rh(1450)$}
    & \multicolumn{1}{c|}{$\!\equiv\!           \rh(1700)$}
    \\ \hhline{|-||---|}
  \mbox{$M_V$}/[\MeV[]]
    &  769.3,\PM 0.8
    & 1465  ,\PM25   
    & 1700  ,\PM20   \\
  \mbox{$\Gatot_V$}/[\MeV[]]
    & 150.9,\PM 3.0
    & 310  ,\PM60  
    & 235  ,\PM50   \\[.25ex]
  \mbox{$\Gaee_V$}/[\keV[]]
    & 6.77,\PM0.32
    & \bm{1.},\bm{63} & \bm{1.},\bm{07} \\
  \mbox{$f_V$}/[\GeV[]]
    & 0.,152\,6
    & \bm{-\,0.},\bm{103}  & \bm{+\,0.},\bm{090\,3} \\[.25ex]
  \mbox{$X_{V}$}/\mbox{${}_{\!,1}$}
    & 4.,48\mbox{$\times\!10^{-5}$}
    & 5.,2\mbox{$\times\!10^{-7}$}  & 6.,0\mbox{$\times\!10^{-7}$} \\
  \mbox{$X_{V}$}/\mbox{${}_{\!,2}$}
    & ,0  & 12.,5  & 55\!/6, \!\cong\! 9.17 \\
  \mbox{$X_{V}$}/\mbox{${}_{\!,3}$}
    & ,1  & 0.,8   & 0.,8  \\[.25ex]
  \mbox{$B_{V\to}$}/\mbox{${}_{\pi^+\pi^-}$\footnote{
    \label{B_X}\vspace*{-1ex}Es ist~$B_{\pi^+\pi^-} \!=\! X_3\!/(1 \!+\! X_2)$ und~$B_{2\pi^+2\pi^-} \!=\! X_2X_3\!/(1 \!+\! X_2)$ aufgrund der Gln.~(\ref{branchings}$'$),~(\ref{branchings}$''$).
  }}
    & ,1  & 8\!/135, \!\cong\! .0593  & 24\!/305, \!\cong\! .0787 \\
  \mbox{$B_{V\to}$}/\mbox{${}_{2\pi^+2\pi^-}$\citeFN{B_X}}
    & ,0  & 20\!/27, \!\cong\! .741   & 44\!/61, \!\cong\! .721 \\ 
  \hhline{:=:b:===:}
  \end{tabular}
  \end{center}
\vspace*{-3.5ex}
\caption[Charakteristik der Vektormesonen~\protect$\rh(770)$,~\protect$\rh(1450)$ und~\protect$\rh(1700)$]{
  Charakteristik der Vektormesonen~$\rh$,~$\rh'$,~$\rh^\dbprime$.   Die Kopplungen~$f_V$ an den elektromagnetischen Strom f"ur~$V \!\equiv\! \rh',\rh^\dbprime$ in Fettdruck folgen unmittelbar aus dem Zahlenwert~$f_\iZS \!=\! -0.137\GeV$, vgl.\@ die Gln.~(\ref{Ansatz}$'$),~(\ref{Ansatz}$''$), der berechnet wird auf Basis der Lichtkegelwellenfunktion~$\ket{2S}$ und sich bezieht auf~$Q^2 \!\equiv\! 0$, vgl.\@ Tabl.~\ref{Tabl:Wfn-Parameter}.   Die Massen sind aktualisiert auf die neueste Ausgabe der Particle Data Group, vgl.\@ Ref.~\cite{PDG00}, die totalen Breiten belassen als die Zahlenwerte von Ref.~\cite{Donnachie87a} bzw.~\cite{PDG98}.   Die Zahlenwerte f"ur~$X_1$,~$X_2$ f"ur~$\rh'$,~$\rh^\dbprime$ stammen von Donnachie, Mirzaie, vgl.\@ Ref.~\cite{Donnachie87a}; sie werden dort extrahiert aus den $\pi^+\pi^-$-, $2\pi^+2\pi^-$-Massespektren f"ur~$e^+e^-$-Annihilation und Photoproduktion.   Im Rahmen der angegebenen Genauigkeit setzen wir~$B_{\rh\to\pi^+\pi^-} \!=\! 1$ und sch"atzen ab das Verzweigungsverh"altnis in zwei oder vier geladene Pionen~$X_3$ f"ur~$\rh'$,~$\rh^\dbprime$ "ubereinstimmend als~$80\,\%$.   Die~$B_{V\to\pi^+\pi^-}$,~$B_{V\to2\pi^+2\pi^-}$ f"ur~$V \!\equiv\! \rh',\rh^\dbprime$ sind berechnet aus den~$X_{V\!,2}$,~$X_{V\!,3}$, vgl.\@ Fu"sn.\,\FN{B_X}.
\vspace*{-.5ex}
}
\label{Tabl:Charakt_rh,rh',rh''}
\end{minipage}
\end{table} 
\renewcommand{\thefootnote}{\thechapter.\arabic{footnote}}
In Tabelle~\ref{Tabl:Charakt_rh,rh',rh''} geben wir an, im Sinne einer zusammenfassenden Gegen"uberstellung, wesentliche Gr"o"sen, durch die charakterisiert und angebunden sind an observable Gr"o"sen die physikalischen Zust"ande~$\rh(770)$,~$\rh(1450)$,~$\rh(1700)$.
Auf Basis der dort angegebenen Zahlenwerte berechnen wir einen expliziten Zahlenwert f"ur den Mischungswinkel~$\Th$.
Diesen Punkt, wie auch die Diskussion der zentralen Gr"o"sen in Fettdruck: der Kopplungen~$f_V$ und der daraus berechneten elektronischen Zerfallsbreiten~$\Gaee_V$ f"ur~$V \!\equiv\! \rh',\rh^\dbprime$, stellen wir f"ur einen Moment zur"uck.
Wir kommentieren zun"achst wie folgt die anderen Zahlenwerte. \\
\indent
Die Massen~$M_V$ sind im Sinne von Definiertheit aktualisiert auf die Zahlenwerte der neuesten Ausgabe der Particle Data Group, vgl.\@ Ref.~\cite{PDG00}; Diskrepanz zu Ref.~\cite{Kulzinger98} bzw.~\cite{PDG98} besteht nur f"ur~$M_\irh$, der Unterschied ist nicht relevant.
F"ur~$V \!\equiv\! \rh$ werden Zahlenwerte f"ur~$X_{V\!,i}$ bestimmt auf Basis der Particle Data Group, Ref.~\cite{PDG00}:
Zerfall des~$\rh(770)$ in die Pion-Endzust"ande~$f \!\equiv\! \pi^+\pi^-,\, 2\pi^+2\pi^-$ wird in Ref~\cite{PDG00} angegeben als~$\sim100\,\%$ beziehungsweise~$(1.8 \!\pm\! 0.9)\!\times\!10^{-5}\,\%$, so da"s wir in der betrachteten Genauigkeit die Verzweigungsverh"altnis\-se gleich Eins und Null setzen:~$B_{\rh\to\pi^+\pi^-} \!=\! 1$ und~$B_{\rh\to2\pi^+2\pi^-} \!=\! 0$.
$X_2$,~$X_3$ folgen so unmittelbar,~$X_1$ aus der elektronischen und totalen Zerfallsbreite.
F"ur~$V \!\equiv\! \rh',\rh^\dbprime$ geben Donnachie, Mirzaie in Ref.~\cite{Donnachie87a} Zahlenwerte an f"ur~$X_{V\!,1}$,~$X_{V\!,2}$, die sie extrahieren aus den experimentellen $\pi^+\pi^-$- und $2\pi^+2\pi^-$-Massespektren im Rho-Kanal~[$I^G(J^{P\!C}) \!=\! 1^+(1^{--})$, vgl.\@ Fu"sn.~\refg{FN:Parity}] f"ur Elektron-Positron-Annihilation und f"ur Photoproduktion elastisch am Proton.
Die totalen Zerfallsbreiten~$\Gatot_V$ f"ur~$V \!\equiv\! \rh',\rh^\dbprime$ sind aus Gr"unden der Konsistenz belassen als deren Werte, vgl.\@ ebenda, die "ubernommen sind in Ref.~\cite{PDG98}; Unterschied zu der neuesten Ausgabe der Particle Data Group besteht nur f"ur~$\Gatot_\irhpp$ und ist von keiner Relevanz.
Die Gr"o"se~$X_{V\!,2}$ f"ur~$V \!\equiv\! \rh',\rh^\dbprime$, das hei"st die Summe der Verzweigungsverh"altnisse in zwei oder vier geladene Pionen, sch"atzen wir ab auf Basis der Particle Data Group f"ur beide Anregungen "ubereinstimmend als~$80\,\%$.
F"ur Vollst"andigkeit sind in den letzten Zeilen von Tabelle~\ref{Tabl:Charakt_rh,rh',rh''} festgehalten die Verzweigungsverh"altnisse~$B_{\pi^+\pi^-}$,~$B_{2\pi^+2\pi^-}$, wie sie folgen aus den Zahlenwerten f"ur~$X_2$,~$X_3$. \\
\indent
Wir wenden uns zu der Diskussion der Gr"o"sen in Fettdruck: den Kopplungen an den elektromagnetischen Strom~$f_\irhp$,~$f_\irhpp$\vspace{-.25ex} und den daraus berechneten, vgl.\@ Gl.~(\ref{APP:fV_Gall},) elektronischen Zerfallsbreiten~$\Gaee_\irhp$,~$\Gaee_\irhpp$.
Die Kopplungen folgen aus~$f_\iZS$ aufgrund des Ansatzes f"ur den Zusammenhang der Zust"ande~$\rh(1450)$,~$\rh(1700)$ mit dem unphysikalischen $2S$-Vektormeson, vgl.\@ die Gln.~(\ref{Ansatz}$'$),~(\ref{Ansatz}$''$).
Zugrundegelegt ist der explizite Zahlenwert~$f_\iZS \!=\! -0.137\GeV$, der berechnet wird auf Basis der  Lichtkegelwellenfunktion~$\ket{2S}$ und sich bezieht auf~$Q^2 \!\equiv\! 0$, vgl.\@ Tabl.~\ref{Tabl:Wfn-Parameter}.
Dabei ist die Kopplung~$f_V$ generell festgelegt bis auf eine komplexe Phase, diese ist {\it zun"achst Konvention\/}:
Der Ansatz unserer $1S$-Lichtkegelwellenfunktion ist gew"ahlt als positiv am Ursprung, unsere $2S$-Lichtkegelwellenfunktion entsprechende negativ.
Proportionalit"at von~$f_V$ zu der Wellenfunktion f"ur verschwindender Separation des Quark-Antiquark-Paares induziert daher f"ur~$f_\iES$ positives, f"ur~$f_\iZS$ negatives Vorseichen.
Diese Vorzeichen werden {\it physikalisch\/} durch folgende "Uberlegung.
Lepton-Antilepton-Annihilation in einen Zustand~$V$ ist in gleicher Weise dominiert durch das Verhalten der Wellenfunktion von~$V$ f"ur kleine Separation.
Das experimentell beobachtete destruktive Interferenzmuster im~$\pi^+\pi^-$-Massespektrums f"ur Elektron-Positron-Annihilation im Bereich einer invarianten Masse des Pion-Endzustands von \mbox{$M \!\cong\! 1.6\GeV$}, vgl.\@ Abb.~\refg{Fig:eebar2pis}, bestimmt daher die relativen Vorzeichen der Kopplungen~$f_V$, \mbox{$V \!\equiv\! \rh,\rh',\rh^\dbprime$}, zu~$[+,-,+]$.
Mit der Konvention der Kopplung~$f_\irh \!=\! f_\iES$ als {\it positiv\/} und der Kopplung~$f_\iZS$ als {\it negativ}, ist die Forderung, da"s \mbox{$f_\irhp \!=\! \cos\Th\, f_\iZS$} {\it negativ\/} und \mbox{$f_\irhpp \!=\! -\sin\Th\, f_\iZS$} {\it positiv} sind, genau dann erf"ullt, wenn der Mischungswinkel~$\Th$ gew"ahlt wird im {\it ersten Quadranten}.
Damit folgt unmittelbar:
\vspace*{-.5ex}
\begin{align} \label{Th-explizit}
\Th\; =\;  0.229&\;\pi\;
      =\; 41.2^{\;\circ}
    \\[-4.5ex]\nn
\end{align}
vgl.\@ Gl.~(\ref{Th}) auf Basis der Zahlenwerte von Tabl.~\ref{Tabl:Charakt_rh,rh',rh''}.

\vspace*{-.5ex}
\bigskip\noindent
Abschlie"send die folgende generelle Bemerkung betreffend unseren Ansatz f"ur die Zust"an\-de~$\rh(1450)$,~$\rh(1700)$ als Mischung einer~$2S$-Komponente und eines Restes, der nur in h"oherer Ordnung an den elektromagnetischen Strom koppelt, etwa eine $2D$- und/oder Hybrid-Komponenete; vgl.\@ die Gln.~(\ref{Ansatz}$'$),~(\ref{Ansatz}$'$):
Clegg, Donnachie finden in Ref.~\cite{Clegg94}, vgl.\@ auch Ref.~\cite{Donnachie87a}, in einer Analyse der~$2\pi^+2\pi^-$-Massespektren f"ur $e^+e^-$-Annihilation [und diffraktive Photoproduktion], wichtige Verzweigung der produzierten Rho-Zust"ande~[$1^+(1^{--})$, vgl.\@ Fu"sn.~\refg{FN:Parity}] in~$\pi\,a_1$, mit~$\pi \!\equiv\! \pi^0,\, 1^-(0^{-+})$ und~$a_1 \!\equiv\! a_1(1260),\, 1^-(1^{++})$.
Close, Page diskutieren gluonische Anregungen von Mesonen: {\it Hybride}, in einem Modell, das basiert auf dem Aufbrechen chromo-elektrischer gluonischer Strings in einer Harmonischen-Oszillator-Approximation, vgl.\@ die Refn.~\cite{Close94,Close97,Page98}.
Sie postulieren f"ur den niedrigst-liegenden Hybrid-Zustand mit den Quantenzahlen des~$\rh(770)$:~$1^+(1^{--})$, als dominanten Zerfallskanal~$\pi\,a_1$ "uber eine~$S$-Partialwelle mit einer Breite von etwa~$150\MeV$.
Demgegen"uber unterdr"uckt sei der Kanal~$\pi\,h_1$, mit~$h_1 \!\equiv\! h_1(1170),\, 0^-(1^{+-})$.
Dazu in Kontrast pr"aferiere der Zerfall eines~$2D$- oder~$2S$-Zustands den Zerfall in~$\pi\,h_1$ gegen"uber dem in~$\pi\,a_1$.
Der gro"se~$\pi\,a_1$-Beitrag, den Clegg, Donnachie konstatieren, ist daher starkes Indiz, da"s in der Wellenfunktion des~$\rh(1450)$ zum einen zumindest eine Hybrid-Komponente anwesend sein sollte.
Close, Page stellen weiter fest~-- auf Basis der Analyse von Clegg, Donnachie~-- da"s zum anderen, aufgrund des nichtverschwindenden $\pi\pi$-Kanals, auch eine radial angeregte Komponente im~$\rh(1450)$ anwesend sein k"onnte.
Diese~$2S$-Komponente k"onnte dann sowohl den beobachteten~$\pi\om$-Beitrag erkl"aren,~$\om \!\equiv\! \om(782),\, 0^-(1^{--})$, wie auch die Verschiebung des physikalischen~$\rh(1450)$-(Eigen)Zustands hin zu dem niedrigeren Massewert relativ zu~$\rh(1700)$.
Eine analoge Analyse der Zerfallskan"ale des~$\rh(1700)$ fordert nicht streng eine Hybrid-Beimischung, l"a"st eine solche aber zu.
Vgl.\@ auch Ref.~\cite{Donnachie99}.

%
\section[Diffraktive Streuquerschnitte: Numerische Analyse]{%
       \!Diffraktive Streuquerschnitte: Numerische Analyse}

Im vorangehenden Kapitel~\ref{Kap:GROUND} wurde diskutiert Leptoproduktion elastisch am Proton der Grundzustand-Vektormesonen~$\rh(770)$,~$\om(782)$,~$\ph(1020)$ und~$\Jps(3097)$.
Vor dem Hintergrund der effektiven Quarkmasse, wie diskutiert in~\ref{Sect:Photon-Wfn_Q2klgl2} beh"alt diese Analyse ihre vollst"andige G"ultigkeit f"ur~$Q^2$ gr"o"ser als~\mbox{$Q_{u\!/\!d,0}^2 \!=\! 1.05\GeV^2$} f"ur~$\rh$,~$\om$ und f"ur~$Q^2$ gr"o"ser als~\mbox{$Q_{s,0}^2 \!=\! 1.6\GeV^2$} f"ur~$\ph$.
Die von~$Q^2$ abh"angende Quarkmasse~$\meff$ figuriert im Sinne von Mimikry von chirale Symmetriebrechung und Confinement der Gestalt, da"s Colour-Dipole beschr"ankt sind auf transversale Ausdehnungen von der Gr"o"senordnung dieser Skalen,~$\sim1\!/\meff(Q^2)$.
Die laufende Quarkmasse der schweren Flavour beschr"ankt von vornherein auf kleinere Ausdehnungen; die grosse Masse des Charm-Quarks beschr"ankt auf Ausdehnungen~$\sim1\!/m_c$ und garantiert effektiiv G"ultigkeit der Analyse f"ur~$\Jps$ bis hinunter zu kleinsten~$Q^2$ und f"ur Photoproduktion. \\
\indent
Die Analyse dort geschieht im Rahmen des Modells des Stochastischen Vakuums~(\DREI[]{M}{S}{V}), das ausf"uhrlich vorgestellt und diskutiert ist in Kapitel~\ref{Kap:VAKUUM} bez"uglich seiner Konstruktion und Annahmen, in Kapitel~\ref{Kap:ANALYT} bez"uglich seiner formalen Anwendung auf diffraktive Hochenergiestreuung und in Kapitel~\ref{Kap:GROUND} bez"uglich seiner expliziten Anwendung auf die genannte Klasse von Reaktionen. \\
\indent
Ein wesentliches Charakteristikum des \DREI{M}{S}{V} ist, da"s es erkl"art als Konsequenz desselben Mechanismus nichtpreturbative Ph"anomene der Nieder- und Hochenergiephysik.
Wir bezeichnen diesen Mechanismus als Strin-String-Mechanismus in dem Sinne, als er zu interpretieren ist als das Ausbilden und Wechselwirken gluonischer Strings zwischen den Quarkkonstituenten hadronischer Zust"ande.
Er ist in gleichem Ma"se verantwortlich f"ur Confinement von statischen Colour-Ladungen zu Colour-Singuletts und die Wechselwirkung schnell bewegter Colour-Dipole in diffraktiver Streuung. \\
\indent
Diese in diesem Sinne nichtperturbative Wechselwirkung auf Quark-Gluon-Niveau ist formal subsumiert in der Funktion~$\tTll$, die mit der $T$-Amplitude f"ur die Streuung zweier Wegner-Wilson-Loops oder Colour-Dipole zusammenh"angt durch Fourier-Transformation bez"uglich des Impaktvektors~$\rb{b}$ der Dipole.
F"ur gro"se invariante Schwerpunktenergie~$\surd s$ und kleinem invariantem Impuls\-transfer~$\surd-t \!=\! \tfbB$, bis auf Korrekturen~$\sim s^{-2}$, vgl.\@ Gl.~(\ref{delta}), ist die $T$-Amplitude f"ur Hadron-Hadron-Streuung~$T\hh$~-- konkret f"ur Photo-/Leptoproduktion $\ga^{\scriptscriptstyle({\D\ast})}$ des Vektormesons~$V$ elastisch am Proton~$p$ gegeben durch die Gln.~(\ref{Thh_Tellp}),~(\ref{Tellp_tTll}):
\vspace*{-.5ex}
\begin{align}
&T\hh\;
  =\; \int \frac{d\zet d^2\rb{r}}{4\pi}\; \vv
        \big[\ps_V^{\D\dagger}\,\ps_\iga\big]\!(\zet, \rb{r})\vv
        T\ellp(\zet, \rb{r}; \tfbB)
    \label{E:Thh_Tellp} \\[1ex]
&T\ellp(\zet, \rb{r}; \tfbB)
    \label{E:Tellp_tTll} \\[-.5ex]
  &=\; 2\pi\, \int_0^{\infty} b\, db\, J_0(\tfbB\, b)\;
        \int \frac{d\zet_\snbr{-} d^2\rb{r}_\snbr{-}}{4\pi} \vv
        \big| \ps_p (\zet_\snbr{-}, \rb{r}_\snbr{-}) \big|^2 \vv
          \tTll(\zet, \rb{r}; \zet_\snbr{-}, \rb{r}_\snbr{-}; \rb{b})
    \nn
    \\[-4.5ex]\nn
\end{align}
mit~$T\ellp$ der $T$-Amplitude f"ur die $T\hh$-inh"arente Streuung des Dipols~$\{\zet,\rb{r}\}$ am Proton.%
\FOOT{
  Die Proton-Wellenfunktion~$\ps_p$, die den Dipols~$\{\zet_\snbr{-}, \rb{r}_\snbr{-}\}$ verteilt, wird angesetzt wie in Gl.~(\ref{Proton-wfn2_ps}).
}
Dabei ist~$T\ellp$ "uber~$\tTll$ \DREI[]{M}{S}{V}-spezifisch aber Photo-/Leptoproduktion-nichtspezifisch.

Die Produktions-Amplitude~$T\hh \!\equiv\! T[\ga^{\scriptscriptstyle({\D\ast})}p \!\to\! Vp]$ ist wesentlich bestimmt durch den "Uberlapp der Lichtkegelwellenfunktionen, der nicht abh"angt vom Azimut von~$\rb{r}$.
Es gilt f"ur Amplitude und differentiellen Wirkungsquerschnitt:
\vspace*{-.5ex}
\begin{alignat}{2}
&T_\la\![\ga^{\scriptscriptstyle({\D\ast})}p \!\to\! Vp]&\;
  &=\; \int_0^{1\!/\!2}\zz d\zet\; \int_0^{\infty}\zz rdr\vv
         \big[\overlap{\la}\big]\!(\zet,r)\vv
         \pT\ellp(\zet,r,\tfbB)
    \label{T_Lepto} \\[.5ex]
&d\si_\la\![\ga^{\scriptscriptstyle({\D\ast})}p \!\to\! Vp]\!/dt&\;
  &=\; 1 \big/ (16\pi\, s^2)\;\cdot\;
       \big| T_\la\![\ga^{\scriptscriptstyle({\D\ast})}p \!\to\! Vp] \big|^2
    \label{dsigmadt_Lepto}
    \\[-4.5ex]\nn
\end{alignat}
f"ur Polarisationen~$\la \!\equiv\! L,T$,%
\FOOT{
  Bzgl.\@ der Untredr"uckung der Beitr"age mit Helizit"ats"anderung~$\la \!\to\! la'$ um zwei (und mehr) Einheiten, vgl. die Diskussion in Anschlu"s an Gl.~(\ref{Tellp-L3-Zerlegung}) auf Seite~\pageref{Tellp-L3-Zerlegung}.
}
vgl.\@ die Gl.~(\ref{dsigmadt-L,T}),~(\ref{T_Lepto,Compton}).
Dabei ist~$\pT\ellp$ gegen"uber~$T\ellp$ gemittelt "uber s"amtliche azimutalen Orientierungen des Dipols~$\{\zet,\rb{r}\}$, der streut an dem festen Proton-Target; vgl.\@ die Definition in Gl.~(\ref{pT_ellp}) und unten in Gl.~(\ref{E:pT_ellp}). \\
\indent
Die $T$-Amplitude f"ur Photo- und Leptoproduktion nach Gl.~(\ref{T_Lepto}), und damit Gl.~(\ref{dsigmadt_Lepto}), ist im kinematischen Bereich~gro"ser~$\surd s$ und kleiner~$\surd-t$, in ihrer G"ultigkeit allein beschr"ankt durch die G"ultigkeit der zugrundegelegten Lichtkegelwellenfunktionen%
~\vspace*{-.5ex}\mbox{$\ps_\iga \!\equiv\! \ps_{\iga(Q^2,\la)}^{h,\bar h}(\zet,\rb{r})$} und%
~\vspace*{-.25ex}\mbox{$\ps_V \!\equiv\! \ps_{V(\la)}^{h,\bar h}(\zet,\rb{r})$} f"ur Photon beziehungsweise Vektormeson.
Nach Diskussion und Konstruktion in den vorangehenden Abschnitten~\ref{Sect:Photon-Wfn_Q2klgl2} und~\ref{Sect:Rho-Anregungen} verf"ugen wir in diesem Kapitel
"uber eine explizite Parametrisierung der Photon-Lichtkegelwellenfunktion f"ur beliebig kleines bis verschwindendes~$Q^2$
und "uber explizite Parametrisierungen nichtperturbativer Lichtkegelwellenfunktionen in demselben Sinne f"ur das Systems~$\rh(770)$,~$\rh(1450)$, $\rh(1700)$ der niedrigst-liegenden~$1^+(1^{--})$-Rho-Mesonen.
Auf dieser Basis dehnen wir zum einen die Analyse von~$\rh(770)$ im vorangehenden Kapitel~\ref{Kap:GROUND} aus auf~$Q^2$ unterhalb der Skala~\vspace*{-.5ex}\mbox{$Q_{u\!/\!d,0}^2 \!=\! 1.05\GeV^2$} auf den~$Q^2$-Bereich~$0 \!-\! 20\GeV^2$ und diskutieren zum anderen neu in diesem Bereich die h"oheren Zust"ande~$\rh(1450)$,~$\rh(1700)$. \\
\indent
Dazu rekapitulieren wir zun"achst kurz, in welcher Weise die Wellenfunktionen beziehungsweise deren "Uberlapp Einflu"s nimmt auf die Streuquerschnitte.
Dies wird exemplifiziert an der angeregten $2S$-Wellenfunktion unter Betonung des Bereichs kleiner bis verschwindender~$Q^2$~-- das hei"st wenn Dipole gro"ser transversaler Ausdehnung involviert sind.
Diese generelle Diskussion setzt uns in die Lage zu {\it verstehen}, wie explizit die Streuquerschnitte generiert werden, die wir daran anschlie"send angeben und diskutieren.
\vspace*{-1ex}

\subsection{Generelle Diskussion}

In Abbildung~\refg{Fig:si-ellp-tot_zet} ist aufgetragen der totale Wirkungsquerschnitt~$\si^{\rm tot}\ellp(\zet, r^2)$ f"ur die Streuung eines Wegner-Wilson-Loops, das hei"st Dipols~$\{\zet,\rb{r}\} \!\equiv\! \{\zet,r,\th\}$, gemittelt "uber seine s"amtlichen azimutalen Orientierungen an dem festen Proton-Target.
Dieser Querschnitt~$\si^{\rm tot}\ellp$ ist aufgrund des optischen Theorems verkn"upft mit der $T$-Amplitude~$\pT\ellp$ in Vorw"artsrichtung und ist, da diese rein imagin"ar ist, proprotional zu der entsprechenden Mittelung~$\pT\ellp$ von~$T\ellp$; diese h"angt weiter ab von der invarianten Schwerpunktenergie~$\surd s$ nur kinematisch, das hei"st durch den Faktor~$s$.
Formal gilt:
\vspace*{-.5ex}
\begin{align}
&\si^{\rm tot}\ellp(\zet, r^2)\;
  =\; \frac{1}{s}\, {\rm Im}\vv
      \pT\ellp(\zet, r, \tfbB \!\equiv\! 0)
    \label{E:sitot_ellp}
    \\[.5ex]
&\text{mit}\qquad
 \pT\ellp(\zet, r, \tfbB)\;
  =\; \int_0^{2\pi}\! \frac{d\th}{2\pi}\vv
        T\ellp(\zet, r, \th, \tfbB)
    \label{E:pT_ellp}
    \\[-4.5ex]\nn
\end{align}
vgl.\@ die Gln.~(\ref{sitot_ellp}),~(\ref{pT_ellp}).
Der totale Wirkungsquerschnitt~$\si^{\rm tot}\ellp$ subsumiert vollst"andig auf Niveau von Dipol-Proton-Streuung die nichtperturbetive \DREI[]{M}{S}{V}-spezifische Quark-Gluon-Wechselwirkung, die basiert auf dem diskutierten String-String-Mechanismus.

Die Funktionen~$\si^{\rm tot}\ellp$,~$\pT\ellp$ h"angen nur marginal ab von~$\zet$, vgl.\@ Abb.~\refg{Fig:si-ellp-tot_zet}.
F"ur festes~$\zet$~[und~$\tfbB$] zeigen sie Potenzverhalten bez"uglich der Quark-Antiquark-Separation~$r$ des Dipols, der streut am Proton:~$\si^{\rm tot}\ellp,\,\pT\ellp \!\propto\! r^{2n}$ mit~$n \!\equiv\! 1$ im Limes verschwindender Separation%
\FOOT{
  analytisches Resultat
};
f"ur~$\zet \!=\! 1\!/\!2$ beginnt im Bereich~$r \grgl 1 \!-\! 2\,a$ der Exponenet~$n$ langsam abzufallen: auf typische Werte~$n \!=\! 0.9 \!-\! 0.85$ f"ur mittlere~$r$ und~$n \!\cong\! 0.75$ f"ur~$r \!=\! 2\fm\, [\cong\! 5.8\,a]$.
Wesentliches Charakteristikum ist das stetige Ansteigen mit~$r$ der Funktionen~$\si^{\rm tot}\ellp$,~$\pT\ellp$, insbesondere, da"s sie nicht abs"attigen.
Dies ist unmittelbare Konsequenz dessen, da"s in der Wechselwirkung involviert sind~-- in~conclusio sogar dominant involviert sind~-- die gluonischen Strings zwischen den Quarkkonstituenten.
Das stetiges Ansteigen r"uhrt allein aus deren nichtlokalen Wechselwirkung im~$C$-Term~[konfinierend]; es ist nicht zu erreichen durch eine lokale Wechselwirkung in dem Sinne, da"s diese nur abh"angt von den Parton-Abst"anden, wie sie repr"asentiert der~$N\!C$-Term~[nicht-konfinierend].%
\FOOT{
  Wir rekapitulieren:   Das \DREI{M}{S}{V} konstatiert allgemein, da"s Pr"asenz eines $C$-Terms "aquivalent ist der Verletzung der Bianchi-Identit"at der Feldst"arken in Termen der {\sl partiellen\/} Ableitung; dies ist der Fall in einer nichtabelschen Theorie, deren Bianchi-Identit"at sich bezieht auf die {\sl kovariante\/} Ableitung, und explizit in einer abelschen Theorie mit magnetischen Monopolen.   Der $C$-Term der Quantenchromodynamik folgt aus Gitterrechnungen, die bestimmen~$\vka \!\cong\! 0.74$, als dominant.   Vgl.\@ die Diskussion der Abbn.~\ref{Fig:int-ampl_NC,C},~\ref{Fig:int-ampl-collinear_NC,C}.
}
Demgegen"uber generieren Modelle auf Basis perturbativer Quark-Gluon-Wechselwirkung, die lokal abh"angen von den Parton-Abst"anden allein, nur Terme von~$N\!C$-Typ.

\begin{figure}
\begin{minipage}{\linewidth}
  \begin{center}
  \setlength{\unitlength}{.9mm}\begin{picture}(120,74.36)   
    \put(0,0){\epsfxsize108mm \epsffile{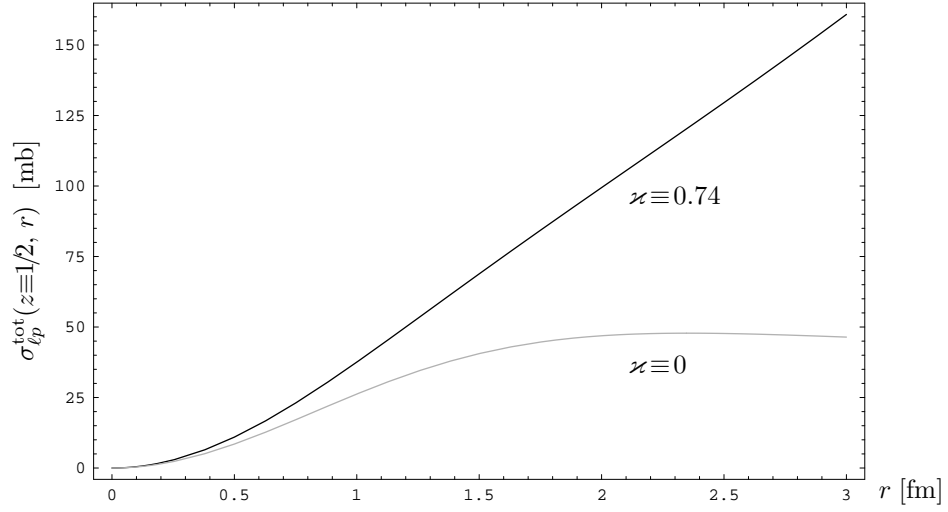}}
    \put(122,0.5){\normalsize$r\;[\fm[]]$}
    \put( -6, 0){\yaxis[66.92mm]{\normalsize%
                               $\si\ellp^{\rm tot}(\zet\zz\equiv\zz1\!/\!2,\, r)\vv[\mbarn[]]$}}
    \put( 85,44){\normalsize$\vka \!\equiv\! 0.74$}
    \put( 85,19){\normalsize$\vka \!\equiv\! 0$}
  \end{picture}
  \end{center}
\vspace*{-4.5ex}
\caption[\protect\vspace*{-.5ex}Dipol-Proton-totaler Wirkungsquerschnitt~\protect$\si\ellp^{\rm tot}|_{\D\vka}(\zet\zz\equiv\zz1\!/\!2,\, r)$,\,~\protect$\vka \!\equiv\! 0 \;{\rm vs.}\; 0.74$\!]{
  Verhalten mit~$r$ des totalen Wirkungsquerschnitts~$\si\ellp^{\rm tot}$ der Loop(Dipol)-Proton-Streuung f"ur festes~$\zet \!=\! 1\!/\!2$.   Die graue Kurve mit~$\vka \!\equiv\! 0$ stellt dar den reinen nicht-konfinierenden~$N\!C$-Term, die schwarze Kurve mit~$\vka \!\equiv\! 0.74$ den vollen physikalischen Verlauf, der weitgehend dominiert ist durch den konfinierenden~$C$-Term.  Austausch perturbativer Gluonen generiert einen Term von~$N\!C$-Typ entsprechend der grauen Kurve.
\vspace*{-1.5ex}
}
\label{Fig:si-ellp-tot_vka}
\end{minipage}
\end{figure}
In Abbildung~\ref{Fig:si-ellp-tot_vka} ist f"ur den Dipol-Proton-totalen Wirkungsquerschnitt~$\si\ellp^{\rm tot}$ gegen"ubergestellt das Verhalten auf Basis eines reinen~$N\!C$-Terms dem physikalischen Verhalten mit dominantem~$C$-Term.
Die graue Kurve,~$\vka \!\equiv\! 0$, zeigt die Abs"attigung des~$N\!C$-Terms bereits f"ur mittlere Dipolausdehnungen~$r$, die schwarze Kurve,~$\vka \!\equiv\! 0.74$, dagegen das starke stetige Ansteigen des \vspace*{-.25ex}physikalischen~$\si\ellp^{\rm tot}$, das basiert auf der nichtlokalen String-String-Wechselwirkung des dominanten~$C$-Terms. \\
\indent
Zu Funktionen \vspace*{-.125ex}analog~$\si^{\rm tot}\ellp$,~$\pT\ellp$ gelangen Autoren auch auf Basis perturbativer Quark-Gluon-Wechselwirkung.
Allerdings r"uhrt etwa in den Refn.~\cite{Kopeliovich91,Nemchik96,Nemchik96a,Nikolaev97} im Bereich~$r \klgl 2\fm$ nur etwa die H"alfte des Streuquerschnitts aus dem Austausch zweier perturbativer Gluonen~-- denn dei entsprechenden Funktionen~$\si^{\rm tot}\ellp$,~$\pT\ellp$ s"atiigen ab bei~$r \!\cong\! 1\fm$~--, f"ur die andere H"alfte ist notwendig ein nichtperturbativer Dipol-Proton-Wirkungsquerschnitt, der bestimmt ist durch Anpassung an verwandte experimentelle Querschnitte.
Dies liegt im Rahmen dessen, was wir erwarten auf Basis von Abbildung~\ref{Fig:si-ellp-tot_vka}. \\
\indent
Zur Differenzierung unseres Zugangs von perturbativen Zug"angen wie geschildert, sind daher transversal m"oglichst ausgedehnte Dipole in die Analyse zu involvieren.
F"ur Ausdehnungen~$r$ im Bereich~\mbox{$1 \!-\! 2\fm$} erwarten wir sich die differierenden Postulate signfikant zu manifestieren.
In unserer Analyse ist dies insbesondere f"ur kleine Virtualit"aten~$Q$ des Photons in der Tat der Fall; f"ur Photoproduktion sind sogar wesentlich involviert Dipole mit transversalen Ausdehnungen~$r$ bis~$2.5 \!-\! 2.8\fm$.
Wir diskutieren dies im folgenden. \\
\indent
Dazu betrachten wir zun"achst, wie die Streuquerschnitte f"ur Photo-/Leptoproduktion generiert werden auf Basis von Gl.~(\ref{T_Lepto}) f"ur die $T$-Amplitude.
Ihrzufolge sind wesentlich zwei Gr"o"sen: zum einen die gerade diskutierte (azimutal gemittelte) Dipol-Proton-Amplitude~$\pT\ellp$, zum anderen der "Uberlapp der Photon-Lichtkegelwellenfunktion mit der des Vektormesons.
Beide Gr"o"sen treten in ein subtiles Wechselspiel miteinander entsprechend, wie geschildert in Kapitel~\ref{Kap:GROUND} f"ur den~$1S$-Zustand.
Der Fall des $2S$-Zustands ist aber insofern komplexer, als dessen Wellenfunktion einen Knoten besitzt, der f"ur den "Uberlapp mit der Photon-Wellenfunktion einen Nulldurchgang induziert.
Beitr"age gro"ser und kleiner Dipole treten auf mit entgegengesetztem Vorzeichen, kompensieren und k"urzen einander zum Teil.

\begin{figure}
\begin{minipage}{\linewidth}
  \vspace*{4.5mm}
  \begin{center}
  \setlength{\unitlength}{.9mm}\begin{picture}(120,73.52)   
    \put(0,0){\epsfxsize108mm \epsffile{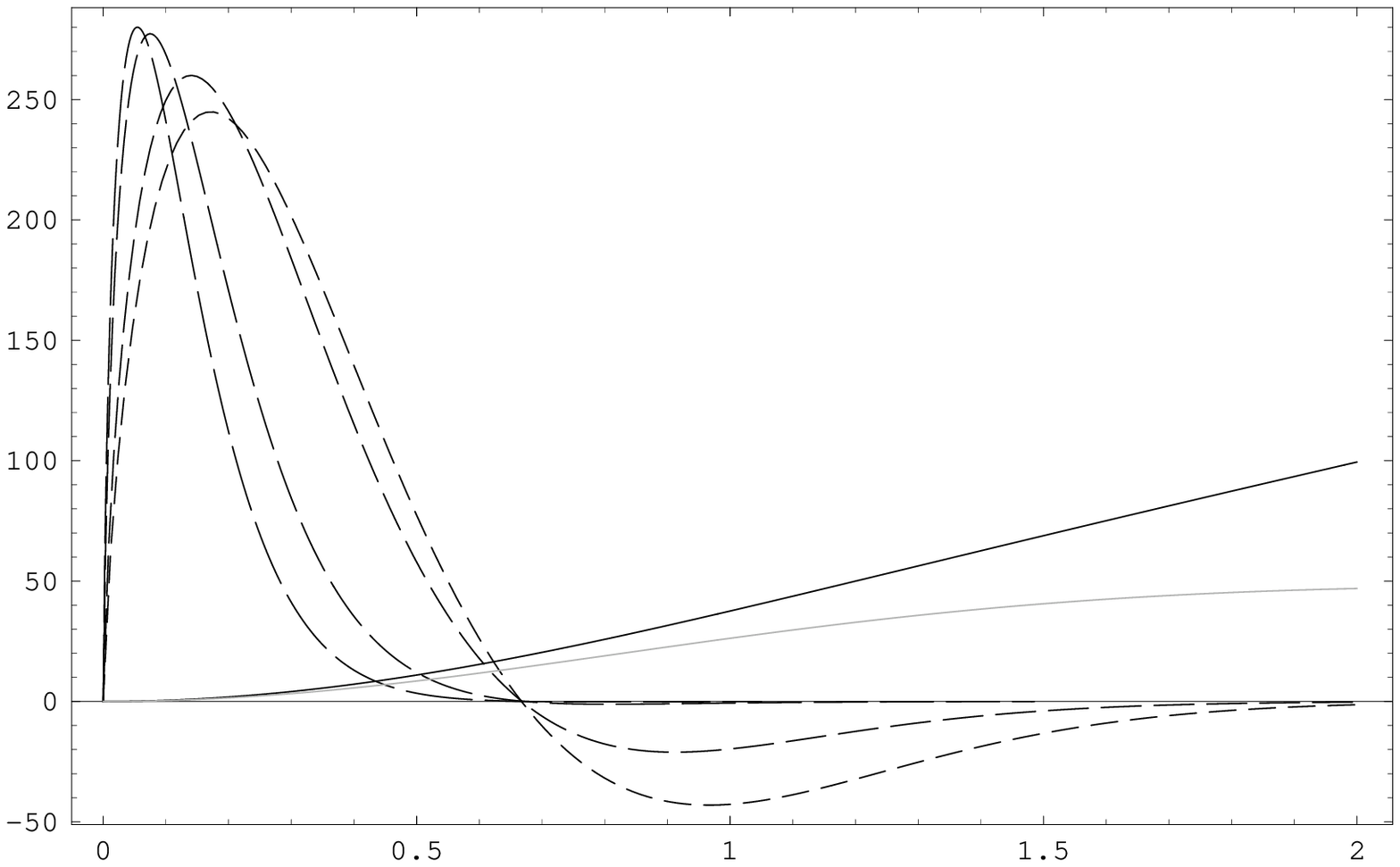}}
    \put(-18,74.5){\normalsize (a)\quad Longitudinal,~$L$\,:}
    \put(121, 0.25){\normalsize$r\;[\fm[]]$}
    \put(-14, 0){\yaxis[66.17mm]{\normalsize$\si\ellp^{\rm tot}\big|_{\D\vka}%
                                             (\zet\zz\equiv\zz1\!/\!2,\, r)%
                                             \vv[\mbarn[]]$}}
    \put( -8, 0){\yaxis[66.17mm]{\normalsize$\&\vv \roverlap[\iZS]{L}/(2\pi)\vv[10^{-3}\fm^{-1}]$}}
    \put(40  ,25){\normalsize$1\GeV^2$}
    \put(33  ,25){\normalsize$2$}
    \put(32.5,20){\normalsize$10$}
    \put(13  ,20){\normalsize$Q^2 \!=\! 20$}
    \put(90,33){\normalsize$\vka \!\equiv\! 0.74$}
    \put(90,24){\normalsize$\vka \!\equiv\! 0$}
  \end{picture}\\[3ex]
  \setlength{\unitlength}{.9mm}\begin{picture}(120,73.33)   
    \put(0,0){\epsfxsize108mm \epsffile{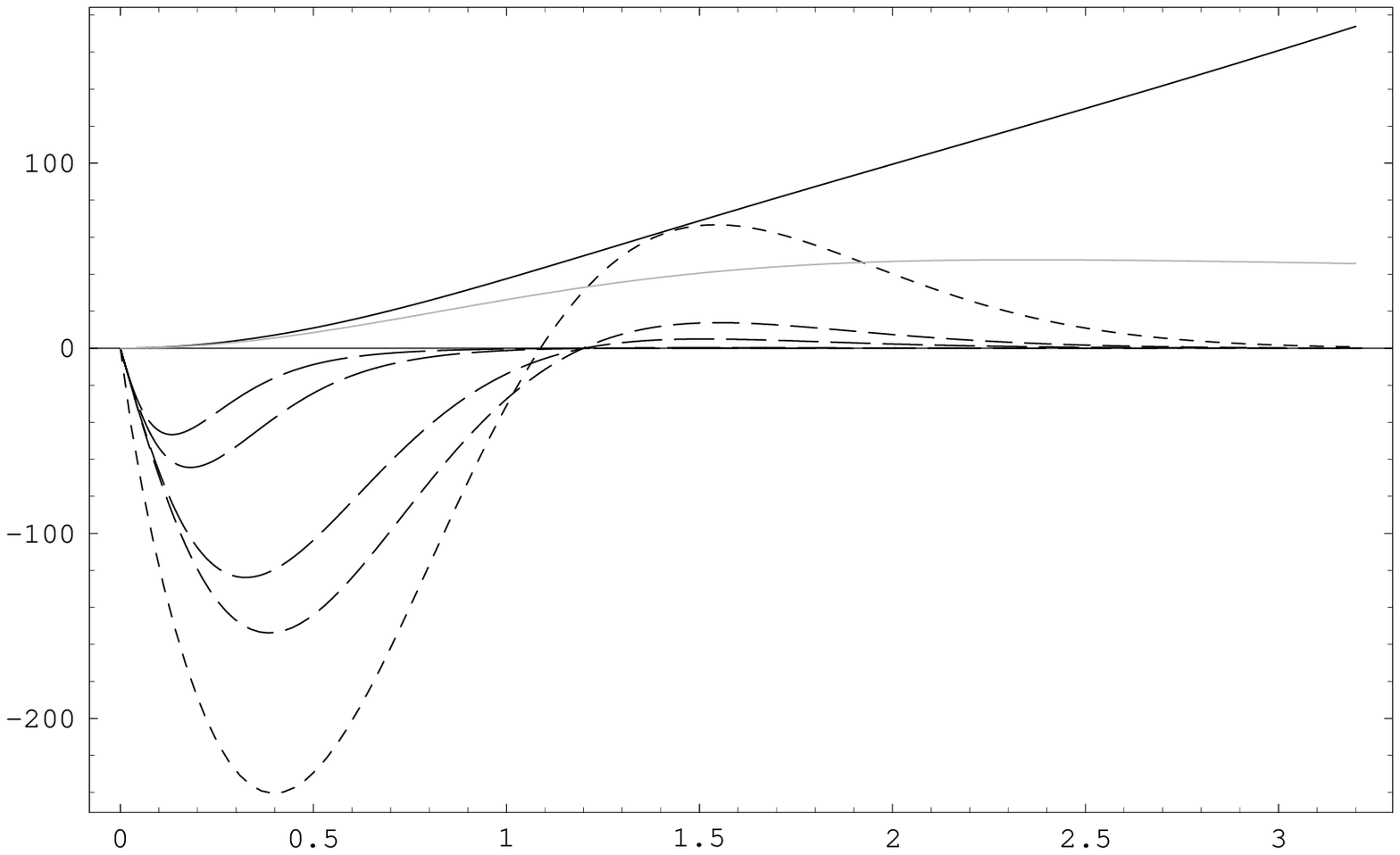}}
    \put(-18,74.5){\normalsize (b)\quad Transversal,~$T$\,:}
    \put(121, 0.25){\normalsize$r\;[\fm[]]$}
    \put(-14, 0){\yaxis[66mm]{\normalsize$\si\ellp^{\rm tot}\big|_{\D\vka}%
                                          (\zet\zz\equiv\zz1\!/\!2,\, r)%
                                          \vv[\mbarn[]]$}}
    \put( -8, 0){\yaxis[66mm]{\normalsize$\&\vv \roverlap[\iZS]{T}/(2\pi)\vv[10^{-3}\fm^{-1}]$}}
    \put(32, 9){\normalsize$Q^2 \!=\! 0\GeV^2$}
    \put(24,14){\normalsize$1$}
    \put(20,24.5){\normalsize$2$}
    \put(17,29.5){\normalsize$10$}
    \put(14,39){\normalsize$20$}
    \put(90,68){\normalsize$\vka \!\equiv\! 0.74$}
    \put(90,52){\normalsize$\vka \!\equiv\! 0$}
  \end{picture}
  \end{center}
\vspace*{-4.5ex}
\caption[Effektiver "Uberlapp~\protect\mbox{$[\roverlap[\iZS]{\la}](r)$},~\protect$\la \!\equiv\! L,T$, f"ur~\protect$Q^2 \!\equiv\! 0,1,2,10,20\GeV^2$\zz\protect\\
         \& Dipol-Proton-Querschnitt~\protect$\si\ellp^{\rm tot}|_{\D\vka}(\zet\zz\equiv\zz1\!/\!2,\, r)$,\,~\protect$\vka \!\equiv\! 0 \;{\rm vs.}\; 0.74$]{
  Als Funktion von~$r$: effektiver "Uberlapp~$[\roverlap[\iZS]{\la}](r)$,~$\la \!\equiv\! L,T$, f"ur~$Q^2 \!\equiv\! 0,1,2,10,20\GeV^2$~[gestrichelt, zunehmende Strichl"ange] und Dipol-Proton-totaler Wirkungsquerschnitt~$\si\ellp^{\rm tot}|_{\D\vka}(\zet\zz\equiv\zz1\!/\!2,\, r)$,~$\vka \!\equiv\! 0 \;{\rm vs.}\; 0.74$~[durchgezogen, grau bzw.\@ schwarz].~Die Produktionsamplitude~$T_\la^{\tfbB\equiv0}[\ga^{\scriptscriptstyle({\D\ast})}p \!\to\! 2S\,p]$ in Vorw"artsrichtung wird {\it mit dem Auge abge\-sch"atzt\/} durch Integration "uber~$r$ des Produkts dieser Gr"o"sen, vgl.\@ Gl.~(\ref{T_Lepto-approx-prop}).
\vspace*{-1ex}
}
\label{Fig:overlap_2S}
\end{minipage}
\end{figure}
In Abbildung~\ref{Fig:overlap_2S} ist dargestellt dieses Wechselspiel explizit.
Dort ist aufgetragen zum einen der Dipol-Proton-totale Wirkungsquerschnitt~$\si\ellp^{\rm tot}(\zet\zz\equiv\zz1\!/\!2,\, r)$ f"ur~$\vka \!\equiv\! 0$ versus~$0.74$ von Abbildung~\ref{Fig:si-ellp-tot_vka}, zum anderen f"ur den~$2S$-Zustand der in Hinblick auf Gl.~(\ref{T_Lepto}) wie folgt definierte~{\it effektive "Uberlapp\/}:
\vspace*{-.75ex}
\begin{align} 
\big[\roverlap{\la}\big]\!(r)\;
  =\; \int_0^{1\!/\!2}\zz d\zet\vv
        r\; \big[\overlap{\la}\big]\!(\zet,r)
    \\[-4.75ex]\nn
\end{align}
dies also der mit dem~$r$ des Integrationsma"ses multiplizierte und "uber~$\zet$ gemittelte "Uberlapp.%
\FOOT{
  Die entsprechende Gr"o"se in Ref.~\cite{Kulzinger98} ist zus"atzlich dividiert durch~$2\pi$, vgl.\@ die Ordinate in Abb.~\ref{Fig:overlap_2S}.
}
Der Dipol-Proton-Querschnitt~$\si\ellp^{\rm tot}(\zet, r)$ ist proportional~$\pT\ellp(\zet, r,\, \tfbB\zz\equiv\zz0)$, vgl.\@ Gl.~(\ref{E:sitot_ellp}).
Diese Gr"o"se genommen f"ur~$\zet \!\equiv\! 1\!/\!2$, das hei"st ihre~$\zet$-Abh"angigkeit vernachl"assigt~-- sie ist in Abb.~\ref{Fig:si-ellp-tot_vka} dokumentiert als nur marginal~-- gilt nach Gl.~(\ref{T_Lepto}) f"ur die $T$-Amplitude $T_\la\![\ga^{\scriptscriptstyle({\D\ast})}p \!\to\! Vp]$ f"ur Photo-/Leptoproduktion in Vorw"artsrichtung {\it approximativ\/}:
\vspace*{-.5ex}
\begin{align} \label{T_Lepto-approx-prop}
T_\la^{\tfbB\equiv0}[\ga^{\scriptscriptstyle({\D\ast})}p \!\to\! Vp]\vv
  \cong\vv \iIM\,s\; \int_0^{\infty}\zz dr\vv
            \big[\roverlap[V]{\la}\big]\!(r)\vv
            \si\ellp^{\rm tot}(\zet\zz\equiv\zz1\!/\!2,\, r)
    \\[-4.5ex]\nn
\end{align}
\vspace*{-.5ex}So k"onnen wir~$T_\la^{\tfbB\equiv0}\![\ga^{\scriptscriptstyle({\D\ast})}p \!\to\! 2S\,p]$\; in Abbildung~\ref{Fig:overlap_2S} {\it mit dem Auge absch"atzen\/} auf Basis der dort aufgetragenen Gr"o"sen~$[\roverlap[\iZS]{\la}]\!(r)$ und~$\si\ellp^{\rm tot}(\zet\zz\equiv\zz1\!/\!2,\, r)$.
Sei ausdr"ucklich betont, da"s wir an keiner Stelle unserer Arbeit mit dem approximativen Ausdruck von Gl.~(\ref{T_Lepto-approx-prop}) arbeiten; er diene lediglich der Diskussion von Abbildung~\ref{Fig:overlap_2S}, der wir uns hiermit zuwenden.

Zun"achst finden wir f"ur den dargestellten Fall des $2S$-Zustands, da"s dessen effektiver "Uberlapp mit dem Photon~$[\roverlap[\iZS]{\la}]\!(r)$~[gestrichelte Kurven] noch deutlich nichtverschwindende Werte annimmt f"ur Quark-Antiquark-Separationen~$r$ des streuenden Dipols, f"ur die sich die Dipol-Proton-Querschnitte~$\si\ellp^{\rm tot}(\zet\zz\equiv\zz1\!/\!2,\, r)$ der verschiedenen Modelle,~$\vka \!\equiv\! 0$ und~$\vka \!\equiv\! 0.74$~[durchgezogene Kurven, grau bzw.\@ schwarz] bereits deutlich unterscheiden.
Wir testen daher in der Tat einen Bereich von Distanzen~$r$, der signifikant differierende Postulate f"ur Streuquerschnitte erwarten l"a"st gegen"uber rein perturbativen Modellen~[grau].

Wir finden weiter, da"s f"ur ansteigendes~$Q^2$ der effektive "Uberlapp immer mehr hin zu kleineren Dipolen verschoben wird.
F"ur kleine~$Q^2 \klgl 2\GeV^2$ und st"arker f"ur transversale als f"ur longitudinale Polarisation%
\FOOT{
  \label{FN:Skalen-der-Abszissen}Die Skalen der Abszissen sind nicht identisch!
}
existiert ein deutlich nichtverschwindender effektiver "Uberlapp {\it rechts des Nulldurchgangs\/} f"ur gro"se Dipole.
Im Aufbau der (Vorw"arts)Amplitude $T_\la^{\tfbB\equiv0}[\ga^{\scriptscriptstyle({\D\ast})}p \!\to\! 2S\,p]$ wird dieser stark gewichtet gegen"uber dem {\it links des Nulldurchgangs\/} f"ur kleine Dipole, der entgegengesetztes Vorzeichen tr"agt, durch den stark ansteigenden Dipol-Proton-Querschnitt; dies in erheblich gr"o"serem Umfang durch die physikalische Kurve f"ur~$\vka \!\equiv\! 0.74$ als durch die nichtkonfinierende f"ur~$\vka \!\equiv\! 0$.
In dieser Weise~-- werden wir sehen~-- sind f"ur kleine~$Q^2$ die Beitr"age gro"ser Dipole in der Lage die Beitr"age kleiner Dipole vollst"andig zu k"urzen und das Vorzeichen von~$T_\la\![\ga^{\scriptscriptstyle({\D\ast})}p \!\to\! 2S\,p]$ herumzudrehen.
Dies ist ein ganz wesentlicher Punkt, den wir noch ausf"uhrlich diskutieren werden.
\begin{figure}
\begin{minipage}{\linewidth}
  \begin{center}
  \vspace*{4.5mm}
  \setlength{\unitlength}{1mm}\begin{picture}(120,74.25)   
    \put(0,0){\epsfxsize120mm \epsffile{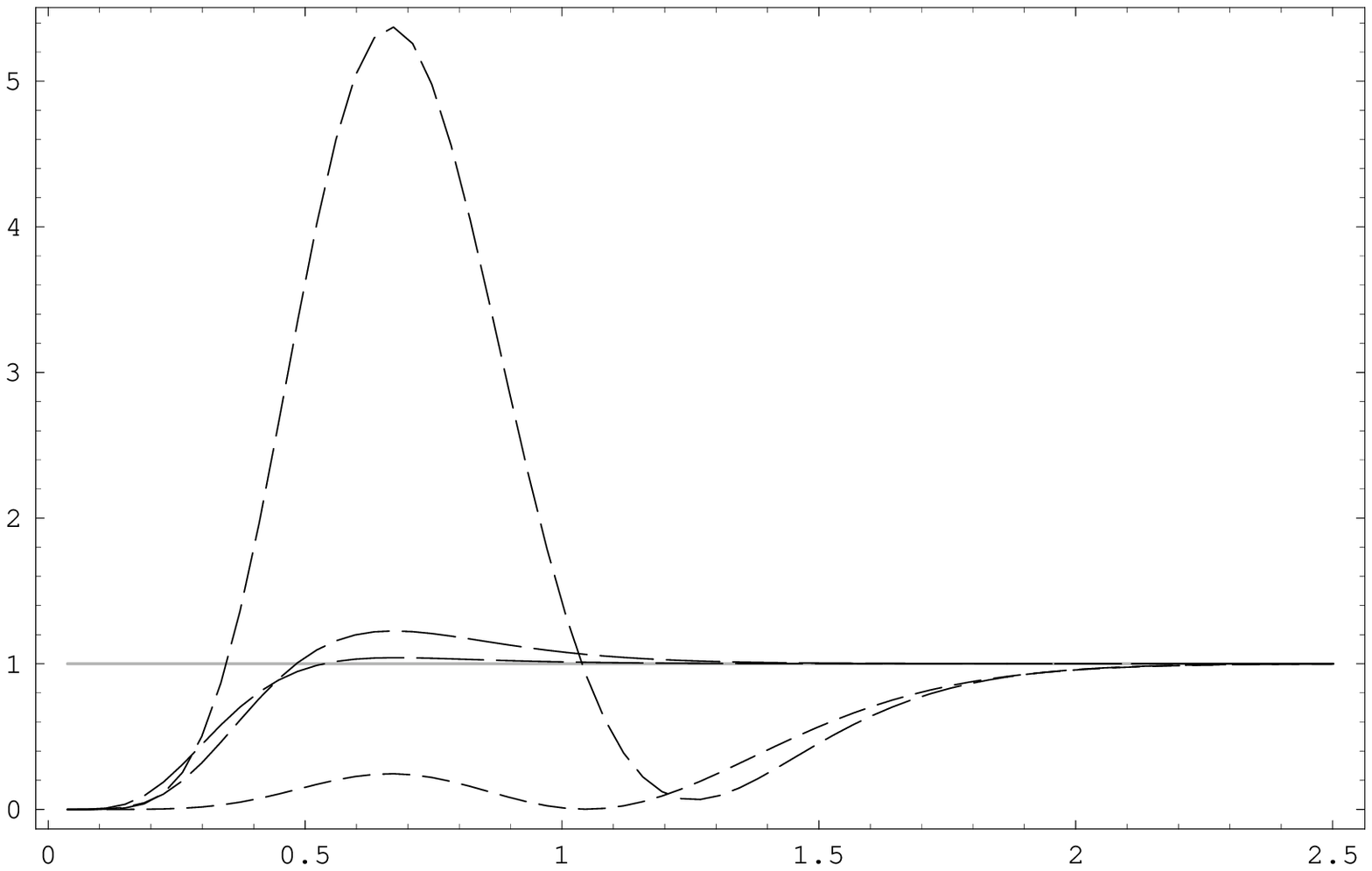}}
    \put(-10,75  ){\normalsize (a)\quad Longitudinal,~$L$\,:}
    \put(  9,69  ){\normalsize$2S$\,:}
    \put(117,-2.5){\normalsize$r_{\rm cut}\;[\fm[]]$}
    \put( -6, 0  ){\yaxis[74.25mm]{\normalsize%
                    $\si_{V,L}(r_{\rm cut})\big/\si_{V,L}(r_{\rm cut} \!=\! \infty)$}}
    \put(31, 9){\normalsize$Q^2 \!=\!  \phantom{0 }1\GeV^2$}
    \put(10,50){\normalsize$Q^2 \!=\!  2\GeV^2$}
    \put(31,21){\normalsize$Q^2 \!=\! 10$}
    \put(39,14){\normalsize$          20\GeV^2$}
    \put(63,38){\epsfxsize56mm \epsffile{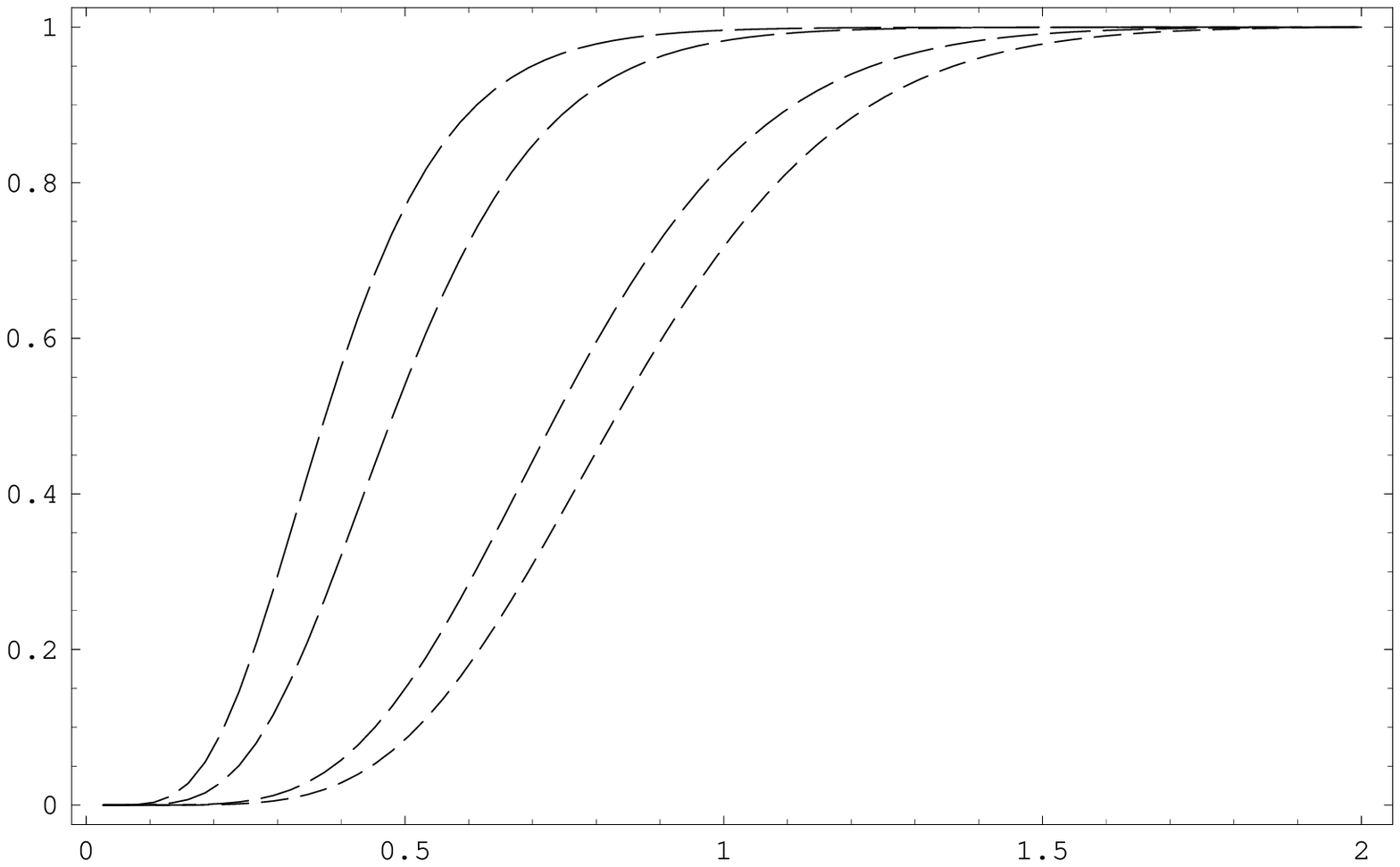}}
    \put(67,69){\normalsize$1S$\,:}
    \put(85.5,41){\small$\phantom{,0}Q^2 \!\equiv\! 20,10,2,1\GeV^2$}
  \end{picture}\\[3ex]
  \setlength{\unitlength}{1mm}\begin{picture}(120,74.3)   
    \put(0,0){\epsfxsize120mm \epsffile{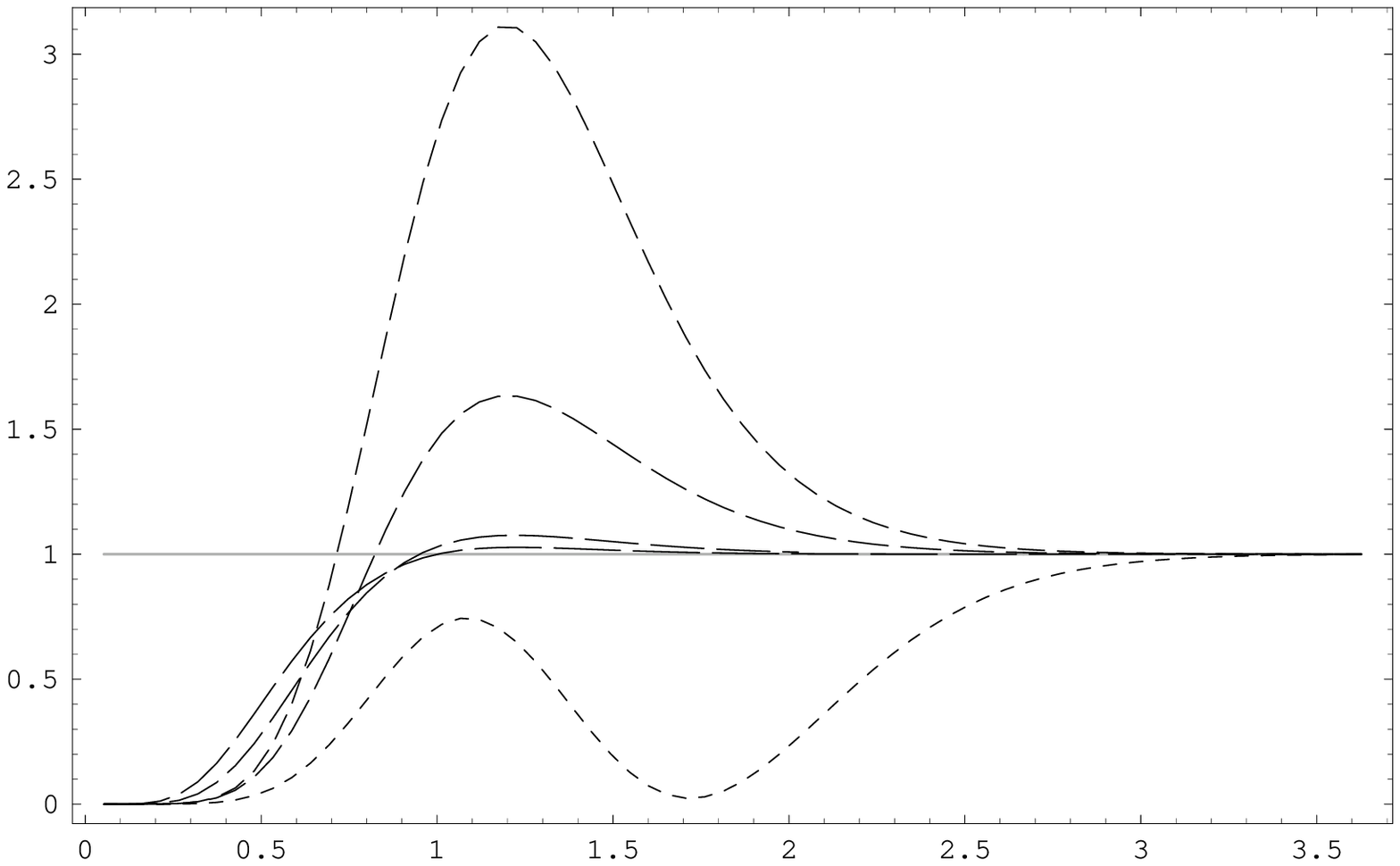}}
    \put(-10,75.25){\normalsize (b)\quad Transversal,~$T$\,:}
    \put(  9,69   ){\normalsize$2S$\,:}
    \put(117,0    ){\normalsize$r_{\rm cut}\;[\fm[]]$}
    \put( -6,0    ){\yaxis[74.3mm]{\normalsize%
                    $\si_{V,T}(r_{\rm cut})\big/\si_{V,T}(r_{\rm cut} \!=\! \infty)$}}
    \put(78,15){\normalsize$Q^2 \!=\!  0$}
    \put(15,50){\normalsize$Q^2 \!=\!  1\GeV^2$}
    \put(37,41){\normalsize$Q^2 \!=\!  2\GeV^2$}
    \put(37,29){\normalsize$Q^2 \!=\! 10$}
    \put(45,23){\normalsize$          20\GeV^2$}
    \put(63,38){\epsfxsize56mm \epsffile{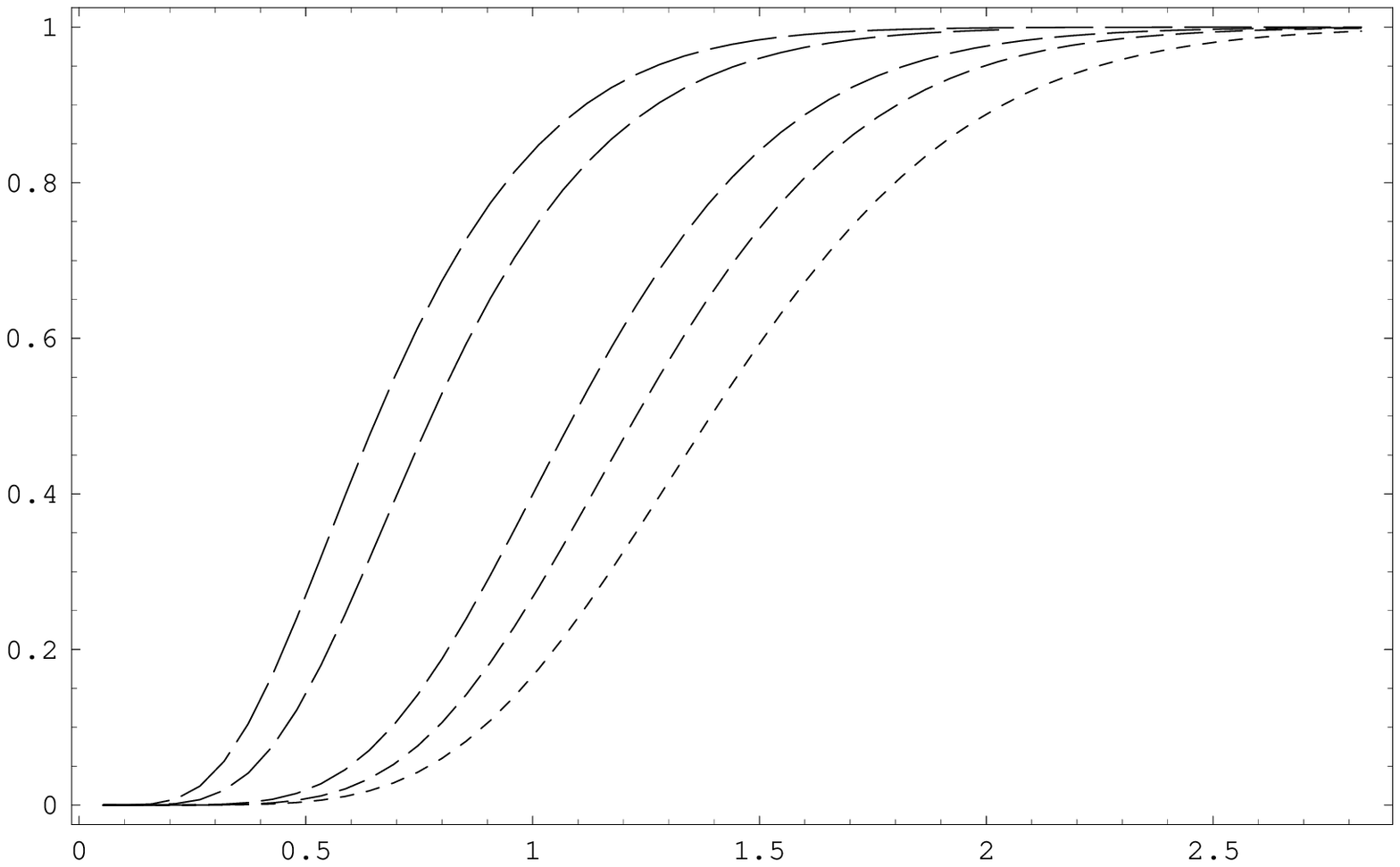}}
    \put(67,69){\normalsize$1S$\,:}
    \put(85.5,41){\small$Q^2 \!\equiv\! 20,10,2,1,0\GeV^2$}
  \end{picture}
  \end{center}
\vspace*{-4.5ex}
\caption[Verh"altnis~\protect$\si_{V,\la}(r_{\rm cut})/\si_{V,\la}(\infty)$ f"ur,~\protect$V \!\equiv\! 1S,2S$-Produktion,~$\la \!\equiv\! L,T$]{
  Integrierter Wirkungsquerschnitt als Funktion des Cut-off~$r_{\rm cut}$ der transversalen Ausdehnung der Dipole; normiert auf den vollen Wert.   Klein: $1S$-, gro"s: $2S$-Pro\-duktion.   Die Kurven beziehen sich auf~$Q^2 \!=\! 0,1,2,10,20\GeV^2$~[zunehmende Strichl"ange], in~(a) auf longitudinale, in~(b) auf transversale Polarisation.   Aufgrund des Knotens der $2S$-Wellenfunktion sind ausgedehntere Dipole von gr"o"serer Bedeutung f"ur kleine~$Q^2$, aufgrund des $\zet$-Endpunkte-Beitrags f"ur transversale mehr als f"ur longitudinale Polarisation.
\vspace*{-1ex}
}
\label{Fig:Rcut_2S[1S]}
\end{minipage}
\end{figure}
\\\indent
Das so weit auf Basis von Abbildung~\ref{Fig:overlap_2S} diskutierte subtile Wechselspiel von effektivem "Uberlapp~\mbox{$[\roverlap[\iZS]{\la}]\!(r)$} und Dipol-Proton-Querschnitt~\mbox{$\si\ellp^{\rm tot}(\zet\zz\equiv\zz1\!/\!2,\, r)$} manifestiert sich weiter und zum Tel noch deutlicher in Abbildung~\ref{Fig:Rcut_2S[1S]}.
Dort ist dokumentiert in den gro"sen Abbildungen~\mbox{f"ur den $2S$-,} in den kleinen f"ur den $1S$-Zustand, Dipole welcher transversaler Ausdehnung~-- abh"angig vom Wert von~$Q^2$~-- die Wirkungsquerschnitte f"ur~\mbox{Photo-/Lepto}\-produktion effektiv bestimmen.
Explizit berechnen wir~$\si_{V,\la}(r_{\rm cut})/\si_{V,\la}(r_{\rm cut}\zz=\zz\infty)$, das hei"st die integrierten Produktionsquerschnitte f"ur die Zust"ande~$V \!\equiv\! 1S$ und~$2S$ auf Basis der Gln.~(\ref{E:sitot_ellp}),~(\ref{E:pT_ellp}), indem wir durch Einf"uhrung des Cut-off~$r_{\rm cut}$ die Integration "uber die transversale Ausdehnung der Dipole beschr"anken auf Werte~$r \!\le\! r_{\rm cut}$; wir skalieren auf den vollen Wert mit entferntem Cut-off.%
\FOOT{
  F"ur Vergleichbarkeit geben wir auch an~$\si_{\iES,\la}(r_{\rm cut})/\si_{\iES,\la}(r_{\rm cut}\zz=\zz\infty)$, das hei"st f"ur den $1S$-Zustand; vgl.\@ auch Abb.~\refg{Fig:dsdtLT-rcut} bzgl.\@ {\sl differentieller Wirkungsquerschnitte in Vorw"artsrichtung}.
} \\
%
\indent
Wir betrachten zun"achst Photoproduktion~$Q^2 \!\equiv\! 0$, die Kurve in kurzen Strichen in Abbildung~\ref{Fig:Rcut_2S[1S]}(b).
Nach Abbildung~\ref{Fig:overlap_2S}(b) besitzt der effektive "Uberlapp der Photon- und der Vektormeson-Lichtkegelwellenfunktion einen Nulldurchgang f"ur~$r \!\cong\! 1.1\fm$.
Entsprechend finden wir hier, da"s der Wirkungsquerschnitt stetig ansteigt f"ur~$r \klgl 1.1\fm$, das hei"st er ist dominiert durch kleine Dipole {\it links des Nulldurchgangs\/}.
Werden gr"o"sere Dipole miteinbezogen, das hei"st {\it rechts des Nulldurchgangs\/}, deren Beitr"age daher auftreten mit entgegengesetztem, positivem Vorzeichen, vermindert sich der Betrag der Amplitude~$T_\la^{\tfbB\equiv0}[\ga^{\scriptscriptstyle({\D\ast})}p \!\to\! 2S\,p]$ und damit der Wirkungsquerschnitt.%
\FOOT{
  Die Amplitude~$T_\la^{\tfbB}$ \vspace*{-.25ex}f"allt ab wie~$\exp-B_0\tfbQ$, das hei"st ihr Vorzeichen und Betrag in Vorw"artsrichtung sind von besonderer Bedeutung auch f"ur~$-t \!\cong\! \tfbQ$-integrierte Querschnitte.
}
F"ur~$r \!\cong\! 1.7\fm$ tritt vollst"andige K"urzung ein, das hei"st Kompensation der Beitr"age kleiner durch die gro"ser Dipole.
Werden noch gr"o"sere Dipole miteinbezogen steigt der Wirkungsquerschnitt wieder an, bis er f"ur etwa~$r \!\cong\! 3\fm$ seinen vollen asymptotischen Wert erreicht.
Wie bereits angemerkt r"uhren dabei wesentliche Beitr"age von Dipolen mit transversaler Ausdehnung im Bereich~$r \!=\! 2.5 \!-\! 2.8\fm$. \\
\indent
Generell tragen gr"o"sere Dipole mehr bei zu den $2S$- als zu den $1S$-Wirkungsquerschnitten; f"ur feste Werte von~$Q^2$ generell mehr f"ur transversale als f"ur longitudinale Polarisation.\citeFN{FN:Skalen-der-Abszissen} \\
\indent
Variation von~$Q^2$ "andert die Position des Nulldurchgangs des effektiven "Uberlapps der Wellenfunktionen; dies spiegelt sich wider in der starken $Q^2$-Abh"angigkeit der Kurven.
Der Einflu"s der Dipole gro"ser Ausdehnung geht mehr und mehr zur"uck f"ur wachsendes~$Q^2$.
F"ur longitudinale Polarisation besitzt der effektive "Uberlapp einen festen Nulldurchgang unabh"angig von~$Q^2$, vgl.\@ Abb.~\ref{Fig:overlap_2S}(a); dies r"uhrt her von der Struktur der Lichtkegelwellenfunktion~$\ket{2S}$, die durch ein Polynom in~$r^2$ bestimmt ist.%
\FOOT{
  \label{FN:2SParameter}In Hinsicht auf die $2S$-Parameter~$\om_{\iZS,\la}$~${\cal N}_{\iZS,\la}$,~$A_\la$ beziehen sich die Abbildungen~\ref{Fig:overlap_2S},~\ref{Fig:Rcut_2S[1S]} auf den Satz von Ref.~\cite{Kulzinger98}: f"ur longitudinale Polarisation sind diese unabh"angig von~$Q^2$, ebenso die Oszillatorparameter beider Polarisationen.   F"ur den {\sl Referenz-Satz\/} h"angen sie dagegen s"amtlich, wenn auch nur leicht, ab von~$Q^2$.   Entsprechend unserer Diskussion des Einflu"ses auf~$A_T$ bezieht sich die einzige nicht von vornherein irrelevante Diskrepanz auf die Position des "`longitudinalen Knotens"' f"ur transversale Polarisation.   Wir erwarten daher kaum eine quantitative "Anderung des diskutierten Bildes.
}
Infolgedessen vermindert sich der effektive "Uberlapp der Wellenfunktionen f"ur wachsendes~$Q^2$ allein aufgrund der globalen D"ampfung durch die modifizierte Besselfunktion zweiter Art~${\rm K}_0(\ep r)$ der Photon-Wellenfunktion.
Das Verhalten von~$\si_{\iZS,L}(r_{\rm cut})/\si_{\iZS,L}(r_{\rm cut}\zz=\zz\infty)$ f"ur~$Q^2 \!\equiv\! 1$ und~$2\GeV^2$, vgl.\@ Abb.~\ref{Fig:Rcut_2S[1S]}(a), ist entsprechend dem f"ur Photoproduktion: Dominanz kleiner Dipole, vollst"andige Kompensation durch gr"o"sere, schlie"slich Dominanz gro"ser Dipole. \\
\indent
F"ur transversale Polarisation besitzt die Lichtkegelwellenfunktion~$\ket{2S}$ zwei verschiedene Polynome in~$r^2$, die durch die effektive Quarkmasse~$\meff[](Q^2)$ abh"angig von~$Q^2$ gegeneinander gewichtet sind, vgl.\@ Fu"sn.\,\FN{FN:2SParameter}.
Dies resultiert effektiv darin, vgl.\@ Abb.~\ref{Fig:overlap_2S}(b), da"s der Nulldurchgang des effektiven "Uberlapps von Photon- und $2S$-Wellenfunktion f"ur ansteigendes~$Q^2$ schnell zu gr"o"seren Dipolausdehnungen~$r$ verschoben wird.
F"ur gr"o"sere~$Q^2$ bewirken die Besselfunktionen~${\rm K}_0(\ep r)$,~${\rm K}_1(\ep r)$ eine starke Unterdr"uckung des effektiven "Uberlapps rechts des Nulldurchgangs.
Bezogen auf die Wirkungsquerschnitte, vgl.\@ Abb.~\ref{Fig:Rcut_2S[1S]}(b), tritt bereits f"ur~$Q^2 \!\equiv\! 1\GeV^2$ Kompensation der Beitr"age kleiner Dipole, links des Nulldurchgangs durch die gro"ser Dipole, rechts des Nulldurchgangs nicht mehr vollst"andig ein:
Von dem einmal durch kleine Dipole erreichten Maximum fallen die Kurven f"ur~$Q^2 \!>\! 1\GeV^2$ monoton ab auf den vollen asymptotischen Wert, das hei"st der einmal erreichte Wirkungsquerschnitt wird durch gr"o"sere Dipole vermindert, ohne da"s deren Beitr"age so gro"s w"urden, sie vollst"andig k"urzen und das Vorzeichen von~$T_\la^{\tfbB\equiv0}[\ga^{\scriptscriptstyle({\D\ast})}p \!\to\! 2S\,p]$ umdrehen zu k"onnen. \\
\indent
Vergleich gegen"uber den grau eingezeichneten Hilfslinien zeigt aber, da"s f"ur beide Polarisationen und s"amtliche betrachteten Werte von~$Q^2$ ein Nulldurchgang existieren mu"s, auch wenn dieser aus Abbildung~\ref{Fig:overlap_2S} nicht mehr abgelesen werden kann.
Denn s"amtliche Wirkungsquerschnitte besitzen ein Maximum, das "uber dem asymptotischen vollen Wert liegt, Verminderung der Amplitude durch gro"se Dipole ist globales Ph"anomen.
Demgegen"uber dokumentieren die kleinen Abbildungen f"ur den $1S$-Zustand stetiges Ansteigen der $T$-Amplitude bis zum vollen Asymptotischen Wert des Wirkungsquerschnitts. \\
\indent
Dem festen Nulldurchgang f"ur longitudinale polarisation stehen gegen"uber zwei gegeneinander gewichtete Polynome f"ur transversale polarisation.
Dies f"uhrt zu einem "`Verwaschungseffekt"', der Grund daf"ur ist, da"s die absolute Gr"o"se des intermedi"ar erreichten Maximums fast um einen Faktor Zwei gr"o"ser ist f"ur longitudinale als f"ur transversale Polarisation; vgl.\@ die Skala der Ordinate in Abb.~\ref{Fig:Rcut_2S[1S]}(a) gegen"uber~(b). \\
\enlargethispage{1.5ex}
\indent
Auf Basis dieser generellen Diskussion spezifizieren wir im folgenden die numerische Analyse im Sinne der Diskussion expliziter Streuquerschnitte.
\vspace*{-.75ex}

%
\subsection{Spezifische Diskussion}

Wir diskutieren die $\pi^+\pi^-$-Massespektren f"ur $e^+e^-$-Annihilation und Photoproduktion in Hinblick auf die unterschiedlichen Interferenzmuster im Bereich einer invarianten Masse~$M$ von etwa~$1.6\GeV$.
In Anhang~\refg{APPSubsect:Photo,l+l-Annih} leiten wir her%
\FOOT{
  \label{FN:}vgl.\@ ebenda Anh.~\ref{APPSubsect:Photo,l+l-Annih} bzgl.\@ aller formalen Details
}
die Formeln f"ur Lepton-An\-tilepton-Annihilation in die Pion-Endzust"ande~\mbox{$f \!\equiv\! \pi^+\pi^-,\, 2\pi^+2\pi^-$}, vgl.\@ Gl.~(\ref{APP:si_ll-to-f}):
\vspace*{-.5ex}
\begin{align} \label{si_ll-to-f}
\si_f(M)\;
  =\; \frac{4\pi e^4}{3}\;
        \Big| {\T\sum}_{V\!=\rh,\rh',\rh^\dbprime} \frac{f_V}{M_V}\;
        \frac{c_{V\!,f}\cdot \sqrt{M_V\Gatot_V\!/\pi}}{%
              M^2 \!-\! M_V^2 \!+\! \iIM\, M_V\Gatot_V}\;
        \sqrt{B_{V\to f}}\; \Big|^2
    \\[-4.5ex]\nn
\end{align}
und f"ur Photoproduktion elastisch am Proton, vgl.\@ Gl.~(\ref{APP:si_gap-to-fp}):
\vspace*{-.5ex}
\begin{align} \label{si_gap-to-fp}
&\frac{1}{2M}\frac{d\si_{f\!,\la}}{dM}\Big|_{Q^2}(M)
    \\
&\phantom{\si_f(M)\;}
  =\; \int_{-\infty}^0 dt\; \frac{1}{16\pi\, s^2}\;
        \bigg| {\T\sum}_{V\!=\rh,\rh',\rh^\dbprime}\vv
        T_{V\!,\la}(s,t)\vv
        \frac{c_{V\!,f}\cdot \sqrt{M_V\Gatot_V\!/\pi}}{%
              M^2 \!-\! M_V^2 \!+\! \iIM\, M_V\Gatot_V}\vv
        \sqrt{B_{V\!\to f}}\; \bigg|^2
    \nn
    \\[-4.5ex]\nn
\end{align}
mit~$M$ der invarianten Masse des Endzustands~$f$.
Dabei seien abk"urzend notiert f"ur die $1^+(1^{--})$-/Rho-Kanal-Querschnitte~\;\mbox{$\si_f \!\equiv\! \si[l^+l^- \!\to\! V \!\to\! f]$},~\;\mbox{$d\si_{f\!,\la} \!\equiv\! d\si_\la[\ga^{\scriptscriptstyle({\D\ast})}p \!\to\! Vp]$} und f"ur die physikalischen $T$-Amplituden~\mbox{$T_{V\!,\la} \!\equiv\! T_\la\![\ga^{\scriptscriptstyle({\D\ast})}p \!\to\! Vp]$} f"ur~$V \!\equiv\! \rh,\rh',\rh^\dbprime$; diese folgen unmittelbar auf Basis der Gln.~(\ref{Ansatz})-(\ref{Ansatz}$''$) aus den $1S$-, $2S$-Amplituden~\mbox{$T_\la\![\ga^{\scriptscriptstyle({\D\ast})}p \!\to\! 1S\,p]$}, \mbox{$T_\la\![\ga^{\scriptscriptstyle({\D\ast})}p \!\to\! 2S\,p]$}.
Die Massen, totalen Zerfallsbreiten, Verzweigungsverh"altnisse~$M_V$,~$\Gatot_V$, $B_{V\!\to f}$ sind diskutiert in Zusammenhang mit Tabelle~\ref{Tabl:Charakt_rh,rh',rh''} und angegeben dort; bzgl.\@ der Normierungsparameter~$c_{V\!,f}$ vgl.\@ Gl.~(\ref{APP:si_gap-to-fp}) und Tabl.~\ref{Tabl-App:cVla} bzgl.\@ expliziter Zahlenwerte.

In den Abbildungen~\ref{Fig:photo2pis},~\ref{Fig:eebar2pis} ist dargestellt unser Resultat f"ur die $\pi^+\pi^-$-Massespektren auf Basis dieser Gln.~(\ref{si_ll-to-f}),~(\ref{si_gap-to-fp}).
Das $e^+e^-$-Spektrum ist perturbativ bestimmt durch kleine Quark-Antiquark-Abst"ande, das hei"st dominiert durch Dipole kleiner (transversaler) Ausdehnung~$r$.
Die relativen Vorzeichen der Kopplungen~$f_V$ f"ur~$V \!\equiv\! \rh,\rh',\rh^\dbprime$ sind~$[+,-,+]$, das Interferenzmuster im Bereich~$M \!\cong\! 1.6\GeV$ entsprechend {\it destruktiv\/}.
Resultat unserer Diskussion der Abbildungen~\ref{Fig:overlap_2S}(b),~\ref{Fig:Rcut_2S[1S]}(b) ist unater anderem, da"s f"ur Photoprpduktion \mbox{$Q^2 \!\equiv\! 0$} die Beitr"age gro"ser Dipole rechts des Nulldurchgangs die kleiner Dipole links des Dipols vollst"andig k"urzen und das Vorzeichen der Amplitude~\mbox{$T_\la\![\ga^{\scriptscriptstyle({\D\ast})}p \!\to\! 2S\,p]$} umdrehen.%
\FOOT{
  F"ur kleine Dipole sind die Vorzeichen von~$f_V\!/M_V$ und~$T_{V,\la}$ in den Gln.~(\ref{si_ll-to-f}),~(\ref{si_gap-to-fp}) identisch, da in derselben Weise determiniert durch das Vorzeichen der $2S$-Wellenfunktion am Ursprung.
}
Dieser Vorzeichenwechsel "ubertr"agt sich entsprechend auf die physikalischen Produktions-Amplituden~\mbox{$T_\la\![\ga^{\scriptscriptstyle({\D\ast})}p \!\to\! Vp]$} f"ur~$V \!\equiv\! \rh',\rh^\dbprime$ und induziert "uber deren koh"arente Addition in Gl.~(\ref{si_gap-to-fp}) zu der konstruktiven Interferenz im $\pi^+\pi^-$-Massespektrum f"ur Photoproduktion.
Da"s wir dieses experimentelle Interferenzmuster reproduzieren ist definitive Konsequenz der Dominanz der Beitr"age von Dipolen gro"ser transversaler Ausdehnung~-- wir haben gesehen, da"s de facto noch Ausdehnungen im Bereich~$r \!=\! 2.5 \!-\! 2.8\fm$ signifikant beitragen.
Diese Dominanz wiederum ist ganz wesentlich verkn"upft mit dem starken Ansteigen mit~$r$~der Dipol-Proton-Amplitude~$\pT\ellp$ f"ur~$\vka \!\equiv\! 0.74$ versus f"ur~$\vka \!\equiv\! 0$ und damit letztlich Konsequenz des \DREI[]{M}{S}{V}-spezifischen Mechanismus von Ausbildung und Wechselwirkung gluonischer Strings. \\
\indent
F"ur Leptoproduktion von~$f \!\equiv\! \pi^+\pi^-$ entsprechend Gl.~(\ref{si_gap-to-fp}) f"ur~\mbox{$Q^2 \!>\! 0$} verm"ogen die Beitr"age gro"ser Dipole rechts des Nulldurchgangs die kleiner Dipole links des Nulldurchgangs bald nicht mehr volllst"andig zu k"urzen, vgl.\@ die Diskussion oben, und so das Vorzeichen von \mbox{$T_\la\![\ga^{\scriptscriptstyle({\D\ast})}p \!\to\! 2S\,p]$} nicht mehr umzudrehen; wir beobachten ein Interferenzmuster entsprechend dem in $e^+e^-$-Annihilation.

\vspace*{-1ex}
\bigskip\noindent
Auf Basis der Gln.~(\ref{T_Lepto}),~(\ref{dsigmadt_Lepto}) werden berechnet in Abh"angigkeit von der Virtualit"at~$Q$ des Photons integrierte Wirkungsquerschnitte \mbox{$\si_{V,\la} \!\equiv\! \si_\la\![\ga^{\scriptscriptstyle({\D\ast})}p \!\to\! Vp]$} f"ur die Zust"ande~$V \!\equiv\! \rh(770)$ und~$2S$, f"ur longitudinale und transversale Polarisation.
Sei dabei im folgenden angesprochen mit~$\rh(770)$ der Vektormeson-Grundzustand, der bisher generisch als~$1S$-Zustand bezeichnet ist.
Die Querschnitte~$\si_{V,\la}$ als Funktion von~$Q^2$ sind angegeben graphisch in Abbildung~\ref{Fig:sigma_rh770,2S}(a) f"ur~$V \!\equiv\! \rh(770)$ und in~\ref{Fig:sigma_rh770,2S}(b) f"ur~$V \!\equiv\! 2S$.
F"ur~$V \!\equiv\! \rh(770)$ sind explizite Zahlenwerte angegeben und verglichen mit experimentellen Daten f"ur Leptoproduktion im Bereich~$0 \!<\! Q^2 \!<\! 20\GeV^2$ in Tabelle~\ref{Tabl:Q>0-sigma_rh770} und f"ur Photoproduktion,~$Q^2 \!\equiv\! 0$, in Tabelle~\ref{Tabl:Q=0_rh,om,ph,Jps}.
\begin{figure}
\begin{minipage}{\linewidth}
  \begin{center}
  \vspace*{4.5mm}
  \setlength{\unitlength}{.9mm}\begin{picture}(120,67.2)   
    \put(0,0){\epsfxsize108mm \epsffile{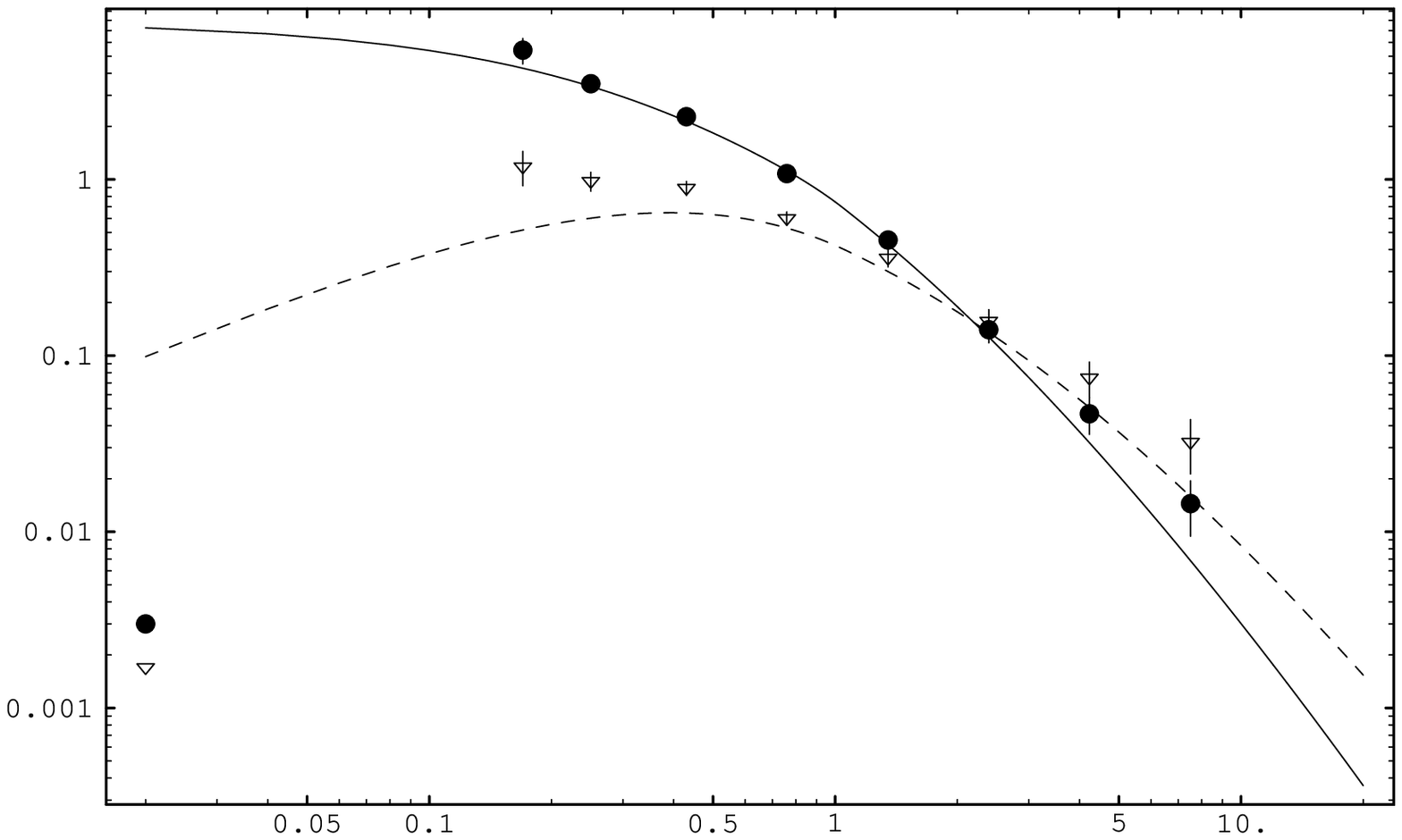}}
    \put(-10,68){\normalsize (a)\quad$\rh(770)$\,:}
    \put(114, 0.5){\normalsize$Q^2\;[\GeV[]^2]$}
    \put( -6, 0  ){\yaxis[60.48mm]{\normalsize$\si_{\irh,\la}(Q^2)\vv[\microbarn[]]$}}
    \put(15,16){\normalsize$85\,\%$\vv \VIER[]{E}{6}{6}{5}-Transversal}
    \put(15,12){\normalsize$85\,\%$\vv \VIER[]{E}{6}{6}{5}-Longitudinal}
    \put(15,58){\normalsize$\la \!\equiv\! T$,~Transversal}
    \put(15,35){\normalsize$\la \!\equiv\! L$,~Longitudinal}
  \end{picture}\\[2.25ex]
  \setlength{\unitlength}{.9mm}\begin{picture}(120,67.2)   
    \put(0,0){\epsfxsize108mm \epsffile{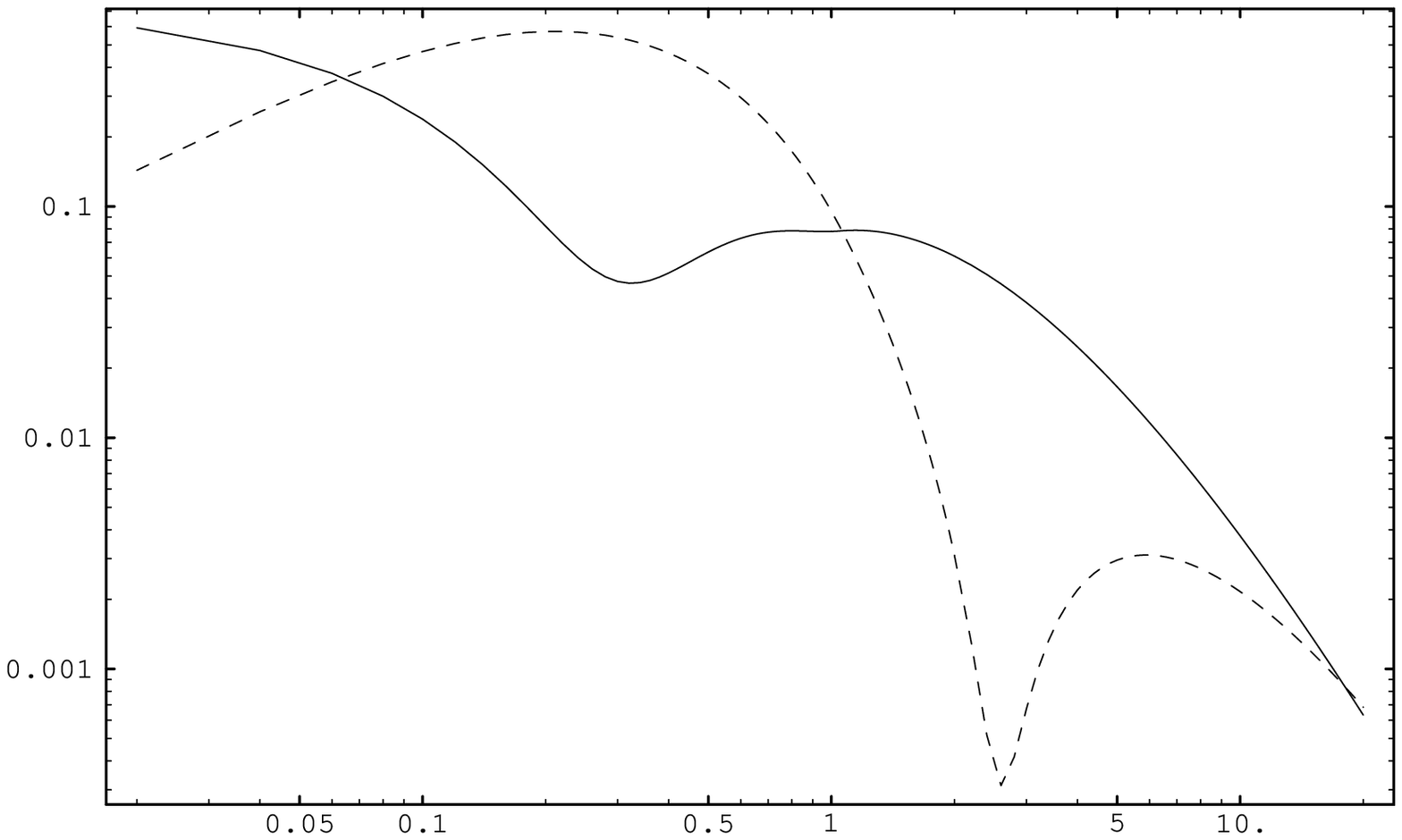}}
    \put(-10,68){\normalsize (b)\quad$2S$\,:}
    \put(114, 0.5){\normalsize$Q^2\;[\GeV[]^2]$}
    \put( -6, 0  ){\yaxis[60.48mm]{\normalsize$\si_{\iZS,\la}(Q^2)\vv[\microbarn[]]$}}
    \put(25,40){\normalsize$\la \!\equiv\! T$,~Transversal}
    \put(61,60){\normalsize$\la \!\equiv\! L$,~Longitudinal}
  \end{picture}
  \end{center}
\vspace*{-4.5ex}
\caption[Integrierte Wirkungsquerschnitte~\protect$\si_{V,\la}(Q^2)$ f"ur~\protect$V \!\equiv\! \rh(770),2S$,~\protect$\la \!\equiv\! L,T$]{
  Integrierte Wirkungsquerschnitte~$\si_{V,\la}$,~$V \!\equiv\! \rh(770),2S$,~$\la \!\equiv\! L,T$, als Funktion von~$Q^2$.   (a) Produktion des~$\rh(770)$-, (b) des~$2S$-Zustands; gestrichelte Kurve f"ur longitudinale, durchgezogene f"ur transversale Polarisation.   Die experimentellen Daten f"ur~$\rh(770)$ stammen von \VIER[]{E}{6}{6}{5}, vgl.\@ Ref.~\cite{Adams97}; sie schlie"sen ein einen Regge-Anteil, den wir abgesch"atzen als etwa~$15\,\%$ und subtrahieren.   Vgl.\@ auch Tabl.~\ref{Tabl:Q>0-sigma_rh770}.
}
\label{Fig:sigma_rh770,2S}
\end{minipage}
\end{figure}
\\\indent
Zun"achst existieren experimentelle Daten nur f"ur~$\rh(770)$.
Fermilab-\VIER[]{E}{6}{6}{5}, vgl.\@ Ref.~\cite{Adams97}, gibt Zahlenwerte an separat f"ur longitudinale und transversale Polarisation.
\DREI{N}{M}{C} ver"offentlicht in Ref.~\cite{Arneodo94} Zahlenwerte f"ur die Gr"o"se~$\si \!=\! \ep \si_L \!+\! \si_T$, mit~$\ep$ der Rate longitudinaler Photonen, die typischerweise im Bereich~$0.5 \!-\! 0.8$ liegt; vgl.\@ die Gln.~(\ref{dsigmadt-exp}),~(\ref{sigma-exp}) und~(\ref{Epsilon}).
Dabei ist zu beachten, da"s die experimentellen Daten einen Regge-Anteil einschlie"sen, das hei"st einen nicht-diffraktiven "`harten"' Anteil, den unser Zugang in der vorgestellten Form nicht diskutiert.
Dieser Anteil ist daher zu subtrahieren.
F"ur kleine~$Q^2$ wird er abgesch"atzt durch Vergleich mit der Parametrisierung des Compton-Streuquerschnitts~$\si^{\rm tot}[\ga p]$ nach Donnachie, Landshoff in den Refn.~\cite{Donnachie92,PDG00}; vgl.\@ auch Gl.~(\ref{T_Lepto,Compton}$'$).
Diese gibt f"ur die betrachtete inveriante Schwerpunktenergie~$\surd s \!=\! 20\GeV$ einen Regge-Anteil an von etwa~$7\,\%$.
Der Wirkungsquerschnitt f"ur die Photo-/Leptoproduktion des Vektormesons~$V$ ist hierzu in Kontrast determiniert durch das (Absolut)Quadrat der entsprechenden $T$-Amplitude, so da"s hier "aquivalent erwartet wird ein Beitrag von~$15\,\%$ des Reggeons.
Wir geben explizite Zahlenwerte experimenteller Daten im folgenden immer an {\it nicht-modifiziert}, den Regge-Anteil einschlie"send. \\
\indent
F"ur unmittelbare Vergleichbarkeit weichen wir hierin nur ab in der Auftragung in Abbildung~\ref{Fig:sigma_rh770,2S}(a), der wir uns nun zuwenden.
Dort ist angegeben im Bereich von~\mbox{$0.02 \!-\! 20\GeV^2$} f"ur~$Q^2$ unser Postulat f"ur~$\si_{\irh,\la}(Q^2)$ und ihm gegen"ubergestellt~$85\,\%$ der \VIER[]{E}{6}{6}{5}-Daten als deren Pomeron-Anteil, wie gerade abgesch"atzt.
Die doppel-logarithmische Auftragung setzt die Betonung auf kleine Werte von~$Q^2$.
F"ur transversale Polarisation ist der niedrigste Datenpunkt bei~$Q^2 \!\cong\! 0.08\GeV^2$ konsistent mit unserem Postulat, die Punkte bis etwa~$4.8\GeV^2$ in guter "Ubereinstimmung.
Wir betonen dabei, da"s die Parameter unseres Zugangs vollst"andig festgelegt sind durch "Uberlegungen, die {\it nicht\/} Bezug nehmen auf Photo- und Leptoproduktion; wir postulieren absolute Zahlenwerte und finden "Ubereinstimmung mit dem Experiment.
F"ur longitudinale Polarisation stellen wir eine Abweichung fest f"ur kleine~$Q^2$ unterhalb~$0.8\GeV^2$.
F"ur longitudinale und transversale Polarisation, f"ur mittlere Werte von~$Q^2$~-- die beiden Datenpunkte bei etwa~$4.8$ und~$7.8\GeV^2$~-- liegt Fermilab-\VIER{E}{6}{6}{5} um etwa~$20 \!-\! 30\,\%$ "uber unserem Postulat.
Tabelle~\ref{Tabl:Q>0-sigma_rh770} dokumentiert, da"s die \VIER[]{E}{6}{6}{5}-Daten und deren Extrapolation zu h"oheren~$Q^2$ allerdings dieselbe dieselbe Diskrepanz zeigen im Vergleich zu den Daten von \DREI[]{N}{M}{C}.
Mit diesen stellen wir gute "Ubereinstimmung fest "uber den gesamten betrachteten $Q^2$-Bereich.
%
%
%
\begin{sidewaystable}
\begin{minipage}{\linewidth}
\renewcommand{\thefootnote}{\thempfootnote}
\begin{center}
  \begin{tabular}{|f{-1}||g{8}|g{10}||g{9}|g{10}||c||g{8}|g{13}||h{4}|} \hline
  \multicolumn{9}{|c|}{Leptoproduktion,~$0 \!<\! Q^2 \!<\! 20\GeV^2$, des $\rh(770)$:
                       ~Theorie versus Experiment} \\
  \hhline{:=:t:==:t:==:t:=:t:==:t:=:}
  \multicolumn{1}{|c||}{$Q^2\;[\!\!\GeV^2]$}
  & \multicolumn{2}{c||}{$\si_{\irh,L}\;[\microbarn[]]$}
  & \multicolumn{2}{c||}{$\si_{\irh,T}\;[\microbarn[]]$}
  & \multicolumn{1}{c||}{$\ep$}
  & \multicolumn{2}{c||}{$\ep\, \si_{\irh,L} \!+\! \si_{\irh,T}\;[\microbarn[]]$}
  & \multicolumn{1}{c|}{}
  \\
  & \multicolumn{1}{c|}{th.\footnote{
    \label{FN:PomeronLepto}Pomeron-Anteil
  }}
  & \multicolumn{1}{c||}{exp.}  
  & \multicolumn{1}{c|}{th.\citeFN{FN:PomeronLepto}}   
  & \multicolumn{1}{c||}{exp.}  
  & \multicolumn{1}{c||}{exp.} 
  & \multicolumn{1}{c|}{th.\citeFN{FN:PomeronLepto}}
  & \multicolumn{1}{c||}{exp.}
  & \multicolumn{1}{c|}{exp.\mbox{$^{\mbox{\scriptsize Ref.}}$}}
  \\ \hhline{|-||--||--||-||--||-|}
  0.17 & 0.,517 & 1.,39 \PM0.26  & 4.,28  & 6.,37 \PM0.89  & 0.76 & 4.,67  & 7.,42 \PM0.91  &
    \ExpB\\
  0.25 & 0.,603 & 1.,15 \PM0.12  & 3.,37  & 4.,11 \PM0.23  & 0.80 & 3.,85  & 5.,03 \PM0.25  &
    \ExpB\\
  0.43 & 0.,645 & 1.,051\PM0.081 & 2.,14  & 2.,67 \PM0.13  & 0.81 & 2.,66  & 3.,52 \PM0.15  &
    \ExpB\\
  0.76 & 0.,530 & 0.,708\PM0.052 & 1.,12  & 1.,269\PM0.073 & 0.81 & 1.,55  & 1.,84 \PM0.084 &
    \ExpB\\
  1.35 & 0.,300 & 0.,422\PM0.040 & 0.,426 & 0.,533\PM0.045 & 0.81 & 0.,669 & 0.,875\PM0.055 &
    \ExpB\\
  2.39 & 0.,135 & 0.,185\PM0.025 & 0.,127 & 0.,165\PM0.022 & 0.81 & 0.,237 & 0.,315\PM0.030 &
    \ExpB\\
  2.5  & 126.,   \milli & \NON & 115.,    \milli & \NON & 0.50 & 178.,   \milli
     & (170.,  \PM31)  \milli & \ExpC\\
  3.5  &  71.,7 \milli & \NON &  51.,6  \milli & \NON & 0.66 &  98.,9 \milli
     &  (60.,  \PM10)  \milli & \ExpC\\
  4.23 & 50.,7\milli & (88.,\PM17)\milli  &  32.,0  \milli & (55.,\PM11)\milli
   & 0.81 & 73.,1\milli & (126.,  \PM18)  \milli & \ExpB\\
  4.5  &  45.,1 \milli & \NON &  27.,3  \milli & \NON & 0.66 &  57.,1 \milli
     &  (65.,  \PM11)  \milli & \ExpC\\
  5.5  &  30.,4 \milli & \NON &  16.,1  \milli & \NON & 0.72 &  38.,0 \milli
     &  (41.,  \PM \phantom{0}7)  \milli & \ExpC\\
  6.9  &  19.,0 \milli & \NON &   8.,68 \milli & \NON & 0.76 &  23.,1 \milli
     &  (23.,  \PM \phantom{0}3)  \milli & \ExpC\\
  7.51 & 15.,8\milli & (38.,\PM11)\milli  &   6.,85 \milli & (17.,\PM 5)\milli 
   & 0.81 & 19.,7\milli &  (47.,8\PM10.2)\milli & \ExpB\\
  8.8  &  11.,1 \milli & \NON &   4.,36 \milli & \NON & 0.78 &  13.,1 \milli
     &  (15.,\phantom{0}  \PM \phantom{0}2\phantom{.0})  \milli & \ExpC\\
  11.9 &   5.,55\milli & \NON &   1.,80 \milli & \NON & 0.82 &   6.,35\milli
     &   (5.,8\PM \phantom{0}0.9)\milli & \ExpC\\
  16.9 &   2.,36\milli & \NON &   0.,617\milli & \NON & 0.81 &   2.,53\milli
     &   (2.,6\PM \phantom{0}0.7)\milli & \ExpC\\
  \hhline{:=:b:==:b:==:b:=:b:==:b:=:}
  \end{tabular}
\vspace*{-3ex}
\caption[Leptoproduktion,~\protect$0 \!<\! Q^2 \!<\! 20\GeV^2$, von~\protect$\rh(770)$: Theorie vs.~Experiment]{
  Integrierte Wirkungsquerschnitte~$\si_{\irh,\la}(Q^2)$,~$\la \!\equiv\! L,T$, f"ur Photo- und Leptoproduktion des $\rh(770)$-Vektormesons.   Unser Postulat~[th.] in Gegen"uberstellung  mit experimentellen Daten~[exp.] von \VIER{E}{6}{6}{5} f"ur~$\si_{\irh,L}$,~$\si_{\irh,T}$, das hei"st f"ur longitudinale und transversale Polarisation separat, vgl.\@ Ref.~\cite{Adams97}, und von \DREI{N}{M}{C} f"ur die Gr"o"se~$\ep\, \si_{\irh,L} \!+\! \si_{\irh,T}$, vgl.\@ Ref.~\cite{Arneodo94}.   Die experimentellen Daten schlie"sen ein einen Regge-Anteil, den wir f"ur die betrachtete Energie~$\surd s \!=\! 20\GeV$ absch"atzen als etwa~$15\,\%$.
\vspace*{-.5ex}
}
\label{Tabl:Q>0-sigma_rh770}
\end{center}
\end{minipage}
\end{sidewaystable}
\renewcommand{\thefootnote}{\thechapter.\arabic{footnote}}

\enlargethispage{1.125ex}
In Abbildung~\ref{Fig:sigma_rh770,2S}(b) ist aufgetragen f"ur demselben Bereich von~\mbox{$Q^2 \!=\! 0.02 \!-\! 20\GeV^2$} unser Postulat f"ur~$\si_{\iZS,\la}(Q^2)$.
Die Kurven f"ur longitudinale und transversale Polarisation unterscheiden sich qualitativ aufgrund der unterschiedliche Struktur der $2S$-Lichtkegelwellenfunk\-tion in Hinsicht auf ihre Knoten.
F"ur longitudinale Polarisation wird der "Uberlapp der Wellenfunktionen f"ur Photon und $2S$-Vektormeson repr"asentiert durch einen einzigen Term, der, und folglich der integrierte Wirkungsquerschnitt~$\si_{\iZS,L}(Q^2)$, f"ur~$Q^2 \!\cong\! 2.5\GeV^2$ eine reelle Nullstelle besitzt.
F"ur transversale Polarisation hingegen existieren zwei Terme: der relativistische mit Bahndrehimpuls~$l_3 \!=\! 1$ und antiparallelen Quark-Spins und der Term mit~$l_3 \!=\! 0$ und parallelen Quark-Spins.
Die Nullstellen der entsprechenden Polynome werden angenommen~-- wie bereits diskutiert in Zusammenhang mit den Abbildungen~\ref{Fig:overlap_2S}(b) und~\ref{Fig:Rcut_2S[1S]}(b)~-- f"ur unterschiedliche transversale Quark-Antiqurk-Separatioenn.
Entsprechend des "`Verwaschungseffektes"' bez"uglich der absoluten H"ohen der intermedi"aren Maxima in Abbildung~\ref{Fig:Rcut_2S[1S]} stellen wir hier in demselben Sinne fest, da"s~-- statt eines definitiven Nulldurchgangs f"ur einen festen Wert von~$Q^2$ f"ur longitudinale Polarisation~-- f"ur transversale Polarisation lediglich ein Minimum existiert f"ur~$Q^2 \!\cong\! 0.3\GeV^2$ und~-- statt eines ausgepr"agten Maximums~-- lediglich ein Plateau bei~$Q^2 \!\cong\! 1 \!-\! 1.1\GeV^2$.
Mit wachsendem Wert~$Q^2$ vermindern sich die absoluten Werte der Wirkungsquerschnitte~$\si_{\iZS,\la}(Q^2)$ f"ur longitudinale wie f"ur transversale Polarisation,~$\la \!\equiv\! L,T$, aufgrund der zunehmenden Unterdr"uckung des "Uberlapps der Wellenfunktionen durch die Funktionen~${\rm K}_0(\vep r)$,~${\rm K}_1(\vep r)$.
Die transversale Ausdehnung der effektiv beitragenden Dipole verschiebt sich zu immer kleineren Separationen~$r$, mehr und mehr zum inneren Teil des "Uberlapps, der negativ ist aufgrund des Vorzeichens der $2S$-Wellenfunktion.
Asymptotisch f"ur gro"se~$Q^2$, vgl.\@ die Gln.~(\ref{Q2-Asymptotik}),~(\ref{Q2-Asymptotik}$'$), dominiert der Wirkungsquerschnitt f"ur longitudinale Polarisation "uber den f"ur transversale Polarisation aufgrund eines zus"atzlichen Faktors~$Q^2$.
Wir stellen fest, da"s diese Asymptotik selbst f"ur~$Q^2 \!=\! 20\GeV^2$ noch nicht erreicht ist.
%
%
%
\begin{table}
\begin{minipage}{\linewidth}
\renewcommand{\thefootnote}{\thempfootnote}
\begin{center}
  \begin{tabular}{|g{7}||g{-1}|g{7}||g{-1}|g{8}||g{-1}|g{5}|} \hline
  \multicolumn{7}{|c|}{Photoproduktion,~$Q^2 \!\equiv\! 0$,~von~$\rh(770)$~--%
                       ~und~$\om(782)$,~$\ph(1020)$,~$\Jps(3097)$:} \\
  \multicolumn{7}{|c|}{Theorie versus Experiment}
  \\ \hhline{:=:t:==:t:==:t:==:}
  \multicolumn{1}{|c||}{}
    & \multicolumn{2}{c||}{$\si\![\ga p \!\to\! 1S\,p]$}
    & \multicolumn{2}{c||}{$d\si\![\ga p \!\to\! 1S\,p]/dt|_{\tfbQ\equiv0}$}
    & \multicolumn{2}{c|}{$B_0\![\ga p \!\to\! 1S\,p]$}
  \\
  \multicolumn{1}{|c||}{}
    & \multicolumn{2}{c||}{$[\microbarn[]]$}
    & \multicolumn{2}{c||}{$[\microbarn[]\GeV^{-2}]$}
    & \multicolumn{2}{c|}{$[\GeV[]^{-2}]$}
  \\[.25ex]
  \multicolumn{1}{|c||}{$1S$}
  & \multicolumn{1}{c|}{th.\footnote{
    \label{FN_0:Pomeron}Pomeron-Anteil.
  }}
    & \multicolumn{1}{c||}{exp.}
    & \multicolumn{1}{c|}{th.\citeFN{FN_0:Pomeron}}   
    & \multicolumn{1}{c||}{exp.}
    & \multicolumn{1}{c|}{th.\citeFN{FN_0:Pomeron}}   
    & \multicolumn{1}{c|}{exp.}
  \\ \hhline{|-||--||--||--|}
  \mbox{$\rh$},\mbox{$(770)$\footnote{
    \label{FN:1S-Parameter}Bzgl.\@ der Parameter~$\om_{\iES,T}(Q^2)$ und~${\cal N}_{\iES,T}(Q^2)$ f"ur~$Q^2 \!\equiv\! 0$ vgl.\@ unten Tabl.~\ref{Tabl:Photo1SParameter}.
    }}
    &  7.,88  &  9.4 ,\PM 1.1\mbox{\footnote{
    \label{FN:si_rh}Ref.~\cite{Aston82}: Gemittelt "uber Photon-Energien~\mbox{$E_\iga \!=\! 20 \!-\! 70\GeV$}.
    }}
    & 83.,3    & 84.4 ,\PM 27.\mbox{\footnote{
    \label{FN:a,b_rh}Ref.~\cite{Aston82}: Parameter~$a$,~$b$ der Anpassung\;~\mbox{$d\si\!/dt(\tfbQ) \!=\! a\exp\{-b\tfbQ\}$} im Bereich~\mbox{$\tfbQ \!=\! 0.06 \!-\! 0.3\GeV^2$}; gemittelt "uber Photon-Energien \mbox{$E_\iga \!=\! 20 \!-\! 30\GeV$}.
    }}
    & 13.,3    &  7.6 ,\PM 1.2\citeFN{FN:a,b_rh}
  \\ \hhline{|-||--||--||--|}
  \mbox{$\om$},\mbox{$(782)$\citeFN{FN:1S-Parameter}}
    &  0.,871\mbox{\footnote{
    \label{FN:ratio-om/rh}F"ur die Verh"altnisse von~$\rh(770)$- zu~$\om(782)$-Produktion erhalten wir beidesmal~$\cong\! 11.1\!/100$, vgl.\@ Gl.~(\ref{fV2ratio-om/rh}).
    }}
    &  1.2 ,\PM 0.3\mbox{\footnote{
    \label{FN:si_om}Ref.~\cite{Aston82}: Gemittelt "uber Photon-Energien~\mbox{$E_\iga \!=\! 20 \!-\! 45\GeV$}.
    }}
    &  9.,27\citeFN{FN:ratio-om/rh}
    &  9.4 ,\PM 1.3\mbox{\footnote{
    \label{FN:a,b_om}Ref.~\cite{Aston82}: Parameter~$a$,~$b$ der Anpassung\;~\mbox{$d\si\!/dt(\tfbQ) \!=\! a\exp\{-b\tfbQ \!+\! c(\tfbQ)^2\}$}, mit~\mbox{$c \!=\! (3.4 \!\pm\! 2.6)\GeV^{-4}$}, im Bereich~\mbox{$\tfbQ \!=\! 0.06 \!-\! 0.8\GeV^2$}; gemittelt "uber Photon-Energien \mbox{$E_\iga \!=\! 20 \!-\! 45\GeV$}.
    }}
    & 13.,4    &  8.3 ,\PM 1.3\citeFN{FN:a,b_om}
  \\
  \mbox{$\ph$},\mbox{$(1020)$\citeFN{FN:1S-Parameter}}
    &  0.,730  &  0.96 ,\PM 0.40\mbox{\footnote{
    \label{FN:si_ph}Ref.~\cite{Derrick96}: Integriert "uber den Bereich~\mbox{$\tfbQ \!<\! 0.5\GeV^2$}, durchschnittliche Energie~\mbox{$\surd s \!\cong\! 70\GeV$}.
    }}
    &  6.,69   &  7.2 ,\PM 3.9\mbox{\footnote{
    \label{FN:a,b_ph}Ref.~\cite{Derrick96}: Parameter~$a$,~$b$ der Anpassung\;~\mbox{$d\si\!/dt(\tfbQ) \!=\! a\exp\{-b\tfbQ\}$} im Bereich~\mbox{$\tfbQ \!=\! 0.1 \!-\! 0.5\GeV^2$}.
    }}
    & 11.,5    &  7.3 ,\PM 1.8\citeFN{FN:a,b_ph}
  \\ 
  \mbox{$\Jps$},\mbox{$(3097)$\citeFN{FN:1S-Parameter}}
    &  0.,0211  &  0.018 ,\PM 0.002\mbox{\footnote{
    \label{FN:si_Jps}Ref.~\cite{Binkley82} f"ur durchschnittliche Photon-Energie~\mbox{$\vev{E_\iga} \!\cong\! 150\GeV$}.   Konsistent mit Ref.~\cite{Arneodo94a}: Extrapolation nach~\mbox{$Q^2 \!\equiv\! 0$} durch Anpassen des Propagator-Terms~\mbox{$[1 \!+\! Q^2\!/\!M_0^2]^{-2}$} an die~$Q^2$-Abh"angigkeit, ergibt: \mbox{$\si\![\ga N \!\to\! \Jps N] \!=\! (18 \!\pm\! 3)\nbarn$},~\mbox{$M_0 \!=\! (3.2 \!\pm\! 0.6)\GeV$} f"ur~\mbox{$\vev{E_\iga} \!\cong\! 150\GeV$}.
    }}
    &  0.,150
    &  0.080 ,\PM 0.013\mbox{\footnote{
    \label{FN:a,b_Jps}Ref.~\cite{Binkley82}: Parameter~$a$,~$b$ der Anpassung\;~\mbox{$d\si\!/dt(\tfbQ) \!=\! a\exp\{-b\tfbQ \!+\! c(\tfbQ)^2\}$}, mit~\mbox{$c \!=\! (2.9 \!\pm\! 1.3)\GeV^{-4}$}, im Bereich~\mbox{$\tfbQ \!=\! 0.0 \!-\! 1.0\GeV^2$}; durchschnittliche Photon-Energie~\mbox{$\vev{E_\iga} \!=\! 150\GeV$}.
    }%
    $^{\!,}$\footnote{
    \label{FN:th-vs-exp_Jps}Extrapolation der Daten zu~$Q^2 \!\equiv\! 0$, vgl.\@ Abb.~\refg{Fig:dsigmadt_Jps}, l"a"st einen signifikant gr"o"seren experimentellen Wert f"ur~$d\si_\la\!/dt|_{\tfbQ\equiv0}$ erwarten, \vspace*{-1.125ex}der "ubereinstimmt mit unserem Postulat.
    }}
    &  9.,26    &  5.6 ,\PM 1.2\citeFN{FN:a,b_Jps}
  \\ \hhline{:=:b:==:b:==:b:==:}
  \end{tabular}
  \end{center}
\vspace*{-3.5ex}
\caption[Photoproduktion von~\protect$\rh(770)$~-- und~\protect\mbox{$\om(782)$,~$\ph(1020)$,~$\Jps(3097)$}: Theorie vs. Experiment]{
  Photoproduktion von~$\rh(770)$~-- und~$\om(782)$,~$\ph(1020)$,~$\Jps(3097)$, Theorie~\mbox{versus} Experiment.   Die experimentellen $\rh(770)$-Daten stammen von \FUNF{O}{M}{E}{G}{A}-\DREI[]{S}{P}{S}, \VIER[]{C}{E}{R}{N}, Ref.~\cite{Aston82}.   Der berechnete slope-Parameter~$B_0$ ist die logarithmische Steigung der Kurve~$d\si_\la\!/dt(\tfbQ)$ im Limes~$\tfbQ \!\to\! 0$, die der String-String-Mechanismus des \DREI{M}{S}{V} postuliert als stark konkav, das hei"st aufw"arts gekr"ummt, vgl.\@ etwa Abb.~\ref{Fig:dsdt_V,la}.   Der experimentelle Wert ist Resultat einer Zwei- oder Drei-Parameter-Anpassung f"ur einen endlichen, teils h"oheren Bereich von~$\tfbQ$.~Daher die Diskrepanz.   Ansonsten finden wir~-- auch f"ur das~$\ph(1020)$~-- gute "Ubereinstimmung.\zz
\vspace*{-.5ex}
}
\label{Tabl:Q=0_rh,om,ph,Jps}
\end{minipage}
\end{table}
\renewcommand{\thefootnote}{\thechapter.\arabic{footnote}}
\begin{table}[b]
\begin{minipage}{\linewidth}
\renewcommand{\thefootnote}{\thempfootnote}
\begin{center}
  \begin{tabular}{|h{-1}||g{-1}|g{-1}|g{-1}|g{-1}|} \hline
  \multicolumn{5}{|c|}{Parameter f"ur~$Q^2 \!\equiv\! 0$ der $1S$-/Grundzustand-Vektormesonen}
  \\ \hhline{:=:t:====:}
  & \mbox{$\rh(7$},\mbox{$70)$}
  & \mbox{$\om(7$},\mbox{$82)$}
  & \mbox{$\ph(1$},\mbox{$020)$}
  & \mbox{$\Jps(3$},\mbox{$097)$}
  \\ \hhline{|-||----|}
  \mbox{$\om_{V,T}$}/[\GeV[]]
    & 0.,213 & 0.,207 & 0.,260 & 0.,573 \\
  \mbox{${\cal N}_{V,T}$}/\mbox{$[\surd\Nc\!\equiv\!3]$}
    & 3.,44  & 3.,43  & 3.,21  & 2.,15  \\
  \hhline{:=:b:====:}
  \end{tabular}
\end{center}
\vspace*{-3.5ex}
\caption[\protect$1S$-/Grundzustand-Vektormesonen: Parameter f"ur~\protect$Q^2 \!\equiv\! 0$]{
  Die Zahlenwerte basieren auf effektiven Quarkmassen von~\mbox{$m_{u\!/\!d,0} \!=\! 0.220\GeV$}, \mbox{$m_{s,0} \!=\! 0.310\GeV$} f"ur \mbox{$Q^2 \!\equiv\! 0$} beziehungsweise der laufenden Charm-Masse~\mbox{$m_c \!=\! 1.3\GeV$}.~Bzgl. der Parameterwerte oberhalb der Schwellen~$Q_{f,0}^2$ vgl.\@ Tabl.~\refg{Tabl:Charakt_rh,om,ph,Jps}; die~$\Jps(3097)$-Parameter sind unabh"angig von~$Q^2$ und daher identisch mit den Werten dort.   Vgl.\@ Text.
}
\label{Tabl:Photo1SParameter}
\end{minipage}
\end{table} 
\renewcommand{\thefootnote}{\thechapter.\arabic{footnote}}
\\\indent
In Tabelle~\ref{Tabl:Q=0_rh,om,ph,Jps} ist konfrontiert mit dem Experiment unser Postulat f"ur Photoproduktion~$Q^2 \!\equiv\! 0$.
Dabei seien zun"achst ignoriert die Eintr"age f"ur die Grundzustand-Vektormesonen $\om(782)$,~$\ph(1020)$,~$\Jps(3097)$, die wir angeben f"ur Vollst"andigkeit in Nachtrag zum vorhergehenden Kapitel~\ref{Kap:GROUND}.
Wir finden gute "Ubereinstimmung, stellen aber fest, da"s insbesondere f"ur den in~$-t \!\cong\! \tfbQ$ differentiellen Wirkungsquerschnitt~$d\si_\la\!/dt(\tfbQ\zz\equiv\zz0)$ in Vorw"artsrichtung die experimentellen Zahlenwerte mit einer erheblichen Unsicherheit behaftet sind.
Wir erinnern daran, da"s ein entsprechender Beitrag des Reggeons von den experimentellenZahlenwerten zu subtrahieren ist.
Der von uns postulierte und angegebene slope-Parameter~$B_0$ ist definiert dessen Limes in Vorw"artsrichtung, der experimentelle Wert dagegen ist Resultat einer Anpassung f"ur einen endlichen Bereich von~$\tfbQ$~[f"ur~$\om(782)$ dar"uberhinaus nur einer von zwei angepa"sten Parametern].
Aufgrund der starken Konkavit"at des differentiellen Wirkungsquerschnitts~$d\si_\la\!/dt(\tfbQ)$ als Postulat des MSV, vgl.\@ etwa Abb.~\ref{Fig:dsdt_V,la}, ist die Diskrepanz erwartungsgem"a"s gro"s.
Wir merken hierzu weiter an, da"s ein starkes Ansteigen von~$d\si_\la\!/dt(\tfbQ)$ hin zu kleinen~$\tfbQ$ experimentell schwierig zu verifizieren ist. \\
\indent
Zur Vollst"andigkeit im Sinne eines Nachtrags zu Kapitel~\ref{Kap:GROUND} geben wir in Tabelle~\ref{Tabl:Q=0_rh,om,ph,Jps} Zahlenwerte an f"ur Photoproduktion der $1S$-Grundzustand-Vektormesonen~$\om(782)$,~$\ph(1020)$, $\Jps(3097)$.
Wir sind in der Lage die Analyse f"ur~$\om$ und~$\ph$ durchzuf"uhren auf Basis der universellen Lichtkegelwellenfunktion des Photons, wie diskutiert zu Beginn des Kapitels in~\ref{Sect:Photon-Wfn_Q2klgl2}, vgl.\@ die Gln.~(\ref{E:Photon-Wfn}),~(\ref{E:Photon-Wfn}$'$).
Dabei werden effektive Quarkmassen~$\meff[u\!/\!d,](Q^2)$ beziehungsweise~$\meff[s,](Q^2)$ zugrundegelegt wie diskutiert dort.
Wir rekapitulieren, da"s~$\meff(Q^2)$ im Bereich~$Q^2 \!=\! 0$ bis~$Q_{f,0}^2$ linear bez"uglich~$Q^2$ interpoliert zwischen der Konstituentenquark-Masse~$\meff(Q^2\zz\equiv\zz0) \!=\! m_{f,0}$ und der laufenden Quarkmasse~$\meff(Q^2\zz\ge\zz Q_{f,0}^2) \!=\! m_f$.
Explizit folgt: \mbox{$Q_{u\!/\!d,0}^2 \!=\! 1.05\GeV^2$}, \mbox{$m_{u\!/\!d,0} \!=\! 0.220\GeV$}, \mbox{$m_{u\!/\!d} \!=\! 0$} beziehungsweise \mbox{$Q_{s,0}^2 \!=\! 1.6\GeV^2$}, \mbox{$m_{s,0} \!=\! 0.310\GeV$}, \mbox{$m_s \!=\! 0.150\GeV$}.
Die Parameter~$\om_{\iES,T}$,~${\cal N}_{\iES,T}$ f"ur~$Q^2 \!\equiv\! 0$ f"ur~$\om(782)$ und~$\ph(1020)$ folgen damit in vollst"andiger Analogie zu der Fixierung der $\rh(770)$-Parameter; vgl.\@ Abschnitt~\ref{Subsect:2SParameter} auf Seite~\pageref{T:SchrittNull}, Schritt~Null.
Die numerischen Zahlenwerte sind zusammengefa"st in Tabelle~\ref{Tabl:Photo1SParameter}.
Dabei sind die~$\Jps(3097)$-Parameter unabh"angig von~$Q^2$ und identisch den Werten in Tabelle~\refg{Tabl:Charakt_rh,om,ph,Jps}; die Werte f"ur~$\rh(770)$ sind entnommen Tabelle~\ref{Tabl:Wfn-Parameter}.
Auf dieser Basis folgen die Streu-Observablen von Tabelle~\ref{Tabl:Q=0_rh,om,ph,Jps}.
\begin{figure}
\begin{minipage}{\linewidth}
  \begin{center}
  \setlength{\unitlength}{.9mm}\begin{picture}(120,69)   
    \put(0,0){\epsfxsize108mm \epsffile{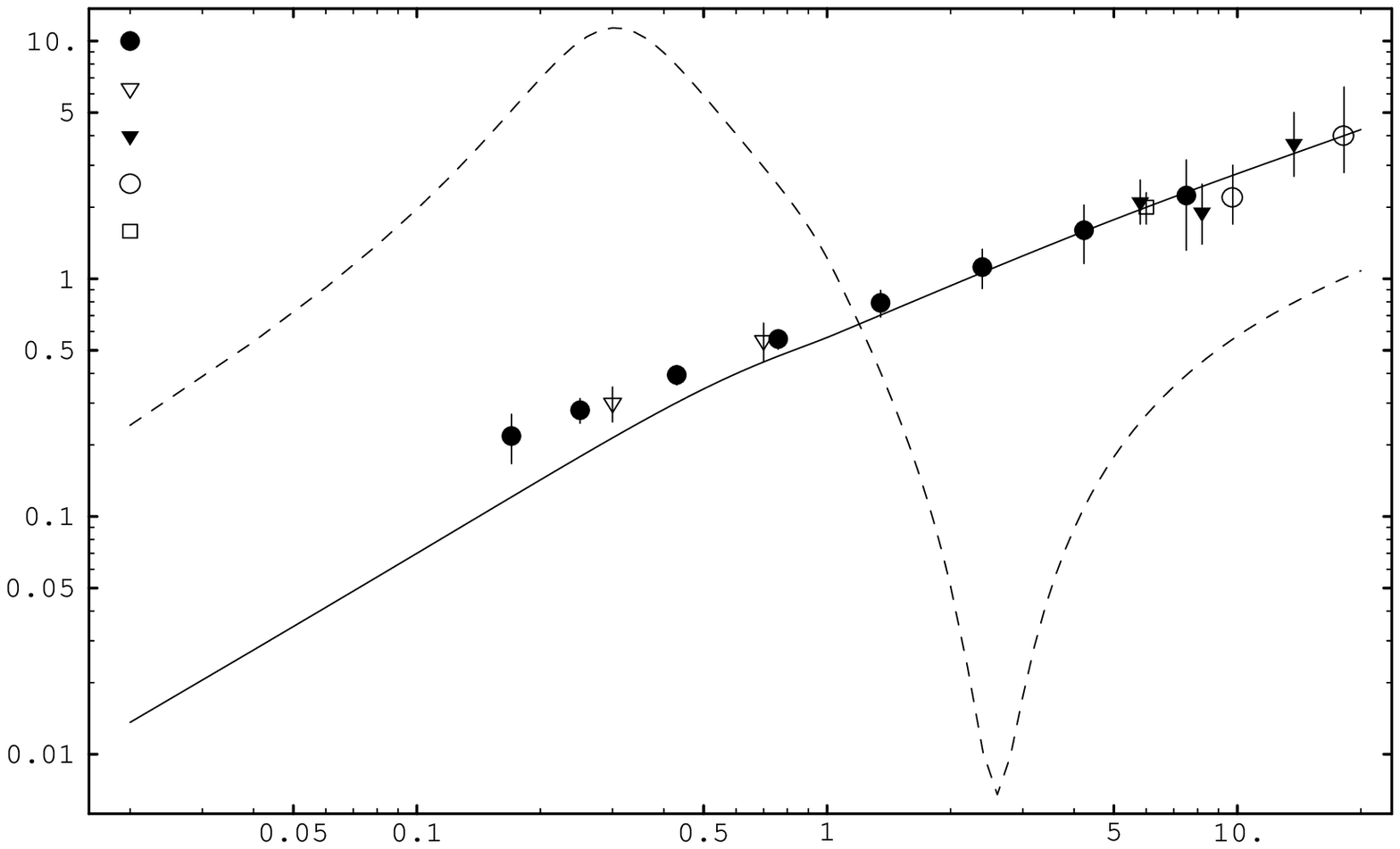}}
    \put(115, 0.5){\normalsize$Q^2\;[\GeV[]^2]$}
    \put( -6, 0  ){\yaxis[62.1mm]{\normalsize%
                                    $\RLT[V,](Q^2) \!=\! \si_{V,L}(Q^2)/\si_{V,T}(Q^2)$}}
    \put(11, 6){\normalsize$V \!\equiv\! \rh(770)$}
    \put(11,31){\normalsize$V \!\equiv\! 2S$}
    \put(13,65   ){\normalsize \VIER[]{E}{6}{6}{5}}
    \put(13,61   ){\normalsize \VIER{Z}{E}{U}{S} \DREI{B}{P}{C} 1995 prel.}
    \put(13,57   ){\normalsize \VIER{Z}{E}{U}{S} 1994 prel.}
    \put(13,53   ){\normalsize \ZWEI{H}{1} 1994 prel.}
    \put(13,49   ){\normalsize \DREI[]{N}{M}{C}}
  \end{picture}
  \end{center}
\vspace*{-5ex}
\caption[\protect$Q^2$-Abh"angigkeit von~\protect\mbox{$\RLT[V,] \!=\! \si_{V,L} \!/\! \si_{V,T}$} f"ur Produktion von~\protect$V \!\equiv\! \rh(770), 2S$]{
  Verh"altnis~$\RLT[V,]$ f"ur longitudinale zu transversale Polarisation als Funktion von~$Q^2$.   Durchgezogene Kurve f"ur~$V \!\equiv\! \rh(770)$-, gestrichelte f"ur~$2S$-Produktion.   Experimentelle Daten liegen vor nur f"ur~$\rh(770)$.   Vgl.\@ auch Abb.~\refg{Fig-G:R_LT}.
\vspace*{-.5ex}
}
\label{Fig:R_LT}
\end{minipage}
\end{figure}
\\\indent
In Abbildung~\ref{Fig:R_LT} ist dargestellt als Funktion von~$Q^2$ das Verh"altnis integrierter Wirkungsquerschnitte, longitudinale zu transversale Polarisation:
%
\begin{align} \label{E:R_LT}
\RLT[V,](Q^2)\; =\; \si_{V,L} / \si_{V,T} \qquad
  V \!\equiv\! \rh(770),2S
\end{align}
vgl.\@ Gl.~(\ref{R_LT}).
Zun"achst merken wir an, da"s sich aufgrund der Fixierung der Parameter des \DREI[]{M}{S}{V}, vgl.\@ Seite~\pageref{T:Parameter}, die postulierten Wirkungsquerschnitte beziehen auf eine invariante Schwerpunktenergie~$\surd s \!=\! 20\GeV$.
F"ur Verh"altnisse von Wirkungsquerschnitten erwarten wir globalere G"ultigkeit, da sich die~$s$-Abh"angigkeit tendenziell herausdividieren sollte.
Explizit f"ur~$\rh(770)$ beschreibt unser Postulat f"ur~$\RLT(Q^2)$~-- die durchgezogene Kurve~-- perfekt den beobachteten starken Anstieg mit~$Q^2$; dieser wird insbesondere beobachtet in Studien zu {\it colour transparency\/} in Kernen, vgl.\@ Ref.~\cite{Adams95}.
Der leichte Knick bei~$Q^2 \!=\! Q_0^2 \!=\! 1.05$ r"uhrt her von der effektiven Quarkmasse~$\meff$, die an dieser Stelle einsetzt abzuweichen von dem verschwindenden laufenden Wert~$m_{u\!/\!d}$.
Die Daten scheinen einen etwas konvexeren Verlauf zu suggerieren.
Ein solcher kann leicht erreicht werden durch eine sensiblere Interpolation zwischen Konstituenten- und laufender Quarkmasse als die zugrundegelegte $Q^2$-lineare Abh"angigkeit.
Wir verweisen auch auf Abbildung~\refg{Fig-G:R_LT} in Hinsicht der experimentellen Status Quo im Moment unserer Ver"offentlichungen Ref.~\cite{Dosch96} und~\cite{Kulzinger98}.
F"ur die Anregungen~$\rh(1450)$,~$\rh(1700)$ existieren keine experimentellen Daten.
Wir geben unser Postulat f"ur~$\RLT(Q^2)$ bez"uglich des $2S$-Zustandes an als gestrichelte Kurve in Abbildung~\ref{Fig:R_LT}.
Wir finden eine starke Abh"angigkeit von~$Q^2$, die sensitiver als die blo"sen Wirkungsquerschnitte in den Abbildungen~\ref{Fig:overlap_2S}-\ref{Fig:sigma_rh770,2S}, die Struktur der Knoten der $2S$-Wellenfunktion testet.
Die reelle Nullstelle des longitudinalen Wirkungsquerschnitts~$\si_{\iZS,L}$ f"ur~$Q^2 \!\cong\! 2.5\GeV^2$, vgl.\@ Abb.~\ref{Fig:sigma_rh770,2S}(b), "ubertr"agt sich unmittelbar auf~$\RLT[\iZS,]$.
Diese Sensitivit"at bezieht sich in erster Linie auf die Position des "`transversalen Knotens"' der $2S$-Lichtkegelwellenfunktion; signifikanter Unterschied bez"uglich der diskutierten verschiedenen Parameters"atze, vgl.\@ Tabl.~\ref{Tabl:Wfn-Parameter}, bezieht sich ausschlie"slich auf~$A_T$, das hei"st auf die Position des "`longitudinalen Knotens"'.
Dennoch erwarten wir~$\RLT(Q^2)$ als den besten Kandidaten, diese zu unterscheiden auf Niveau von Streuquerschnitten.
Wir weisen aber in diesem Zusammenhang darauf hin, da"s die $2S$-Lichtkegelwellenfunktion in Hinsicht auf die Struktur ihrer Knoten eher modelliert als konstruiert ist.
Wir betonen, da"s unser Postulat insbesondere der Gr"o"se~$\RLT[\iZS,]$ nicht zu interpretieren ist im Sinne eines definitiven Postulats der Position der Knoten.
Wir betonen vielmehr, da"s die explizite Struktur der Struktur wesentlichen Einflu"s nimmt auf Observablen der Streuung.
\begin{sidewaysfigure}
\begin{minipage}{\linewidth}
\setlength{\unitlength}{0.943396mm}   
\makebox(220,122.6){
  \begin{picture}(220,122.6)
  \put(3,61.4){   
    \makebox(       110, 61.2){
    \begin{picture}(110, 61.2)
      \put(4,0){\epsfxsize100mm \epsffile{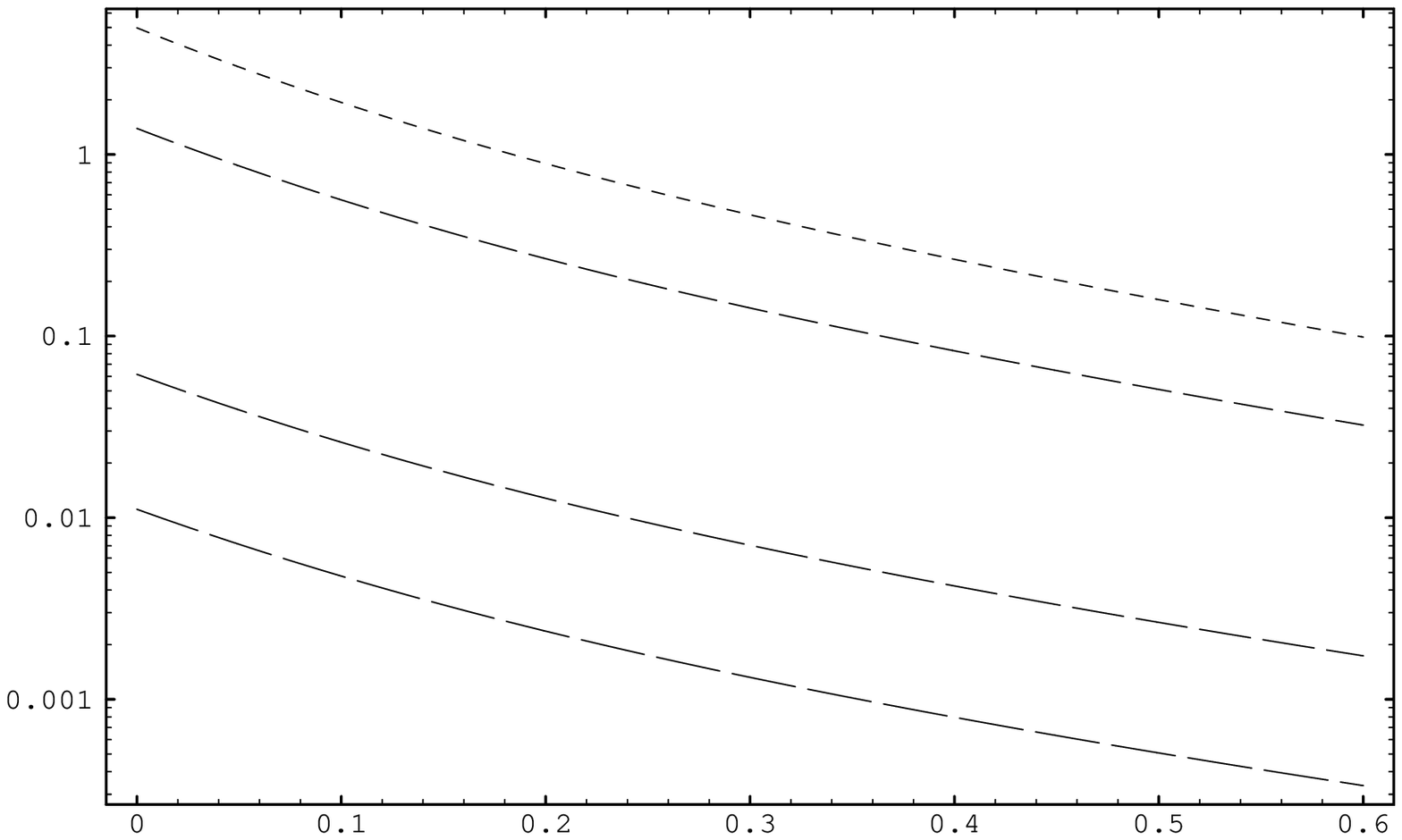}}
      \put(17,4.5){\normalsize (a)\vv $\rh(770)$,~$\la \!\equiv\! L$~-- Longitudinal}
      \put(0,0){\yaxis[57.7358mm]{\normalsize$
                                    d\si_{\irh,L}\!\big/dt\big|_{Q^2}(\tfbQ)%
                                    \vv[\microbarn[]\GeV^{-2}]$}}
      \put(60,46.5){\normalsize$Q^2 \!\equiv\! 0.25\GeV^2$}
      \put(17,42  ){\normalsize$Q^2 \!\equiv\! 2\GeV^2$}
      \put(60,22.2){\normalsize$Q^2 \!\equiv\! 10\GeV^2$}
      \put(20,13  ){\normalsize$Q^2 \!\equiv\! 20\GeV^2$}
    \end{picture}
    }}
    \put(120,61.4){   
    \makebox(       110, 61.2){
    \begin{picture}(110, 61.2)
      \put(4,0){\epsfxsize100mm \epsffile{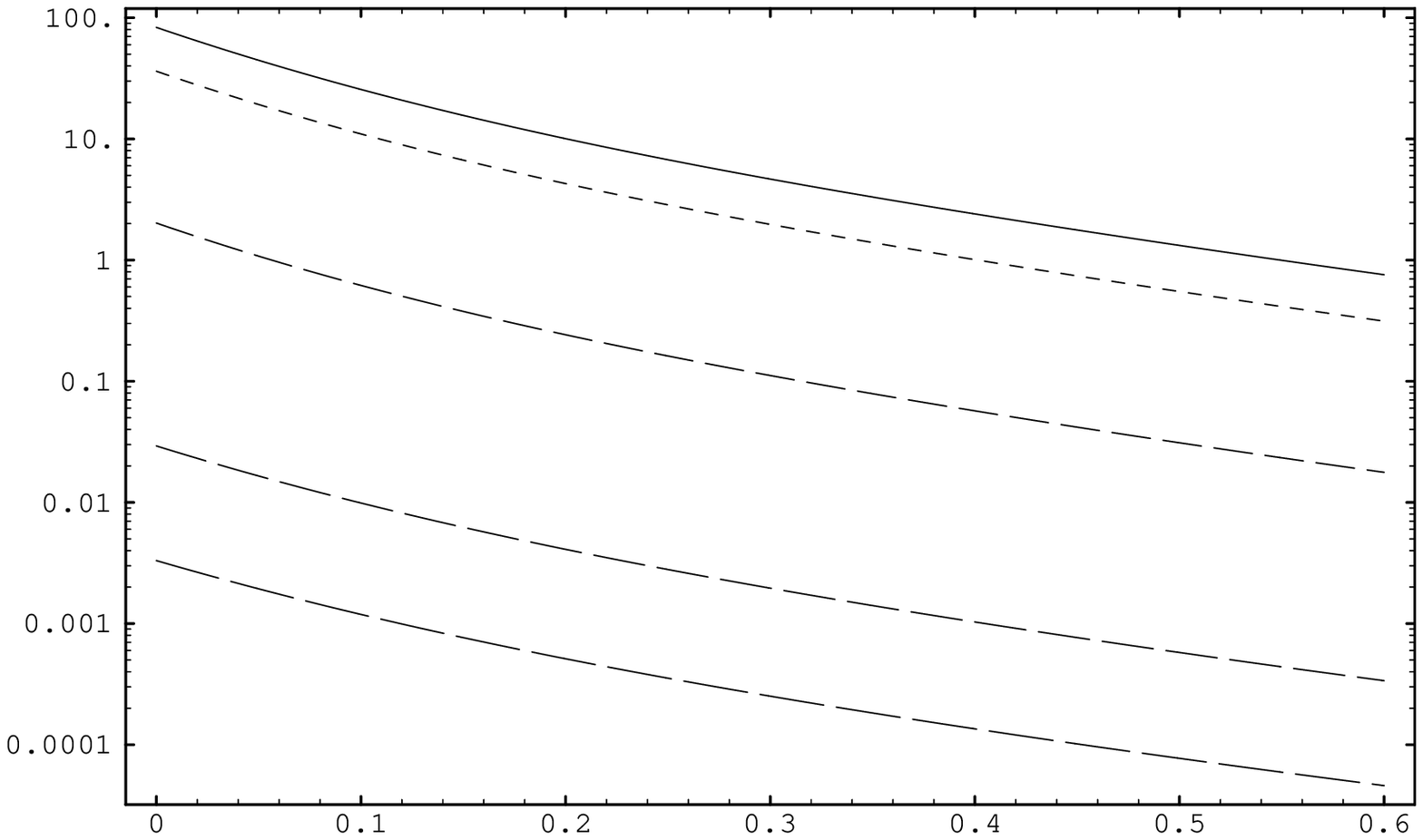}}
      \put(17,4.5){\normalsize (b)\vv $\rh(770)$,~$\la \!\equiv\! T$~-- Transversal}
      \put(0,0){\yaxis[61.2mm]{\normalsize$%
                                    d\si_{\irh,T}\!\big/dt\big|_{Q^2}(\tfbQ)%
                                    \vv[\microbarn[]\GeV^{-2}]$}}
      \put(60,49.5){\normalsize$Q^2 \!\equiv\! 0$}
      \put(60,38  ){\normalsize$Q^2 \!\equiv\! 0.25\GeV^2$}
      \put(20,35.5){\normalsize$Q^2 \!\equiv\! 2\GeV^2$}
      \put(60,19.5){\normalsize$Q^2 \!\equiv\! 10\GeV^2$}
      \put(20,11  ){\normalsize$Q^2 \!\equiv\! 20\GeV^2$}
    \end{picture}
    }}
  \put(3,0){   
    \makebox(       110, 61.4){
    \begin{picture}(110, 61.4)
      \put(4,0){\epsfxsize100mm \epsffile{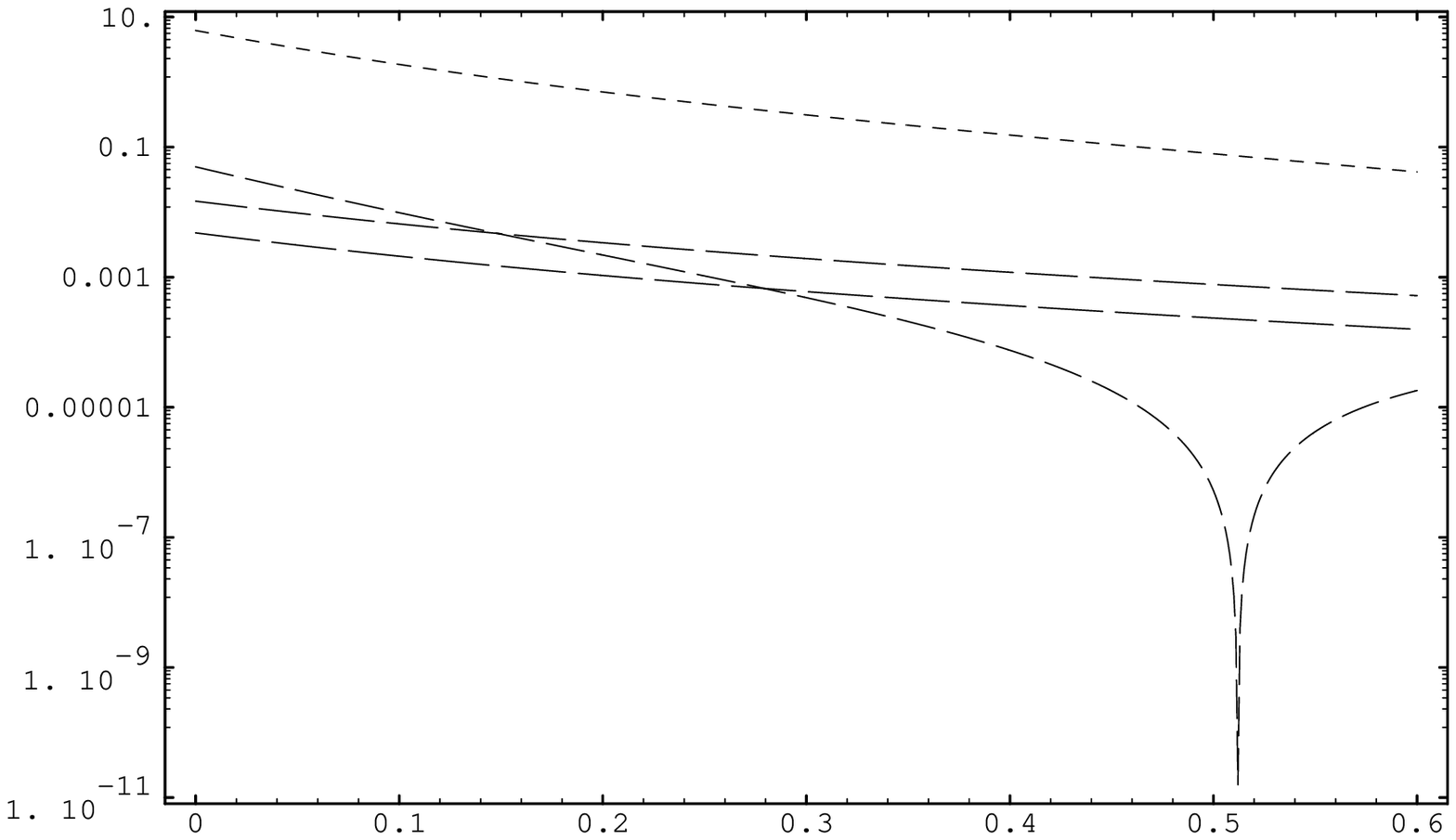}}
      \put(17,4.5){\normalsize (c)\vv $2S$,~$\la \!\equiv\! L$~-- Longitudinal}
      \put(94,-3.5){\normalsize$\tfbQ\;[\GeV[]^2]$}
      \put(0,0){\yaxis[57.9245mm]{\normalsize$
                                    d\si_{\iZS,L}\!\big/dt|_{Q^2}(\tfbQ)%
                                    \vv[\microbarn[]\GeV^{-2}]$}}
      \put(60,54  ){\normalsize$Q^2 \!\equiv\! 0.25\GeV^2$}
      \put(73, 6  ){\normalsize$Q^2 \!\equiv\! 2\GeV^2$}
      \put(60,43.5){\normalsize$Q^2 \!\equiv\! 10\GeV^2$}
      \put(20,38  ){\normalsize$Q^2 \!\equiv\! 20\GeV^2$}
    \end{picture}
    }}
  \put(120,0){   
    \makebox(       110, 61.4){
    \begin{picture}(110, 61.4)
      \put(4,0){\epsfxsize100mm \epsffile{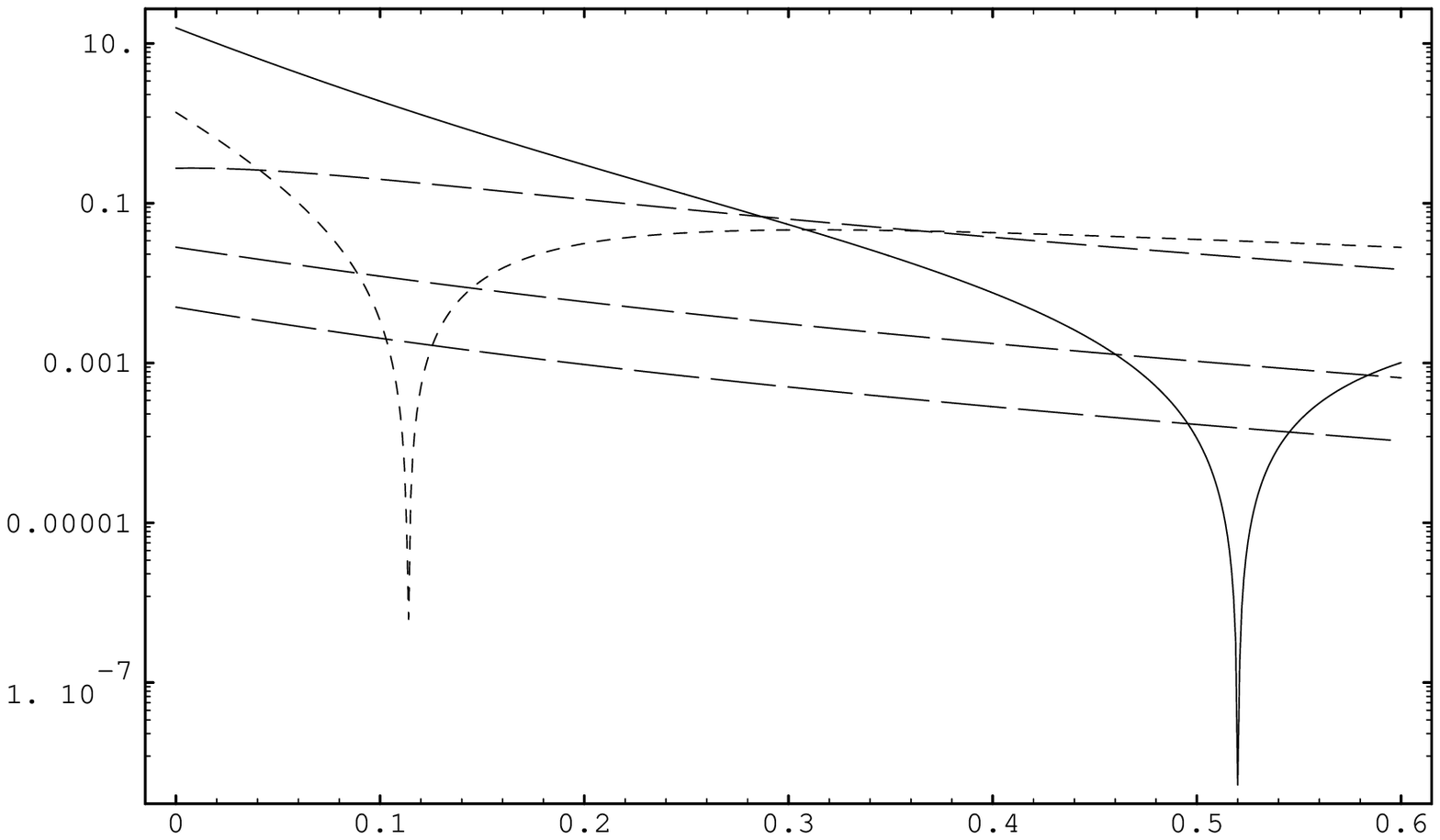}}
      \put(17,4.5){\normalsize (d)\vv $2S$,~$\la \!\equiv\! T$~-- Transversal}
      \put(94,-3.5){\normalsize$\tfbQ\;[\GeV[]^2]$}
      \put(0,0){\yaxis[61.4mm]{\normalsize$%
                                    d\si_{\iZS,T}\!\big/dt\big|_{Q^2}(\tfbQ)%
                                    \vv[\microbarn[]\GeV^{-2}]$}}
      \put(84, 6  ){\normalsize$Q^2 \!\equiv\! 0$}
      \put(37,15  ){\normalsize$Q^2 \!\equiv\! 0.25\GeV^2$}
      \put(90,38  ){\normalsize$Q^2 \!\equiv\! 2\GeV^2$}
      \put(46,38.5){\normalsize$\phantom{Q^2 \!\equiv\! \;} 10\GeV^2$}
      \put(46,30.7){\normalsize$Q^2 \!\equiv\! 20$}
    \end{picture}
    }}
  \end{picture}}
\vspace*{-.5ex}   
\caption[\protect$d\si_{V,\la}\!/dt|_{Q^2}(\tfbQ)$,\,~\protect$V \!\equiv\! \rh(770),2S$,~\protect$\la \!\equiv\! L,T$,\, f"ur~\protect$Q^2 \!\equiv\! 0,\,0.25,\,2,\,10,\,20\GeV^2$\!]{
  In~$\tfbQ \!\cong\! -t$ differentieller Wirkungsquerschnitt~$d\si_{V,\la}\!/dt|_{Q^2}(\tfbQ)$,\vv$V \!\equiv\! \rh(770),2S$,\vv$\la \!\equiv\! L,T$, als Funktion von~$\tfbQ$ f"ur~\mbox{Parameterwerte} \mbox{$Q^2 \!\equiv\! 0$} und~\mbox{$0.25,\,2,\,10,\,20\GeV^2$}~[durchgezogene Kurve bzw.\@ zunehmende Strichl"ange].   Aufgrund der Knoten verschwindet f"ur kleine~$Q^2$ die~Amp\-litude des $2S$-Zustands f"ur spezifische Werte von~$\tfbQ$; in~(d) beobachten wir, wie diese Werte nach innen wandern f"ur~$Q^2 \!\equiv\! 0,\,0.25,\,2\GeV^2$, wenn mehr und mehr kleine Dipole dominieren.   Bzgl.\@ der Kurven f"ur $\rh(770)$-Produktion vgl.\@ die Tabln.~\ref{Tabl:dsigmadtQ2_0}-\ref{Tabl:dsigmadtQ2_20T} in Anh.~\ref{APP:TABLES}
}
\label{Fig:dsdt_V,la}
\end{minipage}
\end{sidewaysfigure}
\\\indent
Dieselbe Bemerkung ist zu beziehen auch auf die in~$\tfbQ \!\cong -t$ differentiellen Streuquerschnitte~$d\si_{V,\la}\!/dt(\tfbQ)$ f"ur den $2S$-Zustand, wie angegeben graphisch in Abbildung~\ref{Fig:dsdt_V,la}(c),(d).
In Abbildung~\ref{Fig:dsdt_V,la} ist dargestellt f"ur eine Reihe von Parameterwerten~$Q^2$ unser Postulat f"ur die differentiellen Wirkungsquerschnitte~$d\si_{V,\la}\!/dt|_{Q^2}(\tfbQ)$; wir geben diese an separat f"ur~$V \!\equiv\! \rh(770),2S$ und~$\la \!\equiv\! L,T$.
Wir wenden uns zun"achst den Gr"o"sen f"ur~$\rh(770)$-Produktion zu: Abbildung~\ref{Fig:dsdt_V,la}(a),(b).
In Anhang~\ref{APP:TABLES} geben wir ohne weiteren Kommentar an als die Tabellen~\ref{Tabl:dsigmadtQ2_0}-\ref{Tabl:dsigmadtQ2_20T} Ausz"uge aus Zahlenlisten, die den Abbildungen f"ur~$\rh(770)$ zugrundeliegen.
Wir finden global, da"s der Verlauf der Kurven abweicht von einem rein exponentiellen Abfallverhalten.
Diese Konkavit"at oder Aufw"arts-Kr"ummung ist besonders ausgepr"agt f"ur kleine Werte von~$\tfbQ$ und spiegelt wider die gro"sen Werte des postulierten slope-Parameters~$B_0$, der definiert ist~-- vgl.\@ Gl.~(\ref{slope-Parameter_Thh})~-- als die logarithmische Steigung im Limes verschwindnden Impulstransfers~$\tfbQ \!\to\! 0$; vgl.\@ die Diskussion in Zusammenhang mit Tabelle~\ref{Tabl:Q=0_rh,om,ph,Jps}.
Bzgl.\@ der Bedeutung von~$B_0$ im \DREI{M}{S}{V} sei verwiesen auf Seite~\pageref{T:slope}, die Diskussion von Gl.~(\ref{slope_R1,R2}). \\
\indent
F"ur~$\rh(770)$-Produktion verlaufen die Kurven f"ur verschiedene Werte von~$Q^2$ im wesentlichen parallel zueinander, vgl.\@ Abb.~\ref{Fig:dsdt_V,la}(a),~(b).
Der Vergleich der Kurven f"ur unterschiedliche Polarisalion, vgl.\@ Abb.~\ref{Fig:dsdt_V,la}(a) in Kontrast zu~(b), zeigt st"arkeren Abfall mit~$\tfbQ$ f"ur transversale als f"ur longitudinale Polarisation~[sechs Dekaden gegen"uber vier].
Dis ist bereits dokumentiert in Abbildung~\refg{Fig:dsigmadt-LT} und erkl"art dort in Termen der effektiven slope-Parameter, die unmittelbar abh"angen von den transversalen Ausdehnungen der effektiv beitragenden Dipolen~-- die aufgrund des~$\zet$-Endpiunkte-Verhaltens tendenziell gr"o"ser sind f"ur transversale als f"ur longitudinale Polarisation. \\
\enlargethispage{1.25ex}
\indent
F"ur die Produktion des $2S$-Zustands, vgl.\@ die Abb.~\ref{Fig:dsdt_V,la}(c),~(d), finden wir generell dasselbe Verhalten, doch ist dieses "uberlagert f"ur kleine Werte von~$Q^2$ durch den Einflu"s der Knoten der $2S$-Wellenfunktion, die f"uhren zu signifikanten Minima und Nullstellen bez"uglich~$\tfbQ$.
Die Werte~$\tfbQ$, an denen diese Minima und Nullstellen~-- "`Dips"'~-- der Amplitude auftreten, wandern nach innen f"ur wachsendes~$Q^2$ und dokumentieren so, da"s Dominanz hin zu immer kleineren Dipolen verschoben wird.
Wie bereits angedeutet h"angt die explizite Position dieser "`Dips"' in hohem Ma"se ab von der Parametrisierung der~\mbox{$2S$-Wel}\-lenfunktion in Hinsicht auf die explizite Position ihrer Knoten.
Wir betrachten daher nicht deren genaue Position wohl aber deren Existenz "uberhaupt als definitives Postulat.
Experimentell ist der starke Abfall in der N"ahe dieser "`Dips"' sicher ein leicht zu detektierendes Signal.
Wir erwarten allerdings eine Auswaschung des Effektes, wenn statt der Wirkungsquerschnitte~\mbox{$d\si_{\iZS,\la}\!/dt$},~$\la \!\equiv\! L,T$, f"ur longitudinale und transversale Polarisation separat deren Superposition~\mbox{$\ep\,d\si_{\iZS,L}\!/dt + d\si_{\iZS,T}\!/dt$} gemessen wird.
Dies erschwert die Detektion nicht unerheblich.
Doch k"onnte aus einer experimentellen Bestimmung der Positionen der "`Dips"' wichtige Information extrahiert werden f"ur die theoretische Konstruktion der $2S$-Lichtkegelwellenfunktion in Hinsicht ihrer Knoten.
\vspace*{-1ex}

\bigskip\noindent
Wir wenden uns erneut zu den~$\pi^+\pi^-$- und~$2\pi^+2\pi^-$-Massespektren f"ur Elektron-Positron-Annihilation und f"ur Photoproduktion elastisch am Proton.
Wir richten besonderen Augenmerk auf einen Bereich der invarianten Masse~$M$ dieser Pion-Endzust"ande von~\mbox{$1 \!-\! 2\GeV$}.
Dieser Bereich ist wesentlich dominiert durch die Interferenz der Resonanzen~\mbox{$\rh'$~[$\equiv\! \rh(1450)$]} und~\mbox{$\rh^\dbprime$~[$\equiv\! \rh(1700)$]} und die beobachteten unterschiedlichen Muster dieser Interferenz~-- destruktiv f"ur Annihilation, konstruktiv f"ur Photoproduktion~-- Charakteristikum von ganz herausragender Bedeutung.
Wir rekapitulieren die groben Z"uge der Diskussion auf Seite~\pageref{si_ll-to-f} in Zusammenhang mit den Gln.~(\ref{si_ll-to-f}),~(\ref{si_gap-to-fp}). \\
\enlargethispage{.625ex}
\indent
In unserer Konvention ist f"ur verschwindende transversale Quark-Antiquark-Se\-paration die $1S$-Wellenfunktion {\it positiv\/}, dagegen die $2S$-Wellenfunktion {\it negativ\/}~\footnote{
  \vspace*{-.5ex}Im Sinne der allgemein "ublichen Konvention, die  Wellenfunktion schmiege sich im Limes gro"ser Separa\-tionen~$|\mskip-.5mu\vec{r}\mskip-.5mu|$ von oben an die $|\mskip-.5mu\vec{r}\mskip-.5mu|$-Achse an, vgl.\@ etwa Godfrey, Isgur in Ref.~\cite{Godfrey85}
}.
Das experimentell beobachtete destruktive Interferenzmuster im $\pi^+\pi^-$-Massespektrum f"ur~$e^+e^-$-Annihila\-tion~-- vgl.\@ Abb.~\ref{Fig:eebar2pis}~-- legt damit aufgrund unseres Ansatzes in Form der Gln.~(\ref{Ansatz})-(\ref{Ansatz}$''$) den Mischungswinkel~$\Th$ fest auf den {\it ersten Quadranten}.
Die $T$-Amplituden von~$\rh'$ und~$\rh^\dbprime$ folgen aus~\mbox{$T_\la\![\ga^{\scriptscriptstyle({\D\ast})}p \!\to\! 2S\,p]$}, vgl.\@ Gl.~(\ref{T_Lepto}), durch Multiplikation von\;~$(\cos\Th)$ beziehungsweise\;~$(-\sin\Th)$ und besitzen entgegengesetztes Vorzeichen.
Und folglich sind die (relativen) Vorzeichen der $T$-Amplituden~\mbox{$T_\la\![\ga^{\scriptscriptstyle({\D\ast})}p \!\to\! Vp]$} f"ur~\mbox{$V \!\equiv\! \rh,\rh',\rh^\dbprime$} im Limes verschwindender Quark-Antiquark-Separationen~$r$ bestimmt zu~$[+,-,+]$.
Lepton-Antilepton-Annihilation ist perturbativ dominiert, durch kleine Quark-Antiquark-Separationen: die Kopplungen~$f_V$ an den elektromagnetischen Strom in erster Ordnung konventioneller St"orungstheorie ist proportional der Wellenfunktion von~$V$ am Ursprung.
Folglich ist~$[+,-,+]$ genau auch das Vorzeichenmuster der Kopplungen~$f_V$ f"ur~$V \!\equiv\! \rh,\rh',\rh^\dbprime$.
Dieses Muster liegt zugrunde der experimentell beobachteten {\it destruktiven\/} Interferenz wenig unterhalb~$M \!\cong\! 1.6\GeV$ im $\pi^+\pi^-$-Massespektrum f"ur~$e^+e^-$-Annihilation.
Vgl.\@ Abb.~\ref{Fig:eebar2pis}.
Die experimentellen Daten dort stammen von den \DREI[]{D}{M}{1}- und \DREI[]{D}{M}{2}-Detektor, \FUNF[]{O}{R}{S}{A}{Y}, und den Detektoren \VIER{O}{L}{Y}{A} und \DREI{C}{M}{D} am \VIER{V}{E}{P}{P}-\ZWEI{2}{M}-Beschleuniger in Nowosibirsk, vgl.\@ Ref.~\cite{Quenzer78,Bisello85} bzw.~\cite{Barkov85}.
Die zwei in Abbildung~\ref{Fig:eebar2pis} angegebenen Kurven sind zum einen~-- durchgezogen~-- die Anpassung von Donnachie, Mirzaie aus Ref.~\cite{Donnachie87a}, der auch entnommen sind die experimentellen Daten, und~-- gestrichelt~-- unser Resultat f"ur das $\pi^+\pi^-$-Spektrum. \\
\indent
Diesem liegt zugrunde~-- wir formulieren unser generelles Konzept f"ur die Berechnung der Massespektren~-- die Parametrisierung der Resonanzen~$\rh$,~$\rh'$,~$\rh^\dbprime$ wie diskutiert und deren Verteilung entsprechend einfacher Breit-Wigner-Distributionen.
Unser Analyse geschieht dahingehend, zu einem allgemeinen Verst"andnis der Ph"anomenologie zugelangen.
Nicht durch subtile Ans"atze und Feinadjustierung bestimmter Parameter in Hinsicht auf globale "Ubereinstimmung ein Verst"andnis zu suggerieren, das wir weder angesichts des status quo von Theorie noch von Experiment besitzen. \\
\indent
So weichen wir allein hier im Fall des $\pi^+\pi^-$-Massespektrums f"ur Elektron-Positron-Annhilation und nur in dem einen Punkt ab, da"s wir die totale Zerfallspreite~$\Gatot_\irh$ des~$\rh(770)$ in allgemein "ublicher Weise ersetzen durch eine Parametrisierung durch ein Polynom zweiten Grades in~$M^2\!/M_\irh^2$.
Sei vewiesen auf Gl.~(\ref{si_ll-to-f}) bzw.\@ auf Anh.~\ref{APPSubsect:Photo,l+l-Annih} bzgl.\@ aller formaler Details.
Wir finden, vgl.\@ Abb.~\ref{Fig:eebar2pis}, da"s wir die beobachtete Interferenz gut reproduzieren auf Basis des Mischungswinkels~$\Th \!\cong\! 41.2$, vgl.\@ Gl.~(\ref{Th-explizit}).
Dieser ist Resultat einer einfachen Rechnung, vgl.\@ Gl.~(\ref{Th}) und Anh.~\ref{APPSubsect:Mischungswinkel}, die benutzt die Zahlenwerte f"ur die Verzweigungsverh"altnisse~$B_{V\to f}$ [in Gestalt der Gr"o"sen~$X_{V\!,i}$], wie sie angegeben werden von Donnachie, Mirzaie in Ref.~\cite{Donnachie87a} und zusammengestellt sind in Tabelle~\ref{Tabl:Charakt_rh,rh',rh''}.
Wir betonen, da"s wir den resultierenden Wert von~$\Th$ in keiner Weise modifizieren oder anpassen.
Das Bild zu dem wir gelangen in Gestalt der gestrichelten Kurve ist konsistent mit der subtileren Anpassung  von Donnachie, Mirzaie in Ref.~\cite{Donnachie87a}~[durchgezogene Linie] und den experimentellen Datenpunkten~-- im Rahmen deren Unsicherheit und in Anbetracht der Details unserer Kurve.
\begin{figure}
\begin{minipage}{\linewidth}
  \begin{center}
  \setlength{\unitlength}{.9mm}\begin{picture}(120,71.8)   
    \put(0,0){\epsfxsize108mm \epsffile{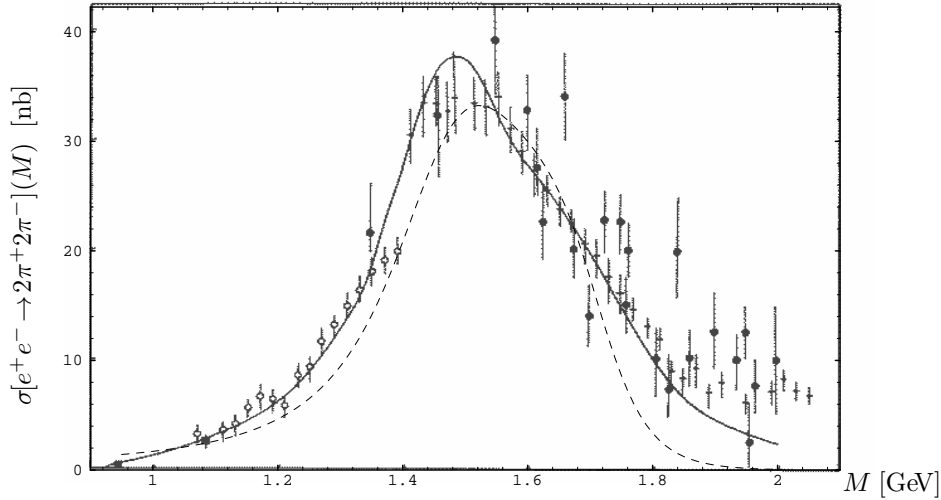}}
    \put(116,-0.5){\normalsize$M\;[\GeV[]]$}
    \put( -7, 0  ){\yaxis[64.62mm]{\normalsize%
                                    $\si\mskip-2mu[e^+e^- \!\to\! 2\pi^+2\pi^-](M)\vv%
                                     [\nbarn[]]$}}
  \end{picture}
  \end{center}
\vspace*{-4.25ex}
\caption[Massespektrum f"ur~\protect$e^+e^-$-Annihilation in~\protect$2\pi^+2\pi^-$]{
  Massespektrum f"ur~\protect$e^+e^-$-Annihilation in~\protect$2\pi^+2\pi^-$.   Die durchgezogene~Kur\-ve ist die Anpassung von Donnachie, Mirzaie, vgl.\@ Ref.~\cite{Donnachie87a}.   Die gestrichelte Kurve ist~Resul\-tat unserer Parametrisierung, vgl.\@ die Gln.~(\ref{si_ll-to-f}),~(\ref{si_gap-to-fp}) bzw.\@ Anh.~\ref{APPSubsect:Photo,l+l-Annih} und Tabl.~\ref{Tabl:Charakt_rh,rh',rh''}.
\vspace*{-.5ex}
}
\label{Fig:eebar4pis}
\end{minipage}
\end{figure}
%
%
Unser Resultat f"ur ~$2\pi^+2\pi^-$-Massespektrum f"ur Elektron-Positron-Annihilation wie angegeben als gestrichelte Kurve in Abbildung~\ref{Fig:eebar4pis} ist zu verstehen in demselben Sinne.
Wir finden in demselben Bereich leicht unterhalb von~$M \!\cong\! 1.6\GeV$ ein Maximum von etwa~$35\nbarn$ und insgesamt wieder ein Bild konsistent mit der subtileren Anpassung von Donnachie, Mirzaie in Ref.~\cite{Donnachie87a}~[durchgezogene Linie] und den experimentellen Datenpunkten, die wieder entnommen sind dieser Referenz und wieder stammen aus Novosibirsk und \FUNF[]{O}{R}{S}{A}{Y}, vgl.\@ Ref.~\cite{Kurdadze81} bzw.~\cite{Cordelier82}.  \\
\indent
F"ur Photoproduktion des $2S$-Zustands~-- vgl.\@ die Abbn.~\ref{Fig:overlap_2S}(b),~\ref{Fig:Rcut_2S[1S]}(b), die Kurven in kurzen Strichen~-- dominieren die Dipole gro"ser transversaler Ausdehnung rechts des Nulldurchgangs bei~$r \!\cong\! 1.1\fm$ "uber die links davon.
Dominanz der Physik gro"ser Abst"ande: der String-String-Mechanismus des~\DREI[]{M}{S}{V}, dreht das Vorzeichen von~\mbox{$T_\la\![\ga^{\scriptscriptstyle({\D\ast})}p \!\to\! 2S\,p]$} um und folglich die Vorzeichen der Amplituden f"ur~$\rh'$ und~$\rh^\dbprime$.
Es folgen die (relativen) Vorzeichen~$[+,+,-]$ der $T$-Amplituden~\mbox{$T_\la\![\ga^{\scriptscriptstyle({\D\ast})}p \!\to\! Vp]$} f"ur~$V \!\equiv\! \rh,\rh',\rh^\dbprime$ im Bereich gro"ser Separationen~$r$, durch den Photoproduktion dominiert ist.
Dies f"uhrt auf die experimentell beobachtete {\it konstruktive\/} Interferenz im Bereich~$M \!\cong\! 1.6\GeV$ des $\pi^+\pi^-$-Massespektrums f"ur Photoproduktion; vgl.\@ Abb.~\ref{Fig:eebar2pis}.
In Gestalt der durchgezogenen Kurve dort finden wir gute Reproduktion dieser Interferenz.
Dabei leigt unserer Berechnung zugrunde~-- auf Basis desselben Zahlenwerts f"ur den Mischungswinkel:~$\Th \!\cong\! 41.2$, vgl.\@ Gl.~(\ref{Th-explizit}),~-- die Parametrisierung der Resonanzen~$\rh$,~$\rh'$,~$\rh^\dbprime$ wie diskutiert, vgl.\@ die Gln.~(\ref{Ansatz})-(\ref{Ansatz}$''$), und deren Verteilung durch einfache unmodifiziert belassene Breit-Wigner-Distributionen.
Im Sinne des ausgef"uhrten allgemeinen Konzepts einfacher Ans"atze verzichten wir darauf insbesondere Modifikationen dieser Distributionen miteinzubeziehen, die herr"uhren von den Effekten nach Ross, Stodolsky oder Drell, S"oding, vgl.\@ Ref.~\cite{Ross66} bzw.~\cite{Drell60,Soeding66}.
Sei verwiesen auf Gl.~(\ref{si_gap-to-fp}) bzw.\@ auf Anh.~\ref{APPSubsect:Photo,l+l-Annih} bzgl.\@ aller formaler Details.
Angesichts dieses Hintergrunds ist die "Ubereinstimmung unseres Postulats mit den experimentellen Datenpunkten in Abbildung~\ref{Fig:eebar2pis} in hohem Ma"se nichttrivial.
Wir beobachten eine Anhebung des Spektrums an etwa derselben Stelle, wie sie indiziert wird von den experimentellen Daten von~\FUNF[]{O}{M}{E}{G}{A}-\DREI[]{S}{P}{S}, \VIER[]{C}{E}{R}{N}, vgl.\@ Ref.~\cite{Aston80}.
Wir merken an, da"s f"ur~$\pi^+\pi^-$-Photoproduktion {\it Vorab-Daten\/} von Fermilab-\VIER{E}{6}{8}{7} vorgestellt sind, vgl.\@ Ref.~\cite{Lebrun97}, die eine Genauigkeit erreichen, die scheinen, die Produktion des $3^{--}$-Vektormesons~$\rh_3(1690)$, das hei"st des hypothetischen $3S$-Zustands dokumentieren zu k"onnen.
F"ur Vergleichbarkeit auf einem solchen Niveau ist aber sicher unser Konzept zu modifizieren und unser Ansatz zu verfeinern.
Dies zeigt sich bereits in unserem Resultat f"ur das $2\pi^+2\pi^-$-Massespektrum f"ur Photoproduktion elastisch am Proton.
%
%
So finden wir in Abbildung~\ref{Fig:eebar4pis}, da"s die von uns postulierte durchgezogene Kurve um fast einen Faktor Zwei "uber den experimentell angegebenen Datenpunkten liegt, die stammen vom \FUNF[]{O}{M}{E}{G}{A}-Spektrometer am \DREI[]{S}{P}{S}, \VIER[]{C}{E}{R}{N}, vgl.\@ Ref.~\cite{Aston81}.
Dabei ist anzumerken, da"s es experimentell eine nicht zu untersch"atzende Herausforderung ist, den vorhandenen Untergrund in diesem Bereich von~$M$ zu subtrahieren und so die resonanten Beitr"age zu identifizieren.
Auch dies k"onnte mit Grund sein f"ur die dokumentierte Diskrepanz.
\vspace*{-1ex}

\bigskip\noindent
In dieser Hinsicht sicherer ist sich zu konzentrieren weniger auf absolute Zahlenwerte als auf die relative Variation~$Q^2$ und auf Verh"altnisse von Wirkungsquerschnitten.
So sind die Minima~-- oder "`Dips"'~-- in den Abbildungen~\ref{Fig:sigma_rh770,2S}(b),~\ref{Fig:R_LT} und~\ref{Fig:dsdt_V,la}(c),(d) klares Postulat unseres Zugangs auf Basis des \DREI[]{M}{S}{V}, wenn auch ihre exakte Position wesentlich abh"angt von den Details der $2S$-Wellenfunktion in Hinblick auf der position ihrer Knoten.
\begin{figure}
\begin{minipage}{\linewidth}
  \begin{center}
  \setlength{\unitlength}{.9mm}\begin{picture}(120,70.8)   
    \put(0,0){\epsfxsize108mm \epsffile{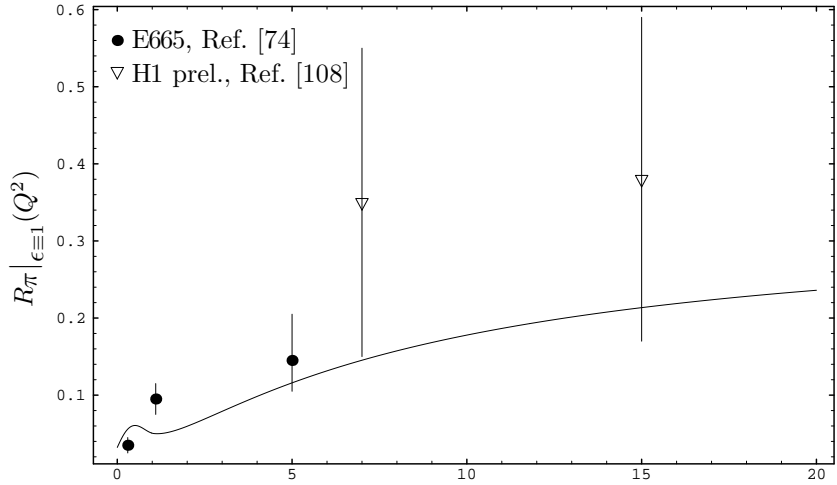}}

    \put( -7, 0  ){\yaxis[64.62mm]{\normalsize$\Rpi\big|_{{\D\ep}\equiv1}(Q^2)$}}
    \put(11,64){\normalsize\VIER[]{E}{6}{6}{5}, Ref.~\cite{Salgado95}}
    \put(11,59){\normalsize\ZWEI{H}{1} prel., Ref.~\cite{H196}}
  \end{picture}
  \end{center}
\vspace*{-4.25ex}
\caption[Verh"altnis~\protect$\Rpi(Q^2)$:~\protect$2\pi^+2\pi^-$-Photoproduktion via~\protect$\rh',\rh^\dbprime$ zu\;~\protect$\pi^+\pi^-$ via~\protect$\rh$]{
  Als Funktion von~$Q^2$ das Verh"altnis~$\Rpi$, das definiert ist als integrierter Wirkungsquerschnitt f"ur $2\pi^+2\pi^-$-Photoproduktion via~$\rh'$,~$\rh^\dbprime$ dividiert durch den Querschnitt f"ur~$\pi^+\pi^-$-Produktion via~\protect$\rh$, vgl.\@ die Gln.~(\ref{R_pi}),~(\ref{R_pi}$'$) und~(\ref{R_pi-siLT}),~(\ref{R_pi-siLT}$'$).   Unser Postulat bezieht sich auf die Rate~$\ep \!\equiv\! 1$ longitudinaler Photonen.   Die experimentellen Datenpunkte stammen von Fermilab-\VIER{E}{6}{6}{5} und \ZWEI[]{H}{1}, vgl.\@ Ref.~\cite{Salgado95} bzw.~\cite{H196}.
\vspace*{-.5ex}
}
\label{Fig:R_pi}
\end{minipage}
\end{figure}
\\\indent
\vspace*{-.25ex}Die Gr"o"se~$\RLT[V,](Q^2)$, vgl.\@ Gl.~(\ref{E:R_LT}), setzt in Beziehung die integrierten Wirkungsquerschnitte f"ur longitudinale und transversale Polarisation.
Analog setzt in Beziehung die integrierten Wirkungsquerschnitte von~$1S$- und $2S$-Zustand die Gr"o"se~\vspace*{-0.375ex}$\Rpi\big|_{\D\ep}(Q^2)$.
Diese ist definiert als das Verh"altnis der integrierten Wirkungsquerschnitte f"ur Photo- und Leptoproduktion von $2\pi^+2\pi^-$ via~\mbox{$\rh',\rh^\dbprime$~[$\equiv\! \rh(1450),\rh(1700)$]} zu~$\pi^+\pi^-$ via~$\rh$~[$\equiv\! \rh(770)$].
Wir definieren formal~\footnote{
  Seien Argumente und Indizes unterdr"uckt, wenn diese folgen aus dem Zusammenhang.
}:
%
\vspace*{-.5ex}
\begin{alignat}{2} \label{R_pi}
&\Rpi\big|_{\D\ep}(Q^2)&\vv
  =\vv &\si^{\{\rh',\rh^\dbprime\}}_{f\equiv2\pi^+2\pi^-}\;
      \Big/\;
      \si^{\{\rh\}}_{f\equiv\pi^+\pi^-}
    \\[1ex]
&&&\text{mit}\qquad
  \si^{\{V\}}_f\big|_{\D\ep}(Q^2)\vv
    =\vv \ep\, \si^{\{V\}}_{f,L}\;
           +\; \si^{\{V\}}_{f,T}
    \tag{\ref{R_pi}$'$}
    \\[-4.5ex]\nn
\end{alignat}
Dabei ist, mit~$d\si\!/dM^2 \!=\! 1\!/(2M)\,(d\si\!/dM)$:
\vspace*{-.5ex}
\begin{align} \label{R_pi-siLT}
&\si^{\{V\}}_{f,\la}(Q^2)\;
  =\; \int_{s_f}^\infty dM^2\vv
        \frac{1}{2M}
        \frac{d\si^{\{V\}}_{f,\la}}{dM}\Big|_{Q^2}
    \\[-.5ex]
&\frac{1}{2M}\frac{d\si^{\{V\}}_{f\!,\la}}{dM}\Big|_{Q^2}(M)\;
    \tag{\ref{R_pi-siLT}$'$} \\[-.75ex]
&\phantom{\si^{\{V\}}_{f,\la}(Q^2)\;}
  =\; \int_{-\infty}^0 dt\; \frac{1}{16\pi\, s^2}\;
        \bigg| {\T\sum}_{V\!=\{V'\}}\vv
        T_{V\!,\la}(s,t)\vv
        \frac{c_{V\!,f}\cdot \sqrt{M_V\Gatot_V\!/\pi}}{%
              M^2 \!-\! M_V^2 \!+\! \iIM\, M_V\Gatot_V}\vv
        \sqrt{B_{V\!\to f}}\; \bigg|^2
    \nn
    \\[-4.5ex]\nn
\end{align}
Zur Berechnung von~\vspace*{-.25ex}$\Rpi$ auf Basis dieser Relatione ist die Summation "uber die Vektormesonen~$\{V'\}$ zu beziehen im Nenner auf~$\rh$ und im Z"ahler auf~$\rh'$,~$\rh^\dbprime$, das hei"st gegen"uber Gl.~(\ref{si_gap-to-fp}) eingeschr"ankt auf entweder~$\rh$ oder~$\rh'$,~$\rh^\dbprime$.
Es ist simultan zu setzen~\vspace*{-.125ex}$f \!\equiv\! \pi^+\pi^-$ mit~\vspace*{-.125ex}$s_f \!=\! (2m_\pi)^2$ beziehungsweise~$f \!\equiv\! 2\pi^+2\pi^-$ mit~$s_f \!=\! (4m_\pi)^2$; bzgl. der~$c_{V\!,f}$ vgl.\@ Gl.~(\ref{APP:cVla}) und \vspace*{-.25ex}Tabl.~\ref{Tabl-App:cVla}. \\
\indent
Numerisch finden wir eine Reduktion des Resultats f"ur~\vspace*{-.25ex}$\Rpi$ auf~$47\!\pm\!18\,\%$ gegen"uber der {\it zero width approximation\/}, das hei"st der Approximation auf Basis verschwindender Breiten der Resonanzen; die Unsicherheit r"uhrt dabei aus der Unsicherheit der Breiten der Resonanzen, vgl.\@ Tabl.~\ref{Tabl:Charakt_rh,rh',rh''}.
\vspace*{-.375ex}Die Gr"o"se~$\Rpi$ ist experimentell zug"anglich.
In Abbildung~\ref{Fig:R_pi} geben wir an unser Postulat in Konfrontation mit experimentellen Daten von Fermilab-\VIER{E}{6}{6}{5} und \ZWEI[]{H}{1}, vgl.\@ Ref.~\cite{Salgado95} bzw.~\cite{H196}.
Wir finden zun"achst, bei einem Wert von~$Q^2$ von etwa~$1\GeV^2$ eine leichte Fluktuation.
Diese r"uhrt her von dem Wechselspiel der Wirkungsquerschnitte f"ur longitudinale und transversale Polarisation.
Diese wiederum sind dominiert durch die unterschiedliche Struktur der $2S$-Wellenfunktion in hinsicht auf die Position ihrer Knoten.
Zum einen ist unsere Kenntnis um deren exakte Position nur sehr begrenzt.
Und insbesondere erstreckt sich die transversale Ausdehnung der $2S$-Wellenfunktion weit "uber ihren Nulldurchgang, das hei"st ist der Wirkungsquerschnitt f"ur transversale Polarisation ist in hohem Ma"se sensitiv auf die Struktur ihrer beider Knoten~-- der f"ur Quarkspins~$S \!\equiv\! 0$ und der f"ur~$S \!\equiv\! 0$~-- im Wechselspiel.
Zum anderen ist die Fluktuation nur leicht
Wir halten daher durchaus f"ur denkbar, da"s sie vollst"andig ausgewaschen sein k"onnte.
Was wir dagegen betrachten als definitives Postulat ist das Verhalten von~$\Rpi$ f"ur gr"o"sere Werte von~$Q^2$.
Der Anstieg von~$\Rpi(Q^2)$ sollte die tats"achliche Physik wiederspiegeln.
Wir finden dies best"atigt in Konfrontation mit den beiden experimentellen Datenpunkten von~\ZWEI[]{H}{1} f"ur~$Q^2 \!\cong\! 7$ und~$15\GeV^2$.
Bleibt anzumerken, da"s diese noch mit einer erheblichen Unsicherheit behaftet sind und stammen aus einer {\it Vorab-Ver"offentlichung\/} von~\ZWEI[]{H}{1}, vgl.\@ Ref.~\cite{H196}.
\vspace{-1ex}

\bigskip\noindent
Wir fassen dieses Kapitel zusammen wie folgt.
Wir untersuchen die $1^+(1^{--})$-/Rho-Resonan\-zen~$\rh,\rh',\rh^\dbprime$~[$\!\equiv\! \rh(770),\rh(1450),\rh(1700)$] f"ur Photo- und Leptoproduktion.
Das~\mbox{$\rh$-Me}\-son wird beschrieben als $1S$-Zustand bezogen auf das Verhalten der Quark-Antiquark-Wellenfunktion im nichtrelativistischen Limes.
Die $\rh'$-, $\rh^\dbprime$-Mesonen werden aufgefa"st als Mischung zweier Anteile: eines $2S$-Zustands, der wesentlich bestimmt die Querschnitte f"ur Photo- und Leptoproduktion, und ein diesbez"uglich inerter Rest, der zusammenfa"st ein $2D$-Zustand und gluonische Anregungen: Hybride.
Dieser Ansatz ist suggeriert durch die Analyse deren Zerfallskan"ale, insbesondere den Rho-Kanal-$\pi^+\pi^-$- und $2\pi^+2\pi^-$-Massespektren f"ur Elektron-Positron-Annihilation und Photoproduktion elastisch am Proton.
Der Mischungswinkel folgt unmittelbar und bestimmt die beiden Anteile als in etwa gleich gro"s.
Eine solche Hybrid-Beimischung wird gefordert durch die Charakteristik des~$\rh'$ und ist konsistent mit der des~$\rh^\dbprime$.
Auf Basis dieses Ansatzes werden konstruiert Lichtkegelwellenfunktionen f"ur die Vektormesonen.
Die Formulierung einer universellen Lichtkegelwellenfunktion f"ur das Photon in Hinblick auf beliebig kleine bis verschwindende Virtualit"at macht zug"anglich nichtperturbativ dominierte Photo- und Leptoproduktion von~$\rh,\rh',\rh^\dbprime$ im Rahmen des Modells des Stochastischen Vakuums~(\DREI[]{M}{S}{V}).
In diesem Zugang werden die Wirkungsquerschnitte wesentlich determiniert durch einen \DREI[]{M}{S}{V}-spezifischen String-String-Mechanismus, der verantwortlich ist f"ur das stetige und starke Ansteigen des zugrundeliegenden Dipol-Proton-Querschnitts mit der transversalen Ausdehnung des Dipols.
Wesentliche Beitr"age r"uhren her von dem Bereich~\mbox{$1 \!-\! 2\fm$}, f"ur Photoproduktion bis~\mbox{$2.5 \!-\! 2.8\fm$}.
Die Parameter des \DREI[]{M}{S}{V} sind allgemeiner Natur und fixiert nich in Zusammenhang der untersuchten Reaktionen.
In Anbetracht dessen gelangen wir zu guter bis sehr guter "Ubereinstimmung mit dem Experimente dort, wo Daten verf"ugbar sind, und zu definitiven Postulaten sonst.
Wir ziehen ausf"uhrlich Resum\'ee unserer Arbeit \vspace*{1.5ex}im folgenden.
\theendnotes

%% file: RESUMEE.tex
\chapter*{\huge \vspace*{-.75ex}Resum\'ee}
\chaptermark{}
\addcontentsline{toc}{chapter}{\numberline{}Resum\'ee}
\lhead[\fancyplain{}{\sc\thepage}]%
      {\fancyplain{}{\sc{Resum\'ee}}}
\rhead[\fancyplain{}{\sc{Resum\'ee}}]%
      {\fancyplain{}{\sc\thepage}}

Die vorliegende Arbeit~-- Hochenergiestreuung im nichtperturbativen Vakuum der Quantenchromodynamik~-- ist zentral orientiert auf den Bereich {\it nichtperturbativer\/} Quantenchromodynamik~(\DREI{Q}{C}{D}), f"ur den im Gegensatz zur Diagrammatik perturbativer Quantenchromodynamik kein etablierter Zugang existiert. \\
\indent
Etabliert ist lediglich, da"s Zusammenhang besteht zwischen dem nichtperturbativen Bereich der Theorie und der Nichttrivialit"at der Struktur ihres Vakuums.
Das Modell des Stochastischen Vakuums~(\DREI{M}{S}{V}) von Dosch und Simonov kn"upft an genau an diese Struktur.
Es liegt wesentlich zugrunde unserer Arbeit.
Observable sind im Formalismus des Funktionalintegrals gegeben durch eichinvariante Vakuumerwartungswerte von Funktionalen im Eichfeld~$A$, die in diesem Sinne "uber s"amtliche Raumzeit-Konfigurationen von~$A(x)$ zu mitteln sind~-- gewichtet mit dem Wirkungsfunktional mit Yang-Mills- und Eichfixierungs-Lagrangedichte und der Fermion- und Geist-Determinante.
Das~\DREI{M}{S}{V} basiert wesentlich auf der Annahme, die dominanten Konfigurationen von~$A(x)$ im nichtperturbativen Bereich k"onnen approximiert werden durch einen stochastischen Proze"s~-- das Vakuum-Eichfeld fluktuiert stochastisch.
Formal ist die Annahme nicht zu beziehen auf das nicht-physikalische Eichfeld~$A$, sondern auf den paralleltransportierten Feldst"arkentensor~$F[A]$.
Die Annahme eines stochastischen Prozesses ist dann gleichbedeutend damit, da"s f"ur observable Vakuumerwartungswerte von Funktionalen im paralleltransportierten Feldst"arkentensor eine konvergente Entwicklung in Kumulanten existiert.
Mit~\mbox{\,$F^{(i)} \!\equiv\! F_{\mu_i\nu_i}\!(x_i; x_0,{\cal C}_{x_{\!0}\!x_{\!i}})$} dem Feldst"arkentensor, dessen Colour-Gehalt entlang der Kurven%
  ~\mbox{${\cal C}_{x_{\!0}\!x_{\!i}}$} und%
  ~\mbox{${\cal C}_{x_{\!0}\!x_{\!i}}^{\vv-1} \!\equiv\! {\cal C}_{x_{\!i}\!x_{\!0}}$} paralleltransportiert ist an den Referenzpunkt~$x_0$, sei die Kumulante \mbox{$n$-ter} Ordnung bezeichnet durch~\mbox{\,$\cum{ g^n F^{(1)} \cdots F^{(n)} }$}; sie ist charakterisiert durch ihre {\it Cluster-Eigenschaft\/} wie folgt.
Ihr Wert ist maximal~-- gleich%
  ~\vspace*{-.125ex}\mbox{\,$\cum{ g^n F^{(1)} \cdots F^{(n)} }|_{x_1\equiv x_2\ldots\equiv x_n}$}, dem "`n-ten"' Parameter des stochastischen Prozesses~--, wenn s"amtliche~$n$ Raumzeit-Punkte~$x_i$ identisch und die~$F^{(i)}$ auf diesen Punkt "`zusammengeclustert"' sind.
Ihr Wert f"allt ab, wenn nur ein Weltpunkt~$x_{\hat\imath}$ diesen Cluster verl"a"st, exponentiell mit dem Abstand, wenn dieser gr"o"ser ist als die Korrelationsl"ange~$a$, dem "`0-ten"' Parmeter des Prozesses.%
\FOOT{
  Ein stochastischer Proze"s mit~$n$ unabh"angigen Kumulanten besitzt daher~\mbox{$n \!+\! 1$} Parameter.
}
Dieses Bild ist beeindruckend intuitiv auf einer Raumzeit mit definiter Metrik, in der das Verschwinden des invarianten (Vierer)Abstands zweier Weltpunkte gleichbedeutend ist mit ihrer r"aumlichen und zeitlichen Koinzidenz:
Die Colour-elektrischen und -magnetischen Felder sind maximal korreliert, wenn sie lokalisiert sind an demselben Weltpunkt, sie sind nicht mehr korreliert, wenn sie wesentlich weiter entfernt sind als ein gewisser Abstand~$a$.
Das~\DREI{M}{S}{V} ist urspr"unglich formuliert in der Euklidischen Raumzeit, die folgt aus der physikalischen Minkowskischen Raumzeit durch analytische Fortsetzung deren Zeitkomponente zu imagin"aren Werten.
Ohne den stochastischen Proze"s in irgendeiner Weise spezifizieren zu m"ussen, postuliert das~\DREI{M}{S}{V} bereits {\it lineares Confinement\/} f"ur ein statisches Quark-Antiquark-Paar.
Unter Spezifizierung des Prozesses als Gau"s'sch kann explizit berechnet werden die Stringspannung des gluonischen Flu"sschlauches, der sich ausbildet zwischen dem Quark und Antiquark,~-- und dieser detailliert ausgemessen werden.
Ein Quark-Antiquark-Paar ist dabei formal repr"asentiert durch den Vakuumerwartungswert {\it eines\/} Wegner-Wilson-Loops, der mithilfe des nichtabelschen Stokes'schen Satzes Funktional ist des paralleltransportierten Feldst"arkentensors~$F[A]$. \\
\indent
Die \mbox{$T$}-Amplitude f"ur die Streuung zweier Quark-Antiquark-Paare ist formal repr"asentiert durch den Vakuumerwartungswert zweier Wegner-Wilson-Loops.
Wir betrachten daher {\it Streuung\/} auch in Hinblick darauf, den String-Mechanismus, den das~\DREI{M}{S}{V} postuliert, zu identifizieren als den dominierenden Mechanismus in der Bestimmung von Vakuumerwartungswerten {\it zweier\/} Wegner-Wilson-Loops.
Wir betrachten {\it Hochenergie\/}streuung, da die {$T$-Amp}\-litude in Termen zweier Wegner-Wilson-Loops notwendig voraussetzt, da"s sich die sie repr"asentierenden (Anti)Quarks fundamental nicht "andern "uber die Dauer der Streuung~-- in dem Sinne, da"s Abstrahlung von Gluonen weitestgehend ausgeschlossen sei.
Dies ist gegeben f"ur gro"se Werte von~$s$, dem Quadrat der invarianten Schwerpunktenergie der Streuung.
Gro"se~$s$ wiederum implizieren, da"s die streuenden Wegner-Wilson-Loops in ihrem Schwerpunktsystem (nahezu) lichtartig sind, das hei"st (nahezu) parallel verlaufen zu den Achsen~$x^\pm$ des Lichtkegels.
Der Wegner Wilson-Loop eines statischen Quark-Antiquark-Paares verl"auft dagegen parallel zur Zeitachse~$x^0$ desselben Schwerpunktsystems.
Insofern untersucht {\it Hochenergie\-streuung\/} einen essentiell anderen, essentiell anders repr"asentierten Bereich nichtperturbativer~\DREI{Q}{C}{D}.
Identifizierung des \mbox{\DREI{M}{S}{V}-spe}\-zifischen String-Mechanismus wiese massiv hin auf dessen fundamentale Bedeutung.
\vspace*{-.5ex}

\bigskip\noindent
{\bf Kapitel~\ref{Kap:VAKUUM}}\quad
  gibt zun"achst an Observable als eichinvariante Eichfeld-Funktionalintegrale.
Dies geschieht unter expliziter Herleitung von%
  ~\vspace*{-.125ex}\mbox{\,$\HaarDmu$}, dem {\it Haarschen Eichfeldma"s\/} der~\DREI{Q}{C}{D}.
Dies ist die Gr"o"se, die das~\DREI{M}{S}{V} approximiert durch die Annahme, ihm liege zugrunde im nichtperturbativen Bereich ein stochastischer Proze"s.
Im Anschlu"s werden eingef"uhrt die wesentlichen mathematischen Hilfsmittel, auf denen die Konstituierung des~\DREI{M}{S}{V} basiert.
So werden eingef"uhrt {\it Konnektoren\/}~$\Ph$, mit deren Hilfe explizit konstruiert werden paralleltransportierte Feldst"arken als die eigentlich relevanten Gr"o"sen.
Es wird formuliert der {\it Nichtabelsche Stokes'sche Satz\/}, durch den Wegner-Wilson-Loops, die a~priori Funktionale sind des Eichfeldes~$A$, dargestellt werden als manifest eichinvariante Funktionale des paralleltransportierten Feldst"arkentensors~$F$.
Es wird ferner explizit formuliert der Begriff der {\it Entwicklung in Kumulanten\/}.
Auf Basis dieses Formalismus werden abschlie"send ausf"uhrlich diskutiert die Annahmen des~\DREI{M}{S}{V} und dieses explizit konstituiert.
\vspace*{-.5ex}

\bigskip\noindent
{\bf Kapitel~\ref{Kap:ANALYT}}\quad
  stellt her den \vspace*{-.125ex}Zusammenhang zu Hochenergiestreuung.
Diesem Zusammenhang liegt zugrunde die \mbox{$T$-Amp}\-litude%
  ~\vspace*{-.125ex}\mbox{\,$\tTll \!\equiv\! \tTll^{(s,\rb{b})}$} f"ur die Streuung der Wegner-Wilson-Loops%
  ~$W\Dmfp$,~$W\Dmfm$; es ist~$\rb{b}$ ihr (transversaler) Impaktvektor, Fourier-konjugiert zum invarianten Impuls"ubertrag.
Wir zeichnen nach die Herleitung Nachtmanns einer \vspace*{-.125ex}\mbox{\,$s \!\to\! \infty$-asymp}\-totischen Formel f"ur%
  ~\vspace*{-.125ex}\mbox{$\tTll$}.
Diese ist im wesentlichen gegeben als der Vakuumerwartungswert zweier~-- als Konsequenz des Limes~-- {\it exakt\/} lichtartiger Wegner-Wilson-Loops; Energieabh"angigkeit besteht nurmehr "uber den kinematischen Faktor~$s$, der verschwindet in den Wirkungsquerschnitten.
Wir argumentieren vor dem Hintergrund von Arbeiten von Verlinde, Verlinde, da"s die Formel Nachtmanns in ihrer G"ultigkeit hin zu {\it gro"sen aber endlichen Werten von\/}~$s$ erweitert werden kann dadurch, da"s die klassischen Parton-Trajektorien mit kleiner zeitartiger Komponente~-- die {\it nahezu\/} lichtartig sind~-- nicht ersetzt werden durch ihre exakt lichtartigen~\mbox{\,$s \!\to\! \infty$-Limi}\-tes.
Diese Argumentation wird zus"atzlich untermauert durch die Arbeiten Meggiolaros f"ur Quark-Quark-Streuung, die basieren \vspace*{-.125ex}auf diesen "Uberlegungen. \\
\indent
W"ahrend die in diesem Sinne exakt lichtartige \vspace*{-.125ex}\mbox{$T$-Amp}\-litude a~priori nicht analytisch fortgesetzt werden kann in die ins Euklidische fortgesetzte Minkowski-Raumzeit, ist dies m"oglich f"ur die nahezu lichtartige \vspace*{-.125ex}\mbox{$T$-Amp}\-litude.
Dies machte Hochenergiestreuung prinzipiell zug"anglich der Auswertung durch eine im Euklidischen formulierten Theorie.
Dies w"are in erster Linie die modellunabh"angige Formulierung von~\DREI{Q}{C}{D} als Gittereichtheorie; wir erinnern ferner daran, da"s die Annahmen des~\DREI{M}{S}{V} eigentlich stringent sind nur in seiner Euklidischen Formulierung:
Maximal korreliert~-- mit verschwindendem invariantem Abstand~-- sind im Minkowskischen alle lichtartig separierten Weltpunkte, die r"aumlich und zeitlich im allgemeinen nicht koinzidieren, sondern im Gegeenteil beliebig gro"sen Abstand haben. \\
\indent
Nach Einf"uhrung in die zugrundeliegende Kinematik rekapitulieren wir daher explizit die Konstruktion Nachtmanns der \mbox{\,$s \!\to\! \infty$-asymp}\-totischen Formel f"ur die \mbox{$T$-Amp}\-litude%
  ~\vspace*{-.125ex}\mbox{$\tTll$} f"ur Loop-Loop-Streuung, eingebunden in die Formel f"ur die \mbox{$T$-Amp}\-litude%
  ~\vspace*{-.125ex}\mbox{$T\hh \!\equiv\! T\hh^{(s,t)}$} f"ur Hadron-Hadron-Streuung; dies geschieht in Hinblick auf die klassischen nahezu lichtartigen Parton-Trajektorien, die konstituieren nahezu lichtartige Wegner-Wilson-Loops.
Diese Konstruktion wird durchgef"uhrt auf Partonniveau: \vspace*{-.125ex}$\tTll$, und angebunden "uber hadronische Lichtkegelwellenfunktionen an Hadronniveau: \vspace*{-.125ex}$T\hh$. \\
\indent
Die resultierende Formel f"ur~\vspace*{-.25ex}$\tTll$ wird dann ausgewertet auf Basis des~\DREI{M}{S}{V} und mit differentialgeometrischen Methoden von suggestiver Anschauung.
Explizit werden die streuenden Wegner-Wilson-Loops~$W\Dmfp$,~$W\Dmfm$ aufgefa"st als zwei initial ruhende Loops unter {\it aktiven Lorentz-Boosts\/} mit Geschwindigkeiten~\mbox{\,$\be\Dmfp, \be\Dmfm \!\to\! 1$} aufeinander zu.
Die Beta-Parameter werden dabei genau in der Weise gew"ahlt~-- \oE~$\be\Dmfm$ als Funktion von~$\be\Dmfp$~--, da"s das System in demselben initialen Schwerpunktsystem in Ruhe bleibt.
Es ist daher das Quadrat der invarianten Schwerpunktenergie~$s$ Funktion von~$\be\Dmfp$[und~$\be\Dmfm$], und umgekehrt~$\be\Dmfp$[und~$\be\Dmfm$] Funktion von~$s$.
Ferner sind in eindeutiger Weise gegeben durch%
  ~\mbox{\,$\be\Dmfp \!=:\! \tanh\ps\Dmfp$} und%
  ~\mbox{\,$\be\Dmfm \!=:\! \tanh\ps\Dmfm$} die {\it hyperbolischen Winkel\/}%
  ~\vspace*{-.125ex}\mbox{\,$\ps\Dmfp, \ps\Dmfm \!\in\!  (0,\infty)$} der Loops~$W\Dmfp$,~$W\Dmfm$ gegen die Zeitachse~$x^0$ des Systems~$I\idx{0}$, in dem ihr gemeinsamer Schwerpunkt ruht.
Es ist~\mbox{\,$\ps \!:=\! \ps\Dmfp \!+\! \ps\Dmfm$} der hyperbolische Winkel zwischen den Wegner-Wilson-Loops im Minkowski-Diagramm mit orthogonalen Achsen bez"uglich~$I\idx{0}$. \\
\indent
Die Auswertung der \mbox{$T$-Amp}\-litude~$\tTll$ auf Basis der nahezu lichtartigen Wegner-Wilson-Loops~$W\Dmfp$ und~$W\Dmfm$ geschieht nun genau bez"uglich Koordinatenlinien~$\bm\mfp$ und~$\bm\mfm$, die definiert sind durch deren Richtungen.
Die entsprechenden Komponenten~$x^\mfp$,~$x^\mfm$ n"ahern sich per~constructionem im Limes%
  ~\mbox{\,$s \!\to\! \infty$} mehr und mehr an den Komponenten~$x^+$,~$x^-$ bez"uglich Lichtkegelkoordinaten, ohne f"ur endliche Werte von~$s$~-- ergo%
  ~\mbox{\,$\be\Dmfp,\be\Dmfm \!<\! 1$} und%
  ~\mbox{\,$\ps\Dmfp,\ps\Dmfm \!<\! 1$}~-- exakt in diese "uberzugehen.
Konsequenz ist, da"s sich die Auswertung gegen"uber dem \mbox{$s \!\to\! \infty$-asymp}\-totischen Fall~-- den etwa diskutiert unsere Diplomarbeit~-- genau darin "andert, da"s die Licht\-kegel-Koordinatenlinien%
  ~\vspace*{-.125ex}\mbox{\,$\bar\mu \!\in\! \{+,-,1,2\}$} "ubergehen in die Koordinatenlinien%
  ~\mbox{\,$\tilde\mu \!\in\! \{\mfp,\mfm,1,2\}$}.
Dabei ist zu beachten, da"s die longitudinalen Komponenten%
  ~\vspace*{-.125ex}\mbox{\,$g_{\mfp\mfp} \!\equiv\! g_{\mfm\mfm}$} und%
  ~\mbox{\,$g_{\mfp\mfm} \!\equiv\! g_{\mfm\mfp}$} des metrischen Tensors%
  ~\mbox{\,$\tilde{g} \!\equiv\! \big(g_{\tilde\mu\tilde\nu}\big)$} "uber~$\ps\Dmfp$,~$\ps\Dmfm$ beziehungsweise~$\be\Dmfp$,~$\be\Dmfm$ Funktionen sind von~$s$.
Es gilt per~constructionem%
  ~\vspace*{-.125ex}\mbox{\,$g_{\mfp\mfp} \!\to\! g_{++}$} und%
  ~\vspace*{-.125ex}\mbox{\,$g_{\mfp\mfm} \!\to\! g_{+-}$} f"ur die diagonale und au"serdiagonale Komponente im Limes~\mbox{\,$s \!\to\! \infty$}.
Terme, die im asymptotischen Fall identisch verschwinden, da sie multipliziert sind mit%
  ~\vspace*{-.125ex}\mbox{$g_{++} \!\equiv\! g_{--} \!\equiv\! 0$}, sind f"ur endliche Werte von~$s$ multipliziert mit~\mbox{$g_{\mfp\mfp} \!\equiv\! g_{\mfm\mfm} \!\nequiv\! 0$} treten also zus"atzlich auf. \\
\indent
Wir finden zun"achst, da"s 
die relevanten Funktionen faktorisieren in Form%
  ~\mbox{\,$g_{\imath\jmath} \!\big/\! g_{+-}
                               \!\cdot\! \tilde{X}\idx{\imath\jmath}$} mit%
  ~\mbox{\,$\imath,\jmath \!\in\! \{\mfp,\mfm\}$}, wobei die Funktionen%
  ~\mbox{$\tilde{X}\idx{\imath\jmath}$} vollst"andig bestimmt sind durch die Geometrie in der \mbox{$x^1\!x^2$-Trans}\-versalebene, die nicht abh"angt von~$s$, in die aber projiziert ist die longitudinale Dynamik der Streuung in der Form, da"s der effektive Impaktvektor~$\rb{b}$ der Streuung abh"angt von den \mbox{$\zet$-An}\-teilen, die die (Anti)Quarks der Wegner-Wilson-Loops~$W\Dmfp$,~$W\Dmfm$ an deren gesamten Lichtkegelimpuls tragen.
Die \mbox{$s$-Ab}\-h"angigkeit ist vollst"andig absorbiert in den longitudinalen Komponenten%
  ~$g_{\imath\jmath}$ des metrischen Tensors.
Dieses Resultat ist konsistent damit, da"s bereits auf Basis allgemeiner "Uberlegungen, unabh"angig von den Details der Wechselwirkung, Verlinde, Verlinde zeigen, da"s Hochenergiestreuung in~$d$ Dimensionen faktorisiert in einen~\mbox{$s$-ab}\-h"angigen longitudinalen Anteil und in einen \mbox{$s$-unab}\-h"angigen Anteil, der beschrieben wird durch eine {\it effektive Quantenfeldtheorie\/} in den~\mbox{$(d \!-\! 2)$} nicht-longitudinalen Dimensionen. \\
\indent
Die mit~$g_{\mfp\mfm}$ multiplizierte Funktion%
  ~\vspace*{-.125ex}\mbox{$\tilde{X}\idx{\mfp\mskip-2mu\mfm}$} ist dieselbe wie im asymptotischen Fall, demgegen"uber f"ur endliche Werte von~$s$ zus"atzlich auftreten die mit~$g_{\mfp\mfp}$,~$g_{\mfm\mfm}$ multiplizierten Funktionen%
  ~\vspace*{-.125ex}\mbox{$\tilde{X}\idx{\mfp\mskip-2mu\mfp}$},~\mbox{$\tilde{X}\idx{\mfm\mskip-2mu\mfm}$}.
Sie sind offensichtlich bestimmt durch die $\bm\mfp$- beziehungsweise \vspace*{-.125ex}\mbox{$\bm\mfm$-Sei}\-te der Streuung allein.
Bereits aufgrund der manifesten Eichinvarianz der Loop-Loop-Amplitude~\vspace*{-.125ex}$\tTll$ folgt, da"s diese Funktionen bestimmt sein m"ussen durch den respektiven Vakuumerwartungswert des einzelnen Wegner-Wilson-Loops:~$\vac{W\Dmfp}$,~$\vac{W\Dmfm}$.
Dies wird explizit verifiziert.
Die Funktionen%
  ~\vspace*{-.25ex}\mbox{$\tilde{X}\idx{\imath\jmath}$} sind gegeben als die Summe eines Anteils bez"uglich der konfinierenden Struktur%
  ~\vspace*{-.125ex}\mbox{\,$t\oC{}_{\zzzz \tilde\mu\tilde\nu\tilde\rh\tilde\si}$} und eines Anteils bez"uglich der nicht-konfinierenden Struktur%
  ~\vspace*{-.125ex}\mbox{\,$t\oNC{}_{\zzzz \tilde\mu\tilde\nu\tilde\rh\tilde\si}$} des fundamentalen Korrelations-Lorentztensors:%
  ~\vspace*{-.25ex}\mbox{\,$\tilde{X}\idx{\imath\jmath} \!=\!
     \vka \tilde{X}\idx{\imath\jmath}\oC \!+\! (1 \!-\! \vka) \tilde{X}\idx{\imath\jmath}\oNC$} mit~$\vka \!\cong\! 0.74$ dem gewichtenden Parameter.
Wir gelangen zu expliziten Darstellungen f"ur die bekannten Funktionen%
  ~\vspace*{-.125ex}\mbox{$\tilde{X}\oC\idx{\mfp\mskip-2mu\mfm}$},%
   \mbox{$\tilde{X}\oNC\idx{\mfp\mskip-2mu\mfm}$} und die neu neu auftretenden Funktionen%
  ~\vspace*{-.125ex}\mbox{$\tilde{X}\oC\idx{\mfp\mskip-2mu\mfp}$},%
   \mbox{$\tilde{X}\oNC\idx{\mfp\mskip-2mu\mfp}$} und%
  ~\vspace*{-.125ex}\mbox{$\tilde{X}\oC\idx{\mfm\mskip-2mu\mfm}$},%
   \mbox{$\tilde{X}\oNC\idx{\mfm\mskip-2mu\mfm}$}~-- f"ur allgemeine Korrelationsfunktionen wie auch f"ur die Korrelationsfunktionen des bekannten expliziten Ansatzes.
F"ur die konfinierenden Funktionen identifizieren wir in der Tat den postulierten String-Mechanismus.
So ist die \vspace*{-.125ex}bekannte Funktion%
  ~\vspace*{-.125ex}\mbox{$\tilde{X}\oC\idx{\mfp\mskip-2mu\mfm}$} bestimmt durch die {\it Wechselwirkung\/} der gluonischen Strings, die sich ausbilden in den Wegner-Wilson-Loops~$W\Dmfp$,~$W\Dmfm$.
Die Funktionen%
  ~\vspace*{-.125ex}\mbox{$\tilde{X}\oC\idx{\mfp\mskip-2mu\mfp}$} und%
   \mbox{$\tilde{X}\oC\idx{\mfm\mskip-2mu\mfm}$} sind respektive bestimmt durch den gluonischen String von%
  ~$W\Dmfp$ und~\vspace*{-.125ex}$W\Dmfm$.
Wir zeigen, da"s diese Funktionen zu interpretieren sind als die {\it dynamische Stringspannung\/} der \vspace*{-.125ex}Wegner-Wilson-Loops. \\
\indent
F"ur die Loop-Loop-Amplitude~\vspace*{-.125ex}$\tTll$ konstatieren wir noch unabh"angig von der expliziten Gestalt der Funktionen%
  ~\vspace*{-.125ex}\mbox{$\tilde{X}\idx{\imath\jmath}$} einen leichten Anstieg mit~$s$~-- konsistent mit dem experimentell beobachteten Anstieg hadronischer Wirkungsquerschnitte.
Explizite Verifizierung erfordert explizite Auswertung der angegebenen Loop-Loop-Amplitude f"ur gro"se aber endliche Werte von~$s$ und die Angabe hadronischer Wellenfunktionen~-- die gleichfalls abh"angen sollten von~$s$.
Beides ginge weit hinaus "uber den Rahmen der vorliegenden Arbeit. \\
\indent
Wir beschr"anken uns darauf~-- abschlie"send Kapitel~\ref{Kap:ANALYT}~-- explizit durchzuf"uhren die analytische Fortsetzung der nahezu lichtartigen \mbox{$T$-Amp}\-litude in die ins Euklidische fortgesetzte Minkowski-Raumzeit.
Dies geschieht dadurch, da"s analog wie im Minkowskischen Koordinatenlinien%
  ~\mbox{\,$\xE[\tilde\mu] \!\in\! \{\mfp,\mfm,1,2\}$} konstruiert werden im Euklidischen, deren longitudinalen Komponenten definiert sind durch die Richtungen der Wegner-Wilson-Loops in der Euklidischen \mbox{$\xE^4\!\xE^3$-Ebe}\-ne, die entspricht der Minkowskischen \mbox{$x^0\!x^3$-Boost-Ebe}\-ne.
Die Loops schlie"sen ein im Euklidischen den {\it Winkel\/}~$\th$, im Minkowskischen den {\it hyperbolischen Winkel\/}~\vspace*{-.125ex}$\ps$.
Resultat ist, da"s die \mbox{$T$-Amp}\-lituden%
  ~\vspace*{-.125ex}\mbox{\,$\xE[(\tTll)] \!\equiv\! \xE[(\tTll^{(s,\rb{b})})]$} und%
  ~\mbox{\,$\tTll \!\equiv\! \tTll^{(s,\rb{b})}$} f"ur die Streuung der Wegner-Wilson-Loops%
  ~\vspace*{-.125ex}$\xE[W]{}\Dmfp$,~$\xE[W]{}\Dmfm$ in Euklidischer beziehungsweise%
  ~$W\Dmfp$,~$W\Dmfm$ in Minkowskischer Raumzeit in einfacher Weise zusammenh"angen durch die analytische Fortsetzung der (hyperbolischen) Winkel wie%
  ~\vspace*{-.125ex}\mbox{\,$-\iIM\,\th \!\leftrightarrow\! \ps$}.
Zu derselben Fortsetzungsvorschrift gelangt Meggiolaro f"ur Quar-Quark-Streuung.%
\FOOT{
  Vgl.\@ Fu"sn.\,\FNg{FN:th-Meggiolaro}: Meggiolaro definiert den Winkel~$\th$ mit {\sl umgekehrtem\/} Vorzeichen. 
}
Wir schlie"sen Kapitel~\vspace*{-.125ex}\ref{Kap:ANALYT} mit der expliziten Angabe dieses Zusammenhangs in Form:%
  ~\vspace*{-.25ex}\mbox{\,$\xE[(\tTll)][\th] \!=\! \tTll[\ps \!\to\! -\iIM\th]$}, umgekehrt:%
  ~\mbox{\,$\tTll[\ps] \!=\! \xE[(\tTll)][\th \!\to\! \iIM\ps]$}, und der ausf"uhrlichen Diskussion der Relevanz dieses Resultats.
\vspace*{-.5ex}

\bigskip\noindent
{\bf Kapitel~\ref{Kap:GROUND}}\quad
-- wie Kapitel~\ref{Kap:EXCITED}~-- stehen chronologisch {\it vor\/} der Herleitung der nahezu lichtartigen \mbox{$T$-Amp}\-litude~$\tTll$ f"ur Loop-Loop-Streuung in Kapitel~\ref{Kap:ANALYT}.
Beiden Kapiteln liegt zugrunde noch die bekannte asymptotische Formel f"ur~\vspace*{-.125ex}$\tTll$ im Limes~\mbox{$s \!\to\! \infty$}.
Beide diskutieren explizite Streureaktionen physikalischer Hadronen im betrachteten \vspace*{-.125ex}kinematischen Bereich, f"ur Definiertheit~\mbox{\,$-t \!<\! 1\GeV^2$} und~\mbox{\,$\surd s \!=\! 20\GeV$} der Fixierungspunkt der Parameter des~\DREI{M}{S}{V}. \\
\indent
Unter "Ubergang der Koordinatenlinien%
  ~\mbox{\,$\tilde\mu \!\in\! \{\mfp,\mfm,1,2\} \to \bar\mu \!\in\! \{+,-,1,2\}$}, das hei"st Substitution%
  ~\mbox{\,$\mfp \!\to\! +$},~\mbox{\,$\mfm \!\to\! -$} bez"uglich der Lorentz-Indizes kann Kapitel~\ref{Kap:ANALYT} gelesen werden als Herleitung der \mbox{$s \!\to\! \infty$-asymp}\-totischen exakt lichtartigen \mbox{$T$-Amp}\-litude~$\tTll$.
In diesem Sinne gehen wir "uber durch%
  ~\mbox{\,$g_{\mfp\mfp} \!\big/\! g_{+-} \!\to\! 0$} und%
  ~\mbox{\,$g_{\mfp\mfm} \!\big/\! g_{+-} \!\to\! 1$} zu der \mbox{$s \!\to\! \infty$-asymp}\-totischen \mbox{$T$-Amp}\-litude f"ur die Streuung exakt lichtartiger Wegner-Wilson-Loops~$W\Dmfp$,~$W\Dmfm$, die wir auffassen als die \mbox{$T$-Amp}\-litude f"ur die Streuung der entsprechenden Quark-Antiquark-Dipole und an den Anfang stellen unserer Analysen in Kapitel~\ref{Kap:GROUND} und~\ref{Kap:EXCITED}. \\
\indent
Der "Ubergang zu Hadronen geschieht, wie dargestellt in Kapitel~\ref{Kap:ANALYT}:
Hadronen werden aufgefa"st als Superposition von Quark-Antiquark-Dipolen, die verteilt sind gem"a"s quantentheoretischer Quark-Antiquark-Wellenfunktionen, cum grano salis {\it Lichtkegelwellenfunktionen\/}%
  ~\mbox{\,$\ps_i(\zet_i,\rb{k}_i)$} im Sinne der perturbativen Theorie auf dem Lichtkegel.
Ihre relevanten Raumzeit-Paramete~$\zet_i$ und~$\rb{k}_i$ sind bestimmt durch die Forderung ihrer Kovarianz unter Lorentz-Transformationen; dabei ist~\vspace*{-.125ex}$\zet$ der Anteil des Quarks am Lichtkegelimpuls des Dipols:%
  ~\mbox{\,$p_i^\pm \!=\! \zet_i P_i^\pm$} bei Propagation in die \mbox{$x^\pm$-Rich}\-tung, und%
  ~\vspace*{-.125ex}\mbox{\,$\rb{k}_i \!=\!
    \frac{1}{2}\big(\rb{p}_i \!-\! \rbb{p}_i\big) \!+\! \big(\frac{1}{2} \!-\! \zet_i\big)\rb{P}_i$} der transversalen Lichtkegel"`relativ"'impuls.
Die Konstruktion der involvierten Lichtkegelwellenfunktionen wird explizit durchgef"uhrt unmittelbar in \vspace*{-.125ex}Anschlu"s an die Diskussion der \DREI{M}{S}{V}-spezifischen Dipol-Dipol-, also der Loop-Loop-Amplitude~\vspace*{-.125ex}$\tTll$. \\
\indent
Die Diskussion von~\vspace*{-.125ex}$\tTll$ f"ur~\mbox{\,$s \!\to\! \infty$} reduziert sich auf die Funktionen%
  ~\mbox{$\tilde{X}\oC\idx{\mfp\mskip-2mu\mfm}$},%
   \mbox{$\tilde{X}\oNC\idx{\mfp\mskip-2mu\mfm}$}, diese wiederum sind vollst"andig bestimmt durch die Geometrie der Streuung in der \mbox{$x^1\!x^2$-Trans}\-versalebene~-- also identisch f"ur endliche Werte von~$s$.
Wesentliche Feststellung ist, da"s in die Transversalebene projiziert wird die longitudinale Dynamik der Streuung in der Form, da"s der relevante Impaktvektor~$\rb{b}$ abh"angt von den Anteilen~$\zet_i$ der respektiven Quarks der Dipole an deren gesamten Lichtkegelimpuls.
In diesem Zusammenhang verifizieren wir a~posteriori Annahme~(3) des~\DREI{M}{S}{V}, da"s "`bei m"oglichst symmetrischer Wahl"' die Amplitude nur schwach abh"angt von der Position des Referenzpunkts~$x_0$ und, damit in Zusammenhang, nur schwach von der Wahl der Mantelfl"achen der verallgemeinerten Pyramiden, "uber die die paralleltransportierten Feldst"arken in Kapitel~\ref{Kap:ANALYT} integriert werden. \\
\indent
Die Funktionen%
  ~\vspace*{-.125ex}\mbox{$\tilde{X}\oC\idx{\mfp\mskip-2mu\mfm}$} und%
  ~\mbox{$\tilde{X}\oNC\idx{\mfp\mskip-2mu\mfm}$} zeigen wesentlich unterschiedliches Verhalten.
So sind in%
  ~\vspace*{-.125ex}\mbox{$\tilde{X}\oC\idx{\mfp\mskip-2mu\mfm}$} s"amtliche Weltpunkte zwischen den (Anti)Quarks entlang der transversal projizierten Pyramiden-Mantelfl"achen miteinander korreliert, dagegen in%
  ~\mbox{$\tilde{X}\oNC\idx{\mfp\mskip-2mu\mfm}$} nur die (Anti)Quarks selbst.
Dies f"uhrt zu der Interpretation, da"s die $C$-Funktion dominiert ist durch die Wechselwirkung gluonischer Strings, in der $N\!C$-Funktionen dagegen keine Strings ausgebildet werden.
Diese Feststellung wird weiter herausgearbeitet und verifiziert f"ur Hochenergiestreuung die Begriffe "`konfinierend"' und "`nicht-konfinierend"' f"ur die Funktionen der respektiven Strukturen des Korrelations-Lorentztensors.
Wichtiges Resultat in diesem \vspace*{-.125ex}Zusammenhang ist, da"s der totale Wirkungsquerschnitt~\vspace*{-.125ex}\mbox{\,$\si^{\rm tot}\ellp(\zet, r^2)$} f"ur die Streuung eines Dipols am Proton ansteigt mit der transversalen Ausdehnung~$r$ des Dipols aufgrund des String-String-Mechanismus der $C$-Funktion, wohingegen er abs"attigt, wenn die $C$-Funktion ausgeschaltet wird und "ubrig bleibt nur die $N\!C$-Funktion.
Perturbative Modelle auf Basis konventioneller Quark-Quark-Wechselwirkung f"uhren zu Quark-additiven $N\!C$-Funktionen.
Unterschiede sollten sich daher signifikant manifestieren f"ur Dipole mit \vspace*{-.125ex}gro"ser transversaler Ausdehnung. \\
\indent
Die in dieser Weise verstandenen Amplituden f"ur Dipol-Dipol- und Dipol-Proton-Streu\-ung liegen zugrunde der Analyse von Kapitel~\ref{Kap:GROUND}: exklusive (Photo- und) Leptoproduk\-tion%
  ~\mbox{\,$\ga^{\scriptscriptstyle({\D\ast})}p \!\to\! V p$} der Grund\-zustand-Vektormesonen%
  ~\vspace*{-.125ex}\mbox{$V \!\equiv\! \rh(770), \om(782), \ph(1020), \Jps(3097)$} mit respektivem Fla\-vour-Gehalt%
  ~\vspace*{-.125ex}\mbox{\,$(u\bar{u} \!-\! d\bar{d})\!\big/\!\surd2$},%
  ~\mbox{\,$(u\bar{u} \!-\! d\bar{d})\!\big/\!\surd2$},%
  ~\mbox{\,$s\bar{s}$},%
  ~\mbox{\,$c\bar{c}$}.%
\FOOT{
  F"ur~$charm$ kann betrachtet werden Photoproduktion; s.u.\@ die Lichtkegelwellenfunktion des Photons.
}
%
\\\indent\enlargethispage{.25ex}
Es werden Lichtkegelwellenfunktionen%
  ~\vspace*{-.125ex}\mbox{\,$\ps(\zet,\rb{k})$} f"ur die Verteilung von Quark-Antiquark-Dipolen konstruiert im Sinne der perturbativen Theorie auf dem Lichtkegel.
F"ur das {\it Proton\/} wird der einfache Ansatz gemacht, der Lichtkegelimpuls werde zu gleichen Teilen getragen von Quark und Diquark:~\mbox{\,$\zet_2 \!\equiv\! 1\!/\!2$}, ihr transversaler Relativimpuls~$\rb{k}_2$ sei Gau"s'sch verteilt, das hei"st gem"a"s der $1S$-Wellenfunktion des transversalen Harmonischen Oszillators. \\
\indent
Die Lichtkegelwellenfunktion des {\it Photons\/} folgt unmittelbar aus der perturbativen Theorie.
Parameter sind Virtualit"at~$Q$ und Helizit"at~$\la$ des Photons.
Die \mbox{$T$-Amp}\-litude wesentlich bestimmende Gr"o"se ist die transversale Ausdehnung~$r$ der Dipole, die bestimmt wird in einem subtilen Wechselspiel der Parameter:
Sie tritt auf in Form des Produkts~\mbox{$\vep r$} als Argument der modifizierten Besselfunktionen zweiter Art~$K_0$ und~$K_1$; deren \vspace*{-.125ex}exponentieller Abfall f"ur gro"ses Argument induziert den Zusammenhang~\mbox{\,$r \!\sim\! \vep^{-1}$} mit%
  ~\vspace*{-.125ex}\mbox{\,$\vep \!\equiv\! \sqrt{\zbz Q^2 \!+\! m_f{}^{\!2}}$} und~$m_f$ der laufenden renormierten Quarkmasse mit Flavour~$f$.
F"ur die leichten Quark-Flavour~$u$,~$d$ mit~\mbox{\,$m_u \!\cong\! m_d \!\cong\! 0$} ist~$r$ gro"s f"ur kleine Virtualit"aten~$Q$ und an den Endpunkten des \mbox{$\zet$-Inter}\-valls~\mbox{\,$[0,1]$}; die \mbox{$\zet$-End}\-punkte sind unterdr"uckt durch einen expliziten Faktor~$\zbz$ in der Wellenfunktion f"ur longitudinale Polarisation, nicht aber in der f"ur transversale Polarisation.
F"ur longitudinale Polarisation tritt auf ein Term~\mbox{$\propto\! K_0(\vep r)$}, f"ur transversale die Terme~\mbox{$\propto\! K_1(\vep r)$} und~\mbox{$\propto\! m_f\,K_0(\vep r)$}.
Um zu gro"se Dipole zu vermeiden~-- die physikalisch ausgeschlossen sind durch die Manifestierung der chiralen Symmetriebrechung und dem Ph"anomen von Confinement~-- beschr"anken wir uns f"ur die Flavour~$up$,~$down$,~$strange$ auf Werte von~$Q^2$ gr"o"ser als~\mbox{\,$1 \!-\! 2\GeV^2$}; da%
  ~\mbox{$m_c \!\cong\! 1.3\GeV$} kann f"ur~$charm$ Photoproduktion betrachtet werden.
Da unser Zugang nicht einschlie"st die Streuung perturbativer Onia, das hei"st "`perturbativ kleiner"' Dipole, beschr"anken wir uns konservativ auf Werte von~$Q^2$ nicht gr"o"ser als~$10\GeV^2$. \\
\indent
Die Lichtkegelwellenfunktion der {\it Vektormesonen\/} wird konstruiert analog zu der des Photons.
Es wird "ubernommen der Anteil, der die Helizit"aten der \vspace*{-.125ex}(Anti)Quarks beschreibt.
Ersetzt wird der Photon-spezifische Energienenner%
  ~\vspace*{-.125ex}\mbox{\,$(\rb{k}^2 \!+\! \vep^2)^{-1}$} durch die $1S$-Wellenfunktion des transversalen Harmonischen Oszillators~\vspace*{-.125ex}\mbox{\,$\exp\frac{1}{2}\om^2r^2$} mit~$\om$ dem Oszillatorparameter, die $\zet$-Abh"anggikeit wird repr"asentiert nach Stech, Wirbel, Bauer.
Die zwei Parameter des Ansatzes werden fixiert durch die Forderung von Normierung und Reproduktion des experimentellen Zahlenwerts f"ur die Kopplung an den elektromagnetischen Strom.
Die eigentlich relevante transversale Ausdehnung der Dipole wird hier in "ahnlicher Weise bestimmt wie im \vspace*{-.125ex}Photon, wobei die \mbox{Besselfunktionen entsprechend ersetzt sind durch die Gau"s'schen Funktionen}. \\
\indent
Auf Basis dieser hadronischen Wellenfunktionen werden diskutiert \mbox{$t$-diffe}\-rentielle Wirkungsquerschnitte%
  ~\vspace*{-.125ex}\mbox{\,$d\si\!\big/\!dt$} als Funktion von~$t$ und ihr Integral~$\si$ im Bereich~\mbox{\,$-t \!=\! 0 \!-\! 0.6\GeV^2$}, beide Gr"o"sen im betrachteten Bereich von~\vspace*{-.125ex}$Q^2$ zwischen~\mbox{\,$1 \!-\! 2$} und~$10\GeV^2$ und f"ur longitudinale und transversale Polarisation~\vspace*{-.125ex}\mbox{\,$\la \!\equiv\! L,T$} von Photon/Vektormeson. \\
\indent
Wir analysieren zun"achst~-- am Beispiel von \mbox{$\rh(770)$-Pro}\-duktion als Prototyp mit der gr"o"sten transversalen Ausdehnung~-- die allgemeine \vspace*{-.125ex}Charakteristik unserer Postulate.
Dies involviert die \mbox{$Q^2 \!\to\! 0$}- und die \mbox{$Q^2 \!\to\! \infty$-Asymp}\-totik.
Wir finden, da"s der betrachtete Bereich von~$Q^2$ intermedi"ar ist, das hei"st weder bezeichnet werden kann als asymptotisch klein noch als asymptotisch gro"s; dasselbe Resultat konstatieren wir f"ur die experimentellen Wirkungsquerschnitte.
Wir diskutieren die Wirkungsquerschnitte hin auf die transversale Ausdehnung~$r$ der Dipole, die sie wesentlich bestimmen.
F"ur longitudinale Polarisation tragen wesentlich bei noch \mbox{Dipole mit~$r$ bis~$1.5\fm$, f"ur transversale Polarisation sogar bis "uber~$2\fm$}. \\
\indent
In der expliziten Gegen"uberstellung der von uns postulierten Wirkungsquerschnitte und der experimentellen Daten finden wir f"ur alle betrachteten Observablen gute bis sehr gute "Ubereinstimmung~-- mit genau einer Ausnahme.
Wir verzichten darauf, die Reihe der Observablen an dieser Stelle noch einmal aufzuz"ahlen und verweisen auf die Abbildungen und deren Diskussion im Haupttext.
Wir gehen nur ein auf die eine Diskrepanz.
Diese bezieht sich auf den \mbox{$t$-inte}\-grierten Wirkungsquerschnitt f"ur \mbox{$\ph(1020)$-Pro}\-duktion.
Wir verifizieren das von \DREI{N}{M}{C} beobachtete Skalierungsverhalten wie~$1\!\big/\!Q^4$, postulieren aber absolut etwa das Doppelte des experimentellen Wirkungsquerschnitts.
Wir merken an im Haupttext, da"s wir mit demselben funktionalen Ansatz arbeiten f"ur die Lichtkegelwellenfunktionen s"amtlicher Vektormesonen; Modifizierung des \mbox{$\ph$-An}\-satzes hin zu einer st"arker um~\mbox{$\zet \!=\! 1\!\big/\!2$} gepeakten Funktion f"uhrte zu einer Verringerung des postulieretn Wirkungsquerschnitts ohne die \mbox{$1\!\big/\!Q^4$-Ska}\-lierung zu ver"andern.
Wir nehmen Abstand von einem solchen {\it fine tuning\/}.
Tats"achlich postulieren wir in Kapitel~\ref{Kap:EXCITED} f"ur Photoproduktion von~$\ph(1020)$ einen Zahlenwert, der innerhalb des experimentellen Fehlerbereichs liegt und sogar kleiner als der zentrale Wert.
Wir sehen dies als Argument daf"ur, da"s auch die Daten f"ur Leptoproduktion mit einem gr"o"seren Fehler behaftet sein k"onnten als angegeben. \\
\indent
Wir betonen, da"s die gute bis sehr gute "Ubereinstimmung unserer Postulate mit den experimentellen Daten zu sehen ist vor dem Hintergrund, da"s das \DREI{M}{S}{V} vollst"andig bestimmt ist durch drei Parameter: die Korrelationsl"ange~$a$, das Gluonkondensat~$\vac{g^2FF}$ und der Proton-Radius bei der Referenzenergie von~\mbox{\,$\surd s \!=\! 20\GeV$}.
Diese Parameter sind a~priori festgelegt innerhalb sehr eng beschr"ankter Intervalle.
Sie sind dar"uberhinaus fixiert durch Forderungen, die allenfalls in mittelbarem Zusammenhang stehen mit den hier postulierten Wirkungsquerschnitten: durch die Forderung von Konsistenz der \DREI{M}{S}{V}-Stringspannung eines statischen Quark-Antiquark-Paares mit dem Wert numerischer Gittersimulationen und der Forderung von Reproduktion des totalen Wirkungsquerschnitts und des {\it slope\/}-Parameters in Proton-Proton-Streuung f"ur die Referenzenergie von~\mbox{\,$\surd s \!=\! 20\GeV$}.
\vspace*{-.5ex}

\bigskip\noindent 
{\bf Kapitel~\ref{Kap:EXCITED}}\quad
liegt zugrunde~-- wie Kapitel~\ref{Kap:GROUND}~-- die \vspace*{-.125ex}\mbox{\,$s \!\to\! \infty$-asymp}\-totische Formel f"ur die Loop-Loop-Amplitude~$\tTll$.
Wir untersuchen exklusive Photo- und Leptoproduk\-tion%
  ~\vspace*{-.125ex}\mbox{\,$\ga^{\scriptscriptstyle({\D\ast})}p \!\to\! V p$} der%
  ~\vspace*{-.125ex}\mbox{$1^+(1^{--})$-/Rho-Vek}\-tormesonen%
  ~\vspace*{-.125ex}\mbox{$V \!\equiv\! \rh(770), \rh(1450), \rh(1700)$}.
Gegen"uber der Analyse in Kapitel~\ref{Kap:EXCITED} ist zum einen der Bereich von~$Q^2$ erweitert zu kleineren bis verschwindenden Werten, zum anderen betrachtet h"ohere Rho-Anregungen.
Beides impliziert gr"o"sere transversale Ausdehnungen der involvierten Dipole.
Der \DREI{M}{S}{V}-spezifische String-String-Mechanismus~-- die Ausbildung und Wechselwirkung gluonischer Flu"sschl"auche zwischen den Quark-Konstituenten~-- wird noch weitgehender und effektiver untersucht.
Konzeptionell erfordert diese Analyse zum einen die Kenntnis der Lichtkegelwellenfunktion des Photons auch f"ur kleinere bis verschwindende~$Q^2$, zum anderen die der h"oheren Rho-Anregungen. \\
\indent
Dosch, Gousset, Pirner diskutieren, da"s sehr weitgehende Analogie besteht zwischen dem transversalen (zweidimensionalen) Harmonischen Oszillator und der Photon-Lichtkegelwel\-lenfunktion der perturbativen Theorie auf dem Lichtkegel.
F"ur den Harmonischen Oszillator sind die Greenfunktionen analytisch bekannt.
Wir f"uhren den Gang der Argumentation ausf"uhrlich aus im Haupttext.
Resultat ist, da"s~-- "ubertragen auf die Lichtkegelfunktion des Photons~-- diese im stark wechselwirkenden Bereich in einfacher Weise approximiert werden kann durch einen {\it shift\/} von~$Q^2$, der in definierter Weise abh"angt von~$Q^2$,~-- besser als in Form der Superposition zahlreicher Residuen im Sinne des \vspace*{-.125ex}Vektormeson-Dominanz-Modells.
Mit~\vspace*{-.125ex}\mbox{\,$\vep \!\equiv\! \surd\zbz Q^2 \!+\! m_f{}^{\!2}
     \to \surd\zbz Q^2 \!+\! \meff[f,]^2(Q^2)$} wird dieser \mbox{$Q^2$-ab}\-h"angige {\it shift\/} absorbiert in einer effektiven Quarkmasse.
Diese Funktion berechnen Dosch, Gousset, Pirner f"ur~$up/down$ und~$strange$.
Resultat ist, da"s~$\meff[f,](Q^2)$ oberhalb einer Schwelle~$Q_{f,0}^2$ identisch ist der laufenden Quarkmasse~$m_f$ und von~$Q_{f,0}^2$ an~-- im wesentlichen~-- linear ansteigt auf den Wert~\mbox{\,$m_{f,0} \!>\! m_f$} einer Konstituentenmasse f"ur~\mbox{\,$Q^2 \!\equiv\! 0$}.
Wir legen unserer Analyse zugrunde diese universelle Lichtkegelwellenfunktion des Photons mit den expliziten Werten~\mbox{\,$Q_{u\!/\!d,0}^2 \!\equiv\! 1.05\GeV^2$},~\mbox{\,$m_{u\!/\!d,0} \!\equiv\! 0.220\GeV$} und~\mbox{\,$m_{u\!/\!d} \!\equiv\! 0$}.%
\FOOT{
  der Photoproduktion von~$\ph(1020)$ die Werte~\mbox{\,$Q_{s,0}^2 \!\equiv\! 1.6\GeV^2$},~\mbox{\,$m_{s,0} \!\equiv\! 0.310\GeV$} und~\mbox{\,$m_s \!\equiv\! 0.150\GeV$}
}
\\\indent
In Bezug auf die Lichtkegelwellenfunktionen der Rho-Anregungen~$\rh(1450)$,~$\rh(1700)$ diskutieren wir zun"achst deren Massespektren in Photoproduktion am Proton und \mbox{$e^+e^-$-Anni}\-hilation.
Im Bereich der invarianten Masse von~\mbox{\,$1.6\GeV$} wird beobachtet respektive eine konstruktive und destruktive Interferenz; dies ist nicht zu erkl"aren auf Basis nur einer Resonanz.
Die Analyse der Spektren und ph"anomenologischen Zerfallskan"ale zeigt, da"s zwei Resonanzen stark koppeln an den elektromagnetischen Strom.
Wir machen den Ansatz eines (Quark-Antiquark-)$2S$-Zustands und eines "`Rests"', dessen Kopplung an den elektromagnetischen Strom unterdr"uckt ist~-- etwa der $2D$- oder hybride Zust"ande wie Quark-Antiquark-Gluon.
Kopplung an den elektromagnetische Strom und Mischungswinkel folgen aus den Spektren.
In den betrachteten Produktionsprozessen tr"agt per definitionem der~"`Rest"' nur vernachl"assigbar bei.
Es gen"ugt, eine \mbox{Wellenfunktion anzugeben f"ur den $2S$-Zustand}. \\
\indent
Dies geschieht vollst"andig analog zum (Quark-Antiquark-)$1S$-Zustand des~$\rh(770)$.
Dabei wird der Anregung Rechnung getragen durch einen "`transversalen"' Knoten in Form der $2S$-Wellenfunktion des transversalen Harmonischen Oszillators und einem "`longitudinalen"' Knoten.
Der Ansatz involviert zwei Parameter, die fixiert werden durch die Forderung von Normierung und Orthogonalit"at bez"uglich der $1S$-Wellenfunktionen.
Die Oszillatorparameter~$\om_{2S,L}$,~$\om_{2S,T}$ werden zun"achst "ubernommen von der $1S$-Funktion, dann in der Weise minimal ge"andert, da"s folgen identische Kopplungen an den elektromagnetischen Strom f"ur longitudinale und transversale Polarisation. \\
\indent
Auf dieser Basis werden explizit berechnet \mbox{$t$-dif}\-ferentielle Wirkungsquerschnitte%
  ~\vspace*{-.125ex}\mbox{\,$d\si\!\big/\!dt$} als Funktion von~$t$ und ihr Integral~$\si$ im Bereich~\mbox{\,$-t \!=\! 0 \!-\! 0.6\GeV^2$}.
Wir beziehen uns hier auf Werte von~$Q^2$ von~\mbox{\,$0 \!-\! 20\GeV^2$} und \mbox{diskutieren longitudinale und transversale Polarisation}. \\
\indent
Zun"achst diskutieren wir wieder die allgemeine Charkteristik unserer Postulate.
Im effektiven "Uberlapp der Photon- und $2S$-Lichtkegelwellenfunktion als Funktion der transversalen Ausdehnung~$r$ der korrespondierenden Dipole identifizieren wir ein subtiles Wechselspiel zwischen den Beitr"agen rechts und links des Knotens, die eingehen mit entgegengesetztem Vorzeichen.
So dominieren f"ur kleiner werdende Werte von~$Q^2$ die "au"seren positiven Beitr"age "uber die inneren negativen.
Die \mbox{$T$-Amp}\-litude in Photoproduktion~--\mbox{\,$Q^2 \!\equiv\! 0$}, gro"se Dipole~-- hat umgekehrtes Vorzeichen wie die \mbox{$T$-Amp}\-litude in \mbox{$e^+e^-$-Anni}\-hilation~-- kleine Dipole.
Dies erkl"art die Umkehr des Interferenzmusters in den Massespektren.
Wesentlich ist, da"s das Wechselspiel rechts-versus-links-des-Knotens darauf basiert, da"s der Dipol-Proton-Wirkungsquerschnitt nicht abs"attigt, sondern ansteigt mit~\vspace*{-.125ex}$r$, und dies wiederum "uber die $C$-Funktion%
  ~\vspace*{-.125ex}\mbox{$\tilde{X}\oC\idx{\mfp\mskip-2mu\mfm}$} unmittelbare Konsequenz ist des \DREI{M}{S}{V}-spezifischen String-String-Mechanismus.
Perturbative Modelle auf Basis Quark-additiver Wechselwirkung f"uhren auf eine abs"attigende $N\!C$-Funktion allein und erkl"aren daher nicht die Rho-Massespektren. \\
\indent
Wir finden ferner, da"s wesentlich beitragen zu den Wirkungsquerschnitten Dipole mit transversaler Ausdehnung~$r$ bis~$2\fm$ f"ur longitudinale und bis~$3\fm$ f"ur transversale Polarisation.
Dies ist sicherlich ein Bereich, in dem die Beschreibung durch perturbative~\DREI{Q}{C}{D} versagt und die allgemein konstatierte gute bis sehr gute "Ubereinstimmung unserer Postulate mit den experimentellen Daten aufgefa"st werden kann als beeindruckende Verifizierung des~\DREI{M}{S}{V} und unseres Zugangs im allgemeinen.
Diese "Ubereinstimmung ist ausf"uhrlich dokumentiert in den Abbildungen und deren Diskussion im Haupttext, so da"s wir sie hier nicht noch einmal rekapitulieren.
Wir schlie"sen \vspace*{20.25ex}Resum\'ee und Arbeit.
\theendnotes

%% file: APP_ALGEBRA.tex
\lhead[\fancyplain{}{\sc\thepage}]
      {\fancyplain{}{\sc\rightmark}}
\rhead[\fancyplain{}{\sc{{\footnotesize Anhang~\thechapter:} Algebra}}]
      {\fancyplain{}{\sc\thepage}}
\psfull
\chapter[Algebra]{%
   \huge Algebra}
\label{APP:Algebra}   

In diesem Anhang geben wir grundlegende Definitionen und Relationen an, die die Algebra bez"uglich Gr"o"sen mit Lorentz-, Dirac- beziehungsweise Eichgruppenindizes betrifft.
Wir legen keinen Wert auf Vollst"andigkeit.
Absicht aber ist, die im Haupttext benutzten Definitionen und Relationen festzuhalten.

\section{Lorentz-Algebra}
\label{APP:Lorentz-Algebra}

Wir geben zun"achst an unsere generellen Konventionen und Definitionen bez"uglich kontra- und kovarianter Komponenten von Tensoren, bez"uglich metrischem Tensor, Epsilon-Pseudo\-tensor und Basen des Minkowski-Raumes.
Wir halten dann fest unsere Konventionen in Zusammenhang mit Lichtkegelkoordinaten.

\subsection{Generelle Konventionen und Definitionen}
\label{APP-Subsect:KonvDef}

Ein Lorentz-Vektor~$x$ sei definiert als Spaltenvektor seiner {\it kontravarianten Komponenten\/}:
\vspace*{-.5ex}
\begin{align} \label{APP:kontrav}
x\;
  =\; \big(x^0,x^1,x^2,x^3\big){}^{\T t}\;
  \equiv\; \big(x^\mu\big)\qquad
  \mu \in \{0,1,2,3\}
    \\[-4.5ex]\nn
\end{align}
Das Lorentz-invariante {\it Vierer-Skalarprodukt\/} zweier Vektoren~$x$,~$y$ ist gegeben durch
\vspace*{-.5ex}
\begin{align} 
x\cdot y\;
  =\; g_{\mu\nu}\, x^\mu\, y^\nu\;
  =\; x_\mu\; y^\mu
    \\[-4.5ex]\nn
\end{align}
mit Definition {\it kovarianter Komponenten\/} durch
\vspace*{-.5ex}
\begin{align} \label{APP:kovariant}
x_\mu\;
  :=\; g_{\mu\nu}\, x^\nu
    \\[-4.5ex]\nn
\end{align}
Dabei ist~$g$ der {\it metrische Tensor\/}, der definiert ist durch seine kovarianten Komponenten:
\vspace*{-.5ex}
\begin{align} \label{APP:g}
g\; \equiv\; \big(g_{\mu\nu}\big)\;
  =\; {\rm diag}[+1,-1,-1,-1]
    \\[-4.5ex]\nn
\end{align}
Seine kontravarianten Komponenten~$g^{\mu\nu}$ sind definiert durch~$x^\mu \!=:\! g^{\mu\nu}x_\nu$, das hei"st durch die inverse Relation zu Gl.~(\ref{APP:kovariant}); daraus folgt:\FOOT{
  Sei die zweite und letzte Identit"at verstanden als suggestive Notation.
}
%
\vspace*{-.5ex}
\begin{align} 
\de^\mu_\nu\vv
  =\; g^{\mu\si}\, g_{\si\nu}\;
  =\; g^\mu{}_\nu\vv
  =\; g_{\nu\si}\, g^{\si\mu}\;
  =\; g_\nu{}^\mu
    \\[-4.5ex]\nn
\end{align}
das hei"st die Matrix der kontravarianten Komponenten~$\big(g^{\mu\nu}\big)$ spielt die Rolle der Inversen~$g^{-1}$ und ist identisch mit~$g$ aufgrund deren expliziten Gestalt, vgl.\@ Gl.~(\ref{APP:g}):
\vspace*{-.5ex}
\begin{align} 
g^{-1}\;
  \equiv\; \big(g^{\mu\nu}\big)\;
  =\; {\rm diag}[+1,-1,-1,-1]\;
  =\; g
    \\[-3.5ex]\nn
\end{align}

Der {\it Epsilon-Pseudotensor\/} ist allgemein definiert als das Signum der Permutation seiner Indizes,~-- durch seine kontravarianten Komponenten,~\mbox{$\ep \!\equiv\! (\ep^{\mu\nu\rh\si})$ mit~$\ep^{0123} \!\equiv\! 1$}, wie folgt:
%
\begin{align} \label{APP:epTensor-kontrav}
&\ep^{\mu_0\cdots\mu_3}\;
  \equiv\; \pmatrixZD{\mu_0}{\cdots}{\mu_3}{0}{\cdots}{3}\; \ep^{0123}\qquad
  \Longrightarrow\quad
  \ep_{\mu_0\cdots\mu_3}
  \equiv \pmatrixZD{\mu_0}{\cdots}{\mu_3}{0}{\cdots}{3}\; \ep_{0123}
    \\[.25ex]
&\text{\it per definitionem:}\qquad
  \ep^{0123}\;
  \equiv\; +1\qquad
  \text{\it folglich:}\qquad
  \ep_{0123} \equiv -1
    \tag{\ref{APP:epTensor-kontrav}$'$}
\end{align}
Dabei ist
\vspace*{-.25ex}
\begin{align} \label{APP:SignumPermutation}
\pmatrixZD{a_1}{\cdots}{a_s}{b_1}{\cdots}{b_s}\;
  :=\; {\rm sign}(\si^{\D\mskip-1mu\star})\qquad
  \text{$\si^{\D\mskip-1mu\star} \!\in\! S_s$\vv
        mit\vv $\si^{\D\mskip-1mu\star}(a_i) =  b_i,\;\forall i=1,\ldots s$}
    \\[-4.25ex]\nn
\end{align}
Standard-Notation f"ur das Signum der Indexpermutation~\mbox{$\si^{\D\mskip-1mu\star}\!: a_i \!\to\! b_i$}; vgl.\@ etwa Ref.~\cite{Klingmann93}. \\
\indent
Wir notieren kurz Symmetrisierung und Antisymmetrisierung von Tensorindizes, indem wir diese schreiben in geschweiften beziehungsweise eckigen Klammern:
\vspace*{-.5ex}
\begin{alignat}{2} \label{APP:(Anti)Symmetrisierung}
&T^{\{\mu_1\cdots\mu_s\}}&\;
  &:=\;
    \frac{1}{s!}\;
      {\T\sum}_{\T \si \!\in\! S_s}\vv
      T^{\mu_{\si(1)}\cdots\mu_{\si(s)}}
    \\[.5ex]
&T^{[\mu_1\cdots\mu_s]}&\;
  &:=\;
    \frac{1}{s!}\;
      {\T\sum}_{\T \si \!\in\! S_s}\vv
      {\rm sign}(\si)\vv
      T^{\mu_{\si(1)}\cdots\mu_{\si(s)}}
    \tag{\ref{APP:(Anti)Symmetrisierung}$'$}
    \\[-4.5ex]\nn
\end{alignat}
Von der (Anti-)Symmetrisierung ausgenommene Indizes seien notiert zwischen vertikalen Strichen, etwa:~$T^{\la[\mu|\nu\rh|\si]} \!=\! 1\!/2!\, (T^{\la\mu\nu\rh\si} \!-\! T^{\la\si\nu\rh\mu})$. \\
\indent
Eine {\it Basis\/} des Minkowski-Raumes ist gegeben durch vier linear unabh"angigge Lorentz-Vektoren~$e_{(\mu)} \!\equiv\! \big(e_{(\mu)}{}^\nu\big)$,~$\mu \!\in\! \{0,1,2,3\}$.%
~Das Vierbein~$\{e_{(\mu)}\}$ ist {\it orthonormierte Basis\/}, falls:\zz
\vspace*{-.5ex}
\begin{align} \label{APP:e_(mu)-Norm}
&e_{(\mu)}\cdot e_{(\nu)}\;
  =\; g_{\mu\nu}
    \\[.5ex]
  &\text{d.h.}\qquad
  e_{(0)}\cdot e_{(0)} = +1\qquad
  e_{(1)}\cdot e_{(1)} = e_{(2)}\cdot e_{(2)} = e_{(3)}\cdot e_{(3)} = -1
    \nn
    \\[-4.5ex]\nn
\end{align}
Seien weiter definiert Vektoren~$e^{(\mu)} \!\equiv\! \big(e^{(\mu)\nu}\big)$ durch
\vspace*{-1.25ex}
\begin{align} \label{APP:e^(mu)-Def}
e^{(\mu)}\;
  :=\; g^{\mu\nu}\, e_{(\nu)}
    \\[-4.25ex]\nn
\end{align}
Wegen~$0 \!\ne\! \det \big(g^{\mu\nu}\big) \!=\! \det g^{-1} \!=\! (\det g)^{-1}$ bilden diese wieder eine Basis des~\mbox{Minkowski-Rau}\-mes.\FOOT{
  Unsere Notation deute an, da"s unter Lorentz-Transformatioenen die~$e_{(\mu)}$ wie kovariante, die~$e^{(\mu)}$ wie kontravariante Tensorkomponenten transformieren; vgl.\@ Anh.~\ref{APP:Boosts}.
}
Mit~\mbox{$e^{(\mu)}\cdot e^{(\nu)} \!=\! g_{\rh\si}\, e^{(\mu)\rh}\, e^{(\nu)\si} \!=\! g_{\rh\si}\, g^{\mu\al}\, e_{(\al)}{}^\rh\, g^{\nu\be}\, e_{(\nu)}{}^\si \!=\! g^{\mu\al}\, g^{\nu\be}\, e_{(\al)}\cdot e_{(\be)}$} folgt
\vspace*{-.5ex}
\begin{align} 
e^{(\mu)}\cdot e^{(\nu)}\;
  =\; g^{\mu\nu}
    \\[-4.5ex]\nn
\end{align}
unmittelbar aus Gl.~(\ref{APP:e_(mu)-Norm}),
das hei"st mit~$\{e_{(\mu)}\}$ ist~$\{e^{(\mu)}\}$ orthonormiert.
F"ur einen beliebigen Lorentz-Vektor~$x$ existieren eindeutige Zerlegungen:
\vspace*{-.5ex}
\begin{align} \label{APP:x-Zerlegung}
x\;
  =\; x^\mu\, e_{(\mu)}\;
  =\; x_\mu\, e^{(\mu)}
    \\[-4.5ex]\nn
\end{align}
Die Komponenten~$x^\mu$,~$x_\mu$ bez"uglich~$\{e_{(\mu)}\}$,~$\{e^{(\mu)}\}$ folgen umgekehrt als Projektionen:
\vspace*{-.5ex}
\begin{align} \label{APP:x-Projektion:x^mu,x_mu}
x^\mu\;
  =\; x\cdot e^{(\mu)}\qquad
  \text{bzw.}\qquad
x_\mu\;
  =\; x\cdot e_{(\mu)}
    \\[-4.5ex]\nn
\end{align}
Diese Gleichungen sind Tensorgleichungen, die gelten in einem beliebigen (inertialen)~Koordinatensystem.
In diesem Sinne werden konstruiert Vierbeine~\mbox{$\{\bar{e}_{(\bar{\mu})}\}$},~\mbox{$\bar{\mu} \!\in\! \{+,-,1,2\}$}, und~\mbox{$\{\tilde{e}_{(\tilde{\mu})}\}$},~\mbox{$\tilde{\mu} \!\in\! \{\mfp,\mfm,1,2\}$} deren Vektoren verlaufen in Richtung der Lichtkegel-Koordinaten\-linien bzw.\@ der Weltlinien zweier aktiv geboosteter Teilchen.
Vgl.\@ Anh.~\ref{APP:LC-Koord} bzw.\@ Anh.~\ref{APP:Boosts}. \\
\indent
Es sind realisiert {\it karthesische\/} Koordinaten durch explizite Definition der~$e_{(\mu)} \!\equiv\! (e_{(\mu)}{}^\nu)$ als die Einheitsvektoren in Richtung der (kontravarianten) karthesischen Koordinatenachsen:
\begin{samepage}
%
\begin{align} \label{APP:e_(mu)-karthesisch}
e_{(\mu)}{}^\nu\;
  :=\; g_\mu{}^\nu\;
  =\; \de_\mu^\nu
\end{align}
Mit~$e^{(\mu)\nu} \!=\! g^{\mu\rh}\, e_{(\rh)}{}^\nu$ folgt daraus unmittelbar:
\vspace*{-.5ex}
\begin{align} 
e^{(\mu)\nu}\;
  =\; g^{\mu\nu}
    \\[-4.5ex]\nn
\end{align}
Es sind~$x^\mu$,~$x_\mu$ die konventionellen karthesischen kontra-/kovarianten Komponneten von~$x$.
\end{samepage}

\subsection{Lichtkegelkoordinaten}
\label{APP:LC-Koord}

Sei~$x \!\equiv\! (x^\mu)$, $\mu \!\in\! \{0,1,2,3\}$, ein beliebiger Lorentz-Vektor nach Gl.~(\ref{APP:kontrav}).
Seine Darstellung in Lichtkegelkoordinaten ist analog definiert durch seine {\it kontravarianten\/} Komponenten:\FOOT{
  Tensoren und Tensorkomponeten seien generell bezeichnet im Sinne~\mbox{$T \!\equiv\! \big(T^{\mu_1\cdots\mu_s}\big)$} versus~\mbox{$\bar{T} \!\equiv\! \big(T^{{\bar\mu}_1\cdots{\bar\mu}_s}\big)$}.
}
%
\begin{align} \label{APP:kontravLC}
{\bar x}\;
  =\; \big(x^+,x^-,x^1,x^2\big){}^{\T t}\;
  \equiv\; \big(x^{\bar\mu}\big)\qquad
  {\bar\mu} \in \{+,-,1,2\}
\end{align}
Seien diese definiert durch:
%
\begin{align} \label{APP:kontravLC_Def}
x^\pm\;
  :=\; \al\, (x^0 \!\pm\! x^3)\qquad
x^{\bar i}\;
  :=\; x^i\quad i \in \{1,2\}
\end{align}
in Matrixnotation~$\bar{x} \!=\! \mathbb{L}\, x$, explizit:%
\FOOT{
  \label{APP-FN:longKomp}Seien hier und im folgenden ausgeschrieben nur die nichttrivialen longitudinalen Komponenten.
}
%
\vspace*{-.5ex}
\begin{align}
x^{\bar\mu}\;
  =\; \mathbb{L}^{\bar\mu}{}_\nu\; x^\nu\qquad
  \text{mit}\qquad
\mathbb{L}
  \equiv \big(\mathbb{L}^{\bar\mu}{}_\nu\big)
  = \al\, \pmatrixZZ{1}{1}{1}{-1}\qquad
\det\mathbb{L}
  = -2\al^2
    \tag{\ref{APP:kontravLC_Def}$'$}
    \\[-4.5ex]\nn
\end{align}
und~$\al \!\in\! \bbbr^+$ einer noch offenen Normierung, die wir in praxi setzen:~$\al \!\equiv\! 1\!/\surd2$ [in der Literatur nicht einheitlich, h"aufig~$1$ oder~$1\!/\!\surd2$].
Invertiert gilt:
\vspace*{-.5ex}
\begin{align} \label{APP:kontravLC_Def^-1}
x^0\;
  =\; \frac{1}{2\al}\, \big(x^+ \!+\! x^-\big)\qquad
x^3\;
  =\; \frac{1}{2\al}\, \big(x^+ \!-\! x^-\big)\qquad
x^i\;
  =\; x^{\bar i}\quad
  i \in \{1,2\}
    \\[-4.5ex]\nn
\end{align}
mit der kontragredienten Matrix:\citeFN{APP-FN:longKomp}
\vspace*{-.5ex}
\begin{align}
x^\mu\;
  =\; \mathbb{L}_{\bar\nu}{}^\mu\; x^{\bar\nu}\qquad
  \text{mit}\qquad
\big(\mathbb{L}_{\bar\mu}{}^\nu\big)
  \equiv \big(g_{\bar\mu\bar\rh}\,
    \mathbb{L}^{\bar\rh}{}_\si\,
    g^{\si\nu}\big)
  = \mathbb{L}^{-1\T\:t}
  = \frac{1}{2\al}\, \pmatrixZZ{1}{1}{1}{-1}
    \tag{\ref{APP:kontravLC_Def^-1}$'$}
    \\[-4.5ex]\nn
\end{align}
Der metrische Tensor ist dabei definiert durch~\mbox{$\bar{g} \!\equiv\! \big(g_{\bar{\mu}\bar{\nu}}\big)$}, das hei"st durch die Matrix seiner kovarianten Komponenten.%
\FOOT{
  Die folgenden Definitionen, Konventionen, Zusammenh"ange stehen in vollst"andiger Analogie zu~\ref{APP-Subsect:KonvDef}.
}
Es wird gefordert Invarianz des Vierer-Skalarprodukts, das hei"st formal f"ur~$x$,~$y$ zwei beliebige Lorentz-Vektoren:
\vspace*{-1ex}
\begin{align} \label{APP:xy=!xybar}
x\cdot y\;
  =\; g_{\mu\nu}\, x^\mu\, y^\nu\;
\stackrel{\D!}{=}\; {\bar x}\cdot {\bar y}\;
  =\; g_{{\bar\mu}{\bar\nu}}\, x^{\bar\mu}\, y^{\bar\nu}
    \\[-4.5ex]\nn
\end{align}
F"ur~\mbox{$\bar{g} \!\equiv\! \big(g_{\bar{\mu}\bar{\nu}}\big)$} folgen die Komponenten:
\vspace*{-.25ex}
\begin{align} \label{APP:gbar}
&g_{++}
  = g_{--}
  = 0\qquad
g_{+-}
  = g_{-+}
  = 1/(2\al^2)
  = -\, [\det\mathbb{L}]^{-1}
    \\[.25ex]
&g_{\bar{i}\bar{j}}
  = g_{ij}
  = -\de_{ij}\quad
      i \in \{1,2\}
    \tag{\ref{APP:gbar}$'$}
    \\[-4.25ex]\nn
\end{align}
Dann sind {\it kovariante\/} Lichtkegelkomponenten~\mbox{$x_{\bar\mu} \!=\! \mathbb{L}_{\bar\mu}{}^\nu\; x_\nu$} gegeben durch:
%
\vspace*{-.25ex}
\begin{align} \label{APP:kovariantLC}
x_{\bar\mu}\;
  =\; g_{{\bar\mu}{\bar\nu}}\, x^{\bar\nu}
    \\[-4.25ex]\nn
\end{align}
Durch~\mbox{$x^{\bar\mu} \!=:\! g^{\bar{\mu}\bar{\nu}}\,x_{\bar\nu}$}, das hei"st durch die zu Gl.~(\ref{APP:kovariantLC}) inverse Relation sind definiert die kontravarianten Komponenten des metrischen Tensors.
Es folgt:
\vspace*{-.25ex}
\begin{align} \label{APP:gbar^-1}
&g^{++}
  = g^{--}
  = 0\qquad
g^{+-} 
  = g^{-+}
  = 2\al^2
  = - \det\mathbb{L}
    \\[.25ex]
&g^{\bar{i}\bar{j}}\;
  = g^{ij}
  = -\de^{ij}\quad
      i \in \{1,2\}
    \tag{\ref{APP:gbar^-1}$'$}
    \\[-4.25ex]\nn
\end{align}
Ihre Matrix spielt die Rolle der Inversen zu~$\bar{g}$, und wir schreiben:~$\bar{g}^{-1}\! \equiv\! \big(g^{\bar{\mu}\bar{\nu}}\big)$. \\
\indent
Der Epsilon-Pseudotensor:~\mbox{$\bar\ep \!\equiv\! \big(\ep^{\bar\mu\bar\nu\bar\rh\bar\si}\big)$} mit~\mbox{$\bar\mu,\bar\nu,\bar\rh,\bar\si \!\in\! \{+,-,1,2\}$}, ist definiert durch seine kontravarianten Komponenten, die folgen mithilfe der Leibnitz'schen Determinantenformel:
%
\begin{align} \label{APP:epTensorLC-kontrav}
&\ep^{\bar\mu\bar\nu\bar\rh\bar\si}\;
  =\; \mathbb{L}^{\bar\mu}{}_\al\,
      \mathbb{L}^{\bar\nu}{}_\be\,
      \mathbb{L}^{\bar\rh}{}_\ga\,
      \mathbb{L}^{\bar\si}{}_\de\,
      \ep^{\al\be\ga\de}\vv
  =\; \det\big(\mathbb{L}^{\bar\mu}{}_\al\big)\cdot
      \pmatrixZV{\bar\mu}{\bar\nu}{\bar\rh}{\bar\si}
                {+}{-}{1}{2}\;
       \ep^{0123}
    \\[.25ex]
&\text{mit}\qquad
  \det\big(\mathbb{L}^{\bar\mu}{}_\al\big)
  = \det\mathbb{L}
  = -\, 2\al^2
  = -\, g^{+-}
    \tag{\ref{APP:epTensorLC-kontrav}$'$}
\end{align}
entsprechend die kovarianten Komponenten:
%
\begin{align} \label{APP:epTensorLC-kov}
&\ep_{\bar\mu\bar\nu\bar\rh\bar\si}\;
  =\; \mathbb{L}_{\bar\mu}{}^\al\,
      \mathbb{L}_{\bar\nu}{}^\be\,
      \mathbb{L}_{\bar\rh}{}^\ga\,
      \mathbb{L}_{\bar\si}{}^\de\,
      \ep_{\al\be\ga\de}\vv
  =\; \det\big(\mathbb{L}_{\bar\mu}{}^\al\big)\cdot
       \pmatrixZV{\bar\mu}{\bar\nu}{\bar\rh}{\bar\si}
                 {+}{-}{1}{2}\;
       \ep_{0123}
    \\[.25ex]
&\text{mit}\qquad
  \det\big(\mathbb{L}_{\bar\mu}{}^\al\big)
  = [\det\mathbb{L}]^{-1}
  = -\, 1/(2\al^2)
  = -\, g_{+-}
    \tag{\ref{APP:epTensorLC-kov}$'$}
\end{align}
Unsere Konvention f"ur~\mbox{$\ep \!\equiv\! \big(\ep^{\mu\nu\rh\si}\big)$} ist~\mbox{$\ep^{0123} \!\equiv\! +1 \Rightarrow \ep_{0123} \!=\! -1$}, das Klammersymbol steht f"ur das Signum der Indexpermutation~\mbox{$(\bar\mu,\bar\nu,\bar\rh,\bar\si) \!\to\! (+,-,1,2)$}; vgl.\@ Gl.~(\ref{APP:epTensor-kontrav}$'$) bzw.~(\ref{APP:SignumPermutation}). \\
\indent
Seien~\mbox{$e_{(\bar{\mu})} \!\equiv\! \big(e_{(\bar{\mu})}{}^{\bar\nu}\big)$},~\mbox{$\bar{\mu},\bar{\nu} \!\in\! \{+,-,1,2\}$}, vier linear unabh"angige Lorentz-Vektoren mit
\vspace*{-.5ex}
\begin{align} \label{APP:e_(mu)LC-Norm}
e_{(\bar{\mu})}\cdot e_{(\bar{\nu})}\;
  =\; g_{\bar{\mu}\bar{\nu}}
    \\[-4.5ex]\nn
\end{align}
Dann ist das Vierbein~$\{e_{(\bar{\mu})}\}$ orthonormale Basis des Minkowski-Raumes, vgl.\@ Gl.~(\ref{APP:e_(mu)-Norm}); seien die Vektoren~$e_{(\bar{\mu})}$ f"ur~\mbox{$\bar\mu \!\in\! \{+,-,1,2\}$} ferner gew"ahlt als als die Einheitsvektoren in Richtung der (kontravarianten) Lichtkegel-Koordinatenlinien~$\{+,-,1,2\}$:
\vspace*{-.5ex}
\begin{align} \label{APP:e_(mu)LC-karthesisch}
e_{(\bar{\mu})}{}^{\bar\nu}\;
  :=\; g_{\bar\mu}{}^{\bar\nu}\;
  =\; \de_{\bar\mu}^{\bar\nu}
    \\[-4.5ex]\nn
\end{align}
Seien weiter definiert "`kontravariante"' Vektoren~\mbox{$e^{(\bar{\mu})} \!\equiv\! \big(e^{(\bar{\mu})\bar\nu}\big)$} durch:
\vspace*{-.5ex}
\begin{align} \label{APP:e^(mu)LC-Def}
e^{(\bar{\mu})}\;
  :=\; g^{\bar{\mu}\bar{\nu}}\, e_{(\bar{\nu})}\;
    \\[-4.25ex]\nn
\end{align}
Mit~\mbox{$e^{(\bar\mu)}\cdot e^{(\bar\nu)} \!=\! g^{\bar\mu\bar\al}\, g^{\bar\nu\bar\be}\; e_{(\bar\al)}\cdot e_{(\bar\be)}$} folgt unmittelbar aus Gl.~(\ref{APP:e_(mu)LC-Norm}):
\vspace*{-.25ex}
\begin{align} 
e^{(\bar{\mu})}\cdot e^{(\bar{\nu})}\;
  =\; g^{\bar{\mu}\bar{\nu}}
    \\[-4.25ex]\nn
\end{align}
und mit~$e^{(\bar\mu)\bar\nu} \!=\! g^{\bar\mu\bar\si}\, e_{(\bar\si)}{}^{\bar\nu}$ weiter aus Gl.~(\ref{APP:e_(mu)LC-karthesisch}):
\vspace*{-.25ex}
\begin{align} 
e^{(\bar{\mu})\bar\nu}\;
  =\; g^{\bar{\mu}\bar{\nu}}
    \\[-4.5ex]\nn
\end{align}
Dann besitzt ein beliebiger Lorentz-Vektor~$x$ eindeutige Zerlegungen:
\vspace*{-.5ex}
\begin{align} \label{APP:xbar-Zerlegung}
\bar{x}\;
  =\; x^{\bar\mu}\, e_{(\bar{\mu})}\;
  =\; x_{\bar\mu}\, e^{(\bar{\mu})}
    \\[-4.5ex]\nn
\end{align}
mit Komponenten:
\vspace*{-.5ex}
\begin{align} \label{APP:xbar-Projektion:x^mu,x_mu}
x^{\bar\mu}\;
  =\; \bar{x}\cdot e^{(\bar\mu)}\qquad
  \text{bzw.}\qquad
x_{\bar\mu}\;
  =\; \bar{x}\cdot e_{(\bar\mu)}\qquad
    \\[-4.5ex]\nn
\end{align}
Aufgrund der Definition der~$e_{(\bar\mu)}$ als die Einheitsvektoren in Richtung der (kontravarianten) Lichtkegel-Koordinatenachsen~-- Gl.~(\ref{APP:e_(mu)LC-karthesisch})~-- sind~$x^{\bar\mu}$,~$x_{\bar\mu}$ genau die konventionellen kontra- beziehungsweise kovarianten Lichtkegelkomponenten der Gln.~(\ref{APP:kontravLC_Def}),~(\ref{APP:kontravLC_Def}$'$) und~(\ref{APP:kovariantLC}).

\section[Dirac-Algebra]{%
         Dirac-Algebra~\bffootnote}
\label{APP:Dirac-Algebra}
\footnotetext{
  Bzgl.\@ der Dirac-Algebra orientieren wir uns in Notation und Konvention an den Refn.~\cite{Itzykson88,Nachtmann92}.
}
%

Definierende Relation der Dirac- oder Clifford-Algebra ist:
%
\begin{align} 
\{\ga^\mu,\ga^\nu\}\;
  \equiv\; \ga^\mu\,\ga^\nu\; +\; \ga^\nu\,\ga^\mu\;
  =\; 2\, g^{\mu\nu}\vv \bbbone
\end{align}
mit~$\mu,\nu \!\in\! \{0,1,2,3\}$.
Es ist~$\bbbone$ die Eins des Dirac-Raums von~$4\!\times\!4$-Matrizen und~$\ga^\mu$ vier Matrizen mit~$\ga^0$ hermitesch und unit"ar,~$\ga^i$ antihermitesch und antiunit"ar:
%
\begin{alignat}{2} \label{APP:ga_hermitesch/unitaer}
 \ga^0\; &=\; \ga^{0\D\dagger}&\; &=\; (\ga^0)^{-1}
    \\
-\ga^i\; &=\; \ga^{i\D\dagger}&\; &=\; (\ga^i)^{-1}
    \tag{\ref{APP:ga_hermitesch/unitaer}$'$}
    \\[-3.5ex]\nn
\end{alignat}
Weiter ist definiert~\mbox{$\ga_5 \!=\! \ga^5 \!=\! \iIM\ga^0\ga^1\ga^2\ga^3 \!=\! {\rm-i}/4!\, \ep_{\mu\nu\rh\si} \ga^\mu \ga^\nu \ga^\rh \ga^\si$} mit~$\ga_5^2 \!=\! \bbbone$ und~$\{\ga_5,\ga^\mu\} \!=\! 0$.
Die~$\ga^\mu$ sind spurlos; allgemein folgt f"ur Produkte von Gamma-Matrizen mit~$s$ ungerade:
\vspace*{-.5ex}
\begin{align} \label{APP:tr-ga_s=odd}
\tr\, \ga^{\mu_1}\, \cdots\, \ga^{\mu_s}\; =\; 0 \qquad
  \text{$\forall\; s$~ungerade}
    \\[-4.5ex]\nn
\end{align}
F"ur~$s$ gerade folgt:
\vspace*{-.5ex}
\begin{align} \label{APP:tr-ga_s=024}
&\tr\, \bbbone\; =\; 4 \\
  &\tr\, \ga^{\mu_1} \ga^{\mu_2}\; =\; 4\,g^{\mu_1\mu_2}
    \tag{\ref{APP:tr-ga_s=024}$'$} \\
  &\tr\, \ga^{\mu_1} \ga^{\mu_2} \ga^{\mu_3} \ga^{\mu_4}\;
        =\; 4\, \big( g^{\mu_1\mu_2}\, g^{\mu_3\mu_4}
                  - g^{\mu_1\mu_3}\, g^{\mu_2\mu_4}
                  + g^{\mu_1\mu_4}\, g^{\mu_2\mu_3} \big)
    \tag{\ref{APP:tr-ga_s=024}$''$}
    \\[-4.5ex]\nn
\end{align}
Hermitesche Konjugation wird vermittelt durch
\vspace*{-.5ex}
\begin{align} 
\ga^0\, \ga^\mu\, \ga^0\; =\; \ga^{\mu\D\dagger}
    \\[-4.5ex]\nn
\end{align}
Ladungskonjugation durch
\vspace*{-.5ex}
\begin{align} 
C^{-1}\, \ga^\mu\, C\; =\; -\ga^{\mu{\T t}} \qquad
  -C = C^{\T t} = C^{\D\dagger} = C^{-1}
    \\[-4.5ex]\nn
\end{align}
das hei"st~$C$ ist antisymmetrisch, antihermitesch und unit"ar.
Wir bezeichnen
\vspace*{-.5ex}
\begin{align} 
\si^0\; =\; \bbbone\;
  =\; \pmatrixZZ{1}{0}{0}{1}\qquad
    \\[-4.5ex]\nn
\end{align}
und definieren die Pauli-Matrizen durch
\vspace*{-.5ex}
\begin{align} \label{APP:PauliMatrizen_si^i}
\si^1\;
  =\; \pmatrixZZ{0}{1}{1}{0}\qquad
\si^2\;
  =\; \pmatrixZZ{0}{\rm-i}{\iIM}{0}\qquad
\si^3\;
  =\; \pmatrixZZ{1}{0}{0}{-1}
    \\[-4.5ex]\nn
\end{align}
Die Dirac-Darstellung ist damit definiert durch
\vspace*{-.5ex}
\begin{alignat}{4} \label{APP:DiracDrst}
&\ga_{\rm Dirac}^0\;&
  &=\;& \si^3\, &\otimes\,& \bbbone\;
  &=\; \pmatrixZZ{\bbbone}{0}  {0}{-\bbbone}
    \\
&\ga_{\rm Dirac}^i\;&
  &=\;& {\iIM}\, \si^2\, &\otimes\,& \si^i\;
  &=\; \pmatrixZZ{0}{\si^i}  {-\si^i}{0}
    \tag{\ref{APP:DiracDrst}$'$}
    \\[-4.5ex]\nn
\end{alignat}
und~$\ga_5 \!=\! \si^1 \otimes \bbbone$.
Die Gamma-Matrizen einer beliebigen anderen Darstellung folgen aus
\begin{samepage}
\vspace*{-.5ex}
\begin{align} 
\ga^\mu\;
  =\; U\, \ga_{\rm Dirac}^\mu\, U^{\D\dagger}
    \\[-4.5ex]\nn
\end{align}
Die Dirac-Spinoren~$u$,~$v$ sind Funktionen des {\it On~mass-shell\/}-Impulses~$p$, f"ur den also gilt: \mbox{$p_{0+} \!=\! \surd\vec{p}^{\,2} \!+\! m^2$}, und l"osen die freie Dirac-Gleichung:
\vspace*{-.5ex}
\begin{alignat}{5} \label{APP:uv_DiracGl}
&(\ga^\mu p_\mu - m)&\; u_s\!(p)\; &=&\; 0 \qquad
  &\text{bzw.}&\qquad
  &({\iIM}\ga^\mu \paR_\mu - m)\; u_s\!(p)\, \efn{{\rm-i}\,p x}&\; &=\; 0
    \\[.5ex]
&(\ga^\mu p_\mu + m)&\; v_s\!(p)\; &=&\; 0 \qquad
  &\text{bzw.}&\qquad
  &({\iIM}\ga^\mu \paR_\mu + m)\; v_s\!(p)\,  \efn{{\iIM}\,p x}&\; &=\; 0
    \tag{\ref{APP:uv_DiracGl}$'$}
    \\[-4.5ex]\nn
\end{alignat}
Dirac-Konjugation impliziert Multiplikation mit~$\ga^0$; konjugierte Dirac-Spinoren sind definiert durch:
\vspace*{-.5ex}
\begin{align} \label{APP:uv-konj}
\overline{u}\; =\; u^{\D\dagger}\, \ga^0
    \\[.5ex]
\overline{v}\; =\; v^{\D\dagger}\, \ga^0
    \tag{\ref{APP:uv-konj}$'$}
    \\[-4.5ex]\nn
\end{align}
Sie l"osen die hermitesch konjugierte Dirac-Gleichung:
\end{samepage}
\vspace*{-.5ex}
\begin{alignat}{6} \label{APP:uv-konj_DiracGl}
&\overline{u}_s(p)\; (\ga^\mu p_\mu - m)&\; &=&\; 0 \qquad
  &\text{bzw.}&\qquad
  &\overline{u}_s(p)\,  \efn{{\iIM}\,p x}&\; &({\iIM}\ga^\mu \paL_\mu + m)&\; &=\; 0
    \\[.5ex]
&\overline{v}_s(p)\; (\ga^\mu p_\mu + m)&\; &=&\; 0 \qquad
  &\text{bzw.}&\qquad
  &\overline{v}_s(p)\, \efn{{\rm-i}\,p x}&\; &({\iIM}\ga^\mu \paL_\mu + m)&\; &=\; 0
    \tag{\ref{APP:uv-konj_DiracGl}$'$}
    \\[-4.5ex]\nn
\end{alignat}
F"ur die Projektoren~$\La_{(\pm)}$ auf Zust"ande positiver beziehungsweise negativer Energie folgt:
\vspace*{-.5ex}
\begin{alignat}{5} \label{APP:Projektoren-La_+-}
&\La_{(+)}&\; &\equiv&\;  &{\T\sum}_{s = \pm 1\!/\!2}\; u_s\!(p)\, \overline{u}_s\!(p)&\;
                &=&\;  &{\iIM}\, \ga^\mu p_\mu + m
    \\[.5ex]
&\La_{(-)}&\; &\equiv&\; -&{\T\sum}_{s = \pm 1\!/\!2}\; v_s\!(p)\, \overline{v}_s\!(p)&\;
                &=&\; -&{\iIM}\, \ga^\mu p_\mu + m
    \tag{\ref{APP:Projektoren-La_+-}$'$}
    \\[-4.5ex]\nn
\end{alignat}
Die Dirac-Spinoren $u$,~$v$ erf"ullen die Orthogonalit"atsrelationen
\vspace*{-.5ex}
\begin{alignat}{4} \label{APP:uv-Spinoren_Eins}
&\overline{u}_s\!(p)\, u_{s'}\!(p)&\;
  &=&\; -&\overline{v}_s\!(p)\, v_{s'}\!(p)&\; &=\; 2m\, \de_{ss'}
    \\[.5ex]
&\overline{v}_s\!(p)\, u_{s'}\!(p)&\;
  &=&\;  &\overline{u}_s\!(p)\, v_{s'}\!(p)&\; &=\; 0
    \tag{\ref{APP:uv-Spinoren_Eins}$'$}
    \\[-4.5ex]\nn
\end{alignat}
und die Relationen
\vspace*{-.5ex}
\begin{alignat}{2} \label{APP:uv-Spinoren_ga^mu}
&\overline{u}_s\!(p)\, \ga^\mu\, u_{s'}\!(p)&\; &=\; 2p^\mu\, \de_{ss'}
    \\[.5ex]
&\overline{v}_s\!(p)\, \ga^\mu\, v_{s'}\!(p)&\; &=\; 2p^\mu\, \de_{ss'}
    \tag{\ref{APP:uv-Spinoren_ga^mu}$'$}
    \\[-4.5ex]\nn
\end{alignat}
bei Einschub einer Gamma-Matrix.

\section[\protect$\suNc$-Eichalgebra]{%
         \bm{\suNc}-Eichalgebra~\bffootnote}
\label{APP:suNc-Eichalgebra}
\footnotetext{
  Allgemeine nichtabelsche Eichgruppen~$\mathfrak{G}$ wie~$\SUNc$,~$SO(\Nc)$,~$Sp(\Nc)$,~$G_2$,~$E_6$,~$F_4$,~$E_7$ werden diskutiert in Ref.~\cite{Cvitanovic76}.   Zur Einf"uhrung in die Gruppen- und Darstellungstheorie der~$\SUNc$ bzgl.\@ allgemeinen~$\Nc$ sei verwiesen auf Ref.~\cite{Pokorski87}, vgl.\@. auch die Refn.~\cite{Peskin99,Pascual84},~-- bzgl.~$\Nc \!\equiv\! 2,3$ auf die Refn.~\cite{Nachtmann92,Field89}.
}
%

Eichtransformationen werden vermittelt durch speziell-unit"are $\Nc \!\times\! \Nc$-Matrizen.
Diese sind aufzufassen als Elemente der zugrundeliegenden Liegruppe~$\SUNc$; deren Dimension ist:
\vspace*{-.5ex}
\begin{align} 
\dimNc\; =\; \Nc^2 - 1
    \\[-4.5ex]\nn
\end{align}
das hei"st eine Eichtransformation ist determiniert durch~$\dimNc$ reelle Parameter.
Sei
\vspace*{-.5ex}
\begin{align} \label{APP:Generatoren}
\{ T^a \}\qquad a = 1, 2, \ldots \dimNc
    \\[-4.5ex]\nn
\end{align}
eine Basis der assoziierten Liealgebra~$\suNc$, das hei"st {\it per definitionem\/} gilt:
\vspace*{-.5ex}
\begin{align} \label{APP:Ta-Kommutator_pre}
[ T^a, T^b ]\; =\; {\iIM}\, f_{abc}\, T^c
    \\[-4.5ex]\nn
\end{align}
mit~$f_{abc}$ den {\it Strukturkonstanten\/} der~$\suNc$, die o.E.d.A.\@ gew"ahlt seien als {\it vollst"andig antisymmetrisch\/} und {\it reell}.
Die~$T^a$ sind {\it spurlose hermitesche Operatoren\/}:
\begin{samepage}
\vspace*{-.5ex}
\begin{alignat}{2} \label{APP:Ta_spurlos/hermitesch}
&\tr T^a&\;
  &=\; 0
    \\
&T^{a\D\dagger}&\;
  &=\; T^a \qquad \forall a = 1, 2, \ldots \dimNc
    \tag{\ref{APP:Ta_spurlos/hermitesch}$'$}
    \\[-4.5ex]\nn
\end{alignat}
und beziehen sich allgemein auf eine Darstellung~$\Drst{R}$ der~$\SUNc$; sie werden bezeichnet als deren {\it Erzeugende\/} oder {\it Generatoren}.
Ein Index $\Drst{R}$ zur Notation einer allgemein gedachten Darstellung~$\Drst{R}$ sei unterdr"uckt, wenn klar aus dem Zusammenhang.

Bez"uglich des Begriffs der {\it Darstellung\/} sei verwiesen auf Ref.~\cite{Nachtmann92}.
Wir fassen zusammen:
Eine Darstellung~$\Drst{R}$ der Dimension~$\dimDrst{R}$ ist eine lineare Abbildung der Eins-Zusammenhang\-komponente der Gruppe:~$\mf{E}\big|\SUNc$, in einen $\dimDrst{R} \!\times\! \dimDrst{R}$-dimensionalen linearen Raum, unter der die Gruppenrelation erhalten ist:
\vspace*{-.5ex}
\begin{align} \label{APP:Gruppenrelation-erhaltend}
\Drst{R}(U) \cdot \Drst{R}(V)\; =\; \Drst{R}(U \cdot V)\qquad
  \forall\; U,V \in \mf{E}\big|\SUNc
    \\[-4.5ex]\nn
\end{align}
Aufgrund der {\it Liestruktur\/} der~$\SUNc$ gen"ugt es,~Ele\-mente der Gruppe zu betrachten, die sich nur {\it infinitesimal\/} von der Eins unterscheiden, das hei"st Elemente
\vspace*{-.5ex}
\begin{align} \label{APP:infGruppenelement}
U\; =\; \bbbone\; +\; {\iIM}\, \de\vph_a\, T^a
    \\[-4ex]\nn
\end{align}
\end{samepage}%
mit~$\dimNc$ reellen infinitesimalen Parametern~$\de\vph_a$ und spurlosen hermiteschen $\Nc \!\times\! \Nc$-Matri\-zen~$T^a$, die der Kommutatorrelation Gl.~(\ref{APP:Ta-Kommutator_pre}) gen"ugen.
Es zeigt sich, da"s eine Darstellung aufgrund von Gl.~(\ref{APP:Gruppenrelation-erhaltend}) diese Kommutatorrelation erh"alt und sie vollst"andig definiert ist durch die Abbildungsvorschrift in infinitesimaler Form:
\vspace*{-.5ex}
\begin{align} \label{APP:Drst-Abb}
\Drst{R}:\vv \bbbone\; +\; {\iIM}\, \de\vph_a\, T^a\vv
  \longrightarrow\vv \bbbOne{R}\; +\; {\iIM}\, \de\vph_a\, T_\Drst{R}^a
    \\[-4.5ex]\nn
\end{align}
mit~$T_\Drst{R}^a$ den Generatoren der Darstellung~$\Drst{R}$.
Diese Vorschrift von~$\Drst{R}$~-- aus der Liegruppe in die Liealgebra~-- gilt f"ur ein beliebiges Gruppenelement~$U \!=\! \bbbone \!+\! {\iIM}\de\vph_a T^a$ nach Gl.~(\ref{APP:infGruppenelement}), das hei"st f"ur beliebige Konstanten~$\de\vph_a$, so da"s f"ur jedes feste~$a \!=\! 1,2,\ldots\dimNc$ die Matrix~$T^a$ auf den Opera\-tor~$T_\Drst{R}^a$ abgebildet wird.

Sei im folgenden eine Darstellung~$\Drst{R}$ aufgefa"st als {\it Matrix\/}darstellung, das hei"st die sie erzeugenden Operatoren~$T_\Drst{R}^a$ als~$\dimDrst{R} \!\times\! \dimDrst{R}$-Matrizen, die den Gln.~(\ref{APP:Ta_spurlos/hermitesch}),~(\ref{APP:Ta_spurlos/hermitesch}$'$) gen"ugen, in Komponenten:
\vspace*{-.5ex}
\begin{align} \label{APP:Ta_spurlos/hermitesch_Komp}
&T^a_{\al\al}\;
  =\; 0
    \\
&(T^{a\D\dagger})_{\al\be}\;
  =\; T^a_{\be\al}{}^{\D\zz\ast}\; =\; T^a_{\al\be}\qquad
  a  = 1,2, \ldots \dimNc\qquad
  \al,\, \be = 1,2, \ldots \dimDrst{R}
    \tag{\ref{APP:Ta_spurlos/hermitesch_Komp}$'$}
    \\[-4.5ex]\nn
\end{align}
Aus dieser Gestalt folgt unmittelbar die Abh"angigkeit der Darstellung~$\Drst{R}$ von~$\dimDrst{R}^2 \!-\! 1$ reellen Parametern.

Zwei Darstellungen sind {\it physikalisch\/} von besonderer Relevanz:
Die~$T^a$ auf der linken Seite von Gl.~(\ref{APP:Drst-Abb}) k"onnen formal identifiziert werden mit den Generatoren~$T_\Drst{F}^a$ einer Darstellung~$\Drst{F}$ mit Dimension~$\dimDrst{F} \!=\! \Nc$.
Diese Darstellung ist insofern ausgezeichnet und wird bezeichnet als definierende oder {\it fundamentale Darstellung}.
Die Anzahl ihrer Parameter ist mit~$\dimDrst{R}^2 \!-\! 1 \!=\! \Nc^2 \!-\! 1$ identisch der Anzahl~$\dimNc$ der Parameter der Gruppe~$\SUNc$, also {\it minimal\/}~-- also $Bild\,\Drst{F}$ der gesamte Darstellungsraum, nicht nur ein Unterraum.
Die zweite physikalisch besonders relevante Darstellung ist die {\it adjungierte Darstellung\/}~$\Drst{A}$, die bestimmt ist dadurch, da"s ihre Generatoren entsprechend
\vspace*{-.5ex}
\begin{align} \label{APP:TaA_fabc}
(T_\Drst{A}^a)_{bc}\;
  =\; {\rm-i}\, f_{abc}
    \\[-4.5ex]\nn
\end{align}
definiert sind in Zusammenhang mit den Strukturkonstanten~$f_{abc}$ der Algebra.
Dies definiert in der Tat eine Darstellung:
Die Matrizen~$T_\Drst{A}^a$ erf"llen die Forderung von Spurfreiheit und Hermitezit"at, das hei"st die Gln.~(\ref{APP:Ta_spurlos/hermitesch}),~(\ref{APP:Ta_spurlos/hermitesch}$'$), in Komponenten die Gln.~(\ref{APP:Ta_spurlos/hermitesch_Komp}),~(\ref{APP:Ta_spurlos/hermitesch_Komp}$'$)~-- aufgrund der Wahl der~$f_{abc}$ als voll antisymmetrisch und reell, vgl.\@ die Bem.\@ zu Gl.~(\ref{APP:Ta-Kommutator_pre}) und unten Gl.~(\ref{APP:f_antisym,reell}).
Und sie erf"ullen die Kommutatorrelation Gl.~(\ref{APP:Ta-Kommutator_pre})~-- aufgrund der allgemeinen Jacobi-Identit"at, s.u.\@ die Gln.~(\ref{APP:Jacobi-Kommutator}),~(\ref{APP:Jacobi-Kommutator}$'$).
Die~$T_\Drst{A}^a$ sind~\mbox{$\dimNc \!\times\! \dimNc$-Ma}\-trizen, so da"s die adjungierte Darstellung die Dimension~$\dimDrst{A} \!\equiv\! \dimNc \!=\! \Nc^2 \!-\! 1$ besitzt.

Der fundamentalen Darstellung und adjungierten Darstellung geh"oren die {\it Materie-Spi\-norfelder\/}~$\ps$ beziehungsweise die {\it Eich-Vektorfelder\/}~$A$ an:
\vspace*{-.5ex}
\begin{alignat}{5} \label{APP:Quarks/Gluonen-DrstF/A}
&\ps_n&\; &\in\; \Drst{F}& \qquad
  &n = 1,2,\ldots \dimDrst{F}& \qquad
  &\text{mit}&\qquad
  &\dimDrst{F} = \Nc
    \\
&A_a&\; &\in\; \Drst{A}& \qquad
  &a = 1,2,\ldots \dimDrst{A}& \qquad
  &\hspace*{0em}\sim&\qquad
  &\dimDrst{A} \equiv \dimNc = \Nc^2 - 1
    \tag{\ref{APP:Quarks/Gluonen-DrstF/A}$'$}
    \\[-4.5ex]\nn
\end{alignat}
das hei"st physikalisch die Quarks beziehungsweise Gluonen.
\vspace*{-.5ex}

\bigskip\noindent
Wir kommen zur"uck auf die Kommutatorrelation Gl.~(\ref{APP:Ta-Kommutator_pre}):
\vspace*{-.5ex}
\begin{align} \label{APP:Ta-Kommutator}
[ T^a, T^b ]\;
  =\; {\iIM}\, f_{abc}\, T^c
    \\[-3.5ex]\nn
\end{align}
Sie gilt zun"achst f"ur die Matrizen~$T^a \!\in\! \SUNc$ auf der linken Seite von Gl.~(\ref{APP:Drst-Abb}), die identifiziert werden mit den Generatoren~$T_\Drst{F}^a$ der fundamentalen Darstellung~$\Drst{F}$; sie "ubertr"agt sich aber unmittelbar auf die rechte Seite, also von~$\Drst{F}$ auf eine beliebige Darstellung~$\Drst{R}$ der~$\SUNc$.
Gl.~(\ref{APP:Ta-Kommutator}) wird daher bezeichnet als definierende oder {\it fundamentale Kommutatorrelation}. \\
Wie wir leicht nachrechnen, folgt aus Gl.~(\ref{APP:Ta-Kommutator})~-- in dem Sinne, da"s ihre Gestalt diese Wahl bereits impliziert:
\vspace*{-.5ex}
\begin{align} \label{APP:f_antisym,reell}
f_{abc} \qquad
  \text{voll antisymmetrisch,}\quad
  \text{reell}
    \\[-4.5ex]\nn
\end{align}
und hieraus
\vspace*{-.5ex}
\begin{align} \label{APP:f-kontrahiert}
f_{aab}\; =\; 0
    \\[-4.5ex]\nn
\end{align}
bei Kontraktion zweier Indizes.

Entsprechend Gl.~(\ref{APP:Ta-Kommutator}) kann f"ur die Matrizen auf der linken Seite von Gl.~(\ref{APP:Drst-Abb}), das hei"st f"ur die fundamentale Darstellung die Antikommutatorrelation
\vspace*{-.5ex}
\begin{align}
\{ T_\Drst{F}^a, T_\Drst{F}^b \}\;
  =\; m_\Drst{F}\, \de^{ab}\, \bbbOne{F}\; +\; d_{abc}\, T_\Drst{F}^c
    \tag{\ref{APP:Ta-Kommutator}$'$}
    \\[-4.5ex]\nn
\end{align}
angegeben werden~-- Konsequenz daraus, da"s~$Bild\,\Drst{F}$ der gesamte Darstellungsraum ist.

Wir sehen leicht, da"s sich Gl.~(\ref{APP:Ta-Kommutator}$'$) unter einer Abbildung entsprechend Gl.~(\ref{APP:Drst-Abb}) {\it nicht\/} "ubertr"agt auf eine andere Darstellung:
F"ur jede Darstellung~$\Drst{R}$ ungleich der fundamentalen Darstellung~$\Drst{F}$ gilt:~$\dimDrst{R}^2 \!-\! 1 > \dimNc$,~$\forall \Drst{R} \!\neq\! \Drst{F}$; das hei"st ist die Anzahl ihrer Parameter~$\dimDrst{R}^2 \!-\! 1$ {\it gr"o"ser\/} als die Anzahl~$\dimNc \!=\! \Nc^2 \!-\! 1$ der Parameter der Gruppe, das hei"st als die Anzahl Generatoren.
Die Wahl von~$\dimNc \!<\! \dimDrst{R}^2 \!-\! 1$ Generatoren im Darstellungsraum von~$\Drst{R}$ w"ahlt also willk"urlich nur einen Unterraum aus.
Die Eigenschaft einer Darstellung~$\Drst{R}$, die Gruppenrelation zu erhalten, vgl.\@ Gl.~(\ref{APP:Gruppenrelation-erhaltend}), garantiert, da"s der Antikommutator zweier Elemente des Darstellungsraumes wieder ein Element des Darstellungsraumes ist, aber im allgemeinen {\it nicht\/}, da"s der Antikommutator zweier Elemente eines willk"urlich gew"ahlten Unterraums wieder Element dieses Unterraums ist. \\
Aus Gl.~(\ref{APP:Ta-Kommutator}$'$) folgt~-- da im Sinne von Gl.~(\ref{APP:f_antisym,reell}) o.E.d.A.\@ bereits implizit gew"ahlt:
%
\vspace*{-.5ex}
\begin{align}
d_{abc} \qquad
  \text{voll symmetrisch,}\quad
  \text{reell}
    \tag{\ref{APP:f_antisym,reell}$'$}
    \\[-4.5ex]\nn
\end{align}
Gl.~(\ref{APP:f-kontrahiert}), die folgt aufgrund der vollst"andigen Antisymmetrie der Strukturkonstanten~$f_{abc}$, gilt aufgrund der Spurfreiheit der Generatoren, vgl.\@ die Gln.~(\ref{APP:Ta_spurlos/hermitesch}),~(\ref{APP:Ta_spurlos/hermitesch_Komp}), analog auch f"ur die Konstanten~$d_{abc}$:
\vspace*{-.5ex}
\begin{align}
d_{aab}\; =\; 0
    \tag{\ref{APP:f-kontrahiert}$'$}
    \\[-4.5ex]\nn
\end{align}
vgl.\@ dazu Bem.\,2 zu Gl.~(\ref{APP:fd_(Anti)Kommutator}$'$) auf Seite\,\pageref{APP-T:f-kontrahiert'}.
Die Relation
\vspace*{-.5ex}
\begin{align} 
d_{aeg}\, f_{beg}\; =\; 0
    \\[-4.5ex]\nn
\end{align}
schlie"slich ist Konsequenz der Gln.~(\ref{APP:f_antisym,reell}),~(\ref{APP:f_antisym,reell}$'$).
\vspace*{-.5ex}

\bigskip\noindent
Zur {\it Normierung\/} der Generatoren~$T^a$ einer Darstellung~$\Drst{R}$, vgl.\@ Gl.~(\ref{APP:Generatoren}), betrachten wir die Spur des Produkts zweier Generatoren; mithilfe der Gln.~(\ref{APP:Ta_spurlos/hermitesch}$'$),~(\ref{APP:Ta_spurlos/hermitesch_Komp}$'$) finden wir
\begin{samepage}
\vspace*{-.5ex}
\begin{align} 
\tr\, T^a\, T^b\;
  =\; T^a_{\al\be}\, T^b_{\be\al}\;
  =\; T^a_{\al\be}\, T^b_{\al\be}{}^{\D\zz\ast}
    \\[-4.5ex]\nn
\end{align}
als eine bez"uglich des Indexpaares~$(a,b)$ hermitesche Matrix.
Diese kann folglich auf Diagonalgestalt mit reellen Eintr"agen gebracht werden; bei Absorption entsprechender Faktoren in die~$T^a$ ist diese Matrix proportional zur Einheitsmatrix.
F"ur die~$T^a$ kann daher die Normierung
\end{samepage}
\vspace*{-.5ex}
\begin{align} \label{APP:Ta-Normierung}
\tr\, T^a\, T^b\;
  =\; \normDrst{R}\, \de^{ab} \qquad
  \text{$\normDrst{R}$~reell}
    \\[-4.5ex]\nn
\end{align}
gefordert werden.
Die "ubliche Konvention
\vspace*{-.5ex}
\begin{align}
\normDrst{F}\; &=\; 1/2
    \label{APP:nDrstF-Konv} \\
\normDrst{A}\; &=\; \Nc
    \label{APP:nDrstA-Konv}
    \\[-4.5ex]\nn
\end{align}
liegt auch unserer Arbeit zugrunde.
\begin{samepage}
Durch Spurbildung folgt aus Gl.~(\ref{APP:Ta-Kommutator}$'$)
\vspace*{-.5ex}
\begin{align} \label{APP:nF_mF,dimF}
2\, \normDrst{F}\; =\; m_\Drst{F}\, \dimDrst{F}
    \\[-4.5ex]\nn
\end{align}
als Relation zur Bestimmung von~$m_\Drst{F}$ in Abh"angigkeit von~$\normDrst{F}$.
\end{samepage}

Der {\it quadratische Casimir-Operator\/} einer Darstellung~$\Drst{R}$ ist definiert als die Summe der quadrierten Generatoren; er vertauscht daher mit einem beliebigen Operator, das hei"st ist proportional zur Eins und bestimmt durch die Proportionalit"atskonstante~$\csDrst{R}$:
\vspace*{-.5ex}
\begin{align} \label{APP:CasimirOperator}
T^a\, T^a\;
  =\; \csDrst{R}\, \bbbOne{R}
    \\[-4.5ex]\nn
\end{align}
Die Spur dieser Relation mithilfe von Gl.~(\ref{APP:Ta-Normierung}) lautet:
\vspace*{-.5ex}
\begin{align} \label{APP:csDrst*dimDrst_nDrst*dimNc}
\csDrst{R}\, \dimDrst{R}\;
  =\; \normDrst{R}\, \dimNc
    \\[-4.5ex]\nn
\end{align}
so da"s mithilfe Gl.~(\ref{APP:nF_mF,dimF}) folgt
\vspace*{-.5ex}
\begin{align} \label{APP:csF_mF,dimNc}
2\, \csDrst{F}\;
  =\; m_\Drst{F}\, \dimNc
    \\[-4.5ex]\nn
\end{align}
als Relation zur Bestimmung von~$m_\Drst{F}$ in Abh"angigkeit von~$\csDrst{F}$.
Die Konstanten~$\normDrst{R}$~-- daher oft bezeichnet mit~$\cssDrst{R}$~-- und~$\csDrst{R}$ sind die einzigen Invarianten der Darstellung~$\Drst{R}$.
\vspace*{-.5ex}

\bigskip\noindent
Wir skizzieren die Herleitung wichtiger Relationen.
Die Spuren der mit einem Generator multiplizierten (Anti)Kommutatorrelationen, vgl.\@ die Gln.~(\ref{APP:Ta-Kommutator}),~(\ref{APP:Ta-Kommutator}$'$), lauten:
\vspace*{-.5ex}
\begin{alignat}{2} \label{APP:fd_(Anti)Kommutator}
f_{abc}\;
  &=&\; \frac{1}{{\iIM} \normDrst{R}}\;
          &\tr{}  [T^a,T^b]T^c
    \\
d_{abc}\;
  &=&\; \frac{1}{\normDrst{F}}\;
          &\tr \{T_\Drst{F}^a,T_\Drst{F}^b\}T_\Drst{F}^c
    \tag{\ref{APP:fd_(Anti)Kommutator}$'$}
    \\[-4.5ex]\nn
\end{alignat}
dabei gilt Gl.~(\ref{APP:fd_(Anti)Kommutator}) in einer beliebigen, Gl.~(\ref{APP:fd_(Anti)Kommutator}$'$) nur in der fundamentalen Darstellung.

Zu diesen Gleichungen zwei Bemerkungen.
Bem.\,1: \label{APP-T:fff-f}
F"ur die adjungierte Darstellung~$\Drst{A}$ mit der expliziten Gestalt~$(T_\Drst{A}^a)_{\!bc} = {\rm-i} f_{abc}$ f"ur ihre Generatoren, vgl.\@ Gl.~(\ref{APP:TaA_fabc}),~-- und mit $\csDrst{A} \!=\! \normDrst{A} \!/\! \dimDrst{A} \!\cdot\! \dimNc \!=\! \normDrst{A}$, vgl.\@ Gl.~(\ref{APP:csDrst*dimDrst_nDrst*dimNc}),~-- schreibt sich Gl.~(\ref{APP:fd_(Anti)Kommutator}):
\vspace*{-.5ex}
\begin{align} \label{APP:fff_f-pre}
f_{aeg}\, f_{bgh}\, f_{che}\;
  =\; \frac{1}{2}\, \csDrst{A}\, f_{abc}\;
  =\; \frac{1}{2}\, \normDrst{A}\, f_{abc}\;
    \\[-4.5ex]\nn
\end{align}
Bem.\,2: \label{APP-T:f-kontrahiert'}
Kontraktion zweier Indizes der Gl.~(\ref{APP:fd_(Anti)Kommutator}$'$) ergibt~$d_{aab} \!=\! 0$, das hei"st Gl.~(\ref{APP:f-kontrahiert}$'$).
Dabei gehen ein: Gl.~(\ref{APP:CasimirOperator}), Spurfreiheit der Generatoren, vgl.\@ die Gln.~(\ref{APP:Ta_spurlos/hermitesch}),~(\ref{APP:Ta_spurlos/hermitesch_Komp}), und Invarianz der Spurbildung unter zyklischer Vertauschung der Faktoren des Arguments.

Zur"uck zu den Gln.~(\ref{APP:fd_(Anti)Kommutator}),~(\ref{APP:fd_(Anti)Kommutator}$'$).
Diese aufgel"ost nach den Spuren und addiert f"ur die fundamentale Darstellung~$\Drst{F}$ folgt:
\begin{samepage}
\vspace*{-.5ex}
\begin{align} 
\tr T_\Drst{F}^a T_\Drst{F}^b T_\Drst{F}^c\;
  =\; \frac{1}{2} \normDrst{F}\, (d_{abc} + {\iIM}\, f_{abc})
    \\[-4.5ex]\nn
\end{align}
Diese Relation folgt alternativ aus
\end{samepage}
\vspace*{-.5ex}
\begin{align} \label{APP:TaTb_mDrst,df}
T^a T^b\;
  &=\; \frac{1}{2}\,
       \big[\, \{T^a, T^b\} + [T^a, T^b]\, \big]
    \\
  &\underset{\D\Drst{F}}{=}\;
       \frac{1}{2}\,
       \Big[\, m_\Drst{F}\, \de^{ab}\, \bbbOne{F} + (d_{abc} + {\iIM}\, f_{abc})\, T^c\, \Big]
    \tag{\ref{APP:TaTb_mDrst,df}$'$}
    \\[-4.5ex]\nn
\end{align}
durch Multiplikation mit einem Generator und anschlie"sende Spurbildung.

Gl.~(\ref{APP:CasimirOperator}) gilt f"ur eine beliebige Darstellung~$\Drst{R}$.
F"ur die adjungierte Darstellung~$\Drst{A}$ mit der expliziten Gestalt~$(T_\Drst{A}^a)_{\!bc} = {\rm-i} f_{abc}$ ihrer Generatoren, vgl.\@ Gl.~(\ref{APP:TaA_fabc}), schreibt sie sich:
\vspace*{-.5ex}
\begin{align} \label{APP:ff-de}
f_{aeg}\, f_{beg}\;
  =\; \csDrst{A}\, \de_{ab}\;
  =\; \normDrst{A}\, \de_{ab}
    \\[-4.5ex]\nn
\end{align}
mit der letzten Identit"at wegen~$\csDrst{A} \!=\! \normDrst{A} \!/\! \dimDrst{A} \!\cdot\! \dimNc \!=\! \normDrst{A}$, vgl.\@ Gl.~(\ref{APP:csDrst*dimDrst_nDrst*dimNc}).
Analog mu"s~$d_{aeg}\, d_{beg}$ proportional zu~$\de_{ab}$ sein; wir definieren durch
\vspace*{-.5ex}
\begin{align}
d_{aeg}\, d_{beg}\;
  =\; \al_1\, \de_{ab} \qquad
  \text{bzgl.~$\al_1$ s.u.\@ Gl.~(\ref{APP:Konstante-al1})}
    \tag{\ref{APP:ff-de}$'$}
    \\[-4.5ex]\nn
\end{align}
die Konstante~$\al_1$, die wir im folgenden berechnen, s.u.\@ Gl.~(\ref{APP:Konstante-al1}).

Wir kontrahieren mit~$f_{abc}$ die Kommutatorrelation nach Gl.~(\ref{APP:Ta-Kommutator}) und finden mithilfe Gl.~(\ref{APP:ff-de}) f"ur eine beliebige Darstellung~$\Drst{R}$:
\vspace*{-.5ex}
\begin{align} \label{APP:fabcTbTc_csDrst}
f_{abc}\, T^b T^c\;
  =\; \frac{{\iIM}}{2}\, \csDrst{A}\, T^a
  =\; \frac{{\iIM}}{2}\, \normDrst{A}\, T^a
    \\[-4.5ex]\nn
\end{align}
F"ur die adjungierte Darstellung~$\Drst{A}$, durch Einsetzen von~$(T_\Drst{A}^a)_{\!bc} = {\rm-i} f_{abc}$, vgl.\@ Gl.~(\ref{APP:TaA_fabc}), folgt wieder Gl.~(\ref{APP:fff_f-pre}) aus Bem.\,1, Seite\,\pageref{APP-T:fff-f}.

Analog kontrahieren wir mit~$d_{abc}$ die Antikommutatorrelation nach Gl.~(\ref{APP:Ta-Kommutator}$'$) und finden f"ur die fundamentale Darstellung~$\Drst{F}$:
\vspace*{-.5ex}
\begin{align}
d_{abc}\, T_\Drst{F}^b T_\Drst{F}^c\;
  =\; \frac{1}{2}\, \al_1\, T_\Drst{F}^a
    \tag{\ref{APP:fabcTbTc_csDrst}$'$}
    \\[-4.5ex]\nn
\end{align}
mit~$\al_1$ der Konstanten aus Gl.~(\ref{APP:ff-de}$'$). \\
\indent
Wir bestimmen diese.
Einerseits ergibt Multiplikation von Gl.~(\ref{APP:TaTb_mDrst,df}$'$) mit~$T_\Drst{F}^b$:
\vspace*{-.5ex}
\begin{align} 
T_\Drst{F}^a T_\Drst{F}^b T_\Drst{F}^b\;
  =\; \frac{1}{2}\,
        \Big[\, m_\Drst{F}\, T_\Drst{F}^a
              + (d_{abc} - {\iIM}\, f_{abc})\, T_\Drst{F}^b T_\Drst{F}^c\, \Big]
    \\[-4.5ex]\nn
\end{align}
und mithilfe von~$m_\Drst{F} \!=\! 2 \csDrst{F} \!/\! \dimNc$, vgl.\@ Gl.~(\ref{APP:csF_mF,dimNc}), und den Gln.~(\ref{APP:fabcTbTc_csDrst}),~(\ref{APP:fabcTbTc_csDrst}$'$):
\vspace*{-.5ex}
\begin{align} \label{APP:TaTbTb-1}
T_\Drst{F}^a T_\Drst{F}^b T_\Drst{F}^b\;
  =\; \frac{1}{2}\,
        \Big[\, \frac{2}{\dimNc}\, \csDrst{F}
              + \frac{1}{2} (\al_1 + \csDrst{A})\, \Big]\, T_\Drst{F}^a
    \\[-4.5ex]\nn
\end{align}
Andererseits gilt mithilfe Gl.~(\ref{APP:CasimirOperator}) f"ur eine beliebige Darstellung~$\Drst{R}$:
\vspace*{-.5ex}
\begin{align} \label{APP:TaTbTb-2}
T^a T^b T^b\;
  =\; \csDrst{R}\, T^a
    \\[-4.5ex]\nn
\end{align}
Gleichsetzen der Gln.~(\ref{APP:TaTbTb-1}),~(\ref{APP:TaTbTb-2}) f"ur die fundamentale Darstellung~$\Drst{F}$ ergibt:
\vspace*{-.5ex}
\begin{align} \label{APP:Konstante-al1}
\al_1\;
  &=\; 4\, \Big[ 1 \!-\! \frac{1}{\dimNc} \Big]\, \csDrst{F} - \csDrst{A}
    \\
  &=\; 4\, \normDrst{F}\, \frac{1}{\dimDrst{F}}\, (\dimNc \!-\! 1) - \normDrst{A}\;
   =\; 2\, \normDrst{F}\cdot \frac{1}{\Nc}\, (\Nc^2 \!-\! 4)
    \nn
    \\[-4.5ex]\nn
\end{align}
Seien die vollst"andigen Kontraktionen der Konstanten~$f_{abc}$ und~$d_{abc}$ bezeichnet mit~$f^2$ beziehungsweise~$d^2$:
\vspace*{-.5ex}
\begin{align} \label{APP:f2-Def}
f^2 \;&\equiv\; f_{abc}\, f_{abc}
    \\[.5ex]
d^2 \;&\equiv\; d_{abc}\, d_{abc}
    \tag{\ref{APP:f2-Def}$'$}
    \\[-4.5ex]\nn
\end{align}
F"ur diese gilt aufgrund Gl.~(\ref{APP:ff-de}) beziehungsweise aufgrund Gl.~(\ref{APP:ff-de}$'$) mit~(\ref{APP:Konstante-al1}):
\vspace*{-.5ex}
\begin{align} \label{APP:f2-explizit}
f^2\;
  &=\; \csDrst{A}\, \dimNc
    \\[.5ex]
  &=\; \normDrst{A}\, \dimNc\;
   =\; 2\, \normDrst{F}\cdot \Nc (\Nc^2 \!-\! 1)
    \nn \\[1ex]
d^2\;
  &=\; 4\, [ \dimNc \!-\! 1 ]\, \csDrst{F} - \csDrst{A}\, \dimNc
    \tag{\ref{APP:f2-explizit}$'$} \\[.5ex]
  &=\; 4\, \normDrst{F}\, \frac{1}{\dimDrst{F}}\,
         (\dimNc \!-\! 1)\, \dimNc - \normDrst{A}\, \dimNc\;
   =\; 2\, \normDrst{F}\cdot \frac{1}{\Nc}\, (\Nc^2 \!-\! 4)(\Nc^2 \!-\! 1)
    \nn
    \\[-4.5ex]\nn
\end{align}
Das hei"st in der Summe
\vspace*{-.5ex}
\begin{align} \label{APP:d2+f2}
d^2 + f^2\;
  &=\; 4\, [ \dimNc \!-\! 1 ]\, \csDrst{F}
    \\[.5ex]
  &=\; 4\, \normDrst{F}\, \frac{1}{\dimDrst{F}}\, (\dimNc \!-\! 1)\, \dimNc\;
   =\; 4\, \normDrst{F}\cdot \frac{1}{\Nc}\, (\Nc^2 \!-\! 2)(\Nc^2 \!-\! 1)
    \nn
    \\[-4.5ex]\nn
\end{align}
wird gerade der Term mit~$\csDrst{A}$ beziehungsweise~$\normDrst{A}$ gek"urzt. \\
Die zweiten Zeilen in Gl.~(\ref{APP:Konstante-al1}), in den Gln.~(\ref{APP:f2-explizit}),~(\ref{APP:f2-explizit}$'$) und in Gl.~(\ref{APP:d2+f2}) gehen entsprechend der Relation~$\csDrst{R} \!=\! \normDrst{R} \dimNc \!/\! \dimDrst{R}$, vgl.\@ Gl.~(\ref{APP:csDrst*dimDrst_nDrst*dimNc}), von den quadratischen Casimir-Faktoren "uber zu den Normierungen.
Die letzten Identit"aten benutzen~$\normDrst{A} \!=\! 2 \normDrst{F} \dimDrst{F}$ mit~$\dimDrst{F} \!=\! \Nc$.
Wir zeigen unten, da"s~$\normDrst{A}$ von~$\normDrst{F}$ "uber diese Relation abh"angt, vgl.\@ Gl.~(\ref{APP:nA_nF-explizit}); im Moment verstehen wir sie als Konsequenz der Konvention~$\normDrst{F} \!=\! 1\!/\!2$,~$\normDrst{A} \!=\! \Nc$, vgl.\@ die Gln.~(\ref{APP:nDrstF-Konv}),~(\ref{APP:nDrstA-Konv}).

Da die Generatoren der adjungierten Darstellung wie~$(T_\Drst{A}^a)_{\!bc} = {\rm-i} f_{abc}$, vgl.\@ Gl.~(\ref{APP:TaA_fabc}), "uber die {\it darstellungsunabh"angigen\/} Strukturkonstanten der~$\suNc$ definiert sind, erwarten wir, da"s die Normierung~$\normDrst{R}$ einer beliebigen Darstellung~$\Drst{R}$, vgl.\@ Gl.~(\ref{APP:Ta-Normierung}), nicht unabh"angig von der Normierung der adjungierten Darstellung~$\normDrst{A}$ ist.
So folgt aus der Wahl der Normierung einer Darstellung~-- o.E.d.A.\@ aus der Wahl~$\normDrst{F} \!=\! 1\!/\!2$, vgl.\@ Gl.~(\ref{APP:nDrstF-Konv}), f"ur die fundamentale Darstellung~$\Drst{F}$~-- die Normierung einer beliebigen anderen Darstellung.

Wir geben mit
\vspace*{-.5ex}
\begin{align} \label{APP:AnsatzR}
&T_{\al\be}^a\, T_{\ga\de}^a\;
  =\; \be_1^\Drst{R}\cdot
        \Big(\,                        \de_{\al\de}\, \de_{\be\ga}
                + \be_2^\Drst{R}\cdot \de_{\al\be}\, \de_{\ga\de}
                + \be_3^\Drst{R}\cdot \de_{\al\ga}\, \de_{\be\de}\, \Big)
    \\[.5ex]
&\qquad\al,\be,\ga,\de = 1,2,\ldots \dimDrst{R} \qquad
   \forall\Drst{R} \setminus \Drst{A}
    \nn
    \\[-4.5ex]\nn
\end{align}
eine explizite Formel an, die als Zerlegung in elementare, hier also: {\it Kronecker\/}-Tensorkom\-ponenten mit entsprechenden Konstanten~$\be_i^\Drst{R}$ f"ur eine beliebige Darstellung~$\Drst{R} \!\setminus\! \Drst{A}$ gilt.
Diese Konstanten sind reell aufgrund der Hermitezit"at der Generatoren; aufgrund ihrer Spurfreiheit und Normierung gem"a"s Gl.~(\ref{APP:Ta-Normierung}) gilt:
\vspace*{-.5ex}
\begin{align} \label{APP:be1,2,3_R}
\be_1^\Drst{R}\;
   =\; \normDrst{R}\qquad
\be_2^\Drst{R}\;
   =\; \frac{\dimDrst{R}(\dimDrst{R} \!-\! 1) - \dimNc}{\dimDrst{R}(\dimDrst{R} \!-\! 1)}\qquad
\be_3^\Drst{R}\;
   =\; - \frac{(\dimDrst{R}^2 \!-\! 1) - \dimNc}{\dimDrst{R} \!-\! 1}
    \\[-4.5ex]\nn
\end{align}
insbesondere f"ur die fundamentale Darstellung~$\Drst{F}$:
\vspace*{-.5ex}
\begin{align} \label{APP:be1,2,3_F}
\be_1^\Drst{F}\;
  =\; \normDrst{F} \qquad
\be_2^\Drst{F}\;
  =\; - \frac{1}{\Nc} \qquad
\be_3^\Drst{F}\;
  =\; 0
    \\[-4.5ex]\nn
\end{align}
Kontrahiert der Zusammenhang der Strukturkonstanten~$f_{abc}$ und der Spur dreier Generatoren~$T^a$, vgl.\@ Gl.~(\ref{APP:fd_(Anti)Kommutator}), und gleichgesetzt mit der Kontraktion~$f^2$ nach Gl.~(\ref{APP:f2-explizit}) folgt f"ur eine beliebige Darstellung~$\Drst{R}$ die Identit"at
%
\vspace*{-.5ex}
\begin{align} 
\normDrst{A}\, \dimNc\;
  =\; - \frac{1}{\normDrst{R}^2}\cdot \tr{} [T^a,T^b]T^c \cdot \tr{} [T^a,T^b]T^c
    \\[-4.5ex]\nn
\end{align}
Auf der rechten Seite dieser Identit"at eingesetzt Gl.~(\ref{APP:AnsatzR}), folgt eine Relation zwischen der Normierung~$\normDrst{R}$ einer beliebigen Darstellung~$\Drst{R} \!\setminus\! \Drst{A}$ und~$\normDrst{A}$.
Resultat ist:
\begin{samepage}
\vspace*{-.5ex}
\begin{align} \label{APP:nA_nR-explizit}
&\normDrst{A}\, \dimNc\;
  =\; 2\, \normDrst{R}\, \dimDrst{R}\, (\dimDrst{R}^2 \!-\! 1)\cdot
         f_\Drst{R}(\dimDrst{R},\dimNc)
    \\[1ex]
  &\text{mit}\qquad
  f_\Drst{R}(\dimDrst{R},\dimNc)\;
  =\; 1\; +\; \be_3^\Drst{R}\cdot
                \Big[\, 3 (\dimDrst{R} \!+\! 1)^{\!-1}\, (1 \!-\! \be_3^\Drst{R})
                      - (\be_3^\Drst{R})^2\,
                \Big]
    \tag{\ref{APP:nA_nR-explizit}$'$}
    \\[-4.5ex]\nn
\end{align}
und~$\be_3^\Drst{R}$ wie in Gl.~(\ref{APP:be1,2,3_R}).
F"ur die fundamentale Darstellung~$\Drst{F}$ ist~$f_\Drst{F} \!\equiv\! 1$ wegen~$\be_3^\Drst{F} \!\equiv\! 0$, und es folgt unmittelbar:
\end{samepage}
\vspace*{-.5ex}
\begin{align} \label{APP:nA_nF-explizit}
\normDrst{A}\;
  =\; 2\, \normDrst{F}\, \dimDrst{F}
    \\[-4.5ex]\nn
\end{align}
Aus~$\normDrst{F}$ folgt also unmittelbar~$\normDrst{A}$~-- die Konvention~$\normDrst{F} \!=\! 1\!/\!2$, Gl.~(\ref{APP:nDrstF-Konv}), impliziert bereits~$\normDrst{A} \!=\! \Nc$, Gl.~(\ref{APP:nDrstA-Konv})~-- und mittelbar~$\normDrst{R}$,~$\forall \Drst{R}$.

Analog zu Gl.~(\ref{APP:AnsatzR}) existiert eine Zerlegung in elementare Tensorkomponenten f"ur die adjungierte Darstellung~$\Drst{A}$.
Ihre Gestalt ist komplizierter, da sie a~priori au"ser Kroneckersymbolen~$\de_{ab}$ auch Konstanten~$d_{abc}$ und Strukturkonstanten~$d_{abc}$ involviert.
Um sie entwickeln zu k"onnen, geben wir zun"achst die unabh"angigen {\it Jacobi-Identit"aten\/} an:
\vspace*{-.5ex}
\begin{align} \label{APP:Jacobi-Kommutator}
0\;
  &=\; \big[[T^a,T^b],T^c\big] + \big[[T^b,T^c],T^a\big] + \big[[T^c,T^a],T^b\big]
    \\[.5ex]
  &\Rightarrow\quad
  0\; =\; f_{abe}\, f_{cde} - f_{ace}\, f_{bde} + f_{ade}\, f_{bce}
    \tag{\ref{APP:Jacobi-Kommutator}$'$}
    \\[-4.5ex]\nn
\end{align}
und:
\vspace*{-.5ex}
\begin{align} \label{APP:Jacobi-Antikommutator}
0\;
  &=\; \big[\{T^a,T^b\},T^c\big] + \big[\{T^b,T^c\},T^a\big] + \big[\{T^c,T^a\},T^b\big]
    \\[.5ex]
  &\underset{\D\Drst{F}}{\Rightarrow}\quad
  0\; =\; f_{abe}\, d_{cde} + f_{ace}\, d_{bde} + f_{ade}\, d_{bce}
    \tag{\ref{APP:Jacobi-Antikommutator}$'$}
    \\[-4.5ex]\nn
\end{align}
Dabei gelten die Gln.~(\ref{APP:Jacobi-Kommutator}),~(\ref{APP:Jacobi-Antikommutator}) f"ur beliebige Matrizen~$T^a$.
Die Gln.~(\ref{APP:Jacobi-Kommutator}$'$),~(\ref{APP:Jacobi-Antikommutator}$'$) folgen durch Einsetzen der expliziten Kommutatorrelation in einer beliebigen Darstellung beziehungsweise der expliziten Kommutator- und Antikommutatorrelation in der fundamentalen Darstellung, vgl. die Gln.~(\ref{APP:Ta-Kommutator}),~(\ref{APP:Ta-Kommutator}$'$).

Aus den Jacobi-Identit"aten~-- den Gln.~(\ref{APP:Jacobi-Kommutator}$'$),~(\ref{APP:Jacobi-Antikommutator}$'$)~-- wiederum folgen:
\vspace*{-.5ex}
\begin{alignat}{2}
&f_{aeg}\, f_{bgh}\, f_{che}&\;
  &=\; \phantom{-} \frac{1}{2}\, \normDrst{A}\; f_{abc}
    \label{APP:fff_f}
    \\[.5ex]
&d_{aeg}\, f_{bgh}\, f_{che}&\;
  &=\; - \frac{1}{2}\, \normDrst{A}\; d_{abc}
    \label{APP:dff_d} \\[.5ex]
&d_{aeg}\, d_{bgh}\, f_{che}&\;
  &=\; - \frac{1}{2}\, \al_1\; f_{abc}\qquad
  \text{bzgl.~$\al_1$ vgl.\@ Gl.~(\ref{APP:Konstante-al1})}
    \label{APP:ddf_f}
    \\[-4.5ex]\nn
\end{alignat}
und zwar: 
Gl.~(\ref{APP:fff_f}) durch Kontraktion von Gl.~(\ref{APP:Jacobi-Kommutator}$'$) mit~$f_{cdg}$,
Gl.~(\ref{APP:dff_d}) durch Kontraktion von Gl.~(\ref{APP:Jacobi-Antikommutator}$'$) mit~$f_{abg}$
und Gl.~(\ref{APP:ddf_f}) durch Kontraktion von Gl.~(\ref{APP:Jacobi-Antikommutator}$'$) mit~$d_{cdg}$.
Dabei ist die letzte dieser Relationen nur Rekapitulation von oben, vgl.\@ Gl.~(\ref{APP:fff_f-pre}) und die Bem.\@ zu Gl.~(\ref{APP:fabcTbTc_csDrst}). \\
Die analoge Kontraktion dreier Konstanten~$d_{abc}$ ist voll symmetrisch, so da"s gilt:
%
\begin{align} 
d_{aeg}\, d_{bgh}\, d_{che}\;
  =\; \hspace*{1.5em}
      \al'_1\; d_{abc} \label{APP:ddd_d} \qquad
  \text{bzgl.~$\al'_1$ s.u.\@ Gl.~(\ref{APP:Konstante-al'1})}
\end{align}
mit~$\al'_1$ einer noch zu bestimmenden Konstanten.

\begin{samepage}
Wir geben nun eine Formel analog zu Gl.~(\ref{APP:AnsatzR}) f"ur die adjungierte Darstellung~$\Drst{A}$ an.
Mit der expliziten Gestalt der Generatoren wie~$(T_\Drst{A}^a)_{\!bc} = -i f_{abc}$, vgl.\@ Gl.~(\ref{APP:TaA_fabc}), steht auf der linken Seite (bis auf ein Vorzeichen)~$f_{abe} f_{cde}$, auf der rechten Seite~-- mit jeweils vier freien Indizes~$a,b,c,d$~-- zun"achst beliebige Kombinationen von Kroneckersymbolen~$\de_{ab}$, Konstanten~$d_{abc}$ und Strukturkonstanten~$f_{abc}$.
Aufgrund der Antisymmetrie unter der Vertauschung~$a \!\leftrightarrow\! b$ und~$c \!\leftrightarrow\! d$ einzeln und der Symmetrie unter~$a \!\leftrightarrow\! c$ und~$b \!\leftrightarrow\! d$ gleichzeitig vereinfacht sich die Struktur erheblich:\FOOT{
   Aufgrund dieser Symmetrien und der Jacobi-Identit"at f"ur die Strukturkonstanten~$f_{abc}$, vgl.\@ Gl.~(\ref{APP:Jacobi-Kommutator}$'$), gen"ugt es o.E.d.A. sich rechts auf Kroneckersymbole und Konstanten~$d_{abc}$ zu beschr"anken.
}
%
\vspace*{-.5ex}
\begin{align} \label{APP:AnsatzA}
&f_{abe} f_{cde}\;
  =\; \ga_1\cdot \Big(\, \de_{ac} \de_{bd} - \de_{ad} \de_{bc}\, \Big)
    + \ga_2\cdot \Big(\, d_{ace}  d_{bde}  - d_{ade}  d_{bce}\,  \Big)
    \\[.5ex]
&\qquad
  a,b,c,d,e = 1,2,\ldots \dimDrst{A} \!\equiv\! \dimNc
    \nn
    \\[-4.5ex]\nn
\end{align}
mit Konstanten~$\ga_1$,~$\ga_2$; vgl.\@ unten die Gln.~(\ref{ga1ga2-explizit}),~(\ref{ga1ga2-explizit}$'$).
\end{samepage}

Wir bestimmen die Konstanten~$\ga_1$,~$\ga_2$ wie folgt.
Kontraktion von Gl.~(\ref{APP:AnsatzA}) mit~$\de_{bd}$ ergibt mithilfe der Gln.~(\ref{APP:ff-de}),~(\ref{APP:ff-de}$'$):
\vspace*{-.5ex}
\begin{align} \label{APP:ga1ga2-1}
\normDrst{A}\;
  =\; \ga_1\, (\dimNc \!-\! 1) - \ga_2\, \al_1
    \\[-4.5ex]\nn
\end{align}
Analoge Kontraktion von Gl.~(\ref{APP:AnsatzA}) mit~$f_{cdg}$ ergibt mithilfe der Gln.~(\ref{APP:ff-de}),~(\ref{APP:ddf_f}):
\vspace*{-.5ex}
\begin{align} \label{APP:ga1ga2-2}
\normDrst{A}\;
  =\; 2\, \ga_1 + \ga_2\, \al_1
    \\[-4.5ex]\nn
\end{align}
Und aus den Gln.~(\ref{APP:ga1ga2-1}),~(\ref{APP:ga1ga2-2}) zusammen:
\vspace*{-.5ex}
\begin{alignat}{3} \label{ga1ga2-explizit}
&\ga_1&\;
  &=\; 2\, \normDrst{A}\, \frac{1}{\dimNc \!+\! 1}&\;
  &=\; 4\, \normDrst{F}\cdot \frac{1}{\Nc}
    \\
&\ga_2&\;
  &=\; \frac{\normDrst{A}}{\al_1}\, \frac{\dimNc \!-\! 3}{\dimNc \!+\! 1}&\;
  &=\; \frac{2\, \Nc\, \normDrst{F}}{ 2\, \normDrst{F}\, \Nc^{-1}\, (\Nc^2 \!-\! 4)}\,
         \frac{\Nc^2 \!-\! 4}{\Nc^2}\;
   =\; 1
    \tag{\ref{ga1ga2-explizit}$'$}
    \\[-4.5ex]\nn
\end{alignat}
Dabei impliziert die zweite Identit"at die Relationen~$\dimNc \!=\! \Nc^2 \!-\! 1$ und~$\normDrst{A} \!=\! 2\dimDrst{F} \normDrst{F}$,~$\dimDrst{F} \!=\! \Nc$, vgl.\@ Gl.~(\ref{APP:nA_nF-explizit}).
Die Konstante~$\al_1$, vgl.\@ Gl.~(\ref{APP:Konstante-al1}), ist in diesem Sinne in der Gestalt
\vspace*{-.5ex}
\begin{align} \label{APP:Konstante-al1_nF,Nc}
\al_1\;
  =\; 4\, \normDrst{F}\, \frac{1}{\dimDrst{F}}\, (\dimNc \!-\! 1) - 2\,
            \dimDrst{F}\, \normDrst{F}\;
  =\; 2\, \normDrst{F}\cdot \frac{1}{\Nc}\, (\Nc^2 \!-\! 4)
    \\[-4.5ex]\nn
\end{align}
eingesetzt.
Wir finden~$\ga_2 \!\equiv\! 1$ unabh"angig von der Konvention f"ur~$\normDrst{F}$ und f"ur alle~$\Nc$.

Die Formel nach Gl.~(\ref{APP:AnsatzA}) ist vollst"andig bestimmt.
Sie setzt uns in die Lage, die Konstante~$\al'_1$ aus Gl.~(\ref{APP:ddd_d}) zu bestimmen.
Dazu kontrahieren wir Gl.~(\ref{APP:AnsatzA}) mit~$d_{bdg}$, das hei"st die Indizes~$a,c,g$ sind frei.
Auf der linken Seite tritt die Kontraktion~$f_{abe}f_{ced}d_{gbd}$ auf, die mithilfe Gl.~(\ref{APP:dff_d}) auf einen Term proportional zu~$d_{acg}$ reduziert wird; nach der Kontraktion~$d_{ade}d_{ceb}d_{gbd}$ auf der rechten Seite l"osen wir auf.
Wir finden:
\vspace*{-.5ex}
\begin{align} 
&d_{aeg}\, d_{bgh}\, d_{che}\;
  =\; \al'_1\; d_{abc}
    \\[.5ex]
  &\text{mit}\qquad \al'_1\;
  =\; \al_1 - \ga_2^{-1}\, \Big( \ga_1 + \frac{1}{2}\, \normDrst{A} \Big)
    \\[-4.5ex]\nn
\end{align}
und explizit:
\vspace*{-.5ex}
\begin{align} \label{APP:Konstante-al'1}
\al'_1\;
  =\; \al_1\cdot \frac{1}{2}\, \frac{\dimNc \!-\! 11}{\dimNc \!-\! 3}
  =\; \normDrst{F}\cdot \frac{1}{\Nc}\, (\Nc^2 \!-\! 12)
    \\[-4.5ex]\nn
\end{align}
Die erste Identit"at in Gl.~(\ref{APP:Konstante-al'1}) ist unmittelbare Konsequenz der allgemeinen, ersten Ausdr"ucke f"ur~$\ga_1$,~$\ga_2$, vgl.\@ die Gln.~(\ref{ga1ga2-explizit}),~(\ref{ga1ga2-explizit}$'$).
Die zweite Identit"at folgt mithilfe~$\al_1$ in der Gestalt von Gl.~(\ref{APP:Konstante-al1_nF,Nc}), das hei"st aufgrund der Relation~$\normDrst{A} \!=\! 2\dimDrst{F} \normDrst{F}$,~$\dimDrst{F} \!=\! \Nc$. \\
\indent
Wir schlie"sen hiermit die Diskussion der~$\suNc$-Eichalgebra bez"uglich allgemeinem~$\Nc$ und betrachten explizit~$\Nc \!\equiv\! 2$ und~$\Nc \!\equiv\! 3$.
\vspace*{-.5ex}

\paragraph{\bm{\SUNc} mit~\bm{\Nc \!\equiv\! 2}} besitzt Gruppendimension~\mbox{$\dimNc \!=\! \Nc^2 \!-\! 1 \!=\! 3$}.
Die Strukturkonstanten sind gegeben durch
\vspace*{-.5ex}
\begin{align} \label{APP:StrukturkonstantenNc=2}
f_{abc}\; 
  =\; \ep^{abc}\qquad
  \text{mit}\qquad
  f_{123}\; =\; \ep^{123}\; =\; 1
    \\[-4.5ex]\nn
\end{align}
das hei"st durch die kontravarianten Komponenten des dreidimensionalen Epsilon-Pseudoten\-sors in der Definition von Gl.~(\ref{APP:epTensor-kontrav}).
Die Konstanten~$d_{abc}$ verschwinden identisch:
\vspace*{-.5ex}
\begin{align} \label{APP:Konstanten-d_abc-Nc=2}
d_{abc}\; \equiv\; 0
    \\[-4.5ex]\nn
\end{align}
Die Dimension der fundamentalen Darstellung ist~$\dimDrst{F} \!=\! \Nc \!=\! 2$ und eine explizite Realisierung gegeben "uber
\vspace*{-.5ex}
\begin{align} \label{APP:GeneratorenNc=2}
T_\Drst{F}^a\;
  =\; \frac{1}{2}\, \si^a\qquad a \!=\! 1,2,3
    \\[-4.5ex]\nn
\end{align}
durch die drei $2 \!\times\! 2$-Pauli-Matrizen~$\si^a$, vgl.\@ Gl.~(\ref{APP:PauliMatrizen_si^i}).
Die Relationen dieses Anhangs~\ref{APP:suNc-Eichalgebra} explizit f"ur~$\Nc \!\equiv\! 2$ folgen unmittelbar durch Einsetzen der Gln.~(\ref{APP:StrukturkonstantenNc=2})-(\ref{APP:GeneratorenNc=2}).
\vspace*{-.5ex}

\paragraph{\bm{\SUNc} mit~\bm{\Nc \!\equiv\! 3}} besitzt Gruppendimension~\mbox{$\dimNc \!=\! \Nc^2 \!-\! 1 \!=\! 8$}.
Explizite Zahlenwerte f"ur die Strukturkonstanten~$f_{abc}$ sind gegeben durch:%
\FOOT{
  \label{APP-FN:f,d-explizit}s"amtliche unabh"angige nichtverschwindende Zahlenwerte
}
%
\vspace*{-.5ex}
\begin{alignat}{4} 
&f_{123}&\;
  &=&\; &1& &
    \\
&f_{147}&\;
  &=&\; - &f_{157}&\;
   =\;    &f_{246}\;
   =\;     f_{257}\;
   =\;     f_{345}\;
   =\;  -  f_{367}\;
   =\;     1/2
    \nn \\
&f_{458}&\;
  &=&\;   &f_{678}&\;
   =\;    &\sqrt3/2
    \nn
    \\[-4.5ex]\nn
\end{alignat}
und f"ur die Konstanten~$d_{abc}$ durch:%
\citeFN{APP-FN:f,d-explizit}
%
\vspace*{-.5ex}
\begin{alignat}{7} 
&d_{118}\;
   =\;     d_{228}&\;
  &=&\;   &d_{338}&\;
  &=&\; - &d_{888}&\;
  &=&\;   &1/\sqrt3
    \\
&d_{146}\;
   =\;     d_{157}&\;
  &=&\; - &d_{247}&\;
  &=&\;   &d_{256}&\;
  &=&\;   &d_{344}\;
   =\;     d_{355}\;
   =\;  -  d_{366}\;
   =\;  -  d_{377}\;
   =\;     1/2
    \nn \\
&d_{448}\;
   =\;     d_{558}&\;
  &=&\;   &d_{668}&\;
  &=&\;   &d_{778}&\;
  &=&\; - &1/(2\sqrt3)
    \nn
    \\[-4.5ex]\nn
\end{alignat}
vgl.\@ etwa  Nachtmann, Ref.~\cite{Nachtmann92}, Tab.~C-1.
Die Dimension der fundamentalen Darstellung ist~$\dimDrst{F} \!=\! \Nc \!=\! 3$ und eine explizite Realisierung gegeben "uber
\vspace*{-.5ex}
\begin{align} \label{APP:GeneratorenNc=3}
T_\Drst{F}^a\;
  =\; \frac{1}{2}\, \la^a\qquad
  a \!=\! 1,2,\ldots 8
    \\[-4.5ex]\nn
\end{align}
durch die acht $3 \!\times\! 3$-Gell-Mann-Matrizen~$\la^a$:
\vspace*{-.5ex}
\begin{alignat}{3} 
\la_1\; &=\; \pmatrixDD{0}{1}{0}{1}{0}{0}{0}{0}{0}&\qquad
\la_4\; &=\; \pmatrixDD{0}{0}{1}{0}{0}{0}{1}{0}{0}&
\la_6\; &=\; \pmatrixDD{0}{0}{0}{0}{0}{1}{0}{1}{0}
    \\[.5ex]
\la_2\; &=\; \pmatrixDD{0}{\rm-i}{0}{\iIM}{0}{0}{0}{0}{0}&\qquad
\la_5\; &=\; \pmatrixDD{0}{0}{\rm-i}{0}{0}{0}{\iIM}{0}{0}&
\la_7\; &=\; \pmatrixDD{0}{0}{0}{0}{0}{\rm-i}{0}{\iIM}{0}
    \nn \\[.5ex]
\la_3\; &=\vv {\rm diag}[1,-1,0]&\qquad
\la_8\; &=\vv 1\!\surd3\cdot {\rm diag}[1,1,-2]&&
    \nn
    \\[-4.5ex]\nn
\end{alignat}
vgl.\@ Nachtmann, Ref.~\cite{Nachtmann92}, die Gln.~(C-44).\FOOT{
  \label{APP-FN:DrstF_Nc-allg}Wir bemerken, da"s die Konstruktion dieser Matrizen in Anlehnung an die Pauli-Matrizen geschieht und deren Verallgemeinerung auf gr"o"sere~$\Nc$ suggeriert.   Wir formulieren Matrizen wie folgt f"ur~$\Nc \!\ge\! 2$, also eingeschlossen die Pauli-Matrizen selbst:

Zum einen $\Nc \!-\! 1$ diagonale Matrizen~$\De^{(j)} \!=\! n_j\, {\rm diag}[1,\ldots,1,-j,0,\ldots,0]$, wobei~$j \!=\! 1,\ldots,\Nc \!-\! 1$, mit~$j$~Ein\-tr"agen Einsen,~$\Nc \!-\! (j \!+\! 1)$ Eintr"agen Nullen und~$n_j \!=\! \sqrt{\normDrst{F}/j(j\!+\!1)}$, vgl. die Pauli-Matrix~$\si^3$ nach Gl.~(\ref{APP:PauliMatrizen_si^i}). \\
Zum anderen~$\Nc(\Nc\!-\!1)$ nicht-diagonale Matrizen~$\Si^{(i,k)}$, wobei~$i \!=\! 1,2$ und~$k \!=\! 1,2,\ldots, 1\!/\!2\cdot\Nc(\Nc\!-\!1)$,~in Anlehnung an die nicht-diagonalen Pauli-Matrizen~$\si^i$, vgl. Gl.~(\ref{APP:PauliMatrizen_si^i}).
Die~$\Si^{(i,k)}$ werden folgenderma"sen konstruiert:
Zun"achst wird das Element der Matrix~$\Si^{(i,k)}$, mit~$i$,~$k$ fest, an der Stelle~$(\al_k,\be_k)$, mit \mbox{$1 \!\le\! \al_k \!<\! \be_k \!\le\! \Nc$} fest, identifiziert mit~$(\si^i)_{12}$, das~Ele\-ment an der Stelle~$(\be_k,\al_k)$ mit~$(\si^i)_{21}$, alle anderen identisch Null gesetzt.
Dann wird diese Matrix mit dem Faktor~$n_{i,k} \!=\! \sqrt{\normDrst{F}/2}$ versehen.
Da~\mbox{$1\!/\!2 \!\cdot\! \Nc(\Nc\!-\!1)$} verschiedene Paare~$\{\al_k,\be_k\}$, mit~$1 \!\le\! \al_k \!<\! \be_k \!\le\! \Nc$,~existie\-ren (Anzahl der Elemente einer~$\Nc \!\times\! \Nc$-Matrix "uber der Hauptdiagonalen), folgen aus dieser Vorschrift in eindeutiger Weise~$\Nc(\Nc\!-\!1)$ Matrizen~$\Si^{(i,k)}$.

Die nach dieser Vorschrift konstruierten~$\dimNc \!=\! \Nc^2 \!-\! 1$ Matrizen~$\De^{(j)}$,~$\Si^{(i,k)}$ sind spurlos, hermitesch und normiert im Sinne von Gl.~(\ref{APP:Ta-Normierung}).
Wir "uberlegen uns leicht, da"s sie linear unabh"angig sind und folglich nach Gl.~(\ref{APP:Generatoren}) eine Basis bilden der fundamentalen Darstellung~$\Drst{F}$ der Liealgebra~$\suNc$; wir identifizieren diese Matrizen daher mit deren Generatoren~$T_\Drst{F}^a$.
}

Abschlie"send geben wir eine Relation an, die speziell f"ur~$\Nc \!\equiv\! 3$ gilt; ihre Existenz folgt durch Abz"ahlen irreduzibler Tensorstrukturen mit vier freien Indizes~$a,b,c,d$:
\vspace*{-.5ex}
\begin{align} \label{APP:dd_dede}
d_{abe}\, d_{cde} + d_{ace}\, d_{bde} + d_{ade}\, d_{bce}\;
  =\; \frac{1}{3}\,
      \Big[\, \de_{ab}\, \de_{cd} + \de_{ac}\, \de_{bd} + \de_{ad}\, \de_{bc}\, \Big]
    \\[-4.5ex]\nn
\end{align}
Sei verwiesen auf Ref.~\cite{Pascual84}, Gl.~(A.26), und die Argumentation dort.
\vspace*{5ex}
\theendnotes

%% file: APP_KINEMATIK.tex
\lhead[\fancyplain{}{\sc\thepage}]
      {\fancyplain{}{\sc\rightmark}}
\rhead[\fancyplain{}{\sc{{\footnotesize Anhang~\thechapter:} Kinematik bei gro"sen~$s$}}]
      {\fancyplain{}{\sc\thepage}}
\psfull
\chapter[Kinematik f"ur gro"se~\protect\boldmath$s$]{%
   \huge Kinematik f"ur gro"se~\protect\boldmath$s$}
\label{APP:Kinematik}

In Abschnitt~\ref{Sect:Kinematik} wird diskutiert Streuung
\vspace*{-.5ex}
\begin{align} \label{APP:2hto2h}
h^1(P_1) \;+\; h^2(P_2)\;
  \longrightarrow\;
  h^{1'}(P_{1'}) \;+\; h^{2'}(P_{2'})
    \\[-4.5ex]\nn
\end{align}
Dieser Anhang formalisiert und vertieft diese Diskussion in Hinblick auf die Kinematik f"ur gro"se~$s$.
Zun"achst skizzieren wir die Herleitung wichtiger, im Haupttext nur zitierter Relationen, die zugrunde liegen den dann angegebenen Reihen-Entwicklungen der relevanten Gr"o"sen f"ur gro"se invariante Schwerpunktenergie, das hei"st f"ur kleinen Parameter
\vspace*{-.5ex}
\begin{align} \label{APP:kappa}
\ka\; \equiv\; \frac{1}{s}
    \\[-4.5ex]\nn
\end{align}
Wir diskutieren den allgemeinen Fall, der Erhaltung des Gesamtimpulses impliziert, aber noch nicht spezifiziert das Bezugsystem~-- und den Fall, der dieses konkretisiert als das Schwerpunktsystem bez"uglich der drei Raumrichtungen.

\section[Zusammenhang von~\protect$s$ und~\protect$\vep_i$ f"ur~\protect$i \!=\! 1,1'$]{%
         Zusammenhang von~\bm{s} und~\bm{\vep_i} f"ur~\bm{i \!=\! 1,1'}}
\label{APP-Sect:vepi-s}

Im Haupttext sind alle relevanten Gr"o"sen und damit ihre Abh"angigkeit von der invarianten Schwerpunktenergie~$\surd s$ ausgedr"uckt durch die Gr"o"sen~$\vep_i$ f"ur~$i \!=\! 1,1',2,2'$:
so die Impulskomponenten~$P_i^+$,~$P^-_i$ und~$P_i^0$,~$P_i^3$ und das Quadrat des invarianten Impulstransfers~$-t$.

Mit~$\vep_2$ ist das Produkt~$\vep \!\equiv\! \vep_1\vep_2$ bestimmt in Termen der Funktion~$\vep_1$, vgl.\@ die Gln.~(\ref{vep2_vrh_vep1}), (\ref{vep2_vrh_vep1}$'$) und~(\ref{vep1vep2-vrh}).
Es bleibt daher zu bestimmen~$\vep_1$ als Funktion von~$s$.
Durch Indexsubstitution~$1 \!\to\! 1'$ und~$2 \!\to\! 2'$ im Sinne von Gl.~(\ref{Subst1<->2,1'<->2'}) folgen unmittelbar~$\vep_{1'}$,~$\vep' \!\equiv\! \vep_{1'}\vep_{2'}$ und~$\vep_{2'}$.
Wir bestimmen zun"achst umgekehrt~$s$ als Funktion von~$\vep_1$.

\subsection[Funktion~\protect$s(\vep_i)$]{%
            Funktion~\bm{s(\vep_i)} f"ur~\bm{i \!=\! 1,1'}}

Wir leiten her die Gln.~(\ref{s_vep1}),~(\ref{s_vep1}$'$).
Statt unmittelbar~$s$, bestimmen wir zun"achst die~Vari\-able~$\si$ als Funktion von~$\vep_1$; $s$ selbst folgt "uber deren Definition:
\vspace*{-.5ex}
\begin{align} \label{APP:s_vep1--1}
\si(s)\; =\;
    \frac{1}{2}\, \Big[\, s\; -\;
      \big[\rb{M}_1^2 + \rb{M}_2^2 - \big(\rb{P}_1 \!+\! \rb{P}_2\big)^2\big]\,
      \Big]
    \\[-4.5ex]\nn
\end{align}
vgl.\@ Gl.~(\ref{vep1vep2_si_s}$'$).
Aufgel"ost nach~$\si$ folgt aus Gl.~(\ref{Mandelstam-st}):
\vspace*{-.5ex}
\begin{alignat}{3} \label{APP:s_vep1-0}
2\,\si\;
  &=&\; \big(\vep_1\vep_2\big)^{-1}&\;
        \big\{\rb{M}_1^2\, \rb{M}_2^2\; +&\; \big(\vep_1\vep_2\big)^2&\big\}
    \\[.5ex]
  &=&\;    \big(\vep|\isi\big)^{-1}&\;
        \big\{\rb{M}_1^2\, \rb{M}_2^2\; +&\;    \big(\vep|\isi\big)^2&\big\}
    \tag{\ref{APP:s_vep1-0}$'$} \\[.5ex]
  &=&\;   \big(\vep|\ivrh\big)^{-1}&\;
        \big\{\rb{M}_1^2\, \rb{M}_2^2\; +&\;    \big(\vep|\ivrh\big)^2&\big\}
    \tag{\ref{APP:s_vep1-0}$'$}
    \\[-4ex]\nn
\end{alignat}
Dabei folgt die letzte Zeile durch Identifizieren von~$\vep|\isi$ mit~$\vep|\ivrh$, vgl.\@ die Gln.~(\ref{vep2_vrh_vep1}),~(\ref{vep1vep2_si_s}).
Partiell eingesetzt hierf"ur der Ausdruck in Termen von~$\vrh$ nach Gl.~(\ref{vep1vep2-vrh}), folgt hieraus:
\vspace*{-.5ex}
\begin{align} 
2\,\si\;
  &=\; \big(\vep|\ivrh\big)^{-1}\,
        \Big\{ \rb{M}_1^2\, \rb{M}_2^2\;
        +\; \vep_1^2\; \Big[-2\,\vrh\;
              \Big(\, -\vrh\; +\; \sqrt{\vrh^2 + \rb{M}_2^2}\, \Big)\;
              +\; \rb{M}_2^2 \Big]
        \Big\}
    \\[-4.5ex]\nn
\end{align}
das hei"st:
\vspace*{-.5ex}
\begin{align} \label{APP:s_vep1-1}
2\,\si\;
  &=\; \big(\vep|\ivrh\big)^{-1}\,
        \big\{\vep_1\, \big(\rb{M}_1^2\, \vep_1^{-1} + \vep_1\big)\, \rb{M}_2^2\;
          -\; 2\,\vrh\; \vep_1\, \big(\vep|\ivrh\big)
        \big\}
    \\[-4.5ex]\nn
\end{align}
Im ersten Term wird benutzt~$\big(\vep|\ivrh\big)^{-1}\,\vep_1 \!=\! \vep_2^{-1}$, im zweiten gek"urzt~$\vep|\ivrh$; es folgt:
\vspace*{-.5ex}
\begin{align} \label{APP:s_vep1-2}
2\,\si\;
  &=\; \rb{M}_2^2\, \vep_2^{-1}\cdot \big(\rb{M}_1^2\, \vep_1^{-1} + \vep_1\big)\;
         -\; 2\,\vrh\, \vep_1
    \\[-4.5ex]\nn
\end{align}
Mithilfe der expliziten Darstellung von~$\vrh$ wie folgt:
\vspace*{-.5ex}
\begin{align} \label{APP:s_vep1-3}
\vrh\;
  =\; \frac{1}{2}\, \big(\rb{M}_1^2\; \vep_1^{-1} - \vep_1\big)\;
        -\; (P_1^3 \!+\! P_2^3)^2
    \\[-4.5ex]\nn
\end{align}
vgl.\@ Gl.~(\ref{vep2_vrh_vep1}$'$), schreiben wir Gl.~(\ref{Bezugsystem}) um in die Form:
\vspace*{-.5ex}
\begin{align} \label{APP:s_vep1-4}
\rb{M}_2^2\, \vep_2^{-1}\; =\; \vep_2 + 2\,\vrh
    \\[-4.5ex]\nn
\end{align}
Diese Relation eingesetzt in Gl.~(\ref{APP:s_vep1-2}) folgt unmittelbar:
\vspace*{-.5ex}
\begin{align} \label{APP:s_vep1-5}
2\,\si\;
  &=\; \big(\vep_2 + 2\,\vrh\big)\cdot \big(\rb{M}_1^2\, \vep_1^{-1} + \vep_1\big)\;
         -\; 2\,\vrh\, \vep_1
    \\[.5ex]
  &=\; \vep_2\cdot \big(\rb{M}_1^2\, \vep_1^{-1} + \vep_1\big)\;
         +\; 2\,\vrh\cdot \rb{M}_1^2\, \vep_1^{-1}
    \tag{\ref{APP:s_vep1-5}$'$}
    \\[-4.5ex]\nn
\end{align}
Schlie"slich geschrieben~$\vep_2$ in Termen von~$\vrh$, vgl.\@ Gl.~(\ref{vep2_vrh_vep1}), folgt mithilfe Gl.~(\ref{APP:s_vep1--1}):
\vspace*{-.5ex} 
\begin{align} \label{APP:s_vep1}
&s(\vep_1)\;
  =\; \big[\rb{M}_1^2 + \rb{M}_2^2 - \big(\rb{P}_1 \!+\! \rb{P}_2\big)^2\big]\; +\; 2\,\si
    \\[1ex]
  &\text{mit}\qquad
  2\,\si(\vep_1)\;
    =\; \sqrt{\vrh^2 + \rb{M}_2^2}\cdot \big(\rb{M}_1^2\; \vep_1^{-1} + \vep_1\big)\;
                          +\; \vrh\cdot \big(\rb{M}_1^2\; \vep_1^{-1} - \vep_1\big)
    \tag{\ref{APP:s_vep1}$'$}
    \\[-4.5ex]\nn
\end{align}
Dabei ist~$\vrh \!\equiv\! \vrh(\vep_1)$ nach Gl.~(\ref{APP:s_vep1-3}) aufzufassen als Funktion von~$\vep_1$.
Dies ist genau die~Dar\-stellung im Haupttext, vgl.\@ die Gln.~(\ref{s_vep1}),~(\ref{s_vep1}$'$).

Seien in Gl.~(\ref{APP:s_vep1}$'$) die~$\vep_1$ umgeschrieben als~$P_1^+$ mithilfe der Relation~$\vep_1 \!\equiv\! \al\rb{M}_1^2\!/P_1^+$, vgl.\@ Gl.~(\ref{epsilons}),~-- das hei"st explizit gemacht Faktoren~$\al\rb{M}_1^2$.
Dies ist die Darstellung von~$s$ als Funktion von~$P_1^+$, wobei die Abh"angigkeit von~$P_2^-$ eliminiert ist.

\subsection[Funktion~\protect$\vep_i(s)$]{%
            Funktion~\bm{\vep_i(s)}, f"ur~\bm{i \!=\! 1,1'}}

Wir leiten her die Gln.~(\ref{vep1_s})-(\ref{vep1_s}$''$).
Durch Gl.~(\ref{vep1vep2-vrh}) ist~$\vep|\ivrh$~definiert in Termen von~$\vrh$; wir schreiben diese Relation um in die Form:
\vspace*{-.5ex}
\begin{align} \label{APP:vep1_s-0}
\vep|\ivrh\;
  =\; -\vep_1\,\vrh\;
        +\; {\rm sign}(\vep_1)\sqrt{\vep_1^2\, \left(\vrh^2 + \rb{M}_2^2\right)}
    \\[-4.5ex]\nn
\end{align}
Aufgel"ost nach dem Wurzelausdruck und die Gleichung quadriert, folgt:
\vspace*{-.5ex}
\begin{align} \label{APP:vep1_s-1}
0\;
  =\; \vep_1^2\, \rb{M}_2^2\;
        -\; 2\, \big(\vep|\ivrh\big)\; \big(\vep_1\, \vrh\big)\;
        -\; \big(\vep|\ivrh\big)^2
    \\[-4.5ex]\nn
\end{align}
Aufgrund der expliziten Darstellung von~$\vrh$, vgl.\@ die Gln.~(\ref{vep2_vrh_vep1}$'$),~(\ref{APP:s_vep1-3}), gilt andererseits:\zz
\vspace*{-.5ex}
\begin{align} \label{APP:vep1_s-2}
\vep_1\,\vrh\;
  =\; \frac{1}{2}\, \big(\rb{M}_1^2 - \vep_1^2\big)\;
        -\; \vep_1\, (P_1^3 \!+\! P_2^3)^2
    \\[-4.5ex]\nn
\end{align}
Dies eingesetzt in Gl.~(\ref{APP:vep1_s-1}) folgt die in~$\vep_1$ quadratische Gleichung:
\vspace*{-.5ex}
\begin{align} \label{APP:vep1_s-3}
0\;
  &=\; \vep_1^2\cdot \big(\rb{M}_2^2 + \vep|\ivrh\big)\;
        +\; 2\, \vep_1\cdot \vep|\ivrh\, (P_1^3 \!+\! P_2^3)\;
        -\; \vep|\ivrh\, \big(\rb{M}_1^2 + \vep|\ivrh\big)
    \\[.5ex]
  &=\; \vep|\ivrh\; \be\vv \left\{
        \left[\vep_1\big/\be + (P_1^3 \!+\! P_2^3)\right]^2\;
        -\; \left[(P_1^3 \!+\! P_2^3)^2 + \ga\right]
      \right\}
    \tag{\ref{APP:vep1_s-3}$'$}
    \\[-4.5ex]\nn
\end{align}
\vspace*{-.375ex}mit abk"urzend~\mbox{$\be \!\stackrel{\D!}{=}\! \vep|\ivrh\!\big/(\rb{M}_2^2 \!+\! \vep|\ivrh)$} und~\mbox{$\ga \!\stackrel{\D!}{=}\! (\rb{M}_2^2 \!+\! \vep|\ivrh)\!\big/\be$}.
Wir lesen unmittelbar ab die Nullstellen der geschweiften Klammer:
\vspace*{-.5ex}
\begin{align} \label{APP:vep1_s-4}
\vep_1\big/\be\;
  =\; -(P_1^3 \!+\! P_2^3)\;
        \pm\; \sqrt{(P_1^3 \!+\! P_2^3)^2 + \ga}
    \\[-4.5ex]\nn
\end{align}
Wir identifizieren~$\vep|\ivrh \!\equiv\! \vep|\isi$; diese Funktion, abk"urzend notiert durch~$\vep \!\equiv\! \vep|\isi$, fassen wir auf mithilfe der Gln.~(\ref{vep1vep2_si_s}),~(\ref{vep1vep2_si_s}$'$) als Funktion von~$\si(s)$, also letztlich von~$s$.
Wir finden zusammenfassend:%
\FOOT{
  Das Vorzeichen des Wurzelausdrucks folgt aus der Analyse der Vorzeichen der involvierten Gr"o"sen im Limes~"`$s$~gro"s"', das hei"st~"`Epsilons klein"': dabei ist~${\rm sign}(\vep_i) \!\equiv\! {\rm sign}(\rb{M}_i^2)$, vgl.\@ die Gln.~(\ref{epsilons}),~(\ref{epsilons}$'$), und~der Radiant reell positiv.   Entsprechend die Gln.~(\ref{vep2_vrh_vep1}) und~(\ref{vep1vep2_si_s}).
}
%
\vspace*{-.5ex} 
\begin{align} \label{APP:vep1_s}
&\vep_1(s)\;
  =\; \be\, \Big[-\big(P_1^3 \!+\! P_2^3\big)\;
        +\; \sqrt{\big(P_1^3 \!+\! P_2^3\big)^2 + \ga}\Big]
    \\[.5ex]
  &\text{mit}\qquad
  \be(\vep)\;
    =\; \vep\big/\big(\rb{M}_2^2 + \vep\big)
    \tag{\ref{APP:vep1_s}$'$} \\
  &\phantom{\text{mit}\qquad}
  \ga(\vep)\;
    =\; \big(\rb{M}_1^2 + \vep\big) \big(\rb{M}_2^2 + \vep\big)\big/\vep
    \tag{\ref{APP:vep1_s}$''$}
    \\[-4.5ex]\nn
\end{align}
Dies ist genau die Darstellung im Haupttext, vgl.\@ die Gln.~(\ref{vep1_s})-(\ref{vep1_s}$''$).

\section[Entwicklung f"ur gro"se~\protect$s$]{%
         Entwicklung f"ur gro"se~\bm{s}}
\label{APP-Sect:Entwicklungen}

Sei zugrundegelegt~$P \!\equiv\! 0$, vgl.\@ Gl.~(\ref{ViererDifferenzimpuls}), das hei"st impliziert, da"s die der Streuung zugrundeliegende Wechselwirkung den Gesamtimpuls erh"alt.
Dann wird das Bezugsystem bestimmt durch die drei einlaufenden Impulse, die wir abk"urzend notieren in der Form
\vspace*{-.5ex}
\begin{alignat}{2} \label{APP:Piin}
&\rb{P}\iin&\; &\equiv\; \rb{P}_1 + \rb{P}_2
    \\
&P\iin^3&\; &\equiv\; P_1^3 + P_2 ^3
    \tag{\ref{APP:Piin}$'$}
    \\[-4.5ex]\nn
\end{alignat}
Wir betrachten die F"alle:
\begin{align} \label{APP:Bzgsystem}
\hspace*{-2em}
\text{(i)}\quad
  &\rb{P}\iin,\vv P\iin^3\vv
  \text{-- das hei"st Bezugsystem -- nicht spezifiziert}
    \\
\hspace*{-2em}
\text{(ii)}\quad
  &\rb{P}\iin \equiv \bm{0},\vv P\iin^3 \equiv 0:\vv
  \text{Schwerpunktsystem bzgl.\@ der drei Raumrichtungen}
    \tag{\ref{APP:Bzgsystem}$'$}
\end{align}
und entwickeln Fall~(ii) bis h"ohere Ordnung gegen"uber Fall~(i). \\
\indent
Aufgrund der bestehenden Symmetrien beschr"anken wir uns o.E.d.A., explizit anzugeben nur die Entwicklungen f"ur die Gr"o"sen mit~{\bf Index~\bm{1}}.
Die entsprechenden Entwicklungen der Gr"o"sen mit~{\bf Index~\bm{1'}} folgen durch Indexsubstitution~{\bf\bm{1 \!\to\! 1'} und~\bm{2 \!\to\! 2'}} im Sinne von Gl.~(\ref{Subst1->1',2->2'}), die der Gr"o"sen mit~{\bf Index~\bm{1'}} und~\bm{2'} weiter durch Indexsubstitution~{\bf\bm{1 \!\leftrightarrow\! 2} und~\bm{1' \!\leftrightarrow\! 2'}} im Sinne von Gl.~(\ref{Subst1<->2,1'<->2'}).

\subsection[(Inverse) Epsilon~\protect$\vep_i$,~\protect$\vep_i^{-1}$%
             ~bzw. Impulskomponenten~\protect$P_i^+$,~\protect$P_i^-$]{%
            (Inverse) Epsilon~\bm{\vep_i},~\bm{\vep_i^{-1}}
              bzw.\@ Impulskomponenten~\bm{P_i^+},~\bm{P_i^-}}

Mit den~$\vep_i$,~$\vep_i^{-1}$ unmittelbar in Zusammenhang stehen die Lichtkegelkomponenten der Impulse~$P_i^+$,~$P_i^-$.
Es gilt:
\begin{alignat}{2}
P_i^+ &= \al\, \rb{M}_i^2\vv \vep_i^{-1}&
  \qquad\qquad&\text{f"ur\;~$i \!=\! 1,1'$}
    \label{APP:h-Impuls+} \\
P_i^- &= \al\, \rb{M}_i^2\vv \vep_i^{-1}&
  \qquad\qquad&\text{f"ur\;~$i \!=\! 2,2'$}
    \tag{\ref{APP:h-Impuls+}$'$}
    \\[-6ex]\nn
\intertext{\vspace{-1ex}und:}
P_i^- &= \al\, \vep_i&
  \qquad\qquad&\text{f"ur\;~$i \!=\! 1,1'$}
    \label{APP:h-Impuls-} \\
P_i^+ &= \al\, \vep_i&
  \qquad\qquad&\text{f"ur\;~$i \!=\! 2,2'$}
    \tag{\ref{APP:h-Impuls-}$'$}
\end{alignat}
vgl.\@ die Gln.~(\ref{epsilons}),~(\ref{epsilons}$'$) bzw.~(\ref{h-Impuls-}),~(\ref{h-Impuls-}$'$).
Entsprechend f"ur die Reihen-Entwicklungen.

\subsubsection{(i)\vv Nicht-spezifiziertes Bezugsystem:}

%
\begin{align} \label{APP:vep1}
&\vep_1(\ka)
    \\[.5ex]
&=\;
  \rb{M}_1^2
    \vv {\sqrt{\ka}}
    \nn \\[.5ex]
&\phantom{=\;}
  - \rb{M}_1^2\,P\iin^3
    \vv \ka
    \nn \\[.5ex]
&\phantom{=\;}
  + {\frac{
      \rb{M}_1^2
    }{2}} \,
  \Big\{
      2\,\rb{M}_2^2 + {{P\iin^3}^2} - 
        \rb{P}\iin^2
  \Big\}
    \vv \ka^{3\!/\!2}
    \nn \\[.5ex]
&\phantom{=\;}
  + \rb{M}_1^2\,P\iin^3\,
  \Big\{
      -\rb{M}_2^2 + 
      \rb{P}\iin^2
  \Big\}
    \vv {{\ka}^2}
    \nn \\[.5ex]
&\phantom{=\;}
  + {\frac{
      \rb{M}_1^2
    }{8}} \,
  \Big\{
      8\,\rb{M}_1^2\,\rb{M}_2^2 + 
        8\,{{\rb{M}_2^2}^2} + 
        4\,\rb{M}_2^2\,{{P\iin^3}^2} - 
        {{P\iin^3}^4}
  - 6\,\Big[
      2\,\rb{M}_2^2 + 
           {{P\iin^3}^2}
  \Big] \, \rb{P}\iin^2
    \nn \\
&\phantom{=\; \quad}
  + 3\,{{\left( \rb{P}\iin^2 \right) }^2}
  \Big\}
    \vv \ka^{5\!/\!2}
    \nn \\[.5ex]
&\phantom{=\;}
  - \rb{M}_1^2\,P\iin^3\,
  \Big\{
      \rb{M}_2^2\,
      \left( \rb{M}_1^2 + \rb{M}_2^2 \right)  - 
      2\,\rb{M}_2^2\,
      \rb{P}\iin^2 + 
      {{\left( \rb{P}\iin^2 \right)}^2}
  \Big\}
    \vv {{\ka}^3}
    \nn \\[.5ex]
&\phantom{=\;}
  + {\frac{
      \rb{M}_1^2
    }{16}} \,
  \Big\{
      16\,{{\rb{M}_1^2}^2}\,\rb{M}_2^2 + 
        16\,{{\rb{M}_2^2}^3} + 
        8\,{{\rb{M}_2^2}^2}\,{{P\iin^3}^2} - 
        2\,\rb{M}_2^2\,{{P\iin^3}^4} + 
        {{P\iin^3}^6}
    \nn \\
&\phantom{=\; \quad}
    + 8\,\rb{M}_1^2\,\rb{M}_2^2\,
    \Big[ 6\,\rb{M}_2^2 + 
           {{P\iin^3}^2}
    \Big]
    - 5\,\Big[ 8\,\rb{M}_1^2\,\rb{M}_2^2 + 
           8\,{{\rb{M}_2^2}^2} + 
           4\,\rb{M}_2^2\,{{P\iin^3}^2} - 
           {{P\iin^3}^4}
    \Big] \,
         \rb{P}\iin^2
    \nn \\
&\phantom{=\; \quad}
    + 15\,\left( 2\,\rb{M}_2^2 + 
           {{P\iin^3}^2} \right) \,
         {{\left( \rb{P}\iin^2
               \right) }^2} - 
        5\,{{\left( \rb{P}\iin^2 \right) }^3}
  \Big\}
    \vv \ka^{7\!/\!2}
    \nn \\[.5ex]
&\phantom{=\;}
  + {{{\cal O}(\ka^4)}}
    \nn
\end{align}
%

%
\begin{align} \label{APP:vep1^-1}
&\vep_1^{-1}(\ka)
    \\[.5ex]
&=\;
  {\frac{1}{\rb{M}_1^2}}
    \vv \frac{1}{\sqrt{\ka}}
    \nn \\[.5ex]
&\phantom{=\;}
  + {\frac{P\iin^3}{\rb{M}_1^2}}
    \nn \\[.5ex]
&\phantom{=\;}
  + {\frac{1
  }{2\,\rb{M}_1^2}}\,
  \Big\{
    -2\,\rb{M}_2^2 + 
        {{P\iin^3}^2} + 
        \rb{P}\iin^2
  \Big\}
    \vv {\sqrt{\ka}}
    \nn \\[.5ex]
&\phantom{=\;}
  - {\frac{\rb{M}_2^2\,P\iin^3
  }{\rb{M}_1^2}}
    \vv \ka
    \nn \\[.5ex]
&\phantom{=\;}
  - {\frac{1
  }{8\,\rb{M}_1^2}}\,
  \Big\{
    8\,\rb{M}_1^2\,\rb{M}_2^2 + 
        4\,\rb{M}_2^2\,{{P\iin^3}^2} + 
        {{P\iin^3}^4} + 
    \Big[ -4\,\rb{M}_2^2 + 
           2\,{{P\iin^3}^2}
    \Big] \,
         \rb{P}\iin^2 + 
        {{\left( \rb{P}\iin^2
              \right) }^2}
  \Big\}
  \vv \ka^{3\!/\!2}
    \nn \\[.5ex]
&\phantom{=\;}
  + {\frac{\rb{M}_2^2\,P\iin^3
  }{\rb{M}_1^2}}\,
  \Big\{
      -\rb{M}_1^2 + 
        \rb{P}\iin^2
  \Big\}
    \vv {{\ka}^2} 
    \nn \\[.5ex]
&\phantom{=\;}
  + {\frac{1
  }{16\,
      \rb{M}_1^2}}\,
  \Big\{
    -16\,{{\rb{M}_1^2}^2}\,
         \rb{M}_2^2 + 
        2\,\rb{M}_2^2\,{{P\iin^3}^4} + 
        {{P\iin^3}^6} - 
        8\,\rb{M}_1^2\,\rb{M}_2^2\,
    \Big[ 2\,\rb{M}_2^2 + 
           {{P\iin^3}^2}
    \Big]
    \nn \\
&\phantom{=\; \quad}
    + 3\,\Big[ 8\,\rb{M}_1^2\,\rb{M}_2^2 + 
           4\,\rb{M}_2^2\,{{P\iin^3}^2} + 
           {{P\iin^3}^4}
    \Big] \,
         \rb{P}\iin^2
    + \Big[ -6\,\rb{M}_2^2 + 
           3\,{{P\iin^3}^2}
    \Big] \,
         {{\left( \rb{P}\iin^2
               \right) }^2}
    \nn \\
&\phantom{=\; \quad}
    + {{\left( \rb{P}\iin^2
              \right) }^3}
  \Big\}
  \vv \ka^{5\!/\!2}
    \nn \\[.5ex]
&\phantom{=\;}
  - {\frac{\rb{M}_2^2\,P\iin^3\,
  }{\rb{M}_1^2}}\,
  \Big\{
    \rb{M}_1^2\,
         \left( \rb{M}_1^2 + \rb{M}_2^2 \right) 
          - 2\,\rb{M}_1^2\,
         \rb{P}\iin^2 + 
        {{\left( \rb{P}\iin^2
              \right) }^2}
  \Big\}
  \vv {{\ka}^3}
    \nn
\end{align}
\begin{align}
&\phantom{=\;}
  - {\frac{1
  }{128\,
      \rb{M}_1^2}}\,
  \Big\{
    128\,{{\rb{M}_1^2}^3}\,
         \rb{M}_2^2 + 
        64\,{{\rb{M}_1^2}^2}\,\rb{M}_2^2\,
    \Big[ 6\,\rb{M}_2^2 + 
           {{P\iin^3}^2}
    \Big]  + 
        {{P\iin^3}^6}\,
    \Big[ 8\,\rb{M}_2^2 + 
           5\,{{P\iin^3}^2}
    \Big]
    \nn \\
&\phantom{=\; \quad}
    + 16\,\rb{M}_1^2\,\rb{M}_2^2\,
    \Big[ 8\,{{\rb{M}_2^2}^2} + 
           4\,\rb{M}_2^2\,{{P\iin^3}^2} - 
           {{P\iin^3}^4}
    \Big]
    \nn \\
&\phantom{=\; \quad}
    - 20\,
    \Big[ 16\,{{\rb{M}_1^2}^2}\,
            \rb{M}_2^2 + 
           8\,\rb{M}_1^2\,\rb{M}_2^2\,
            \left( 2\,\rb{M}_2^2 + 
              {{P\iin^3}^2} \right)  - 
           {{P\iin^3}^4}\,
            \left( 2\,\rb{M}_2^2 + 
              {{P\iin^3}^2} \right)
    \Big] \,
         \rb{P}\iin^2
    \nn \\
&\phantom{=\; \quad}
    + 30\,
    \Big[ 8\,\rb{M}_1^2\,\rb{M}_2^2 + 
           4\,\rb{M}_2^2\,{{P\iin^3}^2} + 
           {{P\iin^3}^4}
    \Big] \,
         {{\left( \rb{P}\iin^2
               \right) }^2}
    + 20\,
    \Big[ -2\,\rb{M}_2^2 + 
           {{P\iin^3}^2}
    \Big] \,
         {{\left( \rb{P}\iin^2
               \right) }^3}
    \nn \\
&\phantom{=\; \quad}
    + 5\,{{\left( \rb{P}\iin^2 \right) }^4}
  \Big\}
  \vv \ka^{7\!/\!2}
    \nn \\[.5ex]
&\phantom{=\;}
  + {{{\cal O}(\ka^4)}}
    \nn
\end{align}

\subsubsection{(ii)\vv Schwerpunktsystem,~\bm{\rb{P}\iin \!\equiv\! \bm{0}} und~\bm{P\iin^3 \!\equiv\! 0}:}
%
\begin{align} \label{APP:vep1-P3IN,PtrIN=0}
\vep_1(\ka)\;
&=\;
  \rb{M}_1^2
    \vv {\sqrt{\ka}}
    \\[.5ex]
&\phantom{=\;}
  + \rb{M}_1^2\,\rb{M}_2^2
    \vv \ka^{3\!/\!2}
    \nn \\[.5ex]
&\phantom{=\;}
  + \rb{M}_1^2\,\rb{M}_2^2\,
    \left( \rb{M}_1^2 + \rb{M}_2^2 \right)
    \vv \ka^{5\!/\!2}
    \nn \\[.5ex]
&\phantom{=\;}
  + \rb{M}_1^2\,\rb{M}_2^2\,
    \left( {{\rb{M}_1^2}^2} + 
      3\,\rb{M}_1^2\,\rb{M}_2^2 + 
      {{\rb{M}_2^2}^2} \right)
    \vv \ka^{7\!/\!2}
    \nn \\[.5ex]
&\phantom{=\;}
  + \rb{M}_1^2\,\rb{M}_2^2\,
    \left( {\rb{M}_1^2} + {\rb{M}_2^2} \right)\,
    \left( {{\rb{M}_1^2}^2} + 
      5\,\rb{M}_1^2\,\rb{M}_2^2 + 
      {{\rb{M}_2^2}^2} \right)
    \vv \ka^{9\!/\!2}
    \nn \\[.5ex]
&\phantom{=\;}
  + {{{\cal O}(\ka^{11\!/\!2})}}
    \nn
\end{align}
%

%
\begin{align} \label{APP:vep1^-1-P3IN,PtrIN=0}
\vep_1^{-1}(\ka)\;
&=\;
  {\frac{1}{\rb{M}_1^2}}
    \vv \frac{1}{\sqrt{\ka}}
    \\[.5ex]
&\phantom{=\;}
  - {\frac{\rb{M}_2^2}{
      \rb{M}_1^2}}
    \vv {\sqrt{\ka}}
    \nn \\[.5ex]
&\phantom{=\;}
  - \rb{M}_2^2
    \vv \ka^{3\!/\!2}
    \nn \\[.5ex]
&\phantom{=\;}
  - \rb{M}_2^2\,\left( \rb{M}_1^2 + 
      \rb{M}_2^2 \right)
    \vv \ka^{5\!/\!2}
    \nn \\[.5ex]
&\phantom{=\;}
  - \rb{M}_2^2\,
    \left( {{\rb{M}_1^2}^2} + 
      3\,\rb{M}_1^2\,\rb{M}_2^2 + 
      {{\rb{M}_2^2}^2} \right)
    \vv \ka^{7\!/\!2}
    \nn \\[.5ex]
&\phantom{=\;}
  - \rb{M}_2^2\,
    \left( {\rb{M}_1^2} + {\rb{M}_2^2} \right)\,
    \left( {{\rb{M}_1^2}^2} + 
      5\,\rb{M}_1^2\,\rb{M}_2^2 + 
      {{\rb{M}_2^2}^2} \right)
    \vv \ka^{9\!/\!2}
    \nn \\[.5ex]
&\phantom{=\;}
  + {{{\cal O}(\ka^{11\!/\!2})}}
    \nn
\end{align}
\subsection[Impulskomponenten~\protect$P_i^0$,~\protect$P_i^3$]{%
            Impulskomponenten~\bm{P_i^0},~\bm{P_i^3}}

Die longitudinalen Impulskomponenten~$P_i^0$,~$P_i^3$~-- das hei"st Energie und Impuls in \mbox{$x^3$-Rich}\-tung~-- h"angen zusammen mit~$\vep_i$,~$\vep_i^{-1}$ wie folgt:%
\FOOT{
  Null-Komponente oberes, Drei-Komponente unteres Vorzeichen.
}
%
\vspace*{-.5ex}
\begin{alignat}{2} \label{APP:P03_vep}
P_1^{\stackrel{\scriptstyle0}{3}}\;
  &=\; \big(\rb{M}_1^2\; \vep_1^{-1} \pm \vep_1 \big)\!\big/2\al&
  \qquad\qquad
  \text{\hspace*{-2pt}und:}\qquad
  P_2^{\stackrel{\scriptstyle0}{3}}\;
  &=\; \big(\vep_2 \pm \rb{M}_2^2\; \vep_2^{-1}\big)\!\big/2\al
    \\[.5ex]
P_{1'}^{\stackrel{\scriptstyle0}{3}}\;
  &=\; \big(\rb{M}_{1'}^2\; \vep_{1'}^{-1} \pm \vep_{1'}\big)\!\big/2\al&
  \qquad\qquad
  P_{2'}^{\stackrel{\scriptstyle0}{3}}\;
  &=\; \big(\vep_{2'} \pm \rb{M}_{2'}^2\; \vep_{2'}^{-1}\big)\!\big/2\al
    \tag{\ref{APP:P03_vep}$'$}
    \\[-4.5ex]\nn
\end{alignat}
vgl.\@ die Gln.~(\ref{P03_vep}),~(\ref{P03_vep}$'$).
Entsprechend die Reihen-Entwicklungen.

\subsubsection{(i)\vv Nicht-spezifiziertes Bezugsystem:}

%
\begin{align} \label{APP:SYMM-P1^0}
&P_1^0(\ka)
    \\[.5ex]
&=\;
  {\frac{1}{2}}
    \vv \frac{1}{\sqrt{\ka}}
    \nn \\[.5ex]
&\phantom{=\;}
  + {\frac{P\iin^3}{2}}
    \nn \\[.5ex]
&\phantom{=\;}
  + {\frac{1
  }{4}}\,
  \Big\{
    2\,(\rb{M}_1^2 \!-\! \rb{M}_2^2) + {{P\iin^3}^2} + 
        {\rb{P}\iin^2}
  \Big\}
    \vv {\sqrt{\ka}}
    \nn \\[.5ex]
&\phantom{=\;}
  - {\frac{P\iin^3\,
  }{2}}\,
  (\rb{M}_1^2 \!+\! \rb{M}_2^2)
    \vv \ka
    \nn \\[.5ex]
&\phantom{=\;}
  + {\frac{1
  }{16}}\,
  \Big\{
    4\,(\rb{M}_1^2 \!-\! \rb{M}_2^2)\,
    \Big[ {{P\iin^3}^2} - 
           {\rb{P}\iin^2}
    \Big]
    - {{\Big[ {{P\iin^3}^2} + 
             {\rb{P}\iin^2}
    \Big] }^2}
  \Big\}
    \vv \ka^{3\!/\!2}
    \nn \\[.5ex]
&\phantom{=\;}
  + {\frac{P\iin^3
  }{4}}\,
  \Big\{
    {{(\rb{M}_1^2 \!-\! \rb{M}_2^2)}^2} - 
        (\rb{M}_1^2 \!+\! \rb{M}_2^2)\,
    \Big[ (\rb{M}_1^2 \!+\! \rb{M}_2^2) - 
           2\,{\rb{P}\iin^2}
    \Big]
  \Big\}
    \vv {{\ka}^2}
    \nn \\[.5ex]
&\phantom{=\;}
  + {\frac{1
  }{32}}\,
  \Big\{
    {{\Big[ {{P\iin^3}^2} + 
             {\rb{P}\iin^2}
    \Big] }^3} - 
        2\,(\rb{M}_1^2 \!-\! \rb{M}_2^2)\,
    \Big[ {{P\iin^3}^4} + 
           6\,{{P\iin^3}^2}\,{\rb{P}\iin^2} - 
           3\,{{\rb{P}\iin^2}^2}
    \Big]
  \Big\}
    \vv \ka^{5\!/\!2}
    \nn \\[.5ex]
&\phantom{=\;}
  + {\frac{P\iin^3
  }{4}}\,
  \Big\{
    {{(\rb{M}_1^2 \!-\! \rb{M}_2^2)}^2}\,
    \Big[ (\rb{M}_1^2 \!+\! \rb{M}_2^2) - 2\,{\rb{P}\iin^2}
    \Big]
    \nn \\
&\phantom{=\; \quad}
    - (\rb{M}_1^2 \!+\! \rb{M}_2^2)\,
    \Big[ {{(\rb{M}_1^2 \!+\! \rb{M}_2^2)}^2} - 
           2\,(\rb{M}_1^2 \!+\! \rb{M}_2^2)\,{\rb{P}\iin^2} + 
           2\,{{\rb{P}\iin^2}^2}
    \Big]
  \Big\}
    \vv {{\ka}^3}
    \nn \\[.5ex]
&\phantom{=\;}
  + {\frac{1
  }{256}}\,
  \Big\{
    -5\,{{\Big[ {{P\iin^3}^2} + 
              {\rb{P}\iin^2}
    \Big] }^4}
    \nn \\
&\phantom{=\; \quad}
    + 8\,(\rb{M}_1^2 \!-\! \rb{M}_2^2)\,
    \Big[ {{P\iin^3}^6} + 
           5\,{{P\iin^3}^4}\,{\rb{P}\iin^2} + 
           15\,{{P\iin^3}^2}\,
            {{\rb{P}\iin^2}^2} - 5\,{{\rb{P}\iin^2}^3}
    \Big]
  \Big\}
    \vv \ka^{7\!/\!2}
    \nn \\[.5ex]
&\phantom{=\;}
  + {{{\cal O}(\ka^4)}}
    \nn
\end{align}
%
%

%
\begin{align} \label{APP:SYMM-P1^3}
&P_1^3(\ka)
    \\[.5ex]
&=\;
  {\frac{1}{2}}
    \vv \frac{1}{\sqrt{\ka}}
    \nn \\[.5ex]
&\phantom{=\;}
  + {\frac{P\iin^3}{2}}
    \nn \\[.5ex]
&\phantom{=\;}
  + {\frac{1
  }{4}}\,
  \Big\{
    -2\,(\rb{M}_1^2 \!+\! \rb{M}_2^2) + {{P\iin^3}^2} + 
        {\rb{P}\iin^2}
  \Big\}
    \vv {\sqrt{\ka}}
    \nn \\[.5ex]
&\phantom{=\;}
  + {\frac{P\iin^3
  }{2}}\,
  (\rb{M}_1^2 \!-\! \rb{M}_2^2)
    \vv \ka
    \nn \\[.5ex]
&\phantom{=\;}
  + {\frac{1
  }{16}}\,
  \Big\{
    4\,{{(\rb{M}_1^2 \!-\! \rb{M}_2^2)}^2} - 
        4\,{{(\rb{M}_1^2 \!+\! \rb{M}_2^2)}^2} - 
        4\,(\rb{M}_1^2 \!+\! \rb{M}_2^2)\,
    \Big[ {{P\iin^3}^2} - 
           {\rb{P}\iin^2}
    \Big]
    \nn \\
&\phantom{=\; \quad}
    - {{\Big[ {{P\iin^3}^2} + 
             {\rb{P}\iin^2}
    \Big] }^2}
  \Big\}
    \vv \ka^{3\!/\!2}
    \nn \\[.5ex]
&\phantom{=\;}
  - {\frac{P\iin^3\,
      {\rb{P}\iin^2}
  }{2}}\,
  (\rb{M}_1^2 \!-\! \rb{M}_2^2)
    \vv {{\ka}^2}
    \nn \\[.5ex]
&\phantom{=\;}
  + {\frac{1
  }{32}}\,
  \Big\{
    -8\,{{(\rb{M}_1^2 \!+\! \rb{M}_2^2)}^3} - 
        4\,{{(\rb{M}_1^2 \!+\! \rb{M}_2^2)}^2}\,
    \Big[ {{P\iin^3}^2} - 
           3\,{\rb{P}\iin^2}
    \Big]
    \nn \\
&\phantom{=\; \quad}
    + 4\,{{(\rb{M}_1^2 \!-\! \rb{M}_2^2)}^2}\,
    \Big[ 2\,(\rb{M}_1^2 \!+\! \rb{M}_2^2) + {{P\iin^3}^2} - 
           3\,{\rb{P}\iin^2}
    \Big]
    + {{\Big[ {{P\iin^3}^2} + 
             {\rb{P}\iin^2}
    \Big] }^3}
    \nn
\end{align}
\begin{align}
&\phantom{=\; \quad}
    + 2\,(\rb{M}_1^2 \!+\! \rb{M}_2^2)\,
    \Big[ {{P\iin^3}^4} + 
           6\,{{P\iin^3}^2}\,{\rb{P}\iin^2} - 
           3\,{{\rb{P}\iin^2}^2}
    \Big]
  \Big\}
    \vv \ka^{5\!/\!2}
    \nn \\[.5ex]
&\phantom{=\;}
  + {\frac{P\iin^3\,
      {{\rb{P}\iin^2}^2}
  }{2}}\,
  (\rb{M}_1^2 \!-\! \rb{M}_2^2)
    \vv {{\ka}^3}
    \nn \\[.5ex]
&\phantom{=\;}
  + {\frac{1
  }{256}}\,
  \Big\{
    -16\,{{(\rb{M}_1^2 \!-\! \rb{M}_2^2)}^4} - 
        80\,{{(\rb{M}_1^2 \!+\! \rb{M}_2^2)}^4} - 
        32\,{{(\rb{M}_1^2 \!+\! \rb{M}_2^2)}^3}\,
    \Big[ {{P\iin^3}^2} - 
           5\,{\rb{P}\iin^2}
    \Big]
    \nn \\
&\phantom{=\; \quad}
    - 5\,{{
    \Big[ {{P\iin^3}^2} + 
              {\rb{P}\iin^2}
    \Big] }^4} + 
        8\,{{(\rb{M}_1^2 \!+\! \rb{M}_2^2)}^2}\,
    \Big[ {{P\iin^3}^4} + 
           10\,{{P\iin^3}^2}\,
            {\rb{P}\iin^2} - 
           15\,{{\rb{P}\iin^2}^2}
    \Big]
    \nn \\
&\phantom{=\; \quad}
    - 8\,(\rb{M}_1^2 \!+\! \rb{M}_2^2)\,
    \Big[ {{P\iin^3}^6} + 
           5\,{{P\iin^3}^4}\,{\rb{P}\iin^2} + 
           15\,{{P\iin^3}^2}\,
            {{\rb{P}\iin^2}^2} - 5\,{{\rb{P}\iin^2}^3}
    \Big]
    \nn \\
&\phantom{=\; \quad}
    + 8\,{{(\rb{M}_1^2 \!-\! \rb{M}_2^2)}^2}\,
    \Big[ 12\,{{(\rb{M}_1^2 \!+\! \rb{M}_2^2)}^2} - 
           {{P\iin^3}^4} - 
           10\,{{P\iin^3}^2}\,
            {\rb{P}\iin^2} + 
           15\,{{\rb{P}\iin^2}^2}
    \nn \\
&\phantom{=\; \quad}
    + 4\,(\rb{M}_1^2 \!+\! \rb{M}_2^2)\,
    \Big( {{P\iin^3}^2} - 
              5\,{\rb{P}\iin^2}
    \Big)
    \Big]
  \Big\}
    \vv \ka^{7\!/\!2}
    \nn \\[.5ex]
&\phantom{=\;}
  + {{{\cal O}(\ka^4)}}
    \nn
\end{align}

\subsubsection{(ii)\vv Schwerpunktsystem,~\bm{\rb{P}\iin \!\equiv\! \bm{0}} und~\bm{P\iin^3 \!\equiv\! 0}:}

%
\vspace*{-.5ex}
\begin{align} \label{APP:P1^0-P3IN,PtrIN=0}
P_1^0(\ka)\;
&=\;
  {\frac{1}{2}}
    \vv \frac{1}{\sqrt{\ka}}
    \\[.5ex]
&\phantom{=\;}
  + {\frac{1
  }{2}}\,
  \left( \rb{M}_1^2 - \rb{M}_2^2 \right)
    \vv {\sqrt{\ka}}
    \nn \\[.5ex]
&\phantom{=\;}
  + {{{\cal O}(\ka^{11\!/\!2})}}
    \nn
    \\[-5ex]\nn
\end{align}
%

%
\vspace*{-.5ex}
\begin{align} \label{APP:P1^3-P3IN,PtrIN=0}
P_1^3(\ka)\;
&=\;
  {\frac{1}{2}}
    \vv \frac{1}{\sqrt{\ka}}
    \\[.5ex]
&\phantom{=\;}
  - {\frac{1
  }{2}}\,
  \left( \rb{M}_1^2 + \rb{M}_2^2 \right)
    \vv {\sqrt{\ka}}
    \nn \\[.5ex]
&\phantom{=\;}
  - \rb{M}_1^2\,\rb{M}_2^2
    \vv \ka^{3\!/\!2}
    \nn \\[.5ex]
&\phantom{=\;}
  - \rb{M}_1^2\,\rb{M}_2^2\,
      \left( \rb{M}_1^2 + \rb{M}_2^2 \right)
    \vv \ka^{5\!/\!2}
    \nn \\[.5ex]
&\phantom{=\;}
  - \rb{M}_1^2\,\rb{M}_2^2\,
      \left( {{\rb{M}_1^2}^2} + 
      3\,\rb{M}_1^2\,\rb{M}_2^2 + 
      {{\rb{M}_2^2}^2} \right)
    \vv \ka^{7\!/\!2}
    \nn \\[.5ex]
&\phantom{=\;}
  - \rb{M}_1^2\,\rb{M}_2^2\,
    \left( {\rb{M}_1^2} + {\rb{M}_2^2} \right)\,
    \left( {{\rb{M}_1^2}^2} + 
      5\,\rb{M}_1^2\,\rb{M}_2^2 + 
      {{\rb{M}_2^2}^2} \right)
    \vv \ka^{9\!/\!2}
    \nn \\[.5ex]
&\phantom{=\;}
  + {{{\cal O}(\ka^{11\!/\!2})}}
    \nn
\end{align}
\subsection[Dreier-Impuls-Betrag~\protect$|\vec{P}_i|$]{%
            Dreier-Impuls-Betrag~\bm{|\vec{P}_i|}}

Der Betrag des Dreier-Impulses~$|\vec{P}_i|$ h"angt ab von~$\vep_i$,~$\vep_i^{-1}$ "uber~$P_i^3$ in der Form:
\vspace*{-.5ex}
\begin{align}
\big|\vec{P}_i\big|\;
  =\; \sqrt{\rb{P}_i^2 \!+\! {P_i^3}^2}
    \\[-4.5ex]\nn
\end{align}
Daraus folgen Reihen-Entwicklungen wie folgt.

\subsubsection{(i)\vv Nicht-spezifiziertes Bezugsystem:}

%
\vspace{-.5ex}
\begin{align} \label{APP:P1vec}
&\big|\vec{P}_1(\ka)\big|\;
    \\[.5ex]
&=\; {\frac{1
  }{2}}
    \vv \frac{1}{\sqrt{\ka}}
    \nn
    \\[-4.5ex]\nn
\end{align}
\vspace*{-4.5ex}
\begin{align}
&\phantom{=\;}
  + {\frac{P\iin^3
  }{2}}
    \nn \\[.5ex]
&\phantom{=\;}
  + {\frac{1
  }{4}}\,
  \Big\{
    -2\,(\rb{M}_1^2 \!+\! \rb{M}_2^2) + 
        4\,{\rb{P}_1^2} + {{P\iin^3}^2} + 
        {\rb{P}\iin^2}
  \Big\}
    \vv {\sqrt{\ka}}
    \nn \\[.5ex]
&\phantom{=\;}
  + {\frac{P\iin^3
  }{2}}\,
  \Big\{
    (\rb{M}_1^2 \!-\! \rb{M}_2^2) - 2\,{\rb{P}_1^2}
  \Big\}
    \vv \ka
    \nn \\[.5ex]
&\phantom{=\;}
  + {\frac{1
  }{16}}\,
  \Big\{
    4\,{{(\rb{M}_1^2 \!-\! \rb{M}_2^2)}^2} - 
        4\,{{(\rb{M}_1^2 \!+\! \rb{M}_2^2)}^2} + 
        16\,(\rb{M}_1^2 \!+\! \rb{M}_2^2)\,{\rb{P}_1^2} - 
        16\,{{\rb{P}_1^2}^2}
    \nn \\
&\phantom{=\; \quad}
  - 4\,(\rb{M}_1^2 \!+\! \rb{M}_2^2)\,{{P\iin^3}^2} +
        8\,{\rb{P}_1^2}\,{{P\iin^3}^2} - 
        {{P\iin^3}^4} + 
        4\,(\rb{M}_1^2 \!+\! \rb{M}_2^2)\,{\rb{P}\iin^2} - 
        8\,{\rb{P}_1^2}\,{\rb{P}\iin^2}
    \nn \\
&\phantom{=\; \quad}
  - 2\,{{P\iin^3}^2}\,{\rb{P}\iin^2} - 
        {{\rb{P}\iin^2}^2}
  \Big\}
    \vv \ka^{3\!/\!2}
    \nn \\[.5ex]
&\phantom{=\;}
  - {\frac{P\iin^3
  }{2}}\,
  \Big\{
    (\rb{M}_1^2 \!-\! \rb{M}_2^2)\,
         \left( 2\,{\rb{P}_1^2} + 
           {\rb{P}\iin^2} \right)  - 
        2\,{\rb{P}_1^2}\,
         \left( -2\,(\rb{M}_1^2 \!+\! \rb{M}_2^2) + 
           3\,{\rb{P}_1^2} + {\rb{P}\iin^2}
            \right)
  \Big\}
    \vv {{\ka}^2}
    \nn \\[.5ex]
&\phantom{=\;}
  + {\frac{1
  }{32}}\,
  \Big\{
    -8\,{{(\rb{M}_1^2 \!+\! \rb{M}_2^2)}^3} + 
        64\,{{\rb{P}_1^2}^3} + 
        64\,(\rb{M}_1^2 \!-\! \rb{M}_2^2)\,{\rb{P}_1^2}\,
         {{P\iin^3}^2} - 
        144\,{{\rb{P}_1^2}^2}\,{{P\iin^3}^2}
    \nn \\
&\phantom{=\; \quad}
  - 4\,{\rb{P}_1^2}\,{{P\iin^3}^4} + 
        {{P\iin^3}^6} + 
        48\,{{\rb{P}_1^2}^2}\,{\rb{P}\iin^2} - 
        24\,{\rb{P}_1^2}\,{{P\iin^3}^2}\,
         {\rb{P}\iin^2} +
        3\,{{P\iin^3}^4}\,{\rb{P}\iin^2} + 
        12\,{\rb{P}_1^2}\,{{\rb{P}\iin^2}^2}
    \nn \\
&\phantom{=\; \quad}
  + 3\,{{P\iin^3}^2}\,{{\rb{P}\iin^2}^2} + 
        {{\rb{P}\iin^2}^3} +
        4\,{{(\rb{M}_1^2 \!-\! \rb{M}_2^2)}^2}\,
    \Big[ 2\,(\rb{M}_1^2 \!+\! \rb{M}_2^2) - 4\,{\rb{P}_1^2} + 
           {{P\iin^3}^2} - 3\,{\rb{P}\iin^2}
    \Big]
    \nn \\
&\phantom{=\; \quad}
  + 4\,{{(\rb{M}_1^2 \!+\! \rb{M}_2^2)}^2}\,
    \Big[ 12\,{\rb{P}_1^2} - 
           {{P\iin^3}^2} + 3\,{\rb{P}\iin^2}
    \Big]
    \nn \\
&\phantom{=\; \quad}
  - 2\,(\rb{M}_1^2 \!+\! \rb{M}_2^2)\,
    \Big[ 48\,{{\rb{P}_1^2}^2} - 
           40\,{\rb{P}_1^2}\,{{P\iin^3}^2}
    - {{P\iin^3}^4} + 
           24\,{\rb{P}_1^2}\,
            {\rb{P}\iin^2} - 
           6\,{{P\iin^3}^2}\,{\rb{P}\iin^2}
    \nn \\
&\phantom{=\; \qquad}
    + 3\,{{\rb{P}\iin^2}^2}
    \Big]
  \Big\}
    \vv \ka^{5\!/\!2}
    \nn \\[.5ex]
&\phantom{=\;}
  - {\frac{P\iin^3
  }{2}}\,
  \Big\{ (\rb{M}_1^2 \!-\! \rb{M}_2^2)\,
    \Big[ 4\,(\rb{M}_1^2 \!+\! \rb{M}_2^2)\,
            {\rb{P}_1^2} - 
           6\,{{\rb{P}_1^2}^2} + 
           4\,{\rb{P}_1^2}\,{{P\iin^3}^2}
    - 4\,{\rb{P}_1^2}\,{\rb{P}\iin^2} - 
           {{\rb{P}\iin^2}^2}
    \Big]
    \nn \\
&\phantom{=\; \quad}
  + 2\,{\rb{P}_1^2}\,
    \Big[ -{{(\rb{M}_1^2 \!-\! \rb{M}_2^2)}^2} +
           4\,{{(\rb{M}_1^2 \!+\! \rb{M}_2^2)}^2} + 
           10\,{{\rb{P}_1^2}^2} - 
           4\,{\rb{P}_1^2}\,{{P\iin^3}^2} + 
           6\,{\rb{P}_1^2}\,{\rb{P}\iin^2} + 
           {{\rb{P}\iin^2}^2}
    \nn \\
&\phantom{=\; \qquad}
    - 2\,(\rb{M}_1^2 \!+\! \rb{M}_2^2)\,
    \Big( 6\,{\rb{P}_1^2} - 
              {{P\iin^3}^2} + 
              2\,{\rb{P}\iin^2}
    \Big)
    \Big] 
  \Big\}
    \vv {{\ka}^3}
    \nn \\[.5ex]
&\phantom{=\;}
  + {\frac{1
  }{256}}\,
  \Big\{
    -16\,{{(\rb{M}_1^2 \!-\! \rb{M}_2^2)}^4} - 
        80\,{{(\rb{M}_1^2 \!+\! \rb{M}_2^2)}^4} - 
        1280\,{{\rb{P}_1^2}^4} + 
        6400\,{{\rb{P}_1^2}^3}\,{{P\iin^3}^2}
    \nn \\
&\phantom{=\; \quad}
  - 480\,{{\rb{P}_1^2}^2}\,{{P\iin^3}^4} + 
        16\,{\rb{P}_1^2}\,{{P\iin^3}^6} - 
        5\,{{P\iin^3}^8} - 
        1280\,{{\rb{P}_1^2}^3}\,{\rb{P}\iin^2} + 
        2880\,{{\rb{P}_1^2}^2}\,{{P\iin^3}^2}\,
         {\rb{P}\iin^2}
    \nn \\
&\phantom{=\; \quad}
  + 80\,{\rb{P}_1^2}\,{{P\iin^3}^4}\,
         {\rb{P}\iin^2} - 
        20\,{{P\iin^3}^6}\,{\rb{P}\iin^2} - 
        480\,{{\rb{P}_1^2}^2}\,{{\rb{P}\iin^2}^2} + 
        240\,{\rb{P}_1^2}\,{{P\iin^3}^2}\,
         {{\rb{P}\iin^2}^2} - 
        30\,{{P\iin^3}^4}\,{{\rb{P}\iin^2}^2}
    \nn \\
&\phantom{=\; \quad}
  - 80\,{\rb{P}_1^2}\,{{\rb{P}\iin^2}^3} - 
        20\,{{P\iin^3}^2}\,{{\rb{P}\iin^2}^3} - 
        5\,{{\rb{P}\iin^2}^4}
    \nn \\
&\phantom{=\; \quad}
  + 256\,(\rb{M}_1^2 \!-\! \rb{M}_2^2)\,{\rb{P}_1^2}\,
         {{P\iin^3}^2}\,
    \Big[ 6\,(\rb{M}_1^2 \!+\! \rb{M}_2^2) - 
           12\,{\rb{P}_1^2} + 
           {{P\iin^3}^2} - 5\,{\rb{P}\iin^2}
    \Big]
    \nn \\
&\phantom{=\; \quad}
  + 32\,{{(\rb{M}_1^2 \!+\! \rb{M}_2^2)}^3}\,
    \Big[ 20\,{\rb{P}_1^2} - 
           {{P\iin^3}^2} + 5\,{\rb{P}\iin^2}
    \Big]
  - 8\,{{(\rb{M}_1^2 \!+\! \rb{M}_2^2)}^2}\,
    \Big[ 240\,{{\rb{P}_1^2}^2} - 
           216\,{\rb{P}_1^2}\,
            {{P\iin^3}^2}
    \nn \\
&\phantom{=\; \qquad}
    - {{P\iin^3}^4} + 
           120\,{\rb{P}_1^2}\,
            {\rb{P}\iin^2} - 
           10\,{{P\iin^3}^2}\,
            {\rb{P}\iin^2} +
           15\,{{\rb{P}\iin^2}^2}
    \Big]
    \nn \\
&\phantom{=\; \quad}
  + 8\,(\rb{M}_1^2 \!+\! \rb{M}_2^2)\,
    \Big[ 320\,{{\rb{P}_1^2}^3} - 
           816\,{{\rb{P}_1^2}^2}\,
            {{P\iin^3}^2} + 
           28\,{\rb{P}_1^2}\,{{P\iin^3}^4} - 
           {{P\iin^3}^6} + 
           240\,{{\rb{P}_1^2}^2}\,
            {\rb{P}\iin^2}
    \nn \\
&\phantom{=\; \qquad}
    - 200\,{\rb{P}_1^2}\,{{P\iin^3}^2}\,
            {\rb{P}\iin^2} - 
           5\,{{P\iin^3}^4}\,{\rb{P}\iin^2} + 
           60\,{\rb{P}_1^2}\,
            {{\rb{P}\iin^2}^2} - 
           15\,{{P\iin^3}^2}\,
            {{\rb{P}\iin^2}^2} + 5\,{{\rb{P}\iin^2}^3}
    \Big]
    \nn \\
&\phantom{=\; \quad}
  + 8\,{{(\rb{M}_1^2 \!-\! \rb{M}_2^2)}^2}\,
    \Big[ 12\,{{(\rb{M}_1^2 \!+\! \rb{M}_2^2)}^2} + 
           48\,{{\rb{P}_1^2}^2} - 
           8\,{\rb{P}_1^2}\,{{P\iin^3}^2} - 
           {{P\iin^3}^4} + 
           40\,{\rb{P}_1^2}\,
            {\rb{P}\iin^2}
    \nn \\
&\phantom{=\; \qquad}
    - 10\,{{P\iin^3}^2}\,
            {\rb{P}\iin^2} + 
           15\,{{\rb{P}\iin^2}^2} - 
           4\,(\rb{M}_1^2 \!+\! \rb{M}_2^2)\,
    \Big( 12\,{\rb{P}_1^2} - 
              {{P\iin^3}^2} + 
              5\,{\rb{P}\iin^2}
    \Big)
    \Big] 
  \Big\}
    \vv \ka^{7\!/\!2}
    \nn \\[.5ex]
&\phantom{=\;}
  + {{{\cal O}(\ka^4)}}
    \nn
    \\[-4.5ex]\nn
\end{align}

\subsubsection{(ii)\vv Schwerpunktsystem,~\bm{\rb{P}\iin \!\equiv\! \bm{0}} und~\bm{P\iin^3 \!\equiv\! 0}:}

%
\begin{align} \label{APP:P1vec-P3IN,PtrIN=0}
&\big|\vec{P}_1(\ka)\big|
    \\[.5ex]
&=\;
  {\frac{1
  }{2}}
    \vv \frac{1}{\sqrt{\ka}}
    \nn \\[.5ex]
&\phantom{=\;}
  -{\frac{1
  }{2}}\,
  \Big[
  (\rb{M}_1^2 \!+\! \rb{M}_2^2) - 
     2\, {\rb{P}_1^2}
  \Big]
    \vv {\sqrt{\ka}}
    \nn \\[.5ex]
&\phantom{=\;}
  + {\frac{1
  }{4}}\,
  \Big\{
    {{(\rb{M}_1^2 \!-\! \rb{M}_2^2)}^2} - 
  {{\left[ (\rb{M}_1^2 \!+\! \rb{M}_2^2) - 2\,{\rb{P}_1^2}
  \right] }^2}
  \Big\}
    \vv \ka^{3\!/\!2}
    \nn \\[.5ex]
&\phantom{=\;}
  + {\frac{1
  }{4}}\,
  \left[
    (\rb{M}_1^2 \!+\! \rb{M}_2^2) - 2\,{\rb{P}_1^2}
  \right]\,
  \Big\{
    {{(\rb{M}_1^2 \!-\! \rb{M}_2^2)}^2} - 
        {{\left[ (\rb{M}_1^2 \!+\! \rb{M}_2^2) - 2\,{\rb{P}_1^2}
              \right] }^2}
  \Big\}
    \vv \ka^{5\!/\!2}
    \nn \\[.5ex]
&\phantom{=\;}
  + {\frac{1
  }{16}}\,
  \Big\{
    -{{(\rb{M}_1^2 \!-\! \rb{M}_2^2)}^4} + 
        6\,{{(\rb{M}_1^2 \!-\! \rb{M}_2^2)}^2}\,
         {{\left[ (\rb{M}_1^2 \!+\! \rb{M}_2^2) - 
              2\,{\rb{P}_1^2} \right] }^2}
    \nn \\
&\phantom{=\; \quad}
  - 5\,{{\left[ (\rb{M}_1^2 \!+\! \rb{M}_2^2) - 
              2\,{\rb{P}_1^2} \right] }^4}
  \Big\}
    \vv \ka^{7\!/\!2}
    \nn \\[.5ex]
&\phantom{=\;}
  - {\frac{1
  }{16}}
  \left[
    (\rb{M}_1^2 \!+\! \rb{M}_2^2) - 2\,{\rb{P}_1^2}
  \right]\,
  \Big\{
    3\,{{(\rb{M}_1^2 \!-\! \rb{M}_2^2)}^4} - 
        10\,{{(\rb{M}_1^2 \!-\! \rb{M}_2^2)}^2}\,
         {{\left[ (\rb{M}_1^2 \!+\! \rb{M}_2^2) - 
              2\,{\rb{P}_1^2} \right] }^2}
    \nn \\
&\phantom{=\; \quad}
  + 7\,{{\left[ (\rb{M}_1^2 \!+\! \rb{M}_2^2) - 
              2\,{\rb{P}_1^2} \right] }^4}
  \Big\}
    \vv \ka^{9\!/\!2}
    \nn \\[.5ex]
&\phantom{=\;}
  + {{{\cal O}(\ka^{11\!/\!2})}}
    \nn
\end{align}
\noindent
Aus~\mbox{$\rb{P}_1 \!+\! \rb{P}_2 \!\equiv\! \rb{P}\iin \!\equiv\! \bm{0}$} folgt~\mbox{$\rb{M}_1^2 \!-\! \rb{M}_2^2 \!\equiv\! M_1^2 \!-\! M_2^2$} und~\mbox{$\rb{M}_1^2 \!+\! \rb{M}_2^2 \!-\! 2\,\rb{P}_1^2 \!\equiv\! M_1^2 \!+\! M_2^2$}, das hei"st in Termen der {\it invarianten\/} Massen:
%
\begin{align}
&\hspace*{-6pt}
\big|\vec{P}_1(\ka)\big|
    \tag{\ref{APP:P1vec-P3IN,PtrIN=0}$'$} \\[.5ex]
&=\;
  {\frac{1}{2}}
    \vv \frac{1}{\sqrt{\ka}}
    \nn \\[.5ex]
&\phantom{=\;}
  - {\frac{1
  }{2}}\,
  \left( M_1^2 + M_2^2 \right)
    \vv {\sqrt{\ka}}
    \nn \\[.5ex]
&\phantom{=\;}
  - M_1^2\,M_2^2
    \vv \ka^{3\!/\!2}
    \nn \\[.5ex]
&\phantom{=\;}
  - M_1^2\,M_2^2\,
      \left( M_1^2 + M_2^2 \right)
    \vv \ka^{5\!/\!2}
    \nn \\[.5ex]
&\phantom{=\;}
  - M_1^2\,M_2^2\,
      \left( {{M_1^2}^2} + 
      3\,M_1^2\,M_2^2 + 
      {{M_2^2}^2} \right)
    \vv \ka^{7\!/\!2}
    \nn \\[.5ex]
&\phantom{=\;}
  - M_1^2\,M_2^2\,
    \left( {M_1^2} + {M_2^2} \right)\,
    \left( {{M_1^2}^2} + 
      5\,M_1^2\,M_2^2 + 
      {{M_2^2}^2} \right)
    \vv \ka^{9\!/\!2}
    \nn \\[.5ex]
&\phantom{=\;}
  + {{{\cal O}(\ka^{11\!/\!2})}}
    \nn
\end{align}
vgl.\@ Gl.~(\ref{APP:P1^3-P3IN,PtrIN=0}).

\subsection[Quadrat des invarianten Impulstransfers~\protect$-t$]{%
            Quadrat des invarianten Impulstransfers~\bm{-t}}

Das Quadrat des invarianten Impulstransfers~$-t$ wird ausgedr"uckt durch~$\vep_i$,~$\vep_i^{-1}$, in symmetrisierter Form:
\vspace*{-.5ex}
\begin{align} \label{APP:Mandelstam-t-symm}
\hspace*{-6pt}
t\;
&=\; -(\rb{P}_{1'} \!-\! \rb{P}_1)^2
    \\
&\phantom{=\;} +\; \frac{1}{2}\, \Big[
           \rb{M}_1^2\,    \big(1 - \vep_1^{-1}\, \vep_{1'}\big)
         + \rb{M}_{1'}^2\, \big(1 - \vep_1\,      \vep_{1'}^{-1}\big)
         + \rb{M}_2^2\,    \big(1 - \vep_2^{-1}\, \vep_{2'}\big)
         + \rb{M}_{2'}^2\, \big(1 - \vep_2\,      \vep_{2'}^{-1}\big)
         \Big]
    \nn
    \\[-4.5ex]\nn
\end{align}
vgl.\@ Gl.~(\ref{Mandelstam-t-symm}).
Diese Formel impliziert die Reihen-Entwicklung wie folgt.

\subsubsection{(i)\vv Nicht-spezifiziertes Bezugsystem:}

%
\begin{align} \label{APP:t-symm}
&t(\ka)
    \\[.5ex]
&=\;
  -(\rb{P}_{1'} \!-\! \rb{P}_1)^2
    \nn \\[.5ex]
&\phantom{=\;}
  + {\frac{1
  }{4}}\,
  \Big\{
    -4\,(\rb{M}_1^2 \!-\! \rb{M}_{1'}^2)\,(\rb{M}_2^2 \!-\! \rb{M}_{2'}^2) -
  (\rb{M}_1^2 \!-\! \rb{M}_{1'}^2)\,{\rb{P}\iin^2} - 
        (\rb{M}_2^2 \!-\! \rb{M}_{2'}^2)\,{\rb{P}\iin^2}
  \Big\}
    \vv \ka
    \nn \\[.5ex]
&\phantom{=\;}
  + {\frac{1
  }{4}}\,
  P\iin^3\,{\rb{P}\iin^2}\,
  \Big\{
    (\rb{M}_1^2 \!-\! \rb{M}_{1'}^2) - (\rb{M}_2^2 \!-\! \rb{M}_{2'}^2)
  \Big\}
    \vv \ka^{3\!/\!2}
    \nn \\[.5ex]
&\phantom{=\;}
  + {\frac{1
  }{16}}\,
  \Big\{
    -8\,{{(\rb{M}_1^2 \!-\! \rb{M}_{1'}^2)}^2}\,(\rb{M}_2^2 \!+\! \rb{M}_{2'}^2)
    + \Big[ 6\,(\rb{M}_2^2 \!-\! \rb{M}_{2'}^2) - (\rb{M}_2^2 \!+\! \rb{M}_{2'}^2)
    \Big] \,
      {{\rb{P}\iin^2}^2}
    \nn \\
&\phantom{=\; \quad}
  - 2\,(\rb{M}_1^2 \!-\! \rb{M}_{1'}^2)\,
    \Big[ 4\,(\rb{M}_1^2 \!+\! \rb{M}_{1'}^2)\,(\rb{M}_2^2 \!-\! \rb{M}_{2'}^2) + 
           4\,(\rb{M}_2^2 \!-\! \rb{M}_{2'}^2)\,(\rb{M}_2^2 \!+\! \rb{M}_{2'}^2)
    \nn \\
&\phantom{=\; \qquad}
    - 12\,(\rb{M}_2^2 \!-\! \rb{M}_{2'}^2)\,{\rb{P}\iin^2} + 
           4\,(\rb{M}_2^2 \!+\! \rb{M}_{2'}^2)\,{\rb{P}\iin^2} - 
           3\,{{\rb{P}\iin^2}^2}
    \Big]
    \nn \\
&\phantom{=\; \quad}
  - (\rb{M}_1^2 \!+\! \rb{M}_{1'}^2)\,
    \Big[ 8\,{{(\rb{M}_2^2 \!-\! \rb{M}_{2'}^2)}^2} + 
           8\,(\rb{M}_2^2 \!-\! \rb{M}_{2'}^2)\,{\rb{P}\iin^2} + 
           {{\rb{P}\iin^2}^2}
    \Big]
  \Big\}
    \vv \ka^2
    \nn \\[.5ex]
&\phantom{=\;}
  + {\frac{1
  }{16}}\,
  P\iin^3\,{\rb{P}\iin^2}\,
  \Big\{
    2\,(\rb{M}_2^2 \!-\! \rb{M}_{2'}^2)\,{{P\iin^3}^2} + 
        9\,(\rb{M}_2^2 \!-\! \rb{M}_{2'}^2)\,{\rb{P}\iin^2} - 
        2\,(\rb{M}_2^2 \!+\! \rb{M}_{2'}^2)\,{\rb{P}\iin^2}
    \nn \\
&\phantom{=\; \quad}
  + 2\,(\rb{M}_1^2 \!+\! \rb{M}_{1'}^2)\,
    \Big[ 2\,(\rb{M}_2^2 \!-\! \rb{M}_{2'}^2) + 
           {\rb{P}\iin^2}
    \Big]
    \nn \\
&\phantom{=\; \quad}
  - (\rb{M}_1^2 \!-\! \rb{M}_{1'}^2)\,
    \Big[ 4\,(\rb{M}_2^2 \!+\! \rb{M}_{2'}^2) + 
           2\,{{P\iin^3}^2} + 
           9\,{\rb{P}\iin^2}
    \Big]
  \Big\}
    \vv \ka^{5\!/\!2}
    \nn \\[.5ex]
&\phantom{=\;}
  + {\frac{1
  }{32}}\,
  \Big\{
    -8\,{{(\rb{M}_1^2 \!-\! \rb{M}_{1'}^2)}^3}\,(\rb{M}_2^2 \!-\! \rb{M}_{2'}^2) - 
        2\,(\rb{M}_2^2 \!+\! \rb{M}_{2'}^2)\,{{P\iin^3}^2}\,
         {{\rb{P}\iin^2}^2} -
        19\,(\rb{M}_2^2 \!-\! \rb{M}_{2'}^2)\,{{\rb{P}\iin^2}^3}
    \nn \\
&\phantom{=\; \quad}
    + 6\,(\rb{M}_2^2 \!+\! \rb{M}_{2'}^2)\,{{\rb{P}\iin^2}^3} -
        8\,{{(\rb{M}_1^2 \!+\! \rb{M}_{1'}^2)}^2}\,(\rb{M}_2^2 \!-\! \rb{M}_{2'}^2)\,
    \Big[ 2\,(\rb{M}_2^2 \!-\! \rb{M}_{2'}^2) + 
           {\rb{P}\iin^2}
    \Big]
    \nn \\
&\phantom{=\; \quad}
  - 2\,(\rb{M}_1^2 \!+\! \rb{M}_{1'}^2)\,
    \Big[ 8\,{{(\rb{M}_2^2 \!-\! \rb{M}_{2'}^2)}^2}\,(\rb{M}_2^2 \!+\! \rb{M}_{2'}^2) - 
           24\,{{(\rb{M}_2^2 \!-\! \rb{M}_{2'}^2)}^2}\,{\rb{P}\iin^2}
    \nn \\
&\phantom{=\; \qquad}
    + 12\,(\rb{M}_2^2 \!-\! \rb{M}_{2'}^2)\,(\rb{M}_2^2 \!+\! \rb{M}_{2'}^2)\,
            {\rb{P}\iin^2} - 
           24\,(\rb{M}_2^2 \!-\! \rb{M}_{2'}^2)\,{{\rb{P}\iin^2}^2} + 
           4\,(\rb{M}_2^2 \!+\! \rb{M}_{2'}^2)\,{{\rb{P}\iin^2}^2}
    \nn \\
&\phantom{=\; \qquad}
    + {{P\iin^3}^2}\,{{\rb{P}\iin^2}^2} - 
           3\,{{\rb{P}\iin^2}^3}
    \Big]
    \nn \\
&\phantom{=\; \quad}
  - 16\,{{(\rb{M}_1^2 \!-\! \rb{M}_{1'}^2)}^2}\,
    \Big[ {{(\rb{M}_2^2 \!-\! \rb{M}_{2'}^2)}^2} + 
           (\rb{M}_2^2 \!-\! \rb{M}_{2'}^2)\,{\rb{P}\iin^2}
    + (\rb{M}_2^2 \!+\! \rb{M}_{2'}^2)\,
    \Big( (\rb{M}_1^2 \!+\! \rb{M}_{1'}^2)
    \nn \\
&\phantom{=\; \qquad}
      + (\rb{M}_2^2 \!+\! \rb{M}_{2'}^2) - 
              3\,{\rb{P}\iin^2}
    \Big)
    \Big]
    \nn \\
&\phantom{=\; \quad}
  - (\rb{M}_1^2 \!-\! \rb{M}_{1'}^2)\,
    \Big[ 8\,{{(\rb{M}_1^2 \!+\! \rb{M}_{1'}^2)}^2}\,
            (\rb{M}_2^2 \!-\! \rb{M}_{2'}^2) + 8\,{{(\rb{M}_2^2 \!-\! \rb{M}_{2'}^2)}^3}
    \nn \\
&\phantom{=\; \qquad}
    + 8\,(\rb{M}_2^2 \!-\! \rb{M}_{2'}^2)\,{{(\rb{M}_2^2 \!+\! \rb{M}_{2'}^2)}^2} + 
           16\,{{(\rb{M}_2^2 \!-\! \rb{M}_{2'}^2)}^2}\,{\rb{P}\iin^2}
    \nn \\
&\phantom{=\; \qquad}
    - 48\,(\rb{M}_2^2 \!-\! \rb{M}_{2'}^2)\,(\rb{M}_2^2 \!+\! \rb{M}_{2'}^2)\,
            {\rb{P}\iin^2} +
           8\,{{(\rb{M}_2^2 \!+\! \rb{M}_{2'}^2)}^2}\,{\rb{P}\iin^2} + 
           84\,(\rb{M}_2^2 \!-\! \rb{M}_{2'}^2)\,{{\rb{P}\iin^2}^2}
    \nn \\
&\phantom{=\; \qquad}
    - 48\,(\rb{M}_2^2 \!+\! \rb{M}_{2'}^2)\,{{\rb{P}\iin^2}^2} + 
           19\,{{\rb{P}\iin^2}^3} +
           24\,(\rb{M}_1^2 \!+\! \rb{M}_{1'}^2)\,
    \Big( 2\,(\rb{M}_2^2 \!-\! \rb{M}_{2'}^2)\,(\rb{M}_2^2 \!+\! \rb{M}_{2'}^2)
    \nn \\
&\phantom{=\; \qqquad}
      - 2\,(\rb{M}_2^2 \!-\! \rb{M}_{2'}^2)\,{\rb{P}\iin^2} +
           (\rb{M}_2^2 \!+\! \rb{M}_{2'}^2)\,{\rb{P}\iin^2}
    \Big) 
    \Big]
  \Big\}
    \vv \ka^3
    \nn \\[.5ex]
&\phantom{=\;}
  + {\frac{1
  }{32}}\,
  P\iin^3\,{\rb{P}\iin^2}\,
  \Big\{
    -4\,{{(\rb{M}_1^2 \!-\! \rb{M}_{1'}^2)}^2}\,(\rb{M}_2^2 \!-\! \rb{M}_{2'}^2) + 
        4\,{{(\rb{M}_1^2 \!+\! \rb{M}_{1'}^2)}^2}\,(\rb{M}_2^2 \!-\! \rb{M}_{2'}^2)
    \nn \\[.5ex]
&\phantom{=\; \quad}
    - 3\,(\rb{M}_2^2 \!-\! \rb{M}_{2'}^2)\,{{P\iin^3}^4} - 
        15\,(\rb{M}_2^2 \!-\! \rb{M}_{2'}^2)\,{{P\iin^3}^2}\,
         {\rb{P}\iin^2} + 
        2\,(\rb{M}_2^2 \!+\! \rb{M}_{2'}^2)\,{{P\iin^3}^2}\,
         {\rb{P}\iin^2}
    \nn \\
&\phantom{=\; \quad}
    - 36\,(\rb{M}_2^2 \!-\! \rb{M}_{2'}^2)\,{{\rb{P}\iin^2}^2} + 
        15\,(\rb{M}_2^2 \!+\! \rb{M}_{2'}^2)\,{{\rb{P}\iin^2}^2}
    \nn \\
&\phantom{=\; \quad}
    - (\rb{M}_1^2 \!+\! \rb{M}_{1'}^2)\,
    \left[ 2\,(\rb{M}_2^2 \!-\! \rb{M}_{2'}^2) + 
           {\rb{P}\iin^2}
    \right] \,
    \Big[ 2\,{{P\iin^3}^2} + 
           15\,{\rb{P}\iin^2}
    \Big]
    \nn \\
&\phantom{=\; \quad}
  + (\rb{M}_1^2 \!-\! \rb{M}_{1'}^2)\,
    \Big[ 4\,{{(\rb{M}_2^2 \!-\! \rb{M}_{2'}^2)}^2} - 
           4\,{{(\rb{M}_2^2 \!+\! \rb{M}_{2'}^2)}^2} + 
           4\,(\rb{M}_2^2 \!+\! \rb{M}_{2'}^2)\,{{P\iin^3}^2} + 
           3\,{{P\iin^3}^4}
    \nn
    \\[-7.25ex]\nn
\end{align}
\begin{align}
&\phantom{=\; \qquad}
    + 30\,(\rb{M}_2^2 \!+\! \rb{M}_{2'}^2)\,{\rb{P}\iin^2} + 
           15\,{{P\iin^3}^2}\,
            {\rb{P}\iin^2} + 
           36\,{{\rb{P}\iin^2}^2}
    \Big]
  \Big\}
    \vv \ka^{7\!/\!2}
    \nn \\[.5ex]
&\phantom{=\;}
  + {{{\cal O}(\ka^4)}}
    \nn
\end{align}
\subsubsection{(ii)\vv Schwerpunktsystem,~\bm{\rb{P}\iin \!\equiv\! \bm{0}} und~\bm{P\iin^3 \!\equiv\! 0}:}

%
\vspace*{-.5ex}
\begin{align} \label{APP:t-P3IN,PtrIN=0}
&t(\ka)
    \\[.5ex]
&=\;
  -(\rb{P}_{1'} \!-\! \rb{P}_1)^2
    \nn \\[.5ex]
&\phantom{=\;}
  - (\rb{M}_1^2 \!-\! \rb{M}_{1'}^2)\,(\rb{M}_2^2 \!-\! \rb{M}_{2'}^2)\,\ka
    \nn \\[.5ex]
&\phantom{=\;}
  - {\frac{1
  }{2}}\,
  \Big[
    (\rb{M}_1^2 \!-\! \rb{M}_{1'}^2) + (\rb{M}_2^2 \!-\! \rb{M}_{2'}^2)
  \Big]
    \nn \\[-.5ex]
&\phantom{=\; - {\frac{1}{2}}}
  \times \Big\{
    (\rb{M}_1^2 \!+\! \rb{M}_{1'}^2)\,(\rb{M}_2^2 \!-\! \rb{M}_{2'}^2) + 
        (\rb{M}_1^2 \!-\! \rb{M}_{1'}^2)\,(\rb{M}_2^2 \!+\! \rb{M}_{2'}^2)
  \Big\}
    \vv {{\ka}^2}
    \nn \\[.5ex]
&\phantom{=\;}
  + {\frac{1
  }{4}}\,
  \Big\{
    -\Big[ {{(\rb{M}_1^2 \!-\! \rb{M}_{1'}^2)}^3}\,
           (\rb{M}_2^2 \!-\! \rb{M}_{2'}^2)
    \Big]
    \nn \\
&\phantom{=\; \quad}
  - 2\,(\rb{M}_1^2 \!+\! \rb{M}_{1'}^2)\,{{(\rb{M}_2^2 \!-\! \rb{M}_{2'}^2)}^2}\,
    \Big[
      (\rb{M}_1^2 \!+\! \rb{M}_{1'}^2) + (\rb{M}_2^2 \!+\! \rb{M}_{2'}^2)
    \Big]
    \nn \\
&\phantom{=\; \quad}
  - (\rb{M}_1^2 \!-\! \rb{M}_{1'}^2)\,(\rb{M}_2^2 \!-\! \rb{M}_{2'}^2)\,
    \Big[ {{(\rb{M}_1^2 \!+\! \rb{M}_{1'}^2)}^2} + {{(\rb{M}_2^2 \!-\! \rb{M}_{2'}^2)}^2}
    \nn \\
&\phantom{=\; \qquad}
    + 6\,(\rb{M}_1^2 \!+\! \rb{M}_{1'}^2)\,(\rb{M}_2^2 \!+\! \rb{M}_{2'}^2) + 
           {{(\rb{M}_2^2 \!+\! \rb{M}_{2'}^2)}^2}
    \Big]
    \nn \\
&\phantom{=\; \quad}
  - 2\,{{(\rb{M}_1^2 \!-\! \rb{M}_{1'}^2)}^2}\,
    \Big[ {{(\rb{M}_2^2 \!-\! \rb{M}_{2'}^2)}^2} + 
           (\rb{M}_2^2 \!+\! \rb{M}_{2'}^2)\,
    \Big(
      (\rb{M}_1^2 \!+\! \rb{M}_{1'}^2) + (\rb{M}_2^2 \!+\! \rb{M}_{2'}^2)
    \Big)
    \Big]
  \Big\}
    \vv {{\ka}^3}
    \nn \\[.5ex]
&\phantom{=\;}
  + {\frac{1
  }{8}}\,
  \Big\{
    -\Big[ {{(\rb{M}_1^2 \!-\! \rb{M}_{1'}^2)}^4}\,
           (\rb{M}_2^2 \!+\! \rb{M}_{2'}^2)
    \Big]
    \nn \\
&\phantom{=\; \quad}
  - 3\,{{(\rb{M}_1^2 \!-\! \rb{M}_{1'}^2)}^3}\,(\rb{M}_2^2 \!-\! \rb{M}_{2'}^2)\,
    \Big[ (\rb{M}_1^2 \!+\! \rb{M}_{1'}^2) + 3\,(\rb{M}_2^2 \!+\! \rb{M}_{2'}^2)
    \Big]
    \nn \\
&\phantom{=\; \quad}
  - (\rb{M}_1^2 \!+\! \rb{M}_{1'}^2)\,{{(\rb{M}_2^2 \!-\! \rb{M}_{2'}^2)}^2}\,
    \Big[ 3\,{{(\rb{M}_1^2 \!+\! \rb{M}_{1'}^2)}^2} + 
           {{(\rb{M}_2^2 \!-\! \rb{M}_{2'}^2)}^2}
    \nn \\
&\phantom{=\; \qquad}
    + 10\,(\rb{M}_1^2 \!+\! \rb{M}_{1'}^2)\,(\rb{M}_2^2 \!+\! \rb{M}_{2'}^2) + 
           3\,{{(\rb{M}_2^2 \!+\! \rb{M}_{2'}^2)}^2}
    \Big]
    \nn \\
&\phantom{=\; \quad}
  - {{(\rb{M}_1^2 \!-\! \rb{M}_{1'}^2)}^2}\,
    \Big[ 11\,(\rb{M}_1^2 \!+\! \rb{M}_{1'}^2)\,{{(\rb{M}_2^2 \!-\! \rb{M}_{2'}^2)}^2} + 
           3\,{{(\rb{M}_1^2 \!+\! \rb{M}_{1'}^2)}^2}\,(\rb{M}_2^2 \!+\! \rb{M}_{2'}^2)
    \nn \\
&\phantom{=\; \qquad}
    + 11\,{{(\rb{M}_2^2 \!-\! \rb{M}_{2'}^2)}^2}\,(\rb{M}_2^2 \!+\! \rb{M}_{2'}^2) + 
           10\,(\rb{M}_1^2 \!+\! \rb{M}_{1'}^2)\,{{(\rb{M}_2^2 \!+\! \rb{M}_{2'}^2)}^2} + 
           3\,{{(\rb{M}_2^2 \!+\! \rb{M}_{2'}^2)}^3}
    \Big]
    \nn \\
&\phantom{=\; \quad}
  - (\rb{M}_1^2 \!-\! \rb{M}_{1'}^2)\,(\rb{M}_2^2 \!-\! \rb{M}_{2'}^2)\,
    \Big[ {{(\rb{M}_1^2 \!+\! \rb{M}_{1'}^2)}^3} + 
           15\,{{(\rb{M}_1^2 \!+\! \rb{M}_{1'}^2)}^2}\,(\rb{M}_2^2 \!+\! \rb{M}_{2'}^2)
    \nn \\
&\phantom{=\; \qquad}
    + 3\,{{(\rb{M}_2^2 \!-\! \rb{M}_{2'}^2)}^2}\,(\rb{M}_2^2 \!+\! \rb{M}_{2'}^2) + 
           {{(\rb{M}_2^2 \!+\! \rb{M}_{2'}^2)}^3}
    \nn \\
&\phantom{=\; \qquad}
    + 3\,(\rb{M}_1^2 \!+\! \rb{M}_{1'}^2)\,
    \Big( 3\,{{(\rb{M}_2^2 \!-\! \rb{M}_{2'}^2)}^2} + 
              5\,{{(\rb{M}_2^2 \!+\! \rb{M}_{2'}^2)}^2}
    \Big)
    \Big]
  \Big\}
    \vv {{\ka}^4}
    \nn \\[.5ex]
&\phantom{=\;}
  + {\frac{1
  }{16}}\,
  \Big\{
    -\Big[ {{(\rb{M}_1^2 \!-\! \rb{M}_{1'}^2)}^5}\,
           (\rb{M}_2^2 \!-\! \rb{M}_{2'}^2)
    \Big]
    \nn \\
&\phantom{=\; \quad}
  - 2\,{{(\rb{M}_1^2 \!-\! \rb{M}_{1'}^2)}^3}\,(\rb{M}_2^2 \!-\! \rb{M}_{2'}^2)\,
    \Big[ 3\,{{(\rb{M}_1^2 \!+\! \rb{M}_{1'}^2)}^2} + 
           7\,{{(\rb{M}_2^2 \!-\! \rb{M}_{2'}^2)}^2}
    \nn \\
&\phantom{=\; \qquad}
    + 26\,(\rb{M}_1^2 \!+\! \rb{M}_{1'}^2)\,(\rb{M}_2^2 \!+\! \rb{M}_{2'}^2) + 
           19\,{{(\rb{M}_2^2 \!+\! \rb{M}_{2'}^2)}^2}
    \Big]
    \nn \\
&\phantom{=\; \quad}
  - 4\,(\rb{M}_1^2 \!+\! \rb{M}_{1'}^2)\,{{(\rb{M}_2^2 \!-\! \rb{M}_{2'}^2)}^2}\,
    \Big[ {{(\rb{M}_1^2 \!+\! \rb{M}_{1'}^2)}^3} + 
           2\,(\rb{M}_1^2 \!+\! \rb{M}_{1'}^2)\,{{(\rb{M}_2^2 \!-\! \rb{M}_{2'}^2)}^2}
    \nn \\
&\phantom{=\; \qquad}
    + 7\,{{(\rb{M}_1^2 \!+\! \rb{M}_{1'}^2)}^2}\,(\rb{M}_2^2 \!+\! \rb{M}_{2'}^2) + 
           {{(\rb{M}_2^2 \!-\! \rb{M}_{2'}^2)}^2}\,(\rb{M}_2^2 \!+\! \rb{M}_{2'}^2)
    \nn \\
&\phantom{=\; \qquad}
    + 7\,(\rb{M}_1^2 \!+\! \rb{M}_{1'}^2)\,{{(\rb{M}_2^2 \!+\! \rb{M}_{2'}^2)}^2} + 
           {{(\rb{M}_2^2 \!+\! \rb{M}_{2'}^2)}^3}
    \Big]
    \nn \\
&\phantom{=\; \quad}
  - 4\,{{(\rb{M}_1^2 \!-\! \rb{M}_{1'}^2)}^4}\,
    \Big[ 2\,{{(\rb{M}_2^2 \!-\! \rb{M}_{2'}^2)}^2}
      + (\rb{M}_2^2 \!+\! \rb{M}_{2'}^2)\,
    \Big(
      (\rb{M}_1^2 \!+\! \rb{M}_{1'}^2) + 2\,(\rb{M}_2^2 \!+\! \rb{M}_{2'}^2)
    \Big)
    \Big]
    \nn
    \\[-7.25ex]\nn
\end{align}
\begin{align}
&\phantom{=\; \quad}
  - (\rb{M}_1^2 \!-\! \rb{M}_{1'}^2)\,(\rb{M}_2^2 \!-\! \rb{M}_{2'}^2)\,
    \Big[ {{(\rb{M}_1^2 \!+\! \rb{M}_{1'}^2)}^4} + {{(\rb{M}_2^2 \!-\! \rb{M}_{2'}^2)}^4}
    \nn \\
&\phantom{=\; \qquad}
    + 28\,{{(\rb{M}_1^2 \!+\! \rb{M}_{1'}^2)}^3}\,(\rb{M}_2^2 \!+\! \rb{M}_{2'}^2) + 
           6\,{{(\rb{M}_2^2 \!-\! \rb{M}_{2'}^2)}^2}\,{{(\rb{M}_2^2 \!+\! \rb{M}_{2'}^2)}^2} + 
           {{(\rb{M}_2^2 \!+\! \rb{M}_{2'}^2)}^4}
    \nn \\
&\phantom{=\; \qquad}
    + 2\,{{(\rb{M}_1^2 \!+\! \rb{M}_{1'}^2)}^2}\,
    \Big( 19\,{{(\rb{M}_2^2 \!-\! \rb{M}_{2'}^2)}^2} + 
              35\,{{(\rb{M}_2^2 \!+\! \rb{M}_{2'}^2)}^2}
    \Big)
    \nn \\
&\phantom{=\; \qquad}
    + 4\,(\rb{M}_1^2 \!+\! \rb{M}_{1'}^2)\,
    \Big( 13\,{{(\rb{M}_2^2 \!-\! \rb{M}_{2'}^2)}^2}\,(\rb{M}_2^2 \!+\! \rb{M}_{2'}^2) + 
              7\,{{(\rb{M}_2^2 \!+\! \rb{M}_{2'}^2)}^3}
    \Big)
    \Big]
    \nn \\
&\phantom{=\; \quad}
  - 4\,{{(\rb{M}_1^2 \!-\! \rb{M}_{1'}^2)}^2}\,
    \Big[ 2\,{{(\rb{M}_2^2 \!-\! \rb{M}_{2'}^2)}^4} + 
           {{(\rb{M}_1^2 \!+\! \rb{M}_{1'}^2)}^3}\,(\rb{M}_2^2 \!+\! \rb{M}_{2'}^2)
    \nn \\
&\phantom{=\; \qquad}
    + 8\,{{(\rb{M}_2^2 \!-\! \rb{M}_{2'}^2)}^2}\,{{(\rb{M}_2^2 \!+\! \rb{M}_{2'}^2)}^2} + 
           {{(\rb{M}_2^2 \!+\! \rb{M}_{2'}^2)}^4}
    \nn \\
&\phantom{=\; \qquad}
    + {{(\rb{M}_1^2 \!+\! \rb{M}_{1'}^2)}^2}\,
    \Big( 8\,{{(\rb{M}_2^2 \!-\! \rb{M}_{2'}^2)}^2} + 
              7\,{{(\rb{M}_2^2 \!+\! \rb{M}_{2'}^2)}^2}
    \Big)
    \nn \\
&\phantom{=\; \qquad}
    + (\rb{M}_1^2 \!+\! \rb{M}_{1'}^2)\,
    \Big( 26\,{{(\rb{M}_2^2 \!-\! \rb{M}_{2'}^2)}^2}\,(\rb{M}_2^2 \!+\! \rb{M}_{2'}^2) + 
              7\,{{(\rb{M}_2^2 \!+\! \rb{M}_{2'}^2)}^3}
    \Big)
    \Big]
  \Big\}
    \vv {{\ka}^5}
    \nn \\[.5ex]
&\phantom{=\;}
  + {{{\cal O}(\ka^6)}}
    \nn
\end{align}
\theendnotes

%% file: APP_STREUUNG.tex
\lhead[\fancyplain{}{\sc\thepage}]
      {\fancyplain{}{\sc\rightmark}}
\rhead[\fancyplain{}{\sc{{\footnotesize Anhang~\thechapter:} Streuung der Zust"ande $h^i$}}]
      {\fancyplain{}{\sc\thepage}}
\psfull
\chapter[Streuung der Zust"ande~\protect\bm{h^i}]{%
   \huge Streuung der Zust"ande~\bm{h^i}}
\label{APP:Streuung}

In diesem Anhang leiten wir her Identit"aten die Streuung der Zust"ande~\mbox{\,$\ket{h^i(P_i)}$} betreffend, die wir ben"otigen im Haupttext, dort aber nur zitieren k"onnen.

\section[Skalarprodukt~$\protect
                \bracket{h^{\!i'}}{h^i}$]{%
         Skalarprodukt~\bm{\bracket{h^{\!i'}}{h^i}}}
\label{APP:Skalarprodukt}

Wir berechnen das Skalarprodukt~$\bracket{h^{i'}(P_{i'})}{h^i(P_i)}$ f"ur feste~$i,\,i'$, beide entweder~$\in\! \{1,1'\}$ oder~\mbox{$\in\! \{2,2'\}$}.
F"ur~$i' \!=\! i$ hei"st dies die Norm des Zustandes~$\ket{h^i(P_i)}$, vgl.\@ Gl.~(\ref{h_Norm}), f"ur~$i' \!\neq\! i$ der "Uberlapp zweier verschiedener, entweder "`plus"'- oder "`minus"'-Zust"ande, vgl.\@ Gl.~(\ref{h_bracket}).

Wir schreiben mithilfe von Gl.~(\ref{h-ket})
%
\begin{align} \label{APP:h_bracket}
&\bracket{h^{i'}(P_{i'})}{h^i(P_i)}
    \\[.5ex]
&=\; \frac{\de_{n_{i'}\!{\bar n}_{i'}}}{\sqrt{N_{\rm\!c}}}\vv
        \frac{\de_{n_i\!{\bar n}_i}}{\sqrt{N_{\rm\!c}}}\vv
        \int \frac{d^2\rb{k}_{i'}}{(2\pi)^2}
        \int_0^1 \frac{d\zet_{i'}}{2\pi}\vv
        \int \frac{d^2\rb{k}_i}{(2\pi)^2}
        \int_0^1 \frac{d\zet_i}{2\pi}\vv
        {\tilde\vph}_{s_{i'}\mskip-1mu\bar{s}_{i'}}^{{i'}\D\dagger}(\zet_{i'}, \rb{k}_{i'})\,
        {\tilde\vph}_{s_i\mskip-1mu\bar{s}_i}^i (\zet_i, \rb{k}_i)
    \nn \\[.5ex]
&\phantom{=\;}
   \times\bracket{q_{s_{i'},n_{i'}}(\zet_{i'}, \rb{k}_{i'})\;
     \bar{q}_{{\bar s}_{i'},{\bar n}_{i'}}(\bzet_{i'}, \rbb{k}_{i'})}{
     q_{s_i,n_i}(\zet_i, \rb{k}_i)\;
     \bar{q}_{{\bar s}_i,{\bar n}_i}(\bzet_i, \rbb{k}_i)}
    \nn
\end{align}
Wir dr"ucken das Vier-Quark-Matrixelement aus durch Erzeugungs- und Vernichtungsoperatoren und erhalten mithilfe der Antikommutatorrelationen, vgl.\@ Gl.~(\ref{Antikommutator}), unmittelbar
%
\begin{align} \label{APP:4Q-Matrixelement}
&\bracket{q_{s_{i'},n_{i'}}(\zet_{i'}, \rb{k}_{i'})\;
          \bar{q}_{{\bar s}_{i'},{\bar n}_{i'}}(\bzet_{i'}, \rbb{k}_{i'})}{
          q_{s_i,n_i}(\zet_i, \rb{k}_i)\;
          \bar{q}_{{\bar s}_i,{\bar n}_i}(\bzet_i, \rbb{k}_i)}\;
  =\; \de({i'},i)\, \de({\bar i}',{\bar i})
    \\[1ex]
&\text{mit}\qquad
  \de({i'},i)\;
  =\; \de_{s_{i'}\!s_i} \de_{n_{i'}\!n_i}\; (2\pi)^3\,
        2\sqrt{(p_{i'})_{0\!+}(p_i)_{0\!+}}\;
        \de(\vec{p}_{i'} \!-\! \vec{p}_i)
    \tag{\ref{APP:4Q-Matrixelement}$'$}
\end{align}
die Definition von~$\de({i'},i)$ nach Gl.~(\ref{Antikommutator}); bzgl.~$\de({\bar i}',{\bar i})$ vgl.\@ Gl.~(\ref{Antikommutator}$'$).
Wir betrachten weiter den Anteil des Konfigurationsraumes, das hei"st ohne die Kronecker-Symbole in den Eichgruppen- und Spinindizes.
In Lichtkegelkoordinaten gilt
%
\begin{align} \label{APP:Delta_LC-Koord}
&\de({i'},i)\big|_{\text{Konfig.}}
    \nn \\[1ex]
&=\; (2\pi)^3\,
        2\sqrt{(p_{i'}^+ \!+\! p_{i'}^-)(p_i^+ \!+\! p_i^-)}\vv
        \de\big((p_{i'}^+ \!+\! p_{i'}^-) - (p_i^+ \!+\! p_i^-)\big)\vv
        \de(\rb{p}_{i'} \!-\! \rb{p}_i)
    \\[1ex]
&=\; (2\pi)^3\,
        2\sqrt{\zet_{i'} P_{i'}^+
           \bigg[ 1 \!+\! \frac{\al^2\rb{m}_{i'}^2}{(\zet_{i'}P_{i'}^+)^2} \bigg]\cdot
               \zet_i P_i^+
           \bigg[ 1 \!+\! \frac{\al^2\rb{m}_i^2}{(\zet_iP_i^+)^2} \bigg]}
        \tag{\ref{APP:Delta_LC-Koord}$'$}
    \\[.5ex]
&\phantom{=\;}
  \times\de\bigg(
    \zet_{i'} P_{i'}^+
      \bigg[ 1 \!+\! \frac{\al^2\rb{m}_{i'}^2}{(\zet_{i'} P_{i'}^+)^2} \bigg]
  - \zet_i P_i^+
      \bigg[ 1 \!+\! \frac{\al^2\rb{m}_{i'}^2}{(\zet_i P_i^+)^2} \bigg]
    \bigg)\vv
    \de\big((\zet_{i'} \rb{P}_{i'} \!+\! \rb{k}_{i'}) - (\zet_i \rb{P}_i \!+\! \rb{k}_i)\big)
    \nn \\[1ex]
&\underset{\text{$s \!\to\! \infty$}}{\sim}\;
    (2\pi)^3\,
      \sqrt{\zet_{i'} P_{i'}^+\, \zet_i P_i^+}\vv   
      \de(\zet_{i'} P_{i'}^+ \!-\! \zet_i P_i^+)\vv
      \de((\zet_{i'} \rb{P}_{i'} \!+\! \rb{k}_{i'}) \!-\! (\zet_i \rb{P}_i \!+\! \rb{k}_i))
    \tag{\ref{APP:Delta_LC-Koord}$''$}
\end{align}
mit~$\rb{m}$ der transversalen Quarkmasse entsprechend Gl.~(\ref{transversaleMasse}).
Dabei sind wir in Gl.~(\ref{APP:Delta_LC-Koord}$''$) zu den "`gro"sen"' Impulskomponenten "ubergegangen und haben pr"azisiert den Fall~$i \!\equiv\! 1$ (\mbox{$i' \!\in\!\{1,\,1'\}$}).
Im Fall~$i \!\equiv\! 2$ (\mbox{$i' \!\in\!\{2,\,2'\}$}) sind die "`gro"sen"' Komponenten der Impulse die Minus-Komponen\-ten und es ist zu ersetzen~$P^+ \!\to\! P^-$.
Gl.~(\ref{APP:Delta_LC-Koord}$''$) folgt f"ur~\mbox{\,$s \!\to\! \infty$}.

Mithilfe der Sustitution nach Gl.~(\ref{Q-AQ-Substitution}):~\mbox{\,$\zet \!\to\! \bzet$} und~\mbox{\,$\rb{k} \!\to\! -\rb{k}$}, gelangen wir zu dem analogen Aus\-druck~\mbox{\,$\de({\bar i}',{\bar i})$} f"ur Antiquarks.

F"ur das Vier-Quark-Matrixelement nach Gl.~(\ref{APP:4Q-Matrixelement}) folgt weiter
\vspace*{-.25ex}
\begin{align} 
&\bracket{q_{s_{i'},n_{i'}}(\zet_{i'}, \rb{k}_{i'})\;
          \bar{q}_{{\bar s}_{i'},{\bar n}_{i'}}(\bzet_{i'}, \rbb{k}_{i'})}{
          q_{s_i,n_i}(\zet_i, \rb{k}_i)\;
          \bar{q}_{{\bar s}_i,{\bar n}_i}(\bzet_i, \rbb{k}_i)}
    \\
&=\; \de_{s_{i'}\!s_i} \de_{n_{i'}\!n_i}\;
        \de_{{\bar s}_{i'}\!{\bar s}_i} \de_{{\bar n}_{i'}\!{\bar n}_i}\; (2\pi)^6\,
        \sqrt{\zet_{i'} P_{i'}^+\, \zet_i P_i^+}\;      
        \sqrt{\bzet_{i'} P_{i'}^+\, \bzet_i P_i^+}
    \nn \\
&\phantom{=\;}\times
  \de\big(\zet_{i'} P_{i'}^+ \!-\! \zet_i P_i^+\big)\vv
  \de\big((\zet_{i'} \rb{P}_{i'} \!+\! \rb{k}_{i'})
                \!-\! (\zet_i \rb{P}_i \!+\! \rb{k}_i)\big)
    \nn \\
&\phantom{=\;}\times
  \de\big([P_{i'}^+ \!-\! P_i^+]
        - (\zet_{i'} P_{i'}^+ \!-\! \zet_i P_i^+)\big)\vv
  \de\big([\rb{k}_{i'} \!-\! \rb{k}_i]
        - [(\zet_{i'} \rb{P}_{i'} \!+\! \rb{k}_{i'}) \!-\!
           (\zet_i \rb{P}_i \!+\! \rb{k}_i)]\big)
    \nn
    \\[-4.25ex]\nn
\end{align}
wobei wir~$\bzet \!=\! 1 \!-\! z$ benutzt haben.
Im Sinne von Distributionen verschwinden aufgrund der vorletzten Zeile die Terme in den eckigen Klammern der letzten Zeile; es folgen die Delta-Distributionen~$\de(P_{i'}^+ \!-\! P_i^+)$ und~$\de(\rb{k}_{i'} \!-\! \rb{k}_i)$.
Diese wiederum haben Konsequenzen f"ur die Delta-Distributionen der vorletzten Zeile.
Wir erhalten:
\vspace*{-.25ex}
\begin{align} \label{APP:4Q-Matrixelement_2}
&\bracket{q_{s_{i'},n_{i'}}(\zet_{i'}, \rb{k}_{i'})\;
          \bar{q}_{{\bar s}_{i'},{\bar n}_{i'}}(\bzet_{i'}, \rbb{k}_{i'})}{
          q_{s_i,n_i}(\zet_i, \rb{k}_i)\;
          \bar{q}_{{\bar s}_i,{\bar n}_i}(\bzet_i, \rbb{k}_i)}
    \nn \\[.5ex]
&=\; \de_{s_{i'}\!s_i} \de_{n_{i'}\!n_i}\;
        \de_{{\bar s}_{i'}\!{\bar s}_i} \de_{{\bar n}_{i'}\!{\bar n}_i}\; (2\pi)^6\,
        \sqrt{\zet_{i'} P_{i'}^+\, \zet_i P_i^+}\;      
        \sqrt{\bzet_{i'} P_{i'}^+\, \bzet_i P_i^+}
    \\
&\phantom{=\;}\times
  (P_{i'}^+)^{-1}\;
  \de(\zet_{i'} \!-\! \zet_i)\vv
  \de(\rb{P}_{i'} \!-\! \rb{P}_i)\vv
  \de(P_{i'}^+ \!-\! P_i^+)\vv
  \de(\rb{k}_{i'} \!-\! \rb{k}_i)
    \nn \\[.5ex]
&=\; (2\pi)^3\; 2(P_{i'})_{0\!+}\;
  \de(\vec{P}_{i'} \!-\! \vec{P}_i)\vv
  \de_{s_{i'}\!s_i} \de_{n_{i'}\!n_i}\;
    \de_{{\bar s}_{i'}\!{\bar s}_i} \de_{{\bar n}_{i'}\!{\bar n}_i}
    \tag{\ref{APP:4Q-Matrixelement_2}$'$} \\
&\phantom{=\;}\times
  2\zet_{i'}\bzet_{i'}\; (2\pi)^3\;
  \de(\zet_{i'} \!-\! \zet_i)\vv
  \de(\rb{k}_{i'} \!-\! \rb{k}_i)
    \nn 
    \\[-4.25ex]\nn
\end{align}
wobei wir in Gl.~(\ref{APP:4Q-Matrixelement_2}$'$) benutzt haben:~$P_{i'}^+\de(P_{i'}^+ \!-\! P_i^+) \sim (P_{i'})_{0\!+}\de(P_{i'}^3 \!-\! P_i^3)$ im Limes~$s \!\to\! \infty$ mit den nullten beziehungsweise dritten Komponenten.

Eingesetzt in Gl.~(\ref{APP:h_bracket}) folgt unmittelbar
\vspace*{-.5ex}
\begin{align} \label{APP:h_bracket_1}
&\bracket{h^{i'}(P_{i'})}{h^i(P_i)}
  \\
&=\; (2\pi)^3\, 2(P_{i'})_{0\!+}\, \de(\vec{P}_{i'} \!-\! \vec{P}_i)\vv
        \int \frac{d^2\rb{k}_{i'}}{(2\pi)^2}
        \int_0^1 \frac{d\zet_{i'}}{2\pi}\vv 2\zet_{i'}\bar{\zet}_{i'}\vv
        {\tilde\vph}_{s_{i'}\mskip-1mu\bar{s}_{i'}}^{{i'}\D\dagger}(\zet_{i'}, \rb{k}_{i'})\,
        {\tilde\vph}_{s_{i'}\mskip-1mu\bar{s}_{i'}}^i (\zet_{i'}, \rb{k}_{i'})
    \nn
    \\[-4.5ex]\nn
\end{align}
Wir haben diese Gleichung streng hergeleitet f"ur Indizes~\mbox{$i \!=\! 1$} und~\mbox{$i' \!\in\!\{1,\,1'\}$}.
F"ur Indizes \mbox{$i \!=\! 2$} und~\mbox{$i' \!\in\!\{2,\,2'\}$} sind die "`gro"sen"' Impulskomponeneten, "uber die die Herleitung geschieht, die Minus-Komponenten.
Resultat ist wieder Gl.~(\ref{APP:h_bracket_1}).

Seien die Lichtkegelwellenfunktionen~\mbox{\,${\tilde\vph}_{s\mskip-1mu\bar{s}}^i$} normiert wie in Gl.~(\ref{vph-k_Norm}), ihre Fourier-Trans\-formierten (bzgl.\@ des transversalen Vektors) definiert durch Gl.~(\ref{vph-x_vph-k}).
Dann ist Gl.~(\ref{APP:h_bracket_1}) f"ur Indizes~\mbox{\,$i' \!=\! i$} genau Gl.~(\ref{h_Norm}): die Normierungsbedingung der Zust"ande~\mbox{\,$\ket{h^i(P_i)}$}, f"ur Indizes~\mbox{\,$i' \!\neq\! i$}~-- beide entweder~\mbox{\,$\in\!\{1,1'\}$} oder~\mbox{\,$\in\!\{2,2'\}$}~-- genau Gl.~(\ref{h_bracket}): deren "Uberlapp.
\vspace*{-.5ex}

\section[Disconnected Terme:~\protect$\de(i',i)$
           als Integral~\protect$\int d_{i'\mskip-1mu i}$]{%
         Disconnected Terme:~\bm{\de(i',i)}
           als Integral~\bm{\int d_{i'\mskip-1mu i}}}
\label{APP:discon}

Wir zeigen explizit die Umformung der {\it disconnected\/}, das hei"st nichtzusammenh"angenden Terme~$\de(i',i)$ der Darstellung nach Gl.~(\ref{S-Element2h_calM}) in die nach Gl.~(\ref{S-Element2h_WW-V}). 

Diese aus dem \DREI{L}{S}{Z}-Formalismus heraus auftretenden Terme, vgl.\@ die Gln.\,(\ref{S-Element_calM}),\,(\ref{S-Element2h_calM}$''$), stehen dann in derselben Form wie die Einsen~\mbox{\,$\bbbone$}, \vspace*{-.375ex}die zu subtrahieren sind von den Wegner-Wilson-Linien~\mbox{\,$V\idx{+}$},~\mbox{\,$V\idx{+}^{\D\dagger}$} und~\mbox{\,$V\idx{-}$},~\mbox{\,$V\idx{-}^{\D\dagger}$}; vgl.\@ die Gln.~(\ref{calM-Q_WW-V}),~(\ref{calM-Q_WW-V}$'$) und~(\ref{calM-AQ_WW-V}),~(\ref{calM-AQ_WW-V}$'$).
Die gemeinsame Form erm"oglicht die simultane Behandlung.

Wir betrachten den Ausdruck f"ur~\mbox{\,$\de(i',i)$} nach Gl.~(\ref{APP:Delta_LC-Koord})~-- pr"azisiert auf Indizes der "`plus"'-Zust"ande,~\mbox{\,$(i,\,i') \!=\! (1,\,1')$}:
\vspace*{-.5ex}
\begin{align} 
&\de({i'},i)\big|_{\text{Konfig.}}
    \nn \\[-.25ex]
&=\; (2\pi)^3\,
        2\sqrt{(p_{i'}^+ \!+\! p_{i'}^-)(p_i^+ \!+\! p_i^-)}\;
        \de\big((p_{i'}^+ \!+\! p_{i'}^-) - (p_i^+ \!+\! p_i^-)\big)\;
        \de(\rb{p}_{i'} \!-\! \rb{p}_i)
    \\[-4.5ex]\nn
\end{align}
Im Limes~\mbox{\,$s \!\to\! \infty$} folgt durch Darstellung der Delta-Distributionen durch Fourier-Integrale:
\vspace*{-.5ex}
\begin{align} \label{APP:Delta-Konfig-lim}
&\de({i'},i)\big|_{\text{Konfig.}}
    \nn \\[-.25ex]
&=\; (2\pi)^3\,
        2\sqrt{p_{i'}^+p_i^+}\;
        \de\big((p_{i'}^+ \!+\! p_{i'}^-) - (p_i^+ \!+\! p_i^-)\big)\;
        \de(\rb{p}_{i'} \!-\! \rb{p}_i)
    \\[-.25ex]
&=\; 2\sqrt{p_{i'}^+p_i^+}\;
        \int d(g_{+-} x^-) d^2\rb{x}\vv
        \exp\iIM\big[ (p_{1'}^+ \!-\! p_1^+) (g_{+-} x^-)
                    - (\rb{p}_{1'} \!-\! \rb{p}_1) \!\cdot\! \rb{x} \big]
    \tag{\ref{APP:Delta-Konfig-lim}$'$}
    \\[-4.5ex]\nn
\end{align}
Vergleich mit der Definition des Integrationsma"ses~\mbox{\,$d_{i'\mskip-1mu i}$} in Gl.~(\ref{di'i}) zeigt
\begin{samepage}
\vspace*{-.5ex}
\begin{align} 
\de(i',i)\;
  =\; \de_{s_i'\!s_i} \de_{n_{i'}\!n_i}\vv
        \int d_{i'\mskip-1mu i}(x^-,\rb{x})
    \\[-4.5ex]\nn
\end{align}
Zun"achst nur f"ur die Indizes~\mbox{\,$(i',\,i) \!=\! (1',\,1)$}, folgt diese Relation in analoger Herleitung f"ur die anderen im $S$-Matrixelement relevanten Kombinationen~\mbox{\,$(2',\,2)$} und~\mbox{\,$(\bar1',\,\bar1)$},~\mbox{\,$(\bar2',\,\bar2)$}.
In suggestiver Schreibweise mit Eisen~$\bbbone$ der Eichgruppe gilt f"ur Teilchen:
\vspace*{-.5ex}
\begin{alignat}{3} \label{APP:Delta_diStrichi-Mass}
&\de(1',1)\;&
  &=\; \de_{s_{1'}\!s_1}\;&
        &\int d_{1'\mskip-1mu 1}\!(x^-, \rb{x})\vv
          \bbbone_{n_{1'}\!n_1}
    \\[-.5ex]
&\de(2',2)\;&
  &=\; \de_{s_{2'}\!s_2}\;&
        &\int d_{2'\mskip-1mu 2}\!(y^+, \rb{y})\vv
          \bbbone_{n_{2'}\!n_2}
    \tag{\ref{APP:Delta_diStrichi-Mass}$'$}
    \\[-4.5ex]\nn
\end{alignat}
und f"ur Antiteilchen:
\vspace*{-.5ex}
\begin{alignat}{3} \label{APP:Delta_diStrichi-Mass-bar}
&\de(\bar1',\bar1)\;&
  &=\; \de_{\bar{s}_{1'}\!\bar{s}_1}\;&
        &\int d_{{\bar1}'\mskip-1mu {\bar1}}\!({\bar x}^-,\rbb{x})\vv
          \bbbone_{\bar{n}_{1'}\!\bar{n}_1}
    \\[-.5ex]
&\de(\bar2',\bar2)\;&
  &=\; \de_{\bar{s}_{2'}\!\bar{s}_2}\;&
        &\int d_{{\bar2}'\mskip-1mu {\bar2}}\!({\bar y}^+,\rbb{y})\vv
          \bbbone_{\bar{n}_{2'}\!\bar{n}_2}
    \tag{\ref{APP:Delta_diStrichi-Mass-bar}$'$}
    \\[-4.5ex]\nn
\end{alignat}
Damit ist gezeigt die Darstellung der aus dem \DREI{L}{S}{Z}-Formalismus herr"uhrenden disconnected Termen~\mbox{\,$\de(i',i)$} in Gl.~(\ref{S-Element2h_WW-V}) als Einsen~$\bbbone$ unter den Integralen~\mbox{\,$d_{i'\mskip-1mu i}$},~-- \vspace*{-.5ex}"aquivalent zu den~Ein\-sen~\mbox{\,$\bbbone$}, die auftreten als Summanden der Wegner-Wilson-Linien~$V\idx{+}$,~$V\idx{+}^{\D\dagger}$ und~$V\idx{-}$,~$V\idx{-}^{\D\dagger}$.
Sie sind somit zu verstehen und zu behandeln im gleichen Sinne.
\vspace*{-.5ex}

\section[\protect$S$-Matrixelement
           \protect$\bracket{h^{\!2'}\, h^{\!1'}, \IN}{%
                       \,S\, \bracketM h^1\, h^2, \IN}$]{%
         \bm{S}-Matrixelement
           \bm{\bracket{h^{\!2'}\, h^{\!1'}, \IN}{%
                  \,S\, \bracketM h^1\, h^2, \IN}}}
\label{APP:S-Element2h}

Wir leiten explizit her Gl.~(\ref{S-Element2h_WW-V_APP}) aus Gl.~(\ref{S-Element2h_WW-V}).
Dies impliziert die Identifizierung der relevanten Variablen, von denen allein abh"angt der Vakuumerwartungswert der vier \vspace*{-.5ex}Wegner-Wilson-Linien~\mbox{\,$\vac{V\idx{+} V\idx{+}^{\D\dagger} V\idx{-} V\idx{-}^{\D\dagger}}$}. \\
\indent
Wir betrachten das Differential
\end{samepage}
\vspace*{-.5ex}
\begin{align} \label{APP:4di'i}
&d_{1'\mskip-1mu 1}            \!(      x^-,  \rb{x}) \;
  d_{{\bar1}'\mskip-1mu {\bar1}}\!({\bar x}^-,\rbb{x}) \;
  d_{2'\mskip-1mu 2}            \!(      y^+,  \rb{y}) \;
  d_{{\bar2}'\mskip-1mu {\bar2}}\!({\bar y}^+,\rbb{y})
    \\
&=\; 2\sqrt{p_{1'}^+p_1^+}\;
        2\sqrt{{\bar p}_{1'}^+{\bar p}_1^+}\;
        2\sqrt{p_{2'}^+p_2^+}\;
        2\sqrt{{\bar p}_{2'}^+{\bar p}_2^+}
    \nn \\
&\phantom{=\;}\times
  (g_{+-})^4\vv
    dx^-         d^2 \rb{x} \vv
    d{\bar x}^-  d^2\rbb{x} \vv
    dy^+         d^2 \rb{y} \vv
    d{\bar y}^+  d^2\rbb{y}
    \nn \\
&\phantom{=\;}\times
  \begin{aligned}[t]
    \exp \iIM \Big[
      &g_{+-}(p_{1'}^+ \!-\! p_1^+) x^-
         - (\rb{p}_{1'} \!-\! \rb{p}_1) \!\cdot\! \rb{x}
    \nn \\
  +\; &g_{+-}({\bar p}_{1'}^+ \!-\! {\bar p}_1^+) {\bar x}^-
         - (\rbb{p}_{1'} \!-\! \rbb{p}_1) \!\cdot\! \rbb{x}
    \nn \\
  +\; &g_{+-}(p_{2'}^- \!-\! p_2^-) y^+
         - (\rb{p}_{2'} \!-\! \rb{p}_2) \!\cdot\! \rb{y}
    \nn \\
  +\; &g_{+-}({\bar p}_{2'}^- \!-\! {\bar p}_2^-) {\bar y}^+
         - (\rbb{p}_{2'} \!-\! \rbb{p}_2) \!\cdot\! \rbb{y}
    \Big]
    \nn
  \end{aligned}
    \nn
    \\[-4.5ex]\nn
\end{align}
vgl.\@ die Definition in den Gln.~(\ref{di'i}),~(\ref{di'i}$'$).
Dieses Differential wirke auf eine Funktion mit gleicher Abh"angigkeit von~$x$,~${\bar x}$ und~$y$,~${\bar y}$ wie
\begin{align} \label{APP:vac_4WW-V}
&\vac{\, V\idx{+} V\idx{+}^{\D\dagger} V\idx{-} V\idx{-}^{\D\dagger}\, } \\
&\hspace*{\equalindent} \equiv\;
        \vacL\,
        \trDrst{F}\big[ V\idx{+}(\infty,x^-,        \rb{x})\,
                        V\idx{+}^{\D\dagger}(\infty,{\bar x}^-,\rbb{x}) \big]\;
        \trDrst{F}\big[ V\idx{-}(y^+,       \infty, \rb{y})\,
                        V\idx{-}^{\D\dagger}({\bar y}^+,\infty,\rbb{y}) \big]\,
        \vacR \nn
\end{align}
Wir beziehen uns gegen"uber Gl.~(\ref{S-Element2h_WW-V}) auf den nichttrivialen Ausdruck in Bezug auf die Abh"an\-gigkeit von~$x$,~${\bar x}$ und~$y$,~${\bar y}$ und lassen weg die Einsen bez"uglich der Eichgruppe.

Wir betrachten separat das Exponential in Gl.~(\ref{APP:4di'i}) und dr"ucken die Impulse der (Anti)Quarks aus durch die der Zust"ande~\mbox{\,$h^i(P_i)$} und die Bruchteile~\mbox{\,$\zet_i$} der Quarks an den "`gro"sen"' Lichtkegelimpulsen, vgl.\@ die Gln.~(\ref{Q_Impulse}),~(\ref{Q_Impulse}$'$) und~(\ref{Q-AQ-Substitution}):
\vspace*{-.5ex}
\begin{align} \label{APP:4di'i_Exp}
\exp\;\iIM\, \Big[\; \cdots\; \Big]\;
&=\; \exp\;\iIM\, \Big[\vv
  \begin{aligned}[t]
  &g_{+-}\; \Big(
       (P_{1'}^+ \!-\! P_1^+) {\bar x}^-
     + (P_{2'}^- \!-\! P_2^-) {\bar y}^+
    \Big)
    \\[-.25ex]
  +\; &g_{+-}\; \Big(
     + (\zet_{1'} P_{1'}^+ \!-\! \zet_1 P_1^+) X^-
     + (\zet_{2'} P_{2'}^- \!-\! \zet_2 P_2^-) Y^+
  \Big)
    \\
  -\; &\Big(
       (\rb{k}_{1'} \!-\! \rb{k}_1)\cdot \rb{X}
     + (\rb{k}_{2'} \!-\! \rb{k}_2)\cdot \rb{Y}
     \Big)
    \\
  -\; &\Big(
       (\zet_{1'} \rb{P}_{1'} \!-\! \zet_1 \rb{P}_1)\cdot \rb{x}
     + (\bzet_{1'} \rb{P}_{1'} \!-\! \bzet_1 \rb{P}_1)\cdot \rbb{x}
    \\[-.25ex]
  &\phantom{\,}
     + (\zet_{2'} \rb{P}_{2'} \!-\! \zet_2 \rb{P}_2)\cdot \rb{y}
     + (\bzet_{2'} \rb{P}_{2'} \!-\! \bzet_2 \rb{P}_2)\cdot \rbb{y}
    \Big)\vv \Big]
  \end{aligned}
    \\[-4.5ex]\nn
\end{align}
unter Definition durch
\vspace*{-.25ex}
\begin{align} 
X\;
  =\; x\; -\; {\bar x}\qquad
  \text{und}\qquad
Y\;
  =\; y\; -\; {\bar y}
    \\[-4.25ex]\nn
\end{align}
der Vierer-Differenzvektoren~$X$,~$Y$.
\vspace*{-.5ex}

\paragraph{Longitudinal.}
Wir betrachten zun"achst die Koordinatentransformation der longitudinalen Komponenten.
Sei sie vermittelt durch die~$4\!\times\!4$-Matrix~${\bf A}_L$:
\begin{samepage}
\vspace*{-.5ex}
\begin{align} \label{APP:bfA_L}
\begin{pmatrix} x^- \\ {\bar x}^- \\ y^+ \\ {\bar y}^+ \end{pmatrix}
        \vv\begin{CD}@>{\D\vv{\bf A}_L\vv}>>\end{CD}\vv
  \begin{pmatrix} x^- \\ X^- \\ y^+ \\ Y^+ \end{pmatrix} \qqquad
{\bf A}_L\; =\;
\begin{pmatrix} 1 & -1 & 0 &  0 \\
                0 &  1 & 0 &  0 \\
                0 &  0 & 1 & -1 \\
                0 &  0 & 0 &  1 \end{pmatrix}
    \\[-4.5ex]\nn
\end{align}
Als obere Dreieckmatrix folgt f"ur die Jacobi-Determinante der Transformation unmittelbar
\vspace*{-.5ex}
\begin{align} \label{APP:bfA_L_Det}
\frac{\pa(X^-,x^-,Y^+,y^+)}{\pa(x^-,{\bar x}^-,y^+,{\bar y}^+)}\;
  =\; \det\,{\bf A}_L\;
  =\; 1
    \\[-4.5ex]\nn
\end{align}
Die neuen Koordinaten sind insbesondere {\it linear unabh"angig\/}. \\
\indent
Translationsinvarianz des Vakuumerwartungswertes~\mbox{\,$\vac{\;\cdot\;}$} impliziert f"ur Gl.~(\ref{APP:vac_4WW-V})
\vspace*{-.75ex}
\begin{align} \label{APP:vac_4WW-V_L}
&\vac{\, V\idx{+} V\idx{+}^{\D\dagger} V\idx{-} V\idx{-}^{\D\dagger}\, }
    \\[-.5ex]
&\equiv\;
        \vacL\,
        \trDrst{F}\big[ V\idx{+}            (\infty,X^-, \rb{x})\,
                        V\idx{+}^{\D\dagger}(\infty,0,  \rbb{x}) \big]\vv
        \trDrst{F}\big[ V\idx{-}            (Y^+,\infty, \rb{y})\,
                        V\idx{-}^{\D\dagger}(0,  \infty,\rbb{y}) \big]\,
        \vacR \nn
    \\[-4.75ex]\nn
\end{align}
Damit ist gezeigt, da"s~\mbox{\,$\vac{\, V\idx{+} V\idx{+}^{\D\dagger} V\idx{-} V\idx{-}^{\D\dagger}\, }$} longitudinal bez"uglich der neuen Variablen abh"angt nur von~\mbox{\,$X^-$} und~\mbox{\,$Y^+$}.
Nach Substitution
\vspace*{-.25ex}
\begin{align} 
{X'}^-\;
  =\; P_1^+\, X^-\qquad
  \text{und}\qquad
{Y'}^+\;
  =\; P_2^-\, Y^+
    \\[-4.25ex]\nn
\end{align}
treten auf in Gl.~(\ref{APP:vac_4WW-V_L}) statt~\mbox{\,$X^-$},~\mbox{\,$Y^+$} respektive die Variablen~\mbox{\,$(P_1^+)^{-1}\,{X'}^-$},~\mbox{\,$(P_2^-)^{-1}\,{Y'}^+$}.
Diese verschwinden im Limes~\mbox{\,$s \!\to\! \infty$}, f"ur den gilt~\mbox{\,$P_1^+,\,P_2^- \to\infty$}.
Es folgt
\vspace*{-.75ex}
\begin{align} \label{APP:vac_4WW-V_L1}
&\vac{\, V\idx{+} V\idx{+}^{\D\dagger} V\idx{-} V\idx{-}^{\D\dagger}\, }
    \\[-.5ex]
&\equiv\;
        \vacL\,
        \trDrst{F}\big[ V\idx{+}            (\infty,0, \rb{x})\,
                        V\idx{+}^{\D\dagger}(\infty,0,\rbb{x}) \big]\;
        \trDrst{F}\big[ V\idx{-}            (0,\infty, \rb{y})\,
                        V\idx{-}^{\D\dagger}(0,\infty,\rbb{y}) \big]\,
        \vacR \nn
    \\[-4.5ex]\nn
\end{align}
unabh"angig von den longitudinalen Variablen.
\end{samepage}

Konsequenz ist, da"s die Terme der Exponentialfunktion in Gl.~(\ref{APP:4di'i_Exp}) mit~$X^-$,~$x^-$ und~$Y^+$,~$y^+$ ausintegriert werden k"onnen.
Die erste Zeile ergibt die Delta-Distributionen
\begin{align} 
2\pi(g_{+-})^{-1}\cdot \de(P_{1'}^+ \!-\! P_1^+) \qquad
        \text{und}\qquad
  2\pi(g_{+-})^{-1}\cdot \de(P_{2'}^- \!-\! P_2^-)
\end{align}
und hiermit die zweite Zeile:
\begin{align} \label{APP:Delta-DistrL_zet}
2\pi\,(g_{+-}\,P_1^+)^{-1}\cdot \de(\zet_{1'} \!-\! \zet_1) \qquad
        \text{und}\qquad
  2\pi\,(g_{+-}\,P_2^-)^{-1}\cdot \de(\zet_{2'} \!-\! \zet_2) 
\end{align}

\paragraph{Transversal.}
Die Delta-Distributionen in Gl.~(\ref{APP:Delta-DistrL_zet}) haben Konsequenzen f"ur die transversalen Komponenten.
Die Terme der Exponentialfunktion von Gl.~(\ref{APP:4di'i_Exp}) in der dritten bis f"unften Zeile schreiben sich:
\vspace*{-.5ex}
\begin{align} \label{APP:4di'i_ExpT}
&\exp \big[\; \cdots\; \big]\,\big|_T
    \nn \\
&=\; \exp \; \iIM\, \big[
       (\rb{k}_{1'} \!-\! \rb{k}_1)\cdot \rb{X}
     + (\rb{k}_{2'} \!-\! \rb{k}_2)\cdot \rb{Y}\;
     +\; (\rb{P}_{1'} \!-\! \rb{P}_1)\cdot \rb{X}_\zet
       + (\rb{P}_{2'} \!-\! \rb{P}_2)\cdot \rb{Y}_\zet
        \big]
    \\[-.25ex]
&=\; \exp \; \iIM\, \big[
       (\rb{k}_{1'} \!-\! \rb{k}_1)\cdot \rb{X}
     + (\rb{k}_{2'} \!-\! \rb{k}_2)\cdot \rb{Y}\;
     +\; \tfb\cdot (\rb{X}_\zet \!-\! \rb{Y}_\zet)
       - \rb{P}\cdot \rb{Y}_\zet
        \big]
    \tag{\ref{APP:4di'i_ExpT}$'$}
    \\[-4.75ex]\nn
\end{align}
Dabei definieren wir $\zet_i$-gewichtete Differenzvektoren
\begin{align} 
X_\zet\;
  =\; \zet_1\, x\; +\; \bzet_1\, {\bar x}\qquad
  \text{und}\qquad
Y_\zet\;
  =\; \zet_2\, y\; +\; \bzet_2\, {\bar y}
\end{align}
und in Gl.~(\ref{APP:4di'i_ExpT}$'$) zus"atzlich die Vierer-Impulse
\begin{align} \label{APP:P,ta}
P\;
  =\; (P_{1'} + P_{2'}) - (P_1 + P_2)\qquad
  \text{und}\qquad
\tf\;
  =\; P_{1'} - P_1
\end{align}
Es ist $P$ der Differenzimpuls zwischen den ein- und auslaufenden Zust"anden, wie definiert in Gl.~(\ref{ViererDifferenzimpuls}), $\tf$ der Vierer-Impuls"ubertrag, dessen Quadrat identisch der Invarianten~$t$ nach Mandelstam ist:~$\tfQ \!=\! t$, vgl.\@ Gl.~(\ref{Mandelstam_Def}).

In Gl.~(\ref{APP:4di'i_ExpT}$'$) werden identifiziert~$\rb{X}$,~$\rb{Y}$ und \mbox{$\rb{X}_\zet \!-\! \rb{Y}_\zet$},~$\rb{Y}_\zet$ als die relevanten transversalen Variablen.
Sie werden vermittelt durch die~$8\!\times\!8$-Matrix~${\bf A}_T$:
\begin{samepage}
\vspace*{-.5ex}
\begin{align} \label{APP:bfA_T}
\begin{pmatrix} \rb{x} \\ \rbb{x} \\ \rb{y} \\ \rbb{y} \end{pmatrix}
        \vv\begin{CD}@>{\D\vv{\bf A}_T\vv}>>\end{CD}\vv
  \begin{pmatrix} \rb{X} \\ \rb{Y} \\ \rb{X}_\zet \!-\! \rb{Y}_\zet \\ \rb{Y}_\zet
  \end{pmatrix} \qqquad
{\bf A}_T\; =\;
  \begin{pmatrix} \bbbone       & -\bbbone       & \bm{0}         & \bm{0}          \\
                  \bm{0}        & \bm{0}         & \bbbone        & \bm{0}          \\
                  \zet_1\bbbone & \bzet_1\bbbone & -\zet_2\bbbone & -\bzet_1\bbbone \\
                  \bm{0}        & \bm{0}         &  \zet_2\bbbone &  \bzet_1\bbbone
  \end{pmatrix}
    \\[-4.5ex]\nn
\end{align}
in Termen von~$2\!\times\!2$-Matrizen.
Die Jacobi-Determinante der Transformation ist
%
\begin{align} \label{APP:bfA_T_Det}
\frac{\pa(\rb{X}, \rb{Y},\rb{X}_\zet \!-\!\rb{Y}_\zet,\rb{Y}_\zet)}{
      \pa(\rb{x},\rbb{x},\rb{y},\rbb{y})}\;
 &=\; \det\,{\bf A}_T
    \\[-.5ex]
 &=\; \det
        \begin{pmatrix} \zet_1\bbbone & \bzet_1\bbbone \\
                        \bbbone       & -\bbbone
        \end{pmatrix}\cdot
      \det
        \begin{pmatrix} \zet_2\bbbone & \bzet_2\bbbone \\
                        \bbbone       & -\bbbone
        \end{pmatrix}
  =\; 1
    \nn
\end{align}
Die neuen Koordinaten sind insbesondere {\it linear unabh"angig\/}.

F"ur die Inverse der Matrix~${\bf A}_T$ folgt:
\begin{samepage}
%
\begin{align} 
{{\bf A}_T}^{-1}\; =\;
  \begin{pmatrix} \bzet_1\bbbone & \bm{0}         & \vv\bbbone\vv & \vv\bbbone\vv \\
                  -\zet_1\bbbone & \bm{0}         & \bbbone       & \bbbone       \\
                  \bm{0}         & \bzet_2\bbbone & \bm{0}        & \bbbone       \\
                  \bm{0}         & -\zet_2\bbbone & \bm{0}        & \bbbone
  \end{pmatrix}
\end{align}
\end{samepage}%
Der Vierer-Impakt~$b$ der Streuung ist definiert als der zum Vierer-Impuls"ubertrag~$\ta$~-- vgl.\@ Gl.~(\ref{APP:P,ta})~-- Fourier-konjugierte Vektor; aus Gl.~(\ref{APP:4di'i_ExpT}$'$) lesen wir ab:
%
\begin{align} \label{APP:Vierer-b}
b\;\;
  =\; \phantom{(}
      X_\zet - Y_\zet\quad
      \phantom{)/2}
  =\; \phantom{(}
        (\zet_1\, x + \bzet_1\, \bar{x})
      - (\zet_2\, y + \bzet_2\, \bar{y})
\end{align}
und definieren
\end{samepage}%
%
\begin{align} \label{APP:Vierer-om}
\om\;
  =\; (X_\zet + Y_\zet)/2\quad
  =\; \big((\zet_1\, x + \bzet_1\, \bar{x})
         + (\zet_2\, y + \bzet_2\, \bar{y})\big)\big/2
    \\[-4.5ex]\nn
\end{align}
Die Transversalprojektionen der (Anti-)Quark-Koordinaten in den Ortsraum folgen mithilfe von~${{\bf A}_T}^{-1}$; es folgt f"ur die "`plus"'-Zust"ande~(Index 1):
\vspace*{-.75ex}
\begin{align} \label{APP:Q-AQ-Pos_rb1}
&\rb{x}\;
  =\; \phantom{-}
      \bzet_1\, \rb{X}\; +\; \rb{b}/2\; +\; \rbG{\om}
    \\[-.5ex]
&\rbb{x}\;
  =\; -\zet_1\, \rb{X}\; +\; \rb{b}/2\; +\; \rbG{\om} 
    \tag{\ref{APP:Q-AQ-Pos_rb1}$'$}
    \\[-4.75ex]\nn
\end{align}
f"ur "`minus"'-Zust"ande~(Index 2):
\vspace*{-.75ex}
\begin{align} \label{APP:Q-AQ-Pos_rb2}
&\rb{y}\;
  =\; \phantom{-}
      \bzet_2\, \rb{Y}\; -\; \rb{b}/2\; +\; \rbG{\om}
    \\[-.5ex]
&\rbb{y}\;
  =\; -\zet_2\, \rb{Y}\; -\; \rb{b}/2\; +\; \rbG{\om}
    \tag{\ref{APP:Q-AQ-Pos_rb2}$'$}
    \\[-4.75ex]\nn
\end{align}
Mit~\mbox{\,$%
  (\rb{X}_\zet \!-\! \rb{Y}_\zet) \!+\! \rb{Y}_\zet
    \!=\! \rb{b}/2 \!+\! \rbG{\om}$},%
   ~\mbox{\,$%
  \rb{Y}_\zet
    \!=\! -\rb{b}/2 \!+\! \rbG{\om}$} folgt~$\rbG{\om}$ als global additiver Term, der o.E.d.A.\@ weggelassen wird%
\FOOT{
  \label{APP-FN:om=0}Dies entspricht Wahl von~$\om \!=\! (X_\zet \!+\! Y_\zet)\!/2$ als Koordinatenursprung; i.a.\@ wird nicht a~priori~\mbox{$\om \!\equiv\! 0$} gesetzt.
}:
%
\vspace*{-.5ex}
\begin{align} \label{APP:vac_4WW-V_red}
\vac{\, V\idx{+} V\idx{+}^{\D\dagger} V\idx{-} V\idx{-}^{\D\dagger}\, }\;
  \equiv\;
  \vacL\,
 &\trDrst{F}\big[
    V\idx{+}            (\infty, 0, \bzet_1\rb{X} \!+\! \rb{b}\!/\!2)\,
    V\idx{+}^{\D\dagger}(\infty, 0, -\zet_1\rb{X} \!+\! \rb{b}\!/\!2) \big]
    \\[-.5ex]
  \times
 &\trDrst{F}\big[
    V\idx{-}            (0, \infty, \bzet_2\rb{Y} \!-\! \rb{b}\!/\!2)\,
    V\idx{-}^{\D\dagger}(0, \infty, -\zet_2\rb{Y} \!-\! \rb{b}\!/\!2) \big]\,
  \vacR
    \nn
    \\[-4.5ex]\nn
\end{align}
aufgrund der Translationsinvarianz des Vakuumerwartungswertes~$\vac{\;\cdot\;}$.
\vspace*{.5ex}

\bigskip\noindent
Wir fassen das Resultat der Koordinatentransformationen von Gl.~(\ref{APP:bfA_L}) und~(\ref{APP:bfA_T}) zusammen.
Im Limes~\mbox{\,$s \!\to\! \infty$} gilt~\mbox{\,$P_{1'}^+ \!-\! P_1^+ \!=\! P^+$} und~\mbox{\,$P_{2'}^- \!-\! P_2^- \!=\! P^-$}, vgl.\@ die Gln.~(\ref{APP:4di'i_Exp}),~(\ref{APP:4di'i_ExpT}), so da"s f"ur das Differential nach Gl.~(\ref{APP:4di'i}) folgt
%
\begin{align} \label{APP:4di'i_1}
&d_{1'\mskip-1mu 1}            \!(      x^-,  \rb{x})\;
  d_{{\bar1}'\mskip-1mu {\bar1}}\!({\bar x}^-,\rbb{x})\;
  d_{2'\mskip-1mu 2}            \!(      y^+,  \rb{y})\;
  d_{{\bar2}'\mskip-1mu {\bar2}}\!({\bar y}^+,\rbb{y})
    \\[.5ex]
&\underset{\text{$s \!\to\! \infty$}}{\sim}\;
 \begin{aligned}[t]
  &2\sqrt{\zet_{1'} P_{1'}^+ \zet_1 P_1^+}\vv
   2\sqrt{\bzet_{1'} P_{1'}^+ \bzet_1 P_1^+}\vv
   2\sqrt{\zet_{2'} P_{2'}^- \zet_2 P_2^-}\vv
   2\sqrt{\bzet_{2'} P_{2'}^+ \bzet_2 P_2^-}
    \\[.25ex]
   &\times\, \big(g_{+-}\big)^4\vv
    dX^-\, dY^+\, d{\bar x}^-\, d{\bar y}^+ \vv
    d^2\rb{X}\, d^2\rb{Y}\, d^2(\rb{X}_\zet \!-\! \rb{Y}_\zet)\, d^2\rb{Y}_\zet
    \\[.25ex]
   &\times\,
   \begin{aligned}[t]
    \exp\; \iIM\, \big[ g_{+-}
        &\big( P_1^+ (\zet_{1'} \!-\! \zet_1) X^- + P_2^- (\zet_{2'} \!-\! \zet_2) Y^+
              + P^+ {\bar x}^- + P^- {\bar y}^+ \big)
    \\[.25ex]
    -\, &\big( (\rb{k}_{1'} \!-\! \rb{k}_1)\cdot \rb{X}
              + (\rb{k}_{2'} \!-\! \rb{k}_2)\cdot \rb{Y}
              - \tfb\cdot(\rb{X}_\zet \!-\! \rb{Y}_\zet)
              + \rb{P}\cdot\rb{Y}_\zet \big)
    \big]
   \end{aligned}
 \end{aligned}
    \nn
\end{align}
Dieses ist zu beziehen auf eine Funktion, die abh"angt {\it longitudinal\/} \vspace*{-.375ex}von~\mbox{\,$X^-$},~\mbox{\,$Y^+$},~\mbox{${\bar x}^-$},~\mbox{\,${\bar y}^+$} und {\it transversal\/} von~\mbox{\,$\rb{X}$},~\mbox{\,$\rb{Y}$},~\mbox{\,$\rb{X}_\zet \!-\! \rb{Y}_\zet$},~\mbox{\,$\rb{Y}_\zet$} in derselbe Weise wie~\mbox{\,$\vac{\,V\idx{+} V\idx{+}^{\D\dagger} V\idx{-} V\idx{-}^{\D\dagger}\,}$} in Gl.~(\ref{APP:vac_4WW-V_red}),~-- einschlie"slich der Eichgruppen-Einsen im urspr"unglichen Vakuumerwartungswert. \\
\indent
Die Integrationen der longitudinalen Variablen sind daher nur zu beziehen auf das Exponential:
Integration von~\mbox{\,$X^-,\,Y^+$} ergibt Gl.~(\ref{APP:Delta-DistrL_zet}): Delta-Distributionen in~\mbox{$(\zet_{1'} \!-\! \zet_1)$}, respektive in~\mbox{\,$\zet_{2'} \!-\! \zet_2$}.
Integration von~\mbox{${\bar x}^-,\, {\bar y}^+$} ergibt Erhaltung des {\it longitudinalen\/} Gesamtimpulses:~Delta-Distribu\-tionen in~$P^+$,~$P^-$.
Integration "uber~\mbox{\,$\rb{Y}_\zet$} schlie"slich ist im selben Sinne frei und ergibt Erhaltung des {\it transversalen\/} Gesamtimpulses:~\mbox{\,$\de(\rb{P})$}.

Wir benutzen die Implikationen dieser Delta-Distributionen f"ur die Wurzelausdr"ucke, setzen~\mbox{\,$\rb{b} \!=\! \rb{X}_\zet - \rb{Y}_\zet$} nach Definition in Gl.~(\ref{APP:Vierer-b}) und erhalten f"ur Gl.~(\ref{APP:4di'i_1}) weiter
\begin{samepage}
%
\begin{align} \label{APP:4di'i_2}
&d_{1'\mskip-1mu 1}            \!(      x^-,  \rb{x}) \;
  d_{{\bar1}'\mskip-1mu {\bar1}}\!({\bar x}^-,\rbb{x}) \;
  d_{2'\mskip-1mu 2}            \!(      y^+,  \rb{y}) \;
  d_{{\bar2}'\mskip-1mu {\bar2}}\!({\bar y}^+,\rbb{y})
    \\[.5ex]
&\underset{\text{$s \!\to\! \infty$}}{\sim}\,
 \begin{aligned}[t]
  &2^4\vv \zet_1\bzet_1(P_1^+)^2\vv \zet_2\bzet_2(P_2^-)^2\vv 
                (g_{+-})^4
    \\[.25ex]
  &\times\,
       (2\pi)^6\; (g_{+-})^{-4}\vv (P_1^+P_2^-)^{-1}\vv
         \de(\zet_{1'} \!-\! \zet_1)\, \de(\zet_{2'} \!-\! \zet_2)\vv
         \de(P^+)\, \de(P^-)\, \de(\rb{P})
    \\[.25ex]
  &\times\,
       d^2\rb{b} \exp\big[-\iIM\, \tfb\cdot \rb{b}\big] \vv
       d^2\rb{X} \exp\big[-\iIM\, (\rb{k}_{1'} \!-\! \rb{k}_1)\cdot \rb{X}\big] \vv
       d^2\rb{Y} \exp\big[-\iIM\, (\rb{k}_{2'} \!-\! \rb{k}_2)\cdot \rb{Y}\big]
 \end{aligned}
    \nn
\end{align}
Es gilt im Limes~\mbox{\,$s \!\to\! \infty$}:
\end{samepage}
\vspace*{-.5ex}
\begin{alignat}{3}
&2\, P_1^+P_2^-&
  &\;\underset{\text{$s \!\to\! \infty$}}{\sim}\vv&
  &(g_{+-})^{-1}\vv s
    \label{APP:Mandelstam-s_lim} \\
&\de(P^+)\, \de(P^-)&
  &\;\underset{\text{$s \!\to\! \infty$}}{\sim}\vv&
  &g_{+-}\vv \de(P^0)\, \de(P^3)
    \label{APP:de-P+P-_de-P0P3}
    \\[-4.25ex]\nn
\end{alignat}
Folglich:
%
\begin{align} \label{APP:4di'i_red}
&d_{1'\mskip-1mu 1}            \!(      x^-,  \rb{x})\vv
  d_{{\bar1}'\mskip-1mu {\bar1}}\!({\bar x}^-,\rbb{x})\vv
  d_{2'\mskip-1mu 2}            \!(      y^+,  \rb{y})\vv
  d_{{\bar2}'\mskip-1mu {\bar2}}\!({\bar y}^+,\rbb{y})
    \\[1ex]
&\underset{\text{$s \!\to\! \infty$}}{\sim}\;
  \begin{aligned}[t]
    \iIM\, &(2\pi)^4\; \de(P)
    \\
  &\times\,
    (-2\iIM\,s)\vv
      d^2\rb{b} \exp\big[-\iIM\, \tfb\cdot \rb{b}\big]
    \nn \\[-.5ex]
  &\times\,
    2\pi\, \de(\zet_{1'} \!-\! \zet_1)\vv
      2\zet_1\bzet_1\vv
      d^2\rb{X}\vv
      \exp\big[-\iIM\, (\rb{k}_{1'} \!-\! \rb{k}_1)\cdot \rb{X}\big]
    \nn \\
  &\times\,
    2\pi\, \de(\zet_{2'} \!-\! \zet_2)\vv
      2\zet_2\bzet_2\vv
      d^2\rb{Y}\vv
      \exp\big[-\iIM\, (\rb{k}_{2'} \!-\! \rb{k}_2)\cdot \rb{Y}\big]
  \end{aligned}
    \nn
\end{align}
in Termen der relevanten Transversalvektoren~$\rb{b}$ und~$\rb{X}$,~$\rb{Y}$.
\vspace*{-.5ex}

\bigskip\noindent
\vspace*{-.5ex}In Gl.~(\ref{S-Element2h_calM}) steht das Differential~\mbox{\,$d_{1'\mskip-1mu 1} d_{{\bar1}'\mskip-1mu {\bar1}} d_{2'\mskip-1mu 2} d_{{\bar2}'\mskip-1mu {\bar2}}$} des Vakuumerwartungswerts der vier Wegner-Wilson-Linien~\mbox{\,$\vac{\, V\idx{+} V\idx{+}^{\D\dagger} V\idx{-} V\idx{-}^{\D\dagger}\, }$}~-- einschlie"slich der Einsen der Eichgruppe~-- unter den Integralen%
\FOOT{
  also insgesamt von Dimension Zw"olf
}
%
\vspace*{-.5ex}
\begin{align} 
{\T\prod}_{i=1,2}\vv
\int d{\tilde\vph}_{s_{i'}\!{\bar s}_{i'}}^{i'\D\dagger} (\zet_{i'}, \rb{k}_{i'})\;
\int d{\tilde\vph}_{s_i\!{\bar s}_i}^i (\zet_i, \rb{k}_i)
    \\[-4.5ex]\nn
\end{align}
die implizieren die Lichtkegelwellenfunktionen der hadronischen Zust"ande; zur Definition der~\mbox{$d{\tilde\vph}_{s_i\!{\bar s}_i}^i (\zet_i, \rb{k}_i)$} vgl.\@ Gl.~(\ref{h-ket_Mass}).
F"ur~\mbox{\,$i \!=\! 1,\,2$} und~$\rb{Z}$ summarisch f"ur~$\rb{X}$ oder~$\rb{Y}$ gilt:
\begin{samepage}
%
\begin{align} \label{APP:FT-to-dvphi'i}
&\int d{\tilde\vph}_{s_{i'}\!{\bar s}_{i'}}^{i'\D\dagger} (\zet_{i'}, \rb{k}_{i'})\vv
  \int d{\tilde\vph}_{s_i\!{\bar s}_i}^i (\zet_i, \rb{k}_i)
    \nn \\[-.75ex]
  &\phantom{\int d{\tilde\vph}_{s_{i'}\!{\bar s}_{i'}}^{i'\D\dagger}}
   \times\,
        2\pi\, \de(\zet_{1'} \!-\! \zet_1)\; 2\zet_1\bzet_1\vv
        \int d^2\rb{Z}\; \exp\big[-\iIM\, (\rb{k}_{1'} \!-\! \rb{k}_1)\cdot \rb{Z}\big]
    \nn \\[-1.375ex]
&=\;
  \begin{alignedat}[t]{5}
  \int d^2\rb{Z} \int_0^1 \frac{d\zet_i}{2\pi}\;
        &\bigg( \int \frac{d^2\rb{k}_{i'}}{(2\pi)^2}&\;
                &\exp\big[\iIM\, \rb{k}_{i'}\cdot \rb{Z}\big]&\;
                &\sqrt{2\zet_i\bzet_i}&\vv
                &{\tilde\vph}_{s_i\!{\bar s}_i}^{i'} (\zet_i, \rb{k}_{i'})&
        &\bigg) \Big.^\sizeq{\Large}{\!\dagger}
    \\[-.25ex]
  \times\,
        &\bigg( \int \frac{d^2\rb{k}_i}{(2\pi)^2}&\;
                &\exp\big[\iIM\, \rb{k}_i\cdot \rb{Z}\big]&\;
                &\sqrt{2\zet_i\bzet_i}&\vv
                &{\tilde\vph}_{s_i\!{\bar s}_i}^i (\zet_i, \rb{k}_i)&
        &\bigg)
  \end{alignedat}
    \\
&=\;
  \int d^2\rb{Z} \int_0^1 \frac{d\zet_i}{2\pi}\;
        \vph_{s_i\!{\bar s}_i}^{i'\D\dagger} (\zet_i, \rb{Z})\vv
        \vph_{s_i\!{\bar s}_i}^i (\zet_i, \rb{Z})
    \tag{\ref{APP:FT-to-dvphi'i}$'$}
    \\[-4.5ex]\nn
\end{align}
unter Absorption von Faktoren~\mbox{$\surd2\zet_i\bzet_i$} in die transversal Fourier-transformierten Lichtkegelwellenfunktionen~\mbox{$\vph_{s_i\!{\bar s}_i}^i$}, vgl.\@ deren Definition in Gl.~(\ref{vph-x_vph-k}). \\
\indent
Wir definieren das Integrationsma"s
%
\begin{align} \label{APP:dvphi'i}
d\vph_{i',i} (\zet, \rb{Z})\;
  \equiv\; d^2\rb{Z}\vv
             \frac{d\zet}{2\pi}\vv
             \vph_{s\mskip-1mu{\bar s}}^{i'\D\dagger} (\zet, \rb{Z})\vv
             \vph_{s\mskip-1mu{\bar s}}^i (\zet, \rb{Z})\qqquad
  i = 1,2
    \\[-6ex]\nn
\end{align}
das gewichtet entsprechend dem "Uberlapp~\mbox{\,$\vph_{s\mskip-1mu{\bar s}}{}^{\zz i'\D\dagger} \vph_{s\mskip-1mu{\bar s}}^i$} der Lichtkegelwellenfunktionen; Summation "uber die Spins von Quark und Antiquark sei impliziert.
Dann gilt f"ur das~\mbox{$S$-Matrix}\-element in~conclusio:
%
\begin{align} \label{APP:S-Element2h_WW-V}
&\hspace*{-10pt}
 \bracket{\, h^{2'}\!(P_{2'})\, h^{1'}\!(P_{1'}),\,\IN \,}{\,
             S\, \bracketM\,
             h^1(P_1)\, h^2(P_2),\,\IN \,}
    \\[.75ex]
&\hspace*{-10pt}
 \underset{\text{$s \!\to\! \infty$}}{\sim}\;
        \iIM\, (2\pi)^4\; \de(P)
    \nn \\[-1ex]
&\hspace*{-10pt}
 \phantom{P_2}\times
        -\, 2\iIM\,s\; \int d^2\rb{b}\;
             \efn{\T-\iIM\,\tfb \!\cdot\! \rb{b}}\vv
             \int d\vph_{1',1} (\zet_1, \rb{X})\;
             \int d\vph_{2',2} (\zet_2, \rb{Y})
    \nn \\[-1.5ex]
&\hspace*{-10pt}
 \phantom{P_2}\times\,
 \hspace*{-5pt}
  \begin{aligned}[t]
  \vacL\,
    &\trDrst{F}\!\big[
      Z\idx{2}^{-1} \big(
        V\idx{+}             (\infty, 0, \bzet_1 \rb{X} \!+\! \rb{b}\!/2)
        - \deZ\,\bbbOne{F} \big)\;
      Z\idx{2}^{-1} \big(
        V\idx{+}^{\D\dagger} (\infty, 0, -\zet_1 \rb{X} \!+\! \rb{b}\!/2)
        - \deZ\,\bbbOne{F} \big)
    \big]
    \nn \\[-.5ex]
  \times
    &\trDrst{F}\!\big[
      Z\idx{2}^{-1} \big(
        V\idx{-}             (0, \infty, \bzet_2 \rb{Y} \!-\! \rb{b}\!/2)
        - \deZ\,\bbbOne{F} \big)\;
      Z\idx{2}^{-1} \big(
        V\idx{-}^{\D\dagger} (0, \infty, -\zet_2 \rb{Y} \!-\! \rb{b}\!/2)
        - \deZ\,\bbbOne{F} \big)
    \big]\,
  \vacR
    \nn
  \end{aligned}
    \nn
    \\[-3.5ex]\nn
\end{align}
Dies ist genau Gl.~(\ref{S-Element2h_WW-V_APP}).
\end{samepage}
\vspace*{4.5ex}

%% file: APP_BOOSTS-F.tex
\lhead[\fancyplain{}{\sc\thepage}]
      {\fancyplain{}{\sc\rightmark}}
\rhead[\fancyplain{}{\sc{{\footnotesize Anhang~\thechapter:} Aktive Lorentz-Boosts}}]
      {\fancyplain{}{\sc\thepage}}
\psfull
\chapter[Aktive Lorentz-Boosts]{%
   \huge Aktive Lorentz-Boosts~\bffootnote}
\label{APP:Boosts}
\footnotetext{
  Diesem Anhang zugrunde liegen die Konventionen bzgl.\@ der Lorentz-Algebra von Anh.~\ref{APP:Lorentz-Algebra}, speziell bzgl.\@ Lichtkegelkoordinaten von Anh.~\ref{APP:LC-Koord}.   Bzgl.\@ der einf"uhrenden Abschnitte vgl.\@ Ref.~\cite{Nachtmann92}, insbes.\@ Ref.~\cite{Sexl92}.
}

Zwei Inertialsysteme~$I$ und~$I'$ k"onnen differieren bez"uglich zehn Parametern:
bez"uglich einer Verschiebung zeitlich um~$a^0$, r"aumlich um~$\vec{a}$, einer Drehung um~$\vec{\al}$ und einer gleichf"ormigen Bewegung mit Geschwindigkeit~$\vec{v}$. \\
\indent
Wir betrachten spezielle orthochrone Lorentz-Transformationen: Elemente der \vspace*{-.25ex}Eins-Zu\-sammenhangkomponente der \vspace*{-.25ex}Lorentz-Gruppe~${\cal L}_+^{\bm{\scriptstyle\uparrow}}$,~-- charakterisiert durch~\mbox{$(a^0,\vec{a}^{\T\,t}) \!\equiv\! 0$}~und f"ur kontravariante Komponenten vermittelt durch pseudo-orthogonale \vspace*{-.25ex}Matrizen \mbox{$\La \!\equiv\! \big(\La^\mu{}_\nu\big)$}, \mbox{$\mu,\nu \!\in\! \{0,1,2,3\}$}, im Sinne~\mbox{$\La^{\T t}\, g\, \La \!=\! g$}, das hei"st~\mbox{$g_{\mu\nu} \La^\mu{}_\rh \La^\nu{}_\si \!=\! g_{\rh\si}$}, mit~\mbox{$\La^0{}_0 \!>\! 0$},~\mbox{$\det\La \!\equiv\! +1$}.
\vspace*{-.5ex}

\section{Lorentz-Boosts}
\label{APP-Sect:Lorentz-Boosts}

Als {\it Lorentz-Boost\/} wird bezeichnet die reine Geschwindigkeitstransformation, die keine Drehung enth"alt:~$\vec{\al} \!\equiv\! \vec{0}$.
Sei~$\vec{\be} \!=\! \vec{v}\!/c$ die Boost-, das hei"st Dreier-Geschwindigkeit von~$I'$ gemessen in~$I$.
Dann besteht zwischen den ungestrichenen und gestrichenen Komponneten eines beliebigen Lorentz-(Spalten)Vektors~$x \!\equiv\! (x^\mu) \!\equiv\! (x^{\mu'})$ der Zusammenhang:%
\FOOT{
  Es ist~$\vec{\be}$ der dreidimensionale Spalten-Vektor,~$\bbbone$ die~$3 \!\times\! 3$-Matrix und mit~$t$ bezeichnet Transposition.
}
%
\begin{align}
&x^{\mu'}\;
  =\; \La^\mu{}_\nu\; x^\nu\qquad
    \mu,\,\nu \in \{0,1,2,3\}
    \label{APP:LT} \\
&\La
  =\; \big(\La^\mu{}_\nu\big)
  =\; \pmatrixZZ{\ga}{-\ga\, \vec{\be}^{\;\T t}}{-\ga\, \vec{\be}}
        {\bbbone\; +\; \frac{\D\ga^2}{\D\ga \!+\! 1}\, \vec{\be}\, \vec{\be}^{\;\T t}}
    \label{APP:LT-La}
    \\[-4.75ex]\nn
\end{align}
mit
\vspace*{-1ex}
\begin{align} \label{APP:gamma,beta}
\ga \equiv \ga(\be)\;
  =\; 1\big/\sqrt{1 - \be^2}\qquad
  \be\; =\; |\vec{\be}|
    \\[-4.5ex]\nn
\end{align}
Sei im folgenden die~$x^3$-Achse gew"ahlt in Boost-Richtung:~\mbox{$\vec{\be} \!=\! \be\,(0,0,1)^{\T t}$}, so da"s gilt:%
\FOOT{
  \label{APP-FN:only-longitudinal}Es sind ausgeschrieben nur die longitudinalen Komponenten; die transversalen transformieren trivial.
}
%
\vspace*{-.5ex}
\begin{align} \label{APP:LT-La3}
\La\;
  =\; \big(\La^\mu{}_\nu\big)\;
  =\; \pmatrixZZ{\ga}{-\ga\, \be}{-\ga\, \be}{\ga}\;
  =\; \pmatrixZZ{\cosh\ps}{-\sinh\ps}{-\sinh\ps}{\cosh\ps}
    \\[-4.5ex]\nn
\end{align}
[bzgl.~$\ps$ vgl.\@ die Gln.~(\ref{APP:coshps}),~(\ref{APP:sinhps})].
Mit Konvention des Beta-Parameters~$\be$ als positiv~-- das hei"st~$\be \!\in\! [0,1)$~-- wird Umkehrung der Boost-Richtung realisiert durch Substitution~\mbox{$\be \!\to\! -\be$}. \\
\indent
Wir bedienen uns des {\it Minkowski-Diagramms\/} als wichtigem Instrument f"ur Argumentation und Anschauung.
Wir rekapitulieren:
Gl.~(\ref{APP:LT}) mithilfe~(\ref{APP:LT-La3}) lautet explizit:
\vspace*{-.5ex}
\begin{alignat}{2} \label{APP:LT3-explizit}
&x^{0'}&\;
  &=\; \ga\, \big(x^0 - \be\, x^3\big)
    \\
&x^{3'}&\;
  &=\; \ga\, \big(-\be\, x^0 + x^3\big)
    \tag{\ref{APP:LT3-explizit}$'$}
    \\[-4ex]\nn
\end{alignat}
\vspace*{-.375ex}Die~$x^{3'}$-Achse ist charakterisiert durch verschwindende Zeitkomponente in~$I'$:~\mbox{$x^{0'} \!\stackrel{\D!}{=}\! 0$}, das hei"st~\mbox{$x^0\big(x^3\big) \!=\! \be\, x^3$}, entsprechend die~$x^{0'}$-Achse durch~\mbox{$x^{3'} \!\stackrel{\D!}{=}\! 0$}, das hei"st~\mbox{$x^0\big(x^3\big) \!=\! \be^{-1}\, x^3$}.
\begin{figure}
\vspace*{-1ex}
\begin{minipage}{\linewidth}
  \begin{center}
  \setlength{\unitlength}{1mm}\begin{picture}(100,100) 
    \put(0,0){\epsfxsize100mm \epsffile{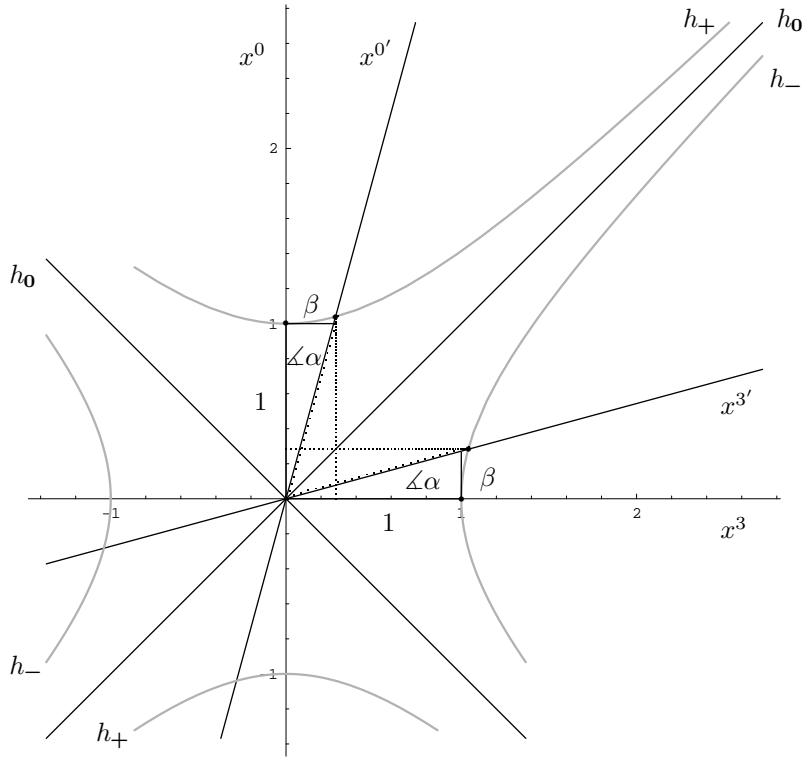}}
    \put(28,92){\normalsize$x^0$}
    \put(44,92){\normalsize$x^{0'}$}
    \put(92,29){\normalsize$x^3$}
    \put(92,45.5){\normalsize$x^{3'}$}
    \put( 9  , 2){\normalsize$h\idx{+}$}
    \put(-2.5,63){\normalsize$h\idx{0}$}
    \put(-2.5,11){\normalsize$h\idx{-}$}
    \put(87  ,97){\normalsize$h\idx{+}$}
    \put(99.5,97){\normalsize$h\idx{0}$}
    \put(98.5,89){\normalsize$h\idx{-}$}
    \put(36.5,59){\normalsize$\be$}
    \put(34.25,52){\normalsize$\measuredangle\al$}   
    \put(30,46){\normalsize$1$}
    \put(60,36){\normalsize$\be$}
    \put(50,35.5){\normalsize$\measuredangle\al$}
    \put(47,30){\normalsize$1$}
    \put(34.23,57.65){\circle*{.8}}
    \put(40.83,58.5 ){\circle*{.8}}
    \linethickness{.3pt}
      \qbezier[25](34.43,34.25)(37.73,46.375)(41.03,58.5)   
    \linethickness{.3pt}
      \qbezier[50](40.83,34.25)(40.83,50    )(40.83,58.5)
    \put(58.5 ,40.85){\circle*{.8}}
    \put(57.55,34.25){\circle*{.8}}
    \linethickness{.3pt}
      \qbezier[25](34.23,34.55)(46.365,37.85)(58.5,41.15)   
    \linethickness{.3pt}
      \qbezier[50](34.23,40.85)(50,    40.85)(58.5,40.85)
  \end{picture}
  \end{center}
\vspace*{-4ex}
\caption[Minkowski-Diagramm: Lorentz-Boost]{
  Minkowski-Diagramm: Allgemeiner Lorentz-Boost.   Sei~$I$ Inertialsystem mit Achsen~$x^0$,~$x^3$.   Relativ zu~$I$ und gemessen in~$I$ bewege sich gleichf"ormig mit Geschwindigkeit~$\be \!\in\! [0,1)$ in positive $x^3$-Richtung das Bezugsystem~$I'$, das ergo auch Inertialsystem ist.   Seine Achsen~$x^{0'}$,~$x^{3'}$ besitzen dann bez"uglich der Achse~$x^3$ Steigung~$\be^{-1}$ bzw.~$\be$:   Sie sind hineingedreht in den 1.\@ Quadranten um den Winkel~$-\al$ bzw.~$\al$ mit Definition~$\tan\al \!:=\! \be$.   Die Einsen der Achsen folgen als deren Schnittpunkte mit den zeit- bzw.\@ raumartigen Hyperboloiden~[$h\idx{+}$ bzw.~$h\idx{-}$, grau]:   Der Eins-Abschnitt auf~$x^{0'}$ ist gr"o"ser als der auf~$x^0$~-- Zeitdilatation, der Eins-Abschnitt auf~$x^{3'}$ ist gr"o"ser bzgl.~$x^3$~-- L"angenkontraktion~[weit vs.\@ eng punktiert].    F"ur~$\be \!\to\! 1$ folgt~$\al \!\to\! 45^\circ$, das hei"st die Achsen approximieren simultan die erste Winkelhalbierende~-- den "`Lichtkegel"'~$h\idx{0}$~--, und ihre Einsen gemessen in~$I$ gehen gegen Unendlich.   Umkehrung der Boost-Richtung in negative $x^3$-Richtung geschieht durch Substitution~$\be \!\to\! -\be$, so da"s~$\al \!\to\! -\al$.   Vgl.\@ Text und Fu"sn.\,\ref{APP-FN:be->-be}.
\vspace*{-.5ex}
}
\label{Fig:MinkowskiGENERAL}
\end{minipage}
\end{figure}
Im Minkowski-Diagramm mit $x^0$- der vertikalen und~$x^3$- der horizontalen Achse~-- vgl.\@ Abb.~\ref{Fig:MinkowskiGENERAL}~-- sind dies Geraden der Steigung~$\be$ und~$\be^{-1}$, die gegen"uber den Achsen~$x^0$ und~$x^3$~\mbox{verdreht sind} in den ersten Quadranten, um den Winkel~$-\al$ beziehungsweise~$\al$, der definiert ist durch
\begin{samepage}
\vspace*{-.5ex}
\begin{align} \label{APP:al_be}
\tan\, \al\;
  :=\; \be\qquad
    \\[-4.5ex]\nn
\end{align}
mit~$\al \!\in\! [0,\pi\!/4)$ f"ur~$\be \!\in\! [0,1)$.
F"ur wachsende Boost-Geschwindigkeit:~$\be \!\to\! 1$, gilt~$\al \!\to\! 45^\circ$, das hei"st die positiven Abschnitte der Achsen~$x^{0'}$ und~$x^{3'}$ n"ahern sich simultan an dem positiven Abschnitt der ersten Winkelhalbierenden.%
\FOOT{
  \label{APP-FN:be->-be}Zur Umkehrung des Boosts in negative $x^3$-Richtung ist zu substituieren~$\be \!\to\! -\be$, folglich~$\al \!\to\! -\al$:   Die Achsen~$x^{0'}$ und~$x^{3'}$ sind hineingedreht in den 2.\@ bzw.\@ 4.\@ Quadranten; f"ur~\mbox{$\be \!\to\! 1$} n"ahern sich so ihre positiven Abschnitte simultan an dem positiven bzw.\@ negativen Abschnitt der zweiten Winkelhalbierenden.
}
Der Grenzwert selbst ist daher singul"ar.
\end{samepage}

Werde~$x \!\equiv\! (x^\mu) \!\equiv\! (x^{\bar\mu})$ aufgefa"st als Differenzvektor von Ereignissen in~\mbox{$(x^0,x^3) \!\simeq\! (x^{0'},x^{3'})$} und~$(0,0)$.
Dann ist sein Lorentz-Quadrat invariant {\it per~definitionem\/}:
%
\begin{align} 
(x^0)^2 - (x^3)^2\;
  =\; (x^{0'})^2 - (x^{3'})^2\;
  =:\; c  
\end{align}
Sei betrachtet~$x^0$ als Funktion von~$x^3$, das hei"st:~\mbox{$x^0\big(x^3\big) \!=\! \pm\, \sqrt{c + (x^3)^2}$}.
Das Vorzeichen der Invarianten~$c$ bestimmt drei Klassen von Ereignissen bez"uglich~$(0,0)$.
Seien entsprechend f"ur~\mbox{$c \!=\! +1$, $-1$} und~$0$ definiert die Hyperboloide:
%
\begin{alignat}{2} \label{APP:Hyperboloide-h_pm1,0}
&h\idx{+}(x^3)&\:
  &:=\; \pm\, \sqrt{+1 + (x^3)^2}
    \\[.5ex]
&h\idx{-}(x^3)&\:
  &:=\; \pm\, \sqrt{-1 + (x^3)^2}
    \tag{\ref{APP:Hyperboloide-h_pm1,0}$'$} \\[.5ex]
&h\idx{0}(x^3)&\:
  &:=\; \pm\,  \sqrt{0 + (x^3)^2}\;
  =\; \pm\, \big|x^3\big|
    \tag{\ref{APP:Hyperboloide-h_pm1,0}$''$}
\end{alignat}
Dann liegen die von~$(0,0)$ mit "`Abstand Eins"' separierten Weltpunkte~$x\!/\!\sqrt{|c|}$ auf~$h\idx{+}$ bei zeitartiger, auf~$h\idx{-}$ bei raumartiger Separation~[$c \!>\! 0$ bzw.~$c \!<\! 0$]; die mit "`Abstand Null"'~-- lichtartig~-- separierten Weltpunkte liegen auf~$h\idx{0}$~[$c \!=\! 0$].
Die Einsen der $x^{0'}$- und $x^{3'}$-Achsen sind folglich gegeben als deren Schnittpunkte mit~$h\idx{+}$ beziehungsweise~$h\idx{-}$.
Konsequenz sind {\it Zeitdilatation\/} und {\it L"angenkontraktion\/} in einem relativ bewegten Bezugsystem, vgl.\@ Abb.~\ref{Fig:MinkowskiGENERAL}:
Die Eins-Abschnitte von~$x^{0'}$ und $x^{3'}$ in~$I'$ sind l"anger als Eins in~$I$ f"ur alle~\mbox{$\be \!>\! 0$}~[vgl.\@ die vertikale bzw.\@ horizontale {\it eng\/} versus entsprechende {\it weit\/} gepunktete Strecke im Sinne achsenparalleler Projektionen].~--
Dabei hei"st L"angenmessung {\it per~definitionem\/} Bestimmung der Endpunke {\it zu derselben Zeit\/}, das hei"st:~\mbox{$x^0 \!=\! const.$} in~$I$ und~\mbox{$x^{0'} \!=\! const.$} in~$I'$.
\vspace*{-.5ex}

\bigskip\noindent
Neben der graphischen Darstellung von Lorentz-Boosts im Minkowski-Diagramm ist von Vorteil ihre Formulierung in Lichtkegelkoordinaten.
Gl.~(\ref{APP:LT}) schreibt sich "aquivalent:
%
\begin{align} \label{APP:LT-bar}
x^{\bar{\mu}'}\;
  =\; \La^{\bar\mu}{}_{\bar\nu}\; x^{\bar\nu}\qquad
  \bar{\mu},\, \bar{\nu} \in \{+,-,1,2\}
\end{align}
Die Komponenten der Matrix~\mbox{$\bar{\La} \!\equiv\! \big(\La^{\bar\mu}{}_{\bar\nu}\big)$}, \mbox{$\bar{\mu},\bar{\nu} \!\in\! \{+,-,1,2\}$}, werden berechnet wie folgt:
\vspace*{-.25ex}
\begin{align} \label{APP:LT-La-bar0}
x^{\pm'}\;
  &=\; \al\, \big(x^{0'} \pm x^{3'}\big)\;
   =\; \al\, \big( \ga(x^0 - \be\, x^3)\;
             \pm\; \ga(-\be\, x^0 + x^3) \big)
    \\[.25ex]
  &=\; \ga(1 \mp \be)\cdot \al\, \big(x^0 \pm x^3\big)
    \tag{\ref{APP:LT-La-bar0}$'$} \\[.25ex]
  &=\; \ga(1 \mp \be)\cdot x^\pm\;
   =\; \exp\mp\ps\cdot x^\pm
    \tag{\ref{APP:LT-La-bar0}$''$}
    \\[-4.25ex]\nn
\end{align}
die letzte Identi"at aufgrund~\mbox{$\ga(1 \!\mp\! \be) \!=\! \big[(1 \!+\! \be)\!/\!(1 \!-\! \be)\big]{}^{\mp1\!/\!2}$} und Definition:
%
\begin{align} \label{APP:ps-hyperbolWinkel}
&\ps\;
  :=\; \frac{1}{2}\; \ln\, \frac{1 \!+\! \be}{1 \!-\! \be}\;
   =\; {\rm artanh}\, \be
    \\
&\text{d.h.}\qquad
\be\;
    =\; \tanh\, \ps\quad
    =\; \iIM\, \tan -{\iIM}\ps
    \tag{\ref{APP:ps-hyperbolWinkel}$'$}
\end{align}
mit~\mbox{$\ps \!\in\! [0,+\infty)$} f"ur~\mbox{$\be \!\in\! [0,1)$}.
In Hinblick auf Gl.~(\ref{APP:al_be}) wird~$\ps$ bezeichnet als der~{\it hyper\-bolische Winkel\/} des Lorentz-Boosts.
Aus\;~$\be \!=\! \tanh\ps$ folgt, vgl.\@ Gl.~(\ref{APP:gamma,beta}):
\begin{samepage}
\vspace*{-.25ex}
\begin{alignat}{3}
&\ga&\;
  &=\; 1\!/\sqrt{1 \!-\! \be^2}\;
   =\; [1 \!-\! \tanh^2\ps]^{-1\!/\!2}&\;
  &=\; \cosh\ps
    \label{APP:coshps}
    \\[-5.25ex]\nn
\intertext{\vspace*{-1.25ex}folglich:}
&\ga\, \be&\;
  &=\; \cosh\ps\, \tanh\ps&\;
  &=\; \sinh\ps
    \label{APP:sinhps}
    \\[-4.25ex]\nn
\end{alignat}
Zusammenfassend gilt:
\vspace*{-.5ex}
\begin{align} \label{APP:LT-Labar3}
\bar{\La}\;
  \equiv\; \big(\La^{\bar\mu}{}_{\bar\nu}\big)\;
  =\; \pmatrixZZ{\ga(1 - \be)}{0}{0}{\ga(1 + \be)}\;
  =\; \pmatrixZZ{\exp-\ps}{0}{0}{\exp+\ps}
    \\[-4.5ex]\nn
\end{align}
Die Matrix~\mbox{$\bar{\La} \!\equiv\! \big(\La^{\bar\mu}{}_{\bar\nu}\big)$} ist diagonal~-- insofern Lorentz-Boosts besonders einfach in Lichtkegelkoordinaten.
Unter Umkehrung der Boost-Richtung,~$\be \!\to\! -\be$, vertauschen~$x^+$ und~$x^-$ ihre Rollen.
Mit~\mbox{$\det\bar{\La} \!=\! 0$} f"ur~\mbox{$\pm\be \!=\! 1$} ist dokumentiert die Singularit"at des Grenzwerts.
\end{samepage}

\section{Aktive versus passive Transformation}
\label{APP-Sect:aktiv-passiv}

Der soweit diskutierten {\it passiven\/} Auffassung von Lorentz-Transformationen steht gegen"uber die {\it aktive\/} Auffassung, die unserer Diskussion im Haupttext zugrunde liegt.
Wir konfrontieren beide Auffassungen, beziehen uns dabei weiterhin o.E.d.A.\@ auf Vierer-Vektoren.

Sei~$\{e_{(\mu)}\}$ eine orthogonale Basis des Minkowski-Raumes, das hei"st ihre Vektoren normiert gem"a"s:
\vspace*{-.5ex}
\begin{align} \label{APP:MinkowskiBasis}
e_{(\mu)}\cdot e_{(\nu)}\;
  =\; g_{\mu\nu}\qquad
  \mu,\,\nu \in \{0,1,2,3\}
    \\[-4.5ex]\nn
\end{align}
mit~$\big(g_{\mu\nu}\big) \!=\!  {\rm diag}[+1,-1,-1,-1]$ dem metrischen Tensor, vgl.\@ die Gln.~(\ref{APP:g}),~(\ref{APP:e_(mu)-Norm}).
\vspace*{-.5ex}

\paragraph{\label{APP-Para:passiv}Passive Transformation}
hei"st~-- wir rekapitulieren Gl.~(\ref{APP:LT})~--, ein und denselben Vektor~$x$ zu betrachten bez"uglich zweier Inertialsysteme~$I$ und~$I'$.
Bez"uglich deren Basen~$\{e_{(\mu)}\}$ und~$\{e'_{(\mu)}\}$ besitze er Komponenten~$x^\mu$ beziehungsweise~$x^{\mu'}$, wir schreiben: \mbox{$x \!\equiv\! (x^\mu) \!\equiv\! (x^{\bar\mu})$}.
Es gilt:
\vspace*{-1.25ex}
\begin{align} \label{APP:LT-passiv}
x\;
  =\; x^\mu\, e_{(\mu)}\vv
  \stackrel{\D!}{=}\; x^{\mu'}\, e'_{(\mu)}\;
  =\; \La^\mu{}_\nu\, x^\nu\, e'_{(\mu)}
     \\[-4.5ex]\nn
\end{align}
mit der letzten Identit"at aufgrund~$x^{\mu'} \!=\! \La^\mu{}_\nu\, x^\nu$, vgl.\@ Gl.~(\ref{APP:LT}).
Folglich:
\vspace*{-.5ex}
\begin{align} \label{APP:e_e'}
e_{(\nu)}\;
  =\; \La^\mu{}_\nu\, e'_{(\mu)}
    \\[-4.5ex]\nn
\end{align}
Sei die Matrix~$\big(\La_\mu{}^\nu\big)$ definiert durch die hierzu inverse Relation:
\vspace*{-.5ex}
\begin{align} \label{APP:e'_e}
e'_{(\mu)}\;
  =:\; \La_\mu{}^\nu\, e_{(\nu)}
    \\[-4.5ex]\nn
\end{align}
Aus den Gln.~(\ref{APP:e_e'}),~(\ref{APP:e'_e}) folgt unmittelbar:
\vspace*{-.5ex}
\begin{align} \label{APP:La-kontragredient}
\La^\mu{}_\rh\, \La_\nu{}^\rh\;
  =\; \La_\rh{}^\mu\, \La^\rh{}_\nu\;
  =\; \de^\mu_\nu
    \\[-4.5ex]\nn
\end{align}
das hei"st die Matrizen~$\big(\La^\mu{}_\nu\big)$ und~$\big(\La_\mu{}^\nu\big)$ sind kontragredient~-- die eine die transponierte Inverse der anderen.
\vspace*{-.5ex}

\paragraph{\label{APP-Para:aktiv}Aktive Transformation}
hei"st die Abbildung des gesamten Minkow\-ski-Raums in sich unter Erhaltung des Vierer-Skalarprodukts:%
\FOOT{
  In Gl.~(\ref{APP:LT-passiv}) "andern sich die Komponenten desselben Vektors, in Gl.~(\ref{APP:LT-aktiv}) der Vektor selbst; entsprechend steht der Strich zur Notation.
}
%
\vspace*{-.5ex}
\begin{alignat}{3}
&x&\; &\longrightarrow\; x'&\;
  &=\; \La\; x\qquad\qquad
  \text{unter}\qquad
  x'{}^2\;
    =\; x^2
    \label{APP:LT-aktiv}
    \\[-6ex]\nn
\intertext{\vspace*{-2ex}Dies impliziert die Abbildung der Vektoren der Basis~$\{e_{(\mu)}\}$ wie folgt:}
&e_{(\mu)}&\; &\longrightarrow\; e'_{(\mu)}&\;
  &=\; \La\; e_{(\mu)}
    \label{APP:e'=La-e}
    \\[-4.5ex]\nn
\end{alignat}
mit~$e'_{(\mu)}\cdot e'_{(\nu)} \!=\! e_{(\mu)}\cdot e_{(\nu)} \!=\! g_{\mu\nu}$, vgl.\@ Gl.~(\ref{APP:MinkowskiBasis}):
Die so-definierten~$e'_{(\mu)}$ bilden zum einen eine Basis~$\{e'_{(\mu)}\}$ und besitzen zum anderen eine Zerlegung bez"uglich der alten Basis:
\begin{samepage}
\vspace*{-.5ex}
\begin{align} \label{APP:e'-nach-e}
e'_{(\mu)}\;
  =\; \La_\mu{}^\nu\, e_{(\nu)}
    \\[-4.5ex]\nn
\end{align}
vgl.\@ Gl.~(\ref{APP:e'_e}).
Folglich gilt:
\vspace*{-.5ex}
\begin{align} 
x'\;
  =\; \La\, x\;
  =\; \La\; x^\mu\, e_{(\mu)}\;
  =\; x^\mu\; \La\, e_{(\mu)}\;
  =\; x^\mu\; \La_\mu{}^\nu\, e_{(\nu)}\vv
  \stackrel{\D!}{=}\; x'{}^\mu\; e_{(\mu)}
    \\[-4.5ex]\nn
\end{align}
mit der vorletzten Identit"at aufgrund der Gln.~(\ref{APP:e'=La-e}),~(\ref{APP:e'-nach-e}).
F"ur die Komponenten von~$x'$ bez"uglich der alten Basis~$\{e_{(\mu)}\}$ folgt daher:
\vspace*{-.5ex}
\begin{align} \label{APP:x'_x}
x'{}^\mu\;
  =\; \La_\nu{}^\mu\, x^\nu
    \\[-4.5ex]\nn
\end{align}
\end{samepage}
und invertiert:
\vspace*{-.5ex}
\begin{align} \label{APP:x'_x-inv}
x^\mu\;
  =\; \La^\mu{}_\nu\, x'{}^\nu
    \\[-4ex]\nn
\end{align}
Gl.~(\ref{APP:x'_x}) steht in direktem Kontrast zu Gl.~(\ref{APP:LT}):
Die {\it passive\/} Transformation wird vermittelt durch~$\La \!\equiv\! \big(\La^\mu{}_\nu\big)$, die {\it aktive\/} Transformation durch die transponierte inverse Matrix:~$\La^{\!-1\T\:t} \!\equiv\! \big(\La^\mu{}_\nu\big)$.
De~facto dreht die Boost-Geschwindigkeit gerade ihr Vorzeichen um.
Es gilt, vgl.\@ Gl.~(\ref{APP:LT-La3}) und die Gln.~(\ref{APP:coshps}),~(\ref{APP:sinhps}):
%
\begin{align} \label{APP:LT-La-tInv-3}
&x'{}^\mu\;
  =\; \La_\nu{}^\mu\; x^\nu
    \\[-.25ex]
&\text{mit}\qquad
  \big(\La_\mu{}^\nu\big)
  = \pmatrixZZ{\ga}{+\ga\, \be}{+\ga\, \be}{\ga}
  = \pmatrixZZ{\cosh\ps}{\sinh\ps}{\sinh\ps}{\cosh\ps}\vv
  \equiv\vv \La^{\!-1\T\:t}
    \tag{\ref{APP:LT-La-tInv-3}$'$}
    \\[-4.5ex]\nn
\end{align}
"aquivalent in Lichtkegelkoordinaten, vgl.\@ Gl.~(\ref{APP:LT-Labar3}):
%
\begin{align} \label{APP:LT-Labar-tInv-3}
&x'{}^{\bar\mu}\;
  =\; \La_{\bar\nu}{}^{\bar\mu}\; x^{\bar\nu}
    \\[-.25ex]
&\text{mit}\qquad
  \big(\La_{\bar\mu}{}^{\bar\nu}\big)
  = \pmatrixZZ{\exp+\ps}{0}{0}{\exp-\ps}\vv
  \equiv\vv \bar\La^{\!-1\T\:t}
    \tag{\ref{APP:LT-Labar-tInv-3}$'$}
    \\[-4.5ex]\nn
\end{align}
mit Definition\;~$\tanh\ps \!:=\! \be$\; nach Gl.~(\ref{APP:ps-hyperbolWinkel}).

%
\section[Streuung im Bild aktiver Lorentz-Boosts]{%
         Streuung im Bild aktiver Lorentz-Boosts}
\label{APP-Sect:Streuung_aktiveBoosts}

\begin{figure}
\begin{minipage}{\linewidth}
  \begin{center}
  \setlength{\unitlength}{1mm}\begin{picture}(100,100) 
    \put(0,0){\epsfxsize100mm \epsffile{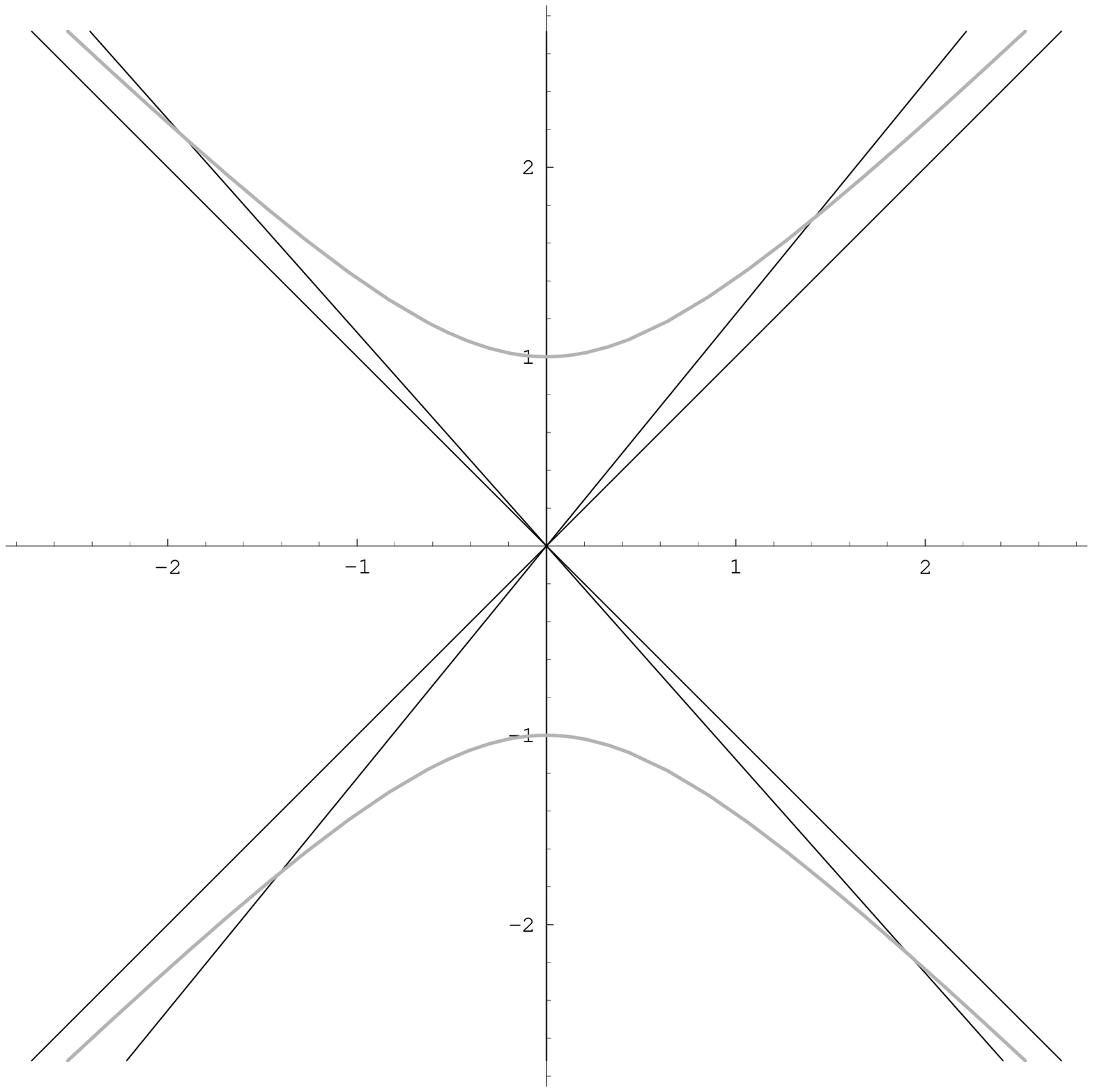}}
    \put(44,94){\normalsize$x^0$}
    \put(94,44){\normalsize$x^3$}
    \put(32,77){\normalsize$h\idx{+}$}
    \put(62,23){\normalsize$h\idx{+}$}
    \put( 3,86.5){\normalsize$\subset\! h\idx{0}$}
    \put(89,81.5){\normalsize$\subset\! h\idx{0}$}
    \put( 3,90){\normalsize${\cal C}\idx{-}$}
    \put(17,90){\normalsize${\cal C}\Dmfm$}
    \put(52,90){\normalsize${\cal C}\idx[\mfm]{0}$}
    \put(52,85){\normalsize$\equiv\! {\cal C}\idx[\mfp]{0}$}
    \put(73,85){\normalsize${\cal C}\Dmfp$}
    \put(89,85){\normalsize${\cal C}\idx{+}$}
    \put(16.35,88){\circle*{.8}}
    \put(50,67.5){\circle*{.8}}
    \put(74.7,80.3){\circle*{.8}}
    \put(25.25,19.75){\circle*{.8}}
    \put(50,32.5){\circle*{.8}}
    \put(83.35,12.2){\circle*{.8}}
  \end{picture}
  \end{center}
\vspace*{-4.5ex}
\caption[Minkowski-Diagramm: Weltlinien unter aktiven Lorentz-Boosts]{
  Minkowski-Diagramm: Weltlinien unter aktiven Lorentz-Boosts~-- head-to-head.   Teilchen~$\mfp$ und~$\mfm$ seien zun"achst gemeinsam in Ruhe bzgl.~\mbox{$I\idx{0} \!\equiv\! I\idx[\mfp]{0} \!\equiv\! I\idx[\mfm]{0}$}: ihre initialen Weltlinien~${\cal C}\idx[\mfp]{0}$,~${\cal C}\idx[\mfm]{0}$ liegen auf der Achse~$x^0$ von~$I\idx{0}$ [vgl.\@ Identit"aten~$(\ast)$].   Werde Teilchen~$\mfp$ aktiv geboostet mit~$\be\Dmfp$, Teilchen~$\mfm$ mit~$-\be\Dmfm$ in $x^3$-Richtung;~$\be\Dmfp, \be\Dmfm \!\in\! [0,1)$.   Dann ruhen sie in Inertialsystemen~$I\Dmfp$,~$I\Dmfm$, die sich aufeinander zu bewegen mit Geschwindigkeit~$\be\Dmfp$ in positive bzw.\@ mit~$-\be\Dmfm$ in negative $x^3$-Richtung relativ zu~$I\idx{0}$.   Ihre Weltlinien~${\cal C}\Dmfp$,~${\cal C}\Dmfm$ fallen zusammen mit den Zeitachsen von~$I\Dmfp$,~$I\Dmfm$.   F"ur~$\be\Dmfp,\, \be\Dmfm \!\to\! 1$ gehen diese "uber in~${\cal C}\idx{+}$,~${\cal C}\idx{-}$, die auf dem Lichtkegel~$h\idx{0}$ liegen; ihre L"angen~[vgl.\@ die Abschnitte zwischen den~Schnittpunkten mit~$h\idx{+}$] wachsen gemessen in~$I\idx{0}$ im Sinne der Zeitdilatation an gegen Unendlich.
}
\label{Fig:MinkowskiTRAJECTORY}
\end{minipage}
\end{figure}

Wir formulieren Streuung auf Basis aktiver Lorentz-Boosts.
Dies f"uhrt auf Koordinaten, die formal vereinfachen und interpretatorisch suggestiv sind. \\
\indent
Das zugrundeliegende Bild sei zun"achst grob umrissen. 
Wir betrachten zwei Teilchen~$\mfp$ und~$\mfm$~-- "`pus"' und~"`minus"'~-- unter simultanen aktiven Lorentz-Boosts head-to-head.
Das hei"st sie ruhen in Inertialsystemen~$I\Dmfp$ und~$I\Dmfm$, die sich mit Geschwindigkeit~$\be\Dmfp$ in positive beziehungsweise mit~$\be\Dmfm$ in negative $x^3$-Richtung aufeinander zu bewegen:%
~\mbox{Ihre Weltlinien} ${\cal C}\Dmfp$,~${\cal C}\Dmfm$ liegen auf deren Zeitachsen.%
\FOOT{
  \label{APP-FN:ausRuhe}Wir betrachten physikalische Teilchen mit nichtverschwindenden Massequadraten:~$M\Dmfp^2,M\Dmfm^2 \!>\! 0$, deren Weltlinien folglich innerhalb des Lichtkegels liegen, so da"s garantiert ist die Existenz (momentaner) inertialer Ruhsysteme, wie sie unsere Konstruktion voraussetzt.
}
Die invariante Schwerpunktenergie der Streuung~$\surd s$ ist Funktion der Boost-Parameter~$\be\Dmfp$,~$\be\Dmfm$; dabei gilt~$\surd s \!\to\! \infty$, falls~\mbox{$\be\Dmfp \!\to\! 1$ und/oder~$\be\Dmfm \!\to\! 1$}.
Es folgt~$\be\Dmfm$ als Funktion von~$\be\Dmfp$ durch die Forderung, da"s der Schwerpunkt beider Teilchen ruht in einem Inertialsystem~$I\idx{0}$~-- vor wie nach den Boosts.
Seien~$I\idx[\mfp]{0}$ und~$I\idx[\mfm]{0}$ die entsprechenden Ruhsysteme vor den Boosts.
Dann nehmen wir o.E.d.A.\@ an:~\mbox{$I\idx{0} \!\equiv\! I\idx[\mfp]{0} \!\equiv\! I\idx[\mfm]{0}$}, das hei"st, da"s die Boosts erfolgen "`aus der gemeinsamen Ruhe bzgl.\@ $x^3$"', das hei"st die longitudinalen Projektionen ihrer initialen Weltlinien~${\cal C}\idx[\mfp]{0}$,~${\cal C}\idx[\mfm]{0}$ liegen gemeinsam auf der Zeitachse von~$I\idx{0}$.
Vgl.\@ Abb.~\ref{Fig:MinkowskiTRAJECTORY}.
Wir formalisieren dieses Bild wie folgt.
\vspace*{-.5ex}

\bigskip\noindent
Seien~$P\Dimath \!\equiv\! (P\Dimath^\mu)$ mit~$\imath \!=\! \mfp,\mfm$ die Vierer-Impulse der Teilchen in den urspr"unglichen Inertialsystemen~$I\idx[\mfp]{0}$,~$I\idx[\mfm]{0}$:
\begin{samepage}
\vspace*{-1ex}
\begin{align} \label{APP:P_i}
&P\Dmfp\!
  \equiv\! (P\Dmfp^\mu)
  = \begin{pmatrix}P\Dmfp^0\\ \rb{P}\\ P^3
        \end{pmatrix}\vv
  \stackrel{({\D\ast})}{=}\vv
      \begin{pmatrix}P\Dmfp^0\\ \rb{P}\\ 0
        \end{pmatrix}\qquad
P\Dmfm\!
  \equiv\! (P\Dmfm^\mu)
  = \begin{pmatrix}P\Dmfm^0\\ -\rb{P}\\ -P^3
        \end{pmatrix}\vv
  \stackrel{({\D\ast})}{=}\vv
      \begin{pmatrix}P\Dmfm^0\\ -\rb{P}\\ 0
        \end{pmatrix}
    \\[-.875ex]
&\text{mit}\qquad
  P\Dimath^0
  = \sqrt{M\Dimath^2 \!+\! \rb{P}^2 \!+\!(P^3)^2}\vv
  \stackrel{({\D\ast})}{=}\vv
    \sqrt{M\Dimath^2 \!+\! \rb{P}^2}\qquad
  M\Dimath^2 > 0\qquad
  \imath = \mfp,\mfm
    \label{APP:P_i^0}
    \\[-4.5ex]\nn
\end{align}
"aquivalent~$\bar{P}\Dimath \!\equiv\! (P\Dmfp^{\bar\mu})$ in Lichtkegelkoordinaten:
\vspace*{-.5ex}
\begin{align} \label{APP:Pbar_i}
&\bar{P}\Dmfp\;
  \equiv\; (P\Dmfp^{\bar\mu})\;
  =\; \begin{pmatrix}P\Dmfp^+\\ P\Dmfp^-\\ \rb{P}
        \end{pmatrix}\qquad
  \text{und}\qquad
\bar{P}\Dmfm\;
  \equiv\; (P\Dmfm^{\bar\mu})\;
  =\; \begin{pmatrix}P\Dmfm^+\\ P\Dmfm^-\\ -\rb{P}
        \end{pmatrix}
    \\[-.875ex]
&\text{mit}\qquad
  P\Dimath^\pm
    = \al\, (P\Dimath^0 \pm P\Dimath^3)\vv
    \stackrel{({\D\ast})}{=}\vv \al\, P\Dimath^0\qquad
  \imath = \mfp,\mfm
    \label{APP:Pbar_i^pm-Def}
    \\[-4ex]\nn
\end{align}
vgl.\@  Gl.~(\ref{APP:kontravLC}) und~(\ref{APP:kontravLC_Def}),~(\ref{APP:kontravLC_Def}$'$).
Dabei gelten die Identit"aten~$(\ast)$ aufgrund Identifizierung der Systeme:~\mbox{$I\idx{0} \!\equiv\! I\idx[\mfp]{0} \!\equiv\! I\idx[\mfm]{0}$}, im Sinne eines {\it gemeinsamen\/} initialen Ruhsystems.
\end{samepage}
\\
\indent
Es gilt f"ur~\mbox{$s\idx{0} \!=\! (P\Dmfp \!+\! P\Dmfm)^2$}, das Quadrat der initialen invarianten Schwerpunktenergie:
%
\begin{align} \label{APP:s0_Pimath}
s\idx{0}\;
  &=\;
    M\Dmfp^2 + M\Dmfm^2
    + 2\, \big\{P\Dmfp^0 P\Dmfm^0 + \rb{P}^2 + (P^3)^2\big\}
    \\[.5ex]
  &=\;
    M\Dmfp^2 + M\Dmfm^2
      + 2\,
        \big\{g_{+-}\, \big[P\Dmfp^+ P\Dmfm^- \!+\! P\Dmfp^- P\Dmfm^+\big]
               + \rb{P}^2\big\}
    \tag{\ref{APP:s0_Pimath}$'$}
\end{align}
mit~$g_{+-} \!=\! 1\!/2\al^2$ nach Gl.~(\ref{APP:gbar}). \\
\indent
Die aktiv transformierten Vierer-Impulse~$P'\Dimath \!\equiv\! (P\Dimath^{\prime\nu}$) und~$\bar{P}'\Dimath \!\equiv\! (P\Dimath^{\prime\bar\nu})$ f"ur~$\imath \!=\! \mfp,\mfm$ folgen mithilfe der Gln.~(\ref{APP:LT-La-tInv-3}),~(\ref{APP:LT-La-tInv-3}$'$) bzw.~(\ref{APP:LT-Labar-tInv-3}),~(\ref{APP:LT-Labar-tInv-3}$'$):
%
\begin{align} \label{APP:P'_i}
&P\Dimath^{\prime\mu}\;
  =\; (\La\Dimath)_\nu{}^\mu\; P\Dimath^\nu\qquad\qquad
        \mu,\nu \in \{0,1,2,3\}\qquad
        \imath = \mfp, \mfm
    \\[.75ex]
&\text{mit}\qquad
  \begin{alignedat}[t]{3}
    &\big((\La\Dmfp)_\mu{}^\nu\big)&\;
      &=\; \pmatrixZZ{\ga\Dmfp}{+\ga\Dmfp\, \be\Dmfp}{+\ga\Dmfp\, \be\Dmfp}{\ga\Dmfp}&\;
      &=\; \pmatrixZZ{\cosh\ps\Dmfp}{+\sinh\ps\Dmfp}{+\sinh\ps\Dmfp}{\cosh\ps\Dmfp}
    \\[.5ex]
    &\big((\La\Dmfm)_\mu{}^\nu\big)&\;
      &=\; \pmatrixZZ{\ga\Dmfm}{-\ga\Dmfm\, \be\Dmfm}{-\ga\Dmfm\, \be\Dmfm}{\ga\Dmfm}&\;
      &=\; \pmatrixZZ{\cosh\ps\Dmfm}{-\sinh\ps\Dmfm}{-\sinh\ps\Dmfm}{\cosh\ps\Dmfm}
  \end{alignedat}
    \tag{\ref{APP:P'_i}$'$}
\end{align}
und
\vspace*{-.5ex}
\begin{align} \label{APP:gamma,beta_imath}
\ga\Dimath\;
  \equiv\; \ga(\be\Dimath)\;
  =\; 1\!/\sqrt{1 \!-\! \be\Dimath^2}\qquad
        \imath = \mfp, \mfm
    \\[-4.5ex]\nn
\end{align}
vgl.\@ Gl.~(\ref{APP:gamma,beta})~-- und "aquivalent in Lichtkegelkoordinaten:
\vspace*{-.5ex}
\begin{align} \label{APP:Pbar'_i}
&P\Dimath^{\prime\bar\mu}\;
  =\; (\La\Dimath)_{\bar\nu}{}^{\bar\mu}\; P\Dimath^{\bar\nu}\qquad\qquad
        \bar{\mu},\bar{\nu} \in \{+,-,1,2\}\qquad
        \imath = \mfp, \mfm
    \\[.5ex]
&\text{mit}\qquad
  \begin{alignedat}[t]{2}
  &\big((\La\Dmfp)_{\bar\mu}{}^{\bar\nu}\big)&\;
    &=\; \pmatrixZZ{\exp+\ps\Dmfp}{0}{0}{\exp-\ps\Dmfp}
    \\
  &\big((\La\Dmfm)_{\bar\mu}{}^{\bar\nu}\big)&\;
    &=\; \pmatrixZZ{\exp-\ps\Dmfm}{0}{0}{\exp+\ps\Dmfm}
  \end{alignedat}
    \tag{\ref{APP:Pbar'_i}$'$}
    \\[-4.5ex]\nn
\end{align}
und Definition
\vspace*{-.5ex}
\begin{align} \label{APP:psp,psm-Def}
\tanh\ps\Dimath\;
  =\; \be\Dimath\qquad
  \imath = \mfp, \mfm
    \\[-4.5ex]\nn
\end{align}
so da"s~$\ps\Dimath \in [0,+\infty)$ wegen~$\be\Dimath \!\in\! [0,1)$ f"ur~$\imath \!=\! \mfp, \mfm$ {\it per~definitionem\/}, vgl.\@ Gl.~(\ref{APP:ps-hyperbolWinkel}). \\
\indent
Es gilt f"ur~$s \!=\! (P'\Dmfp + P'\Dmfm)^2$, das Quadrat der invarianten Schwerpunktenergie nach der aktiven Transformation:
\vspace*{-.5ex}
\begin{align} \label{APP:s_Pimath}
s\;
&=\; 
  \begin{aligned}[t]
  s\idx{0}\; +\; 2\, \Big\{
      \big(\ga\Dmfp\ga\Dmfm\, (1 \!+\! \be\Dmfp\be\Dmfm) \!-\! 1\big)\;
      &\big[P\Dmfp^0 P\Dmfm^0 + (P^3)^2\big]
    \\
     +\; \ga\Dmfp\ga\Dmfm\, (\be\Dmfp \!+\! \be\Dmfm)\;
      &\big[P\Dmfp^0 + P\Dmfm^0\big] P^3
    \Big\}
  \end{aligned}
    \\[.5ex]
&=\; 
  s\idx{0}\; +\; \frac{1}{\al^2}\, \Big\{
    \big(\efn{\D\ps} \!-\! 1)\; P\Dmfp^+ P\Dmfm^-\;
    +\; \big(\efn{\D-\ps} \!-\! 1\big)\; P\Dmfp^- P\Dmfm^+
    \Big\}
    \tag{\ref{APP:s_Pimath}$'$} \\[.5ex]
&=\; 
  \begin{aligned}[t]
  s\idx{0}\; +\; \frac{1}{\al^2}\, \Big\{
      \big(\cosh\ps - 1\big)\;
      &\big[P\Dmfp^+ P\Dmfm^- + P\Dmfp^- P\Dmfm^+\big]
    \\
     +\; \sinh\ps\;
      &\big[P\Dmfp^+ P\Dmfm^- - P\Dmfp^- P\Dmfm^+\big] P^3
    \Big\}
  \end{aligned}
    \tag{\ref{APP:s_Pimath}$''$}
    \\[-4.5ex]\nn
\end{align}
unter Definition
\vspace*{-.5ex}
\begin{align} \label{APP:ps=psp+psm}
\ps\;
  :=\; \ps\Dmfp + \ps\Dmfm
    \\[-4.5ex]\nn
\end{align}
mit~$\ps \in [0,+\infty)$ nach Gl.~(\ref{APP:psp,psm-Def}); bzgl.~$s\idx{0}$ vgl.\@ die Gln.~(\ref{APP:s0_Pimath}),~(\ref{APP:s0_Pimath}$'$).
Unter Identifizierung der initialen Systeme:~\mbox{$I\idx{0} \!\equiv\! I\idx[\mfp]{0} \!\equiv\! I\idx[\mfm]{0}$}, das hei"st Boost aus dem gemeinsamen Ruhsystem~-- die Identit"aten~$(\ast)$~--, vereinfacht sich Gl.~(\ref{APP:s_Pimath}) und~(\ref{APP:s_Pimath}$''$) respektive wie folgt:
%
\begin{align} \label{APP:s*_Pimath}
s\;
&=\; 
  s\idx{0}\; +\;
      2\, \big(\ga\Dmfp\ga\Dmfm\, (1 \!+\! \be\Dmfp\be\Dmfm) \!-\! 1\big)\;
      P\Dmfp^0 P\Dmfm^0
    \\[.5ex]
&=\; 
  s\idx{0}\; +\;
      2\, \big(\cosh\ps - 1\big)\;
      P\Dmfp^0 P\Dmfm^0
    \tag{\ref{APP:s*_Pimath}$'$}
    \\[-4.5ex]\nn
\end{align}
mit~$s\idx{0} \!=\! (P\Dmfp^0 \!+\! P\Dmfm^0)^2$ und~\mbox{$P\Dimath^0 \!=\! \sqrt{M\Dimath^2 \!+\! \rb{P}^2}$}.
[Gesetzt~\mbox{$\be\Dmfp \!\stackrel{\D!}{=}\! 0$ und~$\be\Dmfm \!\stackrel{\D!}{=}\! 0$}, folgt wieder~$s \!=\! s\idx{0}$.] \\
\indent
Der Schwerpunkt des Systems beider Teilchen ruhe in~$I\idx{0}$ {\it nach wie vor\/} den Boosts, das hei"st sie geschehen in dieser Weise {\it simultan\/}.
Dadurch ist~$\be\Dmfm$ determiniert als Funktion von~$\be\Dmfp$.
Wir beschr"anken uns darauf den Zusammnehang zu bestimmen unter Identifizierung der initialen Systeme:~\mbox{$I\idx{0} \!\equiv\! I\idx[\mfp]{0} \!\equiv\! I\idx[\mfm]{0}$}; auf Basis der Identit"aten~$(\ast)$ wird gefordert:
\vspace*{-.5ex}
\begin{align} \label{APP:CMS-Forderung}
&P'\Dmfp{}^3\;
  + P'\Dmfm{}^3\;
  \stackrel{\D!}{=}\; 0
    \\
&\text{d.h.}\qquad
  \ga\Dmfp\, \be\Dmfp\; P\Dmfp^0\;
  +\; \ga\Dmfm\, \be\Dmfm\; P\Dmfm^0\;
  \stackrel{\D!}{=}\; 0
    \tag{\ref{APP:CMS-Forderung}$'$}
    \\[-4ex]\nn
\end{align}
vgl.\@ die Gln.~(\ref{APP:P'_i}),~(\ref{APP:P'_i}$'$) und~(\ref{APP:gamma,beta_imath}).
Daraus folgt unmittelbar:
\begin{samepage}
\vspace*{-.5ex}
\begin{align} \label{APP:be2-mfm,1-be2-mfm}
\be\Dmfm^2\;
  =\; \frac{(1 \!-\! \be\Dmfp^2)}{%
            (1 \!-\! \be\Dmfp^2) + \mu\, \be\Dmfp^2}\qquad
  \text{und}\qquad
1 - \be\Dmfm^2\;
  =\; \frac{\mu\, \be\Dmfp^2}{%
            (1 \!-\! \be\Dmfp^2) + \mu\, \be\Dmfp^2}
    \\[-4.5ex]\nn
\end{align}
mit
\end{samepage}
\vspace*{-.5ex}
\begin{align} \label{APP:mu_P0mfp,P0mfm}
\mu\;
  :=\; (P\Dmfp^0)^2\! \big/ (P\Dmfm^0)^2
    \\[-4ex]\nn
\end{align}
Und daraus:
\vspace*{-.5ex}
\begin{align} \label{APP:be-mfm,ga-mfm}
\be\Dmfm\;
  &=\; \be\Dmfp\cdot
      \frac{\sqrt\mu}{%
            \sqrt{(1 \!-\! \be\Dmfp^2) + \mu\, \be\Dmfp^2}}
    \\[.5ex]
\ga\Dmfm\;
  &=\; \ga\Dmfp\cdot
      \sqrt{(1 \!-\! \be\Dmfp^2) + \mu\, \be\Dmfp^2}
    \tag{\ref{APP:be-mfm,ga-mfm}$'$}
    \\[-4.5ex]\nn
\end{align}
Aus\;~$\tanh\ps\Dimath \!=\! \be\Dimath$ f"ur~$\imath \!=\! \mfp, \mfm$, vgl.\@ Gl.~(\ref{APP:psp,psm-Def}), folgt unmittelbar:
%
\begin{alignat}{3}
&\ga\Dimath&\;
  &=\; \cosh\ps\Dimath&&
    \label{APP:coshps-imath} \\[.25ex]
&\ga\Dimath\, \be\Dimath&\;
  &=\; \sinh\ps\Dimath&&\qquad
  \imath = \mfp,\mfm
    \label{APP:sinhps-imath}
\end{alignat}
vgl.\@ die Gln.~(\ref{APP:coshps}),~(\ref{APP:sinhps}), und die Gln.~(\ref{APP:be-mfm,ga-mfm}),~(\ref{APP:be-mfm,ga-mfm}$'$) nehmen die einfache Form an:
\vspace*{-1ex}
\begin{alignat}{3} \label{APP:be-mfm,ga-mfm_ps}
&\be\Dmfm&\;
  &=\; \frac{\sqrt\mu\, \sinh^2\ps\Dmfp}{\sqrt{1 + \mu\, \sinh^2\ps\Dmfp}}&\vv
  &\stackrel{\D!}{=}\vv \tanh\ps\Dmfm
    \\[-.5ex]
&\ga\Dmfm&\;
  &=\; \sqrt{1 + \mu\, \sinh^2\ps\Dmfp}&\vv
  &\stackrel{\D!}{=}\vv \cosh\ps\Dmfm
    \tag{\ref{APP:be-mfm,ga-mfm_ps}$'$}
    \\[-4.5ex]\nn
\end{alignat}
Aus dem Additionstheorem f"ur hyperbolische Funktionen folgt die Zerlegung:
\vspace*{-.5ex}
\begin{align} \label{APP:cosh-ps_ps-mfp}
\cosh\ps\;
  &=\; \cosh(\ps\Dmfp \!+\! \ps\Dmfm)
    \\[.5ex]
  &=\; \cosh\ps\Dmfp\, \cosh\ps\Dmfm\;
       +\; \sinh\ps\Dmfp\, \sinh\ps\Dmfm
    \tag{\ref{APP:cosh-ps_ps-mfp}$'$} \\[.5ex]
  &=\; \big(\cosh\ps\Dmfp\;
       +\; \sinh\ps\Dmfp\, \tanh\ps\Dmfm\big)\, \cosh\ps\Dmfm
    \tag{\ref{APP:cosh-ps_ps-mfp}$''$} \\
  &=\; \cosh\ps\Dmfp\cdot \sqrt{1 + \mu\,\sinh^2\ps\Dmfp}\;
       +\; \sinh\ps\Dmfp\cdot \sqrt{\mu}\,\sinh\ps\Dmfp
    \tag{\ref{APP:cosh-ps_ps-mfp}$'''$}
    \\[-4.5ex]\nn
\end{align}
mit der letzten Identit"at aufgrund der Gln.~(\ref{APP:be-mfm,ga-mfm_ps}),~(\ref{APP:be-mfm,ga-mfm_ps}$'$).

Der Zusammenhang zwischen~$s$ und~$\cosh\ps$ nach  Gl.~(\ref{APP:s*_Pimath}$'$) ist damit ein Zusammenhang zwischen~$s$ und~$\cosh\ps\Dmfp$.
Zusammengefassend gilt:%
\FOOT{
  \label{APP-FN:cosh2-1=sinh2}Wir dr"ucken diese Relationen aus durch nur eine hyperbolische Funktion, vgl.\@ die Gln.~(\ref{APP:cosh-ps_ps-mfp}$'''$),~(\ref{APP:s*_ps-mfp}$'$).
}
%
\vspace*{-.5ex}
\begin{align} \label{APP:s*_ps-mfp}
&s\;
  =\; s\idx{0}\; +\;
      2\, \big(\cosh\ps - 1\big)\;
      P\Dmfp^0 P\Dmfm^0
    \\[.5ex]
&\text{mit}\qquad
  \cosh\ps\;
  =\; \cosh\ps\Dmfp\cdot \sqrt{1 + \mu\,\big(\cosh^2\ps\Dmfp \!-\! 1\big)}\;
      +\; \sqrt\mu\,\big(\cosh^2\ps\Dmfp \!-\! 1\big)
    \tag{\ref{APP:s*_ps-mfp}$'$}
    \\[-4.5ex]\nn
\end{align}
und invertiert:\citeFN{APP-FN:cosh2-1=sinh2}
\vspace*{-.5ex}
\begin{align} \label{APP:ps-mfp_s*}
&\cosh\ps\Dmfp\;
  =\; \bigg[1\; -\; \frac{\cosh^2\ps - 1}{(\cosh\ps + \sqrt\mu)^2}\bigg]^{-1\!/\!2}
    \\[.5ex]
&\text{mit}\qquad
  \cosh\ps\;
  =\; 1\; +\; \frac{s - s\idx{0}}{2\, P\Dmfp^0 P\Dmfm^0}
    \tag{\ref{APP:ps-mfp_s*}$'$}
    \\[-4.5ex]\nn
\end{align}
Dabei ist~\mbox{$s\idx{0} \!=\! M\Dmfp^2 \!+\! M\Dmfm^2 \!+\! 2(P\Dmfp^0 P\Dmfm^0 \!+\! \rb{P}^2)$} und~\mbox{$\mu \!=\! (P\Dmfp^0)^2\!\big/ (P\Dmfm^0)^2$} mit~\mbox{$P\Dimath^0 \!=\! \sqrt{M\Dimath^2 \!+\! \rb{P}^2}$},~\mbox{$M\Dimath^2 \!>\! 0$}, f"ur~$\imath \!=\! \mfp,\mfm$.
Die Parameter dieser Relationen:~$M\Dmfp^2$,~$M\Dmfm^2$ und~$\rb{P}^2$~[mit~\mbox{$\rb{P} \!\equiv\! \rb{P}\Dmfp \!\equiv\! -\rb{P}\Dmfm$} per~def.], sind Teilchen-spezifisch beziehungsweise subsumieren deren transversale Kinematik.
\vspace*{-.5ex}

\bigskip\noindent
"Uber die angegebenen Relationen stehen~$s$ und~$\be\Dmfp$,~$\be\Dmfm$ in eineindeutigem Zusammenhang.
Es ist daher "aquivalent, zwei aufeinander zu laufende Teilchen im Schwerpunktsystem zu betrachten in Termen des Quadrats ihrer invarianten Schwerpunktenergie~$s$ oder in Termen der Beta-Parameter~$\be\Dmfp$ und~$\be\Dmfm$ entsprechender aktiver Lorentz-Boosts. \\
\indent
Wir gehen den zweiten Weg und definieren auf Basis von~$\be\Dmfp$,~$\be\Dmfm$ beziehungsweise~$\ps\Dmfp$,~$\ps\Dmfm$ geeignete longitudinale Koordinaten~$x^\mfp$,~$x^\mfm$.
Der Limes~$s \!\to\! \infty$ wird "aquivalent betrachtet in Form~$\be\Dmfp,\be\Dmfm \!\to\! 1$ beziehungsweise~$\ps\Dmfp,\ps\Dmfm \!\to\! \infty$.%
\FOOT{
  F"ur~$s \!\to\! \infty$ folgt~$\ps,\ps\Dmfp \!\to\! \infty$, aus Gl.~(\ref{APP:s*_ps-mfp}$'$) die Asymptotik:~~\mbox{$\cosh\ps \!\sim\! \sqrt\mu\, [2\cosh^2\ps\Dmfp \!-\! 1] \!=\! \sqrt\mu\, \cosh 2\ps\Dmfp$}, und~-- hiermit konsistent~-- aus Gl.~(\ref{APP:ps-mfp_s*}) die Asymptotik:~~\mbox{$\cosh\ps\Dmfp \!\sim\! (2\sqrt\mu)^{-1\!/\!2} \sqrt{\cosh\ps \!+\! 1} \!=\! \mu^{-1\!/\!4}\, \cosh\ps\!/\!2$}.
}

%
\section[Minkowskische Koordinaten:%
           ~\protect{$\{e_{(\tilde\mu)}, x^{\tilde\mu}\}$};~-- versus%
           ~\protect{$\{e_{(\bar\mu)},   x^{\bar\mu}\}$} und%
           ~\protect{$\{e_{(\mu)},       x^\mu\}$}]{%
         Minkowskische Koordinaten:
           ~\bm{\{e_{(\tilde\mu)}, x^{\tilde\mu}\}};~\bm{-}\\ versus%
           ~\bm{\{e_{(\bar\mu)},   x^{\bar\mu}\}} und%
           ~\bm{\{e_{(\mu)},       x^\mu\}}~\bffootnote}
\label{APP-Sect:Minkowski}

\footnotetext{
  \label{APP-FN:mf<->bm(mf)}Seien~$\mfp$,~$\mfm$ Lorentz-Indizes,~\bm{\mfp},~\bm{\mfm} dagegen Notation der Teilchen und deren (momentanen) Ruhsysteme.
}
%

\paragraph{Koordinatenlinien~\bm{\tilde\mu \in \{\mfp,\mfm,1,2\}}.}
Seien f"ur feste~$\be\Dmfp$,~$\be\Dmfm$ beziehungsweise~$\ps\Dmfp,\ps\Dmfm$%
\FOOT{
  Aufgefa"st~$\be\Dmfp$ und~$\be\Dmfm \!\equiv\! \be\Dmfm(\be\Dmfp)$ als die Funktionswerte f"ur festes~$s$.
}
betrachtet die Inertialsysteme~$I\Dmfp$,~$I\Dmfm$, die folgen aus dem initialen System~$I\idx{0}$ durch aktiven Lorentz-Boosts mit~$\La\Dmfp$ beziehungsweise~$\La\Dmfm$.
Seien~$e_{(\mu)} \!\equiv\! \big(e_{(\mu)}{}^\nu\big)$,~\mbox{$\mu,\nu \!\in\! \{0,1,2,3\}$}, die (karthesischen) Basisvektoren von~$I\idx{0}$, f"ur die gilt:
\vspace*{-.75ex}
\begin{align} \label{APP:e-kov-karthesisch}
&e_{(\mu)}\cdot e_{(\mu)}\;
  =\; g_{\mu\nu}
    \\[.25ex]
&\text{mit}\qquad
  e_{(\mu)}{}^\nu\; =\; g_\mu{}^\nu\; =\; \de_\mu^\nu\qquad
  \mu,\nu \in \{0,1,2,3\}
    \tag{\ref{APP:e-kov-karthesisch}$'$}
    \\[-4.5ex]\nn
\end{align}
Unter den aktiven Boosts geht diese Basis~$\{e_{(\mu)}\}$ "uber in die Basen~$\{e'_{(\mu)}\}$,~$\{e^\dbprime_{(\mu)}\}$ der Syste\-me~$I\Dmfp$,~$I\Dmfm$; deren Vektoren besitzen Zerlegungen nach der alten Basis, vgl.\@ Gl.~(\ref{APP:e'-nach-e}):
\vspace*{-1ex}
\begin{alignat}{3} \label{APP:e'-nach-e_Imf}
&e'_{(\mu)}&\;
  &=\; (\La\Dmfp)_\mu{}^\nu&\; &e_{(\nu)}
    \\[.25ex]
&e^\dbprime_{(\mu)}&\;
  &=\; (\La\Dmfm)_\mu{}^\nu&\; &e_{(\nu)}\qquad
  \mu \in \{0,1,2,3\}
    \tag{\ref{APP:e'-nach-e_Imf}$'$}
    \\[-4.5ex]\nn
\end{alignat}
im Sinne der Zerlegung
\vspace*{-.75ex}
\begin{align} \label{APP:x-Zerlegung-kontra}
x\;
  =\; x^\mu\, e_{(\mu)}\qquad
  \mu \in \{0,1,2,3\}
    \\[-4.5ex]\nn
\end{align}
des beliebigen Lorentz-Vektors~$x$ in Komponenten~$x^\mu$ entlang der Achsen $e_{(\mu)}$,~$\mu \!\in\! \{0,1,2,3\}$, von~$I\idx{0}$.
Die Zeitachsen~$e'_{(0)}$,~$e^\dbprime_{(0)}$ der Systeme~$I\Dmfp$,~$I\Dmfm$ sind explizit gegeben durch:
\vspace*{-1ex}
\begin{alignat}{3} \label{APP:Imfp,Imfm-Zeitachsen}
&e'_{(0)}&\;
  &=\;& \ga\Dmfp\;
        &\big(e_{(0)}\;
         +\; \be\Dmfp\; e_{(3)}\big)
    \\[.25ex]
&e^\dbprime_{(0)}&\;
  &=\;& \ga\Dmfm\;
        &\big(e_{(0)}\;
         -\; \be\Dmfm\;  e_{(3)}\big)
    \tag{\ref{APP:Imfp,Imfm-Zeitachsen}$'$}
    \\[-4.5ex]\nn
\end{alignat}
vgl.\@ die Gln.~(\ref{APP:e'-nach-e_Imf}),~(\ref{APP:e'-nach-e_Imf}$'$) und~(\ref{APP:P'_i}$'$). \\
\indent
Dann ist ein Koordinatensystem definiert durch die Lorentz-Vektoren:
\vspace*{-.5ex}
\begin{alignat}{3} \label{APP:etilde_e}
&e_{(\mfp)}&\;
  &:=\;& \vrh\;\cdot \ga\Dmfp\;
        &\big(e_{(0)}\;
          +\; \be\Dmfp\; e_{(3)}\big)
    \\[.25ex]
&e_{(\mfm)}&\;
  &:=\;& \vrh\;\cdot \ga\Dmfm\;
        &\big(e_{(0)}\;
          -\; \be\Dmfm\; e_{(3)}\big)
    \tag{\ref{APP:etilde_e}$'$}
    \\[-4.5ex]\nn
\end{alignat}
\vspace*{-5ex}
\begin{align}
\text{und}\qquad
e_{(\tilde{i})}\;
  :=\; e_{(i)}\qquad
    i \in \{1,2\}
    \tag{\ref{APP:etilde_e}$''$}
    \\[-4.5ex]\nn
\end{align}
mit~\mbox{$\vrh \!\in\! \bbbr^+$} einer geeignet zu w"ahlenden Konstanten.
Die Vektoren~$e_{(\mfp)}$,~$e_{(\mfm)}$ sind definiert bis auf die Konstante~$\vrh$ genau als die Zeitachsen~$e'_{(0)}$,~$e^\dbprime_{(0)}$ der Systeme~$I\Dmfp$,~$I\Dmfm$.
Die Weltlinien~${\cal C}\Dmfp$,~${\cal C}\Dmfm$ der Teilchen~$\mfp$ und~$\mfm$~-- vgl.\@ Abb.~\ref{Fig:MinkowskiTRAJECTORY}~-- besitzen folglich longitudinal nur eine $x^\mfp$-~beziehungsweise~$x^\mfm$-Komponente entsprechend der Zerlegung
\vspace*{-1ex}
\begin{align} \label{APP:xtilde-Zerlegung-kontra}
x\;
  =\; x^{\tilde\mu}\, e_{(\tilde\mu)}\qquad
    \tilde\mu \in \{\mfp,\mfm,1,2\}
    \\[-4.5ex]\nn
\end{align}
im Sinne von Gl.~(\ref{APP:x-Zerlegung-kontra}). \\
\indent
F"ur nichtverschwindende Boosts:~$\be\Dmfp,\be\Dmfm \!>\! 0$, sind die Vektoren~$e_{(\tilde\mu)}$,~$\tilde\mu \!\in\! \{\mfp,\mfm,1,2\}$ linear unabh"angig,~$\{e_{(\tilde\mu)}\}$ Basis des Minkowski-Raumes und die Zerlegung entsprechend Gl.~(\ref{APP:xtilde-Zerlegung-kontra}) wohldefiniert.
In Termen der Vektoren~$e_{(\tilde\mu)}$ und den Komponenten~$x^{\tilde\mu}$ mit~$\tilde\mu \!\in\! \{\mfp,\mfm,1,2\}$ geschieht im Haupttext die Auswertung und analytische Fortsetzung der $T$-Streuamplitude. \\
\indent
Wir formalisieren f"ur den beliebigen Lorentz-Vektor~$x$ den Zusammenhang zwischen~dem Koordinatensystem~$I\idx{0}$ mit Komponenten~$x^\mu$ bez"uglich der Basis~$\{e_{(\mu)}\}$,~\mbox{$\mu \!\in\! \{0,1,2,3\}$} nach Gl.~(\ref{APP:e_(mu)-Norm})ff.\@, zwischen Lichtkegelkomponenten~$x^{\bar\mu}$ bez"uglich der Basis~$\{e_{(\bar\mu)}\}$,~\mbox{$\bar\mu \!\in\! \{+,-,1,2\}$} nach Gl.~(\ref{APP:e_(mu)LC-Norm})ff. und dem Koordinatensystem mit Komponenten~$x^{\tilde\mu}$ bez"uglich der Basis $\{e_{(\tilde\mu)}\}$, \mbox{$\tilde\mu \!\in\! \{\mfp,\mfm,1,2\}$} nach Gl.~(\ref{APP:etilde_e})-(\ref{APP:etilde_e}$''$).
Wir schreiben in diesem Sinne f"ur ein und denselben Vektor bez"uglich der verschiedenen Basen:~$x \!\equiv\! (x^\mu)$,~$\bar{x} \!\equiv\! (x^{\bar\mu})$ und~$\tilde{x} \!\equiv\! (x^{\tilde\mu})$. \\
\indent
Prim"ares Interesse gilt dem Hochenergielimes~\mbox{$s \!\to\! \infty$}, das hei"st~\mbox{$\be\Dmfp,\be\Dmfm \!\to\! 1$} beziehungsweise \mbox{$\ps\Dmfp,\ps\Dmfm \!\to\! \infty$}, in dem die physikalischen Weltlinien~${\cal C}\Dmfp$,~${\cal C}\Dmfm$ "ubergehen in~${\cal C}\idx{+}$,~${\cal C}\idx{-}$ und die Zeitachsen~$e'_{(0)}$,~$e^\dbprime_{(0)}$, ergo die Vektoren~$e_{(\mfp)}$,~$e_{(\mfm)}$ zeigen in Richtung~$e_{(+)}$,~$e_{(-)}$.
Wir legen daher Betonung auf den Zusammenhang der Systeme%
~$\{x^{\tilde\mu}, e_{(\tilde\mu)}\}$,~\mbox{$\tilde\mu \!\in\! \{\mfp,\mfm,1,2\}$},
und~$\{x^{\bar\mu}, e_{(\bar\mu)}\}$, \mbox{$\bar\mu \!\in\! \{+,-,1,2\}$}; Konventionen werden getroffen in Hinblick eines einfachen "Ubergangs.
\vspace*{-.5ex}

\bigskip\noindent
Wir betrachten die Vektoren~$e_{(\tilde\mu)}$,~$\tilde\mu \!\in\! \{\mfp,\mfm,1,2\}$ nach den Gln.~(\ref{APP:etilde_e})-(\ref{APP:etilde_e}$''$).
In Gegen\-"uberstellung mit Gl.~(\ref{APP:x-Zerlegung-kontra}) lesen wir wie folgt ab die Komponenten von~$e_{(\mfp)}$,~$e_{(\mfm)}$ bez"uglich der Basis~$\{e_{(\mu)}\}$,~$\mu \!\in\! \{0,1,2,3\}$:
\vspace*{-.5ex}
\begin{alignat}{3}
&e_{(\mfp)}{}^0&\;
  &=\; \vrh\;\cdot \ga\Dmfp&\;
  &=\; \vrh\;\cdot \cosh\ps\Dmfp
    \label{APP:emfp^03} \\[.5ex]
&e_{(\mfp)}{}^3&\;
  &=\; \vrh\;\cdot \ga\Dmfp\, \be\Dmfp&\;
  &=\; \vrh\;\cdot \sinh\ps\Dmfp
    \tag{\ref{APP:emfp^03}$'$}
    \\[-6ex]\nn
\intertext{\vspace*{-1.5ex}und:}
&e_{(\mfm)}{}^0&\;
  &=\; \vrh\;\cdot \ga\Dmfm&\;
  &=\; \vrh\;\cdot \cosh\ps\Dmfm
    \label{APP:emfm^03} \\[.5ex]
&e_{(\mfm)}{}^3&\;
  &=\; \vrh\;\cdot \ga\Dmfp\, (-\be\Dmfm)&\;
  &=\; \vrh\;\cdot (-\sinh\ps\Dmfm)
    \tag{\ref{APP:emfm^03}$'$}
    \\[-4.5ex]\nn
\end{alignat}
Gem"a"s~$x^\pm \!=\! \al\,(x^0 \!\pm\! x^3)$, vgl.\@ Gln~(\ref{APP:kontravLC_Def}), folgen die Lichtkegelkomponenten:
\vspace*{-.5ex}
\begin{alignat}{5} \label{APP:emfpm^+-}
&e_{(\mfp)}{}^\pm&\;
  &=&\; &\vrh \al\, \big(\cosh\ps\Dmfp \pm \sinh\ps\Dmfp\big)&\;
  &=&\; &\vrh \al\; \efn{\D\pm\ps\Dmfp}
    \\[.5ex]
&e_{(\mfm)}{}^\pm&\;
  &=&\; &\vrh \al\, \big(\cosh\ps\Dmfm \mp \sinh\ps\Dmfm\big)&\;
  &=&\; &\vrh \al\; \efn{\D\mp\ps\Dmfm}
    \tag{\ref{APP:emfpm^+-}$'$}
    \\[-4.5ex]\nn
\end{alignat}
Die Vektoren der Basis~$\{e_{(\bar\mu)}\}$,~$\bar\mu \!\in\! \{+,-,1,2\}$ zeigen in Richtung der Lichtkegel-Koordina\-tenlinien; sie erf"ullen:
\vspace*{-.25ex}
\begin{align} \label{APP:ebar-kov-karthesisch}
&e_{(\bar\mu)}\cdot e_{(\bar\nu)}\;
  =\; g_{\bar\mu\bar\nu}
    \\[.5ex]
&\text{mit}\qquad
  e_{(\bar\mu)}{}^{\bar\nu}\; =\; g_{\bar\mu}{}^{\bar\nu}\; =\; \de_{\bar\mu}^{\bar\nu}\qquad
  \bar\mu,\bar\nu \in \{+,-,1,2\}
    \tag{\ref{APP:ebar-kov-karthesisch}$'$}
    \\[-4.25ex]\nn
\end{align}
vgl.\@ die Gln.~(\ref{APP:e_(mu)LC-Norm}),~(\ref{APP:e_(mu)LC-karthesisch}) und~(\ref{APP:gbar}),~(\ref{APP:gbar}$'$).
Die Zerlegung
%
\begin{align} 
x\;
  =\; x^{\bar\mu}\, e_{(\bar\mu)}\qquad
    \bar\mu \in \{+,-,1,2\}
\end{align}
im Sinne der Gln.~(\ref{APP:x-Zerlegung-kontra}),~(\ref{APP:xtilde-Zerlegung-kontra}) lautet f"ur die Vektoren~$e_{(\tilde\mu)}$,~$\tilde\mu \!\in\! \{\mfp,\mfm,1,2\}$:
\vspace*{-.25ex}
\begin{alignat}{4} \label{APP:etilde_ebar}
&e_{(\mfp)}&\;
  &:=\;& \vrh \al\;
         \big( \efn{\D\ps\Dmfp}\; &e_{(+)}&\;
        &+\; \efn{\D-\ps\Dmfp}\; e_{(-)} \big)
    \\[.5ex]
&e_{(\mfm)}&\;
  &:=\;& \vrh \al\;
         \big( \efn{\D-\ps\Dmfm}\; &e_{(+)}&\;
        &+\; \phantom{-}\efn{\D\ps\Dmfm}\; e_{(-)} \big)
    \tag{\ref{APP:etilde_ebar}$'$}
    \\[-4.5ex]\nn
\end{alignat}
\vspace*{-4ex}
\begin{align}
\text{und}\qquad
e_{(\tilde{i})}\;
  :=\; e_{(\bar{i})}\qquad
    i \in \{1,2\}
    \tag{\ref{APP:etilde_ebar}$''$}
    \\[-4.25ex]\nn
\end{align}
vgl.\@ die Gln.~(\ref{APP:etilde_e})-(\ref{APP:etilde_e}$''$).
Im Limes~$\ps\Dmfp, \ps\Dmfm \!\to\! \infty$ zeigen~$e_{(\mfp)}$,~$e_{(\mfm)}$ in Richtung~$e_{(+)}$,~$e_{(-)}$.\!
\vspace*{-.5ex}

\paragraph{Basisvektoren~\bm{e^{(\tilde\mu)}},~\bm{e_{(\tilde\mu)}}
           und induzierte Metrik~\bm{\tilde{g} \!\equiv\! \big(g_{\tilde\mu\tilde\nu}\big)},%
           ~\bm{\tilde\mu \in \{\mfp,\mfm,1,2\}}.}
Die Gln.~(\ref{APP:etilde_ebar})-(\ref{APP:etilde_ebar}$''$) lauten in Matrixform:%
\FOOT{
  \label{APP-FN:only-longitudinal-Rep}Explizit ausgeschrieben seien nur die nichttrivialen longitudinalen Komponenten, vgl.\@ Fu"sn.\,\FN{APP-FN:only-longitudinal}.
}
\begin{samepage}
%
\begin{align} \label{APP:etilde_ebarM-kov}
e_{(\tilde\mu)}\;
  =\; \mathbb{S}_{\tilde\mu}{}^{\bar\nu}\; e_{(\bar\nu)}\qquad
    \tilde\mu \in \{\mfp,\mfm,1,2\},\vv
    \bar\nu \in \{+,-,1,2\}
\end{align}
vgl.\@  die Gln.~(\ref{APP:e'-nach-e}) und~(\ref{APP:e'-nach-e_Imf}),~(\ref{APP:e'-nach-e_Imf}$'$); es ist:
\end{samepage}
\vspace*{-.5ex}
\begin{align} \label{APP:Stilde_mu^nu}
&\big(\mathbb{S}_{\tilde\mu}{}^{\bar\nu}\big)\;
  =\; \vrh \al\,
      \pmatrixZZ{\efn{\D\ps\Dmfp}}{\efn{\D-\ps\Dmfp}}
                {\efn{\D-\ps\Dmfm}}{\efn{\D\ps\Dmfm}}
    \\[.5ex]
&\text{mit}\qquad
  \det\! \big(\mathbb{S}_{\tilde\mu}{}^{\bar\nu}\big)\;
  =\; (g_{+-})^{-1}\, \vrh^2\, \sinh\ps
    \tag{\ref{APP:Stilde_mu^nu}$'$}
    \\[-4ex]\nn
\end{align}
ferner:
\vspace*{-.5ex}
\begin{align} \label{APP:al,g+-,detL,detg}
&2\al^2\;
  =\; g^{+-}\;
  =\; (g_{+-})^{-1}\;
  =\; -\det \mathbb{L}
    \\
&\text{und}\qquad
  g_{+-}\;
  =\; \sqrt{-\det\bar{g}}
    \tag{\ref{APP:al,g+-,detL,detg}$'$}
    \\[-4.5ex]\nn
\end{align}
mit~$\mathbb{L} \!\equiv\! \big(\mathbb{L}^{\bar\mu}{}_\nu\big)$ und~$\bar{g} \!\equiv\! \big(g_{\bar\mu\bar\nu}\big)$; vgl.\@ die Gln.~(\ref{APP:gbar}),~(\ref{APP:gbar^-1}) bzw.~\mbox{\,$\det g \!=\! -(g_{+-})^2$}. \\
\indent
Die~$e_{(\tilde\mu)}$,~$\tilde\mu \!\in\! \{\mfp,\mfm,1,2\}$ repr"asentieren Koordinatenlinien, die induzieren einen metrischen Tensor~\mbox{$\tilde{g} \!\equiv\! \big(g_{\tilde\mu\tilde\nu}\big)$} durch
\vspace*{-.25ex}
\begin{align} \label{APP:etilde-kov-karthesisch}
&e_{(\tilde\mu)}\cdot e_{(\tilde\nu)}\;
  =\; g_{\tilde\mu\tilde\nu}
    \\[.5ex]
&\text{mit}\qquad
e_{(\tilde\mu)}{}^{\tilde\nu}\;
  =\; g_{\tilde\mu}{}^{\tilde\nu}\; =\; \de_{\tilde\mu}^{\tilde\nu}\qquad
  \tilde\mu,\tilde\nu \in \{\mfp,\mfm,1,2\}
    \tag{\ref{APP:etilde-kov-karthesisch}$'$}
    \\[-4.25ex]\nn
\end{align}
vgl.\@ die Gln.~(\ref{APP:e-kov-karthesisch})~(\ref{APP:e-kov-karthesisch}$'$) und~(\ref{APP:ebar-kov-karthesisch}),~(\ref{APP:ebar-kov-karthesisch}$'$).
F"ur dessen Komponenten~\mbox{\,$g_{\tilde\mu\tilde\nu}$} folgt aus
\vspace*{-.5ex}
\begin{align} \label{APP:emu-dot-enu-kov}
e_{(\tilde\mu)}\cdot e_{(\tilde\nu)}\;
   =\; \mathbb{S}_{\tilde\mu}{}^{\bar\al}\;
       \mathbb{S}_{\tilde\nu}{}^{\bar\be}\;
         e_{(\bar\al)}\cdot e_{(\bar\be)}
    \\[-4.5ex]\nn
\end{align}
vgl.\@ Gln~(\ref{APP:etilde_ebarM-kov}), mithilfe der Gln.~(\ref{APP:ebar-kov-karthesisch}),~(\ref{APP:etilde-kov-karthesisch}) unmittelbar:
\vspace*{-.5ex}
\begin{align} \label{APP:S-pseudo-orth}
g_{\tilde\mu\tilde\nu}\;
  =\; \mathbb{S}_{\tilde\mu}{}^{\bar\al}\;
       \mathbb{S}_{\tilde\nu}{}^{\bar\be}\;
         g_{\bar\al\bar\be}
    \\[-4.5ex]\nn
\end{align}
In Form~\mbox{\,$g_{\tilde\mu\tilde\nu} \!=\!
  g_{+-}\, \big(\mathbb{S}_{\tilde\mu}{}^{+}\, \mathbb{S}_{\tilde\nu}{}^{-}
          \!+\! \mathbb{S}_{\tilde\mu}{}^{-}\, \mathbb{S}_{\tilde\nu}{}^{+}\big)$} folgt aus der expliziten Gestalt der Matrix~$\big(\mathbb{S}_{\tilde\mu}{}^{\bar\nu}\big)$, vgl.\@ Gl.~(\ref{APP:Stilde_mu^nu}):
\vspace*{-.25ex}
\begin{alignat}{3} \label{APP:gtilde-kov}
&g_{\mfp\mfp}&\;
  &=\; g_{\mfm\mfm}&\;
  &=\; \vrh^2
    \\[.5ex]
&g_{\mfp\mfm}&\;
  &=\; g_{\mfm\mfp}&\;
  &=\; \vrh^2\cdot \cosh\ps
    \tag{\ref{APP:gtilde-kov}$'$}
    \\[-4.5ex]\nn
\end{alignat}
\vspace*{-4ex}
\begin{align}
\text{und}\qquad
g_{\tilde{i}\tilde{j}}\;
  =\; g_{\bar{i}\bar{j}}\;
  =\; -\de_{ij}\qquad
  i \in \{1,2\}
    \tag{\ref{APP:gtilde-kov}$''$}
    \\[-4.25ex]\nn
\end{align}
mit~$\ps \!=\! \ps\Dmfp \!+\! \ps\Dmfm$, vgl.\@ Gl.~(\ref{APP:ps=psp+psm}). \\
\indent
Dies impliziert~\mbox{\,$\det \tilde{g} \!=\! -(\vrh^2\, \sinh\ps)^2$} f"ur die Determinante des metrischen Tensors und umgekehrt:
\vspace*{-.5ex}
\begin{align} \label{APP:det-gtilde}
\vrh^2\,\sinh\ps\;
  =\; \sqrt{-\det\tilde{g}}
    \\[-4.5ex]\nn
\end{align}
Wir nehmen im folgenden an, die Lorentz-Boosts seien nicht\-trivial:~\mbox{\,$\ps\Dmfp,\ps\Dmfm \!>\! 0$}, so da"s die inverse Metrik~\mbox{\,$\tilde{g}^{-1} \!\equiv\! \big(g^{\tilde\mu\tilde\nu}\big)$} existiert.
F"ur diese folgt unmittelbar:
\vspace*{-.5ex}
\begin{alignat}{3} \label{APP:gtilde-kontrav}
&g^{\mfp\mfp}&\;
  &=\; g^{\mfm\mfm}&\;
  &=\; \phantom{-\,} g_{\mfp\mfp}\; \big/ \det\tilde{g}
    \\[.5ex]
&g^{\mfp\mfm}&\;
  &=\; g^{\mfm\mfp}&\;
  &=\; -\, g_{\mfp\mfm}\; \big/ \det\tilde{g}
    \tag{\ref{APP:gtilde-kontrav}$'$}
    \\[-4.5ex]\nn
\end{alignat}
\vspace*{-4ex}
\begin{align}
\text{und}\qquad
g^{\tilde{i}\tilde{j}}\;
  =\; g^{\bar{i}\bar{j}}\;
  =\; -\de^{ij}\qquad
  i \in \{1,2\}
    \tag{\ref{APP:gtilde-kontrav}$''$}
    \\[-4.5ex]\nn
\end{align}
und explizite Ausdr"ucke mithilfe der Gln.~(\ref{APP:gtilde-kov}),~(\ref{APP:gtilde-kov}$'$). \\
\indent
Seien definiert Lorentz-Vektoren~$e^{(\tilde\mu)}$,~\mbox{$\tilde\mu \!\in\! \{\mfp,\mfm,1,2\}$} durch:
\vspace*{-.5ex}
\begin{align} \label{APP:etilde^(mu)-Def}
&e^{(\tilde\mu)}\;
  :=\; g^{\tilde\mu\tilde\nu}\; e_{(\tilde\nu)}
    \\[.25ex]
&\Longrightarrow\qquad
  e_{(\tilde\mu)}\;
  =\; g_{\tilde\mu\tilde\nu}\; e^{(\tilde\nu)}\qquad
    \tilde\mu \in \{\mfp,\mfm,1,2\}
    \tag{\ref{APP:etilde^(mu)-Def}$'$}
    \\[-4.5ex]\nn
\end{align}
vgl.\@ Gl.~(\ref{APP:e^(mu)-Def}) und~(\ref{APP:e^(mu)LC-Def}).
Dann gilt:
\begin{samepage}
\vspace*{-.5ex}
\begin{align} \label{APP:etilde_ebarM-kontrav}
e^{(\tilde\mu)}\;
  =\; \mathbb{S}^{\tilde\mu}{}_{\bar\nu}\; e^{(\bar\nu)}\qquad
    \tilde\mu \in \{\mfp,\mfm,1,2\},\vv
    \bar\nu \in \{+,-,1,2\}
    \\[-4.5ex]\nn
\end{align}
unter Definition
\vspace*{-.5ex}
\begin{align} \label{APP:Stilde^mu_nu-Def}
\mathbb{S}^{\tilde\mu}{}_{\bar\nu}\;
  :=\; g^{\tilde\mu\tilde\rh}\vv
       \mathbb{S}_{\tilde\rh}{}^{\bar\si}\vv
       g_{\bar\si\bar\nu}
    \\[-4.5ex]\nn
\end{align}
vgl.\@ die Gln.~(\ref{APP:etilde_ebarM-kov}) und~(\ref{APP:Stilde_mu^nu}),~(\ref{APP:Stilde_mu^nu}$'$).
\end{samepage}

Die Matrix~$\big(\mathbb{S}^{\tilde\mu}{}_{\bar\nu}\big)$ wird bestimmt wie folgt.
Kontraktion von Gl.~(\ref{APP:Stilde^mu_nu-Def}) mit~\mbox{\,$g_{\tilde\al\tilde\mu}$} ergibt mithilfe von Gl.~(\ref{APP:S-pseudo-orth})~[Umbenenung~$\al \!\to\! \mu$]:
\vspace*{-.5ex}
\begin{align} \label{APP:S,S-1t-pre}
&g_{\tilde\mu\tilde\rh}\;
  \mathbb{S}^{\tilde\rh}{}_{\bar\nu}\;
  =\; \mathbb{S}_{\tilde\mu}{}^{\bar\si}\;
        g_{\bar\si\bar\nu}
    \\
&\Longleftrightarrow\qquad
\mathbb{S}_{\tilde\mu}{}^{\bar\si}\;
  \mathbb{S}_{\tilde\rh}{}^{\bar\be}\;
  \mathbb{S}^{\tilde\rh}{}_{\bar\nu}\;
    g_{\bar\si\bar\be}\;
  =\; \mathbb{S}_{\tilde\mu}{}^{\bar\si}\;
        g_{\bar\si\bar\nu}\qquad
  \text{d.h.}\qquad
  \mathbb{S}_{\tilde\rh}{}^{\bar\be}\, \mathbb{S}^{\tilde\rh}{}_{\bar\nu}\;
    =\; \de^\be_\nu
    \tag{\ref{APP:S,S-1t-pre}$'$}
    \\[-4.5ex]\nn
\end{align}
und weiter~\mbox{%
  \,$\mathbb{S}^{\tilde\mu}{}_{\bar\be}\, \mathbb{S}_{\tilde\rh}{}^{\bar\be} \!=\! \de^\be_\nu$} durch Kontraktion mit~\mbox{\,$\mathbb{S}^{\tilde\mu}{}_{\bar\be}$}.
Zusammenfassend:
\vspace*{-.25ex}
\begin{align} \label{APP:S-kontragredient}
&\mathbb{S}_{\tilde\rh}{}^{\bar\mu}\;
  \mathbb{S}^{\tilde\rh}{}_{\bar\nu}\;
  =\; \de^{\bar\mu}_{\bar\nu}\qquad
 \mathbb{S}^{\tilde\mu}{}_{\bar\rh}\;
  \mathbb{S}_{\tilde\nu}{}^{\bar\rh}\;
  =\; \de^{\tilde\mu}_{\tilde\nu}
    \\
&\text{d.h.}\qquad
 \big(\mathbb{S}^{\tilde\mu}{}_{\bar\nu}\big)^{-1}\;
  =\; \big(\mathbb{S}_{\tilde\mu}{}^{\bar\nu}\big){}^{\T t}
    \tag{\ref{APP:S-kontragredient}$'$}
    \\[-4.25ex]\nn
\end{align}
Die Matrizen~$\big(\mathbb{S}^{\tilde\mu}{}_{\bar\nu}\big)$ und~$\big(\mathbb{S}_{\tilde\mu}{}^{\bar\nu}\big)$ sind {\it kontragredient\/}: die eine die transponierte Inverse der anderen.
Wir bezeichnen:
\vspace*{-.5ex}
\begin{align} \label{APP:S,S^-1t}
&\mathbb{S}\;
  \equiv\; \big(\mathbb{S}^{\tilde\mu}{}_{\bar\nu}\big)\qquad
    \\[-.5ex]
&\Longrightarrow\qquad
 \big(\mathbb{S}_{\tilde\mu}{}^{\bar\nu}\big)\;
  \equiv\; \mathbb{S}^{-1\T\:t}\qquad
    \tilde\mu \in \{\mfp,\mfm,1,2\},\vv
    \bar\nu \in \{+,-,1,2\}
    \tag{\ref{APP:S,S^-1t}$'$}
    \\[-4.5ex]\nn
\end{align}
Durch Kontraktion von Gl.(\ref{APP:S,S-1t-pre}) mit~$\mathbb{S}^{\tilde\mu}{}_{\bar\rh}$ folgt mithilfe von Gl.~(\ref{APP:S-kontragredient}):
\vspace*{-.25ex}
\begin{align} 
g_{\tilde\mu\tilde\nu}\;
  \mathbb{S}^{\tilde\mu}{}_{\bar\rh}\;
  \mathbb{S}^{\tilde\nu}{}_{\bar\si}\;
  =\; g_{\bar\rh\bar\si}\qquad
  \text{d.h.}\qquad
\mathbb{S}^{\,\T t}\; \tilde{g}\; \mathbb{S}\;
  =\; \bar{g}
    \\[-4.25ex]\nn
\end{align}
das hei"st~\mbox{\,$\mathbb{S} \!\equiv\! \big(\mathbb{S}^{\tilde\mu}{}_{\bar\nu}\big)$} ist {\it pseudo-orthogonal\/}.~--
Vgl.\@ die Matrizen~$\La \!\equiv\! \big(\La^\mu{}_\nu\big)$ und~$\La^{\!-1\T\:t} \!\equiv\! \big(\La_\mu{}^\nu\big)$ f"ur Lorentz-Boosts, Gl.~(\ref{APP:La-kontragredient}), und die Matrizen~$\mathbb{L} \!\equiv\! \big(\mathbb{L}^{\bar\mu}{}_\nu\big)$ und~$\mathbb{L}^{-1\T\:t} \!\equiv\! \big(\mathbb{L}_{\bar\mu}{}^\nu\big)$ f"ur die Transformation auf Lichtkegelkoordinaten, Gl.~(\ref{APP:La-kontragredient}).
Gl.~(\ref{APP:S-kontragredient}$'$) impliziert ferner:
\vspace*{-.5ex}
\begin{align} \label{APP:detS,detS^-1t}
&\det \mathbb{S}
  \equiv \det\! \big(\mathbb{S}^{\tilde\mu}{}_{\bar\nu}\big)\;
  =\; \big[
        \det\! \big( \mathbb{S}{}^{-1\T\:t} \big)
      \big]^{-1}\;
  \equiv\;
      \big[
        \det\! \big(\mathbb{S}_{\tilde\mu}{}^{\bar\nu}\big)
      \big]^{-1}
    \\
&=\; g_{+-}\, [\vrh^2\, \sinh\ps]^{-1}\;
  =\; [-\det \mathbb{L}\;\cdot \vrh^2\, \sinh\ps]^{-1}\;
  =\; \sqrt{-\det\bar{g}}\; \Big/ \sqrt{-\det\tilde{g}}
    \nn
    \\[-4.5ex]\nn
\end{align}
vgl.\@ Gl.~(\ref{APP:Stilde_mu^nu}$'$), bzgl.\@ der letzten Identit"at die Gln.~(\ref{APP:al,g+-,detL,detg}),~(\ref{APP:al,g+-,detL,detg}$'$) und~(\ref{APP:det-gtilde}). \\
\indent
Die Matrix~\mbox{\,$\mathbb{S} \!\equiv\! \big(\mathbb{S}^{\tilde\mu}{}_{\bar\nu}\big)$} folgt aus~\mbox{\,$\mathbb{S}^{-1\T\:t} \!\equiv\! \big(\mathbb{S}_{\tilde\mu}{}^{\bar\nu}\big)$} in Form von Gl.~(\ref{APP:Stilde_mu^nu}) durch Invertieren und Tranponieren oder "aquivalent durch Ausmultiplizieren von Gl.~(\ref{APP:Stilde^mu_nu-Def}) in Form:
\vspace*{-.5ex}
\begin{align} \label{APP:Stilde^mu_nu-prepre}
\big(\mathbb{S}^{\tilde\mu}{}_{\bar\nu}\big)\;
  =\; \frac{1}{\det\tilde{g}}\,
        \pmatrixZZ{g_{\mfm\mfm}}{-g_{\mfp\mfm}}
                 {-g_{\mfm\mfp}}{g_{\mfp\mfp}}\cdot
      \vrh \al\,
        \pmatrixZZ{\efn{\D\ps\Dmfp}}{\efn{\D-\ps\Dmfp}}
                  {\efn{\D-\ps\Dmfm}}{\efn{\D\ps\Dmfm}}\cdot
      \pmatrixZZ{0}{g_{+-}}{g_{-+}}{0}
    \\[-4.5ex]\nn
\end{align}
mithilfe der Relationen:
\vspace*{-.5ex}
\begin{alignat}{4} \label{APP:exp-expcosh}
&\efn{\D\pm\ps\Dmfp}&\;
  &-\;& \efn{\D\mp\ps\Dmfm}\; \cosh\ps\;
  &=\;& \pm\, \efn{\D\mp\ps\Dmfm}\; \sinh\ps&
    \\[.5ex]
&\efn{\D\pm\ps\Dmfm}&\;
  &-\;& \efn{\D\mp\ps\Dmfp}\; \cosh\ps\;
  &=\;& \pm\, \efn{\D\mp\ps\Dmfp}\; \sinh\ps&\qquad
    \ps = \ps\Dmfp \!+\! \ps\Dmfm
    \tag{\ref{APP:exp-expcosh}$'$}
    \\[-4.5ex]\nn
\end{alignat}
In Gl.~(\ref{APP:Stilde^mu_nu-prepre}) ist~\mbox{\,$\tilde{g}^{-1} \!\equiv\! \big(g^{\tilde\mu\tilde\nu}\big)$} im Sinne der Inversen ausgedr"uckt durch die Komponenten des metrischen Tensors~\mbox{\,$\tilde{g} \!\equiv\! \big(g_{\tilde\mu\tilde\nu}\big)$}, vgl.\@ die Gln.~(\ref{APP:gtilde-kontrav})-(\ref{APP:gtilde-kontrav}$''$).
Wir haben:
\begin{samepage}
\vspace*{-.5ex}
\begin{alignat}{3} \label{APP:Stilde^mu_nu}
&\mathbb{S}&\;
  &\equiv\; \big(\mathbb{S}^{\tilde\mu}{}_{\bar\nu}\big)&\;
  &=\; \det\mathbb{S}\cdot
       \vrh \al\,
       \pmatrixZZ{\efn{\D\ps\Dmfm}}{-\efn{\D-\ps\Dmfm}}
                 {-\efn{\D-\ps\Dmfp}}{\efn{\D\ps\Dmfp}}
    \\[-.5ex]
&\mathbb{S}{}^{-1\T\:t}&\;
  &\equiv\; \big(\mathbb{S}_{\tilde\mu}{}^{\bar\nu}\big)&\;
  &=\; \vrh \al\,
      \pmatrixZZ{\efn{\D\ps\Dmfp}}{\efn{\D-\ps\Dmfp}}
                {\efn{\D-\ps\Dmfm}}{\efn{\D\ps\Dmfm}}
    \tag{\ref{APP:Stilde^mu_nu}$'$}
    \\[-4.5ex]\nn
\end{alignat}
vgl.\@ Gl.~(\ref{APP:Stilde_mu^nu}), mit~\mbox{\,$\det\mathbb{S} \!\equiv\! \det\! \big(\mathbb{S}_{\tilde\mu}{}^{\bar\nu}\big)$}, explizit nach Gl.~(\ref{APP:detS,detS^-1t}).
\vspace*{-.5ex}

\paragraph{Komponenten~\bm{x^{\tilde\mu}},~\bm{x_{\tilde\mu}}%
           ~mit~\bm{\tilde\mu \in \{\mfp,\mfm,1,2\}}.}
F"ur einen beliebigen Lorentz-Vektor~$x$ sind die Komponenten~$x^{\tilde\mu}$,~$x_{\tilde\mu}$ bez"uglich der Vektoren~$e^{(\tilde\mu)}$,~$e_{(\tilde\mu)}$ mit~$\tilde\mu \!\in\! \{\mfp,\mfm,1,2\}$ genau die Projektionen im Sinne des Skalarprodukts~-- vgl.\@ die Gln.~(\ref{APP:x-Projektion:x^mu,x_mu}),~(\ref{APP:xbar-Projektion:x^mu,x_mu}):
\end{samepage}%
\vspace*{-.5ex}
\begin{alignat}{2} \label{APP:xtilde-Projektion}
&x^{\tilde\mu}\;
  =\; x\cdot e^{(\tilde\mu)}
    \\
&\text{und}\qquad
  x_{\tilde\mu}\;
  =\; x\cdot e_{(\tilde\mu)}\qquad\qquad
  \mu \in \{\mfp,\mfm,1,2\}
    \tag{\ref{APP:xtilde-Projektion}$'$}
    \\[-4.5ex]\nn
\end{alignat}
Eingesetzt die Gln.~(\ref{APP:etilde_ebarM-kov}),~(\ref{APP:etilde_ebarM-kontrav}) folgt unmittelbar f"ur kontravariante Komponenten:
\vspace*{-.5ex}
\begin{align} \label{APP:xtilde_xbar-kontrav}
&x^{\tilde\mu}\;
  =\; \mathbb{S}^{\tilde\mu}{}_{\bar\nu}\; x^{\bar\nu}
    \\[-.25ex]
&\text{mit}\qquad
  \mathbb{S}\;
  \equiv\; \big(\mathbb{S}^{\tilde\mu}{}_{\bar\nu}\big)\;
  =\; \det\mathbb{S}\cdot
      \vrh \al\,
      \pmatrixZZ{\efn{\D\ps\Dmfm}}{-\efn{\D-\ps\Dmfm}}
                {-\efn{\D-\ps\Dmfp}}{\efn{\D\ps\Dmfp}}
    \tag{\ref{APP:xtilde_xbar-kontrav}$'$}
    \\[-4.5ex]\nn
\end{align}
und~\mbox{\,$\det\mathbb{S}$} nach Gl.~(\ref{APP:detS,detS^-1t})~-- und f"ur kovariante Komponenten:
\vspace*{-.5ex}
\begin{align} \label{APP:xtilde_xbar-kov}
&x_{\tilde\mu}\;
  =\; \mathbb{S}_{\tilde\mu}{}^{\bar\nu}\; x_{\bar\nu}
    \\[-.25ex]
&\text{mit}\qquad
  \mathbb{S}^{-1\T\:t}\;
  \equiv\; \big(\mathbb{S}_{\tilde\mu}{}^{\bar\nu}\big)\;
  =\; \vrh \al\,
      \pmatrixZZ{\efn{\D\ps\Dmfp}}{\efn{\D-\ps\Dmfp}}
                {\efn{\D-\ps\Dmfm}}{\efn{\D\ps\Dmfm}}
    \tag{\ref{APP:xtilde_xbar-kov}$'$}
    \\[-4.5ex]\nn
\end{align}
vgl.\@ die Gln.~(\ref{APP:Stilde^mu_nu}),~(\ref{APP:Stilde^mu_nu}$'$). \\
\indent
F"ur Vollst"andigkeit halten wir fest den entsprechenden Zusammenhang mit den konventionellen karthesischen Koodinaten; er lautet f"ur kontravariante Komponenten:
\vspace*{-.5ex}
\begin{align} \label{APP:xtilde_x-kontrav}
&x^{\tilde\mu}\;
  =\; \mathbb{S}^{\tilde\mu}{}_{\bar\rh}\,
      \mathbb{L}^{\bar\rh}{}_\nu\; x^\nu\;
  =\; (\mathbb{S}\, \mathbb{L})^{\tilde\mu}{}_\nu\; x^\nu
    \\[.5ex]
&\text{mit}\quad
 \begin{aligned}[t]
  \big(\mathbb{S}^{\tilde\mu}{}_{\bar\rh}\,
       \mathbb{L}^{\bar\rh}{}_{\nu}\big)\,
  &=\, \mathbb{S}\, \mathbb{L}\,
  \equiv\, \big((\mathbb{S}\, \mathbb{L})^{\tilde\mu}{}_\nu\big)
    \\[-.25ex]
  &=\, \det\mathbb{S}\, \det\mathbb{L}\cdot
         (-\vrh)\,
         \pmatrixZZ{\sinh\ps\Dmfm}{\cosh\ps\Dmfm}{\sinh\ps\Dmfp}{-\cosh\ps\Dmfp}
 \end{aligned}
    \tag{\ref{APP:xtilde_x-kontrav}$'$}
    \\[-4.5ex]\nn
\end{align}
und f"ur~kovariante Komponenten:
\vspace*{-.5ex}
\begin{align} \label{APP:xtilde_x-kov}
&x_{\tilde\mu}\;
  =\; \mathbb{S}_{\tilde\mu}{}^{\bar\rh}\,
      \mathbb{L}_{\bar\rh}{}^\nu\; x_\nu\;
  =\; (\mathbb{S}\, \mathbb{L})_{\tilde\mu}{}^\nu\; x_\nu
    \\[1ex]
&\text{mit}\quad
 \begin{aligned}[t]
  \big(\mathbb{S}_{\tilde\mu}{}^{\bar\rh}\,
       \mathbb{L}_{\bar\rh}{}^{\nu}\big)\,
  &=\, \mathbb{S}^{-1\T\:t}\, \mathbb{L}^{-1\T\:t}\,
   =\, \big(\mathbb{L}^{-1}\, \mathbb{S}^{-1}\big){}^{\T\:t}\,
   =\, (\mathbb{S}\, \mathbb{L})^{-1\T\:t}\,
  \equiv\, \big((\mathbb{S}\, \mathbb{L})_{\tilde\mu}{}^\nu\big)
    \\
  &=\, \vrh\,
         \pmatrixZZ{\cosh\ps\Dmfp}{\sinh\ps\Dmfp}{\cosh\ps\Dmfm}{-\sinh\ps\Dmfm}
 \end{aligned}
    \tag{\ref{APP:xtilde_x-kov}$'$}
    \\[-4.5ex]\nn
\end{align}
Bzgl.\@ der Determinanten~\mbox{\,$\det\mathbb{S}$},~\mbox{$\det\mathbb{L}$\,} vgl.\@ Gl.~(\ref{APP:detS,detS^-1t}) beziehungsweise:
\vspace*{-.5ex}
\begin{align} \label{APP:detL,detL^-1t}
&\det \mathbb{L}
  \equiv \det\! \big(\mathbb{L}^{\bar\mu}{}_\nu\big)\;
  =\; \big[
        \det\! \big( \mathbb{L}{}^{-1\T\:t} \big)
      \big]^{-1}
  \equiv\; \big[
        \det\! \big(\mathbb{L}_{\bar\mu}{}^\nu\big)
      \big]^{-1}
    \\[.25ex]
&=\; [-\, g_{+-}]^{-1}\;
  =\; -\, g^{+-}
  =\; -\, 2\al^2
    \nn
    \\[-4.5ex]\nn
\end{align}
vgl.\@ Gl.~(\ref{APP:kontravLC_Def}$'$) und die Gln.~(\ref{APP:gbar}),~(\ref{APP:gbar^-1}).
Wir verf"ugen mit den Gln.~(\ref{APP:xtilde_x-kontrav}),~(\ref{APP:xtilde_x-kontrav}$'$) und~(\ref{APP:xtilde_x-kov}),~(\ref{APP:xtilde_x-kov}$'$) "uber die allgemeinen Transformationsformeln f"ur kontra- beziehungsweise kovariante Lorentz-Tensorkomponenten.
\vspace*{-.5ex}

\paragraph{Epsilon-Pseudotensor%
           ~\bm{\tilde\ep \!\equiv\! \big(\ep^{\tilde\mu\tilde\nu\tilde\rh\tilde\si}\big)}%
           ~mit~\bm{\tilde\mu,\tilde\nu,\tilde\rh,\tilde\si \!\in\! \{\mfp,\mfm,1,2\}}.}
Der Epsilon-Pseudo\-tensor:~\mbox{$\tilde\ep \!\equiv\! \big(\ep^{\tilde\mu\tilde\nu\tilde\rh\tilde\si}\big)$} mit~\mbox{$\tilde\mu,\tilde\nu,\tilde\rh,\tilde\si \!\in\! \{\mfp,\mfm,1,2\}$}, ist definiert durch seine kontravarianten Komponenten.
Diese folgen aus den Gln.~(\ref{APP:xtilde_xbar-kontrav}),~(\ref{APP:xtilde_xbar-kontrav}$'$) mithilfe der Leibnitz'schen Determinantenformel und den Gln.~(\ref{APP:epTensorLC-kontrav}),~(\ref{APP:epTensorLC-kontrav}$'$):
\begin{samepage}
\vspace*{-.5ex}
\begin{align} \label{APP:epTensor-tilde-kontrav}
\ep^{\tilde\mu\tilde\nu\tilde\rh\tilde\si}\;
  &=\; \mathbb{S}^{\tilde\mu}{}_{\bar\al}\,
       \mathbb{S}^{\tilde\nu}{}_{\bar\be}\,
       \mathbb{S}^{\tilde\rh}{}_{\bar\ga}\,
       \mathbb{S}^{\tilde\si}{}_{\bar\de}\;
         \ep^{\bar\al\bar\be\bar\ga\bar\de}
    \\
  &=\; \det\big(\mathbb{S}^{\tilde\mu}{}_{\bar\al}\big)\,
      \det\big(\mathbb{L}^{\bar\mu}{}_\al\big)\cdot
      \pmatrixZV{\tilde\mu}{\tilde\nu}{\tilde\rh}{\tilde\si}
                {\mfp}{\mfm}{1}{2}\;
       \ep^{0123}
    \\[-4.5ex]\nn
\end{align}
entsprechend die kovarianten Komponenten mithilfe der Gln.~(\ref{APP:epTensorLC-kov}),~(\ref{APP:epTensorLC-kov}$'$):
\vspace*{-.5ex}
\begin{align} \label{APP:epTensor-tilde-kov}
\ep_{\tilde\mu\tilde\nu\tilde\rh\tilde\si}\;
  &=\; \mathbb{S}_{\tilde\mu}{}^{\bar\al}\,
       \mathbb{S}_{\tilde\nu}{}^{\bar\be}\,
       \mathbb{S}_{\tilde\rh}{}^{\bar\ga}\,
       \mathbb{S}_{\tilde\si}{}^{\bar\de}\;
         \ep_{\bar\al\bar\be\bar\ga\bar\de}
    \\
  &=\; \det\big(\mathbb{S}_{\tilde\mu}{}^{\bar\al}\big)\,
      \det\big(\mathbb{L}_{\bar\mu}{}^\al\big)\cdot
      \pmatrixZV{\tilde\mu}{\tilde\nu}{\tilde\rh}{\tilde\si}
                 {\mfp}{\mfm}{1}{2}\;
       \ep_{0123}
    \\[-4.5ex]\nn
\end{align}
bzgl.\@ der Determinanten~\mbox{\,$\det\big(\mathbb{S}^{\tilde\mu}{}_{\bar\al}\big)$},~\mbox{\,$\det\big(\mathbb{S}_{\tilde\mu}{}^{\bar\al}\big)$} und~\mbox{\,$\det\big(\mathbb{L}^{\bar\mu}{}_\al\big)$},~\mbox{\,$\det\big(\mathbb{L}_{\bar\mu}{}^\al\big)$} sei verwiesen auf Gl.~(\ref{APP:detS,detS^-1t}) bzw.~(\ref{APP:detL,detL^-1t}).
Unsere Konvention f"ur~\mbox{$\ep \!\equiv\! \big(\ep^{\mu\nu\rh\si}\big)$} ist
\end{samepage}
\vspace*{-.5ex}
\begin{align} \label{APP:ep^0123,ep_0123}
\ep^{0123}\;
  \equiv\; +1\qquad
  \Longrightarrow\qquad
  \ep_{0123}\; =\; -1
    \\[-4.5ex]\nn
\end{align}
vgl.\@ Gl.~(\ref{APP:epTensor-kontrav}$'$).
Das Signum der Indexpermutation~\mbox{$(\tilde\mu,\tilde\nu,\tilde\rh,\tilde\si) \!\to\! (\mfp,\mfm,1,2)$} ist notiert durch das Klammersymbol entsprechend Gl.~(\ref{APP:SignumPermutation}).
\vspace*{-.75ex}

\paragraph{Fixierung von~\bm{\vrh}.}
Die positiv-reelle Konstante~$\vrh$ wird bestimmt durch die Forderung von {\it L"angentreue\/} der Abbildung zwischen den Koordinaten~$\bar\mu$ und~$\tilde\mu$, da"s sie ergo die Determinante der Metrik invariant l"a"st oder "aquivalent~-- vgl.\@ Gl.~(\ref{APP:detS,detS^-1t})~-- die Determinante der Transformationsmatrix identisch Eins ist:
\vspace*{-.5ex}
\begin{alignat}{2} \label{APP:vrh-Forderung}
&\det \tilde{g}\;
    \equiv\; \det\! \big(g_{\tilde\mu\tilde\nu}\big)\vv
  \stackrel{\D!}{=}\vv
  \det \bar{g}\;
    \equiv\; \det\! \big(g_{\bar\mu\bar\nu}\big)&&
    \\[-.5ex]
&\Longleftrightarrow\qquad
 \det \mathbb{S}\;
    \equiv\; \det\! \big(\mathbb{S}^{\tilde\mu}{}_{\bar\nu}\big)\vv
  \stackrel{\D!}{=}\vv 1&
 &\text{f"ur alle\vv $\ps \in (0,\infty)$}
    \tag{\ref{APP:vrh-Forderung}$'$}
    \\[-4.5ex]\nn
\end{alignat}
Dann impliziert Gl.~(\ref{APP:Stilde_mu^nu}$'$) f"ur~\mbox{$\vrh \!\in\! \bbbr^+$}, mit~$\ps \!\in\! (0,\infty)$:
\vspace*{-.5ex}
\begin{align} \label{APP:vrh}
&\vrh^2\vv
  =\; g_{+-}\, \sinh^{-1}\!\ps
    \\[-.5ex]
&\Longleftrightarrow\qquad
  \vrh\;
  \equiv\; \vrh(\ps)\vv
  =\; \big( g_{+-}\, \sinh^{-1}\!\ps \big)^{1\!/\!2}
    \tag{\ref{APP:vrh}$'$}
    \\[-4.5ex]\nn
\end{align}
F"ur die Komponenten des metrischen Tensors~\mbox{\,$\tilde{g} \!\equiv\! \big(g_{\tilde\mu\tilde\nu}\big)$} folgt:
\vspace*{-.5ex}
\begin{align}
&g_{\mfp\mfp}\;
  \equiv\; g_{\mfm\mfm}\;
  =\; \vrh^2\quad
  =\; g_{+-}\, \sinh^{-1}\!\ps
    \label{APP:g_pp-expl} \\[-.5ex]
 &\hspace*{140pt}
  \longrightarrow\vv
      g_{++}\; \equiv\; g_{--}\;
        \equiv\; 0\qquad
   \text{f"ur\vv $\ps \to \infty$}
    \tag{\ref{APP:g_pp-expl}$'$} \\[.75ex]
&g_{\mfp\mfm}\;
  \equiv\; g_{\mfm\mfp}\;
  =\; \vrh^2\, \cosh\ps\quad
  =\; g_{+-}\, \tanh^{-1}\!\ps
    \label{APP:g_pm-expl} \\[-.5ex]
 &\hspace*{164pt}
  \longrightarrow\vv
      g_{+-}\; \equiv\; g_{-+}\qquad
  \text{f"ur\vv $\ps \to \infty$}
    \tag{\ref{APP:g_pm-expl}$'$}
    \\[-5ex]\nn
\end{align}
vgl.\@ \vspace*{-.25ex}Gl.~(\ref{APP:gtilde-kov}) bzw.~(\ref{APP:gtilde-kov}$'$). \\
\indent
Im Hochenergielimes:%
  ~\vspace*{-.125ex}\mbox{$\ps\Dmfp,\ps\Dmfm \!\to\! \infty \Rightarrow \ps \!\to\! \infty$}, gehen {\it per~constructionem\/} die Koordinatenlinien ineinander "uber entsprechend \mbox{\,$\bm{\mfp} \!\to\! \bm{+}$},~\mbox{\,$\bm{\mfm} \!\to\! \bm{-}$},~-- entsprechend die Metriken
\vspace*{-.5ex}
\begin{align} 
\tilde{g}\;
    \equiv\; \big(g_{\tilde\mu\tilde\nu}\big)\vv
  \longrightarrow\vv
  \bar{g}\;
    \equiv\; \big(g_{\bar\mu\bar\nu}\big)\qquad
  \text{f"ur\vv $\ps \to \infty$}
    \\[-4.5ex]\nn
\end{align}
wie verifiziert durch die Gln.~(\ref{APP:g_pp-expl}$'$),~(\ref{APP:g_pm-expl}$'$).
Die \vspace*{-.125ex}Diskrepanz zwischen Hochenergielimes und endlichen Lorentz-Boosts:%
  ~\vspace*{-.125ex}\mbox{\,$\ps\Dmfp,\ps\Dmfm \!<\! \infty \Rightarrow \ps \!<\! \infty$}, ist subsumiert in Faktoren hyperboli\-scher Funktionen und im Nichtverschwinden der {\it Diagonalelemente\/}~$g_{\mfp\mfp}$, $g_{\mfm\mfm}$ der Metrik.\zz
\vspace*{-.75ex}

\paragraph{"`L"angen"' der Weltlinien~\bm{{\cal C}\Dmfp},~\bm{{\cal C}\Dmfm}:%
           ~Eigenzeiten~\bm{T\Dmfp},~\bm{T\Dmfm}.}
\label{APP-T:Tmfp,Tmfm}Die "`L"angen"' der Weltlinien ${\cal C}\Dmfp$,~${\cal C}\Dmfm$~-- die Eigenzeiten~$T\Dmfp$,~$T\Dmfm$ der Propagation der Teilchen~$\mfp$,~$\mfm$ in den Systemen~$I\Dmfp$,~$I\Dmfm$~-- h"angen ab von den~\mbox{Lorentz-Boosts}, das hei"st deren Parameter~$\ps\Dmfp$,~$\ps\Dmfm$. \\
\indent
Wir betrachten das geboostete Teilchen~$\mfp$, im System~$I\Dmfp$.
Es propagiert nach Konstruktion~-- vgl.\@ Abb.~\ref{Fig:MinkowskiTRAJECTORY}~-- mit konstanter Geschwindigkeit "uber die Eigenzeit~$\De\ta \!=\! T\Dmfp$, so da"s seine Weltlinie~${\cal C}\Dmfp$ genau die geradlinige Verbindung sei zwischen den Weltpunkten~$-\tilde\ze$ und~$\tilde\ze$ mit~\mbox{$\pm\tilde\ze \!\equiv\! (\pm\ze^{\tilde\mu}) \!=\! (\pm\ze\Dmfp^\mfp,0,0,0)^{\T t}$}.
Wir haben die explizite Parametrisierung:%
\FOOT{
  Seien~$\mfp$,~$\mfm$ Lorentz-Indizes,~\bm{\mfp},~\bm{\mfm} Notation der Teilchen und deren Ruhsysteme, vgl.\@ Fu"sn.\,\FN{APP-FN:mf<->bm(mf)}.   Seien~ferner in diesem Paragraphen o.E.d.A.\@ die konstanten Koordinaten identisch Null gesetzt.
}
\begin{samepage}
%
\begin{align} \label{APP:Cmfp-Paramet}
{\cal C}\Dmfp\;
  \equiv\; {\cal C}\Dmfp(\la)\;
  =\; \big\{\tilde{\ze}\big|\vv \tilde{\ze}(\la)
              = \ze^{\tilde\mu}\!(\la)\; e_{(\tilde\mu)}
              = (2\la \!-\! 1)\, \ze\Dmfp^\mfp\; e_{(\mfp)},\vv
            \la\!:\, 0 \to 1\big\}
\end{align}
mit~$\la$ dem Kurvenparameter.
Wir berechnen den Zusammenhang von Eigenzeit~$T\Dmfp$ und Parameter~$\ze\Dmfp$ der Weltlinien-Endpunkte.
Aus Gl.~(\ref{APP:Cmfp-Paramet}) folgt unmittelbar:
\vspace*{-.5ex}
\begin{align} 
\frac{d\ze^{\tilde\mu}}{d\la}(\la)\;
  =\; \de^{\tilde\mu}_\mfp\vv 2\, \ze\Dmfp^\mfp
    \\[-4.5ex]\nn
\end{align}
und hiermit f"ur die Eigenzeit~-- dem invarianten Linienelement entlang der Weltlinie:
\end{samepage}
\vspace*{-.5ex}
\begin{align} \label{APP:Tmfp_zeDmfp^mfp-pre}
T\Dmfp\;
  &=\; \int_0^{T\Dmfp} d\ta\quad
   \equiv\; \int_{{\cal C}\Dmfp} ds\;
   =\; \int_0^1 d\la\; \frac{ds}{d\la}\;
   =\; \int_0^1 d\la\;
         \sqrt{g_{\tilde\mu\tilde\nu}\;
               \frac{d\ze^{\tilde\mu}}{d\la}\;
               \frac{d\ze^{\tilde\nu}}{d\la}}\;
    \\[-.5ex]
  &=\; \int_0^1 d\la\;
         \sqrt{g_{\mfp\mfp}\;
               (2\, \ze\Dmfp^\mfp)^2}
  =\; \sqrt{g_{\mfp\mfp}}\cdot 2\, \ze\Dmfp^\mfp
    \tag{\ref{APP:Tmfp_zeDmfp^mfp-pre}$'$}
    \\[-4ex]\nn
\end{align}
umgekehrt die Weltlinien-Endpunkte als Funktion der Eigenzeit der Propagation:%
\FOOT{
  \label{APP-FN:Subst-mfp->mfm}Der Zusammenhang f"ur Teilchen~$\mfm$ folgt in vollst"andiger Analogie durch Substitution~\mbox{$\bm{\mfp} \!\to\! \bm{\mfm}$} und~\mbox{$\mfp \!\to\! \mfm$}.
}
%
\vspace*{-.5ex}
\begin{alignat}{2} \label{APP:Tmfp_zeDmfp^mfp}
&\ze\Dmfp^\mfp&\;
  &=\; 1\!/\!\sqrt{g_{\mfp\mfp}}\, \cdot\, T\Dmfp/2
    \\
&\ze\Dmfm^\mfm&\;
  &=\; 1\!/\!\sqrt{g_{\mfm\mfm}}\, \cdot\, T\Dmfm/2
    \tag{\ref{APP:Tmfp_zeDmfp^mfp}$'$}
    \\
&\text{mit}&&\qquad 
  g_{\mfp\mfp}\;
  =\; g_{\mfm\mfm}\;
  =\; \vrh^2\;
  =\; g_{+-}\, \sinh^{-1}\!\ps
    \label{APP:g_mfpmfp}
    \\[-4.5ex]\nn
\end{alignat}
vgl.\@ die Gln.~(\ref{APP:gtilde-kov}),~(\ref{APP:vrh}).
Gemessen im gemeinsamen initialen Ruhsystem~\mbox{$I\idx{0} \!\equiv\! I\idx[\mfp]{0} \!\equiv\! I\idx[\mfm]{0}$} wandern die Weltlinien-Endpunkte~$(\pm\ze\Dmfp,0,0,0)^{\T t}$,~$(0,\pm\ze\Dmfm,0,0)^{\T t}$ der geboosteten Teilchen im Sinne der Zeitdilatation wie~\mbox{$\sim\! \sinh^{1\!/\!2}\!\ps$} ins Unendliche.
Analog folgt:
\vspace*{-.5ex}
\begin{align}
&T\idx{0}\;
  =\; \sqrt{g_{00}}\cdot 2\, \ze\idx{0}^0
    \\
&\Longleftrightarrow\qquad
  \ze\idx{0}^0\;
  =\; 1\!/\!\sqrt{g_{00}}\, \cdot\, T\idx{0}/2\qquad\qquad
  \text{mit}\qquad
  g_{00} \equiv 1
    \label{APP:Endpkt^0_T0}
    \\[-4.5ex]\nn
\end{align}
f"ur~\mbox{\,$T\idx{0}$} die Eigenzeit der "`Propagation"' in~$I\idx{0}$ und~\mbox{\,$(\pm\ze\idx{0}^0,0,0,0)^{\T t}$} die Weltlinien-Endpunkte.
\vspace*{-.5ex}

\section[Euklidische Koordinaten:%
           ~\protect{$\{\eE{}_{(\tilde\mu)}, \xE^{\tilde\mu}\}$};~-- versus%
           ~\protect{$\{\eE{}_{(\mu)},       \xE^\mu\}$}]{%
         Euklidische Koordinaten:
           ~\bm{\{\eE{}_{(\tilde\mu)}, \xE^{\tilde\mu}\}};~\bm{-}\\ versus%
           ~\bm{\{\eE{}_{(\mu)},       \xE^\mu\}}~\citeFNbf{APP-FN:mf<->bm(mf)}}
\label{APP-Sect:Euklid}

Analog wie auf der physikalischen Minkowskischen Raumzeit werden definiert Koordinaten auf deren Analytischen Fortsetzung ins Euklidische.
Wir entwickeln diese Definitionen.
\vspace*{-.5ex}

\subsection[Analytische Fortsetzung:%
              ~\protect\mbox{\,$\{\eE{}_{(\mu)}, \xE^\mu\}$},%
              ~\protect\mbox{$\mu \!\in\! \{0,1,2,3\}$}
              nach~\protect\mbox{\,$\{e_{(\mu)}, x^\mu\}$},
               \protect\mbox{$\mu \!\in\! \{1,2,3,4\}$}]{%
            Analytische Fortsetzung:%
              ~\bm{\{\eE{}_{(\mu)}, \xE^\mu\}}
              nach~\bm{\{e_{(\mu)}, x^\mu\}}~\bffootnote}
%
\footnotetext{
  Das Skript~"`E"' stellt klar den Raumzeit-Index~$\mu$ als Euklidisch:~\mbox{\,$\mu \!\in\! \{1,2,3,4\}$} statt~\mbox{\,$\in\! \{0,1,2,3\}$}.
}
%

Analytische Fortsetzung der Minkowski-Raumzeit in eine Euklidische sei verstanden als formale Substitution der pseudo-Riemannschen Metrik durch eine Riemannsche im Sinne:
\vspace*{-.5ex}
\begin{align} \label{APP:g->deE}
&g\;
  \equiv\; (g_{\mu\nu})\;
  =\; {\rm diag}[+1,-1,-1,-1]
    \\[-.75ex]
&\phantom{{\iIM} x^0\;}
  \stackrel{\D!}{\longrightarrow}\;
      {\rm diag}[-1,-1,-1,-1]\;
  =\; -(\xE[\de]{}_{\mu\nu})\;
  \equiv\; -\xE[\de]\;
  \equiv\; \xE[g]
    \nn
    \\[-4.5ex]\nn
\end{align}
F"ur zeitartige Lorentz-Komponeneten impliziert dies Konsequenzen, die folgen aus der Forderung von Invarianz des Vierer-Skalarprodukts beliebiger Vektoren~$a$,~$b$:%
\FOOT{
  Unsere Konvention sei: \mbox{$+{\iIM}x^0 \!\to\! \xE^4$}; Euklidische Gr"o"sen tragen konsequent das Skript~"`E"', ihre~(Lo\-rentz)Indizes~$\mu$ beziehen sich dann auf die Werte~\mbox{$\in\! \{1,2,3,4\}$}.
}
%
\vspace*{-.5ex}
\begin{align} \label{APP:xy->xyE}
&a\cdot b\;
  =\; g_{\mu\nu}\, a^\mu\, b^\nu\;
  =\; a^0\, b^0 - \de_{ij}\, a^i\, b^j
  =\; -\, \de_{00}\, ({\iIM} a^0)\, ({\iIM} b^0) - \de_{ij}\; a^i\; b^j
    \\[-1ex]
&\phantom{{\iIM} x^0\;}
  \stackrel{\D!}{\longrightarrow}\;
  \begin{aligned}[t]
  &\xE[a]\cdot \xE[b]\;
    =\; \xE[g]{}_{\mu\nu}\, \xE[a]^\mu\, \xE[b]^\nu\;
    =\; -\, \xE[\de]{}_{\mu\nu}\, \xE[a]^\mu\, \xE[b]^\nu
    \\[-.5ex]
  &\hspace*{115pt}
   =\; -\, \xE[\de]{}_{44}\; \xE[a]^4\; \xE[b]^4
          - \xE[\de]{}_{ij}\; \xE[a]^i\; \xE[b]^j
  \end{aligned}
    \nn
    \\[-5ex]\nn
\end{align}
vgl.\@ Gl.~(\ref{APP:xy=!xybar}).~--
Bezogen auf Weltpunkte~$x$ impliziert Substitution der Metrik~\mbox{$g \!\to\! \xE[g] \!=\! -\xE[\de]$} bei Forderung von Invarianz des Skalarprodukts~-- vgl.\@ die Gln.~(\ref{APP:g->deE}),~(\ref{APP:xy->xyE}):
\begin{samepage}
\vspace*{-.75ex}
\begin{alignat}{2} \label{APP:x->xE}
&{\iIM} x^0&\;
  &\longrightarrow\; \xE^4
    \\
&x^i&\;
  &\longrightarrow\; \xE^i\qquad
  i \in \{1,2,3\}
    \tag{\ref{APP:x->xE}$'$}
    \\[-4.5ex]\nn
\end{alignat}
das hei"st Identifizierung von~\vspace*{-.125ex}$\xE^4$ mit~$\iIM x^0$ und trivial von~$\xE^i$ mit~$x^i$ f"ur~$i \!\in\! \{1,2,3\}$. \\
\indent
Durch Gl.~(\ref{APP:xy->xyE}) ist das Vierer-Skalarprodukt definiert als {\it negativ definit\/}:
\vspace{-.5ex}
\begin{align} \label{APP:xcdoty-E}
\xE[a]\cdot \xE[a]\;
  =\; -\, \xE[a]^\mu\, \xE[a]^\mu\vv
  \le\vv 0
    \\[-4.5ex]\nn
\end{align}
Kovariante Komponenten seien allgemein definiert durch:
\vspace*{-.75ex}
\begin{align} \label{APP:xE-kov}
\xE[a]{}_\mu\;
  :=\; \xE[g]{}_{\mu\nu}\, \xE[a]^\nu\;
   =\; -\xE[\de]{}_{\mu\nu}\, \xE[a]^\nu
    \\[-4ex]\nn
\end{align}
\end{samepage}%
Sie differieren durch ein {\it Signum\/} von den kontravarianten Komponenten:~\mbox{$\xE[a]{}_\mu \!=\! -\xE[a]^\mu$}.
Es sind daher auch bez"uglich Euklidischer Gr"o"sen streng zu unterscheiden kontra- und~ko\-variante (Lorentz-)Tensorkomponenten.~--
Diese Konventionen geschehen in vollst"andiger Analogie zum Minkowskischen; sie reduzieren unsere explizite Rechnung auf ein Minimum. \\
\indent
\vspace*{-.25ex}Der Euklidische Epsilon-Pseudotensor~\mbox{$\xE[\ep] \!\equiv\! \big(\xE[\ep]^{\mu\nu\rh\si}\big)$} sei definiert als das Signum der Permutation seiner Indizes~-- durch seine kontravrianten Komponenten wie folgt:
\vspace*{-.25ex}
\begin{align} \label{APP:epE-Def}
&\xE[\ep]^{\mu_1\cdots\mu_4}\;
  \equiv\; \pmatrixZD{\mu_1}{\cdots}{\mu_4}{1}{\cdots}{4}\; \ep^{1234} 
    \\[-.25ex]
&\text{\sl per definitionem:}\qquad
  \xE[\ep]^{1234}\;
  \equiv\; +1
    \tag{\ref{APP:epE-Def}$'$}
    \\[-4.25ex]\nn
\end{align}
und {\it folglich:\/}%
  ~\vspace*{-.125ex}\mbox{$\xE[\ep]{}_{1234} \!=\!
    \xE[g]{}_{1\mu}\, \xE[g]{}_{2\nu}\, \xE[g]{}_{3\rh}\, \xE[g]{}_{4\si}\,
      \xE[\ep]^{\mu\nu\rh\si} \!=\!
    (-1)^4\, \xE[\ep]^{1234} \!\equiv\! +1$};
vgl.\@ die Gln.~(\ref{APP:epTensor-kontrav}),~(\ref{APP:SignumPermutation}). \\
\indent
Mit~\mbox{\,$\ep_{0123} \!\equiv\! -1$} folgt durch Permutation der Indizes:
\vspace*{-.5ex}
\begin{align} 
\ep_{\mu\nu\rh\si}\;
  =\; \pmatrixZV{\mu}{\nu}{\rh}{\si}  {0}{1}{2}{3}\; \ep_{0123}\;
  =\; (-1)^{1+3}\, \pmatrixZV{\mu}{\nu}{\rh}{\si}  {1}{2}{3}{0}
    \\[-4.5ex]\nn
\end{align}
und damit f"ur beliebige Lorentz-Vektoren~\mbox{\,$a,b,c,d$}:
\vspace*{-.5ex}
\begin{align} \label{APP:ep->epE}
&\ep_{\mu\nu\rh\si}\;
  a^\mu\; b^\nu\; c^\rh\; d^\si\;
  =\; - \iIM\cdot \pmatrixZV{\mu}{\nu}{\rh}{\si}  {1}{2}{3}{0}\vv
        \iIM\cdot a^\mu\; b^\nu\; c^\rh\; d^\si
    \\[-.5ex]
&\phantom{{\iIM} x^0\;}
  \longrightarrow\;
  - \iIM\cdot \pmatrixZV{\mu}{\nu}{\rh}{\si}  {1}{2}{3}{4}\vv
      \xE[a]^\mu\; \xE[b]^\nu\; \xE[c]^\rh\; \xE[d]^\si\;
  =\; - \iIM\cdot \xE[\ep]{}_{\mu\nu\rh\si}\;
           \xE[a]^\mu\; \xE[b]^\nu\; \xE[c]^\rh\; \xE[d]^\si
    \nn
    \\[-4.5ex]\nn
\end{align}
aufgrund der Gln.~(\ref{APP:x->xE}),~(\ref{APP:x->xE}$'$), da genau ein Index den Wert%
  ~\vspace*{-.25ex}\mbox{"`$0$"'} annimmt. \\
\indent
Lorentz-Tensorgleichungen involvieren o.E.\@ Vektorfunktionen%
  ~\vspace*{-.125ex}\mbox{$a^\mu\!(x)$}, den metrischen Tensor~$g$ und den Epsilon-Pseudotensor~$\ep$; es folgt als allgemeine Substitutionsvorschrift f"ur den "Ubergang von Minkowskischer zu Euklidischer Raumzeit~-- vgl.\@ die Gln.~(\ref{APP:xy->xyE}),~(\ref{APP:ep->epE}):
\vspace*{-.5ex}
\begin{alignat}{3} \label{APP:TensorGln:M->E}
&a^\mu\!(x)\;&
  &\longrightarrow\vv
      \phantom{-\iIM\; } \xE[a]^\mu\!(\xE)&&
    \\[-.125ex]
&g_{\mu\nu}\;&
  &\longrightarrow\vv
      \phantom{-\iIM\; } \xE[g]{}_{\mu\nu}\;&
     &=\; -\, \xE[\de]{}_{\mu\nu}
    \tag{\ref{APP:TensorGln:M->E}$'$} \\[-.125ex]
&\ep_{\mu\nu\rh\si}\;&
  &\longrightarrow\vv
      -\iIM\; \xE[\ep]{}_{\mu\nu\rh\si}&&
    \tag{\ref{APP:TensorGln:M->E}$''$}
    \\[-4.125ex]\nn
\end{alignat}
mit Euklidischen Vektor-Komponenten~\vspace*{-.25ex}\mbox{\,$\xE[a]^\mu$} im Sinne der Gln.~(\ref{APP:x->xE}),~(\ref{APP:x->xE}$'$). \\
\indent
Die zugeh"origen Basisvektoren~\vspace*{-.125ex}$\eE{}_{(\mu)} \!\equiv\! (\eE{}_{(\mu)}{}^\nu)$ sind analog definiert wie im Minkowskischen, vgl.\@ die Gln.~(\ref{APP:e-kov-karthesisch}),~(\ref{APP:e-kov-karthesisch}$'$):
\vspace*{-.5ex}
\begin{align} \label{APP:eE-kov-karthesisch}
&\eE{}_{(\mu)}\cdot \eE{}_{(\mu)}\;
  =\; \xE[g]{}_{\mu\nu}
    \\
&\text{mit}\qquad
  \eE{}_{(\mu)}{}^\nu\;
  =\; \xE[g]{}_\mu{}^\nu\;
  =\; \xE[\de]{}_\mu^\nu\qquad
  \mu,\nu \in \{1,2,3,4\}
    \tag{\ref{APP:eE-kov-karthesisch}$'$}
    \\[-4.5ex]\nn
\end{align}
Analog gelten die weiteren Relationen von Anh.~\ref{APP-Subsect:KonvDef}~-- unter Anbringung des Skripts~"`E"' f"ur~"`Euklidisch"' und Bezug der (Lorentz)Indizes auf die Werte~$\in\! \{1,2,3,4\}$.
\vspace*{-.5ex}

\subsection[Aktive~\protect$O(4)$-Drehungen:\:%
              ~\protect\mbox{$\{\eE{}_{(\tilde\mu)}, \xE^{\tilde\mu}\}$}
              versus~\protect\mbox{$\{e_{(\tilde\mu)}, x^{\tilde\mu}\}$},%
              ~\protect\mbox{$\tilde\mu \!\in\! \{\mfp,\mfm,1,2\}$}]{%
            \bm{O(4)}-Drehungen:%
              ~\bm{\{\eE{}_{(\tilde\mu)}, \xE^{\tilde\mu}\}}
              versus~\bm{\{e_{(\tilde\mu)}, x^{\tilde\mu}\}},%
              ~\bm{\tilde\mu \!\in\! \{\mfp,\mfm,1,2\}}}
\label{Subsect:AktiveO(4)Drehungen}

Die Konstruktion des Koordinatensystems~\mbox{$\{\eE{}_{(\tilde\mu)}, \xE^{\tilde\mu}\}$} im Euklidischen wie folgt geschieht~in vollst"andiger Analogie zu der von~\mbox{$\{e_{(\tilde\mu)}, x^{\tilde\mu}\}$} im Minkowskischen, vgl.\@ Abschn.~\ref{APP-Sect:Minkowski}.
Sei~zu\-n"achst der Zusammenhang hergestellt zwischen~$O(4)$-Drehungen und Lorentz-Boosts.
\vspace*{-.5ex}

\begin{figure}
\vspace*{-1ex}
\begin{minipage}{\linewidth}
  \begin{center}
  \setlength{\unitlength}{1mm}\begin{picture}(100,100) 
    \put(0,0){\epsfxsize100mm \epsffile{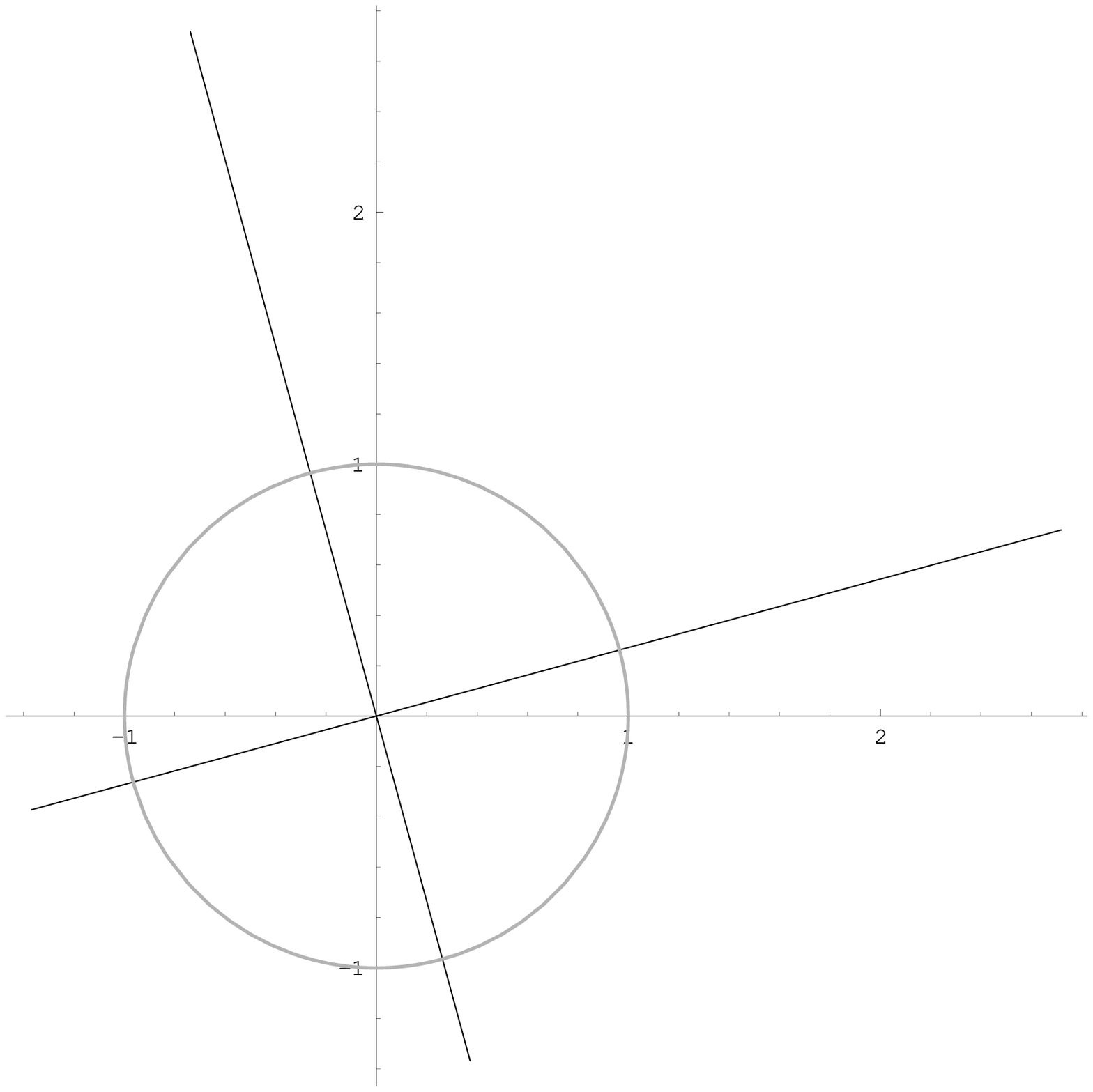}}
    \put(27.5,92){\normalsize$\xE^4$}
    \put(11,92){\normalsize$\xE^{4'}$}
    \put(92,29){\normalsize$\xE^3$}
    \put(92,45.5){\normalsize$\xE^{3'}$}
    \put(29.55,52){\normalsize$\measuredangle\th$}   
    \put(50,35.5){\normalsize$\measuredangle\th$}
    \put(28.12,56.77){\circle*{.8}}
    \put(34.23,57.65){\circle*{.8}}
    \put(56.66,40.48){\circle*{.8}}
    \put(57.55,34.25){\circle*{.8}}
  \end{picture}
  \end{center}
\vspace*{-4ex}
\caption[Euklidische Raumzeit-Diagramm: \protect$O(4)$-Drehung]{
  Euklidische Raumzeit: Allgemeine $O(4)$-Drehung.   Seien~$\xE^4$,~$\xE^3$ die karthe\-sischen Achsen des Euklidischen Bezugsystems~$\xE[I]$.   Das System~$\xE[I]'$ mit Achsen~$\xE^{4'}$,~$\xE^{3'}$ geht hervor aus~$\xE[I]$ wie dargestellt durch eine~$O(4)$-Drehung mit Winkel~$\th$ mit~\mbox{$0 \!\le\! |\th| \!<\! \pi\!/\!4$}.   Die Einsen der Achsen bei definiter Metrik sind bestimmt durch eine Sph"are~-- statt durch Hyperboloide bei indefiniter Metrik.   Vgl.\@ Abb.~\ref{Fig:MinkowskiGENERAL}.
\vspace*{-.5ex}
}
\label{Fig:EuklidGENERAL}
\end{minipage}
\end{figure}
\paragraph{(Aktive) \bm{O(4)}-Drehungen versus (aktive) Lorentz-Boosts.}
Wir betrachten nach Abbildung~\ref{Fig:EuklidGENERAL} die~$O(4)$-Drehung in der~$\xE^4\!\xE^3$-Ebene um den Winkel~\mbox{\,$0 \!<\! \th \!<\! \pi\!/\!4$}.
F"ur~$\xE$ ein beliebiger Vektor der Raumzeit gilt in {\it passiver\/} Auffassung, vgl.\@ oben Seite\,\pageref{APP-Para:passiv}:%
\FOOT{
  \label{APP-FN:4vor3}\vspace*{-.25ex}Wir schreiben aus nur die nichttrivialen longitudinalen Komponenten, vgl.\@ Fu"sn.\,\FN{APP-FN:only-longitudinal} und~\FN{APP-FN:only-longitudinal-Rep},~-- f"ur Kommensurabilit"at die Vier- vor der Drei-Komponente:~\mbox{$\xM[(0,3)] \!\leftrightarrow\! \xM[(+,-)] \!\leftrightarrow\! \xM[(\mfp,\mfm)]$} versus~\mbox{$\xE[(4,3)] \!\leftrightarrow\! \xE[(\mfp,\mfm)]$}.
}
%
\vspace*{-.5ex}
\begin{align} \label{APP:allgO(4)-passiv}
&\xE^{\mu'}\;
  =\; \xE[\La]{}^\mu{}_\nu\; \xE^\nu\qquad
    \mu,\,\nu \in \{1,2,3,4\}
    \\[-.25ex]
&\text{mit}\qquad
  \xE[\La]\;
  \equiv\; \big(\xE[\La]{}^\mu{}_\nu\big)\;
  =\; \pmatrixZZ{\cos\th}{-\sin\th}{\sin\th}{\cos\th}
    \tag{\ref{APP:allgO(4)-passiv}$'$}
    \\[-4.25ex]\nn
\end{align}
explizit:
\vspace*{-.5ex}
\begin{align} \label{APP:O(4)-passiv0}
\pmatrixZE{\xE^{4'}}{\xE^{3'}}\;
  =\; \pmatrixZZ{\cos\th}{-\sin\th}{\sin\th}{\cos\th}\;
      \pmatrixZE{\xE^4}{\xE^3}
    \\[-4.5ex]\nn
\end{align}
Die Gleichung der~$\xE^{4'}$-Achse:~\mbox{$\xE^{3'} \!=\! 0$}, ist gegeben durch \mbox{$\xE^4 \!=\! -[\tan\th]^{-1}\: \xE^3$}, die der~$\xE^{3'}$-Achse: \mbox{$\xE^{4'} \!=\! 0$}, entsprechend durch~\mbox{$\xE^4 \!=\! \tan\th\: \xE^3$}.
Mit
\vspace*{-.5ex}
\begin{alignat}{6} \label{APP:cosh-cos,sinh-sin}
&\cosh z&\;
  &=\;& &\cos(-\iIM z)&\quad
  &\Longleftrightarrow\quad
 \cos z'&\;
  &=\;& &\cosh(\iIM z')
    \\[.25ex]
&\sinh z&\;
  &=\;& \iIM\, &\sin(-\iIM z)&\quad
  &\Longleftrightarrow\quad
 \sin z'&\;
  &=\;& -\iIM\, &\sinh(\iIM z')\qquad
    \forall z,z' \in \bbbc
    \tag{\ref{APP:cosh-cos,sinh-sin}$'$}
    \\[-4.5ex]\nn
\end{alignat}
folgt aus Gl.~(\ref{APP:O(4)-passiv0}) unmittelbar:
\vspace*{-.5ex}
\begin{align} \label{APP:allgO(4)-passiv1}
\pmatrixZE{\xE^{4'}}{\xE^{3'}}\;
  &=\; \pmatrixZZ{\cosh\, \iIM\th}{\iIM\: \sinh\, \iIM\th}
                {-\iIM\: \sinh\, \iIM\th}{\cosh\, \iIM\th}\;
      \pmatrixZE{\xE^4}{\xE^3}
    \\[-.25ex]
  &=\; \pmatrixZZ{\iIM\: \cosh\, \iIM\th}{\iIM\: \sinh\, \iIM\th}
                {\sinh\, \iIM\th}{\cosh\, \iIM\th}\;
      \pmatrixZE{-\iIM\,\xE^4}{\xE^3}
    \tag{\ref{APP:allgO(4)-passiv1}$'$}
    \\[-4.5ex]\nn
\end{align}
Multiplikation der ersten Zeile mit dem Faktor~\mbox{\,$-\iIM$} ergibt weiter:
\vspace*{-.5ex}
\begin{align} \label{APP:allgO(4)-passiv2}
\pmatrixZE{-\iIM\,\xE^{4'}}{\xE^{3'}}\;
  =\; \pmatrixZZ{\cosh\, \iIM\th}{\sinh\, \iIM\th}
                 {\sinh\, \iIM\th}{\cosh\, \iIM\th}\;
      \pmatrixZE{-\iIM\,\xE^4}{\xE^3}
    \\[-4.5ex]\nn
\end{align}
In Konfrontation mit
\vspace*{-.5ex}
\begin{align} \label{APP:allgO(4)-passiv-M}
\pmatrixZE{x^{0'}}{x^{3'}}\;
  =\; \pmatrixZZ{\cosh\ps}{-\sinh\ps}
                {-\sinh\ps}{\cosh\ps}
      \pmatrixZE{x^0}{x^3}
    \\[-4.5ex]\nn
\end{align}
vgl.\@ die Gln.~(\ref{APP:LT}),~(\ref{APP:LT-La3}),~-- folgt mit~\mbox{${\iIM} x^0 \!\leftrightarrow\! \xE^4$} und~\mbox{$x^i \!\leftrightarrow\! \xE^i$},~\mbox{$i \!\in\! \{1,2,3\}$}, vgl.\@ die Gln.\,(\ref{APP:x->xE}), (\ref{APP:x->xE}$'$), und~\mbox{\,$\cosh z \!=\! \cosh-z$},~\mbox{\,$\sinh z \!=\! -\sinh-z$},~$\forall z \!\in\! \bbbc$, der Zusammenhang:~\mbox{\,$-\iIM\,\th \!\leftrightarrow\! \ps$}. \\
\indent
Es gilt anlog in {\it aktiver\/} Auffassung, vgl.\@ oben Seite\,\pageref{APP-Para:aktiv}:
\vspace*{-.5ex}
\begin{align} \label{APP:allgO(4)-aktiv}
&\xE^{\prime\mu}\;
  =\; \xE[\La]{}_\nu{}^\mu{}\; \xE^\nu\qquad
    \mu,\,\nu \in \{1,2,3,4\}
    \\[-.25ex]
&\text{mit}\qquad
  \xE[\La]^{\!-1\T\:t}\;
  \equiv\; \big(\xE[\La]{}_\nu{}^\mu{}\big)\;
  =\; \pmatrixZZ{\cos\th}{-\sin\th}{\sin\th}{\cos\th}
    \tag{\ref{APP:allgO(4)-aktiv}$'$}
    \\[-4.5ex]\nn
\end{align}
explizit:
\vspace*{-.5ex}
\begin{align} \label{APP:O(4)-aktiv0}
\pmatrixZE{\xE^{\prime4}}{\xE^{\prime3}}\;
  =\; \pmatrixZZ{\cos\th}{\sin\th}{-\sin\th}{\cos\th}\;
      \pmatrixZE{\xE^4}{\xE^3}
    \\[-4.5ex]\nn
\end{align}
An Stelle von Gl.~(\ref{APP:allgO(4)-passiv2}) tritt:%
\FOOT{
  Substitution~$\th \!\to\! -\th$
}
%
\vspace*{-.5ex}
\begin{align} \label{APP:allgO(4)-aktiv2}
\pmatrixZE{-\iIM\,\xE^{\prime4}}{\xE^{\prime3}}\;
  =\; \pmatrixZZ{\cosh\, -\iIM\th}{\sinh\, -\iIM\th}
                {\sinh\, -\iIM\th}{\cosh\, -\iIM\th}\;
      \pmatrixZE{-\iIM\,\xE^4}{\xE^3}
    \\[-4.5ex]\nn
\end{align}
In Konfrontation mit
\vspace*{-.5ex}
\begin{align} \label{APP:allgO(4)-aktiv-M}
\pmatrixZE{x^{\prime0}}{x^{\prime3}}\;
  =\; \pmatrixZZ{\cosh\ps}{\sinh\ps}
                {\sinh\ps}{\cosh\ps}
      \pmatrixZE{x^0}{x^3}\;
    \\[-4.5ex]\nn
\end{align}
vgl.\@ die Gln.~(\ref{APP:LT-La-tInv-3}),~(\ref{APP:LT-La-tInv-3}$'$),~-- folgt wie oben der Zusammenhang:~\mbox{\,$-\iIM\,\th \!\leftrightarrow\! \ps$}.
\vspace*{-.5ex}

\paragraph{Koordinatenlinien~\bm{\xE[\tilde\mu] \in \{\mfp,\mfm,1,2\}}.}
Im Minkowskischen werden betrachtet aktive Lo\-rentz-Boosts~$\La\Dmfp$,~$\La\Dmfm$ gegeneinander, die induzieren Basisvektoren%
~\mbox{\,$e'_{(\mu)} \!=\! (\La\Dmfp)_\mu{}^\nu\; e_{(\nu)}$}
und \mbox{\,$e^\dbprime_{(\mu)} \!=\! (\La\Dmfm)_\mu{}^\nu\; e_{(\nu)}$}, mit~\mbox{$\mu,\nu \!\in\! \{0,1,2,3\}$}, vgl.\@ die Gln.~(\ref{APP:e'-nach-e_Imf}), (\ref{APP:e'-nach-e_Imf}$'$) und Gl.~(\ref{APP:P'_i}$'$).
Im Euklidischen treten an ihre Stelle respektive: aktiv zu verstehende~$O(4)$-Drehungen~$\xE[\La\Dmfp]$,~$\xE[\La\Dmfm]$ und Vektoren~$\eE'{}_{(\mu)}$,~$\eE^\dbprime{}_{(\mu)}$ wie folgt:
\begin{samepage}
\vspace*{-.5ex}
\begin{align} \label{APP:e'-nach-e_ImfE}
&\eE'{}_{(\mu)}\;
  =\; (\xE[\La\Dmfp])_\mu{}^\nu\; \eE{}_{(\nu)}\qquad
    \mu,\,\nu \in \{1,2,3,4\}
    \\[-.25ex]
&\text{mit}\qquad
  \big((\xE[\La\Dmfp])_\mu{}^\nu\big)\;
    =\; \pmatrixZZ{\cos\th\Dmfp}{-\sin\th\Dmfp}
                  {\sin\th\Dmfp}{\cos\th\Dmfp}
    \tag{\ref{APP:e'-nach-e_ImfE}$'$}
    \\[-5ex]\nn
\end{align}
und
\vspace*{-.5ex}
\begin{align} \label{APP:e''-nach-e_ImfE}
&\eE^\dbprime{}_{(\mu)}\;
  =\; (\xE[\La\Dmfm])_\mu{}^\nu\; \eE{}_{(\nu)}\qquad
    \mu,\,\nu \in \{1,2,3,4\}
    \\[-.25ex]
&\text{mit}\qquad
  \big((\xE[\La\Dmfm])_\mu{}^\nu\big)\;
    =\; \pmatrixZZ{\cos\th\Dmfm}{\sin\th\Dmfm}
                 {-\sin\th\Dmfm}{\cos\th\Dmfm}
    \tag{\ref{APP:e''-nach-e_ImfE}$'$}
    \\[-4.5ex]\nn
\end{align}
Seien analog diese Drehungen aufgefa"st im Sinne {\it nichtverschwindender~$O(4)$-Transforma\-tionen gegeneinander\/}, die Drehwinkel definiert als {\it positiv\/}:%
\FOOT{
  \label{APP-FN:OrientierungM->E}Mit der Vorzeichenkonvention der Gln.~(\ref{APP:e'-nach-e_ImfE}$'$),~(\ref{APP:e''-nach-e_ImfE}$'$) vermitteln~\mbox{$\xE[\La\Dmfp]$},~\mbox{$\xE[\La\Dmfm]$} bzgl.\@ Abb.~\ref{Fig:EuklidGENERAL} Drehungen ergo in {\sl mathematisch positiven\/} bzw.\@ {\sl negativen\/} Sinne; die Darstellung der Weltlinien im Euklidischen ist gegeben durch Abb.~\ref{Fig:MinkowskiTRAJECTORY} unter Substitution der Achsen~\mbox{\,$x^0 \!\to\! \xE^4$},~\mbox{\,$x^3 \!\to\! \xE^3$} und Trajektorien~\mbox{\,${\cal C}\Dmfm \!\to\! \eE[{\cal C}\Dmfp]$}, \mbox{\,${\cal C}\Dmfp \!\to\! \eE[{\cal C}\Dmfm]$}.   Notation der Vier- vor der Drei-Komponente, vgl.\@ Fu"sn.\,\FN{APP-FN:4vor3}, entspricht einem {\sl Linkssystem\/}.
}
%
\vspace*{-.5ex}
\begin{alignat}{2}
&0\;
  <\; \th\Dimath\;
  <\; \pi/4&&\qquad
  \imath = \mfp, \mfm
    \label{APP:thp,thm}
    \\[-6ex]\nn
\intertext{\vspace*{-2ex}Dann gilt, vgl.\@ oben:}   
&-\, \iIM\,\th\Dimath\vv
  \longleftrightarrow\vv \ps\Dimath&&\qquad
  \imath = \mfp, \mfm
    \label{APP:thp,thm-psp,psm}
    \\[-4.5ex]\nn
\end{alignat}
\end{samepage}%
als Zusammenhang im Sinne analytischer Fortsetzung mit~\mbox{$0 \!<\! \ps\Dimath \!<\! \infty$},~\mbox{$\imath \!=\! \mfp,\mfm$}, den hyperbolischen Winkeln der Lorentz-Boosts; vgl.\@ Gl.~(\ref{APP:psp,psm-Def}).
Wir merken an den Zusammenhang der Drehwinkel~$\th\Dimath$ im Euklidischen und~$\al$,~\mbox{\,$\tan\al \!:=\! \be$}, im Minkowskischen; es gilt~\mbox{\,$0 \!<\! \al \!<\! \pi\!/\!4$} f"ur~\mbox{\,$0 \!<\! \be \!<\! 1$} und~\mbox{\,$\al \!\to\! \pi\!/\!4$} im Limes~\mbox{\,$\be \!\to\! 1$}.
Vgl.\@ die Gln.~(\ref{APP:al_be}),~(\ref{APP:ps-hyperbolWinkel}$'$) und Abb.~\ref{Fig:MinkowskiGENERAL}. \\
%
\indent
Es werden longitudinale Koordinaten~defi\-niert bez"uglich der Richtungen der transformierten Weltlinien, ergo bez"uglich der Zeit-, das hei"st der Vier-Achsen~\mbox{$\eE^\prime{}_{(4)}$},~\mbox{$\eE^\dbprime{}_{(4)}$}:
\vspace*{-.25ex}
\begin{alignat}{5} \label{APP:etilde_eE}
&\eE{}_{(\mfp)}&\;
  &:=\;& \xE[\vrh]\;
          \big(\cos\th\Dmfp\; &\eE{}_{(4)}&\;
         &-\;& \sin\th\Dmfp\; &\eE{}_{(3)}\big)
    \\[.5ex]
&\eE{}_{(\mfm)}&\;
  &:=\;& \xE[\vrh]\;
          \big(\cos\th\Dmfm\; &\eE{}_{(4)}&\;
         &+\;& \sin\th\Dmfm\; &\eE{}_{(3)}\big)
    \tag{\ref{APP:etilde_eE}$'$}
    \\[-4.5ex]\nn
\end{alignat}
\vspace*{-3.5ex}
\begin{align}
\text{und}\qquad
\eE{}_{(\tilde{i})}\;
  :=\; \eE{}_{(i)}\qquad
    i \in \{1,2\}
    \tag{\ref{APP:etilde_eE}$''$}
    \\[-4.25ex]\nn
\end{align}
mit~\mbox{$\xE[\vrh] \!\in\! \bbbr^+$} einer geeignet zu w"ahlenden Konstanten; vgl.\@ die Gln.~(\ref{APP:etilde_e})-(\ref{APP:etilde_e}$''$). \\
\indent
Die Vektoren~$\eE{}_{(\tilde\mu)}$,~$\tilde\mu \!\in\! \{\mfp,\mfm,1,2\}$, repr"asentieren Koordinatenlinien, diebez"uglich induziert wird ein metrischer Tensor~\mbox{$\xE[\tilde{g}] \!\equiv\! \big(\xE[g]{}_{\tilde\mu\tilde\nu}\big)$} durch:
\vspace*{-.5ex}
\begin{align} \label{APP:etilde-kov-karthesischE}
&\eE{}_{(\tilde\mu)}\cdot \eE{}_{(\tilde\nu)}\;
  =\; \xE[g]{}_{\tilde\mu\tilde\nu}
    \\[.25ex]
&\text{mit}\qquad
\eE{}_{(\tilde\mu)}{}^{\tilde\nu}\;
  =\; \xE[g]{}_{\tilde\mu}{}^{\tilde\nu}\;
  =\; \xE[\de]{}_{\tilde\mu}^{\tilde\nu}\qquad
  \tilde\mu,\tilde\nu \in \{\mfp,\mfm,1,2\}
    \tag{\ref{APP:etilde-kov-karthesischE}$'$}
\end{align}
vgl.\@ die Vektoren~$e_{(\tilde\mu)}$,~$\tilde\mu \!\in\! \{\mfp,\mfm,1,2\}$ im Minkowskischen, die Gln.~(\ref{APP:etilde-kov-karthesisch}), (\ref{APP:etilde-kov-karthesisch}$'$).
\vspace*{-.5ex}

\paragraph{Basisvektoren~\bm{\eE^{(\tilde\mu)}},~\bm{\eE{}_{(\tilde\mu)}}
           und induzierte
           Metrik~\bm{\xE[\tilde{g}] \!\equiv\! \big(\xE[g]{}_{\tilde\mu\tilde\nu}\big)},%
           ~\bm{\xE[\tilde\mu] \in \{\mfp,\mfm,1,2\}}.}
Die Gln.~(\ref{APP:etilde_eE})-(\ref{APP:etilde_eE}$''$) lauten in Ma\-trixform:%
\FOOT{
  Im Minkowskischen geschieht die Transformation "uber Lichtkegelkoordinaten: zun"achst~$\mathbb{L}$, dann~$\mathbb{S}$; sei hier die Matrix der unmittelbaren Transformation suggestiv bezeichnet mit beiden Buchstaben:~\mbox{$\slE \!\equiv\! \xE[(\mathbb{SL})]$}.
}
%
\vspace*{-.5ex}
\begin{align} \label{APP:etilde_ebarM-kovE}
\eE{}_{(\tilde\mu)}\;
  =\; \slE{}_{\tilde\mu}{}^\nu\; \eE{}_{(\nu)}\qquad
    \tilde\mu \in \{\mfp,\mfm,1,2\},\vv
    \nu \in \{1,2,3,4\}
    \\[-4.5ex]\nn
\end{align}
dabei ist
\vspace*{-.5ex}
\begin{align} \label{APP:Stilde_mu^nuE}
&\big(\slE{}_{\tilde\mu}{}^\nu\big)\;
  =\; \xE[\vrh]\,
      \pmatrixZZ{\cos\th\Dmfp}{-\sin\th\Dmfp}
                {\cos\th\Dmfm}{\sin\th\Dmfm}
    \\[.25ex]
&\text{mit}\qquad
  \det\! \big(\slE{}_{\tilde\mu}{}^\nu\big)\;
  =\; \xE[\vrh]^2\; \sin\th
    \tag{\ref{APP:Stilde_mu^nuE}$'$}
    \\[-4.5ex]\nn
\end{align}
unter Definition
\vspace*{-.5ex}
\begin{align} \label{APP:th=thp+thm}
\th\;
  :=\; \th\Dmfp + \th\Dmfm
    \\[-4.5ex]\nn
\end{align}
entsprechend~$\ps \!=\! \ps\Dmfp \!+\! \ps\Dmfm$ nach Gl.~(\ref{APP:ps=psp+psm}). \\
\indent
Analog zu Gl.~(\ref{APP:emu-dot-enu-kov}) im Minkowskischen gilt f"ur die Transformation des (Vierer)Skalar\-produkts der Basisvektoren:
%
\begin{align} 
\eE{}_{(\tilde\mu)}\cdot \eE{}_{(\tilde\nu)}\;
   =\; \slE{}_{\tilde\mu}{}^\al\;
       \slE{}_{\tilde\nu}{}^\be\;
         \eE{}_{(\al)}\cdot \eE{}_{(\be)}
\end{align}
Mithilfe von Gl.~(\ref{APP:eE-kov-karthesisch}) und~(\ref{APP:etilde-kov-karthesischE}) gilt daher:
\begin{samepage}
%
\begin{align} \label{APP:S-pseudo-orthE}
\xE[g]{}_{\tilde\mu\tilde\nu}\;
  =\; \slE{}_{\tilde\mu}{}^\al\;
      \slE{}_{\tilde\nu}{}^\be\;
         \xE[g]{}_{\al\be}
\end{align}
In Form~\mbox{\,$\xE[g]{}_{\tilde\mu\tilde\nu} \!=\!
  -\, \big(\slE{}_{\tilde\mu}{}^{4}\, \slE{}_{\tilde\nu}{}^{4}
     \!+\! \slE{}_{\tilde\mu}{}^{3}\, \slE{}_{\tilde\nu}{}^{3}\big)$} folgen aus der expliziten Gestalt der Matrix~$\big(\slE{}_{\tilde\mu}{}^\nu\big)$ nach Gl.~(\ref{APP:Stilde_mu^nuE}) die Komponeneten~\mbox{\,$\xE[g]{}_{\tilde\mu\tilde\nu}$} des Metrischen Tensors::
\vspace*{-.5ex}
\begin{alignat}{3} \label{APP:gtilde-kovE}
&\xE[g]{}_{\mfp\mfp}&\;
  &=\; \xE[g]{}_{\mfm\mfm}&\;
  &=\; -\xE[\vrh]^2
    \\[.25ex]
&\xE[g]{}_{\mfp\mfm}&\;
  &=\; \xE[g]{}_{\mfm\mfp}&\;
  &=\; -\xE[\vrh]^2\cdot \cos\th
    \tag{\ref{APP:gtilde-kovE}$'$}
    \\[-4.5ex]\nn
\end{alignat}
\vspace*{-4ex}
\begin{align}
\text{und}\qquad
\xE[g]{}_{\tilde{i}\tilde{j}}\;
  =\; \xE[g]{}_{ij}\;
  =\; -\de_{ij}\qquad
  i \in \{1,2\}
    \tag{\ref{APP:gtilde-kovE}$''$}
    \\[-4.5ex]\nn
\end{align}
Vgl.\@ die Gln.~(\ref{APP:gtilde-kov})-(\ref{APP:gtilde-kov}$''$) im Minkowskischen.
Als die Determinante des metrischen Tensors folgt:
\vspace*{-.5ex}
\begin{align} \label{APP:det-gtildeE}
\det \xE[\tilde{g}]\;
  =\; (\xE[\vrh]^2\; \sin\th)^2
    \\[-4.5ex]\nn
\end{align}
\end{samepage}%
und umgekehrt:~\mbox{\,$\xE[\vrh]^2\,\sin\th \!=\! \sqrt{\det\xE[\tilde{g}]}$\,}; vgl.\@ Gl.~(\ref{APP:det-gtilde}).
Die inverse Metrik~\mbox{$\xE[\tilde{g}]^{-1} \!\equiv\! \big(\xE[g]^{\tilde\mu\tilde\nu}\big)$} exis\-tiert, falls nicht beide Drehwinkel verschwinden, wie bereits angenommen ist mit Gl.~(\ref{APP:thp,thm}).
Wir finden dann:
\vspace*{-.5ex}
\begin{alignat}{3} \label{APP:gtilde-kontravE}
&\xE[g]^{\mfp\mfp}&\;
  &=\; \xE[g]^{\mfm\mfm}&\;
  &=\; \phantom{-\,} \xE[g]{}_{\mfp\mfp}\; \big/ \det\xE[\tilde{g}]
    \\[.5ex]
&\xE[g]^{\mfp\mfm}&\;
  &=\; \xE[g]^{\mfm\mfp}&\;
  &=\; -\, \xE[g]{}_{\mfp\mfm}\; \big/  \det\xE[\tilde{g}]
    \tag{\ref{APP:gtilde-kontravE}$'$}
    \\[-4.5ex]\nn
\end{alignat}
\vspace*{-4ex}
\begin{align}
\text{und}\qquad
\xE[g]^{\tilde{i}\tilde{j}}\;
  =\; \xE[g]^{ij}\;
  =\; -\de^{ij}\qquad
  i \in \{1,2\}
    \tag{\ref{APP:gtilde-kontravE}$''$}
    \\[-4.5ex]\nn
\end{align}
vgl.\@ die Gln.~(\ref{APP:gtilde-kontrav}),~(\ref{APP:gtilde-kontrav}$'$). \\
\indent
Analog zu den Lorentz-Vektoren~$e^{(\tilde\mu)}$,~\mbox{$\tilde\mu \!\in\! \{\mfp,\mfm,1,2\}$} im Minkowskischen~-- vgl.\@ die Gln. (\ref{APP:etilde^(mu)-Def}),~(\ref{APP:etilde^(mu)-Def}$'$)~-- seien definiert im Euklidischen die Vektoren~$\eE^{(\tilde\mu)}$,~\mbox{$\tilde\mu \!\in\! \{\mfp,\mfm,1,2\}$}, durch:
\vspace*{-.5ex}
\begin{align} \label{APP:etilde^(mu)-DefE}
&\eE^{(\tilde\mu)}\;
  :=\; \xE[g]^{\tilde\mu\tilde\nu}\; \eE{}_{(\tilde\nu)}
    \\[.25ex]
&\Longrightarrow\qquad
  \eE{}_{(\tilde\mu)}\;
  =\; \xE[g]{}_{\tilde\mu\tilde\nu}\; \eE^{(\tilde\nu)}\qquad
    \tilde\mu \in \{\mfp,\mfm,1,2\}
    \tag{\ref{APP:etilde^(mu)-DefE}$'$}
    \\[-4.5ex]\nn
\end{align}
Dann gilt:
%
\begin{align} \label{APP:etilde_ebarM-kontravE}
\eE^{(\tilde\mu)}\;
  =\; \slE{}^{\tilde\mu}{}_\nu\; \eE^{(\nu)}\qquad
    \tilde\mu \in \{\mfp,\mfm,1,2\},\vv
    \nu \in \{1,2,3,4\}
\end{align}
unter Definition
\vspace*{-.5ex}
\begin{align} \label{APP:Stilde^mu_nu-DefE}
\slE{}^{\tilde\mu}{}_\nu\;
  :=\; \xE[g]^{\tilde\mu\tilde\rh}\vv
       \slE{}_{\tilde\rh}{}^\si\vv
       \xE[g]{}_{\si\nu}
    \\[-4ex]\nn
\end{align}
vgl.\@ die Gln.~(\ref{APP:etilde_ebarM-kontrav}),~(\ref{APP:Stilde^mu_nu-Def}). \\
\indent
In vollst"andiger Analogie zum Minkowskischen~-- diskutieren wir und geben wir explizit an die Transformationsmatrix~\mbox{$\big(\slE{}^{\tilde\mu}{}_\nu\big)$}.
Kontraktion von Gl.~(\ref{APP:Stilde^mu_nu-DefE}) mit~$\xE[g]{}_{\tilde\al\tilde\mu}$ ergibt~[Umbenennung~\mbox{$\al \!\to\! \mu$}]:
\vspace*{-.5ex}
\begin{align} \label{APP:S,S-1t-preE}
&\xE[g]{}_{\tilde\mu\tilde\rh}\;
  \slE{}^{\tilde\rh}{}_\nu\;
  =\; \slE{}_{\tilde\mu}{}^\si\;
        \xE[g]{}_{\si\nu}
    \\
&\begin{aligned}[t]
  \Longleftrightarrow\qquad
 \slE{}_{\tilde\mu}{}^\si\;
 &\slE{}_{\tilde\rh}{}^\be\;
  \slE{}^{\tilde\rh}{}_\nu\;
    \xE[g]{}_{\si\be}\;
  =\; \slE{}_{\tilde\mu}{}^\si\;
        \xE[g]{}_{\si\nu}
    \\
  \text{d.h.}\qquad
 &\slE{}_{\tilde\rh}{}^\be\, \mathbb{S}^{\tilde\rh}{}_\nu\;
    =\; \xE[\de]{}^\be_\nu
 \end{aligned}
    \tag{\ref{APP:S,S-1t-preE}$'$}
    \\[-4.5ex]\nn
\end{align}
und weiter~\mbox{%
  \,$\slE{}^{\tilde\mu}{}_\be\, \slE{}_{\tilde\rh}{}^\be \!=\! \xE[\de]{}^\be_\nu$} durch Kontraktion mit~\mbox{\,$\slE{}^{\tilde\mu}{}_{\bar\be}$}.
Zusammenfassend:
\vspace*{-.25ex}
\begin{align} \label{APP:S-kontragredientE}
&\slE{}_{\tilde\rh}{}^\mu\;
  \slE{}^{\tilde\rh}{}_\nu\;
  =\; \xE[\de]{}^\mu_\nu\qquad
 \slE{}^{\tilde\mu}{}_\rh\;
  \slE{}_{\tilde\nu}{}^\rh\;
  =\; \xE[\de]{}^{\tilde\mu}_{\tilde\nu}
    \\[.25ex]
&\text{d.h.}\qquad
 \big(\slE{}^{\tilde\mu}{}_\nu\big)^{-1}\;
  =\; \big(\slE{}_{\tilde\mu}{}^\nu\big){}^{\T t}
    \tag{\ref{APP:S-kontragredientE}$'$}
    \\[-4.25ex]\nn
\end{align}
Die Matrizen~$\big(\slE{}^{\tilde\mu}{}_\nu\big)$ und~$\big(\slE{}_{\tilde\mu}{}^\nu\big)$ sind {\it kontragredient\/}: die eine die transponierte Inverse der anderen.
Wir bezeichnen:
\vspace*{-.5ex}
\begin{align} \label{APP:S,S^-1tE}
&\slE{}\;
  \equiv\; \big(\slE{}^{\tilde\mu}{}_\nu\big)\qquad
    \\[.25ex]
&\Longrightarrow\qquad
 \big(\slE{}_{\tilde\mu}{}^\nu\big)\;
  \equiv\; \slE{}^{-1\T\:t}\qquad
    \tilde\mu \in \{\mfp,\mfm,1,2\},\vv
    \bar\nu \in \{1,2,3,4\}
    \tag{\ref{APP:S,S^-1tE}$'$}
    \\[-4.5ex]\nn
\end{align}
Durch Kontraktion von Gl.(\ref{APP:S,S-1t-preE}) mit~$\slE{}^{\tilde\mu}{}_\rh$ folgt mithilfe von Gl.~(\ref{APP:S-kontragredient}):
%
\begin{align} 
\xE[g]{}_{\tilde\mu\tilde\nu}\;
  \slE{}^{\tilde\mu}{}_\rh\;
  \slE{}^{\tilde\nu}{}_\si\;
  =\; \xE[g]{}_{\rh\si}\qquad
  \text{d.h.}\qquad
\slE^{\,\T t}\vv \xE[\tilde{g}]\vv \slE\;
  =\; \xE[g]
\end{align}
das hei"st~\mbox{\,$\slE \!\equiv\! \big(\slE{}^{\tilde\mu}{}_\nu\big)$} ist {\it pseudo-orthogonal\/}.~--
Vgl.\@ die Matrizen~\mbox{\,$\mathbb{S} \!\equiv\! \big(\mathbb{S}^{\tilde\mu}{}_{\bar\nu}\big)$}, \mbox{\,$\mathbb{S}^{-1\T\:t} \!\equiv\! \big(\mathbb{S}_{\tilde\mu}{}^{\bar\nu}\big)$} und~\mbox{\,$\mathbb{L} \!\equiv\! \big(\mathbb{L}^{\bar\mu}{}_\nu\big)$}, \mbox{\,$\mathbb{L}^{-1\T\:t} \!\equiv\! \big(\mathbb{L}_{\bar\mu}{}^\nu\big)$}~-- folglich~\mbox{\,$\mathbb{SL} \!\equiv\! \big(\mathbb{S}^{\tilde\mu}{}_{\bar\rh}\, \mathbb{L}^{\bar\rh}{}_\nu\big)$},\ldots~-- im Minkowskischen. \\
\indent
F"ur die Determinante von~$\slE$ impliziert Gl.~(\ref{APP:S-kontragredientE}$'$) ferner:
\begin{samepage}
\vspace*{-.5ex}
\begin{alignat}{2} \label{APP:detSL-detSL^-1t-E}
&\det\slE
  \equiv \det\big(\slE{}^{\tilde\mu}{}_\al\big)\;
  =\; \big[ \det\! \big( \slE{}^{-1\T\:t} \big)
            \equiv
            \det\! \big(\slE{}_{\tilde\mu}{}^\al\big)
      \big]^{-1}&&
    \\
&=\; [\xE[\vrh]^2\; \sin\th]^{-1}\;
  =\; \sqrt{\det\xE[g]}\: \Big/\; \sqrt{\det\xE[\tilde{g}]}&
  &\det\xE[g] \!\equiv\! 1
    \nn
    \\[-4.5ex]\nn
\end{alignat}
\end{samepage}%
vgl.\@ Gl.~(\ref{APP:Stilde_mu^nuE}$'$) bzw.~(\ref{APP:det-gtildeE}).
Die Matrix~\mbox{\,$\slE \!\equiv\! \big(\slE{}^{\tilde\mu}{}_\nu\big)$} folgt durch explizites Invertieren und Transponieren von~\mbox{\,$\slE^{-1\T\:t} \!\equiv\! \big(\slE{}_{\tilde\mu}{}^\nu\big)$} nach Gl.~(\ref{APP:Stilde_mu^nuE}).
"Aquivalent durch explizites Ausmultiplizieren von Gl.~(\ref{APP:Stilde^mu_nu-DefE}) in Form:
\vspace*{-.5ex}
\begin{align} \label{APP:Stilde^mu_nu-prepreE}
\big(\slE{}^{\tilde\mu}{}_{\bar\nu}\big)\;
\vspace*{-.5ex}
  =\; \frac{1}{\det \xE[\tilde{g}]}\,
        \pmatrixZZ{\xE[g]{}_{\mfm\mfm}}{-\xE[g]{}_{\mfp\mfm}}
                 {-\xE[g]{}_{\mfm\mfp}}{\xE[g]{}_{\mfp\mfp}}\cdot
      \xE[\vrh]\,
      \pmatrixZZ{\cos\th\Dmfp}{-\sin\th\Dmfp}
                {\cos\th\Dmfm}{\sin\th\Dmfm}\cdot
      \pmatrixZZ{\xE[g]{}_{44}}{0}{0}{\xE[g]{}_{33}}
    \\[-5ex]\nn
\end{align}
analog zu Gl.~(\ref{APP:Stilde^mu_nu-prepre}) im Minkowskischen.
Durch Addition und Subtraktion der Relationen von Gl.~(\ref{APP:exp-expcosh}) folgt mithilfe der Gln.~(\ref{APP:cosh-cos,sinh-sin}),~(\ref{APP:cosh-cos,sinh-sin}$'$):
\vspace*{-.5ex}
\begin{alignat}{4} \label{APP:exp-expcoshE}
&\cos\th\Dmfp&\;
  &-\;& \cos\th\Dmfm\; \cos\th\;
  &=\;& \sin\th\Dmfm\; \sin\th&
    \\[.25ex]
&\sin\th\Dmfp&\;
  &+\;& \sin\th\Dmfm\; \cos\th\;
  &=\;& \cos\th\Dmfm\; \sin\th&\qquad
    \th = \th\Dmfp + \th\Dmfm
    \tag{\ref{APP:exp-expcoshE}$'$}
    \\[-4.5ex]\nn
\end{alignat}
analog~\mbox{$\mfp \!\leftrightarrow\! \mfm$} aus Gl.~(\ref{APP:exp-expcosh}$'$).
Mithilfe dieser Relationen f"uhrt Gl.~(\ref{APP:Stilde^mu_nu-prepreE}) auf einen expliziten Ausdruck f"ur die Matrix~\mbox{\,$\slE \!\equiv\! \big(\slE{}^{\tilde\mu}{}_\nu\big)$}.
Zusammen mit Gl.~(\ref{APP:Stilde_mu^nuE}) gilt:
\vspace*{-.5ex}
\begin{alignat}{3} \label{APP:Stilde^mu_nu-E}
&\slE&\;
  &=\; \big(\slE{}^{\tilde\mu}{}_\nu\big)&\;
  &=\; \det\slE\cdot
       \xE[\vrh]\,
       \pmatrixZZ{\sin\th\Dmfm}{-\cos\th\Dmfm}
                 {\sin\th\Dmfp}{\cos\th\Dmfp}
    \\[-.5ex]
&\slE{}^{-1\T\:t}&\;
  &=\; \big(\slE{}_{\tilde\mu}{}^\nu\big)&\;
  &=\; \xE[\vrh]\,
       \pmatrixZZ{\cos\th\Dmfp}{-\sin\th\Dmfp}
                 {\cos\th\Dmfm}{\sin\th\Dmfm}
    \tag{\ref{APP:Stilde^mu_nu-E}$'$}
    \\[-4.5ex]\nn
\end{alignat}
Sei verwiesen auf Gl.~(\ref{APP:detS,detS^-1t}) und die Gln.~(\ref{APP:Stilde^mu_nu}),~(\ref{APP:Stilde^mu_nu}$'$) im MInkowskischen.
\vspace*{-.5ex}

\paragraph{Komponenten~\bm{\xE^{\tilde\mu}},~\bm{\xE{}_{\tilde\mu}}%
           ~mit~\bm{\tilde\mu \in \{\mfp,\mfm,1,2\}}.}
Das Vierer-Skalarprodukt eines beliebigen Vektors~$\xE$ mit~$\eE{}_{(\mu)}$,~$\eE{}^{(\mu)}$ oder~$\eE{}_{(\mu)}$,~$\eE{}^{(\mu)}$ ist genau die Projektion auf dessen entsprechende ko-~bezie\-hungsweise kontravariante Komponente:
\vspace*{-.5ex}
\begin{alignat}{2}
&\xE^\mu\;
  =\; \xE\cdot \eE^{(\mu)}&&
    \label{APP:xE-Projektion} \\
&\text{und}\qquad
  \xE{}_\mu\;
  =\; \xE\cdot \eE{}_{(\mu)}&\qquad\qquad
  &\mu \in \{1,2,3,4\}
    \nn
    \\[-6.5ex]\nn
\intertext{\vspace*{-1.75ex}beziehungsweise:}
&\xE^{\tilde\mu}\;
  =\; \xE\cdot \eE^{(\tilde\mu)}&&
    \label{APP:xtildeE-Projektion} \\
&\text{und}\qquad
  \xE{}_{\tilde\mu}\;
  =\; \xE\cdot \eE{}_{(\tilde\mu)}&\qquad\qquad
  &\mu \in \{\mfp,\mfm,1,2\}
    \nn
    \\[-4.5ex]\nn
\end{alignat}
vgl.\@ die Gln.~(\ref{APP:xtilde-Projektion}),~(\ref{APP:xtilde-Projektion}$'$).
Das Skalarprodukt von~$\xE$ mit den Gln.~(\ref{APP:etilde_ebarM-kovE}),~(\ref{APP:etilde_ebarM-kontravE}) stellt daher unmittelbar dar die Transformationsformeln f"ur die Komponenten.
F"ur kontravariante Komponenten:
\vspace*{-.5ex}
\begin{align} \label{APP:xtilde_x-kontravE}
\xE^{\tilde\mu}\;
  =\; \slE{}^{\tilde\mu}{}_\nu\; \xE^\nu
    \\[-4.5ex]\nn
\end{align}
f"ur kovariante Komponenten:
\vspace*{-.5ex}
\begin{align} \label{APP:xtilde_x-kovE}
\xE{}_{\tilde\mu}\;
  =\; \slE{}_{\tilde\mu}{}^\nu\; \xE{}_\nu\;
    \\[-4.5ex]\nn
\end{align}
\begin{samepage}
bzgl.\@ der Matrixelemente%
~\mbox{\,$\slE{}^{\tilde\mu}{}_\nu$},%
~\mbox{\,$\slE{}_{\tilde\mu}{}^\nu$} vgl.\@ die Gln.~(\ref{APP:Stilde^mu_nu-E}),~(\ref{APP:Stilde^mu_nu-E}$'$).
Wir verf"ugen mit den Gln.~(\ref{APP:xtilde_x-kontravE}),~(\ref{APP:xtilde_x-kovE}) "uber die allgemeinen Transformationsformeln f"ur kontra- und kovariante (Lorentz-)Tensorkomponenten im Euklidischen; sie sind zu kontrastieren mit den Minkowskischen Formeln, vgl.\@ die Gln.~(\ref{APP:xtilde_x-kontrav}),~(\ref{APP:xtilde_x-kontrav}$'$) bzw.~(\ref{APP:xtilde_x-kov}),~(\ref{APP:xtilde_x-kov}$'$).
\vspace*{-.5ex}

\paragraph{Epsilon-Pseudotensor%
           ~\bm{\xE[\tilde\ep]
              \!\equiv\! \big(\xE[\ep]^{\tilde\mu\tilde\nu\tilde\rh\tilde\si}\big)}%
           ~mit~\bm{\tilde\mu,\tilde\nu,\tilde\rh,\tilde\si \!\in\! \{\mfp,\mfm,1,2\}}.}
Der Epsilon-Pseu\-dotensor auf der ins Euklidische fortgesetzten Raumzeit:~\mbox{\,$\xE[\tilde\ep] \!\equiv\! \big(\xE[\ep]^{\tilde\mu\tilde\nu\tilde\rh\tilde\si}\big)$},~\mbox{\,$\tilde\mu,\tilde\nu,\tilde\rh,\tilde\si \!\in\! \{\mfp,\mfm,1,2\}$}, wird berechnet in vollst"an\-diger Analogie wie auf der Minkowskischen.
Wir finden f"ur seine kontravarianten Komponenten~-- vgl.\@ die Gln.~(\ref{APP:epTensor-tilde-kontrav})-(\ref{APP:epTensor-tilde-kontrav}$''$):
\end{samepage}
\vspace*{-.5ex}
\begin{align} \label{APP:epTensor-tilde-kontravE}
\xE[\ep]^{\tilde\mu\tilde\nu\tilde\rh\tilde\si}\;
  &=\; \slE{}^{\tilde\mu}{}_\al\,
       \slE{}^{\tilde\nu}{}_\be\,
       \slE{}^{\tilde\rh}{}_\ga\,
       \slE{}^{\tilde\si}{}_\de\;
         \xE[\ep]^{\al\be\ga\de}
    \\
  &=\; \det\big(\slE{}^{\tilde\mu}{}_\al\big)\cdot
      \pmatrixZV{\tilde\mu}{\tilde\nu}{\tilde\rh}{\tilde\si}
                {\mfp}{\mfm}{1}{2}\;
       \xE[\ep]^{1234}
    \\[-4.5ex]\nn
\end{align}
und f"ur seine kovarianten Komponenten~-- vgl.\@ die Gln.~(\ref{APP:epTensor-tilde-kov})-(\ref{APP:epTensor-tilde-kov}$''$):
\vspace*{-.5ex}
\begin{align} \label{APP:epTensor-tilde-kovE}
\xE[\ep]{}_{\tilde\mu\tilde\nu\tilde\rh\tilde\si}\;
  &=\; \slE{}_{\tilde\mu}{}^\al\,
       \slE{}_{\tilde\nu}{}^\be\,
       \slE{}_{\tilde\rh}{}^\ga\,
       \slE{}_{\tilde\si}{}^\de\;
         \xE[\ep]{}_{\al\be\ga\de}
    \\
  &=\; \det\big(\slE{}_{\tilde\mu}{}^\al\big)\cdot
      \pmatrixZV{\tilde\mu}{\tilde\nu}{\tilde\rh}{\tilde\si}
                 {\mfp}{\mfm}{1}{2}\;
       \xE[\ep]{}_{1234}
    \\[-4.5ex]\nn
\end{align}
bzgl.\@ der Determinanten~\mbox{\,$\det\big(\slE{}^{\tilde\mu}{}_\al\big)$\;} und~\mbox{\,$\det\big(\slE{}_{\tilde\mu}{}^\al\big)$\;} vgl.\@ die Gln.~(\ref{APP:detSL-detSL^-1t-E}),~(\ref{APP:detSL-detSL^-1t-E}$'$).
Unsere Konvention f"ur~\mbox{$\xE[\ep] \!\equiv\! \big(\xE[\ep]^{\mu\nu\rh\si}\big)$} ist
%
\begin{align} 
\xE[\ep]^{1234}\;
  \equiv\; +1\qquad
  \Longrightarrow\qquad
  \xE[\ep]{}_{1234}\; =\; +1
\end{align}
vgl.\@ Gl.~(\ref{APP:epE-Def}$'$).
Das Signum der Indexpermutation~\mbox{$(\tilde\mu,\tilde\nu,\tilde\rh,\tilde\si) \!\to\! (\mfp,\mfm,1,2)$} ist notiert durch das Klammersymbol entsprechend Gl.~(\ref{APP:SignumPermutation}).
\vspace*{-.5ex}

\paragraph{Fixierung von~\bm{\xE[\vrh]}.}
Analog zu~$\vrh$ im Minkowskischen wird die positiv-reelle Konstante~$\xE[\vrh]$ bestimmt durch die Forderung, da"s die Determinante der Metrik invariant ist unter der Transformation, oder "aquivalent~-- vgl.\@ Gl.~(\ref{APP:detSL-detSL^-1t-E})~-- die Determinante der Transformationsmatrix identisch Eins ist:
\vspace*{-.5ex}
\begin{alignat}{2} \label{APP:vrhE-Forderung}
&\det \xE[\tilde{g}]\;
    \equiv\; \det\! \big(\xE[g]{}_{\tilde\mu\tilde\nu}\big)\vv
  \stackrel{\D!}{=}\vv
  \det \xE[g]\;
    \equiv\; \det\! \big(\xE[g]{}_{\mu\nu}\big)&&
    \\[-.5ex]
&\Longleftrightarrow\qquad
 \det \slE\;
    \equiv\; \det\! \big(\slE{}^{\tilde\mu}{}_\nu\big)\vv
  \stackrel{\D!}{=}\vv 1&
 &\text{f"ur alle\vv $\th \in (0,\pi\!/\!2)$}
    \tag{\ref{APP:vrhE-Forderung}$'$}
\end{alignat}
Dann impliziert Gl.~(\ref{APP:Stilde_mu^nuE}$'$) f"ur~\mbox{$\vrh \!\in\! \bbbr^+$}, mit~$\ps \!\in\! (0,\infty)$:
\vspace*{-.5ex}
\begin{align} \label{APP:vrhE}
&\xE[\vrh]^2\vv
  =\; \sin^{-1}\!\th
    \\[-.25ex]
&\Longleftrightarrow\qquad
  \xE[\vrh]\;
  \equiv\; \xE[\vrh](\th)\vv
  =\; \sin^{-1\!/\!2}\!\th
    \tag{\ref{APP:vrhE}$'$}
    \\[-4.5ex]\nn
\end{align}
F"ur die Komponenten des metrischen Tensors~\mbox{\,$\xE[\tilde{g}] \!\equiv\! \big(\xE[g]{}_{\tilde\mu\tilde\nu}\big)$} folgt:
\vspace*{-.5ex}
\begin{align} \label{APP:gE_pp-expl}
&\xE[g]{}_{\mfp\mfp}\;
  =\; \xE[g]{}_{\mfm\mfm}\;
  =\; -\, \xE[\vrh]^2\quad
  =\; -\, \sin^{-1}\!\th
    \\[-.5ex]
  &\hspace*{128.5pt}
   \longrightarrow\vv
      \xE[g]{}_{44}\; =\; \xE[g]{}_{33}\;
        \equiv\; -\, 1\qquad
  \text{f"ur\vv $\th \to \pi\!/\!2$}
    \tag{\ref{APP:gE_pp-expl}$'$} \\[.75ex]
&\xE[g]{}_{\mfp\mfm}\;
  =\; \xE[g]{}_{\mfm\mfp}\;
  =\; -\, \xE[\vrh]^2\, \cos\th\quad
  =\; -\, \tan^{-1}\!\th
    \label{APP:gE_pm-expl} \\[-.5ex]
  &\hspace*{138pt}
   \longrightarrow\vv
      \xE[g]{}_{43}\; =\; \xE[g]{}_{34}\;
        \equiv\; 0\qquad
  \text{f"ur\vv $\th \to \pi\!/\!2$}
    \tag{\ref{APP:gE_pm-expl}$'$}
    \\[-5ex]\nn
\end{align}
vgl.\@ Gl.~(\ref{APP:gtilde-kovE}) bzw.~(\ref{APP:gtilde-kovE}$'$). \\
\indent
Im Hochenergielimes~-- realisiert durch die maximalen Drehwinkel:~\mbox{$\th\Dmfp,\th\Dmfm \!\to\! \pi\!/\!4 \Rightarrow\! \th \!\to\! \pi\!/\!2$}, vgl.\@ Fu"sn.\,\FN{APP-FN:OrientierungM->E},~-- sind {\it per~constructionem\/} die Koordinatenlinien~\bm{\mfp},~\bm{\mfm} orthogonal wie die Linien~\bm{4},~\bm{3} [diese sind nur gedreht um~$+\pi\!/\!4$]; dies imliziert f"ur die Metriken
%
\begin{align} 
\xE[\tilde{g}]\;
    \equiv\; \big(\xE[g]{}_{\tilde\mu\tilde\nu}\big)\vv
  \longrightarrow\vv
  \xE[\bar{g}]\;
    \equiv\; \big(\xE[g]{}_{\bar\mu\bar\nu}\big)\qquad
  \qquad\;
  \text{f"ur\vv $\th \to \pi\!/\!2$}
\end{align}
wie verifiziert durch die Gln.~(\ref{APP:gE_pp-expl}$'$),~(\ref{APP:gE_pm-expl}$'$).
Die Diskrepanz zwischen Hochenergielimes und endlichen $O(4)$-Drehungen:~\mbox{\,$\th\Dmfp,\th\Dmfm \!<\! \pi\!/\!4 \Rightarrow \th \!<\! \pi\!/\!2$}, ist subsumiert in simplen Faktoren von Winkelfunktionen, insbesondere in den nichtverschwindenden {\it Au"serdiagonalelementen\/}~$\xE[g]{}_{\mfp\mfm}$, $\xE[g]{}_{\mfm\mfp}$ der Metrik.
\vspace*{-.5ex}

\bigskip\noindent
Wir verf"ugen "uber die formalen wie interpretatorischen Zusammenh"ange, um die Diskussion der $T$-Amplitude im Haupttext~-- ihre Auswertung in Minkowskischer und ihre analytische Fortsetzung in Euklidische Raumzeit~-- konzise und pr"agnant durchzuf"uhren.
Wir schlie"sen mit einem Kompendium ferner ben"otigter Zusammenh"ange, die~-- die analytische Fortsetzung der Korrelationsfunktionen und die Lorentz-induzierte Wahl der Raumzeit-Parameter von Lichtkegelwellenfunktionen~-- respektive in unmittelbarer und mittelbarer Beziehung stehen zu dem entwickelten Formalismus.
\vspace*{7.5ex}

\section[Kompendium]{%
         \vspace*{-.375ex}Kompendium}

%
\subsection[Wick-Drehung der Korrelationsfunktionen]{%
            \vspace*{-.125ex}Wick-Drehung der Korrelationsfunktionen~\bffootnote}
\label{APP-Subsect:Wick}
\footnotetext{
  Bzgl.\@ der expliziten Darstellungen der Korrelationsfunktionen sei verwiesen auf Anhang~\ref{APP:CLTFN}.
}
%

Die Korrelationsfunktionen im Minkowskischen~--%
  ~\vspace*{-.25ex}$D\uC$,~$D\uNC$ bzw.~$F\oC$,~$F\oNC$~-- sind konstruiert mithilfe des {\it"`$\nu$-verallgemeinerten"'\/} Feynman-Propagators~$D_\nu$, mit%
  ~\vspace*{-.125ex}\mbox{\,$\De_{\mskip-1mu F} \!\equiv\! D_1|_{\la_1\equiv1}$} dem~{\it kon\-ventionellen\/}; vgl.\@ Kap.~\ref{Abschn:ANN-KONST} auf Seite\,\pageref{xi}ff., insbes.\@ Anh.~\ref{APP:CLTFN}.
Sie zeigen {\it per~constructionem\/}~das\-selbe Verhalten in der komplexen~\mbox{$k^0$-Ebe}\-ne in Bezug auf Position von Singularit"aten,~assozi\-ierter Epsilon-Vorschrift~\mbox{$\ep \!\to\! 0\!+$} und Abfall f"ur~\mbox{$|k^0| \!\to\! \infty$}; sei daher summarisch diskutiert
\vspace*{-.5ex}
\begin{align} \label{APP:F(xi)}
F(\xi^2)\;
  &=\; \int \frac{d^4k}{(2\pi)^4}\vv
         \efn{\D-\iIM k \!\cdot\! \xi}\vv
         \tilde{F}(k^2)\qquad\qquad
  F \equiv D_\nu, \De_{\mskip-1mu F}, \ldots
    \\[-.25ex]
  &=\; \int \frac{d^3\vec{k}}{(2\pi)^3}\vv
        \efn{\D\iIM \vec{k} \!\cdot\! \vec\xi}\vv
        \bigg\{ \int_{-\infty}^\infty \frac{dk^0}{2\pi}\vv
          \efn{\D-\iIM k^0\xi^0}\vv
          \tilde{F}\big((k^0)^2 \!-\! \vec{k}{}^2\big)
        \bigg\}
    \tag{\ref{APP:F(xi)}$'$}
    \\[-4.5ex]\nn
\end{align}
\begin{figure}
\vspace*{-1ex}
\begin{minipage}{\linewidth}
  \begin{center}
  \setlength{\unitlength}{1mm}\begin{picture}(100,100)
    \put(0,0){\epsfxsize100mm \epsffile{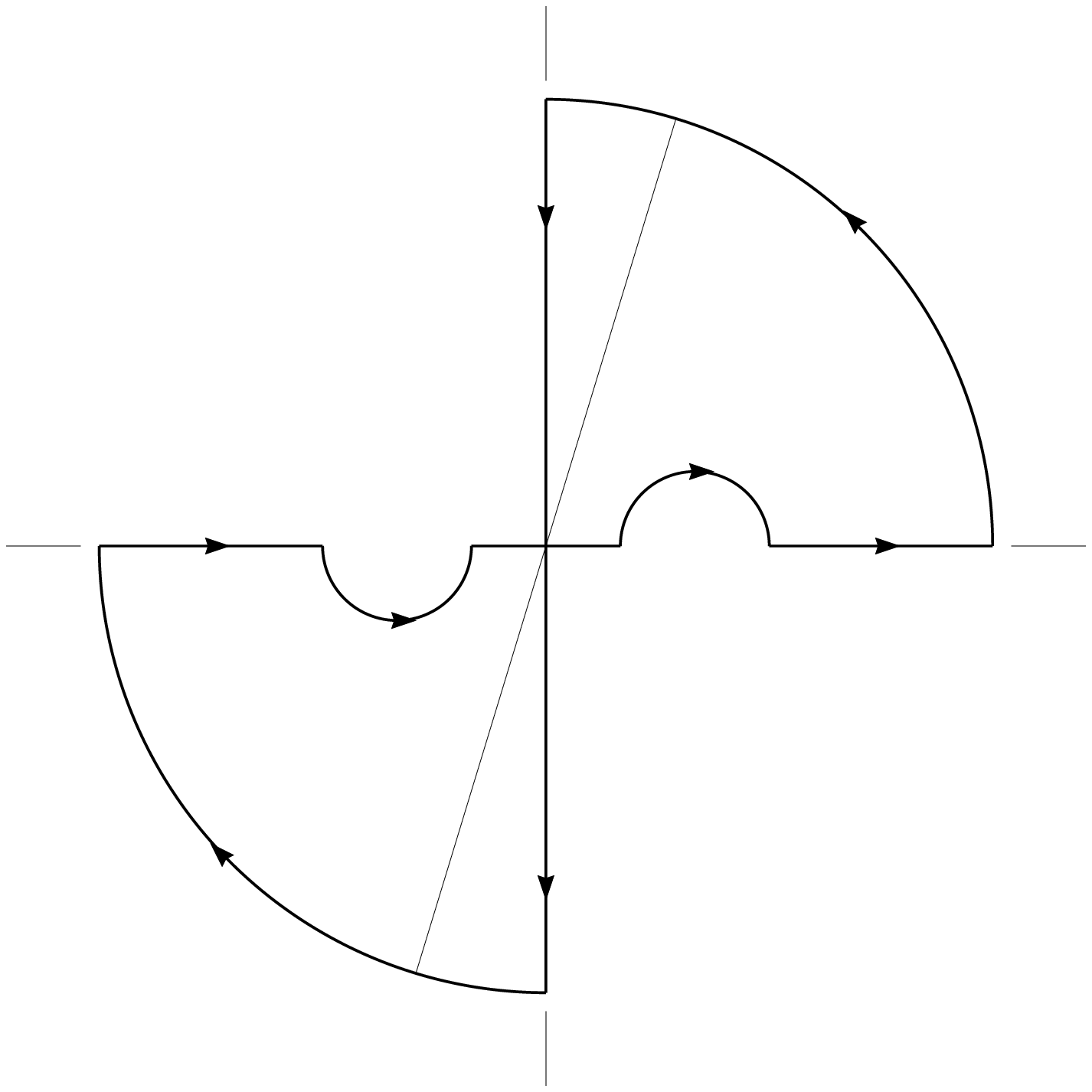}}
    \put(96,43){\normalsize${\rm Re}\,k^0$}
    \put(80,53){\normalsize${\cal C}_\Re$}
    \put(64,50){\circle*{1}}
    \put(55,43){\normalsize$+\surd m^2 \!+\! \vec{k}{}^2$}
    \put(39,96){\normalsize${\rm Im}\,k^0$}
    \put(40,80){\normalsize$-{\cal C}_\Im$}   
    \put(36.5,50){\circle*{1}}
    \put(27.5,55){\normalsize$-\surd m^2 \!+\! \vec{k}{}^2$}
    \put(67,18){\normalsize{\large$\int_{{\cal C}_\Re}$}\;+\;{\large$\int_{-{\cal C}_\Im}$}%
                              \;=\vv0}
    \put(67,11){\normalsize$\Rightarrow\vv${\large$\int_{{\cal C}_\Re}$}%
                              \;=\;$-$\;{\large$\int_{-{\cal C}_\Im}$}%
                              \;=\;{\large$\int_{{\cal C}_\Im}$}}
    \put(57.5,66){\normalsize$R$}
    \put(40.5,34){\normalsize$R$}
  \end{picture}
  \end{center}
\vspace*{-5ex}
\caption[Wick-Drehung der Koorelationsfunktionen in komplexer~$k^0$-Ebene]{
  Wick-Drehung in komplexer~\,$k^0$-Ebene.   Eine Funktion von~\mbox{\,$k^0 \!\in\! \bbbc$} sei zu integrieren "uber die reelle Achse; sei gegeben eine Epsilon-Vorschrift, die im dargestellten Sinne definiert die Behandlung der Pole bei~\mbox{\,$k^0 \!=\! \pm\surd m^2 \!+\! \vec{k}{}^2$}, und besitze die Funktion keine weiteren Pole im 1.\@ und 3.\@ Quadranten.   Dann verschwindet identisch ihr Integral entlang der dargestellten geschlossenen Kurve.   Falle sie ferner gen"ugend schnell ab f"ur gro"sen Absolutbetrag von~\,$k^0$, so da"s die Beitr"age "uber die Kreisb"ogen mit Radius~\,$R$ verschwinden im Limes~\mbox{\,$R \!\to\! \infty$}.   Dann ist ihr Integral "uber die reelle Achse gleich dem "uber die imagin"are:   Der "Ubergang von~\mbox{\,${\cal C}_\Re$} zu~\mbox{\,${\cal C}_\Im$} als Integrationsweg wird bezeichnet als Wick-Drehung.
\vspace*{-.5ex}
}
\label{Fig:WickRotation}
\end{minipage}
\end{figure}
\noindent
explizit die geschweifte Klammer in der Form
\vspace*{-.5ex}
\begin{align} 
f\big((\xi^0)^2\big)\;
  \equiv\; \int_{\D{\cal C}_\Re} \frac{dk^0}{2\pi}\vv
        \efn{\D-\iIM k^0\xi^0}\vv
        \tilde{f}\big((k^0)^2\big)\qquad\qquad
  k^0 \in \bbbc
    \\[-4.5ex]\nn
\end{align}
mit~\mbox{\,$\tilde{f}\big((k^0)^2\big) \!:=\! \tilde{F}\big((k^0)^2 \!-\! \vec{k}{}^2\big)$}.
Aufgrund der Eigenschaften der Funktion~\mbox{$\tilde{F}$} bez"uglich der Singularit"aten bei%
  ~\mbox{\,$k^0 \!=\! \pm\surd m^2 \!+\! \vec{k}{}^2$} mit der assoziierten Epsilon-Vorschrift und dem Abfall\-verhalten f"ur%
  ~\mbox{$|k^0| \!\to\! \infty$} kann bekannterma"sen~-- vgl.\@ Abb.~\ref{Fig:WickRotation}~-- die Integration "uber die~\mbox{\it reelle\/} \mbox{$k^0$-Ach}\-se ersetzt werden durch die Integration "uber die {\it imagin"are\/}~\mbox{$k^0$-Ach}\-se,~\mbox{\,${\cal C}_\Re \!\to\! {\cal C}_\Im$}:
\vspace*{-.5ex}
\begin{align} \label{APP:ftilde_CIm-0}
f\big((\xi^0)^2\big)\;
  &=\; \int_{\D{\cal C}_\Im} \frac{dk^0}{2\pi}\vv
        \efn{\D-\iIM k^0\xi^0}\vv
        \xE[\tilde{f}]\big((k^0)^2\big)
    \\[.25ex]
  &=\; \int_{-\iIM\infty}^{\iIM\infty} \frac{dk^0}{2\pi}\vv
        \efn{\D-\iIM k^0\xi^0}\vv
        \xE[\tilde{f}]\big((k^0)^2\big)
    \tag{\ref{APP:ftilde_CIm-0}$'$}
    \\[-4.5ex]\nn
\end{align}
Die Funktion unter dem Integral mit Argument auf der imagin"aren \mbox{$k^0$-Ach}\-se ist insofern die analytische Fortsetzung der Funktion%
  ~\mbox{$\tilde{f}$} und daher mit dem Skript~"`E"' indiziert:~\mbox{$\xE[\tilde{f}]$}.
Unter Substitution~\mbox{\,$\iIM k^0 \!\to\! \xE[k]^4 \!=\! \iIM k^0$}~-- vgl.\@ Gl.~(\ref{APP:x->xE})~-- ergo:~\mbox{\,$dk^0 \!=\! -\iIM d\xE[k]^4$}, folgt:
\vspace*{-.5ex}
\begin{align} \label{APP:ftilde_CIm-1}
f\big((\xi^0)^2\big)\;
  &=\; -\iIM\cdot \int_{\infty}^{-\infty} \frac{d\xE[k]^4}{2\pi}\vv
        \efn{\D\iIM \xE[k]^4 (\iIM\xi^0)}\vv
        \xE[\tilde{f}]\big((-\iIM\xE[k]^4)^2\big)
    \\[.25ex]
  &=\; \iIM\cdot \int_{-\infty}^{\infty} \frac{d\xE[k]^4}{2\pi}\vv
        \efn{\D\iIM \xE[k]^4 (\iIM\xi^0)}\vv
        \xE[\tilde{f}]\big(\!-\!(\xE[k]^4)^2\big)
    \tag{\ref{APP:ftilde_CIm-1}$'$}
    \\[-4.5ex]\nn
\end{align}
Mit%
  ~\mbox{\,$\xE[\tilde{f}]\big(\!-\!(\xE[k]^4)^2\big)
    \!=\! \xE[\tilde{F}]\big(\!-\!(\xE[k]^4)^2 \!-\! \vec{k}^2\big)$} folgt f"ur~$F(\xi^2)$ nach Gl.~(\ref{APP:F(xi)}$'$):
%
\begin{align} 
F(\xi^2)\;
  =\; \int \frac{d^3\xE[\vec{k}]}{(2\pi)^3}\vv
        \efn{\D\iIM \vec{k} \!\cdot\! \vec\xi}\;
        \bigg\{\, \iIM\cdot \int_{-\infty}^\infty \frac{d\xE[k]^4}{2\pi}\vv
          \efn{\D\iIM \xE[k]^4 (\iIM\xi^0)}\vv
          \xE[\tilde{F}]\big(\!-\!(\xE[k]^4)^2 \!-\! \vec{k}^2\big)
        \,\bigg\}
\end{align}
"Ubergang gem"a"s~\mbox{\,$k^i \!\to\! \xE[k]^i$} und%
  ~\mbox{\,$\iIM\xi^0 \!\to\! \xE[\xi]^4$},%
  ~\mbox{\,$\xi^i \!\to\! \xE[\xi]^i$} zu den analytisch fortgesetzten Komponenten~-- vgl.\@ die Gln.~(\ref{APP:x->xE}),~(\ref{APP:x->xE}$'$)~-- f"uhrt unmittelbar auf:
\vspace*{-.5ex}
\begin{align} \label{APP:F(xi)E-1}
F(\xi^2 \!\to\! \xE[\xi]^2)\;
  &=\; \iIM\cdot \int \frac{d^4\xE[k]}{(2\pi)^4}\vv
         \efn{\D\iIM \xE[k]^\mu \xE[\xi]^\mu}\vv
         \xE[\tilde{F}](-\xE[k]^\nu\xE[k]^\nu)
    \\[.25ex]
  &=\; \iIM\cdot \int \frac{d^4\xE[k]}{(2\pi)^4}\vv
         \efn{\D-\iIM \xE[k] \!\cdot\! \xE[\xi]}\vv
         \xE[\tilde{F}](\xE[k]^2)\qquad
   \equiv\; \iIM\cdot \xE[F](\xE[\xi]^2)
    \tag{\ref{APP:F(xi)E-1}$'$}
    \\[-4.5ex]\nn
\end{align}
die zweite Zeile mit%
  ~\mbox{\,$\xE[k]^2 \!=\! \xE[k] \!\cdot\! \xE[k] \!=\! -\xE[k]^\mu\xE[k]^\mu$} negativ definit und~\mbox{\,$\xE[k] \!\cdot\! \xE[\xi] \!=\! -\xE[k]^\mu\xE[\xi]^\mu$} nach Gl.~(\ref{APP:xcdoty-E}) und Identifizierung der Fourier-Transformierten in die Euklidische Raumzeit.
Es gilt:%
\FOOT{
  Derselbe Zusammenhang folgt bei analytischer Fortsetzung~\mbox{\,$\iIM x^0 \!\to\! -\xE^4$} mit {\sl umgekehrtem\/} Vorzeichen.
}
\begin{samepage}
\vspace*{.25ex}
\begin{align} \label{APP:F(xi)-FE(xiE)}
\xE[F](\xE[\xi]^2)\;
  =\; -\iIM\cdot F\big(\xi^2 \!\to\! \xE[\xi]^2\big)
    \\[-3.5ex]\nn
\end{align}
F"ur respektive~\vspace*{-.125ex}\mbox{$F \!\equiv\! D\uC, D\uNC, F\oC, F\oNC$} ist dies genau die Vorschrift zur Fortsetzung der Korrelationsfunktionen von Minkowskischer zu Euklidischer Raumzeit.
\vspace*{-.5ex}

\subsection[Raumzeit-Parameter~\protect$\{\zet,\rb{k}\}$ der Lichtkegelwellenfunktionen
               unter speziellen orthochronen Lorentz-Transformationen]{%
            Raumzeit-Parameter~\bm{\{}$\zet$\bm{,\rb{k}\}} der Lichtkegelwellenfunktionen
               unter speziellen orthochronen Lorentz-Transformationen}
\label{APP-Subsect:LCWFN-Kovarianz}

Unsere Arbeit wesentlich zugrunde liegt das Konzept hadronischer Lichtkegelwellenfunktionen.
Behauptung ist%
\FOOT{
  Wir beziehen uns in Notation auf Abschnitt~\ref{Subsect:HadronniveauI}.
},
da"s diese bei Abh"angigkeit {\it longitudinal\/} von~$\zet$, dem Anteil des Quarks am gesamten Lichtkegelimpuls [$\bzet \!\equiv\! 1 \!-\! \zet$ der des Antiquarks],~-- ferner nur abh"angen von dem {\it transversalen\/} Vektor~$\rb{k}$.
Die Definition dieses Vektors~$\rb{k}$ hat in der Weise zu erfolgen, da"s unter beliebigen speziellen orthochronen Lorentz-Transformationen~$\La$ garantiert ist Kovarianz der involvierten Raumzeit-Parameter.%
\FOOT{
  \label{APP-FN:La-be-al}In expliziter Diskussion von speziell-orthochronen Lorentz-Boosts~\mbox{\,$\La(\vec\be)$} und reinen Drehungen~\mbox{\,$\La(\vec\al)$} zeigen wir dar"uberhinaus:   Es existiert keine Transformation~$\La$~-- da darstellbar als~\mbox{\,$\La(\vec\al) \!\circ\! \La(\vec\be)$} oder~\mbox{\,$\La(\vec\be) \!\circ\! \La(\vec\al)$} mit geeignetem Boost- und Drehvektor~--, f"ur die eine beliebig andere Definition von~$\rb{k}$ nicht im Widerspruch st"unde mit Lorentz-Kovarianz.
}
\end{samepage}

Sei betrachtet%
\FOOT{
  o.E.d.A.\@ ein {\sl Quark\/} mit Geschwindigkeit~\mbox{$\be \!\to\! 1$} in {\sl positive\/}~$x^3$-Richtung.~-- Gefordert~\mbox{\,$\rb{p} \!+\! \rbb{p} \!=\! \rb{P}$}, folgt das {\sl Antiquark\/} durch~\mbox{\,$p \!\to\! \bar{p},\, \zet \!\to\! \bzet,\, \rb{k} \!\to\! -\rb{k}$} und~\mbox{\,$\la \!\to\! -\la$}, entsprechend in {\sl negative\/}~$x^3$-Richtung durch~\mbox{\,$+ \!\leftrightarrow\! -$} bzgl.\@ Lorentz-Indizes.   Gl.~(\ref{APP:LCWFN-momentum_i}) ist die allgemeinste Definition des Vektors~\mbox{\,$\rb{k}$} bis auf Vielfache:~\mbox{\,$\rb{k}' \!=\! c\,\rb{k}, c \!\in\! \bbbr$}.
}
%
\vspace*{-.25ex}
\begin{alignat}{2}
&p^+&\;
  &=\; \zet\, P^+
    \label{APP:LCWFN-momentum_+} \\
&\rb{p}&\;
  &=\; (\zet + \la)\; \rb{P} + \rb{k}
    \label{APP:LCWFN-momentum_i}
    \\[-4.25ex]\nn
\end{alignat}
mit~\mbox{\,$p^+, P^+$} {\it gro"s\/} und formal~\mbox{\,$k^\pm \!\equiv\! 0$}.
Gefordert~\mbox{\,$\rb{p} \!+\! \rbb{p} \!=\! \rb{P}$}, impliziert Gl.~(\ref{APP:LCWFN-momentum_i}):
\vspace*{-.25ex}
\begin{align} 
\rb{k}\;
  =\; \frac{1}{2}\, \big(\rb{p} - \rbb{p}\big)
        + \Big(\frac{1}{2} \!-\! \big(\zet \!+\! \la\big)\Big)\, \rb{P}
    \\[-4.25ex]\nn
\end{align}
F"ur~\mbox{\,$\la \!\ne\! 1\!/\!2 \!-\! \zet$} differiert~$2\rb{k}$ von~\mbox{\,$(\rb{p} \!-\! \rbb{p})$}, dem Relativimpuls der Quarks.
Wir zeigen:
\vspace*{-.25ex}
\begin{align} \label{APP:la=0}
\la\; \equiv\; 0
    \\[-4.25ex]\nn
\end{align}
folgt notwendig aus der Forderung von Lorentz-Kovarianz; insbes.\@ also~\mbox{\,$\la \!\ne\! 1\!/\!2 \!-\! \zet$}, vgl.\@ Nachtmann in Ref.~\cite{Nachtmann96}. \\
\indent
Spezielle orthochrone Lorentz-Transformationen werden vermittelt durch Matrizen:
\vspace*{-.5ex}
\begin{align} \label{APP:spez-orthLT}
&\La\; \equiv\; \big(\La^\mu{}_\nu\big)\qquad
  \mu,\nu \in \{0,1,2,3\}
    \\[.25ex]
&\text{mit}\qquad
 \begin{alignedat}[t]{2}
 &\La^0{}_0\; >\; 0\quad[\text{\sl orthochron\/}]\qquad\;
  \det\La\; \equiv\; +1&&\quad[\text{\sl speziell\/}]
    \\[-.75ex]
 &\La^{\T t}\, g\, \La\; =\; g\qquad
  \text{d.h.}\quad
    g_{\mu\nu}\, \La^\mu{}_\rh\, \La^\nu{}_\si\;
    =\; g_{\rh\si}&&\quad[\text{\sl pseudo-orthogonal\/}]
 \end{alignedat}
    \tag{\ref{APP:spez-orthLT}$'$}
    \\[-4.5ex]\nn
\end{align}
Sei~\mbox{\,$\La \!\equiv\! \big(\La^\mu{}_\nu\big)$} verstanden als Lorentz-Boost oder Drehung, zun"achst aber nicht weiter explizit gemacht.
Wir betrachten f"ur den beliebigen Lorentz-Vektor~\mbox{\,$x \!\equiv\! (x^\mu)$},~\mbox{\,$\mu \!\in\! \{0,1,2,3\}$}, die~-- o.E.d.A.\@ passive~-- Transformation:
\vspace*{-.25ex}
\begin{align} 
x^{\mu'}\;
  =\; \La^\mu{}_\nu\, x^\nu
    \\[-4.25ex]\nn
\end{align}
in Lichtkegelkoordinaten:
\vspace*{-.5ex}
\begin{align} \label{APP:spez-orthLTbar-Tf}
&x^{\bar\mu'}\;
  =\; \La^{\bar\mu}{}_{\bar\nu}\, x^{\bar\nu}
    \\
&\bar\La\; \equiv\; \big(\La^{\bar\mu}{}_{\bar\nu}\big)\qquad
    \bar\mu,\bar\nu \in \{+,-,1,2\}
    \tag{\ref{APP:spez-orthLTbar-Tf}$'$}
    \\[-4.5ex]\nn
\end{align}
Wir beziehen diese Transformationsformeln auf die Vektorkomponenten~\mbox{\,$p^+ \!\to\! p^{+'} \!=\! \La^+{}_{\bar\mu}p^{\bar\mu}$} und~\mbox{\,$p^i \!\to\! p^{i'} \!=\! \La^i{}_{\bar\mu}p^{\bar\mu}$}, vgl.\@ die Gln.~(\ref{APP:LCWFN-momentum_+}),~(\ref{APP:LCWFN-momentum_i}).
Es gilt f"ur~\mbox{\,$p^{+'}$}, mit~\mbox{\,$i,j \!\in\! \{1,2\}$}:
%
\begin{align} \label{APP:p^+'-pre}
p^{+'}\;
  &=\; \La^+{}_+\, p^+
      + \La^+{}_-\, p^-
      + \La^+{}_i\, p^i
    \\
  &\sim\; \La^+{}_+\, p^+
      + \La^+{}_i\, p^i
    \tag{\ref{APP:p^+'-pre}$'$} \\
  &\phantom{\sim\; }
   =\; \zet\, \big[
         \La^+{}_+\, P^+
       + \La^+{}_i\, P^i \big]
       + \La^+{}_i\, k^i\;
       + \la\; \La^+{}_i\, P^i
    \tag{\ref{APP:p^+'-pre}$''$} \\
  &\phantom{\sim\; }
   \sim\; \zet\, \big[
         \La^+{}_+\, P^+
       + \La^+{}_-\, P^-
       + \La^+{}_i\, P^i \big]
       + \La^+{}_i\, k^i\;
       + \la\; \La^+{}_i\, P^i
    \tag{\ref{APP:p^+'-pre}$'''$}
    \\[-5ex]\nn
\end{align}
ergo:
\begin{samepage}
\vspace*{-.25ex}
\begin{align} \label{APP:p^+'}
p^{+'}\;
  \sim\; \big(\zet\, P^{+'}
       + k^{+'}\big)\vv
       + \la\; \La^+{}_i\, P^i\qquad
  i \in \{1,2\}
\end{align}
mit~\mbox{\,$k^{+'} \!=\! \La^+{}_{\bar\mu}\, k^{\bar\mu} \!=\! \La^+{}_i\, k^i$} wegen~$k^\pm \!\equiv\! 0$.
Es folgt die dritte Zeile~-- Gl.~(\ref{APP:p^+'-pre}$''$)~--~mit\-hilfe Gl.~(\ref{APP:LCWFN-momentum_i}).
Approximative Geichheit~"`$\sim$"' ist zu verstehen unter Vernachl"assigung der initialen kleinen Komponenten~$p^-$ und~$P^-$, explizit:
\vspace*{-.25ex}
\begin{align} 
p^- = 1 \!\big/\! p^+\;
  \sim\; 0\qquad
  \text{und}\qquad
P^- = 1 \!\big/\! P^+\;
  \sim\; 0
    \\[-4ex]\nn
\end{align}
Es gilt f"ur~\mbox{\,$p^{i'}$} analog zu Gl.~(\ref{APP:p^+'-pre}):~\mbox{\,$p^{i'} \!=\! \La^i{}_+\, p^+ \!+\! \La^i{}_-\, p^- \!+\! \La^i{}_j\, p^j$}, ergo~-- analog den Gln.~(\ref{APP:p^+'-pre}$'$)-(\ref{APP:p^+'-pre}$'''$):
%
\begin{align} \label{APP:p^i'}
p^{i'}\;\,
  &\sim\; \big(\zet\, P^{i'}
       + k^{i'}\big)\vv
       + \la\; \La^i{}_j\, P^j\qquad
  i,j \in \{1,2\}
\end{align}
\end{samepage}%
mit~\mbox{$k^{i'} \!=\! \La^i{}_{\bar\mu}\, k^{\bar\mu} \!=\! \La^i{}_j\, k^j$} wegen~\mbox{$k^\pm \!\equiv\! 0$}.
Allgemein sind~\mbox{$\La^+{}_iP^i$} und~\mbox{$\La^i{}_jP^j$},~\mbox{$i,j \!\in\! \{1,2\}$}, nicht (simultan) Null f"ur~\mbox{\,$\rb{P} \!\ne\! \bm{0}$} und beliebige Matrix~\mbox{\,$\bar\La \!\equiv\! \big(\La^{\bar\mu}{}_{\bar\nu}\big)$}~-- dies bereits, ohne f"ur~\mbox{\,$\La \!\equiv\! \big(\La^\mu{}_\nu\big)$} zu fordern die Relationen nach Gl.~(\ref{APP:spez-orthLT}$'$).
In Konfrontation der Gln.(\ref{APP:p^+'}),(\ref{APP:p^i'}) mit~(\ref{APP:LCWFN-momentum_+}),~(\ref{APP:LCWFN-momentum_i}) konstatieren wir, da"s Kovarianz bez"uglich {\it Lorentz-Transforma\-tionen im weitesten Sinne\/} unmittelbar fordert Gl.~(\ref{APP:la=0}):~\mbox{\,$\la \!\equiv\! 0$}.
Quod erat demonstrandum.
\vspace*{-.5ex}

\bigskip\noindent
F"ur Vollst"andigkeit~-- und in Hinblick auf Fu"snote\,\FN{APP-FN:La-be-al}~-- betrachten wir explizit {\it Lorentz-Boosts\/}~\mbox{\,$\La \!\equiv\! \La(\vec\be)$} und {\it reine Drehungen\/} \mbox{\,$\La \!\equiv\! \La(\vec\al)$} mit~\mbox{\,$\vec\be$} der (Dreier-)Boost-Geschwindigkeit respektive~\mbox{\,$\vec\al$} dem (Dreier-)Drehvektor.
Als Verkettung~\mbox{\,$\La(\vec\be) \!\circ\! \La(\vec\al)$} oder~\mbox{\,$\La(\vec\al) \!\circ\! \La(\vec\be)$} mit geeigenten Boost- und Drehvektoren ist darstellbar eine beliebige spezielle orthochrone Lorentz-Transformation. \\
\indent
In vollst"andiger Analogie zu den Gln.(\ref{APP:LT-La-bar0})-(\ref{APP:LT-La-bar0}$''$)~-- unmittelbar durch~\mbox{\,$x^\pm \!=\! \al(x^0 \!\pm\! x^3)$} und \mbox{\,$x_\pm \!=\! g_{+-}\al(x^0 \!\mp\! x^3)$} mit~\mbox{\,$g_{+-}\al \!=\! 1\!/\!2\al$}~-- werden die Komponenten~\mbox{\,$\La^{\bar\mu}{}_{\bar\nu}$},~\mbox{\,$\bar\mu,\bar\nu \!\in\! \{+,-,1,2\}$} der Transformationsmatrix ausgedr"uckt durch die Komponenten~\mbox{\,$\La^\mu{}_\nu$},~\mbox{\,$\mu,\nu \!\in\! \{0,1,2,3\}$}:
\vspace*{-.5ex}
\begin{alignat}{3}
&\La^\pm{}_+&\;
  &=\;& \frac{1}{2}\, \Big(
         \big(&\La^0{}_0 \pm \La^3{}_0\big) + \big(\La^0{}_3 \pm \La^3{}_3\big)
       \Big)
    \label{APP:spez-orthLTbar-pm} \\
&\La^\pm{}_-&\;
  &=\;& \frac{1}{2}\, \Big(
         \big(&\La^0{}_0 \pm \La^3{}_0\big) - \big(\La^0{}_3 \pm \La^3{}_3\big)
       \Big)
    \tag{\ref{APP:spez-orthLTbar-pm}$'$} \\
&\La^\pm{}_{\bar{i}}&\;
  &=\;& \al\, \big(&\La^0{}_i \pm \La^3{}_i\big)
    \tag{\ref{APP:spez-orthLTbar-pm}$''$}
    \\[-6ex]\nn
\intertext{\vspace*{-2ex}und}
&\La^{\bar{i}}{}_\pm&\;
  &=\;& \frac{1}{2\al}\, \big(&\La^i{}_0 \pm \La^i{}_3\big)
    \label{APP:spez-orthLTbar-i} \\
&\La^{\bar{i}}{}_{\bar{j}}&\;
  &=\;& \La^i{}_j&
    \tag{\ref{APP:spez-orthLTbar-i}$'$}
    \\[-4.5ex]\nn
\end{alignat}
mit~$\al$ der Normierung der Lichtkegelkoordinaten, vgl.\@ die Gln.~(\ref{APP:kontravLC_Def}),~(\ref{APP:kontravLC_Def}$'$).
\vspace{-.5ex}

\paragraph{Lorentz-Boosts} sind vollst"andig bestimmt durch den Boost-(Dreier)Vektor
%
\begin{align} 
\vec\be\;
  =:\; \be\, \hat\be\qquad
  \text{mit}\qquad
  |\hat\be|\; \equiv\! \hat\be^i\, \hat\be^i\; =\; 1\quad
  i \in \{1,2,3\}
\end{align}
Es sind~\mbox{\,$\be$} und~\mbox{\,$\hat\be$} respektive Betrag und Richtung der Boost-Geschwindigkeit.
Lorentz-Boosts werden vermittelt durch Matrizen~\mbox{\,$\La(\vec\be) \!\equiv\! \big(\La^\mu{}_\nu\big)$}.
Diese sind explizit gegeben durch Gl.~(\ref{APP:LT-La})~-- vgl.\@ etwa Ref.~\cite{Sexl92}.
Wir rekapitulieren in Komponenten:
%
\begin{alignat}{3} \label{APP:LT-La-Komponenten}
&\La^0{}_0&\;
  &=\;& &\ga
    \\[.25ex]
&\La^i{}_0&\;
  &=\;& -\, &\ga \be\, \hat\be^i
    \tag{\ref{APP:LT-La-Komponenten}$'$} \\[.25ex]
&\La^0{}_j&\;
  &=\;& -\, &\ga \be\, \hat\be^j
    \tag{\ref{APP:LT-La-Komponenten}$''$} \\[.25ex]
&\La^i{}_j&\;
  &=\;& &\de^i_j + (\ga \!-\! 1)\, \hat\be^i\, \hat\be^j
    \tag{\ref{APP:LT-La-Komponenten}$'''$}
\end{alignat}
die letzte Zeile mit der Identit"at~\mbox{\,$\ga^2\be^2\!/(\ga^2 \!-\! 1) = \ga \!-\! 1$}. \\
\indent
Umgekehrt f"ur gegebene Boost-Transformationsmatrix~\mbox{\,$\La \!\equiv\! \big(\La^\mu{}_\nu\big)$}, die charakterisiert ist durch Gl.~(\ref{APP:spez-orthLT}$'$), folgt aus den Relationen
\begin{samepage}
%
\begin{align} \label{APP:Boost-Betrag,Richtung}
&\tr \La\;
  =\; \La^\mu{}_\mu\;
  =\; 2\, (\ga + 1)
    \\
&\hat\be^i\;
  =\; -\, \frac{1}{\ga\be}\; \La^i{}_0
    \tag{\ref{APP:Boost-Betrag,Richtung}$'$}
\end{align}
\vspace*{-.25ex}zun"achst {\it Betrag\/}~$\be$, dann {\it Richtung\/}~$\hat\be$ des Boosts. \\
\indent
In Hinblick auf Gl.~(\ref{APP:p^+'}) und~(\ref{APP:p^i'}) definieren wir durch die Komponenten
%
\begin{align} 
n^{\bar\mu}\;
  =\; \La^{\bar\mu}{}_i\, P^i\qquad
  \bar\mu \in \{+,-,1,2\}
\end{align}
den Vierer-Vektor~\mbox{\,$\bar{n} \!=\! \bar{n}[\rb{P}] \!\equiv\! (n^{\bar\mu}[\rb{P}])$} als Funktion von~$\rb{P} \!=\! (P^1,P^2)^{\T t}$.
Falls f"ur nichtverschwindenden Vektor~$\rb{P}$ simultan gilt~\mbox{\,$n^+[\rb{P}] \!=\! 0$} und~\mbox{\,$\rb{n}[\rb{P}] \!=\! 0$}, mit~\mbox{\,$\rb{n} \!=\! (n^1,n^2)^{\T t}$}, ist Kovarianz vertr"aglich mit~\mbox{\,$\la \!\ne\! 0$}~-- f"ur die Transformation~$\La$, die vermittelt wird durch die Komponenten~\mbox{\,$\big(\La^{\bar\mu}{}_{\bar\nu}\big)$}.
Wir zeigen, da"s dies f"ur keine Transformation~$\La$ der Fall ist.
\end{samepage}

Es ist~\mbox{$\La^+{}_i$},~\mbox{\,$i \!\in\! \{1,2\}$}, explizit gegeben durch:
%
\begin{align} \label{APP:La+i-Boost}
\La^+{}_i\;
  =\; \al\, \big[-\, \ga\be + (\ga \!-\! 1)\, \hat\be^3\big]\, \hat\be^i\qquad
  i \in \{1,2\}
\end{align}
vgl.\@ Gl.~(\ref{APP:spez-orthLTbar-pm}$''$) und~(\ref{APP:LT-La-Komponenten}$''$),~(\ref{APP:LT-La-Komponenten}$'''$). \\
\indent
Wir diskutieren das Verschwinden von~\mbox{\,$n^+[\rb{P}] \!=\! \La^+{}_iP^i$}.
Es gilt~\mbox{\,$n^+[\rb{P}] \!=\! 0$}, falls~{\it (i)\/} die ek\-kige Klammer in Gl.~(\ref{APP:La+i-Boost}) identisch verschwindet.
F"ur \vspace*{-.25ex}\mbox{\,$\hat\be^3 \!=\! \ga\be\!\big/\!(\ga \!-\! 1) \!=\! [(\ga \!+\! 1)\!\big/\!(\ga \!-\! 1)]^{1\!/\!2}$} ist dies formal der Fall,~-- steht aber, da gr"o"ser Eins, im Widerspruch dazu, da"s~\mbox{\,$\hat\be$} Einheitsvektor ist.
Da die eckige Klammer in Gl.~(\ref{APP:La+i-Boost}) ergo ungleich Null ist, verschwindet~\mbox{\,$n^+[\rb{P}]$} genau dann, wenn~{\it (ii)} gilt:~\mbox{\,$\hat\be^iP^i \!=\! 0$}~[etwa wenn~\mbox{\,$\hat\be^i \!\equiv\! 0$} f"ur~\mbox{\,$i \!\in\! \{1,2\}$}, das hei"st~\mbox{\,$\hat\be$} vollst"andig longitudinal ist:~\mbox{\,$\hat\be^3 \!\equiv\! 1$}]. \\
\indent
F"ur~\mbox{$\La^i{}_j$},~\mbox{\,$i,j \!\in\! \{1,2\}$}, folgt der explizite Ausdruck, vgl.\@ die Gln.~(\ref{APP:spez-orthLTbar-i}$'$),~(\ref{APP:LT-La-Komponenten}$'''$):
\vspace*{-.5ex}
\begin{align} \label{APP:Laij-Boost}
\La^i{}_j\;
  =\; \de^i_j + (\ga \!-\! 1)\, \hat\be^i\, \hat\be^j\qquad
  i,j \in \{1,2\}
    \\[-4.5ex]\nn
\end{align}
Wir diskutieren das simultane Verschwinden von~\mbox{\,$n^+ \!=\! \La^+{}_iP^i$} und des transversalen Vektors \mbox{$\rb{n} \!\equiv\! (n^i) \!=\! \big(\La^i{}_jP^j\big)$}, mit~\mbox{\,$i,j \!\in\! \{1,2\}$}.
Es bleibt nichts zu zeigen f"ur~{\it (i)}.
F"ur~{\it (ii)} folgt aus~\mbox{\,$\hat\be^iP^i \!=\! 0$} unmittelbar~\mbox{\,$\rb{n}[\rb{P}] \!=\! \rb{P} \!\ne\! \bm{0}$}. \\
\indent
Zusammen:
F"ur nichtverschwindenden Vektor~\mbox{\,$\rb{P}$} kann simultan {\it nicht\/} erreicht werden sowohl~\mbox{$n^+[\rb{P}] \!=\! \La^+{}_iP^i \!=\! 0$} als auch~\mbox{$\rb{n}[\rb{P}] \!=\! \big(\La^i{}_jP^j\big) \!=\! \bm{0}$}, mit~\mbox{\,$i,j \!\in\! \{1,2\}$}.~--
Es existiert {\it kein\/} speziell-orthochroner Lorentz-Boost~\mbox{\,$\La(\vec\be)$}, f"ur den Kovarianz vertr"aglich w"are mit~\mbox{\,$\la \!\ne\! 0$}.
\vspace*{-.5ex}

\paragraph{Reine Drehungen} sind vollst"andig bestimmt durch den (Dreier)Drehvektor
%
\begin{align} 
\vec\al\;
  =:\; \al\, \hat\al\qquad
  \text{mit}\qquad
  |\hat\al|\; \equiv\! \hat\be^i\, \hat\be^i\; =\; 1\quad
  i \in \{1,2,3\}
\end{align}
Es sind~\mbox{\,$\al \!=\! |\vec\al| \!<\! \pi$} und~\mbox{\,$\hat\al$} respektive Winkel und orientierte Achse der Drehung.
Reine Drehungen werden vermittelt durch Matrizen~\mbox{\,$\La(\vec\al) \!\equiv\! \big(\La^\mu{}_\nu\big)$}~-- vgl.\@ etwa Ref.~\cite{Sexl92}:
\vspace*{-.5ex}
\begin{align} \label{APP:Drehung}
&\La(\vec\al)\;
  \equiv\! \big(\La^\mu{}_\nu\big)
  =\; \pmatrixZZ{1}{\vec0^{\D\,t}}{\vec0}{R}\;
    \\[.25ex]
&\text{mit}\qquad
  R\; \equiv\! \big(R^i{}_j\big)\quad
  i,j \in \{1,2,3\}
    \qquad\text{\sl orthogonal\/}
    \tag{\ref{APP:Drehung}$'$}
    \\[-4.5ex]\nn
\end{align}
Explizit in Komponenten, mit~\mbox{\,$i,j \!\in\! \{1,2,3\}$}:
\begin{samepage}
\vspace*{-.5ex}
\begin{alignat}{2} \label{APP:DrehungRij}
&\La^0{}_0&\;
  &=\; 1
    \\[.25ex]
&\La^i{}_0&\;
  &=\; 0
    \tag{\ref{APP:DrehungRij}$'$} \\[.25ex]
&\La^0{}_j&\;
  &=\; 0
    \tag{\ref{APP:DrehungRij}$''$} \\[.25ex]
&\La^i{}_j&\;
  &=\; R^i{}_j\;
   =\; \hat\al^i\, \hat\al^j
       + \big(\de^i_j - \hat\al^i\, \hat\al^j\big)\, \cos\al
       + \ep^{ijk}\, \hat\al^k\, \sin\al
    \tag{\ref{APP:DrehungRij}$'''$}
    \\[-4.5ex]\nn
\end{alignat}
Durch~\mbox{\,$R^i{}_j\, R^k{}_j \!=\! \de^i_k$} ist garantiert~\mbox{\,$RR^{\D t} \!=\! R^{\D t}R \!=\! \bbbone$}, das hei"st Orthogonalit"at von~\mbox{\,$R \!\equiv\! \big(R^i{}_j\big)$}. \\
\indent
Umgekehrt f"ur vorgegebene orthogonale Matrix~\mbox{\,$R \!\equiv\! \big(R^i{}_j\big)$},~\mbox{$i,j \!\in\! \{1,2,3\}$}, gelten die Relationen
\vspace*{-.5ex}
\begin{align} \label{APP:Drehung-Betrag,Richtung}
&\tr R\;
  =\; R^i{}_i\;
  =\; 1 + 2\cos\al
    \\
&\hat\al^i\;
  =\; \frac{1}{2\sin\al}\; \ep^{ijk}\, R^j{}_k
    \tag{\ref{APP:Drehung-Betrag,Richtung}$'$}
    \\[-4.5ex]\nn
\end{align}
aus denen folgt zun"achst der {\it Betrag\/}~$\al$, dann im Sinne der orientierten Achse die {\it Richtung\/}~\mbox{\,$\hat\al$} der Drehung.
\end{samepage}

Es ist~\mbox{$\La^+{}_i$},~\mbox{\,$i \!\in\! \{1,2\}$}, explizit gegeben durch, vgl.\@ Gl.~(\ref{APP:spez-orthLTbar-pm}$''$):
\begin{samepage}
%
\begin{align} 
\La^+{}_i\;
  =\; \al\, \big[\hat\al^3\, \hat\al^i\, (1 \!-\! \cos\al)
               + \hat\al_\perp^i\, \sin\al\big]\qquad
  i \in \{1,2\}
\end{align}
vgl.\@ Gl.~(\ref{APP:spez-orthLTbar-pm}$''$) und~(\ref{APP:DrehungRij}$''$),~(\ref{APP:DrehungRij}$'''$).
Die globale Konstante~$\al$ ist die Normierung der Lichtkegelkoordinaten.
Es ist ferner definiert der Zweier-Vektor~\mbox{\,$\hat\al_\perp \equiv\; (\hat\al_\perp^i)$},~\mbox{\,$i \!\in\! \{1,2\}$} durch:
\vspace*{-.5ex}
\begin{align} 
\hat\al_\perp^i
  :=\; \ep^{3ik}\, \hat\al^k\qquad
  \text{d.h.}\qquad
  \hat\al_\perp\;
  =\; \pmatrixZE{\hat\al^2}{-\, \hat\al^1}\;
  =\; \pmatrixZZ{0}{1}{-1}{0} \pmatrixZE{\hat\al^1}{\hat\al^2}
    \\[-5ex]\nn
\end{align}
\end{samepage}%
das hei"st~\mbox{\,$\hat\al_\perp$} ist {\it per construktionem\/} orthogonal zum transversalen Anteil~\mbox{\,$(\hat\al^1,\hat\al^2)^{\T t}$} von~\mbox{\,$\hat\al$} und Einheitsvektor:~\mbox{\,$1 \!=\! \hat\al_\perp^i\hat\al_\perp^i \!=\! \hat\al^i\hat\al^i$}, genau dann, wenn~\mbox{\,$\hat\al^3 \!\equiv\! 0$}. \\
\indent
Wir diskutieren das Verschwinden von~\mbox{\,$n^+[\rb{P}] \!=\! \La^+{}_iP^i$} in Unterscheidung der folgenden F"alle.
Sei~{\it (i)\/} die Drehachse vollst"andig longitudinal:~\mbox{\,$\hat\al^3 \!\equiv\! 1$}, \mbox{\,$\hat\al^i,\hat\al_\perp^i \!\equiv\! 0$} f"ur~\mbox{\,$i \!\in\! \{1,2\}$}.
Dann folgt unmittelbar~\mbox{\,$n^+[\rb{P}] \!=\! 0$} f"ur einen beliebigen Vektor~\mbox{\,$\rb{P}$}.
Sei umgekehrt~{\it (ii)\/} die Drehachse vollst"andig transversal:~\mbox{\,$\hat\al^3 \!\equiv\! 0$}, \mbox{\,$\hat\al^i,\hat\al_\perp^i \!\ne\! 0$} f"ur mindestens einen Index~\mbox{\,$i \!\in\! \{1,2\}$}.
Dann verschwindet~\mbox{\,$n^+[\rb{P}] \!=\! \al\,\hat\al_\perp^iP^i\sin\al$} genau dann, wenn~\mbox{\,$\rb{P}$} orthogonal ist zu~\mbox{\,$\hat\al_\perp$}, das hei"st (anti)parallel ist zur Drehachse~\mbox{\,$\hat\al$}, folglich: \mbox{$P^i \!=\! \pm|\rb{P}|\hat\al^i$}.
Sei schlie"slich~{\it (iii)\/} der Drehvektor nicht vollst"andig longitudinal, das hei"st \mbox{\,$\hat\al^3 \!\ne\! 0$} und~\mbox{\,$\hat\al^i \!\ne\! 0$} f"ur mindestens einen Index~\mbox{\,$i \!\in\! \{1,2\}$}.
Dann sind die transversalen Vektoren~\mbox{\,$(\hat\al^1,\hat\al^2)^{\T t}$} und~\mbox{\,$(\hat\al^2,-\hat\al^1)^{\T t}$} nichtverschwindend orthogonal, insbesondere linear unabh"angig; es folgt unmittelbar~\mbox{\,$n^+[\rb{P}] \!\ne\! 0$} f"ur einen beliebigen Vektor~\mbox{\,$\rb{P}$}. \\
\indent
F"ur~\mbox{$\La^i{}_j$},~\mbox{\,$i,j \!\in\! \{1,2\}$}, folgt der explizite Ausdruck, vgl.\@ die Gln.~(\ref{APP:spez-orthLTbar-i}$'$),~(\ref{APP:DrehungRij}$'''$):
%
\begin{align} \label{APP:Laij-Drehung}
\La^i{}_j\;
  =\; \hat\al^i\, \hat\al^j
      + \big(\de^i_j - \hat\al^i\, \hat\al^j\big)\, \cos\al
      + \ep^{ij3}\, \hat\al^3\, \sin\al\qquad
  i,j \in \{1,2\}
\end{align}
Wir diskutieren das simultane Verschwinden von~\mbox{\,$n^+ \!=\! \La^+{}_iP^i$} und des transversalen Vektors~\mbox{\,$\rb{n} \!\equiv\! (n^i) \!=\! \big(\La^i{}_jP^j\big)$}, mit~\mbox{\,$i,j \!\in\! \{1,2\}$}.
Es folgt,~{\it (i)}, f"ur vollst"andig longitudinalen Drehvektor:~\mbox{\,$n^i[\rb{P}] \!=\! \ep^{ij3} P^j \hat\al^3 \sin\al \!\ne\! 0$}.
Es folgt,~{\it (ii)}, f"ur vollst"andig transversalen Drehvektor aus \mbox{\,$P^i \!=\! \pm|\rb{P}|\hat\al^i$} unmittelbar~\mbox{\,$\rb{n}[\rb{P}] \!=\! \rb{P} \!\ne\! \bm{0}$}.
F"ur~{\it (iii)} bleibt nichts zu zeigen. \\
\indent
Zusammen:
Es existiert {\it keine\/} reine Drehung~\mbox{\,$\La(\vec\al)$}, f"ur die simultan identisch Null w"aren \mbox{\,$n^+ \!=\! \La^+{}_iP^i$} und \mbox{\,$n^i \!=\! \La^i{}_jP^j$}, mit~\mbox{\,$i \!\in\! \{1,2\}$},~-- das hei"st f"ur die Kovarianz nicht im Widerspruch st"unde mit~\mbox{\,$\la \!\ne\! 0$}.
\vspace*{-.5ex}

\bigskip\noindent
Wir schlie"sen den Anhang mit der Feststellung, da"s f"ur {\it jede nichttriviale\/} spezielle orthochrone Lorentz-Transformation~-- da darstellbar als Verkettung von Lorentz-Boosts und reinen Drehungen~-- die Forderung~\mbox{\,$\la \!\ne\! 0$} im Widerspruch steht mit Kovarianz der Raumzeit-Parameter nach den Gln.~(\ref{APP:LCWFN-momentum_+}),~(\ref{APP:LCWFN-momentum_i}).
Es ist daher zu fordern~Gl.~(\ref{APP:la=0}):~\mbox{\,$\la \!\equiv\! 0$}.
\theendnotes

%% file: APP_INTEGRATE-F.tex
\lhead[\fancyplain{}{\sc\thepage}]
      {\fancyplain{}{\sc\rightmark}}
\rhead[\fancyplain{}{\sc{{\footnotesize Anhang~\thechapter:} Integration}}]
      {\fancyplain{}{\sc\thepage}}
\psfull
\chapter[Integration]{%
   \huge Integration}
\label{APP:Integration}

Wir geben an unsere Konventionen bez"uglich Integration auf allgemeinen (pseudo-)Riemann\-schen Mannigfaltigkeiten, die unter anderem bezogen werden auf die Minkowskische Raumzeit und die Koordinatenlinien~\mbox{\,$\tilde\mu \!\in\! \{\mfp,\mfm,1,2\}$}, wie definiert in Anhang~\ref{APP:Boosts}.
Auf Basis dieses Formalismus geschieht im Haupttext etwa die explizite Auswertung der~\mbox{zentralen Funktio}\-nen~\mbox{\,$\tilde\ch\oC\idx{\imath\jmath}$},~\mbox{\,$\tilde\ch\oNC\idx{\imath\jmath}$},~\mbox{$\imath,\jmath \!\in\! \{\mfp,\mfm\}$}; formalere Schritte dieser Auswertung sind festgehalten hier.
\vspace*{-.5ex}

\section{Integration auf (pseudo-)Riemannschen Mannigfaltigkeiten.}
\label{APP-Sect:Integration}

Sei~\mbox{$G \!\equiv\! G_{d,\si}$} eine (pseudo-)Riemannsche Mannigfaltigkeit der Dimension~\mbox{$d \!\equiv\! \ze \!+\! \rh$} und Signatur~\mbox{$\si \!\equiv\! \ze \!-\! \rh$} mit~$\ze$ der Anzahl zeitartiger und~$\rh$ der Anzahl raumartiger Dimensionen.%
\FOOT{
  \mbox{$G$} hei"st Riemannsch, falls~\mbox{\,$\ze \!\equiv\! 0$} oder~\mbox{\,$\rh \!\equiv\! 0$}, pseudo-Riemannsch sonst.   Es ist~$G_{4,-2}$~-- mit~\mbox{$\ze \!\equiv\! (d \!+\! \si)\!/2 \!=\! 1$} und~\mbox{$\rh \!\equiv\! (d \!-\! \si)\!/2 \!=\! 3$}~-- {\sl lokal isomorph\/} zum Minkowski-Raum,~$G_{n,-n}$ zum gew"ohnlichen Euklidischen~$\bbbr^n$.
}

Der metrische Tensor~\mbox{\,$g \!\equiv\! \big(g_{\mu\nu}\big)$} von~$G$ h"angt allgemein ab vom Weltpunkt~$x$:~\mbox{\,$g \!\equiv\! g(x)$}; es existiert lokal eine Basis, in der er Diagonalform besitzt, o.E.d.A.
\begin{samepage}
%
\begin{align} \label{APP:g(x)-lokal-diag}
g\;
  \equiv\; \big(g_{\mu\nu}\big)\;
  =\; {\rm diag}[+1,\ldots,+1,-1,\ldots,-1]
\end{align}
mit den~$\ze$ ersten Eintr"age $+1$, den~$\rh$ "ubrigen~$-1$. 
Unter allgemeiner Lorentz-Transforma\-tion%
  ~\mbox{$\La \!\equiv\! \big(\La^\mu{}_\nu\big)$} gilt
  ~\mbox{\,$g'_{\mu\nu} \!=\! \La_\mu{}^\rh \La_\nu{}^\si g_{\rh\si}$}, ergo%
  ~\mbox{\,$\det g' \!=\! \det g \big/\! [\det \La]^2$} und%
  ~\mbox{\,${\rm sign}(\det g') \!\equiv\! (-1)^\rh$}. \\
\indent
Der Epsilon-Pseudotensor~$\ep \!\equiv\! (\ep^{\mu_1\cdots\mu_d})$ ist definiert durch seine kontravarianten Komponenten
\vspace*{-.5ex}
\begin{align} \label{APP:epTensor-G}
&\ep^{\mu_1\cdots\mu_d}\;
  \equiv\; \pmatrixZD{\mu_1}{\cdots}{\mu_d}{1}{\cdots}{d}\; \ep^{1\cdots d}\qquad
  \text{\it per def.:}\qquad
  \ep^{1\cdots d}\;
  \equiv\; 1\big/\!\sqrt{(-1)^\rh \det g}
    \\[-4.5ex]\nn
\end{align}
vgl.\@ Gl.~(\ref{APP:epTensor-kontrav}); seine kovarianten Komponenten folgen durch
%
\begin{align} 
\ep_{\mu_1\cdots\mu_d}\;
  =\; (-1)^\rh \det g\vv \ep^{\mu_1\cdots\mu_d}
\end{align}

Sei definiert f"ur~\mbox{\,$s \!\in\! \bbbn$} mit~\mbox{$1 \!\leq\! s \!\leq\! d$} das verallgemeinerte Kronecker-Symbol~\mbox{\,$\de^{\mu_1\cdots\mu_s}_{\, \nu_1\cdots\nu_s}$} als die Determinante konventioneller Kroneckersymbole:%
\FOOT{
  Das konventionelle Kronecker-Symbol~\mbox{\,$\de^\mu_\nu$} transformiert wie ein~\mbox{\,$(1,1)$}-Tensor, das hei"st wie ein Tensor zweiter Stufe mit einem kontra- und einem kovarianten Index,~-- folglich~\mbox{\,$\de^{\mu_1\cdots\mu_s}_{\, \nu_1\cdots\nu_s}$} wie ein~\mbox{\,$(s,s)$}-Tensor.
}
\end{samepage}
\vspace*{-.5ex}
\begin{align} \label{APP:Kronecker_det-Def}
\de^{\mu_1\cdots\mu_s}_{\, \nu_1\cdots\nu_s}\;
  =\; \det \big( \de^{\mu_i}_{\nu_j} \big)\;
  \equiv\; \det
      \begin{pmatrix} \de^{\mu_1}_{\nu_1} & \cdots & \de^{\mu_1}_{\nu_s} \\
                      \vdots & & \vdots \\
                      \de^{\mu_s}_{\nu_1} & \cdots & \de^{\mu_s}_{\nu_s}
      \end{pmatrix}\qqquad
  i,j \in \{1,2,\ldots,s\}
    \\[-4ex]\nn
\end{align}
mithilfe der Leibnitz'schen Determinantenformel:
\vspace*{-.5ex}
\begin{align} \label{APP:Kronecker_det-Leibnitz}
\de^{\mu_1\cdots\mu_s}_{\, \nu_1\cdots\nu_s}\;
  =\; {\T\sum}_{\T \si \!\in\! S_s}\vv {\rm sign}(\si)\vv
        \de^{\mu_1}_{\nu_{\si(1)}}\, \cdots\, \de^{\mu_s}_{\nu_{\si(s)}}
    \\[-4.5ex]\nn
\end{align}
Diese Definition impliziert vollst"andige Antisymmetrie:
%
\begin{align} \label{APP:Kronecker-antisym}
&\de^{\mu_1\cdots\mu_s}_{\, \nu_1\cdots\nu_s}\;
  =\; \de^{[\mu_1 \cdots \mu_s]}_{\;\, \nu_1 \cdots \nu_s}\;
  =\; \de^{\, \mu_1 \cdots \mu_s}_{[\nu_1 \cdots \nu_s]}
\end{align}
bzgl.\@ der Notation vgl.\@ Gl.~(\ref{APP:(Anti)Symmetrisierung}).
Es ist ferner~\mbox{\,$\de^{\mu_1\cdots\mu_s}_{\, \nu_1\cdots\nu_s}$} gleich dem Signum der Indexpermutation~\mbox{$\si^{\D\mskip-1mu\star}\!: \mu_i \!\to\! \nu_i$}, f"ur alle~\mbox{\,$i \!\in\! \{1,2,\ldots,s\}$}, und kann daher~-- vgl.\@ Gl.~(\ref{APP:epTensor-G})~-- ausgedr"uckt werden durch die Komponenten des Epsilon-Pseudotensors:
%
\begin{align} \label{APP:Kronecker_ep}
\de^{\mu_1\cdots\mu_s}_{\, \nu_1\cdots\nu_s}\;
  &=\; \pmatrixZD{\mu_1}{\cdots}{\mu_s}{\nu_1}{\cdots}{\nu_s}
    \\[.5ex]
  &=\; \frac{(-1)^\rh}{(d \!-\! s)!}\; \ep^{\mu_1\cdots\mu_s \rh_1\cdots\rh_{d-s}}\;
                  \ep_{\nu_1\cdots\nu_s \rh_1\cdots\rh_{d-s}}
    \tag{\ref{APP:Kronecker_ep}$'$}
\end{align}
Kontraktion mit dem verallgemeinerten Kronecker-Symbol bewirkt vollst"andige Antisymmetrisierung bez"uglich der kontrahierten Indizes, vgl.\@ Gl.~(\ref{APP:(Anti)Symmetrisierung}$'$) und~(\ref{APP:Kronecker_det-Leibnitz}):
%
\begin{align} \label{APP:ProjectAntisymm}
\frac{1}{s!}\;
  \de^{\mu_1\cdots\mu_s}_{\, \nu_1\cdots\nu_s}\vv
  T^{\nu_1\cdots\nu_s\rh_1\cdots\rh_r}\;
  =\; T^{[\mu_1\cdots\mu_s]\rh_1\cdots\rh_r}
\end{align}

Wir betrachten~$G_s \!\subset\! G$, eine Untermannigfaltigkeit von~$G$ der Dimension~$s$ mit~$1 \!\leq\! s \!\leq\! d$.
Seien~$d_ix$~-- mit~$i \!=\! 1,\ldots s$~-- $s$ linear unabh"angige Differentiale auf~$G_s$ und sei
\begin{align} \label{APP:Differential-i}
d_ix\; =\; \frac{\partial x}{\partial u_i}\; du_i\qquad
  \text{{\it nicht\/} summiert "uber~$i$}
\end{align}
eine Parameterdarstellung.
Dann definieren wir ein $s$-dimensiona\-les Volumenelement~$dV$ auf~$G_s$, durch seine kontravarianten Komponenten:
\begin{align} \label{APP:dV_Gs_Def}
dV^{\mu_1\cdots\mu_s}\;
  &\equiv\; \de^{\mu_1\cdots\mu_s}_{\, \nu_1\cdots\nu_s}\vv
        d_1x^{\nu_1}\, \cdots\, d_sx^{\nu_1} \\[1ex]
  &=\; d_1x^{\mu_1}\,\wedge\, \cdots\, \wedge\, d_sx^{\mu_s}
        \tag{\ref{APP:dV_Gs_Def}$'$}
\end{align}
wobei~Gl.~(\ref{APP:dV_Gs_Def}$'$) verkn"upft mit dem Kalk"ul alternierender Differentialformen.
Das Volumenelement~$dV$ nach Gl.~(\ref{APP:dV_Gs_Def}) folgt in Standardform mithilfe Gl.~(\ref{APP:Kronecker_det-Def}):
\begin{samepage}
\vspace*{-.5ex}
\begin{align} \label{APP:dV_Gs}
dV^{\mu_1\cdots\mu_s}\;
  &=\; \det \big( \de^{\mu_i}_{\nu_j} \big)\vv
         \frac{\partial x^{\nu_1}}{\partial u_1}\, \cdots\,
           \frac{\partial x^{\nu_s}}{\partial u_s}\vv
         du_1\, \cdots\, du_s
    \\[.5ex]
  &=\; \det \Big( \frac{dx^{\mu_i}}{du_j} \Big)\vv
         du_1\, \cdots\, du_s
    \tag{\ref{APP:dV_Gs}$'$} \\[.5ex]
  &=\; \frac{\pa(x^{\mu_1},\ldots x^{\mu_s})}{\pa(u_1,\ldots u_s)}\vv
         du_1\, \cdots\, du_s
    \tag{\ref{APP:dV_Gs}$''$}
    \\[-4.5ex]\nn
\end{align}
mit der konventionellen Jacobi-Determinante in der letzten Zeile.

Wir formulieren den {\bf Stokes'schen Satz} f"ur eine~$(s \!-\! 1)$-Form~$\om$ auf~$G$ im Kalk"ul alternierender Differentialformen:
\vspace*{-.5ex}
\begin{align} \label{APP:Stokes_Diff'formen}
\int_{G_s}\; d\wedge\om\; =\; \int_{\pa G_s}\; \om
    \\[-4.5ex]\nn
\end{align}
und "aquivalent f"ur einen Tensor~$T$ der Stufe~$(s \!-\! 1)$ im Tensorkalk"ul:
\vspace*{-.5ex}
\begin{align} \label{APP:Stokes}
\int_{G_s} T_{\mu_1\cdots\mu_{s-1};\rh}\;
  dV^{\rh \mu_1\cdots\mu_{s-1}}\;
  =\; \int_{\pa G_s} T_{\mu_1\cdots\mu_{s-1}}\;
       dV^{\mu_1\cdots\mu_{s-1}}
    \\[-4.5ex]\nn
\end{align}
\end{samepage}%
Wir arbeiten in der Standard-Kommanotation, in der~"`$;\rh$"'steht f"ur die kovariante Ableitung nach dem Index~$\mu$; sie ist identisch~"`$,\rh$"', der partiellen Ableitung, unter der vollst"andigen Antisymmetrisierung, wie sie gegeben ist unter Kontraktion mit dem Volumenelement; vgl.\@ Gl.~(\ref{APP:ProjectAntisymm}).

Im folgenden geben wir die Formulierung dieses Satzes in Termen dualer Tensoren an.
Wir definieren zun"achst {\bf Dualit"at} in~$G$:
Sei~$T$ ein beliebiger Tensor~$s$-ter Stufe.
Dann folgt der zu~$T$ duale Tensor~$\tilde{T}$ der Stufe~$(d \!-\! s)$ durch Anwenden des Dualit"atsoperators:%
\FOOT{
  \label{FN:ko-kontravar}Durch Kontraktion mit Metriken, hier explizit mit:~$g_{\rh_1\mu_1}\cdots g_{\rh_s\mu_s}$, folgen alle Relationen mit kovarianten durch kontravariante Indizes ersetzt, und umgekehrt.   Aufgrund der allgemeinen Identit"aten~$g_{\mu\nu;\rh} \!\equiv\! 0$ und~$\ep_{\mu_1\cdots\mu_d;\rh} \!\equiv\! 0$ gilt dies weiter, wenn kovariante Ableitungen involviert sind.
}
%
\vspace*{-.5ex}
\begin{align} \label{APP:Ttilde_T-d(s)}
\tilde{T}_{\nu_1\cdots\nu_{d-s}}\;
  =\; d_{(d-s)\, \nu_1\cdots\nu_{d-s} \mu_1\cdots\mu_s}\vv
        T^{\mu_1\cdots\mu_s}
    \\[-4.5ex]\nn
\end{align}
Der Dualit"atsoperator~$d_{(d-s)}$ ist per definitionem Dualit"at proportional dem Epsilon-Pseudo\-tensor, die Proportionalit"atskonstante h"angt ab von~dem Index~$d\!-\!s$ (der Zahl freier Indizes nach Dualisieren):
\vspace*{-.5ex}
\begin{align} \label{APP:d(s)}
d_{(d-s)\, \mu_1\cdots\mu_d}\;
  =\; c_{(d-s)}\vv \ep_{\mu_1\cdots\mu_d}
    \\[-4.5ex]\nn
\end{align}
Wir lassen~$c_{(d-s)}$ zun"achst offen.
Aus der Kontraktion von Gl.~(\ref{APP:Ttilde_T-d(s)}) mit~$\ep^{\mu_1\cdots\mu_s \rh_1\cdots\rh_{d-s}}$ folgt mithilfe Gl.~(\ref{APP:Kronecker_ep}):
\vspace*{-.5ex}
\begin{align} \label{APP:T_Ttilde}
(-1)^\rh\; T^{\mu_1 \cdots \mu_s}\;
  =\; \frac{1}{c_{(d-s)}\cdot s!(d\!-\!s)!}\vv
        \ep^{\nu_1\cdots\nu_{d-s} \mu_1\cdots\mu_s}\vv \tilde{T}_{\nu_1\cdots\nu_{d-s}}
    \\[-4.5ex]\nn
\end{align}
Andererseits f"uhrt zweifaches Dualisieren~-- bis auf das definierte Vorzeichen~$(-1)^\rh$~-- zur"uck auf den urspr"unglichen Tensor; wir haben:
\vspace*{-.5ex}
\begin{align} 
\dbtilde{T}^{\mu_1\cdots\mu_s}\;
  =\; (-1)^\rh\; T^{\mu_1 \cdots \mu_s}
    \\[-4.5ex]\nn
\end{align}
Damit ist Gl.~(\ref{APP:T_Ttilde}), analog zu Gl.~(\ref{APP:Ttilde_T-d(s)}), zu lesen im Sinne
\vspace*{-.5ex}
\begin{align} \label{APP:T_Ttilde-dtilde(s)}
\dbtilde{T}^{\mu_1\cdots\mu_s}\;
  =\; \tilde{d}_{(s)}{}^{\! \mu_1\cdots\mu_s \nu_1\cdots\nu_{d-s}}\vv
        \tilde{T}_{\nu_1\cdots\nu_{d-s}} 
    \\[-4ex]\nn
\end{align}
Der hierdurch definierte Dualit"atsoperator~$\tilde{d}_{(s)}$ tr"agt eine Tilde zur kennzeichnung, da"s er auf duale Tensoren wirkt.
Analog zu Gl.~(\ref{APP:d(s)}) schreiben wir:
\vspace*{-.5ex}
\begin{align} \label{APP:dtilde_(s)}
\tilde{d}_{(s)}{}^{\! \mu_1\cdots\mu_d}\;
  =\; \tilde{c}_{(s)}\vv \ep^{\mu_1\cdots\mu_d}
    \\[-4.5ex]\nn
\end{align}
dabei folgt die Konstante~$\tilde{c}_{(s)}$ in Abh"angigkeit von~$c_{(d-s)}$ aus den Gln.~(\ref{APP:T_Ttilde})-(\ref{APP:dtilde_(s)}):
\begin{samepage}
\vspace*{-.5ex}
\begin{align} \label{APP:ctilde(s)_c(s)}
\tilde{c}_{(s)}\;
  =\; \frac{\sgnd(s)}{c_{(d-s)}\cdot s!(d\!-\!s)!}\qquad
  \text{mit}\qquad
  \sgnd(s)\; =\; (-1)^{s(d-s)}
    \\[-4.5ex]\nn
\end{align}
Dabei ist das Signum genau das Signum der Permutation der~$s$ Indizes~$\mu_i$ vorbei an den~$d\!-\!s$ Indizes~$\nu_j$ im Epsilon-Pseudotensor in Gl.~(\ref{APP:T_Ttilde}).
Wir werden sehen, da"s die symmetrische Definition:~$c_{(s)} \!\equiv\! \tilde{c}_{(s)} \!\equiv\! (-1)^s\!/s!$, {\it nicht\/} die g"unstigste ist.

Aus Gl.~(\ref{APP:Ttilde_T-d(s)}) folgt explizit f"ur das zu~$dV$, vgl.\@ Gl.~(\ref{APP:dV_Gs_Def}), duale Volumenelement:
\vspace*{-.5ex}
\begin{alignat}{2} \label{APP:dVtilde_dV}
d\tilde{V}_{\nu_1\cdots\nu_{d-s}}\;
  &=\; &&c_{(d-s)}\vv
         \ep_{\nu_1\cdots\nu_{d-s} \mu_1\cdots\mu_s}\vv
         dV^{\mu_1\cdots\mu_s}
    \\[.5ex]
  &=\; s!\; &&c_{(d-s)}\vv
         \ep_{\nu_1\cdots\nu_{d-s} \mu_1\cdots\mu_s}\vv
         d_1x^{\mu_1}\, \cdots\, d_sx^{\mu_s}
    \tag{\ref{APP:dVtilde_dV}$'$}
    \\[-4.5ex]\nn
\end{alignat}
Als Verkn"upfung mit dem Differentialformenkalk"ul ist
\vspace*{-.5ex}
\begin{align} 
d\tilde{V}_{\nu_1\cdots\nu_{d-s}}\;
  =\; c_{(d-s)}\vv
        \ep_{\nu_1\cdots\nu_{d-s} \mu_1\cdots\mu_s}\vv
        d_1x^{\mu_1}\,\wedge\, \cdots\, \wedge\, d_sx^{\mu_s}
    \\[-4.5ex]\nn
\end{align}
das Pendant zu Gl.~(\ref{APP:dV_Gs_Def}$'$).
\end{samepage}

Wir betrachten die bez"uglich allgemeiner Koordinatentransformationen skalaren Integranden im Stokes'schen Satz nach Gl.~(\ref{APP:Stokes}).
Wir haben:
\vspace*{-.5ex}
\begin{align} \label{APP:Stokes_Integranden}
&T_{\mu_1\cdots\mu_{s-1};\rh}\; dV^{\rh \mu_1\cdots\mu_{s-1}}
    \\
&=\; (-1)^\rh\vv (s\!-\!1)!\; (d-(s\!-\!1))!\vv \tilde{c}_{(s-1)}\;\cdot\;
       \tilde{c}_{(s)}\, (-1)^{d-1}\vv
       \tilde{T}^{\nu_1\cdots\nu_{d-s}\rh}{}_{;\rh}\vv
       d\tilde{V}_{\nu_1\cdots\mu_{d-s}}
    \nn \\[1ex]
&T_{\mu_1\cdots\mu_{s-1}}\; dV^{\mu_1\cdots\mu_{s-1}}
    \tag{\ref{APP:Stokes_Integranden}$'$}
  \\
&=\; (-1)^\rh\vv (s\!-\!1)!\; (d-(s\!-\!1))!\vv \tilde{c}_{(s-1)}\;\cdot\;
       \tilde{c}_{(s-1)}\vv
       \tilde{T}^{\nu_1\cdots\nu_{d-s}\rh}\vv
       d\tilde{V}_{\nu_1\cdots\mu_{d-s}\rh}
    \nn
    \\[-4.5ex]\nn
\end{align}
Aus Gl.~(\ref{APP:ctilde(s)_c(s)}) folgt die Identit"at
\vspace*{-.5ex}
\begin{align} 
(-1)^{d-1}\; \sgnd(s)\;
  =\; \sgnd(s\!-\!1)
    \\[-4.5ex]\nn
\end{align}
Division der Integranden der Gln.~(\ref{APP:Stokes_Integranden}),~(\ref{APP:Stokes_Integranden}$'$) durch den Ausdruck vor den Punkten und Multiplikation mit~$\sgnd(s\!-\!1)$ ergibt mithilfe von Gl.~(\ref{APP:Stokes}) den Stokes'schen Satz in seiner dualen Formulierung:
\begin{align} \label{APP:Stokes_dual}
&\tilde{c}_{(s)}\; \sgnd(s)\;
  \int_{G_s}
  \tilde{T}^{\nu_1\cdots\nu_{d-s}\rh}{}_{;\rh}\vv
  d\tilde{V}_{\nu_1\cdots\nu_{d-s}}
    \\
&=\vv \tilde{c}_{(s-1)}\; \sgnd(s\!-\!1)\;
        \int_{\pa G_s}
        \tilde{T}^{\nu_1\cdots\nu_{d-s}\rh}\vv
        d\tilde{V}_{\nu_1\cdots\nu_{d-s}\rh}
    \nn
\end{align}
indem wir benutzen die (kovariante) Divergenzfreiheit von metrischem Tensor und Epsilon-Pseudotensor:~$g_{\mu\nu;\rh} \!\equiv\! 0$ und~$\ep_{\mu_1\cdots\mu_d;\rh} \!\equiv\! 0$.
Der Faktor~$\sgnd(s)$ k"urzt gerade das in~$\tilde{c}_{(s)}$ explizit definierte Vorzeichen, vgl.\@ Gl.~(\ref{APP:ctilde(s)_c(s)}).

F"ur den Fall~$s \!\equiv\! d$ wird Gl.~(\ref{APP:Stokes_dual}) zu
\vspace*{-.5ex}
\begin{align} \label{APP:Stokes_duald}
&\tilde{c}_{(d)}\; \sgnd(d)\;
  \int_{G_d} \tilde{T}^{\rh}{}_{;\rh}\; d\tilde{V}\vv
  =\vv \tilde{c}_{(d-1)}\; \sgnd(d\!-\!1)\;
     \int_{\pa G_d} \tilde{T}^{\rh}\; d\tilde{V}_{\rh}
    \\[1ex]
&\text{mit}\qquad
  d\tilde{V}\; \hspace*{.25em}
    =\; d!\; c_{(0)}\vv
        \ep_{\mu_1\cdots\mu_d}\vv
        d_1x^{\mu_1}\, \cdots\, d_dx^{\mu_d}
    \tag{\ref{APP:Stokes_duald}$'$} \\[.5ex]
&\hspace*{3.5em}
  d\tilde{V}_\rh\;
    =\; (d \!-\! 1)!\; c_{(1)}\vv
        \ep_{\rh \mu_1\cdots\mu_{d-1}}\vv
        d_1x^{\mu_1}\, \cdots\, d_{d-1}x^{\mu_{d-1}}
    \tag{\ref{APP:Stokes_duald}$''$}
    \\[-4.5ex]\nn
\end{align}
Die Faktoren vor den Integralen in den Gln.~(\ref{APP:Stokes_dual}) und~(\ref{APP:Stokes_duald}) werden zu Eins~-- und dies genau sei unsere Wahl~--, wenn die Konstanten~$c_{(s)}$ gesetzt werden wie folgt:
\vspace*{-.5ex}
\begin{align} 
c_{(s)}\;
  =\; \frac{1}{s!(d\!-\!s)!}
    \\[-4.5ex]\nn
\end{align}
Denn dies impliziert~$\tilde{c}_{(s)} \!\equiv\! \sgnd(s)$.
Die definierenden Gleichungen bez"uglich Dualit"at~-- die Gln.~(\ref{APP:Ttilde_T-d(s)}) und~(\ref{APP:T_Ttilde-dtilde(s)})~-- lauten in dieser Konvention:
\vspace*{-.5ex}
\begin{align} \label{APP:Ttilde_T-Konv}
\tilde{T}_{\nu_1\cdots\nu_{d-s}}\;
  =\; \frac{1}{s!(d\!-\!s)!}\vv
         \ep_{\nu_1\cdots\nu_{d-s} \mu_1\cdots\mu_s}\vv
         T^{\mu_1\cdots\mu_s}
    \\[-4.5ex]\nn
\end{align}
beziehungsweise:
\vspace*{-.5ex}
\begin{align} \label{APP:T_Ttilde-Konv}
T^{\mu_1\cdots\mu_s}\;
  &=\; (-1)^\rh\; \sgnd(s)\vv
         \ep^{\mu_1\cdots\mu_s \nu_1\cdots\nu_{d-s}}\vv
         \tilde{T}_{\nu_1\cdots\nu_{d-s}}
    \\[.5ex]
  &=\; (-1)^\rh\vv
         \tilde{T}_{\nu_1\cdots\nu_{d-s}}\vv
         \ep^{\nu_1\cdots\nu_{d-s} \mu_1\cdots\mu_s}
    \tag{\ref{APP:T_Ttilde-Konv}$'$}
    \\[-4.5ex]\nn
\end{align}
Speziell f"ur das duale Volumenelement:
\vspace*{-.5ex}
\begin{align} \label{APP:dVtilde_dV-Konv}
&d\tilde{V}_{\nu_1\cdots\nu_{d-s}}\;
  =\; \frac{1}{(d\!-\!s)!}\vv
        \ep_{\nu_1\cdots\nu_{d-s} \mu_1\cdots\mu_s}\vv
        d_1x^{\mu_1}\, \cdots\, d_sx^{\mu_s}
    \\[-4.5ex]\nn
\end{align}
das also impliziert Division durch die Fakult"at der Anzahl freier Indizes.

Den diese Konvention bez"uglichen Stokes'schen Satz in seiner dualen Formulierung halten wir fest wie folgt~-- vgl.\@ Gl.~(\ref{APP:Stokes_dual}):
\begin{align} \label{APP:Stokes_dual-Konv}
&\tilde{c}_{(s)}\; \sgnd(s)\;
  \int_{G_s}
  \tilde{T}^{\nu_1\cdots\nu_{d-s}\rh}{}_{;\rh}\vv
  d\tilde{V}_{\nu_1\cdots\nu_{d-s}}
    \\
&=\vv \tilde{c}_{(s-1)}\; \sgnd(s\!-\!1)\;
        \int_{\pa G_s}
        \tilde{T}^{\nu_1\cdots\nu_{d-s}\rh}\vv
        d\tilde{V}_{\nu_1\cdots\nu_{d-s}\rh}
    \nn
\end{align}
F"ur den Fall~$s \!\equiv\! d$ gilt~-- vgl.\@ die Gln.~(\ref{APP:Stokes_duald})-(\ref{APP:Stokes_duald}$''$):
\vspace*{-.5ex}
\begin{align} \label{APP:Stokes_duald-Konv}
&\tilde{c}_{(d)}\; \sgnd(d)\;
  \int_{G_d} \tilde{T}^{\rh}{}_{;\rh}\; d\tilde{V}\vv
  =\vv \tilde{c}_{(d-1)}\; \sgnd(d\!-\!1)\;
     \int_{\pa G_d} \tilde{T}^{\rh}\; d\tilde{V}_{\rh}
    \\[1ex]
&\text{mit}\qquad
  d\tilde{V}\phantom{{}_\rh}\; 
    =\; \ep_{\mu_1\cdots\mu_d}\vv
        d_1x^{\mu_1}\, \cdots\, d_dx^{\mu_d}
    \tag{\ref{APP:Stokes_duald-Konv}$'$} \\[.5ex]
&\hspace*{3.5em}
  d\tilde{V}_\rh\;
    =\; \ep_{\rh \mu_1\cdots\mu_{d-1}}\vv
        d_1x^{\mu_1}\, \cdots\, d_{d-1}x^{\mu_{d-1}}
    \tag{\ref{APP:Stokes_duald-Konv}$''$}
    \\[-3.25ex]\nn
\end{align}
\section[Integration der Funktionen~\protect$\tilde\ch\oC$~\protect$\tilde\ch\oNC$]{%
         Integration der Funktionen~\bm{\tilde\ch\oC},~\bm{\tilde\ch\oNC}}
\label{APP-Sect:tilde-ch-C,NC}

Wir f"uhren explizit durch formale Schritte in Zusammenhang mit der Integration der Funktionen~$\tilde\ch\oC$,~$\tilde\ch\oNC$, die im Haupttext nur zitiert werden k"onnen.

\subsection[Integration der longitudinalen Parameter~\protect$u\Dmfp$,~\protect$u\Dmfm$:
              Funktional~\protect\mbox{\,$I_L[F]$}]{%
            Integration der longitudinalen Parameter~\bm{u\Dmfp},~\bm{u\Dmfm}:
              Funktional~\bm{\,I_L[F]}}
\label{APP-Subsect:umfp,umfm-Int}

Die Integration der paralleltransportierten Feldst"arken "uber die Mantelfl"achen~\mbox{$S\Dmfp$},~\mbox{$S\Dmfm$} der Pyramiden~\mbox{$P\Dmfp$},~\mbox{$P\Dmfm$} impliziert Integrale der Form~--  vgl.\@ die Gln.~(\ref{IL[F]-Def}),~(\ref{IL[F]-Def}$'$):%
\FOOT{
  \label{APP-INT-FN:mf<->bm(mf)}Seien~$\mfp$,~$\mfm$ Lorentz-Indizes,~\bm{\mfp},~\bm{\mfm} dagegen Notation der Pyramiden und Teilchen; vgl.\@ Fu"sn.\,\FN{APP-FN:mf<->bm(mf)}.   Bzgl.~der Koordinatenlinien~\mbox{$\tilde\mu \!\in\! \{\mfp,\mfm,1,2\}$} als der Richtungen der (Anti)Quark-Weltlinien vgl.\@ Anh.\,\ref{APP-Sect:Minkowski}.   Aus dem Zusammenhang geht hervor, ob sich Gr"o"sen dieses Abschnitts, die versehen sind mit einer Tilde, beziehen auf diese Koordinatenlinien~-- oder auf die Fourier-Transformierte im Impulsraum.
}
\begin{samepage}
%
\begin{align} \label{APP:longInt-0}
I_L[F]\;
  &=\; \int_{V\Dmfp \vee L\Dmfp} \dsiP{\mfp}\;
       \int_{V\Dmfm \vee L\Dmfm} \dsiM{\mfm}\vv
         F(\tilde{x}^2)
    \\[.75ex]
  &=\; \int_{-T\Dmfp\!/\!2}^{T\Dmfp\!/\!2}\;
         \frac{\pa x\Dmfp^\mfp}{\pa u\Dmfp}\; du\Dmfp\;
       \int_{-T\Dmfm\!/\!2}^{T\Dmfm\!/\!2}\;
         \frac{\pa x\Dmfm^\mfm}{\pa u\Dmfm}\; du\Dmfm\vv
         F(\tilde{x}^2)
    \tag{\ref{APP:longInt-0}$'$}
\end{align}
mit~\mbox{\,$\dsiI{\tilde\mu} \!\equiv\! dx\Dimath^{\tilde\mu}$} f"ur~\mbox{$\imath \!=\! \mfp,\mfm$},~\mbox{$\tilde\mu \!\in\! \{\mfp,\mfm,1,2\}$}, vgl.\@ Gl.~(\ref{dx^mu=dsi^mu(x)}), und definierten Funktionen~$F$. \\
\indent
Es ist~\mbox{$\tilde{x}$} der Differenzvektor der Weltpunkte~$\tilde{x}\Dmfp$,~$\tilde{x}\Dmfm$, die Elemente sind entweder der Volumina~\mbox{$V\Dmfp$},~\mbox{$V\Dmfm$} oder der Grundfl"achen~\mbox{$L\Dmfp$},~\mbox{$L\Dmfm$} der Pyramiden.
Im Sinne der expliziten Parametrisierungen der Gln.~(\ref{Vmfp-Parametr}),~(\ref{Lmfp-Parametr}) seien die \mbox{"`\bm{\mfp}"'-Mannig}\-faltigkeiten {\it longitudinal\/}, das hei"st in Richtung~$e_{(\mfp)}$, parametrisiert mit~\mbox{$u\Dmfp: -T\Dmfp/2 \!\to\! T\Dmfp/2$}; entsprechend die \mbox{"`\bm{\mfm}"'-Mannig}\-faltigkeiten in Richtung~$e_{(\mfm)}$ mit~\mbox{$u\Dmfm: -T\Dmfm/2 \!\to\! T\Dmfm/2$}.
Dann gilt~-- ohne Bezug zu nehmen auf die Parametrisierungen der Gln.~(\ref{Vmfp-Parametr}),~(\ref{Lmfp-Parametr}):%
\citeFN{APP-INT-FN:mf<->bm(mf)}
%
\begin{align} \label{APP:WeltpunktPyramide}
&\tilde{x}\;
  =\; \tilde{x}\Dmfp(u\Dmfp) - \tilde{x}\Dmfm(u\Dmfm)
    \\[.25ex]
 &\phantom{x\;}
  =\;     \frac{\pa x\Dmfp^\mfp}{\pa u\Dmfp}\, u\Dmfp\; e_{(\mfp)}\;
      -\; \frac{\pa x\Dmfm^\mfm}{\pa u\Dmfm}\, u\Dmfm\; e_{(\mfm)}\;
      +\; (x\Dmfp^i \!-\! x\Dmfm^i)\, e_{(i)}
    \tag{\ref{APP:WeltpunktPyramide}$'$} \\[.75ex]
&\text{mit}\qquad
  \tilde{x}\Dmfp\; \in\; V\Dmfp \vee L\Dmfp\qquad
  \tilde{x}\Dmfm\; \in\; V\Dmfm \vee L\Dmfm
    \label{xmf-in-V,Lmf}
\end{align}
\end{samepage}%
Dabei impliziert Gl.~(\ref{APP:WeltpunktPyramide}$'$) bereits die Forderung von {\it Planarit"at\/} der Pyramiden~$P\Dmfp$,~$P\Dmfm$ be\-z"uglich ihrer respektiven longitudinalen Richtung~$e_{(\mfp)}$,~$e_{(\mfm)}$~-- und o.E.d.A.\@ {\it Linearit"at\/} der Parametrisierung, das hei"st konstante Parameter-Geschwindigkeiten:
%
\begin{align} \label{APP:dx/du-const}
\frac{\pa x\Dmfp^\mfp}{\pa u\Dmfp}
  \vv\Big._,\vv
  \frac{\pa x\Dmfm^\mfm}{\pa u\Dmfm}\qquad
  \text{konstant bzgl.\@}\quad
  u\Dmfp, u\Dmfm
\end{align}
mit explizit~$s\Dmfp\vrh^{-1}$,~$s\Dmfm\vrh^{-1}$ f"ur die Volumina~$V\Dmfp$,~$V\Dmfm$ und~$\vrh^{-1}$ f"ur die Grundfl"achen~$L\Dmfp$,~$L\Dmfm$ im Falle der Parametrisierungen der Gln.~(\ref{Vmfp-Parametr}),~(\ref{Lmfp-Parametr}).
Diese Faktoren k"onnen als konstant aus den Integralen herausgezogen werden.
"Ubergang zur Fourier-Transformierten~\mbox{$\tilde{F}(k^2)$} von~\mbox{$F(\tilde{x}^2 \mskip-4mu\equiv\mskip-4mu x^2)$} ergibt dann unter Vertauschung der Integrationen:
\vspace*{-.5ex}
\begin{align}
&I_L[F]\;
  =\; \int \frac{dk}{(2\pi)^4}\; \tilde{F}(k^2)\vv
        \frac{\pa x\Dmfp^\mfp}{\pa u\Dmfp}\; \frac{\pa x\Dmfm^\mfm}{\pa u\Dmfm}\vv
        \int_{-T\Dmfp\!/\!2}^{T\Dmfp\!/\!2}\; du\Dmfp\;
        \int_{-T\Dmfm\!/\!2}^{T\Dmfm\!/\!2}\; du\Dmfm\;
        \efn{\D-\iIM k \!\cdot\! x}
    \label{APP:longInt-1} \\[-.375ex]
&\text{mit}\qquad
  k \cdot x\;
  =\; \tilde{k} \cdot \tilde{x}\;
  =\; k_\mfp\; \frac{\pa x\Dmfp^\mfp}{\pa u\Dmfp}\, u\Dmfp\;
      -\; k_\mfm\; \frac{\pa x\Dmfm^\mfm}{\pa u\Dmfm}\, u\Dmfm\;
      +\; k_i\, x^i
    \label{APP:k-dot-x}
    \\[-5ex]\nn
\end{align}
vgl.~\mbox{\,$%
\tilde{k} \!\cdot\! \tilde{x}
  \!=\! k_{\tilde\mu} x^{\tilde\mu}
  \!=\! \mathbb{S}_{\tilde\mu}{}^{\bar\nu} \mathbb{S}^{\tilde\mu}{}_{\bar\rh}\,
          k_{\bar\nu} x^{\bar\rh}
  \!=\! \de^{\bar\nu}_{\bar\rh}\, k_{\bar\nu} x^{\bar\rh}
  \!=\! k_{\bar\mu} x^{\bar\mu}
  \!=\! \bar{k} \!\cdot\! \bar{x}
    \stackrel{\D!}{=}\! \mathbb{L}_{\bar\mu}{}^\nu \mathbb{L}^{\bar\mu}{}_{\rh}\, k_\nu x^\rh
  \!=\! \de^\nu_\rh\, k_\nu x^\rh
  \!=\! k_\mu x^\mu
  \!=\! k \!\cdot\! x$}.
\\
\indent
Im Limes~\mbox{$T\Dmfp,\, T\Dmfm \!\to\! \infty$} gehen die $u\Dmfp$-,$u\Dmfm$-Integrale wie
\vspace*{-.5ex}
\begin{align} \label{APP:T->infty}
\lim_{T\Dimath\to\infty}\vv
  \int_{-T\Dimath\!/\!2}^{T\Dimath\!/\!2}\; du\Dimath\;
    \efn{\D\pm\iIM z\Dimath u\Dimath}\;
  =\; 2\pi\, \de(z\Dimath)\qquad
  z\Dimath = k_\imath\, \frac{\pa x\Dimath^\imath}{\pa u\Dimath}\qqquad
  \imath = \mfp,\mfm
    \\[-4.5ex]\nn
\end{align}
"uber in Delta-Distributionen.
Nachtmann diskutiert in Ref.~\cite{Nachtmann91}~-- vgl.\@ den Begriff des "`Fem\-to-Universums"'~--, da"s sich auf der Zeitskala der Streuung der Zusatnd der Partonen qualitativ nicht "andert: Parton-Annihilations- und -Produktionsprozesse k"onnen vernachl"assigt werden, so da"s die Propagation geschieht auf geradlinigen Trajektorien, deren L"angen, das hei"st die Eigenzeiten~$T\Dmfp$,~$T\Dmfm$ der Propagation, im Vergleich zu den sonst involvierten Skalen approximiert werden k"onnen als unendlich.
In diesem Sinne wird durchgef"uhrt der formale "Ubergang~\mbox{\,$T\Dmfp, T\Dmfm \!\to\! \infty$}%
\FOOT{
  cum grano salis~\mbox{$\vrh^{-1}T\Dmfp, \vrh^{-1}T\Dmfm \!\to\! \infty$}~-- das hei"st {\sl ausschlie"slich\/} des Anstiegs gegen Unendlich, der Konsequenz ist der aktiven Boosts mit~\mbox{$\be\Dmfp, \be\Dmfm \!\to\! 1$}; bzgl.\@ der eigentlichen Entsprechung~\mbox{$\vrh^{-1}T\Dmfp, \vrh^{-1}T\Dmfm \!\leftrightarrow\! T\idx{0}$}, vgl.\@ die Gln.~(\ref{APP:Tmfp_zeDmfp^mfp}),~(\ref{APP:Tmfp_zeDmfp^mfp}$'$) und~(\ref{APP:Endpkt^0_T0}).
};
f"ur~$I_L[F]$ nach Gl.~(\ref{APP:longInt-1}) folgt mithilfe von Gl.~(\ref{APP:T->infty}):
\begin{samepage}
\vspace*{-.5ex}
\begin{align} \label{APP:longInt-2}
I_L[F]\;
  =\; \int \frac{dk}{(2\pi)^2}\; \efn{\D-\iIM\, k_i\, x^i}\vv
        \tilde{F}(k^2)\vv
      \frac{\pa x\Dmfp^\mfp}{\pa u\Dmfp}\;
        \de\bigg(k_\mfp\, \frac{\pa x\Dmfp^\mfp}{\pa u\Dmfp}\bigg)\vv
      \frac{\pa x\Dmfm^\mfm}{\pa u\Dmfm}\;
        \de\bigg(k_\mfm\, \frac{\pa x\Dmfm^\mfm}{\pa u\Dmfm}\bigg)
    \\[-4.5ex]\nn
\end{align}
und mithilfe~\mbox{\,$|\al| \de(z\al) \!=\! \de(z)$} unmittelbar:
\vspace*{-.25ex}
\begin{align} \label{APP:longInt-3}
I_L[F]\;
  =\; \int \frac{dk}{(2\pi)^2}\; \efn{\D-\iIM\, k_i\, x^i}\vv
        \de(k_\mfp)\, \de(k_\mfm)\vv
        \tilde{F}(k^2)
    \\[-4.25ex]\nn
\end{align}
Es liegt nahe, auch im Lorentz-invarianten Vier-Integrationsma"s~\mbox{\,$dk$} von den Komponen\-ten~\mbox{$k_\mu$},~\mbox{$\mu \!\in\! \{0,1,2,3\}$}, "uberzugehen zu den Komponenten~\mbox{$k_{\tilde\mu}$},~\mbox{$\tilde\mu \!\in\! \{\mfp,\mfm,1,2\}$}:
%
\begin{align} 
k_{\tilde\mu}\;
  =\; \mathbb{S}_{\tilde\mu}{}^{\bar\nu}\, k_{\bar\nu}\;
  =\; \mathbb{S}_{\tilde\mu}{}^{\bar\rh}\, \mathbb{L}_{\bar\rh}{}^{\nu}\, k_\nu\vv
  \equiv\; (\mathbb{SL})_{\tilde\mu}{}^\nu\, k_\nu
\end{align}
Bzgl.\@ dieses "Ubergangs~-- "uber Lichtkegelkomponenten~\mbox{$k_{\bar\mu}$}, \mbox{$\bar\mu \!\in\! \{+,-,1,2\}$}~-- vgl.\@ Anh.~\ref{APP-Sect:Minkowski}, die Gln.~(\ref{APP:xtilde_x-kov}),~(\ref{APP:xtilde_x-kov}$'$).
F"ur die Jacobi-Determinante der Transformation folgt:
%
\begin{align} 
&\frac{\pa(k_0,k_1,k_2,k_3)}{\pa(k_\mfp,k_\mfm,k_1,k_2)}\;
  =\; \bigg[\frac{\pa(k_\mfp,k_\mfm,k_1,k_2)}{\pa(k_0,k_1,k_2,k_3)}\bigg]^{-1}
    \\[.125ex]
&\phantom{k_{\tilde\mu}\;}
  =\; \big[\, \det\!
        \big((\mathbb{SL})_{\tilde\mu}{}^\nu\big) \,\big]^{-1}\;
  =\; \det\! \big((\mathbb{SL})^{\tilde\mu}{}_\nu\big)\;
  =\; \det\mathbb{SL}\qquad\vv
  \mathbb{S} \equiv \big(\mathbb{S}^{\tilde\mu}{}_{\bar\nu}\big),\,
  \mathbb{L} \equiv \big(\mathbb{L}^{\bar\mu}{}_\nu\big)
    \nn
\end{align}
ergo:%
\FOOT{
  Das explizite Vorzeichen, das folgt aus dem "Ubergang von kontra- zu kovarianten Komponenten, k"urzt das \mbox{$\det\mathbb{L}$-impli}\-zite Vorzeichen, das ausdr"uckt die Orientierungsumkehr des "Ubergangs~\mbox{$(k^0,k^3) \!\to\! (k^+,k^-)$}.
}
%
\begin{align} \label{APP:dk,d4k}
dk\;
  \equiv\; dk^0\, dk^3\, d^2\rb{k}\;
  =\; -\, \det\mathbb{SL}\vv dk_\mfp\, dk_\mfm\, d^2\rb{k}
\end{align}
mit~\mbox{\,$d^2\rb{k} \!\equiv\! dk^1dk^2 \!=\! dk_1dk_2$}.
\end{samepage}

F"ur Gl.~(\ref{APP:longInt-3}) folgt:
\vspace*{-.5ex}
\begin{align} \label{APP:longInt-4}
I_L[F]\;
  &=\; -\, \det\mathbb{SL}\vv
       \int \frac{dk^1dk^2}{(2\pi)^2}\vv \efn{\D+\iIM\, k^i\, x^i}\;
         \int dk_\mfp\, \de(k_\mfp)\;
         \int dk_\mfm\, \de(k_\mfm)\vv
         \tilde{F}(k^2)
    \\
  &=\; -\, \det\mathbb{SL}\vv
       \int \frac{d^2\rb{k}}{(2\pi)^2}\vv \efn{\D\iIM\, \rb{k} \!\cdot\! \rb{x}}\vv
        \tilde{F}(-\rb{k}^2)
    \tag{\ref{APP:longInt-4}$'$} 
    \\[-4.5ex]\nn
\end{align}
Sei im Sinne der letzten Relationen ein Operator~\mbox{$\projNFAt$} definiert durch
%
\begin{align} \label{APP:projt-Operator}
\projt{f}{\rb{x}}\;
  :=\; \int \frac{d^2\rb{k}}{(2\pi)^2}\vv
        \efn{\D\iIM\, \rb{k} \!\cdot\! \rb{x}}\vv
        \tilde{f}(k)\Big|_{k^0,k^3\equiv0}
\end{align}
Zusammenfassend gilt:
\vspace*{-.5ex}
\begin{align} \label{APP:longInt}
I_L[F]\;
=\; -\, \det\mathbb{SL}\;\cdot\;
          &\projNAt{F}
    \\
  \text{mit}\qquad
  &\projt{F}{\rb{x}}\;
     =\; \iint_{-\infty}^\infty \frac{d^2\rb{k}}{(2\pi)^2}\vv
           \efn{\D\iIM\, \rb{k} \!\cdot\! \rb{x}}\vv
           \tilde{F}(-\rb{k}^2)
    \nn
    \\[-4ex]\nn
\end{align}
vgl.\@ Gl.(\ref{APP:longInt-4}$'$),~-- mit~$\projNFAt$ nach Gl.~(\ref{APP:projt-Operator}). \\
\indent
Wir halten in Worte fest die Wirkung des Operators~$\projNFAt$:
Es ist~$\projt{f}{\rb{x}}$ die Funktion, die folgt aus einer Funktion~$f(x)$ der Raumzeit auf Basis~-- in~$d$ Dimensionen:
\begin{samepage}
\vspace*{-.25ex}
\begin{alignat}{2} \label{APP:FT-allg}
&\tilde{f}(k)\;&
  &=\; \int d^dx\; \efn{\D\iIM\, k \!\cdot\! x}\vv f(x)
    \\[-.5ex]
&f(x)\;&
  &=\; \int  \frac{d^dk}{(2\pi)^d}\; \efn{\D-\iIM\, k \!\cdot\! x}\vv \tilde{f}(k)
    \tag{\ref{APP:FT-allg}$'$}
    \\[-4.25ex]\nn
\end{alignat}
durch sukzessive {\it Fourier-Transformation\/} in den Impulsraum, {\it Identisch-Null-Setzen\/} der longitudinalen Komponenten~$k^0$,~$k^3$ und {\it Fourier-R"uck-Transformation\/}, vgl.\@ Gl.~(\ref{APP:longInt}). \\
\indent
Das Funktional~$\projNAt{F}$~-- und daher~$I_L[F]$~-- h"angt ab nur von~$|\rb{x}|$, dem Betrag des transver\-salen Differenzvektors der integrierten Feldst"arken, vgl.\@ die Gln.(\ref{APP:WeltpunktPyramide})-(\ref{xmf-in-V,Lmf}).
\vspace*{-.5ex}

\subsection[Integration bez"uglich der transversalen Projektionen%
              ~\protect$S\Dmfp^\perp$,~\protect$S\Dmfm^\perp$:
              Funktional~\protect\mbox{\,$I_T[F]$}]{%
            Integration bez"uglich der transversalen Projektionen%
              ~\bm{S\Dmfp^\perp},~\bm{S\Dmfm^\perp}:
              Funktional~\bm{\,I_T[F]}}
\label{APP-Subsect:S_mf^perp-Int}

In Abschnitt~\ref{Subsect:chC} wird die konfinierende Funktion~$\tilde\ch\oC$ ausgedr"uckt durch das transversale Integation repr"asentierende Funktional~$I_T$~-- vgl.\@ die Gln.~(\ref{chC_IT[F]}),~(\ref{IT[F]-Def}):
\vspace*{-.25ex}
\begin{align} \label{APP:chC_IT[F]}
&\tilde\ch\oC\;
  =\; -\, \det\mathbb{SL}\; g_{\mfp\mfm}\;\cdot\;
         \iIM\; \frac{1}{12}\vv
         I_T[F]
    \\[-4.5ex]\nn
\end{align}
unter Definition:%
\FOOT{
  \label{APP-FN:i->tilde-mu}In Abschnitt~\ref{APP-Sect:Integration} ist ausgef"uhrt der Formalismus von Integration auf (pseudo-)Riemannschen~Mannigfal\-tigkeiten; f"ur unmittelbare Ankn"upfung fassen wir die Transversalprojektionen~${\cal S}\Dmfp\Doperp$,~${\cal S}\Dmfm\Doperp$ auf als Untermannigfaltigkeiten der Minkowski-Raumzeit und summieren im Sinne~\mbox{\,$\tilde\mu,\tilde\nu,\tilde\rh \!\in\! \{\mfp,\mfm,1,2\}$} "uber den gesamten~Index\-bereich~-- {\sl o.E.d.A.\@}, da {\sl identisch verschwinden\/} longitudinale Komponenten der ${\cal S}\Dmfp\Doperp$-,~${\cal S}\Dmfm\Doperp$-Fl"achenelemente.
}
%
\vspace*{-.25ex}
\begin{align} \label{APP:IT[F]-Def}
I_T[F]\;
  =\; -\, g_{\tilde\mu\tilde\nu}\;
        \int_{{\cal S}\Dmfp\Doperp} dx\Dmfp^{\tilde\mu}\vv
        \int_{{\cal S}\Dmfm\Doperp} dx\Dmfm^{\tilde\nu}\vv
       &\pa_{\tilde\rh}\! \left[x^{\tilde\rh} F(\tilde{x}^2)\right]
    \\[-4.5ex]\nn
\end{align}
mit~\mbox{$\pa_{\tilde\mu} \!\equiv\! \pa/\pa x^{\tilde\mu}$} und%
  ~\mbox{$\tilde{x} \!\equiv\! (x^{\tilde\mu})$},%
  ~\mbox{$x^{\tilde\mu} \!=\! x\Dmfp^{\tilde\mu} \!-\! x\Dmfm^{\tilde\mu}$},%
  ~--~\mbox{\,$\tilde\mu,\tilde\nu,\tilde\rh,\ldots \!\in\! \{\mfp,\mfm,1,2\}$}. \\
\indent
Wir werten~$I_T$ aus f"ur die allgemeine Lorentz-skalare~Funktion%
  ~\vspace*{-.125ex}\mbox{$F \!\equiv\! F(\tilde{x}^2)$}, die lediglich insofern eingeschr"ankt sei, als sie im Sinn%
  ~\vspace*{-.125ex}\mbox{$F \!\equiv\! F(|\rb{x}|)$} abh"ange nur vom~-- Betrag des~-- trans\-versalen Vektors~\mbox{$\rb{x} \!\equiv\! (x^1,x^2)^{\T t}$}, und die {\it in~praxi\/} zu identifizieren ist wie
\vspace*{-.375ex}
\begin{align} \label{APP:F_projt(1)}
F(|\rb{x}|)\;
  \equiv\; \projt[(1)]{F\oC}{\rb{x}}\;
  \equiv\; -\, \frac{1}{|\rb{x}|} \frac{d}{d|\rb{x}|}\vv
             \projt{F\oC}{\rb{x}}\qquad
  |\rb{x}| \equiv \surd\rb{x}^2,\vv
  \rb{x}^2 \equiv x^ix^i
    \\[-4.25ex]\nn
\end{align}
\end{samepage}%
mit~\vspace*{-.25ex}$F\oC$ die fundamentale konfinierende Korrelationsfunktion, vgl.\@ die Gln.~(\ref{F_projt(1)}),~(\ref{projt(n)-Def}). \\
\indent
Wir betrachten~$I_T$ nach Gl.~(\ref{APP:IT[F]-Def}) als das Integral der einen paralleltransportierten Feldst"arke "uber~${\cal S}\Dmfp\Doperp$, der anderen "uber~${\cal S}\Dmfm\Doperp$.
Deren Positionen sind gegeben durch
%
\begin{align} \label{APP:ximath}
&\tilde{x}\Dimath\;
  \equiv\; \big(\, 0,\, 0,\, \rb{x}\Dimath^{\T t} \,\big){}^{\T t}\qqquad
  \imath = \mfp,\mfm
    \\[.5ex]
&\text{mit}\qquad
  \tilde{x}\Dimath
    \equiv \tilde{x}\Dimath(s\Dimath)\qquad
  \rb{x}\Dimath
    \equiv \rb{x}\Dimath(s\Dimath)\qqquad
  s\Dimath\!:\; 0 \to 1
    \tag{\ref{APP:ximath}$'$}
\end{align}
deren Differenzvektor durch:
%
\begin{align} \label{APP:x=xmfp-xmfm}
&\tilde{x}\;
  \equiv\; \tilde{x}\Dmfp - \tilde{x}\Dmfm
    \\[.25ex]
&\tilde{x}(s\Dmfp,s\Dmfm)\;
  \equiv\; \tilde{x}\Dmfp(s\Dmfp) -\tilde{x}\Dmfm(s\Dmfm)\qquad
 \rb{x}(s\Dmfp,s\Dmfm)\;
  \equiv\; \rb{x}\Dmfp(s\Dmfp) -\rb{x}\Dmfm(s\Dmfm)
    \tag{\ref{APP:x=xmfp-xmfm}$'$}
\end{align}
vgl.\@ Gl.~(\ref{x=ximath-xjmath}).
Es sind~${\cal S}\Dmfp\Doperp$,~${\cal S}\Dmfm\Doperp$ zu verstehen als {\it abschnittweise parametrisiert\/} bez"uglich ihres Quark- und Antiquark-Abschnitts:%
  ~\mbox{\,${\cal S}\Dmfp\Doperp\big|_Q$},%
  ~\mbox{${\cal S}\Dmfp\Doperp\big|_\AQ$} bzw.%
  ~\mbox{\,${\cal S}\Dmfm\Doperp\big|_Q$},%
  ~\mbox{${\cal S}\Dmfm\Doperp\big|_\AQ$}; diese verlaufen f"ur Parameter~\mbox{$s\Dmfp, s\Dmfm\!: 0 \!\to\! 1$} von~$\rbG{\om}$ zum Quark beziehungsweise von~$\rbG{\om}$ zum Antiquark~-- wobei {\it nicht\/} a~priori gefordert werde {\it abschnittweise Geradlinigkeit\/}:%
\FOOT{
  \label{APP-FN:Subst:mfp->mfm}F"ur K"urze geben wir explizit nur Gr"o"sen an der "`$\bf{\mfp}$"'-Pyramide~$P\Dmfp$; bez"uglich der "`$\bf{\mfm}$"'-Pyramide~$P\Dmfm$ ist konsequent zu ersetzen~\mbox{$\zet_1 \!\to\! \zet_2$},~\mbox{$X \!\to\! Y$}~und \mbox{$b \!\to\! -b$}, vgl.\@ Gl.~(\ref{Subst:mfp->mfm}$'$).
}
%
\begin{alignat}{7} \label{APP:calS_mfp^perp}
&{\cal S}\Dmfp\Doperp|_Q(s\Dmfp\!:\, 0 \!\to\! 1):&\qquad
  &\text{\sl Apex}\!:&\vv
    &x\Dmfp(0)&\;
    &=\, \om&\quad\vv
  \longrightarrow\quad\vv
  &Q\!:&\vv
    &\tilde{x}\Dmfp(1)&\;
    &=\, x
    \\[.75ex]
&{\cal S}\Dmfp\Doperp|_\AQ(s\Dmfp\!:\, 0 \!\to\! 1):&\qquad
  &\text{\sl Apex}\!:&\vv
    &x\Dmfp(0)&\;
    &=\, \om&\quad\vv
  \longrightarrow\quad\vv
  &\AQ\!:&\vv
    &\tilde{x}\Dmfm(1)&\;
    &=\, \bar{x}
    \tag{\ref{APP:calS_mfp^perp}$'$}
\end{alignat}
Es sind~$x$,~$\bar{x}$ die Positionen des (Anti)Quarks, explizit:%
\FOOT{
  Der Zusammenhang stellt klar, ob~\mbox{$\rb{x} \!\equiv\! (x^1,x^2)^{\T t}$}~-- vgl.\@ Gl.~(\ref{APP:x=xmfp-xmfm}) vs.~(\ref{APP:x,barx_r_Q,AQ-mfp})~-- identisch~\mbox{\,$\rb{x}\Dmfp \!-\! \rb{x}\Dmfm$}, dem Differenzvektor der paralleltransportierten Feldst"arken auf den Pyramiden~$P\Dmfp$,\,$P\Dmfm$, oder identisch~\mbox{\,$\rbG{\om} \!+\! \bzet_1\rb{X} \!+\! \rb{b}/2$} \mbox{$=\! \rbG{\om} \!+\! \rb{r}\Dmfp^Q$}, der Position des mit~$P\Dmfp$ assoziierten Quarks~$Q$; vgl.\@ Fu"sn.\,\FNg{FN:x=ximath-xjmath,Qmfp}.
}
\begin{samepage}
\vspace*{-.5ex}
\begin{alignat}{4} \label{APP:x,barx_r_Q,AQ-mfp}
&\rb{x}&\;
  &=\; \rbG{\om}\; +&\;
         (\bzet_1 &\rb{X}\; +\; \rb{b}/2)&\quad
  &=\; \rbG{\om}\; +\; \rb{r}\Dmfp^Q
    \\[-.5ex]
&\rbb{x}&\;
  &=\; \rbG{\om}\; +&\;
         (-\zet_1 &\rb{X}\; +\; \rb{b}/2)&\quad
  &=\; \rbG{\om}\; +\; \rb{r}\Dmfp^\AQ
    \tag{\ref{APP:x,barx_r_Q,AQ-mfp}$'$}
    \\[-4.5ex]\nn
\end{alignat}
und~$\rbG{\om}$ der gemeinsame Apex der {\it verallgemeinerten\/} Pyramiden~$P\Dmfp$,~$P\Dmfm$~-- "uber den Wegner-Wilson-Loops~$W\Dmfp$,~$W\Dmfm$~--, der identifiziert wird mit dem gemeinsamen Referenzpunkt~$x_0$, der wiederum gew"ahlt wird~-- "`im Sinne h"ochster Symmetrie"', vgl.\@  Fu"sn.\,\FNg{FN:h"ochsteSymmetrie}~-- als die Mitte des transversalen Impaktvektors~$\rb{b}$.%
\FOOT{
  \vspace*{-.125ex}Im Falle {\sl planer\/} Pyramiden~-- vgl.\@ die expliziten Parametrisierungen in Kap.~\refg{Subsect:surfacesSDmf}~-- sind die Verbindungen Apex-Basis geradlinig und ihre \vspace*{-.375ex}transversalen Projektionen respektive gegeben durch~$r\Dmfp^Q$,~$r\Dmfp^\AQ$ und~$r\Dmfm^Q$,~$r\Dmfm^\AQ$~-- und suggestiv die zweite Darstellung der Gln.~(\ref{APP:x,barx_r_Q,AQ-mfp}),(\ref{APP:x,barx_r_Q,AQ-mfp}$'$).
} \\
%
\indent
Das Funktional~$I_T$ nach Gl.~(\ref{APP:IT[F]-Def}) involviert die Differentiale%
  ~$dx\Dmfp^{\tilde\mu}$,~$dx\Dmfm^{\tilde\mu}$ bez"uglich%
  ~\mbox{$\tilde{x}\Dmfp\mskip-2mu(s\Dmfp)$}, \mbox{$\tilde{x}\Dmfm\mskip-2mu(s\Dmfm)$}: bez"uglich der \bm{\mfp}-,\bm{\mfm}-Koordinatenlinien entlang~${\cal S}\Dmfp\Doperp$,~${\cal S}\Dmfm\Doperp$.
Dann sind durch
\vspace*{-1ex}
\begin{alignat}{3} \label{APP:dsiPM}
&d\Dmfp{x}^{\tilde\mu}\mskip-2mu(s\Dmfp)&\;
  &\equiv\; \frac{dx^{\tilde\mu}}{ds\Dmfp}\mskip-2mu(s\Dmfp)\vv ds\Dmfp&\quad
  &=\; \frac{dx\Dmfp^{\tilde\mu}}{ds\Dmfp}\mskip-2mu(s\Dmfp)\vv ds\Dmfp\;
   \equiv\; dx\Dmfp^{\tilde\mu}\mskip-2mu(s\Dmfp)
    \\[-2.5ex]
&&&&&\hspace*{13em}\text{\bm{\mfp}-Koordinatenlinien}
    \nn \\[-.75ex]
&d\Dmfm{x}^{\tilde\mu}\mskip-2mu(s\Dmfm)&\;
  &\equiv\; \frac{dx^{\tilde\mu}}{ds\Dmfm}\mskip-2mu(s\Dmfm)\vv ds\Dmfm&\quad
  &=\; -\, \frac{dx\Dmfm^{\tilde\mu}}{ds\Dmfm}\mskip-2mu(s\Dmfm)\vv ds\Dmfm\;
   \equiv\; -\, dx\Dmfm^{\tilde\mu}\mskip-2mu(s\Dmfm)
    \tag{\ref{APP:dsiPM}$'$} \\[-2ex]
&&&&&\hspace*{13em}\text{\bm{\mfm}-Koordinatenlinien}
    \nn
    \\[-4.75ex]\nn
\end{alignat}
gegeben die Differentiale%
  ~\mbox{$d\Dimath{\tilde{x}} \!\equiv \! (d\Dimath{x}^{\tilde\mu})$},%
  ~\mbox{$\imath \!\equiv\! \mfp,\mfm$},%
  ~\mbox{$\tilde\mu \!\in\! \{\mfp,\mfm,1,2\}$}, in der Konvention von Abschnitt~\ref{APP-Sect:Integration}, Gl.~(\ref{APP:Differential-i}), die also ankn"upfen an den dort entwickelten Formalismus. \\
\indent
Im Haupttext werden gew"ahlt die Fl"achen~${\cal S}\Dmfp$,~${\cal S}\Dmfm$ als die Mantelfl"achen planer Pyra\-miden~$P\Dmfp$,~$P\Dmfm$
und auf dieser Basis angegeben f"ur die transversalen Projektionen~${\cal S}\Dmfp\Doperp$,~${\cal S}\Dmfm\Doperp$ explizite Parametrisierungen, vgl.\@ Kap.~\ref{Subsect:surfacesSDmf}; wir rekapitulieren Gl.~(\ref{Smfp^perp-Parametr}) in Form:\citeFN{APP-FN:Subst:mfp->mfm}
\end{samepage}
\vspace*{-.25ex}
\begin{align} \label{APP:Smfp^perp-Parametr_Q,AQ}
&{\cal S}\Dmfp\Doperp\;
  \equiv\; {\cal S}\Dmfp\Doperp\big|_Q
     \circ \big({\cal S}\Dmfp\Doperp\big|_\AQ\big)^{-1}
    \\[.25ex]
&\text{mit}\qquad
 \begin{alignedat}[t]{2}
  &{\cal S}\Dmfp\Doperp\big|_Q\;&
    &=\; \Big\{\, \tilde{x}\Dmfp \!\equiv\! (x\Dmfp^{\tilde\mu})\Big|\;
           \tilde{x}\Dmfp(s\Dmfp)
           =\; \big(\om^i\, + s\Dmfp\; r\Dmfp^{Qi}\big)\, \tilde{e}_{(i)}
         \,\Big\}
    \\[-.25ex]
  &{\cal S}\Dmfp\Doperp\big|_\AQ\;&
    &=\; \Big\{\, \tilde{x}\Dmfp \!\equiv\! (x\Dmfp^{\tilde\mu})\Big|\;
           \tilde{x}\Dmfp(s\Dmfp)
           =\; \big(\om^i\, + s\Dmfp\; r\Dmfp^{\AQ i}\big)\, \tilde{e}_{(i)}
        \,\Big\}
 \end{alignedat}
    \tag{\ref{APP:Smfp^perp-Parametr_Q,AQ}$'$}
    \\[-4.25ex]\nn
\end{align}
und abschnittweise
\vspace*{-.25ex}
\begin{align} \label{APP:s-Lauf}
s\Dmfp\!:\;
  0 \to 1
    \\[-4.25ex]\nn
\end{align}
bez"uglich~${\cal S}\Dmfp\Doperp\big|_Q$,~${\cal S}\Dmfp\Doperp\big|_\AQ$.
Vgl.~\mbox{\,$r\Dmfp^Q \!=\! \bzet_1 X \!+\! b/2$},~\mbox{\,$r\Dmfp^\AQ \!=\! -\zet_1 X + b/2$}~-- die Gln.~(\ref{APP:x,barx_r_Q,AQ-mfp}),~(\ref{APP:x,barx_r_Q,AQ-mfp}$'$). \\
\indent
Wir zerlegen~$I_T[F]$ nach Gl.~(\ref{APP:IT[F]-Def}) in vier Summanden bez"uglich der Kombinationen von (Anti)Quark-Abschnitten~\mbox{$(I,J)$}~--~\mbox{$I,J \!\in\! \{Q,\AQ\}$}:
%
\vspace*{-.5ex}
\begin{align}
&I_T[F]\;
  =\vv  \sum\Big._{\!I,J\equiv Q,\AQ}\vv
       {\rm sign}_{I,J}\quad
        I_T[F]\Big|_{(I,J)}
    \label{APP:IT[F]_IT[F]IJ} \\[1ex]
  &\text{mit}\qquad
  I_T[F]\Big|_{(I,J)}\;
  =\; -\, g_{\tilde\mu\tilde\nu}\;
        \int_{{\T\cal S}\Dmfp\Doperp\big|_{\scriptstyle I}} dx\Dmfp^{\tilde\mu}\vv
        \int_{{\T\cal S}\Dmfm\Doperp\big|_{\scriptstyle J}} dx\Dmfm^{\tilde\nu}\vv
        \pa_{\tilde\rh}\! \left[x^{\tilde\rh} F(\tilde{x}^2)\right]
        \vv\Big|_{(I,J)}
    \label{APP:IT[F]IJ-Def}
    \\[-4.5ex]\nn
\end{align}
Orientierungsumkehr der Antiquark-Abschnitte~-- vgl.\@ Gl.~(\ref{APP:Smfp^perp-Parametr_Q,AQ})~-- ist dabei absorbiert im Signum~\mbox{\,${\rm sign}_{I,J} \!=\! (-1)^{I+J}$}, das definiert ist durch Zuweisen der Zahlenwerte~\mbox{$Q \!\equiv\! 0$},~\mbox{$A\!Q \!\equiv\! 1$}; vgl.\@ Gl.~(\ref{signIJ}) bzgl.\@ der nicht-konfinierenden Tensorstruktur.
F"ur~\vspace*{-.25ex}\mbox{\,$I_T[F]\big|_{(I,J)}$} folgt
\begin{samepage}
\vspace*{-.5ex}
\begin{align} \label{APP:IT[F]IJ-0}
I_T[F]\Big|_{(I,J)}\;
  =\; g_{\tilde\mu\tilde\nu}\;
        \int_{{\T\cal S}\Dmfp\Doperp\big|_{\scriptstyle I}} d\Dmfp{x}^{\tilde\mu}\vv
        \int_{{\T\cal S}\Dmfm\Doperp\big|_{\scriptstyle J}} d\Dmfm{x}^{\tilde\nu}\vv
        \pa_{\tilde\rh}\! \left[x^{\tilde\rh} F(\tilde{x}^2)\right]
        \vv\Big|_{(I,J)}
    \\[-4.5ex]\nn
\end{align}
in Termen der Differentiale~\mbox{\,$d\Dmfp{x}^{\tilde\mu}$},~\mbox{\,$d\Dmfm{x}^{\tilde\mu}$}, vgl.\@ die Gln.~(\ref{APP:dsiPM}),~(\ref{APP:dsiPM}$'$); explizit:
\vspace*{-.5ex}
\begin{align} \label{APP:IT[F]IJ-1}
&\hspace*{-5pt}
 I_T[F]\Big|_{(I,J)}
    \nn \\[-.5ex]
  &\hspace*{-5pt}
   =\; -\, g_{\tilde\mu\tilde\nu}\;
        \int_{{\T\cal S}\Dmfp\Doperp\big|_{\scriptstyle I}}
          ds\Dmfp\; \frac{dx\Dmfp^{\tilde\mu}}{ds\Dmfp}\bigg|_I\vv
        \int_{{\T\cal S}\Dmfm\Doperp\big|_{\scriptstyle J}}
          ds\Dmfm\; \frac{dx\Dmfm^{\tilde\mu}}{ds\Dmfm}\bigg|_J\vv
        \pa_{\tilde\rh}\! \left[x^{\tilde\rh} F(\tilde{x}^2)\right]
        \vv\Big|_{(I,J)}
    \\
  &\hspace*{-5pt}
   =\; -\, \ep_{\tilde\mu\tilde\nu}\;
        \int_{{\T\cal S}\Dmfp\Doperp\big|_{\scriptstyle I}}
          ds\Dmfp\; \frac{dx\Dmfp^{\tilde\mu}}{ds\Dmfp}\bigg|_I\vv
        \int_{{\T\cal S}\Dmfm\Doperp\big|_{\scriptstyle J}}
          ds\Dmfm\; \frac{dx\Dmfm^{\tilde\mu}}{ds\Dmfm}\bigg|_J\vv
        R\; \Big|_{(I,J)}\vv
        \pa_{\tilde\rh}\! \left[x^{\tilde\rh} F(\tilde{x}^2)\right]\; \Big|_{(I,J)}
    \tag{\ref{APP:IT[F]IJ-1}$'$}
    \\[-4.5ex]\nn
\end{align}
zu lesen:%
  ~\mbox{\,$x^{\tilde\mu} F(\tilde{x}^2)
    \equiv x^{\tilde\mu}\mskip-2mu(s\Dmfp, s\Dmfm)\;
             F\big(\tilde{x}^2\mskip-2mu(s\Dmfp, s\Dmfm)\big)$}, mit%
  ~\mbox{\,$x^{\tilde\mu}\mskip-2mu(s\Dmfp, s\Dmfm)
    \equiv x\Dmfp^{\tilde\mu}\mskip-2mu(s\Dmfp) \!-\! x\Dmfm^{\tilde\mu}\mskip-2mu(s\Dmfm)$}, vgl.\@ die Gln.~(\ref{APP:x=xmfp-xmfm}),~(\ref{APP:x=xmfp-xmfm}$'$),~-- und wieder in Termen von~\mbox{\,$d\Dmfp{x}^{\tilde\mu}$},~\mbox{\,$d\Dmfm{x}^{\tilde\mu}$}:
\vspace*{-.5ex}
\begin{align} \label{APP:IT[F]IJ-2}
I_T[F]\Big|_{(I,J)}\;
  =\; \ep_{\tilde\mu\tilde\nu}\;
        \int_{{\T\cal S}\Dmfp\Doperp\big|_{\scriptstyle I}} d\Dmfp{x}^{\tilde\mu}\vv
        \int_{{\T\cal S}\Dmfm\Doperp\big|_{\scriptstyle J}} d\Dmfm{x}^{\tilde\nu}\vv
        R\; \Big|_{(I,J)}\vv
        \pa_{\tilde\rh}\! \left[x^{\tilde\rh} F(\tilde{x}^2)\right]\; \Big|_{(I,J)}
    \\[-5.625ex]\nn
\end{align}
F"ur die Komponenten des Epsilon-Pseudotensors des Transversalraums gilt:~\mbox{\,$\ep_{\tilde\mu\tilde\nu} \!\stackrel{\D!}{=}\! \ep_{\mfp\mfm\tilde\mu\tilde\nu}$}, ergo:~\mbox{\,$\ep_{\tilde\mu\tilde\nu} \!=\! \ep^{\tilde\mu\tilde\nu} \!\equiv \de^{\tilde\mu}_1\de^{\tilde\nu}_2 \!-\! \de^{\tilde\nu}_2\de^{\tilde\mu}_1$}.
Der metrische Tensor in Gl.~(\ref{APP:IT[F]IJ-0}) geht "uber in Gl.~(\ref{APP:IT[F]IJ-2}) in den Epsilon-Pseudotensor:~\mbox{$g_{\tilde\mu\tilde\nu} \to \ep_{\tilde\mu\tilde\nu}$}; dabei tritt auf die~-- auf~\mbox{$(I,J)$},~\mbox{$I,J \!\in\! \{Q,\AQ\}$} zu beziehende~-- Funktion~\mbox{$R \!\equiv\! R(s\Dmfp,s\Dmfm)$}, definiert wie%
\FOOT{
  die erste Zeile in suggestiver Schreibweise in Hinblick auf die Gln.~(\ref{APP:IT[F]IJ-0}),~(\ref{APP:IT[F]IJ-2})
}
%
\vspace*{-.25ex}
\begin{align} \label{APP:FunktionR}
&R\;
  \equiv\; g_{\tilde\mu\tilde\nu}\;
         d\Dmfp{x}^{\tilde\mu}\;
         d\Dmfm{x}^{\tilde\nu}\;
         \Big/\;
           \ep_{\tilde\rh\tilde\si}\;
         d\Dmfp{x}^{\tilde\rh}\;
         d\Dmfm{x}^{\tilde\si}
    \\[-.25ex]
&=\; R(s\Dmfp,s\Dmfm)\;
  \equiv\; \bigg[\;
        g_{\tilde\mu\tilde\nu}\;
          \frac{dx\Dmfp^{\tilde\mu}}{d s\Dmfp}\mskip-2mu(s\Dmfp)\;
          \frac{dx\Dmfm^{\tilde\nu}}{d s\Dmfm}\mskip-2mu(s\Dmfm)\;
      \bigg] \bigg/
      \bigg[\;
        \ep_{\tilde\rh\tilde\si}\;
          \frac{dx\Dmfp^{\tilde\rh}}{d s\Dmfp}\mskip-2mu(s\Dmfp)\;
          \frac{dx\Dmfm^{\tilde\si}}{d s\Dmfm}\mskip-2mu(s\Dmfm)\;
      \bigg]
    \tag{\ref{APP:FunktionR}$'$}
    \\[-4.5ex]\nn
\end{align}
durch die Tangentialvektoren der Mannigfaltigkeiten~${\cal S}\Dmfp\Doperp$,~${\cal S}\Dmfm\Doperp$, vgl.\@ die Gln.~(\ref{APP:dsiPM}),~(\ref{APP:dsiPM}$'$). \\
\indent
Es ist definiert das duale Fl"achenelement~\mbox{$d\tilde{V} \!\equiv\! d\tilde{V}\mskip-2mu(s\Dmfp, s\Dmfm)$}~-- in der Konvention von Abschn.~\ref{APP-Sect:Integration}, Gl.~(\ref{APP:Stokes_duald-Konv}$'$):
\vspace*{-.5ex}
\begin{align} \label{APP:dtildeV}
&d\tilde{V}\;
  =\; \ep_{\tilde\mu\tilde\nu}\;
         d\Dmfp{x}^{\tilde\mu}\;
         d\Dmfm{x}^{\tilde\nu}\;
  =\; -\, \ep_{\tilde\mu\tilde\nu}\;
         dx\Dmfp^{\tilde\mu}\;
         dx\Dmfm^{\tilde\nu}
    \\[.25ex]
&\begin{aligned}[t]
 \equiv\; d\tilde{V}\mskip-2mu(s\Dmfp, s\Dmfm)\;
  &=\; \ep_{\tilde\mu\tilde\nu}\vv
          \frac{dx^{\tilde\mu}}{d s\Dmfp}\mskip-2mu(s\Dmfp)\vv
          \frac{dx^{\tilde\nu}}{d s\Dmfm}\mskip-2mu(s\Dmfm)\vv
          ds\Dmfp\vv
          ds\Dmfm
    \\[-.5ex]
  &=\; -\, \ep_{\tilde\mu\tilde\nu}\vv
          \frac{dx\Dmfp^{\tilde\mu}}{d s\Dmfp}\mskip-2mu(s\Dmfp)\vv
          \frac{dx\Dmfm^{\tilde\nu}}{d s\Dmfm}\mskip-2mu(s\Dmfm)\vv
          ds\Dmfp\vv
          ds\Dmfm
 \end{aligned}
    \tag{\ref{APP:dtildeV}$'$}
    \\[-4.5ex]\nn
\end{align}
\end{samepage}%
Dies ist genau das Element in~\vspace*{-.75ex}\mbox{\,$I_T[F]\big|_{(I,J)}$}.

%
%
\begin{figure}
\begin{minipage}{\linewidth}
  \begin{center}
  \vspace*{-.5ex}
  \setlength{\unitlength}{1mm}\begin{picture}(100,60)   
    \put(41,27){\epsfxsize10mm \epsffile{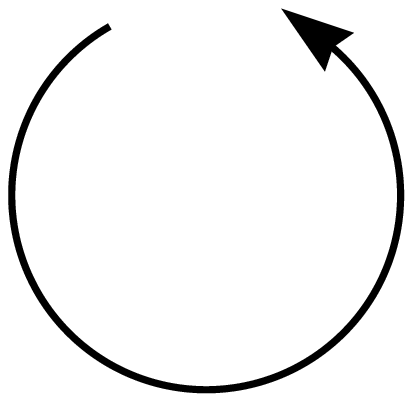}}
    \put(-20,56){\normalsize (Anti)Quark-Abschnitte}
    \put(-20,51.5){\normalsize$I,J \!\in\! \{Q,\AQ\}$:}
    \linethickness{0.3pt}
    \put(13,23){\normalsize$s\Dmfm \!\equiv\! 1$}
    \qbezier(21,33)(28,18)(61, 8)
      \put(32,12){\normalsize${\cal S}\Dmfm\Doperp\big|_{\scriptstyle J}$}
    \put(48, 2){\normalsize$s\Dmfm \!\equiv\! 0$}
    \put(63.5,6.5){\normalsize$s\Dmfp \!\equiv\! 0$}
    \qbezier(57, 6)(62,31)(82,36)
       \put(67,21){\normalsize${\cal S}\Dmfp\Doperp\big|_{\scriptstyle I}$}
    \put(85,32.5){\normalsize$s\Dmfp \!\equiv\! 1$}
    \qbezier(82,34)(49,44)(42,59)
    \qbezier(47,57)(27,52)(22,27)
    \put(55,39){\normalsize\i$\mskip-2mu$v}
    \put(25,47){\normalsize\i}
    \put(38,21){\normalsize\i$\mskip-1mu$\i}
    \put(60,25){\normalsize\i$\mskip-1mu$\i$\mskip-1mu$\i}
    {\thicklines
      \put(36,18.4){\vector(-3,2){0}}
      \put(65,25  ){\vector( 2,3){0}}
      \put(57,44.4){\vector(-3,2){0}}
      \put(30,46  ){\vector( 2,3){0}}}
  \end{picture}
  \end{center}
\vspace*{-5ex}
\caption[Direktes Produkt~\protect\mbox{${\cal S}\Dmfp\Doperp\big|_I \!\otimes\! {\cal S}\Dmfm\Doperp\big|_J$} der Mannigfaltigkeiten~\protect\mbox{${\cal S}\Dmfp\Doperp\big|_I$} und~\protect\mbox{${\cal S}\Dmfm\Doperp\big|_J$}]{
  Der duale Stokes'sche Satz reduziert die Integration "uber das direkte Produkt~\mbox{${\cal S}\Dmfp\Doperp\big|_I \!\otimes\! {\cal S}\Dmfm\Doperp\big|_J$} der Mannigfaltigkeiten \mbox{${\cal S}\Dmfp\Doperp\big|_I$} und~\mbox{${\cal S}\Dmfm\Doperp\big|_J$}~-- mit~\mbox{\,$I,J \!\in\! \{Q,\AQ\}$}~-- auf deren Rand~\mbox{$\pa\big({\cal S}\Dmfp\Doperp\big|_I \!\otimes\! {\cal S}\Dmfm\Doperp\big|_J\big)$}.   Die Pfeile geben an die Orientierung der Faktormannigfaltigkeiten f"ur~\mbox{$s\Dmfp,s\Dmfm\!: 0 \!\to\! 1$}, der Kreispfeil die der Produktmannigfaltigkeit und deren Randes.
\vspace*{-.5ex}
}
\label{Fig:SmfpI-otimes-SmfmJ}
\end{minipage}
\end{figure}
%
%
Das Funktional~\vspace*{-.5ex}\mbox{\,$I_T[F]\big|_{(I,J)}$} ist gegeben in Form von Gl.~(\ref{APP:IT[F]IJ-0}) als direktes Produkt von Linienintegralen entlang der Mannigfaltigkeiten%
  ~${\cal S}\Dmfp\Doperp\big|_{\scriptstyle I}$,~${\cal S}\Dmfm\Doperp\big|_{\scriptstyle J}$~-- und in Form
\begin{samepage}
\vspace*{-.5ex}
\begin{align} \label{APP:IT[F]IJ-3}
I_T[F]\Big|_{(I,J)}\;
  =\vv \iint_{{\T\cal S}\Dmfp\Doperp\big|_{\scriptstyle I}
      \otimes {\T\cal S}\Dmfm\Doperp\big|_{\scriptstyle J}}
        d\tilde{V}\vv
        R\vv
        \pa_{\tilde\mu}\! \left[x^{\tilde\mu} F(\tilde{x}^2)\right]
      \vv\Big|_{(I,J)}
    \\[-4.5ex]\nn
\end{align}
als Fl"achenintegral "uber deren (nichttriviales) direkte Produkt%
  ~\mbox{${\cal S}\Dmfp\Doperp\big|_I \!\otimes\! {\cal S}\Dmfm\Doperp\big|_J$}. \\
\indent
Die Funktion~\mbox{$R(s\Dmfp, s\Dmfm)$} ist konstant bez"uglich~\mbox{$(I,J)$}:
\vspace*{-.75ex}
\begin{align} \label{APP:FunktionR_IJ}
R\big|_{(I,J)}\;
  =\; -\, \rb{r}\Dmfp^I \cdot \rb{r}\Dmfm^J\;
        \Big/\, \big[\vec{r}\Dmfp^{\,I} \times \vec{r}\Dmfm^{\,J}\big]{}^3
    \\[-4.75ex]\nn
\end{align}
mit dem Skalarprodukt der transversalen Vektoren im Z"ahler:%
  ~\mbox{\,$\rb{r}\Dmfp^I \!\cdot\! \rb{r}\Dmfm^J
  \!= \de_{ij} r\Dmfp^{Ii} r\Dmfm^{Jj}$},~\mbox{\;$i,{\mskip-.5mu}j \!\in\! \{1,2\}$}, und der dritten Komponente des Vektorprodukts~-- als Ma"s des transversalen Fl"achenelements~-- im Nenner:%
  ~\mbox{\,$\big[\vec{r}\Dmfp^{\,I} \times \vec{r}\Dmfm^{\,J}\big]{}^3
  \!= \ep_{ij3}\; r\Dmfp^{Ii}\; r\Dmfm^{Jj}
  \!\equiv \ep_{ij}\; r\Dmfp^{Ii}\; r\Dmfm^{Jj}$},~\mbox{\;$i,{\mskip-.5mu}j \!\in\! \{1,2\}$}. \\
\indent
Die Konstante~$R\big|_{(I,J)}$ wird in die eckige Klammer gezogen.
Das Fl"achenintegral "uber die Direkte-Produkt-Mannigfaltigkeit wird reduziert auf das Linienintegral "uber deren Rand
\vspace*{-.5ex}
\begin{alignat}{2} \label{APP:IT[F]_IJ-Stokes}
I_T[F]\Big|_{(I,J)}\;
  &=&\quad \iint_{{\T\cal S}\Dmfp\Doperp\big|_{\scriptstyle I}
          \otimes {\T\cal S}\Dmfm\Doperp\big|_{\scriptstyle J}}\vv
          d\tilde{V}\vv
            \pa_{\tilde\mu}\!
           &\left[R\vv x^{\tilde\mu} F(\tilde{x}^2)\right]
      \vv\Big|_{(I,J)}
    \\[-.5ex]
  &=&\quad \circint_{{\T\pa\big(}{\T\cal S}\Dmfp\Doperp\big|_{\scriptstyle I}
                         \otimes {\T\cal S}\Dmfm\Doperp\big|_{\scriptstyle J}{\T\big)}}\vv
          d\tilde{V}_{\tilde\mu}\,
           &\left[R\vv x^{\tilde\mu} F(\tilde{x}^2)\right]
      \vv\Big|_{(I,J)}
    \tag{\ref{APP:IT[F]_IJ-Stokes}$'$}
    \\[-4.5ex]\nn
\end{alignat}
mithilfe des {\it Stokes'schen Satzes\/} in seiner {\it dualen Formulierung}, vgl.\@ die Gln.~(\ref{APP:Stokes_dual-Konv}), (\ref{APP:Stokes_duald-Konv})-(\ref{APP:Stokes_duald-Konv}$''$),~-- und dem dualen Linienelement%
  ~\mbox{$d\tilde{V}_{\tilde\mu} \!\equiv\! d\tilde{V}_{\tilde\mu}\mskip-2mu(s\Dmfp, s\Dmfm)$} f"ur die~\mbox{\bm{\mfp}- und \bm{\mfm}-Koordinaten}\-linien~[$s\Dmfm$ bzw.~$s\Dmfp \!=\! \text{\sl const.}$]~-- in der Konvention von Abschn.~\ref{APP-Sect:Integration}, Gl.~(\ref{APP:Stokes_duald-Konv}$''$):
\vspace*{-.25ex}
\begin{align} \label{APP:dtildeV_mu}
d\tilde{V}_{\tilde\mu}\mskip-2mu(s\Dmfp, s\Dmfm)\;
  =\; \left\{
  \begin{alignedat}{2}
  &\; \ep_{\tilde\mu\tilde\nu}\;
         d\Dmfp{x}^{\tilde\nu}\mskip-2mu(s\Dmfp)&
    &\qqquad\text{\bm{\mfp}-Koordinatenlinien}
    \\[.25ex]
  &\; \ep_{\tilde\mu\tilde\nu}\;
         d\Dmfm{x}^{\tilde\nu}\mskip-2mu(s\Dmfm)&
    &\qqquad\text{\bm{\mfm}-Koordinatenlinien}
    \end{alignedat}
      \right.
    \\[-4.25ex]\nn
\end{align}
bzgl.%
~\mbox{$d\Dmfp{x}^{\tilde\mu}\mskip-2mu(s\Dmfp) \!\equiv\! dx\Dmfp^{\tilde\mu}\mskip-2mu(s\Dmfp)$},~\mbox{$d\Dmfm{x}^{\tilde\mu}\mskip-2mu(s\Dmfm) \!\equiv\! -dx\Dmfm^{\tilde\mu}\mskip-2mu(s\Dmfm)$} vgl.\@ die Gln.~(\ref{APP:dsiPM}),~(\ref{APP:dsiPM}$'$). \\
\indent
Die Orientierung der Mannigfaltigkeiten%
  ~\vspace*{-.125ex}\mbox{\,${\cal S}\Dmfp\Doperp\big|_I  \!\otimes\!  {\cal S}\Dmfm\Doperp\big|_J$},%
  ~\mbox{\,${\T\pa(}{\cal S}\Dmfp\Doperp\big|_I \!\otimes\! {\cal S}\Dmfm\Doperp\big|_J{\T)}$} ist bestimmt wie folgt:
Definition des dualen Fl"achenelements wie%
  ~\vspace*{-.125ex}\mbox{$d\tilde{V}\mskip-2mu(s\Dmfp, s\Dmfm) \equiv \ep_{\tilde\mu\tilde\nu}\, d\Dmfp{x}^{\tilde\mu}\mskip-2mu(s\Dmfp)\, d\Dmfm{x}^{\tilde\nu}\mskip-2mu(s\Dmfm)$}~-- vgl.\@ die Gln.~(\ref{APP:dtildeV}),~(\ref{APP:dtildeV}$'$)~-- bestimmt%
  ~\vspace*{-.125ex}\mbox{\,$\big(d\Dmfp{x}^{\tilde\mu}\mskip-2mu(s\Dmfp)\big) \!\times\!
            \big(d\Dmfm{x}^{\tilde\nu}\mskip-2mu(s\Dmfm)\big)$} als {\it Rechtssystem\/} und die Orientierung des Randes entsprechend der {\it Rechtsschrauben-Regel\/} wie in Abbildung~\ref{Fig:SmfpI-otimes-SmfmJ}.
\end{samepage}%

In Termen der dort bezeichneten Abschnitte~\mbox{\i},~\mbox{\i$\mskip-1mu$\i},~\mbox{\i$\mskip-1mu$\i$\mskip-1mu$\i} und~\mbox{\i$\mskip-2mu$v} schreibt sich~\mbox{$I_T[F]|_{(I,J)}$} nach Gl.~(\ref{APP:IT[F]_IJ-Stokes}$'$) explizit
\begin{samepage}
%
\begin{align} \label{APP:IT[F]_IJ_i-iv}
I_T[F]\Big|_{(I,J)}
  =\vv \left[\;
         \int_{\mbox{\i} \circlearrowleft}
       + \int_{\mbox{\i$\mskip-1mu$\i} \circlearrowleft}
       + \int_{\mbox{\i$\mskip-1mu$\i$\mskip-1mu$\i} \circlearrowleft}
       + \int_{\mbox{\i$\mskip-2mu$v} \circlearrowleft}
       \right]\vv
         d\tilde{V}_{\tilde\mu}\,
           \left[R\vv x^{\tilde\mu} F(\tilde{x}^2)\right]
       \;\bigg|_{(I,J)}
\end{align}
Die Kreispfeile sind zu verstehen im Sinne des Kreispfeils in Abbildung~\ref{Fig:SmfpI-otimes-SmfmJ}. \\
\indent
In der Summe~\mbox{\,$I_T[F] = \sum_{I,J} {\rm sign}_{I,J} I_T[F]\big|_{(I,J)}$} treten die Integrale "uber die "`inneren Linien"'~-- das hei"st "uber die Abschnitte~\mbox{\i$\mskip-1mu$\i$\circlearrowleft$} und~\mbox{\i$\mskip-1mu$\i$\mskip-1mu$\i$\circlearrowleft$}, charakterisiert respektive durch~\mbox{$s\Dmfp \!\equiv\! 0$} und~\mbox{$s\Dmfm \!\equiv\! 0\vv\text{\sl const.}$}~-- zweimal auf mit jeweils entgegengesetzter Orientierung.
Sie k"urzen sich; "ubrig bleiben die Integrale "uber~\mbox{\i$\circlearrowleft$} und~\mbox{\i$\mskip-2mu$v$\circlearrowleft$}:
%
\begin{align} \label{APP:IT[F]_i-ii}
I_T[F]\;
  =\vv \sum\Big._{\!I,J\equiv Q,\AQ}\vv
       {\rm sign}_{I,J}\vv
       \left[\;
         \int_{\mbox{\i} \circlearrowleft}
       + \int_{\mbox{\i$\mskip-2mu$v} \circlearrowleft}\;
       \right]\vv
          d\tilde{V}_{\tilde\mu}\,
           \left[R\vv x^{\tilde\mu} F(\tilde{x}^2)\right]
       \;\bigg|_{(I,J)}
\end{align}
Die {\it orientierten Abschnitte\/}~\mbox{\i$\circlearrowleft$} und~\mbox{\i$\mskip-2mu$v$\circlearrowleft$} sind charakterisiert bez"uglich der Parameter~$s\Dmfp$,~$s\Dmfm$ respektive durch%
  ~\mbox{\,$\{ s\Dmfp\zz: 1 \!\to\! 0,\, s\Dmfm \!\equiv\! 1 \}$} und durch%
  ~\mbox{\,$\{ s\Dmfp \!\equiv\! 1,\, s\Dmfm\zz: 0 \!\to\! 1 \}$}; vgl.\@ Abb.~\ref{Fig:SmfpI-otimes-SmfmJ}. \\
\indent
Bez"uglich der Kombination~\mbox{$(I,J)$} von (Anti)Quark-Abschnitten~\mbox{$I,J \!\in\! \{Q,\AQ\}$} gilt f"ur die dualen Linienelemente~\mbox{$d\tilde{V}_{\tilde\mu} \!\equiv\! d\tilde{V}_{\tilde\mu}\mskip-2mu(s\Dmfp, s\Dmfm)$}~-- vgl.\@ Gl.~(\ref{APP:dtildeV_mu}):
\vspace*{-.25ex}
\begin{align} \label{APP:dtildeV_mu-explizit}
&d\tilde{V}_{\tilde\mu}\mskip-2mu(s\Dmfp, s\Dmfm)\Big|_{(I,J)}
    \\[-.5ex]
&\qquad=\; \left\{
  \begin{alignedat}{3}
  \; &\ep_{\tilde\mu\tilde\nu}\;
        \frac{dx\Dmfp^{\tilde\nu}}{ds\Dmfp}\mskip-2mu(s\Dmfp)\bigg|_I\;
        ds\Dmfp&\;
  &=\; \phantom{-\,} \ep_{\tilde\mu\tilde\nu}\vv
         r\Dmfp^{I\tilde\nu}\;
         ds\Dmfp&
  &\qqquad\text{\bm{\mfp}-Koordinatenlinien}
    \\[.25ex]
  \; -\, &\ep_{\tilde\mu\tilde\nu}\;
        \frac{dx\Dmfm^{\tilde\nu}}{ds\Dmfm}\mskip-2mu(s\Dmfm)\bigg|_J\;
        ds\Dmfm&\;
  &=\; -\, \ep_{\tilde\mu\tilde\nu}\vv
         r\Dmfm^{J\tilde\nu}\;
         ds\Dmfm&
  &\qqquad\text{\bm{\mfm}-Koordinatenlinien}
  \end{alignedat}
      \right.
    \nn
    \\[-4ex]\nn
\end{align}
und f"ur~\mbox{$x^{\tilde\mu}
  \!\equiv\! x^{\tilde\mu}\mskip-2mu(s\Dmfp, s\Dmfm)$}~-- vgl.\@ die Gln.~(\ref{APP:x=xmfp-xmfm}),~(\ref{APP:x=xmfp-xmfm}$'$) und~(\ref{APP:Smfp^perp-Parametr_Q,AQ}),~(\ref{APP:Smfp^perp-Parametr_Q,AQ}$'$):
\vspace*{-.5ex}
\begin{align} \label{APP:x^mu}
&x^{\tilde\mu}\mskip-2mu(s\Dmfp, s\Dmfm)\Big|_{(I,J)}\;
   =\; x\Dmfp^{\tilde\mu}\mskip-2mu(s\Dmfp)\big|_I\,
       -\, x\Dmfm^{\tilde\mu}\mskip-2mu(s\Dmfm)\big|_J
    \\[-.25ex]
  &=\;     \big( \om^{\tilde\mu} + s\Dmfp\, r\Dmfp^{I\tilde\mu} \big)\,
       -\, \big( \om^{\tilde\mu} + s\Dmfm\, r\Dmfm^{J\tilde\mu} \big)\;
   =\; \phantom{\big(} s\Dmfp\, r\Dmfp^{I\tilde\mu}\,
                   -\, s\Dmfm\, r\Dmfm^{J\tilde\mu}
    \nn
    \\[-4.25ex]\nn
\end{align}
Die Ausdr"ucke der Gln.~(\ref{APP:dtildeV_mu-explizit}),~(\ref{APP:x^mu}) sind zu kontrahieren.
Es gilt:%
\FOOT{
  \label{APP-FN:Nicht-Geradlinigkeit}Je einer der Vektoren~\vspace*{-.25ex}$r\Dmfp^I$,~$r\Dmfm^J$ folgt als Tangentialvektor der Faktor-Mannigfaltigkeiten~${\cal S}\Dmfp\Doperp$,~${\cal S}\Dmfm\Doperp$; sind diese nicht geradlinig, bleibt a~priori stehen die Summe zweier nichttrivialer Terme.
}
%
\vspace*{-.5ex}
\begin{align} 
&\ep_{\tilde\mu\tilde\nu}\;
    x^{\tilde\mu}\mskip-2mu(s\Dmfp, s\Dmfm)\Big|_{(I,J)}\;
    \frac{dx\Dmfp^{\tilde\nu}}{ds\Dmfp}\mskip-2mu(s\Dmfp)\bigg|_I
    \\[-.25ex]
  &=\; \ep_{\tilde\mu\tilde\nu}\;
       \big[ s\Dmfp\, r\Dmfp^{I\tilde\mu}\,
         -\, s\Dmfm\, r\Dmfm^{J\tilde\mu}
       \big]\; r\Dmfp^{I\tilde\nu}\; ds\Dmfp\;
   =\; -\, \ep_{\tilde\mu\tilde\nu}\;
         r\Dmfp^{I\tilde\nu}\, r\Dmfm^{J\tilde\mu}\vv
         s\Dmfm\, ds\Dmfp\;
   =\; + \big[\vec{r}\Dmfp^{\,I} \times \vec{r}\Dmfm^{\,J}\big]{}^3\;
         s\Dmfm\, ds\Dmfp
    \nn
\end{align}
f"ur die \bm{\mfp}- und analog%
  ~\mbox{\,$-\big[\vec{r}\Dmfp^{\,I} \!\times\! \vec{r}\Dmfm^{\,J}\big]{}^3
         s\Dmfp\, ds\Dmfm$} f"ur die \bm{\mfm}-Koordinatenlinien; dabei ist identifi\-ziert:%
  ~\mbox{\,$\ep_{\tilde\mu\tilde\nu} r\Dmfp^{I\tilde\mu} r\Dmfm^{J\tilde\nu}
    \equiv \ep_{ij3}\; r\Dmfp^{Ii}\; r\Dmfm^{Jj}
    = \big[\vec{r}\Dmfp^{\,I} \!\times\! \vec{r}\Dmfm^{\,J}\big]{}^3$},%
  ~\mbox{\,$\tilde\mu,\tilde\nu \!\in\! \{\mfp,\mfm,1,2\}$},%
  ~\mbox{\,$i,j \!\in\! \{1,2\}$},~die dritte Komponente des Vektorprodukts von~$r\Dmfp^I$,~$r\Dmfm^J$~-- wie bereits im Nenner von~\vspace*{-.125ex}\mbox{$R\big|_{(I,J)}$}, vgl.\@ Gl.~(\ref{APP:FunktionR_IJ}).
Es folgt:
\vspace*{-.75ex}
\begin{align} 
d\tilde{V}_{\tilde\mu}
  x^{\tilde\mu}\, (s\Dmfp, s\Dmfm)\; \Big|_{(I,J)}\;
  =\; \left\{
  \begin{alignedat}{2}
  \; +\, &\big[\vec{r}\Dmfp^{\,I} \times \vec{r}\Dmfm^{\,J}\big]{}^3\;
        s\Dmfm\, ds\Dmfp&
    &\qqquad\text{\bm{\mfp}-Koordinatenlinien}
    \\[.5ex]
  \; -\, &\big[\vec{r}\Dmfp^{\,I} \times \vec{r}\Dmfm^{\,J}\big]{}^3\;
        s\Dmfp\, ds\Dmfm&
    &\qqquad\text{\bm{\mfm}-Koordinatenlinien}
    \end{alignedat}
      \right.
    \\[-4.5ex]\nn
\end{align}
und hiermit f"ur~$I_T$ nach Gl.~(\ref{APP:IT[F]_i-ii}):
\vspace*{-.5ex}
\begin{align} 
I_T[F]\;
  &=\vv \sum\Big._{\!I,J\equiv Q,\AQ}\vv
       {\rm sign}_{I,J}\quad
       R\big|_{(I,J)}\vv
       \big[\vec{r}\Dmfp^{\,I} \times \vec{r}\Dmfm^{\,J}\big]{}^3
    \\[-.375ex]
  &\phantom{=\;}\times\,
     \bigg\{\;
         \int_1^0 1\cdot ds\Dmfp\vv
           F\big(\big|s\Dmfp\, \rb{r}\Dmfp^I - 1\cdot \rb{r}\Dmfm^J\big|\big)\;
     -\; \int_0^1 1\cdot ds\Dmfm\vv
           F\big(\big|1\cdot \rb{r}\Dmfp^I - s\Dmfm\, \rb{r}\Dmfm^J\big|\big)
     \;\bigg\}
    \nn
    \\[-4.5ex]\nn
\end{align}
\end{samepage}%
Es ist hier eingesetzt%
  ~\vspace*{-.375ex}\mbox{$\rb{x}\big|_{(I,J)} \equiv s\Dmfp\, \rb{r}\Dmfp^I \!-\! s\Dmfm\, \rb{r}\Dmfm^J$}, vgl.\@ Gl.~(\ref{APP:x^mu}), als das Argument der Funktion
  ~\mbox{$F \!\equiv\! F(|\rb{x}|)$}~-- und f"ur die orientierten Abschnitte%
  ~\mbox{\i$\circlearrowleft$} und~\mbox{\i$\mskip-2mu$v$\circlearrowleft$} respektive%
  ~\mbox{$\{ s\Dmfp\zz: 1 \!\to\! 0,\, s\Dmfm \!\equiv\! 1 \}$} und%
  ~\mbox{$\{ s\Dmfp \!\equiv\! 1,\, s\Dmfm\zz: 0 \!\to\! 1 \}$}. \\
\indent
Ferner eingesetzt%
  ~\mbox{$R\big|_{(I,J)}
            \equiv -\, \rb{r}\Dmfp^I \!\cdot\! \rb{r}\Dmfm^J \big/
                     \big[\vec{r}\Dmfp^{\,I} \!\times\! \vec{r}\Dmfm^{\,J}\big]{}^3$}, vgl.\@ Gl.~(\ref{APP:FunktionR_IJ}), k"urzt sich das Vektor- und bleibt stehen das Skalarprodukt.
F"ur~\mbox{$I_T[F]$} folgt abschlie"send:
%
\begin{align} \label{APP:IT[F]}
&I_T[F]
    \\[-.5ex]
  &\hspace*{16pt}
   =\vv \sum\Big._{\!I,J\equiv Q,\AQ}\vv
       {\rm sign}_{I,J}\quad
       \rb{r}\Dmfp^I \cdot \rb{r}\Dmfm^J\vv
       \int_0^1 ds\;
         \Big\{\;
           F\big(\big|s\, \rb{r}\Dmfp^I - \rb{r}\Dmfm^J\big|\big)\;
       +\; F\big(\big|\rb{r}\Dmfp^I - s\, \rb{r}\Dmfm^J\big|\big)
         \;\Big\}
    \nn
\end{align}
Es gilt f"ur das relevante Parameterintegral~-- vgl.\@ die Gln.~(\ref{APP:IT[F]_IT[F]IJ})-(\ref{APP:IT[F]IJ-1}) versus~(\ref{APP:IT[F]}):%
\FOOT{
  Diese Formel ist "aquivalent zu Gl.~(4.29) in Ref.~\cite{Kulzinger95}, die wir dort nur zitieren.
}
\begin{samepage}
\vspace*{-.5ex}
\begin{align} \label{APP:IJ-Parameterintegral}
\int_0^1 ds\Dmfp\; \int_0^1 ds\Dmfm\vv
  &\pa_i\! \left[x^i F(|\rb{x}|)\right]
     \;\bigg|_{\T\rb{x} \equiv s\Dmfp\, \rb{r}\Dmfp^I \!-\! s\Dmfm\, \rb{r}\Dmfm^J}
    \\[-1ex]
  =\; \int_0^1 ds\;
        &\Big\{\;
           F\big(\big|s\, \rb{r}\Dmfp^I - \rb{r}\Dmfm^J\big|\big)\;
       +\; F\big(\big|\rb{r}\Dmfp^I - s\, \rb{r}\Dmfm^J\big|\big)
         \;\Big\}
    \nn
    \\[-4.5ex]\nn
\end{align}
F"ur~\mbox{$\tilde\ch\oC$} folgt abschlie"send~-- vgl.\@ die Gln.~(\ref{APP:chC_IT[F]}),~(\ref{APP:IT[F]}):
%
\begin{align} \label{APP:chC_IJ}
&\tilde\ch\oC\;
   =\; -\, \det\mathbb{SL}\; g_{\mfp\mfm}\;\cdot\;
         \iIM\; \frac{1}{12}
    \\[-1ex]
  &\hspace*{20pt}\times\,
     \sum\Big._{\!I,J\equiv Q,\AQ}\vv
     {\rm sign}_{I,J}\quad
     \rb{r}\Dmfp^I\cdot \rb{r}\Dmfm^J\vv
       \int_0^1 ds\;
         \Big\{\;
           F\big(\big|s\, \rb{r}\Dmfp^I - \rb{r}\Dmfm^J\big|\big)\;
       +\; F\big(\big|\rb{r}\Dmfp^I - s\, \rb{r}\Dmfm^J\big|\big)
         \;\Big\}
    \nn
\end{align}
Wie betont, ist wie%
  ~\vspace*{-.125ex}\mbox{\,$F(|\rb{x}|) \equiv \projt[(1)]{F\oC}{\rb{x}}$} zu identifizieren mit der fundamentalen Korrela\-tionsfunktion~\mbox{$F\oC$}, vgl.\@ Gl.\,(\ref{APP:F_projt(1)}).
Im Haupttext werden zitiert Gl.~(\ref{APP:IT[F]}) und~(\ref{APP:chC_IJ}).

%
\bigskip\noindent
Wir schlie"sen diesen Anhang mit den folgenden Bemerkungen bez"uglich der G"ultigkeit der Darstellung von~$I_T[F]$ durch Gl.~(\ref{APP:IT[F]})~-- und folglich von~$\tilde\ch\oC$ durch Gl.~(\ref{APP:chC_IJ}). \\
\indent
Zum einen basiert die Herleitung von~$I_T[F]$ entsprechend Gl.~(\ref{APP:IT[F]}) wesentlich auf der Anwendbarkeit des (dualen) Stokes'schen Satzes~-- vgl.\@ Gl.~(\ref{APP:IT[F]_IJ-Stokes}) versus~(\ref{APP:IT[F]_IJ-Stokes}$'$).
Dies setzt offenbar voraus~-- vgl.\@ Gl.~(\ref{APP:IT[F]IJ-3} versus~(\ref{APP:IT[F]_IJ-Stokes})~-- Konstanz der Funktion~\mbox{$R(s\Dmfp, s\Dmfm)$} bez"uglich der Kombinationen~\mbox{$(I,J)$} von (Anti)Quark-Abschnit\-ten~\mbox{$I,J \!\in\! \{Q,\AQ\}$}, vgl.\@ Gl.), das hei"st Geradlinigkeit der Faktor-Mannigfaltigkeiten~${\cal S}\Dmfp\Doperp$,~${\cal S}\Dmfm\Doperp$.
Wir sehen leicht, da"s diese Annahme allerdings nicht notwendig ist:
Bei Nicht-Geradlinigkeit werden~${\cal S}\Dmfp\Doperp$,~${\cal S}\Dmfm\Doperp$ zerlegt in infinitesimal kleine Abschnitte~$I_i$ beziehungsweise~$J_j$; dann ist~\mbox{$R(s\Dmfp, s\Dmfm)$} konstant bez"uglich der Kombinationen~\mbox{$(I_i,J_j)$} von Abschnitten; es folgt%
  ~\vspace*{-.375ex}\mbox{\,$I_T[F]\big|_{(I_i,J_j)}$} in der Form von Gl.~(\ref{APP:IT[F]_IJ-Stokes}) und auf Basis des Stokes'schen Satzes in der Form von Gl.~(\ref{APP:IT[F]_IJ-Stokes}$'$).
In der Summe%
  ~\vspace*{-.375ex}\mbox{\,$I_T[F] = \sum_{I_i,J_j} {\rm sign}_{I_i,J_j} I_T[F]\big|_{(I_i,J_j)}$} k"urzen sich paarweise die Integrale "uber die inneren Linien und es folgt Gl.~(\ref{APP:IT[F]_i-ii}), bezogen auf den~-- nicht-geradlinigen~-- Rand~\mbox{\,${\T\pa(}{\cal S}\Dmfp\Doperp \!\otimes\! {\cal S}\Dmfm\Doperp{\T)}$}. \\
\indent
Zum anderen basiert~$I_T[F]$ entsprechend Gl.~(\ref{APP:IT[F]}) wesentlich auf der Nicht-Trivialit"at der Direkten-Produkt-Mannigfaltigkeit~\mbox{${\cal S}\Dmfp\Doperp \!\otimes\! {\cal S}\Dmfm\Doperp$}, die fordert lineare Unabh"angigkeit der Fak\-tor-Mannigfaltigkeiten~${\cal S}\Dmfp\Doperp$,~${\cal S}\Dmfm\Doperp$.
Sie ist g"ultig f"ur die Funktion%
  ~\mbox{$\tilde\ch\oC \!\equiv\! \tilde\ch\idx{\mfp\mskip-2mu\mfm}\oC (\equiv\! \tilde\ch\idx{\mfm\mfp}\oC)$}.
Sie ist nicht g"ultig f"ur die Funktionen~\mbox{$\tilde\ch\idx{\mfp\mskip-2mu\mfp}\oC$},~\mbox{$\tilde\ch\idx{\mfm\mskip-2mu\mfm}\oC$}~-- die subsumieren Integration beider paralleltransportierter Feldst"arken "uber dieselbe Pyramiden-Mantelfl"ache:
Es ist Bezug zu nehmen auf die Gln.~(\ref{APP:IT[F]_IT[F]IJ}),~(\ref{APP:IT[F]IJ-1}) und entsprechend zu substituieren die Mannigfaltigkeiten~\vspace*{-.25ex}${\cal S}\Dmfp\Doperp$,~${\cal S}\Dmfm\Doperp$.
So gilt f"ur~\mbox{\,$\tilde\ch\idx{\mfp\mskip-2mu\mfp}\oC \propto -\det\mathbb{SL}\, g_{\mfp\mfp}$}~-- mit%
  ~\mbox{$x^{\tilde\mu}\mskip-2mu(s, s')\big|_{(I,J)} = x\Dmfp^{\tilde\mu}\mskip-2mu(s)\big|_I \!-\! x\Dmfp^{\tilde\mu}\mskip-2mu(s')\big|_J = s\, r\Dmfp^{I\tilde\mu} \!-\! s'\, r\Dmfp^{J\tilde\mu}$}:
\vspace*{-.25ex}
\begin{align} \label{APP:IT[F]IJ-mfpmfp}
I_T[F]\Big|_{(I,J)}\;
  &=\; g_{\tilde\mu\tilde\nu}\;
        \int_{{\T\cal S}\Dmfp\Doperp\big|_{\scriptstyle I}} d\Dmfp{x}^{\tilde\mu}\vv
        \int_{{\T\cal S}\Dmfp\Doperp\big|_{\scriptstyle J}} d\Dmfp{x}^{\tilde\nu}\vv
        \pa_{\tilde\rh}\! \left[x^{\tilde\rh} F(\tilde{x}^2)\right]
        \vv\Big|_{(I,J)}
    \\[-.25ex]
  &=\; \rb{r}\Dmfp^I\cdot \rb{r}\Dmfp^J\vv
        \int_0^1 ds\vv
        \int_0^1 ds'\vv
        \pa_{\tilde\rh}\! \left[x^{\tilde\rh} F(\tilde{x}^2)\right]
        \vv\Big|_{(I,J)}
    \tag{\ref{APP:IT[F]IJ-mfpmfp}$'$}
    \\[-4.25ex]\nn
\end{align}
\end{samepage}%
und f"ur%
  ~\vspace*{-.375ex}\mbox{\,$\tilde\ch\idx{\mfm\mskip-2mu\mfm}\oC$} analog:%
  ~\mbox{\,$g_{\mfp\mfp} \!\to\! g_{\mfm\mfm}$} und~\mbox{\,$r\Dmfp \!\to\! r\Dmfm$}.
\theendnotes

%% file: APP_CLTFN.tex
\lhead[\fancyplain{}{\sc\thepage}]
      {\fancyplain{}{\sc\rightmark}}
\rhead[\fancyplain{}{\sc{{\footnotesize Anhang~\thechapter:} Korrelationsfunktionen}}]
      {\fancyplain{}{\sc\thepage}}
\psfull
\chapter[Korrelationsfunktionen]{%
   \huge Korrelationsfunktionen}
\label{APP:CLTFN}

Im Haupttext wird gemacht ein Ansatz f"ur die konfinierende~$C$- und die nicht-konfinierende \mbox{$N\!C$-Struk}\-tur des Lorentz-Korrelationstensors im Impulsraum.
Auf dieser Basis werden berechnet die streu-relevanten Funktionen%
  ~\mbox{$\projt[(1)]{F\oC}{\bm\xi},\, \projt[(2)]{F\oC}{\bm\xi}$} und%
  ~\mbox{$\projt{F\oNC}{\bm\xi}$}. \\
\indent
\vspace*{-.125ex}Wesentlicher Schritt ist die Fourier-Transformation in den Ortsraum einer durch den Ansatz definierten Funktion~\mbox{$\tilde{D}_\nu(k^2, m^2)$}.
Wir verifizieren simultan im Haupttext zitierte Relationen wie die Darstellung des Korrelationstensors im Ortsraum durch nur eine unabh"angige Korrelationsfunktion, indem wir diese Transformation durchf"uhren in dem verallgemeinerten Minkowski-Raum~\mbox{${\cal M}_{d,\si}$} mit Dimension~$d$ und Signatur~$\si$, oder "aquivalent:%
  ~\mbox{$\ze \!\equiv\! (d \!+\! \si)\!/ 2$} der Anzahl zeitartiger und%
  ~\mbox{$\rh \!\equiv\! (d \!-\! \si)\!/ 2$} der Anzahl raumartiger Dimensionen.%
\FOOT{
  \label{APP-FN:calM-d,si}Es ist~${\cal M}_{d,\si}$ der Raum, zu dem der in Anhang~\ref{APP-Sect:Integration} definierte (pseudo-)Riemannsche Raum~$G_{d,\si}$ lokal isoporph ist im Sinne des metrischen Tensors~\mbox{\,$g \!\equiv\! \big(g_{\mu\nu}\big) \!\equiv\! {\rm diag}[+1,\ldots,+1,-1,\ldots,-1]$} nach Gl.~(\ref{APP:g(x)-lokal-diag}); der gew"ohnliche Minkowski-Raum ist~\mbox{\,${\cal M} \!\equiv\! {\cal M}_{4,-2}$} mit~\mbox{$\ze \!\equiv\! 1,\, \rh \!\equiv\! 3$}.   Wir beziehen uns in diesem Anhang generell auf~${\cal M}_{d,\si}$, arbeiten \oE~in den korrespondierenden kartesischen Koordinaten~\mbox{\,$\mu \!\in\! \{1,2,\ldots,d\}$} mit den~$\ze$ ersten zeit-, den~$\rh$ "ubrigen raumartig und pr"aferieren i.a.\@ den Satz von Indizes~\mbox{$(d,\ze)$}, in praxi:~\mbox{$(4,1)$}.
}

%
\section[\protect\mbox{$\tilde{D}_\nu\mskip-2mu(k^2, m^2), D_\nu\mskip-2mu(\xi^2, m^2)$}~--
           "`\protect$\nu$-verallgemeinerter"' Feynman-Propagator
           im verallgemeinerten Minkowski-Raum~\protect${\cal M}_{d,\si}$]{%
         \bm{\tilde{D}_\nu\mskip-2mu(k^2, m^2), D_\nu\mskip-2mu(\xi^2, m^2)}~\bm{-}
           "`\bm{\nu}-verallgemeinerter"'\\
           Feynman-Propagator in~\bm{{\cal M}_{d,\si}}}
\label{APP-Sect:tildeDnu}

In Kapitel~\ref{Abschn:ANN-KONST}, Gl.~(\ref{Dvier_DDkkontrAnsatz}) und~(\ref{DDk}) wird gemacht der Ansatz%
\FOOT{
  Der Index~$\nu$ wird in~praxi~-- {\sl physikalisch\/} suggeriert~-- identisch Vier gew"ahlt.
}
\begin{samepage}
\vspace*{-.5ex}
\begin{align} \label{APP:DDk-Ansatz}
\hspace*{-0pt}
\tilde{D}(k^2)\;
   \equiv\; \tilde{D}\uC&(k^2)\;
   \equiv\; -\, \frac{1}{2}\, k^2\, \tilde{D}^{\D\prime}\uNC(k^2)
    \\[-.25ex]
  &\stackrel{\D!}{=}\;
       6\iIM\, A_\nu\cdot \la_\nu^2 (-k^2)\vv
         \tilde{D}_\nu(k^2, m^2)
         \vv\Big|_\Dstar\qqquad
  \text{mit}\quad
  \mbox{\Large$\star$}\!:\,
    m^2 \equiv 1
    \nn
    \\[-6.25ex]\nn
\end{align}
hier in Termen des "`$\nu$-verallgemeinerten"' Feynman-Propagators~\mbox{\,$\tilde{D}_\nu$}, der definiert ist im Impulsraum durch:%
\FOOT{
  \label{APP-FN:Dnu-konventionell}Der konventionelle Feynman-Propagator folgt f"ur~\mbox{$\nu \!\equiv\! 1$},~\mbox{$\la_1 \!\equiv\! 1$}, ist ergo gegeben durch~\mbox{\,$\tilde\De_{\mskip-1mu F} \!\equiv\! \tilde{D}_1|_{\la_1\equiv1}$}.
}
%
\vspace*{-.25ex}
\begin{align} \label{APP:tildeDnu-Def}
\tilde{D}_\nu(k^2, m^2)\;
  =\; \la_\nu^d\; \frac{1}{(\la_\nu^2 k^2 - m^2 + \iIM\, \ep)^\nu}\qquad
  \la_\nu, m \in \bbbr^+,\vv
  \nu \in \bbbc
    \\[-4.25ex]\nn
\end{align}
Es folgt im Ortsraum:
\vspace*{-.5ex}
\begin{align} \label{APP:Dnu-Def}
\tilde{D}_\nu(\xi^2, m^2)\;
  &\equiv\; \int \frac{d^dk}{(2\pi)^d}\;
         \efn{\T-\iIM\, k \!\cdot\! \xi}\vv \tilde{D}_\nu(k^2, m^2)
    \\[-4.5ex]\nn
\end{align}
\end{samepage}%
mit dem entsprechenden (Lorentz-)invarianten $(d,\ze)$-Skalarprodukt
%
\begin{align} \label{APP:d-Skalarprodukt}
k \!\cdot\! \xi\;
  \equiv\; g_{\mu\nu} k^\mu \xi^\nu\;
  =\; {\T\sum}_{i=1}^\ze\, k^i\, \xi^i
    - {\T\sum}_{i=\ze+1}^d\, k^i\, \xi^i
    \\[-4ex]\nn
\end{align}
Bzgl.\@ der zugrundeliegenden Konventionen vgl.\@ die Gln.~(\ref{APP:FT-allg}),~(\ref{APP:FT-allg}$'$) und Fu"sn.\,\FN{APP-FN:calM-d,si}. \\
\indent
Aus der $\tilde{D}_\nu$ definierenden Gl.~(\ref{APP:tildeDnu-Def}) lesen wir ab:
\vspace*{-.5ex}
\begin{align} \label{APP:tildeDnu-dm2,dk2}
-\, \frac{1}{\la_\nu^2}\, \frac{d}{dk^2}\vv \tilde{D}_\nu(k^2, m^2)\;
  =\; \frac{d}{dm^2}\vv \tilde{D}_\nu(k^2, m^2)
    \\[-4.5ex]\nn
\end{align}
Durch $n$-fache Differentiation folgt ferner:%
\FOOT{
  Im folgenden unterdr"ucken wir jede Bemerkung bez"uglich der G"ultigkeit der angegebenden Relationen, wenn diese~-- wie hier~-- allein bestimmt ist durch die Nicht-Singularit"at der Eulerschen Gamma-Funktion.
}
%
\vspace*{-.25ex}
\begin{align} \label{APP:tildeDnu-dm2-n}
&\hspace*{-8pt}
 \bigg(\frac{d}{dm^2}\bigg)\Big.^{\zz\T n}\vv
  \tilde{D}_\nu(k^2, m^2)\;
  =\; \frac{\Ga(\nu \!+\! n)}{\Ga(\nu)}\vv
        \bigg(\frac{\la_\nu}{\la_{\nu+n}}\bigg)\Big.^{\zz\T d}\vv
        \tilde{D}_{\nu+n}\big((\la_\nu\!/\!\la_{\nu+n}\!\cdot\! k)^2, m^2)
    \\[.25ex]
&\hspace*{-8pt}
 \Longleftrightarrow\quad
  \tilde{D}_\nu(k^2, m^2)\;
  =\; \frac{\Ga(\nu \!-\! n)}{\Ga(\nu)}\vv
        \bigg(\frac{\la_\nu}{\la_{\nu-n}}\bigg)\Big.^{\zz\T d}\vv
        \bigg(\frac{d}{dm^2}\bigg)\Big.^{\zz\T n}\vv
        \tilde{D}_{\nu-n}\big((\la_\nu\!/\!\la_{\nu-n}\!\cdot\! k)^2, m^2\big)
    \tag{\ref{APP:tildeDnu-dm2-n}$'$}
    \\[-4.5ex]\nn
\end{align}
die gestrichene Gleichung durch Aufl"osen nach%
  ~\mbox{$\tilde{D}_{\nu+n}$}, Umbenennung%
  ~\mbox{\,$(\la_\nu\!/\!\la_{\nu+n}\, k) \!\equiv\! \tilde{k} \!\to\! k$} und abschlie"senden "Ubergang%
  ~\mbox{\,$\nu \!\to\! (\nu \!-\! n)$},~-- in Termen von Ableitungen~$d\!/\!dk^2$:
\vspace*{-.25ex}
\begin{align} \label{APP:tildeDnu-dk2-n}
\tilde{D}_\nu(k^2, m^2)\;
  =\; \frac{\Ga(\nu \!-\! n)}{\Ga(\nu)}\vv
        \bigg( \frac{\la_\nu}{\la_{\nu-n}} \bigg)\Big.^{\zz\T d}\vv
        \bigg(\! -\frac{1}{\la_{\nu-n}^2} \frac{d}{dk^2} \bigg)\Big.^{\zz\T n}\vv
        \tilde{D}_{\nu-n}\big((\la_\nu\!/\!\la_{\nu-n}\!\cdot\! k)^2, m^2\big)
    \\[-4.75ex]\nn
\end{align}
vgl.\@ Gl.~(\ref{APP:tildeDnu-dm2,dk2}). \\
\indent
Analog gilt im Ortsraum:%
\FOOT{
  Die Herleitung ist {\sl straight-forward\/} aus der definierenden Gl.~(\ref{APP:tildeDnu-Def}); sie sei hier nicht weiter ausgef"uhrt.
}
\begin{samepage}
\vspace*{-.25ex}
\begin{align} \label{APP:Dnu-dxi2-n}
&D_\nu(\xi^2, m^2)\;
  =\; \frac{\Ga(\nu \!+\! n)}{\Ga(\nu)}\vv
        \bigg(\! -4\, \la_\nu^2\, \frac{d}{d\xi^2}\bigg)\Big.^{\zz\T n}\vv
        D_{\nu+n}\big((\la_{\nu+n}\!/\!\la_\nu\!\cdot\! \xi)^2, m^2)
    \\[.25ex]
&\Longleftrightarrow\quad
  D_{\nu-n}\big((\la_{\nu-n}\!/\!\la_\nu\!\cdot\! \xi)^2, m^2)\;
  =\; \frac{\Ga(\nu)}{\Ga(\nu \!-\! n)}\vv
        \bigg(\! -4\, \la_\nu^2\, \frac{d}{d\xi^2}\bigg)\Big.^{\zz\T n}\vv
        D_\nu\big(\xi^2, m^2\big)
    \tag{\ref{APP:Dnu-dxi2-n}$'$}
    \\[-4.5ex]\nn
\end{align}
die gestrichene Gleichung durch Umbenennung%
  ~\mbox{\,$(\la_{\nu+n}\!/\!\la_\nu\!\cdot\! \xi) \!\equiv\! \tilde\xi \!\to\! \xi$} und wieder abschlie"senden "Ubergang~\mbox{\,$\nu \!\to\! (\nu \!-\! n)$}.
\vspace*{-.5ex}

\section[\protect$C$- und \protect$N\!C$-Korrelationsfunktionen]{%
         \bm{C}- und \bm{N\!C}-Korrelationsfunktionen}

Der Zusammenhang des Korrelationstensors in Orts- und Impulsraum~-- vgl.\@ Gl.~(\ref{Dvier_DDxi0}) versus~(\ref{Dvier_DDk})~-- ist gegeben f"ur die $C$- und $N\!C$-Tensorstruktur respektive durch
\vspace*{-.5ex}
\begin{align} \label{APP:FT_F,D-C}
\pa^2\; F\oC(\xi^2)\;
  \stackrel{\D!}{=}\; D\uC(\xi^2)\;
  \equiv\; \int \frac{d^dk}{(2\pi)^d}\;
         \efn{\T-\iIM\, k \!\cdot\! \xi}\vv
         \tilde{D}\uC(k^2)
    \\[-5.5ex]\nn
\end{align}
und%
\FOOT{
  die zweite Identit"at unter der Annahme, die Randterme verschwinden unter partieller Integration~$\pa\!/\!\pa k^\mu$
}
%
\vspace*{-1ex}
\begin{align} \label{APP:FT_F,D-NC}
&\pa_\mu\, \pa_\nu\; F\oNC(\xi^2)
    \\[-.5ex]
  &\stackrel{\D!}{=}\; \pa_\mu\, \pa_\nu\vv
         \frac{1}{8}\, \int_{-\infty}^{\T\,\xi^2} du\; D\uNC(u)\;
   =\; \int \frac{d^dk}{(2\pi)^d}\;
         \efn{\T-\iIM\, k \!\cdot\! \xi}\vv
         \bigg[ -\frac{1}{2}\, k_\mu\, k_\nu\; \frac{d}{dk^2}\; \tilde{D}\uNC(k^2) \,\bigg]
    \nn
    \\[-4.5ex]\nn
\end{align}
mit~\mbox{\,$\pa_\mu \!\equiv\! \pa/\pa\xi^\mu$} und Definition der eigentlich relevanten Funktionen~$F\oC$,~$F\oNC$ entsprechend der ersten Identit"aten; vgl.\@ Kap.~\ref{Subsect:chNC}, Gl.~(\ref{F,D-NC}) und Kap.~\ref{Subsect:chC}, Gl.~(\ref{F,D-C}).
\vspace*{-.5ex}

\subsection[Zusammenhang~\protect\mbox{\,$\tilde{D}\uC(k^2), F\oC(\xi^2)
              \leftrightarrow \tilde{D}_\nu(k^2, m^2), D_\nu(\xi^2, m^2)$}]{%
            Zusammenhang~\bm{\,\tilde{D}\uC(k^2)\!, F\oC(\xi^2)
              \leftrightarrow \tilde{D}_\nu\mskip-2mu(k^2, m^2)\!, D_\nu\mskip-2mu(\xi^2, m^2)}}
\label{APP-Subsect:D,F_Dnu-C}

Bez"uglich der $C$-Tensorstruktur schreibt sich Gl.~(\ref{APP:DDk-Ansatz}):
\end{samepage}%
%
\begin{align} 
\tilde{D}\uC(k^2)\;
  =\; 6\iIM\, A_\nu\cdot \la_\nu^2 (-k^2)\vv
        \tilde{D}_\nu(k^2,m^2)
        \vv\Big|_\Dstar
    \\[-4.125ex]\nn
\end{align}
Daraus folgt im Ortsraum:
\vspace*{-.5ex}
\begin{align} \label{APP:DuC_Dnu}
&D\uC(\xi^2)\;
  \equiv\; \int \frac{d^dk}{(2\pi)^d}\;
              \efn{\T-\iIM\, k \!\cdot\! \xi}\vv \tilde{D}\uC(k^2)
    \\[-.25ex]
  &\phantom{D\uC}
   =\; \pa^2\vv
       \Big[\,
         6\iIM\, A_\nu\cdot \la_\nu^2\,
         \int \frac{d^dk}{(2\pi)^d}\; \efn{\T-\iIM\, k \!\cdot\! \xi}\vv
         \tilde{D}_\nu(k^2,m^2)
       \,\Big]
         \vv\bigg|_\Dstar
    \tag{\ref{APP:DuC_Dnu}$'$} \\[.25ex]
  &\phantom{D\uC}
   =\; \pa^2\vv
       \Big[\,
         6\iIM\, A_\nu\cdot \la_\nu^2\vv D_\nu(\xi^2,m^2)
       \,\Big]
         \vv\Big|_\Dstar
    \tag{\ref{APP:DuC_Dnu}$''$}
    \\[-5ex]\nn
\end{align}
und f"ur die Funktion~$F\oC$:
\begin{samepage}
%
\begin{align} \label{APP:F^C_Dnu}
F\oC(\xi^2)\;
  =\; 6\iIM\, A_\nu\cdot \la_\nu^2\vv D_\nu(\xi^2,m^2)
        \vv\Big|_\Dstar
\end{align}
vgl.\@ die Gln.~(\ref{APP:FT_F,D-C}),~(\ref{APP:DuC_Dnu}$''$).
\vspace*{-.5ex}

\subsection[Zusammenhang~\protect\mbox{\,$\tilde{D}\uNC(k^2), F\oNC(\xi^2)
              \leftrightarrow \tilde{D}_\nu(k^2, m^2), D_\nu(\xi^2, m^2)$}]{%
            Zusammenhang~\bm{\,\tilde{D}\uNC(k^2)\!, F\oNC(\xi^2)
              \leftrightarrow \tilde{D}_\nu\mskip-2mu(k^2, m^2)\!, D_\nu\mskip-2mu(\xi^2, m^2)}}
\label{APP-Subsect:D,F_Dnu-NC}

Bez"uglich der $N\!C$-Tensorstruktur schreibt sich Gl.~(\ref{APP:DDk-Ansatz}):
%
\begin{align} \label{APP:tildeDuNC_tildeDnu-Ansatz}
-\, \frac{1}{2}\, k^2\, \tilde{D}^{\D\prime}\uNC(k^2)\;
  =\; 6\iIM\, A_\nu\cdot
        \bigg[ -\frac{1}{2}\, k^2\cdot 2\, \la_\nu^2 \,\bigg]\vv
        \tilde{D}_\nu(k^2,m^2)
        \vv\Big|_\Dstar
\end{align}
Es folgt aus Gl.~(\ref{APP:tildeDnu-dk2-n}), mit~\mbox{$n \!\equiv\! 1$}:
%
\begin{align} \label{APP:tildeDuNC-ddk}
\tilde{D}_\nu(k^2, m^2)\;
  =\; \frac{\Ga(\nu \!-\! 1)}{\Ga(\nu)}\vv
        \bigg(\frac{\la_\nu}{\la_{\nu-1}}\bigg)\Big.^{\zz\T d}\vv
        \bigg(-\, \frac{1}{\la_\nu^2}\, \frac{d}{dk^2}\bigg)\vv
        \tilde{D}_{\nu-1}\big((\la_\nu\!/\!\la_{\nu-1}\!\cdot\! k)^2, m^2\big)
\end{align}
und damit f"ur die nicht-differenzierte Funktion \oE:
%
\begin{align} \label{APP:tildeDuNC_tildeDnu}
\tilde{D}\uNC(k^2)\;
  =\; 6\iIM\, A_\nu\cdot
        \bigg[\, (-2)\, \frac{\Ga(\nu \!-\! 1)}{\Ga(\nu)} \,\bigg]\;
        \bigg(\frac{\la_\nu}{\la_{\nu-1}}\bigg)\Big.^{\zz\T d}\vv
        \tilde{D}_{\nu-1}\big((\la_\nu\!/\!\la_{\nu-1}\!\cdot\! k)^2,m^2\big)
        \vv\Big|_\Dstar
\end{align}
aus Gl.~(\ref{APP:tildeDuNC_tildeDnu-Ansatz}). \\
\indent
Es gilt daher im Ortsraum:
%
\begin{align} \label{APP:DuNC_Dnu}
D\uNC&(\xi^2)\;
  \equiv\; \int \frac{d^dk}{(2\pi)^d}\;
              \efn{\T-\iIM\, k \!\cdot\! \xi}\vv \tilde{D}\uNC(k^2)
    \\[-.25ex]
  &=\; 6\iIM\, A_\nu\cdot
         \bigg[\, (-2)\, \frac{\Ga(\nu \!-\! 1)}{\Ga(\nu)} \,\bigg]\;
         \int \frac{d^d\tilde{k}}{(2\pi)^d}\;
           \efn{\T-\iIM\, \tilde{k} \!\cdot\! (\la_{\nu-1}\!/\!\la_\nu\, \xi)}\vv
         \tilde{D}_{\nu-1}(\tilde{k}^2, m^2)
         \vv\bigg|_\Dstar\zz
    \tag{\ref{APP:DuNC_Dnu}$'$} \\[.25ex]
  &=\; 6\iIM\, A_\nu\cdot
         \bigg[\, (-2)\, \frac{\Ga(\nu \!-\! 1)}{\Ga(\nu)} \,\bigg]\;
         D_{\nu-1}\big((\la_{\nu-1}\!/\!\la_\nu\!\cdot\! \xi)^2, m^2\big)
         \vv\Big|_\Dstar
    \tag{\ref{APP:DuNC_Dnu}$''$}
\end{align}
die zweite Identit"at mit Gl.~(\ref{APP:tildeDuNC_tildeDnu}) unter Substitution%
  ~\mbox{\,$k \to \tilde{k} \!\equiv\! (\la_\nu\!/\!\la_{\nu-1}\!\cdot\! k)$} der Integrationsvariablen.
Analog zu Gl.~(\ref{APP:tildeDuNC-ddk}) folgt aus Gl.~(\ref{APP:Dnu-dxi2-n}$'$), mit~\mbox{$n \!\equiv\! 1$}:
\vspace*{-.25ex}
\begin{align} 
D_{\nu-1}\big((\la_{\nu-1}\!/\!\la_\nu\, \xi)^2, m^2)\;
  =\; (-4)\; \frac{\Ga(\nu)}{\Ga(\nu \!-\! 1)}\vv
        \bigg(\la_\nu^2\, \frac{d}{d\xi^2}\bigg)\vv
        D_\nu\big(\xi^2, m^2\big)
    \\[-4.5ex]\nn
\end{align}
und daraus
\vspace*{-.25ex}
\begin{align} \label{APP:DuNC-ddxi2_Dnu}
D\uNC(\xi^2)\;
   =\; 8\cdot
         6\iIM\, A_\nu\cdot \la_\nu^2\vv
         \frac{d}{d\xi^2}\vv
         D_\nu(\xi^2, m^2)
         \vv\Big|_\Dstar
    \\[-4ex]\nn
\end{align}
aus Gl.~(\ref{APP:DuNC_Dnu}$''$). \\
\indent
Die Funktion~$F\oNC$ ist definiert durch:%
\FOOT{
  \label{APP-FN:Int'grenze-infty}In Ref.~\cite{Kulzinger95} bestimmen wir die untere Integrationsgrenze {\sl a~posteriori\/} zu~$-\infty$ durch die Forderung, die Stammfunktion verschwinde identisch f"ur diesen Wert.
}
\end{samepage}%
\vspace*{-.5ex}
\begin{align} \label{APP:F,D-NC}
F\oNC(\xi^2)\;
  =\; \frac{1}{8}\, \int_{-\infty}^{\T\,\xi^2} du\; D\uNC(u)
\end{align}
vgl.\@ die erste Identit"at der Gl.~(\ref{APP:FT_F,D-NC}).
Differenziert nach~$\xi^2$ folgt
\vspace*{-.25ex}
\begin{align} \label{APP:F,D-NC-ddxi2}
\frac{d}{d\xi^2}\vv F\oNC(\xi^2)\;
  =\; \frac{1}{8}\; D\uNC(\xi^2)
    \\[-4.25ex]\nn
\end{align}
Mit Gl.~(\ref{APP:DuNC-ddxi2_Dnu})~\oE:
\vspace*{-.25ex}
\begin{align} \label{APP:F^NC_Dnu}
F\oNC(\xi^2)\;
  =\; 6\iIM\, A_\nu\cdot \la_\nu^2\vv D_\nu(\xi^2,m^2)
        \vv\Big|_\Dstar
    \\[-4ex]\nn
\end{align}
\bigskip\noindent
Es ist~-- vgl.\@ die Gln.~(\ref{APP:F^C_Dnu}),~(\ref{APP:F^NC_Dnu})~--
\vspace*{-.25ex}
\begin{align} \label{APP:F^C,F^NC_Dnu}
F\oC(\xi^2)\;
  \equiv\; F\oNC(\xi^2)\;
  \equiv\; 6\iIM\, A_\nu\cdot \la_\nu^2\vv D_\nu(\xi^2,m^2)
             \vv\Big|_\Dstar
    \\[-4.25ex]\nn
\end{align}
die eine unabh"angige Korrelationsfunktionen beider Tensorstrukturen.
Ihre Darstellung im Impulsraum ist~-- bis auf den Faktor~\mbox{$6\iIM\, A_\nu$}~-- gegeben durch die definierende Gl.~(\ref{APP:tildeDnu-Def}) des "`$\nu$-verallgemeinerten"' Feynman-Propagators~$\tilde{D}_\nu(k^2,m^2)$, ihre Darstellung im Ortsraum durch dessen Fourier-Transformierte~$D_\nu(x^2,m^2)$. \\
\indent
F"ur die Projektionen%
  ~\mbox{\,$\projNAt{F\oC} \!\equiv\! \projNAt{F\oNC}$} folgt~-- bzgl.\@ des Operators~\mbox{$\projNFAt$} vgl.\@ Gl.~(\ref{APP:projt-Operator}), bzgl.~$\tilde{D}_\nu$ Gl.~(\ref{APP:tildeDnu-Def}):%
\FOOT{
  \label{APP-FN:deAnu}Gl.~(\ref{APP:projF^C,NC-0}$''$) ist zu lesen wie folgt:   Wir berechnen das Produkt~\mbox{$(6\iIM A_\nu \zz\cdot\zz D_\nu)$}; da~$A_\nu$ abh"angt von~$d$,~$\ze$, ist zu multiplizieren mit~$\de\!A_\nu$.   Dagegen wird~$\la_\nu$ behandelt als Parameter bez"uglich der festen Werte~$d$,~$\ze$.   \mbox{Diese Kon}\-vention minimiert den expliziten Rechenaufwand und erm"oglicht die suggestive Notation von Gl.~(\ref{APP:projF^C,NC}).
}
\begin{samepage}
\vspace*{-.5ex}
\begin{align} \label{APP:projF^C,NC-0}
&\projt{F\oC}{\bm\xi}\;
   \equiv\; \projt{F\oNC}{\bm\xi}
    \nn \\[-.5ex]
  &\phantom{\projNAt{F\oC}}
   =\; 6\iIM\, A_\nu\cdot \la_\nu^2\vv
         \int \frac{d^{d-2}\rb{k}}{(2\pi)^{d-2}}\;
         \efn{\T-\iIM\, \rb{k} \!\cdot\! \bm{\xi}}\vv
         \la_\nu^d\; \frac{1}{(\la_\nu^2 \rb{k}^2 - m^2 + \iIM\, \ep)^\nu}
         \vv\bigg|_\Dstar
    \\[-.25ex]
  &\phantom{\projNAt{F\oC}}
   =\; \la_\nu^2\;\cdot\;
         6\iIM\, A_\nu\cdot \la_\nu^2\vv
         \Big( D_\nu(\xi^2,m^2)
           \,\Big|_{\Dstar,\, \Dstar\Dstar}
         \Big)
    \tag{\ref{APP:projF^C,NC-0}$'$} \\[-.5ex]
  &\phantom{\projNAt{F\oC}}
   =\; \de\!A_\nu\; \la_\nu^2\;\cdot\,
         \Big(\, 6\iIM\, A_\nu\cdot \la_\nu^2\vv D_\nu(\xi^2,m^2)
           \,\Big|_{\Dstar,\, \Dstar\Dstar}
         \Big)\qqquad
  \de\!A_\nu\, \equiv\, \frac{A_\nu}{A_\nu \big|_{\Dstar,\, \Dstar\Dstar}}
    \tag{\ref{APP:projF^C,NC-0}$''$}
    \\[-4.5ex]\nn
\end{align}
\vspace*{-5.5ex}
\begin{alignat}{2} \label{APP:star,starstar}
\text{mit}\qquad
  &\mbox{\Large$\star$}\!:\quad\vv&
      &m^2 \!\equiv\! 1
    \\[-2ex]
  &\mbox{\Large$\star\star$}\!:\quad\vv&
      &d   \to d'   \!\equiv\!   d \!-\! 2\,,\quad
       \ze \to \ze' \!\equiv\! \ze \!-\! 1\,;\quad
      \text{ergo:}\quad \xi \to \bm{\xi}
    \tag{\ref{APP:star,starstar}$'$}
    \\[-5ex]\nn
\end{alignat}
Die Wirkung der Projektion%
  ~\mbox{$\projNFAt$} besteht darin, da"s gegen"uber vollst"andiger Fourier-Trans\-formation die Integration reduziert ist um eine zeit- und eine raumartige Impulskomponente:%
  ~\mbox{\,$\ze \!\to\! \ze' \!\equiv\! \ze \!-\! 1,\, \rh \!\to\! \rh' \!\equiv\! \rh \!-\! 1$}, das hei"st%
  ~\mbox{\,$d \!\to\! d' \!\equiv\! d \!-\! 2$}; die in diesem Sinne reduzierten~-- "`transversale"'~-- Vektoren sind notiert in \vspace*{-.25ex}(geradem) Fettdruck.
Es folgt:%
\citeFN{APP-FN:deAnu}
\vspace*{-.25ex}
\begin{align} \label{APP:projF^C,NC}
&\projt{F\oC}{\bm\xi}\;
   \equiv\; \projt{F\oNC}{\bm\xi}
    \\[-.625ex]
  &\phantom{\projNAt{F\oC}}
   =\; \de\!A_\nu\; \la_\nu^2\;\cdot\,
        \Big( F\oC(\xi^2)
          \,\Big|_{\Dstar,\, \Dstar\Dstar}
        \Big)\;
   \equiv\; \de\!A_\nu\; \la_\nu^2\;\cdot\,
        \Big( F\oNC(\xi^2)
          \,\Big|_{\Dstar,\, \Dstar\Dstar}
        \Big)
    \nn
    \\[-4.5ex]\nn
\end{align}
die neben%
  ~\vspace*{-.375ex}\mbox{$\projt{F\oNC}{\bm\xi}$} streu-relevanten Funktionen%
  ~\mbox{$\projt[(1)]{F\oC}{\bm\xi},\, \projt[(2)]{F\oC}{\bm\xi}$} durch ein- beziehungsweise zweifache Differentiation%
  ~\vspace*{-.375ex}\mbox{$-\frac{1}{|\bm\xi|}\frac{d}{d|\bm\xi|}$}, vgl.\@ Gl.~(\ref{projt(n)-Def}).
\vspace*{-.5ex}

\bigskip\noindent
Auf dieser Basis folgen explizite Darstellungen unmittelbar aus der expliziten Darstellung im Ortsraum des "`$\nu$-verallgemeinerten"' Feynman-Propagators~$D_\nu(\xi^2,m^2)$ im verallgemeinerten Minkowski-Raum~${\cal M}_{d,\si}$.
Diese wird hergeleitet im folgenden Abschnitt.
\vspace*{-.5ex}

\section[\protect\mbox{$D_\nu\mskip-2mu(\xi^2, m^2)$}:
           explizite Darstellung in~\protect${\cal M}_{d,\si}$]{%
         \vspace*{-0ex}\bm{D_\nu\mskip-2mu(\xi^2, m^2)}:
           explizite Darstellung in~\bm{{\cal M}_{d,\si}}}

Der "`$\nu$-verallgemeinerte"' Feynman-Propagator~\vspace*{-.125ex}\mbox{$D_\nu(\xi^2,m^2)$} in~${\cal M}_{d,\si}$ ist gegeben als die Fou\-rier-Transformierte
\end{samepage}%
\vspace*{-1ex}
\begin{align} \label{APP:Dnu-Def-expl}
D_\nu(\xi^2, m^2)\;
  =\; \int \frac{d^dk}{(2\pi)^d}\;
        \efn{\T-\iIM\, k \!\cdot\! \xi}\vv
        \la_\nu^d\; \frac{1}{(\la_\nu^2 k^2 - m^2 + \iIM\, \ep)^\nu}
\end{align}
%
von~\mbox{$\tilde{D}_\nu(k^2,m^2)$}, vgl.\@ die Gln.~(\ref{APP:tildeDnu-Def}),~(\ref{APP:Dnu-Def}).
Unter Substitution~\mbox{$k \!\to\! k' \!\equiv\! \la_\nu k$} der Integrationsvariablen folgt unmittelbar
\vspace*{-.5ex}
\begin{align} \label{APP:Dnu-Def-expl'}
&D_\nu\big((\la_\nu\xi')^2, m^2\big)\;
  =\; \int \frac{d^dk'}{(2\pi)^d}\;
        \efn{\T-\iIM\, k' \!\cdot\! \xi'}\vv
        \frac{1}{(\la_\nu^2 {k'}^2 - m^2 + \iIM\, \ep)^\nu}
    \\[-.75ex]
 &\phantom{D_\nu}\text{mit}\qquad
  \xi' \equiv \xi/\la_\nu
    \tag{\ref{APP:Dnu-Def-expl'}$'$}
    \\[-4.5ex]\nn
\end{align}
Unter Definition
\vspace*{-.75ex}
\begin{align} \label{APP:la=xi'2}
\la\;
  \equiv\; {\xi'}^2\;
  =\; {\T\sum}_{i=1}^\ze\, \big({\xi'}^i\big)^2
    - {\T\sum}_{i=\ze+1}^d\, \big({\xi'}^i\big)^2
    \\[-4ex]\nn
\end{align}
vgl.\@ Gl.~(\ref{APP:d-Skalarprodukt}), diskutieren wir die F"alle:%
\FOOT{
  \label{APP-FN:(iii)xi'lichtartig}Wir finden f"ur zeit- und raumartigen Vektor~$\xi'$ identische Ausdr"ucke; dessen Limes~\mbox{$\la \!\equiv\! {\xi'}^2 \!\to\! 0\mskip-1mu\pm$} stellt dar den Fall~(iii) lichtartigen Vektors~$\xi'$.
}
%
\vspace*{-.25ex}
\begin{alignat}{4} \label{APP:zeit,raumartig}
&\text{{\bf(i)}}&\quad
  &\text{\bf\bm{\xi'}\vv zeitartig,}&\vv
  &\la \equiv {\xi'}^2 > 0:&\qquad
  &\xi' = \big(\sqrt{\la},\, 0, \ldots0\big)^{\T t}
    \\
&\text{{\bf(ii)}}&\quad
  &\text{\bf\bm{\xi'}\vv raumartig,}&\vv
  &\la \equiv {\xi'}^2 < 0:&\qquad
  &\xi' = \big(0, \ldots0,\, \sqrt{-\la}\big)^{\T t}
    \tag{\ref{APP:zeit,raumartig}$'$}
    \\[-4ex]\nn
\end{alignat}
Nach geeigneter~\mbox{$(d,\ze)$}-(Lorentz-\hspace*{-.625pt})Transformation kann~\oE~angenommen werden die~Dar\-stellung mit nur einer nichtverschwindenden Komponente in Termen der Invarianten~$\la$.
\vspace*{-.5ex}

\subsection{Vorbereitung}

Zur Auswertung der Fourier-Integrale in Gl.~(\ref{APP:Dnu-Def-expl'}) ben"otigen wir zwei Typen elementarer Integrale, die wir~-- voranstellend~-- in geeigneter Form diskutieren.
Durch diese wird dann konstruiert und ausgewertet das relevante multidimensionale Integral.
\vspace*{-1.25ex}

\paragraph{Integral Typ~I:~\bm{\intI{z}{w}{a}}.} Sei definiert
\begin{samepage}
\vspace*{-.25ex}
\begin{align} 
\intI{z}{w}{a}\;
  \equiv\; \int_0^\infty dt\vv \frac{t^{z-1}}{(a + t)^{z+w}}&
    \nn \\[-3.5ex]
  &\qquad
    |\Arg a| < \pi,\vv
    {\rm Re}\, z,\, {\rm Re}\, w > 0
    \\[-4ex]\nn
\end{align}
d.h.\@ $a$ nicht reell-negativ.
Dann folgt durch Substitution~\mbox{\,$t \!\to\! t' \!\equiv\! t/a$}:
%
\begin{align} \label{APP:TypI-0}
&\intI{z}{w}{a}\;
  =\; a^{-w}\vv
        \lim_{R\to\infty}\; I\big({\cal C}_R(-\Arg a)\big)
    \\[-.5ex]
&\text{mit}\quad\vv
  I\big({\cal C}_R(\vph)\big)
  \equiv \int_{{\cal C}_R(\vph)} dt'\;
           \frac{{t'}^{z-1}}{(1 + t')^{z+w}}\qquad
    {\cal C}_R(\vph) = r\efn{\T\iIM\vph},\vv
    r\!: 0 \to R
    \tag{\ref{APP:TypI-0}$'$}
    \\[-4.5ex]\nn
\end{align}
mit~\mbox{$I\big({\cal C}_R(\vph)\big)$} dem Linienintegral in der komplexen $t'$-Ebene entlang des Strahls~\mbox{$\T{\cal C}(-\Arg a)$}, der charakterisiert ist durch konstante Phase%
  ~\mbox{$\vph \!\equiv\! -\Arg a$} und sich erstreckt ins Unendliche im Limes%
  ~\mbox{$R \!\to\! \infty$}; analog ist%
  ~\mbox{$I\big({\cal C}_R(0)\big)$} das Linienintegral entlang der positiven reellen \mbox{$t'$-Ach}\-se.
Sei~$I_{-\Arg a}(R)$ das entsprechende Integral "uber den Kreisbogen, der beide Strahlen verbindet zu einer geschlossenen Kontur.
Dann gilt mithilfe des Residuensatzes:
%
\begin{align} \label{APP:TypI-Res'satz}
I\big({\cal C}_R(-\Arg a)\big)\;
  \equiv\; I\big({\cal C}_R(0)\big)\; -\; I_{-\Arg a}(R)
\end{align}
Wir finden:%
\FOOT{
  Entwicklung des Integranden in Binomialreihe, deren {\sl normale\/} Konvergenz zul"a"st Vertauschen der Limesbildung bez"uglich Integration und Summation;~$i_k$ explizit und Gl.~(\ref{APP:Kreisbogen-Int}$'$) nur f"ur Vollst"andigkeit   
}
%
\begin{align} \label{APP:Kreisbogen-Int}
I_\vph\!(R)\;
  \begin{aligned}[t]
   =\; w^{-1} R^{-w}\; \sum_{k=0}^\infty\; i_k\; R^{-k}&
    \\[-5.5ex]
   &\qquad
    i_k\;
      \equiv\; \frac{(z\!+\!w)_k\, (w)_k}{(w\!+\!1)_k}\;
                 \frac{1}{k!}\,
                 \Big[\, \efn{\T\iIM\vph(w\!+\!k)} \!-\! 1\, \Big]
   \end{aligned}&
    \\[.25ex]
   =\; w^{-1} R^{-w}\;
         \Big[
           \efn{\T\iIM\vph w} \!\cdot\!
             f(\vph) - f(0)
         \Big]\qqquad
  f(\vph)
    \equiv \hyperF{2}{1}\Big(z\!+\!w, w; w\!+\!1; -\frac{\T 1}{\T R}\; \efn{\T\iIM\vph}\Big)&
    \tag{\ref{APP:Kreisbogen-Int}$'$}
\end{align}
\end{samepage}%
mit~\mbox{\,$(\ze)_k \!\equiv\! \Ga(\ze\!+\!k)/\Ga(\ze), (\ze)_0 \!\equiv\! 1$},~\mbox{\,$\ze \!\in\! \bbbc, k \!\in\! \bbbn$}, dem Pochhammer-Symbol und \mbox{\,$\hyperF{2}{1}(\al,\be;\ga;\ze)$} der Gau"s'schen Hypergeometrischen Funktion nach Ref.~\cite{Abramowitz84}, Abramowitz~6.1.22 bzw.~15.1.1.
Das Kreisbogen-Integral~\mbox{$I_{-\Arg a}$} ist nach Gl.~(\ref{APP:Kreisbogen-Int}) Null im Limes~\mbox{$R \!\to\! \infty$} f"ur~\mbox{\,${\rm Re}\,w \!>\! 0$}~[wie auch trivialerweise f"ur~\mbox{$\Arg a \!\equiv\! 0$}]; es folgt~\mbox{\,$\intI{z}{w}{a} \!=\! a^{-w}\, \lim_{R\to\infty} I\big({\cal C}_R(0)\big)$} und explizit:
\vspace*{-.25ex}
\begin{align} \label{APP:TypI}
\intI{z}{w}{a}\;
  &\equiv\; \int_0^\infty dt\vv \frac{t^{z-1}}{(a + t)^{z+w}}\;
   =\; a^{-w}\; \int_0^\infty dt\; \frac{t^{z-1}}{(1 + t)^{z+w}}
    \\[.25ex]
  &\begin{aligned}[t]
   =\; a^{-w}\; {\rm B}(z,w)&
    \nn \\[-2.5ex]
  &\qquad
    |\Arg a| < \pi,\vv
    {\rm Re}\,z,\, {\rm Re}\,w > 0
   \end{aligned}
    \nn
    \\[-4ex]\nn
\end{align}
unter Identifizierung des Integralausdrucks f"ur~\mbox{\,${\rm Re}\,z \!>\! 0,\, {\rm Re}\,w \!>\! 0$} als die Eulersche Beta-Funktion~\mbox{\,${\rm B}(z,w) \equiv \Ga(z)\Ga(w)\big/\Ga(z \!+\! w)$}; vgl.\@ Ref.~\cite{Abramowitz84}, Abramowitz~6.2.1.
\vspace*{-1ex}

\paragraph{Integral Typ~II:~\bm{\intII{\mu}{x}{s}}.} Sei definiert
\vspace*{-.5ex}
\begin{align} 
\intII{\mu}{x}{s}\;
  \equiv\; \int_{-\infty}^\infty \frac{dt}{2\pi}\,
             \frac{\efn{\T\pm\iIM\, tx}}{(s^2 + t^2)^{\mu+\frac{1}{2}}}&
    \\[-2.5ex]
  &\qquad
      {\rm Re}\, \mu > -\frac{1}{2},\;
      x \in \bbbr^+,\;
      |\Arg s| < \frac{\pi}{2}
    \nn
    \\[-4ex]\nn
\end{align}
d.h.\@ $s$ in der rechten Halbebene, $s^2$ nicht reell-negativ.
Der Nenner ist gerade Funktion der Integrationsvariablen, das Integrationsintervall symmetrisch; es folgt:
\vspace*{-.5ex}
\begin{align} \label{APP:TypII}
\hspace*{-15pt}
\intII{\mu}{x}{s}\;
  &\equiv\; \int_{-\infty}^\infty \frac{dt}{2\pi}\vv
              \frac{\efn{\T\pm\iIM\, tx}}{(s^2 + t^2)^{\mu+\frac{1}{2}}}\;
   =\; \frac{1}{\pi}
        \int_0^\infty \frac{dt}{2\pi}\vv
             \frac{\cos(tx)}{(s^2 + t^2)^{\mu+\frac{1}{2}}}
    \\
  &\begin{aligned}[t]
   =\; \frac{(4\pi)^{-1\!/\!2}}{\Ga\big(\mu \!+\! \frac{1}{2}\big)}\,
         \frac{1}{\big(s^2\big)^\mu}\;
         \Big(\frac{1}{2}\Big)^{\!\mu-1}\! (sx)^\mu\, {\rm K}_\mu\!(sx)&
    \nn \\[-2.5ex]
  &\hspace*{-0pt}
    {\rm Re}\, \mu > -\frac{1}{2},\;
    x \in \bbbr^+,\;
    |\Arg s| < \frac{\pi}{2}
   \end{aligned}
    \nn \\[1ex]
  &\hspace*{-10pt}\stackrel{{\rm Re}\, \mu > 0}{=}\;
       (4\pi)^{-1\!/\!2} \frac{\Ga(\mu)}{\Ga\big(\mu \!+\! \frac{1}{2}\big)}\,
         \frac{1}{\big(s^2\big)^\mu}\;
         {\cal K}_\mu\!(sx) 
    \tag{\ref{APP:TypII}$'$}
    \\[-4ex]\nn
\end{align}
Das letzte Integral in Gl.~(\ref{APP:TypII}) ist angegeben in Ref.~\cite{Abramowitz84}, Abamowitz~9.6.25; es ist%
  ~\vspace*{-.125ex}\mbox{${\rm K}_\mu\!(z)$} die modifizierte Besselfunktion zweiter Art mit Index~$\mu$ und%
  ~\mbox{\,${\cal K}_\mu\!(z) \!=\!
    (\frac{1}{2})^{\mu-1} \frac{1}{\Ga(\mu)}\, z^\mu {\rm K}_\mu\!(z)$}, mit%
  ~\vspace*{-.25ex}\mbox{\,${\cal K}_\mu\!(z) \!\to\! 1$} f"ur~\mbox{\,$z \!\to\! 0$}; vgl.\@ Gl.~(\ref{Dvier_DDxi}) auf Seite~\pageref{Dvier_DDxi}.
Herausziehen von~$\Ga(\mu)$ in Gl.~(\ref{APP:TypII}$'$) erfordert Versch"arfung der Forderung an~$\mu$.
Insbesondere gilt~\mbox{\,$\intIINA[+]{\mu} \!\equiv\! \intIINA[-]{\mu}, \forall\mu \!\in\! \bbbc$}.
\vspace*{-1ex}

\paragraph{Multidimensionales Integral~\bm{\intMD{d'}{\ze'}{\nu}{a}}.} In Hinblick auf Gl.~(\ref{APP:Dnu-Def-expl'}) sei definiert
%
\begin{align}
&\begin{aligned}[t]
 \intMD{d'}{\ze'}{\nu}{a}\;
  \equiv\; \bigg(
            {\T\prod}_{i=1}^{d'}\, 
            \int_{-\infty}^\infty \frac{dk_i}{2\pi}
           \bigg)
             \frac{1}{\big(a + {k'}^2\big)^{\T\nu}}&
    \\[-2.5ex]
  &\qquad
    |\Arg a| < \pi,\vv
    {\rm Re}\, \nu > 0
 \end{aligned}
    \label{APP:Multidimensional-Def} \\[-.75ex]
  &\text{mit}\qquad
  {k'}^2\;
  =\; {\T\sum}_{i=1}^{\ze'}\, \big({k'}^i\big)^2
    - {\T\sum}_{i=\ze'+1}^{d'}\, \big({k'}^i\big)^2
    \label{APP:Multidimensional-k'-Def}
\end{align}
d.h.~$\ze'$ Dimensionen zeit- und~\mbox{$\rh' \!\equiv\! (d' \!-\! \ze')$} Dimensionen raumartig; vgl.\@ Gl.~(\ref{APP:d-Skalarprodukt}) und~(\ref{APP:la=xi'2}).
Substituiert~\mbox{\,${k'}^i \!\to\! t_i \!\equiv\! ({k'}^i)^2$}, folgt:
%
\begin{align}
&\begin{aligned}[t]
 \intMD{d}{\ze}{\nu}{a}\;
  =\; \frac{1}{(2\pi)^{d'}}
        \bigg(
          {\T\prod}_{i=1}^{d'}\, 
          \int_0^\infty\! dt_i\; t_i^{-1\!/\!2}
        \bigg)
        \frac{1}{\big(a + t\big)^{\T\nu}}&
    \\[-2.5ex]
  &\qquad
    |\Arg a| < \pi,\vv
    {\rm Re}\, \nu > 0
 \end{aligned}
    \label{APP:Multidimensional-0} \\[-.75ex]
  &\text{mit}\qquad
  t\;
  =\; {\T\sum}_{i=1}^{\ze'}\, t_i
    - {\T\sum}_{i=\ze'+1}^{d'}\, t_i
    \label{APP:Multidimensional-t-Def}
\end{align}
Diese Integrale sind von Typ~I:
%
\begin{align} \label{APP:TypI_nu}
\int_0^\infty\! dt\vv \frac{t^{1\!/\!2 - 1}}{\big(\al + t\big)^{\T\mu}}\;
   \equiv\;
      \intI{\frac{1}{2}}{\mu - \frac{1}{2}}{\al}\;
   =\; \frac{1}{\al^{\T \mu \!-\! \frac{1}{2}}}\;
         {\rm B}\Big(\frac{1}{2}, \mu \!-\! \frac{1}{2}\Big)&
    \\[-1.5ex]
  &\hspace*{-20pt}
    |\Arg \al| < \pi,\vv
    {\rm Re} \Big(\mu \!-\! \frac{1}{2}\Big) > 0
    \nn
    \\[-4ex]\nn
\end{align}
vgl.\@ Gl.~(\ref{APP:TypI}), mit%
  ~\mbox{\,$\mu \!=\! \nu \!-\! \frac{i}{2}$},~\mbox{\,$i \!=\! 1,2,\ldots d'$}.
Damit folgt unmittelbar f"ur die~$\ze'$ zeitartigen Dimensionen:
\vspace*{-.25ex}
\begin{align} \label{APP:Multidimensional-1}
&\hspace*{-10pt}
 \intMD{d'}{\ze'}{\nu}{a}\;
  =\; \frac{1}{(2\pi)^{d'}}
       \bigg[\,
         {\T\prod}_{i=1}^{\ze'}\, 
         {\rm B}\Big(\frac{1}{2},\nu \!-\! \frac{i}{2}\Big)
       \bigg]\!
       \bigg(
         {\T\prod}_{i=\ze'+1}^{d'}\, 
         \int_0^\infty\! dt_i\, t_i^{-1\!/\!2}
       \bigg)
       \frac{1}{\big(a - t'\big)^{\T \nu \!-\! \frac{\ze'}{2}}}
    \\[-4.5ex]\nn
\end{align}
mit~\mbox{\,$- t' \!=\! -{\T\sum}_{i=\ze'+1}^{d'} t_i$}, vgl.\@ Gl.~(\ref{APP:Multidimensional-t-Def}).
Es gilt:
\begin{samepage}
\vspace*{-.25ex}
\begin{align} \label{APP:a->(-a)}
\big(a - t'\big)^{\T-\big(\nu - \frac{\ze'}{2}\big)}\;
  =\; \efn{\T-\!\big(\nu \!-\! \frac{\ze'}{2}\big)\pi\iIM}\;\cdot\;
        \big((-a) + t'\big)^{\T-\big(\nu - \frac{\ze'}{2}\big)}
    \\[-4ex]\nn
\end{align}
Daraus folgt:
\vspace*{-.25ex}
\begin{align} 
&\hspace*{-10pt}
 \intMD{d'}{\ze'}{\nu}{a}\;
    \\[-.75ex]
  &\hspace*{-10pt}
   =\; \efn{\T-(\nu \!-\! \frac{\ze'}{2})\pi\iIM}\cdot
         \frac{1}{(2\pi)^{d'}}
         \bigg[\,
           {\T\prod}_{i=1}^{\ze'}\, 
           {\rm B}(\frac{1}{2},\nu \!-\! \frac{i}{2})
         \bigg]\!
         \bigg(
           {\T\prod}_{i=\ze'+1}^{d'}\, 
           \int_0^\infty\! dt_i\, t_i^{-1\!/\!2}
         \bigg)
         \frac{1}{\big((-a) + t'\big)^{\T \nu \!-\! \frac{\ze'}{2}}}
    \nn
    \\[-4.5ex]\nn
\end{align}
und weiter mithilfe Gl.~(\ref{APP:TypI_nu}) f"ur die~\mbox{$\rh' \!\equiv\! (d' \!-\! \ze')$} "ubrigen raumartigen Dimensionen:
%
\begin{align} 
\intMD{d'}{\ze'}{\nu}{a}\;
  &=\; \efn{\T-\big(\nu \!-\! \frac{\ze'}{2}\big)\pi\iIM}\cdot
         \frac{1}{(2\pi)^{d'}}
         \bigg[\,
             {\T\prod}_{i=1}^{d'}\, 
             {\rm B}(\frac{1}{2},\nu \!-\! \frac{i}{2})
         \bigg]
         \frac{1}{(-a)^{\T \nu \!-\! \frac{\ze'}{2} \!-\! \frac{(d'-\ze')}{2}}}
    \\[-4.25ex]\nn
\end{align}
Mit~\vspace*{-.125ex}\mbox{\,${\T\prod}_{i=1}^{d'} {\rm B}(\frac{1}{2},\nu \!-\! \frac{i}{2})
    = \big(\Ga(\frac{1}{2})\big){}^{d'} \Ga(\nu-\frac{d'}{2}) \big/\Ga(\nu)$} und%
  ~\mbox{\,$\Ga(\frac{1}{2}) \!\equiv\! \pi^{1\!/\!2}$} geben wir abschlie"send an f"ur das multidimensionale Integral~\mbox{\,$\intMD{d'}{\ze'}{\nu}{a}$} nach Gl.~(\ref{APP:Multidimensional-Def}):
\vspace*{-.5ex}
\begin{alignat}{2} \label{APP:Multidimensional}
\intMD{d'}{\ze'}{\nu}{a}\;
  &=\vv& \efn{\T-(\nu \!-\! \frac{\ze'}{2})\pi\iIM}\;\cdot\;
        &(4\pi)^{-d'\!/\!2}\,
         \frac{\Ga(\nu-\frac{d'}{2})}{\Ga(\nu)}\;
         \frac{1}{\big(-a\big)^{\T \nu \!-\! \frac{d'}{2}}}
    \\[-.5ex]
  &=\vv& \efn{\T-\frac{(d'-\ze')}{2}\pi\iIM}\;\cdot\;
        &(4\pi)^{-d'\!/\!2}\,
         \frac{\Ga(\nu-\frac{d'}{2})}{\Ga(\nu)}\;
         \frac{1}{a^{\T \nu \!-\! \frac{d'}{2}}}
    \tag{\ref{APP:Multidimensional}$'$} \\[-1.5ex]
  &&&\hspace*{81pt}
    |\Arg a| < \pi,\vv
    {\rm Re} \Big(\nu \!-\! \frac{d' \!+\! 1}{2}\Big) > 0
    \nn
    \\[-4.5ex]\nn
\end{alignat}
bzgl.\@ der Einschr"ankung f"ur~${\rm Re}\,\nu$ vgl.\@ Gl.~(\ref{APP:TypI_nu}).
Es gilt~-- konsistent mit Gl.~(\ref{APP:a->(-a)}):
\vspace*{-.25ex}
\begin{align} \label{APP:(-a)->a}
\big(-a\big)^{\T-\big(\nu - \frac{d'}{2}\big)}\;
  =\; \efn{\T+\!\big(\nu \!-\! \frac{d'}{2}\big)\pi\iIM}\;\cdot\;
        a^{\T-\big(\nu - \frac{d'}{2}\big)}
    \\[-4ex]\nn
\end{align}
so da"s folgt aus der ersten Darstellung die zweite.%
\FOOT{
  Je nach Vorzeichen von~\mbox{${\rm Re}\,a$} ist zweckm"a"sig zu arbeiten mit Gl.~(\ref{APP:Multidimensional}) oder mit Gl.~(\ref{APP:Multidimensional}$'$).
} \\
%
\indent
F"ur definite Metrik k"onnen in Gl.~(\ref{APP:Multidimensional-Def}) eingef"uhrt werden sph"arische Polarkoordinaten und verifiziert werden die Gln.~(\ref{APP:Multidimensional}),~(\ref{APP:Multidimensional}$'$). \\
\indent
{\bf Seien~(1) s"amtliche Dimensionen zeitartig:~\bm{\,\ze' \!\equiv\! d'}.}
Es ist~\vspace*{-.25ex}\mbox{\,$\ka' \!\equiv\! \surd{k'}^2$} der Betrag~des $d'$-dimensionalen Radialvektors,%
  ~\vspace*{-.25ex}\mbox{\,${k'}^2 \!\equiv\! +{\T\sum}_{i=1}^{d'} \big({k'}^i\big)^2$}, vgl.\@ Gl.~(\ref{APP:Multidimensional-Def}$'$); f"ur~\mbox{\,$\intMD{d'}{d'}{\nu}{a}$} gilt:
\vspace*{-.5ex}
\begin{align} \label{APP:Multidimensional-dd}
\intMD{d'}{d'}{\nu}{a}\;
  &=\; \int \frac{d\Om_{d'}}{(2\pi)^{d'}}\vv
         \int_0^\infty\! d\ka'\,
         \frac{{\ka'}^{d'-1}}{\big(a + {\ka'}^2\big)^{\T\nu}}\;
   =\; \frac{\om_{d'}}{(2\pi)^{d'}}\vv
         \int_0^\infty \frac{dt}{2}\,
         \frac{t^{d'\!/\!2 - 1}}{\big(a + t\big)^{\T\nu}}
    \\[-.25ex]
  &=\; \frac{1}{2} \frac{\om_{d'}}{(2\pi)^{d'}}\;
         \frac{1}{a^{\T \nu \!-\! \frac{d'}{2}}}\;
         {\rm B}\Big(\frac{d'}{2}, \nu \!-\! \frac{d'}{2}\Big)\;
   =\; (4\pi)^{-d'\!/\!2}\;
         \frac{\Ga\big(\nu \!-\! \frac{d'}{2}\big)}{\Ga(\nu)}\;
         \frac{1}{a^{\T \nu \!-\! \frac{d'}{2}}}
    \nn \\
  &\hspace*{-20pt}
   \equiv\; \efn{\T-\big(\nu \!-\! \frac{\ze'}{2}\big)\pi\iIM}\cdot
         (4\pi)^{-d'\!/\!2}\;
         \frac{\Ga\big(\nu \!-\! \frac{d'}{2}\big)}{\Ga(\nu)}\;
         \frac{1}{(-a)^{\T \nu \!-\! \frac{d'}{2}}}
    \tag{\ref{APP:Multidimensional-dd}$'$}
    \\[-4.5ex]\nn
\end{align}
die zweite Identit"at unter Substitution%
  ~\vspace*{-.125ex}\mbox{\,$\ka' \!\to\! t \!\equiv\! {\ka'}^2$}, die folgenden mit Gl.~(\ref{APP:TypI_nu}) und~(\ref{APP:(-a)->a}) und eingesetzt%
  ~\mbox{\,$\int d\Om_{d'} \!\equiv\! \om_{d'}
    \!\equiv\! 2\, \pi^{d'\!/\!2}\!\big/\Ga(d'\!/\!2)$}.%
\FOOT{
  Es ist~\mbox{\,$\om_{d'} \!\equiv\! d' \!\cdot\! v_{\!d'}$} und~\mbox{\,$v_{\!d'} \!=\! \pi^{d'\!/\!2}\!\big/\Ga(d'\!/\!2 \!+\! 1)$} mit~$\om_{d'}$ der Oberfl"ache, $v_{\!d'}$ dem Volumen der \mbox{$d'$-dimensio}\-nalen Einheitskugel; vgl.\@ etwa Ref.~\cite{Forster96}, die Paragraphen~5.7 und~14.9.
}
\end{samepage}%

{\bf Seien~(2) s"amtliche Dimensionen raumartig:~\bm{\,\ze' \!\equiv\! 0}.}
Es ist~\vspace*{-.25ex}\mbox{\,$\ka' \!\equiv\! \surd-{k'}^2$} der Betrag~des $d'$-dimensionalen Radialvektors,%
  ~\vspace*{-.25ex}\mbox{\,${k'}^2 \!\equiv\! -{\T\sum}_{i=1}^{d'} \big({k'}^i\big)^2$}, vgl.\@ Gl.~(\ref{APP:Multidimensional-Def}$'$); f"ur~\mbox{\,$\intMD{d'}{0}{\nu}{a}$} gilt analog:
\vspace*{-.75ex}
\begin{align} \label{APP:Multidimensional-d0}
\intMD{d'}{0}{\nu}{a}\;
  &=\; \efn{\T-\nu\pi\iIM}\cdot
         \int \frac{d\Om_{d'}}{(2\pi)^{d'}}\vv
         \int_0^\infty\! d\ka'\,
         \frac{{\ka'}^{d'-1}}{\big((-a) + {\ka'}^2\big)^{\T\nu}}
    \\[-.25ex]
  &=\; \efn{\T-\nu\pi\iIM}\cdot
         \frac{1}{2} \frac{\om_{d'}}{(2\pi)^{d'}}\;
         \frac{1}{(-a)^{\T \nu \!-\! \frac{d'}{2}}}\;
         {\rm B}\Big(\frac{d'}{2}, \nu \!-\! \frac{d'}{2}\Big)
    \nn \\
  &\hspace*{-20pt}
   \equiv\; \efn{\T-\nu\pi\iIM}\cdot
         (4\pi)^{-d'\!/\!2}\;
         \frac{\Ga\big(\nu \!-\! \frac{d'}{2}\big)}{\Ga(\nu)}\;
         \frac{1}{(-a)^{\T \nu \!-\! \frac{d'}{2}}}
    \tag{\ref{APP:Multidimensional-d0}$'$}
    \\[-4ex]\nn
\end{align}
Wir finden Konsistenz der Gln.~(\ref{APP:Multidimensional-dd}$'$),~(\ref{APP:Multidimensional-d0}$'$) mit Gl.~(\ref{APP:Multidimensional})~-- und in nicht-trivialen Phasenfaktoren manifestiert die Indefinitheit von~\mbox{${\cal M}_{d,\si}$}.
\vspace*{-.5ex}

\subsection{Durchf"uhrung}

Auf Basis der gezeigten Identit"aten wird ausgedr"uckt und ausgewertet das Fourier-Integrale f"ur~\mbox{$\tilde{D}_\nu\big((\la\xi')^2, m^2\big)$}, vgl.\@ Gl.~(\ref{APP:Dnu-Def-expl'}). \\
\indent
{\bf Sei~(i) der Vektor~\bm{\xi'} zeitartig:%
  ~\vspace*{-.375ex}\bm{\,\la \!\equiv\! {\xi'}^2 \!>\! 0}.}
Dann gilt~-- mit~\mbox{\,$t \!\equiv\! {\xi'}^1$}, vgl.\@ Gl.~(\ref{APP:zeit,raumartig}), und dem multidimensionalen Integral%
  ~\mbox{$\intMD{d'}{\ze'}{\nu}{a}$} nach Gl.~(\ref{APP:Multidimensional-Def}):
\begin{samepage}
%
\begin{align}
&D_\nu\big((\la_\nu\xi')^2, m^2\big)\;
  =\; \int_{-\infty}^\infty \frac{dt}{2\pi}\vv
        \efn{\T-\iIM\, t\, x}\vv
        \intMD{d-1}{\ze-1}{\nu}{a}
    \label{APP:Dnu-la>0-0} \\[.5ex]
&\text{mit}\qquad
  x\; \equiv\; \surd\la\qqquad
  a\; \equiv\; s^2 + t^2\qqquad
  -s^2\; \equiv\; m^2 - \iIM\, \ep
   \label{APP:x,a,z_la>0}
\end{align}%
F"ur Indizes~\mbox{$d' \!\equiv\! d \!-\! 1$},~\mbox{$\ze' \!\equiv\! \ze \!-\! 1$} gilt~-- vgl.\@ Gl.~(\ref{APP:Multidimensional}$'$):
\vspace*{-.25ex}
\begin{align} \label{APP:Multidimensional-la>0}
\intMD{d-1}{\ze-1}{\nu}{a}\;
  &=\; \efn{\T-\frac{(d-\ze)}{2}\pi\iIM}\cdot
         (4\pi)^{-(d-1)\!/\!2}\;
         \frac{\Ga(\nu-\frac{d-1}{2})}{\Ga(\nu)}\;
         \frac{1}{a^{\T \nu \!-\! \frac{d-1}{2}}}
    \\
  &=\; \efn{\T-\frac{(d-\ze)}{2}\pi\iIM}\cdot
         (4\pi)^{-(d-1)\!/\!2}\;
         \frac{\Ga(\mu+\frac{1}{2})}{\Ga(\mu+\frac{d}{2})}\;
         \frac{1}{a^{\T \mu \!+\! \frac{1}{2}}}
    \tag{\ref{APP:Multidimensional-la>0}$'$}
    \\[-4.5ex]\nn
\end{align}
die zweite Identit"at unter Definition
\vspace*{-.25ex}
\begin{align} \label{APP:mu-Def}
\mu\;
  \equiv\; \nu - \frac{d}{2}
    \\[-4ex]\nn
\end{align}
Zusammen folgt:
\vspace*{-.5ex}
\begin{align} \label{APP:Dnu-la>0-1}
D_\nu\big((\la_\nu\xi')^2, m^2\big)\;
   =\; \efn{\T-\frac{(d-\ze)}{2}\pi\iIM}\cdot
         (4\pi)^{-(d-1)\!/\!2}\;
         \frac{\Ga(\mu+\frac{1}{2})}{\Ga(\mu+\frac{d}{2})}\;\cdot\;
         \intII[-]{\mu}{x}{s}
    \\[-4ex]\nn
\end{align}
unter Identifizierung des resultierenden Integrals als von Typ~II~-- vgl.\@ Gl.~(\ref{APP:TypII}):
\vspace*{-.5ex}
\begin{align} 
&\int_{-\infty}^\infty \frac{dt}{2\pi}\vv
  \frac{\efn{\T-\iIM\, tx}}{(s^2 + t^2)^{\mu+\frac{1}{2}}}\;
   \equiv\; \intII[-]{\mu}{x}{s}\;
   =\; \frac{(4\pi)^{-1\!/\!2}}{\Ga\big(\mu \!+\! \frac{1}{2}\big)}\,
        \frac{1}{\big(s^2\big)^\mu}\;
        \Big(\frac{1}{2}\Big)^{\!\mu-1}\! (sx)^\mu\, {\rm K}_\mu\!(sx)
    \\[-.5ex]
  &\hspace*{195pt}
    {\rm Re}\, \mu > -\frac{1}{2},\;
    x \in \bbbr^+,\;
    |\Arg s| < \frac{\pi}{2}
    \nn
    \\[-4.5ex]\nn
\end{align}
mit~$s$ bestimmt durch~$s^2$ und der Forderung an~\mbox{\,$\Arg\,s$}.
Es folgt:
\vspace*{-.25ex}
\begin{align} \label{APP:Dnu-la>0-2}
&\hspace*{-10pt}
 D_\nu\big((\la_\nu\xi')^2, m^2\big)\;
  =\; \efn{\T-\big(\mu+\frac{(d-\ze)}{2}\big)\pi\iIM}\cdot
        \frac{(4\pi)^{-d\!/\!2}}{\Ga(\mu+\frac{d}{2})}\,
        \frac{1}{\big(-s^2\big)^\mu}\;
        \Big(\frac{1}{2}\Big)^{\!\mu-1}\! (sx)^\mu\, {\rm K}_\mu\!(sx)
    \\[-.5ex]
  &\hspace*{195pt}
    {\rm Re}\, \mu > -\frac{1}{2},\;
    x \in \bbbr^+,\;
    |\Arg s| < \frac{\pi}{2}
    \nn
    \\[-4.5ex]\nn
\end{align}
\end{samepage}%
umgeformt~\mbox{\,$(s^2)^{-\mu} \!\equiv\! \efn{\T-\mu\pi\iIM} \!\cdot\! (-s^2)^{-\mu}$}, vgl.\@ Gl.~(\ref{APP:(-a)->a}).

Als "`nat"urlicher"' Parameter ist zu identifizieren%
  ~\mbox{\,$\mu \!\equiv\! \nu \!-\! d\!/\!2$}, vgl.\@ Gl.~(\ref{APP:mu-Def}), als "`nat"urliche"' Argumente~$s^2$ und~$sx$.
Mit~\mbox{\,$s^2 \!\equiv\! -(m^2 - \iIM\,\ep)$} und~\mbox{\,$x \!\equiv\! \surd\la$}, vgl.\@ Gl.~(\ref{APP:x,a,z_la>0}), folgt:%
\FOOT{
  \label{APP-FN:cong,surd}Wir schreiben~"`$\cong$"' f"ur "`gleich im Sinne der Epsilon-Vorschrift~\mbox{$\ep \!\to\! 0\!+$}"' unter Absorption von Faktoren~\mbox{$\in\! \bbbr^+$} und verstehen die Notation~"`$\surd$"' standardm"a"sig:~\mbox{\,$\sqrt[n]{z} \!\equiv\! z^{1\!/\!n}, \forall z \!\in\! \bbbc, \forall n \!\in\! \bbbn$}.
}
%
\vspace*{-.5ex}
\begin{align} \label{APP:s2,s-la>0-explizit}
&s^2\;
  =\; -(m^2 - \iIM\, \ep)\;
  \cong\; m^2\, (1 - \iIM\, \ep)\, \efn{\T\pi\iIM}\;
  \cong\; m^2\, \efn{\T-\iIM\,\ep}\, \efn{\T\pi\iIM}\;
  \cong\; m^2\, \efn{\T(1 - \ep)\pi\iIM}
    \\[-.5ex]
  &\Longrightarrow\qquad
  s\;
    \cong\; m\; \efn{\T\frac{1}{2}(1-\ep)\pi\iIM}\qquad
  \text{mit}\qquad
  |\Arg\,s| \equiv \Big|\frac{1}{2}(1 - \ep)\pi\Big| < \frac{\pi}{2}
    \tag{\ref{APP:s2,s-la>0-explizit}$'$}
    \\[-5ex]\nn
\end{align}
ergo:%
\citeFN{APP-FN:cong,surd}
\vspace*{-1ex}
\begin{align} \label{APP:sx-la>0-explizit}
&sx\;
 \begin{aligned}[t]
  &=\; m\; \efn{\T\frac{1}{2}(1-\ep)\pi\iIM}\, \surd\la\;
   =\; m\, \Big[\la\, \efn{\T(1 - \ep)\pi\iIM}\Big]^{1\!/\!2}\;
   =\; m\, \Big[-\la\, \efn{\T-\ep\pi\iIM}\Big]^{1\!/\!2}
    \\[-.5ex]
  &\cong\; m\, \big[-\la\, (1 - \iIM\, \ep)\big]^{1\!/\!2}\;
   \cong\; m\, \big[-\la + \iIM\, \ep\big]^{1\!/\!2}
 \end{aligned}
    \\[.5ex]
  &\Longrightarrow\qquad
  sx\;
    \cong\; m\, \sqrt{-\la + \iIM\, \ep}\qquad
  \text{mit}\qquad
  |\Arg\,sx| \equiv |\Arg\,s|
    \tag{\ref{APP:sx-la>0-explizit}$'$}
\end{align}
wobei benutzt ist%
  ~\mbox{\,$m \!\in\! \bbbr^+$} in Gl.~(\ref{APP:s2,s-la>0-explizit}) und%
  ~\mbox{\,$\la \!\in\! \bbbr^+$} in Gl.~(\ref{APP:sx-la>0-explizit}). \\
\indent
Sei f"ur~$sx$ eingef"uhrt die Notation
%
\begin{align} \label{APP:zem,ze-Def}
&\zem\;
  \equiv\; m\, \sqrt{-\la + \iIM\, \ep}\;
  \equiv\; m\, \ze
    \\
  &\text{und\rmfootnote}\qquad
  \ze\; \equiv\; \zem[1]\;
        =\; \sqrt{-\la + \iIM\, \ep}
    \tag{\ref{APP:zem,ze-Def}$'$}
\end{align}%
\footnotetext{
  in Hinblick auf die Korrelationsfunktionen, f"ur die gilt~\mbox{\,$m^2 \!\equiv\! 1$}, vgl.~"`\mbox{\Large$\star$}"' in Gl.~(\ref{APP:DDk-Ansatz})
}
%
Mit~\mbox{\,$\la \!\equiv\! {\xi'}^2$}, vgl.\@ Gl.~(\ref{APP:la=xi'2}), folgt dann abschlie"send f"ur den "`$\nu$-verallgemeinerten"' Feynman-Propagator~$D_\nu$ mit {\it zeitartigem\/} Argument~$\xi'$:
\vspace*{-.25ex}
\begin{align} \label{APP:Dnu-la>0}
&\hspace*{-12pt}
 D_\nu\big((\la_\nu\xi')^2, m^2\big)\;
  =\; \efn{\T-\big(\mu+\frac{d-\ze}{2}\big)\pi\iIM}\cdot
        \frac{(4\pi)^{-d\!/\!2}}{\Ga\big(\mu+\frac{d}{2}\big)}\,
        \frac{1}{\big(m^2 - \iIM\,\ep\big)^\mu}\;
        \Big(\frac{1}{2}\Big)^{\!\mu-1}\! (\zem)^\mu\, {\rm K}_\mu\!(\zem)
    \\[-.5ex]
  &\hspace*{210pt}
    {\rm Re}\, \mu > -\frac{1}{2}\vv
    \Leftrightarrow\vv
      {\rm Re}\, \nu > \frac{d-1}{2}
    \nn
    \\[-4.25ex]\nn
\end{align}
Die Epsilon-Vorschrift~\mbox{$\ep \!\to\! 0\!+$} ist {\it essentiell}, vgl.\@ Gl.~(\ref{APP:s2,s-la>0-explizit}$'$):
Sie garantiert:%
  ~\mbox{\,$|\Arg\,\zem| \!\equiv\! |\Arg\,s|$} \mbox{$\equiv\! \big|(1 \!-\! \ep)\pi\!/\!2\big| \!<\! \pi\!/\!2$}, das hei"st $\zem$ liegt in der rechten Halbebene, in der~\mbox{\,${\rm K}_\mu\!(\zem)$} gegen Null geht f"ur~\mbox{\,$\big|\zem\big| \!\to\! \infty$}, vgl.\@ Ref.~\cite{Abramowitz84}, Abramowitz~9.6.1. \\
\indent
{\bf Sei~(ii) der Vektor~\bm{\xi'} raumartig:%
  ~\vspace*{-.375ex}\bm{\,\la \!\equiv\! {\xi'}^2 \!<\! 0}.}
Dann gilt~-- mit~\mbox{\,$t \!\equiv\! {\xi'}^d$}, vgl.\@ Gl.~(\ref{APP:zeit,raumartig}$'$), und mit~\mbox{$\intMD{d'}{\ze'}{\nu}{a}$} nach Gl.~(\ref{APP:Multidimensional-Def}):
\begin{samepage}
%
\begin{align}
&D_\nu\big((\la_\nu\xi')^2, m^2\big)\;
  =\; \int_{-\infty}^\infty \frac{dt}{2\pi}\vv
        \efn{\T+\iIM\, t\, x}\vv
        \intMD{d-1}{\ze}{\nu}{a}
    \label{APP:Dnu-la<0-0} \\[.5ex]
&\text{mit}\qquad
  x\; \equiv\; \surd-\!\la\qqquad
  -a\; \equiv\; s^2 + t^2\qqquad
  s^2\; \equiv\; m^2 - \iIM\, \ep
   \label{APP:x,a,z_la<0}
\end{align}
F"ur Indizes~\mbox{$d' \!\equiv\! d \!-\! 1$},~\mbox{$\ze' \!\equiv\! \ze$} [ergo:~\mbox{$\rh' \!\equiv\! \rh \!-\! 1$}] gilt~-- vgl.\@ hier Gl.~(\ref{APP:Multidimensional}):
%
\begin{align} \label{APP:Multidimensional-la<0}
\intMD{d-1}{\ze}{\nu}{a}\;
  &=\; \efn{\T-\big(\nu \!-\! \frac{\ze}{2}\big)\pi\iIM}\cdot
         (4\pi)^{-(d-1)\!/\!2}\;
         \frac{\Ga(\nu-\frac{d-1}{2})}{\Ga(\nu)}\;
         \frac{1}{\big(-a\big)^{\T \nu \!-\! \frac{d-1}{2}}}
    \\
  &=\; \efn{\T-\big(\mu \!+\! \frac{d-\ze}{2}\big)\pi\iIM}\cdot
         (4\pi)^{-(d-1)\!/\!2}\;
         \frac{\Ga(\mu+\frac{1}{2})}{\Ga(\mu+\frac{d}{2})}\;
         \frac{1}{\big(-a\big)^{\T \mu \!+\! \frac{1}{2}}}
    \tag{\ref{APP:Multidimensional-la<0}$'$}
\end{align}
die zweite Identit"at wieder mit%
  ~\mbox{\,$\mu \!\equiv\! \nu \!-\! d\!/\!2$}, vgl.\@ Gl.~(\ref{APP:mu-Def}).
Die Gln.~(\ref{APP:Dnu-la<0-0}),~(\ref{APP:Multidimensional-la<0}$'$) zusammen, folgt unmittelbar:
\vspace*{-.5ex}
\begin{align} \label{APP:Dnu-la<0-1}
D_\nu\big((\la_\nu\xi')^2, m^2\big)\;
   =\; \efn{\T-\big(\mu \!+\! \frac{d-\ze}{2}\big)\pi\iIM}\cdot
         (4\pi)^{-(d-1)\!/\!2}\;
         \frac{\Ga(\mu+\frac{1}{2})}{\Ga(\mu+\frac{d}{2})}\;\cdot\;
         \intII[+]{\mu}{x}{s}
    \\[-4ex]\nn
\end{align}
\end{samepage}%
unter analoger Identifizierung des resultierenden Integrals als von Typ~II.
F"ur dieses wiederum folgt~-- vgl.\@ Gl.~(\ref{APP:TypII}):
\vspace*{-.5ex}
\begin{align} 
&\int_{-\infty}^\infty \frac{dt}{2\pi}\vv
  \frac{\efn{\T+\iIM\, tx}}{(s^2 + t^2)^{\mu+\frac{1}{2}}}\;
   \equiv\; \intII[+]{\mu}{x}{s}\;
   =\; \frac{(4\pi)^{-1\!/\!2}}{\Ga\big(\mu \!+\! \frac{1}{2}\big)}\,
        \frac{1}{\big(s^2\big)^\mu}\;
        \Big(\frac{1}{2}\Big)^{\!\mu-1}\! (sx)^\mu\, {\rm K}_\mu\!(sx)
    \\[-.5ex]
  &\hspace*{195pt}
    {\rm Re}\, \mu > -\frac{1}{2},\;
    x \in \bbbr^+,\;
    |\Arg s| < \frac{\pi}{2}
    \nn
    \\[-4.25ex]\nn
\end{align}
mit~$s$ wieder bestimmt durch~$s^2$ und der Forderung an~\mbox{\,$\Arg\,s$}.
Es folgt:
\vspace*{-.25ex}
\begin{align} \label{APP:Dnu-la<0-2}
&\hspace*{-10pt}
 D_\nu\big((\la_\nu\xi')^2, m^2\big)\;
  =\; \efn{\T-\big(\mu+\frac{d-\ze}{2}\big)\pi\iIM}\cdot
        \frac{(4\pi)^{-d\!/\!2}}{\Ga(\mu+\frac{d}{2})}\,
        \frac{1}{\big(s^2\big)^\mu}\;
        \Big(\frac{1}{2}\Big)^{\!\mu-1}\! (sx)^\mu\, {\rm K}_\mu\!(sx)
    \\[-.5ex]
  &\hspace*{195pt}
    {\rm Re}\, \mu > -\frac{1}{2},\;
    x \in \bbbr^+,\;
    |\Arg s| < \frac{\pi}{2}
    \nn
    \\[-4.25ex]\nn
\end{align}
Da~$s^2$ in Termen~$m^2$ mit entgegengesetztem Vorzeichen definiert sind f"ur zeit- und f"ur raumartiges~$\xi'$~-- vgl.\@ Gl.~(\ref{APP:x,a,z_la>0}) versus~(\ref{APP:x,a,z_la<0})~-- finden wir, da"s sich die Darstellungen f"ur~$D_\nu$~-- vgl.\@ Gl.~(\ref{APP:Dnu-la>0-2}) versus~(\ref{APP:Dnu-la<0-2})~-- unterscheiden nur in~$sx$. \\
\indent
Mit~\mbox{\,$s^2 \!\equiv\! +(m^2 - \iIM\,\ep)$} und~\mbox{\,$x \!\equiv\! \surd-\!\la$}, vgl.\@ Gl.~(\ref{APP:x,a,z_la<0}), folgt f"ur~$s$ und das "`nat"urliche"' Argument~$sx$ analog:%
\citeFN{APP-FN:cong,surd}
\vspace*{-.5ex}
\begin{align} \label{APP:s2,s-la<0-explizit}
&s^2\;
  =\; m^2 - \iIM\, \ep\;
  \cong\; m^2\, (1 - \iIM\, \ep)\;
  \cong\; m^2\, \efn{\T-\iIM\,\ep}\;
  \cong\; m^2\, \efn{\T-\ep\pi\iIM}
    \\[-.5ex]
  &\Longrightarrow\qquad
  s\;
    \cong\; m\; \efn{\T-\frac{1}{2}\ep\pi\iIM}\qquad
  \text{mit}\qquad
  |\Arg\,s|\; \equiv\! \Big|-\frac{1}{2}\ep\pi\Big| < \frac{\pi}{2}
    \tag{\ref{APP:s2,s-la<0-explizit}$'$}
    \\[-5ex]\nn
\end{align}
ergo:
\vspace*{-1ex}
\begin{align} \label{APP:sx-la<0-explizit}
&sx\;
 \begin{aligned}[t]
  &=\; m\; \efn{\T-\frac{1}{2}\ep\pi\iIM}\, \surd-\!\la\;
   =\; m\, \Big[-\la\, \efn{\T-\ep\pi\iIM}\Big]^{1\!/\!2}\;
   \cong\; m\, \big[-\la\, (1 - \iIM\, \ep)\big]^{1\!/\!2}
    \\[-.5ex]
  &\cong\; m\, \big[-\la - \iIM\, \ep\big]^{1\!/\!2}\;
   \cong\; m\, \big[-\la\big]^{1\!/\!2}\;
   \cong\; m\, \big[-\la + \iIM\, \ep\big]^{1\!/\!2}
 \end{aligned}
    \\[.5ex]
  &\Longrightarrow\qquad
  sx\;
    \cong\; m\, \sqrt{-\la + \iIM\, \ep}\qquad
  \text{mit}\qquad
  |\Arg\,sx| \equiv |\Arg\,s|
    \tag{\ref{APP:sx-la<0-explizit}$'$}
    \\[-4ex]\nn
\end{align}
wobei benutzt ist%
  ~\mbox{\,$m \!\in\! \bbbr^+$} in Gl.~(\ref{APP:s2,s-la<0-explizit}) und%
  ~\mbox{\,$-\la \!\in\! \bbbr^+$} in Gl.~(\ref{APP:sx-la<0-explizit}).
Aus Gl.~(\ref{APP:s2,s-la<0-explizit}$'$) lesen wir ab {\it Redundanz\/} der Epsilon-Vorschrift%
  ~\mbox{\,$\ep \!\to\! 0\!+$}; sie kann daher in Gl.~(\ref{APP:sx-la<0-explizit}) zun"achst weggelassen, dann~-- f"ur formale "Ubereinstimmung der~\mbox{\,$\zem \!\equiv\! sx$} f"ur~\mbox{\,$\la \!\gtrless\! 0$}, vgl.\@ die Gln.~(\ref{APP:sx-la>0-explizit}$'$),~(\ref{APP:zem,ze-Def}) und~(\ref{APP:sx-la<0-explizit}$'$)~-- von Hand eingef"uhrt werden mit umgekehrtem Vorzeichen. \\
\indent
In Termen von%
  ~\mbox{\,$\zem \!\equiv\! m\, \sqrt{-\la \!+\! \iIM\, \ep}$}, vgl.\@ Gl.~(\ref{APP:zem,ze-Def}), mit%
  ~\mbox{\,$\la \!\equiv\! {\xi'}^2$}, vgl.\@ Gl.~(\ref{APP:la=xi'2}), folgt dann abschlie"send f"ur den "`$\nu$-verallgemeinerten"' Feynman-Propagator~$D_\nu$ mit {\it raumartigem\/} Argument~$\xi'$~-- identisch Gl.~(\ref{APP:Dnu-la>0}):
\begin{samepage}
\vspace*{-.5ex}
\begin{align} \label{APP:Dnu-la<0}
&\hspace*{-12pt}
 D_\nu\big((\la_\nu\xi')^2, m^2\big)\;
  =\; \efn{\T-\big(\mu+\frac{d-\ze}{2}\big)\pi\iIM}\cdot
        \frac{(4\pi)^{-d\!/\!2}}{\Ga\big(\mu+\frac{d}{2}\big)}\,
        \frac{1}{\big(m^2 - \iIM\,\ep\big)^\mu}\;
        \Big(\frac{1}{2}\Big)^{\!\mu-1}\! (\zem)^\mu\, {\rm K}_\mu\!(\zem)
    \\[-.25ex]
  &\hspace*{210pt}
    {\rm Re}\, \mu > -\frac{1}{2}\vv
    \Leftrightarrow\vv
      {\rm Re}\, \nu > \frac{d-1}{2}
    \nn
    \\[-4.25ex]\nn
\end{align}
Die Epsilon-Vorschrift~\mbox{\,$\ep \!\to\! 0\!+$} ist {\it redundant\/} und eingef"uhrt von Hand f"ur formale "Ubereinstimmung der~\mbox{\,$\zem \!\equiv\! sx$} f"ur~\mbox{\,$\la \!\gtrless\! 0$}:
Mit~\mbox{\,$|\Arg\,\zem| \!\equiv\! |\Arg\,s| \!\equiv\! \big|-\ep\pi\!/\!2\big| \!\cong\! \ep \!<\! \pi\!/\!2$} liegt~$\zem$ immer in der rechten Halbebene, in der~\mbox{\,${\rm K}_\mu\!(\zem)$} gegen Null geht f"ur~\mbox{\,$\big|\zem\big| \!\to\! \infty$}. \\
\indent
{\bf Sei~(iii) der Vektor~\bm{\xi'} lichtartig:%
  ~\vspace*{-.375ex}\bm{\,\la \!\equiv\! {\xi'}^2 \!=\! 0}.}
Dann ist der "`$\nu$-verallgemeinerte"' Feynman-Propagator~\mbox{\,$D_\nu\big((\la_\nu\xi')^2, m^2\big)$} gegeben als der identische Limes
  von Gl.~(\ref{APP:Dnu-la>0}) unter~\mbox{\,$\la \!\to\! 0\!+$} und
  von Gl.~(\ref{APP:Dnu-la<0}) unter~\mbox{\,$\la \!\to\! 0\!-$}, vgl.\@ Fu"sn.\,\FN{APP-FN:(iii)xi'lichtartig}.
Dieser folgt explizit mit
\vspace*{-.25ex}
\begin{align} \label{APP:BesselK-Asymptotik,Re-mu>0}
{\rm K}_\mu\!(z)\vv
  \underset{z \to 0}{\sim}\vv
    2^{\mu-1}\; \Ga(\mu)\; z^{-\mu}\qqquad
  {\rm Re}\, \mu > 0
    \\[-4.5ex]\nn
\end{align}
mit~\mbox{\,${\rm K}_{-\mu} \!\equiv\! {\rm K}_\mu$},~\mbox{\,$\forall \mu \!\in\! \bbbc$}, auch f"ur~\mbox{\,${\rm Re}\, \mu \!<\! 0$}; vgl.\@ Ref~\cite{Abramowitz84}, Abramowitz~9.6.6 bzw.~9.6.9.
\vspace*{-.125ex}

\bigskip\noindent
Mit der Derivationsformel
\end{samepage}%
%
\begin{align} \label{APP:BesselK-Derivation}
\bigg(\! -\, \frac{1}{z} \frac{d}{dz}\bigg)\Big.^{\zz\T n}\;
    \big[z^\mu\, {\rm K}_\mu\!(z)\big]\;
  =\; z^{\mu-n}\, {\rm K}_{\mu-n}\!(z)\qqquad
  \forall \mu \in \bbbc,\vv n \in \bbbn
    \\[-4.75ex]\nn
\end{align}
vgl.\@ Ref.~\cite{Abramowitz84}, Abramowitz~9.6.28,
und mit
\vspace*{-.75ex}
\begin{align} \label{APP:zem-Der}
\zem\; \equiv\; m\, \sqrt{-\la + \iIM\, \ep}\qqquad
  \Longrightarrow\quad
  -\, \frac{1}{\zem}\frac{d}{d\zem}\;
    \equiv\; \frac{2}{m^2} \frac{d}{d\la}
    \\[-4.75ex]\nn
\end{align}
folgt unmittelbar
\vspace*{-.5ex}
\begin{align} \label{APP:BesselK-Derivation_zem,la}
(\zem)^\mu\, {\rm K}_\mu\!(\zem)\;
  =\; \bigg(\frac{2}{m^2} \frac{d}{d\la}\bigg)\Big.^{\zz\T n}\;
    \big[(\zem)^{\mu+n}\, {\rm K}_{\mu+n}\!(\zem)\big]\qqquad
  \forall \mu \in \bbbc
    \\[-4ex]\nn
\end{align}
Daraus~-- auf Basis der expliziten Darstellung von~$D_\nu$, vgl.\@ die Gln.~(\ref{APP:Dnu-la>0}),~(\ref{APP:Dnu-la<0}):
\vspace*{-.25ex}
\begin{align} \label{APP:Dnu-dxi2-n-explizit-0}
D_\nu\big(\la_\nu^2\la, m^2\big)\;
  =\; \frac{\Ga(\nu \!+\! n)}{\Ga(\nu)}\vv
        \bigg(\! -4\, \frac{d}{d\la}\bigg)\Big.^{\zz\T n}\vv
        D_{\nu+n}\big(\la_{\nu+n}^2\la, m^2)
    \\[-4ex]\nn
\end{align}
da die Funktion~\mbox{\,$D_{\nu'}\big(\la_{\nu'}^2\, z, m^2\big)$} nicht abh"angt von~$\la_{\nu'}$ f"ur alle~\mbox{\,$\nu' \!\in\! \bbbc$}; vgl.\@ Gl.~(\ref{APP:Dnu-Def-expl'}).
In Termen von~\mbox{\,$\xi \!\equiv\! \la_\nu\xi'$}, mit~\mbox{\,${\xi'}^2 \!\equiv\! \la$} gilt zum einen%
  ~\mbox{\,$D_\nu\big(\la_\nu^2 \la, m^2\big) \!\equiv\!
            D_\nu\big((\la_\nu\xi')^2, m^2\big) \!\equiv\!
            D_\nu\big(\xi^2, m^2\big)$} und%
  ~\mbox{\,$D_{\nu+n}\big(\la_{\nu+n}^2 \la, m^2) \!\equiv\!
            D_{\nu+n}\big((\la_{\nu+n}\xi')^2, m^2) \!\equiv\!
            D_{\nu+n}\big((\la_{\nu+n}\!/\!\la_\nu\!\cdot\! \xi)^2, m^2)$} und zum anderen
   \mbox{\,$d\!/\!d\la \equiv d/d{\xi'}^2 \equiv \la_\nu^{-2} d/d\xi^2$}.~--
Aus Gl.~(\ref{APP:Dnu-dxi2-n-explizit}) folgt also:
\vspace*{-.25ex}
\begin{align} \label{APP:Dnu-dxi2-n-explizit}
D_\nu(\xi^2, m^2)\;
  =\; \frac{\Ga(\nu \!+\! n)}{\Ga(\nu)}\vv
        \bigg(\! -4\, \la_\nu^2\, \frac{d}{d\xi^2}\bigg)\Big.^{\zz\T n}\vv
        D_{\nu+n}\big((\la_{\nu+n}\!/\!\la_\nu\!\cdot\! \xi)^2, m^2)
    \\[-4ex]\nn
\end{align}
Dies ist genau Gl.~(\ref{APP:Dnu-dxi2-n}), die somit explizit verifiziert ist. \\
\indent
Ihre Herleitung basiert also wesentlich auf der Ableitungsformel f"ur~${\rm K}_\mu$~-- vgl.\@ Gl.~(\ref{APP:BesselK-Derivation})~-- die wiederum g"ultig ist ohne Einschr"ankung an den Index~$\mu$.
Dies "ubertr"agt sich unmittelbar auf Gl.~(\ref{APP:Dnu-dxi2-n-explizit}), die Ableitungsformel f"ur~$D_\nu$~-- und zwar mit genau der Einschr"ankung, da"s wohldefiniert ist der~\mbox{\,$z^\mu{\rm K}_\mu\!(z)$-multi}\-plizierende Faktor, das hei"st {\it cum~grano~salis\/} die Gamma-Funktion:~\mbox{\,$\Ga\big(\mu \!+\! d\!/\!2\big)$}.
Die G"ultigkeit der Darstellung f"ur~$D_\nu$ kann daher ausgedehnt werden auf~\mbox{\,${\rm Re}\,\nu \!\equiv\! {\rm Re}\big(\mu \!+\! d\!/\!2\big) \!>\! 0$}. \\
\indent
Zusammenfassend ist der "`$\nu$-verallgemeinerte"' Feynman-Propagator~$D_\nu$ f"ur zeit-, licht- und raumartiges Argument~$\xi'$~-- ergo~$\xi$~-- geegben durch:
%
\begin{align} \label{APP:Dnu-all-la}
&\hspace*{-0pt}
 \begin{aligned}[t]
  D_\nu\big(\xi^2, m^2\big)\;
  &\equiv\; D_\nu\big((\la_\nu\xi')^2, m^2\big)
    \\[.5ex]
  &=\; \efn{\T-\big(\mu+\frac{(d-\ze)}{2}\big)\pi\iIM}\cdot
        \frac{(4\pi)^{-d\!/\!2}}{\Ga\big(\mu+\frac{d}{2}\big)}\,
        \frac{1}{\big(m^2 - \iIM\,\ep\big)^\mu}\;
        \Big(\frac{1}{2}\Big)^{\!\mu-1}\! (\zem)^\mu\, {\rm K}_\mu\!(\zem)\hspace*{-8pt}
 \end{aligned}
    \\[-.5ex]
  &\hspace*{305pt}
    {\rm Re}\, \nu > 0
    \nn \\[-.5ex]
  &\text{mit}\qquad
    \mu \equiv \nu - d \!/\! 2,\qquad
    \zem \equiv m\, \sqrt{-\la + \iIM\, \ep},\qquad
    \la \equiv {\xi'}^2 \equiv \xi^2 \!/\! \la_\nu^2
    \nn
\end{align}
Wir rekapitulieren:
Die Epsilon-Vorschrift~\mbox{\,$\ep \!\to\! 0\!+$} ist essentiell nur f"ur zeitartiges Argument~$\xi'$,~$\xi$:
F"ur dieses,~\mbox{\,$\la \!>\! 0$} garantiert sie~-- f"ur~\mbox{\,$\la \!<\! 0$} ist dies unmittelbar der Fall~-- da"s~$\zem$ in der rechten Halbebene liegt, in der~\mbox{\,${\rm K}_\mu\!(\zem)$} gegen Null geht f"ur~\mbox{\,$\big|\zem\big| \!\to\! \infty$}.
\vspace*{-.5ex}

\bigskip\noindent
Wir verifizieren Gl.~(\ref{APP:Dnu-all-la}) f"ur den konventionellen Feynman-Propagator im gew"ohnlichen Minkowski-Raum.
Es ist~\mbox{\,$\De_{\mskip-1mu F} \!\equiv\! D_1|_{\la_\nu\equiv1}$}, vgl.\@ Fu"sn.\,\FN{APP-FN:Dnu-konventionell}, und~\mbox{$d \!\equiv\! 4$},~\mbox{$\ze \!\equiv\! 1$}; mit~\mbox{\,$\mu \!\equiv\! -1$} folgt:
\vspace*{-.25ex}
\begin{align} \label{APP:Dnu-konventionell}
&\hspace*{-16pt}
 \De_{\mskip-1mu F}(\xi^2, m^2\big)\;
   \equiv\;
       \int \frac{d^4k}{(2\pi)^4}\;
         \efn{\T-\iIM\, k \!\cdot\! \xi}\vv
         \frac{1}{k^2 - m^2 + \iIM\, \ep}
    \\[.5ex]
  &\hspace*{-16pt}
   \hspace*{11pt}
   =\, \efn{\T-\frac{1}{2}\pi\iIM}\cdot
         \frac{(4\pi)^{-2}}{\Ga(1)}\,
         \big(m^2 \!-\! \iIM\,\ep\big)
         \Big(\frac{1}{2}\Big)^{\!-2}\! (\zem)^{-1}\, {\rm K}_{-1}\!(\zem)\vv
   =\, -\, \frac{\iIM\, m}{4\pi^2}\,
          \frac{{\rm K}_1\!(m\, \sqrt{-\la + \iIM\, \ep})}{\sqrt{-\la + \iIM\, \ep}}
    \tag{\ref{APP:Dnu-konventionell}$'$} \\[.25ex]
  &\hspace*{-16pt}
   \stackrel{\la \equiv \xi^2 \to 0}{\sim}\vv
       -\, \frac{\iIM}{4\pi^2}\,
         \Big[
           \frac{1}{-\la + \iIM\, \ep}\;
           +\; \frac{m^2}{2}\Big(\ln\frac{m\surd}{2} + 1\Big)\;
           +\; m^2\, {\cal O}\big((m\surd)^2\big)
         \Big]
    \tag{\ref{APP:Dnu-konventionell}$''$}
    \\[-4ex]\nn
\end{align}
mit~\mbox{$\surd\, \!\equiv\! \sqrt{-\la + \iIM\, \ep}$}, bzgl.~\mbox{${\rm K}_{-1} \!\equiv\! {\rm K}_1$} vgl.\@ die Bem.\@ zu Gl.\,(\ref{APP:BesselK-Asymptotik,Re-mu>0}).
\mbox{Wir finden "Ubereinstimmung} mit Ref.~\cite{Bogoljubov84}; vgl.\@ ebenda Kap.~18.3 und Anh.~A.V.1\&2, insbes.\@ Gl.~(18.19).%
\FOOT{
  Bogoljubov, \v{S}irkov definieren die kausale Greenfunktion mit {\sl umgekehrtem\/} Vorzeichen:~\mbox{\,$D^c \!\equiv\! -\De_{\mskip-1mu F}$}.
}

%
\section[Parameter~\protect\mbox{$A_\nu,\, \la_\nu$}:
           Fixierung in~\protect\mbox{${\cal M}_{d,\si}$}]{%
         Parameter~\bm{A_\nu,\, \la_\nu}:
           Fixierung in~\bm{{\cal M}_{d,\si}}}

Die bisher freien Parameter~$A_\nu$,~$\la_\nu$ werden fixiert als Konsequenz der Bedeutung der $C$- und $N\!C$-Funktionenen von physikalische {\it Korrelations\/}funktionen.
Die Forderungen sind~-- vgl.\@ Gl.~(\ref{Ala_Bedingg}):
\begin{samepage}
\vspace*{-.5ex}
\begin{alignat}{2} \label{APP:Ala_Bedingg}
1\;
  \stackrel{\D!}{=}\; D \Big|_{\D\xi^2 \!\equiv\! 0}\qquad
  \text{und}\qquad
1\;
  \stackrel{\D!}{=}\; \int_0^{\infty} \! du\; D(-u^2)
    \\[-4.5ex]\nn
\end{alignat}
in Termen des kontrahierten Lorentz-Korrelationstensor%
  ~\mbox{\,$D(\xi^2) \!\equiv\! D_{\mu\nu}{}^{\mu\nu}\!(\xi)$}, vgl.\@ Gl.~(\ref{Dvier_DDkkontr}).
A~priori auf dessen Fourier-Transformierte bezieht sich der Ansatz~-- vgl.\@ Gl.~(\ref{APP:DDk-Ansatz}):
\vspace*{-.5ex}
\begin{align} \label{APP:DDk-Ansatz-D}
\hspace*{-0pt}
&\tilde{D}(k^2)\;
  \stackrel{\D!}{=}\;
        6\iIM\, A_\nu\cdot \la_\nu^2 (-k^2)\vv
          \tilde{D}_\nu(k^2, m^2)\Big|_\Dstar
    \\
&\text{ergo:\rmfootnote}\qquad
  D(\xi^2)\;
    =\; 6\iIM\, A_\nu\cdot \la_\nu^2 \pa^2\vv
          D_\nu(\xi^2, m^2)\Big|_\Dstar
    \tag{\ref{APP:DDk-Ansatz-D}$'$}
    \\[-4.75ex]\nn
\end{align}
mit~\mbox{$\mbox{\Large$\star$}\!:\, m^2 \!\equiv\! 1$}.
\footnotetext{
  konsistent mit~\mbox{\,$D(\xi^2) \!=\!
    D\uC(\xi^2) \!=\!
    \pa^2 F\oC(\xi^2) \!=\!
    \pa^2 \big[6\iIM A_\nu\!\cdot\! \la_\nu^2\, D_\nu(\xi^2,m^2)\big]\big|_\Dstar$, vgl.\@ Gl.\,(\ref{APP:DDk-Ansatz}),\,(\ref{APP:FT_F,D-C}) und~(\ref{APP:F^C,F^NC_Dnu})}
}
\vspace*{-.5ex}

\subsection[\protect\mbox{$D \!\equiv\! D_{\mu\nu}{}^{\mu\nu}$}~--
              Kontrahierter Korrelations-Lorentztensor]{%
            \bm{D \!\equiv\! D_{\mu\nu}{}^{\mu\nu}}~\bm{-}
              Kontrahierter Korrelations-Lorentztensor}

%
Es ist~$D_\nu$ explizit gegben in Gl.~(\ref{APP:Dnu-all-la}), aus der~$D$ explizit folgt durch Ausf"uhren des d'Alembert-Operators~$\pa^2$, das mit~-- vgl.%
  ~\mbox{\,$\la \!\equiv\! {\xi'}^2 \!\equiv\! \xi^2/\la_\nu^2$}
\vspace*{-.25ex}
\begin{align} \label{APP:dAlembert_la}
\la_\nu^2 \pa^2\;
  =\; {\pa'}^2\;
  =\; \Big[\, d + \la \Big(2\, \frac{d}{d\la}\Big)\Big] \Big(2\, \frac{d}{d\la}\Big)\qqquad
  \text{mit}\qquad
  d \equiv g^\mu{}_\mu = {\rm dim}({\cal M}_{d,\si})
    \\[-4ex]\nn
\end{align}
zur"uckgef"uhrt ist auf die Derivationsformel f"ur~$D_\nu$, vgl.\@ Gl.~(\ref{APP:Dnu-dxi2-n-explizit-0}). \\
\indent
Die Wirkung von~$\pa^2$ im Impulsraum besteht dagegen in der Multiplikation mit~$(-k^2)$; wir gehen diesen Weg.
Es ist:
\vspace*{-.25ex}
\begin{align} \label{APP:k2-tildeDnu_tildeDnu-0}
&\la_\nu^2 k^2\vv \tilde{D}_\nu(k^2,m^2)
    \nn \\[-.5ex]
  &\hspace*{20pt}
   =\; \big[ (\la_\nu^2 k^2 - m^2 + \iIM\, \ep)\;
         +\; (m^2 - \iIM\, \ep) \big]\cdot
        \la_\nu^d\; \frac{1}{(\la_\nu^2 k^2 - m^2 + \iIM\, \ep)^\nu}&&
    \\[-.5ex]
  &\hspace*{20pt}
   =\; \frac{\la_\nu^d}{\la_{\nu-1}^d}\cdot
         \la_{\nu-1}^d\; \frac{1}{%
           \big(\la_{\nu-1}^2 (\la_\nu\!/\!\la_{\nu-1}\!\cdot\! k)^2
             - m^2 + \iIM\, \ep\big)^{\nu-1}}&&
    \tag{\ref{APP:k2-tildeDnu_tildeDnu-0}$'$} \\[-.75ex]
       &\phantom{\hspace*{20pt}
         =\; \frac{\la_\nu^d}{\la_{\nu-1}^d}\cdot
         \tilde{D}_{\nu-1}\big((\la_\nu\!/\!\la_{\nu-1}\!\cdot\! k)^2,m^2\big)\vv}
       +\; (m^2 - \iIM\, \ep)\cdot
         \la_\nu^d\; \frac{1}{(\la_\nu^2 k^2 - m^2 + \iIM\, \ep)^\nu}
    \nn
    \\[-6ex]\nn
\end{align}
zusammen:
\vspace*{-.25ex}
\begin{align} \label{APP:k2-tildeDnu_tildeDnu}
&\la_\nu^2 k^2\vv \tilde{D}_\nu(k^2,m^2)
    \\
  &\hspace*{20pt}
   =\; \frac{\la_\nu^d}{\la_{\nu-1}^d}\cdot
         \tilde{D}_{\nu-1}\big((\la_\nu\!/\!\la_{\nu-1}\!\cdot\! k)^2,m^2\big)\;
       +\; (m^2 - \iIM\, \ep)\cdot
         \tilde{D}_\nu(k^2,m^2)
    \nn
    \\[-4.5ex]\nn
\end{align}
Dies Relation ist Fourier-zu-transformieren bez"uglich~$k$.
Im \mbox{$\tilde{D}_{\nu-1}$-Inte}\-gral wird unmittelbar substituiert%
  ~\mbox{\,$k \!\to\! \tilde{k} \!\equiv\! \la_\nu\!/\!\la_{\nu-1}\!\cdot\! k$}; mit ergo%
  ~\mbox{\,$d^d\tilde{k} \!\equiv\! \la_\nu^d\!/\!\la_{\nu-1}^d\!\cdot\! d^dk$} wir absorbiert der Vorfaktor, abschlie"send umbenannt~\mbox{\,$\tilde{k} \!\to\! k$}.
Mit Notation%
  ~\mbox{\,$\tilde\xi \!\equiv\! \la_{\nu-1}\!/\!\la_\nu\!\cdot\! \xi$} folgt:
\vspace*{-.5ex}
\begin{align} \label{APP:pa2-Dnu_Dnu-0}
&-\, \la_\nu^2 \pa^2\vv D_\nu(\xi^2,m^2)
    \nn \\
  &\hspace*{20pt}
   =\; \int \frac{d^dk}{(2\pi)^d}\;
         \efn{\T-\iIM\, k \!\cdot\! \xi}\vv
         \la_\nu^2 k^2\vv \tilde{D}_\nu(k^2,m^2)
    \\
  &\hspace*{20pt}
   =\; \int \frac{d^d\tilde{k}}{(2\pi)^d}\;
       \Big[
         \efn{\T-\iIM\, k \!\cdot\! \tilde\xi}\vv
         \tilde{D}_{\nu-1}(k^2,m^2)\;
       +\; (m^2 - \iIM\, \ep)\cdot
         \efn{\T-\iIM\, k \!\cdot\! \xi}\vv
         \tilde{D}_\nu(k^2,m^2)
       \Big]
    \tag{\ref{APP:pa2-Dnu_Dnu-0}$'$}
    \\[-4.5ex]\nn
\end{align}
zusammen:
\end{samepage}%
\vspace*{-.125ex}
\begin{align} \label{APP:pa2-Dnu_Dnu}
-\, \la_\nu^2 \pa^2\vv D_\nu(\xi^2,m^2)\;
  =\; D_{\nu-1}(\tilde\xi^2,m^2)\;
      +\; (m^2 - \iIM\, \ep)\vv D_\nu(\xi^2,m^2)
\end{align}
Diese Relation eingesetzt in Gl.~(\ref{APP:DDk-Ansatz-D}$'$), folgt:
\vspace*{-.5ex}
\begin{align} \label{APP:D_Dnu-1,Dnu}
D(\xi^2)\;
    =\; -\, 6\iIM\, A_\nu\cdot
          \big[
            D_{\nu-1}\big((\la_{\nu-1}\!/\!\la_\nu\!\cdot\! \xi)^2,m^2\big)\;
            +\; (m^2 - \iIM\, \ep)\vv
                  D_\nu(\xi^2,m^2)
          \big]
        \vv\Big|_\Dstar
    \\[-4.5ex]\nn
\end{align}
Dies ist die geeignetste Darstellung f"ur~$D$~-- in Termen von Funktionen~$D_\nu$ mit expliziter Darstellung nach Gl.~(\ref{APP:Dnu-all-la})~-- zur Fixierung der Parameter~$A_\nu$,~$\la_\nu$ im Sinne von Gl.~(\ref{APP:Ala_Bedingg}).

%
\subsection[Parameter~\protect\mbox{$A_\nu$}]{%
            Parameter~\bm{A_\nu}}

Der globale Faktor~$A_\nu$ ist bestimmt durch Forderung der erste Relation in Gl.~(\ref{APP:Ala_Bedingg}): Normierung auf Eins f"ur verschwindende Raumzeit-Separation, das hei"st maximale Korrelation. \\
\indent
In Hinblick auf die Darstellung von~$D$ in Termen von~$D_\nu$~-- vgl.\@ Gl.~(\ref{APP:D_Dnu-1,Dnu})~-- ben"otigen wir formal deren Grenzwert unter~\mbox{$\xi^2 \!\to\! 0$}.%
\FOOT{
  dies der in Zusammenhang mit Gl.~(\ref{APP:BesselK-Asymptotik,Re-mu>0}) angesprochene Fall~(iii) lichtartigen Vektors~$\xi$
}
Die Raumzeit-Abh"angigkeit der Funktion~$D_\nu$ ist subsumiert in der Funktion~\mbox{\,$z^\mu {\rm K}_\mu\!(z)$}, mit~\mbox{\,$\mu \!\equiv\! \nu \!-\! d\!/\!2$}~-- vgl.\@ Gl.~(\ref{APP:Dnu-all-la}).
Wir betrachten daher deren Limes unter~\mbox{\,$z \!\to\! 0$}, explizit Gl.~(\ref{APP:BesselK-Asymptotik,Re-mu>0}) f"ur~$\mu$, falls~\mbox{\,${\rm Re}\, \mu \!>\! 0$}, und f"ur~$-\mu$, falls~\mbox{\,${\rm Re}\, \mu \!<\! 0$}.
Es gilt:
\begin{samepage}
%
\begin{align} \label{APP:BesselK-Asymptotik}
{\rm K}_\mu\!(z)\vv
  \underset{z \to 0}{\sim}\vv
    2^{\si_{\!\mu}\mu-1}\; \Ga(\si_{\!\mu}\mu)\vv
    z^{-\si_{\!\mu}\mu}\qqquad
  \si_{\!\mu} \equiv {\rm sign}({\rm Re}\, \mu),\vv
  {\rm Re}\, \mu \gtrless 0
\end{align}
indem benutzt ist~\mbox{\,${\rm K}_{-\mu} \!\equiv\! {\rm K}_\mu$},~\mbox{\,$\forall \mu \!\in\! \bbbc$}, im Falle~\mbox{\,${\rm Re}\, \mu \!<\! 0$}.
Multiplikation mit~$z^\mu$ ergibt unmittelbar:
%
\begin{align} \label{APP:z-BesselK-Asymptotik}
z^\mu\, {\rm K}_\mu\!(z)\vv
  \underset{z \to 0}{\sim}\vv
    2^{\si_{\!\mu}\mu-1}\; \Ga(\si_{\!\mu}\mu)\vv
    z^{(1-\si_{\!\mu})\mu}\qqquad
  \si_{\!\mu} \equiv {\rm sign}({\rm Re}\, \mu),\vv
  {\rm Re}\, \mu \gtrless 0
\end{align}
f"ur die $D_\nu$-relevante Funktion. \\
\indent
Wir finden, da"s~\mbox{\,$z^\mu {\rm K}_\mu\!(z)$} f"ur~\mbox{\,$z \!\to\! 0$} geht gegen den endlichen Zahlenwert%
  ~\mbox{\,$2^{\mu-1}\; \Ga(\mu)$} f"ur~\mbox{\,${\rm Re}\, \mu \!>\! 0$} und divergiert wie%
  ~\mbox{\,$z^{2\mu}$} f"ur~\mbox{\,${\rm Re}\, \mu \!<\! 0$}.
Normierbarkeit von~$D_\nu$ f"ur verschwindendes Raumzeit-Argument induziert also mit~\mbox{\,$\mu \!\equiv\! \nu \!-\! d\!/\!2$} wieder die Forderung~\mbox{\,${\rm Re}\, \nu \!>\! d\!/\!2$}.%
\FOOT{
  Wir halten fest:   Der "`$\nu$-verallgemeinerte"' Feynman-Propagator~$D_\nu$ nach Gl.~(\ref{APP:Dnu-all-la}) fordert~\mbox{\,${\rm Re}\, \nu \!>\! 0$}; aufgefa"st als {\sl normierte\/} (Korrelations)Funktion versch"arft~\mbox{\,${\rm Re}\, \nu \!>\! d\!/\!2$}.   So ist der konventionelle Feynman-Propagator im gew"ohnlichen Minkowski-Raum~-- vgl.\@ Gl.~(\ref{APP:Dnu-konventionell}$''$)~-- {\sl wohldefiniert\/}, er mit~\mbox{\,$\nu \!\equiv\! 1 \ngtr d\!/\!2 \!\equiv\! 2$} aber {\sl nicht normierbar\/} f"ur lichtartiges Argument.
}

Aus Gl.~(\ref{APP:Dnu-all-la})~-- mit Gl.~(\ref{APP:z-BesselK-Asymptotik})~-- folgt unmittelbar f"ur~$D_\nu$:
\vspace*{-.5ex}
\begin{align} \label{APP:Dnu_xi2->0}
&\begin{aligned}[t]
 D_\nu(\xi^2, m^2)\vv
  \underset{\xi^2 \to 0}{\sim}\vv
   &D_\nu \Big|_{\D\xi^2 \!\equiv\! 0}
    \\
   &\hspace*{12pt}
    \equiv\; \efn{\T-\big(\nu-\frac{\ze}{2}\big)\pi\iIM}\cdot
      \frac{(4\pi)^{-d\!/\!2}}{\big(m^2 - \iIM\,\ep\big)^{\nu-\frac{d}{2}}}\vv
      \frac{1}{\Ga(\nu)}\,
      \Ga\big(\T\nu \!-\! \frac{d}{2}\big)
 \end{aligned}
    \\[-2.5ex]
  &\hspace*{300pt}
    {\rm Re}\, \nu > \frac{d}{2}
    \nn
    \\[-5ex]\nn
\end{align}
und in Termen von~\mbox{\,$D_\nu \big|_{\D\xi^2 \!\equiv\! 0}$} f"ur~$D_{\nu-1}$:
\vspace*{-.75ex}
\begin{align} \label{APP:Dnu-1_xi2->0}
&D_{\nu-1}(\xi^2, m^2)\vv
  \underset{\xi^2 \to 0}{\sim}\vv
    D_\nu \Big|_{\D\xi^2 \!\equiv\! 0}\;\cdot\;
    \efn{\T\pi\iIM}\cdot
      \big(m^2 - \iIM\,\ep\big)\vv
      \frac{\Ga(\nu)}{\Ga(\nu \!-\! 1)}\,
      \frac{\Ga\big(\nu \!-\! 1 \!-\! \frac{d}{2}\big)}{\Ga\big(\nu \!-\! \frac{d}{2}\big)}
    \\[-.5ex]
  &\hspace*{282pt}
    {\rm Re}\, \nu > \frac{d}{2} + 1
    \nn
    \\[-5.25ex]\nn
\end{align}
Es ist~\mbox{\,${\rm Re}\, \nu > d\!/\!2$} zu fordern wegen~\mbox{\,${\rm Re}\, \mu \!\stackrel{\D!}{>}\! 0$} und~\mbox{\,${\rm Re}\, \nu > d\!/\!2 \!+\! 1$} wegen~\mbox{\,${\rm Re}\, (\mu \!-\! 1) \!\stackrel{\D!}{>}\! 0$}.
F"ur die relevante Summe gilt~-- f"ur~\mbox{\,${\rm Re}\, \nu \!>\! d\!/\!2 \!+\! 1$}:
\end{samepage}%
%
\begin{align} \label{APP:Dnu-1,Dnu-Summe_xi2->0}
&D_{\nu-1}\big((\la_{\nu-1}\!/\!\la_\nu\!\cdot\! \xi)^2,m^2\big)\;
   +\; (m^2 - \iIM\, \ep)\vv D_\nu(\xi^2,m^2)
    \\
  &\hspace*{16pt}
   \underset{\xi^2 \to 0}{\sim}\vv
    D_\nu \Big|_{\D\xi^2 \!\equiv\! 0}\;\cdot\;
    \big(m^2 - \iIM\,\ep\big)\,
    \bigg[
      (-1)\, \big(\nu \!-\! 1\big)\;
         +\; \bigg(\! \nu \!-\! 1 \!-\! \frac{d}{2} \bigg)
    \bigg]\,
    \frac{\Ga\big(\nu \!-\! 1 \!-\! \frac{d}{2}\big)}{\Ga\big(\nu \!-\! \frac{d}{2}\big)}
    \nn
\end{align}
Damit folgt aus Gl.~(\ref{APP:D_Dnu-1,Dnu}):
\begin{samepage}
\vspace*{-2ex}
\begin{alignat}{2} \label{APP:D_Dnu-1,Dnu_xi2->0}
D(\xi^2)\vv
  &\underset{\xi^2 \to 0}{\sim}&\vv
    -\, 6&\iIM\, A_\nu\;\cdot\;
      D_\nu \Big|_{\D\xi^2 \!\equiv\! 0}\;\cdot\;
      \big(m^2 - \iIM\,\ep\big)\,
      \Big(-\frac{d}{2}\Big)\,
      \frac{\Ga\big(\nu \!-\! 1 \!-\! \frac{d}{2}\big)}{\Ga\big(\nu \!-\! \frac{d}{2}\big)}
        \vv\bigg|_\Dstar
    \\
  &&=\vv
    6&\iIM\, A_\nu\cdot
    \efn{\T-\big(\nu-\frac{\ze}{2}\big)\pi\iIM}\cdot
      \frac{(4\pi)^{-d\!/\!2}}{\big(m^2 - \iIM\,\ep\big)^{\nu-1-\frac{d}{2}}}\cdot
      \frac{\frac{d}{2}\, \Ga\big(\nu \!-\! 1 \!-\! \frac{d}{2}\big)}{\Ga(\nu)}
        \vv\bigg|_\Dstar\zzzz
    \tag{\ref{APP:D_Dnu-1,Dnu_xi2->0}$'$}
    \\[-4.5ex]\nn
\end{alignat}
bzgl.\@ der letzten Identit"at vgl.\@ Gl.~(\ref{APP:Dnu_xi2->0}).
Dieser Ausdruck ist entsprechend der erstten Relation in Gl.~(\ref{APP:Ala_Bedingg}) identisch Eins zu setzen; aufgel"ost nach dem Faktor~\mbox{$(-6\iIM\,A_\nu)$}~-- vgl.\@ etwa Gl.~(\ref{APP:D_Dnu-1,Dnu})~--, folgt unmittelbar:
\vspace*{-.5ex}
\begin{align} \label{APP:6iAnu-m2}
-\, 6\iIM\, A_\nu\;
  &=\; -\, \efn{\T\big(\nu-\frac{\ze}{2}\big)\pi\iIM}\cdot
         (4\pi)^{d\!/\!2}\,
         \frac{\Ga(\nu)}{\frac{d}{2}\, \Ga\big(\nu \!-\! 1 \!-\! \frac{d}{2}\big)}\vv
         \big(m^2 - \iIM\,\ep\big)^{(\nu-1)-\frac{d}{2}}
        \vv\bigg|_\Dstar
    \\[-5ex]\nn
\end{align}
Und daraus~-- vgl.\@ Gl.~(\ref{APP:projF^C,NC-0}$''$):
\vspace*{-1.125ex}
\begin{align} \label{APP:deAnu-m2}
\de\!A_\nu\;
   \equiv\; \frac{A_\nu}{A_\nu \big|_{\Dstar,\, \Dstar\Dstar}}\;
   =\; -\, \iIM\, 4\pi\cdot
         \frac{d \!-\! 2}{d}\,
         \frac{\Ga\big(\nu \!-\! \frac{d}{2}\big)}{%
               \Ga\big(\nu \!-\! 1 \!-\! \frac{d}{2}\big)}\vv
         \big(m^2 - \iIM\,\ep\big)^{-1}
        \vv\bigg|_\Dstar
    \\[-5.5ex]\nn
\end{align}
Es gilt
\vspace*{-.75ex}
\begin{align} \label{APP:efn->}
\iIM\, \efn{\T\big(\nu-\frac{\ze}{2}\big)\pi\iIM}\;
   =\; \efn{\T\big(\nu-\frac{\ze-1}{2}\big)\pi\iIM}
    \\[-5ex]\nn
\end{align}
und
\vspace*{-1.5ex}
\begin{align}
&\frac{\Ga(\nu)}{\frac{d}{2}\, \Ga\big(\nu \!-\! 1 \!-\! \frac{d}{2}\big)}\;
  =\; \Ga\Big(\frac{d}{2}\Big)\, \Big(\frac{d}{2} \!+\! 1\Big)\vv
        \pmatrixZE{\nu\!-\!1}{\frac{d}{2}\!+\!1}
    \label{APP:GaFn->} \\[-.5ex]
&\text{mit}\qquad
  \pmatrixZE{\al}{\be}
  \equiv \frac{\al!}{\be!\, (\al-\be)!}
  \equiv \frac{\Ga(\al+1)}{\Ga(\be+1)\, \Ga\big((\al-\be)+1\big)}\qquad
  \al,\be \in \bbbc
    \label{APP:Binomialkoeffizient}
    \\[-4.5ex]\nn
\end{align}
\vspace*{-.125ex}dem Binomialkoeffizienten in Standard-Definition, vgl.\@ Ref.~\cite{Abramowitz84}, Abramowitz~6.1.21.
Eingesetzt in Gl.~(\ref{APP:6iAnu-m2}$'$) sukzessive die Gln.~(\ref{APP:efn->}),~(\ref{APP:GaFn->}), folgt~-- mit~\mbox{$\mbox{\Large$\star$}\!:\, m^2 \!\equiv\! 1$}:
\vspace*{-.5ex}
\begin{align} \label{APP:Anu}
A_\nu\;
  &=\; -\, \efn{\T\big(\nu-\frac{\ze-1}{2}\big)\pi\iIM}\cdot
         (4\pi)^{d\!/\!2}\; \frac{1}{6}\,
         \frac{\Ga(\nu)}{\frac{d}{2}\, \Ga\big(\nu \!-\! 1 \!-\! \frac{d}{2}\big)}
    \\[-.75ex]
  &\begin{aligned}[t]
   =\; -\, \efn{\T\big(\nu-\frac{\ze-1}{2}\big)\pi\iIM}\cdot
         (4\pi)^{d\!/\!2}\; \frac{1}{6}\, 
         \Ga\Big(\frac{d}{2}\Big)\, \Big(\frac{d}{2} \!+\! 1\Big)\;
         \pmatrixZE{\nu\!-\!1}{\frac{d}{2}\!+\!1}&
    \\[-5ex]
  &\qqquad
    {\rm Re}\, \nu > \frac{d}{2} + 1
   \end{aligned}
    \tag{\ref{APP:Anu}$'$}
    \\[-6.5ex]\nn
\end{align}
Und zusammenfassend im gew"ohnlichen Minkowski-Raum mit~\mbox{\,$d \!\equiv\! 4,\, \ze \!\equiv\! 1$}:
\vspace*{-.75ex}
\begin{align} \label{APP:Anu-Minkowski}
&A_\nu\;
  =\; -\, \efn{\T\nu\pi\iIM}\cdot
         \frac{4\pi^2}{3}\,
         \frac{\Ga(\nu)}{\Ga(\nu \!-\! 3)}\;
  =\; -\, \efn{\T\nu\pi\iIM}\cdot 8\pi^2 \pmatrixZE{\nu\!-\!1}{3}
    \\[-1.875ex]
&\de\!A_\nu\;
  = -\, \iIM\, 2\pi\cdot
        (\nu \!-\! 3)
    \tag{\ref{APP:Anu-Minkowski}$'$} \\[-3.5ex]
  &\hspace*{276pt}
    {\rm Re}\, \nu > 3
    \nn
    \\[-8.75ex]\nn
\end{align}
in "Ubereinstimmung mit Gl.~(\ref{A,la-explizit}).
\vspace*{-.5ex}

\subsection[Parameter~\protect\mbox{$\la_\nu$}]{%
            Parameter~\bm{\la_\nu}}
\label{APP-Subsect:la_nu}

Der Parameter~$\la_\nu$ ist bestimmt durch Forderung der zweiten Relation in Gl.~(\ref{APP:Ala_Bedingg}): Normierung auf Eins des Integrals raumartiger Raumzeit-Separationen~$\xi$. \\
\indent
Zur Fixierung des Parameters~$\la_\nu$ ben"otigen wir ein Integral, das wir bezeichnen als vom Typ~III und voranstellen in Form
\vspace*{-.5ex}
\begin{align} \label{APP:TypIII}
&\intIII{\mu}{a}\;
  \equiv\; \int_0^\infty du\; \big(au\big)^\mu\, {\rm K}_\mu\!(au)\quad
  =\; 2^{\mu-1}\, a^{-1}\; \Ga\Big(\frac{1}{2}\Big)\, \Ga\Big(\mu \!+\! \frac{1}{2}\Big)
    \\[-1.25ex]
  &\hspace*{242pt}
    {\rm Re}\, \mu > -\frac{1}{2},\vv
    {\rm Re}\, a > 0
    \nn
    \\[-5.5ex]\nn
\end{align}
\vspace*{-.125ex}vgl.\@ Ref.~\cite{Gradstein81}, Gradstein~6.561.16. \\
\indent
Sei~$D$ dargestellt nach Gl.~(\ref{APP:D_Dnu-1,Dnu}).
In Gl.~(\ref{APP:Dnu-all-la}) ist explizit gegeben~\mbox{$D_\nu(\xi^2,m^2)$}:
\end{samepage}%
\vspace*{-.5ex}
\begin{align} \label{APP:Dnu-all-la_lanu}
&D_\nu\big(\xi^2, m^2\big)\;
  =\; \efn{\T-\big(\nu-\frac{\ze}{2}\big)\pi\iIM}\cdot
        \frac{(4\pi)^{-d\!/\!2}}{\Ga(\nu)}\,
        \frac{1}{\big(m^2 - \iIM\,\ep\big)^\mu}\;
        \Big(\frac{1}{2}\Big)^{\!\mu-1}\! (\zem)^\mu\, {\rm K}_\mu\!(\zem)
    \\[-1.5ex]
  &\hspace*{300pt}
    {\rm Re}\, \nu > 0
    \nn
    \\[-4.5ex]\nn
\end{align}
mit~\mbox{\,$\mu \!\equiv\! \nu \!-\! d\!/\!2$},%
  ~\mbox{\,$\zem \!\equiv\! m\sqrt{-\la \!+\! \iIM\, \ep}$} und%
  ~\mbox{\,$\la \!\equiv\! {\xi'}^2 \equiv \xi^2\!/\!\la_\nu^2$}.
Aus Gl.~(\ref{APP:Dnu-all-la}) folgt analog explizit f"ur%
  ~$D_{\nu-1}$ mit Raumzeit-Argument~\mbox{\,$(\la_{\nu-1}\!/\!\la_\nu\!\cdot\! \xi)^2$}:
\begin{samepage}
\vspace*{-.25ex}
\begin{align} \label{APP:Dnu-1-all-la_lanu}
\hspace*{-9pt} 
 D_{\nu-1}&\big((\la_{\nu-1}\!/\!\la_\nu\!\cdot\! \xi)^2,m^2\big)
    \nn \\
  &=\; \efn{\T-\big((\nu-1)-\frac{\ze}{2}\big)\pi\iIM}\cdot
        \frac{(4\pi)^{-d\!/\!2}}{\Ga(\nu-1)}\,
        \frac{1}{\big(m^2 - \iIM\,\ep\big)^{\mu-1}}\;
        \Big(\frac{1}{2}\Big)^{\!\mu-2}
        (\zem)^{\mu-1}\, {\rm K}_{\mu-1}\!(\zem)
    \\[.25ex]
  &=\; \efn{\T-\big(\nu-\frac{\ze}{2}\big)\pi\iIM}\cdot
        \frac{(4\pi)^{-d\!/\!2}}{\Ga(\nu)}\,
        \frac{1}{\big(m^2 - \iIM\,\ep\big)^\mu}\;
        \Big(\frac{1}{2}\Big)^{\!\mu-1}
    \tag{\ref{APP:Dnu-1-all-la_lanu}$'$} \\[-.25ex]
  &\hspace*{94pt}
    \times\,
        (-1)\cdot \frac{\Ga(\nu)}{\Ga(\nu \!-\! 1)}\cdot
          \big(m^2 - \iIM\,\ep\big)\cdot 2\cdot (\zem)^{\mu-1}\, {\rm K}_{\mu-1}\!(\zem)
    \nn \\[-.75ex]
  &\hspace*{286pt}
    {\rm Re}\, \nu > 1
    \nn
    \\[-4ex]\nn
\end{align}
die Gln.~(\ref{APP:Dnu-1-all-la_lanu}),~(\ref{APP:Dnu-1-all-la_lanu}$'$) ausgedr"uckt durch dieselben Parameter~$\mu$,~$\nu$ wie Gl.~(\ref{APP:Dnu-all-la_lanu}),~insbe\-sondere durch {\it dasselbe Raumzeit-Argument\/}%
  ~\mbox{\,$\zem \!\equiv\! m\sqrt{-\la \!+\! \iIM\, \ep}$},%
  ~\mbox{\,$\la \!\equiv\! \xi^2\!/\!\la_\nu^2 \!\equiv\! (\la_{\nu-1}\!/\!\la_\nu\!\cdot\! \xi)^2\!/\!\la_{\nu-1}^2$}; in Gl.~(\ref{APP:Dnu-1-all-la_lanu}$'$) herausgezogen derselbe Vorfaktor wie in Gl.~(\ref{APP:Dnu-all-la_lanu}). \\
\indent
F"ur die Funktion~$D$ nach Gl.~(\ref{APP:D_Dnu-1,Dnu}) folgt:
%
\begin{align} \label{APP:D_Dnu-1,Dnu-explizit}
\hspace*{-4pt} 
 D&(\xi^2)
    \nn \\[-.5ex]
  &=\; -\, 6\iIM\, A_\nu\cdot
         \big[
           D_{\nu-1}\big((\la_{\nu-1}\!/\!\la_\nu\!\cdot\! \xi)^2,m^2\big)\;
           +\; (m^2 - \iIM\, \ep)\vv D_\nu(\xi^2,m^2)
         \big]
       \vv\Big|_\Dstar
    \\[1ex]
  &=\; -\, 6\iIM\, A_\nu\,\cdot\,
         (-1)\; \efn{\T-\big(\nu-\frac{\ze}{2}\big)\pi\iIM}\cdot
         \frac{(4\pi)^{-d\!/\!2}}{\Ga(\nu)}\,
         \frac{1}{\big(m^2 - \iIM\,\ep\big)^{\mu-1}}\;
         \Big(\frac{1}{2}\Big)^{\!\mu-1}
    \tag{\ref{APP:D_Dnu-1,Dnu-explizit}$'$} \\[-.5ex]
  &\hspace*{95pt}
    \times\,
    \big[
        2 (\nu \!-\! 1)\cdot
          (\zem)^{\mu-1}\, {\rm K}_{\mu-1}\!(\zem)\;
      -\; (\zem)^\mu\, {\rm K}_\mu\!(\zem)
    \big]
       \vv\Big|_\Dstar
    \nn \\[1ex]
  &=\; \frac{1}{\frac{d}{2}\, \Ga\big(\nu \!-\! 1 \!-\! \frac{d}{2}\big)}
         \Big(\frac{1}{2}\Big)^{\!\mu-1}\!
         \big[
           2 (\nu \!-\! 1)\cdot
             (\zem)^{\mu-1}\, {\rm K}_{\mu-1}\!(\zem)\;
         -\; (\zem)^\mu\, {\rm K}_\mu\!(\zem)
         \big]
       \vv\Big|_\Dstar
    \tag{\ref{APP:D_Dnu-1,Dnu-explizit}$''$}
    \\[-4ex]\nn
\end{align}
die letzte Identit"at eingesetzt~\mbox{$(-6\iIM\,A_\nu)$} nach Gl.~(\ref{APP:6iAnu-m2}). \\
\indent
Dieser Ausdruck ist zu integrieren entsprechend der zweiten Relation in Gl.~(\ref{APP:Ala_Bedingg}), aus der wir unmittelbar ablesen bez"uglich der Integrationsvariablen~$u$:
%
\begin{align} \label{APP:Int'var-u}
&-\, u^2\; =\; \xi^2\qqquad
  \text{mit}\qquad
  u\; >\; 0
    \\
&\Longleftrightarrow\qquad
  u^2 \big/ \la_\nu^2\;
  =\; -\, \xi^2 \big/ \la_\nu^2\;
  \equiv\; -\la\;
  >\; 0
    \tag{\ref{APP:Int'var-u}$'$}
\end{align}
Mit~\mbox{\,$\zem = m\, \sqrt{-\la \!+\! \iIM\, \ep} \cong \sqrt{-\la}\, \sqrt{m^2 + \iIM\, \ep}$}, vgl.~\mbox{\,$-\la,\, m^2 \!\in\! \bbbr^+$}, folgt:
%
\begin{align} \label{APP:zem=a-mal-u}
\zem\;
  =\; a \!\cdot\! u\qqquad
  \text{mit\rmfootnote}\qquad
  a\;
    \equiv\; \frac{1}{\la_\nu}\big(m^2 + \iIM\, \ep\big)^{\!1\!/\!2}\;
    \cong\; \frac{m}{\la_\nu}
\end{align}%
\footnotetext{
  Die Epsilon-Vorschrift~\mbox{$\ep \!\to\! 0\!+$} ist redundant f"ur raumartiges~$\xi$~-- vgl.\@ die Bem.\@ zu Gl.~(\ref{APP:Dnu-la<0})~-- und kann \oE~weggelassen werden.
}%
%
Die relevanten Integrale werden identifiziert als von Typ~III:
\vspace*{-.25ex}
\begin{align} \label{APP:int-D_Dnu-1,Dnu-explizit}
\hspace*{-4pt}1\;
  &\stackrel{\D!}{=}\;
     \frac{2^{-(\mu-1)}}{%
           \frac{d}{2}\, \Ga\big(\nu \!-\! 1 \!-\! \frac{d}{2}\big)}\cdot
       \int_0^{\infty} \! du\;
         \big[
           2 (\nu \!-\! 1)\cdot
             (au)^{\mu-1}\, {\rm K}_{\mu-1}\!(au)\;
         -\; (au)^\mu\, {\rm K}_\mu\!(au)
         \big]
       \vv\Big|_\Dstar
    \\[.5ex]
  &=\; \frac{2^{-(\mu-1)}}{%
             \frac{d}{2}\, \Ga\big(\nu \!-\! 1 \!-\! \frac{d}{2}\big)}\cdot
         \big[
           2 (\nu \!-\! 1)\cdot \intIII{\mu-1}{a}\;
         -\;  \intIII{\mu}{a}
         \big]
       \vv\Big|_\Dstar
    \tag{\ref{APP:int-D_Dnu-1,Dnu-explizit}$'$} \\[.5ex]
  &=\; \frac{2^{-(\mu-1)}}{%
             \frac{d}{2}\, \Ga\big(\nu \!-\! 1 \!-\! \frac{d}{2}\big)}\cdot
         2^{\mu-1}\, a^{-1}\; \Ga\Big(\frac{1}{2}\Big)\,
           \Ga\Big(\mu \!-\! 1 \!+\! \frac{1}{2}\Big)
    \tag{\ref{APP:int-D_Dnu-1,Dnu-explizit}$''$} \\[-1.75ex]
  &\hspace*{162pt}
       \times\,
         \Big[
           2 (\nu \!-\! 1)\cdot 2^{-1}\;
         -\; \Big((\mu \!-\! 1) \!+\! \frac{1}{2}\Big)
         \Big]
       \vv\Big|_\Dstar
    \nn
    \\[-4.5ex]\nn
\end{align}
\end{samepage}%
bzgl.~\mbox{\,$\intIII{\mu-1}{a}, \intIII{\mu}{a}$} vgl.\@ Gl.~(\ref{APP:TypIII}).

F"ur Existenz der Integale~\mbox{\,$\intIII{\mu-1}{a}, \intIII{\mu}{a}$} ist nach Gl.~(\ref{APP:TypIII}) zu fordern%
  ~\mbox{\,${\rm Re}(\mu \!-\! 1) \!>\! -1\!/\!2$}, mit%
  ~\mbox{\,$\mu \!\equiv\! \nu \!-\! d\!/\!2$}, ergo%
  ~\mbox{${\rm Re}\, \nu \!>\! (d \!+\! 1)\!/\!2$}; f"ur Existenz der Gamma-Funktion im Nenner strenger%
  ~\mbox{${\rm Re}\, \nu \!>\! d\!/\!2 \!+\! 1$}. \\
\indent
Mit~\mbox{\,$\mu \!\equiv\! \nu \!-\! d\!/\!2$} wird ferner Gl.~(\ref{APP:int-D_Dnu-1,Dnu-explizit}$''$) ausgedr"uckt durch~$\nu$, die eckige Klammer vereinfacht sich zu~\mbox{\,$(d \!-\! 1) \!/\! 2$}; es folgt die $\la_\nu$ determinierende Forderung in Form:
\begin{samepage}
\vspace*{-.25ex}
\begin{align} 
&1\;
  \stackrel{\D!}{=}\;
    \la_\nu\;\cdot\;
      \big(m^2 + \iIM\, \ep\big)^{\!-1\!/\!2}\cdot
      {\T\Ga\Big(\frac{1}{2}\Big)}\,
      \frac{\frac{d-1}{2}\, \Ga\big(\nu \!-\! 1 \!-\! \frac{d-1}{2}\big)}{%
            \frac{d}{2}\; \Ga\big(\nu \!-\! 1 \!-\! \frac{d}{2}\big)}
    \vv\Bigg|_\Dstar
    \\[-4.5ex]\nn
\end{align}
Es folgt:
\vspace*{-.25ex}
\begin{align} \label{APP:lanu}
\la_\nu\;
  =\; \frac{1}{\Ga\big(\frac{1}{2}\big)}\,
      \frac{\frac{d}{2}\; \Ga\big(\nu \!-\! 1 \!-\! \frac{d}{2}\big)}{%
            \frac{d-1}{2}\, \Ga\big(\nu \!-\! 1 \!-\! \frac{d-1}{2}\big)}\cdot
      \big(m^2 + \iIM\, \ep\big)^{\!1\!/\!2}
    \vv\Bigg|_\Dstar\qqquad
    {\rm Re}\, \nu > \frac{d}{2} + 1
    \\[-4ex]\nn
\end{align}
und daraus im gew"ohnlichen Minkowski-Raum mit~\mbox{\,$d \!\equiv\! 4,\, \ze \!\equiv\! 1$}:
%
\begin{align} \label{APP:lanu-Minkowski}
\la_\nu\;
  =\; \frac{4}{3}\,
      \frac{\Ga(\nu \!-\! 3)}{%
            \Ga\big(\frac{1}{2}\big)\,
              \Ga\big((\nu \!-\! 3) \!+\! \frac{1}{2}\big)}\qqquad
    {\rm Re}\, \nu > 3
\end{align}
in "Ubereinstimmung mit Gl.~(\ref{A,la-explizit}$'$)~-- mit%
  ~\mbox{\,$\Ga(\frac{1}{2}) \!\equiv\! \pi^{1\!/\!2}$}, vgl.\@ Ref.~\cite{Abramowitz84}, Abramowitz~6.1.8.
F"ur weitere Umformung geben wir an:
%
\begin{align} 
\Ga\Big(z \!+\! \frac{1}{2}\Big)\;
  =\; 2^{1-2z}\, {\T\Ga\big(\frac{1}{2}\big)}\;
        \frac{\Ga(2z)}{\Ga(z)}\qquad
 \Ga\Big(n \!+\!\frac{1}{2}\Big)\;
  =\; 2^{-n}\, {\T\Ga\big(\frac{1}{2}\big)}\;
        (2n \!-\! 1)!!
\end{align}
f"ur~\mbox{\,$z \!\in\! \bbbc, n \!\in\! \bbbn$}.
Vgl.\@ Ref.~\cite{Abramowitz84}, Abramowitz~6.1.18 und ebenda, die Fu"sn.\@ zu~6.1.49 bzgl.\@ der Doppel-Fakult"at%
  ~\mbox{\,$(2n \!-\! 1)!! \equiv
    {\T\prod}_{n'=1}^n\, (2n' \!-\! 1) \equiv
    1 \!\cdot\! 3 \!\cdot\! 5\cdots (2n \!-\! 1)$}.
\vspace*{-.5ex}

\section[Conclusio:~\protect$C$- und \protect$N\!C$-Korrelationsfunktionen~--
            explizite Darstellung]{%
         \hspace*{-0pt}Conclusio: \bm{C}- und \bm{N\!C}-Korrelationsfunktionen~\bm{-}
            explizite Darstellung}

Der~"`$\nu$-verallgemeinerte"' Feynman-Propagator~\mbox{$D_\nu(\xi^2,m^2)$} in~\mbox{\,${\cal M}_{d,\si}$} ist explizit dargestellt, die Parameter~$A_\la$,~$\la_\nu$ fixiert.
Auf dieser Basis geben wir an explizite Darstellungen der~$C$- und $N\!C$-Korrelationsfunktionen in Termen der Funktionen
\vspace*{-.25ex}
\begin{align} \label{APP:calKmu-Def}
{\cal K}_{\mu'}\!(z)\;
  \equiv\; \Big(\frac{1}{2}\Big)^{\!\mu'-1}\!
             \frac{1}{\Ga(\mu')}\;
             z^{\mu'}\, {\rm K}_{\mu'}\!(z)\vv
  \underset{z \to 0}{\sim}\vv 1\qqquad
  {\rm Re}\, \mu' > 0,\vv
  z \in \bbbc
    \\[-4.5ex]\nn
\end{align}
die {\it per~constructionem\/} normiert sind auf Eins f"ur verschwindendes Argument, vgl.\@ Gl.\,(\ref{APP:z-BesselK-Asymptotik}).
Wir rekapitulieren als die relevanten Gr"o"sen:
%
\begin{align} \label{APP:mu,zem,ze,la}
\mu \equiv \nu - d \!/\! 2,\qquad
    \zem \equiv m\, \ze,\qquad
    \ze  \equiv \sqrt{-\la + \iIM\, \ep},\qquad
    \la  \equiv \xi^2 \!/\! \la_\nu^2
    \\[-4ex]\nn
\end{align}
bzgl.~$\la_\nu$ vgl.\@ Gl.~(\ref{APP:lanu}).

%
\bigskip\noindent
F"ur den kontrahierten Korrelations-Lorentztensor~\vspace*{-.125ex}\mbox{\,$D(\xi^2) \!\equiv\! D_{\mu\nu}{}^{\mu\nu}(\xi^2)$} und die konfinierende Korrelationsfunktion~$D\uC(\xi^2)$ gilt~-- vgl.\@ Gl.~(\ref{APP:D_Dnu-1,Dnu-explizit}$''$) bzw.\@ Gl.~(\ref{APP:DDk-Ansatz}):
\vspace*{-.25ex}
\begin{align} \label{APP:D,D-C_calK}
\hspace*{-4pt}
D(\xi^2)\;
  \equiv\; D\uC(\xi^2)\;
  &=\; \frac{\Ga(\mu \!-\! 1)}{%
             \frac{d}{2}\, \Ga\big(\nu \!-\! 1 \!-\! \frac{d}{2}\big)}
         \big[
           (\nu \!-\! 1)\cdot {\cal K}_{\mu-1}\!(\zem)\;
         -\; (\mu \!-\! 1)\cdot {\cal K}_\mu\!(\zem)
         \big]
       \vv\bigg|_\Dstar
    \\
  &=\; \frac{2}{d}\, (\nu \!-\! 1)\cdot
         {\cal K}_{\nu-1-\frac{d}{2}}\!(\zem)\;
       -\; \frac{2}{d} \Big(\nu \!-\! 1 \!-\! \frac{d}{2}\Big)\cdot
         {\cal K}_{\nu-\frac{d}{2}}\!(\zem)
       \vv\bigg|_\Dstar
    \tag{\ref{APP:D,D-C_calK}$'$}
    \\[-4.5ex]\nn
\end{align}
Wir verifizieren unmittelbar Normierung auf Eins f"ur verschwindendes~$\xi^2$. \\
\indent
F"ur die Funktion~\mbox{\,$6\iIM\, A_\nu\!\cdot\! D_\nu$} gilt~-- vgl.\@ die Gln.~(\ref{APP:Dnu-all-la}),~(\ref{APP:6iAnu-m2}):
\end{samepage}%
\vspace*{-.25ex}
\begin{align} \label{APP:6iAnu-mal-Dnu-0}
6\iIM\, A_\nu\cdot D_\nu\big(\xi^2, m^2\big)\;
  =\; \frac{1}{\frac{d}{2}\, \Ga(\mu \!-\! 1)}\vv
        \Big(\frac{1}{2}\Big)^{\!\mu-1}\! (\zem)^\mu\, {\rm K}_\mu\!(\zem)\;\cdot\;
        \big(m^2 - \iIM\,\ep\big)^{-1}
    \\[-4.5ex]\nn
\end{align}
ergo:
\vspace*{-1ex}
\begin{align} \label{APP:6iAnu-mal-Dnu}
6\iIM\, A_\nu\cdot
    D_\nu\big(\xi^2, m^2\big)
  =\; \frac{2}{d} \Big(\nu \!-\! 1 \!-\! \frac{d}{2}\Big)\cdot
          {\cal K}_{\nu-\frac{d}{2}}\!(\zem)\;\cdot\;
        \big(m^2 - \iIM\,\ep\big)^{-1}
    \\[-4ex]\nn
\end{align}
Diese Formel f"ur Index~\mbox{\,$\nu \!-\! 1$} f"uhrt auf~\mbox{\,$6\iIM\, A_{\nu-1}\!\cdot\! D_{\nu-1}$}, wohingegen in der Formel f"ur~$D$ als Summe von~$D_{\nu-1}$ und~$D_\nu$~-- vgl.\@ Gl.~(\ref{APP:D_Dnu-1,Dnu})~-- eingeht als Summand~\mbox{\,$6\iIM\, A_\nu\!\cdot\! D_{\nu-1}$}. \\
\indent
F"ur~$F\oC$,~$F\oNC$ gilt~-- vgl.\@ Gl.~(\ref{APP:F^C,F^NC_Dnu}) und~(\ref{APP:6iAnu-mal-Dnu}):
\begin{samepage}
\vspace*{-.25ex}
\begin{align} \label{APP:F^C,F^NC}
F\oC(\xi^2)\;
  \equiv\; F\oNC(\xi^2)\;
  =\; \la_\nu^2\cdot
        \frac{2}{d} \Big(\nu \!-\! 1 \!-\! \frac{d}{2}\Big)\cdot
        {\cal K}_{\nu-\frac{d}{2}}\!(\zem)\;\cdot\;
        \big(m^2 - \iIM\,\ep\big)^{-1}
        \vv\bigg|_\Dstar
    \\[-4ex]\nn
\end{align}
Daraus f"ur die Projektionen%
  ~\vspace*{-.25ex}\mbox{$\projNAt{F\oC}$},~\mbox{$\projNAt{F\oNC}$}~-- vgl.\@ Fu"sn.\,\FN{APP-FN:deAnu} und die Gln.~(\ref{APP:projF^C,NC}),~(\ref{APP:deAnu-m2}):
\vspace*{-.25ex}
\begin{align}
&\projt{F\oC}{\bm\xi}\;
  \equiv\; \projt{F\oNC}{\bm\xi}
    \nn \\[-.5ex]
  &\phantom{\projNAt{F\oC}}
   =\; \de\!A_\nu\; \la_\nu^4\;\cdot\,
        \frac{2}{d \!-\! 2} \Big(\nu \!-\! 1 \!-\! \frac{d \!-\! 2}{2}\Big)\cdot
        {\cal K}_{\nu-\frac{d-2}{2}}\!(\bm\zem)\;\cdot\;
        \big(m^2 - \iIM\,\ep\big)^{-1}
        \vv\bigg|_\Dstar
    \label{APP:projF^C,NC-explizit} \\[-.75ex]
  &\phantom{\projNAt{F\oC}}
   =\, -\, \iIM\, 2\pi\cdot\!
         \Big(\frac{2\la_\nu^2}{m^2 \!-\! \iIM\,\ep}\Big)^{\zz2}\;
         \frac{1}{d}\,
         \frac{\Ga\big(\nu \!+\! 1 \!-\! \frac{d}{2}\big)}{%
               \Ga\big(\nu \!-\! 1 \!-\! \frac{d}{2}\big)}\cdot
         {\cal K}_{\nu+1-\frac{d}{2}}\!(\bm\zem)
        \vv\bigg|_\Dstar
    \tag{\ref{APP:projF^C,NC-explizit}$'$} \\[.5ex]
  &\text{mit}\qquad\hspace*{9pt}
    \bm\zem \equiv m\, \bm\ze,\qquad
    \bm\ze  \equiv \sqrt{-\bm\la + \iIM\, \ep},\qquad
    \bm\la  \equiv \bm\xi^2 \!/\! \la_\nu^2
     \label{APP:bm-mu,zem,ze,la}
    \\[-4ex]\nn
\end{align}
In (geradem) Fettdruck sind notiert die Gr"o"sen nach Gl.~(\ref{APP:mu,zem,ze,la}), in denen die zwei ausin\-tegrierten~-- "`longitudinalen"'~[im gew"ohnlichen Minkowski-Raum die~$0$- und $3$-]~-- Komponen\-ten gestrichen sind.
Bez"uglich~$|\bm\xi|$~-- vgl.~$|\rb{x}|$ im Haupttext~-- ist daher zu identifizieren:
\vspace*{-.625ex}
\begin{align} \label{APP:bm-mu,zem,ze,la-|rbx|}
&|\bm\xi|\;
    \equiv\; \sqrt{-\bm\xi^2}\qquad
  \Longleftrightarrow\qquad
  -|\bm\xi|^2\;
    \equiv\; \bm\xi^2
    \\
&\Longleftrightarrow\qquad
  \bm\zem \equiv m\, \bm\ze,\qquad
  \bm\ze  \equiv \sqrt{-\bm\la + \iIM\, \ep},\qquad
  \bm\la  \equiv \bm\xi^2 \!/\! \la_\nu^2
          \equiv -|\bm\xi|^2 \!/\! \la_\nu^2
    \tag{\ref{APP:bm-mu,zem,ze,la-|rbx|}$''$}
    \\[-4.25ex]\nn
\end{align}
Die streu-relevanten Projektionen%
  ~\mbox{$\projt[(n)]{F\oC}{\bm\xi}$} involvieren~\mbox{$n \!=\! 1,2$} Ableitungen%
 ~\vspace*{-.75ex}\mbox{$-\frac{1}{|\bm\xi|}\frac{d}{d|\bm\xi|}$}, vgl.\@ Gl.~(\ref{projt(n)-Def}), f"ur die ergo gilt:
%
\begin{align} \label{APP:rb-x,bm-xi-Diff}
\hspace*{-0pt}
-\frac{1}{|\bm\xi|}\frac{d}{d|\bm\xi|}\;
   =\; 2\, \frac{d}{d\bm\xi^2}\;
   =\; 2\, \frac{d\bm\la}{d\bm\xi^2}\cdot
             \frac{d\bm\zem}{d\bm\la}\cdot
             \frac{d}{d\bm\zem}\;
   =\; \frac{m^2}{\la_\nu^2}\,
         \bigg(\zz -\! \frac{1}{\bm\zem} \frac{d}{d\bm\zem}\bigg)
\end{align}
die letzte Identit"at mit%
  ~\vspace*{-.25ex}\mbox{\,$d\bm\la/d\bm\xi^2 \!=\! \la_\nu^{-2}$} und%
  ~\mbox{\,$d\bm\zem/d\bm\la \!=\! -m^2/2\bm\zem$}, vgl.\@ Gl.~(\ref{APP:bm-mu,zem,ze,la-|rbx|}$''$).
F"ur die Funktionen~\mbox{\,${\cal K}_\mu\!(z)$} nach Gl.~(\ref{APP:calKmu-Def}) gilt ferner die Derivationsformel
%
\begin{align} \label{APP:calK-Derivation}
&\Big(\zz -\! \frac{1}{z} \frac{d}{dz}\Big)^{\zz n}\;
    {\cal K}_\mu\!(z)\;
   =\; \Big(\frac{1}{2}\Big)^{\zz\mu-1}\,
             \frac{1}{\Ga(\mu)}\;
         \Big(\zz -\! \frac{1}{z} \frac{d}{dz}\Big)^{\zz n}\;
         \big[z^\mu\, {\rm K}_\mu\!(z)\big]
    \\
  &\phantom{\Big(\zz -\! \frac{1}{z} \frac{d}{dz}\Big)^{\zz n}\; {\cal K}_\mu\!(z)\;}
   =\; \Big(\frac{1}{2}\Big)^{\zz n}\,
             \frac{\Ga(\mu \!-\! n)}{\Ga(\mu)}\cdot
         \Big(\frac{1}{2}\Big)^{\zz(\mu-n)-1}\,
         \frac{1}{\Ga(\mu \!-\! n)}\; z^{\mu-n}\, {\rm K}_{\mu-n}\!(z)
    \tag{\ref{APP:calK-Derivation}$'$} \\
  &\hspace*{10pt}
   =\; \Big(\frac{1}{2}\Big)^{\zz n}
             \frac{\Ga(\mu \!-\! n)}{\Ga(\mu)}\cdot
         {\cal K}_{\mu-n}\!(z)\qqquad
  \forall{\rm Re}\, \mu > n \in \bbbn_0
    \tag{\ref{APP:calK-Derivation}$''$}
    \\[-4ex]\nn
\end{align}
bzgl.\@ der zweiten Identit"at vgl.\@ Gl.~(\ref{APP:BesselK-Derivation}): die Derivationsformel f"ur die Funktionen~\mbox{\,${\rm K}_\mu\!(z)$}. \\
\indent
Es folgt unmittelbar~-- vgl.\@ die Gln.~(\ref{APP:projF^C,NC-explizit}$'$),~(\ref{APP:rb-x,bm-xi-Diff}) und~(\ref{APP:calK-Derivation}$''$):
\end{samepage}%
\vspace*{-.25ex}
\begin{align} \label{APP:projF^NC_rb-x,bm-xi-Diff}
&\hspace*{-12pt}
 \projt[(n)]{F\oC}{\bm\xi}\;
  \equiv\; \projt[(n)]{F\oNC}{\bm\xi}\;
  =\; \bigg(\zz -\! \frac{1}{|\bm\xi|} \frac{d}{d|\bm\xi|}\bigg)\Big.^{\zz\T n}\vv
             \projt{F\oNC}{\bm\xi}
    \\
  &\hspace*{-8pt}
   =\, -\, \iIM\, 2\pi\cdot\!
         \Big(\frac{2\la_\nu^2}{m^2 \!-\! \iIM\,\ep}\Big)^{\zz2}\,
         \frac{1}{d}
         \frac{\Ga\big(\nu \!+\! 1 \!-\! \frac{d}{2}\big)}{%
               \Ga\big(\nu \!-\! 1 \!-\! \frac{d}{2}\big)}\cdot\!
         \Big(\frac{m^2}{\la_\nu^2}\Big)^{\zz n}\!\cdot\!
         \bigg(\zz -\! \frac{1}{\bm\zem} \frac{d}{d\bm\zem}\bigg)\Big.^{\zz\T n}\vv
         {\cal K}_{\nu + 1 - \frac{d}{2}}\!(\bm\zem)
       \;\bigg|_\Dstar
    \tag{\ref{APP:projF^NC_rb-x,bm-xi-Diff}$'$} \\
  &\hspace*{-8pt}
   =\, -\, \iIM\, 2\pi\cdot\!
         \Big(\frac{2\la_\nu^2}{m^2 \!-\! \iIM\,\ep}\Big)^{\zz2}\,
         \frac{1}{d}
         \frac{\Ga\big(\nu \!+\! 1 \!-\! \frac{d}{2}\big)}{%
               \Ga\big(\nu \!-\! 1 \!-\! \frac{d}{2}\big)}\cdot\!
         \Big(\frac{m^2}{\la_\nu^2}\Big)^{\zz n}\!\cdot\!
         \Big(\frac{1}{2}\Big)^{\zz n}
         \frac{\Ga\big(\nu \!-\! (n \!-\! 1) \!-\! \frac{d}{2}\big)}{%
               \Ga\big(\nu          \!+\! 1  \!-\! \frac{d}{2}\big)}\!\cdot
          {\cal K}_{\nu - (n - 1) - \frac{d}{2}}\!(\bm\zem)
       \;\bigg|_\Dstar
    \tag{\ref{APP:projF^NC_rb-x,bm-xi-Diff}$''$} \\[-1.5ex]
  &\hspace*{-8pt}
   \cong\, -\, \iIM\, 2\pi\cdot\!
         \Big(\frac{2\la_\nu^2}{m^2 \!-\! \iIM\,\ep}\Big)^{\zz2-n}\;
         \frac{1}{d}\,
         \frac{\Ga\big(\nu \!-\! (n \!-\! 1) \!-\! \frac{d}{2}\big)}{%
               \Ga\big(\nu          \!-\! 1  \!-\! \frac{d}{2}\big)}\cdot
         {\cal K}_{\nu - (n \!-\! 1) - \frac{d}{2}}\!(\bm\zem)
       \vv\bigg|_\Dstar
    \tag{\ref{APP:projF^NC_rb-x,bm-xi-Diff}$'''$}
    \\[-4.75ex]\nn
\end{align}
Mit~\mbox{$n \!\equiv\! 0$} verifizieren wir Gl.~(\ref{APP:projF^C,NC-explizit}$'$), mit~\mbox{$n \!\equiv\! 1,2$} folgt
\begin{samepage}
\vspace*{-.25ex}
\begin{align}
&\projt[(1)]{F\oC}{\bm\xi}\;
  =\; -\, \iIM\, 2\pi\cdot
        \la_\nu^2\vv
        \frac{2}{d} \Big(\nu \!-\! 1 \!-\! \frac{d}{2}\Big)\cdot
        {\cal K}_{\nu-\frac{d}{2}}\!(\bm\zem)\;\cdot\;
        \big(m^2 - \iIM\,\ep\big)^{-1}
      \vv\bigg|_\Dstar
    \label{APP:projt(1)} \\[-.75ex]
&\projt[(2)]{F\oC}{\bm\xi}\;
  =\; -\, \iIM\, 2\pi
        \phantom{\la_\nu^2\vv}\cdot
        \frac{1}{d}\cdot
        {\cal K}_{\nu - 1 - \frac{d}{2}}\!(\bm\zem)
      \vv\bigg|_\Dstar
    \label{APP:projt(2)}
    \\[-4.75ex]\nn
\end{align}
f"ur die streu-relevanten konfinierenden Projektionen; vgl.\@ Gl.\,(\ref{APP:6iAnu-mal-Dnu}) bzw.\@ Gl.\,(\ref{APP:D-NC}) unten. \\
\indent
F"ur Vollst"andigkeit geben wir an die nicht-konfinierende Korrelationsfunktion~$D\uNC(\xi^2)$.
Analog zu Gl.~(\ref{APP:rb-x,bm-xi-Diff}) f"ur~$\bm\xi$,~$\bm\zem$ gilt f"ur~$\xi$,~$\zem$:
\vspace*{-.25ex}
\begin{align} \label{APP:xi-zem-Diff}
\hspace*{-0pt}
2\, \frac{d}{d\xi^2}\;
   =\; \frac{m^2}{\la_\nu^2}
         \bigg(\zz -\! \frac{1}{\bm\zem} \frac{d}{d\bm\zem}\bigg)
    \\[-4ex]\nn
\end{align}
so da"s unmittelbar folgt aus Gl.~(\ref{APP:F,D-NC-ddxi2}):
\vspace*{-.25ex}
\begin{align} \label{APP:D-NC-0}
D\uNC&(\xi^2)\;
  =\; 8\; \frac{d}{d\xi^2}\vv F\oNC(\xi^2)
    \\[-.25ex]
  &=\; 4\; \la_\nu^2\;
         \frac{2}{d}\, \Big(\nu \!-\! 1 \!-\! \frac{d}{2}\Big)\cdot
         \frac{m^2}{\la_\nu^2}
         \Big(\zz -\! \frac{1}{\zem} \frac{d}{d\zem}\Big)\vv
         {\cal K}_{\nu-\frac{d}{2}}\!(\zem)\;\cdot\;
         \big(m^2 - \iIM\,\ep\big)^{-1}
       \vv\bigg|_\Dstar
    \tag{\ref{APP:D-NC-0}$'$} \\[-.25ex]
  &=\; 4\; \la_\nu^2\;
         \frac{2}{d}\, \Big(\nu \!-\! 1 \!-\! \frac{d}{2}\Big)\cdot
         \frac{1}{\la_\nu^2}\cdot
         \frac{1}{2}\,
           \frac{\Ga\big(\nu-1-\frac{d}{2}\big)}{%
                 \Ga\big(\nu-\frac{d}{2}\big)}\cdot
         {\cal K}_{\nu-1-\frac{d}{2}}\!(\zem)
       \vv\bigg|_\Dstar
    \tag{\ref{APP:D-NC-0}$''$}
    \\[-4ex]\nn
\end{align}
die zweite Identit"at mit Gl.~(\ref{APP:F^C,F^NC}), die dritte mit Gl.~(\ref{APP:calK-Derivation}$''$) unter K"urzung der Faktoren~\mbox{\,$m^2$} und~\mbox{\,$(m^2 \!-\! \iIM\,\ep)^{-1}$}.
Wir haben daher abschlie"send:
\vspace*{-.5ex}
\begin{align} \label{APP:D-NC}
D\uNC(\xi^2)\;
  =\; \frac{4}{d}\cdot
         {\cal K}_{\nu-1-\frac{d}{2}}\!(\zem)
       \vv\bigg|_\Dstar
    \\[-4.25ex]\nn
\end{align}
f"ur die nicht-konfinierende $D$-Korrelationsfunktion.
\vspace*{-.5ex}

\bigskip\noindent
Der gew"ohnliche Minkowski-Raum~\mbox{${\cal M} \!\equiv\! {\cal M}_{d,2\ze-d}$} ist charakterisiert durch%
  ~\vspace*{-.125ex}\mbox{$d \!\equiv\! 4,\, \ze \!\equiv\! 1$}.
Wir geben an die ${\cal M}$-bez"uglichen $F$-Korrelationsfunktionen~-- vgl.\@ die Gln.~(\ref{APP:F^C,F^NC}),~(\ref{APP:projF^C,NC-explizit}$'$), (\ref{APP:projt(1)}) und~(\ref{APP:projt(2)}):
\vspace*{-.75ex}
\begin{alignat}{3} \label{APP:F^C,NC-Minkowski}
&\hspace*{-4pt}
 F\oC(\xi^2)\;&
  &\equiv\; F\oNC(\xi^2)\;&
  &=\, \phantom{-\, \iIM\, 2\pi\cdot\;\,}
         \la_\nu^2\vv
         \frac{1}{2}\, (\nu \!-\! 3)\cdot
         {\cal K}_{\nu-2}(\ze)
    \\[-.25ex]
&\hspace*{-4pt}
 \projt{F\oC}{\bm\xi}\;&
  &\equiv\; \projt{F\oNC}{\bm\xi}\;&
  &=\, -\, \iIM\, 2\pi\cdot
         \la_\nu^4\vv
         \phantom{\frac{1}{2}\,}
         (\nu \!-\! 3)(\nu \!-\! 2)\cdot
         {\cal K}_{\nu-1}(\bm\ze)\zz
    \tag{\ref{APP:F^C,NC-Minkowski}$'$} \\[-.25ex]
&\hspace*{-4pt}
 \projt[(1)]{F\oC}{\bm\xi}\;&
  &\equiv\; \projt[(1)]{F\oNC}{\bm\xi}\;&
  &=\, -\, \iIM\, 2\pi\cdot
         \la_\nu^2\vv
         \frac{1}{2}\, (\nu \!-\! 3)\cdot
         {\cal K}_{\nu-2}(\bm\ze)
    \tag{\ref{APP:F^C,NC-Minkowski}$''$} \\[-.25ex]
&\hspace*{-4pt}
 \projt[(2)]{F\oC}{\bm\xi}\;&
  &\equiv\; \projt[(2)]{F\oNC}{\bm\xi}\;&
  &=\, -\, \iIM\, 2\pi
         \phantom{\la_\nu^2\vv}\cdot
         \frac{1}{4}\cdot
         {\cal K}_{\nu-3}(\bm\ze)
    \tag{\ref{APP:F^C,NC-Minkowski}$'''$}
    \\[-4.5ex]\nn
\end{alignat}
eingesetzt~\mbox{$\mbox{\Large$\star$}\!:\, m^2 \!\equiv\! 1$}; bzgl.~$\la_\nu$ vgl.\@ Gl.~(\ref{APP:lanu-Minkowski}), bzgl.\@
\vspace*{-.25ex}
\begin{align} \label{APP:F^C,NC-Minkowski-mit}
&\ze  \equiv \sqrt{-\la + \iIM\, \ep},\qquad
     \la  \equiv \xi^2 \!/\! \la_4^2
    \\[.25ex]
  &\bm\ze  \equiv \sqrt{-\bm\la + \iIM\, \ep},\qquad
     \bm\la  \equiv \bm\xi^2 \!/\! \la_\nu^2 = -|\bm\xi|^2 \!/\! \la_\nu^2,\qquad
     \text{mit}\quad
     -|\bm\xi|^2 \equiv \bm\xi^2,\quad
     \bm\xi \equiv \xi\big|_{\Dstar\Dstar}
    \tag{\ref{APP:F^C,NC-Minkowski-mit}$'$}
    \\[-4.25ex]\nn
\end{align}
Gl.~(\ref{APP:mu,zem,ze,la}) bzw.\@ die Gln.~(\ref{APP:bm-mu,zem,ze,la-|rbx|}),~(\ref{APP:bm-mu,zem,ze,la-|rbx|}$'$). \\
\indent
F"ur Vollst"andigkeit geben wir ferner an bez"uglich ${\cal M}$ den kontrahierten Korrelations-Lorentztensor und die $D$-Korrelationsfunktionen:
\vspace*{-.5ex}
\begin{alignat}{2} \label{APP:D,D-C,D-NC-Minkowski}
D(\xi^2)\;
  \equiv\; &D\uC(\xi^2)\vv&
  &=\; \frac{1}{2}\, (\nu \!-\! 1)\cdot
          {\cal K}_{\nu-3}(\ze)\;
       -\; \frac{1}{2}\, (\nu \!-\! 3)\cdot
          {\cal K}_{\nu-2}(\ze)
    \\[-.25ex]
&D\uNC(\xi^2)\;&
  &=\; \phantom{\frac{1}{2}\,}
         {\cal K}_{\nu-3}(\ze)
    \tag{\ref{APP:D,D-C,D-NC-Minkowski}$'$}
    \\[-4.25ex]\nn
\end{alignat}
vgl.\@ Gl.~(\ref{APP:D,D-C_calK}$'$) bzw.~(\ref{APP:D-NC}). \\
\indent
Physikalisch suggeriert ist~\mbox{\,$\nu \!\equiv\! 4$}.
Im Haupttext wird f"ur die Invariante~\vspace*{-.25ex}\mbox{$\xi^2\!\big/\!\la_4^2$} {\it nicht\/} benutzt die Notation~$\la$, aber am Parameter%
  ~\mbox{\,$\la_{\nu\equiv4} \!=\! 8\!\big/\!3\pi$} {\it unterdr"uckt\/} das Skript.
\end{samepage}%
\theendnotes

%% file: APP_LCWFN.tex
\lhead[\fancyplain{}{\sc\thepage}]
      {\fancyplain{}{\sc\rightmark}}
\rhead[\fancyplain{}{\sc{{\footnotesize Anhang~\thechapter:} Lichtkegelwellenfunktionen}}]
      {\fancyplain{}{\sc\thepage}}
\psfull
\chapter[Lichtkegelwellenfunktionen]{%
   \huge Lichtkegelwellenfunktionen}

Wir leiten her die Lichtkegelwellenfunktion des Photons zu niedrigster Ordnung in Lichtkegelst"orungstheorie~(LCPT) und skizzieren ihre Herleitung im Rahmen der kovarianten perturbativen Quantenfeldtheorie~(PT).
Nach ihr modelliert werden die Lichtkegelwellenfunktionen der $1S$-,~$2S$-Vektormeson-Zust"ande~$\ket{1S}$,~$\ket{2S}$, vgl.\@ Kap.~\ref{Kap:GROUND} und/bzw.\@ Kap.~\ref{Kap:EXCITED}; wir fixieren deren Parameter.

Wir schlie"sen den Anhang mit einem Kompendium diverser Relationen in Zusammenhang mit diesen Wellenfunktionen, die benutzt werden im Haupttext.
\vspace*{-1ex}

\section[\vspace*{-.5ex}Photon-Lichtkegelwellenfunktion%
  ~{\protect$\ps^{h,\bar h}_{\iga(Q^2,\la)}(\zet,\rb{r})$}]{%
         Photon-Lichtkegelwellenfunktion%
  ~\bm{\ps^{h,\bar h}_{\iga(Q^2,\la)}(\zet,\rb{r})}}
\label{APPSect:Photon-Wfn}

Die Photon-Wellenfunktion im {\it infinite-momentum-frame\/}~(imf, Bezugsystem, in dem das Photon unendlichen Impuls besitzt) wird diskutiert im Rahmen von LCPT in Ref.~\cite{Bjorken71}.
Wir folgen dieser Diskussion auf Basis der Vorschriften und Konventionen von Ref.~\cite{Lepage80}.

F"ur ein Photon mit Vierer-Impuls~\mbox{$q \!=\! (q^+, q^- \!=\! -\al^2 Q^2 \!/\!q^+, \rb{q} \!=\! \bm{0})^{\T t}$},~$q^+ \zz\to\! \infty$,~[im folgenden $\al \!=\! 1\!/\!\surd2$, d.h.~\mbox{$q^\pm \!=\! (q^0 \!\pm\! q^3)\!/\!\surd2$},~$g_{+-} \!=\! 1$, vgl.\@ Fu"sn.\,\FNg{FN:LC-Impuls}], werden multipliziert:
\renewcommand{\labelitemi}{$\circ$}
\begin{itemize}
\item der Colour-Faktor\quad
  $\surd\Nc$
\item der Flavour-Anteil\quad
  $e_f \de_{\!f\!\bar f}$
\item der Spinor-Term\quad
  $\bar{u}(\zet q^+,\rb{k},h,f)\, \ga^\mu\, \vep_\mu(q,\la)\, v(\bzet q^+,-\rb{k},\bar h,\bar f)$
\item f"ur die Quark- und Antiquark-Linien der Faktor\quad
  $[\sqrt2\zet q^+]^{-1\!/\!2}\, [\sqrt2\bzet q^+]^{-1\!/\!2}$\quad
\item der Lichtkegel-Energienenner\quad
  $-\sqrt2q^+\, [Q^2 + (\rb{k}^2 \!+\! m_f^2)\!/\zbz]^{-1}$
\end{itemize}
Der resultierende Ausdruck
\vspace*{-.5ex}
\begin{align} \label{APP:Ansatz_Photon}
&\tilde{\ps}^{h,\bar h}_{\iga(Q^2,\la)}(\zet,\rb{k})
    \\[-.5ex]
  &=\; \surd\Nc\; e_f \de_{\!f\!\bar f}\vv
        \frac{[-\surd\zbz]}{\zbz Q^2 \!+\! m_f^2 \!+\! \rb{k}^2}\vv
        \bar{u}(\zet q^+,\rb{k},h,f)\; \ga^\mu\, \vep_\mu(q,\la)\;
              v(\bzet q^+,-\rb{k},\bar h,\bar f)
    \nn
    \\[-5ex]\nn
\end{align}
stellt dar die Wahrscheinlichkeitsamplitude daf"ur, da"s das Photon mit Vierer-Impuls~$q$, wie angegeben, mit Virtualit"at~$Q$ und fester Helizit"at~$\la \!\equiv\! 0,\pm1$ fluktuiert in ein Quark-Antiquark-Paar, das charakterisiert ist wie folgt:
Das Quark besitzt Lichtkegelimpuls~$\zet q^+$, transversalen Lichtkegel-Relativimpuls~$\rb{k}$%
\FOOT{
   Vgl.\@ Kap.~\ref{Subsect:HadronniveauI}, die Diskussion in Zusammenhang mit Gl.~(\ref{AQ_Impulse-Forderung}),~-- wie auch Anh.~\ref{APP-Subsect:LCWFN-Kovarianz}.
},
Helizit"at~$h$ und Flavour~$f$, das Antiquark entsprechend~$\bzet q^+$,~$-\rb{k}$,~$\bar h$ und~$\bar f$\/; mit~$m_f$ ist bezeichnet die laufende Quarkmasse.
Dabei sind die Helizit"ats-Vierer-Eigenvektoren~\mbox{$\vep(q,\la) \!=\! (\vep^\mu(q,\la))$} f"ur~$\la \!\equiv\! 0,\pm1$ definiert durch:
\begin{alignat}{5} \label{vep(q,la)-Photon}
&\vep(q,0)&\;
  &=\; \phantom{-1\!/\!\surd2}
       \big(&q^+\zz/\!Q&,&\, Q\!/\!2q^+&,\,& \bm{0}\,&\big){}^{\T t}
    \\
&\vep(q,\pm1)&\;
  &=\; -1\!/\!\surd2
       \big(&         0&,&\,          0&,\,& 1,\, \pm \iIM\,&\big){}^{\T t}
    \tag{\ref{vep(q,la)-Photon}$'$}
\end{alignat}
das hei"st~$\vep^{\mu{\D\ast}}\!(q,\la)\vep_\mu(q,\la)$ ist normiert auf~$\pm1$ f"ur~$\la \!\equiv\! 0$ beziehungsweise~$\pm1$.
Helizit"ats-Dirac-Eigen\-spinoren~$w(h \!\equiv\! \pm1\!/\!2)$ im infinite-momentum-frame sind definiert durch:
\begin{alignat}{6} \label{APP:LC-w}
&w(1\!/\!2)&\;
  &=\; \big(&1\!/\!\surd2&,&\,            0&,&\, -1\!/\!\surd2&,&\,            0&\big)
    \\
&w(-1\!/\!2)&\;
  &=\; \big(&           0&,&\, 1\!/\!\surd2&,&\,             0&,&\, 1\!/\!\surd2&\big)
    \tag{\ref{APP:LC-w}$'$}
\end{alignat}
vgl.\@ Ref.~\cite{Lepage80}.
Bez"uglich dieser Zust"ande folgt f"ur die Spinor-Matrixelemente:%
\FOOT{
  \label{FN-APP:ArgumenteIndizes}Seien Indizes und Argumente der Funktionen nur soweit explizit notiert, wie der Zusammenhang fordert.
}
%
\begin{alignat}{2} \label{APP:ubar-ga-v_explizit}
&\bar{u}\, \ga^+\, v&\;
  &=\; \phantom{-}
       2\sqrt{\zbz}\, q^+\; \de_{h,-\bar{h}}
    \\
&\bar{u}\, \ga^-\, v&\;
  &=\; -\, \frac{\rb{k}^2 \!+\! m_f^2}{\zbz\, q^+}\; \de_{h,-\bar{h}}
    \tag{\ref{APP:ubar-ga-v_explizit}$'$} \\
&\bar{u}\, \ga^i\, v&\;
  &=\; \frac{(1 \!-\! 2\zet)\,k^i \mp \iIM\,\vep^{ij3}k^j}{\zbz}\; \de_{h,-\bar{h}}\;
         \mp\; m\, \frac{\de^{i1} \!\mp\! \iIM\,\de^{i2}}{\zbz}\;  \de_{h,\bar{h}}
    \nn
\intertext{Dabei steht das obere Vorzeichen in der letzten Zeile f"ur~$h \!\equiv\! +1\!/\!2$ und das untere f"ur~$h \!\equiv\! -1\!/\!2$; wir schreiben diese Relation unmittelbar um in die pr"agnante Form:}
&\bar{u}\, \ga^i\, v&\;
  &=\; \frac{(1 \!-\! 2\zet)\,k^i \mp \iIM\,\vep^{ij3}k^j}{\zbz}\; \Hpm\;
         \mp\; m\, \frac{\de^{i1} \!\mp\! \iIM\,\de^{i2}}{\zbz}\;  \Hpp
    \tag{\ref{APP:ubar-ga-v_explizit}$''$}
\end{alignat}
die zu verstehen ist als {\it Summe bez"uglich jeweils der oberen und der unteren Vorzeichen\/} mit
\begin{align}
\Hss{\si}{\bar\si}\;
  \equiv\; \de_{\si,{\rm sign}\,h}\; \de_{\bar\si,{\rm sign}\,\bar h}
\end{align}
als abk"urzende Notation f"ur die Helizit"ats-Konfigurationen der Quarks.

Zum einen, aus den Ausdr"ucken der Gln.~(\ref{APP:ubar-ga-v_explizit}),~(\ref{APP:ubar-ga-v_explizit}$'$) folgt f"ur longitudinale Polarisation,~$\la \!\equiv\! 0$:
\begin{align} \label{ubar-ga-ep-v}
\bar{u}\, \ga^\mu \vep_\mu(q,0)\, v\;
  =\; \frac{\zbz\, Q^2  - (\rb{k}^2 \!+\! m^2)}{\sqrt{\zbz}\, Q}\; \de_{h,-\bar{h}}\;
  =\; 2\sqrt{\zbz}\, Q\; \de_{h,-\bar{h}}
\end{align}
mit der zweiten Identit"at aufgrund der Relation~$-Q^2 \!\equiv\! q^2 \!=\! (\rb{k}^2 \!+\! m_f^2)\!/ \zbz$.%
\FOOT{
  Diese ist Konsequenz der on-mass-shell-Forderungen an das Photon und die Quarks und der Addition derer Impulse zu dem des Photons.
}
\vspace*{-.5ex}Dies eingesetzt in Gl.~(\ref{APP:Ansatz_Photon}), folgt~$\tilde{\ps}^{h,\bar h}_{\iga(Q^2,\la\equiv0)}(\zet,\rb{k})$.

Zum anderen, aus dem Ausdruck von Gl.(\ref{APP:ubar-ga-v_explizit}$''$) folgt f"ur transversale Polarisation,~$\la \!\equiv\! \pm1$:
\begin{align}
\bar{u}\, \ga^\mu \vep_\mu(q,\pm1)\, v\;
  =\; - \frac{1}{\surd\zbz}\;
      \Big\{ \surd2&\;
        \Big[ k\, \efn{\D+ \iIM\,\vph_\rb{k}}\;
          \big( \zet\, \Hss{+}{-} - \bzet\, \Hss{-}{+} \big)\;
          +\; m_f\, \Hss{+}{+}
        \Big]\cdot \de_\la^+
    \tag{\ref{ubar-ga-ep-v}$'$} \\[-1ex]
          + \surd2&\;
        \Big[ k\, \efn{\D- \iIM\,\vph_\rb{k}}\;
          \big( \zet\, \Hss{-}{+} - \bzet\, \Hss{+}{-} \big)\;
          -\; m_f\, \Hss{-}{-}
        \Big]\cdot \de_\la^-\;
      \Big\}
    \nn
\end{align}
mit~$k$,~$\vph_\rb{k}$ Betrag und Azimut von~$\rb{k}$.
Dies eingesetzt in Gl.~(\ref{APP:Ansatz_Photon}), \vspace*{-.375ex}folgt~$\tilde{\ps}^{h,\bar h}_{\iga(Q^2,\la\equiv\pm)}(\zet,\rb{k})$.

Zusammen f"ur~$\la \!\equiv\! 0,\pm1$ gilt in kompakter Schreibweise:
\begin{align} \label{APP:Photon_k}
&\tilde{\ps}^{h,\bar h}_{\iga(Q^2,\la)}(\zet,\rb{k})
    \\
  &=\; \surd\Nc\; e_f \de_{\!f\!\bar f}\;
       \Big\{
         -\, 2\zbz\; Q\vv \de_{h,-\bar h}\;\cdot \de^0_\la
    \nn \\
  &\phantom{=\; \surd\Nc\; e_f \de_{\!f\!\bar f}\; }
   +\, \surd2 \left[
         k\, \efn{\D\pm \iIM\,\vph_\rb{k}}\;
           \big( \zet\, \Hpm - \bzet\, \Hmp \big)\;
         \pm\; m_f\, \Hpp
       \right]\cdot \de^\pm_\la\;
       \Big\}\; \frac{1}{\vep^2 \!+\! k^2}
    \nn
\end{align}
mit den Vorzeichen wie in Gl.~(\ref{ubar-ga-ep-v}$'$) und der Definition:~\mbox{$\vep \!=\! \surd \zbz Q^2 \!+\! m_f^2$}, vgl.\@ Gl.~(\ref{epsilon}).

Sei~$\tilde{f}(\rb{k})$ eine Funktion mit den "ublichen Integrationseigenschaften; dann sei ihre bez"ug\-lich des transversalen Impulses~$\rb{k}$ Fourier-Transformierte~$f(\rb{r})$ definiert durch:
\vspace*{-.5ex}
\begin{align} \label{APP:FT}
\int \frac{d^2\rb{k}}{(2\pi)^2}\; \efn{\D \iIM\,\rb{k} \!\cdot\! \rb{r}}\vv
  \tilde{f}(\rb{k})\;
  \stackrel{\D!}{=}\; f({\bf r})
\end{align}
vgl.\@ Gl.~(\ref{APP:FT-allg}$'$).
Daraus folgt:
\begin{align}
\int \frac{d^2\rb{k}}{(2\pi)^2}\; \efn{\D \iIM\,\rb{k} \!\cdot\! \rb{r}}\vv
  k\, \efn{\D\pm \iIM\,\vph_\rb{k}}\; \tilde{f}(\rb{k})\;
  =\; -\iIM\, (\pa_1 \!\pm\! \iIM\,\pa_2)\; f(\rb{r})\;
  =\; -\iIM\, \efn{\D\pm \iIM\,\vph_\rb{r}}\, \pa_r\; f(r)
    \tag{\ref{APP:FT}$'$}
\end{align}
mit der zweiten Identit"at f"ur den Fall, da"s~$\tilde{f}$ abh"angt nur vom Betrag~$k$ von~$\rb{k}$ und~$f$ folglich nur vom Betrag~$r$ von~$\rb{r}$; bezeichne~$\vph_\rb{r}$ den Azimutwinkel von~$\rb{r}$.

Die bez"uglich~$\rb{k}$ Fourier-Transformierte von Gl.~(\ref{APP:Photon_k}) folgt mithilfe der Gln.~(\ref{APP:FT}),~(\ref{APP:FT}$'$) unmittelbar zu:
\begin{alignat}{2} \label{APP:pre-Photon_r}
&\ps^{h,\bar h}_{\iga(Q^2,\la)}(\zet,\rb{r}) &&
    \\
  &=\; \surd\Nc\; e_f \de_{\!f\!\bar f}\;
         \Big\{
         -\, 2\zbz\; Q\vv \de_{h,-\bar h}\;\cdot \de^0_\la &&
    \nn \\
  &\phantom{=\; \surd\Nc\; e_f \de_{\!f\!\bar f}\; }
     +\, \surd2 \Big[
           \iIM\, \efn{\D\pm \iIM\,\vph_\rb{r}}\;
             \big( \zet\, \Hpm - \bzet\, \Hmp \big)&\, &[-\pa_r]\;
         \pm\; m_f\, \Hpp
         \Big]\cdot \de^\pm_\la\;
         \Big\}
    \nn \\
  &&&\times
     \int \frac{d^2\rb{k}}{(2\pi)^2}\;
       \efn{\D \iIM\,\rb{k} \!\cdot\! \rb{r}}\; \frac{1}{\vep^2 \!+\! k^2}
    \nn
\end{alignat}
Mithilfe
\begin{align}
\int \frac{d^2\rb{k}}{(2\pi)^2}\;
  \efn{\D \iIM\,\rb{k} \!\cdot\! \rb{r}}\; \frac{1}{\vep^2 \!+\! k^2}\;
  =\; \frac{{\rm K}_0(\vep r)}{2\pi}
\end{align}
und~$-d\!/\!d\ze\, {\rm K}_0(\ze) \!=\! {\rm K}_1(\ze)$, mit~${\rm K}_0$,~${\rm K}_1$ den modifizierten Besselfunktion zweiter Art in der Definition von Ref.~\cite{Abramowitz84}, folgt aus Gl.~(\ref{APP:pre-Photon_r}) unmittelbar:
\vspace*{-.5ex}
\begin{alignat}{2} \label{APP:Photon_r}
\hspace*{-1em}
&\ps^{h,\bar h}_{\iga(Q^2,\la)}(\zet,\rb{r})
    \\[-.5ex]
  &=\; \surd\Nc\; e_f \de_{\!f\!\bar f}\;
         \Big\{
         -\, 2\zbz\; Q\vv \de_{h,-\bar h}\vv \frac{{\rm K}_0(\vep r)}{2\pi}\cdot \de^0_\la
    \nn \\
  &\phantom{=\; \surd\Nc\; e_f \de_{\!f\!\bar f}\; }
     +\, \surd2 \Big[
           \iIM\,\vep\, \efn{\D\pm \iIM\,\vph_\rb{r}}\;
             \big( \zet\, \Hpm - \bzet\, \Hmp \big)\;
             \frac{{\rm K}_1(\vep r)}{2\pi}\;
           \pm\; m_f\, \Hpp\;
             \frac{{\rm K}_0(\vep r)}{2\pi} \Big]\cdot \de^\pm_\la\;
         \Big\}
    \nn
    \\[-4.5ex]\nn
\end{alignat}
F"ur Photon-Helizit"at~$\la \!\equiv\! -1$ wird o.E.d.A.\@ abschlie"send ein globales negatives Vorzeichen weggelassen und es folgt f"ur~$\ps^{h,\bar h}_{\iga(Q^2,\la)}(\zet,\rb{r})$ genau die Darstellung der Gln.~(\ref{Photon-Wfn}),~(\ref{Photon-Wfn}$'$) im Haupttext.

\bigskip\noindent
Eine Photon-Wellenfunktion kann perturbativ hergeleitet werden auch im Rahmen der konventionellen, kovarianten Quantenfeldtheorie.
Voraussetzung ist in der Darstellung durch Feynman-Graphen die Interpretation des Photon-Quark-Antiquark-Vertex in Termen einer Photon-Wellenfunktion im Sinne von Lichtkegelst"orungstheorie.
Dies ist m"oglich, wenn die Ordnung bez"uglich~$x^+$ eingeschr"ankt ist auf "`Photon vor Quark-Antiquark"', und de facto der Fall im formalen Limes unendlichen Photon-Impulses in~$x^+$-Richtung:~$q^+ \zz\to\! \infty$. \\
\indent
Zun"achst ist auszuwerten der~$k^-$-Anteils des Quark-Loop-Integrals.
F"ur asymptotische~$q^+$ folgt,~$dk^0dk^3 \!=\! 1\!/\!(2\al^2) dk^+dk^- \!=\! dk^+dk^-$:
\begin{align} 
\frac{dk^+d^2\rb{k}}{(2\pi)^3}\,
  \int_{-\infty}^\infty \frac{dk^-}{2\pi}\,
   &\frac{N(k^+,k^-,\rb{k})}{%
          (k^2 \!-\! m_f^2 \!+\! \iIM\,\ep)([k \!-\! q]^2 \!-\! m_f^2 \!+\! \iIM\,\ep)}
    \\[.5ex]
  \underset{\text{$q^+ \zz\to\! \infty$}}{\sim}\vv
    \frac{d\zet}{4\pi}\frac{d^2\rb{k}}{(2\pi)^2}\,
   &\frac{\iIM\, N(k^+,k^- \zz\equiv\! 0,\rb{k})}{\rb{k}^2 \!+\! m_f^2 \!-\! \zbz q^2}
    \nn
\end{align}
falls~$N$ endlich und ungleich Null f"ur~$k^- \zz\equiv\! 0$.
F"ur~\mbox{$N \!=\! \iIM(\kslash \!+\! m_f)\, [-\iIM e\vepslash(q,\la)]\, \iIM([\kslash \!-\! \qslash] \!+\! m_f)$} folgt explizit:
\vspace*{-.5ex}
\begin{align} 
\hspace*{-.5em}
N &= \iIM\,e\, \big\{
           k \!\cdot\! \vep\, (2\kslash \!-\! \qslash)
           - k \!\cdot\! (k \!-\! q)\, \vepslash
           + \iIM\,\ep_{\mu\nu\rh\si} \ga_5 \ga^\mu k^\nu \vep^\rh q^\si
           + m_f (2k \!\cdot\! \vep \!-\! \vepslash \qslash \!+\! m_f \vepslash) \big\}
    \\[1.5ex]
  &
   \underset{\text{$q^+ \zz\to\! \infty$}}{\sim}\;
     \iIM\,e\, q^+ \ga^-\, \big\{\,
       \de_\la^0\cdot [\rb{k}^2 \!+\! m_f^2 \!+\! \zbz q^2]\!/Q
    \nn \\[-1.5ex]
  &
   \hspace*{-1.125em}
   \phantom{\underset{\text{$q^+ \zz\to\! \infty$}}{\sim}\; ie\, q^+ \ga^-\, \big\{\, }
     + \de_\la^\pm\cdot [(1 \!-\! 2\zet)\, k \!\cdot\! \bm{\vep}(q,\pm1)
                     + \iIM\,\ga_5 \ep^{ij3} k^i \vep^j(q,\pm1)
                     + m_f \vepslash(q,\pm1)]\, \big\}
    \nn
    \\[-4.5ex]\nn
\end{align}
Die Wellenfunktion folgt durch Bilden der Helizit"ats-Matrixelemente~$\bar{w}(h)\, [N\!/\sqrt2q^+]\, w(-\bar h)$ mit den Helizit"ats-Eigenspinoren~$w(h \!\equiv\! \pm1\!/\!2)$ entsprechend den Gln.~(\ref{APP:LC-w}),~(\ref{APP:LC-w}$'$).

\section[\vspace*{-.25ex}Vektormeson-Lichtkegelwellenfunktion%
  ~{\protect$\ps^{h,\bar h}_{V(\la)}(\zet,\rb{r})$}]{%
          \vspace*{-.5ex}Vektormeson-Lichtkegelwellenfunktion%
  ~\bm{\ps^{h,\bar h}_{V(\la)}(\zet,\rb{r})}}
\label{APPSect:Vektormeson-Wfn}

Die Lichtkegelwellenfunktion des Vektormesons~$V$ wird {\it modelliert\/} entsprechend der des Photons.
Ansatzpunkt ist Gl.~(\ref{APP:Photon_k}).
Wir "ubernehmen deren Helizit"atsanteil (geschweifte Klammer) und ersetzen den Energienenner~$[\zbz Q^2 \!+\! m_f^2 \!+\! \rb{k}^2]^{-1}$ durch zun"achst nicht spezifizierte \vspace{-.25ex}skalare Funktion~$\tilde{\ps}_{V(\la)}$ von~$\zet$ und~$k \!\equiv\! |\rb{k}|$:
\vspace*{-.5ex}
\begin{align} \label{VMeson_k}
\tilde{\ps}^{h,\bar h}_{V(\la)}(\zet,\rb{k})\;
  &=\; \Big\{
         4\zbz\, \om_{V,\la}\, \de_{h,-\bar h}\cdot \de^0_\la
    \\
  &\phantom{=\; }
   +\, \left[k\, \efn{\D\pm \iIM\,\vph_\rb{k}}\;
         \big( \zet\, \Hpm - \bzet\, \Hmp \big)\;
         \pm\; m_f\, \Hpp
       \right]\cdot \de^\pm_\la\;
       \Big\}\vv
       \tilde{\ps}_{V(\la)}(\zet,k)
    \nn
    \\[-4.5ex]\nn
\end{align}
Geeignete Faktoren wie etwa~$\om_{V,\la}$ sind aus diesen Funktionen herausgezogen.

Eine Darstellung bez"uglich des transversalen Ortsraums analog zu Gl.~(\ref{APP:pre-Photon_r}) folgt, indem Gl.~(\ref{VMeson_k}) formal Fourier-transformiert wird bez"uglich~$\rb{k}$:
\vspace*{-.5ex}
\begin{align} \label{VMeson_r}
\ps^{h,\bar h}_{V(\la)}(\zet,\rb{r})\;
  &=\; \Big\{
        4\zbz\, \om_{V,\la}\, \de_{h,-\bar h}\cdot \de^0_\la
    \\
  &\phantom{=\; }
   +\, \Big[
         \iIM\, \efn{\D\pm \iIM\,\vph_\rb{r}}\;
           \big( \zet\, \Hpm - \bzet\, \Hmp \big)\, [-\pa_r]\;
         \pm\; m_f\, \Hpp
       \Big]\cdot \de^\pm_\la\;
       \Big\}\;
       \ps_{V(\la)}(\zet,r)
    \nn
    \\[-4.5ex]\nn
\end{align}
Dabei sind die skalaren Funktionen~$\ps_{V(\la)}$ die Fourier-Transformierten von~$\tilde{\ps}_{V(\la)}$:
%
\vspace*{-.5ex}
\begin{align} \label{APP:FT-scalar_k-to-r}
\int \frac{d^2\rb{k}}{(2\pi)^2}\; \efn{\D \iIM\,\rb{k} \!\cdot\! \rb{r}}\vv
  \tilde{\ps}_{V(\la)}(\zet,k)\;
  \stackrel{\D!}{=}\; \ps_{V(\la)}(\zet,r)
    \\[-4.5ex]\nn
\end{align}
nach Gl.~(\ref{APP:FT}).
In den folgenden Abschnitten~\ref{APPSubsect:1S-Vektormeson-Wfn} und~\ref{APPSubsect:2S-Vektormeson-Wfn} werden explizite Ans"atze angegeben f"ur die skalaren Funktionen~$\tilde{\ps}_{V(\la)}$ f"ur~$V \!\equiv\! 1S,2S$, die zu den Darstellungen im Haupttext f"uhren.

\vspace*{-1ex}
\paragraph{\label{APP-T:Charakteristik}Charakteristik.}
Seien zun"achst diskutiert auf Basis der allegmeinen Darstellung durch Gl.~(\ref{VMeson_k}) Charakteristika und Eigenschaften der \vspace*{-.25ex}Lichtkegelwellenfunktion~$\tilde{\ps}^{h,\bar h}_{V(\la)}(\zet,\rb{k})$\vspace*{-.25ex}.
Ziel ist die Formulierung von Relationen f"ur die skalaren Funktionen~$\tilde{\ps}_{V(\la)}$, die es gestatten deren Parameter zu fixieren.

Wichtige Gr"o"se ist die Kopplung~$f_V$ des Vektormesons~$V$ an den elektromagnetischen Strom.
Im folgenden Abschnitt wird sie formal definiert und hergeleitet ihr Zusammenhang mit der leptonischen Zerfallsbreite~$\Gall_V$ von~$V$; vgl.\@ unten die Gln.~(\ref{APP:fV-Def}),~(\ref{APP:fV_Gall}) bzw.~de\-ren Diskussion im Haupttext in Form der Gln.~(\ref{fV-Def}),~(\ref{fV_Gall}).
In Termen der skalaren Funktionen~$\tilde{\ps}_{V(\la)}$ folgt aus Gl.~(\ref{VMeson_k}) explizit f"ur longitudinale und transversale Polarisation:
\begin{alignat}{2} \label{fV_LCWfn}
f_{V,L}\;
  &=\; \hat{e}_V \surd\Nc&\;\cdot\;
         4\, \om_{V,L}\vv
        &\int \frac{d\zet}{4\pi}\frac{d^2\rb{k}}{(2\pi)^2}\vv
           4\zbz\vv
           \tilde{\ps}_{V(\la\equiv0)}(\zet,k)
    \\
f_{V,T}\;
  &=\; \hat{e}_V \surd\Nc&\;\cdot\;
         \frac{4\surd2}{M_V}\vv
        &\int \frac{d\zet}{4\pi}\frac{d^2\rb{k}}{(2\pi)^2}\vv
           \frac{(\zet^2 \!+\! \bzet^2)\, k^2 \!+\! m_f^2}{4\zbz}\vv
           \tilde{\ps}_{V(\la\equiv\pm1)}(\zet,k)
    \tag{\ref{fV_LCWfn}$'$}
\end{alignat}
Dabei ist~$\hat{e}_V$ die effektive Quark-Ladung im Vektormeson~$V$ in Einheiten der Proton-Ladung~$e$; sie ist determiniert durch dessen Flavour-Gehalt beziehungsweise Isospin, vgl.\@ Gl.~(\ref{Flavour-Gehalt}) und bzgl.\@ expliziter Zahlenwerte Tabl.~\refg{Tabl:Charakt_rh,om,ph,Jps}.

Die Vektormeson-Zust"ande~$\ket{1S}$,~$\ket{2S}$,\ldots sind normiert gem"a"s:
\bea 
\bracket{V(q',\la')}{V(q,\la)}\;
  =\; (2\pi)^3\, 2q^+\;
        \de(q^+ \!-\! {q'}^+)\,
        \de_{(2)}(\rb{q} \!-\! \rb{q}')\;
        \de_{\la\la'}
\eea
vgl.\@ Gl.~(\ref{h_bracket}) und die Gln.~(\ref{GROUND:h_bracket}),~(\ref{GROUND:h_bracket}$'$).
Diese Gleichung impliziert insgesamt {\it Orthonormalit"at\/} der Anregungszust"ande eines Vektormesons.
Wir schreiben in diesem Sinne f"ur die Lichtkegelwellenfunktion~$\tilde{\ps}^{h,\bar h}_{V(\la)}(\zet,\rb{k})$, mit~$V,V' \!\equiv\! 1S,2S,\ldots$ und~$\la,\la' \!\equiv\! 0,\pm1$ fest:
\bea \label{APP:OrthoNorm_full}
\de_{VV'}\, \de_{\la\la'}\;
  =\; \int \frac{d\zet}{4\pi}\frac{d^2\rb{k}}{(2\pi)^2}\vv
        {\T\sum}_{h,\bar h}\;
        \tilde{\ps}^{h,\bar h\;{\D\dagger}}_{V'(\la')}(\zet,\rb{k})\,
        \tilde{\ps}^{h,\bar h}_{V(\la)}(\zet,\rb{k})
\eea
wobei Summation "uber die Quark-Helizit"aten~$h,\bar h$ impliziert ist.
In Termen der skalaren Funktionen~$\tilde{\ps}_{V(\la)}(\zet,k)$ lauten diese Relationen explizit bez"uglich Normierung:
\begin{align} \label{APP:Norm_skalar}
1\; &=\; 2\, \om_{V,L}^2\vv
    \int \frac{d\zet}{4\pi}\frac{d^2\rb{k}}{(2\pi)^2}\vv
      \{4\zbz\}^2\vv
      \big| \tilde{\ps}_{V(\la\equiv0)}(\zet,k) \big|^2
    \\
1\; &=\; \int \frac{d\zet}{4\pi}\frac{d^2\rb{k}}{(2\pi)^2}\vv
      \left\{(\zet^2 \!+\! \bzet^2)\, k^2 \!+\! m_f^2\right\}\vv
      \big| \tilde{\ps}_{V(\la\equiv\pm1)}(\zet,k) \big|^2 
    \tag{\ref{APP:Norm_skalar}$'$}
\end{align}
und bez"uglich Orthogonalit"at:
\begin{alignat}{2} \label{APP:OrthoGon_skalar}
0\; &=\;
    \int \frac{d\zet}{4\pi}\frac{d^2\rb{k}}{(2\pi)^2}\vv
      \{4\zbz\}^2\vv
        \tilde{\ps}^{\D\dagger}_{V'(\la)}(\zet,\rb{k})\,
        \tilde{\ps}_{V(\la)}(\zet,\rb{k})&\qquad
  &\text{$\la \!\equiv\! 0$}\hspace*{-1.25em}
    \\
0\; &=\;
    \int \frac{d\zet}{4\pi}\frac{d^2\rb{k}}{(2\pi)^2}\vv
      \left\{(\zet^2 \!+\! \bzet^2)\, k^2 \!+\! m_f^2\right\}\vv
        \tilde{\ps}^{\D\dagger}_{V'(\la)}(\zet,\rb{k})\,
        \tilde{\ps}_{V(\la)}(\zet,\rb{k})&\qquad\;
  &\text{$\la \!\equiv\! \pm1$}\hspace*{-1.25em}
    \tag{\ref{APP:OrthoGon_skalar}$'$}
\end{alignat}
Dabei sind in den Gln.~(\ref{APP:OrthoGon_skalar}),~(\ref{APP:OrthoGon_skalar}$'$) rechts konstante Faktoren weggelassen. \\
\indent
\vspace*{-.25ex}Sei weiter bezeichnet mit~$\vev{\rb{r}^2}_{V\!,\la}$ f"ur definiertes Vektormeson~$V \!\equiv\! 1S,2S,\ldots$ und fester Polarisation~$\la \!\equiv\! L,T$ dessen transversaler rmsq-Radius.
In Termen der Lichtkegelwellenfunk\-tion~$\ps^{h,\bar h}_{V(\la)}(\zet,\rb{r})$ nach Gl.~(\ref{VMeson_r}) ist folglich definiert:
\begin{align} \label{APP:rmsq-Radius}
\vev{\rb{r}^2}_{V\!,\la}\;
  =\; \int \frac{d\zet}{4\pi}\; d^2\rb{r}\vv
        {\T\sum}_{h,\bar h}\;
        \ps^{h,\bar h\;{\D\dagger}}_{V(\la)}(\zet,\rb{r})\;
        \rb{r}^2\;
        \ps^{h,\bar h}_{V(\la)}(\zet,\rb{r})
\end{align}
vgl.\@ Gl.~(\ref{APP:OrthoNorm_full}).

Analog zum Haupttext, vgl.\@ die Gln.~(\ref{Photon-Wfn}),~(\ref{E:Photon-Wfn}) und~(\ref{Vektormeson-Wfn}),~(\ref{E:Vektormeson-Wfn}), wird separiert Colour- und Flavour-Gehalt:
\begin{alignat}{4} \label{APP:Separation}
&\ps_{\iga(Q^2,\la)}^{h,\bar h}&\;
  &\stackrel{\D!}{=}\;
  \surd\Nc\vv e_f \de_{\!f\!\bar f}\vv \ch_{\iga(Q^2,\la)}^{h,\bar h}&
    \qqquad
&\tilde{\ps}_{\iga(Q^2,\la)}^{h,\bar h}&\;
  &\stackrel{\D!}{=}\;
  \surd\Nc\vv e_f \de_{\!f\!\bar f}\vv \tilde{\ch}_{\iga(Q^2,\la)}^{h,\bar h}
    \\
&\ps_{V(\la)}^{h,\bar h}&\;
  &\stackrel{\D!}{=}\;
  1\!/\!\surd\Nc\vv i_{V,f}\vv \ch_{V(\la)}^{h,\bar h}&
    \qqquad
&\tilde{\ps}_{V(\la)}^{h,\bar h}&\;
  &\stackrel{\D!}{=}\;
  1\!/\!\surd\Nc\vv i_{V,f}\vv \tilde{\ch}_{V(\la)}^{h,\bar h}
    \tag{\ref{APP:Separation}$'$}
\end{alignat}
bzgl.~$i_{V,f}$ vgl.\@ Gl.~(\ref{Flavour-Gehalt}).
Darstellungen der angegebenen Relationen in Termen der \vspace*{-.25ex}Funk\-tionen~$\ch_{V(\la)}^{h,\bar h}$,~$\tilde{\ch}_{V(\la)}^{h,\bar h}$ und skalarer Funktionen~$\ch_{V(\la)}$,~$\tilde{\ch}_{V(\la)}$ folgen unmittelbar.

\subsection[\protect$1S$-Vektormeson]{%
                 \bm{1S}-Vektormeson}
\label{APPSubsect:1S-Vektormeson-Wfn}

F"ur die~$1S$-Vektormeson im Sinne der Kapitel~\ref{Kap:GROUND},~\ref{Kap:EXCITED} wird auf Niveau der skalaren Funktionen~$\tilde{\ch}(\zet,k)$ folgender Ansatz gemacht,~$V \!\equiv\! 1S$:
\vspace*{-.5ex}
\begin{align} \label{scalar1S_k-explizit}
\tilde{\ch}_{V\equiv\iES(\la)}(\zet,k)
  &= {\cal N}_{V,\la}\; \sqrt\zbz\;
         \exp \bigg[ -\frac{1}{2}\,
                     \frac{M^2 (\zet \!-\! 1\!/\!2)^2}{\om_{V,\la}^2} \bigg]\cdot
      \frac{2\pi}{\om_{V,\la}^2}\,
      \exp \bigg[ -\frac{1}{2}\, \om_{V,\la}^{-2}\, k^2 \bigg]
    \\[1ex]
  &=\; h_{,\la}(\zet)\cdot \tilde{g}_{V,\la}(k)
    \tag{\ref{scalar1S_k-explizit}$'$}
    \\[-4.5ex]\nn
\end{align}
mit~$M \!\equiv\! M_\iES$ und den Funktionen
\vspace*{-.5ex}
\begin{alignat}{2}
&\tilde{g}_{\iES,\la}(k)&\;
  &=\; \frac{2\pi}{\om_{\iES,\la}^2}\,
         \exp \bigg[ -\frac{1}{2}\, \om_{\iES,\la}^{-2}\, k^2 \bigg]
    \label{APP:tilde-g-Wfn_1S} \\
&h_{\iES,\la}(\zet)&\;
  &=\; {\cal N}_{\iES,\la}\vv [\zbz]^{\D\ta_\la\!/\!2}\vv
         \exp \bigg[ -\frac{1}{2}\,
                     \frac{M^2 (\zet \!-\! 1\!/\!2)^2}{\om_{\iES,\la}^2} \bigg] \qquad
       \text{mit}\qquad
       \ta_\la \equiv 1
    \label{APP:h-Wfn_1S}
    \\[-4.5ex]\nn
\end{alignat}
die f"ur feste Polarisation~$\la \!\equiv\! L,T$ des Vektormesons abh"angen von \mbox{den zwei Parametern} $\om_{V,\la}$,~${\cal N}_{V,\la}$; in Hinblick auf formale Umformungen sind Konstanten~$\ta_L,\ta_T$ eingef"uhrt, die {\it in~praxi\/} identisch Eins gesetzt werden.

Diese Parametrisierung wird vorgeschlagen von Wirbel, Stech, Bauer in Ref.~\cite{Wirbel85}.
Der longitudinale Anteil~$h_{V,\la}(\zet)$, ist symmetrisch bez"uglich des Austauschs von Quark und Antiquark:~$\zet \!\leftrightarrow\! \bzet$ aufgrund~$(\zet \!-\! 1\!/\!2)^2 \!=\! 1\!/\!4 \!-\! \zbz$, und gepeakt f"ur den nicht-relativistischen Fall gleichverteilten Lichtkegelimpulses:~$\zet \!=\! 1\!/\!2$.
Der transversale Anteil~$\tilde{g}_{V,\la}(k)$ ist die~$1S$-Wellenfunktion des transversalen Harmonischen Oszillators im Impulsraum.%
\FOOT{
  \label{FN-APP:moduloFaktoren}bis auf deren Normierungsfaktor
}

Diese Wellenfunktion~$\tilde{g}_{V,\la}$ Fourier-transformiert bez"uglich~$\rb{k}$ ergibt~$g_{V,\la}$, die~$1S$-Wellen\-funktion des transversalen Harmonischen Oszillators im Ortsraum\citeFN{FN-APP:moduloFaktoren}, vgl.\@ Gl.~(\ref{APP:FT-scalar_k-to-r}):
\vspace*{-.5ex}
\begin{align} \label{APP:FT-scalar_1S}
&\int \frac{d^2\rb{k}}{(2\pi)^2}\; \efn{\D \iIM\,\rb{k} \!\cdot\! \rb{r}}\vv
  \tilde{g}_{V,\la}(k)\;
   =\; \int \frac{d^2\rb{k}}{(2\pi)^2}\; \efn{\D \iIM\,\rb{k} \!\cdot\! \rb{r}}\vv
        \frac{2\pi}{\om_{V,\la}^2}\,
        \exp \bigg[ -\frac{1}{2}\, \om_{V,\la}^{-2}\, k^2 \bigg]
    \\
  &=\;   \exp \bigg[ -\frac{1}{2}\, \om_{V,\la}^2\, r^2 \bigg]\;
  \stackrel{\D!}{=}\; g_{V,\la}(r)
    \nn
    \\[-4.5ex]\nn
\end{align}
und f"ur die Fourier-Transformierte~$\ch_{\iES(\la)}(\zet,r)$ von~$\tilde{\ch}_{\iES(\la)}(\zet,k)$ unmittelbar:
\vspace*{-.5ex}
\begin{align}
\ch_{V\equiv\iES(\la)}(\zet,r)\;
  =\; h_{V,\la}(\zet)\cdot g_{V,\la}(r)
    \\[-4.5ex]\nn
\end{align}
Eingesetzt in Gl.~(\ref{VMeson_r}) und ausgef"uhrt die Ableitung~$\pa_r$, folgt f"ur die Lichtkegelwellenfunktionen des~$1S$-Vek\-tormesons~$V$:
\vspace*{-.5ex}
\begin{alignat}{2} \label{APP:1S-Vektormeson-Wfn}
&\hspace*{-1em}
 \ch_{\iES(\la\equiv0)}&\;
  &=\; 4 \zet\bar{\zet}\vv \om_{\iES,L}\vv \de_{h,-\bar h} \cdot
             g_{\iES,L}(r)\; h_{\iES,L}(\zet)
    \\[.5ex]
&\hspace*{-1em}
 \ch_{\iES(\la\equiv+1)}&\;
  &=\; \Big[\,
         \iIM\, \om_{\iES,T}^2r\, \efn{\T +\iIM\,\vph}\;
           \big( \zet\, \de_{h+,\bar h-} - {\bar\zet}\, \de_{h-,\bar h+} \big)\;
     +\; \meff[]\; \de_{h+,\bar h+}\,
                \Big]\cdot g_{\iES,T}(r)\; h_{\iES,T}(\zet) \nn \\[.5ex]
&\hspace*{-1em}
 \ch_{\iES(\la\equiv-1)}&\;
  &=\; \Big[\,
         \iIM\, \om_{\iES,T}^2r\, \efn{\T -\iIM\,\vph}\;
           \big( {\bar\zet}\, \de_{h+,\bar h-} - \zet\, \de_{h-,\bar h+} \big)\;
     +\; \meff[]\; \de_{h-,\bar h-}\,
                \Big]\cdot g_{\iES,T}(r)\; h_{\iES,T}(\zet) \nn
    \\[-4.5ex]\nn
\end{alignat}
\vspace*{-.25ex}mit abk"urzend~$\ch_{V(\la)}$ f"ur~$\ch_{V(\la)}^{h,\bar h}(\zet,\rb{r})$ und identifiziert~$m_f$ mit der effektiven Quarkmasse~$\meff[] \!\equiv\! \meff$.
Dies ist genau die Darstellung im Haupttext, vgl.\@ die Gln.~(\ref{Vektormeson-Wfn}),~(\ref{E:1S-Vektormeson-Wfn}).

\vspace*{-1ex}
\paragraph{\label{APP-T:1S-ParameterFix}Fixierung der Parameter.}
F"ur feste Helizit"at~$\la \!\equiv\! 0,\pm1$ besitzt die so definierte Lichtkegelwellenfunktion des~$1S$-Vektormesons die genannten zwei Parameter~$\om_{\iES,\la}$,~${\cal N}_{\iES,\la}$.
Diese werden fixiert durch die zwei Forderungen an den Zustand~$\ket{1S}$: da"s er die Kopplung~$f_V$ an den elektromagnetischen Strom reproduziert und normiert ist.
Formal sind dies das System der zwei Gln.~(\ref{fV_LCWfn}),~(\ref{APP:Norm_skalar}) [bzw.~(\ref{fV_LCWfn}$'$),~(\ref{APP:Norm_skalar}$'$)].
Seine L"osung ergibt~$\om_{\iES,\la}$,~${\cal N}_{\iES,\la}$ als Funktionen von~$M_\iES$,~$f_\iES$; diese Konstanten sind experimentell wohlbestimmt.
\vspace*{-.5ex}

\bigskip\noindent
Wir formalisieren das Verfahren und verweisen den technisch weniger interessierten Leser auf den folgenden Unterabschnitt~\ref{APPSubsect:2S-Vektormeson-Wfn}.

Sei zun"achst definiert das absolut konvergente bestimmte Integral:
\vspace{-1ex}
\begin{align} \label{APP:I(T,Z)-Def}
I(T,Z)\;
  \stackrel{\D!}{=}\;
    \int_0^1 d\zet\; [\zbz]^{\D T\!/\!2}\;
    \exp\big\{ -Z\!/\!2 (\zet \!-\! 1\!/\!2)^2 \big\} \qquad
  \text{f"ur\vv ${\rm Re}\,T > -2$}
    \\[-4ex]\nn
\end{align}
das insofern von zentraler Bedeutung ist, als alle im folgenden angegebenen Relationen f"ur die Lichtkegelwellenfunktion~$\ket{1S}$,~$\ket{2S}$ vereinfacht sind bis auf Integrale~$I(T,Z)$.%
\FOOT{
  Diese werden ben"otigt f"ur Argumente~$Z \!\in\! \bbbr^+$ und Indizes~$T \!\in\! \bbbn$; zur Berechnung der Kopplungen~$f_V$ f"ur transversale Polarisation zus"atzlich f"ur~$T \!\equiv\! -1$.   Der Parameter~$T$ wird korreliert sein mit den~$\ta_\la$, vgl.\@ die Gln.~(\ref{APP:h-Wfn_1S}),~(\ref{APP:h-Wfn_2S}), so da"s Konvergenz von~$I(T,Z)$,~${\rm Re}\,T \!>\! -2$, induziert die Forderung~${\rm Re}\,\ta_\la \!>\! 0$.
}

Wir haben diese Relationen zum Teil analytisch hergeleitet erst nach Ver"offentlichung von Ref.~\cite{Kulzinger98}, so da"s wir hier Zusammenh"ange exakt angeben, die dort nur approximiert werden konnten.
Diese wiederum suggerieren ein einfaches Schema zur Fixierung der Parameter, das wir nun diskutieren; Abweichungen zu Ref.~\cite{Kulzinger98} werden hervorgehoben diskutiert.~-- 
So kann das dort numerisch approximierte Integral~$I(T,Z)$ analytisch gel"ost werden:
\vspace*{-.5ex}
\begin{align} \label{APP:I(T,Z)-J(T,Z)}
I(T,Z)\;
  &=\; 2 \exp\{-Z\!/\!8\}\cdot
        \int_0^{1\!/\!2} d\zet\; [\zbz]^{\D T\!/\!2}\;
        \exp\{Z\!/\!2\cdot \zbz\}
    \\[.5ex]
  &=\; 2^{1\!-\! T} \exp\{-Z\!/\!8\}\cdot J(T,Z\!/\!8)
    \tag{\ref{APP:I(T,Z)-J(T,Z)}$'$}
    \\[-4.5ex]\nn
\end{align}
indem wir benutzen
\vspace*{-.5ex}
\begin{align} \label{APP:(z-1/2)^2_zbz}
(\zet \!-\! 1\!/\!2)^2\; =\; 1\!/\!4 \!-\! \zbz
    \\[-4.5ex]\nn
\end{align}
und substituieren~$\la \!=\! 4\zbz$,~$d\zet \!=\! d\la\!/\surd1 \!-\! \la$.
Dabei ist das Integral~$J(T,\be)$ in Gl.~(\ref{APP:I(T,Z)-J(T,Z)}$'$) definiert durch:
\begin{align} 
J(T,\be)\;
  &\stackrel{\D!}{=}\;
     \int_0^1 d\la\; [1 \!-\! \la]^{-1\!/\!2}\; \la^{\D T\!/\!2}\; \exp\{\be\cdot \la\}
\end{align}
F"ur dieses folgt, vgl.\@ Ref.~\cite{Gradstein81},~3.383.1:
\begin{align} 
J(T,\be)\;
  =\; {\rm B}(1\!/\!2,T\!/\!2 \!+\! 1)\vv
      \hyperF{1}{1}(T\!/\!2 \!+\! 1; T\!/\!2 \!+\! 3\!/2; \be) \qquad
  \text{f"ur\vv ${\rm Re}\,T > -2$}
\end{align}
mit~${\rm B}(a,b) \!\equiv\! \Ga(a)\Ga(b)\!/\Ga(a \!+\! b)$ der Eulerschen Beta-Funktion und~\mbox{$\hyperF{1}{1}(a;b;z) \!\equiv\! {\rm M}(a,b,z)$}~der verallgemeinerten Hypergeometrischen oder Kummer-Funktion, vgl.\@ auch Ref.~\cite{Abramowitz84}.
Zusammenfassend f"ur~\mbox{$I(T,Z)$,~${\rm Re}\,T > -2$} folgt:
%
\begin{align} \label{APP:I(T,Z)-explizit}
I(T,Z)\;
  =\; 2^{1\!-\! T} \exp\{-Z\!/\!8\}\cdot
        {\rm B}(1\!/\!2,T\!/\!2 \!+\! 1)\vv
        \hyperF{1}{1}(T\!/\!2 \!+\! 1; T\!/\!2 \!+\! 3\!/2; Z\!/\!8)
\end{align}
Differentiation nach~$Z$ f"uhrt auf die ben"otigte erste Ableitung, vgl.\@ die Gln.~(\ref{APP:I(T,Z)-Def}),~(\ref{APP:(z-1/2)^2_zbz}):
\begin{align} \label{APP:d/dz-I(T,Z)}
d/\!dZ\vv I(T,Z)\;
  =\; (1\!/\!2)\, [I(T \!+\! 2,Z) \!-\! (1\!/\!4)\, I(T,Z)]
\end{align}
das hei"st auf Integrale desselben Typs.

Seien weiter definiert die dimensionslosen Gr"o"sen:%
\FOOT{
  \label{FN-APP:Indizes}Indizes sind unterdr"uckt, vgl.\@ Fu"snote~\FN{FN-APP:ArgumenteIndizes}.
}
%
\begin{align} \label{APP:ze,mu}
\ze\; \equiv\; M^2/\om_{\iES,\la}^2 \qquad
  \text{und}\qquad
  \mu\; \equiv\; m_f^2/M^2
\end{align}
so da"s~$m_f^2 \!/\! \om_{\iES,\la}^2 \!=\! \mu\, \ze$, also~$\ze$ die relevante Variable.

In die Gln.~(\ref{fV_LCWfn}),~(\ref{fV_LCWfn}$'$) und~(\ref{APP:Norm_skalar}),~(\ref{APP:Norm_skalar}$'$) werden eingesetzt die expliziten Ans"atze f"ur die skalaren Funktionen~$\tilde{\ps}_{\iES(\la)}(\zet,k)$; es werden analytisch ausgef"uhrt die~$\rb{k}$-Integrationen und die Integrationen bez"uglich~$\zet$ ausgedr"uckt durch Integrale~$I(T,Z)$.
In Hinblick auf deren explizite Darstellung~-- vgl.\@ Gl.~(\ref{APP:I(T,Z)-explizit})~-- gelangen wir zu den also vollst"andig analytischen Ausdr"ucken:
\vspace*{-.5ex}
\begin{align} \label{APP:fV_LCWfn1S-zet}
\left.
\frac{f_{V,L}}{\hat{e}_V M_V} \right|_{V\equiv\iES}
  &=\; \frac{\surd2}{\pi}\; {\cal N}_{\iES,L}\;\cdot
         \ze^{-1\!/\!2}\;\cdot I(\ta_L, \ze)
    \\[1ex]
\left.
\frac{f_{V,T}}{\hat{e}_V M_V} \right|_{V\equiv\iES}
  &=\; \frac{1}{2}\, \frac{\surd2}{\pi}\; {\cal N}_{\iES,T}\;\cdot
         \ze^{-1}\, \big\{
         - 4\;\cdot I(\ta_T, \ze)
         (2 \!+\! \mu\, \ze)\;\cdot I(\ta_T \!-\! 2, \ze)
         \big\}
    \tag{\ref{APP:fV_LCWfn1S-zet}$'$}
    \\[-3.5ex]\nn
\end{align}
Und f"ur die Gln.~(\ref{APP:Norm_skalar}),~(\ref{APP:Norm_skalar}$'$):
\vspace*{-.5ex}
\begin{align} \label{APP:Norm_skalar1S-zet}
\Nc\; &=\; |{\cal N}_{\iES,L}|^2\;\cdot
           I(2\ta_L, 2\ze)
    \\[1ex]
\Nc\; &=\; |{\cal N}_{\iES,T}|^2\;\cdot \big\{
           - 2\cdot  I(2\ta_T \!+\! 2, 2\ze)
           (1 \!+\! \mu\, \ze)\cdot I(2\ta_T, 2\ze)
         \big\}
    \tag{\ref{APP:Norm_skalar1S-zet}$'$}
    \\[-4.5ex]\nn
\end{align}
Die Gln.~(\ref{APP:Norm_skalar1S-zet}),~(\ref{APP:Norm_skalar1S-zet}$'$) werden aufgel"ost nach~${\cal N}_{\iES,L}$,~${\cal N}_{\iES,T}$, diese o.E.d.A. angenommen als reell positiv und eingesetzt in die Gln.~(\ref{APP:fV_LCWfn1S-zet}),~(\ref{APP:fV_LCWfn1S-zet}$'$).
Es folgen implizite Gleichungen f"ur~$\ze$~-- das hei"st f"ur~$\om_{\iES,L}$,~$\om_{\iES,T}$, vgl.\@ Gl.~(\ref{APP:ze,mu})~--  von der Form:
%
\vspace*{-.5ex}
\begin{align} \label{APP:NewtonApprox0}
F(\ze) = 0
    \\[-3.5ex]\nn
\end{align}
mit Funktionen~$F \!=\! F_L,F_T$, die abh"angen von~$M,f_V,\hat{e}_V$ und f"ur transversale Polarisation zus"atzlich von~$m_f$.%
\FOOT{
  F"ur~$Q^2$ unterhalb einer Schwelle~$Q_{f\!,0}^2$ ist~$m_f \!\equiv\! \meff(Q^2)$ und h"angt effektiv ab von der Virtualit"at~$Q$ des Photons.   Dies beeinflu"st die Parameter f"ur transversale polarisation; vgl.\@ die Diskussion im Haupttext.
}
Der Fixpunkt~$\ze \!\equiv\! \ze_\infty$ folgt durch Iteration auf Basis des quadra\-tisch konvergenten {\it Newtonschen Approximationsverfahrens\/}, vgl.\@ etwa die Refn.~\cite{Abramowitz84,Forster88}:
\vspace*{-.5ex}
\begin{align} \label{APP:NewtonApprox}
\ze_{n+1}\; =\; \ze_n - F(\ze_n)/F'(\ze_n) \qquad
  \ze_\infty\; =\; \lim_{n\to\infty} \ze_n
    \\[-4.5ex]\nn
\end{align}
mit~$\ze_0$ einer approximierten L"osung als Anfangswert. \\
\indent
Praktisch, zur Fixierung der Parameter in Ref.~\cite{Kulzinger98}, wird Gl.~(\ref{APP:NewtonApprox}) noch analytisch berechnet in Termen der Integrale~$I(T,Z)$; f"ur die Ableitung~$F'$ wird benutzt Gl.~(\ref{APP:d/dz-I(T,Z)}).
Die Iteration von Gl.~(\ref{APP:NewtonApprox}) geschieht numerisch mit Anfangswert~$\ze_0 \!\cong\! M_V^2 \!/\! (0.4\GeV)^2$, vgl.\@ Gl.~(\ref{APP:ze,mu}).
Die Auswertung der Integrale~$I(T,Z)$ im Rahmen dieser Iteration geschieht numerisch mithilfe des Gau"s'schen Integrationsverfahrens, vgl.\@ die etwa Refn.~\cite{Press86,Press88}.
Eine Genauigkeit von~$10^{-10}$ bez"uglich des Fixpunktes~$\ze$ ist de facto erreicht nach~$n \!=\! 3 \!-\! 4$ Iterationen.

\vspace{-1ex}
\paragraph{Radius-Erwartungswerte~\bm{\vev{\rb{r}^2}_{\iES\!,\la}}.}
Analog zu den angegebenen Gleichungen zur Fixierung der Parameter k"onnen auf Basis von Gl.~(\ref{APP:rmsq-Radius}) formale Relationen in Termen von Integralen~$I(T,Z)$ angegeben werden f"ur den Erwartungswert~$\vev{\rb{r}^2}_{\iES\!,\la}$, der definiert ist als der transversale rmsq-Radius des~$1S$-Vektormesons f"ur feste \vspace*{-.25ex}Polarisation~$\la \!\equiv\! L,T$.

Eingesetzt in Gl.~(\ref{APP:rmsq-Radius}) die Ortsraum-Darstellungen~$\ps^{h,\bar h}_{V\equiv\iES(\la)}(\zet,\rb{r})$, wie angegeben im Haupttext mit den Gln.~(\ref{Vektormeson-Wfn}),~(\ref{E:Vektormeson-Wfn}) bzw.~(\ref{E:Vektormeson-Wfn}),~(\ref{E:1S-Vektormeson-Wfn}), lauten diese Relationen:
\begin{align} \label{APP:rmsq-Radius-1S}
\vev{\rb{r}^2}_{\iES\!,L}\;
  &=\; \om_{\iES,L}^{-2}
    \\[1ex]
\vev{\rb{r}^2}_{\iES\!,T}\;
  &=\; 2\om_{\iES,T}^{-2}\cdot
         \bigg\{
         1 - \frac{1}{2}\, \mu\, \ze\;
             \bigg/
             \bigg[
               (1 \!+\! \mu\, \ze) \!-\! 2\; \frac{I(2\ta_T \!+\! 2, 2\ze)}{I(2\ta_T, 2\ze)}
             \bigg]
         \bigg\}
    \tag{\ref{APP:rmsq-Radius-1S}$'$}
\end{align}
F"ur den Fall verschwindender Quarkmasse,~$\mu \!\equiv\! 0$, vereinfacht sich Gl.~(\ref{APP:rmsq-Radius-1S}$'$) wesentlich.
\vspace*{-.5ex}

\subsection[\protect$2S$-Vektormeson]{%
                 \bm{2S}-Vektormeson}
\label{APPSubsect:2S-Vektormeson-Wfn}

Wir gehen analog vor f"ur das~$2S$-Vektormeson im Sinne von Kapitel~\ref{Kap:EXCITED}.
Es wird folgende Ansatz gemacht f"ur die skalaren Funktionen~$\tilde{\ch}_{V(\la)}$,~$V \!\equiv\! 2S$:
\vspace*{-.5ex}
\begin{align} \label{scalar2S_k-explizit}
\tilde{\ch}_{V\equiv\iZS(\la)}(\zet,k)
  &= {\cal N}_{V,\la}\, \sqrt{\zet{\bar\zet}}\,
         \exp \bigg[ -\frac{1}{2}\,
                     \frac{M^2 (\zet \!-\! 1\!/\!2)^2}{\om_{V,\la}^2} \bigg]\cdot
      \frac{2\pi}{\om_{V,\la}^2}\,
      \exp \bigg[ -\frac{1}{2}\, \om_{V,\la}^{-2}\, k^2 \bigg]
    \\[-.5ex]
      &\phantom{=\; h_{V,\la}(\zet)\cdot \tilde{g}_{V,\la}(k) }\!\times
         \big\{
           \big(\zet{\bar\zet} \!-\! A_\la\big)
         + B_\la\, \big(1 \!-\! \om_{\iZS,\la}^{-2}k^2\big)
         \big\}
    \nn \\[2ex]
  &=\; h_{V,\la}(\zet)\cdot \tilde{g}_{V,\la}(k)\cdot
         \big\{
           \big(\zet{\bar\zet} \!-\! A_\la\big)
         + B_\la\, \big(1 \!-\! \om_{\iZS,\la}^{-2}k^2\big)
         \big\}
    \tag{\ref{scalar2S_k-explizit}$'$}
    \\[-3.5ex]\nn
\end{align}
mit~$B_\la \!\equiv\! \surd2$; vgl.\@ die Gln.~(\ref{scalar1S_k-explizit}),~(\ref{scalar1S_k-explizit}$'$).
Dabei sind die Funktionen~$\tilde{g}_{\iZS,\la}(r)$,~$h_{\iZS,\la}(\zet)$ bez"uglich ihrer funktionalen Abh"angigkeit identisch den Funktionen~$\tilde{g}_{\iES,\la}(r)$,~$h_{\iES,\la}(\zet)$ nach den Gln.~(\ref{APP:tilde-g-Wfn_1S}),~(\ref{APP:h-Wfn_1S}) differieren aber von diesen durch ihre Parameter,~$V \!\equiv\! 2S$:
\vspace*{-.5ex}
\begin{alignat}{2}
&\tilde{g}_{\iZS,\la}(k)&\;
  &=\; \frac{2\pi}{\om_{\iZS,\la}^2}\,
         \exp \bigg[ -\frac{1}{2}\, \om_{\iZS,\la}^{-2}\, k^2 \bigg]
     \\
&h_{\iZS,\la}(\zet)&\;
  &=\; {\cal N}_{\iZS,\la}\vv [\zbz]^{\D\ta'_\la\!/\!2}\vv
         \exp \bigg[ -\frac{1}{2}\,
                     \frac{M^2 (\zet \!-\! 1\!/\!2)^2}{\om_{\iZS,\la}^2} \bigg] \qquad
       \text{mit}\qquad
       \ta'_\la \equiv 1
    \label{APP:h-Wfn_2S}
    \\[-4.5ex]\nn
\end{alignat}
In Hinblick auf formale Umformungen sind wieder Konstanten~$\ta'_\la$ eingef"uhrt, die analog zu~$\ta_\la$ in~praxi identisch Eins gesetzt werden.
Wir betonen, da"s im "`longitudinalen Exponential"'~$h_{\iZS,\la}$ angesetzt ist mit~$M \!\equiv\! M_\iES$ die Masse des Vektormesons im Grundzustand. \\
\indent
Die Polynome im Ansatz von Gl.~(\ref{scalar2S_k-explizit}) stellen genau die funktionale Diskrepanz dar zum~$1S$-Ansatz, vgl.\@ Gl.\,(\ref{scalar1S_k-explizit}$'$); sie vermitteln die "`Knoten"' der Wellenfunktion.
Der~\mbox{$2S$-Zu}\-stand aufgefa"st als {\it radiale\/} Anregung besitzt dabei "`gleichberechtigt"' eine longitudinale und zwei transversale Moden.
Das "`transversale Polynom"' wird gew"ahlt als das Polynom der~\mbox{$2S$-Wel}\-lenfunktion des transversalen Harmonischen Oszillators und gewichtet mit dem Faktor~$B_\la \!\equiv\! \surd2$.%
\FOOT{
  \label{FN-APP:B_la}Die Konstante~$B_\la$ ist genau die Quadratwurzel der Anzahl transversaler gegen"uber longitudinaler Freiheitsgrade; sie wird eingef"uhrt in Hinblick auf formale Umformungen und in~praxi identisch~$\surd2$ gesetzt.
}
Das "`longitudinale Polynom"' f"uhrt den zus"atzlichen Parameter~$A_\la$ ein.
Unser Ansatz f"ur die~$2S$-Lichtkegelwellenfunktion besitzt f"ur feste Polarisation~$\la \!\equiv\! L,T$ die drei Parameter~$\om_{\iZS,\la}$,~${\cal N}_{\iZS,\la}$,~$A_\la$.

Zur Fourier-Transformation des $2S$-Ansatzes bez"uglich~$\rb{k}$ wird ben"otigt die Transformation der~$2S$-Wellenfunktion im Impulsraum~[$\tilde{g}_{\iZS,\la}$ multipliziert mit dem Impulsraum-Polynom] in die~$2S$-Wellenfunktion des Ortsraums, vgl.\@ Fu"snote~\FN{FN-APP:moduloFaktoren}; diese lautet explizit, vgl.\@ Gl.~(\ref{APP:FT-scalar_1S}):
\begin{align} 
&\int \frac{d^2\rb{k}}{(2\pi)^2}\; \efn{\D \iIM\,\rb{k} \!\cdot\! \rb{r}}\vv
  \frac{2\pi}{\om_{V,\la}^2}\,
  \exp \bigg[ -\frac{1}{2}\, \om_{V,\la}^{-2}\, k^2 \bigg]\cdot
  \big(1 \!-\! \om_{\iZS,\la}^{-2}k^2\big)
    \\
  &=\; \exp \bigg[ -\frac{1}{2}\, \om_{V,\la}^2\, r^2 \bigg]\cdot
        \big(\om_{\iZS,\la}^2r^2 \!-\! 1\big)
    \nn
\end{align}
mit Definition entsprechend Gl.~(\ref{APP:FT-scalar_k-to-r}).

Dies eingesetzt in Gl.~(\ref{VMeson_r}) und ausgef"uhrt die Ableitung~$\pa_r$, folgt f"ur die Lichtkegelwellenfunktionen des~$2S$-Vek\-tormesons~$V$:
%
\begin{alignat}{2} 
&\hspace*{-0pt}
 \ch_{\iZS(\la\equiv0)}&\,
  &=\; 4 \zet\bar{\zet}\vv \om_{\iZS,L}\vv \de_{h,-\bar h} \\
  &&&\phantom{=\;\vv +\; \meff[]\; \de_{h+,\bar h+}\! }\times
         \big\{
           \big(\zet{\bar\zet} \!-\! A_\la\big)
         + \surd2\, \big(\om_{\iZS,\la}^2r^2 \!-\! 1\big)
         \big\}\cdot
         g_{\iZS,L}(r)\; h_{\iZS,L}(\zet)
    \nn \\[1ex]
&\hspace*{-0pt}
 \ch_{\iZS(\la\equiv+1)}&\,
  &=\; \Big[
         \iIM\, \om_{\iZS,T}^2r\, \efn{\T +\iIM\,\vph}\;
           \big( \zet\, \de_{h+,\bar h-} - {\bar\zet}\, \de_{h-,\bar h+} \big)
    \nn \\
  &&&\phantom{=\;\vv +\; \meff[]\; \de_{h+,\bar h+}\! }\times
         \big\{
           \big(\zet{\bar\zet} \!-\! A_\la\big)
         + \surd2\, \big(\om_{\iZS,\la}^2r^2 \!-\! 3\big)
         \big\}
    \nn \\
  &&&\phantom{=\;\vv}
      +\; \meff[]\; \de_{h+,\bar h+}\,\cdot
          \big\{
            \big(\zet{\bar\zet} \!-\! A_\la\big)
          + \surd2\, \big(\om_{\iZS,\la}^2r^2 \!-\! 1\big)
          \big\}
       \Big]\cdot
       g_{\iZS,T}(r)\; h_{\iZS,T}(\zet)
    \nn \\[1ex]
&\hspace*{-0pt}
 \ch_{\iZS(\la\equiv-1)}&\,
  &=\; \Big[
         \iIM\, \om_{\iZS,T}^2r\, \efn{\T -\iIM\,\vph}\;
           \big( \bzet\, \de_{h+,\bar h-} - \zet\, \de_{h-,\bar h+} \big)
    \nn \\
  &&&\phantom{=\;\vv +\; \meff[]\; \de_{h+,\bar h+}\! }\times
         \big\{
           \big(\zet{\bar\zet} \!-\! A_\la\big)
         + \surd2\, \big(\om_{\iZS,\la}^2r^2 \!-\! 3\big)
         \big\}
    \nn \\
  &&&\phantom{=\;\vv}
      +\; \meff[]\; \de_{h-,\bar h-}\,\cdot
          \big\{
            \big(\zet{\bar\zet} \!-\! A_\la\big)
          + \surd2\, \big(\om_{\iZS,\la}^2r^2 \!-\! 1\big)
          \big\}
       \Big]\cdot
       g_{\iZS,T}(r)\; h_{\iZS,T}(\zet)
    \nn
\end{alignat}
\vspace*{-.25ex}mit abk"urzend~$\ch_{V(\la)}$ f"ur~$\ch_{V(\la)}^{h,\bar h}(\zet,\rb{r})$.
Dies ist genau die Darstellung im Haupttext, vgl.\@ Gl.~(\ref{2S-Vektormeson-Wfn}).

\vspace{-1ex}
\paragraph{\label{APP-T:2S-ParameterFix}Fixierung der Parameter}
Wir strukturieren analog zu dem Fall des $1S$-Vektormesons:
Der skizzenhaften Schilderung des Verfahrens folge seine explizite technische Formulierung. \\
\indent
Seien zun"achst die Parameter~$\om_{\iES,\la}$,~${\cal N}_{\iES,\la}$ des $1S$-Vektormesons~$\ket{1S}$ fixiert wie diskutiert in~\ref{APPSubsect:1S-Vektormeson-Wfn}.
F"ur feste Helizit"at~$\la \!\equiv\! 0,\pm1$ besitzt die so definierte Lichtkegelwellenfunktion des $2S$-Vektormesons die drei Parameter~$\om_{\iZS,\la}$,~${\cal N}_{\iZS,\la}$,~$A_\la$.\citeFN{FN-APP:B_la}
Diese werden fixiert durch die drei Forderungen an den Zustand~$\ket{2S}$: da"s er die Kopplung~$f_\iZS$ reproduziert, normiert ist und orthogonal zu dem Zustand~$\ket{1S}$; formal also, da"s er erf"ullt das System der drei Gln.~(\ref{fV_LCWfn}),~(\ref{APP:Norm_skalar}),~(\ref{APP:OrthoGon_skalar}) [bzw.~(\ref{fV_LCWfn}$'$),~(\ref{APP:Norm_skalar}$'$),~(\ref{APP:OrthoGon_skalar}$'$)]. \\
\indent
Die L"osung dieses Systems hat anders zu erfolgen als im Fall des $1S$-Vektormesons:
Eingegangen dort sind die experimentell wohldeterminierten Gr"o"sen~$M_\iES$,~$f_\iES$.
Die entsprechenden Gr"o"sen~$M_\iZS$,~$f_\iZS$ sind nicht bekannt aus zwei Gr"unden.
Zun"achst ist der Zustand~$\ket{2S}$ nicht physikalisch.
Dar"uberhinaus ist insbesondere die Kopplung~$f_\iZS$ nicht berechenbar aus denen der physikalischen Zust"ande~$\ket{\rh(1450)}$,~$\ket{\rh(1700)}$, da auch diese nicht bekannt sind. \\
\indent
Wir gehen folgenderma"sen vor.
In dem longitudinalen Anteil~$h_{\iZS,\la}(\zet)$ nach  Wirbel, Stech, Bauer ist bereits angesetzt die Masse~$M \!\equiv\! M_\iES$ des Grundzustandes.
Dies geschieht im Sinne des Harmonischen Oszillators, in dem allen Zust"anden gemeinsam ist derselbe exponentielle Anteil und der jeweilige Zustand bestimmt ist durch das damit multiplizierte Polynom.
Konsequenterweise setzen wir daher in einem {\it ersten Schritt\/} die Oszillatorparameter des $2S$-Zustands an als identisch denen des $1S$-Zustands:
\vspace*{-.5ex}
\begin{align} \label{APP:Parameter1.}
\hspace*{-1em}
  \text{Schritt~Eins:}\qquad
  \om_{\iZS,\la}\;
    \stackrel{\D!}{=}\; \om_{\iES,\la} \qquad
  \text{$\la \!\equiv\! L,T$,\vv fest}
    \\[-4.5ex]\nn
\end{align}
Die verbleibenden zwei Parameter~${\cal N}_{\iZS,\la}$,~$A_\la$ werden bestimmt durch Forderung von Orthonormalit"at bez"uglich des~$1S$-Zustands, das hei"st durch L"osen des Systems der zwei Gleichungen~(\ref{APP:Norm_skalar}),~(\ref{APP:OrthoGon_skalar})~[bzw.~(\ref{APP:Norm_skalar}$'$),~(\ref{APP:OrthoGon_skalar}$'$)]; ${\cal N}_{\iZS,\la}$,~$A_\la$ folgen in Abh"angigkeit von~$M \!\equiv\! M_\iES$ und~$\om_{\iES,\la}$, das hei"st effektiv in Abh"angigkeit von~$M_\iES$,~$f_\iES$.

Die zwei mal drei Parameter sind vollst"andig fixiert.
Mithilfe der Gl.~(\ref{fV_LCWfn}),~(\ref{fV_LCWfn}$'$)] werden~$f_{\iZS,\la}$ berechnet.%
\FOOT{
  \label{FN-APP:M_2S}In Gl.~(\ref{fV_LCWfn}$'$) f"ur~$f_{\iZS,T}$ geht explizit ein die Masse~$M_\iZS$ des nicht-physikalischen $2S$-Vektormesons, die wir im Sinne eines {\sl educated guess\/} als~$M_\iZS \!=\! 1.6\GeV$ ansetzen.
}
Die resultierenden Kopplungen~$f_{\iZS,L}$,~$f_{\iZS,T}$ sind offensichtlich nicht identisch; dies wird aber gefordert in einem {\it zweiten Schritt\/} und realisiert durch Verr"ucken der Oszillatorparameter des $2S$-Zustands gegen"uber denen des $1S$-Zustands, explizit:
\vspace*{-.5ex}
\begin{align} \label{APP:Parameter2.}
\hspace*{-1em}
  \text{Schritt~Zwei:}\qquad
 &f_{\iZS,L}\;
    \stackrel{\D!}{=}\; f_{\iZS,T} \quad
  \Longrightarrow \quad
  \om_{\iZS,\la}\;
    \ne\; \om_{\iES,\la} \qquad
  \text{$\la \!\equiv\! L,T$,\vv fest}
    \\
 &\om_{\iZS,L} \stackrel{\D!}{=} \om_{\iES,L} (1\!-\! \De)
    \tag{\ref{APP:Parameter2.}$'$} \\[-.5ex]
 &\text{\&}\quad
  \om_{\iZS,T} \stackrel{\D!}{=} \om_{\iES,T} (1\!+\! \De) \qquad
  \text{$|\mskip-2mu\De\mskip-2mu|$\vv minimal}
    \nn
    \\[-4.5ex]\nn
\end{align}
Dabei werden die relativen Verr"uckungen gefordert als {\it gleich\/} und {\it minimal\/}; Identit"at der Kopplungen wird erreicht f"ur~$\De \!>\! 0$, das hei"st~$\om_{\iZS,L}$ ist kleiner,~$\om_{\iZS,T}$ gr"o"ser zu w"ahlen.

Auf Basis der so bestimmten Oszillatorparameter~$\om_{\iZS,\la}$ werden neu fixiert die Parameter~${\cal N}_{\iZS,\la}$,~$A_\la$.
Sie folgen in Abh"angigkeit von~$M$ und~$\om_{\iZS,\la}$ anlog wie im ersten Schritt durch L"osen des Systems der zwei Gln.~(\ref{APP:Norm_skalar}),~(\ref{APP:OrthoGon_skalar}) [bzw.~(\ref{APP:Norm_skalar}$'$),~(\ref{APP:OrthoGon_skalar}$'$)].
Die resultierenden Lichtkegelwellenfunktionen~$\tilde{\ps}^{h,\bar h}_{V(\la)}(\zet,\rb{r})$, mit~$V \!\equiv\! 1S,2S$,~$\la \!\equiv\! L,T$  fest, sind daher zusammenfassend {\it orthonormiert\/} und liefern f"ur longitudinale und transversale Polarisation identische Kopplungen~$f_V$; dabei ist~$f_{\iES} \!\equiv\! f_{\iES,{\rm exp}}$ und~$f_\iZS$ Postulat.
\vspace*{-.5ex}

\bigskip\noindent
Wir formalisieren das geschilderte Verfahren.
Dies geschieht in Termen des Integrals~$I(T,Z)$, vgl.\@ Gl.~(\ref{APP:I(T,Z)-Def}), und der~-- analog zu~$\ze$ in Gl.~(\ref{APP:ze,mu})~-- dimensionslos definierten Gr"o"se:\citeFN{FN-APP:Indizes}
\vspace*{-.5ex}
\begin{align} \label{APP:ze'}
  \ze'\; \equiv\; M^2/\om_{\iZS,\la}^2
    \\[-3.5ex]\nn
\end{align}
mit~$m_f^2 \!/\! \om_{\iZS,\la}^2 \!=\! \mu\, \ze'$ wegen~$\mu \!\equiv\! m_f^2/M^2$, vgl.\@ Gl.~(\ref{APP:ze,mu}); damit ist~$\ze'$ die relevante Variable.

Wie im Fall des $1S$-Vektormesons werden in die Gln.~(\ref{fV_LCWfn}),~(\ref{fV_LCWfn}$'$) und~(\ref{APP:Norm_skalar}),~(\ref{APP:Norm_skalar}$'$) zun"achst eingesetzt die expliziten Ans"atze f"ur die skalaren Wellenfunktionen~$\tilde{\ps}_{\iZS(\la)}(\zet,k)$, analytisch ausgef"uhrt die~$\rb{k}$-Integrationen, ausgedr"uckt durch Integrale~$I(T,Z)$ die~$\zet$-Integra\-tionen.
Es folgt, mit abk"urzend\citeFN{FN-APP:Indizes}~\mbox{$i_k \!=\! I(\ta'_\la + k, \ze')$}:
\begin{align} \label{APP:fV_LCWfn2S-zet}
\left.
\frac{f_{V,L}}{\hat{e}_V M_V} \right|_{V\equiv\iZS}
  &=\; \frac{\surd2}{\pi}\; {\cal N}_{\iZS,L}\;\cdot
       {\ze'}^{-1\!/\!2}\, \big\{
         i_2
         - (A_L \!+\! B_L)\cdot i_0
       \big\}
    \\[1.5ex]
\left.
\frac{f_{V,T}}{\hat{e}_V M_V} \right|_{V\equiv\iZS}
  &=\; \frac{1}{2}\, \frac{\surd2}{\pi}\; {\cal N}_{\iZS,T}
    \tag{\ref{APP:fV_LCWfn2S-zet}$'$} \\
  &\phantom{=\; }\; \times
       {\ze'}^{-1}\, \big\{
         -4\cdot i_2
         + [(2 \!+\! \mu\, \ze') \!+\! 4 (A_T \!+\! 3 B_T)]\cdot i_0
    \nn \\
  &\phantom{=\; \times {\ze'}^{-1}\, \big\{ \;}\;
         - [A_T (2 \!+\! \mu\, \ze') \!+\! B_T (6 \!+\! \mu\, \ze')]\cdot i_{-2}
      \big\}
    \nn
\end{align}
und, mit abk"urzend\citeFN{FN-APP:Indizes}~\mbox{$\ip_k \!\equiv\! I(2\ta'_\la \!+\! 2k, 2\ze')$}:
\begin{align} \label{APP:Norm_skalar2S-zet}
\Nc\; &=\; |{\cal N}_{\iZS,L}|^2\cdot \big\{
           \ip_4
           - 2 A_L\cdot \ip_2
           + (A_L^2 \!+\! B_L^2)\cdot \ip_0
         \big\}
    \\[1.5ex]
\Nc\; &=\; |{\cal N}_{\iZS,T}|^2\cdot \big\{
           -2\cdot \ip_3
           + [(1 \!+\! \mu\, \ze') \!+\! 4 (A_T \!+\! B_T)]\cdot \ip_2
    \tag{\ref{APP:Norm_skalar2S-zet}$'$} \\[1ex]
    &\phantom{=\; |{\cal N}_{\iZS,T}|^2\cdot \big\{ \;}
           - 2 [A_T (1 \!+\! \mu\, \ze')
                    \!+\! B_T (1 \!+\! 2 A_T)
                    \!+\! (A_T^2 \!+\! 3 B_T^2)]\cdot \ip_1
    \nn \\[1ex]
    &\phantom{=\; |{\cal N}_{\iZS,T}|^2\cdot \big\{ \;}
           + [A_T^2 (1 \!+\! \mu\, \ze')
                  \!+\! B_T^2 (3 \!+\! \mu\, \ze')
                  \!+\! 2 A_T B_T]\cdot \ip_0
         \big\}
    \nn
\end{align}
vgl.\@ die Gln.~(\ref{APP:fV_LCWfn1S-zet}),~(\ref{APP:fV_LCWfn1S-zet}$'$) bzw.~(\ref{APP:Norm_skalar1S-zet}),~(\ref{APP:Norm_skalar1S-zet}$'$).
Analog f"uhren die Gln.~(\ref{APP:OrthoGon_skalar}),~(\ref{APP:OrthoGon_skalar}$'$), mit abk"urzend\citeFN{FN-APP:Indizes}~\mbox{$\ipp_k \!=\! I(\ta_\la \!+\! \ta'_\la + 2k, \ze \!+\! \ze')$}, auf:
\vspace*{-.5ex}
\begin{alignat}{2} \label{APP:OrthoGon_skalar-zet}
0\; &=\; {\cal N}_{\iZS,L}^{\D\ast}\, {\cal N}_{\iES,L}\;\cdot
         \big\{
           \ipp_1
           - A_L\cdot \ipp_0
         \big\}
    \\[1.5ex]
0\; &=\; {\cal N}_{\iZS,T}^{\D\ast} {\cal N}_{\iES,T}
    \tag{\ref{APP:OrthoGon_skalar-zet}$'$} \\[1ex]
    &\phantom{=\; \;} \times \big\{
           -2\cdot \ipp_2
           + \big[\big(1 \!+\! \mu\, \sqrt{\ze\ze'}\big)
                  \!+\! 2 (A_T \!+\! B_T)\big]\cdot \ipp_1
           - \big[A_T \big(1 \!+\! \mu\, \sqrt{\ze\ze'}\big)
                  \!+\! B_T\big]\cdot \ipp_0
         \big\}
    \nn
    \\[-3.5ex]\nn
\end{alignat}
und daraus unmittelbar:
\vspace{-.5ex}
\begin{align} \label{APP:Ala_ij}
A_L\; &=\; \frac{\ipp_1}{\ipp_0}
    \\[.5ex]
A_T\; &=\; \frac{2\cdot \ipp_2
                 - \big[\big(1 \!+\! \mu\, \sqrt{\ze\ze'}\big)
                        \!+\! 2 B_T\big]\cdot \ipp_1
                 + B_T \ipp_0}{%
               2\cdot \ipp_1
                 - \big(1 \!+\! \mu\, \sqrt{\ze\ze'}\big)\cdot \ipp_0}
    \tag{\ref{APP:Ala_ij}$'$}
    \\[-4.5ex]\nn
\end{align}
F"ur feste~$\ze$,~$\ze'$, das hei"st f"ur feste Oszillatorparameter~$\om_{\iES,\la}$,~$\om_{\iZS,\la}$, sind die~$A_\la$ vollst"andig~be\-stimmt durch die Forderung von Orthogonalit"at der~$1S$- und~$2S$-Lichtkegelwellenfunktionen. \\
\indent
Diese Ausdr"ucke f"ur~$A_\la$ eingesetzt in die Gln.~(\ref{APP:Norm_skalar2S-zet}),~(\ref{APP:Norm_skalar2S-zet}$'$), folgen unmittelbar die Normierungskonstanten~${\cal N}_{\iZS,L}$,~${\cal N}_{\iZS,T}$ in Abh"angigkeit der~$B_\la \!\equiv\! \surd2$ als Funktion der festen~$\ze$,~$\ze'$.
Weiter mithilfe der Gln.~(\ref{APP:fV_LCWfn2S-zet}),~(\ref{APP:fV_LCWfn2S-zet}$'$) folgen schlie"slich die Kopplungen~$f_{\iZS,L}$,~$f_{\iZS,T}$. \\
\indent
\label{APP-T:approx-ort,orth}In Ref.~\cite{Kulzinger98} werden die Parameter~$A_L$,~$A_T$ numerisch approximiert; \vspace*{-.375ex}effektiv wird im Sinne der Gln.~(\ref{APP:OrthoGon_skalar-zet}),~(\ref{APP:OrthoGon_skalar-zet}$'$) implizit gesetzt~$\ze \!\to\! \ze'$.
Konsequenz ist, da"s die Lichtkegelfunktion~$\ket{2S}$ nur approximativ, nicht exakt orthogonal auf~$\ket{1S}$ ist.
Auf Basis der Relationen~(\ref{APP:Ala_ij}),~(\ref{APP:Ala_ij}$'$) und mithilfe des analytischen Ausdrucks f"ur die Integrale~$I(T,Z)$, vgl.\@ Gl.~(\ref{APP:I(T,Z)-explizit}), ist diese N"aherung nicht mehr erforderlich; es folgen f"ur~$A_L$,~$A_T$ unmittelbar Zahlenwerte f"ur beliebige nicht-identische~$\ze$,~$\ze'$.~--
Wir gelangen zu {\it exakter statt approximativer Orthogonalit"at}.
Im Haupttext sind dokumentiert die Konsequenzen beider Ans"atze; sei hier nur angemerkt, da"s die numerische Diskrepanz marginal ist.

Orthonormalit"at und Reproduktion identischer Kopplungen~$f_{\iZS,L}$,~$f_{\iZS,T}$ nach dem diskutierten Verfahren h"angen implizit ab "uber die effektive Quarkmasse~$\meff[]$ von~$Q^2$.
Auf Basis exakter Orthogonalit"at f"ur jedes~$Q^2$ wird in Kapitel~\ref{Subsect:2SParameter} quantitativ diskutiert: zum einen Fixieren der Oszillatorparameter~$\om_{\iZS,\la}$ f"ur~$Q^2 \!\equiv\! 0$ wie geschildert und Einfrieren auf diese Werte und zum anderen ihre laufende Anpassung mit~$Q^2$; das hei"st identische Kopplungen werden generiert nur f"ur~$Q^2 \!\equiv\! 0$ beziehungsweis f"ur jedes~$Q^2$.

In Tabelle~\ref{Tabl-APP:om1ST,De,f2S} sind angegeben Zahlenwerte f"ur~$\om_{\iES,T}(Q^2)$ und~$\De(Q^2)$ in gen"ugender 
Genauigkeit und f"ur das gesamte~$Q^2$-Intervall, in dem eine nichttriviale Abh"angigkeit besteht; der Zahlenwert f"ur den bez"uglich~$Q^2$ konstanten Parameter~$\om_{\iES,L}$ ist angegeben in der Unterschrift.
Die Oszillatorparameter folgen aus den Gln.~(\ref{APP:fV_LCWfn1S-zet}),~(\ref{APP:fV_LCWfn1S-zet}$'$) und~(\ref{APP:Norm_skalar1S-zet}),~(\ref{APP:Norm_skalar1S-zet}$'$), durch numerische Iteration auf Basis des Newtonschen Approximationsverfahrens, vgl.\@ die Gln.~(\ref{APP:NewtonApprox0}),\,(\ref{APP:NewtonApprox}).
Die Funktion~$\De(Q^2)$, durch die determiniert ist die Verr"uckung der~$\om_{\iZS,\la}$ relativ zu den~$\om_{\iES,\la}$, wird bestimmt durch Iteration von Hand in der Weise, da"s f"ur jedes~$Q^2$ die Lichtkegelwellenfunktion~$\ket{2S}$ erf"ullt: Reproduktion identischer Kopplungen~$f_{\iZS,L}$,~$f_{\iZS,T}$, Normiertheit und Orthogonalit"at auf~$\ket{1S}$, vgl.\@ die Gln.~(\ref{APP:fV_LCWfn2S-zet}),~(\ref{APP:fV_LCWfn2S-zet}$'$),~(\ref{APP:Norm_skalar2S-zet}),~(\ref{APP:Norm_skalar2S-zet}$'$) bzw.~(\ref{APP:OrthoGon_skalar-zet}),~(\ref{APP:OrthoGon_skalar-zet}$'$).
Wir gehen hierauf ausf"uhrlich ein im Haupttext in~\ref{Subsect:2SParameter}.
Auf Basis von Tabelle~\ref{Tabl-APP:om1ST,De,f2S} reduziert sich die Bestimmung s"amtlicher "ubriger Parameter auf das Auswerten der angegebenen Relationen in Termen des analytisch gel"osten Integrals~$I(T,Z)$, vgl.\@ Gl.~(\ref{APP:I(T,Z)-explizit})~-- Fixpunkte impliziter Gleichungen sind nicht mehr zu bestimmen.
\begin{table}
\begin{minipage}{\linewidth}
\begin{center}
  \begin{tabular}{|c||c|c||c|c|} \hline
  \multicolumn{5}{|c|}{Parameter~$\om_{\iES,T}$,~$\De$\vv%
                       [und resultierende~$f_{\iZS,L}$,~$f_{\iZS,T}$]}
    \\ \hhline{:=:t:==:t:==:}
  \multicolumn{1}{|c||}{$Q^2\;[\GeV[]^2]$}
    & \multicolumn{1}{c|}{$\om_{\iES,T}\;[\GeV[]]$}
    & \multicolumn{1}{c||}{$\De$}
    & \multicolumn{1}{c|}{$f_{\iZS,L}\;[\GeV[]]$}
    & \multicolumn{1}{c|}{$f_{\iZS,T}\;[\GeV[]]$}
  \\ \hhline{|-||--||--|}
$0.00$ &  \;$0.212\,9547\,611$\; &  \;$0.097\,6137\,4330$\;
    &  \,$-\,0.137\,7811\,726$\, &  \,$-\,0.137\,7811\,726$\, \\
$0.05$ &  $0.213\,8661\,612$ &  $0.098\,9565\,6500$
    &  $-\,0.137\,5566\,129$ &  $-\,0.137\,5566\,129$ \\
$0.10$ &  $0.214\,6983\,169$ &  $0.100\,2242\,9370$
    &  $-\,0.137\,3445\,901$ &  $-\,0.137\,3445\,901$ \\[.5ex]
$0.15$ &  $0.215\,4476\,720$ &  $0.101\,4176\,8150$
    &  $-\,0.137\,1449\,819$ &  $-\,0.137\,1449\,819$ \\
$0.20$ &  $0.216\,1112\,763$ &  $0.102\,5377\,6735$
    &  $-\,0.136\,9576\,179$ &  $-\,0.136\,9576\,179$ \\
$0.25$ &  $0.216\,6869\,190$ &  $0.103\,5858\,6818$
    &  $-\,0.136\,7822\,808$ &  $-\,0.136\,7822\,808$ \\[.5ex]
$0.30$ &  $0.217\,1732\,806$ &  $0.104\,5635\,6470$
    &  $-\,0.136\,6187\,090$ &  $-\,0.136\,6187\,090$ \\
$0.35$ &  $0.217\,5701\,003$ &  $0.105\,4726\,7800$
    &  $-\,0.136\,4666\,004$ &  $-\,0.136\,4666\,004$ \\
$0.40$ &  $0.217\,8783\,607$ &  $0.106\,3152\,3500$
    &  $-\,0.136\,3256\,183$ &  $-\,0.136\,3256\,183$ \\[.5ex]
$0.45$ &  $0.218\,1004\,851$ &  $0.107\,0934\,1830$
    &  $-\,0.136\,1953\,994$ &  $-\,0.136\,1953\,994$ \\
$0.50$ &  $0.218\,2405\,397$ &  $0.107\,8094\,9310$
    &  $-\,0.136\,0755\,668$ &  $-\,0.136\,0755\,668$ \\
$0.55$ &  $0.218\,3044\,271$ &  $0.108\,4656\,9840$
    &  $-\,0.135\,9657\,473$ &  $-\,0.135\,9657\,473$ \\[.5ex]
$0.60$ &  $0.218\,3000\,459$ &  $0.109\,0640\,9450$
    &  $-\,0.135\,8655\,976$ &  $-\,0.135\,8655\,976$ \\
$0.65$ &  $0.218\,2373\,841$ &  $0.109\,6063\,4830$
    &  $-\,0.135\,7748\,401$ &  $-\,0.135\,7748\,401$ \\
$0.70$ &  $0.218\,1284\,961$ &  $0.110\,0934\,6085$
    &  $-\,0.135\,6933\,085$ &  $-\,0.135\,6933\,085$ \\[.5ex]
$0.75$ &  $0.217\,9873\,072$ &  $0.110\,5254\,5573$
    &  $-\,0.135\,6209\,996$ &  $-\,0.135\,6209\,996$ \\
$0.80$ &  $0.217\,8291\,845$ &  $0.110\,9010\,9350$
    &  $-\,0.135\,5581\,221$ &  $-\,0.135\,5581\,221$ \\
$0.85$ &  $0.217\,6702\,266$ &  $0.111\,2177\,2335$
    &  $-\,0.135\,5051\,204$ &  $-\,0.135\,5051\,204$ \\[.5ex]
$0.90$ &  $0.217\,5262\,661$ &  $0.111\,4714\,1320$
    &  $-\,0.135\,4626\,535$ &  $-\,0.135\,4626\,535$ \\
$0.95$ &  $0.217\,4116\,374$ &  $0.111\,6574\,7315$
    &  $-\,0.135\,4315\,072$ &  $-\,0.135\,4315\,072$ \\
$1.00$ &  $0.217\,3378\,524$ &  $0.111\,7713\,5375$
    &  $-\,0.135\,4124\,433$ &  $-\,0.135\,4124\,433$ \\[.5ex]
$1.05$ &  $0.217\,3123\,918$ &  $0.111\,8097\,2285$
    &  $-\,0.135\,4060\,203$ &  $-\,0.135\,4060\,203$ \\
$1.10$ &  $0.217\,3123\,918$ &  $0.111\,8097\,2285$
    &  $-\,0.135\,4060\,203$ &  $-\,0.135\,4060\,203$
  \\ \hhline{:=:b:==:b:==:}
  \end{tabular}
  \end{center}
\vspace*{-3ex}
\caption[\protect$Q^2$-Abh"angigkeit von~\protect$\om_{\iES,T}$,~\protect$\De$~-- und resultierende~\protect$f_{\iZS,L}$,~\protect$f_{\iZS,T}$]{
  \vspace*{-.125ex}Abh"angigkeit von~$Q^2$ der Parameter~$\om_{\iES,T}$,~$\De$~-- und der resultierende Kopplungen~$f_{\iZS,L}$,~$f_{\iZS,T}$.   Auf Basis dieser Zahlewerte f"ur~$\om_{\iES,T}(Q^2)$,~$\De(Q^2)$ und des konstanten \vspace*{-.125ex}Zahlenwerts~\mbox{$\om_{\iES,L} \!\equiv\! 0.329\,9209\,590\GeV$} k"onnen s"amtliche Parameter der Lichtkegelwellenfunktionen~$\ket{1S}$,\,$\ket{2S}$ berechnet werden ohne Fixpunkt-Gleichungen l"osen zu m"ussen.   Die {\it numerische\/} Genauigkeit sei dokumentiert durch die letzten Spalten, die Konstanz f"ur~$Q^2 \!\ge\! 1.05\GeV^2$ durch die letzte Zeile.
\vspace{-1.5ex}
}
\label{Tabl-APP:om1ST,De,f2S}
\end{minipage}
\end{table} 
\vspace{-1.5ex}
\paragraph{\vspace*{-.25ex}Radius-Erwartungswerte~\bm{\vev{\rb{r}^2}_{\iZS\!,\la}}.}
Wir geben die formalen Relationen an f"ur den transversalen rmsq-Radius~$\vev{\rb{r}^2}_{\iZS\!,\la}$ f"ur feste \vspace*{-.625ex}Polarisation~$\la \!\equiv\! L,T$: \\
\indent
Eingesetzt in Gl.~(\ref{APP:rmsq-Radius}) die Ortsraum-Darstellungen~$\ps^{h,\bar h}_{V\equiv\iZS(\la)}(\zet,\rb{r})$, wie angegeben im Haupttext mit den Gln.~(\ref{E:Vektormeson-Wfn}),~(\ref{2S-Vektormeson-Wfn}), und abk"urzend~\mbox{$\ip_k \!=\! I(2\ta'_\la + 2k, 2\ze')$} [wie in den Gln.~(\ref{APP:Norm_skalar2S-zet}),~(\ref{APP:Norm_skalar2S-zet}$'$)], lauten diese Relationen explizit:
\vspace*{-.5ex}
\begin{align} \label{APP:rmsq-Radius-2S}
\vev{\rb{r}^2}_{\iZS\!,L}\;
  &=\; \om_{\iES,L}^{-2}\; |{\cal N}_{\iZS,L}|^2
    \\[.5ex]
    &\phantom{=\; \;} \times \big\{
           \ip_2
           + 2 (-A_L \!+\! B_L)\cdot \ip_1
           + B_L (-2 A_L \!+\! 3 B_L)\cdot \ip_0
         \big\}
    \nn \\[1ex]
\vev{\rb{r}^2}_{\iZS\!,T}\;
  &=\; \om_{\iES,T}^{-2}\; |{\cal N}_{\iZS,T}|^2\;\cdot
       \big\{
         -4\cdot \ip_3
         + [(2 \!+\! \mu\, \ze') \!+\! 8 A_T]\cdot \ip_2
    \tag{\ref{APP:rmsq-Radius-2S}$'$} \\[.5ex]
    &\phantom{=\; \om_{\iES,T}^{-2}\; |{\cal N}_{\iZS,T}|^2\;\cdot \big\{ \;}
         - 2 [A_T (2 \!+\! \mu\, \ze')
              \!+\! 2 A_T^2
              \!-\! B_T\, \mu\, \ze'
              \!+\! 6 B_T^2]\cdot \ip_1
    \nn \\[.5ex]
    &\phantom{=\; \om_{\iES,T}^{-2}\; |{\cal N}_{\iZS,T}|^2\;\cdot \big\{ \;}
         + [A_T^2 (2 \!+\! \mu\, \ze')
              \!-\! 2 A_T B_T\, \mu\, \ze' 
              \!+\! 3 B_T^2 (2 \!+\!  \mu\, \ze')]\cdot \ip_0
         \big\}
    \nn
    \\[-4.5ex]\nn
\end{align}
F"ur den Fall verschwindender Quarkmasse,~$\mu \!\equiv\! 0$, vereinfacht sich Gl.~(\ref{APP:rmsq-Radius-2S}$'$).
\vspace*{-.5ex}

\bigskip\noindent
Der Modellierung der longitudinalen~$2S$-Anregungsmode durch ein Polynom ersten Grades haftet eine gewisse Willk"ur an, so da"s wir unsere Analyse vollst"andig neu durchf"uhren f"ur:
\vspace*{-.25ex}
\begin{align} \label{APP:Pol2}
(\zbz \!-\! A_\la)\vv
  \longrightarrow\vv
  \zbz(\zbz \!-\! A_\la)
    \\[-4ex]\nn
\end{align}
Die Resultate sind zusammenfassend sehr "ahnlich.
Daher beschr"anken wir uns in dieser~Ar\-beit auf das Polynom ersten Grades.
F"ur~Voll\-st"andigkeit seien hier aber~-- ohne~weite\-ren Kommentar~-- festgehalten die relevanten Relationen f"ur das Polynom zweiten Grades. \\
\indent
Mit den Definitionen~$i_k$,~$\ip_k$,~$\ipp_k$ wie eingef"uhrt, gilt f"ur die Kopplung an das Photon:
\vspace*{-.5ex}
\begin{align} \label{APP:fV_LCWfn2S-zet_Pol2}
\left.
\frac{f_{V,L}}{\hat{e}_V M_V} \right|_{V\equiv\iZS}
  &=\; \frac{\surd2}{\pi}\; {\cal N}_{\iZS,L}\;\cdot
       {\ze'}^{-1\!/\!2}\, \big\{
         i_4
         - A_L\cdot i_2
         - B_L\cdot i_0
       \big\}
    \\
\left.
\frac{f_{V,T}}{\hat{e}_V M_V} \right|_{V\equiv\iZS}
  &=\;  \frac{1}{2}\, \frac{\surd2}{\pi}\; {\cal N}_{\iZS,T}
    \tag{\ref{APP:fV_LCWfn2S-zet_Pol2}$'$} \\[-.25ex]
  &\phantom{=\;}\; \times
       {\ze'}^{-1}\, \big\{
         - 4\cdot i_4
         + [2 (1 \!+\! 2 A_T) \!+\! \mu\, \ze']\cdot i_2
    \nn \\
  &\phantom{=\; \times {\ze'}^{-1}\, \big\{ \;}\;
         + [2 (6 B_T \!-\! A_T) \!-\! A_T\; \mu\, \ze'\big]\cdot i_0
         - B_T(6 \!+\! \mu\, \ze')\cdot i_{-2}
       \big\}
    \nn
    \\[-5.5ex]\nn
\end{align}
f"ur Normierung:
\vspace*{-.5ex}
\begin{align} \label{APP:Norm_skalar2S-zet_Pol2}
\hspace*{-10pt}
\Nc\; &=\; |{\cal N}_{\iZS,L}|^2\cdot \big\{
           \ip_4
           - 2 A_L\cdot \ip_3
           + A_L^2\cdot \ip_2
           + B_L^2\cdot \ip_0 
         \big\}
    \\[1.5ex]
\hspace*{-10pt}
\Nc\; &=\; |{\cal N}_{\iZS,T}|^2
    \tag{\ref{APP:Norm_skalar2S-zet_Pol2}$'$} \\[1ex]
    &\phantom{=\; \;} \times \big\{
             -2\cdot \ip_5
             + (1 \!+\! \mu\, \ze' \!+\! 4 A_T)\cdot \ip_4
             - 2 [A_T (1 \!+\! \mu\, \ze') \!-\! 2 B_T \!+\! A_T^2]\cdot \ip_3
    \nn \\[1ex]
    &\phantom{=\; \; \times \big\{ \;}
             + \big[A_T^2 (1 \!+\! \mu\, \ze') \!-\! 2 B_T (1 \!+\! 2 A_T)]\cdot \ip_2
             + 2 B_T (A_T \!-\! 3 B_T)\cdot \ip_1
             + B_T^2 (3 \!+\! \mu\, \ze')\cdot \ip_0
         \big\}
    \nn
    \\[-4.5ex]\nn
\end{align}
f"ur Orthogonalit"at:
\vspace*{-.5ex}
\begin{alignat}{2} \label{APP:OrthoGon_skalar-zet_Pol2}
0\; &=\; {\cal N}_{\iZS,L}^{\D\ast}\, {\cal N}_{\iES,L}\;\cdot
         \big\{
           \ipp_2
           - A_L\cdot \ipp_1
         \big\}
    \\[1.5ex]
0\; &=\; {\cal N}_{\iZS,T}^{\D\ast} {\cal N}_{\iES,T}
    \tag{\ref{APP:OrthoGon_skalar-zet_Pol2}$'$} \\[1ex]
    &\phantom{=\; \;} \times \big\{
             -2\cdot \ipp_3
             + \big[\big(1 \!+\! \mu\, \sqrt{\ze\ze'}\big)
                    \!+\! 2 A_T\big]\cdot \ipp_2
             + \big[2 B_T \!-\! A_T \big(1 \!+\! \mu\, \sqrt{\ze\ze'}\big)\big]\cdot \ipp_1
             - B_T\cdot \ipp_0
         \big\}
    \nn
    \\[-4.5ex]\nn
\end{alignat}
daraus unmittelbar:
\vspace*{-.5ex}
\begin{align} \label{APP:Ala_ij_Pol2}
A_L\; &=\; \frac{\ipp_2}{\ipp_1}
    \\[.5ex]
A_T\; &=\; \frac{2\cdot \ipp_3
                   - \big(1 \!+\! \mu\, \sqrt{\ze\ze'}\big)\cdot \ipp_2
                   - B_T (2\cdot \ipp_1 - \ipp_0)}{%
                 2\cdot \ipp_2
                   - \big(1 \!+\! \mu\, \sqrt{\ze\ze'}\big)\cdot \ipp_1}
    \tag{\ref{APP:Ala_ij_Pol2}$'$}
    \\[-4.5ex]\nn
\end{align}
und f"ur die Quadrate der transversalen rmsq-Radien:
%
\begin{align} \label{APP:rmsq-Radius-2S_Pol2}
\vev{\rb{r}^2}_{\iZS\!,L}\;
  &=\; \om_{\iES,L}^{-2}\; |{\cal N}_{\iZS,L}|^2
    \\[1ex]
    &\phantom{=\; \;} \times \big\{
           \ip_4
           - 2 A_L\cdot \ip_3
           + (A_L^2 \!+\! 2 B_L)\cdot \ip_2
           - 2 A_L B_L\cdot \ip_1
           + 3 B_L^2\cdot \ip_0 
         \big\}
    \nn \\[1.5ex]
\vev{\rb{r}^2}_{\iZS\!,T}\;
  &=\; \om_{\iES,T}^{-2}\; |{\cal N}_{\iZS,T}|^2
    \tag{\ref{APP:rmsq-Radius-2S_Pol2}$'$} \\[1ex]
    &\phantom{=\; \;} \times \big\{
           -4\cdot \ip_5
           + [2 (1 \!+\! 4 A_T) \!+\! \mu\, \ze']\cdot \ip_4
           - 2 [2 A_T (1 \!+\! A_T) \!+\! A_T\; \mu\, \ze']\cdot \ip_3
    \nn \\[1ex]
    &\phantom{=\; \; \times \big\{ \;}
           + [2 A_T^2 \!+\! (A_T^2 \!+\! 2 B_T) \mu\, \ze']\cdot \ip_2
    \nn \\[1ex]
    &\phantom{=\; \; \times \big\{ \;}
           - 2 B_T (6 B_T \!+\! A_T^2 \mu\, \ze')\cdot \ip_1
           + 3 B_T^2 (2 \!+\! \mu\, \ze')\cdot \ip_0
         \big\}
    \nn
\end{align}
Die Relationen f"ur transversale Polarisation h"angen wieder explizit ab von der (effektiven) Quarkmasse.
\vspace*{-.5ex}

\section{Kompendium}

Wir leiten diverse Relationen her, die im Haupttext benutzt werden; bzgl.\@ der Konventionen und Notationen der folgenden beiden Abschnitte, vgl.\@ Anh.~\ref{APP:Dirac-Algebra} und Nachtmann in Ref.~\cite{Nachtmann92}.

\subsection[Leptonische Zerfallsbreite~\protect$\Gall_V$]{%
            Leptonische Zerfallsbreite~\protect\bm{\Gall_V}}

Die Leptonische Zerfallsbreite des Vektormesons~$V$ ist bestimmt durch das $S$-Matrixelement
\begin{align} \label{APP:S-Matrix} 
S\; =\; \bra{l^-(p,r)\, l^+(p',r')}S\ket{V(q,\la)}
\end{align}
Dabei bezeichnen~$p$,~$p'$,~$q$ die Impulse,~$r$,~$r'$,~$\la$ die Spins und Helizit"aten des Leptonpaares.
F"ur das~$T$-Matrixelement, definiert durch~$S =: \iIM\,(2\pi)^4 \de_4(p\!+\!p'\!-\!q)\; T$, folgt daraus:
\begin{align} \label{APP:T-Matrix}
T(s)\; &=\; -e\, \bar{u}_r(p) \ga^\mu v_{r'}(p')\;
         \frac{g_{\mu\nu}}{(p \!+\! p')^2 + \iIM\,\vep}\;
         \bra{\Om}J_{em}^{\nu}(0)\ket{V(q,\lambda)}
    \\
    &=\; -e^2 \bar{u}_r(p) \ga^\mu v_{r'}(p')\;
         \frac{1}{s}\; f_V M_V \vep_{\mu}(q,\la)
    \tag{\ref{APP:T-Matrix}$'$}
\end{align}
mit~$s \!=\! (p \!+\! p')^2$ und~$f_V$ in der Definition:
\begin{align} \label{APP:fV-Def}
\bra{\Om}J_{\rm em}^{\mu}(0)\ket{V(q,\la)}\;
  =\; e f_V\, M_V\, \vep^{\mu}(q,\la)
\end{align}
Das Absolut-Quadrat von Gl.~(\ref{APP:T-Matrix}$'$) wird gemittelt "uber die einlaufenden Spins~$r$,~$r'$ und summiert "uber die auslaufenden Helizit"aten:
\begin{alignat}{2} \label{APP:T-Matrix_sum'}
{\T\sum}'\!|T(s)|^2\;
  &=&\; -\frac{e^4}{3}\;
        &\Big[\frac{f_V M_V}{s}\Big]^2\vv
         {\T\sum}_{r,r'}\; \tr
         \big[\ga^\mu u_r(p) \bar{u}_r(p) \ga_\mu v_{r'}(p') \bar{v}_{r'}(p')\big]
    \\
  &=&\; +\frac{4e^4}{3}\;
        &\Big[\frac{f_V M_V}{s}\Big]^2\vv
         s\cdot \bigg[1 \!+\! \frac{2m_l^2}{s}\bigg]
    \tag{\ref{APP:T-Matrix_sum'}$'$}
\end{alignat}
mit~$m_l$ der Masse des Leptons.
Die Zerfallsrate des Vektormesons in dessen Ruhesystem ist:
\begin{align} \label{APP:dGall}
d\Gall_V\; =\; \frac{1}{2M_V}\; (2\pi)^4 \de_{(4)}(p \!+\! p' \!-\! q)\;
             d^{\mskip-1mu R}\!p\, d^{\mskip-1mu R}\!p'\vv {\T\sum}'\!|T(s)|^2 \qquad
  s \!=\! M_V^2 
\end{align}
mit dem Lorentz-invarianten Ma"s~\mbox{$d^{\mskip-1mu R}\!k \!\equiv\! d^3\vec{k}\!/(2\pi)^32k_{0+} \!=\! d^4k\, \th(k_0)\, \de(k^2\!-\!m^2)/(2\pi)^3$}.
Eingesetzt Gl.~(\ref{APP:T-Matrix_sum'}$'$) und ausgef"uhrt die Phasenraum-Integrationen, folgt:%
\FOOT{
  \label{FN-APP:ml-Faktoren}F"ur Elektronen k"onnen die~$m_l$-abh"angigen Faktoren vernachl"assigt werden.
}
%
\begin{align} \label{APP:fV_Gall}
\Gall_V\;
  =\; \frac{4\pi\al_{\rm em}^2}{3} \frac{f_V^2}{M_V}\cdot
        \bigg[ 1 \!+\! \frac{2m_l^2}{M_V^2} \bigg]
         \sqrt{1 \!-\! \frac{4m_l^2}{M_V^2}}
\end{align}
dabei ist~$\al_{\rm em} \!=\! e^2\!/4\pi \!\cong\! 1\!/137.04$.
Die Gln.~(\ref{APP:fV-Def}),~(\ref{APP:fV_Gall}) werden explizite diskutiert im Haupttext, vgl.\@ die Gln.~(\ref{fV-Def}),~(\ref{fV_Gall}) auf Seite~\pageref{fV-Def}.

\subsection[Photoproduktion und
            \protect$l^+l^-$-Annihilation in~\protect$\pi^+\pi^-\!,\,2\pi^+2\pi^-$]{%
            Photoproduktion und
            \protect\bm{l^+l^-}-Annihilation in~\protect\bm{\pi^+\pi^-\!,\,2\pi^+2\pi^-}}
\label{APPSubsect:Photo,l+l-Annih}

Der differentielle Wirkungsquerschnitt~$d\si_{l^+l^-\to V}$ f"ur die Produktion eines realen Vektormesons~$V$ in~$l^+l^-$-Annihilation wird berechnet auf Basis desselben $S$-Matrixelementes, vgl.\@ Gl.~(\ref{APP:S-Matrix}); mit~${\T\sum}'\!|T(s)|^2$ nach Gl.~(\ref{APP:T-Matrix_sum'}$'$) gilt:
\begin{align} 
d\si_{l^+l^-\to V}\;
  =\; \frac{1}{2w(s,m_l^2,m_l^2)}\; (2\pi)^4 \de_{(4)}(p \!+\! p' \!-\! q)\;
        d^{\mskip-1mu R}\!q\vv {\T\sum}'\!|T(s)|^2
\end{align}
mit~$w(\al,\be,\ga) \!=\! [\al^2 \!+\! \be^2 \!+\! \ga^2 \!-\! 2(\al\be \!+\! \be\ga \!+\! \ga\al)]^{1\!/\!2} \!=\! [\al \!-\! (\sqrt\be \!+\! \sqrt\ga)^2]^{1\!/\!2} [\al \!-\! (\sqrt\be \!-\! \sqrt\ga)^2]^{1\!/\!2}$ der vollst"andig symmetrischen K"allen-Funktion.
Daraus folgt durch Ausf"uhren der Phasenraum-Integrationen:\citeFN{FN-APP:ml-Faktoren}
\vspace*{-1ex}
\begin{align} 
\si_{l^+l^-\to V}\;
  =\; \frac{4\pi e^4}{3} \Big[\frac{f_V}{M_V}\Big]^2\; \de(s - M_V^2)\cdot
        \bigg[ 1 \!+\! \frac{2m_l^2}{M_V^2} \bigg] \Bigg/\zz
         \sqrt{1 \!-\! \frac{4m_l^2}{M_V^2}}
    \\[-4.5ex]\nn
\end{align}

Die Berechnung der Rho-Kanal-Massenspektren f"ur Photoproduktion und~$e^+e^-$-Annihi\-lation in zwei und vier geladene Pionen geschieht durch Distribution der Zust"ande~\mbox{$\rh,\rh',\rh^\dbprime$} \mbox{[$\equiv\! \rh(770),\rh(1450),\rh(1700)$]}, als einfache Breit-Wigner-Resonanzen:
\begin{align} \label{APP:TM}
T_M\; \stackrel{\D!}{=}\;
  T \cdot \frac{c_{V\!,f}\cdot \sqrt{M_V\Gatot_V\!/\pi}}{%
                M^2 \!-\! M_V^2 \!+\! \iIM\, M_V\Gatot_V} \qquad
  f \!\equiv\! \pi^+\pi^-, 2\pi^+2\pi^- 
\end{align}
mit~$M$ der invarianten Masse des Endzustands~$f$.
Die Normierungskonstanten~$c_{V\!,f}$ sind definiert "uber den Zusammenhang der~$T$-Amplituden:
\vspace*{-.5ex}
\begin{align} \label{APP:int-TM}
\int_{s_f}^\infty dM^2\, \big|T_M\big|^2\;
  =\; \big|T\big|^2
    \\[-4.5ex]\nn
\end{align}
mit~$s_f \!\equiv\! (2m_\pi)^2,(4m_\pi)^2$ den kinematischen Schwellen der Reaktionen.
Eingesetzt~$T_M$ nach Gl.~(\ref{APP:TM}) in Gl.~(\ref{APP:int-TM}), kann der Faktor~$|T|^2$ gek"urzt werden; wir erhalten unmittelbar:
\vspace*{-.5ex}
\begin{align} \label{APP:cVla}
c_{V\!,f}\;
  =\; \left[\frac{1}{2} + \frac{1}{\pi}\, \arctan\frac{M_V^2 \!-\! s_f}{M_V}\right]^{-1\!/\!2}
    \\[-4.5ex]\nn
\end{align}
Bzgl.\@ ihrer Abweichung von Eins vgl.\@ Tabl.~\ref{Tabl-App:cVla}.
\begin{table}
\begin{minipage}{\linewidth}
\begin{center}
  \begin{tabular}{|c||g{7}|g{7}|g{7}|} \hline
  \multicolumn{4}{|c|}{Normierungsparameter der Breit-Wigner-Distributionen:}\\[-.375ex]
  \multicolumn{4}{|c|}{Abweichung~von Eins\;~[$c_{V\!,f} \!-\! 1$]}
    \\[.375ex] \hhline{:=:t:===:}
    & \multicolumn{1}{c|}{$\rh$}
    & \multicolumn{1}{c|}{$\rh'$}
    & \multicolumn{1}{c|}{$\rh^\dbprime$}
    \\
  \multicolumn{1}{|c||}{}
    & \multicolumn{1}{c|}{$\!\equiv\! \rh(770)$}
    & \multicolumn{1}{c|}{$\!\equiv\! \rh(1450)$}
    & \multicolumn{1}{c|}{$\!\equiv\! \rh(1700)$}
    \\[.5ex] \hhline{|-||---|}
  $\vv f \!\equiv\! \pi^+\pi^- \vv$\;   
    &\vv 3.743,\centi \vv
    &\vv 3.628,\centi \vv
    &\vv 2.325,\centi \vv\\[.5ex]
  $\vv f \!\equiv\! 2\pi^+2\pi^- \vv$\;
    &\vv 6.942,\centi \vv
    &\vv 4.102,\centi \vv
    &\vv 2.540,\centi \vv\\
  \hhline{:=:b:===:}
  \end{tabular}
  \end{center}
\vspace*{-3ex}   
\caption[Normierungsparameter~\protect$c_{V\!,f}$ der Breit-Wigner-Distributionen]{
  Normierungsparameter~$c_{V\!,f}$ der Breit-Wigner-Distributionen nach Gl.~(\ref{APP:cVla}); Abweichungen von Eins in Prozent~$\centi$.   F"ur die Schwellen~$s_f \!\equiv\! (2m_\pi)^2,(4m_\pi)^2$ liegen zugrunde f"ur die Massen~$M_V$ der Vektormesonen die Zahlenwerte von Tabelle~\refg{Tabl:Charakt_rh,rh',rh''} und~$m_\pi \!=\! 0.139\,570\GeV$ f"ur die Masse~des geladenen Pions, vgl.\@ Ref.~\cite{PDG00}.
\vspace*{-1.375ex}
}
\label{Tabl-App:cVla}
\end{minipage}
\end{table}
F"ur $l^+l^-$-Annihilation in die Endzust"ande \mbox{$f \!\equiv\! \pi^+\pi^-, 2\pi^+2\pi^-$} folgt:
\vspace*{-1ex}
\begin{align} \label{APP:si_ll-to-f}
\si_f(M)\;
  =\; \frac{4\pi e^4}{3}\;
        \Big| {\T\sum}_{V\!=\rh,\rh',\rh^\dbprime} \frac{f_V}{M_V}\;
        \frac{c_{V\!,f}\cdot \sqrt{M_V\Gatot_V\!/\pi}}{%
              M^2 \!-\! M_V^2 \!+\! \iIM\, M_V\Gatot_V}\;
        \sqrt{B_{V\to f}}\; \Big|^2
    \\[-4ex]\nn
\end{align}
mit abk"urzend~$\si_f \!\equiv\! \si[l^+l^- \!\to\! V \!\to\! f]$.
Die Verzweigungsverh"altnisse~$B_{V\to f}$ sind zusammen mit den Massen~$M_V$ und totalen Zerfallsbreiten~$\Gatot_V$ angegeben in Tabl.~\ref{Tabl:Charakt_rh,rh',rh''}.
In dieser Formel wird die totale Zerfallsbreite des~$\rh(770)$ parametrisierte in der Form:
%
\begin{align} \label{APP:Gatot-mod}
\Gatot_\rh\vv
  \longrightarrow\vv
  \Gatot_\rh\cdot
  \big\{1 + c \!\cdot\! \big[M^2\!/\!M_\rh^2 \!-\! 1\big]
          + c' \!\cdot\! \big[M^2\!/\!M_\rh^2 \!-\! 1\big]^2
  \big\}
\end{align}
mit~$\Gatot_\rh$ dem experimentellen Zahlenwert, vgl.\@ Tabl.~\ref{Tabl:Charakt_rh,rh',rh''}, \vspace*{-.25ex}und ersetzt~\mbox{$c_{\rh,f} \!\to\! c^\dbprime \!\cdot c_{\rh,f}$}.%
\FOOT{
  Da~$B_{\rh,2\pi^+2\pi^-} \!=\! 0$, vgl.\@ Tabl.~\ref{Tabl:Charakt_rh,rh',rh''}, betrifft diese Modifikation de facto nur den Zwei-Pion-Endzustand.
}%
~Die Parameter~$c$,~$c'$,~$c^\dbprime$ werden adjustiert%
\FOOT{
  auf die Zahlenwerte~$c \!=\! -0.2954$,~$c' \!=\! 0.02286$,~$c^\dbprime \!=\! 0.8512$
}
durch die Forderung, da"s Gl.~(\ref{APP:si_ll-to-f}) das experimentelle $e^+e^-$-Annihilationspektrum in~$\pi^+\pi^-$ reproduziert, vgl.\@ Abb.~\refg{Fig:eebar2pis}.

Das Massenspektrum f"ur Photoproduktion von~$f \!\equiv\! \pi^+\pi^-, 2\pi^+2\pi^-$ elastisch am Proton folgt analog,~-- dabei sind~$\Gatot_\rh$,~$c_{\rh,f}$ {\it nicht\/} modifiziert:
\vspace*{-.5ex}
\begin{align} \label{APP:si_gap-to-fp}
&\frac{1}{2M}\frac{d\si_{f\!,\la}}{dM}\Big|_{Q^2}(M)
    \\
&=\; \int_{-\infty}^0 dt\; \frac{1}{16\pi\, s^2}\;
        \bigg| {\T\sum}_{V\!=\rh,\rh',\rh^\dbprime}\vv
        T_{V\!,\la}(s,t)\vv
        \frac{c_{V\!,f}\cdot \sqrt{M_V\Gatot_V\!/\pi}}{%
              M^2 \!-\! M_V^2 \!+\! \iIM\, M_V\Gatot_V}\vv
        \sqrt{B_{V\!\to f}}\; \bigg|^2
    \nn
    \\[-4.5ex]\nn
\end{align}
\vspace*{-.25ex}mit abk"urzend~$d\si_{f\!,\la} \!\equiv\! d\si_\la[\ga^{\scriptscriptstyle({\D\ast})}p \!\to\! Vp]$ und~$T_{V\!,\la} \!\equiv\! T_\la[\ga^{\scriptscriptstyle({\D\ast})}p \!\to\! Vp]$ den entsprechenden $T$-Amplituden, die berechnet werden im Rahmen unseres Zugangs auf Basis des \DREI[]{M}{S}{V}.
\vspace*{-.5ex}

\subsection[Mischungswinkel~\protect$\Th$, Kopplung~\protect$f_\iZS$]{%
            Mischungswinkel~\protect\bm{\Th}, Kopplung~\protect\bm{f_\iZS}}
\label{APPSubsect:Mischungswinkel}

Im Haupttext wird f"ur die Vektormesonen~$\rh',\rh^\dbprime$~[$\equiv\! \rh(1450),\rh(1700)$] der Ansatz gemacht:
Sie setzen sich zusammen, bestimmt durch den Mischungswinkel~$\Th$, aus zwei Komponenten, dem $2S$-Quark-Antiquark-Zustand und dem inerten Rest, vgl.\@ die Gln.~(\ref{Ansatz})-(\ref{Ansatz}$''$).
Es werden definiert die Verzweigungsverh"altnisse~\mbox{$X_{V\!,1}, X_{V\!,2}, X_{V\!,3}$} f"ur~$V \!\equiv\! \rh',\rh^\dbprime$ in den Gln.~(\ref{branchings})-(\ref{branchings}$''$).
Wir stellen den folgenden Zusammenhang her.

Aus dem Ansatz f"ur~$\rh',\rh^\dbprime$, vgl.\@ die Gln.~(\ref{Ansatz})-(\ref{Ansatz}$''$), folgt f"ur die Kopplungen:
\begin{alignat}{3} \label{APP:f-rh',rh''_f2S}
&f_{\rh'}&\;
  &=&\;  &\cos\Th\; f_\iZS
    \\
&f_{\rh^\dbprime}&\;
  &=&\; -&\sin\Th\; f_\iZS
    \tag{\ref{APP:f-rh',rh''_f2S}$'$}
\end{alignat}
das hei"st mithilfe Gl.~(\ref{APP:fV_Gall}) gilt f"ur die~$e^+e^-$-Zerfallsbreiten zum einen:
\begin{alignat}{2} \label{APP:Gaee-rh',rh''_Th}
&\Gaee_{\rh'}&\;
  &=\; \frac{4\pi\al^2}{3}\; f_\iZS^2\; \frac{\cos^2\th}{M_{\rh'}}
    \\
&\Gaee_{\rh^\dbprime}&\;
  &=\; \frac{4\pi\al^2}{3}\; f_\iZS^2\; \frac{\sin^2\th}{M_{\rh^\dbprime}}
    \tag{\ref{APP:Gaee-rh',rh''_Th}$'$}
\end{alignat}
Zum anderen gilt in Termen der~$X_{V\!,i}$, vgl.\@ die Gln.~(\ref{branchings})-(\ref{branchings}$''$), f"ur~$V \!\equiv\! \rh',\rh^\dbprime$:
\begin{align} \label{APP:Gaee-rh',rh''_Xi} 
&\Gaee_V\;
  =\; \Gatot_V\cdot  x_V \qquad
  \text{mit}\qquad
  x_V\; \equiv\; X_{V\!,1}(1\!+\!X_{V\!,2})\big / X_{V\!,3}
\end{align}
Gleichsetzen der Relationen~(\ref{APP:Gaee-rh',rh''_Th}),~(\ref{APP:Gaee-rh',rh''_Th}$'$) und~(\ref{APP:Gaee-rh',rh''_Xi}) f"uhrt auf:
\begin{alignat}{2}
&\tan^2\Th&\;
  &=\; M_{\rh'} \Gatot_{\rh'}\, x_{\rh'}\;
      \big/\;
      M_{\rh^\dbprime} \Gatot_{\rh^\dbprime}\, x_{\rh^\dbprime}
    \label{APP:Th} \\[1ex]
&f_\iZS^2&\;
  &=\; \Big[\frac{4\pi\al^2}{3}\Big]^{-1}\, \big\{
        M_{\rh'} \Gatot_{\rh'}\, x_{\rh'}\;
        +\;
        M_{\rh^\dbprime} \Gatot_{\rh^\dbprime}\, x_{\rh^\dbprime}
      \big\}
    \label{APP:f2S}
\end{alignat}
Eingesetzt f"ur die~$X_{V\!,i}$ die Zahlenwerte aus Tabelle~\ref{Tabl:Charakt_rh,rh',rh''} finden wir explizit:
\begin{alignat}{4}
&\Th&\;    &=&\;  0.229&\;\pi&\;
           &=\;    41.2^{\;\circ}
    \tag{\ref{APP:Th}$'$} \\
&f_\iZS&\; &=&\; -0.178&\GeV&&
    \tag{\ref{APP:f2S}$'$}
\end{alignat}
Die hier implizierte Phasenwahl von~$f_\iZS$ und die Wahl von~$\Th$ im ersten Quadranten geschieht dahingehend, das Vorzeichenmuster zu reproduzieren, das sich experimentell manifestiert im~\mbox{$\pi^+\pi^-$-Masse}\-spektrum f"ur $e^+e^-$-Annihilation, vgl.\@ Abb.~\ref{Fig:eebar2pis}.

Die Kopplung~$f_\iZS$ nach~(\ref{APP:f2S}$'$) differiert augenscheinlich von dem Wert von~$-0.137\GeV$, vgl.\@ Tabl.~\refg{Tabl:Wfn-Parameter}, zu dem wir gelangen im Rahmen unseres expliziten Ansatzes f"ur der Lichtkegelwellenfunktionen des~$2S$-Zustandes.
In Hinblick auf die nur geringe Genauigkeit der Zahlenwerte f"ur die~$X_{V\!,i}$, vgl.\@ auch Donnachie, Mirzaie in Ref.~\cite{Donnachie87a}, erscheint es uns legitim, unserer Analyse zugrundezulegen den Zahlenwert~(\ref{APP:Th}$'$) f"ur den Mischungswinkel~$\Th$ und den Wert von~$-0.137\GeV$ aus Tabelle~\ref{Tabl:Wfn-Parameter} f"ur die Kopplung~$f_\iZS$~-- statt durch Fein-Adjustierung globale "Ubereinstimmung der Parameter zu erreichen.
\theendnotes

%% file: APP_TABLES-F.tex
\lhead[\fancyplain{}{\sc\thepage}]
      {\fancyplain{}{\sc{{\footnotesize Anhang~\thechapter:}
                            Tabellen zu Abbildung~\protect\ref{Fig:dsdt_V,la}}}}
\rhead[\fancyplain{}{\sc{{\footnotesize Anhang~\thechapter:}
                            \protect$d\si_{\la\equiv L,T}\!/\!dt|_{Q^2}(\tfbQ)$
                            f"ur~\protect$\rh$-Produktion}}]
      {\fancyplain{}{\sc\thepage}}
\psfull
\chapter[\protect\bm{d\si_{\la\equiv L,T}\!/\!dt|_{Q^2}(\tfbQ)},%
           ~\protect\bm{\rh(770)}-Produktion:
           Tabellen zu Abbildung~\protect\ref{Fig:dsdt_V,la}
 ]{\huge \protect\bm{d\si_{\la\equiv L,T}\!/\!dt|_{Q^2}(\tfbQ)},%
           ~\protect\bm{\rh(770)}-Produktion:\\%
           Tabellen zu Abbildung~\protect\ref{Fig:dsdt_V,la}}
\label{APP:TABLES}

\vspace*{-1ex}
In Abbildung~\ref{Fig:dsdt_V,la} ist aufgetragen als Funktion von~\mbox{\,$\tfbQ = -t \!+\! \tfde \cong -t$}, vgl.\@ Gl.~(\ref{tfbB_-t}), die im Quadrat des invarianten Impulstransfers~\mbox{\,$-t$\,} {\bf differentiellen Wirkungsquerschnitte\/}:
\vspace*{-.5ex}
\begin{align}
\left.\frac{d\si_{V,\la}}{dt}\right|_{Q^2}(\tfbQ)
  \qquad V \equiv \rh(770),2S
  \qquad \la \equiv L,T
    \\[-4.5ex]\nn
\end{align}
{\bf longitudinal\/} und {\bf transversal\/}
f"ur feste Werte~\mbox{\,\bm{Q^2 \equiv 0,\; 0.25,\; 2,\; 10,\; 20\GeV^2}}. \\
\indent
Die Auftragung dort ist logarithmisch "uber viele Dekaden.
F"ur~\bm{\rh(770)}-Produktion geben wir daher an~-- ohne weiteren Kommentar~-- Ausz"uge der $C$-Programm-Listen:
Die erste Zahl ist~\mbox{\,$\tfbQ$} in~\mbox{\,\bm{\GeV[]^2}}, die zweite der zugeh"orige Produktionsquerschnitt in~\mbox{\,\bm{\mbarn[]\,\GeV[]^{-2}}}.
\vfill

\renewcommand{\textfraction}{0}
%

%
\begin{table}[h]
  \begin{minipage}[b]{\linewidth}%
  {\tt%
  \setlength{\unitlength}{1mm}\makebox(140,95){\begin{picture}(140,95)%
  \put( 4.5,0){\makebox(39,95   ){\begin{minipage}{39mm} \input{TABLES-F/Table-H-1a} \end{minipage}}}%
  \put(50.5,0){\makebox(39,90.25){\begin{minipage}{39mm} \input{TABLES-F/Table-H-1b} \end{minipage}}}%
  \put(96.5,0){\makebox(39,90.25){\begin{minipage}{39mm} \input{TABLES-F/Table-H-1c} \end{minipage}}}%
  \end{picture}}%
  }%
  \end{minipage}%

  \vspace*{-4ex}
\caption[\protect$d\si_\la\!/\!dt|_{Q^2}(\tfbQ)$ f"ur~$\rh(770)$:%
   ~\protect$\la \!\equiv\! T$,~transversal,\;\,~\protect$Q^2 \!\equiv\! 0$]{%
   \bf\bm{Q^2 \!\equiv\! 0}, Transversal~\bm{T}
}
\label{Tabl:dsigmadtQ2_0}
\end{table}
\vspace*{-2ex}
\clearpage
\begin{table}[h]
  \begin{minipage}[b]{\linewidth}%
  {\tt%
  \setlength{\unitlength}{1mm}\makebox(140,95){\begin{picture}(140,95)%
  \put( 4.5,0){\makebox(39,95   ){\begin{minipage}{39mm} \input{TABLES-F/Table-H-2a} \end{minipage}}}%
  \put(50.5,0){\makebox(39,90.25){\begin{minipage}{39mm} \input{TABLES-F/Table-H-2b} \end{minipage}}}%
  \put(96.5,0){\makebox(39,90.25){\begin{minipage}{39mm} \input{TABLES-F/Table-H-2c} \end{minipage}}}%
  \end{picture}}%
  }%
  \end{minipage}%

  \vspace*{-4ex}
\caption[\protect$d\si_\la\!/\!dt|_{Q^2}(\tfbQ)$ f"ur~$\rh(770)$:%
   ~\protect$\la \!\equiv\! L$,~longitudinal,~\protect$Q^2 \!\equiv\! 0.25\GeV$]{%
   \bf\bm{Q^2 \!\equiv\! 0.25\GeV^2}, Longitudinal~\bm{L}
}
\label{Tabl:dsigmadtQ2_0.25L}
\end{table}
\vfill
\begin{table}[h]
  \begin{minipage}[b]{\linewidth}%
  {\tt%
  \setlength{\unitlength}{1mm}\makebox(140,95){\begin{picture}(140,95)%
  \put( 4.5,0){\makebox(39,95   ){\begin{minipage}{39mm} \input{TABLES-F/Table-H-3a} \end{minipage}}}%
  \put(50.5,0){\makebox(39,90.25){\begin{minipage}{39mm} \input{TABLES-F/Table-H-3b} \end{minipage}}}%
  \put(96.5,0){\makebox(39,90.25){\begin{minipage}{39mm} \input{TABLES-F/Table-H-3c} \end{minipage}}}%
  \end{picture}}%
  }%
  \end{minipage}%

  \vspace*{-4ex}
\caption[\protect$d\si_\la\!/\!dt|_{Q^2}(\tfbQ)$ f"ur~$\rh(770)$:%
   ~\protect$\la \!\equiv\! T$,~transversal,\;\,~\protect$Q^2 \!\equiv\! 0.25\GeV^2$]{%
   \bf\bm{Q^2 \!\equiv\! 0.25\GeV^2}, Transversal~\bm{T}
}
\label{Tabl:dsigmadtQ2_0.25T}
\end{table}
\clearpage
\begin{table}[h]
  \begin{minipage}[b]{\linewidth}%
  {\tt%
  \setlength{\unitlength}{1mm}\makebox(140,95){\begin{picture}(140,95)%
  \put( 4.5,0){\makebox(39,95   ){\begin{minipage}{39mm} \input{TABLES-F/Table-H-4a} \end{minipage}}}%
  \put(50.5,0){\makebox(39,90.25){\begin{minipage}{39mm} \input{TABLES-F/Table-H-4b} \end{minipage}}}%
  \put(96.5,0){\makebox(39,90.25){\begin{minipage}{39mm} \input{TABLES-F/Table-H-4c} \end{minipage}}}%
  \end{picture}}%
  }%
  \end{minipage}%

  \vspace*{-4ex}
\caption[\protect$d\si_\la\!/\!dt|_{Q^2}(\tfbQ)$ f"ur~$\rh(770)$:%
   ~\protect$\la \!\equiv\! L$,~longitudinal,~\protect$Q^2 \!\equiv\! 2\GeV$]{%
   \bf\bm{Q^2 \!\equiv\! 2\GeV^2}, Longitudinal~\bm{L}
}
\label{Tabl:dsigmadtQ2_2L}
\end{table}
\vfill
\begin{table}[h]
  \begin{minipage}[b]{\linewidth}%
  {\tt%
  \setlength{\unitlength}{1mm}\makebox(140,95){\begin{picture}(140,95)%
  \put( 4.5,0){\makebox(39,95   ){\begin{minipage}{39mm} \input{TABLES-F/Table-H-5a} \end{minipage}}}%
  \put(50.5,0){\makebox(39,90.25){\begin{minipage}{39mm} \input{TABLES-F/Table-H-5b} \end{minipage}}}%
  \put(96.5,0){\makebox(39,90.25){\begin{minipage}{39mm} \input{TABLES-F/Table-H-5c} \end{minipage}}}%
  \end{picture}}%
  }%
  \end{minipage}%

  \vspace*{-4ex}
\caption[\protect$d\si_\la\!/\!dt|_{Q^2}(\tfbQ)$ f"ur~$\rh(770)$:%
   ~\protect$\la \!\equiv\! T$,~transversal,\;\,~\protect$Q^2 \!\equiv\! 2\GeV^2$]{%
   \bf\bm{Q^2 \!\equiv\! 2\GeV^2}, Transversal~\bm{T}
}
\label{Tabl:dsigmadtQ2_2T}
\end{table}
\clearpage
\begin{table}[h]
  \begin{minipage}[b]{\linewidth}%
  {\tt%
  \setlength{\unitlength}{1mm}\makebox(140,95){\begin{picture}(140,95)%
  \put( 4.5,0){\makebox(39,95   ){\begin{minipage}{39mm} \input{TABLES-F/Table-H-6a} \end{minipage}}}%
  \put(50.5,0){\makebox(39,90.25){\begin{minipage}{39mm} \input{TABLES-F/Table-H-6b} \end{minipage}}}%
  \put(96.5,0){\makebox(39,90.25){\begin{minipage}{39mm} \input{TABLES-F/Table-H-6c} \end{minipage}}}%
  \end{picture}}%
  }%
  \end{minipage}%

  \vspace*{-4ex}
\caption[\protect$d\si_\la\!/\!dt|_{Q^2}(\tfbQ)$ f"ur~$\rh(770)$:%
   ~\protect$\la \!\equiv\! L$,~longitudinal,~\protect$Q^2 \!\equiv\! 10\GeV$]{%
   \bf\bm{Q^2 \!\equiv\! 10\GeV^2}, Longitudinal~\bm{L}
}
\label{Tabl:dsigmadtQ2_10L}
\end{table}
\vfill
\begin{table}[h]
  \begin{minipage}[b]{\linewidth}%
  {\tt%
  \setlength{\unitlength}{1mm}\makebox(140,95){\begin{picture}(140,95)%
  \put( 4.5,0){\makebox(39,95   ){\begin{minipage}{39mm} \input{TABLES-F/Table-H-7a} \end{minipage}}}%
  \put(50.5,0){\makebox(39,90.25){\begin{minipage}{39mm} \input{TABLES-F/Table-H-7b} \end{minipage}}}%
  \put(96.5,0){\makebox(39,90.25){\begin{minipage}{39mm} \input{TABLES-F/Table-H-7c} \end{minipage}}}%
  \end{picture}}%
  }%
  \end{minipage}%

  \vspace*{-4ex}
\caption[\protect$d\si_\la\!/\!dt|_{Q^2}(\tfbQ)$ f"ur~$\rh(770)$:%
   ~\protect$\la \!\equiv\! T$,~transversal,\;\,~\protect$Q^2 \!\equiv\! 10\GeV^2$]{%
   \bf\bm{Q^2 \!\equiv\! 10\GeV^2}, Transversal~\bm{T}}
\label{Tabl:dsigmadtQ2_10T}
\end{table}
\clearpage
\begin{table}[h]
  \begin{minipage}[b]{\linewidth}%
  {\tt%
  \setlength{\unitlength}{1mm}\makebox(140,95){\begin{picture}(140,95)%
  \put( 4.5,0){\makebox(39,95   ){\begin{minipage}{39mm} \input{TABLES-F/Table-H-8a} \end{minipage}}}%
  \put(50.5,0){\makebox(39,90.25){\begin{minipage}{39mm} \input{TABLES-F/Table-H-8b} \end{minipage}}}%
  \put(96.5,0){\makebox(39,90.25){\begin{minipage}{39mm} \input{TABLES-F/Table-H-8c} \end{minipage}}}%
  \end{picture}}%
  }%
  \end{minipage}%

  \vspace*{-4ex}
\caption[\protect$d\si_\la\!/\!dt|_{Q^2}(\tfbQ)$ f"ur~$\rh(770)$:%
   ~\protect$\la \!\equiv\! L$,~longitudinal,~\protect$Q^2 \!\equiv\! 20\GeV$]{%
   \bf\bm{Q^2 \!\equiv\! 20\GeV^2}, Longitudinal~\bm{L}
}
\label{Tabl:dsigmadtQ2_20L}
\end{table}
\vfill
\begin{table}[h]
  \begin{minipage}[b]{\linewidth}%
  {\tt%
  \setlength{\unitlength}{1mm}\makebox(140,95){\begin{picture}(140,95)%
  \put( 4.5,0){\makebox(39,95   ){\begin{minipage}{39mm} \input{TABLES-F/Table-H-9a} \end{minipage}}}%
  \put(50.5,0){\makebox(39,90.25){\begin{minipage}{39mm} \input{TABLES-F/Table-H-9b} \end{minipage}}}%
  \put(96.5,0){\makebox(39,90.25){\begin{minipage}{39mm} \input{TABLES-F/Table-H-9c} \end{minipage}}}%
  \end{picture}}%
  }%
  \end{minipage}%

  \vspace*{-4ex}
\caption[\protect$d\si_\la\!/\!dt|_{Q^2}(\tfbQ)$ f"ur~$\rh(770)$:%
   ~\protect$\la \!\equiv\! T$,~transversal,\;\,~\protect$Q^2 \!\equiv\! 20\GeV^2$]{%
   \bf\bm{Q^2 \!\equiv\! 20\GeV^2}, Transversal~\bm{T}
}
\label{Tabl:dsigmadtQ2_20T}
\end{table}

%% file: DANK.tex
\lhead[\fancyplain{}{}]%
      {\fancyplain{}{}}
\rhead[\fancyplain{}{}]%
      {\fancyplain{}{}}
\addtocontents{toc}{\protect\contentsline {chapter}{\numberline {}{\rm Danksagung}}{}}

\setlength{\parskip}{0.3cm}
\begin{center}
{\bf{DANKSAGUNG\vspace*{1cm}}}
\end{center}

\noindent
An erster Stelle m"ochte ich mich ganz herzlich bedanken bei Herrn Dosch.
F"ur die "Uberlassung des interessanten Themas und an dem Interesse an der Entwicklung, die es "uber die Zeit erfahren hat.
F"ur die sch"one Zusammenarbeit, die sehr herzliche Betreuung.
F"ur das stete Interesse zur Diskussion, in der ich sehr viel lernen konnte, von seiner ph"anomenologischen Intuition und seinem Gef"uhl f"ur einfache L"osungen~-- was mich sicher sehr gepr"agt hat in meiner Entwicklung, an Fragen heranzugehen.
F"ur seinen Einsatz, der es mir erm"oglicht hat, Konferenzen zu besuchen und Vortr"age zu halten. 
Nicht zuletzt m"ochte ich mich ganz herzlich bedanken f"ur die sehr herzliche Atmosph"are weit "uber die wissenschaftliche Arbeit hinaus.
F"ur die sch"onen Gespr"ache "uber Musik.
\vspace*{-.5ex}

\bigskip\noindent
Ich danke Herrn Prof.\@ Povh f"ur die "Ubernahme des Korreferats (und die "Ubungsgruppe in meinem ersten Semester, in der er mich begeistert hat f"ur Quarks, Gluonen und Confinement).
\vspace*{-.5ex}

\bigskip\noindent
Ich danke Herrn Prof.\@ Nachtmann, von dessen Arbeiten und Analytizit"at ich viel habe lernen k"onnen; ich danke f"ur die interessierte Diskussion als Zweitbetreuer, die Begutachtung meiner Arbeiten und daf"ur, da"s er als zweiter Theoretiker anwesend ist in der Disputation meiner Arbeit.
\vspace*{-.5ex}

\bigskip\noindent
Ich danke Herrn Prof.\@ Pirner f"ur die Zusammenarbeit und die herzliche Atmosph"are.
\vspace*{-.5ex}

\bigskip\noindent
Ich danke ferner Herrn Prof.\@ Schmidt f"ur die begeisternde Vorlesung "uber Quantenfeldtheorie, die ich bei ihm geh"ort habe, die Diskussion, die Gespr"ache und das stete Interesse~--\\
und ich danke Prof.\@ Donnachie f"ur die Begutachtung meiner Arbeiten.
\vspace*{-.5ex}

\bigskip\noindent
F"ur die finanzielle Unterst"utzung danke ich dem Graduiertenkolleg der \mbox{D}{F}{G} und meinen Eltern.
\vspace*{-.5ex}

\bigskip\noindent
Wichtig f"ur mein Verst"andnis war die st"andige Diskussion mit Post-Docs, Doktoranden und Diplomanden; ich danke vor allen
Edgar Berger, Bastian Bergerhoff, Boris K"ors, Enrico Meggiolaro, Timo Paulus, Michael R"uter.
\vspace*{-.5ex}

\bigskip\noindent
Ich m"ochte mich abschlie"send bedanken f"ur die herzliche Atmosph"are in Arbeitszimmer und Institut "uberhaupt, die sich nicht an einzelnen Namen festmachen l"a"st.